\documentclass[11pt,twoside,a4paper]{cernrep} 
\usepackage{rep_common}
\usepackage{cancel}
\usepackage{placeins}
\usepackage{booktabs}
\usepackage{slashbox}
\usepackage{subfig}
\usepackage{amsmath,amssymb,amsthm}
\usepackage{rotating}
\usepackage{multicol}
\usepackage{tabulary}
\usepackage[colorlinks, urlcolor=blue, linkcolor=blue, citecolor=blue, plainpages=false]{hyperref}
\usepackage{float}
\floatstyle{plaintop}
\restylefloat{table}
\usepackage{xargs}
\usepackage{xstring}

\begin{document}
\newcounter{ncontacts}
\newcommand{\fcontact}[1]{\StrCount{#1}{,}[\tmp]\setcounter{ncontacts}{\tmp}
  Contact\ifthenelse{\value{ncontacts} > 0}{s}{}: #1}
\newcommand{\feditor}[1]{\StrCount{#1}{,}[\tmp]\setcounter{ncontacts}{\tmp}
  Contact Editor\ifthenelse{\value{ncontacts} > 0}{s}{}: #1}
\newcommandx{\asection}[2][1=NONE]{
  \ifthenelse{\equal{#1}{NONE}}
  {\section{#2}}{\section[#2]{#2\footnote{\feditor{#1}}}}}
\newcommandx{\asubsection}[2][1=NONE]{
  \ifthenelse{\equal{#1}{NONE}}
             {\subsection{#2}}{
               \subsection[#2]{#2\footnote{\fcontact{#1}}}}}
\newcommandx{\asubsubsection}[2][1=NONE]{
  \ifthenelse{\equal{#1}{NONE}}
             {\subsubsection{#2}}{
               \subsubsection[#2]{#2\footnote{\fcontact{#1}}}}}
\newcommandx{\asubsubsubsection}[2][1=NONE]{
  \ifthenelse{\equal{#1}{NONE}}
             {\subsubsubsection{#2}}{
               \subsubsubsection[#2]{#2\footnote{\fcontact{#1}}}}}
\newcommand{\main}{.}
\newcommand{\code}[1]{\texttt{#1}}
\newcommand{\ifb}{\mbox{fb$^{-1}$}}
\newcommand{\iab}{\ensuremath{ab^{-1}}\xspace}
\newcommand{\de}{\partial}
\newcommand{\Mb}{\bar{M}}
\newcommand{\Mpl}{M_{\textrm{Pl}}}
\newcommand{\Mp}{M_{\textrm{Pl}}}
\newcommand{\reef}[1]{(\ref{#1})}

\newcommand{\abs}[1]{\left\lvert#1\right\rvert}
\newcommand{\cL}{\mathcal{L}}
\newcommand{\cR}{\mathcal{R}}
\newcommand{\BR}{{\rm BR}}
\newcommand{\blu}{\color{blue}}
\newcommand{\ros}{\color{red}}
\newcommand{\eq}[1]{eq.~(\ref{#1})}
\newcommand{\lag}{\mathcal{L}}
\newcommand{\op}{\mathcal{O}}
\newcommand{\lp}{\left(}
\newcommand{\rp}{\right)}
\newcommand{\nn}{\nonumber}
\newcommand{\vev}[1]{\langle {#1} \rangle}
\newcommand{\pslash}{\!\not\! p}
\newcommand{\be}{\begin{equation}}
\newcommand{\ee}{\end{equation}}
\newcommand{\bea}{\begin{eqnarray}}
\newcommand{\eea}{\end{eqnarray}}
\newcommand{\bc}{\begin{center}}
\newcommand{\ec}{\end{center}}
\newcommand{\ba}{\begin{array}}
\newcommand{\ea}{\end{array}}
\newcommand{\noindentFR}[1]{\vspace{5mm}\noindent {\bf #1}\\}

\newcommand{\deltapt}{p_{T,VV}}
\newcommand{\met}{\cancel{E}_T}
\newcommand{\ptvmin}{p_{T,V}^{(min)}}
\newcommand{\thetastar}{\theta^{*}}
\newcommand{\aq}{a_{q}^{(3)}}
\newcommand{\ptv}{p_{T,V}}
\newcommand{\rf}[1]{{\color{purple} #1}} 
\newcommand{\com}[1]{{\color{red} \sffamily (#1)}} 
\newcommand*{\lHHH}{\ensuremath{\lambda_{HHH}}}
\newcommand*{\kl}{\ensuremath{\kappa_{\lambda}}}
\newcommand{\Br}{{\mathrm{Br}}}

\newcommand\TwoFigBottom{-2}
\newcommand{\cg}{c_{ggh}}
\newcommand{\cgg}{c_{gghh}}
\newcommand{\ctt}{c_{tt}}
\newcommand{\ct}{c_{t}}
\newcommand{\cb}{c_{b}}
\newcommand{\chhh}{c_{hhh}}
\newcommand{\mhh}{m_{hh}}
\newcommand{\pythia}{\textsc{Pythia}\xspace}
\newcommand{\geant}{\textsc{Geant}\xspace}

\newcommand{\as}{\ensuremath{\alpha_\text{s}}\xspace}
\newcommand{\muf}{\ensuremath{\mu_f}\xspace}
\newcommand{\mur}{\ensuremath{\mu_r}\xspace}
\newcommand{\mbottom}{\ensuremath{m_b}\xspace}
\newcommand{\mtop}{\ensuremath{m_t}\xspace}
\newcommand{\mt}{\ensuremath{m_t}\xspace}
\newcommand{\mcharm}{\ensuremath{m_c}\xspace}
\newcommand{\cred}{\color{red}}
\newcommand{\cblue}{\color{blue}}
\newcommand{\cgreen}{\color{X575}}
\newcommand{\tCaa}{\tilde c_{\gamma\gamma}}
\newcommand{\tCza}{\tilde c_{Z\gamma}}
\newcommand{\tCzz}{\tilde c_{ZZ}}
\newcommand{\tCww}{\tilde c_{WW}}
\newcommand{\bbbar}{\ensuremath{\text{b}\overline{\text{b}}}\xspace}
\newcommand{\ttbar}{\ensuremath{\text{t}\overline{\text{t}}}\xspace}
\newcommand{\ttH}{\ensuremath{\ttbar\text{H}}\xspace}
\newcommand{\tH}{\ensuremath{\text{tH}}\xspace}
\newcommand{\Hbb}{\ensuremath{\text{H}\rightarrow\bbbar}\xspace}
\newcommand{\Htobb}{\ensuremath{\text{H}\rightarrow\bbbar}\xspace}
\newcommand{\ttHF}{\ensuremath{\ttbar+\text{HF}}\xspace}
\newcommand{\ww}{\ensuremath{\PW\PW}\xspace}
\newcommand{\hww}{\ensuremath{\PH\to\ww}\xspace}
\newcommand{\tautau}{\ensuremath{\Pgt\Pgt}\xspace}
\newcommand{\tHW}{\ensuremath{\Pqt\PH\PW}\xspace}
\newcommand{\tHq}{\ensuremath{\Pqt\PH\PQq}\xspace}
\newcommand{\zz}{\ensuremath{\PZ\PZ}\xspace}
\newcommand{\bb}{\ensuremath{\PQb\PQb}\xspace}
\newcommand{\hbb}{\hbox{$\PH\to\bb$}\xspace}
\newcommand{\hgg}{\ensuremath{\PH\to\PGg\PGg}\xspace}
\newcommand{\ggH}{\ensuremath{\PGg\PGg \to \PH}\xspace}
\newcommand{\ttjets}{\ensuremath{\PQt\PQt\text{+}\text{jets}}\xspace}
\newcommand{\vbf}{\ensuremath{\mathrm{VBF}}\xspace}
\newcommand{\vh}{\ensuremath{\mathrm{V}\PH}\xspace}
\newcommand{\tth}{\ensuremath{\Pqt\Pqt\PH}\xspace}
\newcommand{\hzz}{\ensuremath{\PH\to\zz}\xspace}
\newcommand{\hzzllll}{\ensuremath{\hzz^{(\ast)}\to 4\ell}\xspace}
\newcommand{\hwwlnln}{\ensuremath{\hww^{(\ast)}\to\ell\Pgn\ell\Pgn}\xspace}
\newcommand{\htt}{\ensuremath{\PH\to\tautau}\xspace}
\newcommand{\hlep}{\ensuremath{\PH\to\text{leptons}}\xspace}
\newcommand{\mumu}{\ensuremath{\PGm\PGm}\xspace}
\newcommand{\hmm}{\ensuremath{\PH\to\mumu}\xspace}
\newcommand{\hzg}{\ensuremath{\PH\to\PZ\PGg}\xspace}
\newcommand{\wh}{\ensuremath{\PW\PH}\xspace}
\newcommand{\zh}{\ensuremath{\PZ\PH}\xspace}
\newcommand{\fbinv}{\mbox{\ensuremath{\,\text{fb}^\text{$-$1}}}\xspace}
\newcommand{\SM}{\ensuremath{\mathrm{SM}}\xspace}
\newcommand{\kZ}{\ensuremath{\kappa_{\mathrm{\PZ}}}\xspace}
\newcommand{\kW}{\ensuremath{\kappa_{\mathrm{\PW}}}\xspace}
\newcommand{\kH}{\ensuremath{\kappa_{\mathrm{\PH}}}\xspace}
\newcommand{\ktau}{\ensuremath{\kappa_{\PGt}}\xspace}
\newcommand{\ktop}{\ensuremath{\kappa_{\Pqt}}\xspace}
\newcommand{\kb}{\ensuremath{\kappa_{\Pqb}}\xspace}
\newcommand{\kmu}{\ensuremath{\kappa_{\Pgm}}\xspace}
\newcommand{\kglu}{\ensuremath{\kappa_{\mathrm{\Pg}}}\xspace}
\newcommand{\kgam}{\ensuremath{\kappa_{\PGg}}\xspace}
\newcommand{\kzgam}{\ensuremath{\kappa_{\PZ\PGg}}\xspace}
\newcommand{\kappat}{\ensuremath{\kappa_{\Pqt}}\xspace}
\newcommand{\kappab}{\ensuremath{\kappa_{\Pqb}}\xspace}
\newcommand{\kappac}{\ensuremath{\kappa_{\Pqc}}\xspace}
\newcommand{\GammaSM}{\ensuremath{\Gamma_{\PH}/\Gamma_{\PH}^{\mathrm{SM}}}\xspace}
\newcommand{\Bbsm}{\ensuremath{\mathrm{B_{BSM}}}\xspace}
\newcommand{\kV}{\ensuremath{\kappa_{\mathrm{V}}}\xspace}
\newcommand{\ggZH}{\ensuremath{\Pg\Pg\PZ\PH}\xspace}
\newcommand{\wip}[1]{\textcolor{red}{\bfseries TODO: #1}\xspace}
\newcommand{\ggh}{\ensuremath{\Pg\Pg\PH}\xspace}
\newcommand{\Hyy}{\hbox{$H\to\gamma\gamma$}}
\newcommand{\HZy}{\hbox{$H \to Z\gamma$}}
\newcommand{\HZZ}{\hbox{$H \to ZZ^* \to 4\ell$}}
\newcommand{\HWW}{\hbox{$H \to WW^* \to e \nu \mu \nu$}}
\newcommand{\pTH}{\ensuremath{{\pT}^{\PH}}\xspace}
\newlength{\UcmsFigWidth}
\setlength{\UcmsFigWidth}{0.49\textwidth}
\newcommand{\UcmsLeft}{upper\xspace}
\newcommand{\UcmsRight}{lower\xspace}
\newcommand{\tHQ}{\ensuremath{tHq}}
\newcommand{\VH}{\ensuremath{VH}}
\newcommand{\mgg}{\ensuremath{m_{\gamma\gamma}}}
\newcommand{\ptgg}{\ensuremath{\pt^{\gamma\gamma}}}

\newcommand{\tril}{\lambda_{3}}
\newcommand{\trilsm}{\tril^{\rm SM}}
\newcommand{\ktre}{\kappa_{3}}
\newcommand{\dsigmah}{\delta \sigma_{\tril}}
\newcommand{\dBR}{\delta {\rm BR}_{\tril}}
\newcommand{\ggF}{gg{\rm F}}
\newcommand{\br}{{\rm BR}}
\newcommand{\muif}{\mu_i^f}
\newcommand{\pnote}[1]{\textbf{[P:} \textit{\color{red} #1}\textbf{]}}
\newcommand{\mptvec}{{/\!\!\!\vec{ p}_T}}
\newcommand{\HH}{\ensuremath{\PH\PH}\xspace}
\newcommand{\bbbb}{\ensuremath{\PQb\PQb\PQb\PQb}\xspace}
\newcommand{\bbgg}{\ensuremath{\PQb\PQb\Pgg\Pgg}\xspace}
\newcommand{\bbtt}{\ensuremath{\PQb\PQb\Pgt\Pgt}\xspace}
\newcommand{\bbWW}{\ensuremath{\PQb\PQb\PW\PW}\xspace}
\newcommand{\bbZZ}{\ensuremath{\PQb\PQb\PZ\PZ}\xspace}
\newcommand{\bbVV}{\ensuremath{\PQb\PQb\text{VV}}\xspace}
\newcommand{\delphes}{\textsc{Delphes}\xspace}
\newcommand{\tauh}{\ensuremath{\Pgt_{\rm{h}}}}
\newcommand{\ptvecmiss}{\ensuremath{{\vec p}_{\mathrm{T}}^{\kern1pt\text{miss}}}\xspace}
\newcommand{\ptmiss}{\ensuremath{\pt^\text{miss}}\xspace}
\newcommand{\TeV}{\ensuremath{\,\text{Te\hspace{-.08em}V}}\xspace}
\newcommand{\GeV}{\ensuremath{\,\text{Ge\hspace{-.08em}V}}\xspace}
\newcommand{\lambdahhh}{\ensuremath{\lambda_{\PH\PH\PH}}\xspace}
\newcommand{\fixme}[1]{
    \GenericWarning{}{There is a FIXME}
    \textcolor{red}{\textbf{FIXME: #1}}
}

\providecommand{\jhugen}{\textsc{JHUGen}\xspace}
\providecommand{\mela}{\textsc{MELA}\xspace}
\providecommand{\hnnlo}{\textsc{hnnlo}\xspace}
\providecommand{\F}{\ensuremath{\mathrm{F}}\xspace}
\providecommand{\V}{\ensuremath{\mathrm{V}}\xspace}
\providecommand{\W}{\ensuremath{\mathrm{W}}\xspace}
\providecommand{\vv}{\ensuremath{\mathrm{vv}}\xspace}
\providecommand{\VV}{\ensuremath{\mathrm{VV}}\xspace}
\providecommand{\ZZ}{\ensuremath{\mathrm{ZZ}}\xspace}
\providecommand{\WW}{\ensuremath{\mathrm{WW}}\xspace}
\providecommand{\HVV}{\ensuremath{\mathrm{HVV}}\xspace}
\providecommand{\HZZ}{\ensuremath{\mathrm{HZZ}}\xspace}
\providecommand{\LC}[1]{\ensuremath{\Lambda_{#1}}\xspace}
\providecommand{\LZGs}{\ensuremath{\LC{1}^{Z\gamma}}\xspace}
\providecommand{\AC}[1]{\ensuremath{a_{#1}}\xspace}
\providecommand{\ai}{\ensuremath{\AC{i}}\xspace}

\providecommand{\fLC}[1]{\ensuremath{f_{\Lambda #1}}\xspace}
\providecommand{\fAC}[1]{\ensuremath{f_{a #1}}\xspace}
\providecommand{\fLZGs}{\ensuremath{\fLC{1}^{Z\gamma}}\xspace}
\providecommand{\fai}{\fAC{i}}

\providecommand{\sLC}[1]{\ensuremath{\tilde{\sigma}_{\Lambda #1}}\xspace}
\providecommand{\sAC}[1]{\ensuremath{\sigma_{#1}}\xspace}
\providecommand{\sLZGs}{\ensuremath{\sLC{1}^{Z\gamma}}\xspace}
\providecommand{\sai}{\sAC{i}}

\providecommand{\pLC}[1]{\ensuremath{\phi_{\Lambda #1}}\xspace}
\providecommand{\pAC}[1]{\ensuremath{\phi_{a #1}}\xspace}
\providecommand{\pLZGs}{\ensuremath{\pLC{1}^{Z\gamma}}\xspace}
\providecommand{\pai}{\pAC{i}}

\providecommand{\cospLC}[1]{\ensuremath{\cos\left(\pLC{#1}\right)}\xspace}
\providecommand{\cospAC}[1]{\ensuremath{\cos\left(\pAC{#1}\right)}\xspace}
\providecommand{\cospLZGs}{\ensuremath{\cos\left(\pLZGs\right)}\xspace}
\providecommand{\cospai}{\cospAC{i}}

\providecommand{\fcospLC}[1]{\ensuremath{\fLC{#1}\cospLC{#1}}\xspace}
\providecommand{\fcospAC}[1]{\ensuremath{\fAC{#1}\cospAC{#1}}\xspace}
\providecommand{\fcospLZGs}{\ensuremath{\fLZGs\cospLZGs}\xspace}
\providecommand{\fcospai}{\fcospAC{i}}

\providecommand{\offshell}{off-shell\xspace}
\providecommand{\onshell}{on-shell\xspace}
\providecommand{\Offshell}{Off-shell\xspace}
\providecommand{\Onshell}{On-shell\xspace}
\providecommand{\abinv} {\mbox{\ensuremath{\,\text{ab}^\text{$-$1}}}\xspace}
\providecommand{\pbinv} {\mbox{\ensuremath{\,\text{pb}^\text{$-$1}}}\xspace}
\providecommand{\fbinv} {\mbox{\ensuremath{\,\text{fb}^\text{$-$1}}}\xspace}
\providecommand{\GH}{\ensuremath{\Gamma_{\PH}}\xspace}
\providecommand{\GHSM}{\ensuremath{\Gamma_{\PH}^{\mathrm{SM}}}\xspace}
\providecommand{\glufu}{\ensuremath{\mathrm{gg}}\xspace}
\providecommand{\kapl}{\ensuremath{\kappa_\lambda}\xspace}

\newcommand{\gsim}{\gtrsim}
\newcommand{\lsim}{\lesssim}
\newcommand{\la}{\left\langle}
\newcommand{\ra}{\right\rangle}
\newcommand{\lc}{\left[}
\newcommand{\rc}{\right]}
\newcommand{\eps}{\varepsilon}
\newcommand{\mcL}{\mathcal{L}}
\newcommand{\mcO}{\mathcal{O}}
\newcommand{\mcP}{\mathcal{P}}
\newcommand{\pyth}{\texttt{Pythia8}~}
\newcommand{\sher}{\texttt{Sherpa}~}
\newcommand{\alpg}{\texttt{ALPGEN}~}
\newcommand{\amcnlo}{{\tt MadGraph5\_aMC@NLO}~}
\newcommand{\cv}{c_V}
\newcommand{\cvv}{c_{2V}}
\newcommand{\ccc}{c_3}
\newcommand{\dcvv}{\delta_{\cvv}}
\newcommand{\dccc}{\delta_{\ccc}}
\newcommand{\xtento}[1]{\ensuremath{\times 10^{#1}}\xspace}

\newcommand{\gHsV}{{\Gamma_H}}
\newcommand{\gHsVSM}{{\Gamma_{H,{\rm SM}}}}
\newcommand{\gHsVtoX}{{\Gamma_{H\to X}}}
\newcommand{\mhsV}{{m_H}}
\newcommand{\n}{\nonumber \\}
\newcommand{\sig}{\langle \sigma_{\rm a} v \rangle}
\newcommand{\rel}{\Omega_\chi h^2}
\newcommand{\MET}{\ensuremath{\slash{\hspace{-2.5mm}E}_T}\xspace}
\newcommand{\dm}{\Delta M}
\newcommand{\mf}{\ensuremath{m_{\tilde f}}\xspace}
\newcommand{\mlsp}{\ensuremath{M_{LSP}}\xspace}
\newcommand{\lsp}{\ensuremath{{\chi^0_1}}\xspace}
\newcommand{\none}{\ensuremath{{\chi^0_1}}\xspace}
\newcommand{\ntwo}{\ensuremath{{\chi^0_2}}\xspace}
\newcommand{\cone}{\ensuremath{{\chi^\pm_1}}\xspace}
\newcommand{\ctwo}{\ensuremath{{\chi^\pm_2}}\xspace}
\newcommand{\ci}{\ensuremath{{\chi^\pm_i}}\xspace}
\newcommand{\mnone}{\ensuremath{M_{\chi^0_1}}\xspace}
\newcommand{\mntwo}{\ensuremath{M_{\chi^0_2}}\xspace}
\newcommand{\mni}{\ensuremath{M_{\chi^0_i}}\xspace}
\newcommand{\mcone}{\ensuremath{M_{\chi^\pm_1}}\xspace}
\newcommand{\mctwo}{\ensuremath{M_{\chi^\pm_2}}\xspace}
\newcommand{\mw}{\ensuremath{m_W}\xspace}
\newcommand{\mz}{\ensuremath{m_Z}\xspace}
\newcommand{\s}{\ensuremath{\tilde}\xspace}
\newcommand{\mh}{\ensuremath{m_{h^0}}\xspace}
\newcommand{\mH}{\ensuremath{m_{H^0}}\xspace}
\newcommand{\mHpm}{\ensuremath{m_{H^\pm}}\xspace}
\newcommand{\ma}{\ensuremath{m_{A}}\xspace}
\newcommand{\non}{\nonumber}	
\newcommand{\ibid}{{\it ibid}.}
\newcommand{\ccto}{\ensuremath{\chi_1^0 \chi_1^0 \rightarrow}\xspace}
\newcommand{\xen}{XENON-100~}
\newcommand{\missET}{\ensuremath{\slash{\hspace{-2.5mm}E}_T}\xspace}
\newcommand{\fbi}{\ensuremath{\,{\rm fb}^{-1}}\xspace}
\newcommand{\abi}{\ensuremath{\,{\rm ab}^{-1}}\xspace}
\newcommand{\fb}{\ensuremath{{\,{\rm fb}}}\xspace}
\newcommand{\mrec}{\ensuremath{m_{\rm rec}}\xspace}
\newcommand{\mee}{\ensuremath{m_{ee}}\xspace}
\newcommand{\oda}[1]{\overset\leftrightarrow{#1}}

\newcommand{\BRHinv}{\mathrm{BR}_\mathrm{inv}}
\newcommand{\MDM}{M_\mathrm{DM}}
\newcommand{\CL}{\ensuremath{~\mathrm{C.L.}}}
\newcommand{\BRNP}{\mathrm{BR}(H\to\mathrm{NP})}
\newcommand{\MHT}{\ensuremath{\mathrm{H}_{\mathrm{T}}^{\mathrm{miss}}}}
\newcommand{\POWHEG}{\textsc{POWHEG}}
\newcommand{\MGvATNLO}{\textsc{aMC@NLO}}
\newcommand{\PYTHIA}{\textsc{PYTHIA}}
\newcommand{\eqn}{equation}
\newcommand{\METvec}{\ensuremath{\vec{\MET}}}
\newcommand{\pt}{\ensuremath{\mathrm{p}_{T}}}
\newcommand{\ptvec}{\ensuremath{\vec{\mathrm{p}_{T}}}}
\newcommand{\BHinv}{\ensuremath{\text{B}(\PH\to \text{inv.})}\xspace}

\newcommand{\yvmkk}{\bar Y^2 \frac{v^2}{m_{KK}^2}}
\newcommand{\hboson}{\ensuremath{\PH}\xspace}
\newcommand{\bquark}{\ensuremath{\Pqb}\xspace}
\newcommand{\cquark}{\ensuremath{\Pqc}\xspace}
\newcommand{\tquark}{\ensuremath{\Pqt}\xspace}
\newcommand{\zboson}{\ensuremath{\PZ}\xspace}
\newcommand{\gluon}{\ensuremath{\Pg}\xspace}
\newcommand{\photon}{\ensuremath{\Pgg}\xspace}
\newcommand{\proton}{\ensuremath{\Pp}\xspace}
\newcommand{\vboson}{\ensuremath{\mathrm{V}}\xspace}
\newcommand{\electron}{\ensuremath{\Pe}\xspace}
\newcommand{\muon}{\ensuremath{\Pgm}\xspace}
\newcommand{\cc}{\ensuremath{\cquark\bar{\cquark}}\xspace}
\newcommand{\hcc}{\ensuremath{\hboson\to\cc}\xspace}
\newcommand{\hzztofourl}{\ensuremath{\hzz^{(\ast)}\to 4\ell}\xspace}
\newcommand{\BRgamgam}{\ensuremath{\mathcal{B}_{\photon\photon}}\xspace}
\newcommand{\BRZZ}{\ensuremath{\mathcal{B}_{\zboson\zboson}}\xspace}
\newcommand{\pth}{\ensuremath{{\pt}^{\hboson}}\xspace}
\newcommand{\msd}{\ensuremath{m_\text{SD}}\xspace}
\newcommand{\pb}{\ensuremath{\,\text{pb}}\xspace}


\newcommand{\emu}{\Pe\Pgm}
\newcommand{\mutau}{\Pgm\Pgt_{\rm{h}}}
\newcommand{\etau}{\Pe\Pgt_{\rm{h}}}
\newcommand{\tauhad}{\ensuremath{\Pgt_{\rm{h}}}}
\newcommand{\ZJets}{\ensuremath{\PZ}/\ensuremath{\Pgg^{*}}{+}\text{jets}\xspace}
\newcommand{\WJets}{\ensuremath{\PW}{+}\text{jets}\xspace}
\newcommand{\ttbarJets}{\ensuremath{\Pqt\Paqt}{+}\text{jets}\xspace}
\newcommand{\ztt}{\ensuremath{\PZ\to\Pgt\Pgt}\xspace}
\newcommand{\pzetavis}{\ensuremath{p_{\zeta}^{\text{vis}}}\xspace}
\newcommand{\pzetamiss}{\ensuremath{p_{\zeta}^{\text{miss}}}\xspace}
\newcommand{\mhmodp}{\ensuremath{m_{\Ph}^{\text{mod+}}}\xspace}
\newcommand{\mA}{\ensuremath{m_{\text{A}}}\xspace}
\newcommand{\e}[1]{\ensuremath{\times 10^{#1}}}
\newcommand{\processbbtt}{\ensuremath{{\Ph}\to{\Pa\Pa}\to2\Pgt2{\Pqb}}\xspace}
\newcommand{\processmmtt}{\ensuremath{{\Ph}\to{\Pa\Pa}\to2\Pgm2{\Pgt}}\xspace}
\newcommand{\mbtt}{\ensuremath{m_{\PQb\Pgt\Pgt}^{\text{vis}}}\xspace}
\newcommand{\gammaDark}{\ensuremath{{\PGg}_{\mathrm{D}}}\xspace}
\newcommand{\nOne}{\ensuremath{\textnormal{n}_{1}}\xspace}
\newcommand{\nDark}{\ensuremath{\textnormal{n}_{\mathrm{D}}}\xspace}
\newcommand{\PhOne}{\ensuremath{\Ph_{1}}\xspace}
\newcommand{\PhTwo}{\ensuremath{\Ph_{2}}\xspace}
\newcommand{\PhOneTwo}{\ensuremath{\Ph_{1,2}}\xspace}
\newcommand{\PaOne}{\ensuremath{\Pa_{1}}\xspace}
\newcommand{\dimuon}{\ensuremath{(\PGm\PGm)}\xspace}
\newcommand{\dimuonOne}{\ensuremath{(\PGm\PGm)_1}\xspace}
\newcommand{\dimuonTwo}{\ensuremath{(\PGm\PGm)_2}\xspace}
\newcommand{\MgammaDark}{\ensuremath{m_{{\PGg}_{\mathrm{D}}}}\xspace}
\newcommand{\MnOne}{\ensuremath{m_{{\textnormal{n}}_{1}}}\xspace}
\newcommand{\MnDark}{\ensuremath{m_{{\textnormal{n}}_{\mathrm{D}}}}\xspace}
\newcommand{\MPhOne}{\ensuremath{m_{\Ph_{1}}}\xspace}
\newcommand{\MPhTwo}{\ensuremath{m_{\Ph_{2}}}\xspace}
\newcommand{\MPhOneTwo}{\ensuremath{m_{\Ph_{1,2}}}\xspace}
\newcommand{\MPhI}{\ensuremath{m_{\Ph_{i}}}\xspace}
\newcommand{\MPa}{\ensuremath{m_{\Pa}}\xspace}
\newcommand{\MPaOne}{\ensuremath{m_{\Pa_{1}}}\xspace}
\newcommand{\Mdimuon}{\ensuremath{m_{(\PGm\PGm)}}\xspace}
\newcommand{\MdimuonOne}{\ensuremath{m_{(\PGm\PGm)_1}}\xspace}
\newcommand{\MdimuonTwo}{\ensuremath{m_{(\PGm\PGm)_2}}\xspace}
\newcommand{\Mmuon}{\ensuremath{m_{\PGm}}\xspace}
\newcommand{\TgammaDark}{\ensuremath{\tau_{{\PGg}_{\mathrm{D}}}}\xspace}
\newcommand{\ratio}{\ensuremath{\overline{r}}\xspace}
\newcommand{\alphaGen}{\ensuremath{\alpha_\text{gen}}\xspace}
\newcommand{\epsilonFull}{\ensuremath{\varepsilon_{\text{full}}}\xspace}
\newcommand{\SOne}{\ensuremath{S_\mathrm{I}(\Mdimuon)}\xspace}
\newcommand{\STwo}{\ensuremath{S_\mathrm{II}(\Mdimuon)}\xspace}
\newcommand{\ASR}{\ensuremath{A_{\mathrm{SR}}}\xspace}
\newcommand{\ACR}{\ensuremath{A_{\mathrm{CR}}}\xspace}
\newcommand{\fSPS}{\ensuremath{f_\mathrm{SPS}}\xspace}
\newcommand{\pp}{\ensuremath{\Pp\Pp}\xspace}
\newcommand{\Hphiphi}{\ensuremath{\PH \to \phi\phi}\xspace}
\newcommand{\GEANTfour}{\textsc{Geant~4}\xspace}
\newcommand{\dtrans}{\ensuremath{d_0}\xspace}
\newcommand{\zo}{\ensuremath{z_o}\xspace}
\newcommand{\HT}{\ensuremath{H_\mathrm{T}}\xspace}
\newcommand{\effLT}{\ensuremath{\epsilon_\mathrm{LT}}\xspace}
\newcommand{\Nbunches}{\ensuremath{N_\mathrm{bunches}}\xspace}
\newcommand{\fLHC}{\ensuremath{f_\mathrm{LHC}}\xspace}
\newcommand{\Nstubs}{\ensuremath{N_\mathrm{stubs}}\xspace}
\newcommand{\SI}{\ensuremath{\mathrm{SI}}\xspace}
\newcommand{\ctau}{\ensuremath{c\tau}\xspace}
\newcommand{\kt}{\ensuremath{k_t}\xspace}
\newcommand{\FASTJET}{\textsc{FastJet}\xspace}
\newcommand{\BSMHphiphi}{\ensuremath{\Phi \to \phi \phi}\xspace}
\newcommand{\tanb}{\ensuremath{\tan\beta}\xspace}
\newcommand{\RunTwo}{Run~2\xspace}
\newcommand{\taulep}{\ensuremath{\tau_{\mathrm{lep}}}\xspace}
\newcommand{\lephad}{\ensuremath{\taulep\thad}\xspace}
\newcommand{\hadhad}{\ensuremath{\thad\thad}\xspace}
\newcommand{\taue}{\ensuremath{\tau_{e}}\xspace}
\newcommand{\taumu}{\ensuremath{\tau_{\mu}}\xspace}
\newcommand{\thad}{\ensuremath{\tau_{\mathrm{had}}}\xspace}
\newcommand{\mTtot}{m_\textnormal{T}^\textnormal{tot}}
\newcommand{\tauhadvis}{\ensuremath{\Pgt_{\text{had-vis}}}\xspace}
\newcommand{\ehad}{\ensuremath{\taue\thad}\xspace}
\newcommand{\muhad}{\ensuremath{\taumu\thad}\xspace}
\def\lra#1{\overset{\text{\scriptsize$\leftrightarrow$}}{#1}}
\def\permille{\ensuremath{{}^\text{o}\mkern-5mu/\mkern-3mu_\text{oo}}}

\makeatletter
\def\@citex[#1]#2{\leavevmode
  \let\@citea\@empty
  \@cite{\@for\@citeb:=#2\do
    {\@citea\def\@citea{,\penalty\@m\ }%
\edef\magic##1{\let##1\expandafter\noexpand\csname bibalias@\@citeb\endcsname}%
\magic\tmp \ifx\tmp\relax\else \let\@citeb\tmp\fi
     \edef\@citeb{\expandafter\@firstofone\@citeb\@empty}%
     \if@filesw\immediate\write\@auxout{\string\citation{\@citeb}}\fi
     \@ifundefined{b@\@citeb}{\hbox{\reset@font\bfseries ?}%
       \G@refundefinedtrue
       \@latex@warning
         {Citation `\@citeb' on page \thepage \space undefined}}%
       {\@cite@ofmt{\csname b@\@citeb\endcsname}}}}{#1}}
\def\bibalias#1#2{\expandafter\def\csname bibalias@#1\endcsname{#2}}
\makeatother
\bibalias{CMS-PAS-HIG-17-030}{Sirunyan:2018two}
\bibalias{ CMS-PAS-HIG-17-030}{Sirunyan:2018two}
\bibalias{geant}{Agostinelli:2002hh}
\bibalias{ geant}{Agostinelli:2002hh}
\bibalias{Collaboration:2296612}{CMS_MTD_TP}
\bibalias{ Collaboration:2296612}{CMS_MTD_TP}
\bibalias{ATLAS-CONF-2018-028}{ATLAS:2018uso}
\bibalias{ ATLAS-CONF-2018-028}{ATLAS:2018uso}
\bibalias{ATLASHiggsJuly2012}{Aad:2012tfa}
\bibalias{ ATLASHiggsJuly2012}{Aad:2012tfa}
\bibalias{ATLAS-CONF-2018-018}{ATLAS:2018bsg}
\bibalias{ ATLAS-CONF-2018-018}{ATLAS:2018bsg}
\bibalias{CMS-PAS-HIG-17-031}{Sirunyan:2018koj}
\bibalias{ CMS-PAS-HIG-17-031}{Sirunyan:2018koj}
\bibalias{bib:hig-17-026}{Sirunyan:2018mvw}
\bibalias{ bib:hig-17-026}{Sirunyan:2018mvw}
\bibalias{CT14}{Dulat:2015mca}
\bibalias{ CT14}{Dulat:2015mca}
\bibalias{MMHT14}{Harland-Lang:2014zoa}
\bibalias{ MMHT14}{Harland-Lang:2014zoa}
\bibalias{CMS_Tracker_TDR}{Klein:2017nke}
\bibalias{ CMS_Tracker_TDR}{Klein:2017nke}
\bibalias{cmstdr-014}{Klein:2017nke}
\bibalias{ cmstdr-014}{Klein:2017nke}
\bibalias{CMSPhase2TrackerTDR}{Klein:2017nke}
\bibalias{ CMSPhase2TrackerTDR}{Klein:2017nke}
\bibalias{CMS-PAS-HIG-16-040}{CMS:2017rli}
\bibalias{ CMS-PAS-HIG-16-040}{CMS:2017rli}
\bibalias{PhysRevD.91.012006}{Aad:2014eva}
\bibalias{ PhysRevD.91.012006}{Aad:2014eva}
\bibalias{HIG16041}{Sirunyan:2017exp}
\bibalias{ HIG16041}{Sirunyan:2017exp}
\bibalias{ATLAS-CONF-2016-112}{ATLAS:2016gld}
\bibalias{ ATLAS-CONF-2016-112}{ATLAS:2016gld}
\bibalias{HIG16044}{Sirunyan:2017elk}
\bibalias{ HIG16044}{Sirunyan:2017elk}
\bibalias{HIG16043}{Sirunyan:2017khh}
\bibalias{ HIG16043}{Sirunyan:2017khh}
\bibalias{HIG-17-019}{CMS-PAS-HIG-17-019}
\bibalias{ HIG-17-019}{CMS-PAS-HIG-17-019}
\bibalias{ATLAS_Pixel_TDR}{ITKPixelTDR}
\bibalias{ ATLAS_Pixel_TDR}{ITKPixelTDR}
\bibalias{Collaboration:2285585}{ITKPixelTDR}
\bibalias{ Collaboration:2285585}{ITKPixelTDR}
\bibalias{ATLAS_TDAQ_TDR}{Collaboration:2285584}
\bibalias{ ATLAS_TDAQ_TDR}{Collaboration:2285584}
\bibalias{pythia}{Sjostrand:2014zea}
\bibalias{ pythia}{Sjostrand:2014zea}
\bibalias{powheg}{Alioli:2008tz}
\bibalias{ powheg}{Alioli:2008tz}
\bibalias{Carvalho2016}{Dall'Osso:2015aia}
\bibalias{ Carvalho2016}{Dall'Osso:2015aia}
\bibalias{Carvalho:2015ttv}{Dall'Osso:2015aia}
\bibalias{ Carvalho:2015ttv}{Dall'Osso:2015aia}
\bibalias{ATLAS-CONF-2018-031}{ATLAS:2018doi}
\bibalias{ ATLAS-CONF-2018-031}{ATLAS:2018doi}
\bibalias{FTR-16-002}{CMS:2017cwx}
\bibalias{ FTR-16-002}{CMS:2017cwx}
\bibalias{CMS-PAS-FTR-16-002}{CMS:2017cwx}
\bibalias{ CMS-PAS-FTR-16-002}{CMS:2017cwx}
\bibalias{CMSHiggsJuly2012}{Chatrchyan:2012xdj}
\bibalias{ CMSHiggsJuly2012}{Chatrchyan:2012xdj}
\bibalias{Goncalves:2018qas}{Goncalves:2018yva}
\bibalias{ Goncalves:2018qas}{Goncalves:2018yva}
\bibalias{CLs_2002}{Read:2002hq}
\bibalias{ CLs_2002}{Read:2002hq}
\bibalias{ATLAS-CONF-2016-024}{ATLAS:2016iqc}
\bibalias{ ATLAS-CONF-2016-024}{ATLAS:2016iqc}
\bibalias{Franceschini:2017xkh}{Franceschini:2017ab}
\bibalias{ Franceschini:2017xkh}{Franceschini:2017ab}
\bibalias{fastjet}{Cacciari:2011ma}
\bibalias{ fastjet}{Cacciari:2011ma}
\bibalias{CMS-PAS-HIG-18-002}{CMS:2018bwq}
\bibalias{ CMS-PAS-HIG-18-002}{CMS:2018bwq}
\bibalias{ATLAS_Strip_TDR}{Collaboration:2017mtb}
\bibalias{ ATLAS_Strip_TDR}{Collaboration:2017mtb}
\bibalias{Collaboration:2257755}{Collaboration:2017mtb}
\bibalias{ Collaboration:2257755}{Collaboration:2017mtb}
\bibalias{CMS-NOTE-2011-005}{ATLAS:2011tau}
\bibalias{ CMS-NOTE-2011-005}{ATLAS:2011tau}
\bibalias{CMSHIGGSJHEP}{Chatrchyan:2013lba}
\bibalias{ CMSHIGGSJHEP}{Chatrchyan:2013lba}
\bibalias{CMS-PAS-HIG-14-040}{CMS:2015udp}
\bibalias{ CMS-PAS-HIG-14-040}{CMS:2015udp}
\bibalias{DAmbrosio:2002vsn}{D'Ambrosio:2002ex}
\bibalias{ DAmbrosio:2002vsn}{D'Ambrosio:2002ex}
\bibalias{CMS-PAS-HIG-17-028}{CMS:2018hhg}
\bibalias{ CMS-PAS-HIG-17-028}{CMS:2018hhg}
\bibalias{HIG-16-042}{CMS-PAS-HIG-16-042}
\bibalias{ HIG-16-042}{CMS-PAS-HIG-16-042}
\bibalias{HIG17010}{Sirunyan:2017dgc}
\bibalias{ HIG17010}{Sirunyan:2017dgc}
\bibalias{ATLASRun2Ditau}{Aaboud:2017sjh}
\bibalias{ ATLASRun2Ditau}{Aaboud:2017sjh}
\bibalias{PhysRevD.90.075004}{Curtin:2013fra}
\bibalias{ PhysRevD.90.075004}{Curtin:2013fra}
\bibalias{bsmh}{Curtin:2013fra}
\bibalias{ bsmh}{Curtin:2013fra}
\bibalias{HIG-17-020}{Sirunyan:2018zut}
\bibalias{ HIG-17-020}{Sirunyan:2018zut}
\bibalias{LEPewkfits}{ALEPH:2010aa}
\bibalias{ LEPewkfits}{ALEPH:2010aa}
\bibalias{MSSMBenchmarks}{Carena:2013ytb}
\bibalias{ MSSMBenchmarks}{Carena:2013ytb}
\bibalias{CMSL1interim}{cmstdr-017}
\bibalias{ CMSL1interim}{cmstdr-017}
\bibalias{CMS_Barrel_TDR}{cmstdr-barrel}
\bibalias{ CMS_Barrel_TDR}{cmstdr-barrel}
\bibalias{Collaboration:2283187}{cmstdr-barrel}
\bibalias{ Collaboration:2283187}{cmstdr-barrel}
\bibalias{CMSPhase2BarrelTDR}{cmstdr-barrel}
\bibalias{ CMSPhase2BarrelTDR}{cmstdr-barrel}
\bibalias{CMS_Muon_TDR}{cmstdr-mu}
\bibalias{ CMS_Muon_TDR}{cmstdr-mu}
\bibalias{Collaboration:2283189}{cmstdr-mu}
\bibalias{ Collaboration:2283189}{cmstdr-mu}
\bibalias{CMSPhase2MuonTDR}{cmstdr-mu}
\bibalias{ CMSPhase2MuonTDR}{cmstdr-mu}
\bibalias{CMS_HGCal_TDR}{cmstdr-ec}
\bibalias{ CMS_HGCal_TDR}{cmstdr-ec}
\bibalias{Collaboration:2293646}{cmstdr-ec}
\bibalias{ Collaboration:2293646}{cmstdr-ec}
\bibalias{CMSPhase2EndcapTDR}{cmstdr-ec}
\bibalias{ CMSPhase2EndcapTDR}{cmstdr-ec}
\bibalias{CMS}{Chatrchyan:2008aa}
\bibalias{ CMS}{Chatrchyan:2008aa}
\bibalias{Chatrchyan:2008zzk}{Chatrchyan:2008aa}
\bibalias{ Chatrchyan:2008zzk}{Chatrchyan:2008aa}
\bibalias{Contardo:2020886}{CMSCollaboration:2015zni}
\bibalias{ Contardo:2020886}{CMSCollaboration:2015zni}
\bibalias{CMSPhase2TP}{CMSCollaboration:2015zni}
\bibalias{ CMSPhase2TP}{CMSCollaboration:2015zni}
\bibalias{Proceedings:2018wrp}{Bahl:2017aev}
\bibalias{ Proceedings:2018wrp}{Bahl:2017aev}
\bibalias{CMS-PAS-HIG-17-009}{CMS:2017xxp}
\bibalias{ CMS-PAS-HIG-17-009}{CMS:2017xxp}
\bibalias{ll1}{Sirunyan:2018pwn}
\bibalias{ ll1}{Sirunyan:2018pwn}

\title{{\normalfont\bfseries\boldmath\huge
\begin{center}
Higgs Physics\\ at the HL-LHC and HE-LHC\\
\begin{normalsize} 
\href{http://lpcc.web.cern.ch/hlhe-lhc-physics-workshop}{Report from Working Group 2 on the Physics of the HL-LHC, and Perspectives at the HE-LHC} 
\end{normalsize}
\end{center}\vspace*{0.2cm}
}}

\newcounter{instituteref}
\newcommand{\iinstitute}[2]{\refstepcounter{instituteref}\label{#1}$^{\ref{#1}}$\href{http://inspirehep.net/record/#1}{#2}}
\newcommand{\iauthor}[3]{\href{http://inspirehep.net/record/#1}{#2}$^{#3}$}
\author{Conveners: \\ 
\iauthor{1064384}{M. Cepeda}{\ref{902725},\ref{905312}},
\iauthor{1059119}{S. Gori}{\ref{1218068}},
\iauthor{1070755}{P. Ilten}{\ref{902671}},
\iauthor{1003971}{M. Kado}{\ref{903100},\ref{902887},\ref{903168}},
\iauthor{1056642}{F. Riva}{\ref{910394}}
\\ \vspace*{4mm} Contributors: \\
\iauthor{1706729}{R. Abdul Khalek}{\ref{903331},\ref{903832}},
\iauthor{1272238}{A. Aboubrahim}{\ref{946076}},
\iauthor{1068305}{J. Alimena}{\ref{1118764}},
\iauthor{1058528}{S. Alioli}{\ref{907960},\ref{907960}},
\iauthor{1034803}{A. Alves}{\ref{912350}},
\iauthor{1063843}{C. Asawatangtrakuldee}{\ref{902770}},
\iauthor{1041900}{A. Azatov}{\ref{904416},\ref{902888}},
\iauthor{1017778}{P. Azzi}{\ref{902884}},
\iauthor{0}{S. Bailey}{\ref{908692}},
\iauthor{1277169}{S. Banerjee}{\ref{908399}},
\iauthor{1017399}{E.L. Barberio}{\ref{1255221}},
\iauthor{1272340}{D. Barducci}{\ref{904416}},
\iauthor{1072893}{G. Barone}{\ref{902689}},
\iauthor{1191766}{M. Bauer}{\ref{908399}},
\iauthor{0}{C. Bautista}{\ref{903185}},
\iauthor{1020248}{P. Bechtle}{\ref{902676}},
\iauthor{1073979}{K. Becker}{\ref{902808}},
\iauthor{1064433}{A. Benaglia}{\ref{909939}},
\iauthor{0}{M. Bengala}{\ref{905303}},
\iauthor{1016657}{N. Berger}{\ref{903421}},
\iauthor{1072897}{C. Bertella}{\ref{1280832}},
\iauthor{1069083}{A. Bethani}{\ref{902984}},
\iauthor{1455824}{A. Betti}{\ref{902676}},
\iauthor{1701380}{A. Biekotter}{\ref{1209215}},
\iauthor{1279838}{F. Bishara}{\ref{902770}},
\iauthor{1016135}{D. Bloch}{\ref{911366}},
\iauthor{1340636}{P. Bokan}{\ref{911852}},
\iauthor{1065538}{O. Bondu}{\ref{910783}},
\iauthor{1058479}{M. Bonvini}{\ref{902887}},
\iauthor{1492000}{L. Borgonovi}{\ref{902878},\ref{902674}},
\iauthor{1262186}{M. Borsato}{\ref{902842}},
\iauthor{1358174}{S. Boselli}{\ref{913314}},
\iauthor{1015470}{S. Braibant-Giacomelli}{\ref{902878},\ref{902674}},
\iauthor{1015081}{G. Buchalla}{\ref{940108}},
\iauthor{1321289}{L. Cadamuro}{\ref{902804}},
\iauthor{1190398}{C. Caillol}{\ref{903349}},
\iauthor{1218056}{A. Calandri}{\ref{902989},\ref{903369}},
\iauthor{1064395}{A. Calderon Tazon}{\ref{907640}},
\iauthor{1014644}{J.M. Campbell}{\ref{902796}},
\iauthor{1053709}{F. Caola}{\ref{908399}},
\iauthor{1686168}{M. Capozi}{\ref{903036}},
\iauthor{984325}{M. Carena}{\ref{902796},\ref{946080}},
\iauthor{1014488}{C.M. Carloni Calame}{\ref{902885}},
\iauthor{1052521}{A. Carmona}{\ref{1280366}},
\iauthor{1048020}{E. Carquin}{\ref{904589}},
\iauthor{1117644}{A. Carvalho Antunes De Oliveira}{\ref{905096}},
\iauthor{1067212}{A. Castaneda Hernandez}{\ref{905672}},
\iauthor{1028975}{O. Cata}{\ref{903203}},
\iauthor{1189874}{A. Celis}{\ref{903038}},
\iauthor{1021757}{A. Cerri}{\ref{903239}},
\iauthor{1014162}{F. Cerutti}{\ref{902953},\ref{903299}},
\iauthor{1071755}{G.S. Chahal}{\ref{902868},\ref{908399}},
\iauthor{1259672}{A. Chakraborty}{\ref{902916}},
\iauthor{1664293}{G. Chaudhary}{\ref{1302872}},
\iauthor{1317597}{X. Chen}{\ref{903370}},
\iauthor{1072913}{A.S. Chisholm}{\ref{902725},\ref{902671}},
\iauthor{1019481}{R. Contino}{\ref{903128}},
\iauthor{0}{A.J. Costa}{\ref{905303}},
\iauthor{1028393}{R. Covarelli}{\ref{902889},\ref{922848}},
\iauthor{1046385}{N. Craig}{\ref{903307}},
\iauthor{1024481}{D. Curtin}{\ref{1084743}},
\iauthor{1422359}{L. D'Eramo}{\ref{903119}},
\iauthor{1293755}{N.P. Dang}{\ref{903537}},
\iauthor{1340629}{P. Das}{\ref{1120892}},
\iauthor{1012276}{S. Dawson}{\ref{902689}},
\iauthor{1096972}{O.A. De Aguiar Francisco}{\ref{902725}},
\iauthor{1021004}{J. de Blas}{\ref{903113},\ref{902884}},
\iauthor{1039860}{S. De Curtis}{\ref{902880}},
\iauthor{1035217}{N. De Filippis}{\ref{902877},\ref{906299}},
\iauthor{1072922}{H. De la Torre}{\ref{903006}},
\iauthor{1706737}{L. de Lima}{\ref{914879}},
\iauthor{1266338}{A. De Wit}{\ref{902770}},
\iauthor{1019444}{C. Delaere}{\ref{910783}},
\iauthor{1315199}{M. Delcourt}{\ref{910783}},
\iauthor{1053232}{M. Delmastro}{\ref{903421}},
\iauthor{1021058}{S. Demers}{\ref{903357}},
\iauthor{1279673}{N. Dev}{\ref{903085}},
\iauthor{1067133}{R. Di Nardo}{\ref{902992}},
\iauthor{1078369}{S. Di Vita}{\ref{922207}},
\iauthor{1210081}{S. Dildick}{\ref{1269381}},
\iauthor{1615019}{L.A.F. do Prado}{\ref{1625414},\ref{903185}},
\iauthor{1057817}{M. Donadelli}{\ref{903186}},
\iauthor{0}{D. Du}{\ref{904187}},
\iauthor{1273362}{G. Durieux}{\ref{903257},\ref{902770}},
\iauthor{1043155}{M. D{\"u}hrssen}{\ref{902725}},
\iauthor{1069650}{O. Eberhardt}{\ref{907907}},
\iauthor{1476947}{K. El  Morabit}{\ref{1693600}},
\iauthor{1221953}{J. Elias-Miro}{\ref{902725}},
\iauthor{1010819}{J. Ellis}{\ref{1277595},\ref{905096},\ref{902725}},
\iauthor{1058345}{C. Englert}{\ref{902823}},
\iauthor{1048347}{R. Essig}{\ref{1228041}},
\iauthor{1498417}{S. Falke}{\ref{903421}},
\iauthor{1071541}{M. Farina}{\ref{1228041}},
\iauthor{1021221}{A. Ferrari}{\ref{903314}},
\iauthor{1020854}{A. Ferroglia}{\ref{910360}},
\iauthor{1067079}{M.C.N. Fiolhais}{\ref{911195}},
\iauthor{1052629}{M. Flechl}{\ref{1441143}},
\iauthor{1069885}{S. Folgueras}{\ref{906106}},
\iauthor{1611747}{E. Fontanesi}{\ref{902878},\ref{902674}},
\iauthor{1062445}{P. Francavilla}{\ref{903119},\ref{902886}},
\iauthor{1052115}{R. Franceschini}{\ref{906528},\ref{907692}},
\iauthor{1040388}{R. Frederix}{\ref{903037}},
\iauthor{1009385}{S. Frixione}{\ref{902881}},
\iauthor{1028676}{G. G\'omez-Ceballos}{\ref{907455}},
\iauthor{1091415}{A. Gabrielli}{\ref{902953},\ref{903299}},
\iauthor{1107853}{S. Gadatsch}{\ref{902725}},
\iauthor{1009009}{M. Gallinaro}{\ref{905303}},
\iauthor{1427143}{A. Gandrakota}{\ref{903404}},
\iauthor{1277998}{J. Gao}{\ref{904740}},
\iauthor{1079088}{F.M. Garay Walls}{\ref{904336}},
\iauthor{1008686}{T. Gehrmann}{\ref{903370}},
\iauthor{1019601}{Y. Gershtein}{\ref{903404}},
\iauthor{1297333}{T. Ghosh}{\ref{945007}},
\iauthor{1008386}{A. Gilbert}{\ref{902725}},
\iauthor{1671457}{R. Glein}{\ref{902748}},
\iauthor{1008202}{E.W.N. Glover}{\ref{908399}},
\iauthor{1319074}{R. Gomez-Ambrosio}{\ref{908399}},
\iauthor{1020595}{R. Gon\c{c}alo}{\ref{905303}},
\iauthor{1078269}{D. Gon\c{c}alves}{\ref{1272953}},
\iauthor{1021199}{M. Gorbahn}{\ref{926895}},
\iauthor{1514504}{E. Gouveia}{\ref{905303}},
\iauthor{1635818}{M. Gouzevitch}{\ref{902974}},
\iauthor{1062192}{P. Govoni}{\ref{909939},\ref{907960}},
\iauthor{1007669}{M. Grazzini}{\ref{903370}},
\iauthor{1615025}{B. Greenberg}{\ref{903404}},
\iauthor{1060980}{K. Grimm}{\ref{903693}},
\iauthor{1007500}{A.V. Gritsan}{\ref{1269385}},
\iauthor{1050762}{A. Grohsjean}{\ref{902770}},
\iauthor{1007486}{C. Grojean}{\ref{902770}},
\iauthor{1274618}{J. Gu}{\ref{1224838}},
\iauthor{1426008}{R. Gugel}{\ref{902808}},
\iauthor{1084007}{R.S. Gupta}{\ref{908399}},
\iauthor{1067001}{C.B. Gwilliam}{\ref{902964}},
\iauthor{1030568}{S. H\"oche}{\ref{903206}},
\iauthor{0}{M. Haacke}{\ref{904336}},
\iauthor{1235870}{Y. Haddad}{\ref{902868}},
\iauthor{1006981}{U. Haisch}{\ref{903036}},
\iauthor{1231547}{G.N. Hamity}{\ref{903196}},
\iauthor{1006825}{T. Han}{\ref{1272953}},
\iauthor{1069568}{L.A. Harland-Lang}{\ref{908692}},
\iauthor{1019568}{R. Harnik}{\ref{902796}},
\iauthor{1006300}{S. Heinemeyer}{\ref{907640},\ref{910767}},
\iauthor{1006298}{G. Heinrich}{\ref{903036}},
\iauthor{1383268}{B. Henning}{\ref{910394}},
\iauthor{1074931}{V. Hirschi}{\ref{903369}},
\iauthor{1005812}{K. Hoepfner}{\ref{910724}},
\iauthor{1068037}{J.M. Hogan}{\ref{1241132},\ref{902692}},
\iauthor{1515880}{S. Homiller}{\ref{910429},\ref{902689}},
\iauthor{1057527}{Y. Huang}{\ref{903123}},
\iauthor{1272381}{A. Huss}{\ref{902725}},
\iauthor{1004354}{S. J\'ez\'equel}{\ref{903421}},
\iauthor{1064540}{Sa. Jain}{\ref{1120892}},
\iauthor{1357945}{S.P. Jones}{\ref{902725}},
\iauthor{1028368}{K. K\"oneke}{\ref{902808}},
\iauthor{1003873}{J. Kalinowski}{\ref{903335}},
\iauthor{1033909}{J.F. Kamenik}{\ref{903408},\ref{903532}},
\iauthor{1032359}{M. Kaplan}{\ref{907455}},
\iauthor{1069015}{A. Karlberg}{\ref{903370}},
\iauthor{1189871}{M. Kaur}{\ref{1302872}},
\iauthor{1676775}{P. Keicher}{\ref{1693600}},
\iauthor{1234959}{M. Kerner}{\ref{903370}},
\iauthor{1003109}{A. Khanov}{\ref{903094}},
\iauthor{1123712}{J. Kieseler}{\ref{902725}},
\iauthor{1407901}{J.H. Kim}{\ref{902912}},
\iauthor{}{M. Kim}{\ref{1294399}},
\iauthor{1396923}{T. Klijnsma}{\ref{903369}},
\iauthor{1273332}{F. Kling}{\ref{903302}},
\iauthor{1020906}{M. Klute}{\ref{907455}},
\iauthor{1060935}{J.R. Komaragiri}{\ref{902658}},
\iauthor{1020007}{K. Kong}{\ref{902912}},
\iauthor{1121767}{J. Kozaczuk}{\ref{1112663}},
\iauthor{0}{P Kozow}{\ref{903335}},
\iauthor{1256558}{C. Krause}{\ref{902796}},
\iauthor{1050782}{S. Lai}{\ref{911852}},
\iauthor{1648705}{J. Langford}{\ref{902868}},
\iauthor{1345396}{B. Le}{\ref{1255221}},
\iauthor{1670004}{L. Lechner}{\ref{1441143}},
\iauthor{1061283}{W.A. Leight}{\ref{902666}},
\iauthor{1066810}{K.J.C. Leney}{\ref{905856}},
\iauthor{1066809}{T. Lenz}{\ref{902676}},
\iauthor{1356922}{C-Q. Li}{\ref{903800}},
\iauthor{1057147}{H. Li}{\ref{904187}},
\iauthor{1074984}{Q. Li}{\ref{903603}},
\iauthor{1061148}{S. Liebler}{\ref{1389990}},
\iauthor{1069465}{J. Lindert}{\ref{908399}},
\iauthor{1393097}{D. Liu}{\ref{902645}},
\iauthor{1336279}{J. Liu}{\ref{902730}},
\iauthor{0}{Y. Liu}{\ref{904118}},
\iauthor{1256188}{Z. Liu}{\ref{902990},\ref{902796}},
\iauthor{1665809}{D. Lombardo}{\ref{910394}},
\iauthor{1069459}{A. Long}{\ref{903156}},
\iauthor{1280606}{K. Long}{\ref{903349}},
\iauthor{999724}{I. Low}{\ref{903083},\ref{902645}},
\iauthor{1056846}{G. Luisoni}{\ref{903036}},
\iauthor{1355155}{L.L. Ma}{\ref{904187}},
\iauthor{1023556}{A.-M. Magnan}{\ref{902868}},
\iauthor{1064163}{D. Majumder}{\ref{902912}},
\iauthor{1662612}{A. Malinauskas}{\ref{908692}},
\iauthor{999053}{F. Maltoni}{\ref{1095325}},
\iauthor{998982}{M.L. Mangano}{\ref{902725}},
\iauthor{1028397}{G. Marchiori}{\ref{903119},\ref{903119}},
\iauthor{1073582}{A.C. Marini}{\ref{907455}},
\iauthor{1049426}{A. Martin}{\ref{903085}},
\iauthor{1051050}{S. Marzani}{\ref{902814},\ref{902881}},
\iauthor{1067362}{A. Massironi}{\ref{902725}},
\iauthor{998430}{K.T. Matchev}{\ref{902804},\ref{1225439}},
\iauthor{1041384}{R.D. Matheus}{\ref{903185}},
\iauthor{998164}{K. Mazumdar}{\ref{1120892}},
\iauthor{1259473}{J. Mazzitelli}{\ref{903370}},
\iauthor{0}{A.E. Mcdougall}{\ref{1255221}},
\iauthor{1025277}{P. Meade}{\ref{910429}},
\iauthor{1055394}{P. Meridiani}{\ref{902887}},
\iauthor{997651}{A.B. Meyer}{\ref{902770}},
\iauthor{1401143}{E. Michielin}{\ref{902884}},
\iauthor{1053883}{P. Milenovic}{\ref{902725},\ref{903925}},
\iauthor{1467369}{V. Milosevic}{\ref{902868}},
\iauthor{1274623}{K. Mimasu}{\ref{1095325}},
\iauthor{1259311}{B. Mistlberger}{\ref{1237813}},
\iauthor{1384622}{M. Mlynarikova}{\ref{902727}},
\iauthor{997010}{M. Mondragon}{\ref{903003}},
\iauthor{1078032}{P.F. Monni}{\ref{902725}},
\iauthor{996990}{G. Montagna}{\ref{903122},\ref{902885}},
\iauthor{1634517}{F. Monti}{\ref{909939},\ref{907960}},
\iauthor{1066789}{M. Moreno Llacer}{\ref{902725}},
\iauthor{996593}{A. Mueck}{\ref{902624}},
\iauthor{1021431}{P.C. Mui\~{n}o}{\ref{905303}},
\iauthor{1081421}{C. Murphy}{\ref{1268258}},
\iauthor{996420}{W.J. Murray}{\ref{903734}},
\iauthor{1062160}{P. Musella}{\ref{903369}},
\iauthor{996070}{M. Narain}{\ref{902692}},
\iauthor{1078009}{R.F. Naranjo Garcia}{\ref{902666}},
\iauthor{996010}{P. Nath}{\ref{946076}},
\iauthor{995832}{M. Neubert}{\ref{911853}},
\iauthor{995696}{O. Nicrosini}{\ref{902885}},
\iauthor{1066672}{K. Nikolopoulos}{\ref{902671}},
\iauthor{995578}{A. Nisati}{\ref{902887},\ref{903168}},
\iauthor{1044763}{J.M. No}{\ref{910767}},
\iauthor{1589937}{M.L. Ojeda}{\ref{903282}},
\iauthor{1074074}{S.A. Olivares Pino}{\ref{904336}},
\iauthor{994850}{A. Onofre}{\ref{905617}},
\iauthor{1063052}{G. Ortona}{\ref{902786}},
\iauthor{1048820}{S. Pagan Griso}{\ref{902953},\ref{903299}},
\iauthor{1274353}{D. Pagani}{\ref{903037}},
\iauthor{1028707}{E. Palencia Cortezon}{\ref{906106}},
\iauthor{1067383}{C. Palmer}{\ref{1118336}},
\iauthor{1265353}{C. Pandini}{\ref{902725}},
\iauthor{1037621}{G. Panico}{\ref{902801},\ref{902880}},
\iauthor{1643518}{L. Panwar}{\ref{902658}},
\iauthor{1061215}{D. Pappadopulo}{\ref{1191473}},
\iauthor{1023777}{M. Park}{\ref{1205048}},
\iauthor{1067799}{R. Patel}{\ref{902748}},
\iauthor{0}{F. Paucar-Velasquez}{\ref{903404}},
\iauthor{1067964}{K. Pedro}{\ref{902796}},
\iauthor{1190393}{L. Pernie}{\ref{1269381}},
\iauthor{1064078}{L. Perrozzi}{\ref{903369}},
\iauthor{993634}{B.A. Petersen}{\ref{902725}},
\iauthor{1066617}{E. Petit}{\ref{902828}},
\iauthor{1056382}{G. Petrucciani}{\ref{902725}},
\iauthor{1066610}{G. Piacquadio}{\ref{903237}},
\iauthor{993440}{F. Piccinini}{\ref{902885}},
\iauthor{993396}{M. Pieri}{\ref{903305}},
\iauthor{993181}{T. Plehn}{\ref{902842}},
\iauthor{991231}{S. Pokorski}{\ref{903335}},
\iauthor{993026}{A. Pomarol}{\ref{907904}},
\iauthor{992997}{E. Ponton}{\ref{903185},\ref{1119124}},
\iauthor{1019193}{S. Pozzorini}{\ref{903370}},
\iauthor{1058764}{S. Prestel}{\ref{902796}},
\iauthor{1058678}{K. Prokofiev}{\ref{906840}},
\iauthor{992324}{M. Ramsey-Musolf}{\ref{945696}},
\iauthor{1058529}{E. Re}{\ref{902725},\ref{908074}},
\iauthor{1222888}{N.P. Readioff}{\ref{902828}},
\iauthor{1214912}{D. Redigolo}{\ref{903259},\ref{903138}},
\iauthor{992031}{L. Reina}{\ref{902803}},
\iauthor{1403062}{E. Reynolds}{\ref{902671},\ref{902671}},
\iauthor{1497192}{M. Riembau}{\ref{910394}},
\iauthor{1040388}{F. Rikkert}{\ref{903037}},
\iauthor{1040385}{T. Robens}{\ref{902678}},
\iauthor{1056069}{R. Roentsch}{\ref{1389986},\ref{902725}},
\iauthor{1019897}{J. Rojo}{\ref{903832}},
\iauthor{1064039}{N. Rompotis}{\ref{902964}},
\iauthor{1041241}{J. Rorie}{\ref{903156}},
\iauthor{991231}{J. Rosiek}{\ref{903335}},
\iauthor{1319409}{J. Roskes}{\ref{1269385}},
\iauthor{1061185}{J.T. Ruderman}{\ref{1191473}},
\iauthor{1239736}{N. Sahoo}{\ref{1120892}},
\iauthor{}{S. Saito}{\ref{1120892}},
\iauthor{1062135}{R. Salerno}{\ref{902786}},
\iauthor{1127850}{P.H. Sales De Bruin}{\ref{903338}},
\iauthor{1066537}{A. Salvucci}{\ref{902855}},
\iauthor{}{K. Sandeep}{\ref{1302872}},
\iauthor{990367}{J. Santiago}{\ref{909079}},
\iauthor{0}{R. Santo}{\ref{905303}},
\iauthor{1019116}{V. Sanz}{\ref{903239}},
\iauthor{1074808}{U. Sarica}{\ref{1269385}},
\iauthor{1029853}{A. Savin}{\ref{903349}},
\iauthor{1023298}{A. Savoy-Navarro}{\ref{1625414},\ref{910133}},
\iauthor{1632393}{S. Sawant}{\ref{1120892}},
\iauthor{1066527}{A.C. Schaffer}{\ref{903100}},
\iauthor{1259433}{M. Schlaffer}{\ref{903342}},
\iauthor{1064622}{A. Schmidt}{\ref{910724}},
\iauthor{1074089}{B. Schneider}{\ref{902796}},
\iauthor{1047939}{R. Schoefbeck}{\ref{1441143}},
\iauthor{1287849}{M. Schr{\"o}der}{\ref{1693600}},
\iauthor{1706140}{M. Scodeggio}{\ref{902666}},
\iauthor{1474586}{E. Scott}{\ref{902868}},
\iauthor{1480952}{L. Scyboz}{\ref{903036}},
\iauthor{1039590}{M. Selvaggi}{\ref{902725}},
\iauthor{1342183}{L. Sestini}{\ref{902884}},
\iauthor{1077844}{H.-S. Shao}{\ref{908583}},
\iauthor{1072204}{A. Shivaji}{\ref{1095325},\ref{911848}},
\iauthor{988661}{L. Silvestrini}{\ref{902887},\ref{902725}},
\iauthor{}{L. Simon}{\ref{902624}},
\iauthor{1024959}{K. Sinha}{\ref{1273509}},
\iauthor{1073494}{Y. Soreq}{\ref{902725},\ref{903257}},
\iauthor{1045921}{M. Spannowsky}{\ref{902782}},
\iauthor{987872}{M. Spira}{\ref{905405}},
\iauthor{1488501}{D. Spitzbart}{\ref{1441143}},
\iauthor{1072178}{E. Stamou}{\ref{946080}},
\iauthor{1021980}{J. Stark}{\ref{902828}},
\iauthor{1055978}{T. Stefaniak}{\ref{902770}},
\iauthor{1065513}{B. Stieger}{\ref{1273761},\ref{903057}},
\iauthor{1439225}{G. Strong}{\ref{905303}},
\iauthor{986929}{M. Szleper}{\ref{1658555}},
\iauthor{1046648}{K. Tackmann}{\ref{902666}},
\iauthor{1051155}{M. Takeuchi}{\ref{1209632}},
\iauthor{1055004}{S. Taroni}{\ref{903085}},
\iauthor{1071093}{M. Testa}{\ref{902807}},
\iauthor{1084006}{A. Thamm}{\ref{902725}},
\iauthor{1361826}{V. Theeuwes}{\ref{902826},\ref{1087875}},
\iauthor{1074097}{L.A. Thomsen}{\ref{903357}},
\iauthor{986030}{S. Tkaczyk}{\ref{902796}},
\iauthor{1064514}{R. Torre}{\ref{902881},\ref{902725}},
\iauthor{1043083}{F. Tramontano}{\ref{903043},\ref{902883}},
\iauthor{1028311}{K.A. Ulmer}{\ref{902748}},
\iauthor{1671474}{T. Vantalon}{\ref{902770}},
\iauthor{1051399}{L. Vecchi}{\ref{1471035}},
\iauthor{1061171}{R. Vega-Morales}{\ref{909079}},
\iauthor{1478448}{E. Venturini}{\ref{904416}},
\iauthor{1039193}{M. Verducci}{\ref{907692},\ref{906528}},
\iauthor{1067845}{C. Vernieri}{\ref{903206}},
\iauthor{1021153}{T. Vickey}{\ref{903196}},
\iauthor{1054164}{M. Vidal Marono}{\ref{910783}},
\iauthor{1054943}{P. Vischia}{\ref{906106}},
\iauthor{1077733}{E. Vryonidou}{\ref{902725}},
\iauthor{1233257}{P. Wagner}{\ref{902676}},
\iauthor{1511804}{V.M. Walbrecht}{\ref{903036}},
\iauthor{984146}{L.-T. Wang}{\ref{902730}},
\iauthor{1070324}{N. Wardle}{\ref{902868}},
\iauthor{1062116}{D.R. Wardrope}{\ref{903311}},
\iauthor{983877}{G. Weiglein}{\ref{902770}},
\iauthor{1315209}{S. Wertz}{\ref{910783}},
\iauthor{983600}{M. Wielers}{\ref{903174}},
\iauthor{1096938}{J.M. Williams}{\ref{907455}},
\iauthor{1064531}{R. Wolf}{\ref{1693600}},
\iauthor{1037622}{A. Wulzer}{\ref{902725}},
\iauthor{1073040}{M. Xiao}{\ref{1269385}},
\iauthor{1080678}{H.T. Yang}{\ref{902953},\ref{903299}},
\iauthor{1046262}{E. Yazgan}{\ref{903123}},
\iauthor{1663425}{Z. Yin}{},
\iauthor{1087415}{T. You}{\ref{902712},\ref{913314}},
\iauthor{1056696}{F. Yu}{\ref{1280366},\ref{902982}},
\iauthor{982365}{G. Zanderighi}{\ref{902725},\ref{903036}},
\iauthor{1074144}{D. Zanzi}{\ref{902725}},
\iauthor{1065776}{M. Zaro}{\ref{903832},\ref{903832}},
\iauthor{1054201}{S.C. Zenz}{\ref{902868},\ref{903146}},
\iauthor{982248}{D. Zerwas}{\ref{903100}},
\iauthor{1510840}{M. Zgubi\v{c}}{\ref{903112}},
\iauthor{0}{J. Zhang}{\ref{904187}},
\iauthor{1077637}{L. Zhang}{\ref{904118}},
\iauthor{1666186}{W. Zhang}{\ref{902692}},
\iauthor{1656814}{X. Zhao}{\ref{910783},\ref{902689}},
\iauthor{1288248}{Y.-M. Zhong}{\ref{945903}}
\vspace*{1cm}} \institute{\small 
\iinstitute{902725}{CERN, Geneva};
\iinstitute{905312}{CIEMAT, Madrid};
\iinstitute{1218068}{UC, Santa Cruz, SCIPP};
\iinstitute{902671}{U. Birmingham, Sch. Phys. Astron.};
\iinstitute{903100}{LAL, Orsay};
\iinstitute{902887}{INFN, Rome 1};
\iinstitute{903168}{U. Rome 1, La Sapienza, Dept. Phys.};
\iinstitute{910394}{U. Geneva, Dept. Theor. Phys.};
\iinstitute{903331}{Vrije U., Amsterdam, Dept. Phys. Astron.};
\iinstitute{903832}{Nikhef, Amsterdam};
\iinstitute{946076}{Northeastern U.};
\iinstitute{1118764}{Ohio State U., Columbus};
\iinstitute{907960}{U. Milan Bicocca, Dept. Phys.};
\iinstitute{912350}{Diadema, Sao Paulo Fed. U.};
\iinstitute{902770}{DESY, Hamburg};
\iinstitute{904416}{SISSA, Trieste};
\iinstitute{902888}{INFN, Trieste};
\iinstitute{902884}{INFN, Padua};
\iinstitute{908692}{U. Oxford, Ctr. Theor. Phys.};
\iinstitute{908399}{Durham U., IPPP};
\iinstitute{1255221}{ARC, CoEPP, Melbourne};
\iinstitute{902689}{Brookhaven Natl. Lab., Dept. Phys.};
\iinstitute{903185}{UNESP Sao Paulo, IFT};
\iinstitute{902676}{U. Bonn, Phys. Inst.};
\iinstitute{902808}{U. Freiburg, Inst. Phys.};
\iinstitute{909939}{INFN, Milan Bicocca};
\iinstitute{905303}{LIP, Lisbon};
\iinstitute{903421}{LAPP, Annecy};
\iinstitute{1280832}{CAS, Beijing};
\iinstitute{902984}{U. Manchester, Sch. Phys. Astron.};
\iinstitute{1209215}{U. Heidelberg, ITP};
\iinstitute{911366}{IPHC, Strasbourg};
\iinstitute{911852}{U. Gottingen, II. Phys. Inst.};
\iinstitute{910783}{Cathol. U. Louvain, CP3};
\iinstitute{902878}{INFN, Bologna};
\iinstitute{902674}{U. Bologna, Dept. Phys.};
\iinstitute{902842}{U. Heidelberg, Phys.Inst.};
\iinstitute{913314}{U. Cambridge};
\iinstitute{940108}{LMU Munich};
\iinstitute{902804}{U. Florida, Gainesville, Dept. Phys.};
\iinstitute{903349}{U. Wisconsin, Madison,  Dept. Phys.};
\iinstitute{902989}{CPPM, Marseille};
\iinstitute{903369}{ETH, Zurich, Dept. Phys.};
\iinstitute{907640}{IFCA, Santander};
\iinstitute{902796}{Fermilab};
\iinstitute{903036}{MPI Phys., Munich};
\iinstitute{946080}{U. Chicago};
\iinstitute{902885}{INFN, Pavia};
\iinstitute{1280366}{U. Mainz, PRISMA};
\iinstitute{904589}{U. Tech. Federico Santa Maria, Valparaiso};
\iinstitute{905096}{NICPB, Tallinn};
\iinstitute{905672}{Sonora U.};
\iinstitute{903203}{U. Siegen, Dept. Phys.};
\iinstitute{903038}{LMU Munich, Dept. Phys.};
\iinstitute{903239}{U. Sussex, Brighton, Dept. Phys. Astron.};
\iinstitute{902953}{LBNL, Berkeley, Div. Phys.};
\iinstitute{903299}{UC, Berkeley, Dept. Phys.};
\iinstitute{902868}{Imperial Coll., London, Dept. Phys.};
\iinstitute{902916}{KEK, Tsukuba};
\iinstitute{1302872}{Panjab U., Chandigarh};
\iinstitute{903370}{U. Zurich, Phys. Inst.};
\iinstitute{903128}{Scuola Normale Superiore, Pisa};
\iinstitute{902889}{INFN, Turin};
\iinstitute{922848}{U. Turin, Dept. Exp. Phys.};
\iinstitute{903307}{UC, Santa Barbara, Dept. Phys.};
\iinstitute{1084743}{U. Toronto};
\iinstitute{903119}{LPNHE, Paris};
\iinstitute{903537}{U. Louisville, Dept. Phys.};
\iinstitute{1120892}{TIFR, Mumbai, DHEP};
\iinstitute{903113}{U. Padua, Dept. Phys.};
\iinstitute{902880}{INFN, Florence};
\iinstitute{902877}{INFN, Bari};
\iinstitute{906299}{Polytech. Bari};
\iinstitute{903006}{Michigan State U., East Lansing, Dept. Phys. Astron.};
\iinstitute{914879}{Federal da Fronteira Sul U.};
\iinstitute{903357}{Yale U., Dept. Phys.};
\iinstitute{903085}{U. Notre Dame, Dept. Phys.};
\iinstitute{902992}{UMass, Amherst, Dept. Phys.};
\iinstitute{922207}{INFN, Italy};
\iinstitute{1269381}{TAMU, College Station};
\iinstitute{1625414}{IRFU, Saclay, DPP};
\iinstitute{903186}{U. Sao Paulo, Inst. Phys.};
\iinstitute{904187}{Shandong U., Jinan};
\iinstitute{903257}{Technion, IIT, Dept. Phys.};
\iinstitute{907907}{IFIC, Valencia};
\iinstitute{1693600}{KIT, Karlsruhe, ETP};
\iinstitute{1277595}{King's Coll. London};
\iinstitute{902823}{U. Glasgow, Sch. Phys. Astron.};
\iinstitute{1228041}{Stony Brook U.};
\iinstitute{903314}{Uppsala U., Dept. Phys. Astron.};
\iinstitute{910360}{CUNY, City Tech.};
\iinstitute{911195}{CUNY, BMCC};
\iinstitute{1441143}{OEAW, Vienna};
\iinstitute{906106}{U. Oviedo, Dept. Phys.};
\iinstitute{902886}{INFN, Pisa};
\iinstitute{906528}{U. Rome 3, Dept. Math. Phys.};
\iinstitute{907692}{INFN, Rome 3};
\iinstitute{903037}{Tech. U., Munich, Dept. Phys.};
\iinstitute{902881}{INFN, Genoa};
\iinstitute{907455}{MIT, Cambridge};
\iinstitute{903404}{Rutgers U., Piscataway, Dept. Phys. Astron.};
\iinstitute{904740}{SJTU, Sch. Phys. Astron., Shanghai};
\iinstitute{904336}{Pontificia U. Catol. Chile, Santiago, Dept. Phys.};
\iinstitute{945007}{U. Hawaii};
\iinstitute{902748}{U. Colorado, Boulder, Dept. Phys.};
\iinstitute{1272953}{U. Pittsburgh};
\iinstitute{926895}{U. Liverpool};
\iinstitute{902974}{IPNL, Lyon};
\iinstitute{903693}{Cal State, East Bay};
\iinstitute{1269385}{Johns Hopkins U.};
\iinstitute{1224838}{UC, Davis};
\iinstitute{902964}{U. Liverpool, Dept. Phys.};
\iinstitute{903206}{SLAC};
\iinstitute{903196}{U. Sheffield, Dept. Phys. Astron.};
\iinstitute{910767}{IFT, Madrid};
\iinstitute{910724}{RWTH Aachen};
\iinstitute{1241132}{Bethel Coll.};
\iinstitute{902692}{Brown U., Dept. Phys.};
\iinstitute{910429}{Stony Brook U., YITP};
\iinstitute{903123}{CAS, IHEP, Beijing};
\iinstitute{903335}{U. Warsaw, Fac. Phys.};
\iinstitute{903408}{J. Stefan Inst., Ljubljana};
\iinstitute{903532}{U. Ljubljana, Fac. Math. Phys.};
\iinstitute{903094}{OKState, Stillwater, Dept. Phys.};
\iinstitute{902912}{U. Kansas, Lawrence, Dept. Phys. Astron.};
\iinstitute{1294399}{POSTECH, Pohang};
\iinstitute{903302}{UC, Irvine, Dept. Phys. Astron.};
\iinstitute{902658}{Indian Inst. Sci., Bangalore};
\iinstitute{1112663}{U. Illinois, Urbana-Champaign};
\iinstitute{902666}{DESY, Zeuthen};
\iinstitute{905856}{Southern Methodist U., Dept. Phys.};
\iinstitute{903800}{USTC, Hefei, DMP};
\iinstitute{903603}{Peking U., Beijing, Sch. Phys.};
\iinstitute{1389990}{KIT, Karlsruhe, TP};
\iinstitute{902645}{Argonne Natl. Lab., HEP Div.};
\iinstitute{902730}{Chicago U., EFI};
\iinstitute{904118}{Nanjing U., Dept. Phys.};
\iinstitute{902990}{U. Maryland, College Park, Dept. Phys.};
\iinstitute{903156}{Rice U., Dept. Phys. Astron.};
\iinstitute{903083}{Northwestern U., Dept. Phys. Astron.};
\iinstitute{1095325}{Cathol. U. Louvain};
\iinstitute{902814}{U. Genoa, Dept. Phys.};
\iinstitute{1225439}{U. Florida, Gainesville};
\iinstitute{903925}{U. Belgrade};
\iinstitute{1237813}{MIT, Cambridge, CTP};
\iinstitute{902727}{Charles U., Prague, Inst. Part. Nucl. Phys.};
\iinstitute{903003}{UNAM, Mexico, IFUNAM};
\iinstitute{903122}{U. Pavia, Dept. Nucl. Theor. Phys.};
\iinstitute{902624}{RWTH, Aachen, Phys. Inst.};
\iinstitute{1268258}{Brookhaven Natl. Lab.};
\iinstitute{903734}{U. Warwick, Dept. Phys.};
\iinstitute{911853}{U. Mainz};
\iinstitute{903282}{U. Toronto, Dept. Phys.};
\iinstitute{905617}{U. Minho, Dept. Math.};
\iinstitute{902786}{LLR, Palaiseau};
\iinstitute{1118336}{Princeton U.};
\iinstitute{902801}{U. Florence, Dept. Phys. Astron.};
\iinstitute{1191473}{New York U.};
\iinstitute{1205048}{Seoultech, Seoul};
\iinstitute{902828}{LPSC, Grenoble};
\iinstitute{903237}{Stony Brook U., Dept. Phys. Astron.};
\iinstitute{903305}{UC, San Diego, Dept. Phys.};
\iinstitute{907904}{U. Barcelona, IFAE};
\iinstitute{1119124}{ICTP-SAIFR, Sao Paulo};
\iinstitute{906840}{HKUST, Hong Kong};
\iinstitute{945696}{UMass, Amherst};
\iinstitute{908074}{LAPTH, Annecy};
\iinstitute{903259}{Tel-Aviv U., Dept. Part. Phys.};
\iinstitute{903138}{IAS, Princeton};
\iinstitute{902803}{Florida State U., Tallahassee, Dept. Phys.};
\iinstitute{902678}{RBT, Zagreb};
\iinstitute{1389986}{KIT, Karlsruhe, Dept. Phys.};
\iinstitute{903338}{U. Washington, Seattle, Dept. Phys.};
\iinstitute{902855}{Chinese U. Hong Kong};
\iinstitute{909079}{U. Granada, CAFPE};
\iinstitute{910133}{APC, Paris};
\iinstitute{903342}{Weizmann Inst. Sci., Rehovot, Fac. Phys.};
\iinstitute{908583}{LPTHE, Paris};
\iinstitute{911848}{IISER, Mohali};
\iinstitute{1273509}{U. Oklahoma, Norman};
\iinstitute{902782}{Durham U., Dept. Phys.};
\iinstitute{905405}{PSI, Villigen};
\iinstitute{1273761}{U. Nebraska, Lincoln};
\iinstitute{903057}{U. Nebraska, Lincoln, Dept. Phys. Astron.};
\iinstitute{1658555}{NCBJ, Warsaw};
\iinstitute{1209632}{U. Tokyo};
\iinstitute{902807}{INFN, LNF, Frascati};
\iinstitute{902826}{U. Gottingen, Inst. Theor. Phys.};
\iinstitute{1087875}{IPhT, Saclay};
\iinstitute{903043}{U. Naples, Dept. Phys. Sci.};
\iinstitute{902883}{INFN, Naples};
\iinstitute{1471035}{EPFL, Lausanne, LPTP};
\iinstitute{903311}{U. Coll. London, Dept. Phys. Astron.};
\iinstitute{903174}{RAL, Didcot};
\iinstitute{902712}{U. Cambridge, Cavendish Lab.};
\iinstitute{902982}{U. Mainz, Inst. Phys.};
\iinstitute{903146}{Queen Mary U. London, Sch. Phys. Astron.};
\iinstitute{903112}{U. Oxford, Part. Phys. Dept.};
\iinstitute{945903}{Boston U.}}

\begin{titlepage}

\vspace*{-1.8cm}

\noindent
\begin{tabular*}{\linewidth}{lc@{\extracolsep{\fill}}r@{\extracolsep{0pt}}}
\vspace*{-1.2cm}\mbox{\!\!\!\includegraphics[width=.14\textwidth]{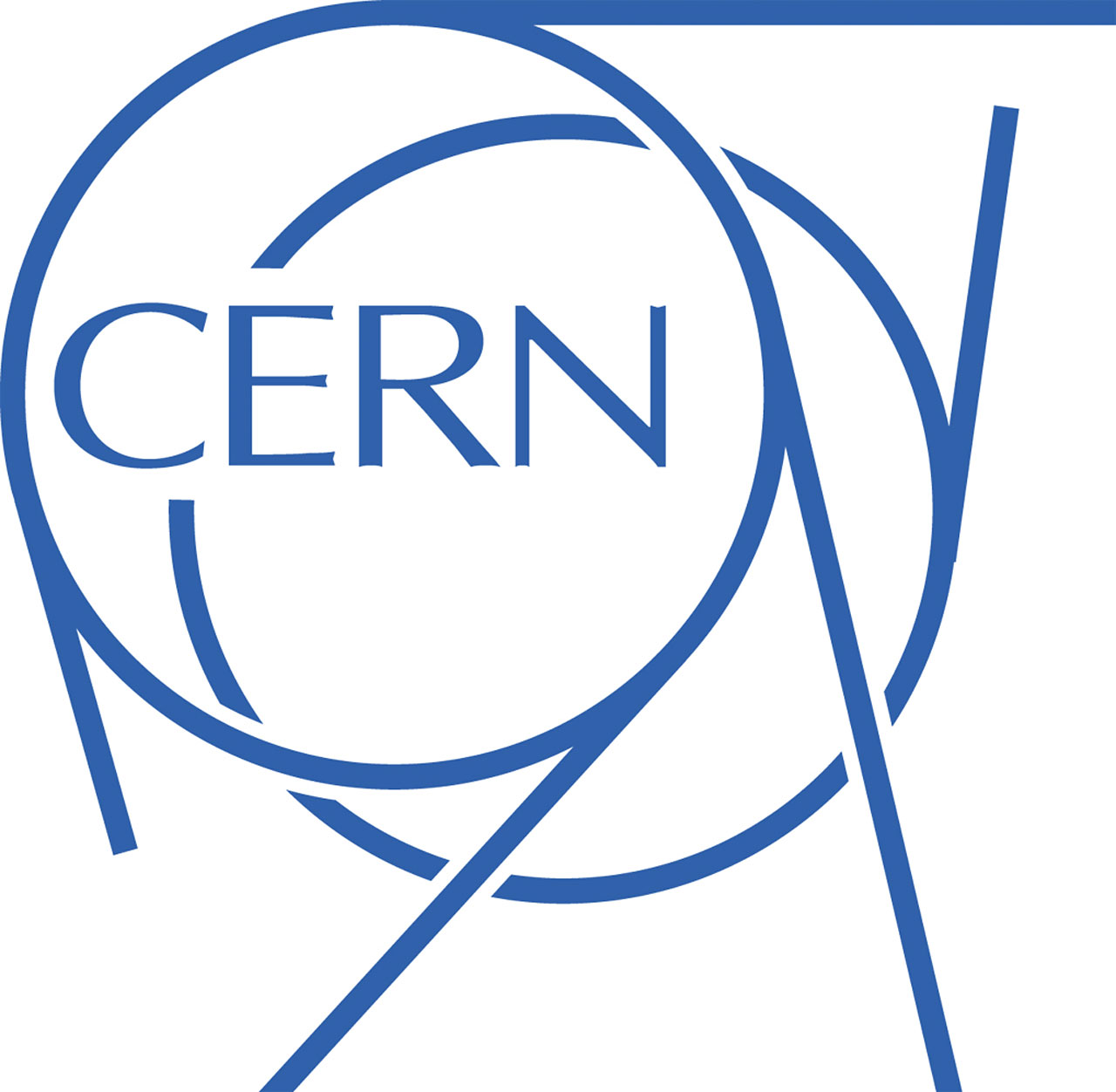}} & & \\
 & & CERN-LPCC-2018-04 \\  
 & & \today \\ 
 & & \\
\hline
\end{tabular*}

\vspace*{0.3cm}
%
%
\maketitle
\vspace{\fill}

\newpage
\begin{abstract}
The discovery of the Higgs boson in 2012, by the
ATLAS and CMS experiments, was a success achieved with only
a percent of the entire dataset foreseen for the
LHC. 
It opened a landscape of possibilities
in the study of Higgs boson properties, Electroweak Symmetry breaking 
and the Standard Model in general, as well as new avenues in probing new  
physics beyond the Standard Model. 
Six years after the discovery, with a conspicuously larger dataset collected during LHC Run 2
 at a  13~\UTeV centre-of-mass energy, the theory and experimental particle physics communities have
started a meticulous exploration of the potential for precision measurements of its properties.
This includes studies of Higgs boson production and
decays processes, the search for rare decays and production modes, high energy observables, and searches for an extended electroweak symmetry breaking
sector. 
This report summarises the potential reach and opportunities in Higgs physics during the High 
Luminosity phase of the LHC, with an expected dataset of pp collisions at 14~\UTeV, 
corresponding to an integrated luminosity of 3~ab$^{-1}$. These studies are performed in light of the 
most recent analyses from LHC collaborations and the latest theoretical developments. 
The potential of an LHC upgrade, colliding
protons at a centre-of-mass energy of 27~\UTeV and producing a dataset 
corresponding to an integrated luminosity of 15~ab$^{-1}$, is also 
discussed. 
\end{abstract}


\vspace*{2.0cm}

\vspace*{2.0cm}


\vspace*{2.0cm}
\vspace{\fill}

\end{titlepage}

\setcounter{tocdepth}{3}
\tableofcontents
\newpage

\section{Introduction}

One of the main goals of the physics program at the Large Hadron
Collider is to elucidate the origin of  electroweak symmetry
breaking.

Relativistic quantum field and gauge theories have been remarkably
successful to describe fundamental particles and their
interactions. 
In this context, the seminal work
of Brout, Englert~\cite{Englert:1964et}, Higgs~\cite{Higgs:1964ia,Higgs:1964pj,Higgs:1966ev} and Guralnik, Hagen and Kibble~\cite{Guralnik:1964eu, Kibble:1967sv}, has provided a consistent mechanism for the generation
of gauge boson masses. 
The Glashow-Weinberg-Salam theory
extended this mechanism proposing a theory of the electroweak
interactions~\cite{Glashow:1961tr,Salam:1968,Weinberg:1967tq}, introducing a doublet of complex scalar
fields, which couples also to fermions, providing them with a  mass  which would otherwise be absent. This is now known as the Standard Model (SM) of particle physics. 
A complete and detailed description of the Higgs mechanism
can be found at~\cite{Tanabashi:2018oca}. A salient prediction of the theory is the
presence of a Higgs boson.
The discovery of the Higgs boson with a mass of 125 \UGeV, during the
first run of the LHC at reduced centre-of-mass energies of 7 \UTeV and
8 \UTeV, is a landmark result that has reshaped the landscape of High
Energy physics~\cite{Aad:2012tfa,Chatrchyan:2012xdj}. 
The mass of the Higgs boson is particularly favourable
as it allows to measure directly a large number of its couplings. It
has also important consequences in terms of probing the self-consistency of the Standard Model both through the global fit of
precision observables and through its interpretation as a measure of the Higgs boson self coupling, allowing to extrapolate the SM at
higher energies and verify the stability of the vacuum.

The existence of the Higgs boson as a light scalar leads to the
hierarchy or naturalness problem, as its mass at the weak scale happens to be particularly sensitive to general larger scales beyond the SM (BSM), therefore apparently requiring a large fine tuning of fundamental parameters.
Addressing the naturalness problem is and has been for decades one of
the main guiding principles for the development of theories beyond the
Standard Model. 
There are two main classes of theories attempting
to address the naturalness problem: the first are weakly coupled theories, 
where the Higgs boson
remains an elementary scalar and its mass is protected by additional
symmetries, as in Supersymmetric theories. 
The second are strongly
coupled solutions, which involve new strong interactions at approximately the \UTeV scale and deliver naturally light composite scalars as Pseudo Nambu-Goldstone bosons. Both approaches can have large effects on the phenomenology
of the Higgs particle and in some cases predict new states that could be 
observed at the LHC.

Other questions of fundamental importance can affect the phenomenology
of the Higgs boson.  The question of the nature of the Electroweak Phase transition is strongly intertwined with Higgs physics where, in many scenarios, a detailed study of the Higgs pair production can reveal the strength of the transition.
Similarly, certain models of Dark Matter involve
potentially large effects on the phenomenology of the Higgs
particle.
These fundamental questions, and many more, can be addressed by the study of the
Higgs boson at the LHC and its high luminosity (HL) and high energy
(HE) upgrades.

Since the discovery, a large campaign of measurements of the properties of the
Higgs boson has started, including exclusive production modes and
differential cross sections. Many new ideas have emerged during the
completion of this program. This chapter presents a reappraised
estimate of the potential of the HL-LHC and the HE-LHC projects to
measure the properties of the Higgs boson, highlighting the
opportunities for measurements of fundamental importance.

Section 2 presents the foreseen program for precision measurements of
the Higgs boson coupling properties through exclusive production modes
and differential cross sections. Section 3 presents the potential to
measure double Higgs production and to constrain the Higgs
trilinear coupling, both through the double Higgs production and
indirect probes from single Higgs boson production. Section 4 is
devoted to a new class of measurements unique to the HL-HE program:
high-energy probes. These include Higgs processes like  associated production  of a Higgs and a $W$ or $Z$ boson,  or vector boson fusion (VBF), for which the centre-of-mass energy is not limited to the Higgs mass, and it extends to
Drell Yan, di-boson processes and vector
boson scattering, which provide a context in which high-energy measurements can be associated with 
 precision observables. Section 5 focuses on measurements of the
Higgs mass and opportunities for the measurement of the Higgs boson
width. Section 6 describes the constraints on the invisible decays of
the Higgs boson and the indirect constraints on the couplings of the
Higgs boson to undetected particles from the measurement of the Higgs
boson couplings, in particular in the framework of Higgs portal
and dark matter models. Section 7 will discuss approaches to constrain light and non
diagonal Higgs Yukawa couplings directly and indirectly. Section 8 is
devoted to a global interpretation of the measurements in the
framework of the Standard Model Effective Field Theory. Section 9 is
devoted to the discussion of the prospects for probing additional Higgs bosons both with a mass above or below 125 \UGeV, and for discovering a wide range of exotic Higgs boson decays.

\subsection{Experimental analysis methods and objects definitions}

Different approaches have been used by the experiments and in theoretical prospect studies, hereafter named projections, to assess the sensitivity in searching for new physics at the HL-LHC and HE-LHC.
For some of the projections, a mix of the approaches described below is used, in order to deliver the most realistic result.
The total integrated luminosity for the HL-LHC dataset is assumed to be $3000$~fb$^{-1}$ at a centre-of-mass energy of $14$~\UTeV. For HE-LHC studies the dataset is assumed to be $15$~ab$^{-1}$ at a centre-of-mass of $27$~\UTeV.
The effect of systematic uncertainties is taken into account based on the studies performed for the existing analyses and using common guidelines for projecting the expected improvements that are foreseen thanks to the large dataset and upgraded detectors, as described in Section~\ref{sec:methods:syst}.

{\bf Detailed-simulations} are used to assess the performance of reconstructed objects in the upgraded detectors and HL-LHC conditions, as described in Sections~\ref{sec:methods:perf},\ref{sec:methods:perf_LHCb}.
For some of the projections, such simulations are directly interfaced to different event generators, parton showering (PS) and hadronisation generators. Monte Carlo (MC) generated events are used for SM and BSM processes, and are employed in the various projections to estimate the expected contributions of each process.

{\bf Extrapolations} of existing results rely on the existent statistical frameworks to estimate the expected sensitivity for the HL-LHC dataset.
The increased centre-of-mass energy and the performance of the upgraded detectors are taken into account for most of the extrapolations using scale factors on the individual processes contributing to the signal regions. Such scale factors are derived from the expected cross sections and from detailed simulation studies.

{\bf Fast-simulations} are employed for some of the projections in order to produce a large number of Monte Carlo events and estimate their reconstruction efficiency for the upgraded detectors. The upgraded CMS detector performance is taken into account encoding the expected performance of the upgraded detector in \delphes~\cite{deFavereau:2013fsa}, including the effects of pile-up interactions. Theoretical contributions use \delphes with the commonly accepted HL-LHC card corresponding to the upgraded ATLAS and CMS detectors.

{\bf Parametric-simulations} are used for some of the projections to allow a full re-optimisation of the analysis selections that profit from the larger available datasets.
Particle-level definitions are used for electrons, photons, muons, taus, jets and missing transverse momentum. These are constructed from stable particles of the MC event record with a lifetime larger than $0.3 \times 10^{-10}$~s within the observable pseudorapidity range. Jets are reconstructed using the anti-$k_t$ algorithm~\cite{Cacciari:2008gp} implemented in the FastJet~\cite{fastjet} library, with a radius parameter of 0.4. All stable final-state particles are used to reconstruct the jets, except the neutrinos, leptons and photons associated to $W$ or $Z$ boson or $\tau$ lepton decays. The effects of an upgraded ATLAS detector are taken into account by applying energy smearing, efficiencies and fake rates to generator level quantities, following parametrisations based on detector performance studies with the detailed simulations. The effect of the high pileup at the HL-LHC is incorporated by overlaying pileup jets onto the hard-scatter events. Jets from pileup are randomly selected as jets to be considered for analysis with $\sim 2\%$ efficiency, based on studies of pile-up jet rejection and current experience.

\subsubsection{ATLAS and CMS performance}
\label{sec:methods:perf}

The expected performance of the upgraded ATLAS and CMS detectors has been studied in detail in the context of the Technical Design Reports
and subsequent studies; the assumptions used for this report and a more detailed description are available in Ref.~\cite{ATLAS_PERF_Note, Collaboration:2650976}. For CMS, the object performance in the central region assumes a barrel calorimeter ageing corresponding to an integrated luminosity of $1000$~fb$^{-1}$.

The triggering system for both experiments will be replaced and its impact on the triggering abilities of each experiment assessed;
new capabilities will be added, and, despite the more challenging conditions, most of the trigger thresholds for common objects are expected
to either remain similar to the current ones or to even decrease~\cite{ATLAS_TDAQ_TDR,CMSL1interim}.

The inner detector is expected to be completely replaced by both experiments, notably extending its coverage to $|\eta|<4.0$.
The performance for reconstructing charged particles has been studied in detail in Ref.~\cite{ATLAS_Pixel_TDR,ATLAS_Strip_TDR,CMS_Tracker_TDR}.

Electrons and photons are reconstructed from energy deposits in the electromagnetic calorimeter and information from the inner tracker\cite{ATLAS_LAr_TDR,CMS_Barrel_TDR,Collaboration:2293646,CMS_MTD_TP}.
Several identification working points have been studied and are employed by the projection studies as most appropriate.

Muons are reconstructed combining muon spectrometer and inner tracker information~\cite{ATLAS_Muon_TDR,CMS_Muon_TDR}.

Jets are reconstructed by clustering energy deposits in the electromagnetic and hadronic calorimeters\cite{ATLAS_Tile_TDR,ATLAS_LAr_TDR,CMS_Barrel_TDR} using the anti-$k_{T}$ algorithm\cite{Cacciari:2008gp}.
B-jets are identified via $b$-tagging algorithms. B-tagging is performed if the jet is within the tracker acceptance ($|\eta|<4.0$).
Multivariate techniques are employed in order to identify $b-$jets and $c-$jets, and were fully re-optimised for the upgraded detectors~\cite{ATLAS_Pixel_TDR,CMS_Tracker_TDR}.
An 70\% $b-$jet efficiency working point is used, unless otherwise noted.

High $p_T$ boosted jets are reconstructed using large-radius anti-$k_{T}$ jets with a distance parameter of 0.8. Various jet substructure variables are employed to identify boosted W/Z/Higgs boson and top quark jets with good discrimination against generic QCD jets. 

Missing transverse energy is reconstructed following similar algorithms as employed in the current data taking.
Its performance has been evaluated for standard processes, such as top pair production~\cite{ATLAS_Pixel_TDR,Contardo:2020886}.

The addition of new precise-timing detectors and its effect on object reconstruction has also been studied in Ref.~\cite{ATLAS_TP_HGTD,CMS_MTD_TP}, although its results are only taken into account in a small subset of the projections in this report.

\subsubsection{LHCb}
\label{sec:methods:perf_LHCb}
The LHCb upgrades are shifted with respect to those of ATLAS and CMS. A first upgrade will happen at the end of Run 2 of the LHC, to run at a luminosity five times larger  ($2\times 10^{33}\text{cm}^{-2}\text{s}^{-1}$) in LHC Run 3 compared to those in Runs 1 and 2, while maintaining or improving the current detector performance. This first upgrade phase (named \mbox{Upgrade~I}) will be followed by by the so-called \mbox{Upgrade~II} phase (planned at the end of Run 4) to run at an even more challenging luminosity of $\sim 2\times 10^{34}\text{cm}^{-2}\text{s}^{-1}$.

The LHCb MC simulation used in this document mainly relies on the \pythia~8 generator~\cite{Sjostrand:2007gs} with a specific LHCb configuration~\cite{LHCb-PROC-2010-056}, using the CTEQ6 leading-order set of parton density functions~\cite{cteq6}. The interaction of the generated particles with the detector, and its response, are implemented using the \geant{} toolkit~\cite{Allison:2006ve,Agostinelli:2002hh}, as described in Ref.~\cite{LHCb-PROC-2011-006}. 

The reconstruction of jets is done using a particle flow algorithm, with the output of this clustered using
the anti-kT algorithm as implemented in FastJet, with a distance parameter of
0.5. Requirements are placed on the candidate jet in order to reduce the background
formed by particles which are either incorrectly reconstructed or produced in additional pp interactions in the same event.

Concerning the increased pile-up, different assumptions are made, but in general the effect is assumed to be similar to the one in Run 2.

\subsubsection{Treatment of systematic uncertainties}
\label{sec:methods:syst}
It is a significant challenge to predict the expected systematic uncertainties of physics results at the end of HL-LHC running.
It is reasonable to anticipate improvements to techniques of determining systematic uncertainties over an additional decade of data-taking.
To estimate the expected performance, experts in the various physics objects and detector systems from ATLAS and CMS have looked at current limitations to
systematic uncertainties in detail to determine which contributions are limited by statistics and where there are more fundamental limitations.
Predictions were made taking into account the increased integrated luminosity and expected potential gains in technique.
These recommendations were then harmonised between the experiments to take advantage of a wider array of expert opinions and to allow the experiments to make sensitivity predictions on equal footing~\cite{ATLAS_PERF_Note,Collaboration:2650976}. For theorists' contributions, a simplified approach is often adopted, loosely inspired by the improvements predicted by experiments. 

General guide-lining principles were defined in assessing the expected systematic uncertainties.
Theoretical uncertainties are assumed to be reduced by a factor of two with respect to the current knowledge, thanks to both
higher-order calculation as well as reduced parton distribution functions (PDF) uncertainties~\cite{Khalek:2018mdn}.
All the uncertainties related to the limited number of simulated events are neglected, under the assumption that sufficiently large simulation samples will be available by the time the HL-LHC becomes operational. For all scenarios, the intrinsic statistical uncertainty in the measurement is reduced by a factor $1/\sqrt{\text{L}}$, where $\text{L}$ is the projection integrated luminosity divided by that of the reference Run~2 analysis.
Systematics driven by intrinsic detector limitations are left unchanged, or revised according to detailed simulation studies of the upgraded detector.
Uncertainties on methods are kept at the same value as in the latest public results available, assuming that the harsher HL-LHC conditions will be compensated by method improvements.

The uncertainty in the integrated luminosity of the data sample is expected to be reduced down to 1\% by a better understanding of the calibration methods and
their stability employed in its determination, and making use of the new capabilities of the upgraded detectors.

In addition to the above scenario (often referred to as ``YR18 systematics uncertainties'' scenario), results are often
compared to the case where the current level of understanding of systematic uncertainties is assumed (``Run 2 systematic uncertainties'')
or to the case of statistical-only uncertainties.

\subsection{Implications for beyond the Standard Model theories}
\subsubsection{Heavy new physics: precision tests and effective field theories}\label{sec:eftintro}
Precision measurements provide an important tool to search for heavy BSM dynamics, associated with mass scales beyond the LHC direct energy reach,  exploiting the fact that such dynamics can still have an impact on processes at smaller energy, via virtual effects.
In this context the well-established framework of effective field theories (EFTs) allows to systematically parametrise BSM effects and how they modify SM processes. Assuming lepton and baryon number conservation, the leading such effects can be captured by dimension-6 operators,
\begin{equation}\label{EFTLAG}
\mathcal{L}_{\rm eff} = \mathcal{L}_{\rm SM} + \frac{1}{\Lambda^2} \sum_{i} c_i \mathcal{O}_i+\cdots
\end{equation}
for dimensionless coefficients $c_i$ and, for simplicity, a common suppression scale $\Lambda$. Table \ref{tab:dim6ops} proposes a  set of operators considered in this report. This set is \emph{redundant}, in the sense that different combinations of operators might lead to the same physical effect; moreover this set is \emph{not complete}, in the sense that there are more  operators at dimension-6 level.
In practical applications we will always be interested in identifying \emph{minimal} (non-redundant) subsets of operators that contribute to a given process; we will also be interested that these operators be complete, at least under some well motivated assumption. For instance, the assumption that new physics only couples to the SM bosons, leads to the \emph{universal} set of operators, from the second panel in table~\ref{tab:dim6ops}. Alternatively, the minimal flavour violation assumption~\cite{DAmbrosio:2002vsn} provides a well-motivated framework to focus on operators with a certain, family-universal, flavour structure; operators with a richer flavour structure will be studied in a dedicated section \ref{sec7}.

\renewcommand{\arraystretch}{1.4}
\begin{table}[h]
\begin{center}
 \caption{A list of dimension-6 SMEFT operators used in this chapter, defined for one family only; operators suppressed in the minimal flavour violation assumption~\cite{DAmbrosio:2002vsn}, have been neglected (in particular dipole-type operators). Some combinations are redundant and can be eliminated as described in the text. }
\label{tab:dim6ops}
{\small
\begin{tabular}{lll}
 \hline\hline
\multicolumn{3}{c}{Higgs-Only Operators}\\
\hline
${\cal O}_H=\frac{1}{2}(\partial^\mu |H|^2)^2$& ${\cal O}_6=\lambda |H|^6$ & \\
${\cal O}_{y_u}   =y_u |H|^2    \bar Q  \widetilde{H} u $ & ${\cal O}_{y_d}   =y_d |H|^2    \bar Q  Hd $ & ${\cal O}_{y_e}   =y_e |H|^2    \bar L  H e $  \\
${\cal O}_{BB}={g}^{\prime 2} |H|^2 B_{\mu\nu}B^{\mu\nu}$ & ${\cal O}_{GG}=g_s^2 |H|^2 G_{\mu\nu}^A G^{A\mu\nu}$ & $\mathcal{O}_{WW} = g^2 |H|^2 W^{I}_{\mu \nu} W^{I\mu\nu}$ \\
\hline\hline
\multicolumn{3}{c}{Universal Operators}\\
\hline
 ${\cal O}_T=\frac{1}{2} (H^\dagger {\lra{D}_\mu} H)^2$  &  $\mathcal{O}_{H D} = (H^\dagger D^\mu H)^*(H^\dagger D_\mu H)$     & ${\cal O}_{3G}= \frac{1}{3!} g_sf_{abc}G^{a\, \nu}_{\mu}G^{b}_{\nu\rho}G^{c\, \rho\mu}$ \\
  ${\cal O}_W=\frac{ig}{2}( H^\dagger  \sigma^a \lra {D^\mu} H  )D^\nu  W_{\mu \nu}^a$ & ${\cal O}_B=\frac{ig'}{2}( H^\dagger  \lra {D^\mu} H  )\partial^\nu  B_{\mu \nu}$  &  $\mathcal{O}_{W\! B} = gg^\prime(H^\dagger \sigma^IH) W^{I}_{\mu\nu} B^{\mu\nu} $\\
   ${\cal O}_{HW}=i g(D^\mu H)^\dagger\sigma^a(D^\nu H)W^a_{\mu\nu}$  &   ${\cal O}_{HB}=i g'(D^\mu H)^\dagger(D^\nu H)B_{\mu\nu}$ & ${\cal O}_{3W}= \frac{1}{3!} g\epsilon_{abc}W^{a\, \nu}_{\mu}W^{b}_{\nu\rho}W^{c\, \rho\mu}$
   \\
     ${\cal O}_{2G}=\frac 12 \left(D^\nu G^a_{\mu\nu}\right)^2$ & ${\cal O}_{2B}=\frac 12 \left(\partial^\rho B_{\mu\nu}\right)^2$ & ${\cal O}_{2W}=\frac 12 \left(D^\rho W^a_{\mu\nu}\right)^2$\\
  \multicolumn{3}{c}{and ${\cal O}_H$, ${\cal O}_6$, ${\cal O}_{BB}$, ${\cal O}_{WW}$, ${\cal O}_{GG}$,  ${\cal O}_y=\sum_\psi{\cal O}_{y_\psi} $ }\\
  \hline\hline
  \multicolumn{3}{c}{Non-Universal Operators that modify $Z/W$ couplings to fermions}\\
  \hline
            ${\cal O}_{HL}=(i H^\dagger {\lra { D_\mu}} H)( \bar L\gamma^\mu L)$     
&         ${\cal O}_{HL}^{(3)}=(i H^\dagger \sigma^a {\lra { D_\mu}} H)( \bar L\sigma^a\gamma^\mu L)$  
&            ${\cal O}_{He} =(i H^\dagger {\lra { D_\mu}} H)( \bar e \gamma^\mu e )$             \\
            ${\cal O}_{HQ}=(i H^\dagger  {\lra { D_\mu}} H)( \bar Q \gamma^\mu Q )$       &
               ${\cal O}_{HQ}^{(3)}=(i H^\dagger \sigma^a {\lra { D_\mu}} H)( \bar Q \sigma^a\gamma^\mu Q )$     &   \\
          ${\cal O}_{Hu} =(i H^\dagger {\lra { D_\mu}} H)( \bar u \gamma^\mu u )$        
 &            ${\cal O}_{Hd} =(i H^\dagger {\lra { D_\mu}} H)( \bar d \gamma^\mu d )$     &      \\ 
  \hline\hline
  \multicolumn{3}{c}{CP-odd operators}\\
  \hline
 $\mathcal{O}_{H \widetilde W} = (H^\dagger H) \widetilde W^{I}_{\mu\nu}W^{I\mu\nu}$    &
 $\mathcal{O}_{H \widetilde B} = (H^\dagger H) \widetilde B_{\mu\nu}B^{\mu\nu}$     &$\mathcal{O}_{ \widetilde W\! B} = ( H^\dagger\sigma ^I H)
\widetilde W_{\mu\nu}^IB^{\mu\nu}$\\
&   ${\cal O}_{3\widetilde W}= \frac{1}{3!} g\epsilon_{abc}W^{a\, \nu}_{\mu}W^{b}_{\nu\rho}\widetilde W^{c\, \rho\mu}$&  \\
\hline\hline
 \end{tabular}
 }
\end{center}
\end{table}
\renewcommand{\arraystretch}{1}

Reduction to a minimal basis is achieved via  integration by parts and  field re-definitions, equivalent in practice to removing combinations proportional to the equations of motion. 
These imply  relations between the operators of table~\ref{tab:dim6ops}; the most important ones being ($Y$ denotes here  hyper-charge)
\begin{gather}
{\cal O}_{HB}={\cal O}_{B}-\frac{1}{4}{\cal O}_{BB}-\frac{1}{4}{\cal O}_{WB}\,,\quad \quad {\cal O}_{HW}={\cal O}_W-\frac{1}{4}{\cal O}_{WW}-\frac{1}{4}{\cal O}_{WB}\label{OpId1}\\
{\cal O}_B= \frac{g^{\prime\, 2}}{2}\sum_\psi Y_\psi {\cal O}_{H\psi}  -\frac{g^{\prime\, 2}}{2}{\cal O}_T\,,\quad\quad  {\cal O}_T={\cal O}_H-2{\cal O}_{HD}\\
{\cal O}_W= \frac{g^2}{2}\big[\left({\cal O}_{y_u}+{\cal O}_{y_d}+{\cal O}_{y_e}+\text{h.c.}\right)-3{\cal O}_{H}+4{\cal O}_{6} +  \frac{1}{2}\sum_{\psi_L} {\cal O}_{H\psi_L}^{(3)}\big]\, ,
 \label{OpIdend}
\end{gather}
and similar expressions for ${\cal O}_{2W}$ and ${\cal O}_{2B}$ in terms of the products of $SU(2)$ and $U(1)$ SM currents.
Eqs.~(\ref{OpId1}-\ref{OpIdend}) can be used to define minimal,  non-redundant operator bases; for instance, in the context of Higgs physics, the operators ${\cal O}_H,{\cal O}_W, {\cal O}_B,{\cal O}_{HW},{\cal O}_{HB}$ are retained  at the expense of  ${\cal O}_{HD}$, ${\cal O}_{WW}$, ${\cal O}_{WB}$, ${\cal O}_{HL}^{(3)}$, ${\cal O}_{HL}$ in what is known as the SILH basis \cite{Giudice:2007fh}, while in the opposite case  we refer to the Warsaw basis \cite{Grzadkowski:2010es}.\footnote{In addition, the SILH basis gives preference to the operators ${\cal O}_{2W}$ and  ${\cal O}_{2B}$, which are more easily found in universal BSM theories, while the Warsaw basis swaps them in terms of four-fermions operators.}

These operators induce two types of effects: some that are proportional to the SM amplitudes and some that  produce  genuinely new amplitudes. The former are  better accessed by high-luminosity experiments in kinematic regions where the SM is the largest. The most interesting example of this class for the LHC are  Higgs couplings measurements in single-Higgs processes. 
The operators in the top panel of table~\ref{tab:dim6ops}  have the form 
$|H|^2 \times {\cal L}_{\textrm{SM}}$, with ${\cal L}_{\textrm{SM}}$ denoting operators in the SM Lagrangian, and  imply small modifications $\propto v^2/\Lambda^2$  of the Higgs couplings to other SM fields, with respect to the SM value.
These are often parametrised as rescalings of the SM rates, $\kappa_{i}^2=\Gamma_{i}/\Gamma^{\textrm{SM}}_{i}$ ($\Gamma^{\textrm{SM}}$ the  Higgs partial width into channel $i$)  \emph{assuming} the same Lorentz structure as that of the SM, i.e. providing an overall energy-independent factor. This is known as the kappa framework~\cite{Heinemeyer:2013tqa}. We discuss Higgs couplings in detail in sections \ref{sec2} and \ref{sec4} .

Among effects associated with new amplitudes, that cannot be put in correspondence with the $\kappa$s, particularly interesting are BSM energy-growing effects. At dimension-6 level we find effects that grow at most quadratically with the energy. This implies a quadratic enhancement of the sensitivity to these effects, as we consider bins at higher and higher energy. 
 This can be contrasted with high-intensity effects, whose sensitivity increases only with the square root of the integrated luminosity, and eventually saturates as systematics become comparable. High-energy effects are the ideal target of the HL and HE LHC programs, as we discuss in section \ref{sec4}.
 In section \ref{sec8}, we combine the results from the various EFT analyses and provide a global perspective on the HL and HE LHC sensitivity to EFT effects. 

Ultimately, the goal of these global fits is to provide a model-independent framework to which large classes of specific models can be matched an analysed. We provide some example in section \ref{sec8}.

\subsubsection{Light new physics: rare processes and new degrees of freedom}\label{sec:BSMintro}
A complementary way to unveil BSM physics affecting the Higgs sector of Nature is the search for very rare processes involving the 125 \UGeV Higgs boson and for extended Higgs sectors. 

The SM predicts several processes involving the Higgs boson to be very rare. Notable examples are the di-Higgs production, as well as the Higgs decays to first and second generation quarks and leptons. The search for these rare processes can unveil the presence of new degrees of freedom. Particularly, measurements of the di-Higgs production cross section (Sec. \ref{sec3}) will give constraints on the Higgs trilinear interaction, therefore providing information on electroweak symmetry breaking and allowing to set constraints on e.g. the nature of the phase transition between the trivial Higgs vacuum and the vacuum we observe at present (Sec. \ref{sec:HH_EWPT}) and on the presence of extended Higgs sectors. The HL and HE stages of the LHC will be crucial to achieve this goal thanks to the relatively sizeable di-Higgs samples that will be produced: $\mathcal O(100~{\rm{K}})$ at HL-LHC and $\mathcal O(2~{\rm{millions}})$ at HE-LHC (compared to the $\mathcal O(6~{\rm{K}})$ di-Higgs produced at Run 1 and 2 LHC). Furthermore, the branching ratios of SM rare Higgs decay modes such as $h\to\mu^+\mu^-$, $h\to Z\gamma$, and $h\to cc$ have been only mildly upper bounded by present LHC searches due in part to the low statistics ($h\to\mu^+\mu^-$, $h\to Z\gamma$) and, in part, to the background limited analyses ($h\to cc$). An important progress on these rare decay modes is expected at the HL and HE-LHC. For example, the HL-LHC will be able to discover and have a $(10-13)\%$ accuracy measurement of the di-muon decay mode (Sec. \ref{Sec:2.3.8}). Knowing the Higgs couplings to light quark and lepton generations will shed light on BSM flavor models and possibly on the SM flavor puzzle (Sec. \ref{sec7}).

Beyond rare SM Higgs processes, BSM models that contain new light degrees of freedom, $X_i$, generically predict rare exotic Higgs, decays $h\to X_i X_j$ or $h\to X_i ~{\rm{SM}}_j$ where ${\rm{SM}}_j$ is a SM particle (Secs. \ref{Sec:6Invisible} and \ref{Sec:9.1Exo}. For a review see e.g. \cite{Curtin:2013fra}). A typical example is the Higgs decaying to light dark matter particles. Thanks to the tiny Higgs width ($\sim 4$ \UMeV), even very feebly coupled new light particles can lead to relatively sizeable Higgs branching ratios that can be probed by the LHC in the future. On the one hand, the HL and HE-LHC will produce huge samples of Higgs bosons from its main production mode, gluon fusion ($\mathcal O(10^8)$ and $\mathcal O(10^9)$, respectively). This can allow the search for super rare and low background signatures. On the other hand, the sample of Higgs bosons produced from sub-leading  production modes in association with other SM particles (e.g. $tth$) will be sizeable, increasing the discovery prospects for rare and more background limited Higgs decay signatures. 
Therefore, the HL/HE-LHC Higgs exotic decay program can be uniquely sensitive to the existence of a broad range of new light weakly coupled particles (on condition that trigger and analysis thresholds will be kept relatively low, to allow capturing this set of soft signatures).

In many BSM theories, electroweak symmetry is broken not only by one Higgs boson, but by several degrees of freedom. Examples are supersymmetric theories, composite Higgs theories, as well as theories of neutral naturalness. Overall, extended Higgs sectors can lead to new interesting signatures that are not contained in the SM. The search for additional Higgs bosons is a high priority for current and future colliders. The ATLAS and CMS collaborations have performed several searches for heavy neutral and charged Higgs bosons during the first two runs of the LHC. At the same time, the LHCb collaboration (as well as ATLAS and CMS) has pursued several searches for new Higgs bosons with a mass below 125 \UGeV.  The reach of all these searches will expand considerably in the future and, especially, at the HL and HE-LHC. In Secs. \ref{sec:Hff}-\ref{Sec:9.4} and \ref{Sec:9.8} of this report, we study the prospects for testing some of the most promising signatures.
Most of the BSM models that predict the existence of an extended Higgs sector, also predict a 125 \UGeV Higgs with the interactions which are generically different from the SM predictions. As we will show in Secs. \ref{Sec:9.5}-\ref{Sec:9.7}, the study of the interplay between new Higgs searches and Higgs coupling measurements will be a powerful tool to probe vast regions of parameter space of BSM theories with an extended electroweak symmetry breaking sector.

\newpage

\asection[S. Alioli, M. D\"uhrssen, P. Milenovic]{Higgs boson precision measurements}\label{sec2}
\asubsection{Introduction}
\label{sec2:introduction}

The large number of events expected in almost all Higgs boson measurement channels for the HL-LHC and HE-LHC will allow very precise measurements of the Higgs boson production cross sections and its couplings to gauge bosons and fermions. In many measurement channels, the expected overall statistical, experimental and theoretical uncertainties will be comparable in size. Therefore, a close interaction between the communities of the experimental and theoretical particle physicists will be needed in order to reach the best possible measurements of the Higgs boson properties.

Experimental sensitivity for the Higgs boson properties measurements is estimated by extrapolating the performance of the existing measurements to the HL-LHC data set, assuming the experiments will have a similar level of detector and triggering performance. Results are presented for two assumptions on the size of the experimental and theoretical systematic uncertainties that will be achievable by the time of HL-LHC (so called conservative and optimistic scenarios). Details on the extrapolation methodology and scenarios will be presented in Section~\ref{sec2:channels}.

Section~\ref{sec2_HXSWG1} provides an overview of theoretical predictions for the Higgs boson production at 14 and 27~\UTeV\ and of the uncertainties that are expected to be reached by the time of the final HL-LHC and HE-LHC measurements. 
These predictions are used as input to sensitivity studies of the ATLAS and CMS Higgs boson cross section and coupling measurements in individual channels that are summarised in Section~\ref{sec2:channels} and for the expectations for differential cross section measurements presented in Section~\ref{sec2:fiducial}.
Section~\ref{sec2:top} puts emphasis on all measurements related to the top Yukawa coupling, as this is the largest Yukawa coupling in the Standard Model with a value close to unity and, hence, of special interest in understanding the Higgs mechanism and its relation to fermions.
The combination of the expected measurements in ATLAS and CMS are presented in Section~\ref{sec2:exp_combination} together with an interpretation in the kappa-model~\cite{LHCHiggsCrossSectionWorkingGroup:2012nn,Heinemeyer:2013tqa} in Section~\ref{sec2:exp_kappa}.

The kappa-framework is closely related to a non-linear EFT, and projections of measurements of EFT coefficients in a non-linear EFT are presented in Section~\ref{sec2:theo_kappa_EFT} together with a translation of these results in terms of composite Higgs scenarios in section~\ref{sec9:CHM}.
Finally, probes of anomalous HVV interactions are discussed in Section~\ref{sec2:anomalous_HVV}.

\asubsection[K. Becker, C. Bertella, M. Bonvini, A. Calderon Tazon,
  J. Campbell, F. Caola, X. Chen, P. Francavilla, S. Frixione,
  R. Frederix, T. Gehrmann, N. Glover, Y. Haddad, V. Hirschi, A. Huss,
  S. Jones, A. Karlberg, M. Kerner, J. Lindert, G. Luisoni,
  G. Marchiori, S. Marzani, A. Massironi, B. Mistlberger, P. Monni,
  M. Moreno Llacer, A. M\"uck, D. Pagani, C. Palmer, C. Pandini, L. Perrozzi,
  S. Pozzorini, E. Re, L. Reina, H.S. Shao, L. Simon, B. Stieger,
  V. Theeuwes, F. Tramontano, M. Zaro]{Theoretical predictions for the
  Higgs boson production}
\label{sec2_HXSWG1}
%
%
%
Cross-section predictions for the high-energy (HE) LHC, and their associated theoretical uncertainties, are discussed and shown in
Section~\ref{sec:he-lhc}. Predictions are computed for a proton-proton collider with a $pp$ centre-of-mass
energy $\sqrt{s}=27$~\UTeV and use a Higgs
boson mass of $m_H=125.09 \pm 0.5$~\UGeV.   All other parameters are taken from YR4~\cite{deFlorian:2016spz},
with exceptions noted where they are important.  
Projections of progress towards a reduction in theoretical uncertainties, on the timescale
of the high-luminosity (HL) LHC (3~ab$^{-1}$ of $pp$ collisions at $\sqrt{s}=14$~\UTeV), are discussed in Section~\ref{sec:hl-lhc}. 
Tables summarising a detailed study of the dependence of the gluon-fusion cross section on the mass of the Higgs boson
are presented in Section~\ref{sec:ggFmhdep}.

\asubsubsection{Cross sections for 13, 14 and 27 \UTeV HE-LHC}
\label{sec:he-lhc}

This section provides updated cross-sections for the LHC operating at
energies of $13$, $14$ and $27$~\UTeV. All predictions~\cite{HXSWGwiki} include the latest 
theoretical input and supersede the older results in YR4~\cite{deFlorian:2016spz}.

\asubsubsubsection{Gluon fusion}
\label{sec:he-lhc-ggF}
In this section we document cross section predictions for a standard
model Higgs boson produced through gluon fusion in 27~\UTeV $pp$ collisions.  To
derive predictions we include contributions based on perturbative
computations of scattering cross sections as studied in
Ref.~\cite{Anastasiou:2016cez}.  We include perturbative QCD
corrections through next-to-next-to-next-to-leading order (N$^3$LO), electroweak (EW) and approximated mixed
QCD-electroweak corrections as well as effects of finite quark
masses. The only modification with respect to YR4~\cite{deFlorian:2016spz} is that we
now include the exact N$^3$LO heavy top effective theory cross section of
Ref.~\cite{Mistlberger:2018etf} instead of its previous approximation. The
result of this modification is only a small change in the central values and
uncertainties. To derive theoretical uncertainties we follow the
prescriptions outlined in Ref.~\cite{Anastasiou:2016cez}.
We use the following inputs:
\begin{equation}
\begin{array}{c  c}
\hline
\text{E}_\text{CM} & 27~\text{\UTeV} \\
\text{m}_\text{t}(\text{m}_\text{t}) & 162.7~\text{\UGeV} \\
\text{m}_\text{b}(\text{m}_\text{b}) & 4.18~\text{\UGeV} \\
\text{m}_\text{c}(3 \text{ \UGeV}) & 0.986~\text{\UGeV} \\
\alpha_\text{S} (m_Z) & 0.118  \\
\text{PDF} & \text{PDF4LHC15\_nnlo\_100}\text{~\cite{Botje:2011sn}}\\
\hline
\end{array}
\end{equation}
All quark masses are treated in the $\overline{\text{MS}}$ scheme. To derive numerical predictions we use the program \texttt{iHixs}~\cite{Dulat:2018rbf}.

Sources of uncertainty for the inclusive Higgs boson production cross section have been assessed recently in refs.~\cite{Anastasiou:2016cez,Harlander:2016hcx,Bonvini:2016frm,deFlorian:2016spz}. 
Several sources of theoretical uncertainties were identified.
\begin{figure}[h]
\begin{center}
    \includegraphics[width=0.7\textwidth]{\main/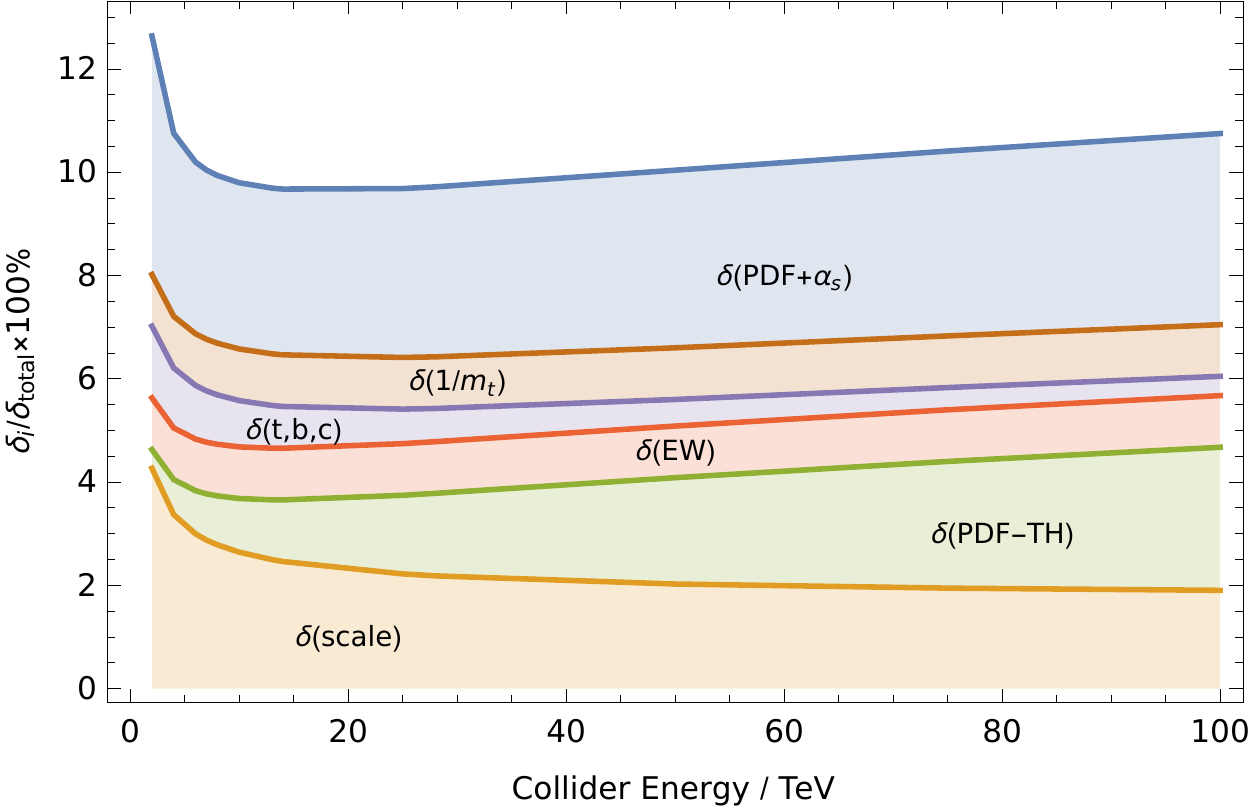}
    \caption{
    The figure shows the linear sum of the different sources of relative uncertainties as a function of the collider energy. 
    Each coloured band represents the size of one particular source of uncertainty as described in the text.
    The component $\delta(\text{PDF}+\alpha_S)$ corresponds to the uncertainties due to our imprecise knowledge of the strong coupling constant and of parton distribution functions combined in quadrature.
        \label{fig:errorplot}}
        \end{center}
\end{figure}
\begin{itemize}
\item Missing higher-order effects of QCD corrections beyond N$^3$LO ($\delta(\text{scale})$).
\item Missing higher-order effects of electroweak and mixed QCD-electroweak corrections at and beyond $\mathcal{O}(\alpha_S \alpha)$ ($\delta(\text{EW})$).
\item Effects due to finite quark masses neglected in QCD corrections beyond NLO  ($\delta(\text{t,b,c})$ and $\delta(1/m_t)$).
\item Mismatch in the perturbative order of the parton distribution functions (PDF) evaluated at NNLO and the perturbative QCD cross sections evaluated at N$^3$LO ($\delta(\text{PDF-TH})$).
\end{itemize}
In the tables the linear sum of the effect of those uncertainties
is referred to as $\delta(\text{theory})$.  In addition, the imprecise
knowledge of the parton distribution functions and of the strong coupling
constant play a dominant role.  The individual size of these
contributions can be seen in fig.~\ref{fig:errorplot} as a function of
the collider energy~\cite{Dulat:2018rbf}.  As can be easily inferred
the relative importance of the different sources of uncertainty is
impacted only mildly by changing the centre of mass energy from $13$
\UTeV to $27$ \UTeV.  Inclusive cross sections for
$m_H=125.09$ \UGeV are given in Table~\ref{tab:results}. As noted above, the
exact treatment of N$^3$LO QCD corrections results in a
small shift in the cross-section at 13~\UTeV, relative to the YR4 result, and
a slight reduction in the overall theoretical uncertainty.
\begin{table}[!h]
\normalsize\setlength{\tabcolsep}{2pt}
\begin{center}
    \begin{tabular}{rrrrr}
        \toprule
        \multicolumn{1}{c}{$\sqrt{s}$}&
        \multicolumn{1}{c}{$\sigma$}&
        \multicolumn{1}{c}{$\delta(\textrm{theory})$}&
        \multicolumn{1}{c}{$\delta(\textrm{PDF})$}&
        \multicolumn{1}{c}{$\delta(\alpha_s)$}\\
        \midrule
13 \UTeV & 48.61 pb & ${}^{+2.08\textrm{pb}}_{-3.15\textrm{pb}}\,\left({}^{+4.27\%}_{-6.49\%}\right) $ & $\pm\,0.89\,\textrm{pb}\,(\pm\,1.85\%)$ & ${}^{+1.24\textrm{pb}}_{-1.26\textrm{pb}}\,\left({}^{+2.59\%}_{-2.62\%}\right) $ \\
14 \UTeV & 54.72 pb & ${}^{+2.35\textrm{pb}}_{-3.54\textrm{pb}}\,\left({}^{+4.28\%}_{-6.46\%}\right) $ & $\pm\,1.00\,\textrm{pb}\,(\pm\,1.85\%)$ & ${}^{+1.40\textrm{pb}}_{-1.41\textrm{pb}}\,\left({}^{+2.60\%}_{-2.62\%}\right) $ \\
27 \UTeV & 146.65 pb & ${}^{+6.65\textrm{pb}}_{-9.44\textrm{pb}}\,\left({}^{+4.53\%}_{-6.43\%}\right) $ & $\pm\,2.81\,\textrm{pb}\,(\pm\,1.95\%)$ & ${}^{+3.88\textrm{pb}}_{-3.82\textrm{pb}}\,\left({}^{+2.69\%}_{-2.64\%}\right) $ \\
   \bottomrule
    \end{tabular}
    \caption{Gluon fusion Higgs boson production cross sections and uncertainties as a function of the $pp$ collider energy.\label{tab:results}}
\end{center}
\end{table}
 
\noindent The dependence of the inclusive gluon-fusion cross-section on the Higgs boson mass at $\sqrt{s}=14$ and 27~\UTeV
is detailed at the end of this note in Section~\ref{sec:ggFmhdep}.

\subsubsubsection*{Impact of threshold and high-energy
  corrections}
Recently, Ref.~\cite{Bonvini:2018ixe} has performed a study of the
effects of simultaneous threshold and high-energy (small Bjorken $x$)
resummations on the inclusive Higgs boson production cross section. In this brief
section we summarise the main conclusions, while the numerical results
will be discussed in the following section. For more details we refer
the reader to Ref.~\cite{Bonvini:2018ixe}:
\begin{enumerate}
\item At different collider energies, it was found that the impact of
  threshold resummation amounts to about $+1\%$ on top of the N$^3$LO
  cross section~\cite{Bonvini:2016frm}. The size of this effect is
  compatible with other estimates of the size of missing higher-order
  corrections.
\item Conversely, the inclusion of small-$x$ resummation was found to
  increase the cross section by about one percent at $13$~\UTeV, and by
  about $3\%-4\%$ at $27$~\UTeV, with respect to the N$^3$LO
  prediction. The correction grows even larger at higher 
  energies, reaching about $+10\%$ for a $100$~\UTeV $pp$ collider. The
  inclusion of high-energy resummation affects differently the
  perturbative coefficient functions and the parton densities.
\begin{itemize}
\item The effect on the coefficient functions is very moderate, and
  remains below the $1\%$ level for different collider energies. This
  indicates that the production of a Higgs boson at present and future
  colliders does not probe very small values of the momentum fraction
  at which the coefficient functions are evaluated.
  In turn, this implies that currently and at future colliders PDFs are probed at intermediate values of x.

\item The parton densities receive a large correction from small-$x$
  resummation. Its effect is twofold: on one hand, the {\it evolution}
  of the gluon density is modified by the inclusion of small-$x$
  effects, and at average values of $x$ probed in Higgs production
  this leads to a moderate effect on the parton densities at $m_H$
  (cf. Fig. 2.2 of Ref.~\cite{Ball:2017otu}). On the other hand, the
  PDFs used in the double-resummed prediction of
  Ref.~\cite{Bonvini:2018ixe} (\texttt{NNPDF31sx\allowbreak\_nnlonllx\allowbreak\_as\allowbreak\_0118}
  \cite{Ball:2017otu}) include small-$x$ data from HERA, which
  Ref.~\cite{Ball:2017otu} observes to require high-energy resummation
  for the fit to be robust. The fixed-order prediction of
  Ref.~\cite{Bonvini:2018ixe} instead uses a PDF set which fits the
  small-$x$ HERA data without including high-energy resummation
  (\verb+NNPDF31sx_nnlo_as_0118+ \cite{Ball:2017otu}).  This results
  in sizeable differences in the parton distribution functions and
  drives the large correction to the N$^3$LO total cross section
  observed in Ref.~\cite{Bonvini:2018ixe}.
\end{itemize}
\end{enumerate}
Summarising, the sizeable corrections to the N$^3$LO prediction due to
high-energy resummation observed in Ref.~\cite{Bonvini:2018ixe} are,
to a large extent, due to the need for high-energy resummation in the
PDF fit which are necessary to get a reliable description of small-$x$
HERA data.  Performing a fit without high energy resummation results
in considerable tension with respect to low $Q^2$ HERA data.  In order
to corroborate these findings, and assess precisely the effect of
high-energy resummation on parton distribution fits, it is important
to make progresses in the theoretical knowledge of small-$x$
dynamics. Furthermore, it would be desirable to include additional
small-$x$ collider data in the fits of parton distributions.  We would
like to encourage the PDF and theory community to further investigate
these effects in view of future high energy colliders.

\subsubsubsection*{Predictions for double-resummed cross
  section}

The setup is the same of the YR4 ($m_H=125$~\UGeV, $m_t=172.5$~\UGeV,
$m_b=4.92$~\UGeV, $m_c=1.51$~\UGeV, $\as(m_Z^2)=0.118$,
$\muf=\mur=m_H/2$), with the only difference being that we do not use
PDF4LHC but the 
\verb+NNPDF31sx_nnlonllx_as_0118+ set of
Ref.~\cite{Ball:2017otu}.  Since these resummed PDFs are available for
a single value of $\as$, we could not compute the $\as$ uncertainty in
our result. The results are collected in Tab.~\ref{tab:xsec1}.

For each value of the collider energy, we give the full
N$^3$LO+N$^3$LL+LL$x$ cross section which includes top, bottom and
charm contributions (as discussed in Ref.~\cite{Bonvini:2018iwt}) and
EW corrections included in the complete factorisation approach, i.e.\
as a $+5\%$ contribution.  The breakdown of the individual terms
contributing to the cross section (the main contribution assuming only
top runs in the loop, the bottom+charm correction, and the EW
correction) is presented in the third column.  In the next columns, we
present various sources of uncertainties,
following Ref.~\cite{Bonvini:2018ixe}:
\begin{itemize}
\item Missing higher-order uncertainty (scale uncertainty)
  $\delta_{\rm scale}^{\rm 42var}$. It is the envelope of standard
  7-point scale variations for each of the sub-leading variations of
  threshold resummed contributions, resulting in a total of 42
  variation.

\item PDF uncertainty $\delta_{\rm PDFs}$. This is the standard NNPDF
  Monte Carlo replica uncertainty, but it does not contain the $\as$
  uncertainty, as previously discussed.

\item Sub-leading small-$x$ logarithms uncertainty
  $\delta_{\rm subl.logs}$. This uncertainty is computed as described
  in Refs.~\cite{Bonvini:2018ixe,Bonvini:2018iwt}, and it likely
  overestimates the effect of sub-leading contributions in the
  coefficient functions. However, as argued in
  Refs.~\cite{Bonvini:2018ixe,Bonvini:2018iwt}, this uncertainty can
  be considered as an estimate of the uncertainty from sub-leading
  contributions in the PDFs. In this respect, this provides an
  alternative to the uncertainty from missing higher-order PDFs
  adopted in YR4, which should thus not be included.
\end{itemize}
Additional uncertainties from missing $1/m_t^2$ effects, missing
bottom+charm effects and sub-leading EW effects should be included
according to the YR4 prescription.  Since the N$^3$LO heavy-top result
is matched to the exact small-$x$ according to the construction of
Ref.~\cite{Bonvini:2018iwt}, the ``truncation of the soft expansion''
uncertainty discussed in YR4 should not be considered.

Finally, in the last column of the table we present the ratio of our
resummed result with a purely fixed-order N$^3$LO cross section
obtained with the same settings but using the NNLO parton distribution
functions 
\verb+NNPDF31sx_nnlo_as_0118+ of Ref.~\cite{Ball:2017otu}.  This is
useful to understand how large the effect of resummation(s) in our
prediction is.  We see in particular that the effect (of small-$x$
resummation) grows with the collider energy, reaching $4.6\%$ at
the HE-LHC.  For any of the scales, approximately $+1\%$ of the effect of
resummations is due to threshold resummation (in the coefficient
functions), while the rest of the effect is due to small-$x$
resummation, which mostly comes from the PDFs (see
Ref.~\cite{Bonvini:2018ixe}) as discussed in the previous subsection.
\begin{table*}[t!]
 \centering
 \begin{tabular}{r cc ccc c}
 \toprule
   \multicolumn{1}{c}{$\sqrt{s}$}
   & $\sigma_\text{N$^3$LO+N$^3$LL+LL$x$}$ & $=\sigma_t+\Delta\sigma_{bc}+\Delta\sigma_{\rm EW}$
   & $\delta_{\rm scale}^{\rm 42var}$ & $\delta_{\rm PDFs}$ & $\delta_{\rm subl.logs}$
   & $\frac{\sigma_\text{N$^3$LO+N$^3$LL+LL$x$}}{\sigma_\text{N$^3$LO}}$ \\
\midrule
  13~\UTeV & $ 48.93$~pb & $( 49.26 -2.66+  2.33)$~pb & ${}_{ -3.8}^{+ 4.0}\%$ & $\pm 1.2\%$ & $\pm 1.8\%$ & $ 1.020$ \\[1ex]
  14~\UTeV & $ 55.22$~pb & $( 55.56 -2.96+  2.63)$~pb & ${}_{ -3.8}^{+ 4.0}\%$ & $\pm 1.1\%$ & $\pm 1.9\%$ & $ 1.023$ \\[1ex]
  27~\UTeV & $ 151.6$~pb & $( 151.6  -7.2+   7.2)$~pb & ${}_{ -4.0}^{+ 4.0}\%$ & $\pm 1.0\%$ & $\pm 2.3\%$ & $ 1.046$ \\
  \bottomrule
 \end{tabular}
 \caption{Values of the N$^3$LO+N$^3$LL+LL$x$ gluon-fusion cross section
   for selected values of the $pp$ collision energy and for a Higgs boson mass $m_H=125$~\UGeV.
   We use the NNPDF31sx PDFs with $\as(m_Z^2)=0.118$, $\mt=173$~\UGeV, $\mbottom=4.92$~\UGeV and $\mcharm=1.51$~\UGeV.}
 \label{tab:xsec1}
\end{table*}

\asubsubsubsection{Vector boson fusion}
\label{sec:he-lhc-VBF}
The vector-boson fusion (VBF) cross sections are computed with the
same settings as in YR4 and reported in Tab.~\ref{tab:vbf_xsec}. The
description of the setup can be found in the YR4 itself. The EW and photon cross sections have
been computed using the
\texttt{LUXqed\_plus\_PDF4LHC\_nnlo\_100}~\cite{Manohar:2016nzj,Manohar:2017eqh}
PDF set and hence the 13 and 14 \UTeV cross sections differ slightly
from those reported in the YR4, where
\texttt{NNPDF23\_nlo\_as\_0118\_qed}~\cite{Ball:2013hta} was used
instead. The QCD cross section was computed at NNLO with
proVBFH~\cite{Cacciari:2015jma,Dreyer:2016oyx}, while the EW and
photon contributions have been computed at NLO with
HAWK~\cite{Ciccolini:2007jr,Ciccolini:2007ec,Denner:2014cla}.

\begin{table}
\centering
\begin{tabular}{ccccccc|c}
\toprule
$\sqrt{s}$~[\UTeV] & $\sigma^{\rm VBF}$~[fb] & $\Delta_{\mathrm{scale}}$~[\%] &
$\Delta_{\mathrm{PDF\oplus\alpha_s}}$~[\%] &
$\sigma_{\rm NNLO}^{\rm DIS}$~[fb] & $\delta_{\rm ELWK}$~[\%] & $\sigma_{\gamma}$~[fb] & $\sigma_{\mbox{\scriptsize $s$-ch}}$~[fb]
\\
\midrule
$13$ & $3766$  &$^{+0.43}_{-0.33}$ &$\pm2.1$ &$3939$  & $-5.3$ & $35.3$ & $1412$ \\
$14$ & $4260$  &$^{+0.45}_{-0.34}$ &$\pm2.1$ &$4460$  & $-5.4$ & $40.7$ & $1555$ \\
$27$ & $11838$ &$^{+0.66}_{-0.36}$ &$\pm2.1$ &$12483$ & $-6.2$ & $129$  & $3495$ \\
\bottomrule
\end{tabular}
\caption{VBF Higgs boson production cross-sections in $pp$ collisions for centre-of-mass energies
up to 27 \UTeV and a Higgs boson mass $m_H=125.09$~\UGeV.  The $s$-channel cross-section is the contribution
from Higgs-strahlung diagrams with hadronic weak-boson decay~\cite{deFlorian:2016spz}.}
\label{tab:vbf_xsec}
\end{table}

We note that the photon induced contribution is more reliably
predicted here than was the case in the YR4 due to the LUXqed
method. In particular the photon PDF should no longer be considered as
a source of uncertainty as in eq. (I.5.7) in the YR4, as it is now
constrained at the percent level. Quantitatively the photon induced
contributions are reduced by about $30\%$ compared to in the YR4.

The s-channel contributions at $13$ and $14$ \UTeV have on the other
hand increased compared to the YR4 results. This is due to the updated
set of
parton distribution functions used for this prediction,
i.e. \texttt{LUXqed\_plus\_PDF4LHC\_nnlo\_100} instead of
\texttt{NNPDF23\_nlo\_as\_0118\_qed}. We also note that the relative
size of the s-channel decreases as the collider energy increases -
from $47\%$ at $7$ \UTeV to $30\%$ at $27$ \UTeV.

\asubsubsubsection{VH production}
\label{sec:he-lhc-VH}
\begin{table}
\centering
\begin{tabular}{ccccc}
\toprule
$\sqrt{s}$~[\UTeV] & $\sigma_{{\rm NNLO~QCD}\otimes{\rm NLO~EW}}$~[pb] & $\Delta_{\mathrm{scale}}$~[\%] &
$\Delta_{\mathrm{PDF\oplus\alpha_s}}$~[\%] \\
\midrule
$13$ & $1.358$ & $^{+0.51}_{-0.51}$ & $1.35$ \\
$14$ & $1.498$ & $^{+0.51}_{-0.51}$ & $1.35$ \\
$27$ & $3.397$ & $^{+0.29}_{-0.72}$ & $1.37$ \\
\bottomrule
\end{tabular}
\caption{Cross-section for the process $p p \to WH$. Both $W^+$ and $W^-$ contributions are included.
The photon contribution is not included. Results are given for a Higgs boson mass $m_H = 125.09~{\rm \UGeV}$.}
\label{tab:wh_xsec}
\end{table}

\begin{table}
\centering
\begin{tabular}{ccccc}
\toprule
$\sqrt{s}$~[\UTeV] & $\sigma_{{\rm NNLO~QCD}\otimes{\rm NLO~EW}}$~[pb] & $\Delta_{\mathrm{scale}}$~[\%] &
$\Delta_{\mathrm{PDF\oplus\alpha_s}}$~[\%] \\
\midrule
$13$ & $0.831$ & $^{+0.74}_{-0.73}$ & $1.79$ \\
$14$ & $0.913$ & $^{+0.64}_{-0.76}$ & $1.78$ \\
$27$ & $1.995$ & $^{+0.43}_{-1.04}$ & $1.84$ \\
\bottomrule
\end{tabular}
\caption{Cross-section for the process $p p \to W^+H$. 
The photon contribution is not included. Results are given for a Higgs boson mass $m_H = 125.09~{\rm \UGeV}$.}
\label{tab:w+h_xsec}
\end{table}

\begin{table}
\centering
\begin{tabular}{ccccc}
\toprule
$\sqrt{s}$~[\UTeV] & $\sigma_{{\rm NNLO~QCD}\otimes{\rm NLO~EW}}$~[pb] & $\Delta_{\mathrm{scale}}$~[\%] &
$\Delta_{\mathrm{PDF\oplus\alpha_s}}$~[\%] \\
\midrule
$13$ & $0.527$ & $^{+0.59}_{-0.63}$ & $2.03$ \\
$14$ & $0.585$ & $^{+0.55}_{-0.68}$ & $1.98$ \\
$27$ & $1.402$ & $^{+0.36}_{-0.93}$ & $2.03$ \\
\bottomrule
\end{tabular}
\caption{Cross-section for the process $p p \to W^-H$. 
The photon contribution is not included. Results are given for a Higgs boson mass $m_H = 125.09~{\rm \UGeV}$.}
\label{tab:w-h_xsec}
\end{table}

\begin{table}
\centering
\begin{tabular}{cccc|c}
\toprule
$\sqrt{s}$~[\UTeV] & $\sigma_{{\rm NNLO~QCD}\otimes{\rm NLO~EW}}$~[pb] & $\Delta_{\mathrm{scale}}$~[\%] &
$\Delta_{\mathrm{PDF\oplus\alpha_s}}$~[\%] & $\sigma_\gamma$\\
\midrule
$13$ & $0.094$ & $^{+0.71}_{-0.70}$ & $1.72$ & $4.1~10^{-3}$\\
$14$ & $0.104$ & $^{+0.61}_{-0.73}$ & $1.70$ & $4.7~10^{-3}$\\
$27$ & $0.232$ & $^{+0.40}_{-0.97}$ & $1.72$ & $1.5~10^{-2}$\\
\bottomrule
\end{tabular}
\caption{Cross-section for the process $p p \to l^+\nu H$. 
The photon contribution is included, and also reported separately in the last column. 
Results are given for a Higgs boson mass $m_H = 125.09~{\rm \UGeV}$.}
\label{tab:l+nuh_xsec}
\end{table}

 \begin{table}
\centering
\begin{tabular}{cccc|c}
\toprule
$\sqrt{s}$~[\UTeV] & $\sigma_{{\rm NNLO~QCD}\otimes{\rm NLO~EW}}$~[pb] & $\Delta_{\mathrm{scale}}$~[\%] &
$\Delta_{\mathrm{PDF\oplus\alpha_s}}$~[\%] & $\sigma_\gamma$\\
\midrule
$13$ & $0.0598$ & $^{+0.57}_{-0.60}$ & $1.94$ & $2.6~10^{-3}$ \\
$14$ & $0.0666$ & $^{+0.52}_{-0.64}$ & $1.89$ & $3.1~10^{-3}$ \\
$27$ & $0.1628$ & $^{+0.34}_{-0.87}$ & $1.90$ & $1.1~10^{-2}$\\
\bottomrule
\end{tabular}
\caption{Cross-section for the process $p p \to l^-\bar\nu H$. 
The photon contribution is included, and also reported separately in the last column. 
Results are given for a Higgs boson mass $m_H = 125.09~{\rm \UGeV}$.}
\label{tab:l-nuh_xsec}
\end{table}

\begin{table}
\centering
\begin{tabular}{cccc}
\toprule
$\sqrt{s}$~[\UTeV] & $\sigma_{{\rm NNLO~QCD}\otimes{\rm NLO~EW}}$~[pb] & $\Delta_{\mathrm{scale}}$~[\%] &
$\Delta_{\mathrm{PDF\oplus\alpha_s}}$~[\%] \\
\midrule
$13$ & $0.880$ & $^{+3.50}_{-2.68}$ & $1.65$ \\
$14$ & $0.981$ & $^{+3.61}_{-2.94}$ & $1.90$ \\
$27$ & $2.463$ & $^{+5.42}_{-4.00}$ & $2.24$ \\
\bottomrule
\end{tabular}
\caption{Cross-section for the process $p p \to ZH$. The predictions for the $gg\to ZH$ channel are computed 
at LO, rescaled by the NLO $K$-factor in the $\mt \to \infty$ limit,
and supplemented by the NLL$_{\rm soft}$ resummation. The photon contribution is
omitted.  Results are given for a Higgs boson mass $m_H = 125.09~{\rm \UGeV}$.}
\label{tab:ZH_xsec}
\end{table}

\begin{table}
\centering
\begin{tabular}{cccc}
\toprule
$\sqrt{s}$~[\UTeV] & $\sigma_{{\rm NNLO~QCD}\otimes{\rm NLO~EW}}$~[pb] & $\Delta_{\mathrm{scale}}$~[\%] &
$\Delta_{\mathrm{PDF\oplus\alpha_s}}$~[\%] \\
\midrule
$13$ & $0.758$ & $^{+0.49}_{-0.61}$ & $1.78$ \\
$14$ & $0.836$ & $^{+0.51}_{-0.62}$ & $1.82$ \\
$27$ & $1.937$ & $^{+0.56}_{-0.74}$ & $2.37$ \\
\bottomrule
\end{tabular}
\caption{Cross-section for the process $p p \to ZH$. The photon 
and $gg\to ZH$ contributions are omitted. Results are given for a Higgs boson mass $m_H = 125.09~{\rm \UGeV}$.}
\label{tab:ZHnogg_xsec}
\end{table}

\begin{table}
\centering
\begin{tabular}{cccc}
\toprule
$\sqrt{s}$~[\UTeV] & $\sigma_{{\rm NNLO~QCD}\otimes{\rm NLO~EW}}$~[pb] & $\Delta_{\mathrm{scale}}$~[\%] &
$\Delta_{\mathrm{PDF\oplus\alpha_s}}$~[\%] \\
\midrule
$13$ & $0.123$ & $^{+24.9}_{-18.8}$ & $4.37$ \\
$14$ & $0.145$ & $^{+24.3}_{-19.6}$ & $7.47$ \\
$27$ & $0.526$ & $^{+25.3}_{-18.5}$ & $5.85$ \\
\bottomrule
\end{tabular}
\caption{Cross-section for the process $g g  \to ZH$. Predictions are computed 
at LO, rescaled by the NLO $K$-factor in the $\mt \to \infty$ limit,
and supplemented by the NLL$_{\rm soft}$ resummation. Results are given for a Higgs boson mass $m_H = 125.09~{\rm \UGeV}$.}
\label{tab:ggZH_xsec}
\end{table}

\begin{table}
\centering
\begin{tabular}{cccc|c}
\toprule
$\sqrt{s}$~[\UTeV] & $\sigma_{{\rm NNLO~QCD}\otimes{\rm NLO~EW}}$~[pb] & $\Delta_{\mathrm{scale}}$~[\%] &
$\Delta_{\mathrm{PDF\oplus\alpha_s}}$~[\%] & $\sigma_{\gamma}$\\
\midrule
$13$ & $2.97~10^{-2}$ & $^{+3.49}_{-2.67}$ & $1.64$ & $1.4~10^{-4}$\\
$14$ & $3.31~10^{-2}$ & $^{+3.59}_{-2.92}$ & $1.89$ & $1.6~10^{-4}$\\
$27$ & $8.32~10^{-2}$ & $^{+5.39}_{-3.97}$ & $1.85$ & $5.4~10^{-4}$\\
\bottomrule
\end{tabular}
\caption{Cross-section for the process $p p   \to l\bar lH$. The photon contribution is included,
and reported separately in the last column.  Results are given for a Higgs boson mass $m_H = 125.09~{\rm \UGeV}$.}
\label{tab:llZH_xsec}
\end{table}

\begin{table}
\centering
\begin{tabular}{cccc}
\toprule
$\sqrt{s}$~[\UTeV] & $\sigma_{{\rm NNLO~QCD}\otimes{\rm NLO~EW}}$~[pb] & $\Delta_{\mathrm{scale}}$~[\%] &
$\Delta_{\mathrm{PDF\oplus\alpha_s}}$~[\%] \\
\midrule
$13$ & $0.177$ & $^{+3.50}_{-2.68}$ & $1.65$ \\
$14$ & $0.197$ & $^{+3.59}_{-2.92}$ & $1.89$ \\
$27$ & $0.496$ & $^{+5.41}_{-3.99}$ & $2.24$ \\
\bottomrule
\end{tabular}
\caption{Cross-section for the process $p p   \to \nu\bar \nu H$. Results are given for a Higgs boson mass $m_H = 125.09~{\rm \UGeV}$.}
\label{tab:nnZH_xsec}
\end{table}

In Tabs.~\ref{tab:wh_xsec}--\ref{tab:nnZH_xsec}
we report the inclusive cross sections for associated production of a Higgs
boson and a weak gauge boson $V=W,Z$, for $pp$ collisions at 13, 14 and 27~\UTeV.
The results have been obtained using HAWK, combining
NNLO QCD and NLO EW
corrections~\cite{Harlander:2018yio,Denner:2014cla,Harlander:2014wda,
Altenkamp:2012sx,Denner:2011id,Brein:2003wg,Ciccolini:2003jy}, by
means of a multiplicative scheme, as described in the YR4 studies (eq.
I.5.15 and I.5.16 of Ref.~\cite{deFlorian:2016spz}).  For ZH production, the
loop-induced $gg\rightarrow ZH$ channel has been computed at NLO+NLL, using a Born-improved Higgs effective field theory (HEFT) approach, and added linearly.

The contribution from photon-induced channels depends on the specific
decay mode of the vector boson, and thus it has been removed from the
total cross-sections, while it is instead included in the total result
for the dedicated cross-sections where decay products are
specified. In the latter cases, the individual photon-induced cross
section is also separately reported.

The results at 27 and 14~\UTeV show a similar pattern of good
perturbative convergence. There are two points that deserve some
specific comment:

\begin{enumerate}
\item As can be evinced from the above tables, photon-induced
  contributions are relatively important in the $p p \to l^\pm \nu H$
  case (where they amount to $\sim 4-7 \%$ of the total cross
  section). For the $pp\to l\bar l H$ case instead, they contribute to
  only $\sim 4-7$ permille.

  We also notice that the relative weight of the photon-induced
  channel is computed more reliably than in the results previously obtained
  for the YR4 study: the changes in the values of $\sigma_\gamma$ from
  the YR4 results (which also had large uncertainties) to those
  presented here are indeed non-negligible, and they are due to the
  fact that the photon PDF is now constrained significantly better, thanks to
  the LUXqed approach~\cite{Manohar:2016nzj,Manohar:2017eqh}. We refer
  the reader to paragraph I.5.2.c of the YR4 for details on how this
  channel was treated previously. For the numbers in the new tables,
  the cross section for $\sigma_\gamma$ was computed using the
  LUXqed\_plus\_PDF4LHC15\_nnlo PDF set. For completeness, we also
  included an update for the 13~\UTeV cross sections using this PDF set.
  
\item As far as the loop-induced $gg\rightarrow ZH$ process is
  concerned, we remind that this channel starts contributing only at
  order $\alpha_S^2$, hence it is part of the NNLO corrections to the
  $pp \rightarrow ZH$ cross section. Nevertheless, due to the gluon
  luminosity, its relative size is important, especially at large
  centre-of-mass energies.
Due to the fact that it is a loop-induced channel, this contribution
is known exactly (i.e. retaining finite values for the top mass) only at
LO.  However, because of its numerical size, and due to the fact that
it contributes to the total cross section with a leading-order-like
scale uncertainty, it is important to compute it at higher order.
Exact NLO corrections to $gg \rightarrow ZH$ are not yet available. The numbers in
the tables are obtained using a Born-improved HEFT approach, which
essentially consists in computing the process at LO exactly, and
rescaling it with the NLO/LO $K$-factor obtained in the
$m_t\to\infty$ limit. NLL threshold effects have also been
included. At order $\alpha_S^3$ there are however many other gluon-gluon initiated sub-processes
that are not yet calculated. It is reasonable
to expect that for VH the correction to the loop induced process will be the
first at order $\alpha_S^3$ to be evaluated in the near future, so that this contribution can provide an order of magnitude estimate of the remaining
perturbative uncertainty
coming from the missing higher orders.
\end{enumerate}

\asubsubsubsection{$t\bar t H$ and $tH$}
\label{sec:he-lhc-ttH}
Cross sections for $t\bar{t}H$ and $tH+\bar{t}H$ production at
$\sqrt{s}=14$ and 27~\UTeV are presented in Tables~\ref{tab:ttH13_xsec}-\ref{tab:ttH27_xsec}
and Tables~\ref{tab:tH13_xsec}-\ref{tab:tH27_xsec} respectively. Results have been obtained using the same setup as
in YR4, and considering three values for $M_H$, namely $M_H=125.09\pm
0.5$~\UGeV. The theoretical uncertainties from renormalisation and
factorisation scale dependence, PDF, and $\alpha_s$ are calculated as
explained in Sec. I.6.2 of YR4~\cite{deFlorian:2016spz}, to which we
refer for full details. $t\bar{t}H$ predictions include NLO
QCD~\cite{Beenakker:2001rj,Beenakker:2002nc,Reina:2001sf,Dawson:2002tg,Dawson:2003zu,Yu:2014cka,Frixione:2015zaa}
and NLO QCD+EW
corrections~\cite{Yu:2014cka,Frixione:2014qaa,Frixione:2015zaa}, while
$tH+\bar{t}H$ predictions are accurate at NLO QCD
only~\cite{Demartin:2015uha}. In both cases, {\tt
MadGraph5\_aMC@NLO}~\cite{Alwall:2014hca,Frederix:2018nkq} has been
employed for the computation of the cross sections. As expected, going
to higher energies greatly enhances both $t\bar{t}H$ and $tH+\bar{t}H$
cross sections.

\begin{table}
\centering
\begin{tabular}{cccccc}%
\toprule
$m_{\rm H}$~[\UGeV] & $\sigma^{\rm NLO}_{\rm QCD+EW}$~[fb] & Scale~[\%] &
$\alpha_s$~[\%] & PDF~[\%] & PDF+${\alpha_s}$~[\%]\\
\midrule
$124.59$ & $512.2$ & $^{+5.8}_{-9.2}$ & $2.0$ & $3.0$ & $3.6$ \\
$125.09$ & $506.5$ & $^{+5.8}_{-9.2}$ & $2.0$ & $3.0$ & $3.6$ \\
$125.59$ & $500.7$ & $^{+5.8}_{-9.2}$ & $2.0$ & $3.0$ & $3.6$ \\
\bottomrule
\end{tabular}%
\caption{NLO QCD+EW cross sections for $t\bar tH$ production at the 13 \UTeV LHC, taken from Ref.~\cite{deFlorian:2016spz}.}
\label{tab:ttH13_xsec}
\end{table}
\begin{table}
\centering
\begin{tabular}{cccccc}
\toprule
$m_{\rm H}$~[\UGeV] & $\sigma^{\rm NLO}_{\rm QCD+EW}$~[fb] & Scale~[\%] &
$\alpha_s$~[\%] & PDF~[\%] & PDF+${\alpha_s}$~[\%]\\
\midrule
$124.59$ & $619.3$ & $^{+6.1}_{-9.2}$ & $1.9$ & $2.9$ & $3.5$ \\
$125.09$ & $612.8$ & $^{+6.0}_{-9.2}$ & $1.9$ & $2.9$ & $3.5$ \\
$125.59$ & $605.6$ & $^{+6.1}_{-9.2}$ & $1.9$ & $2.9$ & $3.5$ \\
\bottomrule
\end{tabular}
\caption{NLO QCD+EW cross sections for $t\bar tH$ production at the 14 \UTeV LHC.}
\label{tab:ttH14_xsec}
\end{table}
\begin{table}
\centering
\begin{tabular}{cccccc}
\toprule
$m_{\rm H}$~[\UGeV] & $\sigma^{\rm NLO}_{\rm QCD+EW}$~[pb] & Scale~[\%] &
$\alpha_s$~[\%] & PDF~[\%] & PDF+${\alpha_s}$~[\%]\\
\midrule
$124.59$ & $2.90$ & $^{+7.9}_{-9.0}$ & $1.8$ & $2.1$ & $2.8$ \\
$125.09$ & $2.86$ & $^{+7.8}_{-9.0}$ & $1.8$ & $2.1$ & $2.8$ \\
$125.59$ & $2.84$ & $^{+7.9}_{-9.0}$ & $1.8$ & $2.1$ & $2.8$ \\
\bottomrule
\end{tabular}
\caption{NLO QCD+EW cross sections for $t\bar tH$ production at a 27~\UTeV proton--proton collider.}
\label{tab:ttH27_xsec}
\end{table}

\begin{table}
\centering
\begin{tabular}{cccccccc}%
\toprule
$m_{\rm H}$~[\UGeV] & $\sigma_{t{\rm H}+\bar t{\rm H}}$~[fb] & Scale+FS~[\%] &
$\alpha_s$~[\%] & PDF~[\%] & PDF+${\alpha_s}$~[\%] & $\sigma_{t{\rm H}}$~[fb] & 
$\sigma_{\bar t{\rm H}}$~[fb]\\
\midrule
$124.59$ & $74.52$ & $^{+6.6}_{-14.7}$ & $1.2$ & $3.5$ & $3.7$ & $49.04$ & $25.49$\\
$125.09$ & $74.26$ & $^{+6.5}_{-14.7}$ & $1.2$ & $3.5$ & $3.7$ & $48.89$ & $25.40$\\
$125.59$ & $74.09$ & $^{+6.5}_{-15.2}$ & $1.2$ & $3.6$ & $3.7$ & $48.75$ & $25.32$\\
\bottomrule
\end{tabular}%
\caption{NLO QCD cross sections for the $t-$channel $tH$ and $\bar t H$
production at the 13 \UTeV LHC, taken from Ref.~\cite{deFlorian:2016spz}.}
\label{tab:tH13_xsec}
\end{table}
\begin{table}
\centering
\begin{tabular}{cccccccc}
\toprule
$m_{\rm H}$~[\UGeV] & $\sigma_{t{\rm H}+\bar t{\rm H}}$~[fb] & Scale+FS~[\%] &
$\alpha_s$~[\%] & PDF~[\%] & PDF+${\alpha_s}$~[\%] & $\sigma_{t{\rm H}}$~[fb] & 
$\sigma_{\bar t{\rm H}}$~[fb]\\
\midrule
$124.59$ & $90.35$ & $^{+6.4}_{-14.6}$ & $1.2$ & $3.4$ & $3.6$ & $59.15$ & $31.21$\\
$125.09$ & $90.12$ & $^{+6.4}_{-14.7}$ & $1.2$ & $3.4$ & $3.6$ & $58.96$ & $31.11$\\
$125.59$ & $89.72$ & $^{+6.4}_{-14.8}$ & $1.2$ & $3.4$ & $3.6$ & $58.70$ & $31.02$\\
\bottomrule
\end{tabular}
\caption{NLO QCD cross sections for the $t-$channel $tH$ and $\bar t H$
production at the 14 \UTeV LHC.}
\label{tab:tH14_xsec}
\end{table}
\begin{table}
\centering
\begin{tabular}{cccccccc}
\toprule
$m_{\rm H}$~[\UGeV] & $\sigma_{t{\rm H}+\bar t{\rm H}}$~[fb] & Scale+FS~[\%] &
$\alpha_s$~[\%] & PDF~[\%] & PDF+${\alpha_s}$~[\%] & $\sigma_{t{\rm H}}$~[fb] & 
$\sigma_{\bar t{\rm H}}$~[fb]\\
\midrule
$124.59$ & $419.0$ & $^{+5.0}_{-12.3}$ & $1.3$ & $2.6$ & $2.9$ & $263.3$ & $155.7$\\
$125.09$ & $417.9$ & $^{+5.0}_{-12.5}$ & $1.3$ & $2.6$ & $2.9$ & $262.8$ & $155.1$\\
$125.59$ & $416.4$ & $^{+5.0}_{-12.6}$ & $1.3$ & $2.6$ & $2.9$ & $261.8$ & $154.7$\\
\bottomrule
\end{tabular}
\caption{NLO QCD cross sections for the $t-$channel $tH$ and $\bar t H$
production at a 27 \UTeV proton--proton collider.}
\label{tab:tH27_xsec}
\end{table}

\asubsubsection{Projections of uncertainty reductions for the HL-LHC}
\label{sec:hl-lhc}

This section discusses improvements to the theoretical predictions that may be
possible on the timescale of the HL-LHC.  Estimates of potential reductions in
current theoretical uncertainties are made where possible and potential limiting
factors identified.

\asubsubsubsection{Gluon fusion}
\label{sec:hl-lhc-ggF}
Improving substantially on any of the current
sources of uncertainty represents a major theoretical challenge that
should be met in accordance with our ability to utilise said precision
and with experimental capabilities. 
The computation of sub-leading mass and EW corrections is
currently being addressed by several groups, and therefore it is
likely to be achieved in the next decade. Although such computations
will allow for a better control over some sources of uncertainty,
their final impact on the full theoretical error is likely to be
moderate as current estimates indicate.
Another source of error that might improve in the forthcoming years is
that related to the parton densities. In particular, the extraction of
N$^3$LO PDFs would lead to the disappearance of the PDF-TH
uncertainty. Similar considerations apply to the error on the strong
coupling constant, that will be reduced due to more accurate
extractions. Overall, the above progress would ultimately lead to a
notable reduction of the uncertainties of Figure~\ref{fig:errorplot}.

It is obvious that the future precision of experimental measurement of Higgs boson properties will challenge the theoretical community.
Achieving a significant improvement of our current theoretical understanding of the Higgs boson and its interactions will inspire us to push the boundaries of our capabilities to predict and extract information.
New ways of utilising quantum field theory in our endeavours have to be explored and our perturbative and non-perturbative understanding of hadron scattering processes has to evolve substantially. 
It is clear that this exciting task can only be mastered by a strong and active collider phenomenology community.
 
\subsubsubsection*{Impact of future precision of parton distribution function}
It is a tantalising question to ask by how much one of the largest sources of uncertainty - the imprecise knowledge of PDFs - would be reduced if already all future LHC data were available.
To this end a study was performed in ref.~\cite{Khalek:2018mdn} (see also Section~\ref{sec2:PDFuncertainties}) that uses simulated future data with accordingly shrunken statistical uncertainties to constrain parton distribution functions.
The authors used pseudo data corresponding to measurements of ATLAS, CMS and LHCb for key precision processes after $3 ab^{-1}$ of integrated luminosity were collected at the High-Luminosity LHC at 14 \UTeV.
They then performed a new fit according to the PDF4LHC15 framework~\cite{Botje:2011sn} and studied the implications of their analysis.
The resulting PDFs are readily available and can be used in order to estimate the impact of this future data on specific observables. 
Three scenarios were considered in this study that assume that experimental systematic uncertainties will shrink at different levels relative to the 8 \UTeV run of the LHC.
Scenario 1, scenario 2 and scenario 3 assume that the future systematic uncertainty will be equal, shrunk by a factor $0.7$ or a factor of $0.4$ w.r.t to the 8 \UTeV run respectively.

Evaluating the inclusive Higgs boson production cross section with this simulated PDFs results in the PDF uncertainties summarised in Tab.~\ref{tab:PDFuncert}.
Note, that the central values stay unchanged and all other uncertainties are not afflicted by the change of PDFs.
\begin{table}[!h]
\normalsize\setlength{\tabcolsep}{2pt}
\begin{center}
    \begin{tabular}{rrrrr}
        \toprule
        \multicolumn{1}{c}{$\textrm{E}_{\textrm{CM}}$}&
        \multicolumn{1}{c}{Current  }&
        \multicolumn{1}{c}{Scenario 1 }&
        \multicolumn{1}{c}{Scenario 2}&
        \multicolumn{1}{c}{Scenario 3} \\
        \midrule
13 \UTeV & $ \pm 1.85 \% $ & $\pm 0.78 \%$ & $\pm 0.69\%$ & $\pm 0.59\%$ \\\midrule
14 \UTeV & $ \pm 1.85 \% $ & $\pm 0.78\%$ & $\pm 0.68\%$ & $\pm 0.58\%$ \\\midrule
27 \UTeV & $ \pm 1.95 \%$  & $\pm 0.81\%$ & $\pm 0.72\%$ & $\pm 0.61\%$ \\\midrule
    \end{tabular}
    \caption{Uncertainty due to imprecise knowledge of PDFs estimated with current and simulated future PDFs for different scenarios and at different collider energies.\label{tab:PDFuncert}}
\end{center}
\end{table}
Even the most pessimistic scenario leads to a reduction of the PDF uncertainty by factor of two. 
However, this projections should be viewed only as a first estimate for the determination of PDFs from future measurements.
Predicting the future development and correlation of systematic experimental uncertainties is non trivial and may differ strongly from observable to observable. 
PDF uncertainties may in the future also be adversely impacted by a more accurate treatment of theoretical uncertainties in the predictions of cross sections that serve as input for PDF extraction.
Data incompatibilities may occur for various reasons.
It is clear that an understanding of the structure of the proton at percent level accuracy is clearly a formidable task and rightly deserves significant research in the future.

\asubsubsubsection{Vector boson fusion}
\label{sec:hl-lhc-VBF}

VBF Higgs boson production is currently known at a very high theoretical accuracy. In
the structure-function approximation, the cross section has been computed fully inclusively at
N$^3$LO accuracy in QCD. Fiducial calculations in the same approximation exist at NNLO accuracy in QCD. The
only contribution which is currently unknown is the contribution from two-loop
diagrams with gluon exchange between the two VBF quark lines. The conceptual
difficulty is that it is a $2 \to 3$ process and that currently there are no methods
available for evaluating two-loop diagrams with more than four external legs. It is
realistic that such methods will become available before the HE-LHC is in
operation. Beyond the VBF approximation, the full NLO corrections in both the
strong and electroweak coupling have been computed. The electroweak
contributions are of the same order as, or in certain phase space regions even
larger than, the NNLO QCD corrections. Taking all of this into account, it has
been estimated that the VBF cross section under typical VBF cuts has an accuracy
at the 1\% level. In order to connect these calculations to experimental
measurements one would ideally need merged 2- and 3-jet NLO samples matched with the parton shower~\cite{Nason:2009ai,Jager:2014vna} (NLOPS level) or even better a fully exclusive generator for VBF matched with the parton shower at  NNLO (NNLOPS) . It is realistic that this
will become available within the next few years and certainly before the
HL-/HE-LHC phases.

\asubsubsubsection{$VH$ production}
\label{sec:hl-lhc-VH}
The Higgs couplings to $W$ and $Z$ bosons are related by $SU(2)_L$ gauge invariance. As such, the measurement of the Higgs associated production with a $W$ or a $Z$ is complementary to the vector boson fusion process, as first considered in $e^+e^−$ colliders in Ref. \cite{Jones:1979bq}.
At the time of writing, the numbers shown in Section~\ref{sec:he-lhc-VH}
are the best estimates
available for the $pp\rightarrow VH$ contribution. As far as the ZH final state is concerned, due to the progress made in the
last couple of years for the computation of top-mass effects at NLO in
Higgs-boson pair production, it is foreseeable that, in the forthcoming years
(definitely in the timescale of HL/HE LHC), an exact NLO result
(including finite-$m_t$ effects) will be available also for $gg\rightarrow ZH$. If
one assumes that a pattern similar to what was found for di-higgs
production~\cite{Borowka:2016ypz} also holds for $gg\rightarrow ZH$, one can expect
that the total NLO/LO $K$-factor will be slightly smaller than in the
HEFT limit (from 1.9--2.0 to $\sim$1.6) and the final scale uncertainty for
the $gg\rightarrow ZH$ cross section will decrease from 18--25\% to about 15\%.\footnote{We stress that these numbers have been obtained as a back-of-the-envelope estimate through a comparison with di-higgs production.}

All the above results have been obtained for a stable Higgs boson. For
the Higgs boson decay to bottom quarks, it is known that higher-order
corrections to the $m_{bb}$ line-shape are relevant, as shown in
Ref.~\cite{Ferrera:2017zex} and also recently confirmed
in Ref.~\cite{Caola:2017xuq}.  Although explicit studies are not available,
one can expect that effects similar to those observed at 13--14~\UTeV in
the region $m_{bb}<m_H$ will persist also at higher energies.

The matching of fixed-order corrections to parton showers (PS) is
available for the $pp\rightarrow VH$ signal processes, at NLO as well
as at NNLO~\cite{Astill:2016hpa,Astill:2018ivh}.
As for Higgs decays to bottom quarks, a fixed-order study~\cite{Caola:2017xuq}
suggests that higher-order corrections to the $m_{bb}$ shape are not always
very well modelled by a LO + parton shower treatment of the $H\to b\bar{b}$ decay.
Event generators as the one developed in Ref.~\cite{Astill:2018ivh}, and improvements thereof
for the treatment of radiation off $b$-quarks~\cite{Buonocore:2017lry},
will allow one to assess this issue in the forthcoming
years. A solid prediction of the $H\to b\bar{b}$ decay, also matched to
parton-showers, can definitely be expected in the timescale of HL/HE LHC.

Furthermore, once the exact $gg\rightarrow ZH$ computation at NLO will
be completed, a NLO matching to parton-shower will be straightforward
to achieve, thereby improving on the currently available more advanced
treatments, where a LO-merging of the exact matrix elements for
$gg\rightarrow ZH$ and $gg\rightarrow ZH+1$-jet is performed.

Finally, as for the VH and VHJ event generators, recently there has been
also the completion of the NLO EW corrections matched to the parton
shower~\cite{Granata:2017iod} showing once again the relevance of the
EW corrections for the distributions for both the fixed order and the
matched predictions.

\asubsubsubsection{$t\bar t H$ and $tH$}
\label{sec:hl-lhc-ttH}
The cross sections for $t\bar t H$ and $tH$ production are known at NLO accuracy in QCD ~\cite{Beenakker:2001rj,Reina:2001sf,Biswas:2012bd} and, in the
case of $t \bar t H$, NLO EW corrections have also been calculated~\cite{Yu:2014cka,Frixione:2014qaa}. The corresponding
theoretical uncertainty is of the order of 10--15\% and is mainly induced by the
residual scale dependence and, to a lesser extent, by PDF uncertainties. A
drastic improvement can only come from the calculation of the NNLO QCD
corrections. Given the ongoing rapid progress in cross section calculations with NNLO accuracy in QCD, it is
foreseeable that NNLO QCD corrections to $t\bar t H$ and $tH$ will become available in the
next decade. In this scenario it is reasonable to expect a factor-two
improvement of the theoretical accuracy.

On the other hand, the extraction of the $t \bar t H$ signal is at the moment mainly
limited by the theoretical uncertainties in the modelling of the background,
mainly $t\bar t b \bar b$ and $t \bar t W+$jets, via Monte Carlo generators. The reliable assessment of
the related uncertainties and their further reduction are the main goals of an
ongoing campaign of theoretical studies within the HXSWG. On a time scale of
5--10 years such background uncertainties may be reduced by a factor two to
three.

\asubsubsection{Predictions for boosted Higgs production}
The HL and HE LHC upgrades would allow for in-depth analyses of high-$p_t$ tail of the Higgs
transverse momentum distribution. This region is particularly interesting as it is
very sensitive to BSM physics in the Higgs sector. For example, measures in the boosted region would 
allow one to lift the degeneracy between $ggH$ and $ttH$ couplings, and more in general to 
probe the internal structure of the $ggH$ interaction. In this section, we report theoretical
predictions for boosted Higgs production. 

\begin{figure}[h]
\begin{center}
    \includegraphics[width=0.45\textwidth]{\main/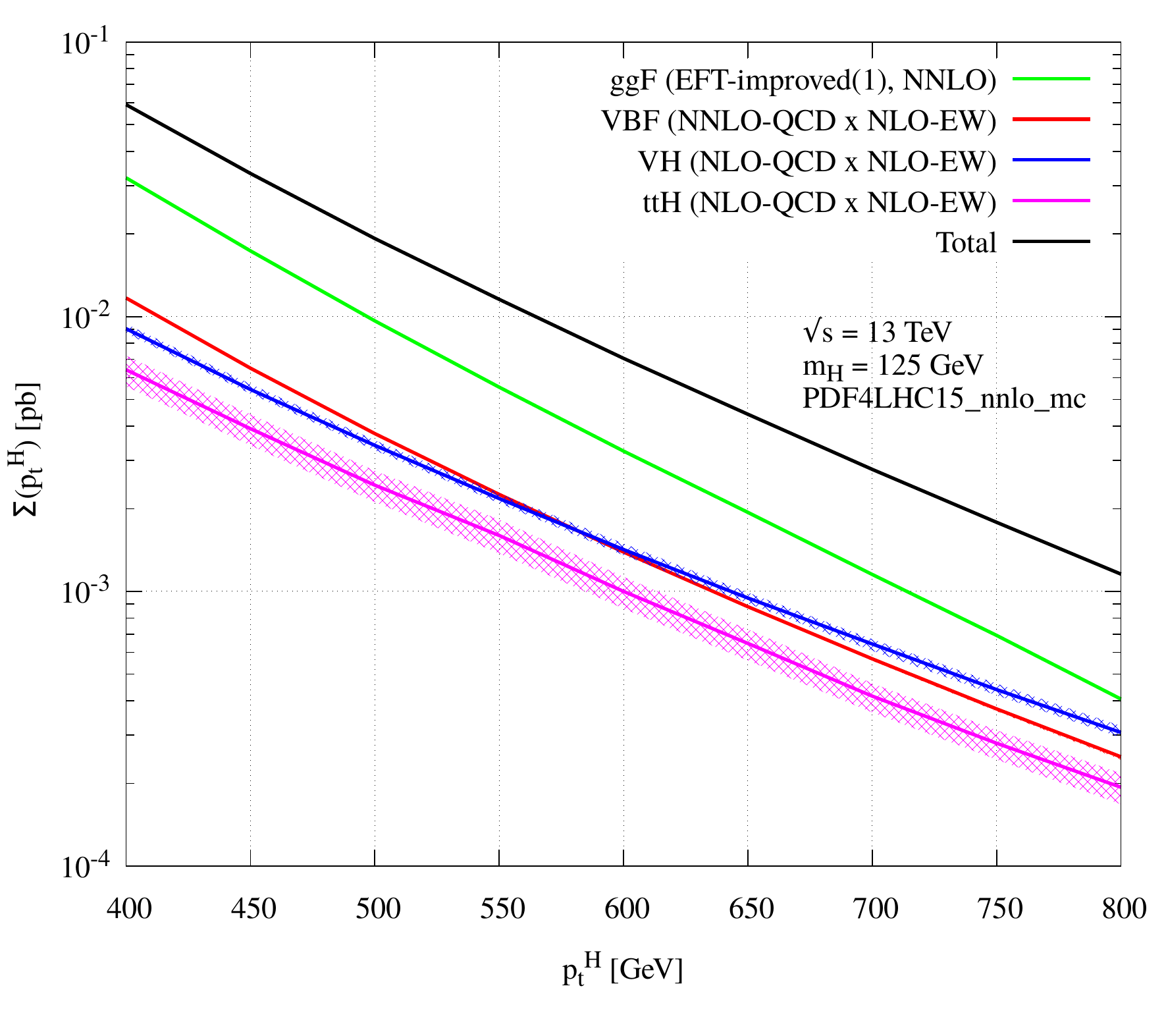}~
    \includegraphics[width=0.45\textwidth]{\main/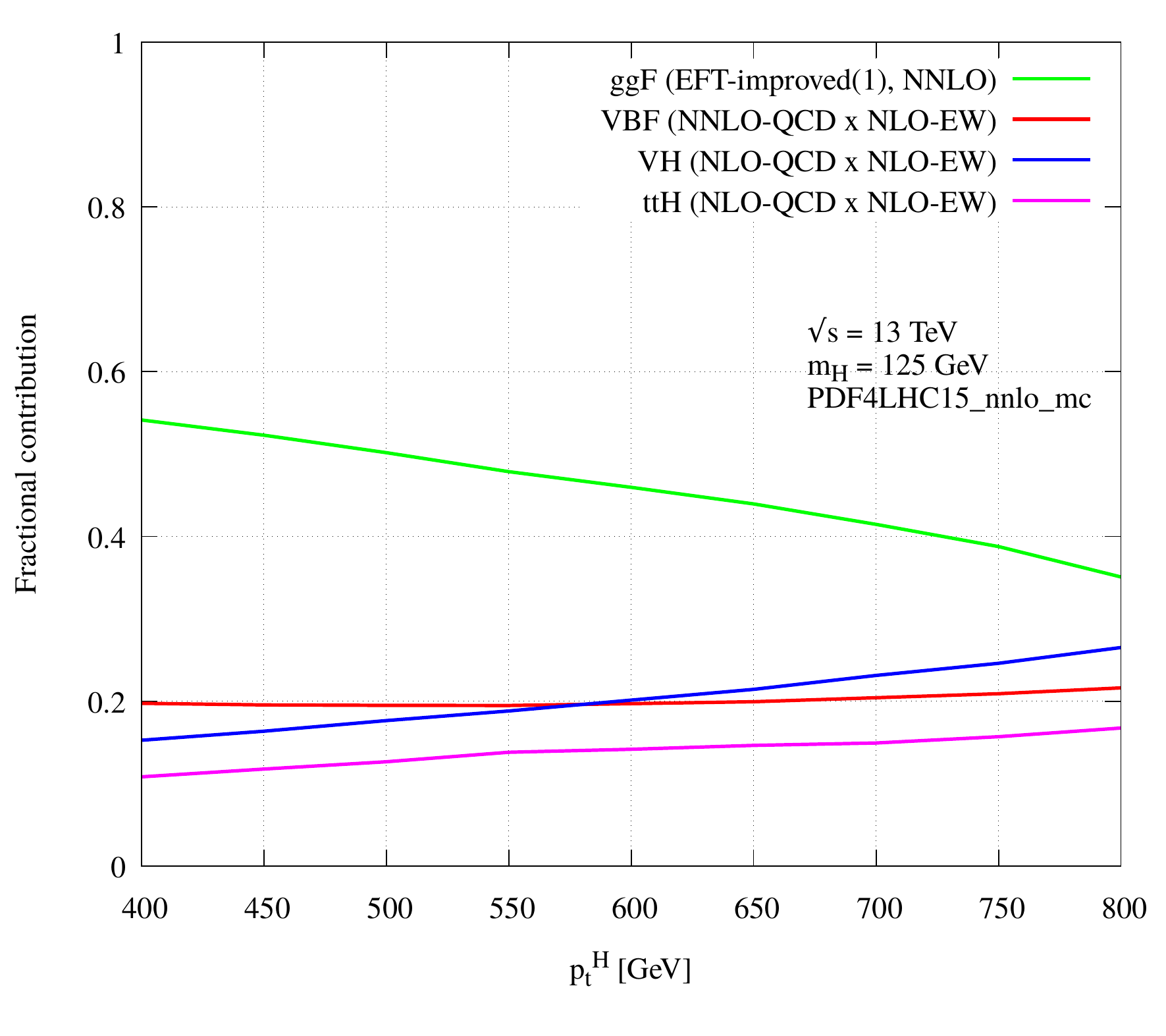}
    \caption{
    Boosted Higgs prediction at the 13-\UTeV LHC. Left: cumulative transverse momentum distribution.
    Right: relative importance of different production mechanisms. See text for details. 
        \label{fig:boosted13}}
        \end{center}
\end{figure}

We first present results for the 13-\UTeV LHC. In Fig.~\ref{fig:boosted13}(left) we show the cumulative Higgs 
transverse momentum distribution, defined as
$$
\Sigma(p_t^H) = \int\limits_{p_t^H}^{\infty} \frac{d\sigma}{d p_t},
$$
for the main production channels.
The $ggF$ prediction is obtained by rescaling the exact NLO with the NNLO $K-$factor in the $m_t\to\infty$
approximation, and it does not contain EW corrections. The VBF and VH predictions include NNLO QCD
and NLO EW corrections, while the $t\bar t H$ prediction includes NLO QCD and EW corrections. In
Fig.~\ref{fig:boosted13}(right), we show the relative importance of the different production mechanisms.\footnote{
The small feature around $p_t\sim 750~{\rm \UGeV}$ in the $ggF$ channel is due to lack of statistics in the
theoretical simulation and it is not a genuine physical feature.}
As it is well known, at high $p_t$ the $ggF$ channel becomes somewhat less dominant. Still, radiative
corrections strongly enhance this channel, which remains the dominant one in the \UTeV region. A very similar picture is expected for the HL-LHC. 

\begin{figure}[h]
\begin{center}
    \includegraphics[width=0.45\textwidth]{\main/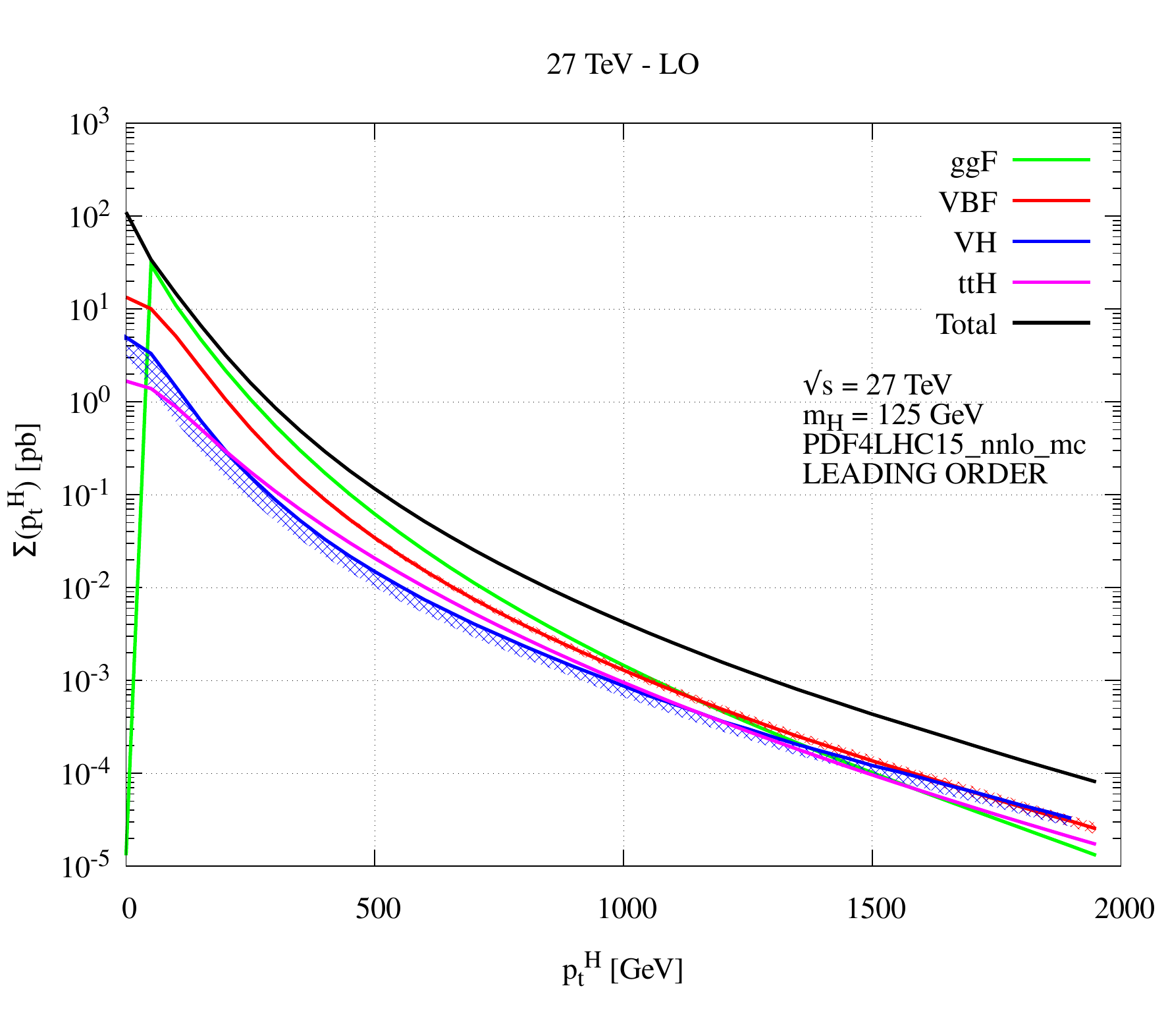}~
    \includegraphics[width=0.45\textwidth]{\main/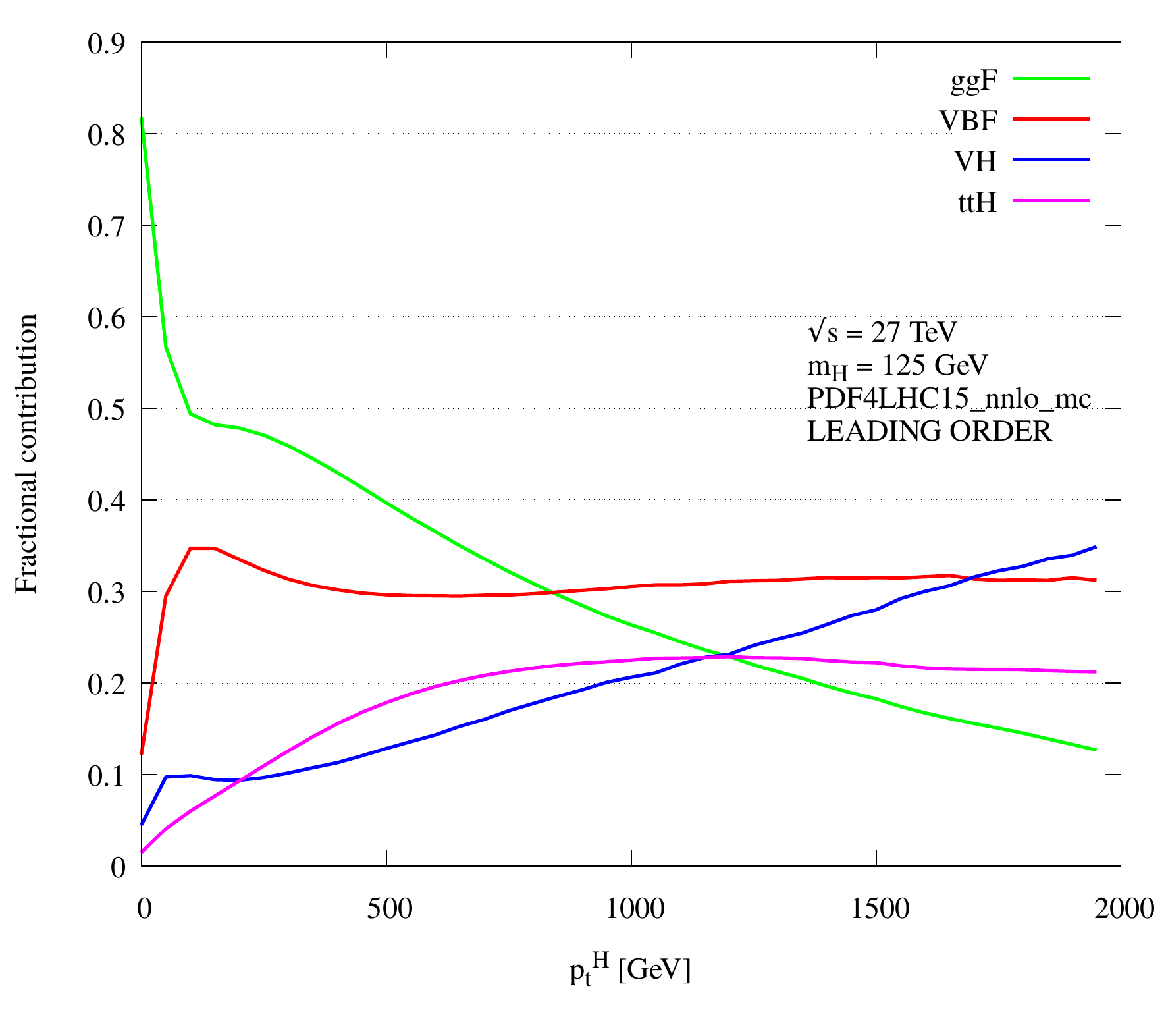}
    \caption{
    LO boosted Higgs prediction at the 27-\UTeV LHC. Left: cumulative transverse momentum distribution.
    Right: relative importance of different production mechanisms. See text for details. 
        \label{fig:boosted27-LO}}
        \end{center}
\end{figure}

\begin{figure}[h]
\begin{center}
    \includegraphics[width=0.45\textwidth]{\main/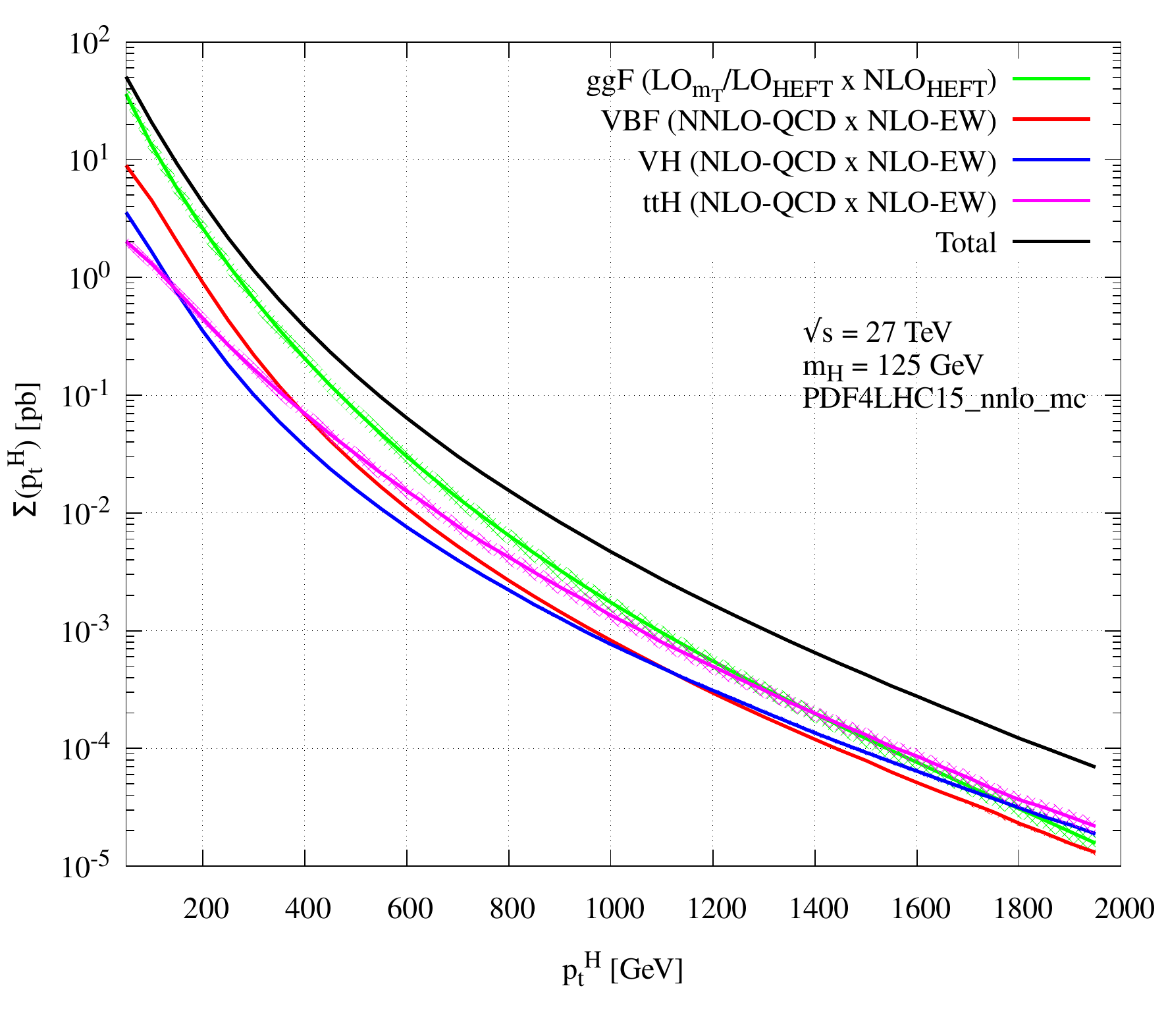}~
    \includegraphics[width=0.45\textwidth]{\main/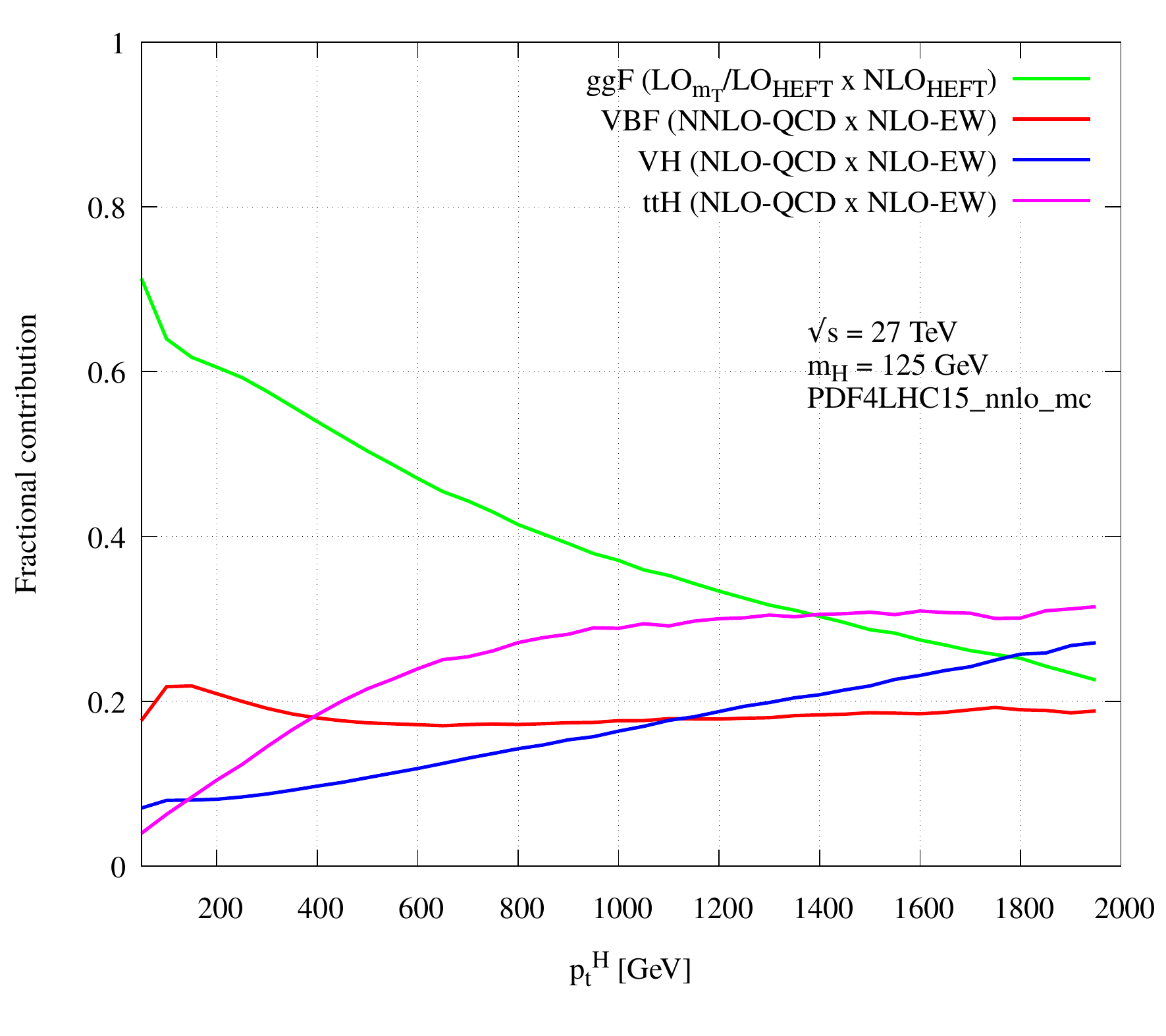}
    \caption{
    Boosted Higgs prediction at the 27-\UTeV LHC, including radiative corrections. Left: cumulative transverse momentum distribution.
    Right: relative importance of different production mechanisms. See text for details. 
        \label{fig:boosted27}}
        \end{center}
\end{figure}

Figs.~\ref{fig:boosted27-LO} and~\ref{fig:boosted27} show similar predictions for the HE-LHC. In Fig.~\ref{fig:boosted27-LO}, all predictions are LO. 
At high $p_t$, the $ggF$ channel become sub-dominant compared to the other ones. VBF becomes the dominant channel around  $p_t\sim 1~{\rm \UTeV}$, 
and VH around $p_t\sim 2~{\rm \UTeV}$. In the \UTeV region, the $t\bar t H$ channel becomes larger than $ggF$. 

This picture is however significantly altered by radiative correction, whose size and impact is very different for different channels. 
This is shown in Fig.~\ref{fig:boosted27}, where predictions include radiative corrections. More precisely, the VBF, VH and $t\bar t H$ 
predictions have the same accuracy as the ones in Fig.~\ref{fig:boosted13}. The $ggF$ prediction contains exact LO mass effects rescaled
by the NLO $K-$factor in the $m_t$ approximation. This is expected to provide an excellent approximation of the exact NLO result. 
Radiative corrections enhance the relative importance of the $ggF$ and $t\bar t H$ channels, which still dominate over VBF well into the
multi-\UTeV region. At large $p_t \sim 1.5~{\rm \UTeV}$, the $t\bar t H$ channel becomes the dominant one. 

Obtaining accurate and precise theoretical predictions in the boosted region is very challenging. Nevertheless, it is natural
to expect progress in the timescale for the HL and HE LHC upgrades. This would allow for a proper scrutiny of the structure of
Higgs interactions in the multi-\UTeV regime. 

\asubsubsection{Dependence of gluon-fusion cross section  at $14$ and $27$ \UTeV on $m_H$}
\label{sec:ggFmhdep}
The dependence of the inclusive gluon-fusion cross-section on the Higgs boson mass is shown in
Tables~\ref{ggF:14TeV} and~\ref{ggF:27TeV}, for $pp$ collisions at $\sqrt{s}=14$ and 27~\UTeV, respectively.
\begin{table}[!h]
\begin{center}
\begin{equation}
\begin{array}{|c|c|c|c|c|c|c|}
 \hline
 m_H  \text{ [\UGeV]}&\text{Cross Section [pb]}&\text{+ $\delta$Th. [\%]}&\text{- $\delta$Th. [\%]}&\pm\delta\text{(PDF+$\alpha_S$) [\%]}& \pm\delta\text{$\alpha_S$ [\%]} & \pm\delta\text{ PDF [\%]}  \\
\hline 
125.09 & 54.72 & 4.29 & -6.46 & 3.20 & 2.61 & 1.85\\
\hline 
124.59 & 55.10 & 4.29 & -6.48 & 3.20 & 2.61 & 1.86\\
\hline 
125.59 & 54.34 & 4.28 & -6.45 & 3.20 & 2.61 & 1.85\\
\hline 
120.00 & 58.85 & 4.37 & -6.61 & 3.23 & 2.63 & 1.87\\
\hline 
120.50 & 58.42 & 4.37 & -6.60 & 3.22 & 2.63 & 1.87\\
\hline 
121.00 & 58.00 & 4.36 & -6.58 & 3.22 & 2.63 & 1.87\\
\hline 
121.50 & 57.56 & 4.35 & -6.57 & 3.22 & 2.62 & 1.86\\
\hline 
122.00 & 57.15 & 4.34 & -6.55 & 3.22 & 2.62 & 1.86\\
\hline 
122.50 & 56.75 & 4.33 & -6.54 & 3.21 & 2.62 & 1.86\\
\hline 
123.00 & 56.35 & 4.32 & -6.52 & 3.21 & 2.62 & 1.86\\
\hline 
123.50 & 55.95 & 4.31 & -6.51 & 3.21 & 2.61 & 1.86\\
\hline 
124.00 & 55.56 & 4.30 & -6.49 & 3.21 & 2.61 & 1.86\\
\hline 
124.10 & 55.48 & 4.30 & -6.49 & 3.20 & 2.61 & 1.86\\
\hline 
124.20 & 55.41 & 4.30 & -6.49 & 3.20 & 2.61 & 1.86\\
\hline 
124.30 & 55.33 & 4.30 & -6.49 & 3.20 & 2.61 & 1.86\\
\hline 
124.40 & 55.25 & 4.30 & -6.48 & 3.20 & 2.61 & 1.86\\
\hline 
124.50 & 55.17 & 4.30 & -6.48 & 3.20 & 2.61 & 1.86\\
\hline 
124.60 & 55.10 & 4.29 & -6.48 & 3.20 & 2.61 & 1.86\\
\hline 
124.70 & 55.02 & 4.29 & -6.47 & 3.20 & 2.61 & 1.86\\
\hline 
124.80 & 54.94 & 4.29 & -6.47 & 3.20 & 2.61 & 1.86\\
\hline 
124.90 & 54.86 & 4.29 & -6.47 & 3.20 & 2.61 & 1.86\\
\hline 
125.00 & 54.79 & 4.29 & -6.47 & 3.20 & 2.61 & 1.86\\
\hline 
125.10 & 54.71 & 4.29 & -6.46 & 3.20 & 2.61 & 1.85\\
\hline 
125.20 & 54.64 & 4.28 & -6.46 & 3.20 & 2.61 & 1.85\\
\hline 
125.30 & 54.56 & 4.28 & -6.46 & 3.20 & 2.61 & 1.85\\
\hline 
125.40 & 54.48 & 4.28 & -6.45 & 3.20 & 2.61 & 1.85\\
\hline 
125.50 & 54.41 & 4.28 & -6.45 & 3.20 & 2.61 & 1.85\\
\hline 
125.60 & 54.33 & 4.28 & -6.45 & 3.20 & 2.61 & 1.85\\
\hline 
125.70 & 54.26 & 4.28 & -6.44 & 3.20 & 2.61 & 1.85\\
\hline 
125.80 & 54.18 & 4.27 & -6.44 & 3.20 & 2.60 & 1.85\\
\hline 
125.90 & 54.11 & 4.27 & -6.44 & 3.20 & 2.60 & 1.85\\
\hline 
126.00 & 54.03 & 4.27 & -6.44 & 3.20 & 2.60 & 1.85\\
\hline 
126.50 & 53.66 & 4.26 & -6.42 & 3.19 & 2.60 & 1.85\\
\hline 
127.00 & 53.29 & 4.25 & -6.41 & 3.19 & 2.60 & 1.85\\
\hline 
127.50 & 52.92 & 4.25 & -6.40 & 3.19 & 2.60 & 1.85\\
\hline 
128.00 & 52.56 & 4.24 & -6.38 & 3.19 & 2.60 & 1.85\\
\hline 
128.50 & 52.20 & 4.23 & -6.37 & 3.18 & 2.59 & 1.85\\
\hline 
129.00 & 51.85 & 4.22 & -6.35 & 3.18 & 2.59 & 1.85\\
\hline 
129.50 & 51.50 & 4.21 & -6.34 & 3.18 & 2.59 & 1.84\\
\hline 
130.00 & 51.15 & 4.20 & -6.33 & 3.18 & 2.59 & 1.84\\
\hline
\end{array}
\nonumber
\end{equation}
\end{center}
\caption{The gluon-fusion cross-section in $pp$ collisions at $\sqrt{s}=14$~\UTeV, for different values of the Higgs boson mass $m_H$. \label{ggF:14TeV}}
\end{table}

\begin{table}[!h]
\begin{center}
\begin{equation}
\begin{array}{|c|c|c|c|c|c|c|}
 \hline
 m_H  \text{ [\UGeV]}&\text{Cross Section [pb]}&\text{+ $\delta$Th. [\%]}&\text{- $\delta$Th. [\%]}&\pm\delta\text{(PDF+$\alpha_S$) [\%]}& \pm\delta\text{$\alpha_S$ [\%]} & \pm\delta\text{ PDF [\%]}  \\
\hline 
125.09 & 146.65 & 4.53 & -6.43 & 3.30 & 2.66 & 1.95\\
\hline 
124.59 & 147.55 & 4.55 & -6.45 & 3.30 & 2.67 & 1.95\\
\hline 
125.59 & 145.75 & 4.52 & -6.42 & 3.30 & 2.66 & 1.95\\
\hline 
120.00 & 156.35 & 4.64 & -6.60 & 3.33 & 2.69 & 1.97\\
\hline 
120.50 & 155.36 & 4.63 & -6.58 & 3.33 & 2.69 & 1.97\\
\hline 
121.00 & 154.36 & 4.62 & -6.56 & 3.33 & 2.69 & 1.97\\
\hline 
121.50 & 153.38 & 4.61 & -6.55 & 3.32 & 2.68 & 1.96\\
\hline 
122.00 & 152.41 & 4.60 & -6.54 & 3.32 & 2.68 & 1.96\\
\hline 
122.50 & 151.45 & 4.59 & -6.52 & 3.32 & 2.68 & 1.96\\
\hline 
123.00 & 150.50 & 4.58 & -6.50 & 3.31 & 2.68 & 1.96\\
\hline 
123.50 & 149.56 & 4.57 & -6.49 & 3.31 & 2.67 & 1.96\\
\hline 
124.00 & 148.64 & 4.56 & -6.47 & 3.31 & 2.67 & 1.95\\
\hline 
124.10 & 148.45 & 4.56 & -6.47 & 3.31 & 2.67 & 1.95\\
\hline 
124.20 & 148.27 & 4.56 & -6.46 & 3.31 & 2.67 & 1.95\\
\hline 
124.30 & 148.08 & 4.55 & -6.46 & 3.31 & 2.67 & 1.95\\
\hline 
124.40 & 147.90 & 4.55 & -6.46 & 3.31 & 2.67 & 1.95\\
\hline 
124.50 & 147.72 & 4.55 & -6.46 & 3.31 & 2.67 & 1.95\\
\hline 
124.60 & 147.53 & 4.55 & -6.45 & 3.30 & 2.67 & 1.95\\
\hline 
124.70 & 147.35 & 4.54 & -6.45 & 3.30 & 2.67 & 1.95\\
\hline 
124.80 & 147.17 & 4.54 & -6.44 & 3.30 & 2.67 & 1.95\\
\hline 
124.90 & 146.99 & 4.54 & -6.44 & 3.30 & 2.67 & 1.95\\
\hline 
125.00 & 146.81 & 4.54 & -6.44 & 3.30 & 2.67 & 1.95\\
\hline 
125.10 & 146.63 & 4.53 & -6.43 & 3.30 & 2.66 & 1.95\\
\hline 
125.20 & 146.45 & 4.53 & -6.43 & 3.30 & 2.66 & 1.95\\
\hline 
125.30 & 146.27 & 4.53 & -6.43 & 3.30 & 2.66 & 1.95\\
\hline 
125.40 & 146.09 & 4.53 & -6.42 & 3.30 & 2.66 & 1.95\\
\hline 
125.50 & 145.91 & 4.52 & -6.42 & 3.30 & 2.66 & 1.95\\
\hline 
125.60 & 145.73 & 4.52 & -6.42 & 3.30 & 2.66 & 1.95\\
\hline 
125.70 & 145.55 & 4.52 & -6.41 & 3.30 & 2.66 & 1.95\\
\hline 
125.80 & 145.37 & 4.52 & -6.41 & 3.30 & 2.66 & 1.95\\
\hline 
125.90 & 145.20 & 4.52 & -6.41 & 3.30 & 2.66 & 1.95\\
\hline 
126.00 & 145.02 & 4.51 & -6.40 & 3.30 & 2.66 & 1.95\\
\hline 
126.50 & 144.14 & 4.50 & -6.39 & 3.29 & 2.66 & 1.94\\
\hline 
127.00 & 143.26 & 4.49 & -6.37 & 3.29 & 2.66 & 1.94\\
\hline 
127.50 & 142.40 & 4.48 & -6.36 & 3.29 & 2.65 & 1.94\\
\hline 
128.00 & 141.54 & 4.48 & -6.34 & 3.28 & 2.65 & 1.94\\
\hline 
128.50 & 140.69 & 4.47 & -6.33 & 3.28 & 2.65 & 1.94\\
\hline 
129.00 & 139.84 & 4.46 & -6.31 & 3.28 & 2.65 & 1.93\\
\hline 
129.50 & 139.00 & 4.46 & -6.30 & 3.27 & 2.64 & 1.93\\
\hline 
130.00 & 138.18 & 4.45 & -6.29 & 3.27 & 2.64 & 1.93\\
 \hline
\end{array}
\nonumber
\end{equation}
\end{center}
\caption{The gluon-fusion cross-section in $pp$ collisions at $\sqrt{s}=27$~\UTeV, for different values of the Higgs boson mass $m_H$. \label{ggF:27TeV}}
\end{table}

\asubsubsection[R. Abdul Khalek, S. Bailey, J. Gao, L.  Harland-Lang,
  J. Rojo]{PDF uncertainty expectations at the HE/HL-LHC}
\label{sec2:PDFuncertainties}
%
{\bf PDFs in the HL-LHC era.}
The detailed understanding of the quark and gluon
structure of the proton, quantified by the parton distribution functions
(PDFs)~\cite{Gao:2017yyd,Rojo:2015acz,Forte:2013wc},
is an essential ingredient for the theoretical predictions at hadron colliders.
PDF uncertainties represent one of the dominant theoretical systematic
errors
both for direct searches of new physics beyond the Standard
Model (BSM)~\cite{Beenakker:2015rna}
as well as in
the profiling of the Higgs boson sector~\cite{deFlorian:2016spz}.
Therefore, improving our knowledge of the proton structure
is an essential task for the high-precision physics
program to be carried out at future runs of the LHC, including
the HL-LHC era.

Modern global PDF fits~\cite{Ball:2017nwa,Dulat:2015mca,
  Harland-Lang:2014zoa,Alekhin:2017kpj}
include a wide range of LHC measurements in
processes such as the production of jets, weak gauge bosons,
and top quark pairs, among others.
Recent breakthroughs in the calculation
of NNLO QCD and NLO QED and electroweak corrections to
most PDF-sensitive processes have been instrumental in
allowing for the full exploitation of the information provided by the LHC measurements.
The impact of high-precision LHC data combined with state-of-the art
perturbative calculations has been quantified for many of the processes
of interest, such as top-quark pair production~\cite{Czakon:2016olj,Guzzi:2014wia},
the transverse momentum spectrum of $Z$ bosons~\cite{Boughezal:2017nla}, 
direct photon production~\cite{dEnterria:2012kvo,Campbell:2018wfu},
$D$ meson production in the forward region~\cite{Gauld:2015yia,Zenaiev:2015rfa,Gauld:2016kpd}, $W$ production
in association with charm quarks~\cite{Aad:2014xca,Chatrchyan:2013mza},
and inclusive jet production~\cite{Currie:2016bfm,Rojo:2014kta}.

From the point of view of PDF determinations, the availability of the
immense data samples at the HL-LHC
will permit a significant extension
of the kinematic coverage of PDF-sensitive measurements
as well as a marked improvement in their statistical and systematic uncertainties.
In this contribution, we summarise the main results
of our PDF projections for the HL-LHC era presented
in~\cite{Khalek:2018mdn}.
The main idea is to quantify the impact of the future HL--LHC
measurements on the proton PDFs and their
uncertainties, with emphasis on
their implications for Higgs physics.
Specifically, we quantify the 
constraints of the HL--LHC pseudo-data
on the PDF4LHC15
set~\cite{Butterworth:2015oua,Gao:2013bia,Carrazza:2015hva,Carrazza:2015aoa}
by means
of the Hessian Profiling
method~\cite{Paukkunen:2014zia} (see also~\cite{Schmidt:2018hvu}).
We choose the PDF4LHC15 set since it broadly represents
the state-of-the-art understanding of
the proton structure.

\begin{figure}[t]
\centering
\includegraphics[width=0.67\textwidth]{\main/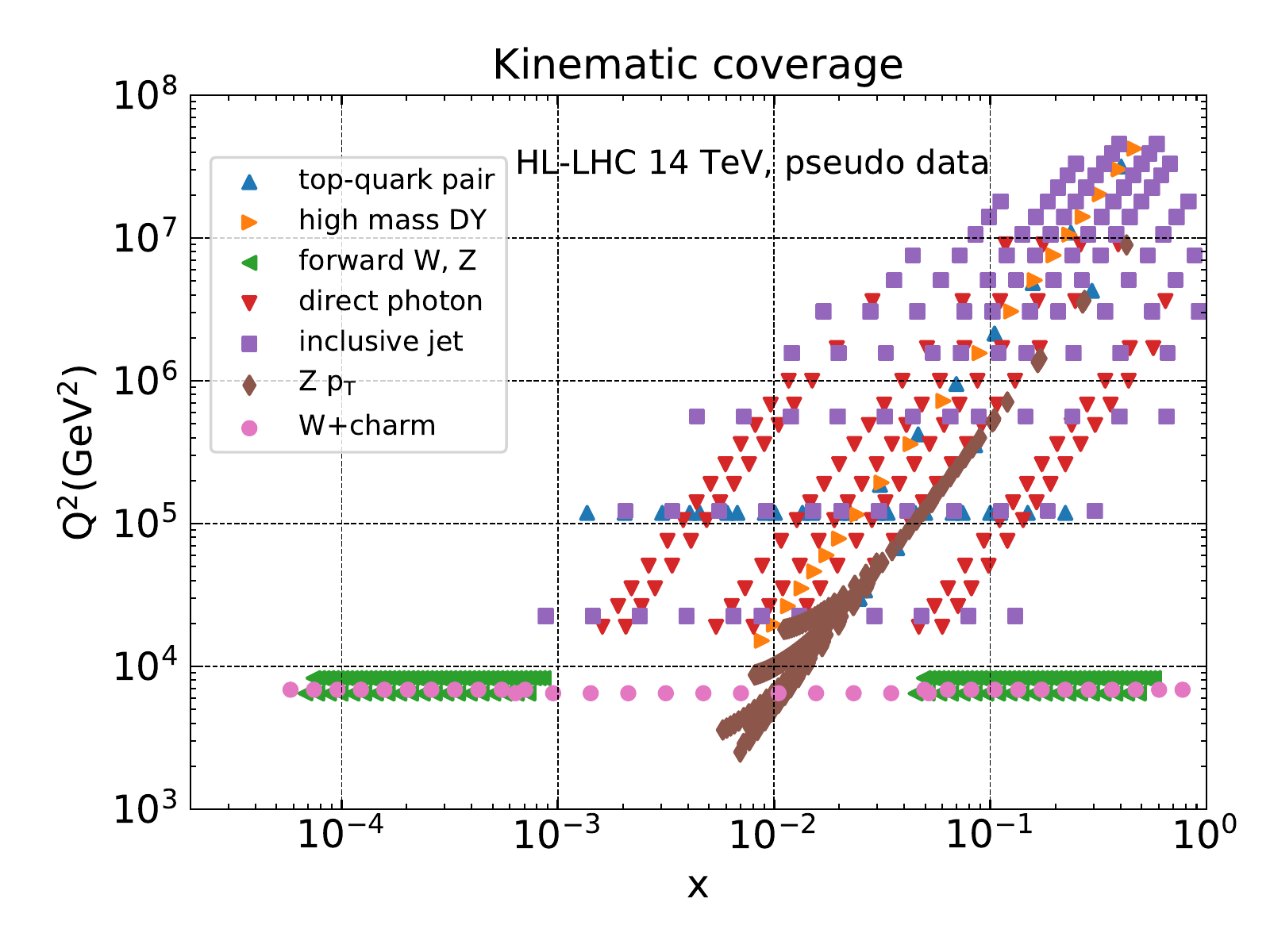}
\caption{\small 
  The kinematic coverage in the $(x,Q^2)$ plane of the
  HL--LHC pseudo-data.
 \label{fig:kinHLLHC}} 
\end{figure}

In Fig.~\ref{fig:kinHLLHC} we show the
kinematic coverage in the $(x,Q^2)$ plane of the
  HL--LHC pseudo-data included in this analysis.
  As indicated there, we have simulated pseudo-data for
  the following processes: top quark pair production,
  high-mass and forward Drell-Yan $W,Z$ production, direct
  photon and inclusive jet production, the transverse momentum
  of $Z$ bosons, and the production of $W$ bosons in association
  with charm quarks.
  The HL--LHC pseudo-data therefore spans a wide region in the kinematic
  plane, namely $6\times 10^{-5} < x < 0.7$ and
  $40~{\rm \UGeV} < Q < 7~{\rm \UTeV}$.
  In particular, one sees that the HL-LHC coverage of the large-$x$ region,
  where current PDF fits exhibit large uncertainties,
  is markedly improved as compared to available LHC
  measurements.
  
\begin{figure}[t]
  \begin{center}
\includegraphics[width=0.44\linewidth,angle=-90]{\main/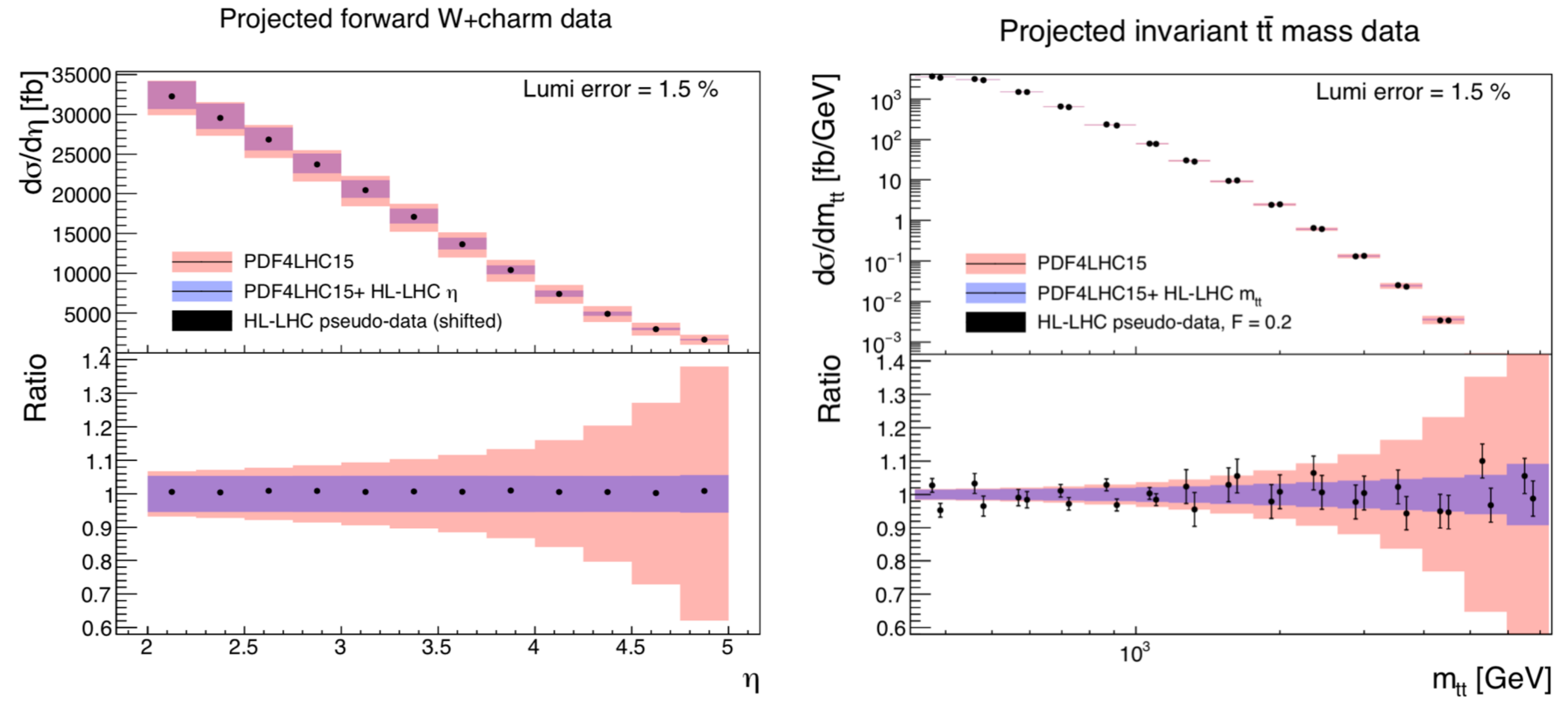}
\caption{\small Comparison between the HL--LHC pseudo-data
  and the theoretical predictions for forward $W+$charm production (left)
  and for the invariant mass $m_{t\bar{t}}$ distribution in top-quark
  pair production (right).
   The theory calculations are shown both before (PDF4LHC15)
   and after profiling.
       \label{fig:resultsPDFs.pdf} }
  \end{center}
\end{figure}

{\bf Results.}
As an illustration of the impact of individual sets
of HL-LHC pseudo-data,
in Fig.~\ref{fig:resultsPDFs.pdf} we show 
the comparison between the HL--LHC projected measurements
and the theoretical predictions for the lepton
rapidity distribution in forward $W+$charm production
  and for the invariant mass $m_{t\bar{t}}$ distribution in top-quark
  pair production. These two particular datasets probe the poorly-known strange quark and
  the gluon at large-$x$, respectively.
   The theory calculations are shown both before (PDF4LHC15)
   and after profiling.
  In the bottom panel, we show the same results normalised
  to the central value of the original theory calculation.
  In both cases we see that the expected precision of the HL-LHC
  measurements is rather higher than the current PDF uncertainties,
  and therefore we observe a marked improvement once they
  are included in PDF4LHC15 via the Hessian profiling.

In this study we have considered three
different scenarios for the
experimental systematic uncertainties of the HL--LHC pseudo-data.
    These scenarios, ranging from more conservative to more optimistic, differ among them in
    the reduction factor
    applied to the systematic errors of the reference
    8 \UTeV or 13 \UTeV measurements, see~\cite{Khalek:2018mdn}
    for more details.
    In particular, in the optimistic scenario we assume a reduction
    of the systematic errors by a factor 2.5 (5) as compared to the
    reference 8 \UTeV (13 \UTeV) measurements, while for
    the conservative scenario we assume no reduction in systematic
    errors with respect to the 8 \UTeV reference.
    Reassuringly, we obtain that the main results of our
    study depend only mildly in the specific assumption for
    the values of this reduction factor.

\begin{figure}[t]
  \begin{center}
    \includegraphics[width=0.49\linewidth]{\main/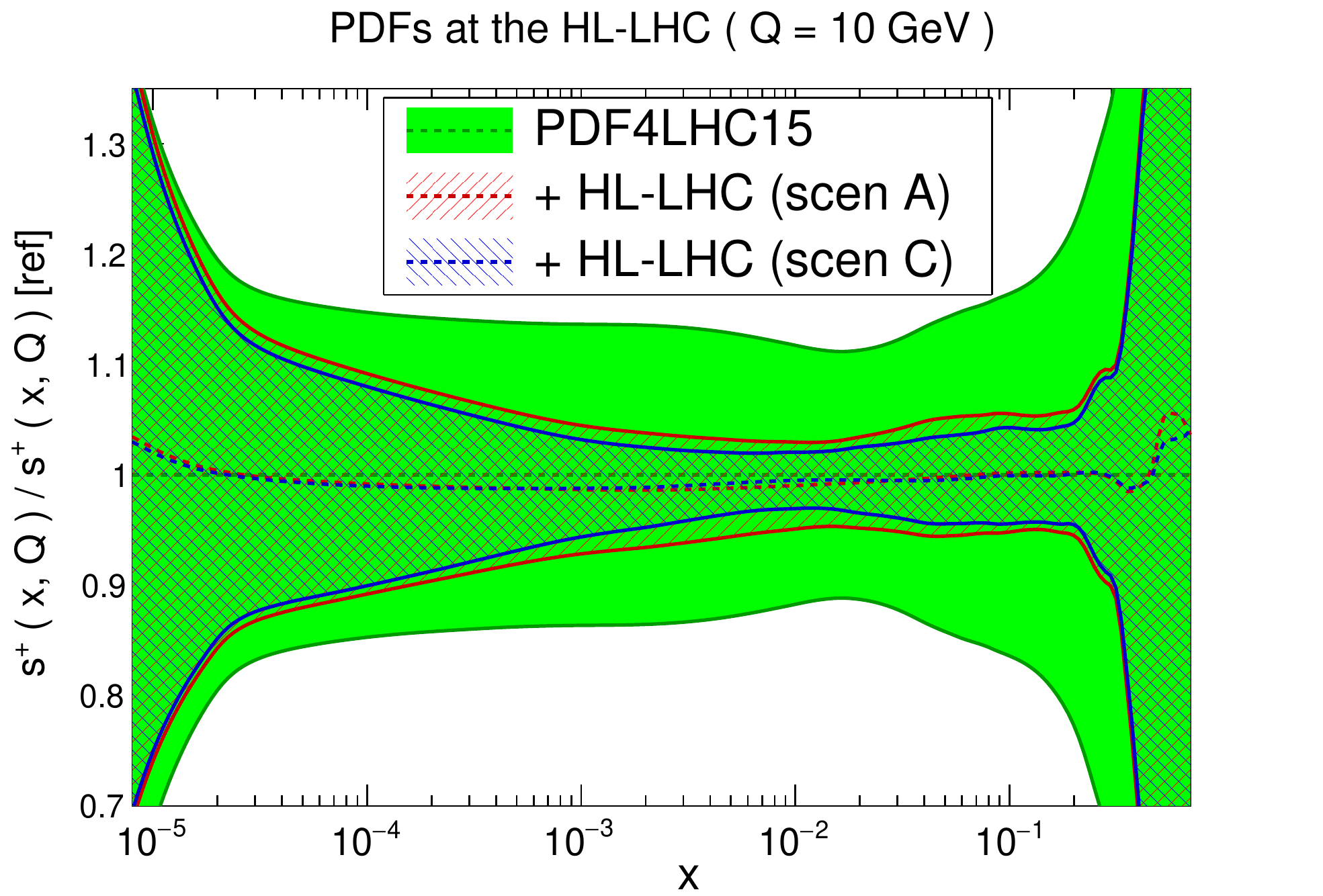}
\includegraphics[width=0.49\linewidth]{\main/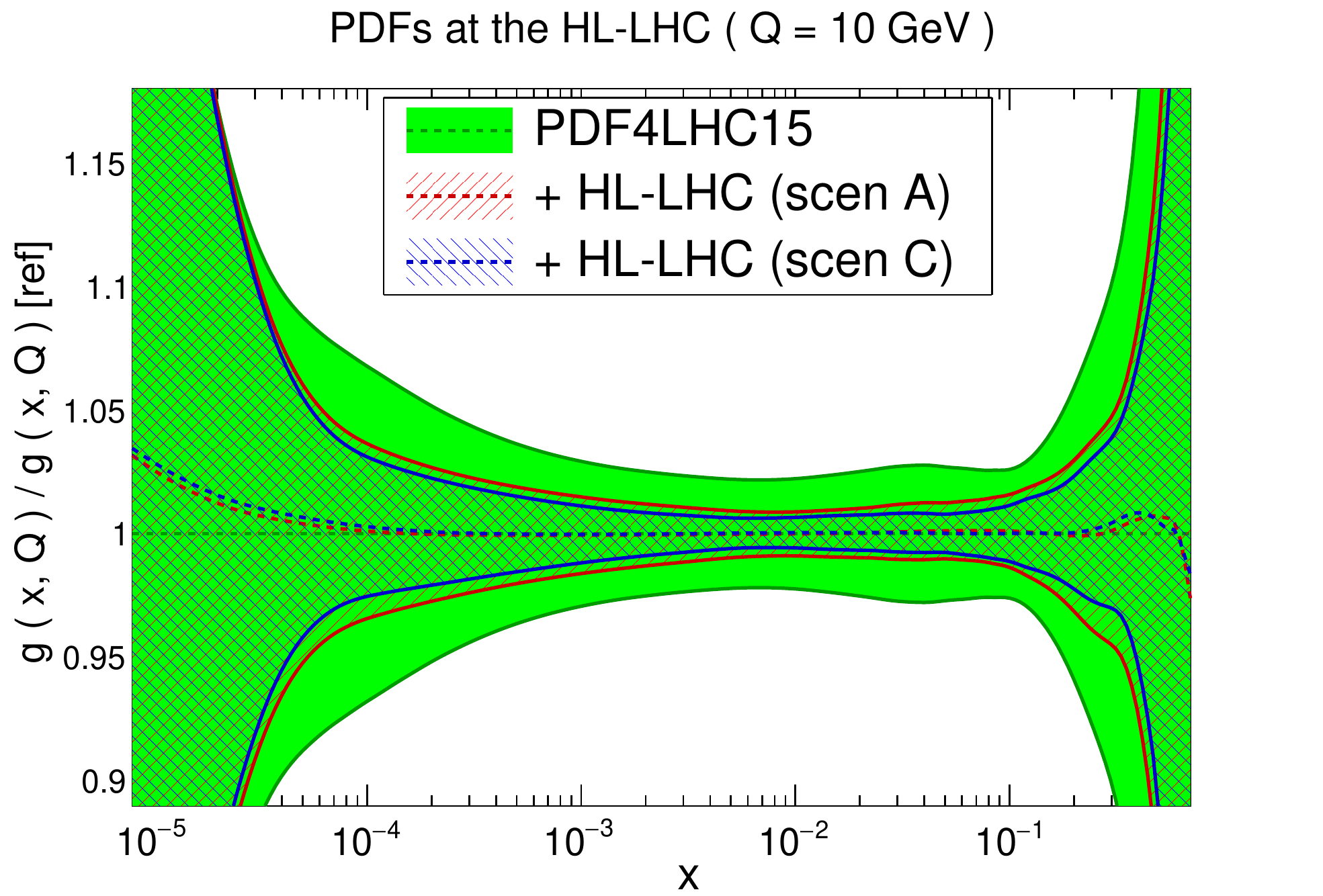}
\caption{\small Comparison of the PDF4LHC15 set with the profiled sets
  with HL--LHC pseudo-data.
  We show the strange (left) and gluon (right) PDFs
  normalised to the central value of
  the baseline.
     \label{fig:PDFratios} }
  \end{center}
\end{figure}

    In Fig.~\ref{fig:PDFratios}  we compare the PDF4LHC15 set
    with the strange quark and gluon PDFs obtained once the entire
    set of HL-LHC pseudo-data summarised in Fig.~\ref{fig:kinHLLHC}
    has been included via profiling.
    We show results both in the conservative (A) and optimistic (C) scenarios
    for the projections of the experimental systematic uncertainties.
    We observe that the
  impact of the HL--LHC pseudo-data is reasonably similar
  in both scenarios.
  This is due to the fact that we have chosen
  those processes which will benefit from a significant improvement in statistics, independent of the specific assumption about the systematic errors. These then tend to lie in kinematic regions where the PDFs themselves are generally less
  well determined.
  We also observe
  a marked reduction of the PDF uncertainties in all cases.
  In the case of the gluon PDF, there is an improvement of
  uncertainties in the complete relevant range of momentum
  fraction $x$.
  This is a direct consequence of the fact that
  we have included several HL--LHC processes
  that have direct sensitivity to the gluon
  content of the proton, including jet, direct photon, and top quark pair
  production, as well as the transverse momentum of $Z$ bosons.
  As we discuss next, this has direct implications for the phenomenology of Higgs
  boson production.
      
{\bf Implications for Higgs physics.}
In Table~\ref{fig:PDFs-HL--LHC-summaryTable} we indicate
the reduction of the PDF uncertainties
in comparison to the PDF4LHC15 baseline for
different initial partonic combinations (that is, a value
of 1 corresponds to no improvement).
  Results are presented for three different bins
  of the invariant mass $M_X$ of the produced system for
  the three initial states relevant for Higgs production:
  gluon-gluon (for $gg\to h$ and $t\bar{t}h$), quark-quark
  (for vector boson fusion) and quark-antiquark
  (for associated $Wh$ and $Zh$ production).
  The values shown outside (inside) the brackets correspond to the
  optimistic (conservative) scenario.
  We can see that for the $M_X$ region relevant for the SM Higgs
  boson production, as well as for related BSM Higgs-like scalars, namely
  $40~{\rm \UGeV}\le M_X\le 1~{\rm \UTeV}$, the HL-LHC pseudo-data leads
  to a reduction by almost a factor four in the optimistic scenario
  in the $gg$ channel, and around a factor three in the $q\bar{q}$
  and $qq$ channels.
  This implies that precision calculations of Higgs production
  at the HL-LHC should be possible with significantly reduced PDF
  uncertainties compared to current state-of-the-art predictions.
  
\begin{table}[t]
  \begin{center}
  \small
   \renewcommand{\arraystretch}{1.70}
\begin{tabular}{c|c|c|c}
\toprule
Ratio to baseline  & $10~{\rm \UGeV}\le M_X\le 40~{\rm \UGeV}$   &
 $40~{\rm \UGeV}\le M_X\le 1~{\rm \UTeV}$&
  $1~{\rm \UTeV}\le M_X\le 6~{\rm \UTeV}$\\
\midrule
gluon-gluon & 0.50 (0.60)  & 0.28 (0.40)  & 0.22 (0.34)     \\
quark-quark &  0.74 (0.79) & 0.37 (0.46) &  0.43 (0.59)     \\
quark-antiquark & 0.71 (0.76) & 0.31 (0.40)  &  0.50 (0.60)     \\
\bottomrule
\end{tabular}
\vspace{+0.4cm}
\caption{\small The reduction of the PDF uncertainties
as compared to the PDF4LHC15 baseline for
different initial partonic combinations in the 
  optimistic (conservative) scenario.
     \label{fig:PDFs-HL--LHC-summaryTable}
 }
  \end{center}
\end{table}

\begin{figure}[t]
  \begin{center}
    \includegraphics[width=0.49\linewidth]{\main/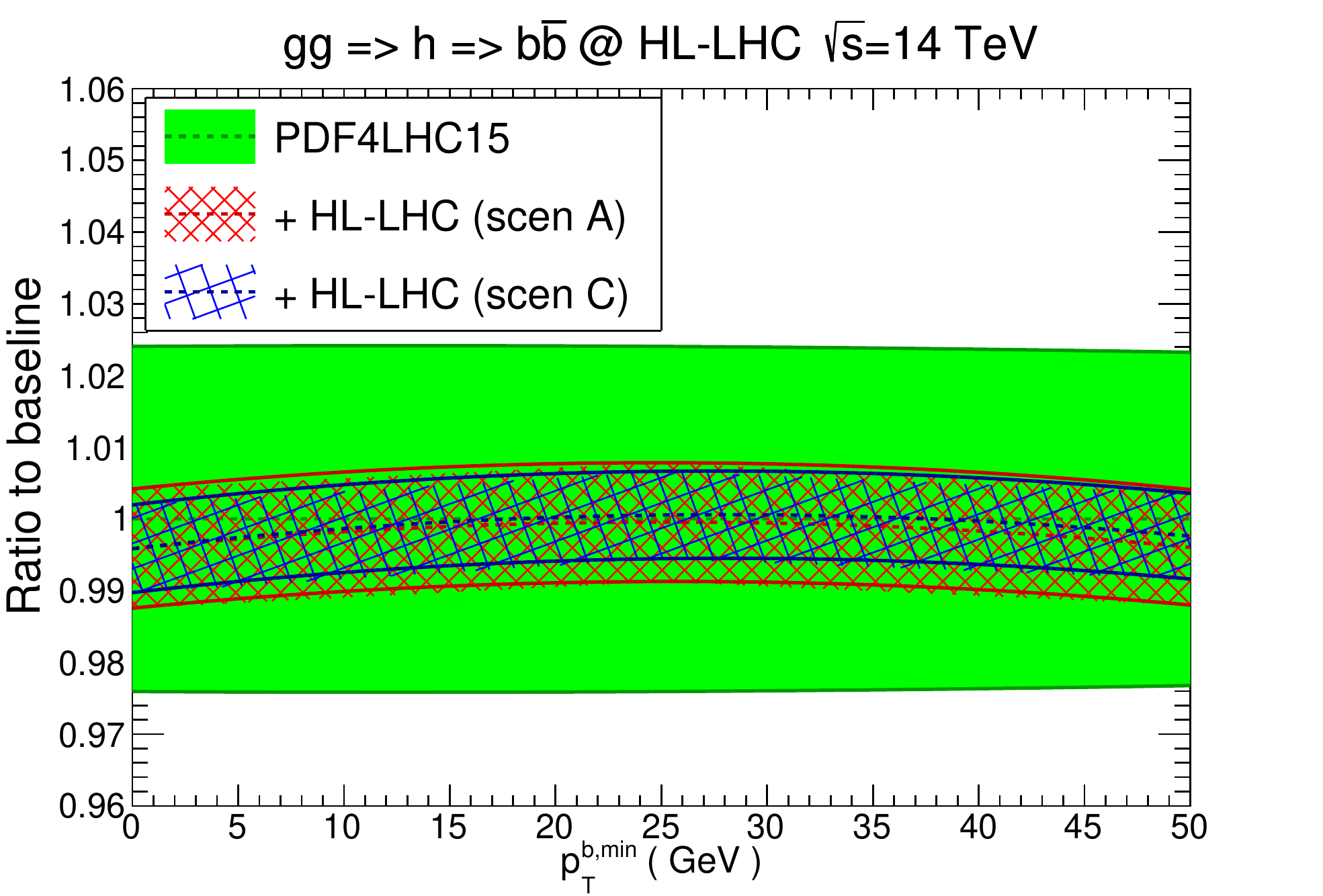}
    \includegraphics[width=0.49\linewidth]{\main/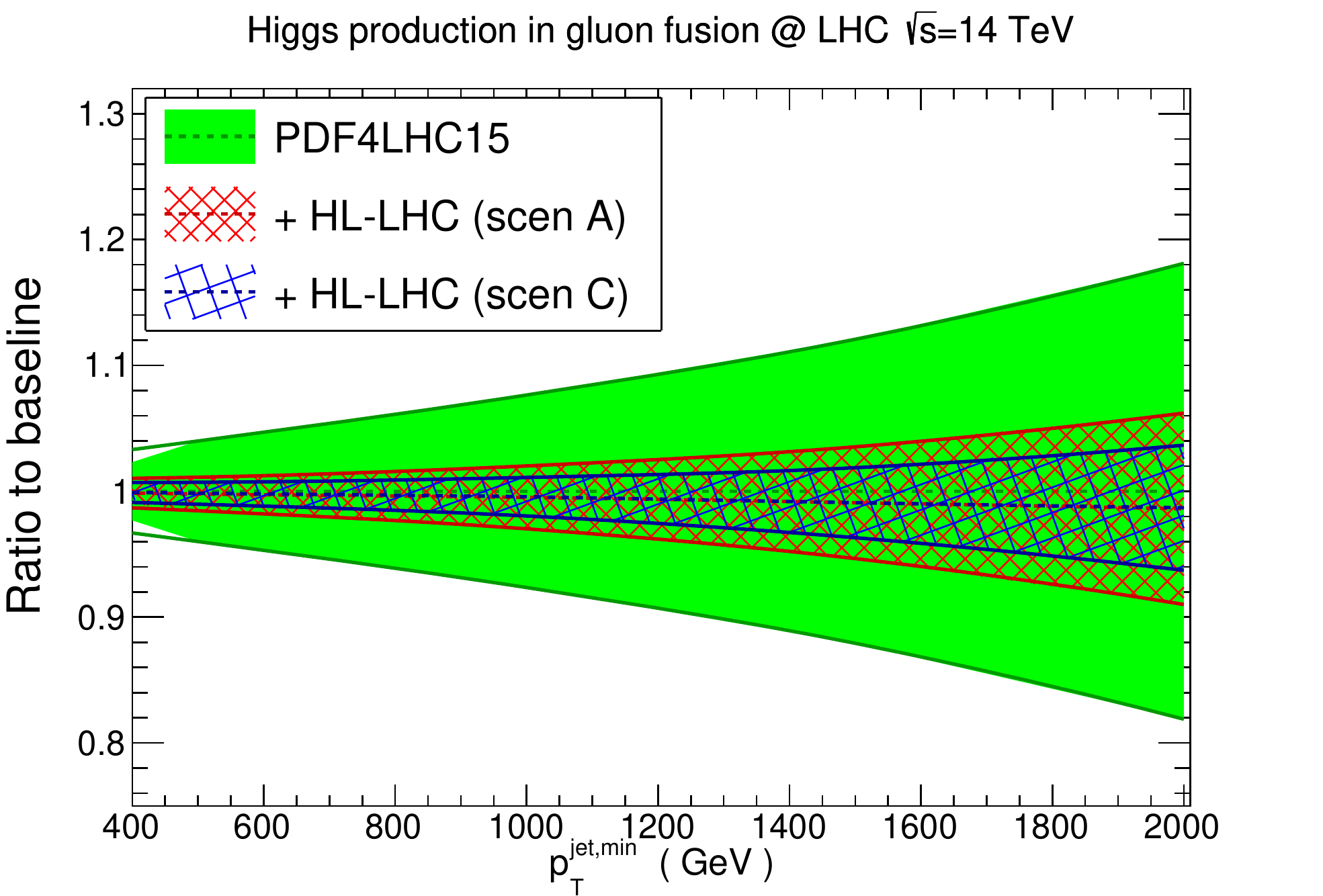}
    \caption{\small Comparison of the predictions for 
    SM Higgs production cross-sections at $\sqrt{s}=14$ \UTeV between the PDF4LHC15
baseline and the profiled PDF sets with HL--LHC pseudo--data.
     \label{fig:MCFMxsects} }
  \end{center}
\end{figure}

To illustrate this improvement, in Fig.~\ref{fig:MCFMxsects}
we present the comparison of the predictions for 
    SM Higgs production at $\sqrt{s}=14$ \UTeV between the PDF4LHC15
    baseline and the profiled PDF sets.
    Specifically, we show
     Higgs boson production in gluon fusion with heavy top
      quark effective theory, both inclusive
      and decaying into $b\bar{b}$ as a function of $p_T^{b,\rm min}$ (left), and
      then in association with a hard jet as a function of its transverse
      momentum $p_{\rm T}^{\rm jet, min}$ (right).
      The calculations have been performed using {\tt MCFM8.2} with leading-order
      matrix elements.
      The marked reduction of PDF uncertainties is consistent with
      the values reported in Table~\ref{fig:PDFs-HL--LHC-summaryTable}.

      Finally, there are two caveats to be added concerning this study.
      First we have only considered a subset of all possible measurements of relevance for PDF fits at HL--LHC.
      Second, possible data incompatibility has not been
      accounted for fully. These may strengthen and weaken, respectively, the constraining powers of future LHC data on PDFs.

The results of this study are made
 publicly available in the {\tt LHAPDF6} format~\cite{Buckley:2014ana},
 for the three
 scenarios that have been considered, and can be
 downloaded from:
 \begin{center}
{\tt  https://data.nnpdf.science/HLLHC\_YR/PDF4LHC15\_nnlo\_hllhc\_scen1.tgz}\\
{\tt https://data.nnpdf.science/HLLHC\_YR/PDF4LHC15\_nnlo\_hllhc\_scen2.tgz}\\
{\tt https://data.nnpdf.science/HLLHC\_YR/PDF4LHC15\_nnlo\_hllhc\_scen3.tgz}
\end{center}


\asubsection[M. Delmastro, N. De Filippis, P. Francavilla, A. Gilbert, S. Jezequel, P. Milenovic, M. Testa]{Overview of experimental analysis for the Higgs boson measurement channels}
\label{sec2:expan}
\label{sec2:channels}
\asubsubsection{Extrapolation assumptions}
\label{sec:HiggsExtrapAss}

The results presented in this Section are based on the extrapolation to an expected integrated luminosity of 3000~fb$^{-1}$ 
of the corresponding ATLAS and CMS Run-2 results. For some of the Higgs decay final states (ATLAS: $WW^*$, $Z\gamma$, $t\bar{t}H$, $\tau\tau$; CMS: $WW^*$, $Z\gamma$, $\gamma\gamma$, $ZZ^*$, $t\bar{t}H$, $\tau\tau$, $b\bar{b}$ and $\mu\mu$) the extrapolation is performed from results obtained with the 2015-2016 36\,$\ifb$ datasets; the remaining final state analyses (ATLAS: $\gamma\gamma$, $ZZ^*$, $b\bar{b}$ and $\mu\mu$) use the results based on the 2015+2016+2017 80\,$\ifb$ data samples. The starting points of the extrapolated results are measurements based on datasets of size $\mathcal{O}(1\%)$ of the expected HL-LHC integrated luminosity. The extrapolations are in this regard very limited with respect to the potential reach of the real HL-LHC analyses, which large statistics will allow to probe corners of the phase space inaccessible at the LHC Run-2.
    
In addition to the increase in integrated luminosity, the ATLAS extrapolations also account for the increase of signal and background cross-sections from $\sqrt{s}$ = 13 \UTeV to 14 \UTeV.  In those cases, the signal yields have been scaled according to the Higgs boson production cross sections values at 13 and 14 \UTeV, as reported in Ref.~\cite{deFlorian:2016spz}. Similarly, the background yields have been scaled according to the parton luminosity ratio between 13 and 14 \UTeV, as reported in Ref.~\cite{Heinemeyer:2013tqa}, by taking into account whether the background process is predominantly quark pair or gluon pair initiated.

Object reconstruction efficiencies, resolutions and fake rates are assumed to be similar in the Run-2 and HL-LHC environments, based on the assumption that the planned upgrades of the ATLAS and CMS detectors will compensate for the effects of the increase of instantaneous luminosity and higher pile-up environment at HL-LHC.
For the systematic uncertainties which include experimental, signal and background components, two scenarios have been considered.
The first scenario (S1) assumes the same values as those used in the published Run-2 analyses.
The second scenario (S2) implements a reduction of the systematic uncertainties according to the improvements expected to be reached at the end of HL-LHC program in twenty years from now: the correction factors follow the recommendations from Ref.~\cite{HLHELHCCommonSystematics}.
In certain analyses some of the systematic uncertainties are treated in a specific way, and this is discussed explicitly in each corresponding section.
In all analyses, the theory uncertainties for signal and background are generally halved, except where more precise extrapolated values have been provided. Details on the projections of theoretical uncertainties are given in Section~\ref{sec:hl-lhc}. The reduction of the theory uncertainties in gluon-fusion Higgs production is for instance associated to a better understanding of the correlation of their components, leading to their sum in quadrature in scenario S2, instead of the linear sum used in S1 (see Section~\ref{sec:hl-lhc-ggF} for details). The uncertainties related to missing higher orders in theory calculations are in particular discussed in Section~\ref{sec2:PDFuncertainties}: these uncertainties are halved in all analyses extrapolation in scenario S2, even though larger improvements are expected in some cases (e.g. gluon-fusion Higgs production).
The uncertainty on the luminosity is set to 1\%.
The uncertainty related to Monte Carlo samples statistics is assumed to be negligible.

The extrapolated results are generally limited by systematic uncertainties. It is worth noting that, despite all efforts to design proper projections, the values of the systematic uncertainties of the Run-2 analyses cannot fully account neither for the HL-LHC data-taking conditions, nor for the level of understanding of the various sources of systematic uncertainties that will be achieved by fully exploiting the large HL-LHC statistic. The systematic models in current Run-2 analyses are in fact designed for the needs of Run-2, and hence lack flexibility and details needed to account for full-fledged HL-LHC analyses. In this sense, these extrapolated uncertainties are to be considered an approximation: future analyses will exploit and gain sensitivity from phase space regions that are not accessible yet, or use analysis techniques that reduce the impact of systematic uncertainties. In many cases one can as well expect that several uncertainties will be be highly constrained with very large luminosity, and therefor updated uncertainty models with greater flexibility will be needed to properly fit the data.

In the following, all analyses segment the selected events according to the objects produced in association with the Higgs boson decay products and their topology, in order to maximise the sensitivity to the main Higgs production modes ($\ggF+b\bar{b}H$, VBF, $VH$ = $qqZH+ggZH+WH$ and top = $t\bar{t}H+tH$)  and to reduce the uncertainties on the respective cross sections. Details on how this segmentation is performed, and on the event selection and categorisation in the various analyses, are found in the Run-2 analysis references quoted in each section.

\asubsubsection[S. Falke]{$H \to \gamma\gamma$}
\label{sec:Hgammagamma}

The measurement of the Higgs boson properties in the \Hyy\ channel is performed using the events that contain two isolated photon candidates passing good quality requirements in the precision regions of the detectors. Events are further segmented according to the objects accompanying the di-photon system, in order to maximise the sensitivity to the main Higgs production modes and to reduce the uncertainties on the respective cross sections, as well as to the Simplified Template Cross Section (STXS, first introduced in Refs. \cite{deFlorian:2016spz,Badger:2016bpw}) in the merged version of Stage-1.
The Higgs production cross sections are measured for a Higgs boson absolute rapidity $|y_H|$ smaller than 2.5, and with further requirements on the objects accompanying the di-photon system (e.g. jet $p_\mathrm{T}$).
The \Hyy\ signal is extracted by means of a combined signal-plus-background fit of the di-photon invariant mass spectra in the various event categories, where both the continuous background and the signal resonance are parametrised by analytic functions. The shape properties of the signal PDF are obtained by Monte Carlo (MC) simulation, and constrained by performance studies of the photon energy scale and resolution. The background PDF is completely determined by the fit on data, with systematic uncertainties attributed to the specific choice of functional form following the procedure described in Ref. \cite{Aad:2012tfa} or using the discrete profiling method \cite{Dauncey:2014xga}. More details on the analyses methods can be found in most recent measurements in the \Hyy\ channel published by ATLAS \cite{ATLAS:2018uso} and CMS \cite{Sirunyan:2018ouh}. 

The performance of the measurement of the Higgs boson properties in the \Hyy\ channel at HL-LHC is extrapolated from the most recent measurements by ATLAS with 80\,$\mathrm{fb}^{-1}$ \cite{ATLAS:2018uso} and by CMS with 36\,$\mathrm{fb}^{-1}$ \cite{Sirunyan:2018ouh}. The main systematic uncertainties affecting the results are the background modelling uncertainty,
missing higher order uncertainties
causing event migrations between the bins, photon isolation efficiencies and jet uncertainties.
%
On top of the common assumptions mentioned in Section~\ref{sec:HiggsExtrapAss}, the 
results of the studies performed by ATLAS include a 10\% increase of the background modelling systematic uncertainties, to account for the potentially worst knowledge of the background composition in each analysis category at HL-LHC: this assumption has anyway negligible impact.
In the Run-2 analyses, a conservative 100\% uncertainty on the heavy flavour resonant background in top-sensitive categories is applied. Measurements by ATLAS and CMS of the heavy flavour content, or the $b$-jet multiplicity, are expected to better constrain these contributions: for the S2 scenario extrapolation, this uncertainty is therefore halved.

Figure~\ref{fig:Hyy_ATLAS_HLLHC_S2} shows the ratio of the extrapolated \Hyy\ ATLAS measurements of the cross sections times branching fraction of the main Higgs production modes to their respective theoretical SM predictions (left), and uncertainties on these measurements for S1, S2, and stat-only scenarios as extrapolated using the \Hyy\ CMS measurements (right). CMS extrapolation is obtained from the simultaneous fit in all production and decay modes, as described in Section \ref{sec:expcomb_prodtimesdecay}. The reduction of the total uncertainty with respect to the 80\,$\mathrm{fb}^{-1}$ results ranges from a factor of about 2 (3) for the S1 (S2) scenario for the $ggH+b\bar{b}H$, VBF, top cross sections, to a factor of about 5(6) for the $VH$ cross section, that remains dominated by the statistical uncertainty.

\begin{figure}
  \centering
  \includegraphics[width=0.56\linewidth]{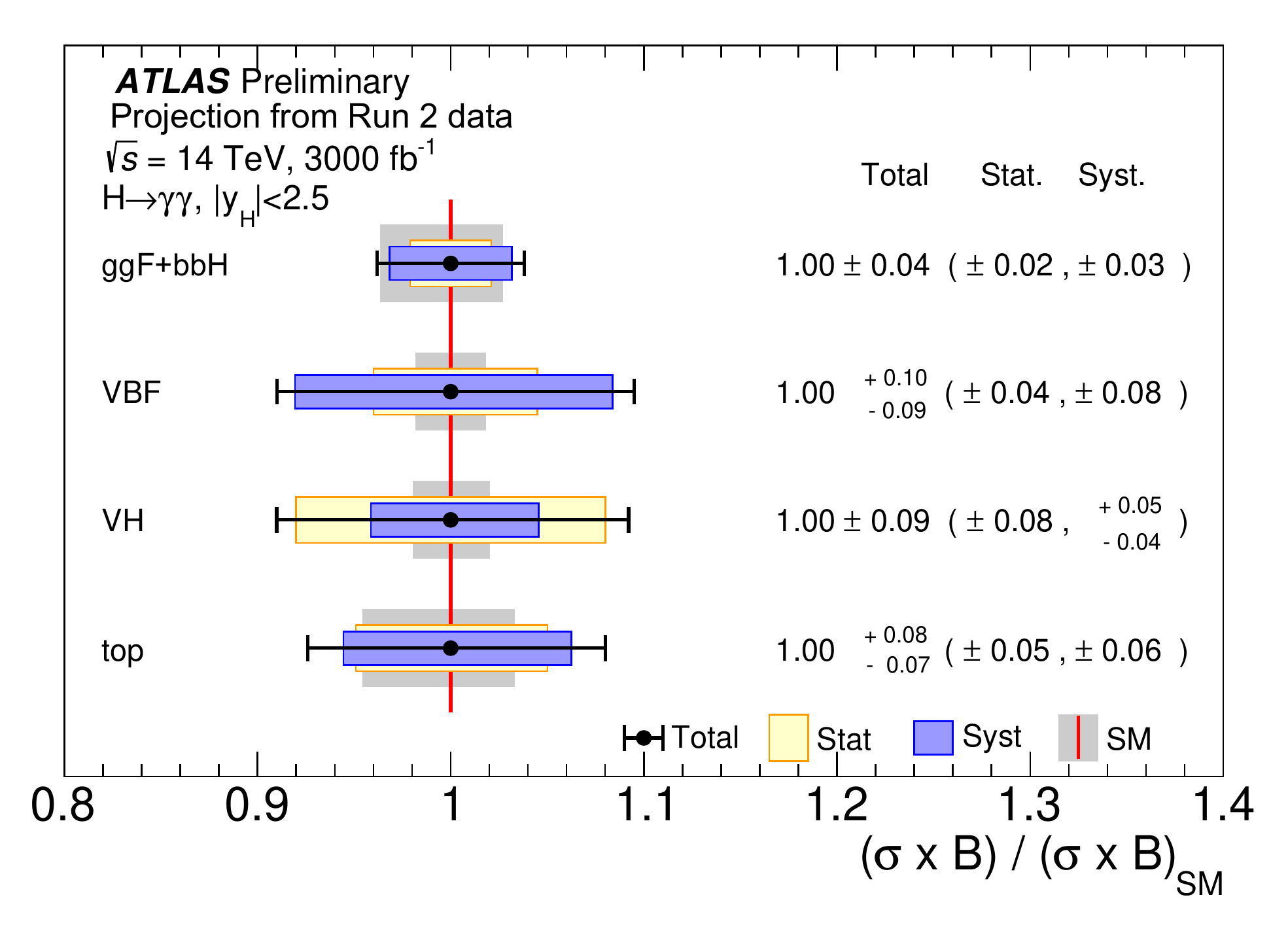}
  \includegraphics[width=0.42\linewidth]{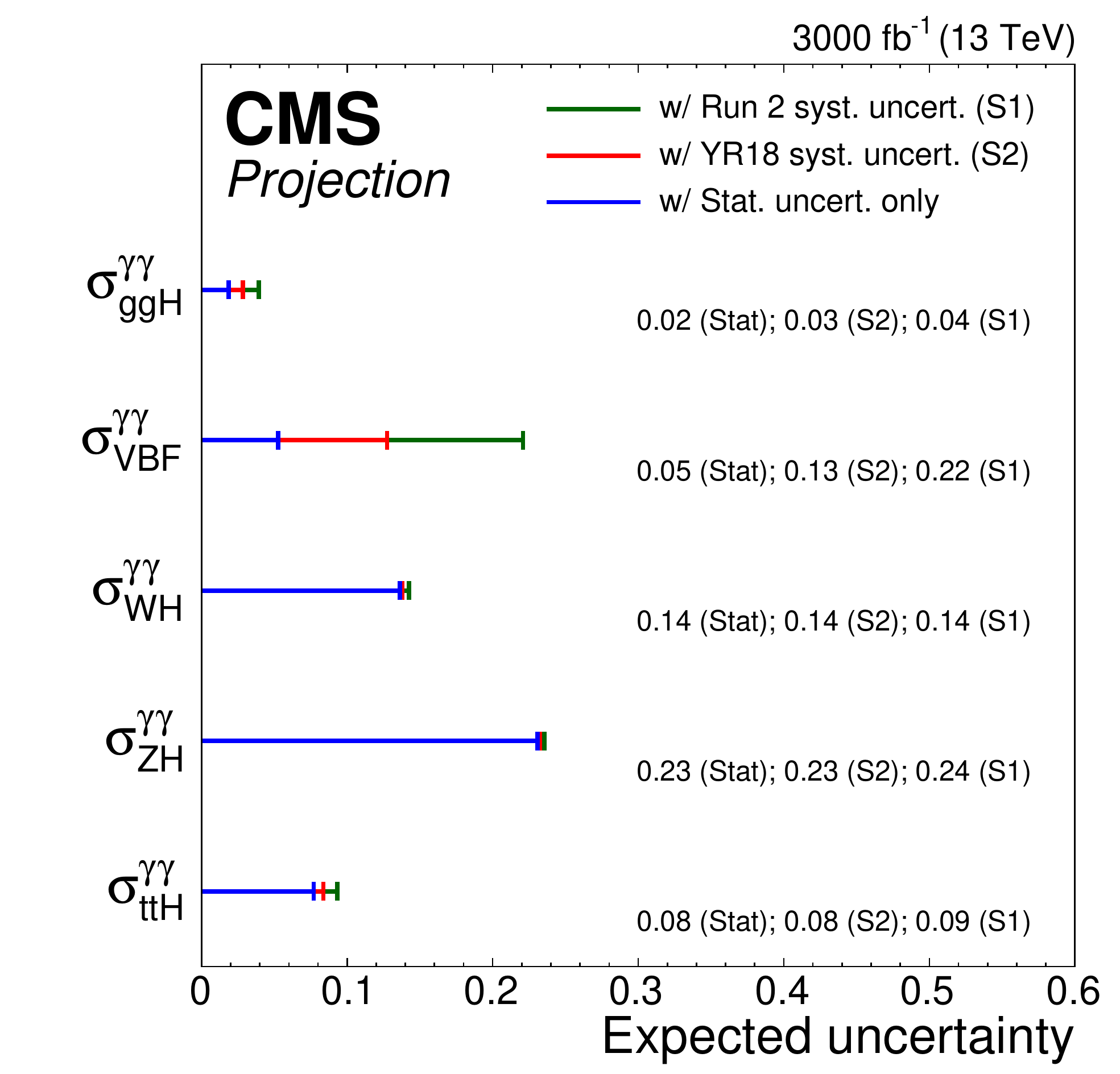}
  \caption{Cross-section times branching fraction measurements of the main Higgs production modes in the \Hyy\ decay channel, as extrapolated at the HL-LHC. In case of ATLAS results (left) the ratios of cross sections to their respective theoretical SM predictions are shown for scenario S2, while in case of CMS results (right) the uncertainties on these measurements are shown for S1, S2, and Stat-only scenarios..}
  \label{fig:Hyy_ATLAS_HLLHC_S2}
\end{figure}

\asubsubsection[ Y. Huang]{$H \to Z\gamma \to \ell\ell\,\gamma$}

Due to the small branching fraction in the SM, the \HZy\ decay has not yet been observed at the LHC. The experimental observed limits at the $95\%$ confidence level are currently 6.6 times the SM prediction for a Higgs boson mass of 125.09 \UGeV by ATLAS and 3.9 times the SM prediction for a Higgs boson mass of 125 \UGeV by CMS, based on the analyses of 36\,$\mathrm{fb}^{-1}$ of $pp$ collision at $\sqrt{s} = 13$ \UTeV described in Ref. \cite{Aaboud:2017uhw, Sirunyan:2018tbk}.

The analyses select events with an isolated photon candidate passing good quality requirements in the precision regions of the detectors, and a di-lepton system with properties compatible with that of the decay of a $Z$ boson. Events are separated according to lepton flavour, the event kinematic properties, and the presence of jets compatible with the VBF production of the Higgs boson, in order to maximise the signal sensitivity. The signal is sought for by means of a combined signal-plus-background fit of the photon-di-lepton invariant mass spectra in various event categories, where both the continuous background and the signal resonance are parametrised by analytic functions. The Run-2 analyses are strongly driven by statistical uncertainty, and the main systematic uncertainties are from the bias associated to the background modelling, based on the MC simulation of some background processes, and on low-statistics data control regions for others.

The extrapolations to HL-LHC are performed with a simple scaling approach, assuming the same signal and background modelling used in the Run-2 analyses. All experimental and systematic uncertainties are considered to remain the same (S1), except the uncertainty associated to the background modelling, which is taken to be negligible. The latter assumption is based on the idea that, thanks to the large HL-LHC statistics and the use of modern functional modelling techniques, the background shape could be constrained exclusively using data with great accuracy, thus dramatically reducing the modelling uncertainty.

The ATLAS expected significance to the SM Higgs boson decaying in $Z\gamma$ is 4.9 $\sigma$ with 3000\,$\mathrm{fb}^{-1}$. Assuming the SM Higgs production cross section and decay branching ratios, the signal strength is expected to be measured with a $\pm0.24$ uncertainty. The cross section times branching ratio for the $pp\rightarrow H \rightarrow Z\gamma$ process is projected to be measured as $1.00\pm0.23$ times the SM prediction. Even at the HL-LHC scenario S1, the analysis sensitivity  to \HZy\ will remain driven by the statistical uncertainty. The dominant source of systematic uncertainty in the extrapolation is that associated to the 
missing higher order uncertainties \cite{ATL-PHYS-PUB-2018-054}.

\asubsubsection[A. Gabrielli, A. Schaffer, V. Walbrecht]{$H \to ZZ^* \to 4\ell$}

The measurement of the Higgs boson properties in the \HZZ\ channel is performed using the events that contain at least two same-flavour opposite-sign di-lepton pairs, chosen from isolated electrons and muons candidates passing good quality requirements in the precision regions of the detectors. Additional constraints on the kinematic properties of the lepton pair associated with the decay of the on-shell $Z$ boson, and on the global topology of the event, helps to improve the signal to background ratio. The four-lepton invariant mass resolution is improved by correcting for the emission of final-state radiation photons by the leptons.
The \HZZ\ signal is extracted from the four-lepton invariant mass spectra in the different event categories, after having evaluated the background components using simulations to constrain their shapes, and data control regions to extrapolate their normalisation in the signal regions. Signal to background sensitivity is in general enhanced using the multivariate and/or matrix-element based techniques. More details on the analyses methods can be found in most recent measurements in the \HZZ\ channel published by ATLAS \cite{ATLAS:2018bsg} and CMS \cite{Sirunyan:2017exp}.

The performance of the measurement of the Higgs boson properties in the \HZZ\ at HL-LHC is extrapolated from the most recent measurements by ATLAS with 80\,$\mathrm{fb}^{-1}$ \cite{ATLAS:2018bsg}, and by CMS with 36\,$\mathrm{fb}^{-1}$ \cite{Sirunyan:2017exp}.
CMS extrapolation is obtained from the simultaneous fit in all production and decay modes, as described in Section \ref{sec:expcomb_prodtimesdecay}.
The dominant systematic uncertainties affecting the extrapolation of the ggH cross section measurement are the lepton reconstruction and identification efficiencies, and pile-up modelling uncertainties. The VBF and VH cross-sections are primarily affected by the uncertainty on the jet energy scale and resolution, and by the
missing higher order uncertainties.
These and
the parton shower modelling primarily affects the extrapolated top cross section.

The VBF, VH and especially top measurements in the \HZZ\ decay channel remain largely dominated by statistical uncertainty when extrapolated to 3000\,$\mathrm{fb}^{-1}$ while the $ggH+b\bar{b}H$ cross section is dominated by systematic uncertainties both in scenario S1 and S2.
Figure~\ref{fig:HZZ_ATLAS_HLLHC_S2} shows the ratio of the extrapolated \HZZ\ ATLAS measurements of the main Higgs boson production modes to their respective theoretical SM predictions in the scenario S2 (left), and uncertainties on these measurements for S1, S2, and stat-only scenarios as extrapolated using the \HZZ\ CMS measurements (right). The ggF and top \HZZ\ measurements at HL-LHC are expected to reach a level of precision comparable to the projected uncertainty on the corresponding theory predictions.

\begin{figure}
  \centering
  \includegraphics[width=0.56\linewidth]{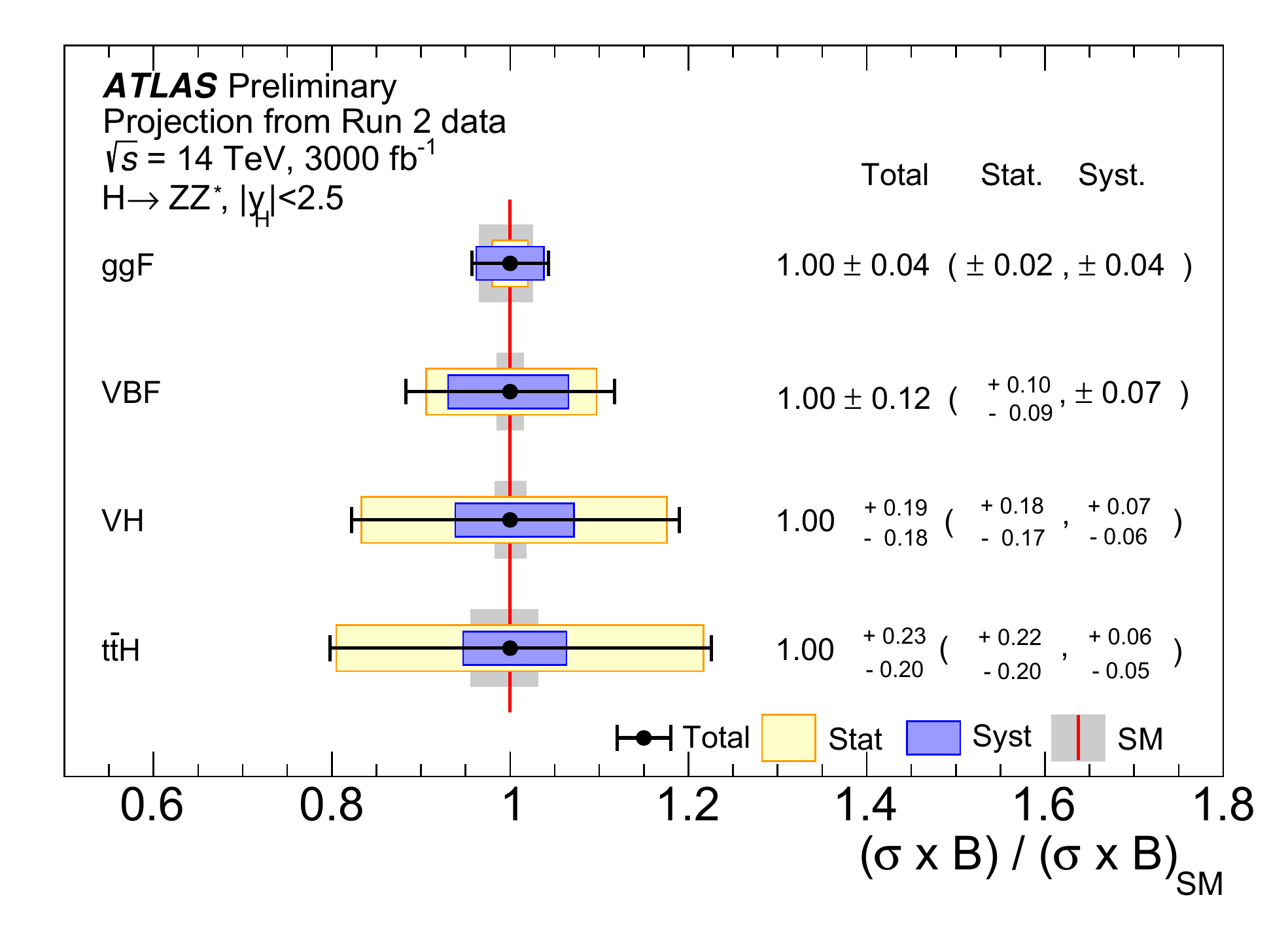}
  \includegraphics[width=0.42\linewidth]{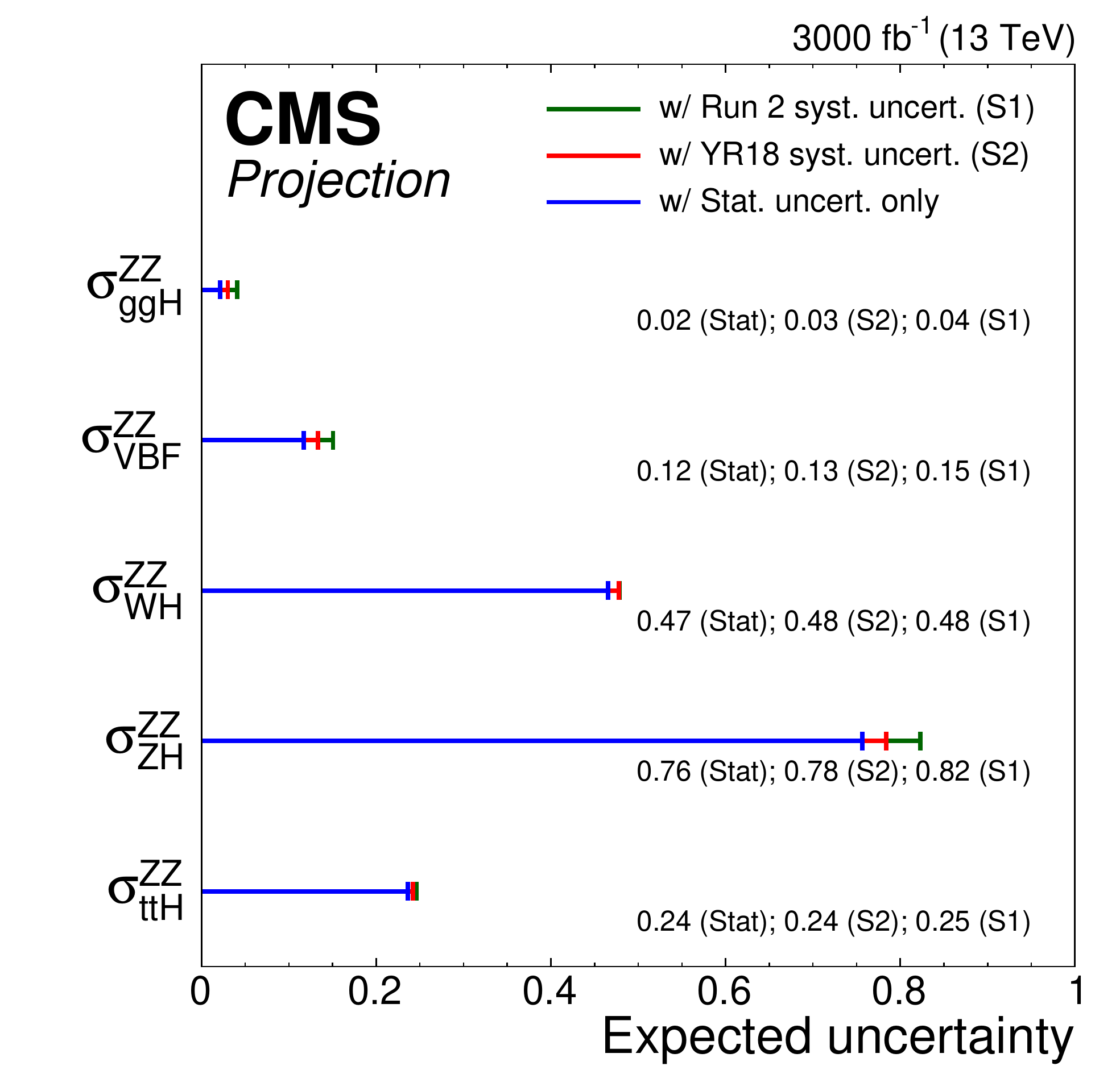}
  \caption{Cross-section times branching fraction measurements of the main Higgs boson production modes in the \HZZ\ decay channel, as extrapolated at the HL-LHC. In case of ATLAS results (left) the ratios of cross sections to their respective theoretical SM predictions are shown for scenario S2, while in case of CMS results (right) the uncertainties on these measurements are shown for S1, S2, and Stat-only scenarios.}
  \label{fig:HZZ_ATLAS_HLLHC_S2}
\end{figure}

\asubsubsection[R. Gugel, K. Koeneke]{$H \to WW^* \to \ell\nu\,\ell\nu$}

The measurement of the Higgs boson properties in the \HWW\ channel is performed using the events that contain two opposite-charged isolated leptons passing good quality requirements in the precision region of the detectors and missing transverse momentum. Additional requirements on the event kinematic properties are applied to reduce the various background components (e.g. requirements on the di-lepton invariant mass, transverse mass of the di-lepton + missing-transverse-energy (MET) system). Events are categorised as a function of the jet multiplicity in order to exploit the different background composition in different categories, and to help extracting the Higgs ggH and VBF production cross sections. The normalisations of the top ($t\bar{t}$ and $W+t$), and $Z\rightarrow\tau\tau$ backgrounds are set using dedicated control regions of the same jet multiplicity as the signal category to which the normalisation is transferred. In case of the (non-resonant) $WW$ background, its normalisation is either determined using dedicated control regions (ATLAS approach) or by using theoretical prediction with corresponding uncertainty on it (CMS approach). More details on the analyses methods can be found in most recent measurements in the \HWW\ channel published by ATLAS \cite{Aaboud:2018jqu} and CMS \cite{Sirunyan:2018egh}.

The performance of the measurements of Higgs boson properties in the \HWW\ channel at HL-LHC is extrapolated from the most recent measurements in this channel performed by ATLAS with 80\,$\mathrm{fb}^{-1}$ \cite{Aaboud:2018jqu} and by CMS with 36\,$\mathrm{fb}^{-1}$ \cite{Sirunyan:2018egh}. These measurements are completely dominated by systematic uncertainties, and their extrapolation to the S2 scenario shows the expected reduction by a factor two. The measurement of the ggH cross section by branching fraction is dominated by theoretical PDF uncertainty, followed by experimental uncertainties affecting the signal acceptance, including uncertainties on the jet energy scale and flavour composition, and lepton mis-identification; the VBF result suffers from similar dominant uncertainties.
Figure~\ref{fig:HWW_ATLAS_HLLHC_S2} shows the ratio of the extrapolated \HWW\ ATLAS measurements of the main Higgs production modes to their respective theoretical SM predictions in scenario S2 (left), and uncertainties on these measurements for S1, S2, and Stat-only scenarios as extrapolated using the \HWW\ CMS measurements (right). CMS extrapolation is obtained from the simultaneous fit in all production and decay modes, as described in Section \ref{sec:expcomb_prodtimesdecay}.

\begin{figure}
  \centering
  \includegraphics[width=0.56\linewidth]{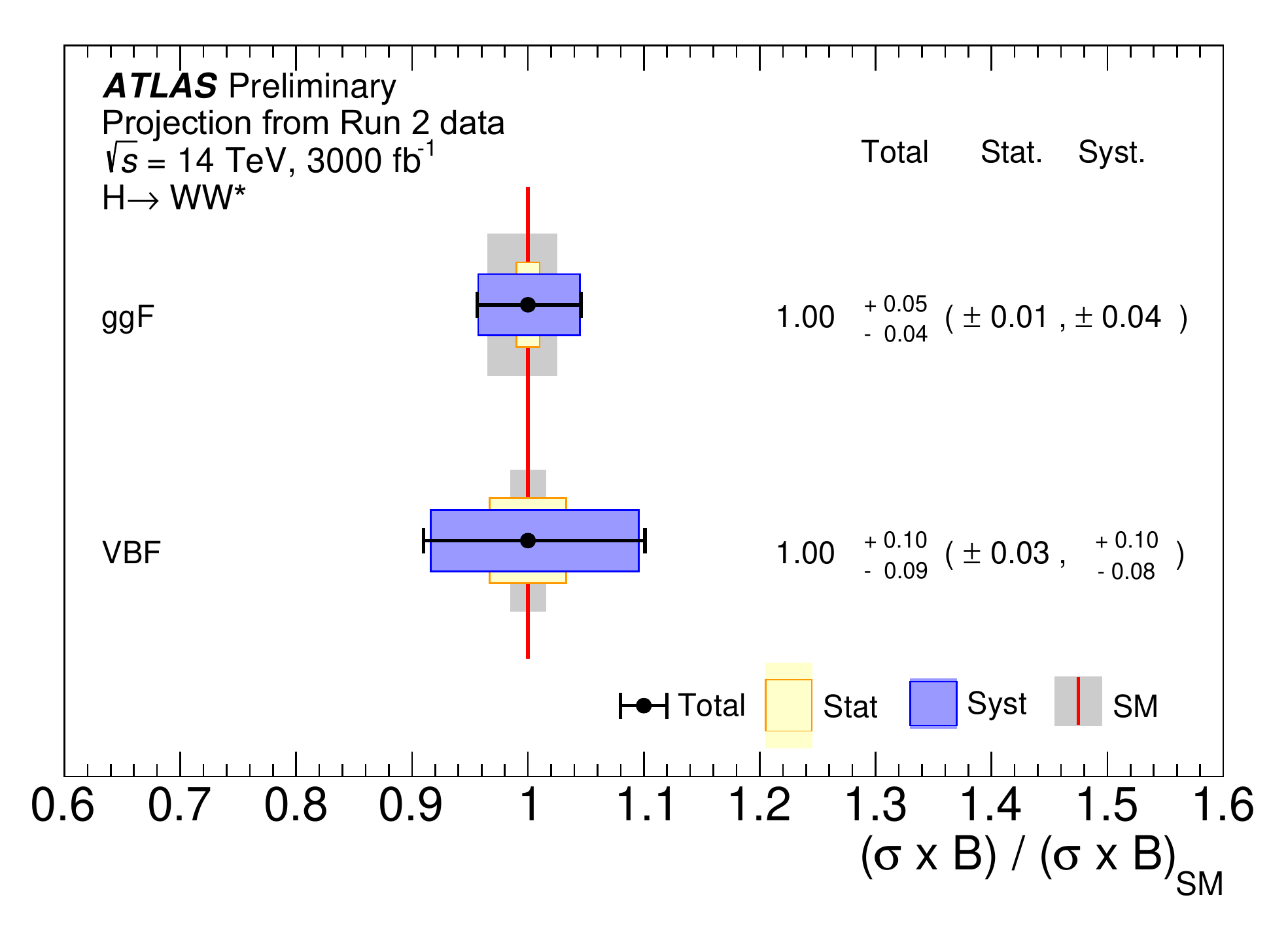}
  \includegraphics[width=0.42\linewidth]{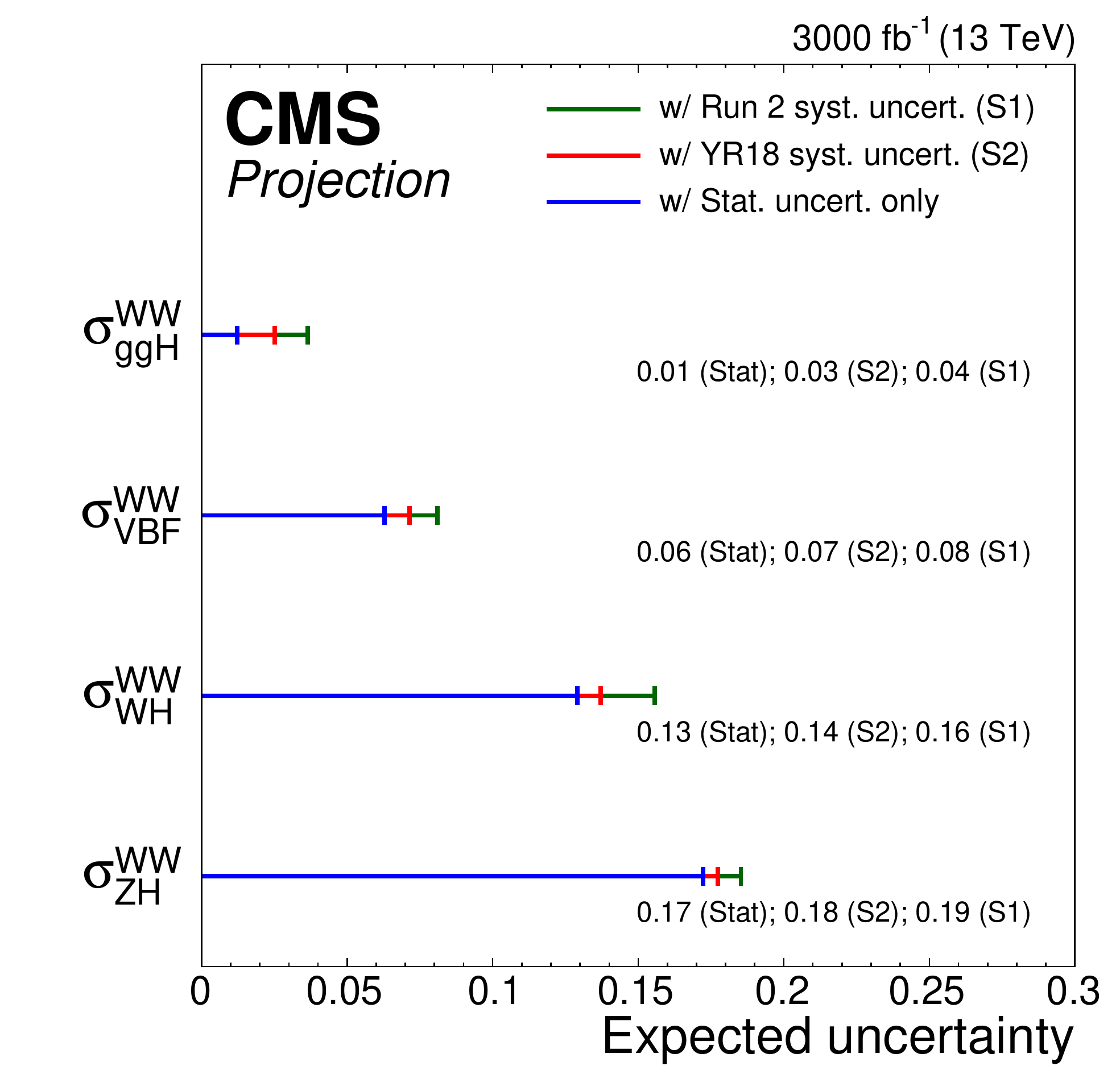}
  \caption{Cross-section times branching fraction measurements of the main Higgs production modes in the \HWW\ decay channel, as extrapolated at the HL-LHC. In case of ATLAS results (left) the ratios of cross sections to their respective theoretical SM predictions are shown for scenario S2, while in case of CMS results (right) the uncertainties on these measurements are shown for S1, S2, and Stat-only scenarios.}
  \label{fig:HWW_ATLAS_HLLHC_S2}
\end{figure}

\asubsubsection[M. Mlynarikova, L. Thomsen]{$H \to \tau^{+}\tau^{-}$}
The measurement of the Higgs boson  in the $H \to \tau^{+}\tau^{-}$ channel considers the leptonic ($\tau_{lep}$) and the hadronic ($\tau_{had}$) decays of the $\tau$ lepton. Three subs-channels ($\tau_{lep}\tau_{lep}$, $\tau_{lep}\tau_{had}$ and $\tau_{had}\tau_{had}$) are defined by requirements on the number of hadronically decaying $\tau$-leptons candidates and leptons (electrons or muons) in the event.
Candidate events are divided into categories using kinematic properties to target cases in which the Higgs boson is produced with a  boost ($\pT > 100$ \UGeV), primarily from gluon  fusion, and cases primarily produced from vector boson fusion, in which the Higgs boson is produced with two jets separated in pseudo-rapidity. Additional requirements are employed to discriminate signal from background. One of the most important variables is the mass of the $\tau\tau$ system, calculated in ATLAS with the Missing Mass Calculator \cite{Elagin:2010aw}, and in CMS with a  dynamical likelihood technique named SVFit \cite{Bianchini:2014vza}.
The normalisation of the dominant backgrounds ($Z\to\ell^{+}\ell^{-}$, $t\bar{t}$, Fake-$\tau_{had}$) is determined using dedicated control regions, or extracted directly in each signal region ($Z\to\tau^{+}\tau^{-}$, the dominant and irreducible background).  More details on the analysis methods can be found in the most recent measurements in the  $H \to \tau^{+}\tau^{-}$ channels published by ATLAS \cite{Aaboud:2018pen} and CMS \cite{Sirunyan:2017khh}.

The performance of the measurements of Higgs boson properties in the $H \to \tau^{+}\tau^{-}$ channel at HL-LHC is extrapolated from the recent measurements in this channel performed by ATLAS \cite{ATLAS-CONF-2018-021} and by CMS \cite{Sirunyan:2017khh} with 36\,$\mathrm{fb}^{-1}$. 
The measurements of the cross sections for the gluon fusion and vector boson fusion production modes are dominated by systematic uncertainties, as can be seen in Table \ref{tab:htt_proj}, which  lists the total expected uncertainties on the cross sections normalised to their SM values as well as  the different contributions from different types of uncertainties. 
The dominant contributions, the experimental and background modelling errors, are due to uncertainties on jet calibration and resolution, on the reconstruction of the $E_{T}^{miss}$, and on the determination of the background normalisation from signal and control regions.

Figure~\ref{fig:htt_proj} shows the ratio of the extrapolated $H \to \tau^{+}\tau^{-}$ ATLAS measurements of the main Higgs production modes to their respective theoretical SM predictions in scenario S2 (left), and uncertainties on these measurements for S1, S2, and Stat-only scenarios as extrapolated using the $H \to \tau^{+}\tau^{-}$ CMS measurements (right).
In case of the ATLAS extrapolation, the SM uncertainties are divided by two compared to their current values, which approximately corresponds to the scaling expected from S2 scenario. The figure shows that at the HL-LHC the measurement will reach a level of precision which is  similar to the theory predictions. These systematic uncertainties are dominated by the theoretical errors on the signal acceptance for the gluon fusion measurement both for S1 and S2. In the measurement of the vector boson fusion cross section, the effects of the experimental errors and uncertainties on the background modelling become more relevant, particularly in S2. 
\begin{table}[th!]
\begin{center}
\renewcommand{\arraystretch}{1.5}
{
\caption{Expected results for the production mode cross-section measurement in the $H\to \tau^{+} \tau^{-}$ channel with 36\fbinv of Run 2 data and at the HL-LHC. Uncertainties are reported relative to the SM cross section at the corresponding centre-of-mass energy. Both scenarios have been considered for the systematic uncertainties in the HL-LHC extrapolation.}
\label{tab:htt_proj}
\begin{tabular}{l | c c c c}
\hline\hline
Experiment, Process & \multicolumn{2}{c}{ATLAS, ggF}& \multicolumn{2}{c}{ATLAS, VBF}\\ 
\hline
Scenario &  S1 & S2 & S1 & S2  \\
Total uncertainty   & $^{+23.1\%}_{-18.5\%}$& $^{+12.3\%}_{-10.8\%}$  & $^{+9.3\%}_{-9.3\%}$ & $^{+8.0\%}_{-7.6\%}$ \\
\hline
Statistical uncert.  & $^{+3.1\%}_{-3.1\%}$ & $^{+3.1\%}_{-3.1\%}$  & $^{+3.4\%}_{-3.4\%}$ & $^{+3.4\%}_{-3.4\%}$\\
Experimental uncert. & $^{+6.0\%}_{-6.2\%}$ & $^{+4.1\%}_{-3.9\%}$  & $^{+5.2\%}_{-5.6\%}$ & $^{+4.9\%}_{-4.5\%}$ \\
Signal theory uncer. & $^{+20.3\%}_{-16.0\%}$ & $^{+10.4\%}_{-9.0\%}$ & $^{+6.3\%}_{-5.3\%}$ & $^{+2.7\%}_{-3.3\%}$\\
Background theory uncer. & $^{+8.0\%}_{-5.5\%}$ & $^{+3.1\%}_{-2.4\%}$ & $^{+3.4\%}_{-3.4\%}$ & $^{+3.8\%}_{-3.8\%}$\\
\hline\hline
\end{tabular}
} 
\end{center}
\end{table}

 \begin{figure}[h!]
\begin{center}
\includegraphics[width=0.56\textwidth]{\main/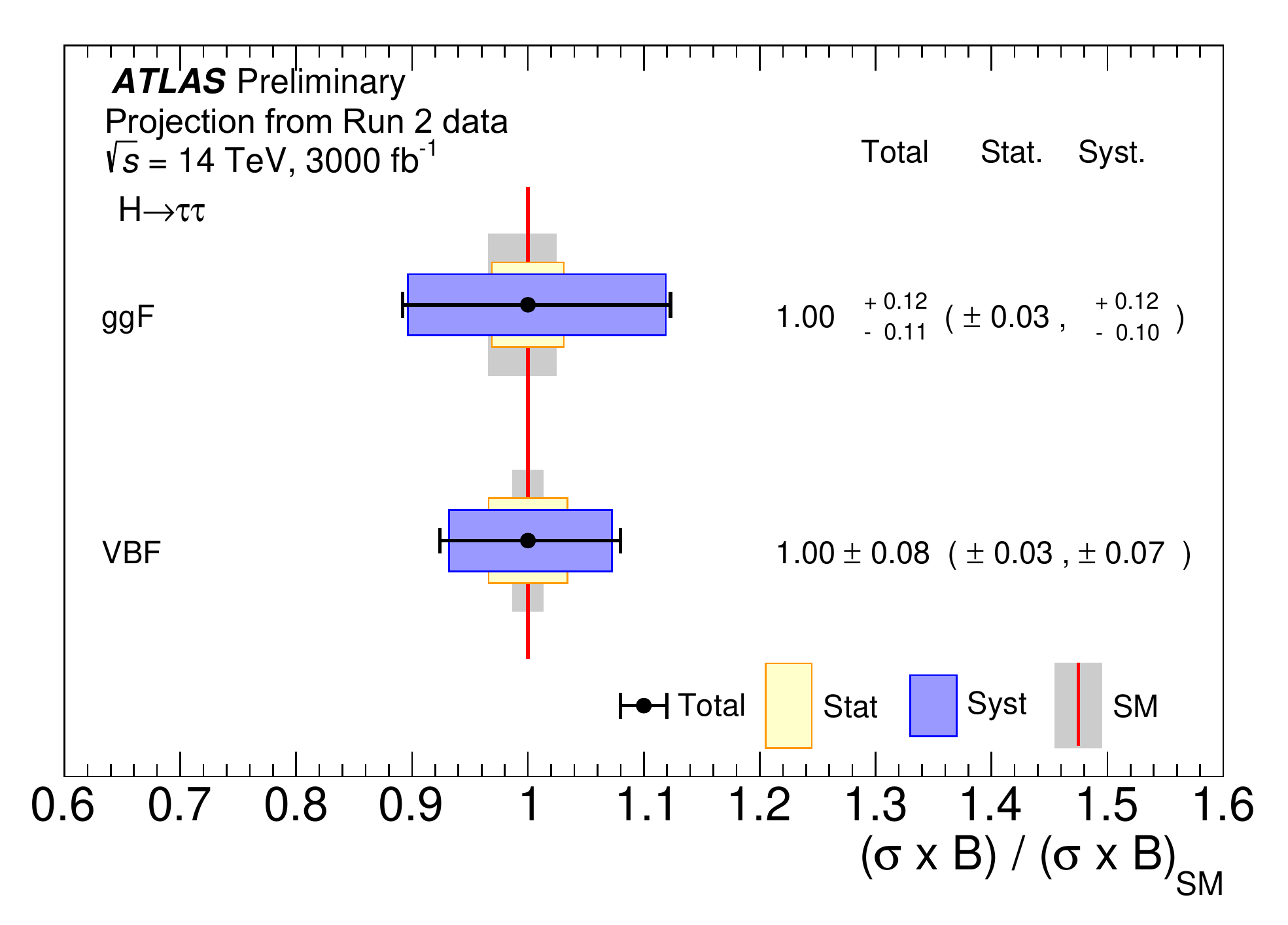}
\includegraphics[width=0.42\textwidth]{\main/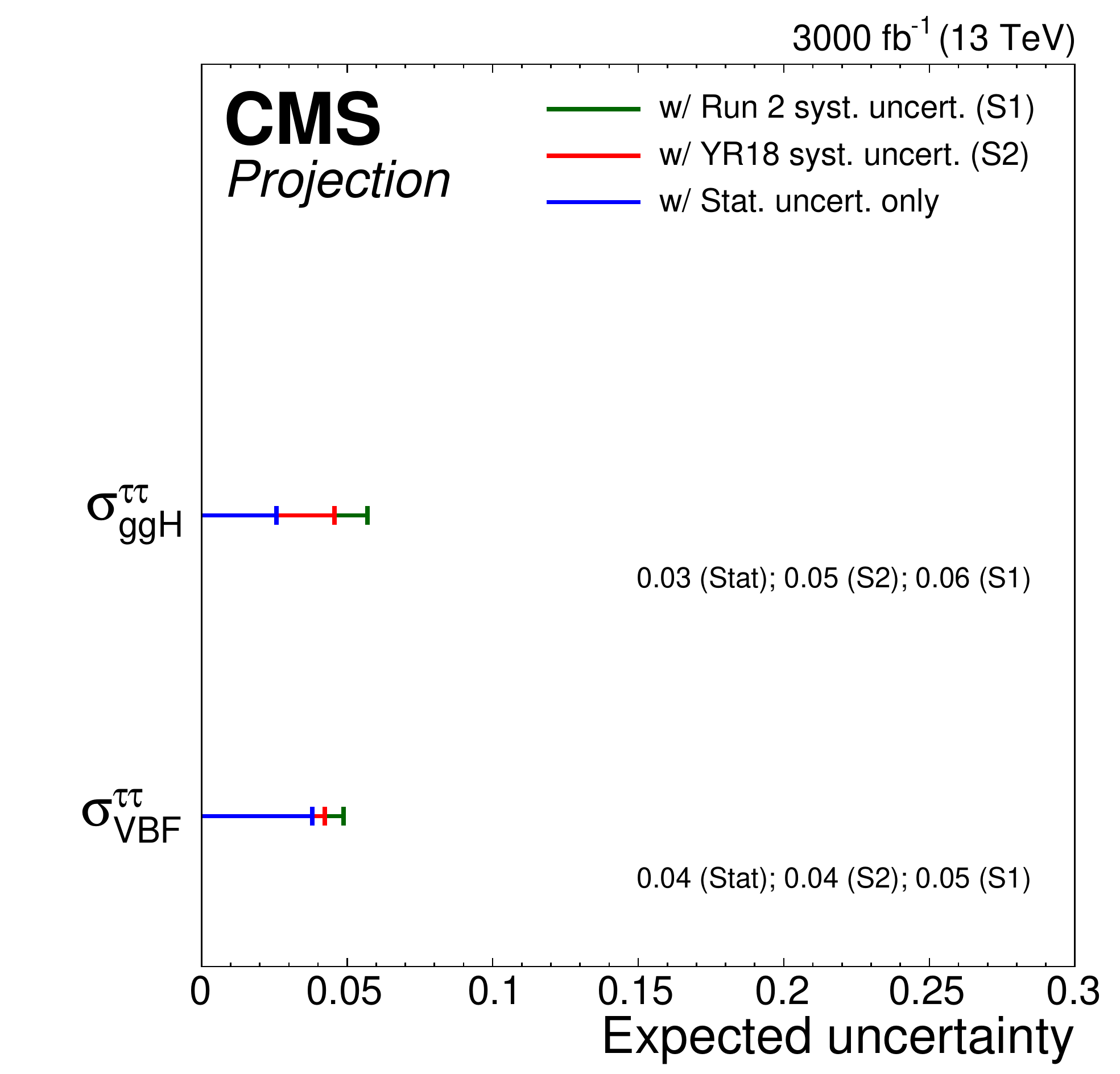}
\end{center}
\caption{(Right) ATLAS comparison, for $H \to \tau^{+}\tau^{+}$ final state applying scenario S2, between the expected precision on production-mode cross section times branching ratio normalised to their SM expectation at HL-LHC and the theoretical uncertainty on the SM prediction. (Left) CMS expected precision on production-mode cross section times branching ratio for $H \to \tau^{+}\tau^{+}$ final state in case of S1, S2, and stat-only scenarios.
}
\label{fig:htt_proj}
\end{figure}

\asubsubsection[L. D'eramo, C. Li, G. Marchiori, A. de Wit]{$H \to b\bar{b}$}

The measurement of the Higgs boson  in the $H \to b\bar{b}$ channel presented here considers the Higgs boson production in association with a vector boson ($V=W/Z$). 
Searches for $H \to b\bar{b}$  in association with a vector boson drove the recent observation  of this decay mode reported by the ATLAS and CMS Collaborations \cite{Aaboud:2018zhk,Sirunyan:2018kst}.
The analyses make use of leptonic decays of the vector boson for triggering and to reduce the multi-jet background: the final states
of the $\vh$ system covered in the analyses always contain two b-jets and either zero, one or two electrons or muons. Both leptons are required to have the same flavour in the two lepton selection.
Major backgrounds arising from SM production of vector boson plus heavy- or light-flavour jets, in addition to $t\bar{t}$ production, are 
controlled and constrained via dedicated control regions. The  b-jet energy resolution is improved by using multivariate energy regression techniques (CMS), or sequential corrections (ATLAS), and a boosted decision tree is used to improve the discrimination 
between signal and background. The distribution of this multivariate discriminator is used as the discriminating
variable in the signal extraction fit. 

The ATLAS and CMS Collaborations have both recently reported the observation of the $H \to b\bar{b}$ decay \cite{Aaboud:2018zhk,Sirunyan:2018kst}.
The studies presented here are performed by extrapolating this most recent ATLAS $H \to b\bar{b}$ measurements using a dataset corresponding to an integrated luminosity of 78.9 \fbinv, and by extrapolating a previous analysis by the CMS Collaboration. In this previous analysis evidence for the $H \to b\bar{b}$ decay in the $\text{VH}$ production mode was reported using a dataset corresponding to an integrated luminosity of 35.9\fbinv \cite{HIG16044}. 



Figure \ref{fig:vhbb_proj_bars} shows the extrapolation of the signal strength uncertainty per-channel (CMS) and per-production mode (ATLAS). The details of the contributions of different sources of uncertainty in scenarios S1 and S2 for the projection of the ATLAS and CMS analyses are shown in Table~\ref{tab:vhbb_uncertbreakdown}.
The large improvement, by a factor 2.5--3, in the uncertainty of the measurement for the $WH$ (1-lepton channel) compared to the Run-2 results (around 45\%) is caused by the integrated luminosity scaling of the uncertainty in the modelling of the $W$ boson $\pT$ distribution for both the collaborations, being the dominant uncertainty in scenario S1.

\begin{figure}[h!]
\begin{center}
\includegraphics[width=0.40\textwidth]{\main/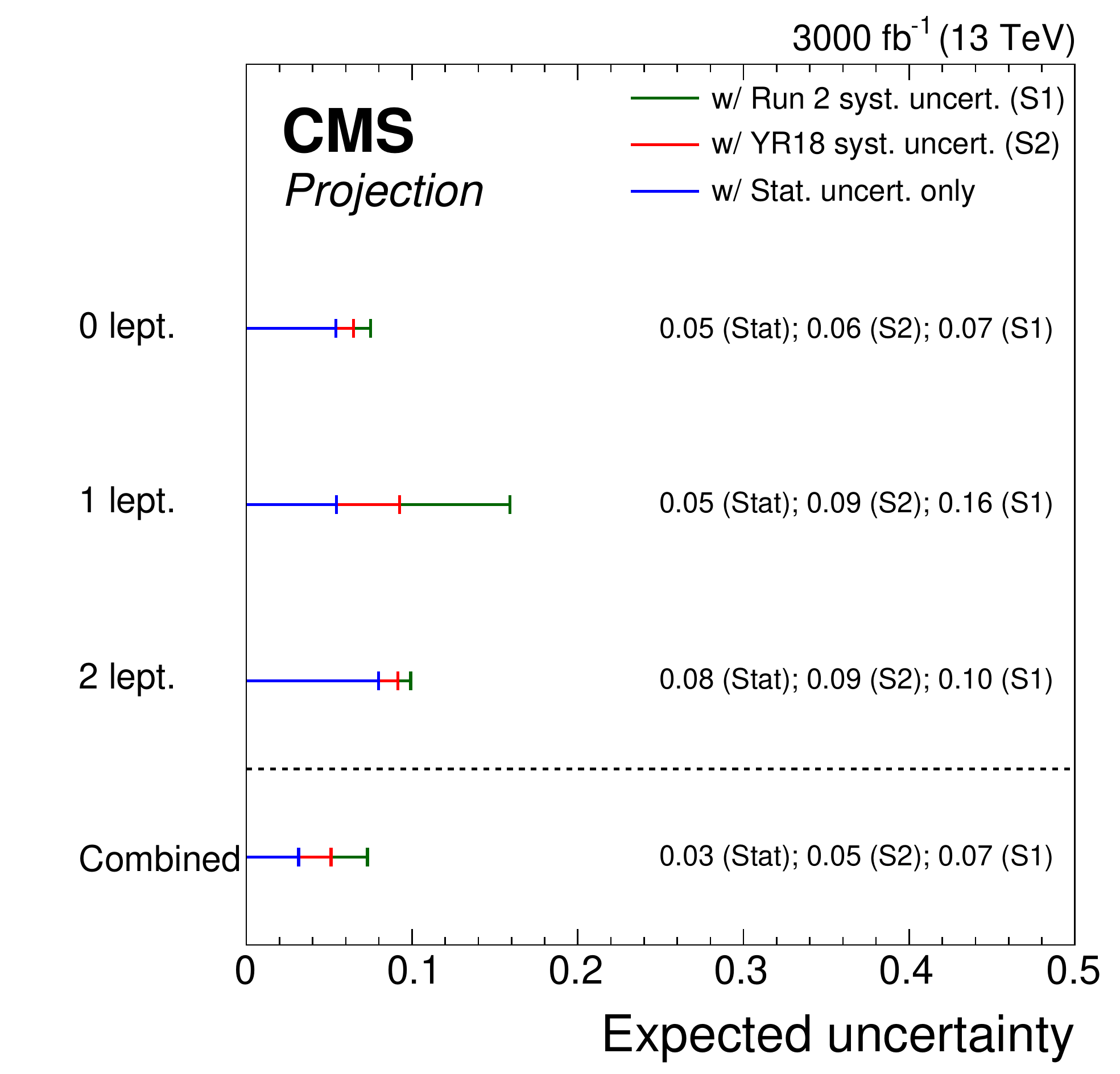}
\includegraphics[width=0.50\textwidth]{\main/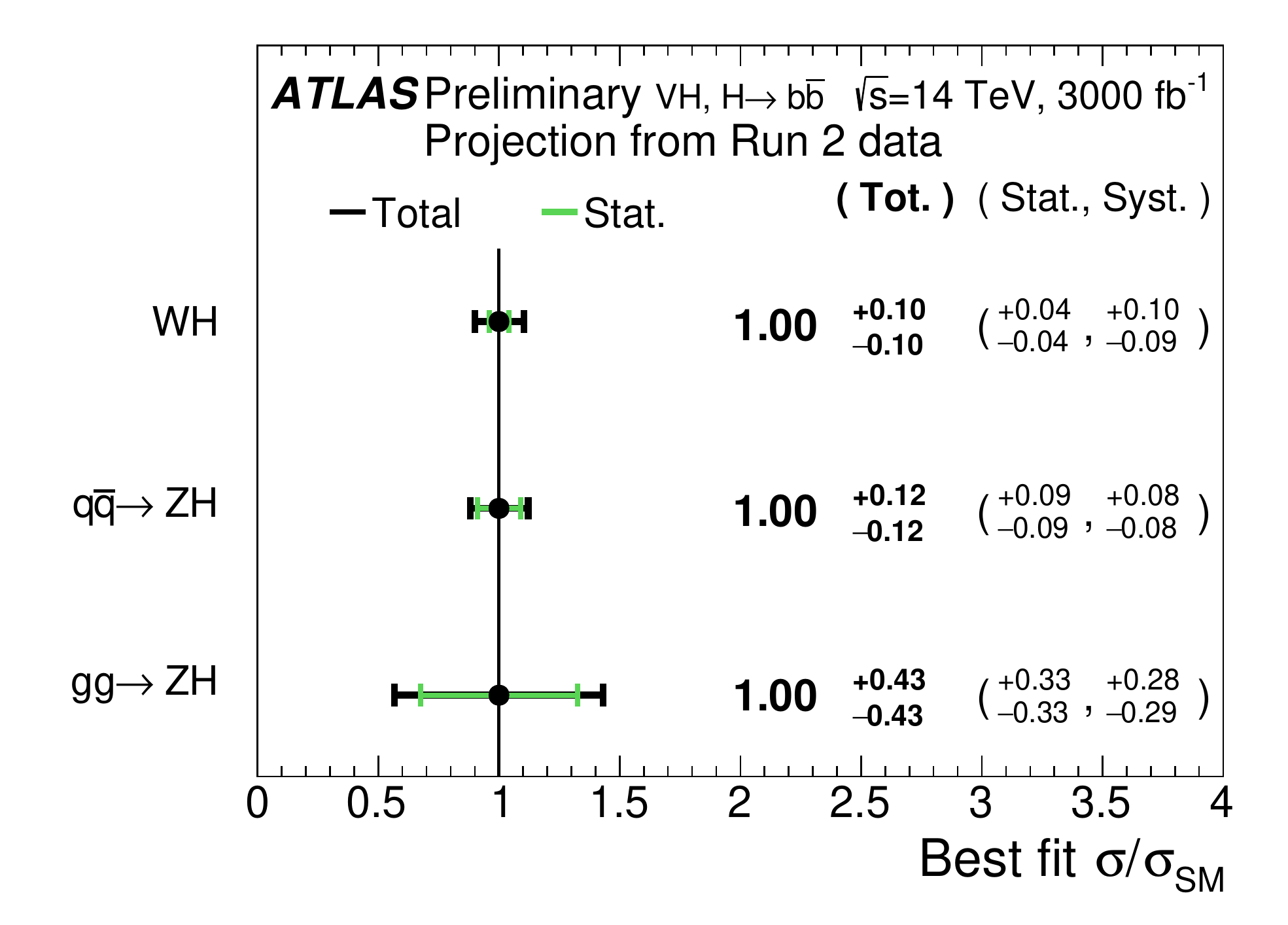}
\end{center}
\caption{Extrapolation of the uncertainties estimated by the CMS collaboration (left) and by the ATLAS collaboration (right) for the $H \to b\bar{b}$ channel. The figure gives the uncertainties per-channel and on the combined signal strengths on the left, and per-production mode on the right. Values are given for the S1 (with Run~2 systematic uncertainties~\cite{HIG16044}) and S2 (with YR18 systematic uncertainties) scenarios, as well as a scenario in which all systematic uncertainties are removed. Only the S2 scenario is presented in the plot by the ATLAS collaboration (S1 is presented in Table \ref{tab:vhbb_uncertbreakdown}).}
\label{fig:vhbb_proj_bars}
\end{figure}
\begin{table}[th!]
\begin{center}
{
\caption{Contributions of particular groups of uncertainties, expressed as percentages, in S1 (with Run~2 systematic uncertainties~\cite{HIG16044}) and S2 (with YR18 systematic uncertainties) for the CMS and ATLAS analyses of the $H \to b\bar{b}$ channel. The total uncertainty is decomposed into four components: signal theory, background theory, experimental and statistical. In the CMS results, the signal theory uncertainty is further split into inclusive and acceptance parts, and the contributions of the b-tagging and JES/JER uncertainties to the experimental component are also given. In the ATLAS results, the contributions of the four groups of uncertainties are presented for $pp\to WH$, $qq\to ZH$ and $gg\to ZH$ separately.}
\label{tab:vhbb_uncertbreakdown}
\subfloat{
\begin{tabular}{l | c c  }
\hline\hline
Experiment & \multicolumn{2}{c}{CMS} \\
Process & \multicolumn{2}{c}{$pp\to VH$}\\ 
\hline
Scenario & S1 & S2 \\
Total uncertainty   & 7.3\%  & 5.1\% \\
\hline
Statistical uncert. & 3.2\% & 3.2\%  \\
Experimental uncert. & 2.6\% & 2.2\% \\
\quad b-tagging & 2.2\%  & 2.0\% \\
\quad JES and JER & 0.7\%  & 0.6\% \\
Signal theory uncer.& 5.4\% & 2.6\% \\
\quad Inclusive & 4.6\% &2.2\%  \\
\quad Acceptance & 2.7\% &1.3\% \\
Background uncert. & 2.8\% & 2.3\%\\
\hline\hline
\end{tabular}
}

\vspace{5mm}
\subfloat{
\renewcommand{\arraystretch}{1.5}
\begin{tabular}{l | c c c c c c }
\hline\hline
Experiment &  \multicolumn{6}{c}{ATLAS}\\
Process  & \multicolumn{2}{c}{$pp\to WH$}& \multicolumn{2}{c}{$qq\to ZH$}& \multicolumn{2}{c}{$gg\to ZH$}\\ 
\hline
Scenario &  S1 & S2  & S1 & S2 & S1 & S2 \\
Total uncertainty   & $^{+14.9\%}_
{-13.8\%}$& $^{+10.4\%}_
{-10.0\%}$ &$^{+13.8\%}_
{-13.2\%}$ & $^{+12.1\%}_
{-11.8\%}$ & $^{+49.8\%}_
{-49.0\%}$ & $^{+43.2\%}_
{-43.3\%}$\\
\hline 
Statistical uncert.  &$^{+4.1\%}_
{-4.1\%}$ & $^{+4.1\%}_
{-4.1\%}$ &$^{+9.0\%}_
{-8.9\%}$ & $^{+9.0\%}_
{-8.9\%}$ & $^{+33.3\%}_
{-33.3\%}$ & $^{+33.3\%}_
{-33.3\%}$ \\
Experimental uncert. & $^{+4.8\%}_
{-4.7\%}$  & $^{+4.4\%}_
{-4.3\%}$ & $^{+6.5\%}_
{-6.3\%}$ & $^{+5.7\%}_
{-5.5\%}$ & $^{+24.9\%}_
{-25.0\%}$ & $^{+20.8\%}_
{-20.4\%}$ \\
Signal theory uncer. & $^{+8.0\%}_
{-7.0\%}$ & $^{+4.6\%}_
{-4.1\%}$ & $^{+6.1\%}_
{-5.5\%}$ & $^{+3.1\%}_
{-2.8\%}$ & $^{+18.1\%}_
{-14.0\%}$ & $^{+9.6\%}_
{-8.0\%}$ \\
Background uncert.  & $^{+10.8\%}_
{-10.0\%}$ & $^{+7.2\%}_
{-6.9\%}$ & $^{+5.4\%}_
{-4.8\%}$ & $^{+4.8\%}_
{-4.6\%}$ & $^{+20.7\%}_
{-21.8\%}$ & $^{+17.7\%}_
{-18.1\%}$ \\
\hline\hline
\end{tabular}
}
} 
\end{center}
\end{table}

Both in scenario S1 and S2 the largest component of the systematic uncertainty is theoretical. This arises from the uncertainty in the gluon-induced $ZH$ ($gg\to ZH$) production cross section due to QCD scale variations. The $gg\to ZH$ process contributes a small fraction of the total $ZH$ process. Despite this, the uncertainty in the production cross section for this process due to QCD scale variations  becomes dominant because it is very large: 25\% for the $gg\to ZH$ process, compared to approximately 0.7\% for the $qq\to ZH$ process \cite{deFlorian:2016spz}. The theoretical uncertainties on the $gg\to ZH$ production are reduced to 15\% in the S2.
An important contribution to the uncertainty is due to category-acceptance uncertainties in the dominant $Z+bb$ and $W+bb$ backgrounds due to QCD scale variations, as well as the uncertainty in the $qq\to ZH$ and $WH$ production
cross section due to QCD scale variations. 
To improve the precision of the measurement it is therefore important to improve these theoretical uncertainties.

In the future, and at the HL-LHC in particular, the b-tagging efficiency may change. The conditions could worsen the efficiency, but
at the same time new detectors and new techniques could also lead to an improvement in the b-tagging efficiency. 
The effect of changes in b-tagging efficiency on the overall signal strength uncertainty has been evaluated by the CMS collaboration, showing that an improvement of 10\% in the b-tagging efficiency leads to a relative improvement in the signal strength uncertainty of up to 6\% \cite{CMS-PAS-FTR-18-011}.
%

\asubsubsection[M. Klute, H. Li, G. Marchiori, A.Marini, M. Verducci, M. Zgubic, J. Zhang]{$H \to \mu^{+}\mu^{-}$}\label{Sec:2.3.8}
The  $H \to \mu^{+}\mu^{-}$ analyses play a crucial role in the determination of the couplings to the second fermion generation. The analyses search for a narrow peak in the di-muon invariant mass over a smoothly falling background, dominated by Drell--Yan and top-pair productions. Events are selected requiring two oppositely charged muons passing loose quality criteria to retain as much signal as possible. The overall sensitivity to this decay mode benefits from multivariate or sequential categorisation techniques that allow separating the two dominant production modes, the vector boson fusion (with the typical presence of a forward-backward jet pair) and the gluon fusion. Additional enhancements in the sensitivity are achieved by a further sub-categorisation based on the muon momentum resolution. More details on the analysis methods can be found in the most recent searches of the  $H \to \mu^{+}\mu^{-}$ channels published by ATLAS \cite{Aaboud:2017ojs} and CMS \cite{Sirunyan:2018hbu}.  

The extrapolation studies presented here by ATLAS Collaboration are based on a previous analysis performed by that collaboration using the 2015--2017 proton-proton collision dataset collected at $\sqrt{s} = 13\,\UTeV$, which corresponds to an integrated luminosity of 79.8\fbinv \cite{Aaboud:2017ojs}. 
In addition to the standard extrapolation procedure, the di-muon signal widths are reduced  by 15-30\% thanks to the improvements expected from the performance of the ATLAS upgrade Inner Tracker (ITk) \cite{Collaboration:2285585}.
In this analysis, the $Z\to\mu^{+}\mu^{-}$ background is fully determined by data, and it is modelled by fitting the di-muon invariant mass $m_{\mu\mu}$ distribution in each category using a Breit--Wigner function convoluted with a Gaussian summed to a smooth function.

Similar studies have been carried out by the CMS Collaboration, based on the analysed data collected during 2016 and corresponding to an integrated luminosity of $36\,\text{fb}^{-1}$~\cite{HIG-17-019}. 
The analysis was optimised to have the overall best sensitivity to a standard model Higgs boson inclusively with respect to the production modes with the data collected in 2016.
In addition to the extrapolation procedure based on the increased luminosity, the di-muon invariant mass width is reduced in order to match the expected increase in performances due to the upgrade in the tracking system~\cite{Klein:2017nke} and displayed in Fig.~\ref{fig:cms_hmm_res_hl}. The di-muon mass resolution plays a crucial role in the analysis performances and in the systematic uncertainty induced by the choice of the background function. 
The CMS experiment~\cite{CMS} benefits from the large $4\,\text{T}$ solenoidal fields that allowed it to achieve down to 1.1\% di-muon mass resolution in 2016 and, with the upgrade projects, the CMS detector will be able to reach in the best category a di-muon mass resolution of 0.65\%~\cite{Klein:2017nke}.

\begin{figure}[h!]
\begin{center}
\includegraphics[width=0.40\textwidth]{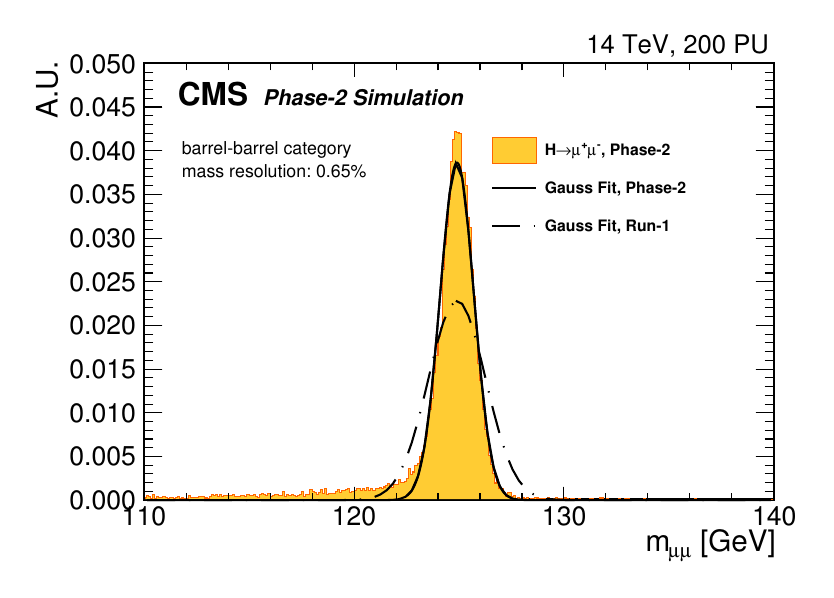}
\end{center}
\caption{The di-muon invariant mass distribution for $H\rightarrow{}\mu^+\mu^-$ decays for muons in the
central region, simulated with the Phase-2 detector.~\protect\cite{Klein:2017nke}.}
\label{fig:cms_hmm_res_hl}
\end{figure}

Table \ref{tab:hmumu_proj} shows the expected precision on the signal strength (ATLAS) and branching fraction (CMS) measurement with 3000\fbinv of HL-LHC data in the scenarios S1 and S2.
In both scenarios, the analysis is limited by the statistical uncertainty, while the leading systematic uncertainty is the bias introduced by the choice of the function describing the background (spurious signal uncertainty), and the uncertainties on the modelling of the signal (their reduction in S2 contributes to an overall improvement of 10\% on the precision of the measurement). Expected uncertainties on signal strength vary from 15 to 13\% (ATLAS) and on the branching fraction vary from 13 to 10\% (CMS), accordingly to the projection scenario. CMS extrapolation is obtained from the simultaneous fit in all production and decay modes, as described in Section \ref{sec:expcomb_prodtimesdecay}.

\begin{table}[th!]
\begin{center}
\renewcommand{\arraystretch}{1.5}
{
\caption{Expected precision on the signal strength measurement in the $H\to\mu^{+}\mu^{-}$ channels with 3000\fbinv of HL-LHC data with the two systematic uncertainties scenarios. For the HL-LHC extrapolation, the improved ITk resolution has been emulated. }
\label{tab:hmumu_proj}
\begin{tabular}{l | c c}
\hline\hline
Experiment & \multicolumn{2}{c}{ATLAS}\\
Process & \multicolumn{2}{c}{Combination}\\ 
\hline
Scenario  & S1 & S2  \\
Total uncertainty   & $^{+15\%}_{-14\%}$  & $^{+13\%}_{-13\%}$  \\
\hline
Statistical uncert. & $^{+12\%}_{-13\%}$  & $^{+12\%}_{-13\%}$  \\
Experimental uncert. & $^{+3\%}_{-3\%}$  & $^{+2\%}_{-2\%}$  \\
Theory uncer. & $^{+8\%}_{-5\%}$ & $^{+5\%}_{-4\%}$  \\
\hline\hline
\end{tabular}

\vspace{5mm}

\label{tab:hmumu_proj_cms}
\begin{tabular}{l | c c}
\hline\hline
Experiment & \multicolumn{2}{c}{CMS}\\
Process & \multicolumn{2}{c}{Combination}\\ 
\hline
Scenario  & S1 & S2  \\
Total uncertainty   & $13\%$  & $10\%$  \\
\hline
Statistical uncert. & $9\%$  & $9\%$  \\
Experimental uncert. & $8\%$  & $2\%$  \\
Theory uncer. & $5\%$  & $3\%$  \\
\hline\hline
\end{tabular}

} 
\end{center}
\end{table}

\asubsection[M. Delmastro, A. Gilbert, T. Klijnsma, J. Langford,
  W. Leight, R. Naranjo Garcia, A. Salvucci, M. Scodeggio,
  K. Tackmann, N. Wardle, C. Vernieri]{Fiducial and differential
  cross-section measurements}
\label{sec2:fiducial}
\asubsubsection[M. Delmastro, T. Klijnsma]{Measurements using $H \to \gamma\gamma$, $H \to ZZ^* \to 4\ell$, (boosted) \Hbb\ decay channels}
\label{sec:diffxs}


\noindent In the context of Higgs boson property measurements, one of the main goals of HL-LHC, differential measurements provide a probe of various Higgs boson properties by looking at distortions of differential distributions.
The $\pTH$ distribution is of particular interest, as potential new physics may reside in the tails of the distribution, which cannot be measured in inclusive measurements~\cite{%
Khachatryan:2016vau,
Aad:2015zhl,
CMS:2018lkl
}.
Differential Higgs boson production cross section measurements are available for a range of observables from both the ATLAS~\cite{%
Aad:2014lwa,
Aad:2014tca,
Aad:2016lvc,
Aaboud:2018xdt,
Aaboud:2017oem,
Aaboud:2018ezd
} and CMS~\cite{%
Khachatryan:2015rxa,
Khachatryan:2015yvw,
Khachatryan:2016vnn,
Sirunyan:2018kta,
Sirunyan:2017exp,
Sirunyan:2018sgc
} Collaborations at $\sqrt{s}$~=~8 and 13~\UTeV.

The most recent $\pTH$ spectra at $\sqrt{s}$~=~13~\UTeV\ from both the ATLAS~\cite{Aaboud:2018ezd} and CMS~\cite{Sirunyan:2018sgc} Collaborations are projected to an integrated luminosity of $3000\fbinv$~\cite{CMS-PAS-FTR-18-011, ATL-PHYS-PUB-2018-040}.
The projection of the $\pTH$ differential cross section measurement by the CMS Collaboration is shown in Fig.~\ref{fig:proj_pth}, for both scenarios S1 and S2. The corresponding total uncertainties are respectively given in Tables~\ref{tab:proj_pth_unc_scen1} and \ref{tab:proj_pth_unc_scen2}. With respect to the uncertainties affecting the measurement based on an integrated luminosity of 35.9\fbinv, the uncertainties at 3000\fbinv in the higher $\pTH$ region are about a factor of ten smaller. This is expected, as the uncertainties in this region remain statistically dominated.
The uncertainties in the lower $\pTH$ region are however no longer statistically dominated, as can been seen by comparing Table~\ref{tab:proj_pth_unc_scen1} with Table~\ref{tab:proj_pth_unc_scen2}, where the reduced systematic uncertainties in S2 yield a reduction in the total uncertainty of up to 25\% compared to S1.

Figure~\ref{fig:ATLAS_proj_differential} shows the ATLAS projections to 3000 fb$^{-1}$ of the differential measurements of $\pTH$, the Higgs rapidity $|y_H|$, the jet multiplicity $N_{\rm jets}$ of jets with $p_{\mathrm{T}} >$ 30 \UGeV and the transverse momentum of the leading jet accompanying the Higgs boson $p_{\mathrm{H}}^{j1}$, as obtained by combining the measurement in the \Hyy\ and \HZZ\ channels, in scenarios S1 and S2. The relative uncertainties affecting the $\pTH$ measurement are given in Tables~\ref{tab:proj_pth_unc_scen1} and \ref{tab:proj_pth_unc_scen2}.
The ATLAS combined $\pTH$ measurement extrapolation exhibits relative uncertainties ranging from about 5\% in the lower $\pTH$ bins to about 9\% in the highest $\pTH$ bin in scenario S1, reducing to uncertainties ranging from $\sim$ 4\% to $\sim$ 8\% in scenario S2.

\begin{figure}
  \begin{center}
    \includegraphics[width=0.49\linewidth]{\main/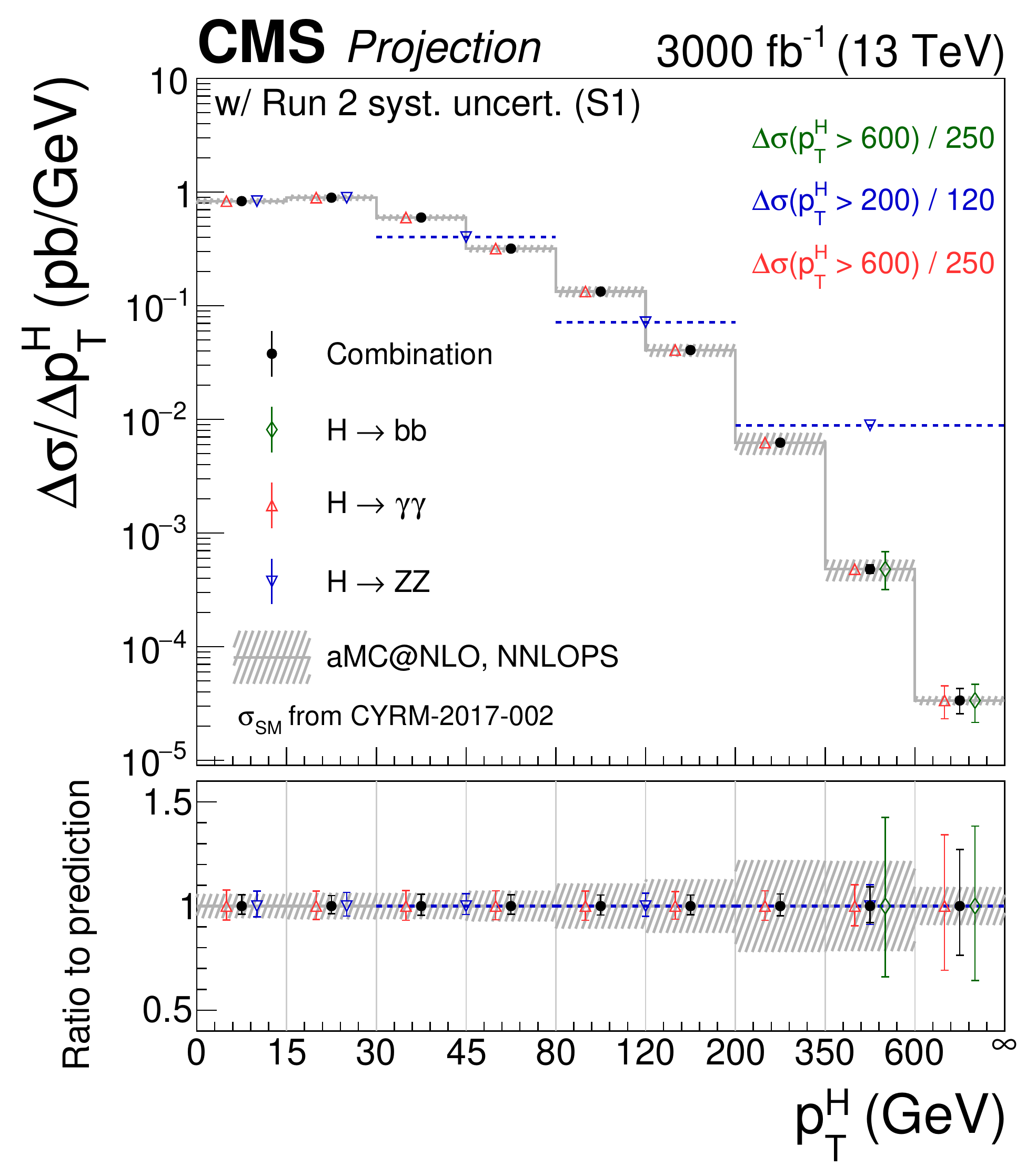}
    \includegraphics[width=0.49\linewidth]{\main/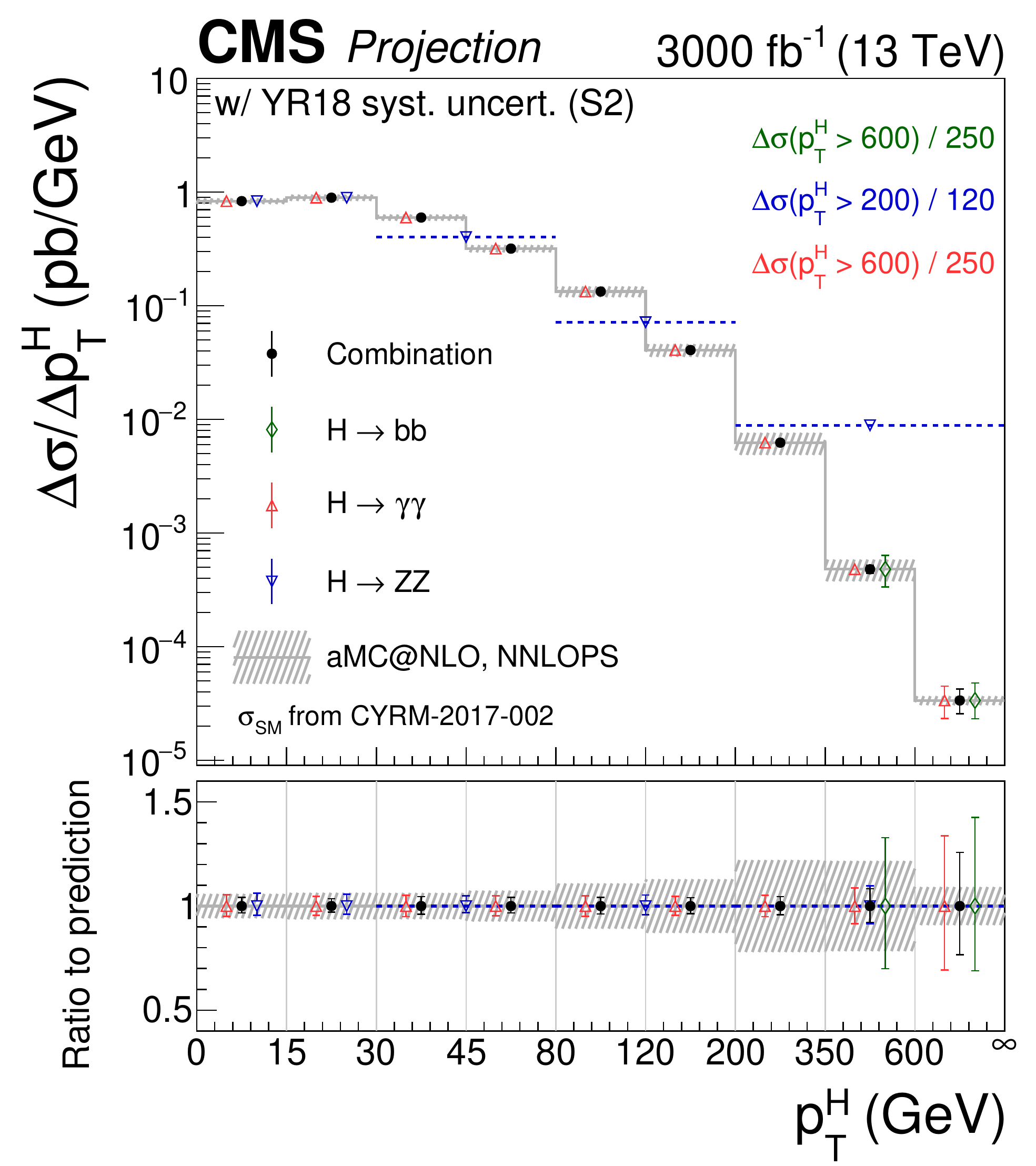}
    \caption{
        Projected differential cross section for $\pTH$ at an integrated luminosity of 3000\fbinv ~\cite{Sirunyan:2018sgc}, under S1 (\UcmsLeft, with Run~2 systematic uncertainties~\cite{CMS-PAS-HIG-17-028}) and S2 (\UcmsRight, with YR18 systematic uncertainties).
        }
    \label{fig:proj_pth}
  \end{center}
\end{figure}

\begin{table}[th]
  \resizebox{\textwidth}{!}{
  \centering
  \begin{tabular}{l|cccccccc|c|c|c|c|c|c|c|c}
    \hline
    \hline
    \multicolumn{17}{c}{3000 fb$^{-1}$ ATLAS}                                                                                                                                                                                                 \\
    \hline
    \hline
    $\pTH$ [\UGeV] & \multicolumn{2}{c|}{0-10}  & \multicolumn{2}{c|}{10-15} & \multicolumn{2}{c|}{15-20} & \multicolumn{2}{c|}{20-30} & 30-45  & 45-60   & 60-80   & 80-120  & 120-200 & 200-350 & \multicolumn{2}{c}{350-1000} \\
    \hline
    $\hgg$       & \multicolumn{2}{c|}{6.5\%} & \multicolumn{4}{c|}{5.9\%} & \multicolumn{2}{c|}{6.2\%} & 6.0\%                      & 6.5\%  & 6.7\%   & 6.0\%   & 5.4\%   & 6.3\%   & \multicolumn{2}{c}{9.5\%}              \\
    \hline
    $\hzz$       & \multicolumn{2}{c|}{9.0\%} & \multicolumn{2}{c|}{8.1\%} & \multicolumn{2}{c|}{8.9\%} & \multicolumn{2}{c|}{6.9\%} & 6.3\%  & 6.8\%   & 6.8\%   & 6.2\%   & 6.7\%   & 13.2\%  & \multicolumn{2}{c}{24.3}     \\
    \hline
    Combination  & \multicolumn{2}{c|}{5.5\%} & \multicolumn{4}{c|}{4.8\%} & \multicolumn{2}{c|}{5.0\%} & 4.7\%                      & 5.0\%  & 5.1\%   & 4.6\%   & 4.4\%   & 5.4\%   & \multicolumn{2}{c}{8.7\%}              \\
    \hline
    \hline
    \multicolumn{17}{c}{3000 fb$^{-1}$ CMS}                                                                                                                                                                                                   \\
    \hline
    \hline
    $\pTH$ [\UGeV] & \multicolumn{4}{c|}{0-15}  & \multicolumn{4}{c|}{15-30} & 30-45                      & \multicolumn{2}{c|}{45-80} & 80-120 & 120-200 & 200-350 & 350-600 & 600-$\infty$                                     \\
    \hline
    $\hgg$       & \multicolumn{4}{c|}{5.1\%} & \multicolumn{4}{c|}{6.8\%} & 7.1\%                      & \multicolumn{2}{c|}{6.9\%} & 7.1\%  & 6.7\%   & 7.1\%   & 9.9\%   & 32.5\%                                           \\     
    \hline
    $\hzz$       & \multicolumn{4}{c|}{5.4\%} & \multicolumn{4}{c|}{5.7\%} & \multicolumn{3}{c|}{5.0\%} & \multicolumn{2}{c|}{5.5\%} & \multicolumn{3}{c}{9.6\%}                                                               \\ 
    \hline
    $\hbb$       & \multicolumn{14}{c|}{none} & 38.2\%                     & 37.1\%                                                                                                                                            \\ 
    \hline
    Combination  & \multicolumn{4}{c|}{4.7\%} & \multicolumn{4}{c|}{4.4\%} & 5.0\%                      & \multicolumn{2}{c|}{4.7\%} & 4.8\%  & 4.7\%   & 5.2\%   & 8.5\%   & 25.4\%                                           \\
    \hline
    \hline
    \multicolumn{17}{c}{6000 fb$^{-1}$}                                                                                                                                                                                                   \\
    \hline
    \hline
    Combination  & \multicolumn{4}{c|}{4.0\%} & \multicolumn{4}{c|}{3.7\%} & 4.0\% & \multicolumn{2}{c|}{3.9\%} & 4.0\% & 4.0\% & 4.3\% & 6.3\% & 18.3\% \\
    \hline
    \hline
  \end{tabular}
  }
  \caption{Relative uncertainties on the projected $\pTH$ spectrum measurements by ATLAS and CMS under S1 at 3000 fb$^{-1}$.  The relative uncertainty of the CMS projection is also given at 6000 fb$^{-1}$ to represent the sensitivity achievable by an eventual ATLAS and CMS combination. }
  \label{tab:proj_pth_unc_scen1}
\end{table}

\begin{table}[th]
  \centering
  \resizebox{\textwidth}{!}{
  \begin{tabular}{l|cccccccc|c|c|c|c|c|c|c|c}
    \multicolumn{1}{l}{} & \multicolumn{1}{l}{} & \multicolumn{1}{l}{} & \multicolumn{1}{l}{} & \multicolumn{1}{l}{} & \multicolumn{1}{l}{} & \multicolumn{1}{l}{} & \multicolumn{1}{l}{} & \multicolumn{1}{l}{} & \multicolumn{1}{l}{} & \multicolumn{1}{l}{} & \multicolumn{1}{l}{} & \multicolumn{1}{l}{} & \multicolumn{1}{l}{} & \multicolumn{1}{l}{} \\[\dimexpr-\normalbaselineskip-\arrayrulewidth] 
    \hline
    \hline
    \multicolumn{17}{c}{3000 fb$^{-1}$ ATLAS}                                                                                                                                                                                                 \\
    \hline
    \hline
    $\pTH$ [\UGeV] & \multicolumn{2}{c|}{0-10}  & \multicolumn{2}{c|}{10-15} & \multicolumn{2}{c|}{15-20} & \multicolumn{2}{c|}{20-30} & 30-45  & 45-60   & 60-80   & 80-120  & 120-200 & 200-350 & \multicolumn{2}{c}{350-1000} \\
    \hline
    $\hgg$       & \multicolumn{2}{c|}{5.3\%} & \multicolumn{4}{c|}{4.6\%} & \multicolumn{2}{c|}{4.9\%} & 4.7\%                      & 5.4\%  & 5.7\%   & 4.9\%   & 4.2\%   & 5.1\%   & \multicolumn{2}{c}{8.7\%}              \\
    \hline
    $\hzz$       & \multicolumn{2}{c|}{8.3\%} & \multicolumn{2}{c|}{7.6\%} & \multicolumn{2}{c|}{8.3\%} & \multicolumn{2}{c|}{6.3\%} & 5.7\%  & 6.2\%   & 6.3\%   & 5.7\%   & 6.4\%   & 13.1\%  & \multicolumn{2}{c}{23.2\%}   \\
    \hline
    Combination  & \multicolumn{2}{c|}{4.5\%} & \multicolumn{4}{c|}{3.8\%} & \multicolumn{2}{c|}{3.9\%} & 3.6\%                      & 4.1\%  & 4.2\%   & 3.7\%   & 3.5\%   & 4.5\%   & \multicolumn{2}{c}{8.2\%}              \\
    \hline
    \hline
    \multicolumn{17}{c}{3000 fb$^{-1}$ CMS}                                                                                                                                                                                                   \\
    \hline
    \hline
    $\pTH$ [\UGeV] & \multicolumn{4}{c|}{0-15}  & \multicolumn{4}{c|}{15-30} & 30-45                      & \multicolumn{2}{c|}{45-80} & 80-120 & 120-200 & 200-350 & 350-600 & 600-$\infty$                                     \\
    \hline
    $\hgg$       & \multicolumn{4}{c|}{5.1\%} & \multicolumn{4}{c|}{4.6\%} & 5.1\%                      & \multicolumn{2}{c|}{4.8\%} & 4.9\%  & 4.5\%   & 5.1\%   & 8.6\%   & 32.2\%                                           \\     
    \hline
    $\hzz$       & \multicolumn{4}{c|}{5.4\%} & \multicolumn{4}{c|}{4.8\%} & \multicolumn{3}{c|}{4.1\%} & \multicolumn{2}{c|}{4.7\%} & \multicolumn{3}{c}{9.1\%}                                                               \\ 
    \hline
    $\hbb$       & \multicolumn{14}{c|}{none} & 31.4\%                     & 36.8\%                                                                                                                                            \\ 
    \hline
    Combination  & \multicolumn{4}{c|}{3.7\%} & \multicolumn{4}{c|}{3.3\%} & 4.2\%                      & \multicolumn{2}{c|}{3.7\%} & 4.0\%  & 3.8\%   & 4.4\%   & 8.0\%   & 24.5\%                                           \\
    \hline
    \hline
    \multicolumn{17}{c}{6000 fb$^{-1}$}                                                                                                                                                                                                   \\
    \hline
    \hline
    Combination  & \multicolumn{4}{c|}{2.9\%} & \multicolumn{4}{c|}{2.6\%} & 3.2\% & \multicolumn{2}{c|}{2.9\%} & 3.0\% & 2.9\% & 3.2\% & 5.8\% & 17.9\% \\
    \hline
    \hline
  \end{tabular}
  }
  \caption{Relative uncertainties on the projected $\pTH$ spectrum measurements by ATLAS and CMS under S2 at 3000 fb$^{-1}$. The relative uncertainty of the CMS projection is also given at 6000 fb$^{-1}$ to represent the sensitivity achievable by an eventual ATLAS and CMS combination.}
  \label{tab:proj_pth_unc_scen2}
\end{table}

\begin{figure}
  \centering
  \subfloat[][]{\includegraphics[width=0.48\textwidth]{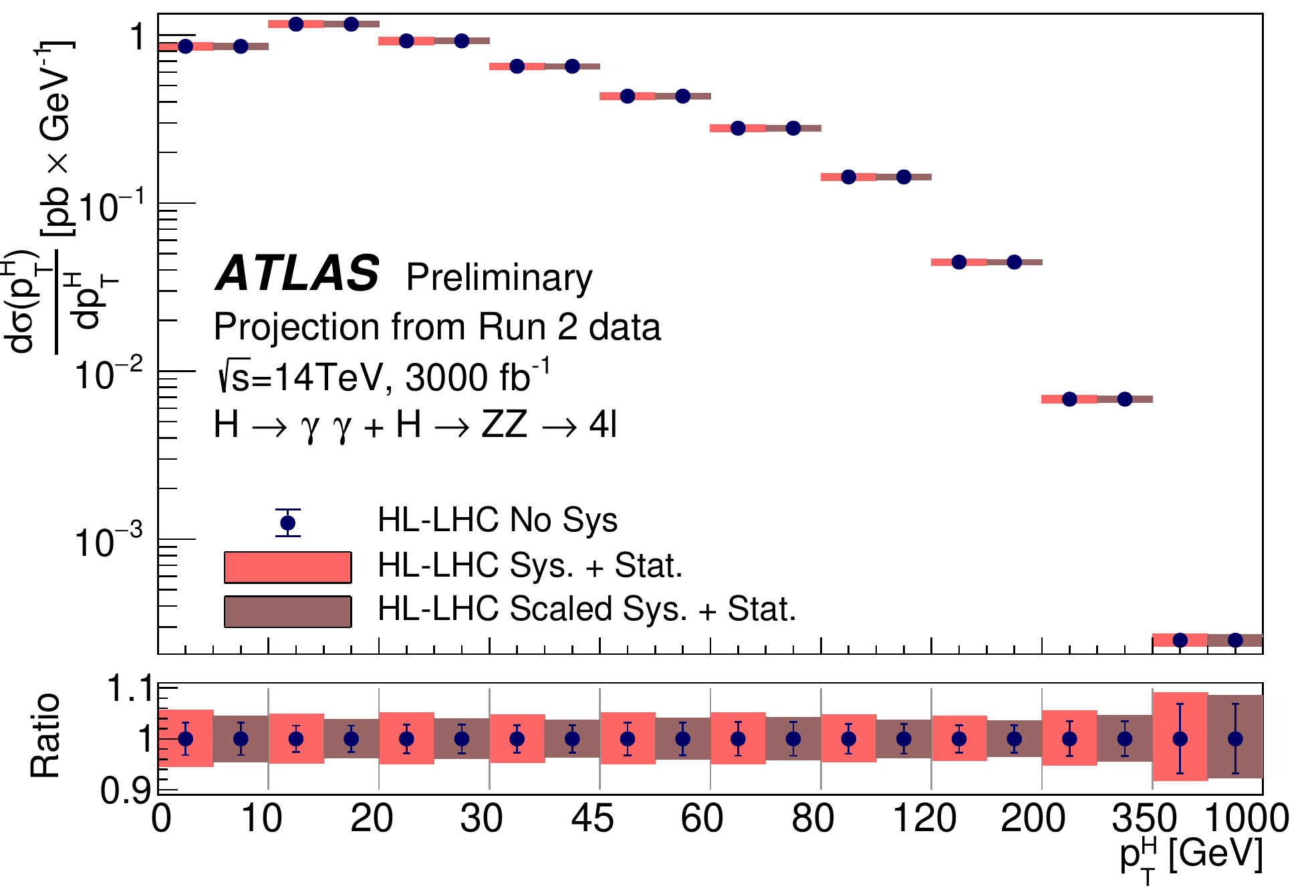}}
  \subfloat[][]{\includegraphics[width=0.48\textwidth]{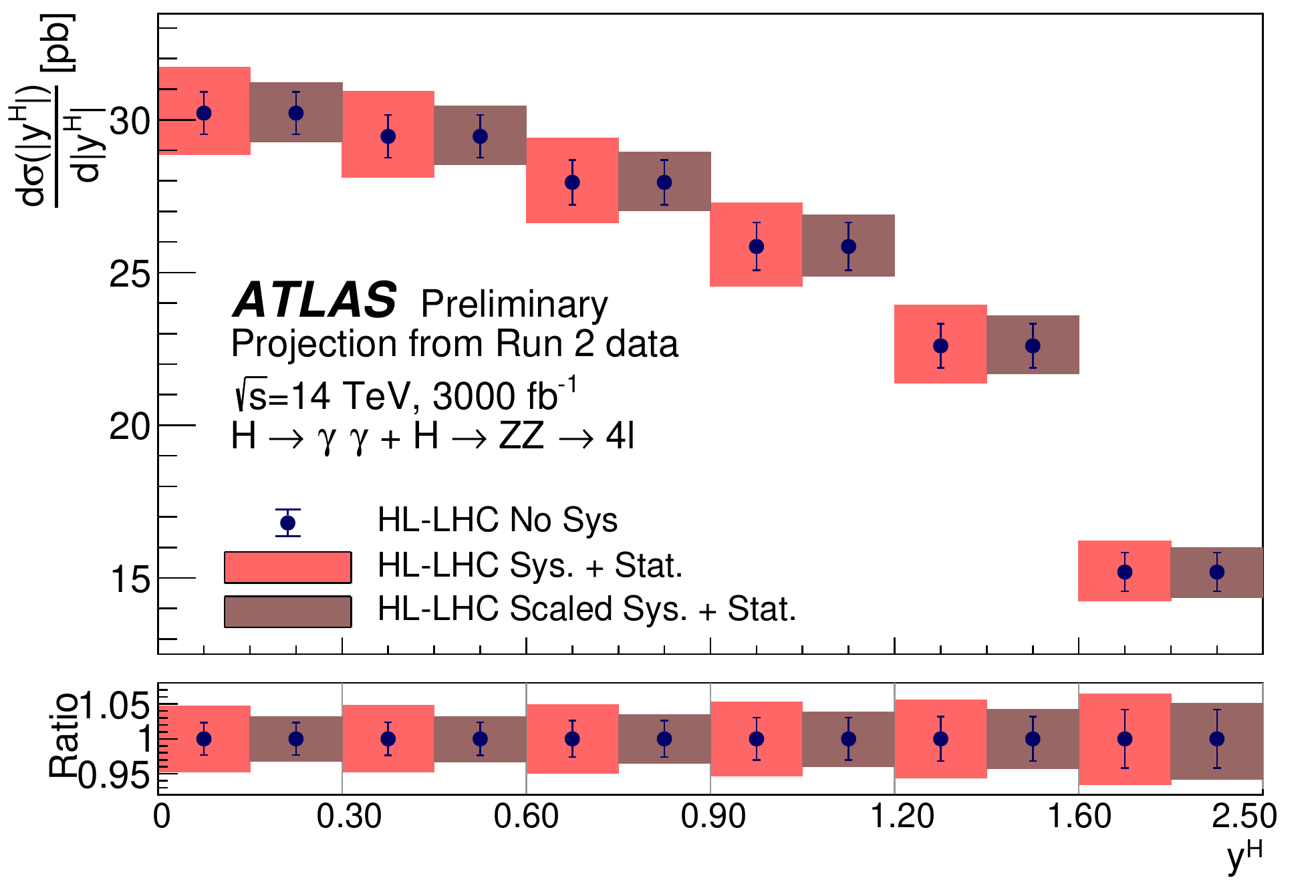}} \\
  \subfloat[][]{\includegraphics[width=0.48\textwidth]{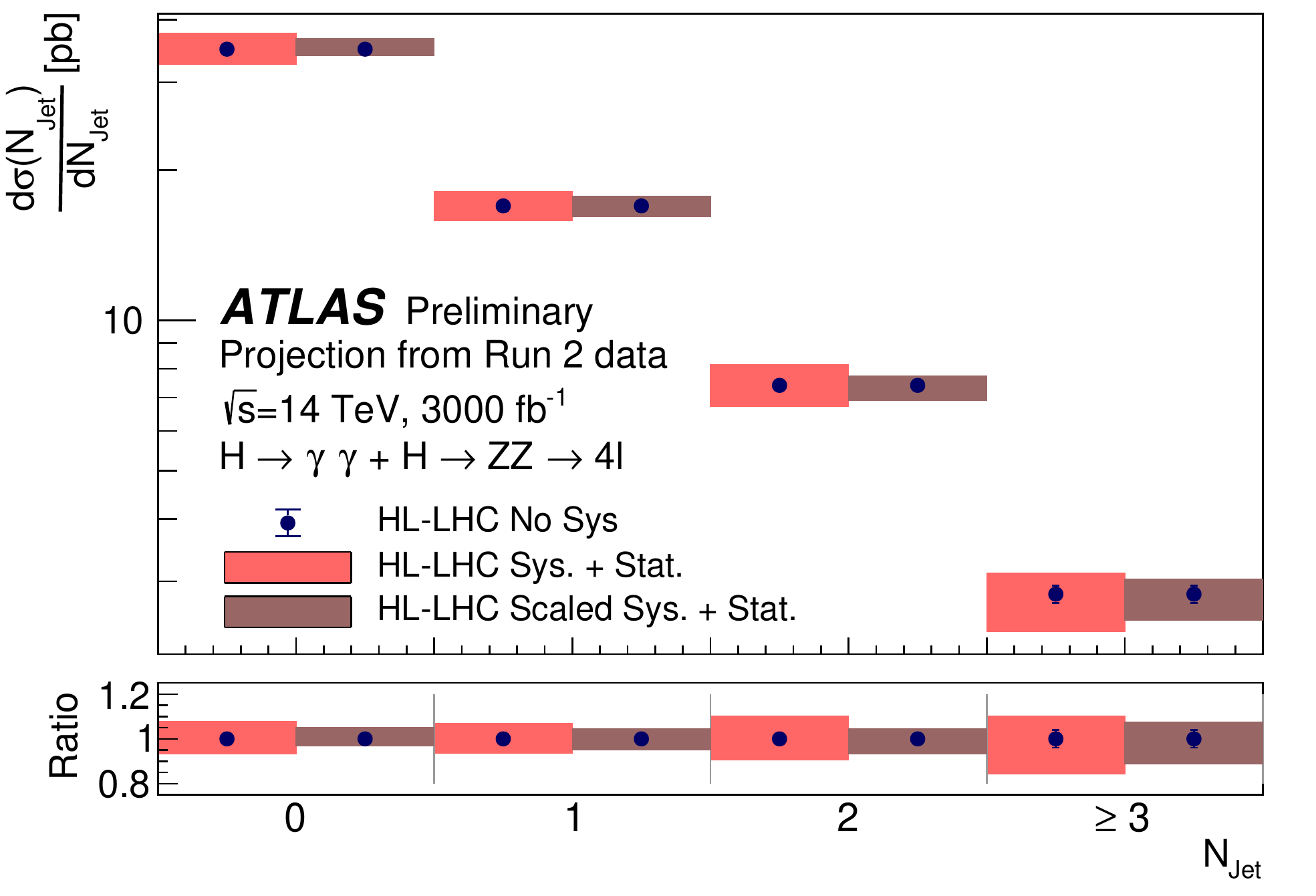}}
  \subfloat[][]{\includegraphics[width=0.48\textwidth]{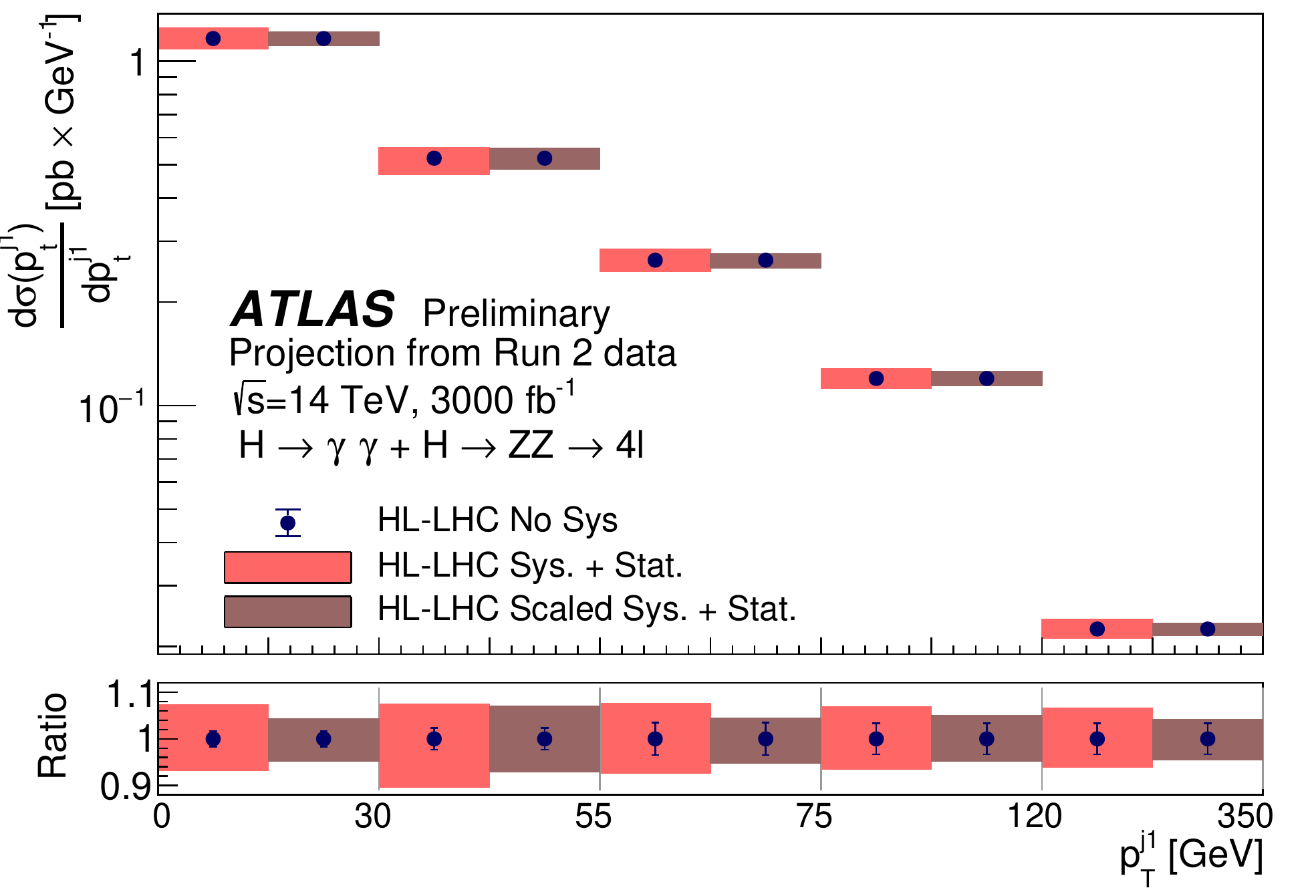}}
  \caption{Differential cross sections measured by ATLAS in the full phase space, extrapolated to the full HL-LHC luminosity for the combination of the \Hyy\ and \HZZ\ decay channels for (a) Higgs boson transverse momentum $\pTH$, (b) Higgs boson rapidity $|y_H|$, (c) number of jets $N_{\mathrm{jets}}$ with  $p_{\mathrm{T}} >$ 30 \UGeV, and (d) the transverse momentum of the leading $p_{\mathrm{H}}^{j1}$. For each point both the statistical (error bar) and total (shaded area) uncertainties are shown. Two scenarios are shown: one with the current Run2 systematic uncertainty (S1) and one with scaled systematic uncertainties (S2).}
   \label{fig:ATLAS_proj_differential}
\end{figure}

Due to a different choice of $\pTH$ binning by ATLAS and CMS, and the lack of a more sophisticated study of the correlation of systematic uncertainties, it was chosen not to combine the projected spectra presented above. Instead, the projections from CMS are scaled to an integrated luminosity of 6000\fbinv, providing a proxy estimate of the overall sensitivity of an eventual combination of measurements by the two experiments. Figure~\ref{fig:proj_pth_6000} shows the CMS projection at 6000\fbinv, with the same systematic scaling as for the projection at 3000\fbinv.  As expected at very high integrated luminosity, the systematic uncertainties dominate the statistical ones.

\begin{figure}
  \centering
  \includegraphics[width=0.49\linewidth]{\main/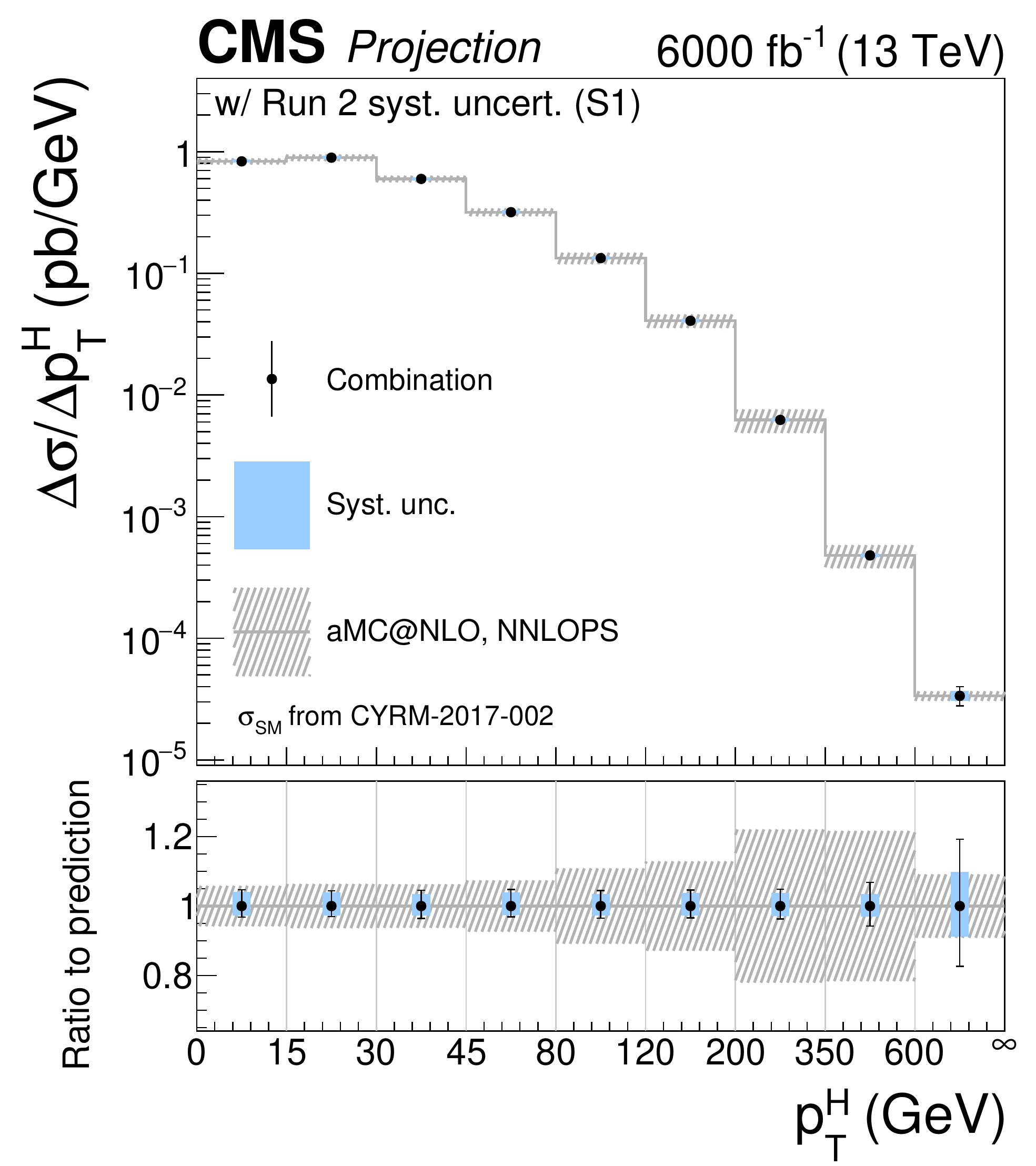}
  \includegraphics[width=0.49\linewidth]{\main/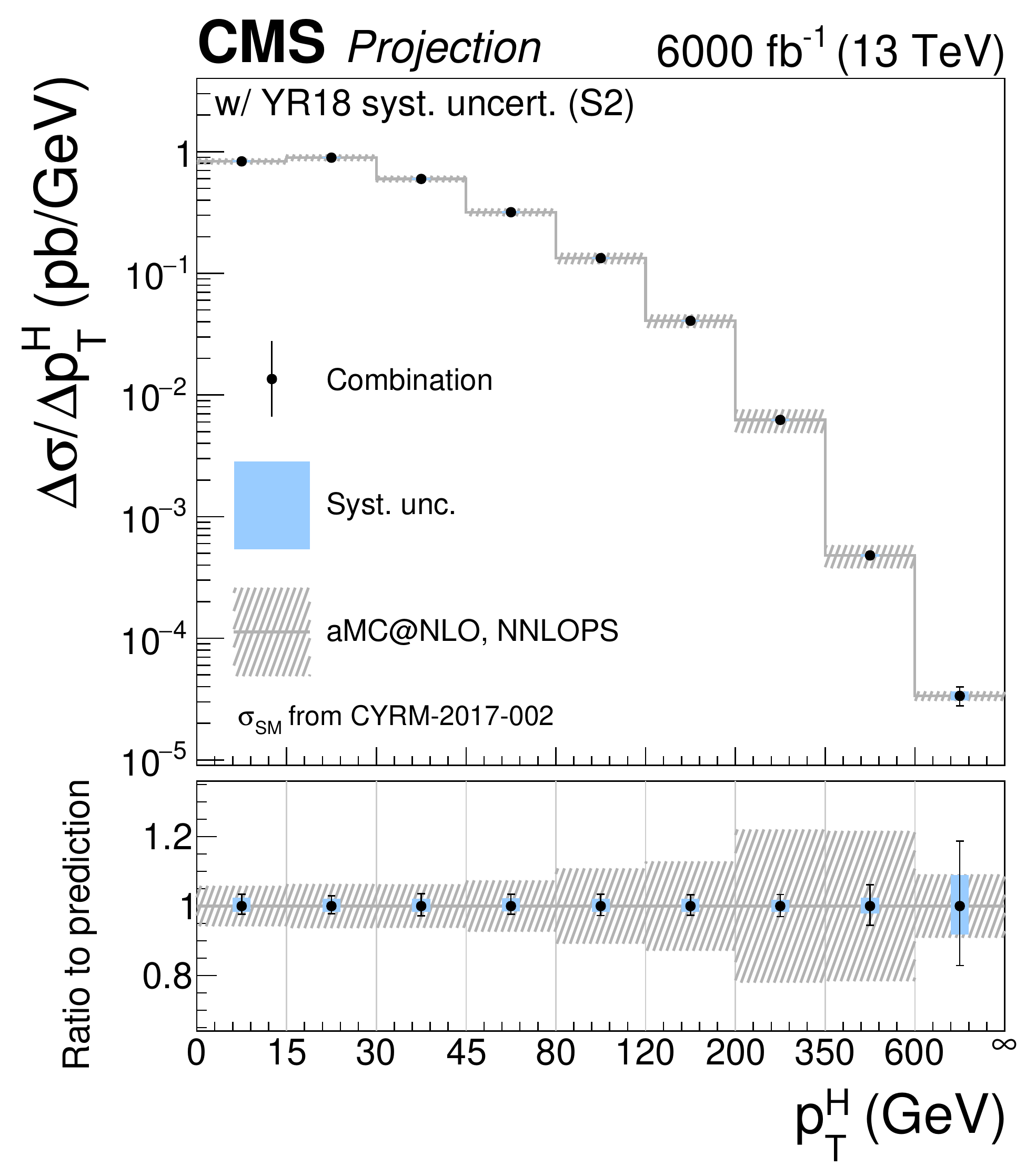}
  \caption{Projected differential cross section for $\pTH$ at an integrated luminosity of 6000\fbinv (representing the sensitivity achievable by an eventual ATLAS and CMS combination), under scenarios S1
    and S2.}
  \label{fig:proj_pth_6000}
\end{figure}

\asubsubsection[N. Wardle, J. Langford]{Measurement of $p_{T}(H)$ spectrum  in \ttH production mode}
\label{sec:ttHdiffxs}

This section describes the strategy for measuring the differential $\pt$ cross section for 
Higgs boson production in association with at least one top quark, and decaying to photons ($\ttH+\tH$, $\hgg$), 
at the High-Luminosity LHC with the CMS Phase-2 detector. The $\hgg$ decay mode provides a final state in which the decay of the Higgs boson can be fully reconstructed, and a direct measurement of the $\pt$ differential cross-section can be made. 

The expected precision of the analysis is determined based on simulated proton-proton ($pp$) events, at a centre of mass energy of 14 \UTeV.
Simulated signal and background events are generated using a combination of {\sc POWHEG}
v2.0~\cite{Alioli:2010xd,Nason:2009ai}, {\sc MADGRAPH5\_AMC@NLO} v2.2.2~\cite{Alwall:2014hca}, {\sc SHERPA} v2.2.5~\cite{Gleisberg:2008ta}, and interfaced with {\sc PYTHIA} v8.205~\cite{Sjostrand:2007gs}. The signal and background events are processed with {\sc DELPHES}~\cite{deFavereau:2013fsa}, using the CMS Phase-2 card, to simulate the response of the upgraded CMS detector to showered particles. Full details of the analysis can be found in Ref.~\cite{CMS-PAS-FTR-18-020}.

\asubsubsubsection{Analysis strategy}

An event selection is applied to the simulated background and signal events following a similar strategy to the CMS Run 2 $\hgg$ strategy \cite{Sirunyan:2018ouh}. The events are required to contain two photons, with $|\eta^\gamma|$~$<$~2.4 excluding the region 1.44~$<$~$|\eta^\gamma|$~$<$~1.57, with an invariant mass satisfying 100~$<$~$m_{\gamma\gamma}$~$<$~180~\UGeV, where the leading-$\pt$ (sub-leading-$\pt$) photon satisfies $\pt^{\gamma}/m_{\gamma\gamma}$~$>$~1/3 (1/4). The two photons are also required to be separated by $\Delta R_{\gamma\gamma}$~$>$~0.4. The photons must also be isolated, which is achieved by requiring that the sum of charged transverse momentum in a cone of radius $\Delta R_{\gamma}$~=~0.4, centred on the photon direction, is less than 0.3$\, \pt^\gamma$. For events where more than one photon pair passes the selection, then the pair with $m_{\gamma\gamma}$ closest to the Higgs boson mass is chosen.

In order to isolate the production of the Higgs boson in association with top quarks, the selection requires all events to have at least one $b-$tagged jet. Such events are separated into two orthogonal categories based on the decay products of the top quark, a hadronic category and a leptonic category. In the hadronic category, events must contain at least 3 jets, clustered using the anti-$k_{T}$ algorithm with a cone size of 0.4, separated by $\Delta R$~$>$~0.4 with respect to both photon candidates. The jets are required to have $\pt>25$ \UGeV and $|\eta|$~$<$~4. In the leptonic category, only 2 jets are required, however, in addition, the events must contain at  least one isolated muon or electron. The muons or electrons must satisfy $\pt$~$>$~20~\UGeV and $|\eta|$~$<$~2.4, excluding the region 1.44~$<$~$|\eta^\gamma|$~$<$~1.57 for electrons. The muons must satisfy an isolation requirement that the sum of all reconstructed particles $\pt$, inside a cone of radius $\Delta R=0.4$, excluding the muon itself, is less than 0.25 times the transverse momentum of the muon. In addition, for electrons, the invariant mass of pairs formed from the electron and either selected photon, $m_{e\gamma}$, is required to be greater than 95 \UGeV to reduce contamination from $Z\rightarrow e^{+}e^{-}$ decays. Events passing the leptonic category selection are excluded from the hadronic selection to maintain orthogonality of the two categories.  
For the signal extraction, boosted decision tree (BDT) classifiers are trained independently in each channel, which distinguish between signal-like and background-like events, using input variables related to the kinematics of the events, such as the lepton and jet momenta and pseudo-rapidities, and the scalar sum of transverse momentum of all final state objects in the event. Events are required to have output BDT values greater than fixed thresholds, which are tuned to provide the best sensitivity to $\kappa_{\lambda}$. The hadronic category is further split into two different regions of BDT output, for events with di-photon transverse momentum ($\pt^{\gamma\gamma}$) less than 350 \UGeV, to reduce the contamination of gluon fusion Higgs boson production. 

Finally, the events are further divided into six bins of $\pt^{\gamma\gamma}$, given in Tab.~\ref{tab:ttHdiff_CMS_ptbins}, making a total of 17 categories. 

\begin{table}[h]

 \centering
 \begin{tabular}{c|c|c|c|c|c|c}
    \hline
    \hline
    \multicolumn{7}{c}{$\pt^{H}$ or $\pt^{\gamma\gamma}$ bin boundaries (\UGeV)}  \\ \hline
    0 & 45 & 80 & 120 & 200 & 350 & $\infty$ \\
    \hline
    \hline
\end{tabular}
\caption{bin boundaries which define the $\pTH$ regions for which the differential cross sections are measured. These also correspond to the bins in which the hadronic and leptonic event categories are sub-divided.}
\label{tab:ttHdiff_CMS_ptbins}
\end{table}

Experimental systematic uncertainties are included in the signal model, which can cause migration both between the different categories and in and out of the fiducial region. The dominant uncertainties are related to the reconstruction and identification efficiencies for photons and b jets as well as the energy scale and resolution of reconstructed jets. 
Furthermore, theoretical uncertainties are included on the rates of $\ggh$ and $\vh$ contamination, which modify both the overall normalisation and the relative contamination between the different categories for these processes. The background estimation follows the same strategy as in the CMS Run 2 $\hgg$ analysis~\cite{Sirunyan:2018ouh}, in that the parameters of the background functions are free to float in the fit, and constrained directly from the data. Therefore the uncertainties on the background will be statistical in nature.
However, the impact of increasing the rate of fake photons in the background component has been studied and was found to reduce the sensitivity to $\kappa_\lambda$ by roughly 10\% in the worst case scenario~\cite{CMS-PAS-FTR-18-020}.  

\asubsubsubsection{Differential cross-section results}

In order to account for resolution effects, the signal events are separated based on the $\pt^H$ at generator level.   Signal and background models are constructed using the simulated events in each category. The signal model accounts for the relative populations of events from the different production processes as well as from different $\pTH$ bins, and the di-photon mass resolution expected from events in each category. The background model is constructed from a fit of smoothly falling functions to the weighted sum of simulated background samples, accounting for the different fake photon rates for each source of background and normalised to the  total background yield expected in 3000 fb$^-1$ of High-Luminosity LHC data. The differential cross-section is determined from a simultaneous maximum likelihood fit to an Asimov data set~\cite{Cowan:2010js} corresponding to $3\abinv$, and assuming SM Higgs boson  production in each category.  Systematic uncertainties are accounted for through the introduction of constrained nuisance parameters in the log-likelihood, which are profiled. 

The results of this fit are given in figure~\ref{fig:ttHdiff_CMS_ptH_xs}. The results shown are unfolded back to a fiducial region which is common to both the hadronic and leptonic selections, and shown using only the hadronic or leptonic categories, and their combination. The theoretical uncertainties displayed on the predicted $\ttH+\tH$ cross section are calculated by modifying the renormalisation and factorisation scales up and down by a factor of 2.

\begin{figure}[htb!]
        \centering
        \includegraphics[width=0.6\textwidth]{\main/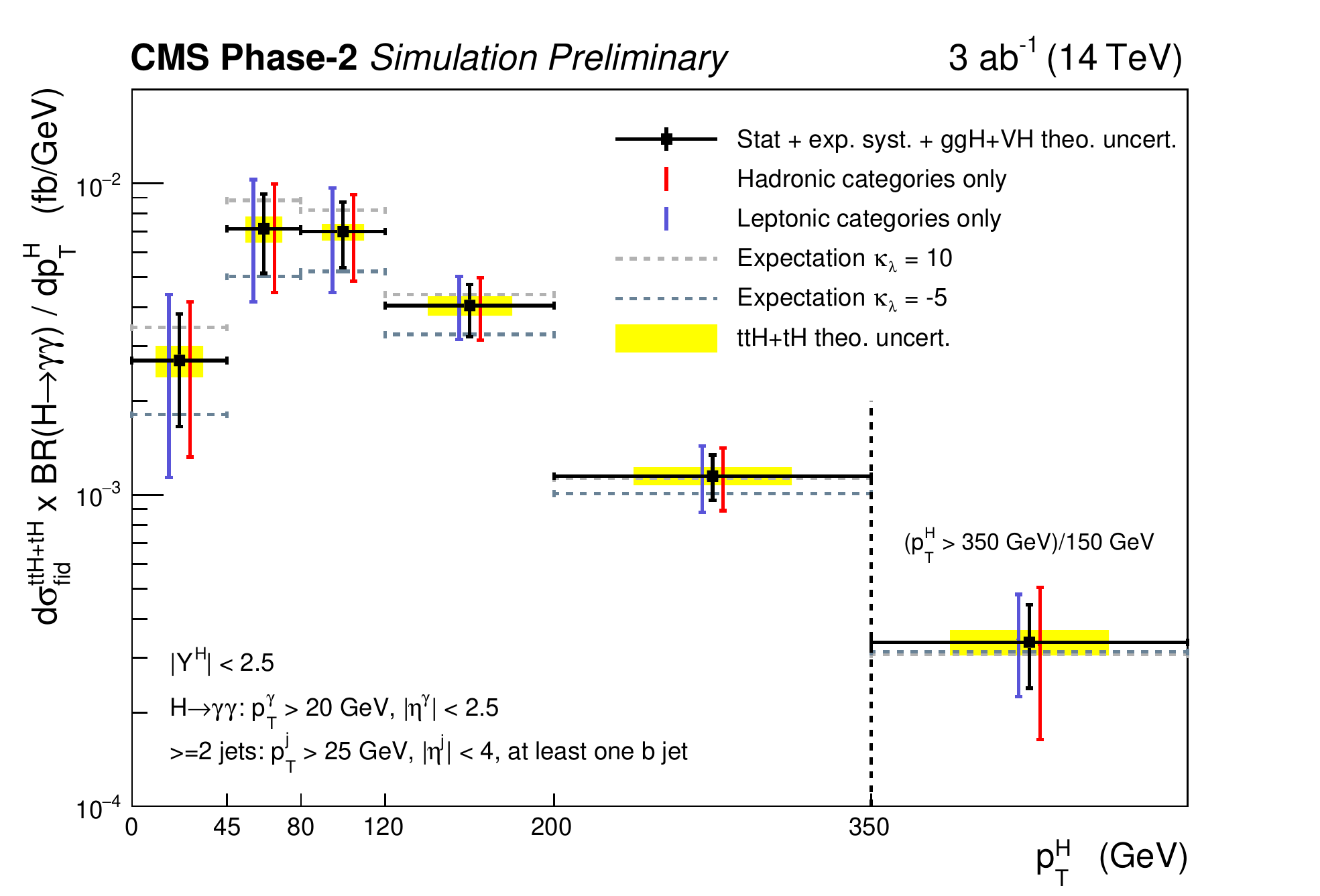}
        \caption{The expected $\pt^H$ differential $\ttH$~+~$\tH$ cross sections times branching ratio, along with their uncertainties ~\cite{CMS-PAS-FTR-18-020}. The error bars on the black points include the statistical uncertainty, the experimental systematic uncertainties and the theoretical uncertainties related to the $\ggh$ and $\vh$ contamination, which is subtracted in the fit.  The cross section for $\pTH > 350$ \UGeV is scaled by the width of the previous bin. The expected $\ttH$~+~$\tH$ cross sections for anomalous values of the Higgs boson self-coupling ($\kappa_\lambda$~=~10 and $\kappa_\lambda$~=~-5) are shown by the horizontal dashed lines.}
        \label{fig:ttHdiff_CMS_ptH_xs}
\end{figure}


\asubsection[A. Calandri,  P. Das, K. El Morabit, S. Folgueras, S. Gadatsch,  A. Gilbert, P. Keicher, T. Klijnsma, K. Mazumdar, M. Schr\"oder]{Direct and indirect probing of top Yukawa coupling}
\label{sec2:top}
\asubsubsection[A. Calandri, M. Schr\"oder]{Measurements in \ttH and \tH production modes}

One of the main targets of the High-Luminosity LHC upgrade is to achieve precision measurements of the Higgs boson properties.
The Yukawa coupling of the Higgs boson to the top quark is expected to be of the order of unity and could be partially sensitive to effects beyond the Standard Model.
Therefore, a direct measurement of the coupling of the Higgs boson to top quarks is extremely important to access possible deviations in the top quark's Yukawa couplings due to couplings to new particles.
Such a measurement can be performed by measuring the rate of the process where the Higgs boson is produced in association with a pair of top quarks (\ttH) or a single top quark (\tH).
Even though the \ttH process is characterised by a small cross section compared to the dominant gluon fusion Higgs boson production (approximately two orders of magnitude smaller), the signature with top quarks in the final state can be exploited to reconstruct the event and gives access to many Higgs boson decay modes.
The SM \tH production cross-section is yet smaller by a factor five, but due to interference effects between diagrams with top-Higgs and W-boson-Higgs couplings, the process allows access to the sign of the top-Higgs Yukawa coupling.
The ATLAS and CMS Collaborations have searched for the \ttH and \tH production with LHC Run 2 data of 2015, 2016, and 2017, and observed the Higgs boson production in association with a top-quark pair~\cite{Aaboud:2018urx,Sirunyan:2018hoz}.
The analyses are sensitive to a large variety of final-state event topologies, H$\rightarrow$WW$^{*}$, H$\rightarrow$ZZ$^{*}$, H$\rightarrow \tau^{+}\tau^{-}$, \Htobb and H$\rightarrow \gamma\gamma$.
Dedicated multivariate analysis techniques, including boosted decision trees and deep neural networks, that combine the information of several discriminating variables, as well as classifiers based on a matrix element method are utilised to identify the signal against the background.

In this Section, projections based on dedicated analyses with 36\fbinv of Run-II data of 2016 are presented, which target the \ttH, \Htobb channel with leptonic decays of the \ttbar system~\cite{Aaboud:2017rss,Sirunyan:2018mvw} and the \ttH multi-lepton final state~\cite{Aaboud:2017jvq}, where the Higgs boson decays into a pair of Z and W vector bosons or into $\tau$ leptons.
Furthermore, results are presented for the projection of a search for \tH production that considers all of the above decay channels~\cite{Sirunyan:2018lzm}.

\asubsubsubsection{Sensitivity to \ttH production in the \bbbar and multi-lepton final states}

The \ttH analyses in the \Htobb final state benefit from the large branching ratio.
At the same time, the relatively poor b jet energy resolution, the large jet combinatorics, and the sizeable background of SM processes with large modelling uncertainties, in particular \ttbar+heavy-flavour jet (\ttHF) production, pose major challenges.
The expected relative precision of the \ttH, \Htobb cross-section measurement reaches the level of 20\%-14\% for the ATLAS analysis and 15\%-11\% for the CMS analysis ~\cite{ATLAS-PHYS-PUB-2018-XY,CMS-PAS-FTR-18-011} corresponding to 7\%-11\% relative uncertainty on the signal strength ($\mu$), depending on the scenario and the assumptions of the \ttHF background modelling, as detailed below.

Table~\ref{tab:tthbb:breakdown:cms} shows a breakdown of the contributing sources of uncertainty in the CMS analysis; their evolution with integrated luminosity is depicted in Fig.~\ref{fig:tthbb:evolution:cms}.
Compared to the result at 35.9\fbinv, the relative contribution of the experimental uncertainties, such as the b-tagging uncertainty, remains approximately the same, while the signal-theory uncertainty component increases and becomes the major uncertainty component, mostly driven by the inclusive cross-section uncertainty on the SM prediction entering $\mu$.
The statistical uncertainty becomes small compared to the systematic components.
A similar behaviour is observed in the ATLAS analysis.
\begin{table}
  \centering
  \caption{
    Breakdown of the contributions to the expected uncertainties on the \ttH signal-strength $\mu$ in the \Htobb channel at different luminosities for the scenarios S1 and S2 at CMS.
    The uncertainties are given in percent relative to $\mu=1$.
    Results with 35.9\fbinv are intended for comparison with the projections to higher luminosities and differ in parts from~\cite{Sirunyan:2018mvw} for consistency with the projected results: uncertainties due to the limited number of Monte Carlo statistics have been omitted and the assumptions in S1/S2 on the theory uncertainties are applied.
  }
  \label{tab:tthbb:breakdown:cms}
  \begin{tabular}{l rrr rrr}
    \hline
    & \multicolumn{2}{c}{S1} & \multicolumn{2}{c}{S2} \\
    Source & \multicolumn{1}{c}{35.9\fbinv} & \multicolumn{1}{c}{3000\fbinv} & \multicolumn{1}{c}{35.9\fbinv} & \multicolumn{1}{c}{3000\fbinv} \\
    \hline
    Total            & 48.7 &11.1 & 46.1 & 7.3 \\
    Stat             & 26.7 & 2.9 & 26.7 & 2.9 \\
    SigTh            & 10.8 & 8.7 &  5.0 & 4.4 \\
    BkgTh            & 28.6 & 4.1 & 25.6 & 3.5 \\
    ~~~\ttHF XS      & 14.6 & 0.8 & 16.5 & 0.7 \\
    Exp              & 17.4 & 4.2 & 16.6 & 2.6 \\
    ~~~Luminosity    &  1.6 & 1.8 &  0.5 & 0.8 \\
    ~~~B tagging     & 12.0 & 2.8 & 10.8 & 1.6 \\
    ~~~JES           & 10.9 & 1.6 & 11.3 & 1.6 \\
    \hline
  \end{tabular}
\end{table}

\begin{figure}
  \centering
  \begin{tabular}{@{}c@{}c@{}}
    \includegraphics[width=0.49\textwidth]{\main/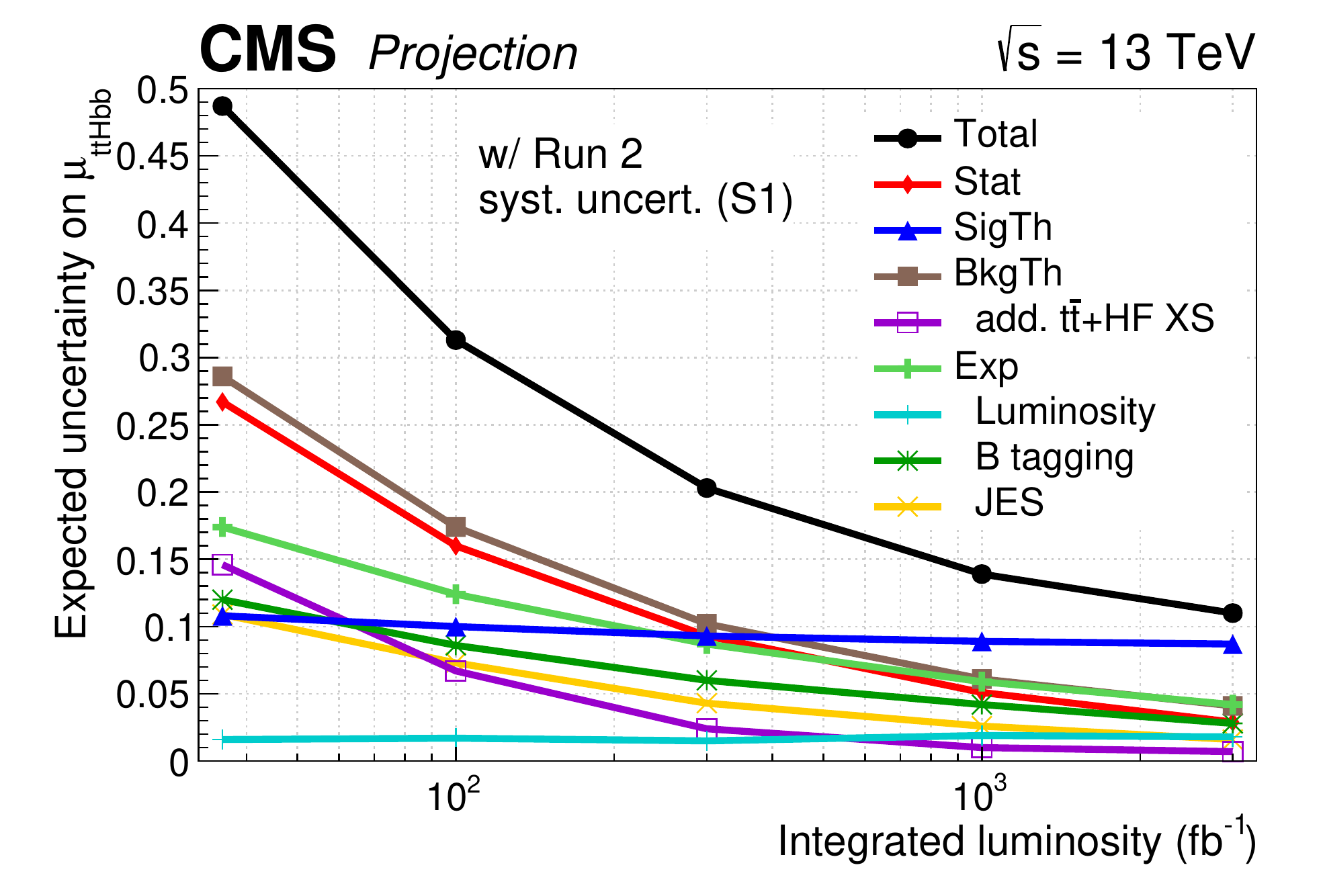} &
    \includegraphics[width=0.49\textwidth]{\main/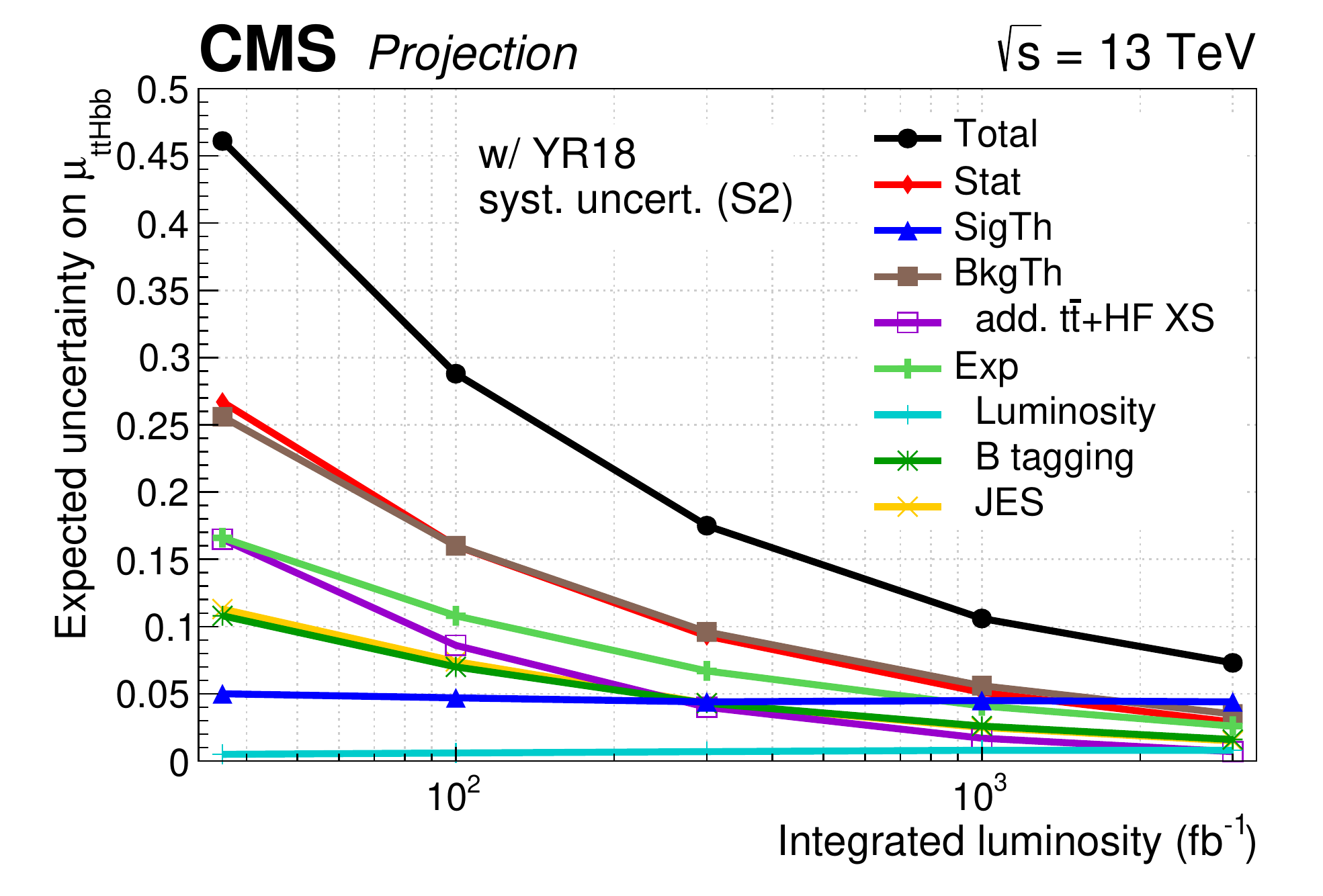} \\
  \end{tabular}
  \caption{
    Expected uncertainties on the \ttH signal strength in the \Htobb channel as a function of the integrated luminosity under the S1 (left) and S2 (right) scenarios at CMS.
    Shown are the total uncertainty (black) and contributions of different groups of uncertainties.
    Results with 35.9\fbinv are intended for comparison with the projections to higher luminosities and differ in parts from~\cite{Sirunyan:2018mvw} for consistency with the projected results: uncertainties due to the limited number of MC events have been omitted and the assumptions in S1/S2 on the theory uncertainties are applied.  
   }
   \label{fig:tthbb:evolution:cms}
\end{figure}

\begin{figure}
  \centering
  \begin{tabular}{@{}c@{}c@{}}
    \includegraphics[width=0.49\textwidth]{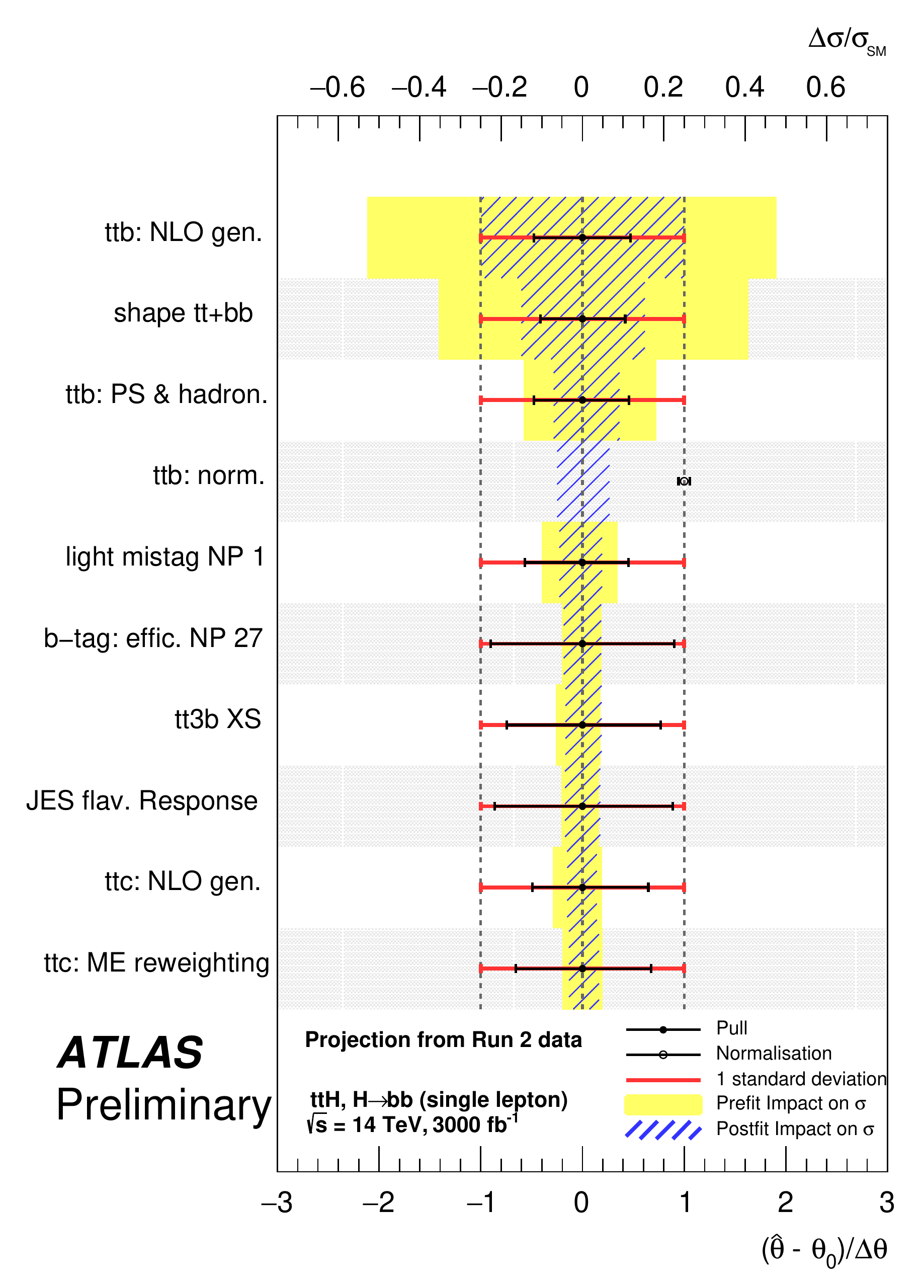} &
    \includegraphics[width=0.49\textwidth]{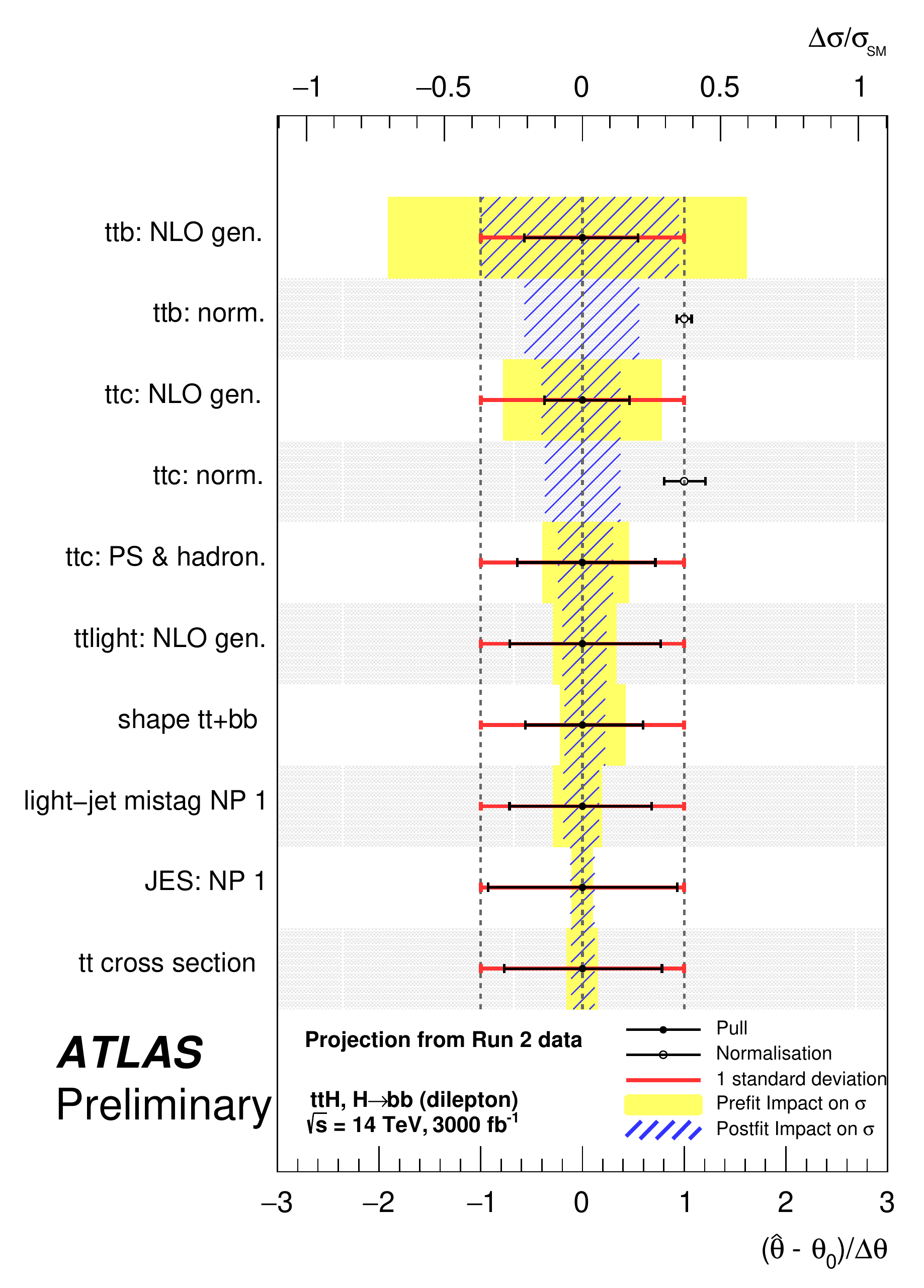} \\
  \end{tabular}  
  \caption{
    Ranking of the ten most significant systematics uncertainties under S2 in the single lepton (a) and di-lepton (b) final states at ATLAS listed in accordance to their post-fit impact on the \ttH cross section.
  }
  \label{fig:tthbb:impacts:atlas}
\end{figure}

In both analyses, a rather sizeable reduction of the uncertainties related to the modelling of the \ttHF background, which relies on MC simulation, is observed.
Relevant nuisance parameters are constrained to a few percent, such as the nuisance parameters describing the difference between four and five-flavour scheme calculations which is treated as a 2-point systematic uncertainty in the ATLAS analysis (Fig.~\ref{fig:tthbb:impacts:atlas}) or the nuisance parameters describing the additional \ttHF cross-section uncertainties in the CMS analysis (Table~\ref{tab:tthbb:breakdown:cms} and Fig.~\ref{fig:tthbb:evolution:cms}).
This is attributed to the increasing power of the profile likelihood fit to constrain the uncertainties.

The results illustrate that the background modelling, which has been designed to work well with 35.9\fbinv of data, will need to be refined at 3000\fbinv, requiring improved simulations or in-situ measurements of the \ttHF processes themselves.
The observed constraints on the \ttHF background model systematics uncertainties shown in Fig.~\ref{fig:tthbb:impacts:atlas} demonstrate that there will be enough data at the HL-LHC to obtain further information about the background beyond the current modelling. The level at which the nuisance parameters are constrained at 3000\fbinv, corresponding to a few percent cross-section uncertainty, demonstrate the level of sensitivity at which the data will be able to distinguish different models and sets a benchmark for the required precision.
Monte Carlo prediction will thus need to improve sufficiently to match the data within the uncertainties expected at 3000\fbinv.

Following the expected improvement by a factor two to three in the theoretical uncertainties on the \ttHF cross-section calculation described in Section~\ref{sec:hl-lhc-ttH}, ATLAS and CMS have also performed the \ttH, \Htobb extrapolation assuming that the reduction of the \ttHF modelling uncertainties is limited to factors of two (in scenario S1) and three (in scenario S2) relative to the uncertainty at 35.9\fbinv.
In this case, the obtained relative \ttHF modelling uncertainties are approximately 23\% (S1) and 15\% (S2) in the ATLAS analysis as reported in Table~\ref{tab:tthbb:breakdown:atlas} and approximately 15\% (S1) and 10\% (S2) in the CMS analysis. These results enter the combined coupling measurement presented in Sections~\ref{sec2:exp_combination} and~\ref{sec2:exp_kappa}. The impact of limiting the constraints of the \ttHF uncertainties on the total uncertainties on the extracted parameters is relatively small, e.g.\ the uncertainty on $\kappa_{\text{t}}$ increases by approximately 10\% and 15\% in CMS and ATLAS, respectively. 
\begin{table}
  \centering\small
\renewcommand{\arraystretch}{1.3}
  \begin{tabular}{c|c|c|cccc|c}
    \hline \hline
    Final state & Scenario & $\Delta_{\textrm {tot}}/\sigma_{\textrm {SM}}$ & $\Delta_{\textrm {stat}}/\sigma_{\textrm {SM}}$ & $\Delta_{\textrm {exp}}/\sigma_{\textrm {SM}}$ & $\Delta_{\textrm {sig}}/\sigma_{\textrm {SM}}$ & $\Delta_{\textrm {bkg}}/\sigma_{\textrm {SM}}$
 & $\Delta\mu_{\textrm {sig}}$ \\
    \hline
    $t\bar{t} H,H \rightarrow$ $b\bar{b}$ & Run 2, 36~\ifb  & $^{+0.61}_{-0.61}$ & $^{+0.22}_{-0.22}$ & $^{+0.27}_{-0.28}$ & $^{+0.10}_{-0.09}$ & $^{+0.47}_{-0.47}$ & $^{+0.15}_{-0.15}$ \\
    (single lepton) & HL-LHC S1   & $^{+0.25}_{-0.20}$      & $^{+0.02}_{-0.02}$     & $^{+0.10}_{-0.10}$      & $^{+0.08}_{-0.06}$ &  $^{+0.22}_{-0.17}$ & $^{+0.10}_{-0.11}$ \\
    & HL-LHC S2  & $^{+0.18}_{-0.15}$      & $^{+0.02}_{-0.02}$     & $^{+0.09}_{-0.09}$      & $^{+0.06}_{-0.05}$ &$^{+0.14}_{-0.11}$  & $^{+0.08}_{-0.07}$ \\
    \hline
    $t\bar{t} H,H \rightarrow$ $b\bar{b}$   & Run 2, 36~\ifb  & $^{+1.06}_{-1.08}$ & $^{+0.51}_{-0.51}$ & $^{+0.32}_{-0.31}$ & $^{+0.11}_{-0.12}$ & $^{+0.90}_{-0.92}$  & $^{+0.14}_{-0.14}$ \\
  (di-lepton) & HL-LHC S1             & $^{+0.32}_{-0.26}$      & $^{+0.06}_{-0.06}$      & $^{+0.13}_{-0.13}$      & $^{+0.08}_{-0.07}$ & $^{+0.27}_{-0.21}$  & $^{+0.11}_{-0.09}$ \\
    & HL-LHC  S2               & $^{+0.23}_{-0.20}$      & $^{+0.06}_{-0.06}$     & $^{+0.11}_{-0.11}$      & $^{+0.06}_{-0.06}$ & $^{+0.17}_{-0.15}$ & $^{+0.08}_{-0.08}$ \\
    \hline\hline
\end{tabular}  
  \caption{
    Breakdown of the contributions to the expected uncertainties on the \ttH cross section in the \Htobb channel at different luminosities for the scenarios S1 and S2 at ATLAS. As discussed in the text, the extrapolation assumes the limitations on the reduction of the \ttHF modelling to a factor 2 and a factor 3 of the Run 2 prior uncertainties (Section \ref{sec:hl-lhc-ttH}).
    Therefore, the additional modelling uncertainty used for the extrapolation is 23\% in S1 and 15\% in S2.
    Uncertainties due to the limited number of Monte Carlo statistics have been omitted and the assumptions in S1/S2 on the theory uncertainties are applied.  
  }
  \label{tab:tthbb:breakdown:atlas}
\end{table}

Studies in the channel of the boosted regime using substructure
techniques, where the Higgs boson has significant transverse momentum, 
have been carried out and reported in~\cite{Costa:2018fdy}. The main background remains the top pair 
production in association with additional heavy flavour quarks, however the boosted 
regime allows to reduce the combinatorial background and therefore the use of 
side-bands in the invariant mass distribution. This study confirms the statistical power 
of the analysis and the side-bands, but would require a more detailed study of the 
background systematic uncertainties to be fully compared with the current projected 
results based on the LHC Run 2 analysis.

In conclusion, \ttH production in the \Htobb final state will provide a powerful channel to probe the top-Higgs Yukawa coupling at the HL-LHC.
The control of the \ttHF background is crucial, and it is expected to benefit from measuring relevant quantities from data, thus mitigating the impact of theoretical uncertainties.

ATLAS performs the extrapolation to HL-LHC also for the ttH multi-lepton (ML) final state~\cite{ATLAS-PHYS-PUB-2018-XY} where the Higgs boson decays into a pair of Z and W vector bosons or into a pair of $\tau$ leptons. Table \ref{tab:tthml:breakdown:atlas} shows the results on the extrapolation to 3000 fb$^{-1}$ under S1 and S2. As shown in the ranking plot in Figure \ref{fig:tthml:impacts:atlas}, in the $\tau$ final state, the dominant uncertainty pertains to the object reconstruction for such a channel. It is also worth noting that the main theoretical systematic uncertainties concerns the modelling of the tt+V background. Finally, fake lepton uncertainties are moderately constrained as well: this is due to the absence of a reduction factor of prior uncertainties for such a source of systematic uncertainties under S1 and S2. 
\begin{table}
  \centering\small
  \renewcommand{\arraystretch}{1.3}
  \begin{tabular}{c|c|c|cccc|c}
    \hline \hline
    Final state & Scenario & $\Delta_{\textrm {tot}}/\sigma_{\textrm {SM}}$ & $\Delta_{\textrm {stat}}/\sigma_{\textrm {SM}}$ & $\Delta_{\textrm {exp}}/\sigma_{\textrm {SM}}$ & $\Delta_{\textrm {sig}}/\sigma_{\textrm {SM}}$ & $\Delta_{\textrm {bkg}}/\sigma_{\textrm {SM}}$
 & $\Delta\mu_{\textrm {sig}}$ \\
    \hline
    $t\bar{t} H,H \rightarrow$ ML   & Run 2, 36~\ifb & $^{+0.40}_{-0.40}$ & $^{+0.33}_{-0.34}$ & $^{+0.15}_{-0.15}$ & $^{+0.10}_{-0.10}$ & $^{+0.13}_{-0.13}$ & $^{+0.13}_{-0.13}$ \\
    (no $\tau$) & HL-LHC S1 & $^{+0.18}_{-0.18}$ & $^{+0.04}_{-0.04}$ & $^{+0.13}_{-0.14}$ & $^{+0.08}_{-0.08}$ & $^{+0.12}_{-0.12}$ & $^{+0.11}_{-0.11}$ \\
    & HL-LHC S2& $^{+0.17}_{-0.17}$      & $^{+0.04}_{-0.04}$     & $^{+0.12}_{-0.13}$      & $^{+0.05}_{-0.05}$ & $^{+0.09}_{-0.09}$ & $^{+0.07}_{-0.07}$ \\
    \hline
    $t\bar{t} H,H \rightarrow$ ML  & Run 2, 36~\ifb  & $^{+0.64}_{-0.64}$ & $^{+0.54}_{-0.54}$ & $^{+0.29}_{-0.29}$ & $^{+0.10}_{-0.09}$ & $^{+0.14}_{-0.13}$  & $^{+0.13}_{-0.13}$ \\
    (with $\tau$) & HL-LHC S1             & $^{+0.27}_{-0.28}$      & $^{+0.07}_{-0.07}$     & $^{+0.23}_{-0.23}$      & $^{+0.09}_{-0.08}$ & $^{+0.12}_{-0.12}$ & $^{+0.11}_{-0.11}$ \\
    & HL-LHC S2           & $^{+0.25}_{-0.25}$      & $^{+0.07}_{-0.07}$     & $^{+0.22}_{-0.22}$      & $^{+0.05}_{-0.05}$ &  $^{+0.07}_{-0.07}$ & $^{+0.07}_{-0.07}$ \\
    \hline\hline
\end{tabular}  
  \caption{
    Breakdown of the contributions to the expected uncertainties on the \ttH cross section in the multi-lepton channel at different luminosities for the scenarios S1 and S2 at ATLAS. Uncertainties due to the limited number of Monte Carlo statistics have been omitted and the assumptions in S1/S2 on the theory uncertainties are applied.
  }
  \label{tab:tthml:breakdown:atlas}
\end{table}
\begin{figure}
  \centering
    \begin{tabular}{@{}c@{}c@{}}
    \includegraphics[width=0.49\textwidth]{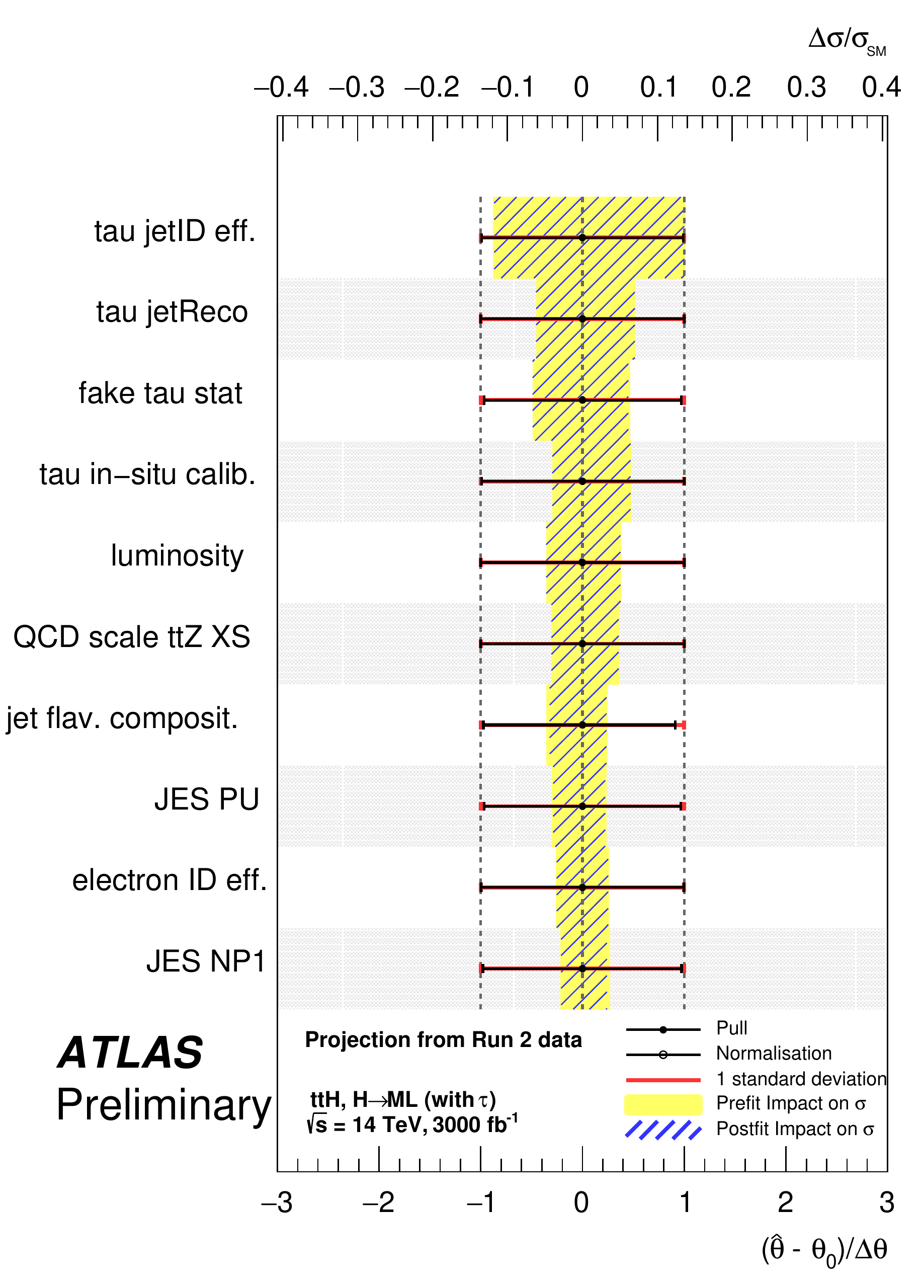} &
    \includegraphics[width=0.49\textwidth]{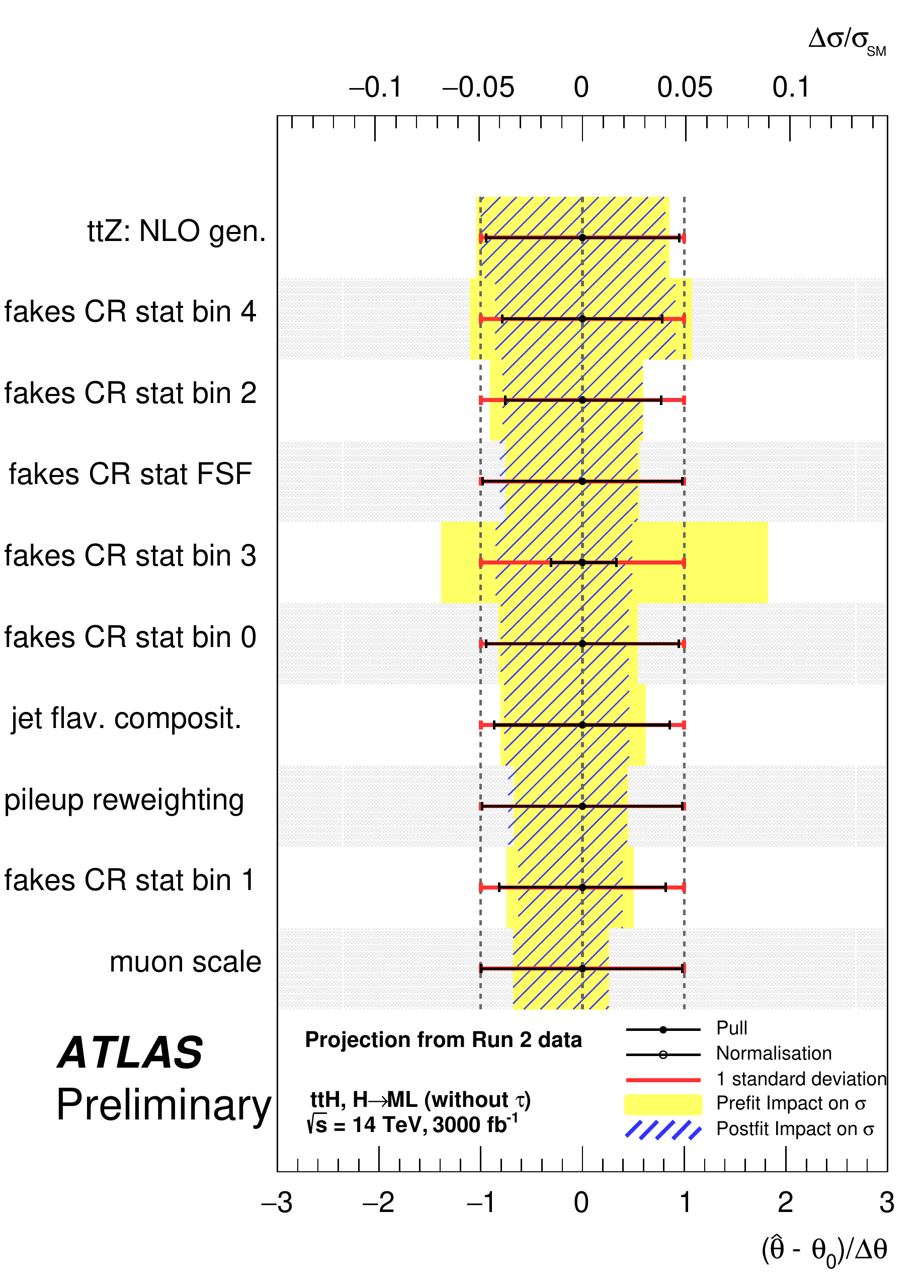} \\
  \end{tabular}  
  \caption{
    Ranking of the ten most significant systematic uncertainties under S2 in the \ttH multi-lepton (ML) final state with (a) and without (b) $\tau$ leptons in the ATLAS analysis listed in accordance to their post-fit impact on the \ttH cross section.
  }
  \label{fig:tthml:impacts:atlas}
\end{figure}

\asubsubsubsection[K. Mazumdar, P. Das]{Sensitivity to \tH production}

The sensitivity to the $\tH$ process at the HL-LHC is determined by extrapolating a combination of Run~2 analyses based on $35.9\fbinv$ of data at $\sqrt{s}= 13 \UTeV$~\cite{Sirunyan:2018lzm}. Two of these analyses are dedicated searches for $\tHq$: one targets a multi-lepton final state in which the Higgs boson decays to $\ww$, $\zz$ or $\tautau$ pairs, and the other targets the $\hbb$ decay. In both analyses the presence of at least one central b tagged jet and an isolated lepton from the top quark decay is required. Furthermore, the presence of a light quark jet at high pseudorapidity, a unique feature of the $\tHq$ production mode, is exploited. Both analyses also rely heavily on multivariate techniques to discriminate the signal against the large $\ttjets$ background. The $\gamma\gamma$ final state is also utilised, via a reinterpretation of the inclusive $\hgg$ analysis~\cite{Sirunyan:2018ouh}. In this analysis the $\mathrm{tHq}$ and $\mathrm{tHW}$ processes primarily contribute to the``$\mathrm{t\bar{t}H}$ leptonic'' and ``$\mathrm{t\bar{t}H}$ hadronic'' event categories, and these are included in the combination.

In Figure~\ref{fig:limit} the variation of the expected upper limits on $\mu_{\tH}$ is shown as a function of the integrated luminosity for the S1 and S2 scenarios. The limits are determined assuming a background-only hypothesis in which the $\tth$ process is considered as following the SM expectation ($\mu_{\tth} = 1$). In order to minimise further assumptions on the rate of $\tth$ production, $\mu_{\tth}$ is treated as a free parameter in the fit.
In the S1 scenario the expected median upper limit on $\mu_{\tH}$ at 3000 \fbinv is determined to be 2.35.
The corresponding value in S2 is 1.51. With the $3000 \fbinv$ dataset and foreseen reduction in systematic uncertainties in S2, the expected upper limit on $\mu_{\tH}$ improves by about a factor of eight with respect to the current exclusion.

\begin{figure}[hbtp]
\begin{center}
\includegraphics[width=0.45\textwidth]{\main/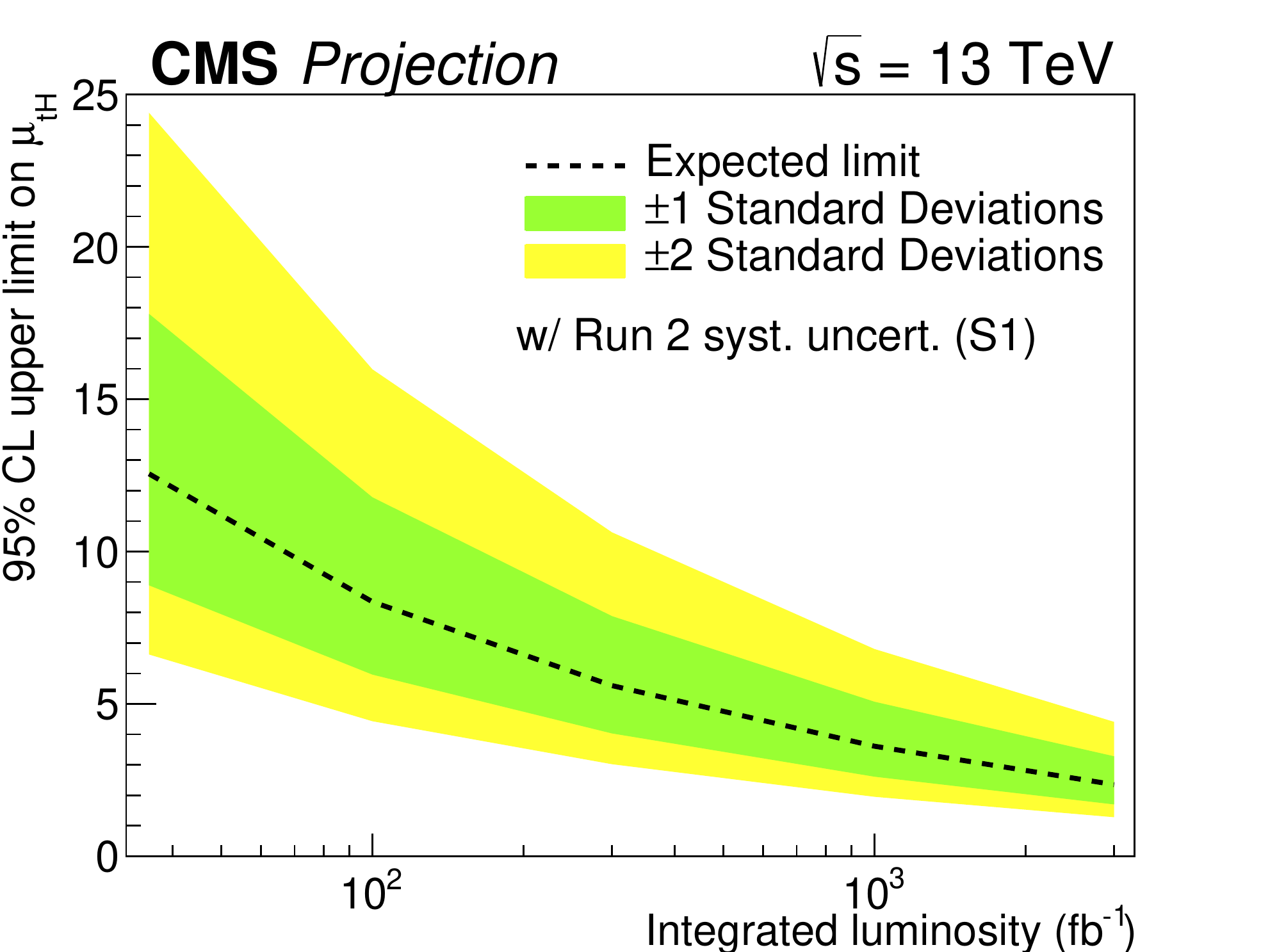} \hspace{1cm}
\includegraphics[width=0.45\textwidth]{\main/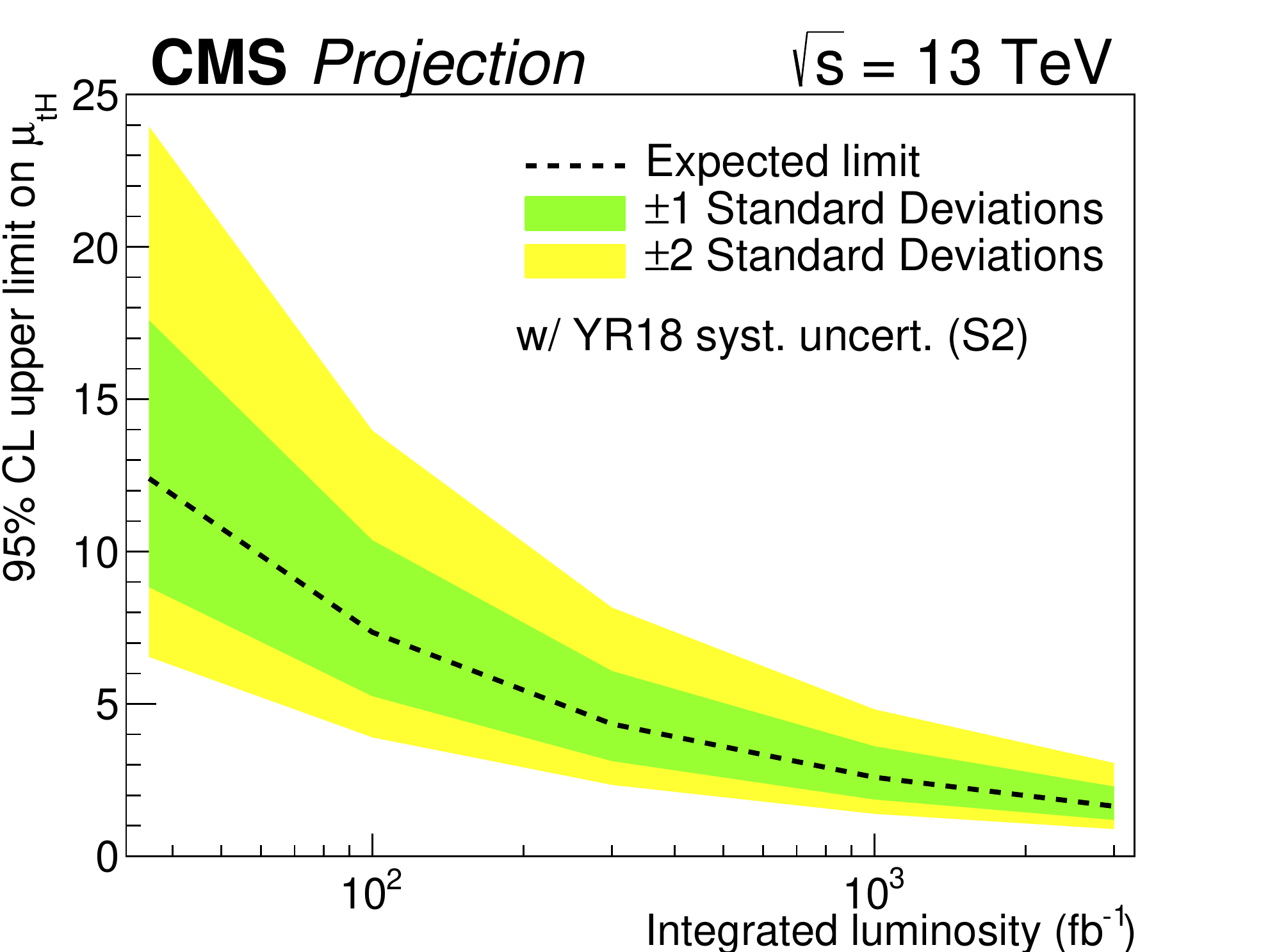}
\end{center}
\caption{The variation of expected upper limit on $\mu_{\tH}$ with integrated luminosity for two projection scenarios S1 (with Run~2 systematic uncertainties~\cite{CMS-PAS-HIG-18-009}) and S2 (with YR18 systematic uncertainties).}
\label{fig:limit}
\end{figure}

The evolution of the expected uncertainty on the measurement of $\mu_{\tH}$, assuming the SM rate, is given in Table~\ref{tab:muunc}. Values are given for two cases of background: one in which $\mu_{\tth}$ is unconstrained in the fit, and one in which it is fixed to the SM value of 1. In the latter case the uncertainties are reduced by around $10\%$ at $3000\fbinv$, indicating that a precise simultaneous measurement of the $\tth$ signal strength will be needed to obtain the optimal sensitivity to the $\tH$ channel. In both cases it is found that the reduced systematic uncertainties in S2 improve the precision by up to $30\%$.

\begin{table}[htbp]
\centering
\caption{The $\pm1\sigma$ uncertainties on expected $\mu_{\tH}$=1 for scenarios S1 (with Run~2 systematic uncertainties~\cite{CMS-PAS-HIG-18-009}) and S2 (with YR18 systematic uncertainties) at all three luminosities, considering also the case when $\mu_{\ttH}$ is fixed at the SM value 1.} \label{tab:muunc}
\begin{tabular}{@{} l c c@{\hskip 0.15in} c c }
 \hline
  &  & $\mu_{\tth}$ floating & $\mu_{\tth}$ fixed \\
  \hline
\multirow{3}{*}{S1} & 35.9 \fbinv  & ${}_{-5.8}^{+6.2}$ & ${}_{-5.4}^{+5.8}$ \\[1pt]
                        & 300 \fbinv & ${}_{-2.8}^{+2.9}$ & ${}_{-2.4}^{+2.5}$ \\[1pt]
                        & 3000 \fbinv & ${}_{-1.2}^{+1.2}$ & ${}_{-1.0}^{+1.1}$ \\[4pt]
\hline
\multirow{3}{*}{S2}  & 35.9 \fbinv  & ${}_{-5.8}^{+6.2}$ & ${}_{-5.3}^{+5.8}$ \\[1pt]
                        & 300 \fbinv & ${}_{-2.2}^{+2.2}$ & ${}_{-2.0}^{+2.0}$ \\[1pt]
                        & 3000 \fbinv & ${}_{-0.9}^{+0.9}$ & ${}_{-0.8}^{+0.8}$ \\[4pt]
 \hline
\end{tabular}
\end{table}

\asubsubsection[T. Klijnsma]{Constraints from differential measurements}
\label{sec:diffxsinterpretation}


Higgs boson couplings can be constrained by fitting theoretical predictions for $\pTH$~\cite{Bishara:2016jga,Grazzini:2017szg,Grazzini:2016paz} to data, exploiting not only the overall normalisation (as is done in inclusive measurements~\cite{%
Khachatryan:2016vau,
Aad:2015zhl,
CMS:2018lkl
}), but also the shape of the distribution.
One of the first constraints on Higgs boson couplings using differential Higgs boson production cross sections was made in Ref.~\cite{Bishara:2016jga}.
The limits $\kappac \in [ -16, 18 ]$ at 95\% CL were found, using data collected by the ATLAS Collaboration at $\sqrt{s}=8$\UTeV~\cite{Aad:2015lha}, corresponding to an integrated luminosity of $20.3$\fbinv.
More recently, the CMS Collaboration performed a similar fit using data~\cite{CMS-PAS-HIG-17-028} collected at $\sqrt{s}=13$\UTeV, corresponding to an integrated luminosity of $36.1$\fbinv.
The limits on $\kappab$ and $\kappac$ are discussed in Section~\ref{sec:HiggsDist}, whereas the interpretation in terms of $\kappat$ and $\cg$, the anomalous direct coupling to the gluon field, is discussed here.
The projected simultaneous limits on $\kappat$ and $\cg$ at $3000\fbinv$ are shown in Fig.~\ref{fig:ktcg_couplingdependentBRs}, assuming branching fractions that scale according to SM predictions.
It is expected to observe the loop in the gluon-fusion production process, which is clear from the fact that heavy top mass limit, given by the point $(\kappat =0, \cg=\sim 1/12)$, is excluded.

\begin{figure}[hbtp]
  \begin{center}
    \includegraphics[width=0.49\linewidth]{\main/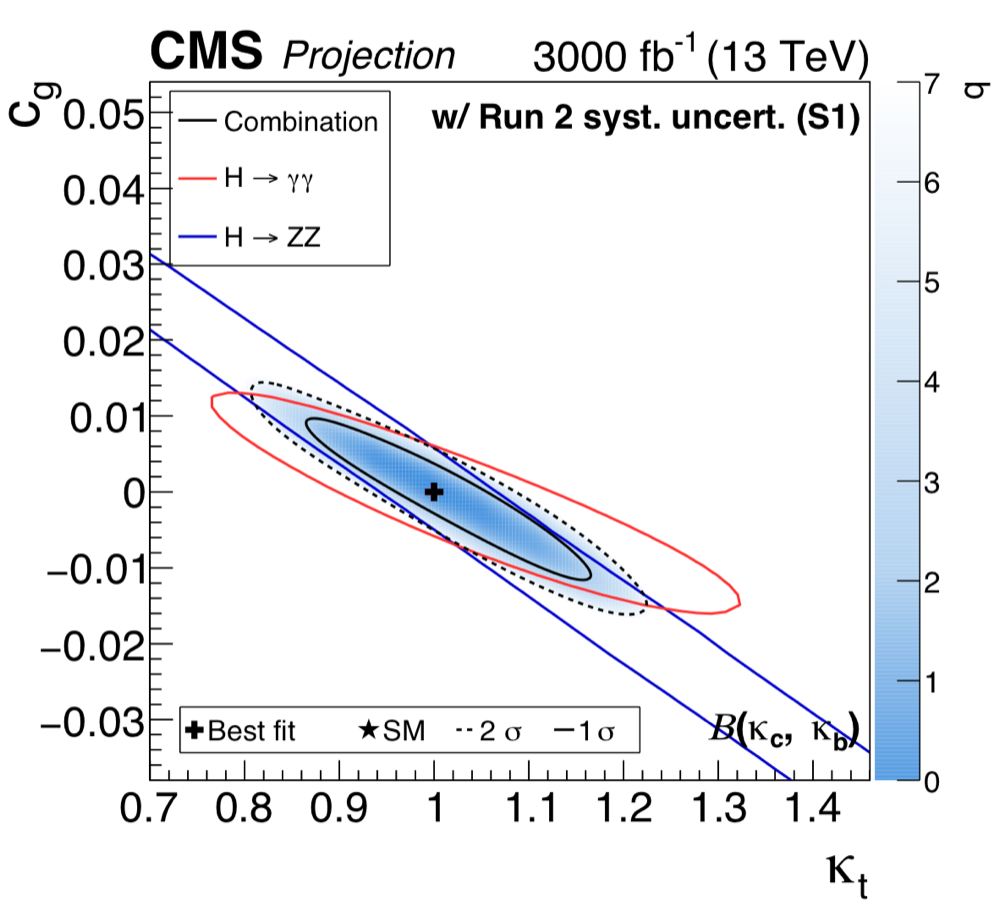}
    \includegraphics[width=0.49\linewidth]{\main/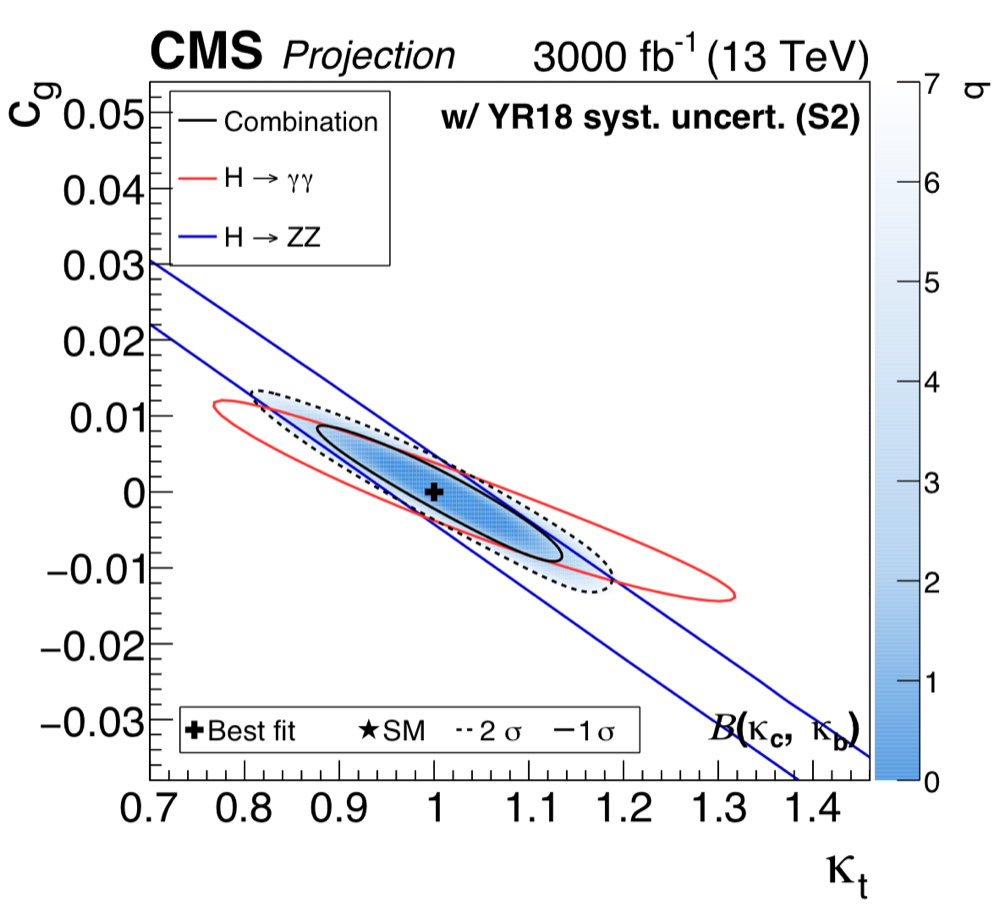}
    \caption{
        Projected simultaneous fit for $\kappat$ and $\cg$, assuming a coupling dependence of the branching fractions for Scenario 1 (left) and Scenario 2 (right).
        The one standard deviation contour is drawn for the combination ($\hgg$, $\hzz$, and $\hbb$), the $\hgg$ channel, and the $\hzz$ channel in black, red, and blue, respectively.
        For the combination the two standard deviation contour is drawn as a black dashed line, and the shading indicates the negative log-likelihood, with the scale shown on the right hand side of the plots.
        }
    \label{fig:ktcg_couplingdependentBRs}
  \end{center}
\end{figure}

In order to determine solely the constraint obtained from the distribution (and not the overall normalisation), the fit is repeated with the branching fractions implemented as nuisance parameters with no prior constraint, effectively profiling the overall normalisation.
With this parametrisation, the sensitivity to the sign of $\kappat$ coming from the $\hgg$ branching fraction is lost.
The fits obtained this way are shown in Fig.~\ref{fig:ktcg_floatingBRs}; although less significantly, the loop is still distinguished from the point-like coupling to the gluon field, using only the information in the shape of the distribution.

\begin{figure}[hbtp]
  \begin{center}
    \includegraphics[width=0.49\linewidth]{\main/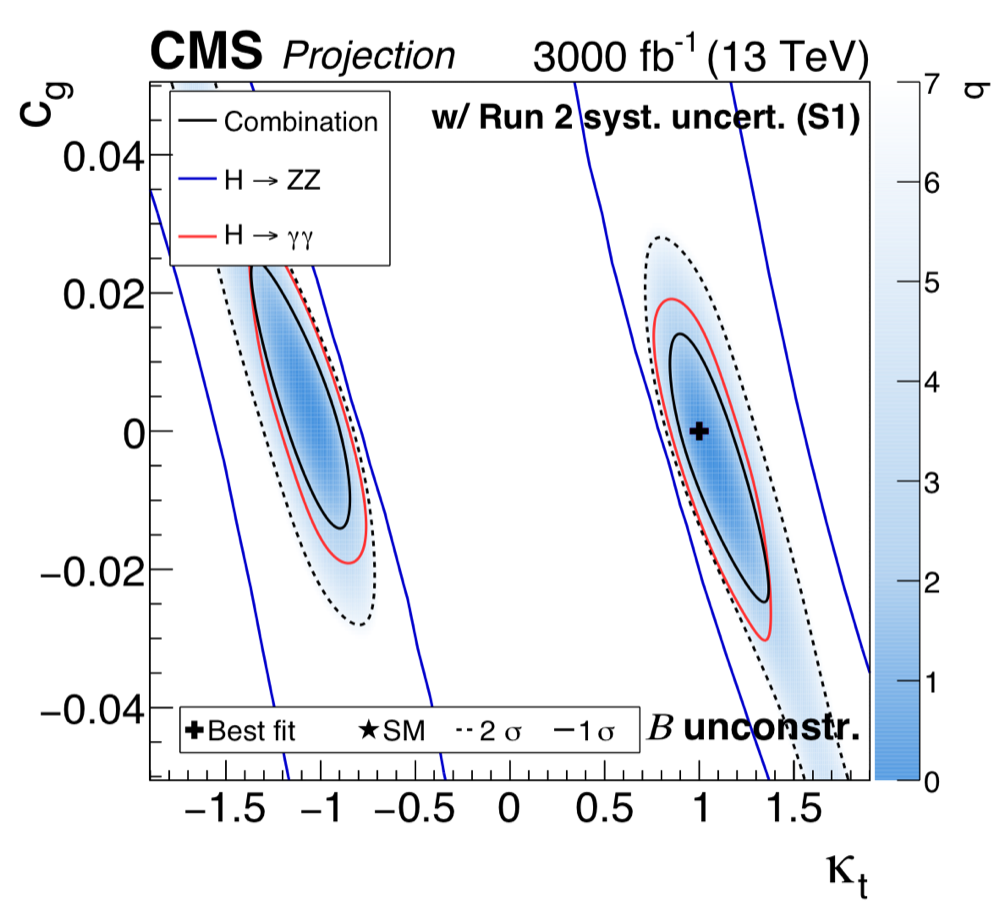}
    \includegraphics[width=0.49\linewidth]{\main/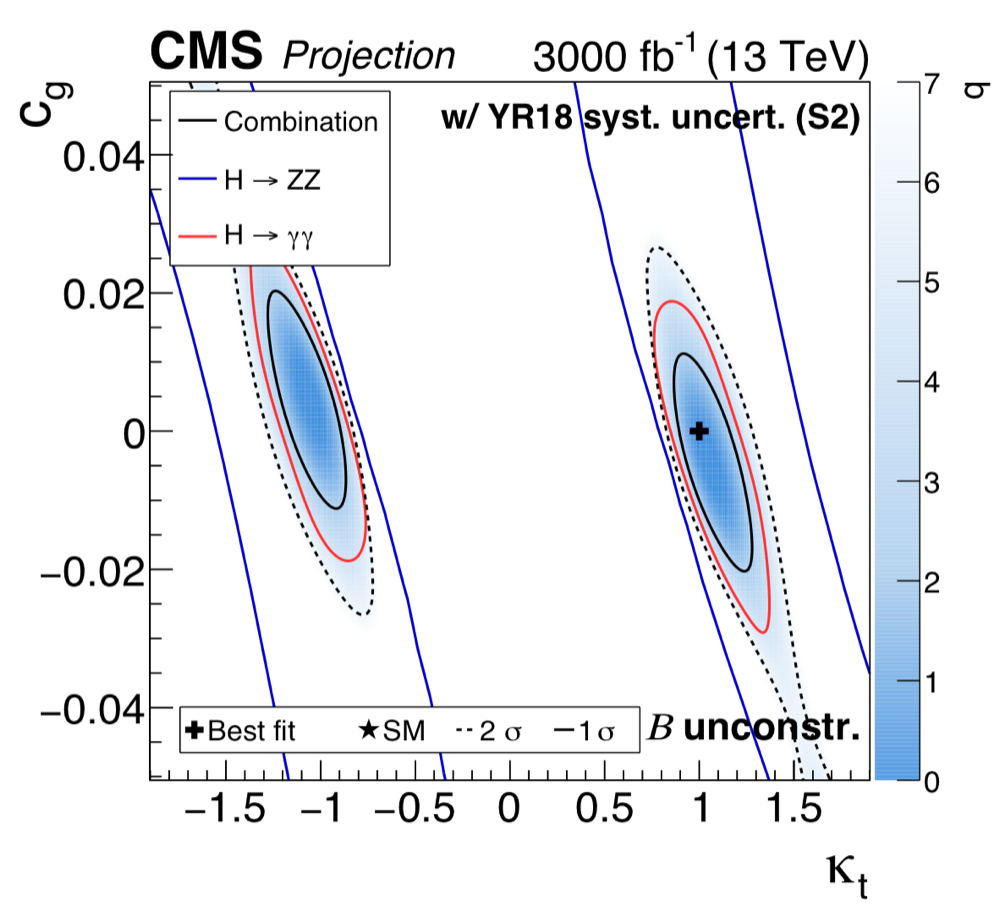}
    \caption{
        Projected simultaneous fit for $\kappat$ and $\cg$ with the branching fractions implemented as nuisance parameters with no prior constraint for Scenario 1 (left) and Scenario 2 (right).
        The one standard deviation contour is drawn for the combination ($\hgg$, $\hzz$, and $\hbb$), the $\hgg$ channel, and the $\hzz$ channel in black, red, and blue, respectively.
        For the combination the two standard deviation contour is drawn as a black dashed line, and the shading indicates the negative log-likelihood, with the scale shown on the right hand side of the plots.
        }
    \label{fig:ktcg_floatingBRs}
  \end{center}
\end{figure}

\asubsection[R. Di Nardo, A. Gilbert, H. Yang, N. Berger,  D. Du,
  M. D\"uhrssen, A. Gilbert, R. Gugel, L. Ma, B. Murray,   P. Milenovic
  ]{Combination of Higgs boson measurement projections}
\label{sec2:exp_combination}
The projections documented in this section~\cite{CMS-PAS-FTR-18-011,ATL-PHYS-PUB-2018-054} are based on the extrapolation of the following analyses:

\begin{itemize}
	\item \hgg, with $\ggh$, $\vbf$, $\vh$ and $\tth$ production~\cite{Sirunyan:2018ouh,ATLAS-CONF-2018-028,Aaboud:2018urx},
	\item \hzzllll, with $\ggh$, $\vbf$, $\vh$ and $\tth$ production~\cite{HIG16041,ATLAS-CONF-2018-018},
	\item \hwwlnln, with $\ggh$, $\vbf$ and $\vh$ production~\cite{HIG-16-042,Aaboud:2018jqu},
	\item \htt, with $\ggh$ and $\vbf$ production~\cite{HIG16043,ATLAS-CONF-2018-021},
	\item \vh production with \hbb decay~\cite{HIG16044,Aaboud:2018zhk},
	\item Boosted H production with \hbb~ decay~\cite{HIG17010},
	\item \tth production with \hlep~\cite{Sirunyan:2018shy,Aaboud:2017jvq},
	\item \tth production with \hbb~\cite{bib:hig-17-026,Sirunyan:2018ygk,Aaboud:2017rss},
	\item \hmm, with $\ggh$ and $\vbf$ production~\cite{HIG-17-019,ATLAS-CONF-2018-026},
	\item \hzg, with \ggh and \vbf production~\cite{Aaboud:2017uhw}.
\end{itemize}

The projected results given in this section are based on the combined measurement of these channels~\cite{Sirunyan:2018koj,ATLAS-CONF-2018-031}. In the following results, the signal model in the $\hmm$ channel is modified to account for the improved di-muon mass resolution in the Phase-2 ATLAS and CMS tracker upgrades~\cite{Klein:2017nke,Collaboration:2285585}. In CMS, it is estimated that the reduced material budget and improved spatial resolution of the upgraded tracker will yield a $40\%$ improvement in the di-muon mass resolution, for example a reduction from $1.1\%$ to $0.65\%$ for muons in the barrel region. In ATLAS, instead, the reduction of the di-muon invariant mass resolution is estimated to be between 15\% and 30\% depending on the analysis categories (forward/central).

In the ATLAS projection the expected signal and background yields in all channels are scaled to account for the increasing cross sections going from $\sqrt{s}=13\TeV$ to $\sqrt{s}=14\TeV$, while no scaling is performed in the CMS projection. The impact of this scaling on the projected sensitivity is found to be small.



Projections are given for three different sets of measurements:
\begin{itemize}
    \item \textbf{Higgs boson production cross sections in different decay channels}: the parameters of interest are the cross sections times branching fractions for ggH, VBF, WH, ZH and ttH production in each relevant decay mode, normalised to their SM predictions.
    \item \textbf{Higgs boson production cross sections}:  the parameters of interest are the production cross sections normalised to the corresponding SM predictions  $\sigma_{i}/\sigma_{i}^{\mathrm{SM}}$  where $i=\ggh$, $\vbf$, $\wh$, $\zh$ and $\tth$, assuming the SM values for the branching fractions. The small contribution from bbH is grouped with \ggh while the \zh process includes \zh production with the gluon-gluon initial state. The overall theoretical uncertainties on the inclusive SM cross section predictions are not included, while the uncertainties on the branching ratios are included as the values are assumed to be given by the SM.
    \item \textbf{Higgs boson branching fractions}: the parameters of interest are the branching fractions normalised to the corresponding SM values $\BR_{f}/\BR_{f}^{\mathrm{SM}}$, where $f = \zz$, $\ww$, $\PGg\PGg$, $\tautau$, $\bb$, $\mumu$, $\PZ\PGg$ assuming the SM cross sections for the production modes. In this case  the uncertainties on the decay branching ratios are not included, while the overall QCD scales and PDF+$\alpha_{S}$ uncertainties on the inclusive production cross sections are included.

\end{itemize}

For each projected measurement the uncertainty is decomposed into four components: statistical, experimental, background theory and signal theory.
The combination is based on the assumption that these components are independent within each experiment. Among them, the statistical and experimental uncertainties are treated as fully uncorrelated between the two experiments, while the signal and background theory uncertainties are assumed to be fully correlated. The combination is performed for each parameter individually using the the BLUE methodology as described in Ref.~\cite{Valassi:2003mu}. This procedure does not take into account correlations that may exist between parameters. These arise when analysis channels are sensitive to more than one production or decay mode and the chosen fit observables do not fully distinguish between these, as well as when the same systematic uncertainties apply to multiple production or decay modes. The effect of including these correlations via a simultaneous combination of all parameters has been verified, utilising the same BLUE methodology and including the complete covariance matrices for both experiments, and is found to have a minor effect on the projection results. Specifically, the effect on the combined statistical and experimental uncertainties is negligible given the reported precision. For the theory uncertainties this procedure can lead to smaller values than in the case where these correlations are neglected. This is a feature of the methodology of Ref.~\cite{Valassi:2003mu}, due to the different approaches concerning the theoretical uncertainties used in the extrapolations by the two experiments, which leads to differences in some of the correlation values. However, it is expected that by the time of the HL-LHC both experiments will employ a more consistent treatment of the theoretical uncertainties, making this reduction largely artificial. This motivates the choice to combine measurement projections independently and neglect such correlations in the following results.

\asubsubsection{Production mode cross-sections in different decay channels}
\label{sec:expcomb_prodtimesdecay}
The expected $\pm 1\sigma$ uncertainties on the production mode cross sections in the different decay channels in S2 are summarised in Figure~\ref{fig:summary_A1_5PD_ggH_VBF} for \ggh and \vbf, in Figure~\ref{fig:summary_A1_5PD_WH_ZH} for \wh and \zh, in Figure~\ref{fig:summary_A1_5PD_ttH}. These are shown for ATLAS, CMS and their combination. Additionally, the numerical values for the ATLAS-CMS combination in scenario S2 are also reported in the figures, with the uncertainty decomposed in three components: statistical, experimental and theory.
There are few cases where the extrapolation is currently available only for one experiment (e.g. only $gg \rightarrow H \rightarrow bb$ in CMS, and only \hzg in ATLAS). In these cases, the combined result is obtained by using the same available extrapolation  for both experiments.
The correlations between the different production mode cross-sections in different decay channels are in general small, with the exception of the \zh and \wh measurements in the \hzz decay (for this reason the \vh cross-section is also reported) and the \ggh and \vbf production mode cross sections in the \hmm decay. 


The numerical values of the expected $\pm 1\sigma$ uncertainties on the per-production-mode cross sections in the different decay channels for the ATLAS and CMS projections are given in Table~\ref{tab:summary_A1_5PD}. The table  gives the breakdown of the uncertainty into four components: statistical, signal theory, background theory and experimental for both scenarios S1 and S2.

\begin{figure}[hbtp]
\centering
\includegraphics[height=0.332\textheight]{\main/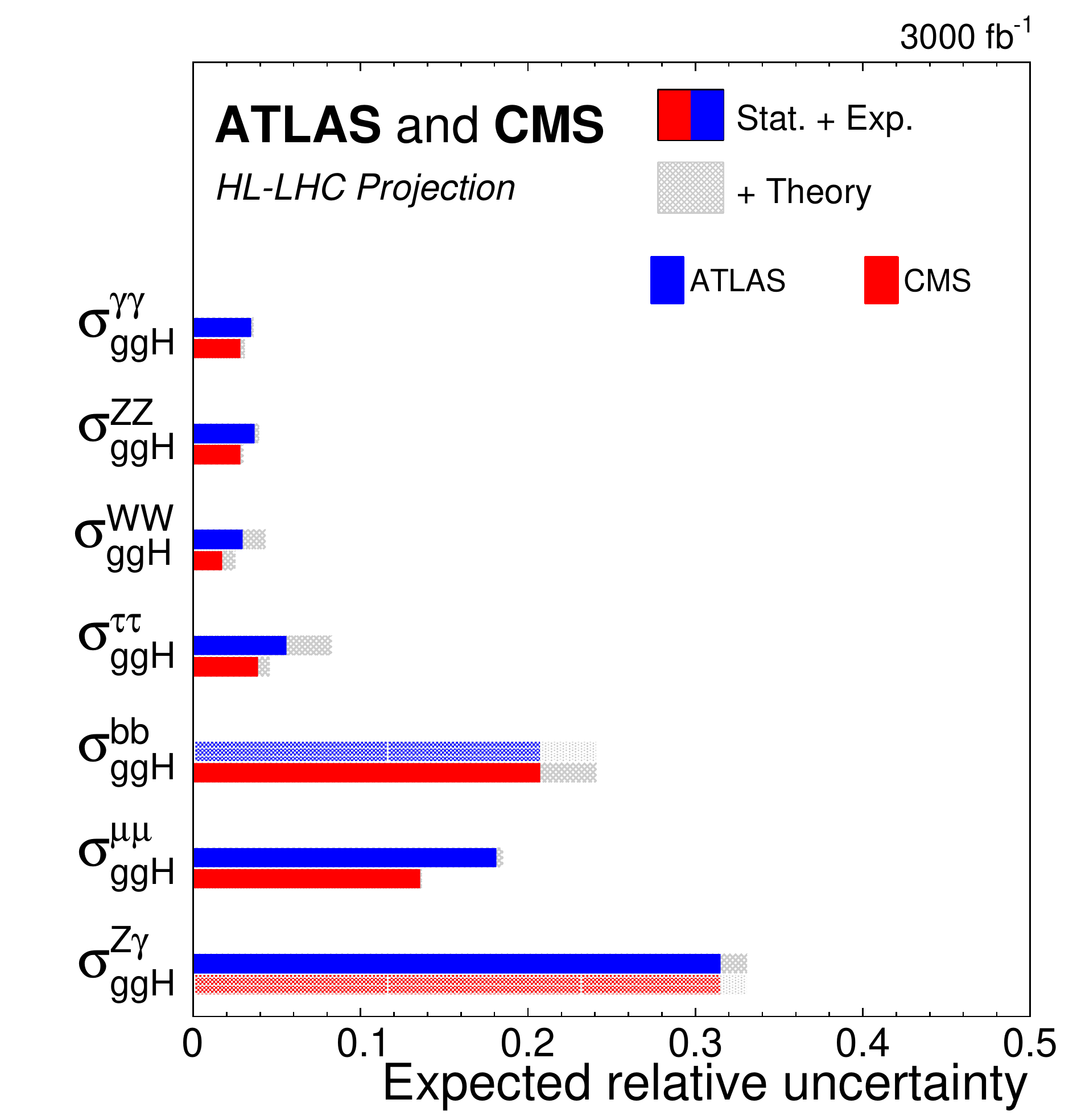}%
\includegraphics[height=0.33\textheight]{\main/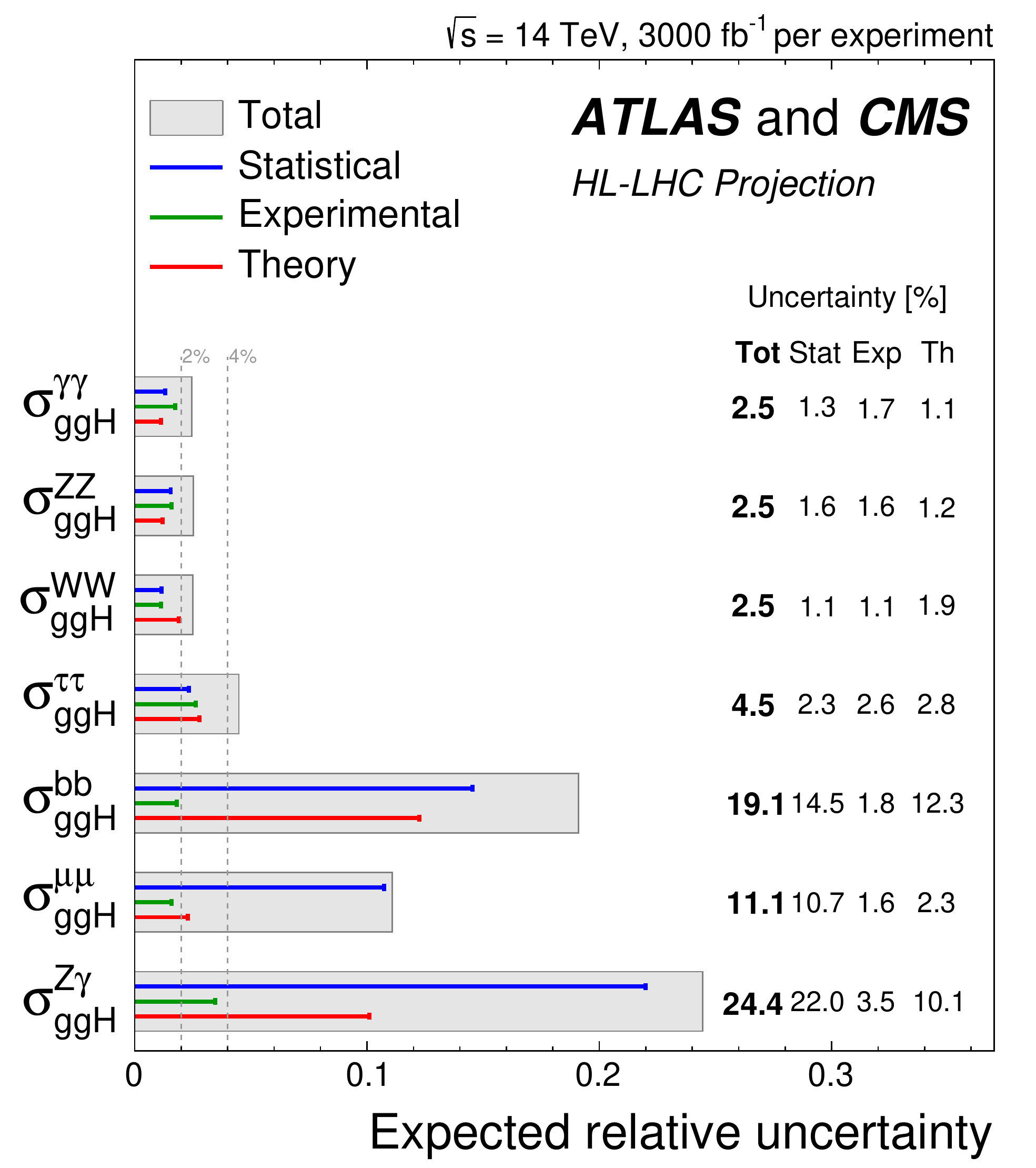}\\
\includegraphics[height=0.332\textheight]{\main/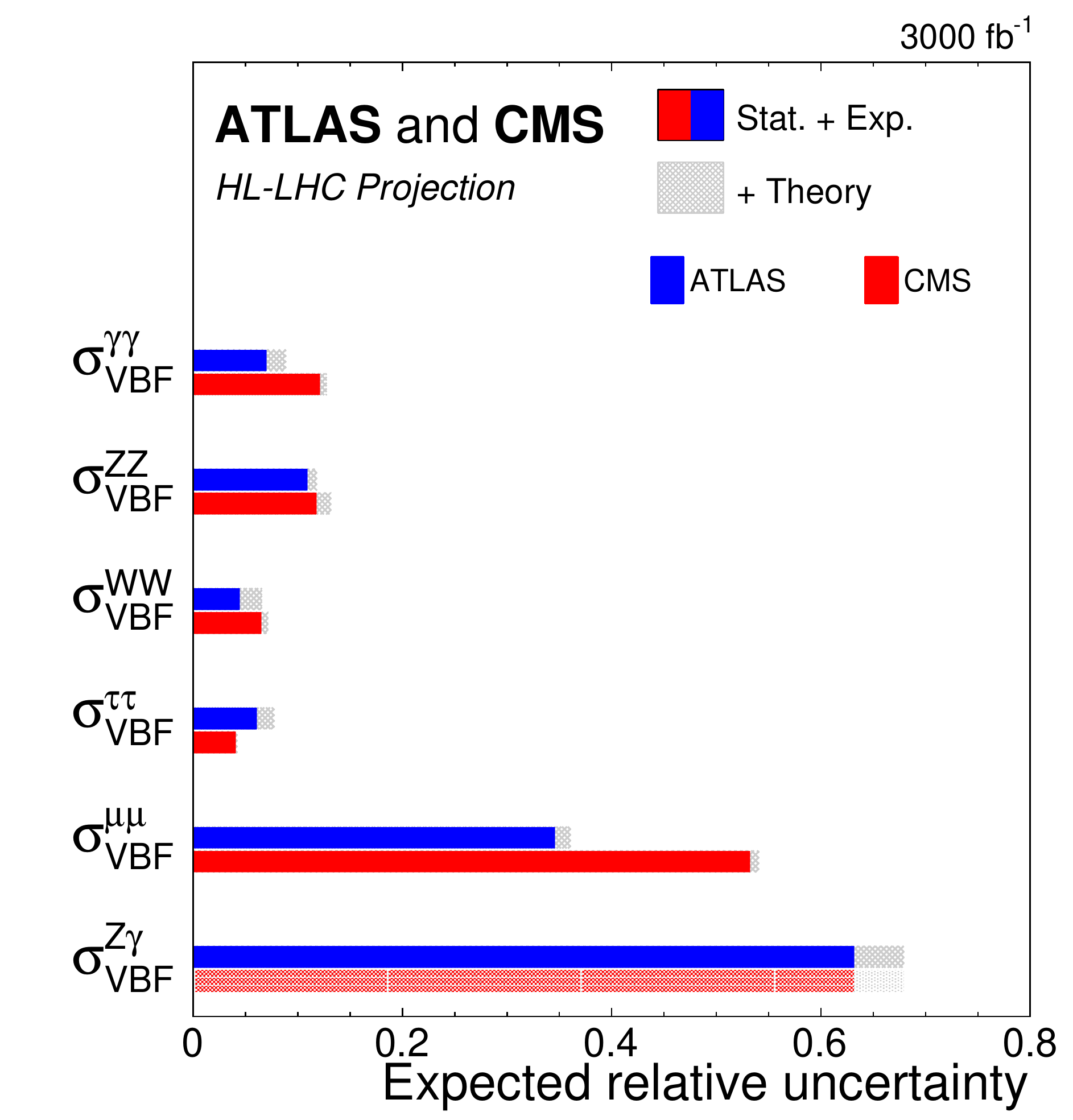}%
\includegraphics[height=0.33\textheight]{\main/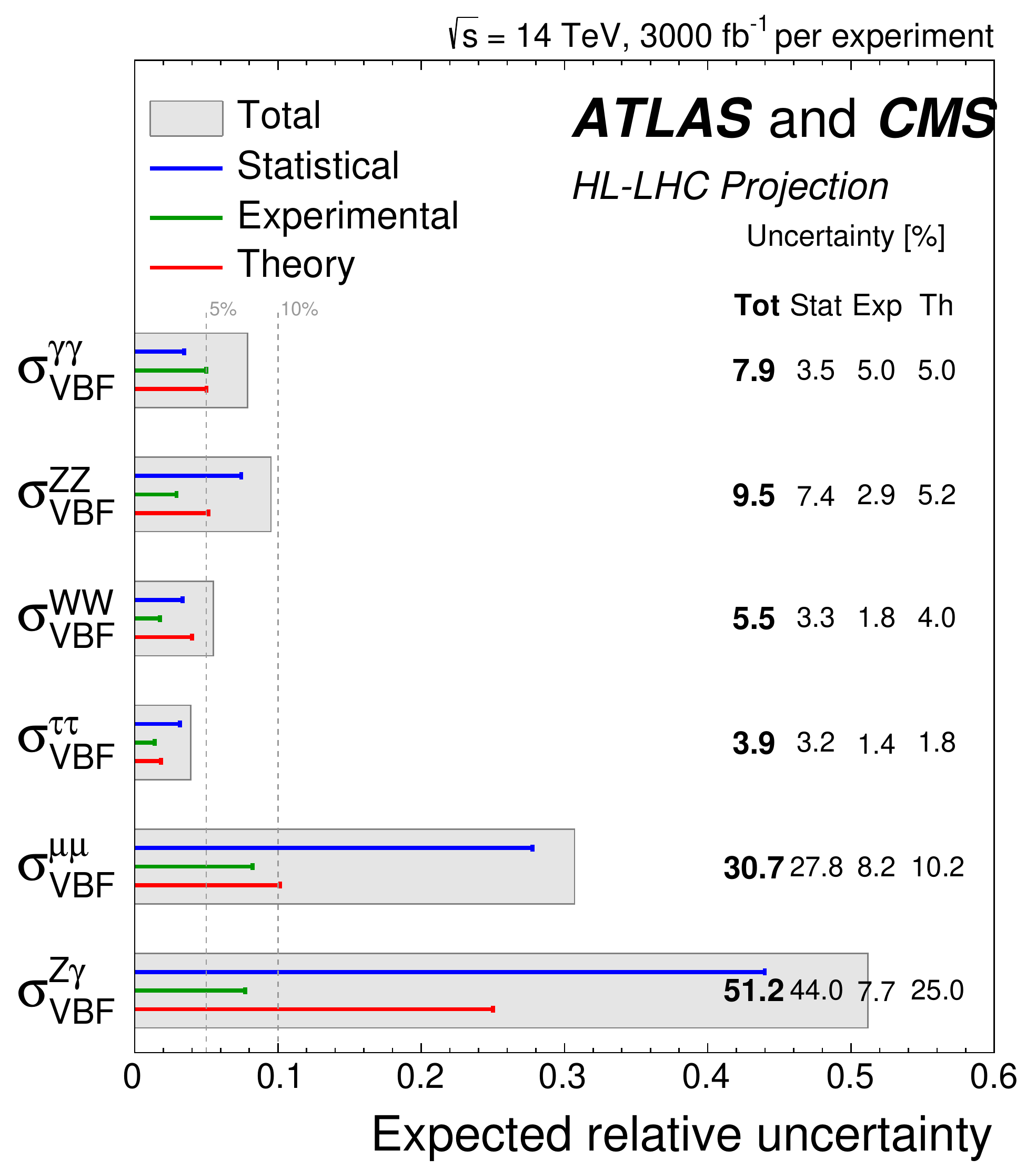}%
\caption{(left) Summary plot showing the total expected $\pm 1\sigma$ uncertainties in S2 (with YR18 systematic uncertainties) on the \ggh (top) and \vbf (bottom) production cross sections in the different decay modes normalised to the SM predictions   for ATLAS(blue)  and CMS (red). The filled coloured box corresponds to the statistical and experimental systematic uncertainties, while the hatched grey area represent the additional contribution to the total uncertainty due to theoretical systematic uncertainties. In the cases where  the extrapolation is performed only by one experiment, same performances are assumed for the other experiment and this is indicated by a  hatched bar.
(right) Summary plot showing the total expected $\pm 1\sigma$  uncertainties in S2 (with YR18 systematic uncertainties) on the \ggh (top) and \vbf (bottom) production cross sections in the different decay modes normalised to the SM predictions for the combination of ATLAS and CMS extrapolations. For each measurement,  the total uncertainty is indicated by a grey box while the statistical, experimental and theory uncertainties are indicated by a blue, green and red line respectively. In addition, the numerical values are also reported.}
\label{fig:summary_A1_5PD_ggH_VBF}
\end{figure}

\begin{figure}[hbtp]
\centering
\includegraphics[height=0.332\textheight]{\main/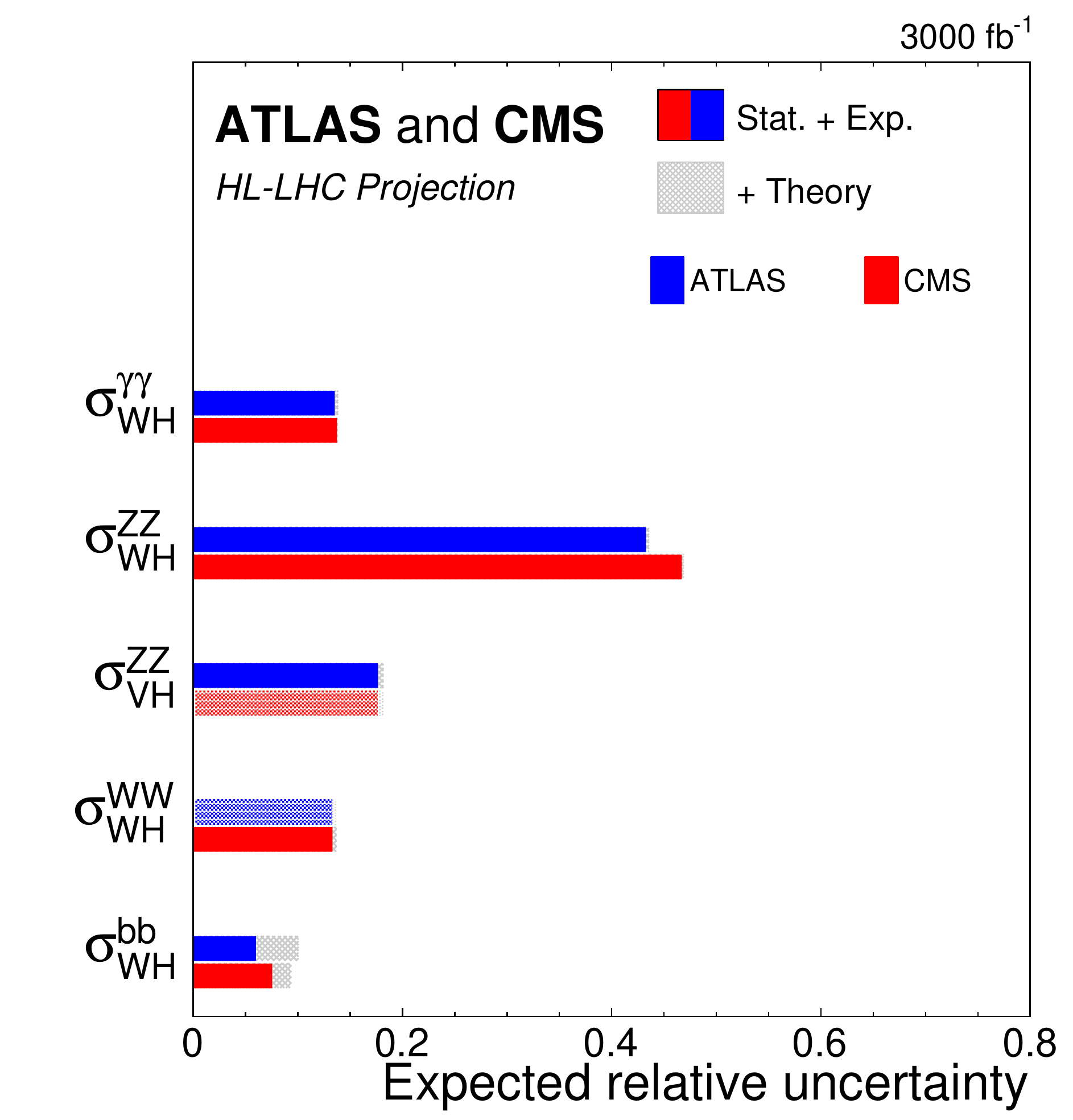}%
\includegraphics[height=0.33\textheight]{\main/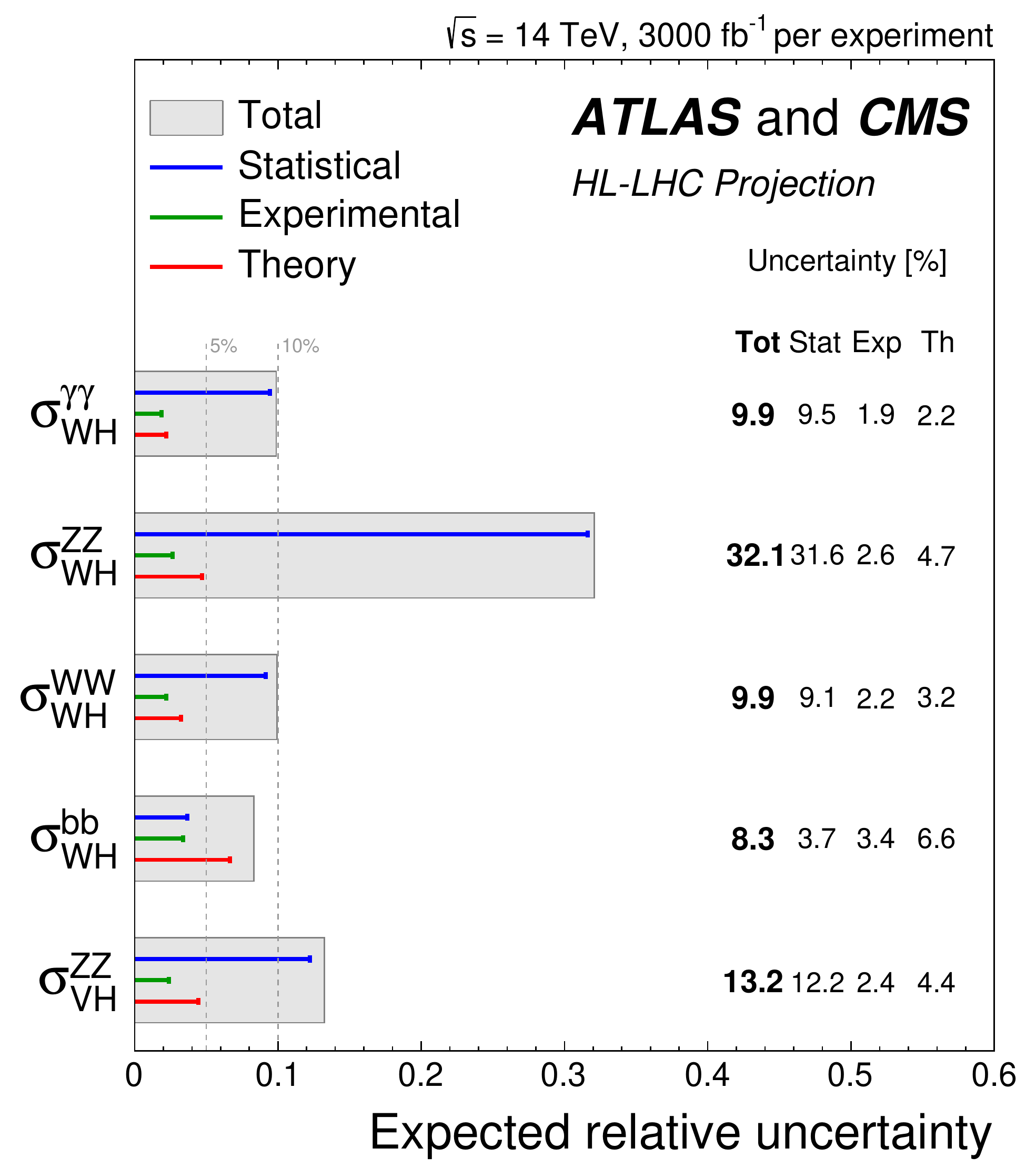}\\
\includegraphics[height=0.332\textheight]{\main/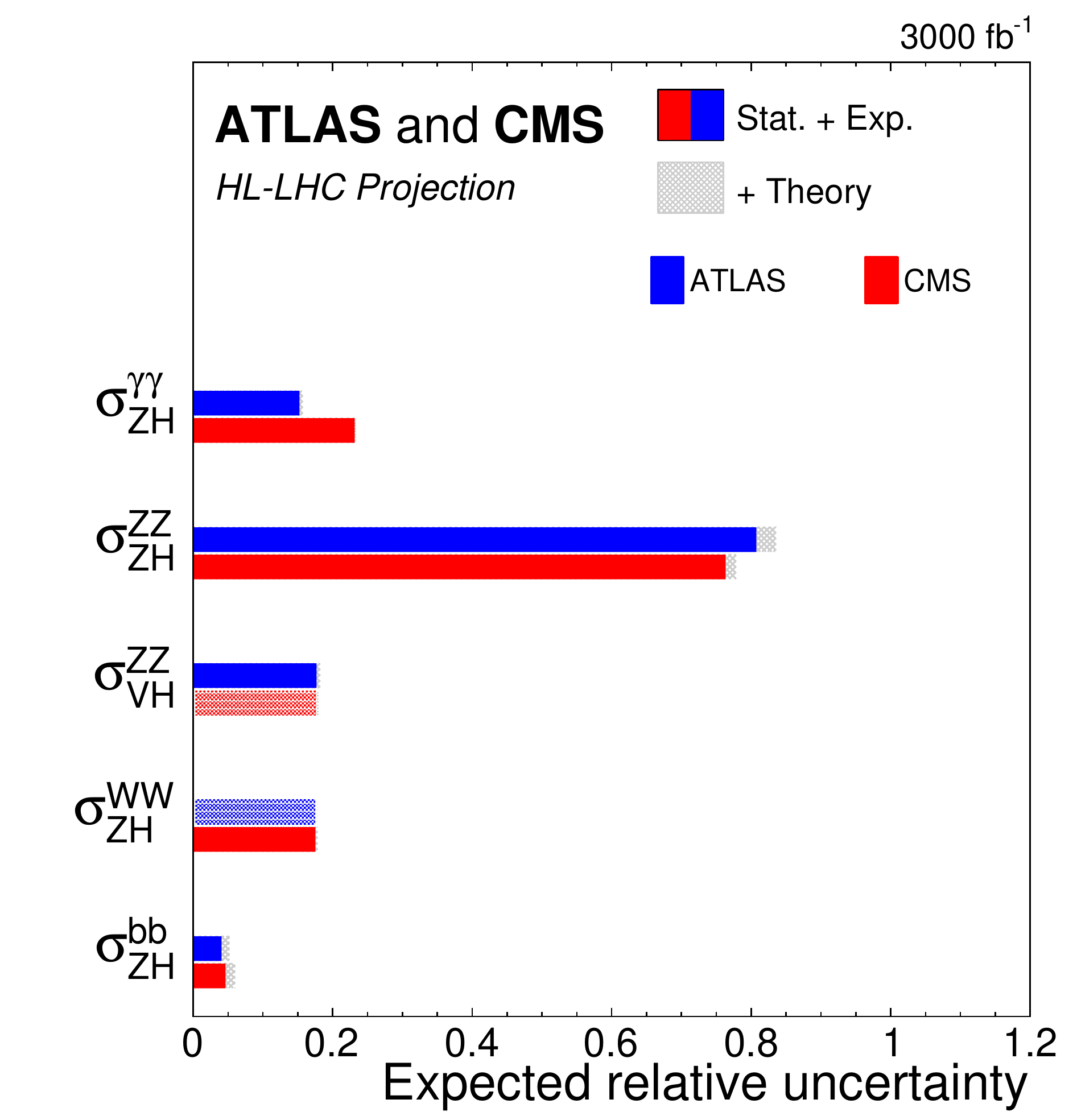}%
\includegraphics[height=0.33\textheight]{\main/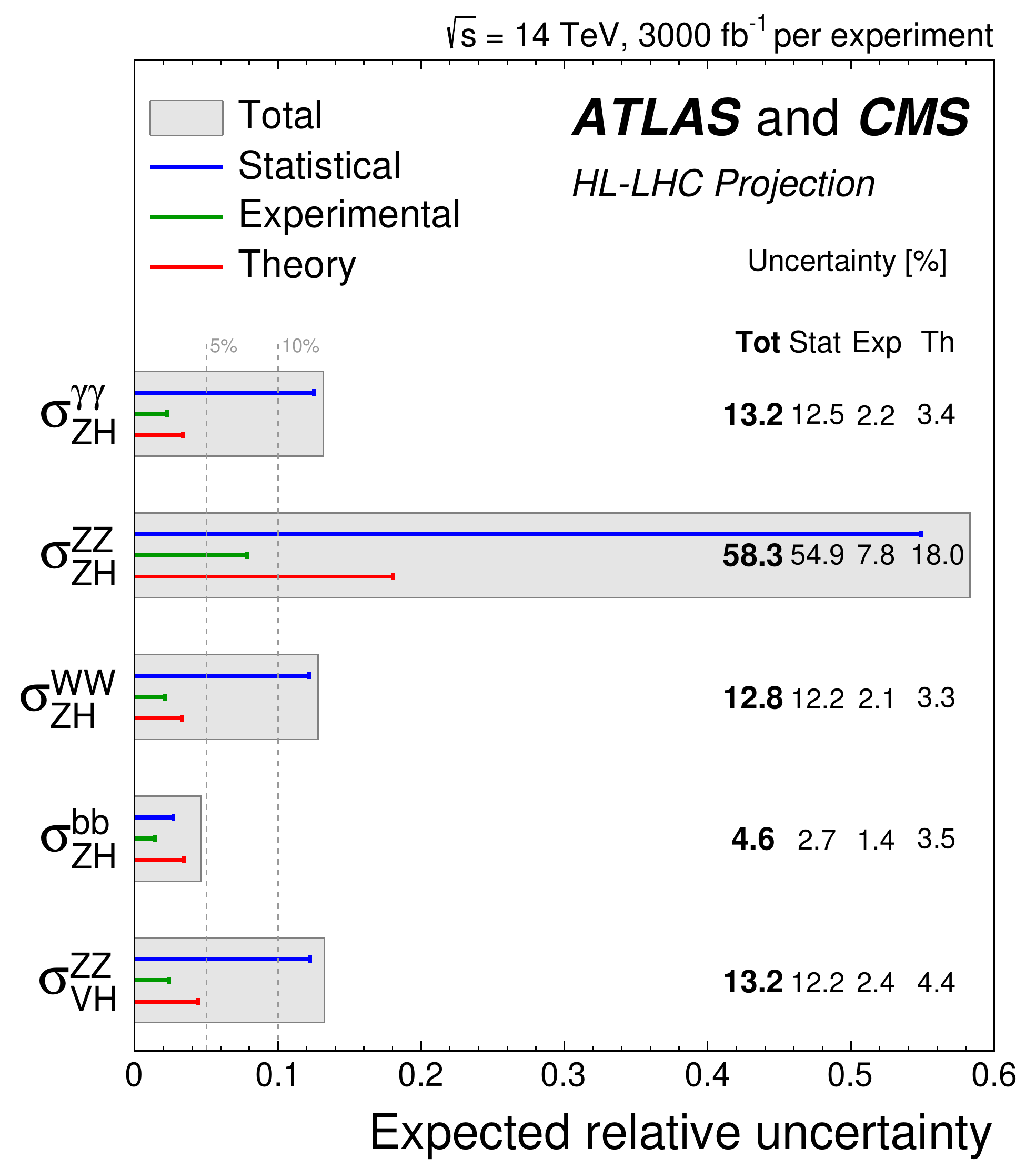}%
\caption{(left) Summary plot showing the total expected $\pm 1\sigma$ uncertainties in S2 (with YR18 systematic uncertainties) on the \wh (top) and \zh (bottom) production cross sections in the different decay modes normalised to the SM predictions   for ATLAS(blue)  and CMS (red). The filled coloured box corresponds to the statistical and experimental systematic uncertainties, while the hatched grey area represent the additional contribution to the total uncertainty due to theoretical systematic uncertainties. In the cases where  the extrapolation is performed only by one experiment, same performances are assumed for the other experiment and this is indicated by a  hatched bar.
(right) Summary plot showing the total expected $\pm 1\sigma$  uncertainties in S2 (with YR18 systematic uncertainties) on the \wh (top) and \zh (bottom) production cross sections in the different decay modes normalised to the SM predictions for the combination of ATLAS and CMS extrapolations. For each measurement,  the total uncertainty is indicated by a grey box while the statistical, experimental and theory uncertainties are indicated by a blue, green and red line respectively. In addition, the numerical values are also reported.}
\label{fig:summary_A1_5PD_WH_ZH}
\end{figure}

\begin{figure}[hbtp]
\centering
\includegraphics[height=0.332\textheight]{\main/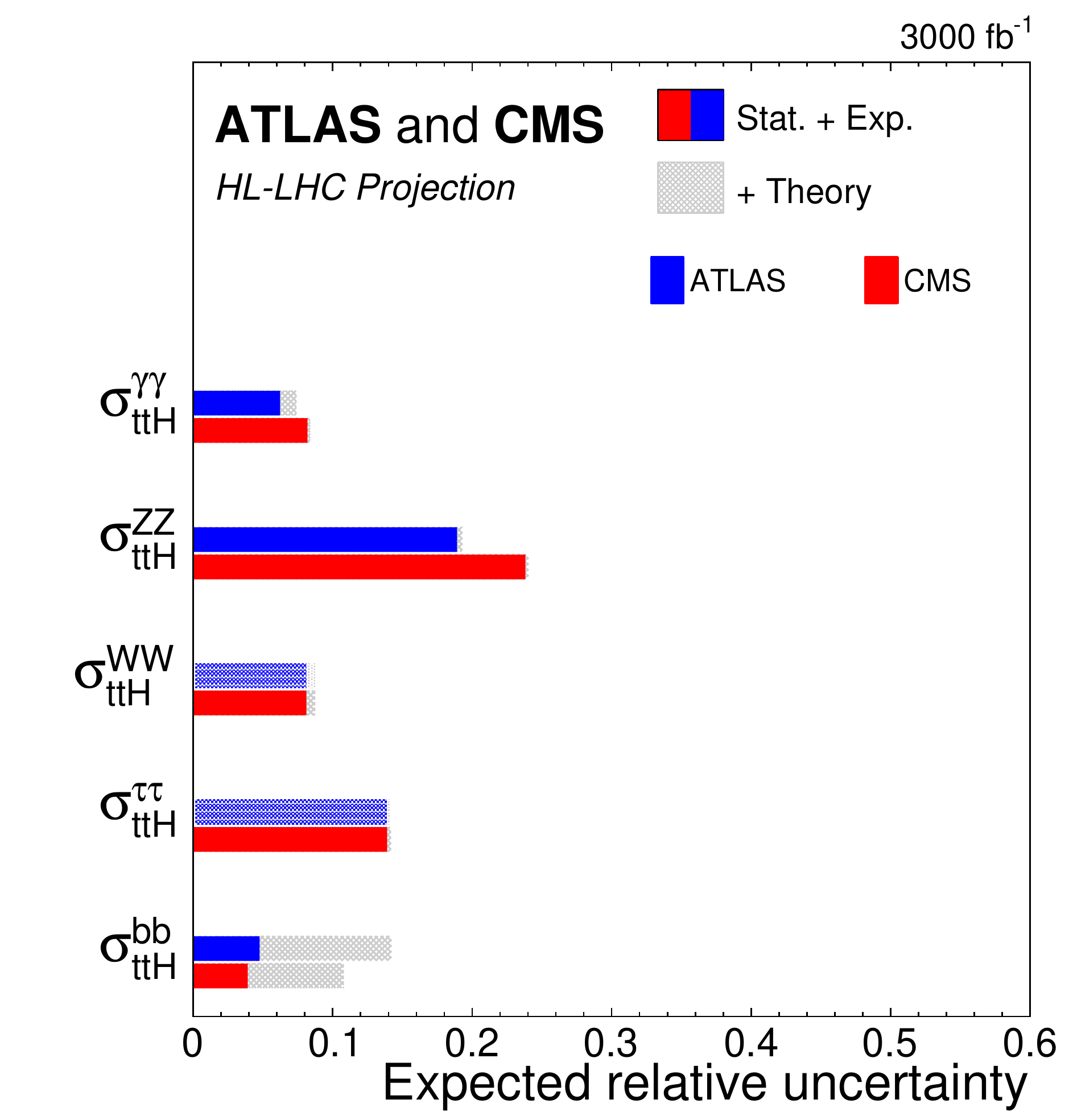}%
\includegraphics[height=0.33\textheight]{\main/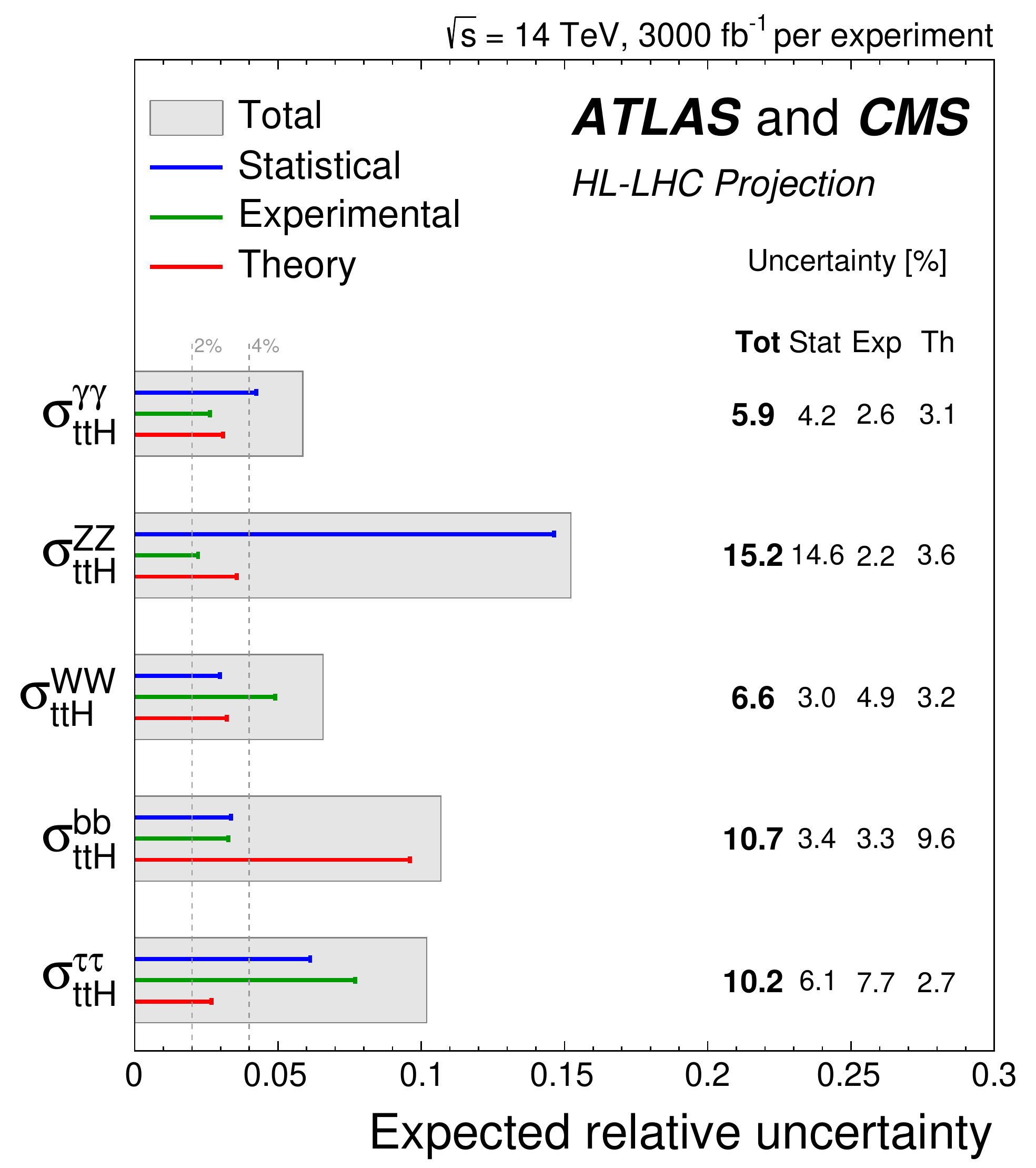}
\caption{(left) Summary plot showing the total expected $\pm 1\sigma$ uncertainties in S2 (with YR18 systematic uncertainties) on the \tth production cross section in the different decay modes normalised to the SM predictions   for ATLAS (blue)  and CMS (red). The filled coloured box corresponds to the statistical and experimental systematic uncertainties, while the hatched grey area represent the additional contribution to the total uncertainty due to theoretical systematic uncertainties. In the cases where  the extrapolation is performed only by one experiment, same performances are assumed for the other experiment and this is indicated by a  hatched bar.
(right) Summary plot showing the total expected $\pm 1\sigma$  uncertainties in S2 (with YR18 systematic uncertainties) on the \tth production cross sections in the different decay modes normalised to the SM predictions for the combination of ATLAS and CMS extrapolations. For each measurement,  the total uncertainty is indicated by a grey box while the statistical, experimental and theory uncertainties are indicated by a blue, green and red line respectively. In addition, the numerical values are also reported.}
\label{fig:summary_A1_5PD_ttH}
\end{figure}

\begin{table}[hbtp]
\centering
\caption{The expected $\pm 1\sigma$ uncertainties, expressed as percentages, on the per-production-mode cross sections in the different decay modes  for ATLAS (left) and CMS (right). Values are given for both S1 (with Run~2 systematic uncertainties~\cite{Sirunyan:2018koj}) and S2 (with YR18 systematic uncertainties). The total uncertainty is decomposed into four components: statistical (Stat), signal theory (SigTh), background theory (BkgTh) and experimental (Exp).}
\scriptsize	
\begin{tabular}{@{} l c c@{\hskip 0.15in} c c c c @{}}
  \hline
   \multicolumn{7}{c}{ATLAS}\\
 \hline
  &  & \multicolumn{5}{c}{3000 $\text{fb}^{-1}$ uncertainty [\%]} \\
  &  & Total & Stat & Exp & SigAcc & BkgTh \\
  \hline
  \multirow{2}{*}{$\sigma_{\mathrm{ggH}}^{\gamma \gamma }$} & S1 &5.2   & 1.7   & 4.7   & 1.1   & 1.2  \\[1pt] 
  & S2 &3.6   & 1.7   & 3.0   & 0.9   & 0.5  \\[4pt] 
  \multirow{2}{*}{$\sigma_{\mathrm{ggH}}^{\mathrm{ZZ}}$} & S1 &4.9   & 2.0   & 3.7   & 1.8   & 1.9  \\[1pt]
  & S2 &3.9   & 2.0   & 3.0   & 1.0   & 1.0  \\[4pt]
  \multirow{2}{*}{$\sigma_{\mathrm{ggH}}^{\mathrm{WW}}$} & S1 &6.0   & 1.2   & 3.2   & 3.7   & 3.4  \\[1pt]
  & S2 &4.3   & 1.2   & 2.7   & 2.1   & 2.4  \\[4pt]
  \multirow{2}{*}{$\sigma_{\mathrm{ggH}}^{\tau \tau }$} & S1 &10.6  & 3.3   & 5.0   & 7.5   & 4.4  \\[1pt]
  & S2 &8.2   & 3.3   & 4.4   & 5.4   & 2.7  \\[4pt]
  \multirow{2}{*}{$\sigma_{\mathrm{ggH}}^{\mu \mu}$} & S1  &19.9  & 17.9  & 2.8   & 8.0   & 0.1  \\[1pt]
  & S2 &18.5  & 17.9  & 2.7   & 3.8   & 0.1  \\[4pt]
  \multirow{2}{*}{$\sigma_{\mathrm{ggH}}^{\mathrm{Z} \gamma }$} & S1 &33.3  & 31.1  & 4.9   & 10.1  & 0.3  \\[1pt]
  & S2 &33.3  & 31.1  & 4.9   & 10.1  & 0.3  \\[4pt]
  \multirow{2}{*}{$\sigma_{\mathrm{VBF}}^{\gamma \gamma }$} & S1  &12.0  & 4.4   & 7.3   & 8.2   & 2.1  \\[1pt]
  & S2 &8.9   & 4.4   & 5.5   & 5.4   & 0.9  \\[4pt]
  \multirow{2}{*}{$\sigma_{\mathrm{VBF}}^{\mathrm{ZZ}}$} & S1 &13.0  & 9.6   & 5.1   & 6.8   & 2.1  \\[1pt]
  & S2 &11.8  & 9.6   & 5.1   & 4.5   & 1.2  \\[4pt]
  \multirow{2}{*}{$\sigma_{\mathrm{VBF}}^{\mathrm{WW}}$} & S1 &10.3  & 3.3   & 3.9   & 7.7   & 4.5  \\[1pt]
  & S2 &6.6   & 3.3   & 2.9   & 4.0   & 2.8  \\[4pt]
  \multirow{2}{*}{$\sigma_{\mathrm{VBF}}^{\tau \tau }$} & S1 &8.7   & 3.7   & 4.1   & 5.5   & 3.8  \\[1pt]
  & S2 &7.8   & 3.7   & 4.8   & 3.2   & 3.6  \\[4pt]
  \multirow{2}{*}{$\sigma_{\mathrm{VBF}}^{\mu \mu }$} & S1 &38.7  & 32.5  & 11.7  & 17.1  & 0.2  \\[1pt]
  & S2 &36.1  & 32.5  & 11.7  & 10.4  & 0.3  \\[4pt]
  \multirow{2}{*}{$\sigma_{\mathrm{VBF}}^{\mathrm{Z} \gamma }$} & S1 &68.2  & 62.2  & 10.9  & 25.0  & 0.5  \\[1pt]
  & S2 &68.2  & 62.2  & 10.9  & 25.0  & 0.5  \\[4pt]
  \multirow{2}{*}{$\sigma_{\mathrm{WH}}^{\gamma \gamma }$} & S1 &14.8  & 13.1  & 5.2   & 4.0   & 1.3  \\[1pt]
  & S2 &13.8  & 13.1  & 3.3   & 2.8   & 0.7  \\[4pt]
  \multirow{2}{*}{$\sigma_{\mathrm{VH}}^{\mathrm{ZZ}}$} & S1 &18.7  & 17.3  & 4.2   & 5.4   & 2.2  \\[1pt]
  & S2 &18.1  & 17.3  & 3.4   & 4.1   & 1.7  \\[4pt]
  \multirow{2}{*}{$\sigma_{\mathrm{WH}}^{\mathrm{bb}}$} & S1 &14.1  & 4.3   & 4.9   & 7.3   & 10.1 \\[1pt]
  & S2 &10.1  & 4.4   & 4.1   & 4.2   & 6.9  \\[4pt]
  \multirow{2}{*}{$\sigma_{\mathrm{ZH}}^{\gamma \gamma }$} & S1 &17.0  & 14.9  & 5.1   & 6.3   & 1.3  \\[1pt]
  & S2 &15.7  & 14.9  & 3.2   & 3.7   & 0.6  \\[4pt]
  \multirow{2}{*}{$\sigma_{\mathrm{ZH}}^{\mathrm{bb}}$} & S1 &7.0   & 3.5   & 2.7   & 4.0   & 3.6  \\[1pt]
  & S2 &5.2   & 3.5   & 2.0   & 2.1   & 2.4  \\[4pt]
  \multirow{2}{*}{$\sigma_{\mathrm{ttH}}^{\gamma \gamma }$} & S1 &10.0  & 4.6   & 5.9   & 6.4   & 1.5  \\[1pt]
  & S2 &7.4   & 4.6   & 4.1   & 3.9   & 0.5  \\[4pt]
  \multirow{2}{*}{$\sigma_{\mathrm{ttH}}^{\mathrm{ZZ}}$} & S1 &20.5  & 18.6  & 4.1   & 7.3   & 1.7  \\[1pt]
  & S2 &19.3  & 18.6  & 3.1   & 3.8   & 0.9  \\[4pt]
  \multirow{2}{*}{$\sigma_{\mathrm{ttH}}^{\mathrm{WW} \tau \tau}$} & S1 &22.1  & 6.3   & 18.2  & 7.0   & 8.1  \\[1pt]
  & S2 &20.2  & 6.3   & 17.9  & 4.3   & 5.1  \\[4pt]
  \multirow{2}{*}{$\sigma_{\mathrm{ttH}}^{\mathrm{bb}}$} & S1 &19.9  & 3.2   & 4.2   & 7.4   & 17.8 \\[1pt]
  & S2 &14.2  & 3.2   & 3.4   & 4.4   & 12.7 \\[4pt]
  \hline
\end{tabular}
\hspace{0.5cm}
\begin{tabular}{@{} l c c@{\hskip 0.15in} c c c c @{}}
 \hline
 \multicolumn{7}{c}{CMS}\\
 \hline
  &  & \multicolumn{5}{c}{3000 $\text{fb}^{-1}$ uncertainty [\%]} \\
  &  & Total & Stat & Exp & SigAcc & BkgTh \\
 \hline
\multirow{2}{*}{$\sigma_{\mathrm{ggH}}^{\gamma \gamma }$} & S1  & 3.9& 1.9 & 3.3 & 0.7 & 1.0  \\[1pt]
                        & S2  & 2.8& 1.9 & 2.1 & 0.8 & 0.9  \\[4pt]
\multirow{2}{*}{$\sigma_{\mathrm{ggH}}^{\mathrm{ZZ}}$} & S1  & 4.1& 2.1 & 2.7 & 1.2 & 1.7  \\[1pt]
                        & S2  & 3.0& 2.1 & 1.8 & 0.8 & 0.7  \\[4pt]
\multirow{2}{*}{$\sigma_{\mathrm{ggH}}^{\mathrm{WW}}$} & S1  & 3.6& 1.2 & 1.5 & 2.9 & 1.0  \\[1pt]
                        & S2  & 2.5& 1.2 & 1.2 & 1.6 & 0.9  \\[4pt]
\multirow{2}{*}{$\sigma_{\mathrm{ggH}}^{\tau \tau }$} & S1  & 5.7& 2.6 & 3.5 & 3.3 & 1.7  \\[1pt]
                        & S2  & 4.6& 2.6 & 2.9 & 2.3 & 0.7  \\[4pt]
\multirow{2}{*}{$\sigma_{\mathrm{ggH}}^{\mathrm{bb}}$} & S1  & 34.3& 20.6 & 10.0 & 23.7 & 3.2  \\[1pt]
                        & S2  & 24.7& 20.6 & 2.6 & 12.2 & 1.5  \\[4pt]
\multirow{2}{*}{$\sigma_{\mathrm{ggH}}^{\mu \mu }$} & S1  & 15.9& 13.4 & 8.0 & 2.6 & 1.9  \\[1pt]
                        & S2  & 13.5& 13.4 & 2.0 & 1.4 & 0.6  \\[4pt]
\multirow{2}{*}{$\sigma_{\mathrm{VBF}}^{\gamma \gamma }$} & S1  & 22.1& 5.2 & 19.9 & 7.9 & 1.3  \\[1pt]
                        & S2  & 12.7& 5.2 & 10.9 & 4.0 & 0.3  \\[4pt]
\multirow{2}{*}{$\sigma_{\mathrm{VBF}}^{\mathrm{ZZ}}$} & S1  & 15.1& 11.7 & 1.8 & 8.8 & 2.4  \\[1pt]
                        & S2  & 13.3& 11.7 & 1.3 & 5.9 & 0.8  \\[4pt]
\multirow{2}{*}{$\sigma_{\mathrm{VBF}}^{\mathrm{WW}}$} & S1  & 8.1& 6.3 & 2.0 & 4.4 & 1.8  \\[1pt]
                        & S2  & 7.2& 6.3 & 1.6 & 2.8 & 1.1  \\[4pt]
\multirow{2}{*}{$\sigma_{\mathrm{VBF}}^{\tau \tau }$} & S1  & 4.9& 3.8 & 2.0 & 2.8 & 1.5  \\[1pt]
                        & S2  & 4.2& 3.8 & 1.3 & 1.2 & 0.4  \\[4pt]
\multirow{2}{*}{$\sigma_{\mathrm{VBF}}^{\mu \mu }$} & S1  & 57.3& 53.2 & 11.3 & 18.0 & 4.5  \\[1pt]
                        & S2  & 54.0& 53.2 & 2.6 & 9.5 & 1.0  \\[4pt]
\multirow{2}{*}{$\sigma_{\mathrm{WH}}^{\gamma \gamma }$} & S1  & 14.3& 13.6 & 3.7 & 2.0 & 1.4  \\[1pt]
                        & S2  & 13.8& 13.6 & 1.7 & 1.5 & 0.2  \\[4pt]
\multirow{2}{*}{$\sigma_{\mathrm{WH}}^{\mathrm{ZZ}}$} & S1  & 47.9& 46.5 & 7.8 & 11.2 & 2.8  \\[1pt]
                        & S2  & 47.8& 46.5 & 3.8 & 4.0 & 0.8  \\[4pt]
\multirow{2}{*}{$\sigma_{\mathrm{WH}}^{\mathrm{WW}}$} & S1  & 15.6& 12.9 & 6.5 & 5.3 & 2.2  \\[1pt]
                        & S2  & 13.7& 12.9 & 3.1 & 2.9 & 1.5  \\[4pt]
\multirow{2}{*}{$\sigma_{\mathrm{WH}}^{\mathrm{bb}}$} & S1  & 16.0& 5.6 & 9.8 & 5.3 & 10.8  \\[1pt]
                        & S2  & 9.4& 5.6 & 5.1 & 2.2 & 5.1  \\[4pt]
\multirow{2}{*}{$\sigma_{\mathrm{ZH}}^{\gamma \gamma }$} & S1  & 23.5& 23.1 & 2.9 & 3.1 & 1.5  \\[1pt]
                        & S2  & 23.2& 23.1 & 1.2 & 2.4 & 0.4  \\[4pt]
\multirow{2}{*}{$\sigma_{\mathrm{ZH}}^{\mathrm{ZZ}}$} & S1  & 82.3& 75.7 & 16.4 & 26.3 & 7.6  \\[1pt]
                        & S2  & 78.4& 75.7 & 9.9 & 15.1 & 1.3  \\[4pt]
\multirow{2}{*}{$\sigma_{\mathrm{ZH}}^{\mathrm{WW}}$} & S1  & 18.5& 17.2 & 3.5 & 5.3 & 2.4  \\[1pt]
                        & S2  & 17.7& 17.2 & 3.0 & 2.8 & 1.7  \\[4pt]
\multirow{2}{*}{$\sigma_{\mathrm{ZH}}^{\mathrm{bb}}$} & S1  & 7.9& 4.2 & 2.3 & 5.6 & 3.1  \\[1pt]
                        & S2  & 6.0& 4.2 & 1.9 & 2.9 & 2.6  \\[4pt]
\multirow{2}{*}{$\sigma_{\mathrm{ttH}}^{\gamma \gamma }$} & S1  & 9.3& 7.7 & 3.9 & 3.5 & 1.0  \\[1pt]
                        & S2  & 8.4& 7.7 & 2.7 & 1.9 & 0.2  \\[4pt]
\multirow{2}{*}{$\sigma_{\mathrm{ttH}}^{\mathrm{ZZ}}$} & S1  & 24.6& 23.6 & 4.2 & 4.9 & 2.5  \\[1pt]
                        & S2  & 24.2& 23.6 & 3.1 & 2.6 & 1.8  \\[4pt]
\multirow{2}{*}{$\sigma_{\mathrm{ttH}}^{\mathrm{WW}}$} & S1  & 11.2& 4.2 & 9.1 & 1.8 & 4.5  \\[1pt]
                        & S2  & 8.7& 4.2 & 6.9 & 1.1 & 3.0  \\[4pt]
\multirow{2}{*}{$\sigma_{\mathrm{ttH}}^{\mathrm{bb}}$} & S1  & 15.9& 2.8 & 3.9 & 0.0 & 15.2  \\[1pt]
                        & S2  & 10.8& 2.8 & 2.7 & 0.1 & 10.0  \\[4pt]
\multirow{2}{*}{$\sigma_{\mathrm{ttH}}^{\tau \tau }$} & S1  & 16.5& 8.7 & 13.1 & 3.4 & 3.5  \\[1pt]
                        & S2  & 14.2& 8.7 & 10.9 & 1.6 & 2.1  \\[4pt]
\hline
\end{tabular}
\label{tab:summary_A1_5PD}
\vspace{0.5cm}
\end{table}




\asubsubsection{Cross sections per-production mode}

The expected $\pm 1\sigma$ relative uncertainties on the per-production-mode cross sections parameters in S2 for ATLAS, CMS and their combination are summarised in Figure~\ref{fig:summary_A1_5P}.
Additionally, the numerical values for the ATLAS-CMS combination  are also given, with the uncertainty decomposed in three components: statistical, experimental and theory.
In scenario S2 the contribution from the statistical, experimental and theoretical uncertainties to the total error for the combined  \ggh and \vbf cross section measurements is similar. For \wh and \zh production cross section measurements, the statistical  and theoretical uncertainty are the dominant one.
Finally, the total uncertainty on the \tth production cross section measurement is dominated by the theoretical uncertainty, which is almost a factor two larger with respect to the other components.
The numerical values of the expected $\pm 1\sigma$ uncertainties on the per-production-mode cross sections for the ATLAS and CMS projections are given in Table~\ref{tab:summary_A1_5P}. The table  gives the breakdown of the uncertainty into four components: statistical, signal theory, background theory and experimental for both scenarios S1 and S2. 

\begin{figure}[hbtp]
\centering
\includegraphics[height=0.332\textheight]{\main/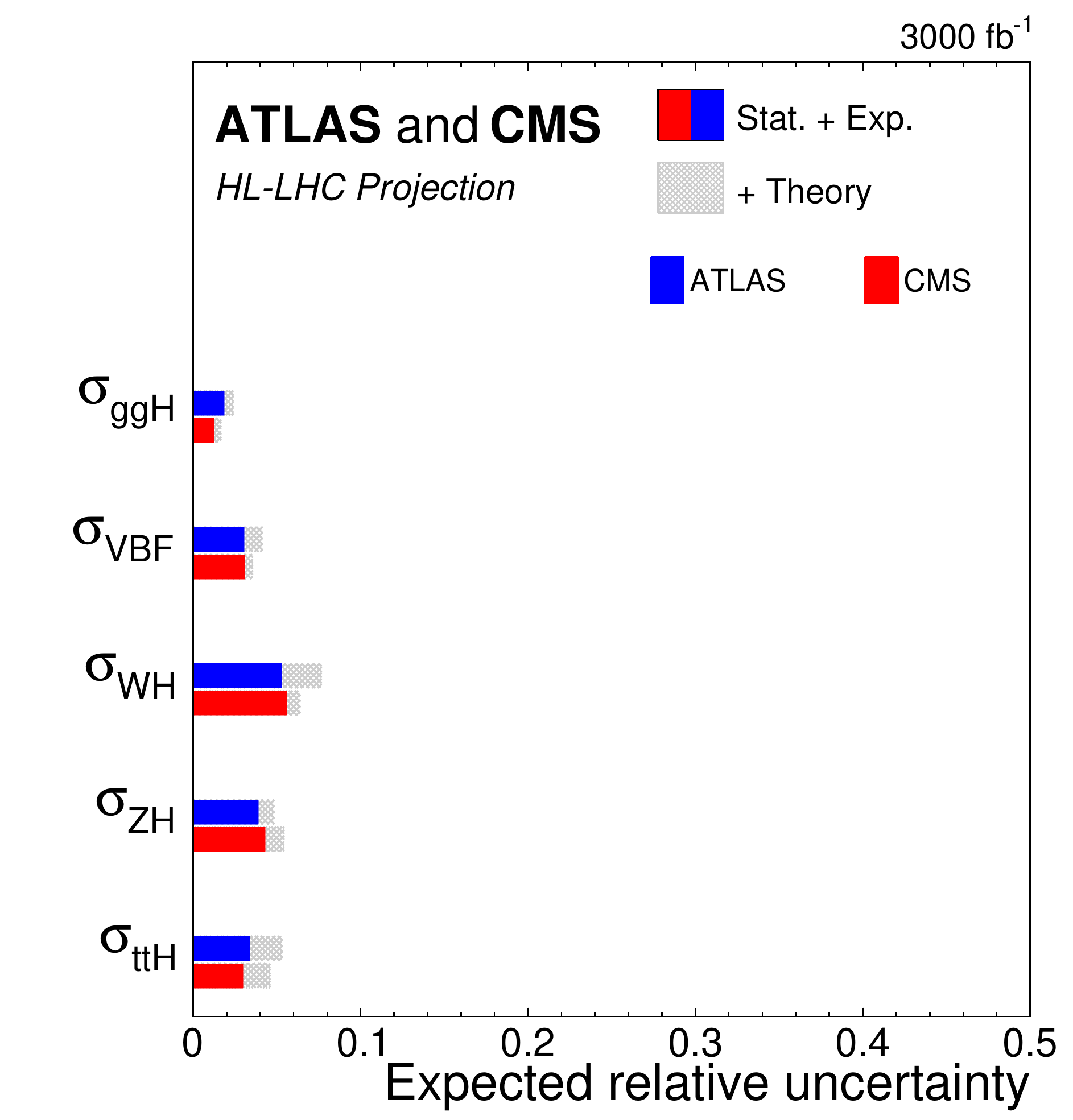}%
\includegraphics[height=0.33\textheight]{\main/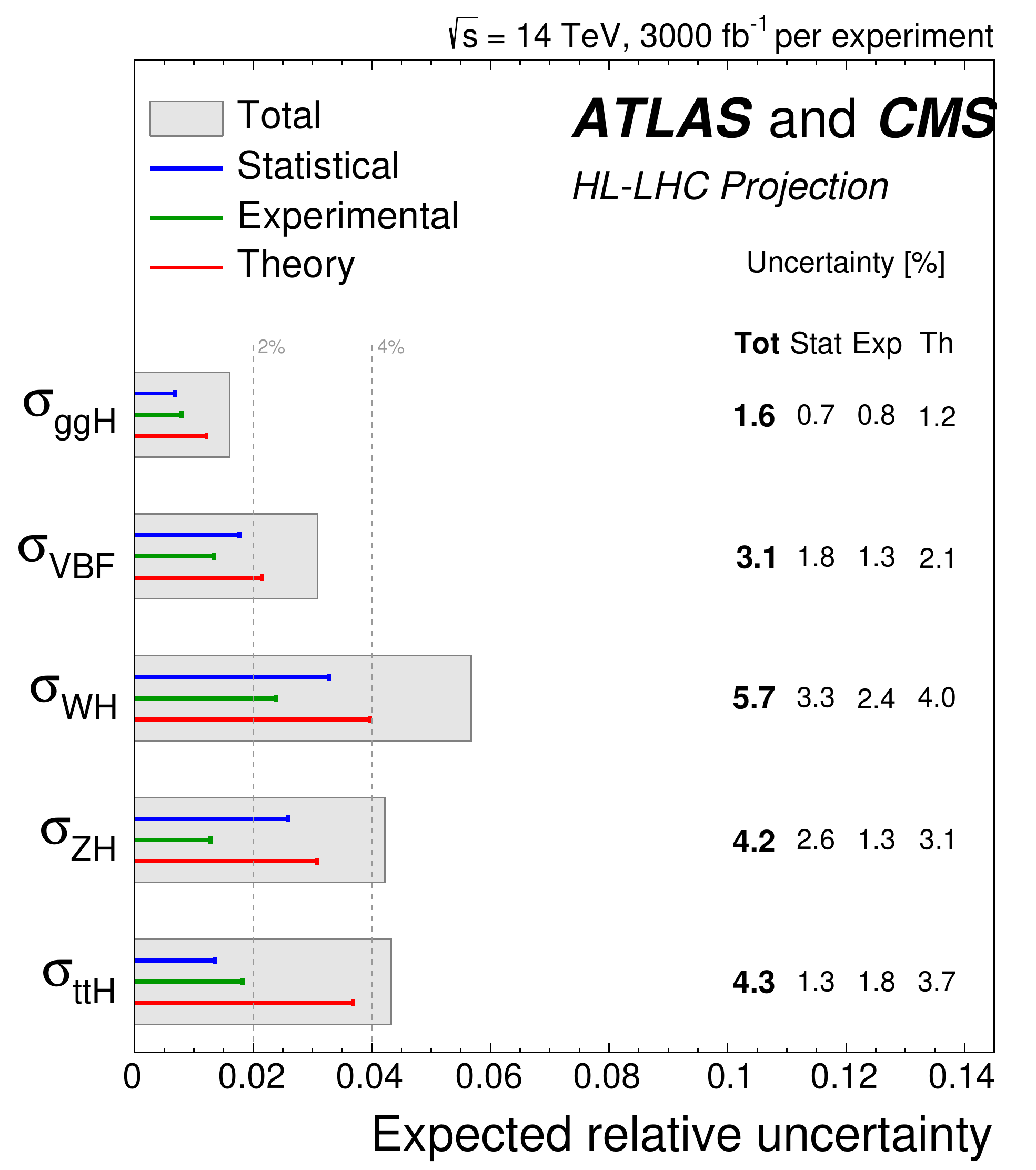}%
\caption{(left) Summary plot showing the total expected $\pm 1\sigma$ uncertainties in S2 (with YR18 systematic uncertainties) on the  per-production-mode cross sections normalised to the SM predictions   for ATLAS (blue)  and CMS (red). The filled coloured box corresponds to the statistical and experimental systematic uncertainties, while the hatched grey area represent the additional contribution to the total uncertainty due to theoretical systematic uncertainties.
(right) Summary plot showing the total expected $\pm 1\sigma$  uncertainties in S2 (with YR18 systematic uncertainties) on the per-production-mode cross sections normalised to the SM predictions for the combination of ATLAS and CMS extrapolations. For each measurement,  the total uncertainty is indicated by a grey box while the statistical, experimental and theory uncertainties are indicated by a blue, green and red line respectively.  In addition, the numerical values are also reported.}
\label{fig:summary_A1_5P}
\end{figure}

\begin{table}[hbtp]
\centering
\caption{The expected $\pm 1\sigma$ uncertainties, expressed as percentages, on the per-production-mode cross sections normalised to the SM values  for  ATLAS (left) and CMS (right). Values are given for both S1 (with Run~2 systematic uncertainties~\cite{Sirunyan:2018koj}) and S2 (with YR18 systematic uncertainties). The total uncertainty is decomposed into four components: statistical (Stat), signal theory (SigTh), background theory (BkgTh) and experimental (Exp).}
\small
\begin{tabular}{@{} l c c@{\hskip 0.15in} c c c c @{}}
  \hline
   \multicolumn{7}{c}{ATLAS}\\
 \hline  
  &  & \multicolumn{5}{c}{3000 $\text{fb}^{-1}$ uncertainty [\%]} \\
  &  & Total & Stat & Exp & SigTh & BkgTh \\
  \hline
  \multirow{2}{*}{$\sigma_{\mathrm{ggH}}$} & S1 &3.5   & 0.8   & 2.1   & 2.1   & 1.6  \\[1pt]
  & S2 &2.4   & 0.8   & 1.7   & 1.2   & 1.0  \\[4pt]
  \multirow{2}{*}{$\sigma_{\mathrm{VBF}}$} & S1 &5.5   & 2.0   & 2.7   & 3.7   & 2.1  \\[1pt]
  & S2 &4.2   & 2.0   & 2.3   & 2.2   & 1.7  \\[4pt]
  \multirow{2}{*}{$\sigma_{\mathrm{WH}}$} & S1 &9.3   & 4.0   & 4.0   & 5.1   & 5.4  \\[1pt]
  & S2 &7.7   & 4.0   & 3.4   & 3.3   & 4.5  \\[4pt]
  \multirow{2}{*}{$\sigma_{\mathrm{ZH}}$} & S1 &6.2   & 3.4   & 2.4   & 3.4   & 3.0  \\[1pt]
  & S2 &4.8   & 3.4   & 1.8   & 2.0   & 2.1  \\[4pt]
  \multirow{2}{*}{$\sigma_{\mathrm{ttH}}$} & S1 &6.7   & 1.9   & 3.1   & 3.7   & 4.3  \\[1pt]
  & S2 &5.3   & 1.9   & 2.8   & 2.4   & 3.3  \\[4pt]
  \hline
\end{tabular}
\hspace{0.5cm}
\begin{tabular}{@{} l c c@{\hskip 0.15in} c c c c @{}}
 \hline
 \multicolumn{7}{c}{CMS}\\
 \hline
  &  & \multicolumn{5}{c}{3000 $\text{fb}^{-1}$ uncertainty [\%]} \\
  &  & Total & Stat & Exp & SigTh & BkgTh \\
 \hline
\multirow{2}{*}{$\sigma_{\mathrm{ggH}}$} & S1  & 2.4& 0.8 & 1.2 & 1.6 & 0.9  \\[1pt]
                        & S2  & 1.7& 0.8 & 0.9 & 0.9 & 0.6  \\[4pt]
\multirow{2}{*}{$\sigma_{\mathrm{VBF}}$} & S1  & 4.1& 2.6 & 2.1 & 2.0 & 1.3  \\[1pt]
                        & S2  & 3.5& 2.6 & 1.6 & 1.8 & 0.3  \\[4pt]
\multirow{2}{*}{$\sigma_{\mathrm{WH}}$} & S1  & 8.1& 4.6 & 5.2 & 2.6 & 3.3  \\[1pt]
                        & S2  & 6.4& 4.6 & 3.2 & 1.5 & 2.7  \\[4pt]
\multirow{2}{*}{$\sigma_{\mathrm{ZH}}$} & S1  & 6.7& 3.9 & 2.1 & 4.3 & 2.5  \\[1pt]
                        & S2  & 5.4& 3.9 & 1.7 & 2.4 & 2.3  \\[4pt]
\multirow{2}{*}{$\sigma_{\mathrm{ttH}}$} & S1  & 5.8& 1.8 & 3.1 & 1.9 & 4.1  \\[1pt]
                        & S2  & 4.6& 1.8 & 2.4 & 1.1 & 3.4  \\[4pt]
\hline
\end{tabular}
\label{tab:summary_A1_5P}
\vspace{0.5cm}
\end{table}





\asubsubsection{Branching ratios per-decay mode}

The expected $\pm 1\sigma$ uncertainties on the per-decay-mode branching ratios normalised to the SM expectations in S2 for ATLAS, CMS and their combination are summarised in Figure~\ref{fig:summary_A1_5D}.
Additionally, the numerical values for the ATLAS-CMS combination  are also reported in the figure, with the uncertainty decomposed in three components: statistical, experimental and theory.
The S2 uncertainties for the combined ATLAS-CMS extrapolation range from $2-4\%$, with the exception of that on $B^{\mu\mu}$ at $8\%$ and on $B^{Z\gamma}$ at $19\%$.
The numerical values in both S1 and S2 for ATLAS and CMS are given in Table~\ref{tab:summary_A1_5D} where the the breakdown of the uncertainty into four components  is provided. 
In projections of both experiments, the S1 uncertainties are up to a factor of 1.5 larger than those in S2, reflecting the larger systematic component. 
 The systematic uncertainties generally dominate in both S1 and S2. In S2 the signal theory uncertainty is the largest, or joint-largest, component for all parameters except $BR^{\mu\mu}$ and $B^{Z\gamma}$, which remain limited by statistics due to the small  branching fractions. 

\begin{figure}[hbtp]
\centering
\includegraphics[height=0.332\textheight]{\main/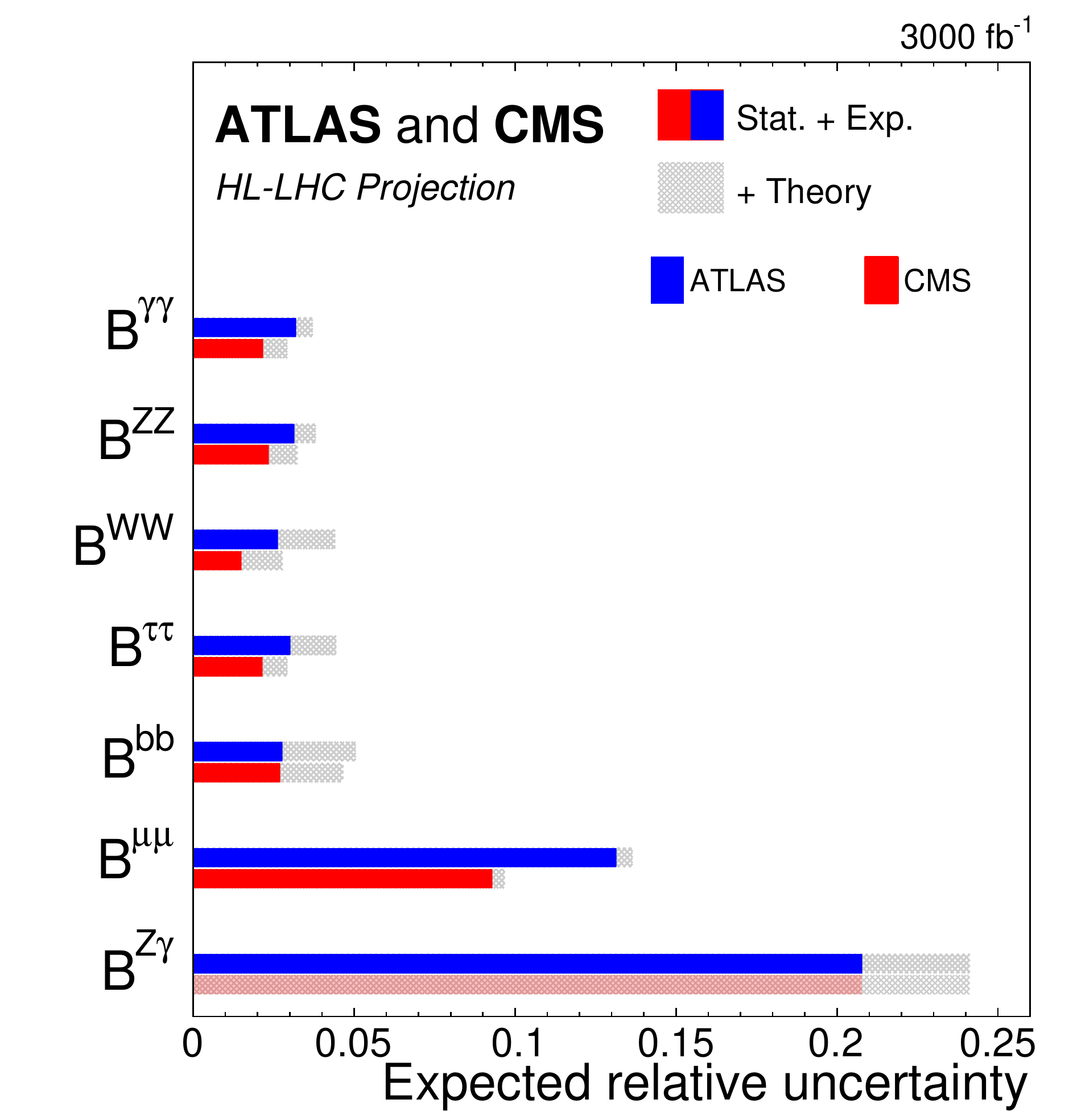}%
\includegraphics[height=0.33\textheight]{\main/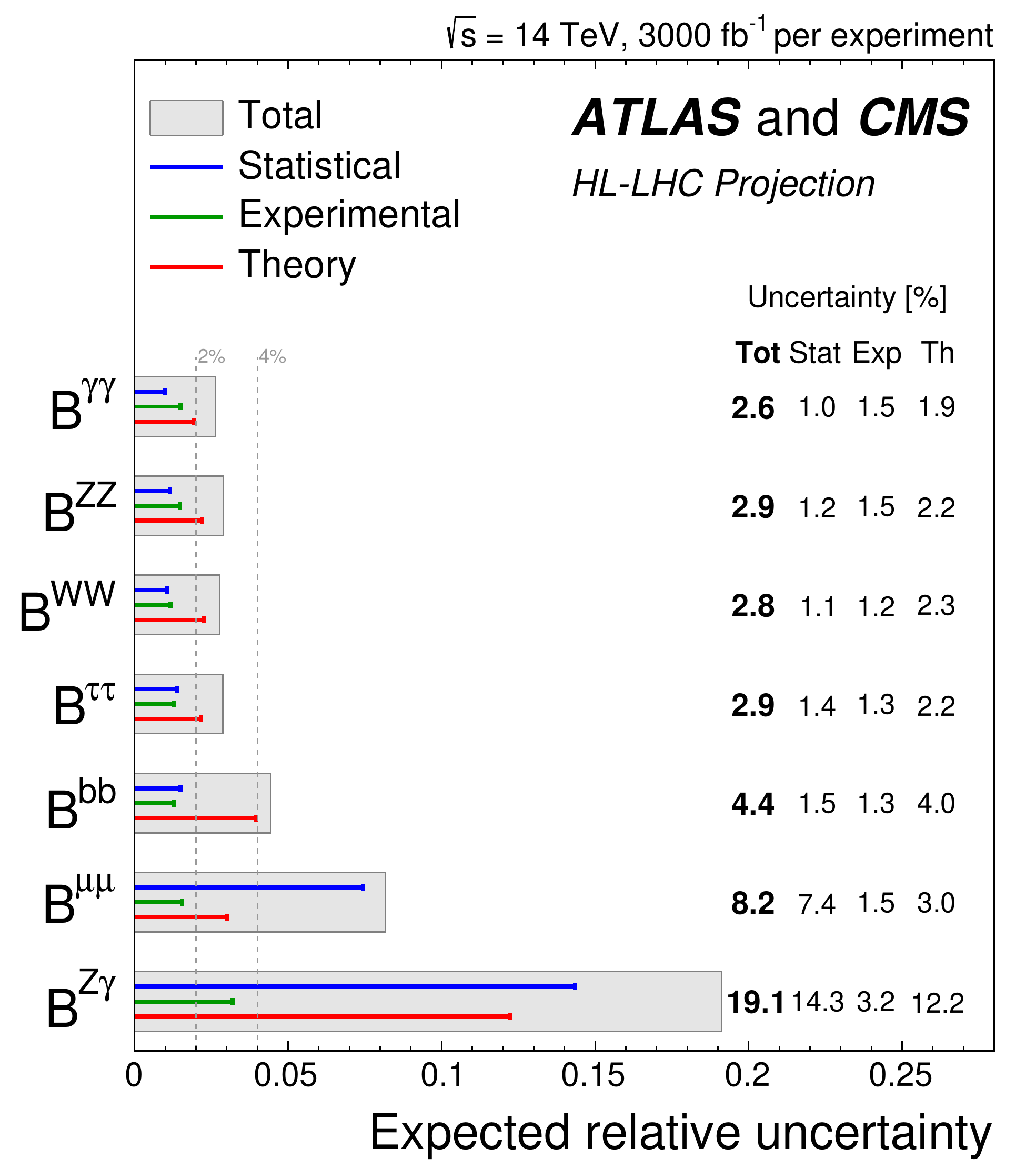}%
\caption{(left) Summary plot showing the total expected $\pm 1\sigma$ uncertainties in S2 (with YR18 systematic uncertainties) on the  per-decay-mode branching ratios normalised to the SM predictions   for ATLAS (blue)  and CMS (red). The filled coloured box corresponds to the statistical and experimental systematic uncertainties, while the hatched grey area represent the additional contribution to the total uncertainty due to theoretical systematic uncertainties.
(right) Summary plot showing the total expected $\pm 1\sigma$  uncertainties in S2 (with YR18 systematic uncertainties) on the per-decay-mode branching ratios normalised to the SM predictions for the combination of ATLAS and CMS extrapolations. For each measurement,  the total uncertainty is indicated by a grey box while the statistical, experimental and theory uncertainties are indicated by a blue, green and red line respectively. In addition, the numerical values are also reported.}
\label{fig:summary_A1_5D}
\end{figure}

\begin{table}[hbtp]
\centering
\caption{The expected $\pm 1\sigma$ relative uncertainties, expressed as percentages, on the Higgs boson branching ratios normalised by the SM expectations for ATLAS (left) and CMS (right). Values are given for both S1 (with Run~2 systematic uncertainties~\cite{Sirunyan:2018koj}) and S2 (with YR18 systematic uncertainties). The total uncertainty is decomposed into four components: statistical (Stat), signal theory (SigTh), background theory (BkgTh) and experimental (Exp).}
\small
\begin{tabular}{@{} l c c@{\hskip 0.15in} c c c c @{}}
  \hline
     \multicolumn{7}{c}{ATLAS}\\
 \hline
  &  & \multicolumn{5}{c}{3000 $\text{fb}^{-1}$ relative uncertainty [\%]} \\
  &  & Total & Stat & Exp & SigTh & BkgTh \\
  \hline
  \multirow{2}{*}{$\mathrm{B}^{\gamma\gamma}$} & S1 &6.0   & 1.2   & 4.7   & 3.3   & 1.4  \\[1pt] 
  & S2 &3.7   & 1.2   & 2.9   & 1.8   & 0.6  \\[4pt]
  \multirow{2}{*}{$\mathrm{B}^{\mathrm{WW}}$} & S1 &5.8   & 1.0   & 2.8   & 4.3   & 2.6  \\[1pt]
  & S2 &4.4   & 1.0   & 2.4   & 3.2   & 1.6  \\[4pt]
  \multirow{2}{*}{$\mathrm{B}^{\mathrm{ZZ}}$} & S1 &5.3   & 1.6   & 3.0   & 3.7   & 1.7  \\[1pt]
  & S2 &3.8   & 1.6   & 2.7   & 1.9   & 1.0  \\[4pt]
  \multirow{2}{*}{$\mathrm{B}^{\mathrm{bb}}$} & S1 &7.6   & 2.0   & 2.4   & 5.0   & 4.7  \\[1pt]
  & S2 &5.0   & 2.0   & 1.9   & 2.8   & 3.2  \\[4pt]
  \multirow{2}{*}{$\mathrm{B}^{\tau\tau }$} & S1 &6.0   & 1.7   & 2.7   & 4.4   & 2.4  \\[1pt]
  & S2 &4.4   & 1.7   & 2.5   & 2.8   & 1.7  \\[4pt]
  \multirow{2}{*}{$\mathrm{B}^{\mu\mu}$} & S1  &14.9  & 12.7  & 3.2   & 6.8   & 0.3  \\[1pt]
  & S2 &13.7  & 12.7  & 3.2   & 3.7   & 0.3  \\[4pt]
  \multirow{2}{*}{$\mathrm{B}^{\mathrm{Z}\gamma}$} & S1 &24.2  & 20.3  & 4.5   & 12.2  & 0.0  \\[1pt]
  & S2 &24.2  & 20.3  & 4.5   & 12.2  & 0.0  \\[4pt]
  \hline
\end{tabular}
\hspace{0.5cm}
\begin{tabular}{@{} l c c@{\hskip 0.15in} c c c c @{}}
 \hline
   \multicolumn{7}{c}{CMS}\\
 \hline
  &  & \multicolumn{5}{c}{3000 $\text{fb}^{-1}$ relative uncertainty [\%]} \\
  &  & Total & Stat & Exp & SigTh & BkgTh \\
 \hline
\multirow{2}{*}{$\mathrm{B}^{\gamma\gamma}$} & S1  & 4.4& 1.3 & 2.6 & 3.3 & 0.3  \\[1pt]
                        & S2  & 3.0& 1.3 & 1.7 & 1.9 & 0.3  \\[4pt]
\multirow{2}{*}{$\mathrm{B}^{\mathrm{WW}}$} & S1  & 4.0& 1.0 & 1.4 & 3.5 & 1.0  \\[1pt]
                        & S2  & 2.8& 1.0 & 1.1 & 2.2 & 0.9  \\[4pt]
\multirow{2}{*}{$\mathrm{B}^{\mathrm{ZZ}}$} & S1  & 5.0& 1.6 & 2.5 & 3.5 & 1.9  \\[1pt]
                        & S2  & 3.2& 1.6 & 1.7 & 2.1 & 0.7  \\[4pt]
\multirow{2}{*}{$\mathrm{B}^{\mathrm{bb}}$} & S1  & 7.0& 2.1 & 2.3 & 5.2 & 3.6  \\[1pt]
                        & S2  & 4.7& 2.1 & 1.7 & 2.4 & 2.9  \\[4pt]
\multirow{2}{*}{$\mathrm{B}^{\tau\tau }$} & S1  & 3.9& 1.6 & 1.9 & 2.6 & 1.5  \\[1pt]
                        & S2  & 2.9& 1.6 & 1.4 & 1.9 & 0.6  \\[4pt]
\multirow{2}{*}{$\mathrm{B}^{\mu\mu}$} & S1  & 12.8& 9.1 & 7.6 & 4.7 & 0.8  \\[1pt]
                        & S2  & 9.6& 9.1 & 1.7 & 2.6 & 0.8  \\[4pt]
\hline
\end{tabular}
\label{tab:summary_A1_5D}
\vspace{0.5cm}
\end{table}

The correlations range up to 40\%, and are largest between modes where the sensitivity is dominated by gluon-fusion production. This reflects the impact of the theory uncertainties affecting the SM prediction of the gluon-fusion production rate.



\asubsection[R. Di Nardo, A. Gilbert, H. Yang, N. Berger,  D. Du,
  M. D\"uhrssen, A. Gilbert, R. Gugel, L. Ma B. Murray,   P. Milenovic]{Kappa interpretation of the combined Higgs boson
  measurement projections}
\label{sec2:exp_kappa}

\subsubsection{Interpretations and results for HL-LHC}

In this section combination results are given for a parametrisation based on the coupling modifier, or $\kappa$-framework~\cite{Heinemeyer:2013tqa}. A set of coupling modifiers, $\vec\kappa$, is introduced to parametrise potential deviations from the SM~predictions of the Higgs boson couplings to SM~bosons and fermions. For a given production process or decay mode $j$, a coupling modifier $\kappa_j$ is defined such that,

\begin{equation}
\label{eq:kappa}
  \kappa_j^2=\sigma_j/\sigma_j^\SM\ \ \ {\rm or}\ \ \  \kappa_j^2=\Gamma^j/\Gamma^j_\SM.
\end{equation}

In the SM, all $\kappa_j$ values are positive and equal to unity. Six coupling modifiers corresponding to the tree-level Higgs boson couplings are defined: $\kW$, $\kZ$, $\ktop$, $\kb$, $\ktau$ and $\kmu$. In addition, the effective coupling modifiers $\kglu$, $\kgam$ and $\kzgam$ are introduced to describe \ggh production, \hgg decay and \hzg decay loop processes. 
The total width of the Higgs boson, relative to the SM prediction, varies with the coupling modifiers as $\GammaSM = \sum_{j}\mathrm{B}^{j}_{\text{SM}}\kappa_{j}^{2} / (1 - \Bbsm)$, where $\mathrm{B}^{j}_{\text{SM}}$ is the SM branching fraction for the $\PH\rightarrow jj$ channel and $\Bbsm$ is the Higgs boson branching fraction to BSM final states. In the results for the $\kappa_j$ parameters presented here $\Bbsm$ is fixed to zero and only decays to SM particles are allowed. Projections are also given for the upper limit on $\Bbsm$ when this restriction is relaxed, in which an additional constraint that $|\kV| < 1$ is imposed. A constraint on $\GammaSM$ is also obtained in this model by treating it as a free parameter in place of one of the other $\kappa$ parameters.

The expected uncertainties for the coupling modifier parametrisation for ATLAS, CMS~\cite{ATL-PHYS-PUB-2018-054,CMS-PAS-FTR-18-011} and their combination for scenario S2 are summarised in Figure~\ref{fig:summary_K2}. The numerical values in both S1 and S2 for ATLAS and CMS are provided in Table~\ref{tab:summary_K2}.
For the combined measurement in S2, the uncertainty components  contribute at a similar level for $\kgam$, $\kW$, $\kZ$ and $\ktau$. The signal theory remains the main component for $\ktop$ and $\kglu$, while $\kmu$ and $\kzgam$ are limited by statistics.

The expected $1\sigma$ uncertainty on $\Bbsm$, for the parametrisation with $\Bbsm\geq0$ and $|\kV|\leq1$, is $0.033$ ($0.049$) in S1 and $0.027$ ($0.032$) in S2 for CMS (ATLAS), where in the latter case the statistical uncertainty is the largest component.
The expected uncertainty for the ATLAS-CMS combination on  $\Bbsm$ is $0.025$ in S2.
The uncertainty on $\GammaSM$, determined for CMS only, is $0.05$ ($0.04$) in S1 (S2).

\begin{figure}[hbtp]
\centering
\includegraphics[height=0.332\textheight]{\main/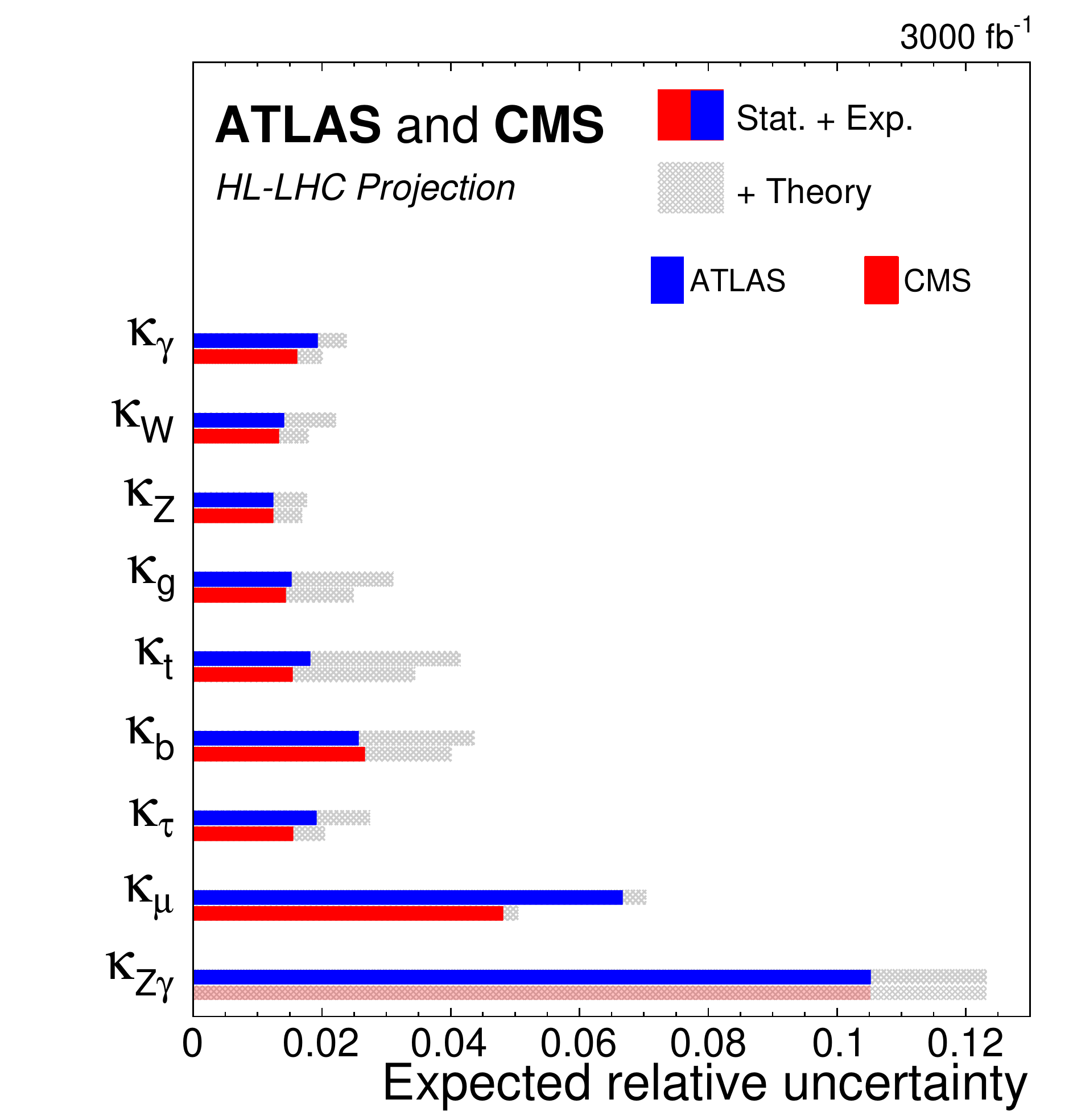}%
\includegraphics[height=0.33\textheight]{\main/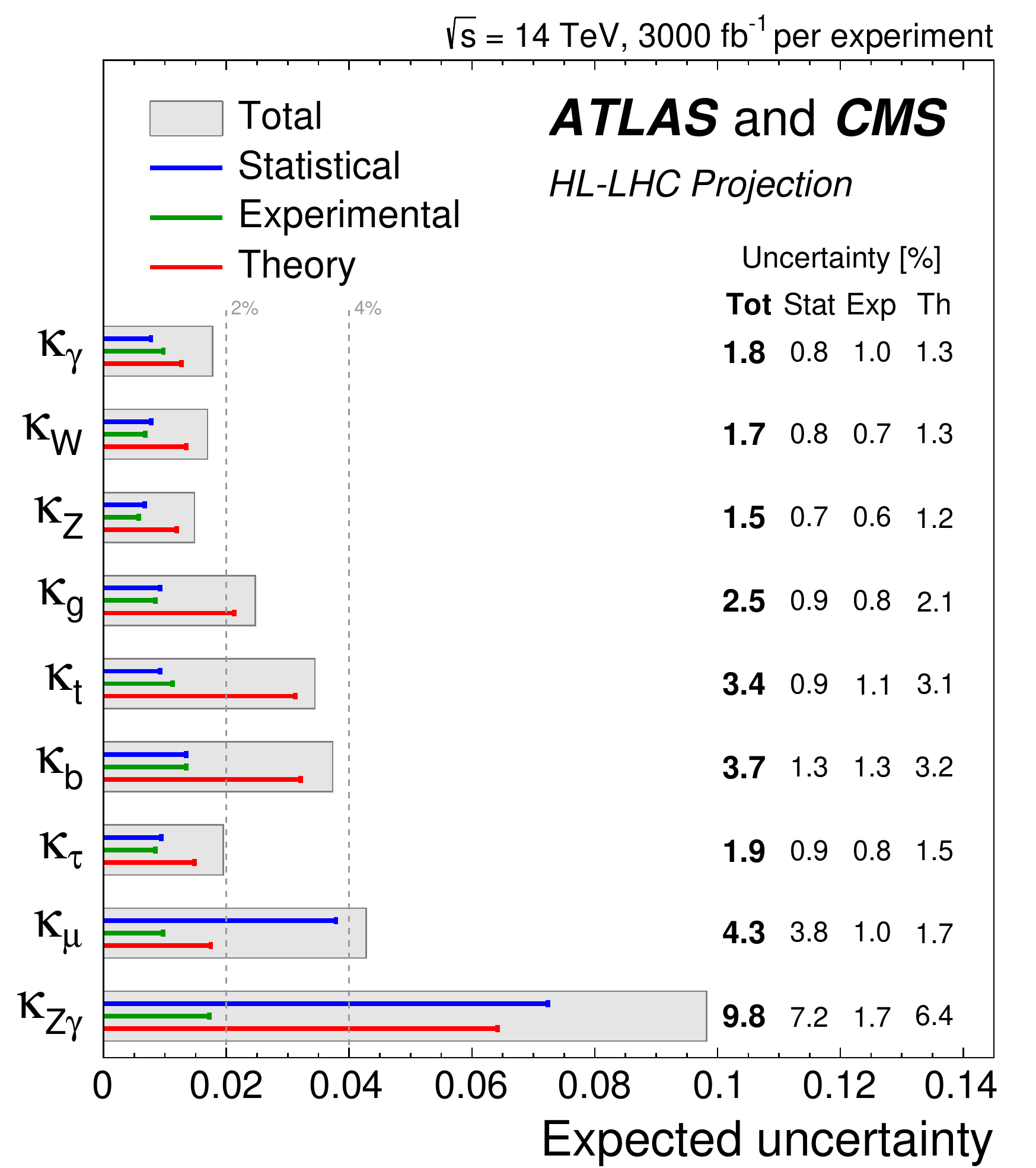}%
\caption{(left) Summary plot showing the total expected $\pm 1\sigma$ uncertainties in S2 (with YR18 systematic uncertainties) on the coupling modifier parameters   for ATLAS (blue)  and CMS (red). The filled coloured box corresponds to the statistical and experimental systematic uncertainties, while the hatched grey area represent the additional contribution to the total uncertainty due to theoretical systematic uncertainties.
(right) Summary plot showing the total expected $\pm 1\sigma$  uncertainties in S2 (with YR18 systematic uncertainties) on the coupling modifier parameters for the combination of ATLAS and CMS extrapolations. For each measurement,  the total uncertainty is indicated by a grey box while the statistical, experimental and theory uncertainties are indicated by a blue, green and red line respectively.}
\label{fig:summary_K2}
\end{figure}

\begin{table}[hbtp]
\centering
\caption{The expected $\pm 1\sigma$ uncertainties, expressed as percentages, on the coupling modifier parameters.
Values are given for both S1 (with Run~2 systematic uncertainties~\cite{Sirunyan:2018koj}) and S2 (with YR18 systematic uncertainties). The total uncertainty is decomposed into four components: statistical (Stat), signal theory (SigTh), background theory (BkgTh) and experimental (Exp).}
\small
\begin{tabular}{@{} l c c@{\hskip 0.15in} c c c c @{}}
  \hline
     \multicolumn{7}{c}{ATLAS}\\
 \hline
  &  & \multicolumn{5}{c}{3000 $\text{fb}^{-1}$ uncertainty [\%]} \\
  &  & Total & Stat & SigTh & BkgTh & Exp \\
  \hline
  \multirow{2}{*}{$\kappa_{\gamma }$} & S1 & 3.7   & 0.9   & 2.2   & 1.4   & 2.5  \\[1pt]
  & S2& 2.4   & 0.9   & 1.1   & 0.9   & 1.7  \\[4pt]
  \multirow{2}{*}{$\kappa_{\mathrm{W}}$} & S1  & 3.1   & 0.8   & 1.9   & 1.9   & 1.3  \\[1pt]
  & S2 & 2.2   & 0.8   & 1.2   & 1.3   & 1.2  \\[4pt]
  \multirow{2}{*}{$\kappa_{\mathrm{Z}}$} & S1 & 2.6   & 0.8   & 1.8   & 1.2   & 1.1  \\[1pt]
  & S2 & 1.8   & 0.8   & 1.0   & 0.8   & 0.9  \\[4pt]
  \multirow{2}{*}{$\kappa_{\mathrm{g}}$} & S1 & 4.2   & 1.0   & 3.2   & 2.2   & 1.4  \\[1pt]
  & S2 & 3.1   & 1.0   & 2.2   & 1.6   & 1.2  \\[4pt]
  \multirow{2}{*}{$\kappa_{\mathrm{t}}$} & S1 & 6.3   & 1.1   & 4.9   & 3.4   & 1.6  \\[1pt]
  & S2 & 4.2   & 1.1   & 2.6   & 2.7   & 1.4  \\[4pt]
  \multirow{2}{*}{$\kappa_{\mathrm{b}}$} & S1 & 6.2   & 1.6   & 3.7   & 4.1   & 2.3  \\[1pt]
  & S2  & 4.4   & 1.6   & 2.1   & 2.8   & 2.0  \\[4pt]
  \multirow{2}{*}{$\kappa_{\tau }$} & S1 & 3.7 & 1.1   & 2.6   & 1.8   & 1.7  \\[1pt]
  & S2 & 2.7   & 1.1   & 1.5   & 1.2   & 1.6  \\[4pt]
  \multirow{2}{*}{$\kappa_{\mu}$} & S1  & 7.7 & 6.4   & 3.6   & 1.4   & 1.9  \\[1pt]
  & S2 & 7.0   & 6.4   & 2.0   & 0.9   & 1.8  \\[4pt]
  \multirow{2}{*}{$\kappa_{\mathrm{Z}\gamma}$} & S1 & 12.7  & 10.2  & 6.9   & 1.4   & 2.5  \\[1pt]
  & S2 & 12.4  & 10.2  & 6.4   & 0.9   & 2.4  \\[4pt]
  \hline
\end{tabular}
\hspace{0.5cm}
\begin{tabular}{@{} l c c@{\hskip 0.15in} c c c c @{}}
 \hline
   \multicolumn{7}{c}{CMS}\\
 \hline
  &  & \multicolumn{5}{c}{3000 $\text{fb}^{-1}$ uncertainty [\%]} \\
  &  & Total & Stat & SigTh & BkgTh & Exp \\
 \hline
\multirow{2}{*}{$\kappa_{\gamma }$} & S1  & 2.9& 1.1 & 1.8 & 1.0 & 1.7  \\[1pt]
                        & S2  & 2.0& 1.1 & 0.9 & 0.8 & 1.2  \\[4pt]
\multirow{2}{*}{$\kappa_{\mathrm{W}}$} & S1  & 2.6& 1.0 & 1.7 & 1.1 & 1.1  \\[1pt]
                        & S2  & 1.8& 1.0 & 0.9 & 0.8 & 0.8  \\[4pt]
\multirow{2}{*}{$\kappa_{\mathrm{Z}}$} & S1  & 2.4& 1.0 & 1.7 & 0.9 & 0.9  \\[1pt]
                        & S2  & 1.7& 1.0 & 0.9 & 0.7 & 0.7  \\[4pt]
\multirow{2}{*}{$\kappa_{\mathrm{g}}$} & S1  & 4.0& 1.1 & 3.4 & 1.3 & 1.2  \\[1pt]
                        & S2  & 2.5& 1.1 & 1.7 & 1.1 & 1.0  \\[4pt]
\multirow{2}{*}{$\kappa_{\mathrm{t}}$} & S1  & 5.5& 1.0 & 4.4 & 2.7 & 1.6  \\[1pt]
                        & S2  & 3.5& 1.0 & 2.2 & 2.1 & 1.2  \\[4pt]
\multirow{2}{*}{$\kappa_{\mathrm{b}}$} & S1  & 6.0& 2.0 & 4.3 & 2.9 & 2.3  \\[1pt]
                        & S2  & 4.0& 2.0 & 2.0 & 2.2 & 1.8  \\[4pt]
\multirow{2}{*}{$\kappa_{\tau }$} & S1  & 2.8& 1.2 & 1.8 & 1.1 & 1.4  \\[1pt]
                        & S2  & 2.0& 1.2 & 1.0 & 0.9 & 1.0  \\[4pt]
\multirow{2}{*}{$\kappa_{\mu}$} & S1  & 6.7& 4.7 & 2.5 & 1.0 & 3.9  \\[1pt]
                        & S2  & 5.0& 4.7 & 1.3 & 0.8 & 1.1  \\[4pt]
\hline
\end{tabular}
\label{tab:summary_K2}
\vspace{0.5cm}
\end{table}



The correlation coefficients between the coupling modifiers are in general larger compared to the one related to the signal strength (up to $+75\%$ ). One reason for this is that the normalisation of any signal process depends on the total width of the Higgs boson, which in turn depends on the values of the other coupling modifiers. The largest correlations involve $\kb$, as this gives the largest contribution to the total width in the SM. Therefore improving the measurement of the $\hbb$ process will improve the sensitivity of many of the other coupling modifiers at the HL-LHC.


Projections have also been determined for a parametrisation based on ratios of the coupling modifiers ($\lambda_{ij} = \kappa_{i}/\kappa_{j}$) together with a reference ratio of coupling modifiers $\kappa_{gZ} = \kglu\kZ/\kH$.
This parametrisation requires no assumption on the total width of the Higgs boson, as its effective modifier $\kH$ has been absorbed into the ratio $\kappa_{gZ}$. The results of this projection for ATLAS, CMS and their combination in S2 are given in Fig~\ref{fig:summary_L1}.
The numerical values in both S1 and S2 for ATLAS and CMS are given in Table~\ref{tab:summary_L1}.
\begin{figure}[hbtp]
\centering
\includegraphics[height=0.332\textheight]{\main/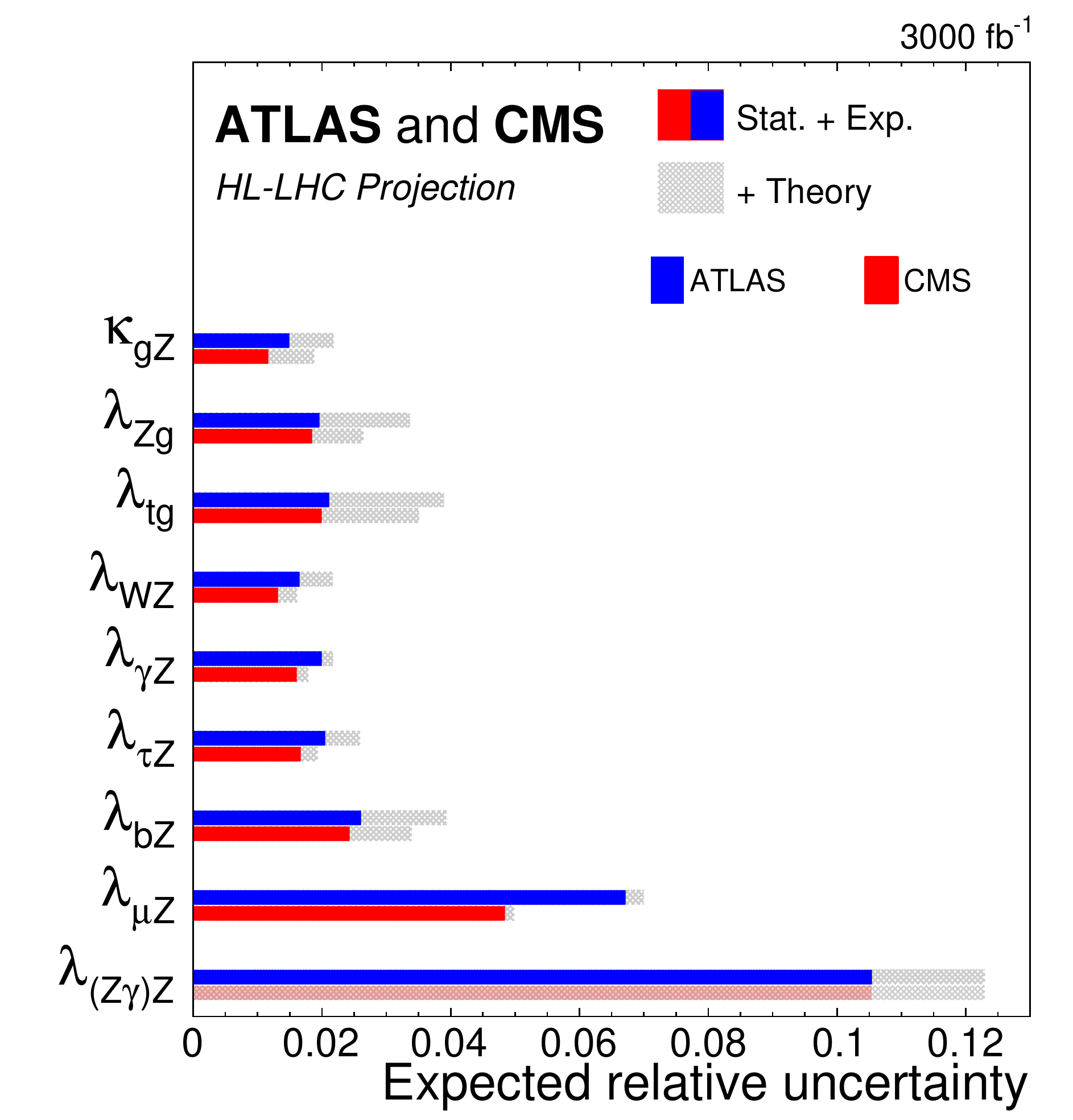}%
\includegraphics[height=0.33\textheight]{\main/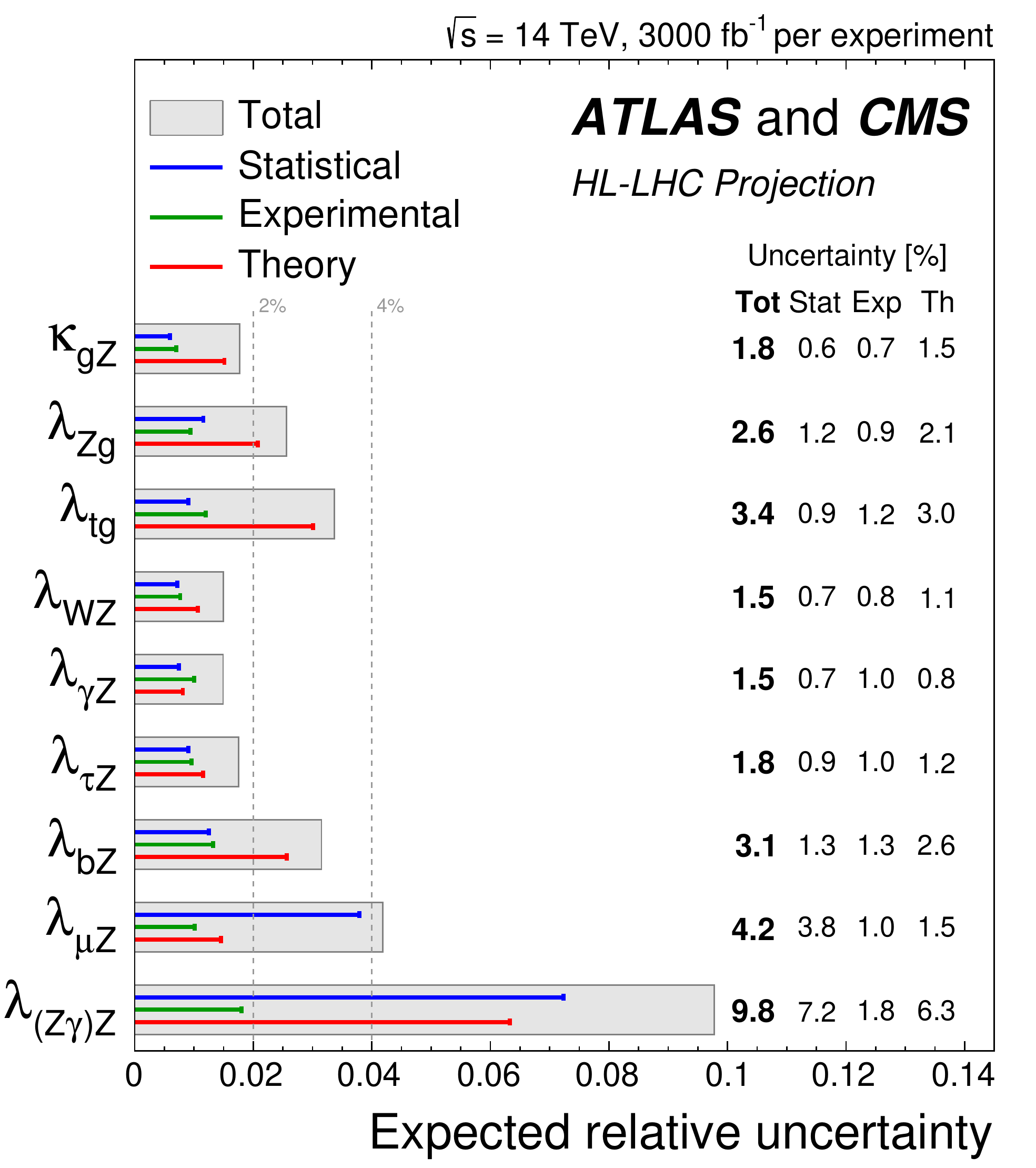}%
\caption{(left) Summary plot showing the total expected $\pm 1\sigma$ uncertainties in S2 (with YR18 systematic uncertainties) on the ratios of coupling modifier parameters  for ATLAS (blue)~\cite{ATL-PHYS-PUB-2018-054} and CMS (red)~\cite{CMS-PAS-FTR-18-011}. The filled coloured box corresponds to the statistical and experimental systematic uncertainties, while the hatched grey area represent the additional contribution to the total uncertainty due to theoretical systematic uncertainties.
(right) Summary plot showing the total expected $\pm 1\sigma$  uncertainties in S2 (with YR18 systematic uncertainties) on the ratios of coupling modifier parameters  for the combination of ATLAS and CMS extrapolations. For each measurement,  the total uncertainty is indicated by a grey box while the statistical, experimental and theory uncertainties are indicated by a blue, green and red line respectively.}
\label{fig:summary_L1}
\end{figure}

\begin{table}[hbtp]
\centering
\caption{The expected $\pm 1\sigma$ uncertainties, expressed as percentages, on the ratios of coupling modifier parameters for ATLAS and CMS~\cite{ATL-PHYS-PUB-2018-054,CMS-PAS-FTR-18-011}.  Values are given for both S1 (with Run~2 systematic uncertainties~\cite{Sirunyan:2018koj}) and S2 (with YR18 systematic uncertainties). The total uncertainty is decomposed into four components: statistical (Stat), signal theory (SigTh), background theory (BkgTh) and experimental (Exp).}
\small
\begin{tabular}{@{} l c c@{\hskip 0.15in} c c c c @{}}
  \hline
     \multicolumn{7}{c}{ATLAS}\\
 \hline
  &  & \multicolumn{5}{c}{3000 $\text{fb}^{-1}$ uncertainty [\%]} \\
  &  & Total & Stat & SigTh & BkgTh & Exp \\
  \hline
  \multirow{2}{*}{$\kappa_{\mathrm{gZ}}$} & S1 & 3.4   & 0.8   & 2.8   & 0.9   & 1.5  \\[1pt] 
  & S2 & 2.2   & 0.8   & 1.5   & 0.5   & 1.3  \\[4pt]  
  \multirow{2}{*}{$\lambda_{\gamma\mathrm{Z}}$} & S1 & 3.1   & 1.0   & 1.5   & 0.6   & 2.4  \\[1pt]
  & S2  & 2.2   & 1.0   & 0.7   & 0.5   & 1.7  \\[4pt]
  \multirow{2}{*}{$\lambda_{\mathrm{WZ}}$} & S1 & 2.7   & 0.9   & 1.5   & 1.3   & 1.5  \\[1pt]
  & S2  & 2.2   & 0.9   & 1.0   & 1.0   & 1.4  \\[4pt]
  \multirow{2}{*}{$\lambda_{\mathrm{Zg}}$} & S1 & 4.5   & 1.3   & 3.7   & 1.6   & 1.6  \\[1pt]
  & S2  & 3.4   & 1.3   & 2.4   & 1.3   & 1.4  \\[4pt]
  \multirow{2}{*}{$\lambda_{\mathrm{tg}}$} & S1 & 6.1   & 1.3   & 5.4   & 1.8   & 1.8  \\[1pt]
  & S2  & 3.9   & 1.3   & 3.0   & 1.3   & 1.7  \\[4pt]
  \multirow{2}{*}{$\lambda_{\mathrm{bZ}}$} & S1 & 5.3   & 1.6   & 3.1   & 3.3   & 2.2  \\[1pt]
  & S2  & 3.9   & 1.6   & 1.8   & 2.3   & 2.0  \\[4pt]
  \multirow{2}{*}{$\lambda_{\tau\mathrm{Z}}$} & S1 & 3.4   & 1.2   & 2.3   & 1.4   & 1.8  \\[1pt]
  & S2  & 2.6   & 1.2   & 1.3   & 1.0   & 1.7  \\[4pt]
  \multirow{2}{*}{$\lambda_{\mu\mathrm{Z}}$} & S1 & 7.7   & 6.4   & 3.6   & 0.9   & 2.1  \\[1pt]
  & S2  & 7.0   & 6.4   & 1.9   & 0.5   & 1.9  \\[4pt]
  \multirow{2}{*}{$\lambda_{\mathrm{Z}\gamma\mathrm{Z}}$} & S1 & 12.7  & 10.2  & 6.9   & 1.0   & 2.6  \\[1pt]
  & S2  & 12.3  & 10.2  & 6.3   & 0.5   & 2.5  \\[4pt]
  \hline
\end{tabular}
\hspace{0.5cm}
\begin{tabular}{@{} l c c@{\hskip 0.15in} c c c c @{}}
 \hline
   \multicolumn{7}{c}{CMS}\\
 \hline
  &  & \multicolumn{5}{c}{3000 $\text{fb}^{-1}$ uncertainty [\%]} \\
  &  & Total & Stat & SigTh & BkgTh & Exp \\
 \hline
\multirow{2}{*}{$\kappa_{\mathrm{gZ}}$} & S1  & 3.2& 0.8 & 2.7 & 0.9 & 1.2  \\[1pt]
                        & S2  & 1.9& 0.8 & 1.4 & 0.4 & 0.8  \\[4pt]
\multirow{2}{*}{$\lambda_{\gamma\mathrm{Z}}$} & S1  & 2.6& 1.0 & 1.1 & 1.0 & 1.8  \\[1pt]
                        & S2  & 1.8& 1.0 & 0.7 & 0.2 & 1.2  \\[4pt]
\multirow{2}{*}{$\lambda_{\mathrm{WZ}}$} & S1  & 2.3& 0.9 & 1.4 & 1.0 & 1.3  \\[1pt]
                        & S2  & 1.6& 0.9 & 0.8 & 0.5 & 0.9  \\[4pt]
\multirow{2}{*}{$\lambda_{\mathrm{Zg}}$} & S1  & 3.9& 1.4 & 3.2 & 1.1 & 1.4  \\[1pt]
                        & S2  & 2.6& 1.4 & 1.8 & 0.7 & 1.1  \\[4pt]
\multirow{2}{*}{$\lambda_{\mathrm{tg}}$} & S1  & 5.8& 1.2 & 5.0 & 1.8 & 1.9  \\[1pt]
                        & S2  & 3.5& 1.2 & 2.5 & 1.3 & 1.6  \\[4pt]
\multirow{2}{*}{$\lambda_{\mathrm{bZ}}$} & S1  & 5.2& 1.7 & 3.4 & 2.6 & 2.3  \\[1pt]
                        & S2  & 3.4& 1.7 & 1.7 & 1.7 & 1.7  \\[4pt]
\multirow{2}{*}{$\lambda_{\tau\mathrm{Z}}$} & S1  & 2.6& 1.2 & 1.2 & 1.2 & 1.6  \\[1pt]
                        & S2  & 1.9& 1.2 & 0.9 & 0.4 & 1.2  \\[4pt]
\multirow{2}{*}{$\lambda_{\mu\mathrm{Z}}$} & S1  & 6.6& 4.7 & 2.2 & 1.1 & 4.0  \\[1pt]
                        & S2  & 5.0& 4.7 & 1.1 & 0.4 & 1.2  \\[4pt]
\hline
\end{tabular}
\label{tab:summary_L1}
\vspace{0.5cm}
\end{table}

\subsubsection{Higgs boson coupling measurements projections estimates for HE-LHC}
\label{sec2:HE-LHC_couplings} 

As discussed above, except for the $\kappa_{\mu}$ and
$\kappa_{Z\gamma}$ coupling modifiers measured directly through the
rare decay channels $H\rightarrow \mu^+\mu^-$ and $H\rightarrow (Z
\rightarrow \ell^+\ell^-)\gamma$, the precision of the measurement of
the Higgs boson couplings at HL-LHC are limited by systematic
uncertainties and in particular those related to the theoretical
predictions and the modelling of the signal and the backgrounds.

Detailed studies on how the main systematic uncertainties will be
reduced with foreseeable theoretical developments and the input of the
large amount of data from the HL-LHC have not been carried out so far,
except for PDF uncertainties. A very approximate estimate can however
be given from the studies made for the projected sensitivities at
HL-LHC, where HE-LHC sensitivities are derived neglecting the
statistical uncertainty taking into account the increase in
centre-of-mass energy of 27~\UTeV and the much larger dataset of
15~ab$^{-1}$ for all measurements, except $H \rightarrow \mu\mu$ and $H \rightarrow Z\gamma$ channel where a simple scaling of the cross sections and luminosities is applied, which is a fair assessment with the current systematic uncertainties and assuming that the experimental performance and
systematic uncertainties are unchanged with respect to the current LHC
experiments. Two scenarios are then assumed for the theoretical and
modelling systematic uncertainties on the signal and backgrounds. The
first (S2) is the foreseen baseline scenario at HL-LHC, and the second (S2$'$) is a scenario where theoretical and modelling systematic
uncertainties are halved, which in many cases would  correspond to uncertainties roughly four times smaller than for current Run 2 analyses. It should be noted that HL-LHC measurements, whose precision is limited by systematic uncertainties, would also improve for S2'. The results of these projections are
reported in Table~\ref{tab:HE-LHC_couplings}. 

\begin{table}[hbtp]
\begin{center}
\begin{tabular}{>{$}c<{$}>{$}c<{$}>{$}c<{$}}
\hline
\mathrm{Coupling}  &   \mathrm{S2}  &  \mathrm{S2'} \\ \hline
k_{\gamma}  &  1.6   &  1.2 \\
k_W    &  1.5   &   1.0 \\
k_Z     & 1.3   &   0.8\\
k_g    &  2.2   &   1.3\\
k_t    &  3.2   &   1.9\\
k_b    &  3.5   &   2.1\\
k_{\tau}  &  1.7   &   1.1\\
k_{\mu}   & 2.2    &  1.7\\
k_{Z{\gamma}} & 6.9  &    4.1 \\\hline
\end{tabular}
\caption{Projected sensitivities of the measurements of Higgs boson couplings at HE-LHC.}
\label{tab:HE-LHC_couplings}
\end{center}
\end{table}

 
\asubsection[J. de Blas, O. Cat\`a, O. Eberhardt, C. Krause]{Higgs couplings precision overview in the Kappa-framework and the nonlinear EFT}
\label{sec2:theo_kappa_EFT}
After the discovery of the Higgs boson at the LHC, the first exploration of the couplings of the new particle at Run I and Run II has achieved an overall precision at the level of ten percent. One of the main goals of Higgs studies at the HL-LHC or HE-LHC will be to push the sensitivity to deviations in the Higgs couplings close to the percent level. 

In this section we study the projected precision that would be possible at such high luminosity and high energy extensions of the LHC from a global fit to modifications of the different single-Higgs couplings. Other important goals of the Higgs physics program at the HL/HE-LHC, such as extending/complementing the studies of the total rates with the information from differential distributions, or getting access to the Higgs trilinear coupling, will be covered in other parts of this document.

In order to study single-Higgs couplings, we introduce a parametrisation, the nonlinear EFT, that transparently connects with the $\kappa$-formalism, but is based on the language of effective field theories (EFTs). 
 We then present a fit to the projected HL/HE-LHC uncertainties both in the $\kappa$-formalism and in the more general nonlinear EFT, discussing the expected sensitivities to deviations on the Higgs couplings at the HL/HE-LHC, and compare with the recent results obtained using current data from \cite{deBlas:2018tjm}. 
 The translation of these results in terms of composite Higgs scenarios will be discussed in section~\ref{sec9:CHM}.


The $\kappa$-formalism was introduced in\cite{LHCHiggsCrossSectionWorkingGroup:2012nn,Heinemeyer:2013tqa} as an interim framework to report on the measurements of the Higgs-boson couplings and characterise the nature of the Higgs boson. The $\kappa_{i}$ are defined as ratios of measured cross sections and decay widths with respect to their SM expectation, {\it i.e.}
\begin{equation}
  \label{eq:kappa.EFT.1}
  \kappa^{2}_{X} = \frac{\sigma(X_i\rightarrow h+X_f)}{\sigma(X_i\rightarrow h+X_f)_{\text{SM}}}, \qquad \kappa^{2}_{Y} = \frac{\Gamma(h\rightarrow Y)}{\Gamma(h\rightarrow Y)_{\text{SM}}},
\end{equation}
so that the SM is recovered for $\kappa_i=1$.
This framework, defined at the level of signal strengths, was appropriate for the observables under study at Run I, which tested deviations in event rates. For Run II and the analyses required at the HL/HE-LHC, differential information is needed and the formalism defined by eq.~\eqref{eq:kappa.EFT.1} has to be extended.
In practice it then becomes more efficient to work directly at the level of Lagrangians. 
Here we will discuss the interpretation of the $\kappa$ factors within the electroweak chiral Lagrangian (EWChL or  HEFT). Within this EFT, the contributions to processes with a single Higgs, in the unitary gauge, are \cite{Buchalla:2015qju,Buchalla:2015wfa,deBlas:2018tjm}
\begin{align}
  \begin{aligned}
    \label{eq:kappa.EFT.2}
    \mathcal{L}_{\text{fit}} &= 2 c_{V} \left(m_{W}^{2}W_{\mu}^{+}W^{-\mu} +\tfrac{1}{2} m^2_Z Z_{\mu}Z^{\mu}\right) \dfrac{h}{v} - \sum_{\psi}c_{\psi} m_{\psi} \bar{\psi} \psi \dfrac{h}{v} \\
 &+ \dfrac{e^{2}}{16\pi^{2}} c_{\gamma} F_{\mu\nu}F^{\mu\nu} \dfrac{h}{v}+ \dfrac{e^{2}}{16\pi^{2}} c_{Z\gamma} Z_{\mu\nu}F^{\mu\nu} \dfrac{h}{v}+\dfrac{g_{s}^{2}}{16\pi^{2}} c_{g}\mathrm{tr}\big[ G_{\mu\nu}G^{\mu\nu}\big]\dfrac{h}{v},
  \end{aligned}
\end{align}
where $m_{i}$ is the mass of particle $i$, $\psi \in \{t, b, c, \tau, \mu\}$, and the $c_{i}$ describe the modifications of the Higgs couplings.
The previous Lagrangian differs from a naive rescaling of Higgs couplings, even though superficially it might seem to be equivalent. In particular, the Standard Model is consistently recovered in eq.~\eqref{eq:kappa.EFT.2} for
\begin{equation}
  \label{eq:kappa.EFT.3}
    c_{i}^{\text{SM}} = \begin{cases} 1 & \text{for } i = V, t, b, c, \tau, \mu\\ 0 & \text{for } i = g, \gamma, Z\gamma. \end{cases}
\end{equation} 
This Lagrangian, taken in isolation, leads to a theory with a parametrically low cutoff: it has therefore to be thought as part of a bigger EFT: the EWChL\cite{Dobado:1989ax,Dobado:1989ue,Dobado:1990zh,Dobado:1990jy,Espriu:1991vm,Herrero:1993nc,Herrero:1994iu,Feruglio:1992wf,Bagger:1993zf,Koulovassilopoulos:1993pw,Burgess:1999ha,Wang:2006im,Grinstein:2007iv,Azatov:2012bz,Alonso:2012px,Buchalla:2012qq,Buchalla:2013rka,Buchalla:2013eza}. This is a bottom-up EFT, constructed with the particle content and symmetries of the SM. These are the same requirements adopted in the construction of the SMEFT. The main difference between both EFTs concerns the Higgs field. In the EWChL, the Higgs boson, $h$, is included as a scalar singlet, with couplings unrelated to the ones of the Goldstone bosons of electroweak symmetry breaking (EWSB). 
Therefore, $h$ is not necessarily part of an SU(2) doublet and consequently (contrary to the SMEFT) the leading-order Lagrangian is already an EFT, leading potentially to $O(1)$ effects in the $\kappa$s and to a potentially low cutoff.
Further details and explanations  are discussed in \cite{Guo:2015isa,Buchalla:2017jlu,Alonso:2017tdy,Buchalla:2013rka,Buchalla:2013eza,Buchalla:2015wfa,Buchalla:2016sop}.

Focusing on the leading effects in single-Higgs processes only, the full EWChL reduces to eq.~\eqref{eq:kappa.EFT.2} which, if needed,  can be extended to describe other processes:   double-Higgs production from gluon fusion, for instance, would require three more operators, corresponding to the interactions $h^{3},\bar{t}th^{2},ggh^{2}$~\cite{Grober:2015cwa,deFlorian:2016spz,Kim:2018uty,Buchalla:2018yce} (see section~\ref{sec:EWChL.double.h}). Since the observed processes are mediated by both tree level and one-loop amplitudes at the first non-vanishing order, operators of leading order in the EFT (first line of eq.~\eqref{eq:kappa.EFT.2}) and next-to-leading order in the EFT (second line of eq.~\eqref{eq:kappa.EFT.2}) have to be included\cite{Buchalla:2015wfa}. Corrections beyond the leading ones, both strong and electroweak, can also be incorporated to arbitrary order in the description of Higgs processes. These corrections involve additional operators, not present in eq.~\eqref{eq:kappa.EFT.2}, but contained in the EWChL.
 
As stated above, the couplings in eq.~\eqref{eq:kappa.EFT.2} can receive a priori large contributions and have to be considered as $\mathcal{O}(1)$ numbers. This is the expectation if new physics contains strongly-coupled new interactions at the electroweak scale. In other scenarios of compositeness, new-physics interactions  can be associated with a larger scale $f$ and progressively decoupled from the SM: it is therefore useful to understand the Wilson coefficients in eq.~\eqref{eq:kappa.EFT.2} as functions of the parameter $\xi = v^{2}/f^{2}$. The scale $f$ could correspond, for example, to the scale of global symmetry breaking in composite Higgs models (see the discussion in sec.~\ref{sec9:CHM}). The SM is then recovered for $\xi=0$. For $\xi\ll 1$, one can perform an expansion in $\xi$ on top of the loop expansion in the EWChL. This yields a double expansion in $\xi$ and $1/16\pi^{2}$ \cite{Buchalla:2014eca}, in the spirit of the strongly-interacting light Higgs (SILH) \cite{Giudice:2007fh}. The expected size of the Wilson coefficients is then
\begin{equation}
  \label{eq:kappa.EFT.4}
    c_{i} =  c_{i}^{\text{SM}} + \mathcal{O}(\xi).
\end{equation}
The mapping of the Wilson coefficients $c_{i}$ to the $\kappa_{i}$ parameters is done using the relations of the signal strengths computed from the Lagrangian in eq.~\eqref{eq:kappa.EFT.2}. The necessary formulas can be found in \cite{Buchalla:2015qju,deBlas:2018tjm}. These relations can be written as
\begin{equation}
\label{eq:kappa.EFT.5}
  \kappa_{i} =  |f_i(c_{j})| \equiv \frac{|{\cal A}_i(c_{j})|}{|{\cal A}_i(c_{j}^{\text{SM}})|}, 
\end{equation}
where ${\cal A}$ is the corresponding transition amplitude of each process (the absolute value on the right hand side is necessary, as the loop functions of the light fermions ($b,\tau,\mu,\dots$) for the $\kappa_{\gamma}$, $\kappa_{g}$ and $\kappa_{Z\gamma}$ are complex).
We can  obtain an approximate inverse of eq.~\eqref{eq:kappa.EFT.5}, to connect both formalisms in the opposite direction, by assuming that all the imaginary parts are negligible.\footnote{This is a good approximation for some of the coefficients in $f_i(c_{j})$, for example for the coefficient of $c_{t}$, or as long as  the Wilson coefficients stay relatively close to the SM value. This is not the case for the coefficients of the light fermion loops, where real and imaginary parts are of similar size.}
With the assumption of vanishing imaginary parts, eq.~\eqref{eq:kappa.EFT.5} becomes
\begin{equation}
  \label{eq:kappa.EFT.6}
  \begin{pmatrix}
    \kappa_{V}\\
    \kappa_{t}\\
    \kappa_{b}\\
    \kappa_{\ell}\\
    \kappa_{g}\\
    \kappa_{\gamma}\\
    \kappa_{Z\gamma}
  \end{pmatrix}
  = 
  \begin{pmatrix}
    1 & 0 & 0 & 0 & 0 & 0 & 0\\
    0 & 1 & 0 & 0 & 0 & 0 & 0\\
    0 & 0 & 1 & 0 & 0 & 0 & 0\\
    0 & 0 & 0 & 1 & 0 & 0 & 0\\
    0 & 1.055 & -0.055 & 0 & 1.389 & 0 & 0\\
    1.261 & -0.268 & 0.004 & 0.004 & 0 & -0.304 & 0\\
    1.059 & -0.060 & 0.001 & 6 \cdot 10^{-5} & 0 & 0 & -0.083
  \end{pmatrix}
  \cdot
  \begin{pmatrix}
    c_{V}\\
    c_{t}\\
    c_{b}\\
    c_{\tau}\\
    c_{g}\\
    c_{\gamma}\\
    c_{Z\gamma}
  \end{pmatrix}.
\end{equation}
These numbers also include the leading QCD corrections of the $h\to \gamma\gamma$ and $gg\to h$ amplitude. An explicit comparison of this approximation and the full formulas shows only negligible numerical differences. The inverse of eq.~\eqref{eq:kappa.EFT.6} is
\begin{equation}
  \label{eq:kappa.EFT.7}
  \begin{pmatrix}
    c_{V}\\
    c_{t}\\
    c_{b}\\
    c_{\tau}\\
    c_{g}\\
    c_{\gamma}\\
    c_{Z\gamma}
  \end{pmatrix}
  = 
  \begin{pmatrix}
    1 & 0 & 0 & 0 & 0 & 0 & 0\\
    0 & 1 & 0 & 0 & 0 & 0 & 0\\
    0 & 0 & 1 & 0 & 0 & 0 & 0\\
    0 & 0 & 0 & 1 & 0 & 0 & 0\\
    0 & -0.759 & 0.040 & 0 & 0.720 & 0 & 0\\
    4.149 & -0.883 & 0.012 & 0.012 & 0 & -3.290 & 0\\
    12.736   & -0.722 & 0.015 & 0.001 & 0 & 0 & -12.029
  \end{pmatrix}
  \cdot
  \begin{pmatrix}
    \kappa_{V}\\
    \kappa_{t}\\
    \kappa_{b}\\
    \kappa_{\ell}\\
    \kappa_{g}\\
    \kappa_{\gamma}\\
    \kappa_{Z\gamma}
  \end{pmatrix}.
\end{equation}
With these relations one can translate the results of a $\kappa_i$ fit into the EWChL~formalism and vice-versa. \\

\noindent
{\bf Fit to HL/HE-LHC Higgs precision data.}
The fits presented in what follows have been performed using the {\tt HEPfit} package~\cite{hepfit,hepfitsite}, and following a Bayesian statistical approach.  
 The prior for the different model parameters both in the EFT and in the $\kappa$ framework are taken as flat, centred around the SM solution, and restricting the ranges to avoid other solutions present due to the parametrisation invariances of the different formalisms.
Since no direct sensitivity to the $H\to c\bar{c}$ channel is available in the HL/HE-LHC inputs used for the fits, we fix the corresponding parameters controlling the $Hc\bar c$ interactions to their SM values ($c_c,\kappa_c=1$).~\footnote{See ~\cite{deBlas:2018tjm} for a discussion of the multiplicities of the different solutions in the fit as well as the effect of letting the charm coupling float in the fits in absence of a significant direct constraint.}

To assess the sensitivity to deviations from the SM, we assume the future measurements are SM-like and include them in the likelihood of the fit assuming Gaussian distributions with standard deviations given by the corresponding projections for the experimental uncertainties. 

The analysis of current constraints has been taken directly from~\cite{deBlas:2018tjm}, which is based on the experimental data from~\cite{Aaltonen:2013ipa,Abazov:2013gmz,Chatrchyan:2013iaa,Chatrchyan:2013vaa,Chatrchyan:2013zna,Aad:2014eha,Aad:2014eva,Aad:2014xzb,ATLAS:2014aga,Chatrchyan:2014nva,Khachatryan:2014ira,Khachatryan:2014jba,Khachatryan:2014qaa,Aad:2015gba,Aad:2015gra,Aad:2015ona,Aad:2015vsa,ATLAS:2016gld,CMS:2016mmc,Khachatryan:2016vau,Aaboud:2017jvq,Aaboud:2017ojs,Aaboud:2017rss,Aaboud:2017uhw,Aaboud:2017vzb,Aaboud:2017xsd,CMS:2017rli,CMS-PAS-HIG-17-007,CMS-PAS-HIG-17-019,Sirunyan:2017elk,Sirunyan:2017exp,Sirunyan:2017khh,Aaboud:2018xdt,ATLAS-CONF-2018-004,Sirunyan:2018egh,Sirunyan:2018mvw,Sirunyan:2018shy,Sirunyan:2018ygk}. 
For the HL-LHC fits we use the corresponding ATLAS and CMS projections presented in this document in sections~\ref{sec2:expan} and \ref{sec2:exp_combination}.  
Since CMS does not currently have any projections
for the $H\to Z\gamma$ channel, we assume measurements with
the same precision as ATLAS.
For the systematics and theory uncertainties we use the 2 scenarios presented in section~\ref{sec:HiggsExtrapAss}: S1, for which the systematics are kept as in current values, and S2, where experimental systematics are reduced with the luminosity and theory errors are reduced.
The correlations between the ATLAS and CMS sets of inputs were not available for these fits and are therefore ignored. We note that the absence of that information can result, in some cases, in somewhat optimistic bounds. This is particularly the case for those couplings whose uncertainty is dominated by theoretical errors (e.g. $\kappa_t$).

Finally, we follow the ATLAS and CMS guidelines to estimate the HE-LHC uncertainties for the different signal strengths.  
Starting from the HL-LHC S2 projections we define the {\it Base} HE-LHC scenario by scaling the statistical uncertainties according to the changes in the production cross section going from 14 \UTeV to 27 \UTeV, as well as the different luminosities (3 ab$^{-1}$ for the HL-LHC and 15 ab$^{-1}$ in the HE-LHC):
\begin{equation}
\delta_{\mathrm{stat}} \mu_{\mbox{\tiny HE-LHC}} = \sqrt{\frac{\sigma_{pp\to H}^{\mathrm{14 \UTeV}}\times 3~\!\mathrm{ab}^{-1}}{\sigma_{pp\to H}^{\mathrm{27 \UTeV}}\times 15~\! \mathrm{ab}^{-1}}} \delta_{\mathrm{stat}} \mu_{\mbox{\tiny HL-LHC}} .
\end{equation}
Other experimental and theory uncertainties are kept as in the HL-LHC S2 case. 
To compare the results with those that would be possible in a scenario where better understanding of theory errors or systematics is achieved, it was suggested to use a more optimistic scenario where such uncertainties are further reduced by a factor of 2. We also include this optimistic scenario in our fits, and we denote it by {\it ``Opt.''}. One should note, however, that such a scenario does not come from a systematic extrapolation of the possible reduction of uncertainties. It is only an educated guess for illustration purposes.
\vspace{1cm}

In Table~\ref{tab:projection.ci} we show the results of the fit for the different scenarios discussed above for the EWChL. The numbers for the HL-LHC and the HE-LHC are reported independently (i.e. the HE-LHC does not includes the HL-LHC results on the Higgs couplings). 
These results are also shown in Figure~\ref{fig:projection.ci}. 
The analogous results for the fit using the $\kappa$ formalism are presented in Table~\ref{tab:projection.kappai} and Figure~\ref{fig:projection.kappai}. In order to put both approaches on an equal theoretical footing, in the $\kappa$ fit we assumed custodial symmetry as well as the absence of extra exotic decays of the Higgs.  As expected from the discussion above, the results show an excellent agreement between the $c_i$ and $\kappa_i$ fits for most of the parameters. The exceptions are the parameters entering in loop-induced processes, $c_{g,\gamma,Z\gamma}$ and $\kappa_{g,\gamma,Z\gamma}$, whose interpretation differs in the two formalisms. In particular, the interplay between $c_{\gamma,Z\gamma}$ and the couplings modifying the SM loops results in a significant difference between the $c_{\gamma,Z\gamma}$ results and the ones for $\kappa_{\gamma,Z\gamma}$.

Focusing our attention on the HL-LHC results, we observe that, even in the conservative S1 scenario, the knowledge of the different EFT parameters, $c_i$, and Higgs-couplings modifiers, $\kappa_i$, will improve by at least a factor of 2-3 with respect to current experimental limits. The improvement is more marked for channels that benefit from very high statistics, such as $H\to \mu^+ \mu^-$, with a precision $\sim 7$ times better than in the current fit,
and $H\to Z \gamma$ where current data does not allow to set
any meaningful bound on $c_{Z\gamma}$, $\kappa_{Z\gamma}$. Further progress is expected at the HE-LHC where, for instance, we foresee
$1\%$ level determinations for the Higgs couplings to vector bosons and $\tau$ leptons.
As one can see by comparing the HL-LHC S2 and HE-LHC Base scenarios, the precision of the interactions associated with the main Higgs couplings 
will be controlled, to a large extent, by systematic and theory errors. 

One must be careful with the interpretation of these results, though. These projections implicitly assume that departures from the SM appear only as modifications in the Higgs couplings or, in other words, that any other interaction entering the relevant Higgs processes is exactly SM-like. However, at the level of precision we observe in the results, close to the $1\%$, this may not be a justified assumption given current bounds on other electroweak interactions that could modify, e.g.~VBF or VH associated production. This comment applies even more to the HE-LHC uncertainties obtained assuming the reduced theory and systematic uncertainties which, in particular, predict a sub-percent precision for the Higgs coupling to vector bosons. We believe this to be too aggressive. A more realistic assessment of the HE-LHC uncertainties would require an equally realistic study of the experimental precision at that machine, as well as the results of a full global fit combining Higgs data with other relevant observables of the EW sector. We refer to section~\ref{sec8} for more details in this regard (in the context of the SMEFT).

\begin{table}[ht!]
\begin{center}
\caption{Comparison of the current and HL/HE-LHC 68\% probability sensitivities to the $c_{i}$ coefficients, as shown in Figure~\ref{fig:projection.ci}.}\label{tab:projection.ci}
\begin{tabular}{  c c c c c c }
  \hline\hline
  & Current limits~\cite{deBlas:2018tjm}  & HL-LHC S1 & HL-LHC S2 & HE-LHC (Base)&HE-LHC (Opt.) \\
  \hline
   $c_{V}$&$1.01\pm0.06$ &$\pm 0.017$&$\pm 0.011$&$\pm 0.009$&$\pm 0.005$\\
  $c_{t}$&$1.04^{+0.09}_{-0.1}$&$\pm 0.040$&$\pm 0.025$&$\pm 0.020$&$\pm 0.010$\\
  $c_{b}$&$0.95\pm0.13$ &$\pm 0.042$&$\pm 0.028$&$\pm 0.023$&$\pm 0.012$ \\
  $c_{\tau}$&$1.02\pm 0.1$ &$\pm 0.023$&$\pm 0.017$& $\pm 0.012$&$\pm 0.007$\\
  $c_{\mu}$&$0.58^{+0.4}_{-0.38} $ &$\pm 0.053$&$\pm 0.042$& $\pm 0.019$&$\pm 0.013$\\
  $c_{g}$&$-0.01^{+0.08}_{-0.07} $ &$\pm 0.032$&$\pm 0.020$& $\pm 0.016$&$\pm 0.008$\\
  $c_{\gamma}$ &$0.05\pm0.2 $&$\pm 0.066$&$\pm 0.045$&$\pm 0.033$&$\pm 0.019$\\
  $c_{Z\gamma}$ &$-$&$\pm 1.061$&$\pm 1.048$&$\pm 0.45$&$\pm 0.314$\\
\hline\hline
\end{tabular}
\end{center}
\end{table}
\begin{figure}[ht]
\includegraphics[width=\textwidth]{\main/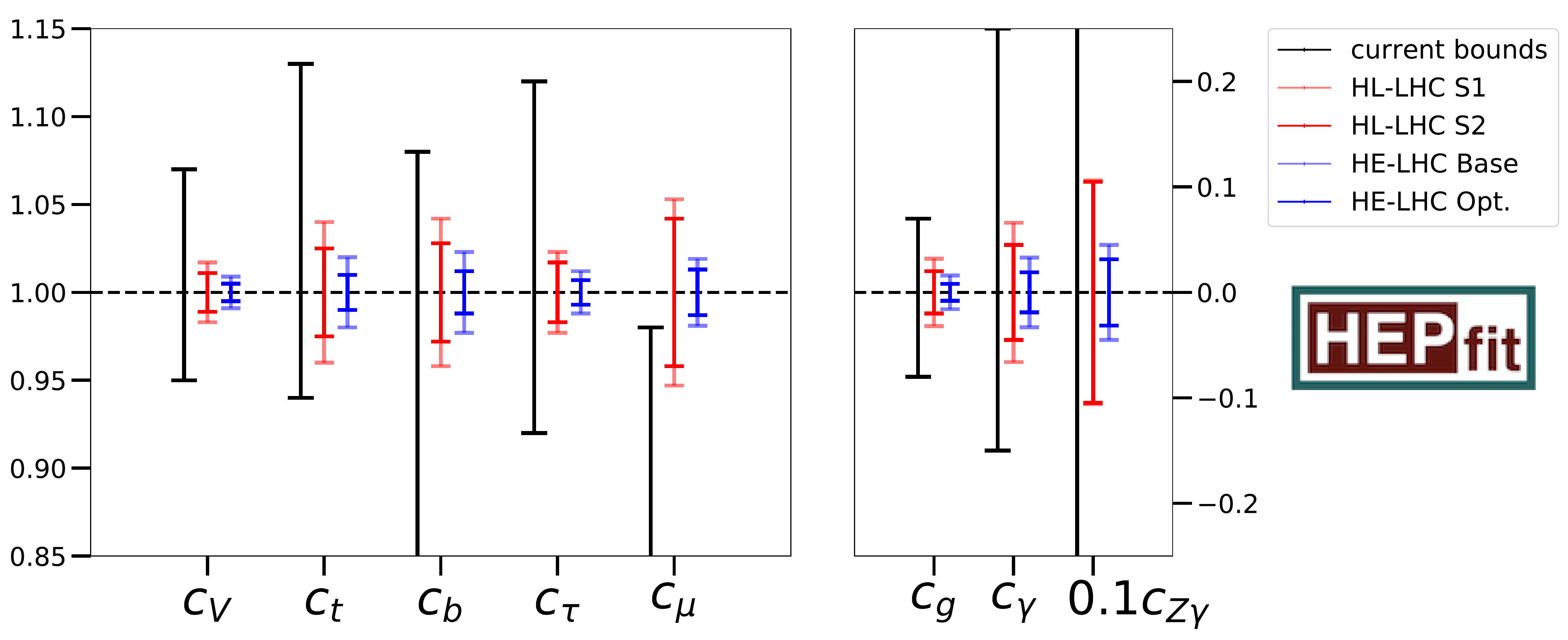}
\caption{Current and HL/HE-LHC constraints on $c_{i}$. The left line of each coupling is the current bound, from Ref.~\cite{deBlas:2018tjm}. The central line is the projection to the HL-LHC, with the S1 scenario in light red and S2 in dark red. The right line is the projection to HE-LHC, with the base scenario in light blue and the optimistic one in dark blue.}\label{fig:projection.ci}
\end{figure}

\begin{table}[ht!]
\begin{center}
\caption{Comparison of the current and HL/HE-LHC 68\% probability sensitivities to the SM Higgs-coupling modifiers $\kappa_{i}$, as shown in Figure~\ref{fig:projection.kappai}.}\label{tab:projection.kappai}
\begin{tabular}{ c c c c c c }
  \hline\hline
  & Current limits~\cite{deBlas:2018tjm}  & HL-LHC S1 & HL-LHC S2 & HE-LHC (Base)&HE-LHC (Opt.) \\
  \hline
   $\kappa_{V}$&$1.01\pm0.06$ &$\pm 0.017$&$\pm 0.011$&$\pm 0.009$&$\pm 0.005$\\
  $\kappa_{t}$&$1.04^{+0.09}_{-0.1}$&$\pm 0.040$&$\pm 0.025$&$\pm 0.020$&$\pm 0.010$\\
  $\kappa_{b}$&$0.94\pm 0.13$ &$\pm 0.042$&$\pm 0.028$&$\pm 0.023$&$\pm 0.012$ \\
  $\kappa_{\tau}$&$1.0\pm 0.1$ &$\pm 0.023$&$\pm 0.017$& $\pm 0.012$&$\pm 0.007$\\
  $\kappa_{\mu}$&$0.58^{+0.4}_{-0.38} $ &$\pm 0.053$&$\pm 0.042$& $\pm 0.019$&$\pm 0.013$\\
  $\kappa_{g}$&$1.02^{+0.08}_{-0.07} $ &$\pm 0.027$&$\pm 0.018$& $\pm 0.015$&$\pm 0.008$\\
  $\kappa_{\gamma}$ &$0.97\pm 0.07 $&$\pm 0.023$&$\pm 0.016$&$\pm 0.012$&$\pm 0.007$\\
  $\kappa_{Z\gamma}$ &$-$&$\pm 0.094$&$\pm 0.093$&$\pm 0.040$&$\pm 0.028$\\
\hline\hline
\end{tabular}
\end{center}
\end{table}
\begin{figure}[ht]
\includegraphics[width=\textwidth]{\main/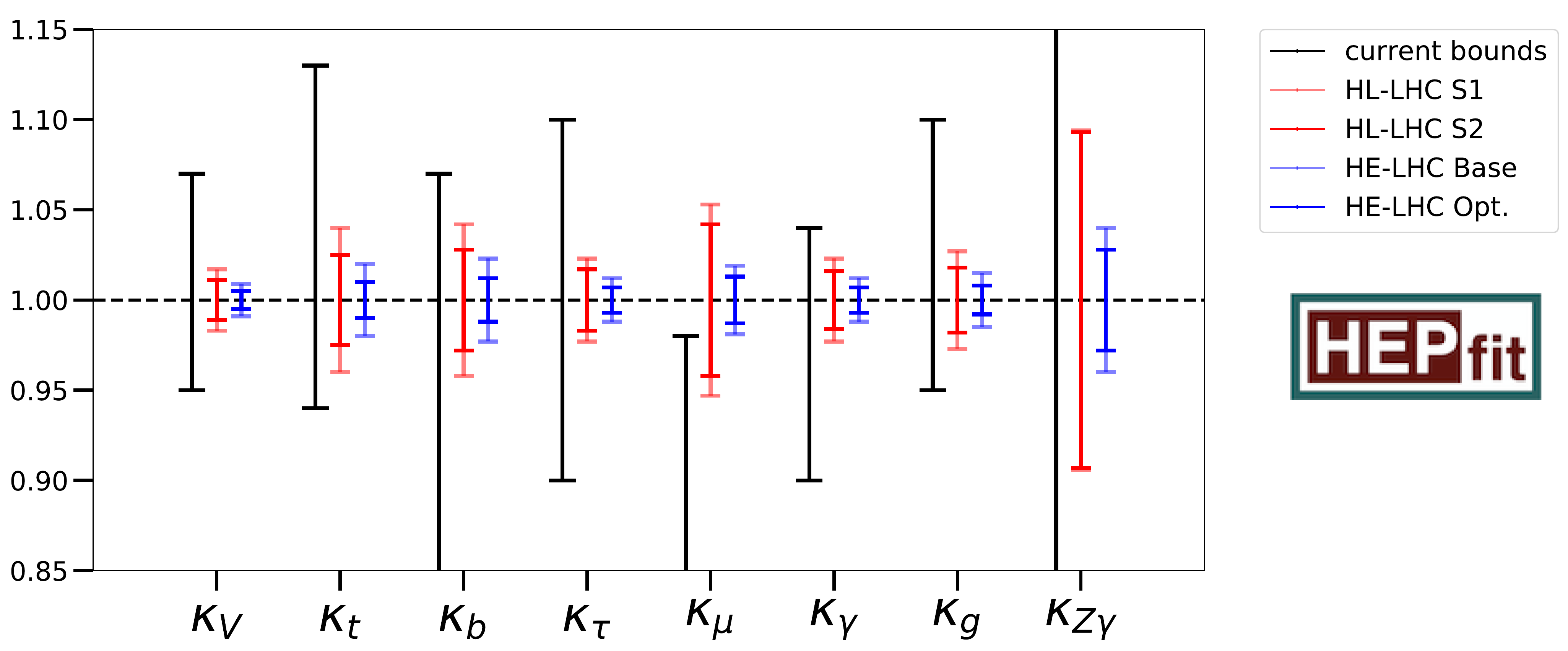}
\caption{Current and future constraints on $\kappa_{i}$. The left line of each $\kappa$ is the current bound, from Ref.~\cite{deBlas:2018tjm}. The central line is the projection to the HL-LHC, with the S1 scenario in light red and S2 in dark red. The right line is the projection to HE-LHC, with the base scenario in light blue and the optimistic one in dark blue.}\label{fig:projection.kappai}
\end{figure}


\asubsection[L. Vecchi]{Interpretation of the Higgs couplings in terms
  of Composite Higgs models}
\label{sec9:CHM}

Composite Higgs (CH) models postulate that the Standard SM Higgs sector is UV-completed by a strongly-coupled dynamics characterised by some scale $m_*$, not too far above the \UTeV. Since, by analogy with QCD, $m_*$ can naturally be small compared to any existing microscopic scale, this framework provides an attractive solution to the hierarchy problem. 

Historically, precision {\emph{indirect}} tests, mainly from EW data, have resulted in important constraints on strongly-coupled extensions of the SM. The discovery of the Higgs boson has removed the uncertainty associated to the value of $m_h$ but otherwise has not improved those bounds qualitatively. On the other hand, {\emph{direct}} access to the Higgs boson properties has had a qualitative impact on CH scenarios: we now know that viable realizations must contain a light scalar resonance $h$ with properties that mimic those of the SM Higgs boson. This observation excludes Higgless solutions to the hierarchy problem (like old-fashioned technicolor), but leaves open a number of options, a representative set of which will be discussed here. Overall, CH scenarios with a Higgs-like resonance continue to offer a very compelling explanation of the weak scale.

In this section we will focus on two representative classes of CH scenarios that predict a light scalar with SM-like couplings:
\begin{itemize}
\item[1)] the Strongly-Interacting Light Higgs (SILH). In this class the exotic strong dynamics generates a light scalar doublet $H$ with the same $SU(2)_w\times U(1)_Y$ charges of the SM Higgs, and it is the latter which spontaneously breaks the EW symmetry~\cite{Kaplan:1983fs, Kaplan:1983sm}. The doublet $H$ may be part of a Nambu-Goldstone multiplet, or simply be an accidentally light scalar. The physical Higgs boson $h$ belonging to the composite doublet behaves as the SM Higgs boson up to corrections induced by higher-dimensional operators suppressed by the strong coupling scale $m_*$. 
\item[2)] the Strongly-Interacting Light Dilaton (SILD). In this class of theories the strong dynamics is assumed to feature the spontaneous breaking of an approximate scale invariance at a scale $f_D$. In such a framework the low energy EFT possesses an approximate Nambu-Goldstone mode, the dilaton, which automatically has couplings aligned along the direction of those of the SM Higgs~\cite{Goldberger:2008zz}. The key difference compared to the SILH is that this is a non-decoupling scenario, in which the new physics threshold is controlled by the EW scale. We interpret the SILD as a representative of CH scenarios based on the EW chiral Lagrangian, in which the EW symmetry is non-linearly realised and the Higgs-like particle $h$ is not embedded in an EW doublet $H$. 
\end{itemize}

The main goal of this section is to review what we can learn about the CH picture from the investigation of the Higgs properties at the HL and HE-LHC. We will focus on modifications of the on-shell couplings, as opposed to off-shell rates like double-Higgs production or $VV\to VV$ scattering. Of course, more direct ways to test the CH hypothesis include the observation of new resonances. Here we however assume that the new resonances are too heavy to be directly accessible and focus on the low energy EFT for the light state $h$.

\begin{table}[t]
\caption{\small List of the dimension-6 operators relevant to our study of modified Higgs couplings in the SILH class. We use the basis of~\cite{Giudice:2007fh}. Here $y_\psi$ are the SM Yukawa couplings and $V=Z,W$. 
\label{tab1}}
\begin{center}
{
\begin{tabular}{c|c|c} 
\rule{0pt}{1.2em}%
Operator name & Operator definition & Main on-shell process \\
\hline
${\cal O}_H$ & $\frac{1}{2}~\partial_\mu(H^\dagger H)\partial^\mu(H^\dagger H)$ & $h\to\psi\bar\psi, VV^*$     \\
${\cal O}_6$ & $\lambda_h(H^\dagger H)^3$ & $h^*\leftrightarrow hh$   \\
${\cal O}_y$ & $\sum_{\psi=u,d,e} y_\psi~\overline\psi_L H\psi_R(H^\dagger H)$ & $h\to\psi\bar\psi$   \\
\hline
%
${\cal O}_{HW}$ & $ig(D^\mu H)^\dagger\sigma^i(D^\nu H)W_{\mu\nu}^i$ & $h\to VV^*,\gamma Z$   \\
${\cal O}_{HB}$ & $ig'(D^\mu H)^\dagger(D^\nu H)B_{\mu\nu}$ & $h\to ZZ^*,\gamma Z$  \\
\hline
${\cal O}_{g}$ & $g_s^2H^\dagger H G^a_{\mu\nu}G^{a\mu\nu}$ & $h\to gg$   \\
${\cal O}_{\gamma}$ & $g'^2H^\dagger HB_{\mu\nu}B^{\mu\nu}$ & $h\to \gamma\gamma,\gamma Z,ZZ$  
\end{tabular}
}
\end{center}
\end{table}

\subsubsection*{The SILH}

The operators that dominantly impact on-shell processes involving $h$ are collected in table \ref{tab1} under the assumption that $H$ is an EW doublet. We do not include operators of higher dimension and those that are severely constrained by precision data, which for this reason are expected to lead to negligible corrections to the rates induced by those in the table.~\footnote{These include ${\cal O}_T=(H^\dagger\overleftrightarrow{D_\mu}H)(H^\dagger\overleftrightarrow{D^\mu}H)/2$, ${\cal O}_W={i}g(H^\dagger\sigma^i\overleftrightarrow{D_\mu} H)(D_\nu W^{\mu\nu})^i/2$, ${\cal O}_B=ig'(H^\dagger\overleftrightarrow{D_\mu} H)(\partial_\nu B^{\mu\nu})/2$, current-current interactions $H^\dagger \overleftrightarrow{D_\mu}H \bar \psi\gamma^\mu \psi$ and $H^\dagger \tau^a\overleftrightarrow{D_\mu}H \bar \psi\gamma^\mu \tau^a\psi$ containing non-universal couplings to the SM fermions $\psi$, and dipole operators. The operator ${\cal O}_T$ is constrained by the EW $\rho$ parameter; whereas ${\cal O}_W+{\cal O}_B$ by the EW $S$ parameter. Current-current operators are constrained by LEP and the non-observation of rare flavor-violating processes. Dipole operators are severely constrained by measurements of electric and magnetic moments. See, e.g.,~\cite{Contino:2013kra} for more details.
} In particular, in realistic SILH scenarios the Higgs coupling to fermions must be aligned to the SM Yukawas. For illustrative purposes, here we further simplify our discussion specialising on realisations in which the SM fermions are all coupled analogously to the strong sector, such that a universal fermionic operator ${\cal O}_y$ is sufficient. (We will comment below on more general scenarios.)

The observables that are mostly affected by the new interactions are shown in the third column of table \ref{tab1}. An estimate of the various Wilson coefficients in concrete CH models reveals that corrections to $h\to WW^*,ZZ,\psi\bar\psi$ are typically dominated by ${\cal O}_{H,y}$~\cite{Giudice:2007fh}. Those to the radiative processes $h\to gg,\gamma\gamma$ are controlled by 1-loop diagrams with an insertion of ${\cal O}_{H,y}$ if the doublet $H$ is a Nambu-Goldstone mode of the strong dynamics, but may also receive important contributions from ${\cal O}_{g,\gamma}$ if $H$ is an accidentally light resonance. For what concerns $h\to\gamma Z$ we find that ${\cal O}_{H,y}$ give contributions parametrically comparable to ${\cal O}_{HW, HB}$. The same is true for ${\cal O}_{\gamma}$, but only when $H$ is not a Nambu-Goldstone mode. Because the sensitivity on $h\to\gamma Z$ is appreciably weaker than $h\to\gamma\gamma$, it makes sense to simplify our discussion by neglecting the impact of ${\cal O}_{HW, HB}$ in our fit. Similarly, we will ignore ${\cal O}_6$ since this operator only controls the very poorly known Higgs self-couplings.

From these considerations follows that the leading on-shell signatures of the Higgs-like state $h$ in SILH scenarios can be characterised by the reduced set of operators 
\begin{eqnarray}\label{SILHvecchi}
\delta{\cal L}_{\rm SILH}=\frac{g_*^2}{m_*^2}\bar c_H{\cal O}_H+\frac{g_*^2}{m_*^2} \bar c_y{\cal O}_y+\frac{g_*^2}{16\pi^2m_*^2}\bar c_g{\cal O}_g+\frac{g_*^2}{16\pi^2m_*^2}\bar c_\gamma{\cal O}_\gamma,
\end{eqnarray}
where $\bar c_{H,y,g,\gamma}$ are expected to be of order unity and $g_*, m_*$ are the typical couplings and mass scale of the new physics. We included a factor of ${g_*^2}/{16\pi^2}$ in front of the last two operators in order to emphasise their radiative nature~\cite{Giudice:2007fh}. We further assumed CP is approximately satisfied by the strong sector.

We can now match the Wilson coefficients appearing in \eqref{SILHvecchi} onto the phenomenological Lagrangian \eqref{eq:kappa.EFT.2} for the light boson $h$, up to additional interactions that are irrelevant to the present analysis. The resulting modified Higgs couplings are collected here:
\begin{eqnarray}\label{4parSILH}
c_V=1-\frac{\bar c_H}{2}\xi,~~~~~c_y=1-\left(\frac{\bar c_H}{2}+\bar c_y\right)\xi,~~~~~c_g=2 \bar c_g\xi,~~~~~c_\gamma=\bar c_\gamma\xi
\end{eqnarray}
where we defined 
\begin{eqnarray}\label{xi}
\xi\equiv\frac{g_*^2v^2}{m_*^2}\equiv\frac{v^2}{f^2},
\end{eqnarray}
with $v=246$ \UGeV. Note that our assumption of SM fermion-universality implies that $c_y$ is a single real number (independent of $\psi=u,d,e$). This leaves us with a total of 4 independent parameters~\eqref{4parSILH}. Our truncation of the EFT to dimension 6 operators is crucially associated to the working hypothesis $\xi\ll1$: operators of higher dimension are suppressed by larger powers of $\xi$. 

\paragraph{Analysis}

The expected reach of the HL ($3$ ab$^{-1}$) and HE ($15$ ab$^{-1}$) LHC on the modified Higgs couplings has been presented in Section \ref{sec2:theo_kappa_EFT}. Here we specialise to scenarios defined by the 4 couplings $c_{V,y,g,\gamma}$.

To better quantify the sensitivity of the future LHC upgrades on CH models we identify the benchmark scenarios shown in table~\ref{bencharkVecchi}. SILH1 is intended to capture the low energy physics of CH models with a Nambu-Goldstone boson $H$, where couplings to the massless gauge bosons are suppressed~\cite{Giudice:2007fh}. SILH2 is expected to mimic scenarios with an accidentally light CH doublet, since in that case one typically expects $|\bar c_{g,\gamma}|$ of order unity. To assess the impact of the fermionic coupling $\bar c_y$ we distinguished between scenarios with $\bar c_y=0$ (a) and $\bar c_y=1$ (b); models in which the SM fermions have different couplings for the various SM representations $\psi=u,d,e$ should lie somewhat in between these two.

In table~\ref{bencharkVecchi} we present, for each benchmark model, the expected HL/HE sensitivity. We treat the systematic uncertainties using the conservative hypothesis S1. Note the significant impact of a non-vanishing $\bar c_{g,\gamma}$ --- especially the coupling to gluons; of the four options with $\bar c_{g,\gamma}=\pm1$ we quote only the most stringent bound for brevity. Making a fair comparison between present LHC data and our projections is not possible because current data slightly favours values $\bar c_{V,y}>1$, a fact that in some benchmark models results in stronger constraints than our projections (even when restricting our fit to the domain $0<\xi<1$). Perhaps a more significant measure of the improvement of the HL/HE-LHC can be obtained if we artificially assume the currently preferred central values are $c_V=c_y=1, c_g=c_\gamma=0$, as in our projections. This way we obtain the $95\%$ CL {\emph{expected limits}} shown in table~\ref{bencharkVecchi}.

\begin{table}[t]
\caption{\small Projected HL/HE sensitivity (95\% CL) on the SILH parameter $\xi=v^2/f^2$ (and on $f$ in \UTeV) in our benchmark scenarios. Systematic uncertainties are treated according to the conservative scenario S1.
\label{bencharkVecchi}}
\begin{center}
{
\begin{tabular}{c|cccc||cc|cc|cc} 
\rule{0pt}{1.2em}%
\multirow{ 2}{*}{} & \multirow{ 2}{*}{$\bar c_H$} & \multirow{ 2}{*}{$\bar c_y$} & \multirow{ 2}{*}{$|\bar c_{g}|$} & \multirow{ 2}{*}{$|\bar c_{\gamma}|$} & \multicolumn{2}{c|}{expected LHC14}  & \multicolumn{2}{c|}{HL-LHC} &  \multicolumn{2}{c}{HE-LHC} \\
&  &   &  &  & $\xi$ & $f~[{\rm \UTeV}]$  & $\xi$  & $f~[{\rm \UTeV}]$ &  $\xi$ & $f~[{\rm \UTeV}]$ \\
\hline
\hline
SILH1a & $1$ & $0$ & $0$ & $0$ & $0.094$ & $0.80$ &      $0.045$ & $1.2$ & $0.022$ & $1.7$ \\
SILH1b &$1$ & $1$ & $0$ & $0$ & $0.064$ & $0.97$ &       $0.021$ & $1.7$ & $0.011$ & $2.3$ \\
\hline
SILH2a &$1$ & $0$ & $1$ & $1$ & $0.019$ & $1.8$ &       $0.012$ & $2.3$ & $0.0062$ & $3.1$\\
SILH2b & $1$ & $1$ & $1$ & $1$ & $0.018$ & $1.8$ &       $0.0096$ & $2.5$ & $0.0050$ & $3.5$ \\
\end{tabular}
}
\end{center}
\end{table}

Recalling our definition \eqref{xi} we can translate the results of table~\ref{bencharkVecchi} into a lower bound on the new physics scale $m_*$ as a function of the size of the typical coupling $g_*$ of the exotic sector. The result is shown in Fig~\ref{SILH}. For presentation purposes, in the figure we only show the reach of the HL-LHC and HE-LHC on the benchmark models SILH1b (black) and SILH2b (red). The constrained region lies on the left of each line. The bounds tend to push the CH scenario towards the SM limit $\xi\to0$, obtained decoupling the new physics scale. We see that in the case the exotic dynamics is maximally strongly coupled ($g_*\sim4\pi$) the LHC will be able to indirectly access mass scales of order $20-30$ \UTeV.

\begin{figure}[t]
\begin{center}
\includegraphics[scale=0.7]{\main/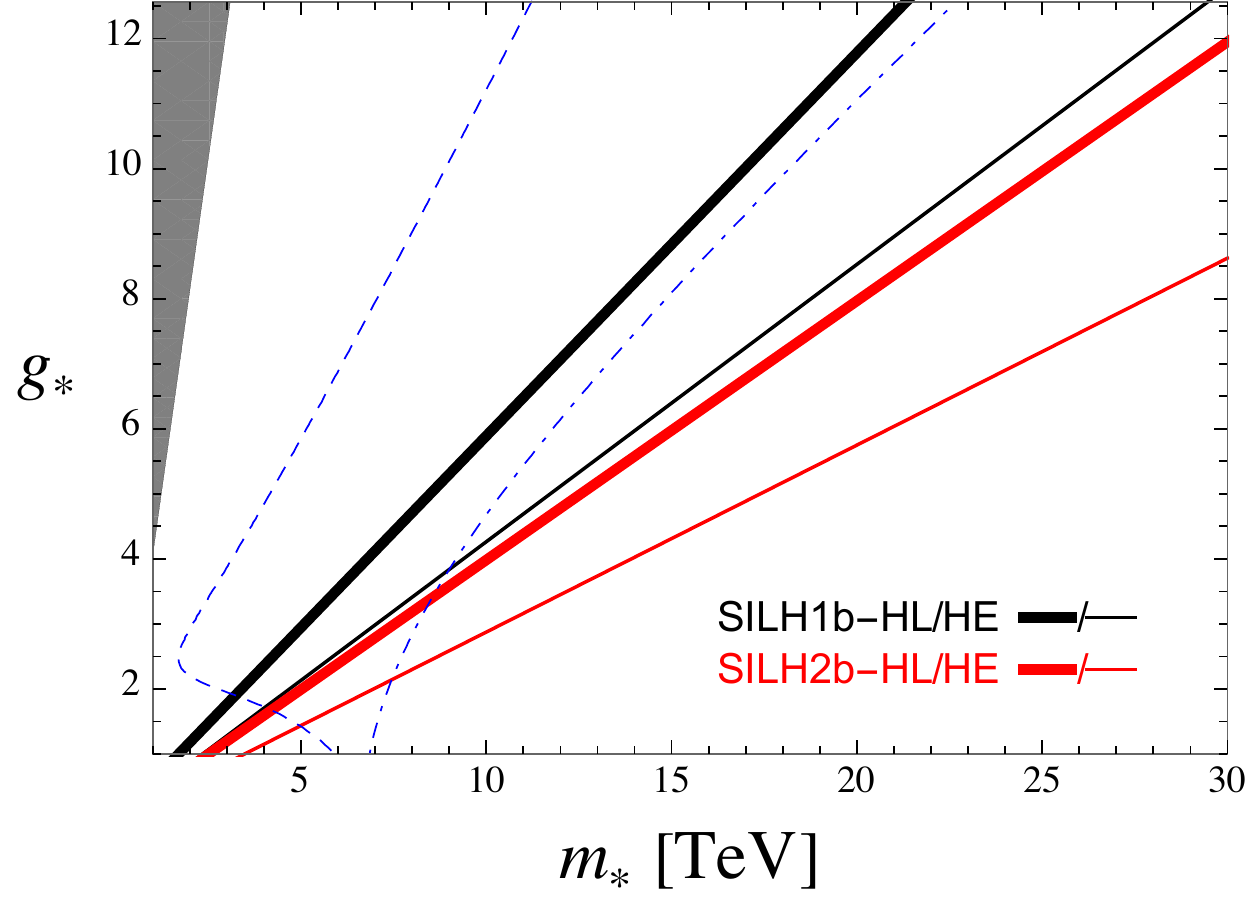}
\caption{The constraints of table~\ref{bencharkVecchi} are interpreted as lower bounds on the new physics scale $m_*$ for a given coupling $g_*$ of the strong dynamics, see \eqref{xi}. The blue lines define lower bounds on $m_*$ from current EW precision tests for two different assumptions on the UV dynamics (see text). The grey region identifies the unphysical regime $\xi>1$. 
}\label{SILH}
\end{center}
\end{figure}

In Fig~\ref{SILH} we also include (blue dashed and dot-dashed lines) the current $95\%$ CL limits derived from precision EW data~\cite{Baak:2014ora} and encoded in the oblique parameters (with $U=0$)
\begin{eqnarray}
&&{\widehat S}=(1-c_V^2)\frac{g^2}{96\pi^2}\ln\frac{m_*}{m_Z}+{\widehat S}_{\rm UV}
\\\nonumber
&&{\widehat T}=-(1-c_V^2)\frac{3g'^2}{32\pi^2}\ln\frac{m_*}{m_Z}+{\widehat T}_{\rm UV}.
\end{eqnarray}
Note that these include 1-loop effects within \eqref{eq:kappa.EFT.2} as well as contributions from heavy particles of mass $\sim m_*$ that we parametrised via $\widehat S_{\rm UV}={m_W^2}/{m_*^2}$ and $\widehat T_{\rm UV}$. The blue dot-dashed line refers to scenarios in which additional violations of custodial symmetry are negligible, $\widehat T_{\rm UV}=0$, whereas the blue dashed line to the more natural expectation $\widehat T_{\rm UV}=\xi 3y_t^2/(16\pi^2)$. Precision EW data already exclude a sizeable portion of parameter space. However, as our plot clearly illustrates, these {\emph{indirect}} bounds significantly depend on unknown physics at the cutoff scale $m_*$. Hence, a {\emph{direct}} probe of the Higgs couplings provides a more robust and model-independent assessment of a given CH scenario.

\subsubsection*{The SILD}

The dominant interactions of the dilaton (still denoted by $h$) to the SM are derived from an EFT with non-linearly realised EW symmetry, where the Nambu-Goldstone bosons eaten by the $W^\pm,Z^0$ are encapsulated into the unitary matrix $\Sigma$, which transforms as $\Sigma\to U_w\Sigma U_Y^\dagger$ under $SU(2)_w\times U(1)_Y$. The powers of the singlet $h$ are fully determined by the approximate conformal symmetry. Neglecting possible (small) sources of custodial symmetry breaking, one identifies the dominant interactions:~\cite{Goldberger:2008zz}
\begin{eqnarray}\label{SILD1}
{\cal L}_{\rm SILD}&=&\frac{v^2}{4}{\rm tr}[D_\mu\Sigma^\dagger D^\mu\Sigma]\left(1+\frac{h}{f_D}\right)^2-\sum_{\psi=u,d,l}m_\psi\bar\psi_L\Sigma\psi_R\left(1+\frac{h}{f_D}\right)^{1+\gamma_\psi}\\\nonumber
&+&\frac{g_s^2}{16\pi^2}\delta_sG^a_{\mu\nu}G^{a\mu\nu}\frac{h}{f_D}+\frac{e^2}{16\pi^2}\delta_e\gamma_{\mu\nu}\gamma^{\mu\nu}\frac{h}{f_D}\\\nonumber
&+&\cdots,
\end{eqnarray}
where the dots refer to operators that impact negligibly our analysis. In the unitary gauge, the first operator in \eqref{SILD1} describes the main coupling between $h$ and the EW vector bosons. The couplings to fermions depend on how the latter interact with the underlying (approximately) conformal dynamics and are therefore model-dependent. A family-universal interaction in \eqref{SILD1} is expected in any UV theory that does not suffer from sizeable flavor-changing effects, which would otherwise be in tension with precision flavor observables. Interactions with the unbroken gauge bosons in the second line of \eqref{SILD1} also depend on the details of the UV dynamics. We do not make any restrictive assumption here (besides an approximate CP symmetry) and instead allow the parameters $\delta_{g,\gamma}$ to acquire any real value. We however included a loop factor to emphasise we expect them to arise at the loop level. Similarly to what we have already stressed above \eqref{SILHvecchi}, novel corrections to $\gamma Z$ are not important to our analysis and can be ignored in a first assessment: the Lagrangian \eqref{SILD1} is enough to capture the dominant on-shell signatures of the SILD scenario as well. From \eqref{SILD} one obtains a Lagrangian like \eqref{eq:kappa.EFT.2} with 
\begin{eqnarray}\label{4parSILD}
c_V=\frac{v}{f_D},~~~~~c_y=(1+\gamma_y)\frac{v}{f_D},~~~~~c_g=2\delta_g\frac{v}{f_D},~~~~~c_\gamma=\delta_\gamma\frac{v}{f_D}.
\end{eqnarray}
Under the simplifying assumption of flavor universality ($\gamma_{u,d,e}=\gamma_y$) our EFT contains only 4 independent real parameters. 

We are now able to draw a few conclusions. First, as anticipated, experimental constraints on $c_V$ force $f_D\simeq v$, see Fig. \ref{SILD}. Therefore the SILD, as any other framework based on the non-linear chiral Lagrangian (i.e. where $h$ is not part of a Higgs doublet), implies the characteristic mass of the new physics lies at the relatively low scale $g_*v\lesssim4\pi v\sim3$ \UTeV. On the one hand this is an exciting possibility because it suggests its UV completion is more likely to be accessible at the LHC. On the other hand, a strong dynamics at such low scales contributing to EW symmetry breaking is in serious tension with EW precision data (e.g. $\widehat S_{\rm UV}\sim m_W^2/m_*^2\sim g^2/(16\pi^2)$ is typically too large as in technicolor). One can only hope the UV theory somehow cures this problem, though no concrete mechanism to achieve this is known.

\begin{figure}[t]
\begin{center}
\includegraphics[scale=0.6]{\main/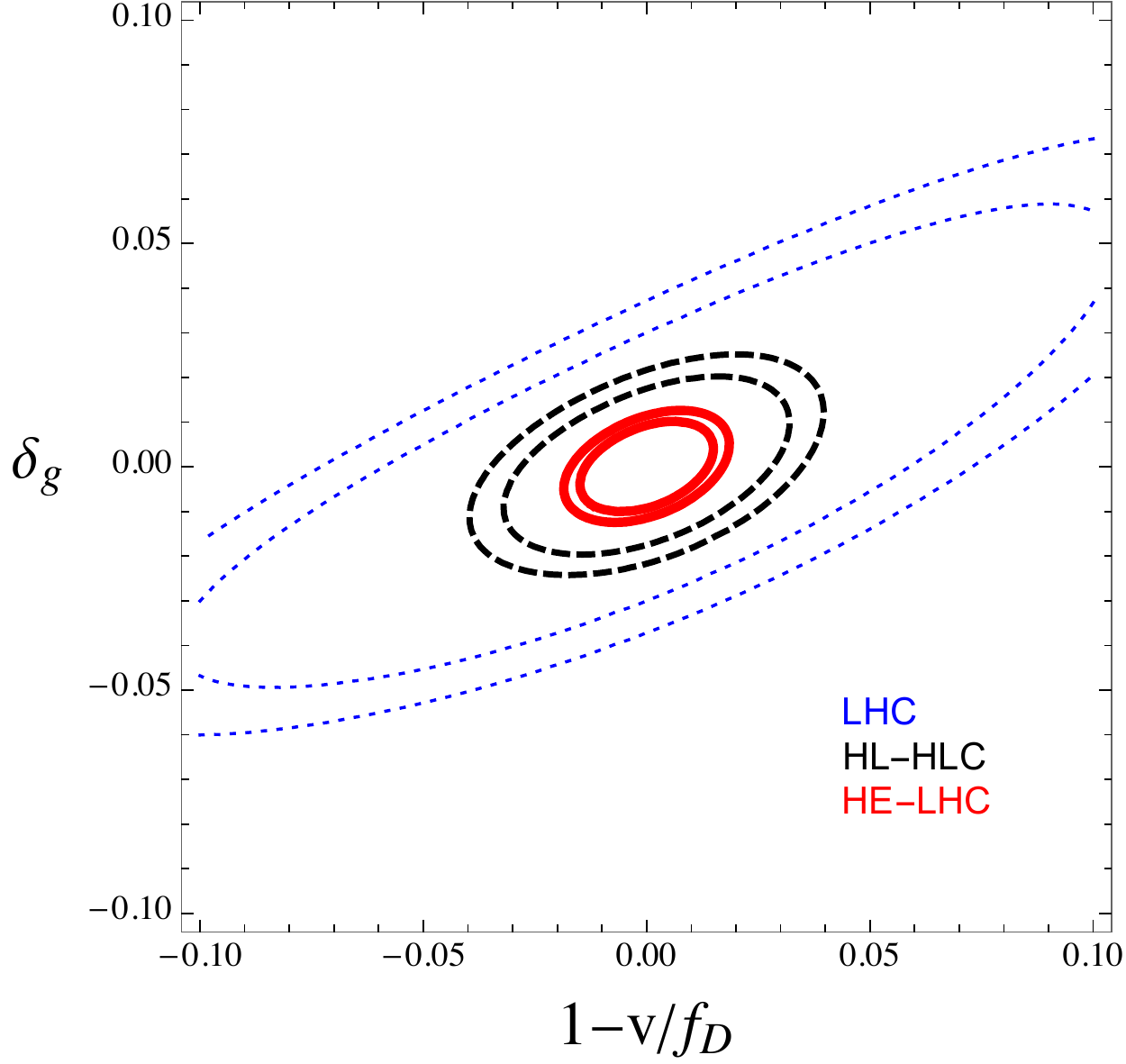}
\caption{Expected sensitivity at the $95\%, 99\%$ CL of the LHC (blue dotted), HL-LHC (black dashed) and HE-LHC (red solid) on SILD scenarios. For definiteness we set $\delta_\gamma=\gamma_y=0$.
}\label{SILD}
\end{center}
\end{figure}

Secondly, explicit realisations of the SILD are characterised by sizeable couplings to the massless gauge bosons, $|\delta_{g,\gamma}|={\cal O}(1)$, and this appears to be in conflict with data. For example, the prototypical SILD scenario --- in which the entire SM is part of the conformal sector --- has $\delta_{g}=+46/3, \delta_\gamma=-34/9$~\cite{Goldberger:2008zz} and is already excluded with large confidence. Scenarios with partially composite SM, $|\delta_{g,\gamma}|\sim0.1$, will be subject to the significant constraints from the HL and HE upgrades, see Fig. \ref{SILD}. Unfortunately, even in this case there is no known symmetry argument that may be invoked to ensure $|\delta_{g,\gamma}|\ll1$. The only way to accommodate such a constraint within explicit strongly-coupled models seems to be via fine tuning.

Overall, the SILD scenario --- and all scenarios based on a non-linear realisation of the EW symmetry --- suffers from a major drawback compared to the SILH: the SM is recovered by {\emph{tuning}} several (often uncorrelated) parameters. This is because the former do not possess a simple decoupling mechanism (analogous to the $\xi\to0$ limit in the SILH) that switches off the new physics corrections to precision data as well as $c_{V,y,g,\gamma}$. The {\emph{simultaneous}} non-observation of new physics at the \UTeV scale and of deviations from the SM in the future LHC upgrades would then unambiguously prove that the Higgs boson must be the missing component of the doublet responsible for EW symmetry breaking. In such a situation the only compelling realisation of the CH paradigm is represented by the SILH class.

\asubsection{Probing of anomalous HVV interactions}
\label{sec2:anomalous_HVV}
\asubsubsection[S. Boselli, C. M. Carloni Calame, G. Montagna,
  O. Nicrosini, F. Piccinini, A. Shivaji, F. Yu, M. M. Llacer]{Probes using differential distributions of CP sensitive
  observables} 
%
 We present prospects for studies on {\it CP}-odd couplings in the
 interactions of the Higgs boson with the electroweak gauge bosons as well
 as in the Yukawa couplings of the Higgs boson with fermions, in
 particular with $\tau^+ \tau^-$ pairs.
 
 \asubsubsubsection{\bf CP-odd $VVH$ couplings}

While a large number of studies assessing the impact of {\it CP}-even  
effective operators on Higgs physics is available in the literature 
(see for instance our analysis in Ref.~\cite{Boselli:2017pef} and the references therein), 
the present analysis is focused on the impact of {\it CP}-odd effective operators on the 
interactions among the Higgs boson and the electroweak bosons. 
In the Higgs basis, the CP-violating (CPV) sector of the BSM Lagrangian affecting $VVH$ couplings 
is given by, 
\begin{eqnarray}
\lag_{\rm CPV} =  \frac{H}{v} \Big[
\tCaa \frac{e^2}{4} A_{\mu\nu}\tilde{A}^{\mu\nu}  
+ \tCza \frac{e\sqrt{g_1^2 + g_2^2}}{2} Z_{\mu\nu}\tilde{A}^{\mu\nu} 
+ \tCzz \frac{g_1^2 + g_2^2}{4} Z_{\mu\nu}\tilde{Z}^{\mu\nu} + \tilde{c}_{WW} \frac{g_2^2}{2}  W^+_{\mu\nu} \tilde W^{-\mu\nu}\Big]
\label{eqn:Hvv}
\end{eqnarray}  
where, $g_1$ and $g_2$ are the $U(1)_Y$  and  $SU(2)_L$ gauge coupling constants. Out of the above four 
parameters, only three  are independent. In particular,
\begin{equation}
 \tCww = \tCzz + 2s_\theta^2 \tCza + s_\theta^4 \tCaa.
\end{equation}

The processes which are sensitive to CP-odd operators are the Higgstrahlung processes ($WH$ and $ZH$), the vector boson fusion (VBF) and the Higgs decay into four charged leptons ($H\to4\ell$). Here we focus on angular observables which are sensitive to CPV effects. Indeed, since the total cross-section is a CP-even quantity,  the $1/\Lambda^2$ effects of CPV operators can affect the shape of some specific kinematic distributions only.

\begin{table*}
        \begin{center}
                \begin{tabular}{lc|ccccc}
                        \multicolumn{2}{c|}{Process}     & Combination & Statistical & Theory (Sig.) & Theory (Bkg.) & Experimental \\ \hline \hline
                        \multirow{5}{*}{${H}\to {ZZ}$}
                        &$\text{ggF}$      & 6.6  & 2.1  & 5.4  & 1.7 & 2.7\\
                        &$\text{VBF}$      & 15.2 & 11.7 & 9.1  & 2.4 & 1.8\\
                        &${WH}$            & 48.0 & 46.5 & 6.2  & 2.8 & 7.8\\
                        &${ZH}$            & 82.5 & 75.7 & 27.0 & 7.6 & 16.4\\
                        &${t\overline tH}$ & 26.9 & 23.6 & 10.9 & 2.5 & 4.2\\ \hline
                \end{tabular}
                \caption{Estimated uncertainties [\%] on the determination of single-Higgs production channels in $H \to 4\ell$ decay mode. These are CMS projections for high-luminosity LHC (14 \UTeV centre of mass energy and 3 ab$^{-1}$ integrated luminosity) in scenario S1 (systematic uncertainties are kept constant with luminosity) taken from Ref.~\cite{CMS-PAS-FTR-18-011}.}\label{tab:one}
        \end{center}
\end{table*}

\asubsubsubsection{\bf Global Fit}

To study the sensitivity on CP-violating parameters $\tCza$ and $\tCzz$ at HL and HE-LHC, we perform a $\chi^2$ fit using, as observable, 
the signal strength ($\mu_{i,f}$) in the Higgs production channel ($i$) and Higgs decay channel ($f$). 
We can build a $\chi^2$  as follows:

\begin{equation}
 \chi^2(\tCza,\tCzz) = \sum_{i,f} \frac{(\mu_{i,f} -\mu_{i,f}^{\rm obs.})^2}{\Delta_{i,f}^2}
\end{equation}
{
The signal strength, $\mu_{i,f}$ is a function of the BSM parameters and it is defined as, 
\begin{eqnarray}
 \mu_{i,f} &=& \mu_i \times \mu_f \\
         &=& \frac{\sigma_i^{\rm BSM}}{\sigma_i^{\rm SM}} \times \frac{{\rm BR}_f^{ \rm BSM}}{{\rm BR}_f^{ \rm SM}}.
\end{eqnarray}

The total uncertainty, $\Delta_{i,f}^2$ includes theoretical, experimental systematic and statistical uncertainties, which 
are added in quadrature.
The one-sigma uncertainties for the high-luminosity (14 \UTeV centre of mass energy and 3 ab$^{-1}$ integrated luminosity) are given in table~\ref{tab:one}. 
Assuming the same acceptance efficiency, we scale the statistical uncertainties at 14 \UTeV and 3 ab$^{-1}$ luminosity appropriately 
to obtain the statistical uncertainties at 27 \UTeV and 15 ab$^{-1}$ 
luminosity. The theoretical and experimental systematic uncertainties are kept unchanged.

When considering kinematic distributions in the fit, we estimate the statistical uncertainty in each bin by scaling 
the overall statistical uncertainty by the fraction of number 
of events in each bin. On the other hand, the theoretical 
and systematic uncertainties are assumed to be the same in all the bins implying 
a very conservative scenario.

Since we are interested in the sensitivity on the CPV parameters that can be reached at HL and HE LHC, due to the present lack of experimental data, we take 
$\mu_{i,f}^{\rm obs.}=1$, implying that the future data would be consistent with the SM hypothesis. In the current analysis, we consider all the single Higgs production channels and Higgs decaying to four charged-leptons, {\it i.e} $i={{\rm ggF, VBF}, ZH, WH, t{\bar t}H}$ and $f=4\ell (2e2\mu, 4e, 4\mu)$. The projected uncertainties in these channels for HL-LHC are given in table~\ref{tab:one}. 
All the results in the following sections are presented taking $M_H=$125 \UGeV.
}\\

{\bf Production signal strengths : Inclusive} \\

{
The first step is to calculate the signal strengths for the relevant production channels in presence of the CP-violating
parameters $\tCza$ and $\tCzz$. We use {\tt Madgraph5\_aMC@NLO}~\cite{Alwall:2014hca} to obtain the inclusive cross sections in presence of these parameters. We have generated the required UFO model file for {\tt Madgraph} using the {\tt FeynRules} package~\cite{Degrande:2011ua,Alloul:2013bka}.  At 14 \UTeV, the production signal strengths are given by, }

\begin{eqnarray}\label{eq:mu14tev}
 \mu_{ZH}^{\rm 14 \UTeV} &=& 1.00 +  0.54~ \tCza^2 + 2.80~ \tCzz^2 + 0.95~ \tCza \tCzz \\
 \mu_{WH}^{\rm 14 \UTeV} &=& 1.00  + 0.84~ \tCza^2 + 3.87~ \tCzz^2 
   + 3.63~\tCza\tCzz \\
 \mu_{\rm VBF}^{\rm 14 \UTeV} &=& 1.00  + 0.25~ \tCza^2 + 0.45~ \tCzz^2  
   + 0.45~\tCza\tCzz
\end{eqnarray}

{ At 27 \UTeV, the corresponding signal strengths are given by,}

\begin{eqnarray}\label{eq:mu27tev}
 \mu_{ZH}^{\rm 27 \UTeV} &=& 1.00 +  0.63~ \tCza^2 + 3.26~ \tCzz^2 + 1.11~ \tCza \tCzz \\
 \mu_{WH}^{\rm 27 \UTeV} &=& 1.00 + 0.98~ \tCza^2 + 4.48~ \tCzz^2 
  + 4.16~\tCza\tCzz \\
 \mu_{\rm VBF}^{\rm 27 \UTeV} &=& 1.00  + 0.32~ \tCza^2 + 0.67~ \tCzz^2  
  + 0.65~\tCza\tCzz
\end{eqnarray}

The BSM predictions for VBF are derived using following cuts, 
\begin{equation}
 p_T(j) > 20~{\rm \UGeV}, |\eta(j)| < 5, \Delta\eta_{jj} > 3, m_{jj} > 130~{\rm \UGeV} \nonumber.
\end{equation}
We find that the $VH$ production modes are more sensitive to $\tCzz$ parameters.
The ggF and $t{\bar t}H$ production channels are unaffected in presence of CP-violating $VVH$ 
couplings. Therefore, 
\begin{eqnarray}
 \mu_{\rm ggF}^{\rm 14 \UTeV} &=& \mu_{\rm ggF}^{\rm 27 \UTeV} =  1.00 \\
 \mu_{t \bar t H}^{\rm 14 \UTeV} &=& \mu_{t\bar t H}^{\rm 27 \UTeV} =  1.00 .
\end{eqnarray}

{ In the present analysis we consider only kinematic distributions of the Higgs decay products, 
in the Higgs rest frame.} \\

{\bf Decay signal strength : Inclusive} \\

Now we turn to the calculation of the signal strength for the decay channel $H \to 4\ell$. This decay 
channel receives contributions from $2e^+2e^-$ ($4e$), $2\mu^+ 2\mu^-$ ($4\mu$) and $e^+e^-\mu^+\mu^-$ ($2e2\mu$) final states.
We use the latest version of the {\tt Hto4l} event generator~\cite{Boselli:2017pef} to obtain the partial decay widths in these 
channels in presence of $\tCza$ and $\tCzz$. Both the $e$ and $\mu$ are treated as massless. The ratio of the partial decay widths in BSM and in SM ($R_\Gamma$) for different channels are given by, 

{
\begin{eqnarray}
 R_{\Gamma}(H \to 2e2\mu) &=& 1  + 1.174~ \tCza^2 + 0.00291~ \tCzz^2  
   + (-0.00762)~\tCza\tCzz \\
 R_{\Gamma}(H \to 4e) 
 &=& R_{\Gamma}(H \to 4\mu) \nonumber \\
 &=& 1  + 1.106~ \tCza^2 + 0.00241~ \tCzz^2 
   + (-0.00595)~\tCza\tCzz.
\end{eqnarray}
}
The above expression for Higgs decay
into $2e2\mu$ and $4e$ are obtained after applying a selection cut of { 4 \UGeV} on the leading and sub-leading lepton pairs
of opposite sign.

In the present analysis, we also assume that the total Higgs decay width remains unchanged in presence of BSM. In this
case, the signal strength for decay is just the ratio of decay widths in BSM and in SM, that is,
\begin{eqnarray}\label{eq:mu4l}
 \mu_{4\ell} &=& \frac{\Gamma^{\rm BSM}_{4\ell}}{\Gamma^{\rm SM}_{4\ell}}  \nn \\
 &=& 1  + 1.138~ \tCza^2 + 
 0.00265~ \tCzz^2  +
 (-0.00674)~ \tCza \tCzz 
\end{eqnarray}
We note that, the dependence of the $4\ell$ decay signal strength on the parameter $\tCzz$ is very weak. \\

{\bf Decay signal strength : Differential} \\

{
We now turn to assessing  the role of kinematic distributions in $H \to 4\ell$ decay channel, 
which are affected by CP-violating $VVH$ couplings, in improving the sensitivity on $\tCza$ and $\tCzz$ at the HL and HE-LHC.
The Higgs rest frame angle $\phi$ between the decay planes of the two intermediate gauge bosons  
is very sensitive to the CP-Violating $VVH$ couplings~\cite{Soni:1993jc,Chang:1993jy,Skjold:1993jd,Buszello:2002uu}. We have considered 50 bins of $\phi$-distribution to perform the fit at differential level. For each bin, we calculate the signal strength ($\mu_{4\ell,j}; j=1\to50$) corresponding 
to Eq.~\ref{eq:mu4l}. Unlike $\mu_{4\ell}$ in Eq.~\ref{eq:mu4l}, $\mu_{4\ell,j}$ is also sensitive to linear terms in $\tCza$ and $\tCzz$.

\asubsubsubsection{\bf Result: HL and HE-LHC Analyses}

\begin{figure}[h!]
\centering
 \includegraphics[scale=0.6]{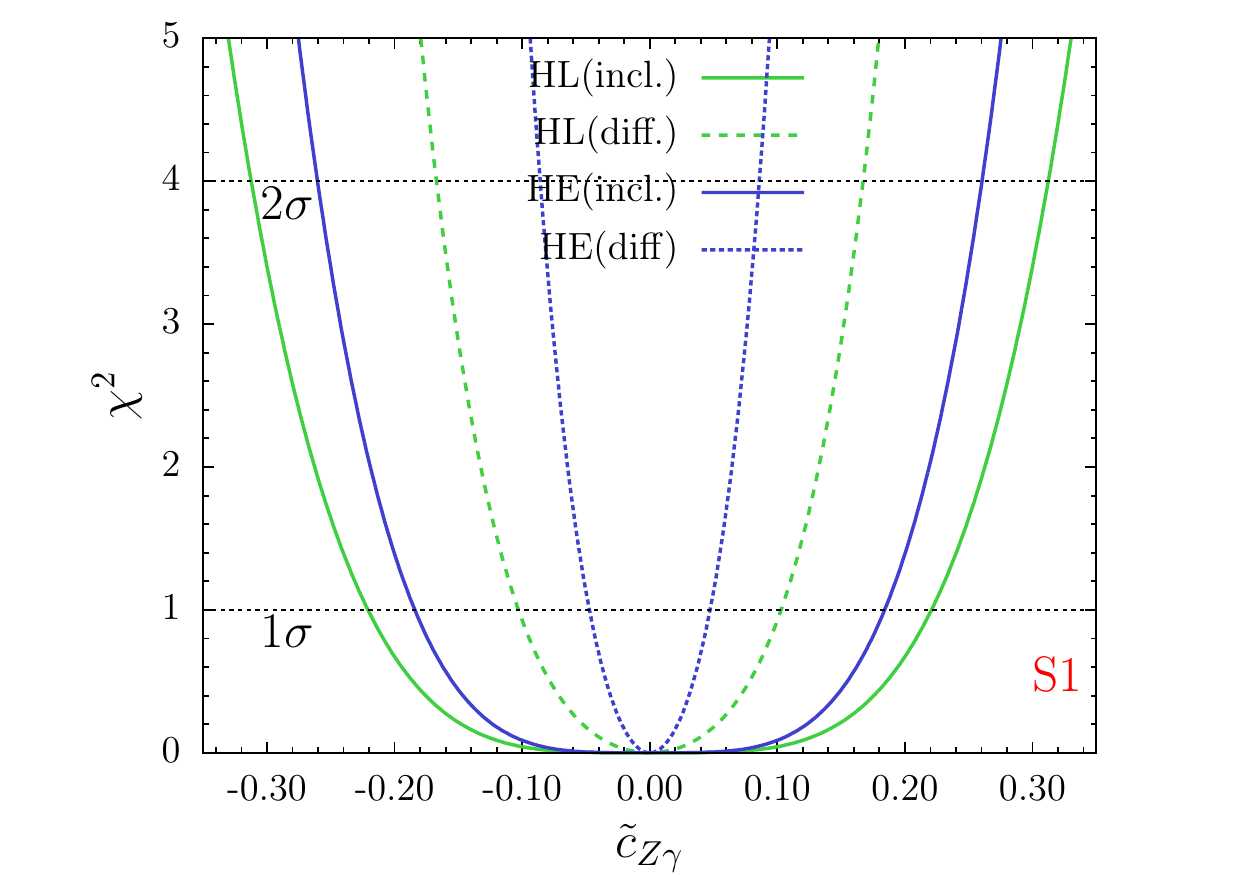}
\includegraphics[scale=0.6]{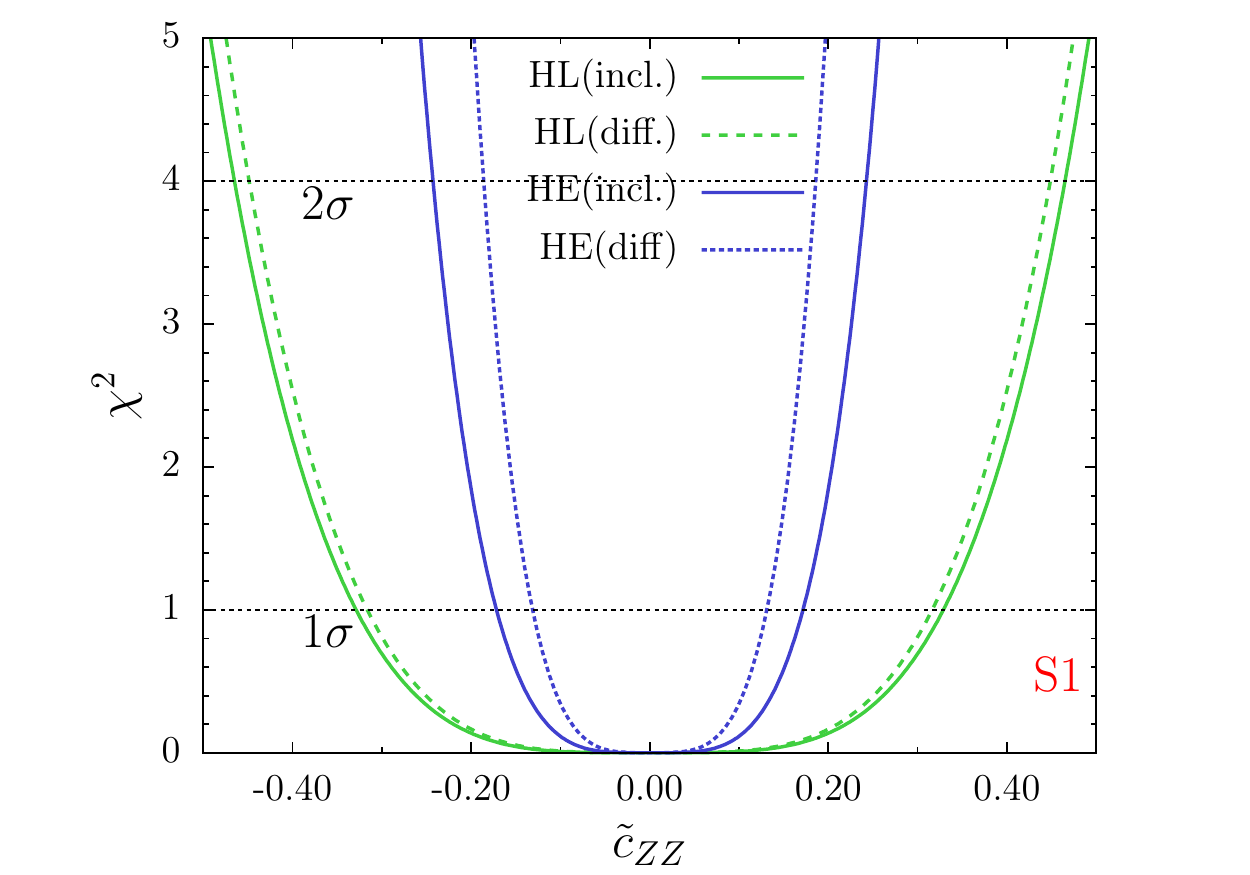}
\caption{ $\chi^2$ dependence on CP-violating parameters taking one parameter non-zero at a time 
at HL-LHC (3 ab$^{-1}$, green) and HE-LHC (15 ab$^{-1}$, blue) for uncertainty scenario S1. The solid lines refer to the fit performed using $H \to 4\ell$ decay width at inclusive level (1 bin) while, the dashed lines refer to the fit obtained using $H \to 4\ell$ decay width at differential level ($\phi$-distribution with 50 bins).}\label{fig:fit1p}
\end{figure}

\begin{figure}
\centering
 \includegraphics[scale=1.2]{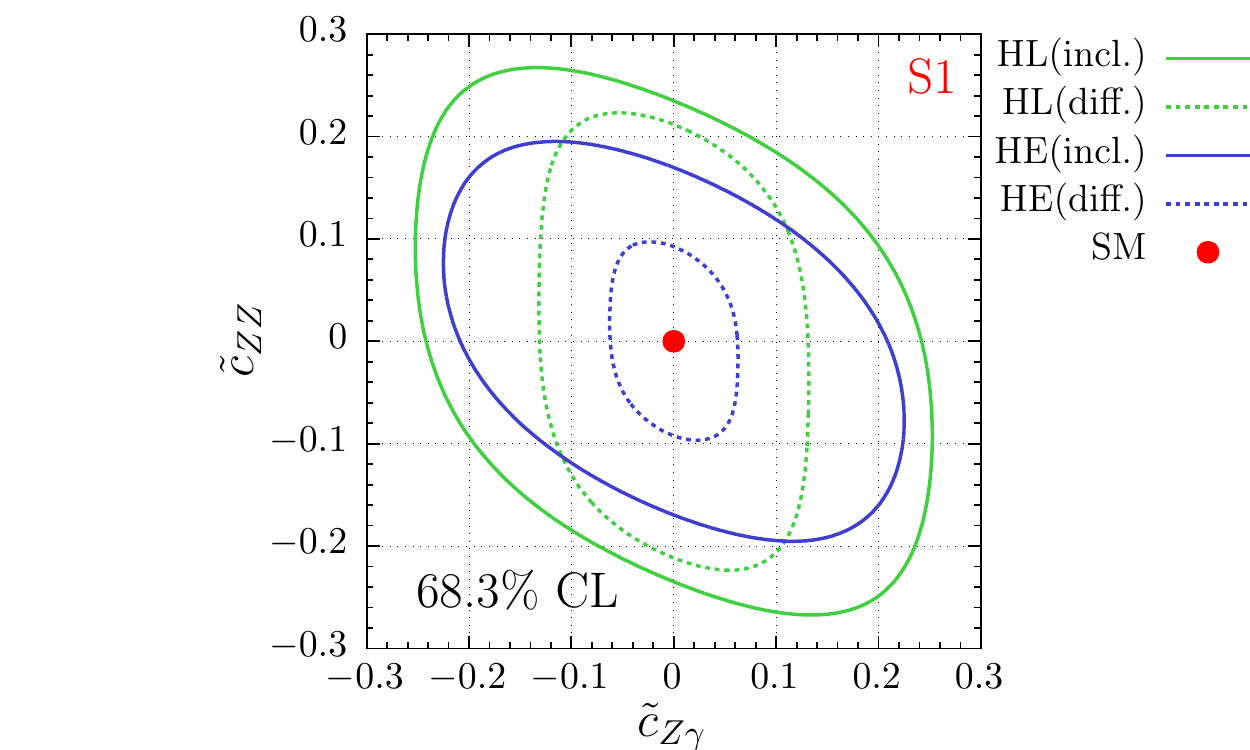}
\caption{ $1\sigma$  reach on $\tCza$ and 
$\tCzz$ at HL-LHC (3 ab$^{-1}$, green) and HE-LHC (15 ab$^{-1}$, blue) for uncertainty scenario S1. The solid lines refer to the fit performed using $H \to 4\ell$ decay width at inclusive level (1 bin) while, the dashed lines refer to the fit obtained using $H \to 4\ell$ decay width at differential level ($\phi$-distribution with 50 bins).  }\label{fig:fit2p}
\end{figure}

The results of the $\chi^2$ fit for CP-violating parameters $\tCza$ and $\tCzz$ are displayed in Fig.~\ref{fig:fit1p} and Fig.~\ref{fig:fit2p}. In these results, 
 {\it incl.} refers to the fit obtained using 
the partial decay width information in the $H \to 4\ell$ channel, while {\it diff.} refers to the fit obtained using 
$\phi$-distribution in $H \to 4\ell$ decay. In Fig.~\ref{fig:fit1p}, we show 
$1\sigma$ and $2\sigma$ bounds on $\tCza$ and $\tCzz$ in a one parameter (1P) 
analysis.  We find that at HL-LHC we are more sensitive to $\tCza$ than to $\tCzz$. At the inclusive level 
we gain better sensitivity on $\tCzz$ than on $\tCza$ when going from HL-LHC 
to HE-LHC. This is mainly due to a stronger dependence of the production signal strength on parameter $\tCzz$. However, due to a stronger dependence of $\mu_{4\ell}$ on $\tCza$ the effect of using $\phi$-distribution in the fit is larger for $\tCza$ than for $\tCzz$.

In Fig.~\ref{fig:fit2p}, we provide $1\sigma$ contour lines in the $\tCza-\tCzz$ plane. We can see that the parameters $\tCza$ and $\tCzz$ are weekly correlated. 
Once again we find that using $\phi$-distribution in the fit improves our 
sensitivity on CP-violating parameters significantly.
The parameter $\tCzz$ is mainly constrained by the production channels $VH$ and VBF.
We have given a summary of $1\sigma$ bounds on $\tCza$ and $\tCzz$ obtained from our analyses for HL and HE-LHC in Table~\ref{tab:tab3}.
}

\begin{table}
 \centering
 \begin{tabular}{l|cc|l}
 \hline
  \backslashbox{Analysis}{Parameter} & $\tCza$ & $\tCzz$ & Case \\
  \hline\hline
    HL-LHC ($4\ell$, incl.) & [-0.22,0.22] & [-0.33,0.33]& 1P \\
                            & [-0.25,0.25] & [-0.27,0.27]& 1P$_{marg.}$ \\
    \hline
    HL-LHC ($4\ell$, diff.) & [-0.10,0.10] & [-0.31,0.31]& 1P \\
                            & [-0.13,0.13] & [-0.22,0.22]& 1P$_{marg.}$ \\
    \hline
    HE-LHC ($4\ell$, incl.) & [-0.18,0.18] & [-0.17,0.17]& 1P \\
                            & [-0.23,0.23] & [-0.20,0.20]& 1P$_{marg.}$ \\
    \hline
    HE-LHC ($4\ell$, diff.) & [-0.05,0.05] & [-0.13,0.13]& 1P  \\
                            & [-0.06,0.06] & [-0.10,0.10]& 1P$_{marg.}$ \\
 \end{tabular}
\caption{ Summary of 1$\sigma$ bounds on $\tCza$ and $\tCzz$ from various analyses considered in our study for uncertainty scenario S1. 1P refers to the case
where only one parameter is non-zero while 1P$_{marg.}$ refers to the case in which the effect of one of the two parameters is marginalised.}\label{tab:tab3}
\end{table}

\asubsubsubsection{$h \to \tau^+ \tau^-$}

The most promising direct probe of CP violation in fermionic Higgs
decays is the $\tau^+ \tau^-$ decay channel, which benefits from a
relatively large $\tau$ Yukawa giving a SM branching fraction of
$6.3\%$. Measuring the CP violating phase in the tau Yukawa requires a measurement of the linear polarisations of both $\tau$ leptons and the azimuthal angle between them. This can be done by analysing tau substructure, namely the angular distribution of the various components of the tau decay products.

The main $\tau$ decay modes studied include $\tau^\pm \to
\rho^\pm (770) \nu$, $\rho^\pm \to \pi^\pm \pi^0$~\cite{Bower:2002zx,
  Desch:2003mw, Desch:2003rw, Harnik:2013aja, Askew:2015mda,
  Jozefowicz:2016kvz} and $\tau^\pm \to \pi^\pm
\nu$~\cite{Berge:2008wi, Berge:2008dr, Berge:2011ij}.  Assuming CPT
symmetry, collider observables for CP violation must be built from
differential distributions based on triple products of three-vectors.
In the first case, $h \to \pi^\pm \pi^0 \pi^\mp \pi^0 \nu \nu$,
angular distributions built only from the outgoing charged and neutral
pions are used to determine the CP properties of the initial $\tau$
Yukawa coupling.  In the second case, $h \to \pi^\pm \pi^\mp \nu \nu$,
there are not enough reconstructible independent momenta to construct an observable sensitive to CP violation, requiring additional kinematic information such as the $\tau$ decay impact parameter.

In the kinematic limit when each outgoing neutrino is taken to be
collinear with its corresponding reconstructed $\rho^\pm$ meson, the
acoplanarity angle, denoted $\Phi$, between the two decay planes
spanned by the $\rho^\pm \to \pi^\pm \pi^0$ decay products is exactly
analogous to the familiar acoplanarity angle from $h \to 4 \ell$
CP-property studies~\cite{Chatrchyan:2012jja, Aad:2013xqa}.  Hence, by measuring the $\tau$ decay products in
the single-prong final state, suppressing the irreducible $Z \to
\tau^+ \tau^-$ and reducible QCD backgrounds, and reconstructing the
acoplanarity angle of $\rho^+$ vs.~$\rho^-$, the differential
distribution in $\Phi$ gives a sinusoidal shape whose maxima and
minima correspond to the CP-phase in the $\tau$ Yukawa coupling.  We can parametrise the $\tau$ Yukawa coupling to include CP violation using the Lagrangian term $\frac{y_\tau}{\sqrt{2}} h \bar{\tau} (\cos \Delta + i \sin \Delta \gamma^5) \tau$, where $y_\tau$ is the magnitude of the tau Yukawa coupling and remains fixed to the SM value for this study.

An optimal observable using the collinear approximation was derived in~\cite{Harnik:2013aja}. Assuming 70\% efficiency for tagging hadronic $\tau$ final states, and
neglecting detector effects, the estimated sensitivity for the
CP-violating phase $\Delta$ of the $\tau$ Yukawa coupling using 3 ab$^{-1}$ at
the HL-LHC is 8.0$^\circ$.  A more sophisticated
analysis~\cite{Askew:2015mda} found that detector resolution effects
on the missing transverse energy distribution degrade the expected
sensitivity considerably, and as such, about 1 ab$^{-1}$ is required
to distinguish a pure scalar coupling (CP phase is zero) from a pure
pseudoscalar coupling (CP phase is 90$^\circ$).

At the HE-LHC, the increased signal cross section for Higgs production
is counterbalanced by the increased background rates, and so the main
expectation is that improvements in sensitivity will be driven by the
increased luminosity and more optimised experimental methodology.
Rescaling with the appropriate luminosity factors, the optimistic
sensitivity to the $\tau$ Yukawa phase from acoplanarity studies is
4-5$^\circ$, while the more conservative estimate is roughly an order
of magnitude bigger.

\asubsubsubsection{$ t\,\bar{t}\,h$}

CP violation in the top quark-Higgs coupling is strongly constrained by EDM measurements and Higgs rate measurements~\cite{Brod:2013cka}. However, these constraints assume that the light quark Yukawa couplings and $hWW$ couplings are set to their SM values. If this is not the case, the constraints on the phase of the top Yukawa coupling are less stringent.
    
Assuming the EDM and Higgs rate  constraints can be avoided, the CP structure of the top quark Yukawa can be probed directly in $pp \to t\bar t h$. Many simple observables, such as $m_{t\bar t h}$ and $p_{T,h}$ are sensitive to the CP structure, but require reconstructing the top quarks and Higgs.

Some $t\bar t h$ observables have been proposed recently that access the CP structure without requiring full event reconstruction. These include the azimuthal angle between the two leptons in a fully leptonic $t/bar{t}$ decay with the additional requirement that the $p_{T,h} > 200\, \text{\UGeV}$~\cite{Buckley:2015vsa}, and the angle between the leptons (again in a fully leptonic $t/\bar t$ system) projected onto the plane perpendicular to the $h$ momentum~\cite{Boudjema:2015nda}. These observables only require that the Higgs is reconstructed and are inspired by the sensitivity of $\Delta \phi_{\ell^+\ell^-}$ to top/anti-top spin correlations in $pp \to t\bar t$~\cite{Mahlon:1995zn}. The sensitivity of both of these observables improves at higher Higgs boost (and therefore higher energy), making them promising targets for the HE-LHC, though no dedicated studies have been carried out to date.

\asubsubsection[K. Prokofiev,  A. Gritsan, P. Milenovic, H. Roskes, U. Sarica]{Experimental constraints on anomalous HVV couplings}
\asubsubsubsection{Experimental constraints from $\hzzllll$ decays}
\label{subsection:ACHVVdecay}

The projections for anomalous coupling measurements from $\hzzllll$ decays at the HL-LHC were studied within the context of the last ECFA HL-LHC Experiments Workshops in 2013 and 2014~\cite{ATL-PHYS-PUB-2013-013}.
The obtained limits are quantified 
in terms of the effective couplings $g_i$  introduced in the invariant amplitude describing the interaction of a spin-0 particle and and two spin-one gauge bosons
introduced in Refs.~\cite{Gao:2010qx} and \cite{Heinemeyer:2013tqa}. The couplings $g_1$, $g_2$ and $g_3$ correspond to the 
interaction with the CP-even and $g_4$ to the interaction with the CP-odd boson, respectively.
The direct measurement of couplings is free of assumptions on the size of the interference effects and its results can be 
expressed in terms of the $(f_{g_i}, \phi_i)$ parametrisation:
$$
f_{g_i} =\frac{|g_i|^2 \sigma _i ^2}{|g_1|^2 \sigma _1 ^2 + |g_2|^2 \sigma _2 ^2 + |g_4|^2 \sigma _4 ^2}; \;\;\; \phi _{g_i} = \arg \left ( \frac{g_i}{g_1} \right ).
$$
In this analysis $g_2$ and $g_4$ are measured separately, assuming the simultaneous presence of
only $g_1$ and of the coupling under study, this corresponds to set $g_2 = 0\; (g_4 = 0)$ in the expression of $f_{g_4}\;
( f_{g_2} )$ above.

The analysis was performed by fitting the observables based on the analytic calculation of Leading Order Matrix Element describing 
 $\hzzllll$ decays in the presence of anomalous couplings. The final fit is based on Monte Carlo modelling of the expected
signal at each bin of the $(\Re (g_i)/g_1; \Im (g_i) /g_1)$ plane, where $g_i$ represents $g_2$ or $g_4$. The irreducible $ZZ$ background 
 was suppressed by using a dedicated Boosted Decision Tree discriminant.  
 
 
Following the event selection and applying the fit methodology described above, the expected exclusion
of the non-Standard Model contributions given the Standard Model data is evaluated for $300$ and $3000~\fbinv$.
Examples of the corresponding exclusion plots are given in Figure~\ref{fig:fgi-atlas-HZZ_anom_couplings}. With a conservative analysis limits of $f_{g_4} < 0.037$ at $95\%$~CL and 
$f_{g_2} < 0.12$ at $95\%$~CL for $3000~\fbinv$ are obtained. This allows a sensitive test of the tensor structure of the $\hzz$ couplings at the HL-LHC.

\begin{figure*}[!htbp]
\centering
\includegraphics[width=0.70\textwidth]{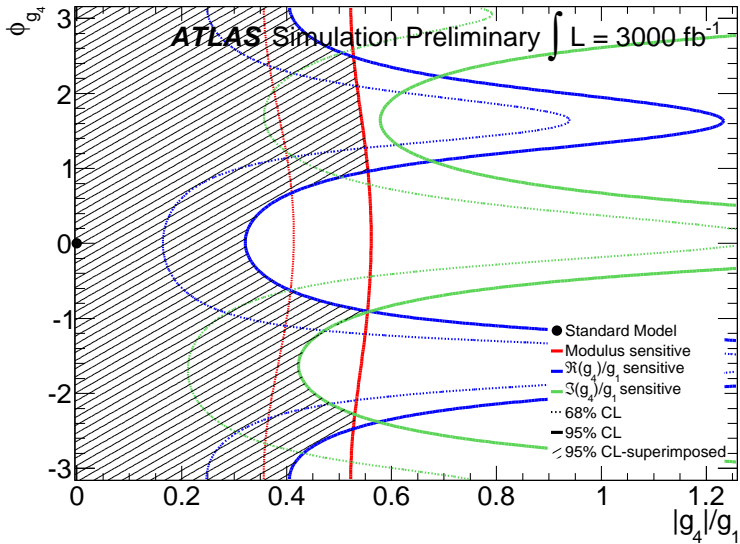}
\caption
{
Results of the $g_4$-sensitive fits projected onto the ($|g_4|/g_1$, $\phi_{g_4}$) plane for $3000~\fbinv$. The shaded area corresponds to the most restrictive exclusion of the three observables. 
}
\label{fig:fgi-atlas-HZZ_anom_couplings}
\end{figure*}

\asubsubsubsection{Experimental constraints from production and decay in $\hzzllll$ channel}
Anomalous contributions in the spin-0 tensor structure of \HVV interactions can be characterised by coefficients $a_2$, $a_3$, $\Lambda_1$, and $\Lambda_Q$ defined in Refs.~\cite{Khachatryan:2014kca, Khachatryan:2015mma}. The $a_2$ and $a_3$ coefficients have one-to-one correspondence with the $g_2$ and $g_4$ coefficients mentioned in Section~\ref{subsection:ACHVVdecay}.
The contribution to the total cross section from these coefficients is parametrised in terms of their fractional contribution to \onshell $\PH\to2\mathrm{e}2\mathrm{\mu}$ decays via the fractions \fai and phases \pai~\cite{Khachatryan:2014kca, Khachatryan:2015mma}.
Constraints on these anomalous contributions can further be improved by including \offshell Higgs boson production. An enhancement of signal events is expected in the presence of either anomalous \HVV couplings or large Higgs boson total width, \GH~\cite{Khachatryan:2015mma,deFlorian:2016spz,CMS-PAS-HIG-18-002}.

In the study from Ref.~\cite{CMS-PAS-FTR-18-011}, only the tensor structure proportional to \AC{3} is considered using either the combination of \onshell and \offshell events or with only \onshell events with $4\ell$ decay, following the techniques described in Refs.~\cite{Khachatryan:2014kca,deFlorian:2016spz,CMS-PAS-HIG-18-002}. Constraints are placed in terms of \fcospai with the assumptions $\pai=0$~or~$\pi$, $\ai^\ZZ=\ai^\WW$, and ${\GH}={\GHSM}$ in the case of limits from the combined \onshell and \offshell likelihood parametrisation.

The projections are shown in Fig.~\ref{fig:fa3-GH-projections} and summarised in Table~\ref{table:fa3-GH-projections}. Systematic uncertainties are found to have a negligible effect on the results for \fcospAC{3} using either \onshell and \offshell events combined or only \onshell events, so only the case when systematic uncertainties are as in Run~2~\cite{CMS-PAS-HIG-18-002}, is shown.

\begin{table*}[!htbp]
\centering
\caption
{
Summary of the 95\%~\CL intervals for \fcospAC{3}, under the assumption $\GH=\GHSM$ for projections at $3000\fbinv$~\cite{CMS-PAS-FTR-18-011}. The constraints are multiplied by $10^{4}$, and the values are given only for the case when systematic uncertainties are as in Run~2~\cite{CMS-PAS-HIG-18-002}.
}
\renewcommand{\arraystretch}{1.25}
\begin{tabular}{cll}
\hline
 Parameter & \multicolumn{1}{c}{Scenario} & \multicolumn{1}{c}{Projected 95\% CL interval}  \\
\hline
 $\fcospAC{3}$ & Only \onshell & $[-1.8, 1.8]\times10^{-4}$  \\ 
 $\fcospAC{3}$ & \Onshell and \offshell & $[-1.6, 1.6]\times10^{-4}$  \\ 
\hline
\end{tabular}
\label{table:fa3-GH-projections}
\end{table*}

\begin{figure*}[!htbp]
\centering
\includegraphics[width=0.90\textwidth]{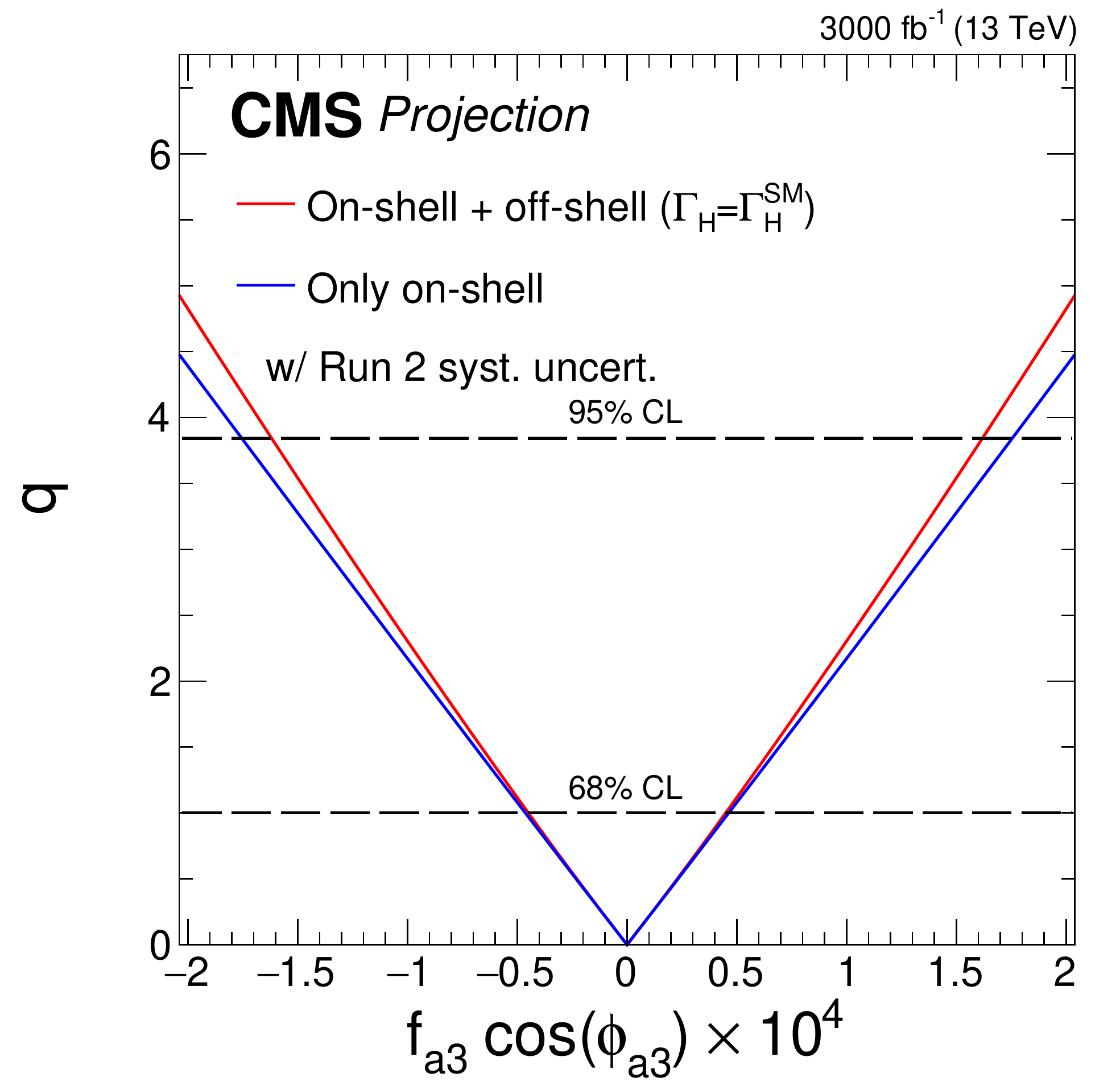}
\caption
{
Likelihood scans for projections on \fcospAC{3} at $3000~\fbinv$~\cite{CMS-PAS-FTR-18-011}. The scans are shown using either the combination of \onshell and \offshell events (red) or only \onshell events (blue). The dashed lines represent the effect of removing all systematic uncertainties.
The dashed horizontal lines indicate the 68\% and 95\% CLs, and the \fcospAC{3} scans assume ${\GH}={\GHSM}$.
}
\label{fig:fa3-GH-projections}
\end{figure*}

\clearpage


\newpage

\asection[L. Cadamuro, E. Petit, M. Riembau]{Di-Higgs production and Higgs self couplings}\label{sec3}

The HL-LHC is expected to be a Higgs boson factory, and the study of the double-Higgs ($HH$) production is one of the key goal of this high-luminosity program. Despite the small production cross-section compared to the single-Higgs boson production, more than 100000 $HH$ pairs should be produced by the HL-LHC per experiment. The trilinear self-interaction of the Higgs boson is described by the coupling strength $\lambda_{HHH}  = \frac{m_H^2}{2v} $, where $m_H$ is the Higgs boson mass, and $v$ the electroweak symmetry breaking vacuum expectation value. Measurements of the Higgs trilinear interaction would provide constraints on the shape of the Higgs potential close to the minimum and would allow to verify the electroweak symmetry breaking mechanism of the SM.
The existence of an extended scalar sector or the presence of new dynamics at higher scales could modify the Higgs boson self-couplings.
In the following the trilinear self-coupling strength, measured relative to the SM expectation is denoted by $\kappa_{\lambda} = c_{hh} = \lambda = \lambda_{HHH}/\lambda_{HHH}^{SM}$.

This section describes the prospects for studies of the Higgs boson pair production at the HL-LHC and HE-LHC and is organised as follows: the state-of-the-art NLO computations of the Higgs boson pair production cross sections is shown in Section~\ref{sec:HH_NLO}. Section~\ref{sec:HH_meas_exp} describes the prospect experimental analyses with the ATLAS and CMS experiments with realistic conditions, while Section~\ref{sec:HH_meas_th} concentrates on alternative methods with phenomenology studies. Studies at the HE-LHC are shown in Section~\ref{sec:HH_HE} with both phenomenological and experimental perspectives. Indirect probes of the trilinear couplings are described in Section~\ref{sec:HH_indirect}, using differential cross-section measurements or global fits. Finally Section~\ref{sec:HH_implications} shows the implications of the trilinear coupling measurements on b-physics and the electroweak phase transition.  

\asubsection{Higgs boson pair production cross section}
\label{sec:HH_NLO}

\asubsubsection{SM Calculation}

\asubsubsubsection[J. Mazzitelli]{HH production via gluon fusion at NNLO}
\label{sec:HH_NNLO}

The fusion of gluons via a heavy-quark (mainly top-quark) loop is the most important production mechanism of Higgs boson pairs at hadron colliders within the SM.
The NLO QCD corrections for this process have been known in the large-$m_t$ limit for some time \cite{Dawson:1998py}, and the NNLO cross section has also been computed within this approximation \cite{deFlorian:2013jea}.
The NLO corrections retaining the full dependence on the top-quark mass have been obtained for the first time in Refs.~\cite{Borowka:2016ehy,Borowka:2016ypz}, and have been recently confirmed by an independent calculation \cite{Baglio:2018lrj}.
On top of this, an improved NNLO prediction --labelled NNLO$_{\mathrm{FTa}}$ for full-theory approximation-- was presented in Ref.~\cite{Grazzini:2018bsd}.
This approximation is obtained by combining one-loop double-real corrections with full $m_t$ dependence with suitably reweighted real-virtual and double-virtual contributions evaluated in the large-$m_t$ limit.
Furthermore, the stability of the QCD perturbative expansion at this order has been confirmed by consistently matching the NNLO$_{\mathrm{FTa}}$ prediction with the next-to-next-to-leading logarithmic terms coming from threshold resummation \cite{deFlorian:2018tah}.

More details on the NLO results with full-$m_t$ dependence (including the effect coming from variations in the Higgs self-coupling and the contributions arising from BSM EFT operators) are provided in the following sections, therefore we focus here on the state-of-the-art NNLO prediction, i.e., the NNLO$_{\mathrm{FTa}}$ result from Ref.~\cite{Grazzini:2018bsd}.

Before focusing on the numerical results, it is worth to stress out that the NNLO cross sections presented here, as well as the NLO predictions for gluon fusion present in the following sections, are computed using the on-shell scheme for the top-quark mass renormalisation .
Some partial results on the uncertainties related to the $m_t$ scheme and scale choice have been presented in Ref.~\cite{Baglio:2018lrj} at NLO, and further studies to gauge the size of their effect on the total cross section and distributions are in progress.
This source of uncertainty is for the moment not considered in the NLO and NNLO predictions.

In Table \ref{table:HH_at_NNLO} we present results for the total cross section at $\sqrt{s} = 14$~\UTeV and 27~\UTeV.
We use the values $m_h = 125$~\UGeV for the Higgs boson mass and $m_t = 173$~\UGeV for the  on-shell top quark mass.
The NNLO PDF4LHC15 sets of parton distribution functions are used, and PDF and $\alpha_S$ uncertainties are also provided.
An estimation of the systematic uncertainty of the approximation due to missing finite-$m_t$ effects is also presented, and it is found to be at the few percent level.
For the renormalisation  and factorisation scales we use the central value $\mu_0 = M_{hh}/2$, which has been shown to provide a better convergence for the fixed order prediction \cite{deFlorian:2018tah}. We obtain the scale uncertainties via the usual 7-point scale variation.

{\renewcommand{\arraystretch}{1.6}
\begin{table}
\begin{center}
\begin{tabular}{|l|c|c|c|c|c|}
\hline
$\sqrt{s}$~[\UTeV] & NNLO$_{\mathrm{FTa}}$~[fb] & $m_t$ unc. & PDF unc. & $\alpha_S$ unc. & PDF$+\alpha_S$ unc. \\
 \hline
$14$ & $36.69^{+2.1\%}_{-4.9\%}$ & $\pm 2.7\%$ & $\pm 2.1\%$ & $\pm 2.1\%$ & $\pm 3.0\%$ \\
\hline
$27$ & $139.9^{+1.3\%}_{-3.9\%}$ & $\pm 3.4\%$ & $\pm 1.7\%$ & $\pm 1.8\%$ & $\pm 2.5\%$ \\
\hline
\end{tabular}
\end{center}
\caption{
Inclusive cross sections for Higgs boson pair production at NNLO$_{\mathrm{FTa}}$ for centre-of-mass energies of 14~\UTeV and 27~\UTeV. 
Scale uncertainties are reported as superscript/subscript.
The estimated uncertainty of the approximation due to finite top-quark mass effects is also presented, as well as the PDF and $\alpha_S$ uncertainties.
}
\label{table:HH_at_NNLO}
\end{table}
}

The NNLO$_{\mathrm{FTa}}$ predictions from Ref.~\cite{Grazzini:2018bsd} are also fully differential in the Higgs boson pair and the associated jet activity.
As an example, we present the Higgs pair invariant mass distribution at 14~\UTeV and 27~\UTeV in Figure~\ref{fig:mhh_NNLO}, together with the corresponding NLO prediction.
We can observe the strong reduction in the size of the scale uncertainties when including the NNLO$_{\mathrm{FTa}}$ corrections, and the sizeable overlap with the NLO uncertainty band (not present between the LO and NLO predictions), suggesting a significant improvement in the perturbative convergence as we move from NLO to NNLO.

\begin{figure}[t!]
\begin{center}
\includegraphics[width=.49\textwidth]{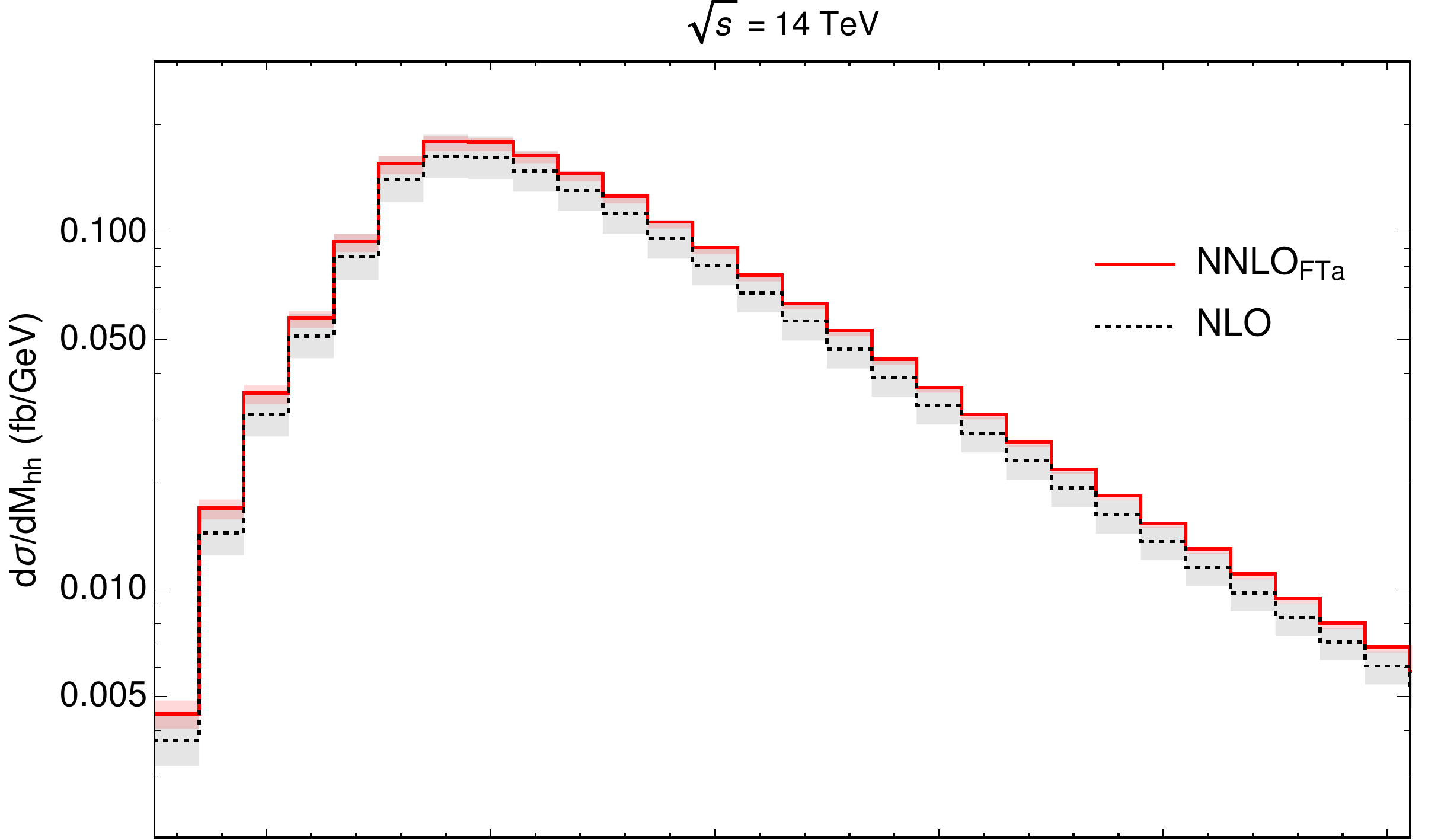}
\hfill
\includegraphics[width=.49\textwidth]{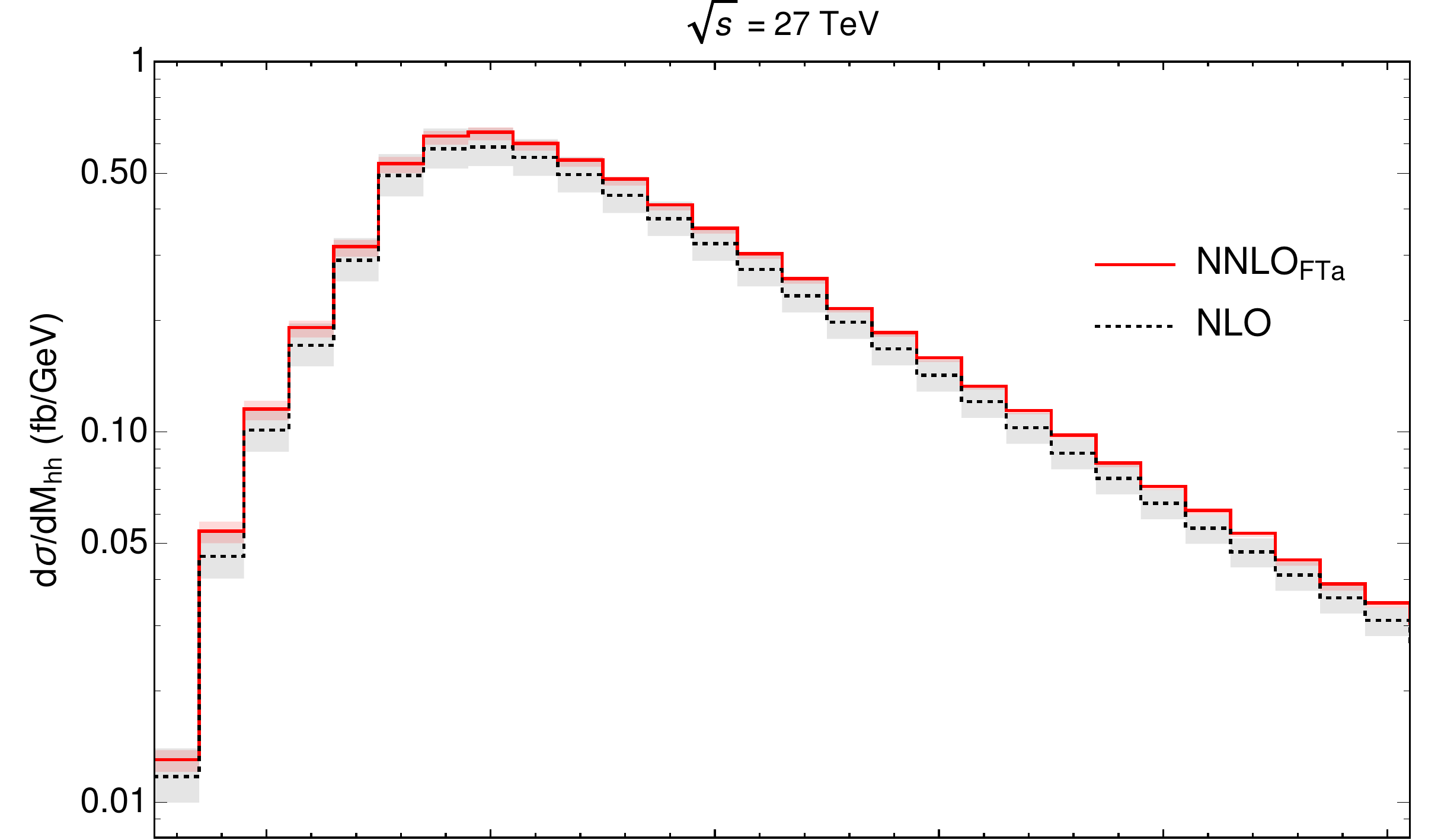}
\\
\includegraphics[width=.49\textwidth]{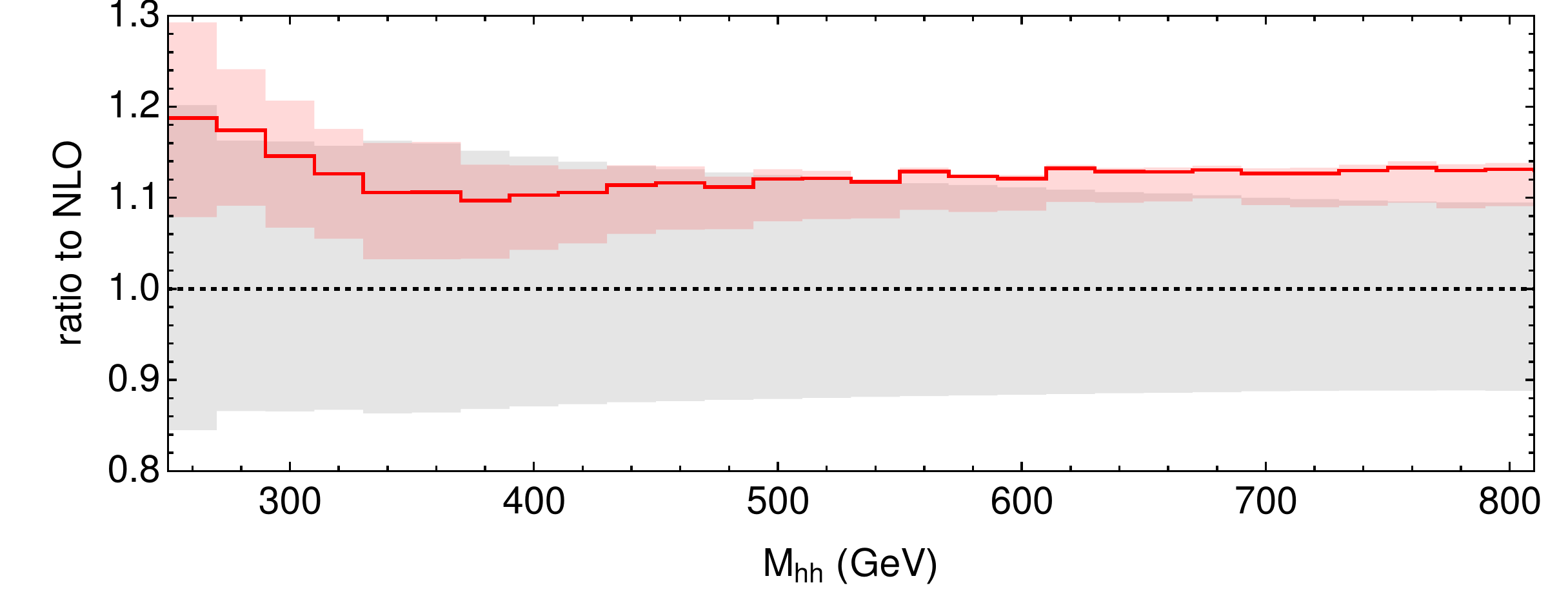}
\hfill
\includegraphics[width=.49\textwidth]{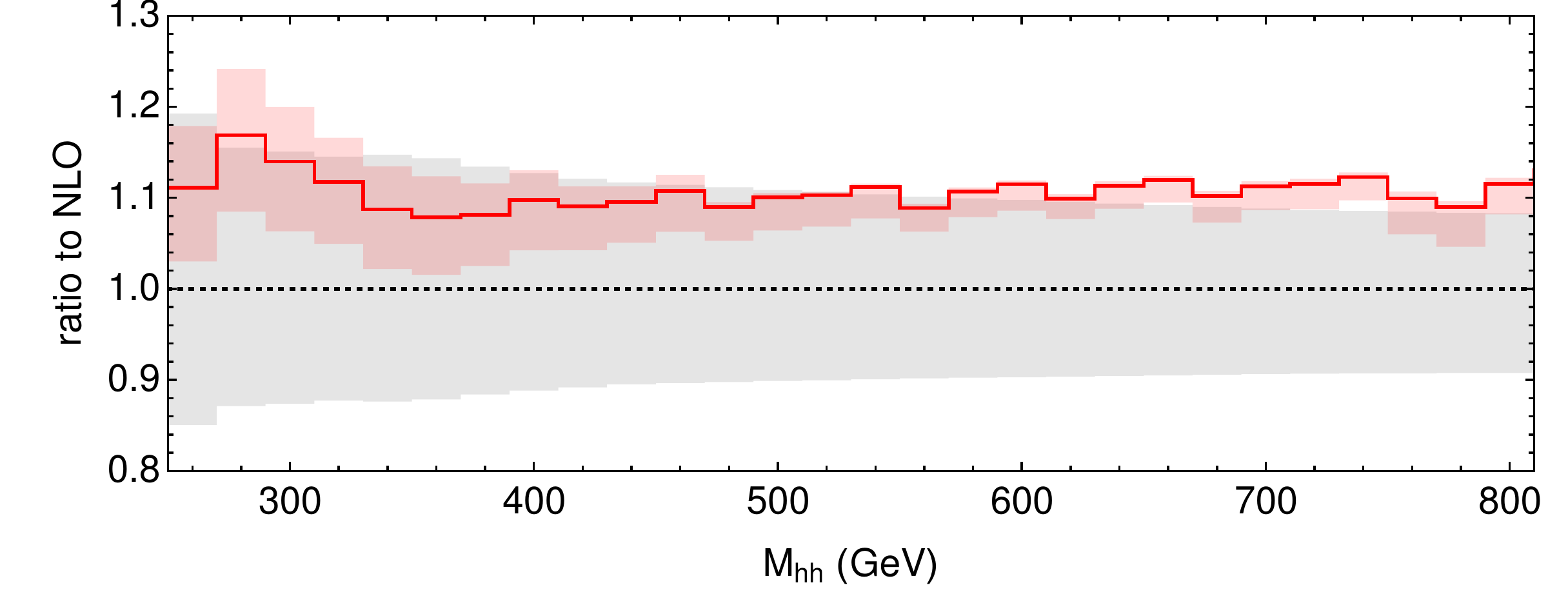}
\end{center}
\vspace{-2ex}
\caption{\label{fig:mhh_NNLO}
Higgs boson pair invariant mass distribution at NNLO$_{\mathrm{FTa}}$, together with the NLO prediction, at $14\,$\UTeV (left) and $27\,$\UTeV (right). The lower panels show the ratio with respect to the NLO prediction, and the filled areas indicate the scale uncertainties.
}
\end{figure}

\asubsubsubsection[E. Vryonidou]{HH production in sub-dominant channels}

\begin{table}[h!]
\renewcommand{\arraystretch}{1.6}
\begin{center}
\resizebox{\textwidth}{!}{%
    \begin{tabular}{|l|c|c|c|c|c|}
        \hline 
$\sqrt{s}$ (\UTeV) & $ZHH$ & $WHH$ & VBF $HH$ & $ttHH$ &$tjHH$ \\ 
 \hline
14 & $0.359^{+1.9\%}_{-1.3\%}\pm 1.7\%$  &  $0.573^{+2.0\%}_{-1.4\%}\pm 1.9\%$  & $1.95^{+1.1\%}_{-1.5\%}\pm 2.0\%$ & $0.948^{+3.9\%}_{-13.5\%}\pm 3.2\%$& $0.0383^{+5.2\%}_{-3.3\%}\pm 4.7\%$ \\ \hline
  27 &  $0.963^{+2.1\%}_{-2.3\%}\pm 1.5\%$ & $1.48^{+2.3\%}_{-2.5\%}\pm 1.7\%$  & $8.21^{+1.1\%}_{-0.7\%}\pm 1.8\%$& $5.27^{+2.0\%}_{-3.7\%}\pm 2.5\%$&  $0.254^{+3.8\%}_{-2.8\%}\pm 3.6\%$ \\ \hline
\end{tabular}}
\end{center} 
\caption{Signal cross section (in fb) for HH production at NLO QCD.}  
\label{tab:HHxsecOthers}
\end{table}
Results  shown in Table~\ref{tab:HHxsecOthers} have been obtained within the {\sc MadGraph5\_aMC@NLO} \cite{Alwall:2014hca} framework, as in Ref. \cite{Frederix:2014hta}. The renormalisation and factorisation scale was set to $m_{HH}/2$ and varied up and down by a factor of two to obtain the scale uncertainties. The 5-flavour PDF4LHC NLO Monte Carlo PDF set was used to obtain the results (LHAPDF set number 90500, \textit {PDF4LHC15\_nlo\_mc}). The $WHH$ results are the sum of the $W^+$ and $W^-$ cross-sections. Similarly $tjHH$ involves both top and anti-top production. 


\asubsubsubsection[G. Heinrich, S. Jones, M. Kerner, G. Luisoni, L. Scyboz]{Probing the Higgs boson self-coupling in di-Higgs
  production with full $m_t$-dependence at NLO QCD}
\label{subsec:varylambda}

In this section we consider the impact of varying the Higgs self-coupling $\lambda$ on the NLO computations of the HH production cross section.
In particular, we announce a version of the {\tt ggHH} code~\cite{Borowka:2016ypz,Heinrich:2017kxx,Borowka:2016ehy} implemented in the {\tt POWHEG-BOX-V2}~\cite{Alioli:2010xd} where variations of  $\lambda$ are accessible to the user in a parton shower Monte Carlo program at full NLO.

\asubsubsubsection{Total cross sections at different values of the trilinear coupling}

In Table \ref{tab:sigmatot} we list total cross sections at 14\,\UTeV and 27\,\UTeV for various values of the trilinear Higgs coupling $\lambda$. 
\begin{table}[htb]
\begin{center}
\begin{tabular}{| c | c | c |c|c|}
\hline
&&&&\\
$\lambda_{\mathrm{BSM}}/\lambda_{\mathrm{SM}}$ & $\sigma_{\rm{NLO}}@14 \mathrm{\UTeV}$\,[fb] & $\sigma_{\rm{NLO}}@27 \mathrm{\UTeV}$\,[fb] &K-fac.@14TeV&K-fac.@27TeV\\
&&&&\\
\hline
1& 32.88$^{+13.5\%}_{-12.5\%}$&127.7$^{+11.5\%}_{-10.4\%}$ &1.66&1.62\\
\hline
2 & 14.91 &  59.10 & 1.58 & 1.52\\
\hline
2.4 & 13.81& 53.67 & 1.65 & 1.60\\
\hline
3& 19.82 & 69.84 & 1.97 & 1.89\\
\hline 
5 & 98.42& 330.61 & 2.21 & 2.18\\
\hline 
0 & 73.84& 275.29& 1.79 & 1.78 \\
\hline 
-1 & 137.69& 504.9 & 1.87 & 1.83\\
\hline
\end{tabular}
\end{center}
\caption{Total cross sections for Higgs boson pair production at full NLO. The given uncertainties are scale uncertainties. 
\label{tab:sigmatot}}
\end{table}

The results have been obtained using the parton distribution functions PDF4LHC15\_nlo\_100\_pdfas~\cite{Butterworth:2015oua,CT14,MMHT14,Ball:2014uwa},
along with the corresponding value for $\alpha_s$ for both the NLO and
the LO calculation.
The masses have been set to $m_h=125$\,\UGeV, $m_t=173$\,\UGeV,
and the top quark width has been set to zero. 
The scale uncertainties are the result of a 7-point scale variation around the central scale $\mu_0 = m_{hh}/2$,
with $\mu_{R,F}=c_{R,F}\,\mu_0$, where 
$c_R,c_F\in \{2,1,0.5\}$, except that the extreme variations $(c_R,c_F)=(2,0.5)$ and $(c_R,c_F)=(0.5,2)$
are omitted. 

Table~\ref{tab:sigmatot} also shows that the K-factors do vary substantially as functions of the trilinear coupling.
This fact is illustrated in Fig.~\ref{fig:Kfacvariation}, where it is demonstrated that the K-factor takes values between 1.57 and 2.16.

\begin{figure}[htb]
  \centering
    \includegraphics[width=0.4\textwidth]{\main/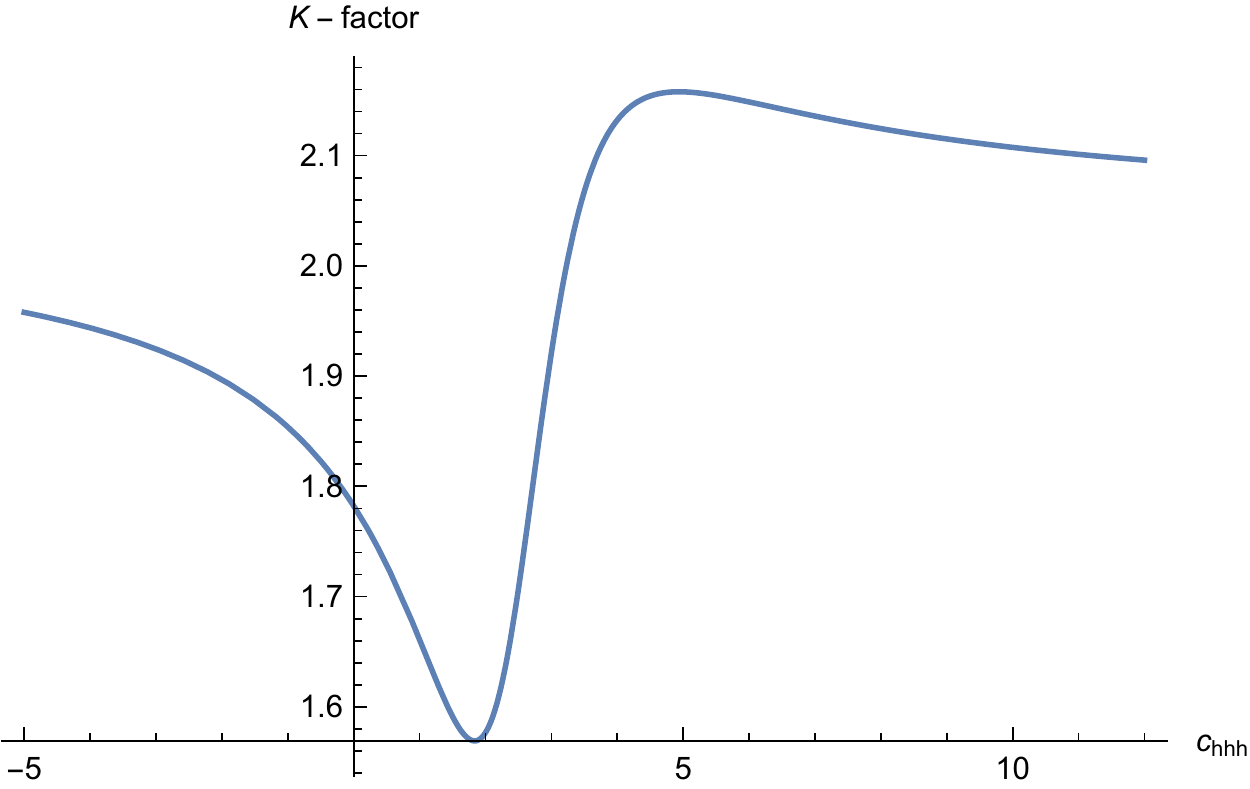}
\caption{Variation of the NLO K-factor with the trilinear coupling, $\sqrt{s}=14$\,\UTeV.}
\label{fig:Kfacvariation}
\end{figure}

\asubsubsubsection{Differential cross sections  at 14 \UTeV and 27 \UTeV}

In Figs.~\ref{fig:lambda_small} and ~\ref{fig:lambda_large} we show the $\mhh$ distribution for various values of $\lambda=\lambda_{\mathrm{BSM}}/\lambda_{\mathrm{SM}}$ at 14 \UTeV.
Figs.~\ref{fig:lambda_small_27} and ~\ref{fig:lambda_large_27} 
 show results for the $\mhh$ distribution at 27 \UTeV. The scale variation band for $\lambda=1$ is also included.
 Note that $\lambda=2.4$ is the value where the cross section as a function of $\lambda$ goes through a minimum, due to maximal destructive interference between diagrams containing the trilinear coupling and diagrams which do not contain Higgs boson self-couplings.
 
\begin{figure}[htb]
  \centering
  \subfloat[]{\label{fig:lambda_small_14}\includegraphics[width=0.6\textwidth]{\main/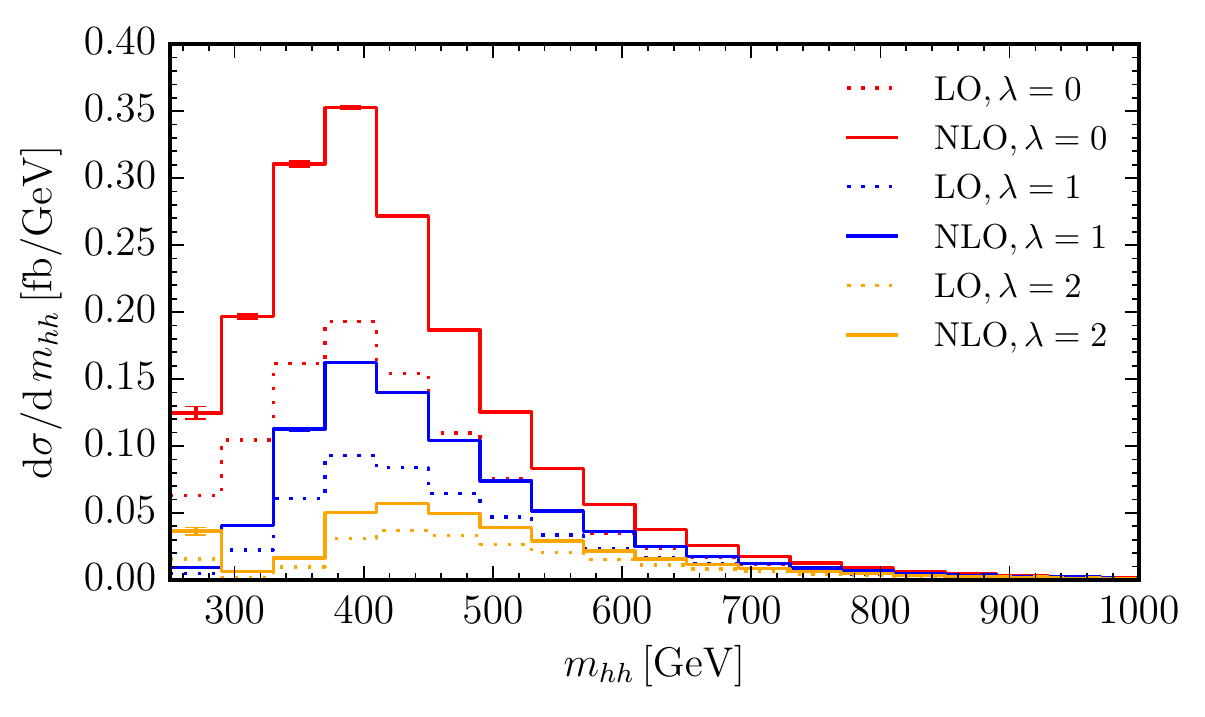}}
 \caption{Higgs boson pair invariant mass distributions for various values of $\lambda$ (relative to $\lambda_{\mathrm{SM}}$)  at 14\,\UTeV.}
\label{fig:lambda_small}
\end{figure}
\begin{figure}[htb]
  \centering
    \includegraphics[width=0.6\textwidth]{\main/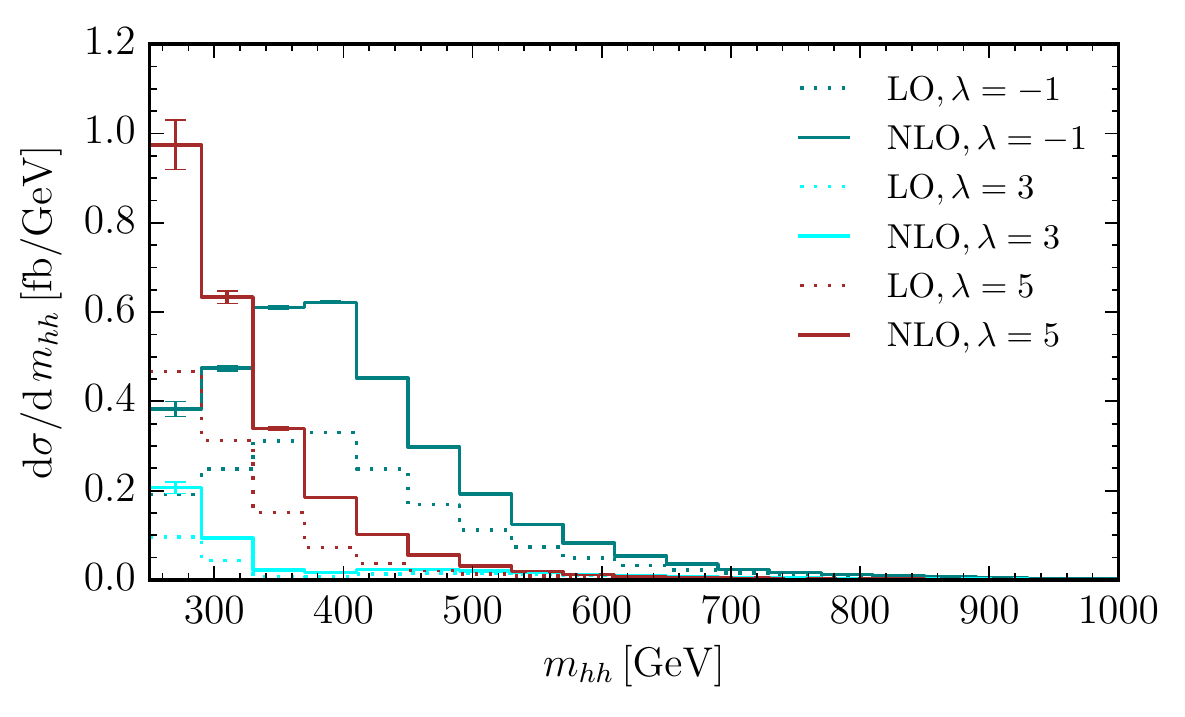}
\caption{Higgs boson pair invariant mass distributions for $\lambda=\lambda_{\mathrm{BSM}}/\lambda_{\mathrm{SM}}=-1,3,5$  at 14\,\UTeV.}
\label{fig:lambda_large}
\end{figure}

\begin{figure}[htb]
  \centering
    \includegraphics[width=0.6\textwidth]{\main/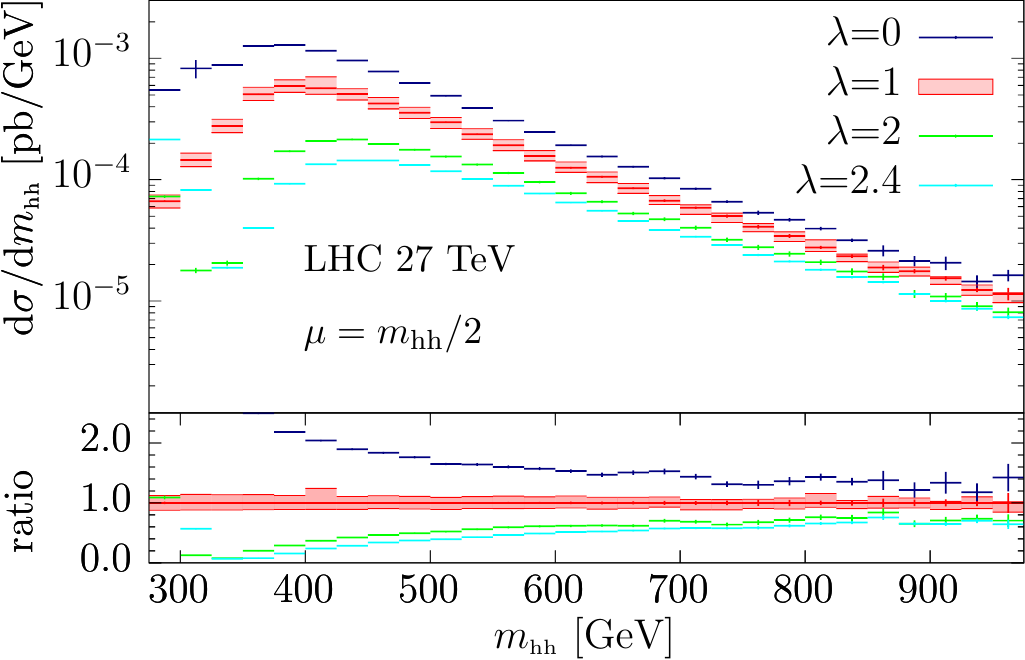}
\caption{Higgs boson pair invariant mass distributions for $\lambda=\lambda_{\mathrm{BSM}}/\lambda_{\mathrm{SM}}=0,1,2,2.4$  at 27\,\UTeV. The scale uncertainties for the SM value of $\chhh$ are shown as a red band.}
\label{fig:lambda_small_27}
\end{figure}

\begin{figure}[htb]
  \centering
    \includegraphics[width=0.6\textwidth]{\main/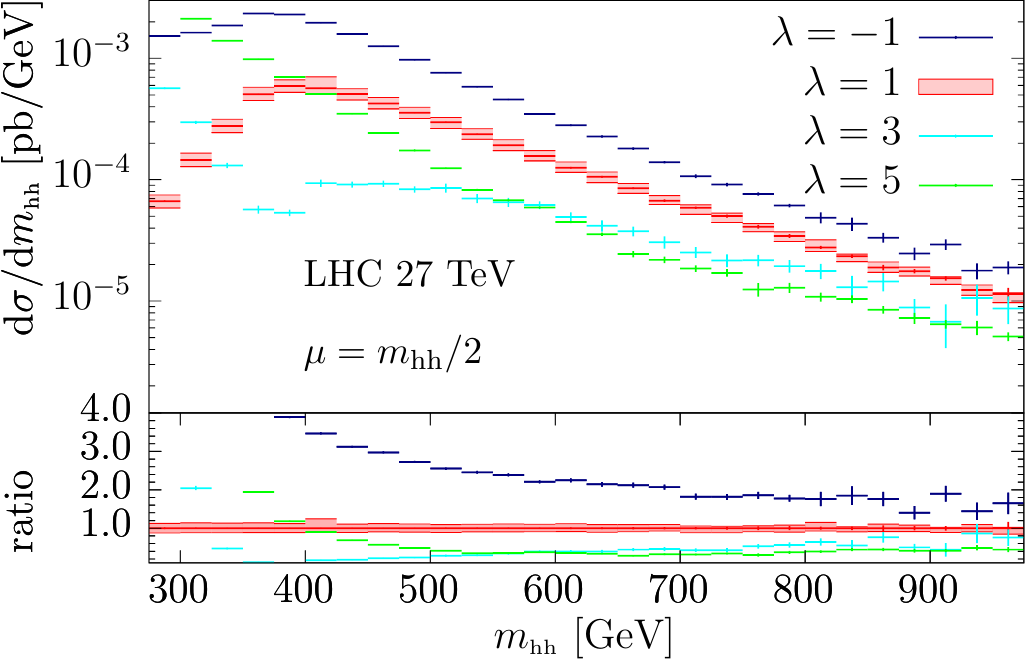}
\caption{Higgs boson pair invariant mass distributions for $\lambda=\lambda_{\mathrm{BSM}}/\lambda_{\mathrm{SM}}=-1,1,3,5$  at 27\,\UTeV. The scale uncertainties for the SM value of $\lambda$ are shown as a red band.}
\label{fig:lambda_large_27}
\end{figure}

\begin{figure}[ht]
\begin{center}
  \includegraphics[width=0.45\textwidth]{\main/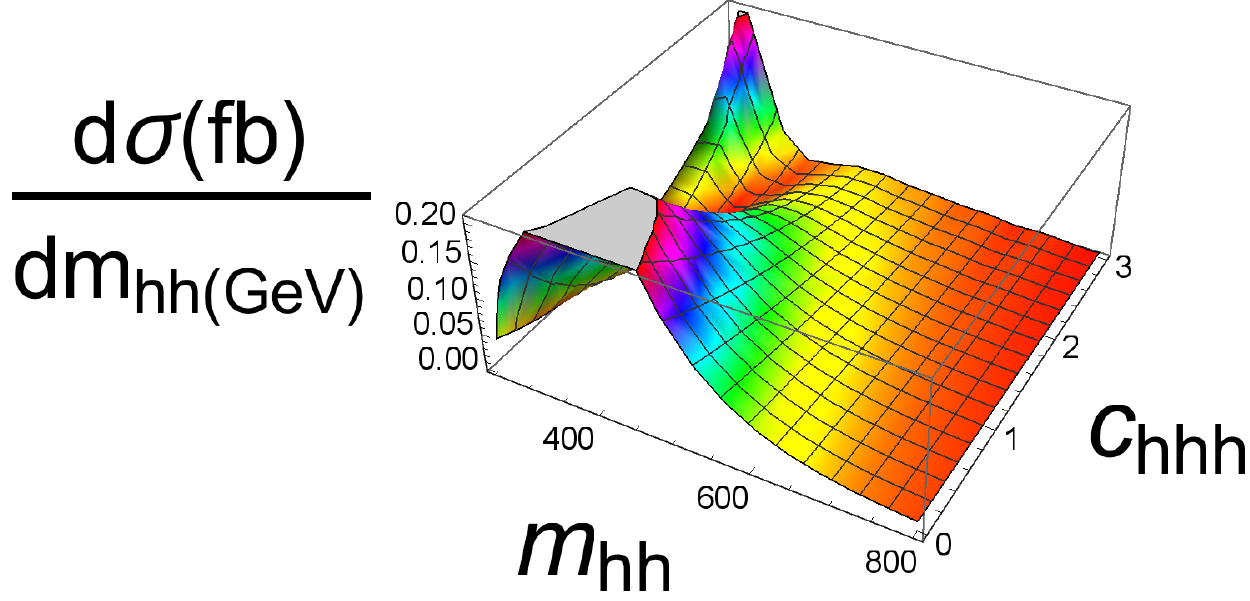}    
\end{center}
\caption{3-dimensional visualisation of the $\mhh$ distribution at
  14\,\UTeV, as a function  $\lambda$.}
\label{fig:chhh_3D}
\end{figure}
Fig.~\ref{fig:chhh_3D} shows the Higgs boson pair invariant mass
distributions at NLO as a function of $\lambda=\chhh$  as a 3-dimensional heat map.
The dip in the distribution around $\chhh=2.4$ is clearly visible.
 



\asubsubsection[G. Buchalla, A. Celis, M. Capozi, G. Heinrich,
  L. Scyboz]{Di-Higgs production in the non-linear EFT with full
  $m_t$-dependence at NLO QCD}

\asubsubsubsection{The Higgs sector in the non-linear EFT framework}
\label{sec:EWChL.double.h}

Below we will describe  the potential impact of physics
beyond the Standard Model through a non-linear Effective Field Theory,
also called the electroweak chiral Lagrangian
including a light Higgs 
boson~\cite{Buchalla:2013rka,Buchalla:2013eza,Buchalla:2017jlu}.
This framework provides us with a consistent EFT
for New Physics in the Higgs sector, where the Higgs field is an electroweak singlet $h$,
independent of the Goldstone matrix $U = \exp(2i\varphi^a T^a/v)$.
The latter transforms as $U\to g_L U g^\dagger_Y$ under the SM gauge group.
The symmetry is non-linearly realised on the Goldstone fields
$\varphi^a$, therefore the name non-linear EFT.
More details about this framework already have been given in
Section~\ref{sec2:theo_kappa_EFT}.
Therefore we restrict ourselves to stating the part of the Lagrangian
relevant for our study of anomalous Higgs couplings:
\begin{align}
{\cal L}\supset 
-m_t\left(c_t\frac{h}{v}+c_{tt}\frac{h^2}{v^2}\right)\,\bar{t}\,t -
c_{hhh} \frac{m_h^2}{2v} h^3+\frac{\alpha_s}{8\pi} \left( c_{ggh} \frac{h}{v}+
c_{gghh}\frac{h^2}{v^2}  \right)\, G^a_{\mu \nu} G^{a,\mu \nu}\;.
\label{eq:ewchl}
\end{align}
To lowest order in the SM $c_t=c_{hhh}=1$ and $c_{tt}=c_{ggh}=c_{gghh}=0$.
In general, all couplings may have arbitrary values of ${\cal O}(1)$.
Note that we have extracted a loop factor from the definition of the
Higgs-gluon couplings.

The leading-order diagrams are shown in Fig.~\ref{fig:hprocess}.
\begin{figure}[htb]
\begin{center}
\includegraphics[width=7.5cm]{\main/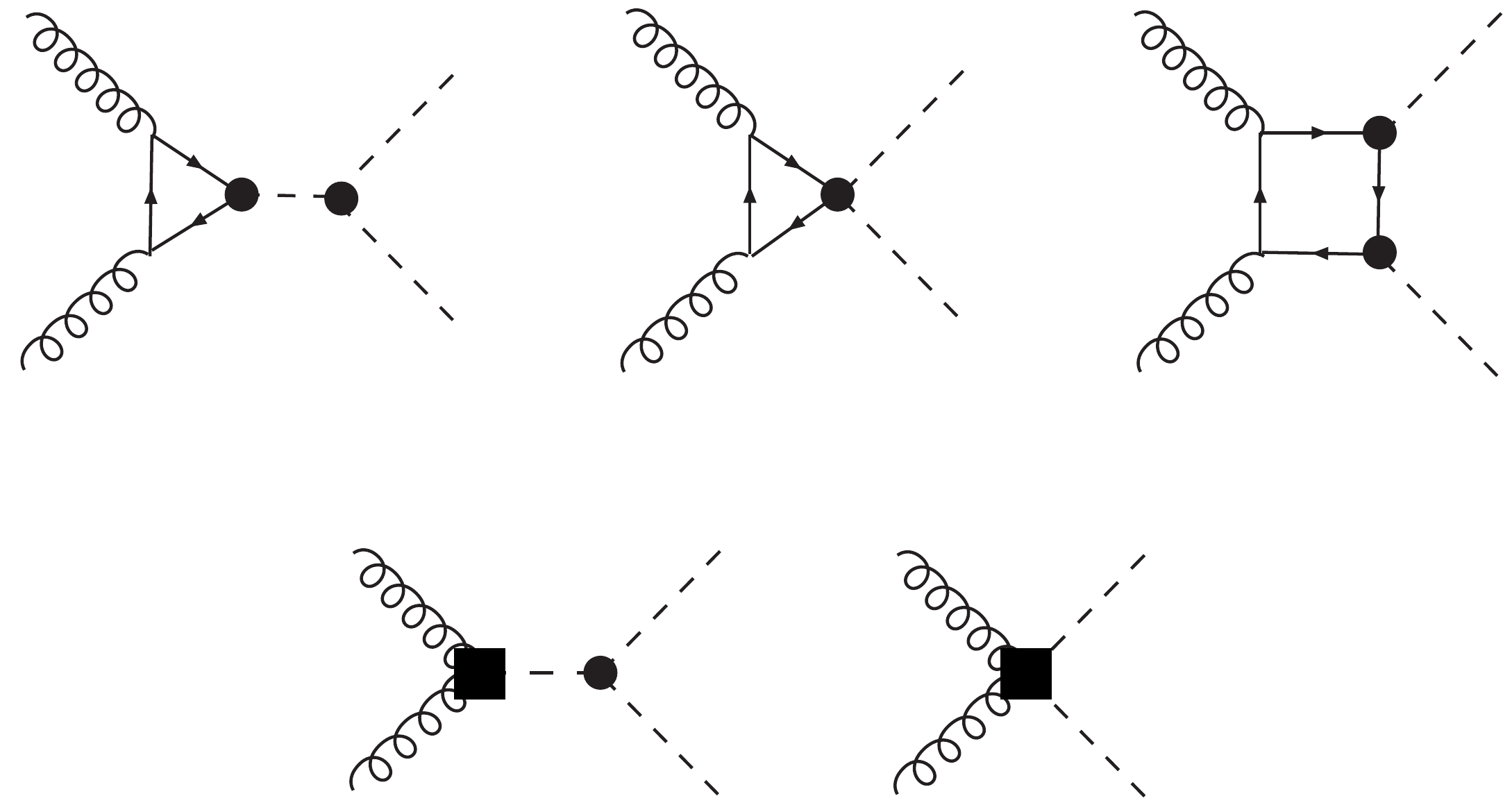}
\end{center}
\caption{Higgs-pair production in gluon fusion at leading order
in the non-linear EFT Lagrangian.}
\label{fig:hprocess}
\end{figure}
Examples for virtual diagrams at NLO are shown in
Fig.~\ref{fig:virt_examples}.
For further details we refer to Ref.~\cite{Buchalla:2018yce}.
\begin{figure}[htb]
\begin{center}
\includegraphics[width=6cm]{\main/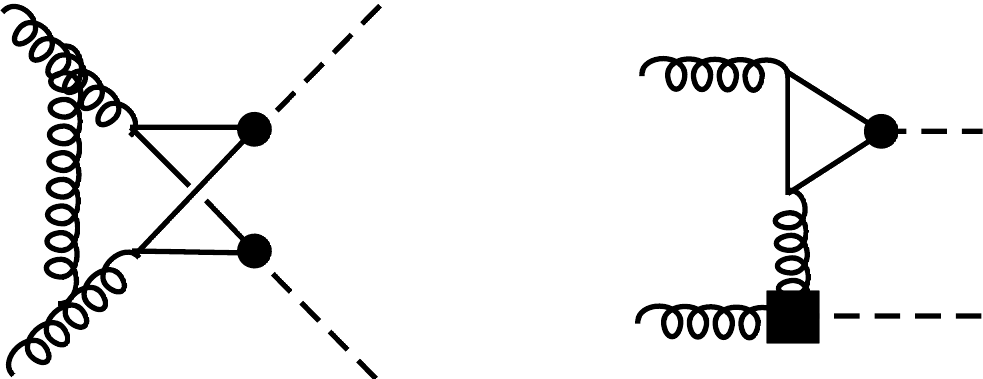}
\end{center}
\caption{Examples of virtual diagrams contributing at NLO QCD.}
\label{fig:virt_examples}
\end{figure}

\asubsubsubsection{Total cross sections for 14 and 27 \UTeV at some benchmark points}

In the following we will show results for some benchmark points, specified in Table~\ref{tab:benchmarks}, 
some of them having been first defined in Refs.~\cite{Carvalho:2015ttv}.
The results at 14\,\UTeV and 27\,\UTeV are given in Table~\ref{sigmatot}.
Note that our conventions for $\cg$ and $\cgg$ differ from the ones in
Ref.~\cite{Carvalho:2015ttv,Carvalho:2016rys}, the relations are $c_{ggh}= \frac{2}{3}c_g $
and $c_{gghh}=-\frac{1}{3}c_{2g}$, where $c_g,c_{2g}$ are the
couplings defined in Refs.~\cite{Carvalho:2015ttv,Carvalho:2016rys}.
We also take into account recent constraints on $\cg$ from
Refs.~\cite{CMS-PAS-HIG-17-031,deBlas:2018tjm}
and the  limits on the Higgs boson pair production cross section
from Refs~\cite{CMS-PAS-HIG-17-030,ATLAS-CONF-2018-043}.
This is why we do not show results for the original benchmark point 5
anymore, as its value for $\cg$ is outside the 2-sigma band of a
combined fit of $\cg,\ct$ from single Higgs production
data~\cite{CMS-PAS-HIG-17-031,deBlas:2018tjm}.
Benchmark point 6 is interesting because its value for $\chhh$ is near
the point where maximal destructive interference takes place between
triangle-type and box-type contributions if the other couplings are SM-like, leading to a total cross
section which is below the SM value.

\begin{table}[htb]
\begin{center}
\begin{tabular}{| c | c  c  c  c  c |}
\hline
Benchmark & $c_{hhh}$ & $c_t$ & $c_{tt}$ & $c_{ggh}$ & $c_{gghh}$ \\
\hline
5a & 1 & 1 & 0 &   $2/15$ & $4/15$\\
\hline
6 & 2.4 & 1 & 0 & $2/15$ & $1/15$  \\
\hline
7 & 5 & 1 & 0 & $2/15$ & $1/15$  \\
\hline
8a & 1 & 1 & 1/2 & $4/15$ & 0\\
\hline
SM & 1 & 1 & 0 & 0 & 0 \\
\hline
\end{tabular}
\end{center}
\caption{Benchmark points used for the distributions shown below.\label{tab:benchmarks}}
\end{table}
\begin{table}[htb]
\begin{center}
\begin{tabular}{|c|c|c|c|c|c|}
\hline
&&&&&\\
Benchmark & $\sigma_{NLO}$ [fb] & K-factor & scale uncert.  &
stat. uncert.   &$\frac{\sigma_{NLO}}{\sigma_{NLO,SM}}$ \\ 
&&&[\%] &[\%] &\\
\hline
$    B_{5a}$ [14 \UTeV] & 38.64 & 1.78& ${+4, \; -12}$ &0.24  &1.17  \\ 
$    B_{5a}$ [27 \UTeV] & 198.64 & 1.75 & ${+2, \; -10}$ & 0.43 & 1.56  \\ 
\hline
$    B_{6}$ [14 \UTeV] &  24.69 & 1.89 & ${+2, \; -11}$ & 2.1 & 0.75 \\ 
$    B_{6}$ [27 \UTeV] &  97.25 &  1.58& ${+1, \; -6}$ &1.6  & 0.76  \\ 
\hline
$   B_7$ [14 \UTeV] & 169.41 & 2.07& ${+9, \; -12}$ & 2.2 & 5.14 \\
$   B_7$ [27 \UTeV] & 598.20 & 2.11 & ${+8, \; -10}$ & 2.0 &  4.68\\
\hline 
$   B_{8a}$ [14 \UTeV] & 41.70 & 2.34& ${+6, \; -9}$ & 0.63 & 1.27 \\
$   B_{8a}$ [27 \UTeV] & 179.52 & 2.33 & ${+4, \; -7}$ & 0.49 & 1.40 \\
\hline 
$   SM $ [14 \UTeV] & 32.95 & 1.66 & ${+14, \; -13}$ & 0.1 & 1\\
$   SM $ [27 \UTeV] & 127.7&  1.62 & ${+12, \; -10}$ & 0.1 & 1\\
\hline
\end{tabular}
\end{center}
\caption{Total cross sections at 14 and 27 \UTeV at NLO (2nd column),
K-factor $\sigma_{NLO}/\sigma_{LO}$ (3rd column),
scale uncertainty (4th column), statistical uncertainty (5th column)
and the ratio to the SM total cross section at NLO (6th column).\label{sigmatot}}
\end{table}
%
Table~\ref{sigmatot} shows that the total cross sections increase by a factor of 3.5-5 when increasing the centre-of-mass energy from 14\,\UTeV to 27\,\UTeV. The increase for $B_{5a}$ is largest because of the large value of $\cgg$, which yields a contribution growing linearly with energy.


\asubsubsubsection{HH  invariant mass distributions at 14 and 27 \UTeV at some benchmark points}

In Figs.~\ref{fig:benchmark7} and \ref{fig:benchmark8a} we show Higgs boson pair invariant mass distributions for the benchmark points 7 and 8a. 
For both of them the shape of the distribution is very different from the SM one, and the K-factor is non-homogeneous over the whole $\mhh$-range. Benchmark point 7 is characterised by a large enhancement of the low $\mhh$ region, induced by the large value of $\chhh$. The lower ratio plot shows the ratio of the two approximations ``Born-improved HEFT" and ``FT$_{\rm{approx}}$" to the full NLO, where the former denotes the $m_t\to\infty$ limit rescaled by the $m_t$-dependent LO, while FT$_{\rm{approx}}$ includes the Born-improved $m_t\to\infty$ limit for the virtual part and the full $m_t$-dependence for the real radiation part. One can see from Fig.~\ref{fig:benchmark7} that these approximations are off by about 20\% even below the $2m_t$ threshold. Therefore one cannot claim that the $m_t\to \infty$ limit works well in the region below $\sim 400$\,\UGeV. As the triangle-type contributions are dominating for $\chhh=5$, their full $m_t$-dependence plays a significant role. 

Benchmark point 8a shows a characteristic dip near $\mhh=2\,m_t$ and an enhancement in the tail compared to the SM. 
As the total cross section for $B_{8a}$ is very similar to the SM one, both at 14\,\UTeV and at 27\,\UTeV, this is an example where the discriminating power of differential information is very important.

\begin{figure}[htb]
\includegraphics[width=0.48\textwidth]{\main/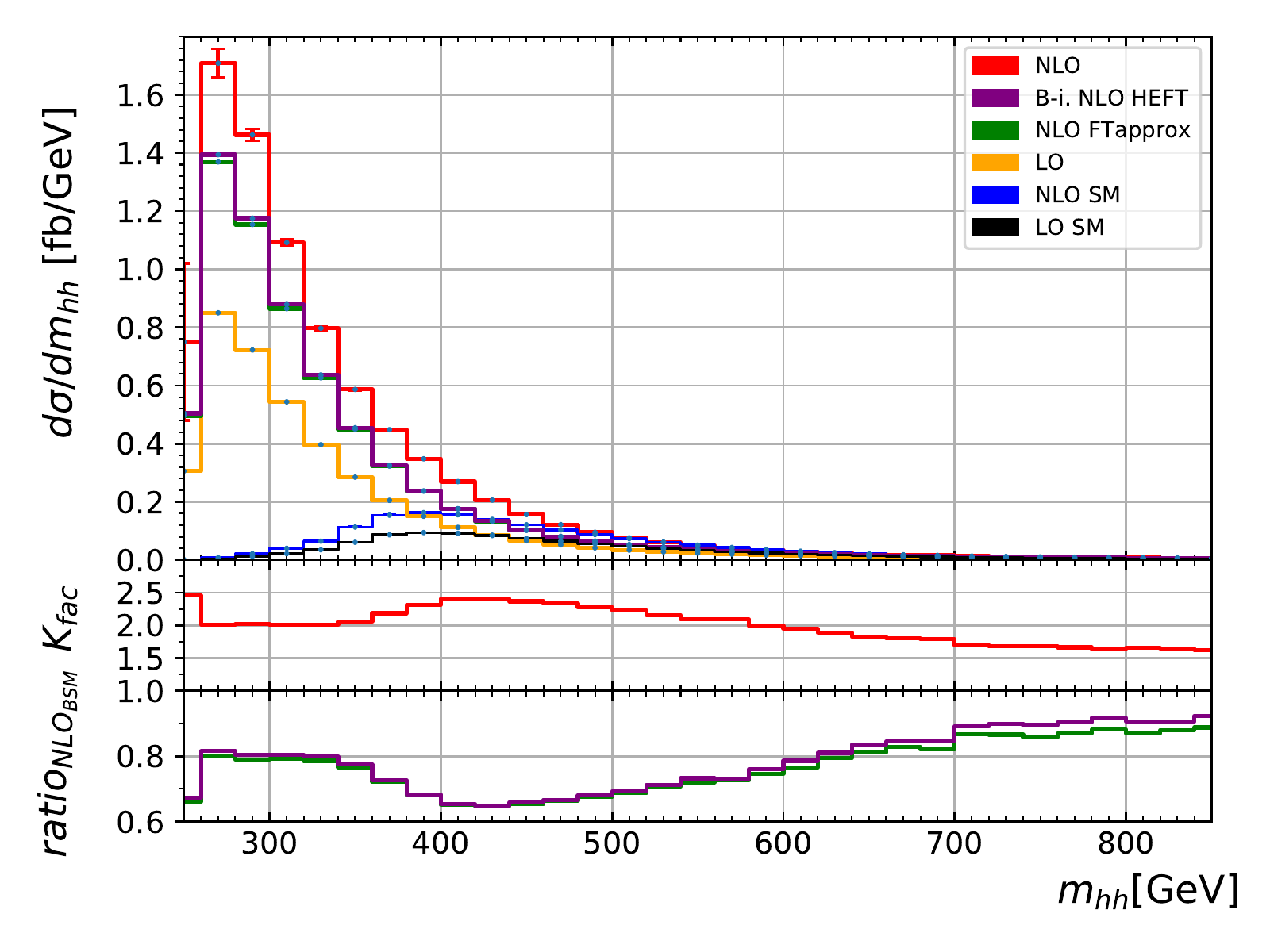}
~
 \includegraphics[width=0.48\textwidth]{\main/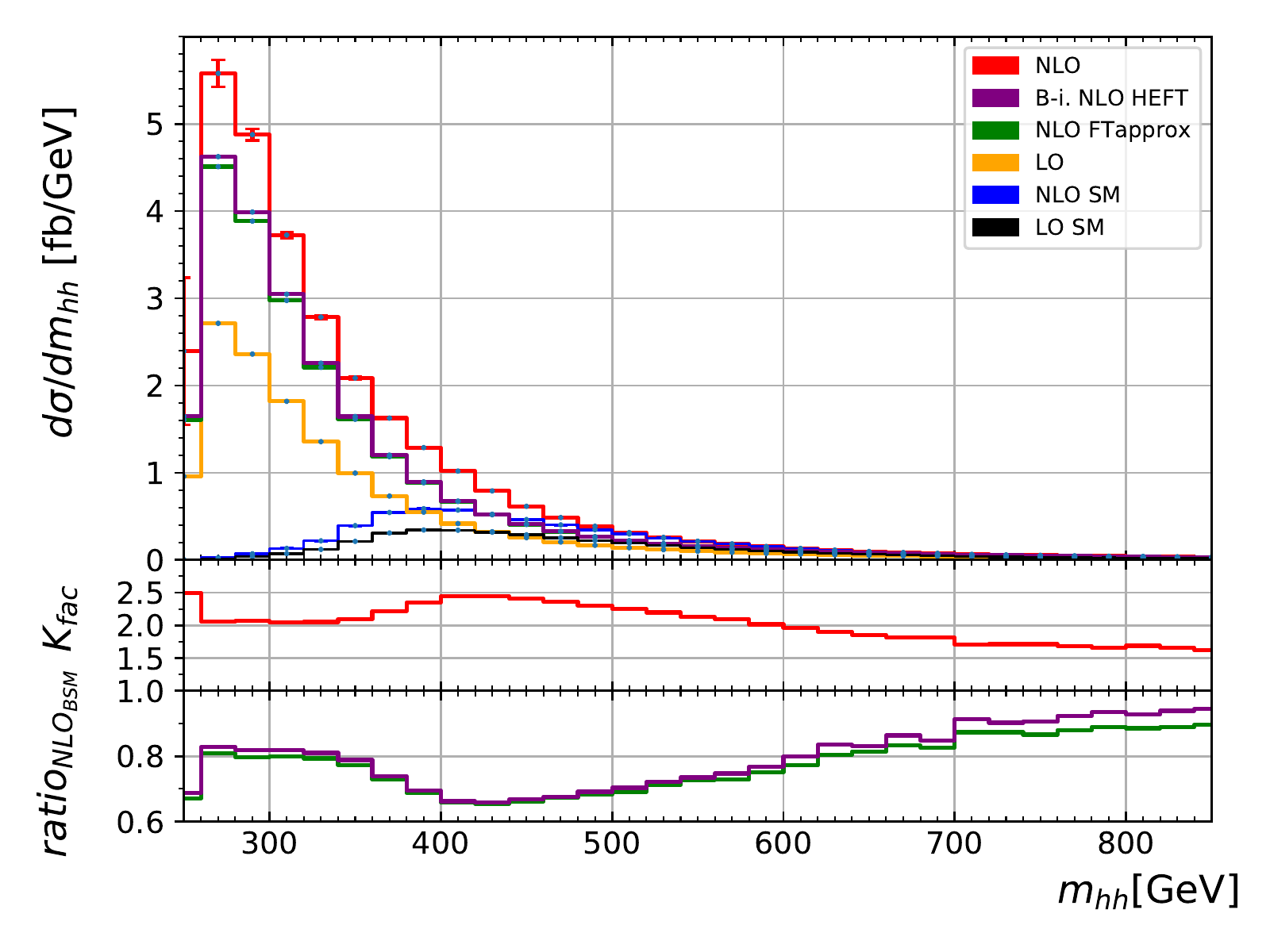}
\caption{Higgs boson pair invariant mass distributions  for benchmark point 7, $\chhh=5,\ct= 1, \ctt=0, \cg=2/15,\cgg=1/15$, at 14\,\UTeV (left) and 27\,\UTeV (right).}
\label{fig:benchmark7}
\end{figure}
\begin{figure}[htb]
  \centering
  \includegraphics[width=0.48\textwidth]{\main/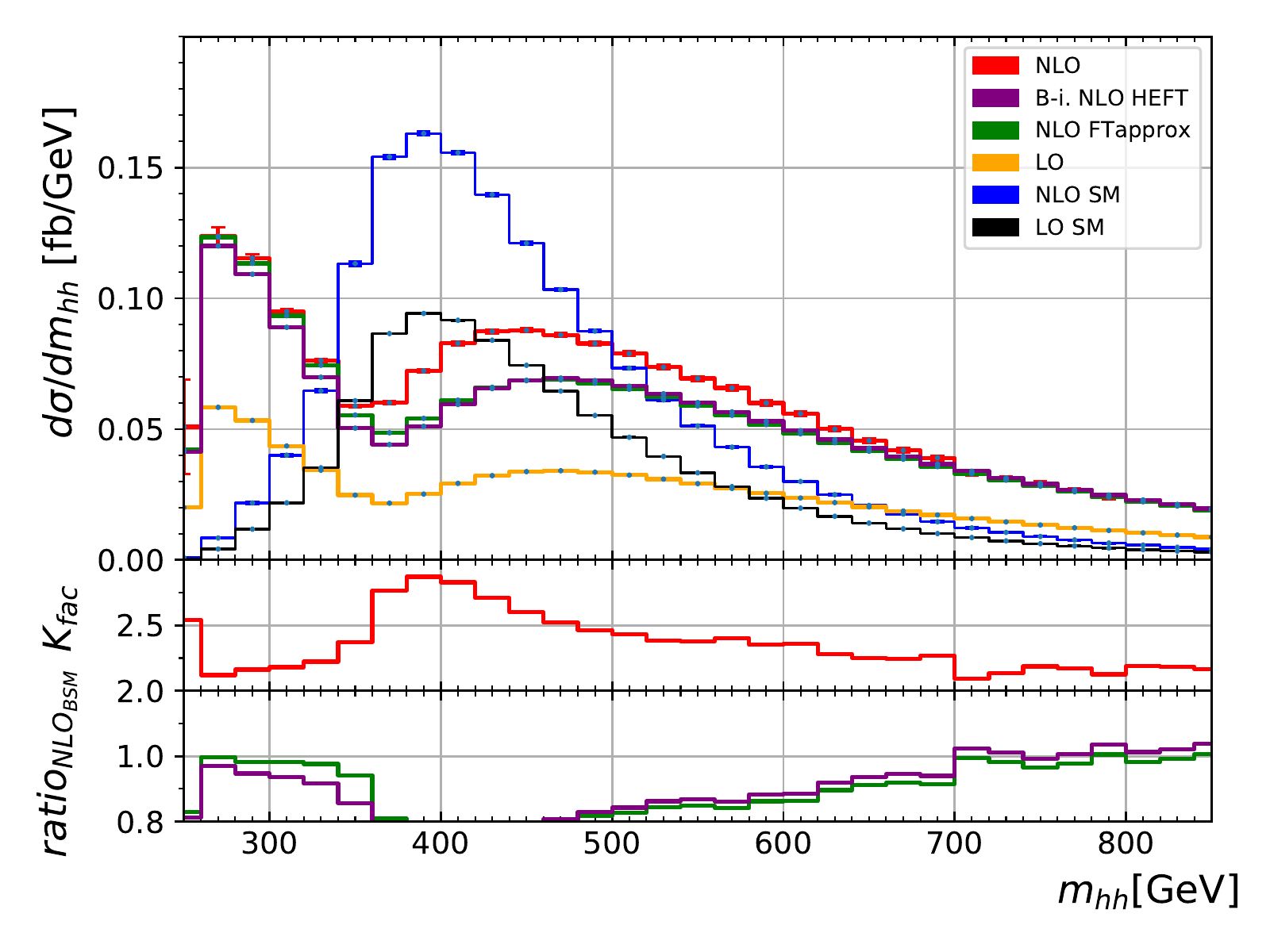} 
~
\includegraphics[width=0.48\textwidth]{\main/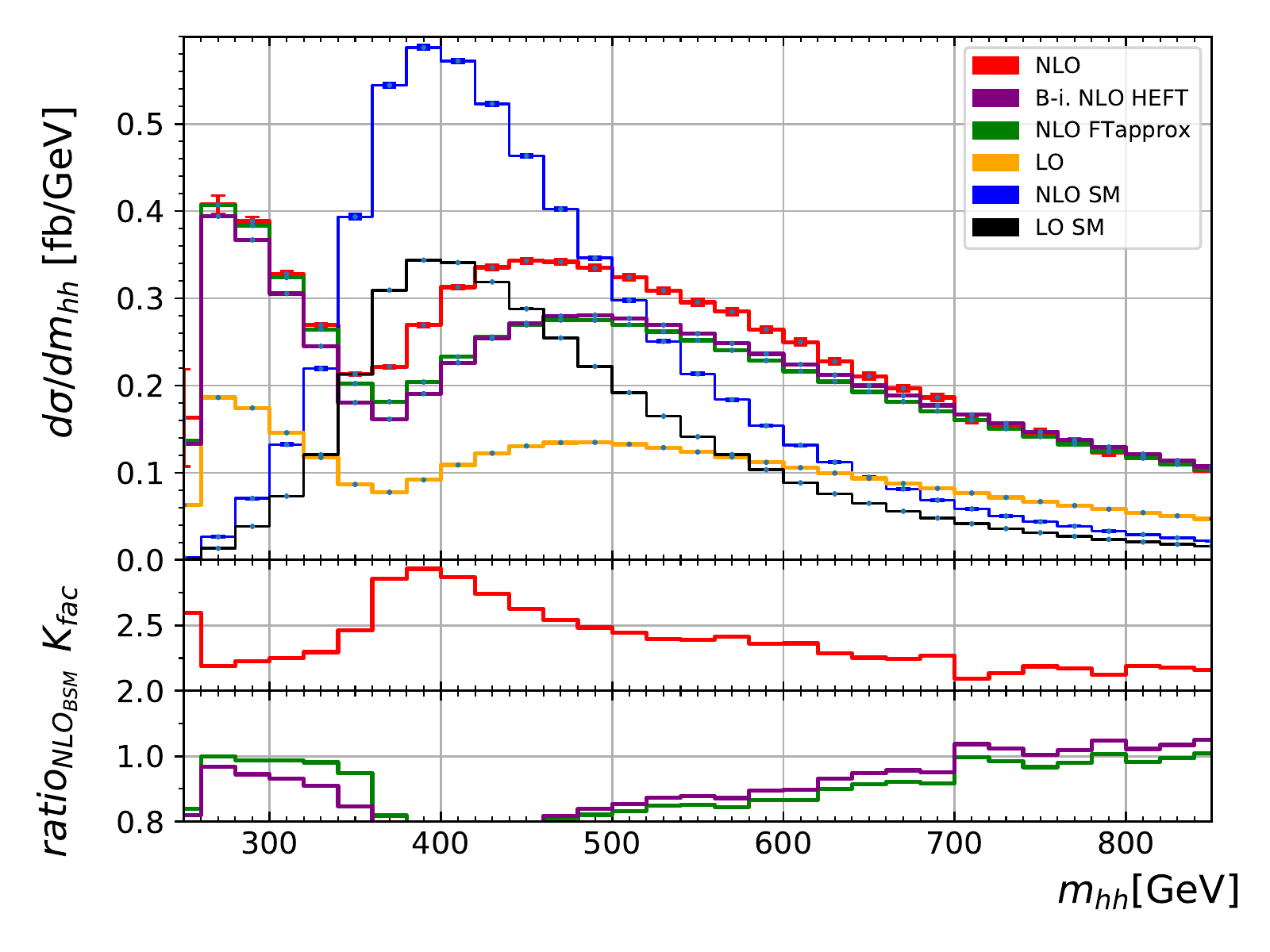}
\caption{Higgs boson pair invariant mass distributions  for benchmark point 8a, $\chhh=1,\ct= 1, \ctt=0.5, \cg=4/15,\cgg=0$, at 14\,\UTeV (left) and 27\,\UTeV (right).}
\label{fig:benchmark8a}
\end{figure}

\asubsubsubsection{Characterising the BSM parameter space}

The total cross section can be written in terms of the 15 coefficients 
$A_1, \ldots, A_{15}$, at LO~\cite{Carvalho:2015ttv,Azatov:2015oxa} and in terms of 23 coefficients at NLO~\cite{Buchalla:2018yce}.
\begin{align}
\label{eq:Acoeffs_all}
&\sigma^{\rm{NLO}}/\sigma^{\rm{NLO}}_{SM}  =\nn\\
& \quad  A_1\, c_t^4 + A_2 \, c_{tt}^2  + A_3\,  c_t^2 \chhh^2  + 
A_4 \, \cg^2 \chhh^2  + A_5\,  \cgg^2  + 
A_6\, c_{tt} c_t^2 + A_7\,  c_t^3 \chhh \nn\\
& + A_8\,  c_{tt} c_t\, \chhh  + A_9\, c_{tt} \cg \chhh + A_{10}\, c_{tt} \cgg + 
A_{11}\,  c_t^2 \cg \chhh + A_{12}\, c_t^2 \cgg \nn\\
& + A_{13}\, c_t \chhh^2 \cg  + A_{14}\, c_t \chhh \cgg +
A_{15}\, \cg \chhh \cgg \nn\\ 
& + A_{16}\, c^3_t \cg + A_{17}\,  c_t c_{tt} \cg 
+ A_{18}\, c_t \cg^2 \chhh + A_{19}\, c_t \cg \cgg 
\nn\\
&+ A_{20}\,  c_t^2 \cg^2 + A_{21}\, c_{tt} \cg^2 
+ A_{22}\, \cg^3 \chhh + A_{23}\, \cg^2 \cgg\,.
\end{align}
Based on our results for $A_1,\ldots, A_{23}$, we produce heat maps for the ratio $\sigma/\sigma_{SM}$, 
varying two of the five parameters, while for the fixed parameters the SM values are used, along with
$\sigma_{SM}^{\rm{LO}}[14\,\mathrm{\UTeV}]=19.85$\,fb,
$\sigma_{SM}^{\rm{NLO}}[14\,\mathrm{\UTeV}]=32.95$\,fb.
The couplings are varied in a range which seems reasonable when taking into account the current constraints on the 
Higgs coupling measurements
as well as recent limits on the di-Higgs production cross
section~\cite{CMS-PAS-HIG-17-030,ATLAS-CONF-2018-043}.

\begin{figure}[ht]
\begin{center}
\includegraphics[width=0.45\textwidth]{\main/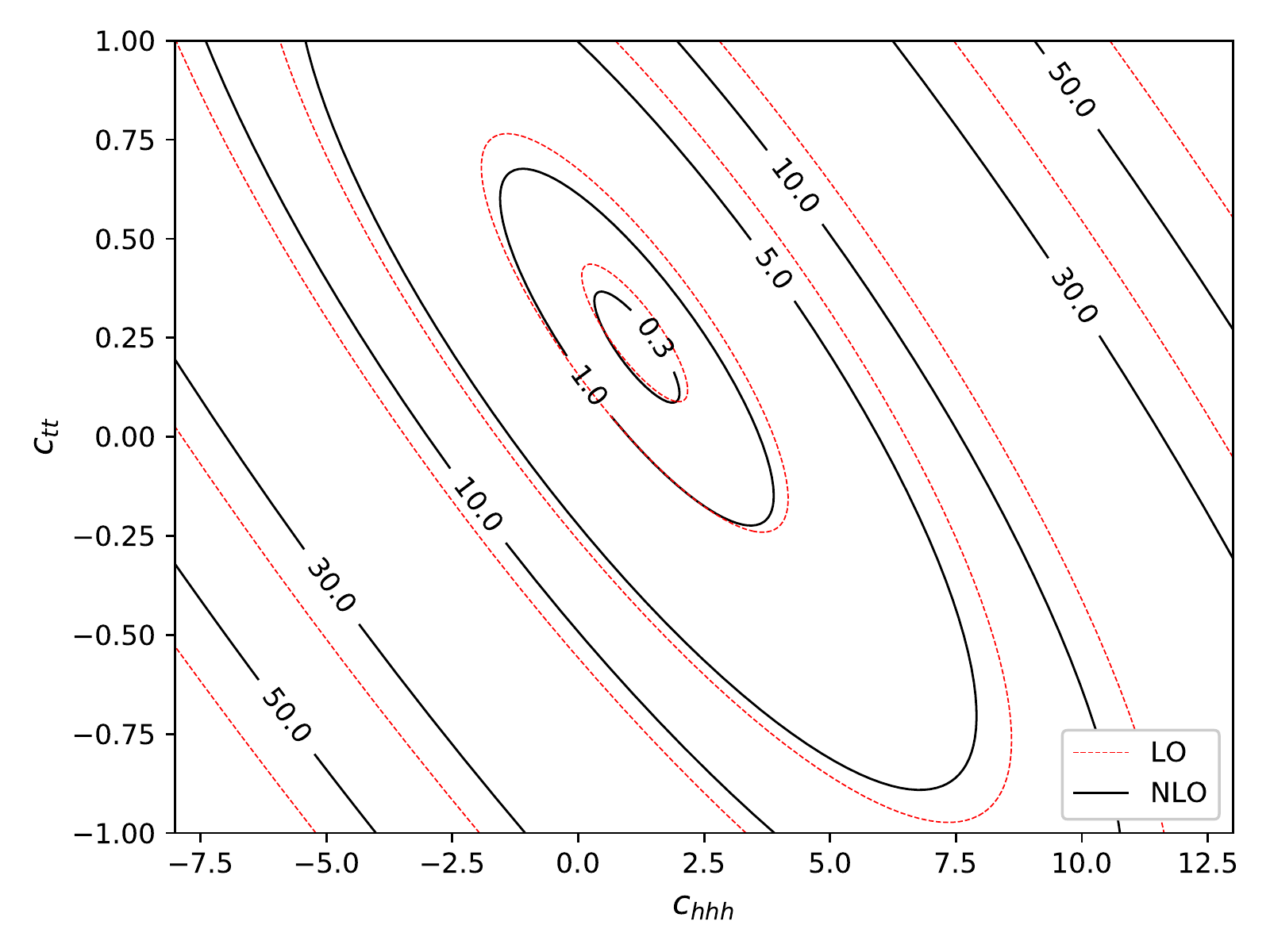}    
~
\includegraphics[width=0.45\textwidth]{\main/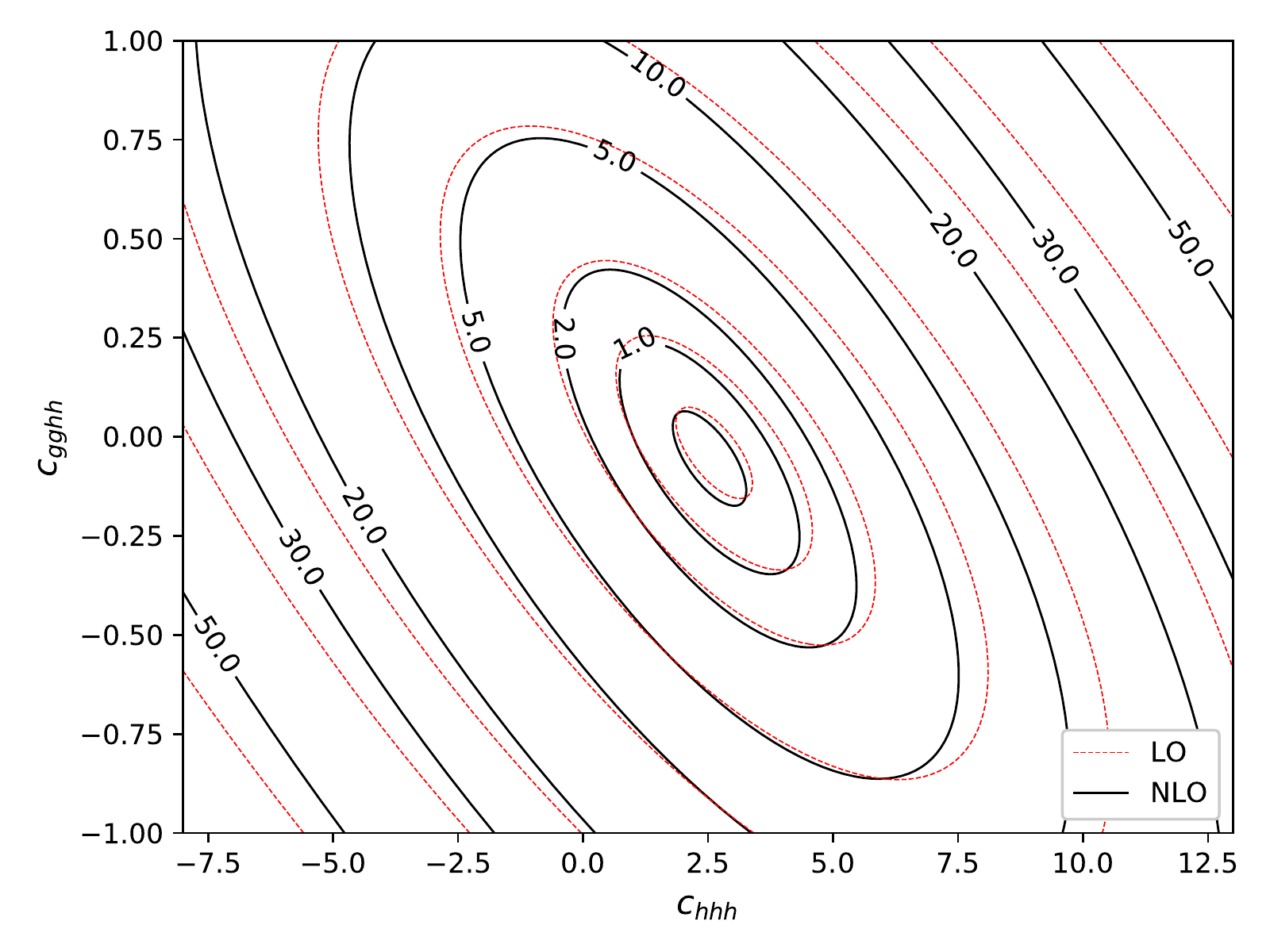}
\end{center}
\caption{Iso-contours of $\sigma/\sigma_{SM}$: (a) $\chhh$ versus $\ctt$ and (b) $\chhh$ versus $\cgg$  at $\sqrt{s}=14$\,\UTeV.}
\label{fig:chhh_ctt_cgg}
\end{figure}

Fig.~\ref{fig:chhh_ctt_cgg} shows variations of the triple Higgs coupling $\chhh$ in combination with $\ctt$ and $\cgg$ at $\sqrt{s}=14$\,\UTeV.
We observe that the deviations from the SM cross section can be substantial.
\begin{figure}[ht]
\begin{center}
\includegraphics[width=9cm]{\main/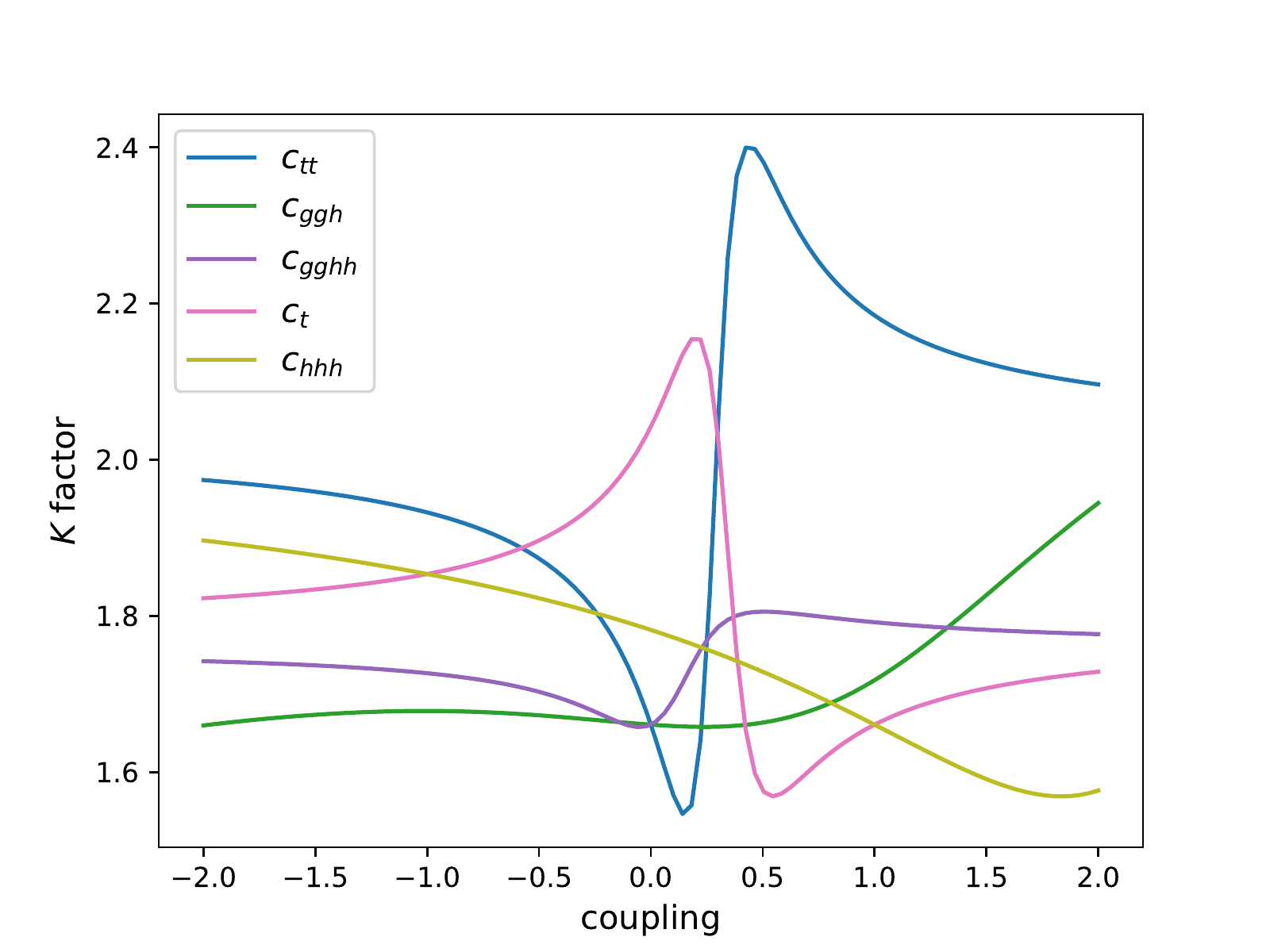}
\end{center}
\caption{K-factors for the total cross section at $\sqrt{s}=14$\,\UTeV as a function of the
  different couplings.}
\label{fig:project_ctt}
\end{figure}
In Fig.~\ref{fig:project_ctt} we show the K-factors as a function of
the coupling parameters, with the others fixed to their SM values. 
It shows that the K-factors exhibit a much stronger dependence on the
coupling parameters once the full top quark mass dependence is taken into account when compared to 
the results in the $m_t\to\infty$ limit~\cite{Grober:2015cwa,deFlorian:2017qfk}.

Fig.~\ref{fig:ctt_3D} shows the Higgs boson pair invariant mass
distributions as a function of (a) $\ctt$ and (b) $\cgg$ as 3-dimensional heat maps.
In case (a) the other couplings are fixed to their SM values.
We can see that large values of $|\ctt|$ lead to a substantial increase of the cross section, in particular at low $\mhh$ values.
\begin{figure}[ht]
\begin{center}
  \includegraphics[width=0.45\textwidth]{\main/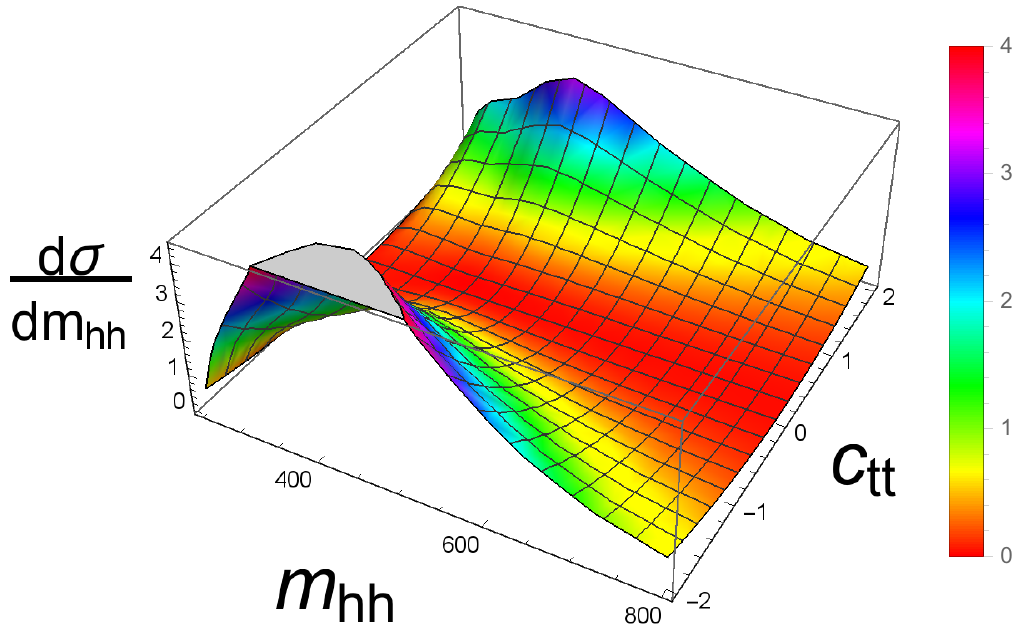}    
~
\includegraphics[width=0.5\textwidth]{\main/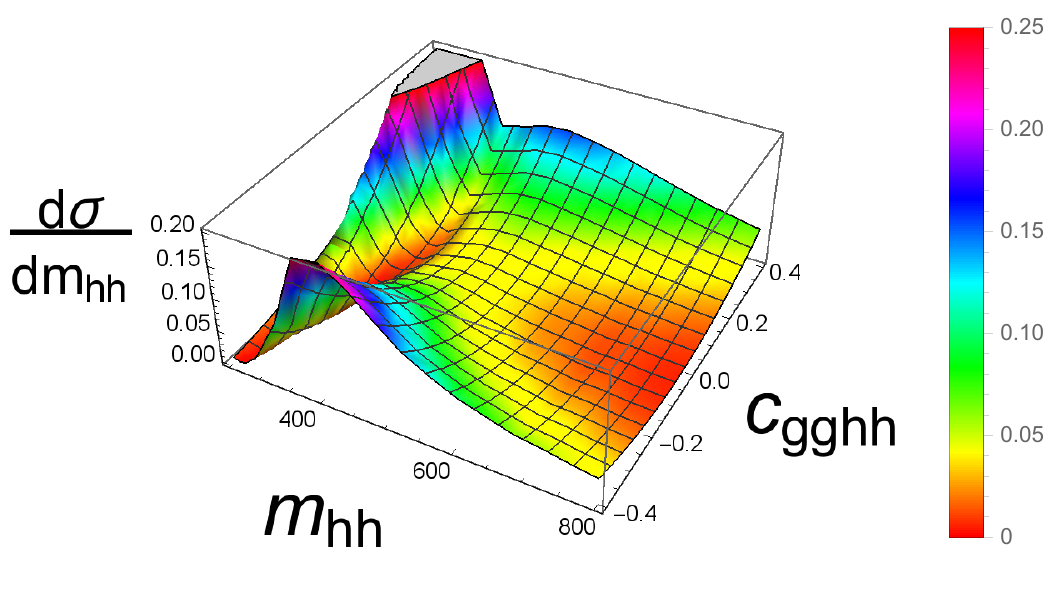}
\end{center}
\caption{3-dimensional visualisation of the $\mhh$ distribution (in units of fb/\UGeV) at 14\,\UTeV as a function of (a) $\ctt$  and (b) $\cgg$.  In case (a) all
other couplings are fixed to their SM values, in case (b) $\chhh= 2.4$.}
\label{fig:ctt_3D}
\end{figure}
In case (b) the other couplings are fixed to their SM values except for $\chhh$, which is fixed to $\chhh=2.4$
in order to demonstrate the following point: varying only $\chhh$, the $\mhh$ distribution shows a dip in the differential cross section just below $\mhh\sim 2m_t$ for $\chhh\sim 2.4$, while the low $\mhh$ region gets enhanced for larger values of $\chhh$, see Section~\ref{subsec:varylambda}.
 However, this pattern can get destroyed by non-zero
 Higgs-gluon contact interactions. While $\cg$ is increasingly well constrained meanwhile,
 $\cgg$ still could be relatively large. We can see from Fig.~\ref{fig:ctt_3D}(b) that the dip is not present for very low (negative) $\cgg$ values and also gets very shallow for values of $\cgg\sim 0.4$.
 Therefore it would be premature to conclude that a dip in the $\mhh$ distribution points to a value of $\chhh$ close to 2.4.

We also point out that the LO and NLO $A_i$ coefficients for both the total cross
section and the $\mhh$ distributions at both 14\,\UTeV and 27\,\UTeV are
available as ancillary files coming with Ref.~\cite{Buchalla:2018yce}. These data files allow to reconstruct the full NLO result for any point in the 5-dimensional parameter space.



\asubsection{Double Higgs measurements and trilinear coupling: experimental prospects}
\label{sec:HH_meas_exp}

The current Run 2 measurements of the Higgs-boson-pair production are performed with approximately 36~$fb^{-1}$ of collision data at a centre-of-mass energy of 13~\UTeV, combining different decay channels~\cite{ATLAS-CONF-2018-043, Sirunyan:2018two}. 
The ATLAS collaboration reports the combined observed (expected) limit on the non-resonant Higgs-boson-pair production cross-section of 6.7 (10.4) times the SM expectation. The ratio of the Higgs boson self-coupling to its SM expectation is observed (expected) to be constrained at 95\% CL to $-5.0<\kappa_{\lambda}<12.1$ ($-5.8<\kappa_{\lambda}<12.0$). 
The reported combined observed (expected) limit on the non-resonant Higgs-boson-pair production cross-section by the CMS collaboration is 22.2 (12.8) times the predicted Standard Model cross-section. The ratio of the Higgs boson self-coupling to its SM expectation is observed (expected) to be constrained at 95\% CL to $-11.8<\kappa_{\lambda}<18.8$ ($-7.1<\kappa_{\lambda}<13.6$).

Only the production of HH pairs through gluon-gluon fusion is considered (the other production mechanisms being more than an order magnitude smaller), with an expected cross-section of 36.69$^{+2.1\%}_{-4.9\%}$ fb at 14 \UTeV as described in Section~\ref{sec:HH_NNLO}. The state of the art NNLO/NNLL calculation with finite top mass effects included at NLO in QCD is used, for a Higgs boson mass of 125 \UGeV. Scale uncertainties are reported as superscript/subscript. The estimated top quark mass uncertainty of the NNLOFTapprox predictions is also computed, together with PDF and $\alpha_{S}$ uncertainties. PDF uncertainties are estimated within the Born-improved approximation. The calculation is performed in the on-shell top quark mass scheme.
The Feynman diagram which exhibits a \lHHH\ dependence interferes destructively with the box diagram that is independent of \lHHH, thus a small increase in the value of \lHHH decreases the expected HH production cross section, and modifies the distributions of event kinematics.

\asubsubsection[A. Bethani, A. Betti, P. Bokan, E. Carquin, M. Donadelli, A. Ferrari, K. Grimm, C. Gwilliam, M. Haacke, S. Lai, K. Leney, T. Lenz, S. Olivares Pino, E. Petit, N. Readioff, P. Sales De Bruin, J. Stark, F. Garay Walls, D. Wardrope, M. Wielers]{Measurements with the ATLAS experiment}
\label{sec:HH_ATLAS}

A direct measurement of the Higgs boson trilinear self-coupling \lHHH\ can be made via the study of Higgs boson pair production. 

The small SM non-resonant HH production cross section means that it is necessary to consider final states where at least one of the two Higgs bosons decays into a final state with a large branching ratio, ie $H \rightarrow b\bar{b}$. The most promising decays channels are $HH \rightarrow b\bar{b}b\bar{b}$, $HH \rightarrow b\bar{b}\tau\tau$ and $HH \rightarrow b\bar{b}\gamma\gamma$ with branching ratios of 33.9, 7.3 and 0.26\% respectively.

The expected performance for the $b\bar{b}b\bar{b}$ and $b\bar{b}\tau\tau$ channels is assessed through extrapolation of measurements performed by the ATLAS detector using 24.3~\fbinv\ and 36.1~\fbinv\ of data, respectively, obtained at $\sqrt{s} = 13$~\TeV\ during Run 2.
The expected performance for the $b\bar{b}\gamma\gamma$ channel is assessed through the use of truth-level Monte Carlo samples. 
These MC samples have been adjusted with parametrised functions to estimate the response of the upgraded ATLAS detector at the HL-LHC, assuming a mean pile-up rate <$\mu$> = 200. 
An 8\% improvement in b-tagging efficiency is expected for all channels as a result of improvements to the inner tracker~\cite{ITKPixelTDR}.
This improvement is factored into the $b\bar{b}b\bar{b}$ and $b\bar{b}\tau\tau$ extrapolations, and it is included in the smearing functions used in the $b\bar{b}\gamma\gamma$ analysis.

A short description of the analysis strategy and of the results is given here, and further details can be found in Ref.~\cite{ATLASHHPUBnote}. The systematic uncertainties used follow the common recommendations for HL-LHC studies~\cite{ATLAS_PERF_Note}.

\paragraph{The $HH \rightarrow b\bar{b}b\bar{b}$ channel}

Projections for this channel were made by extrapolating from the ATLAS Run 2 analysis of 24.3 \fbinv\ of 13~\TeV\ data, described in Ref.~\cite{ITKPixelTDR}. This extrapolation assumes similar detector performance to Run~2. Four central jets with transverse momentum higher than 40~\GeV\ are paired to construct the Higgs boson candidates. Additional requirements are made on Higgs boson mass and transverse momentum, and the pseudorapidity between the two Higgs boson candidates.
The acceptance times efficiency of the full event selection for the SM signal is of 1.6\%, and around 95\% of the background consists of multi-jet events. This dominant background is assessed using data-driven techniques. 

The largest source of systematic uncertainty comes from the ability to model the QCD multi-jet background using control regions in data. A conservative assumption in the extrapolation is made that the systematic uncertainties related to this background are left unchanged. Figure~\ref{fig:ATLAS_HH_4b_a} shows the impact of this uncertainty on the results.

\begin{figure}[!htb]
\centering 
\subfloat[]{\label{fig:ATLAS_HH_4b_a}\includegraphics[width=0.5\textwidth]{\main/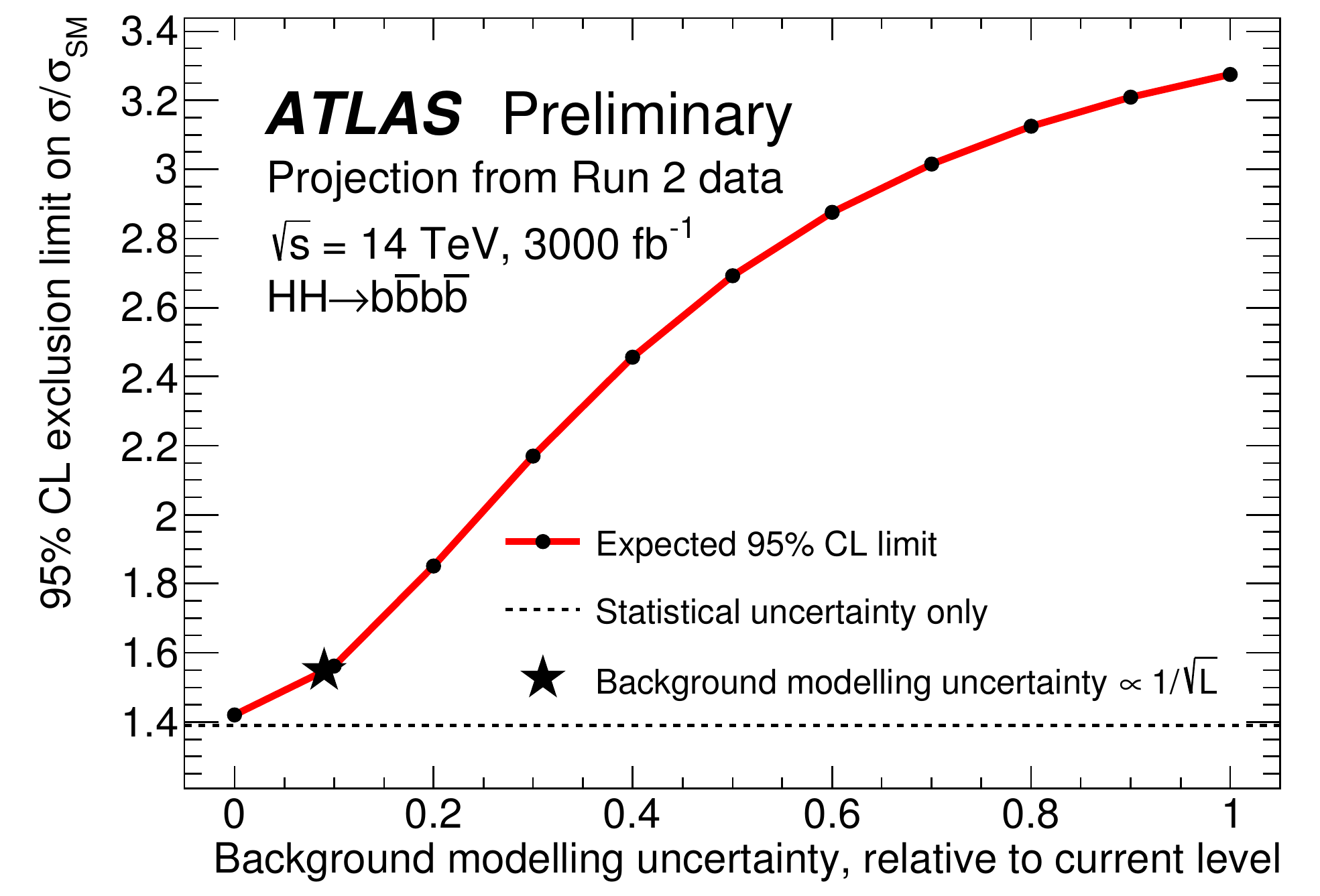}} 
\subfloat[]{\label{fig:ATLAS_HH_4b_b}\includegraphics[width=0.5\textwidth]{\main/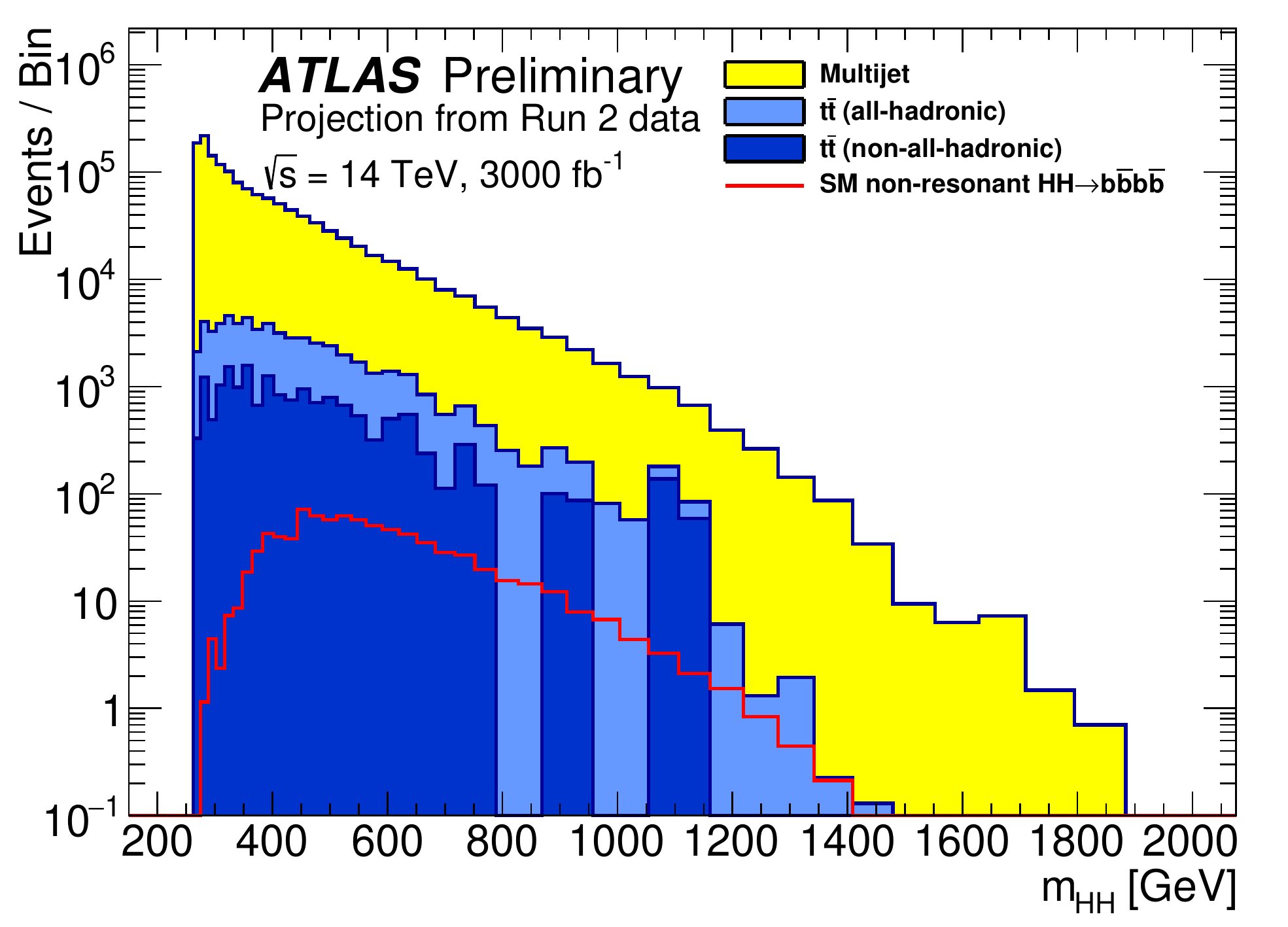}}
\caption{(a) Expected 95\% CL upper limit on $\sigma_{HH}/\sigma_{HH}^{SM}$, as a function of the pre-fit background modelling uncertainties, which are each scaled by a common, constant factor relative to the corresponding uncertainty in the Run~2 analysis (i.e. the uncertainties of the analysis of the 2016 dataset correspond to 1 here). 
The limit achievable assuming that the overall uncertainty scales with luminosity as $1/\sqrt L$ is shown by the star point.
The limit obtained when considering only statistical uncertainties is shown as the dashed line.
The extrapolated sensitivities are calculated assuming a jet \pt\ threshold of 40 \GeV.
(b) Stacked $m_{HH}$ histograms of the \ttbar\ and multi-jet backgrounds extrapolated from 24.3~\fbinv\ (the 2016 dataset) to 3000~\fbinv.
The predicted SM non-resonant Higgs boson pair production signal is shown as the red line.} 
\label{fig:ATLAS_HH_4b} 
\end{figure}

The final analysis discriminant, m$_{HH}$, showed in Figure~\ref{fig:ATLAS_HH_4b_b}, is the invariant mass of the selected four-jet system, after a correction based on the known Higgs boson mass. The significance neglecting the systematic uncertainties is 1.4 standard deviations, while it is 0.61 standard deviations when the current systematic uncertainties are included. 
The high number of pile-up events at the HL-LHC cause difficulties in maintaining high acceptance when triggering on multi-jet final states. Ref.~\cite{Collaboration:2285584} proposes a trigger menu which thresholds corresponding to asking for jets with a transverse energy higher than 75~\GeV. In the scenario without systematic uncertainties this would degrades the sensitivity by 50\% relative to the threshold used by default in this extrapolation.


%
\paragraph{The $HH \rightarrow b\bar{b}\tau\tau$ channel}

Results~\cite{ATLASHHPUBnote} for this channel are computed by extrapolating from the Run 2 analysis of 36.1~\fbinv\ of 13~\TeV\ data~\cite{ATLASrun2HHbbtautau}, which currently sets the world's strongest limit by a single channel on the di-Higgs production. 
The leptonic/hadronic and hadronic/hadronic decay modes of the $\tau$-lepton are considered, the first one being separated in two categories, depending on the trigger used. A multivariate analysis with a Boosted Decision Tree is performed to separate the signal from the background processes. The Run~2 BDT distributions are scaled to the integrated luminosity of 3000~\fbinv, taking into account the change of cross-section with the increased centre-of-mass energy. The binning of those BDT distributions is also redefined to take into account the increased number of events.
A profile-likelihood fit is applied to the BDT score distributions shown in Figures~\ref{fig:ATLAS_HH_bbtautau_a} to~\ref{fig:ATLAS_HH_bbtautau_c}.

\begin{figure}[!htb]
\centering 
\subfloat[$\tau_{lep}\tau_{had}$ channel passing a single-lepton trigger (SLT)]{\label{fig:ATLAS_HH_bbtautau_a}\includegraphics[width=0.5\textwidth]{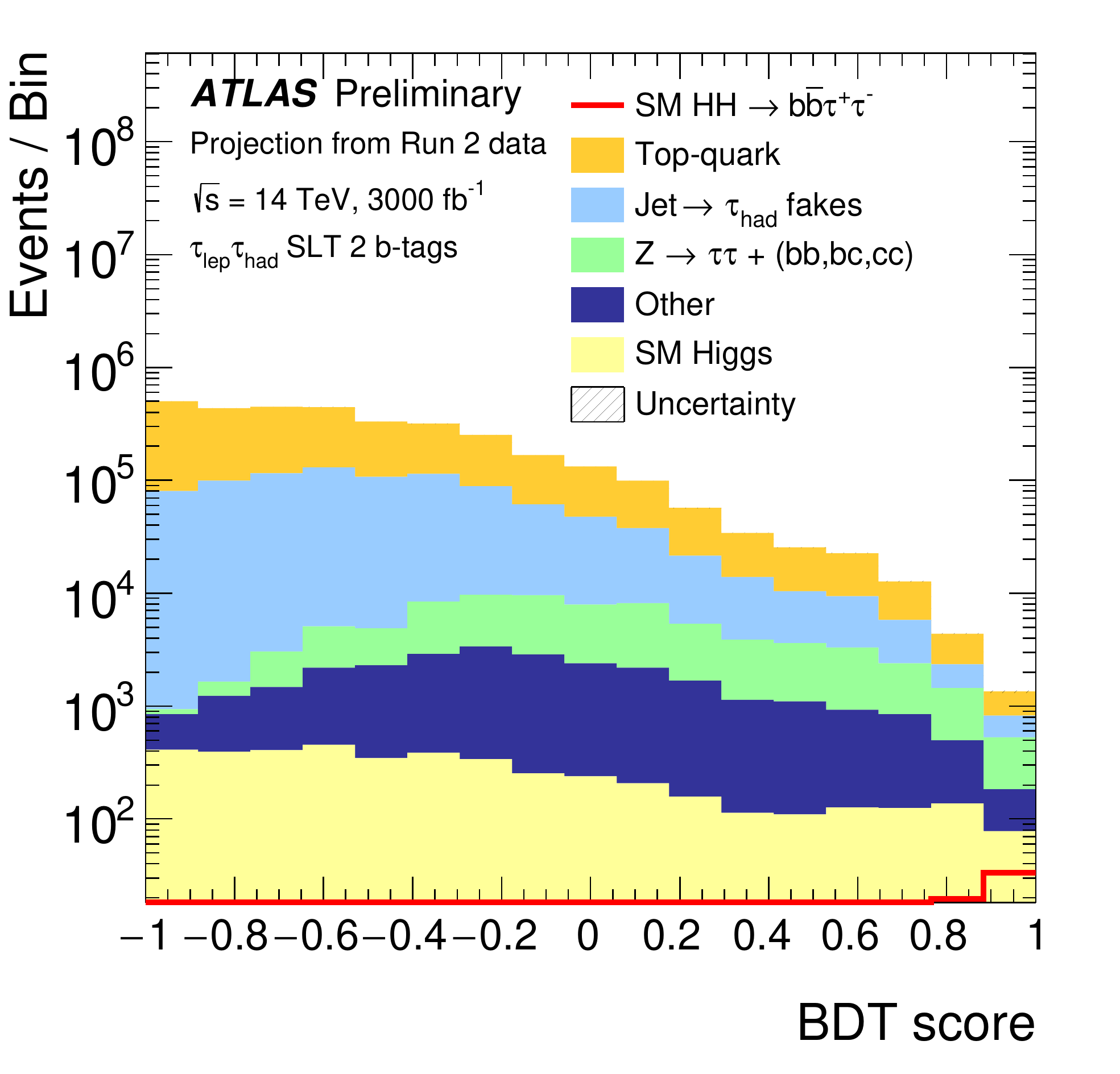}} 
\subfloat[$\tau_{lep}\tau_{had}$ channel passing a lepton-plus-$\tau_{had}$ trigger (LTT)]{\label{fig:ATLAS_HH_bbtautau_b}\includegraphics[width=0.5\textwidth]{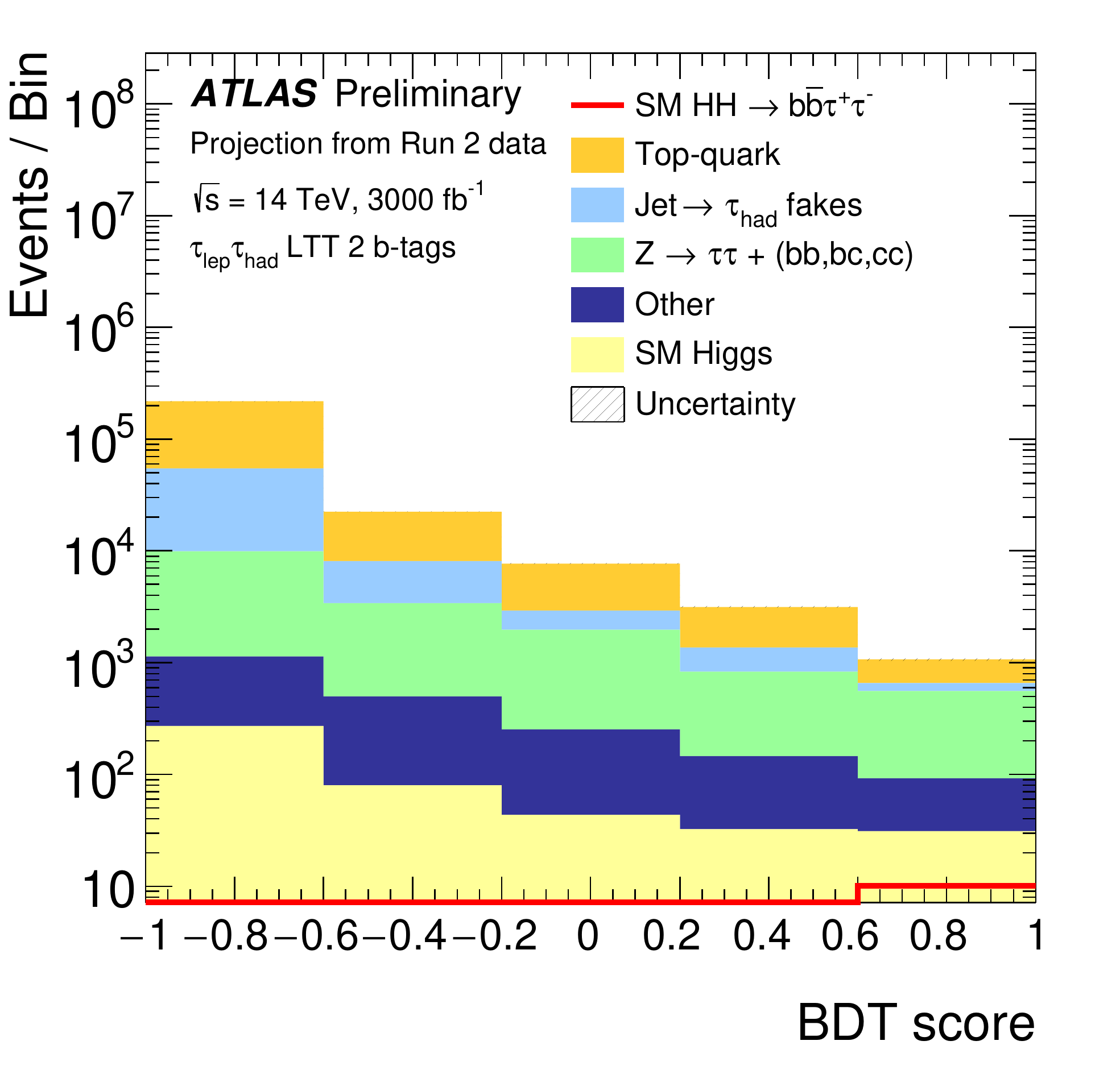}}\\
\subfloat[$\tau_{had}\tau_{had}$ channel]{\label{fig:ATLAS_HH_bbtautau_c}\includegraphics[width=0.5\textwidth]{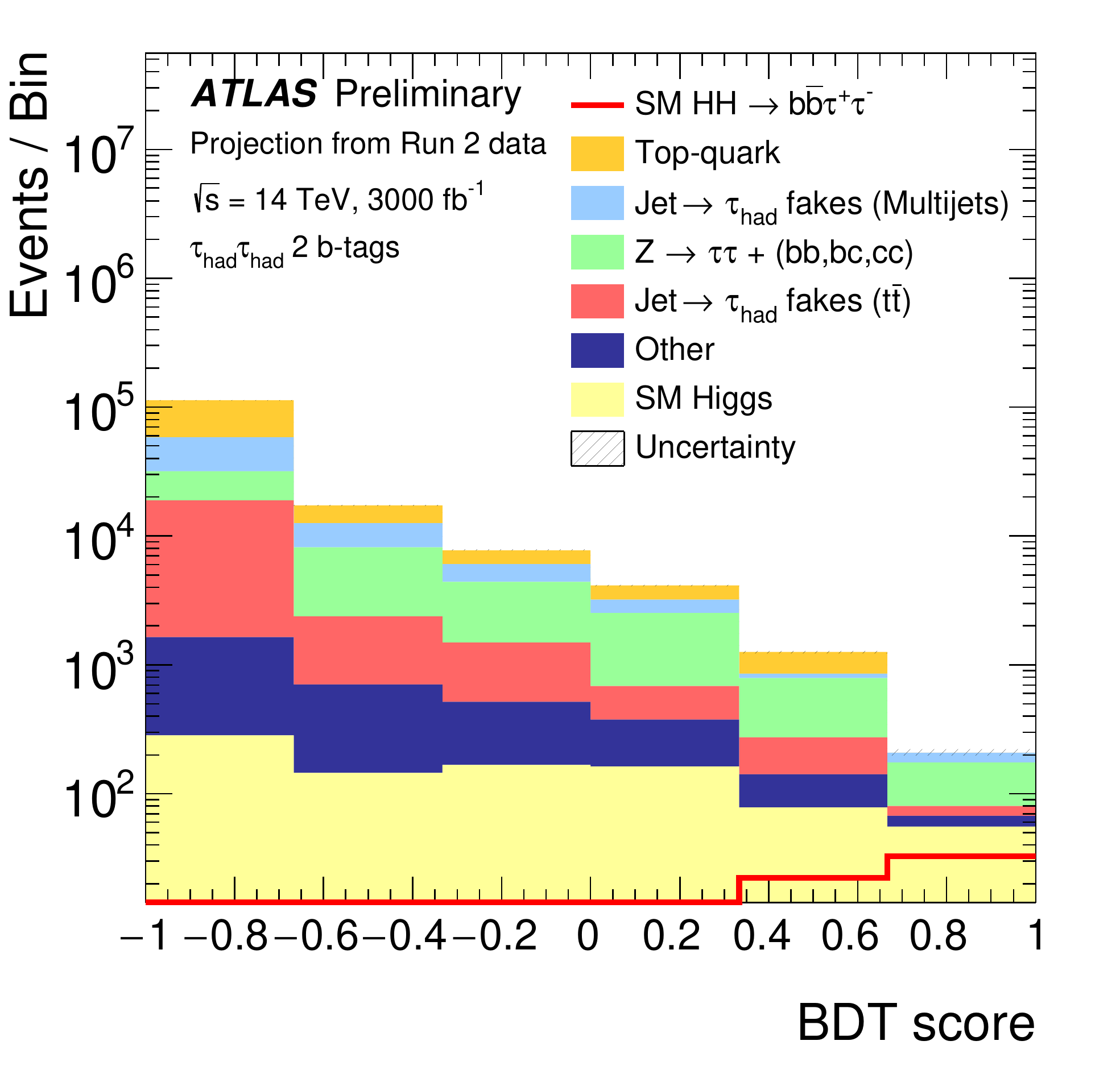}}
\subfloat[]{\label{fig:ATLAS_HH_bbtautau_d}\includegraphics[width=0.5\textwidth]{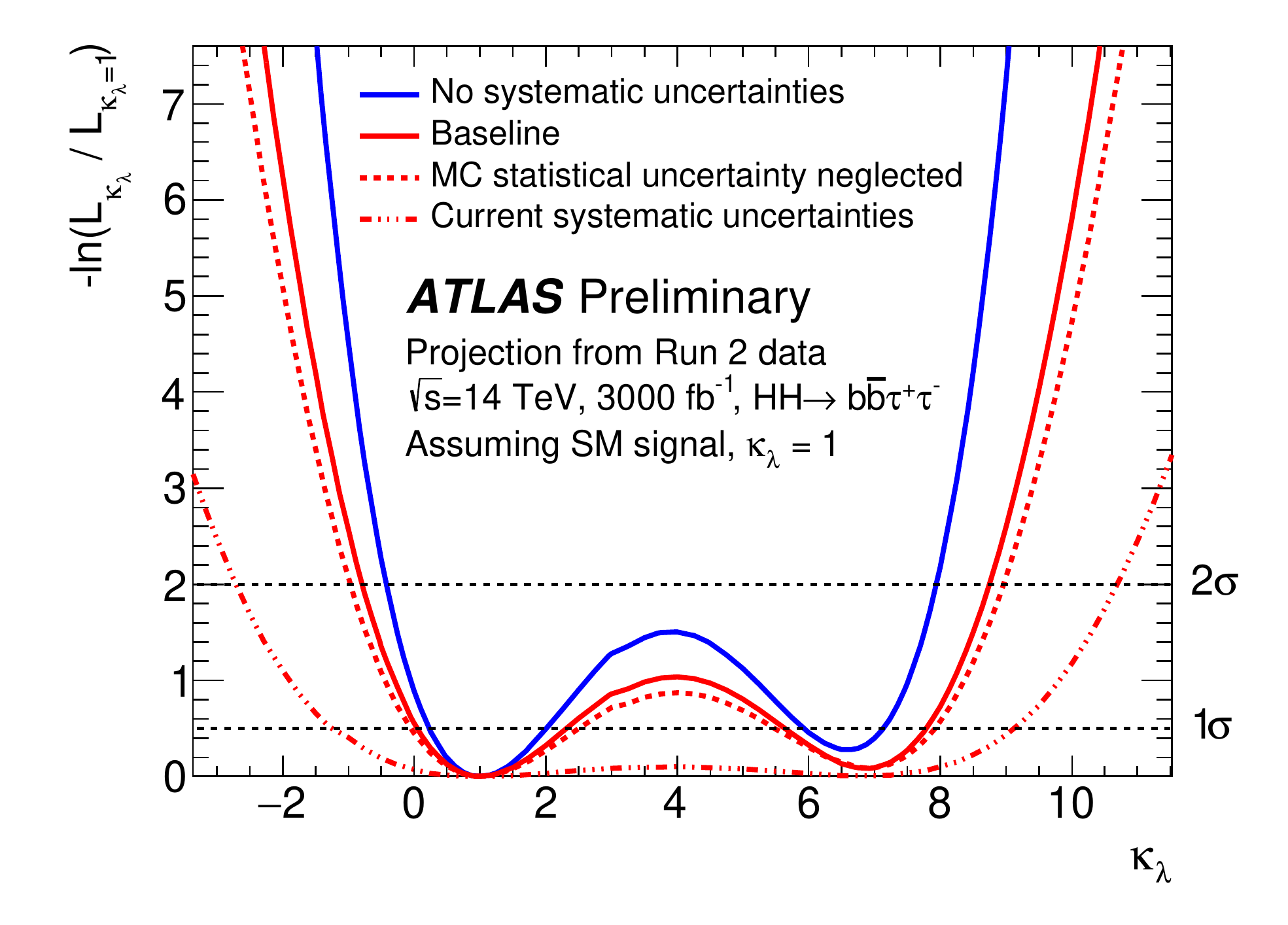}}
\caption{(a), (b), (c) Distributions of the BDT score for the three categories of the analysis, extrapolated to 3000~\fbinv\ of data. The background distributions are shown after the fit based on a background-only Asimov dataset and the signal is scaled to the SM prediction. The hatched bands represent the combined statistical and systematic uncertainty for the baseline scenario. These uncertainty bands are included in the plots for completeness but are very small. (d) Negative natural logarithm of the ratio of the maximum likelihood for $\kappa_{\lambda}$ to the maximum likelihood for $\kappa_{\lambda}$ = 1, obtained from fits to the Asimov dataset that contains the $\kappa_{\lambda}$ = 1 signal. Four different scenarios are considered for the systematic uncertainties.} 
\label{fig:ATLAS_HH_bbtautau} 
\end{figure}

In the Run~2 analysis one of the dominant systematic uncertainty is the one coming from the limited statistics of the MC samples used to estimate the background. In the baseline scenario, following the prescriptions of Ref.~\cite{ATLAS_PERF_Note}, this uncertainty is neglected.
Different scenarios are considered: the one in which the systematic uncertainties remain the same as for the Run~2 analysis (scenario S1 described in Section~\ref{sec:HiggsExtrapAss}); the scenario with the current systematic uncertainties but neglected MC statistical uncertainties and the baseline scenario for systematic uncertainties (scenario S2 described in Section~\ref{sec:HiggsExtrapAss}). The effect on those various scenarios is shown in Figure~\ref{fig:ATLAS_HH_bbtautau_d}.

The expected significance without systematic uncertainties is of 2.5 standard deviations, while it is 2.1 standard deviations when the baseline scenario for the systematic uncertainties is considered.

For the measurement of \kl\ the output score of a BDT trained on the \kl\ = 20 signal is used as the final discriminant. This was shown to provide similar performance with BDTs trained specifically for every \kl\ value, as \kl\ = 20 corresponds to a softer m$_{HH}$ spectrum, which is where the nominal BDT is less sensitive.
The minimum negative-log-likelihood for a SM signal hypothesis is shown in Figure~\ref{fig:ATLAS_HH_bbtautau_d}.


%
\paragraph{The $HH \rightarrow b\bar{b}\gamma\gamma$ channel}

The analysis~\cite{ATLASHHPUBnote} is based on truth level particles convoluted with the detector resolution, efficiencies and fake rates computed for $\mu$ = 200 which were extracted from fully simulated samples using the detector layout described in Ref.~\cite{ITKPixelTDR}. The selection is made using a multivariate analysis with a BDT using the full kinematic information of the event, in particular to reject the continuum and \ttH\ backgrounds. 
The di-photon invariant mass distribution, $\ensuremath{m_{\gamma\gamma}}$, is shown in Figure~\ref{fig:ATLAS_HH_bbyy_a}. The number of signal, single Higgs and continuum background in a 123-127~\GeV\ window is 6.5, 3.2 and 3.7 respectively.

The systematic uncertainties follow the prescriptions of Ref.~\cite{ATLAS_PERF_Note}. Their effect is very small since this channel will still be dominated by statistical uncertainties at the end of the HL-LHC programme.

The sensitivity of the analysis to \kl\ is assessed by using the $\mhh$ distribution for events in a 123 $ < \ensuremath{m_{\gamma\gamma}} $< 127~\GeV. This distribution is shown in Figure~\ref{fig:ATLAS_HH_bbyy_b} for different values of \kl\ and split into eight categories. It should be noted that the BDT was trained on the SM signal only, so the constraints on \kl\ are pessimistic. Using separate BDTs trained on specific values of \kl\ would bring negligible improvements at negative values of \kl, but up to 1$\sigma$ reduction in the expected limit at high positive values of \kl.
The expected significance was evaluated to be 2.1 and 2.0 standard deviations with and without systematic uncertainties included.


\begin{figure}[!htb]
\centering 
\subfloat[]{\label{fig:ATLAS_HH_bbyy_a}\includegraphics[width=0.5\textwidth]{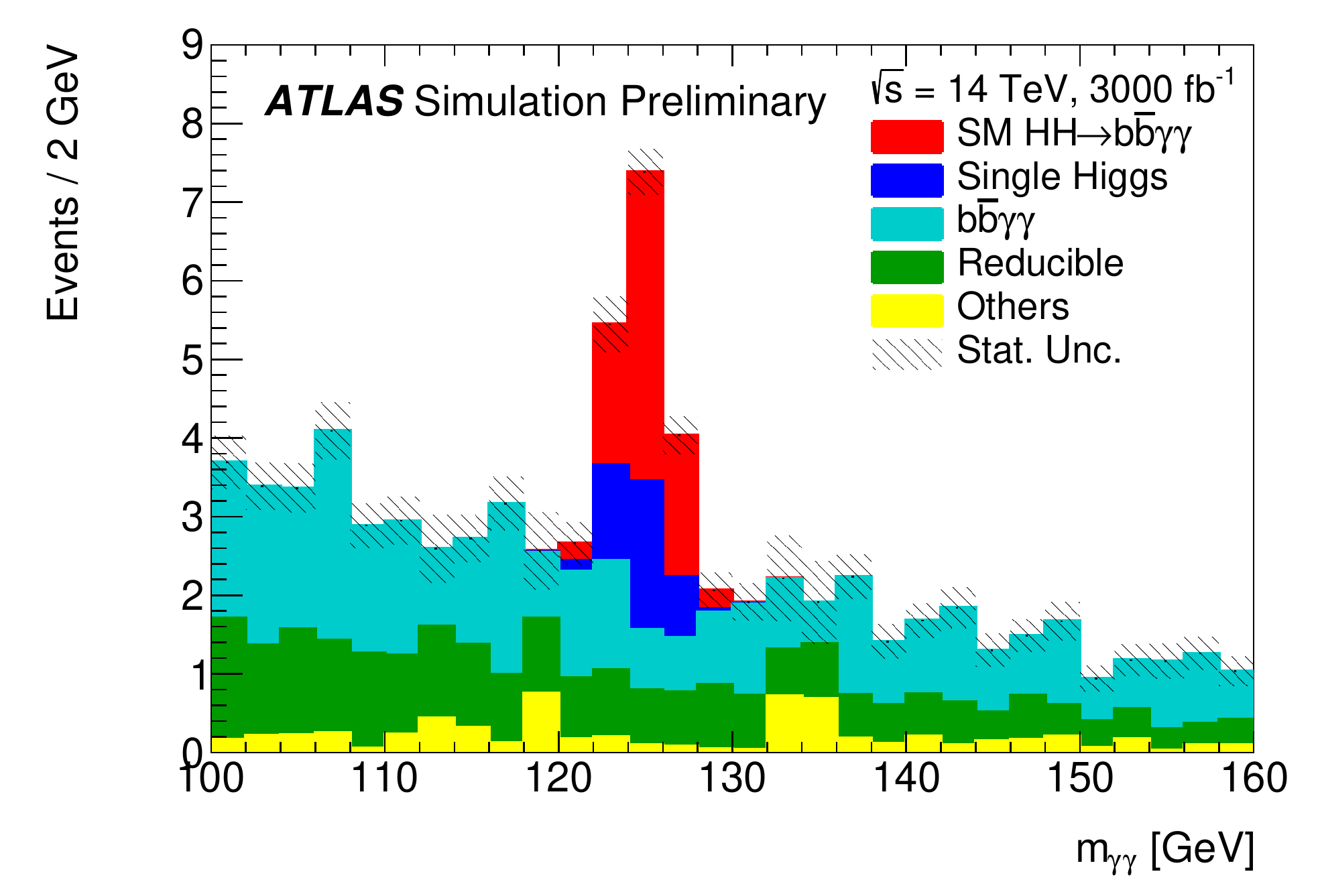}} 
\subfloat[]{\label{fig:ATLAS_HH_bbyy_b}\includegraphics[width=0.5\textwidth]{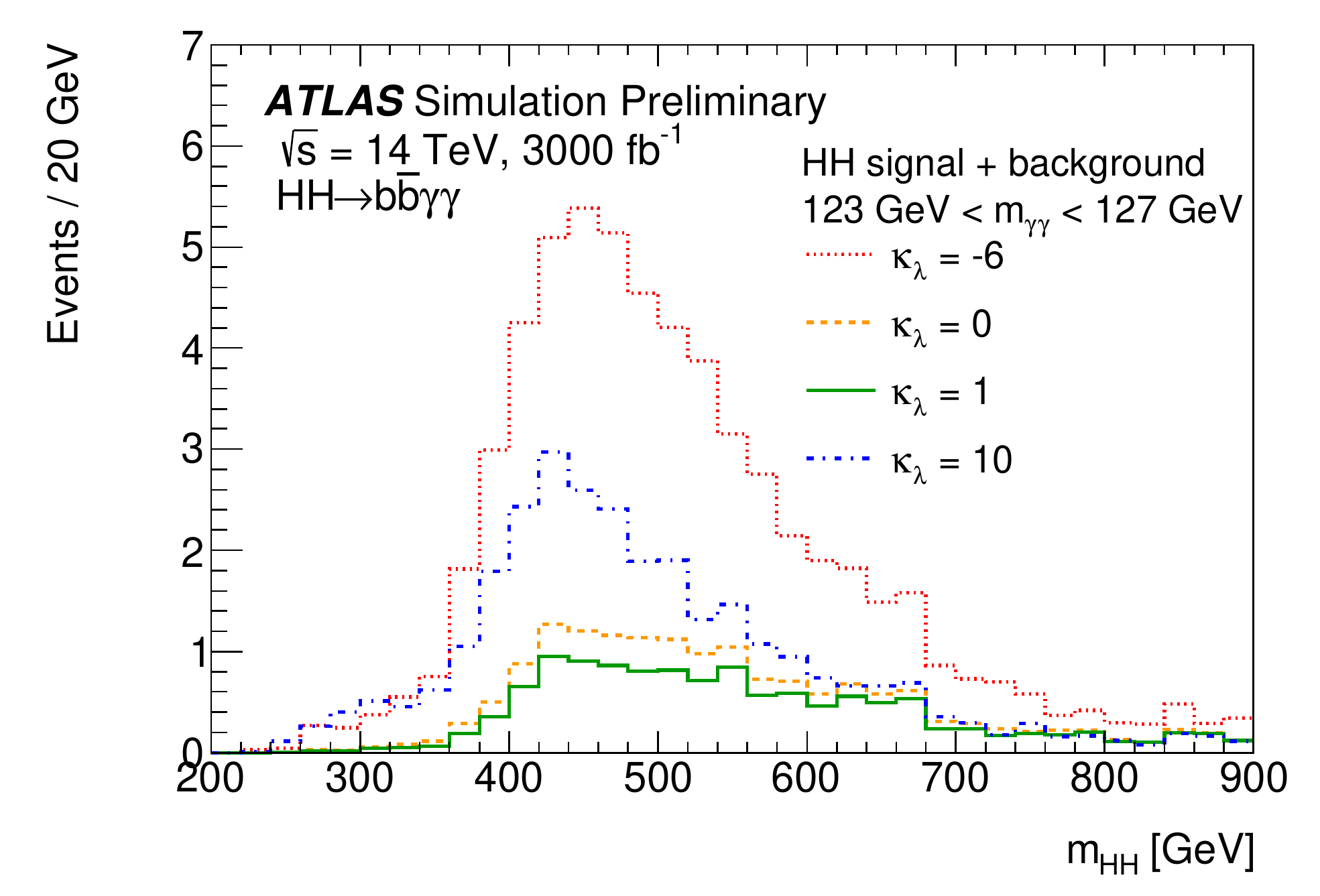}} 
\caption{(a) Distribution of $m_{\gamma\gamma}$ following the BDT response cut. The reducible background processes consist of $c\bar{c}\gamma\gamma$, $jj\gamma\gamma$, $b\bar{b}j\gamma$, $c\bar{c}j\gamma$, and $b\bar{b}jj$ events.  Other background processes come from $Z(b\bar{b})\gamma\gamma$, $t\bar{t}$ and $t\bar{t}\gamma$.
(b) Distributions of $m_{b\bar{b}\gamma\gamma}$ for combined signal and background events passing the BDT selection and the requirement 123 \GeV $ < \ensuremath{m_{\gamma\gamma}} $< 127~\GeV, for various values of $\kappa_{\lambda}$.} 
\label{fig:ATLAS_HH_bbyy} 
\end{figure}

\paragraph{Combined results}

The combination of various channels is realised by constructing a combined likelihood function that takes into account data, models and correlated systematic uncertainties from all channels. 

Setting appropriate nuisance parameters (NP) to be correlated with one another induced a negligible change in the combination results compared to assuming all nuisance parameters are uncorrelated. No strong correlation between any of the NP are found by the fits, with the exception of some correlation between the background models of the $b\bar{b}b\bar{b}$ and $b\bar{b}\tau\tau$ channels. Theoretical uncertainties on the cross-sections have negligible impact on the combined results.

The combined significance is 3.5 and 3.0 standard deviations with and without systematic uncertainties included.
Table \ref{tab:comb_measuredmu} shows the signal strength measured in the individual channels, as well as the combination, when the SM $HH$ signal is injected.

\begin{table}[htb!]
\centering
\begin{tabular}{l  c  c}
\hline
Measured $\mu$  &  \textbf{Statistical-only} & \textbf{Statistical + Systematic}\\
\hline
$HH \rightarrow b\bar{b}b\bar{b}$    & 1.0 $\pm$ 0.6 & 1.0 $\pm$ 1.6 \\
$HH \rightarrow b\bar{b}\tau\tau$      & 1.0 $\pm$ 0.4 & 1.0 $\pm$ 0.5 \\ 
$HH \rightarrow b\bar{b}\gamma\gamma$    & 1.0 $\pm$ 0.6 & 1.0 $\pm$ 0.6 \\ 

Combined   & 1.00 $\pm$ 0.31 & 1.0 $\pm$ 0.4 \\ 
\hline
\end{tabular}
\caption{Signal strength measured in the individual channels and their combination using an Asimov dataset with SM $HH$ signal injected.}
\label{tab:comb_measuredmu}
\end{table}

The combined sensitivity of the three channels to \kl\ is assessed by generating an Asimov dataset containing the background plus SM signal. The ratio of the negative-log-likelihood of the maximum likelihood fit for \kl\ was calculated and shown in Figure~\ref{fig:ATLAS_HH_comb}. A morphing technique~\cite{ATL-PHYS-PUB-2015-047} is used to generate signal distributions of m$_{HH}$ for any arbitrary value of \kl.

\begin{figure}[!htb]
\centering 
\subfloat[Statistical uncertainties only]{\label{fig:ATLAS_HH_comb_a}\includegraphics[width=0.5\textwidth]{\main/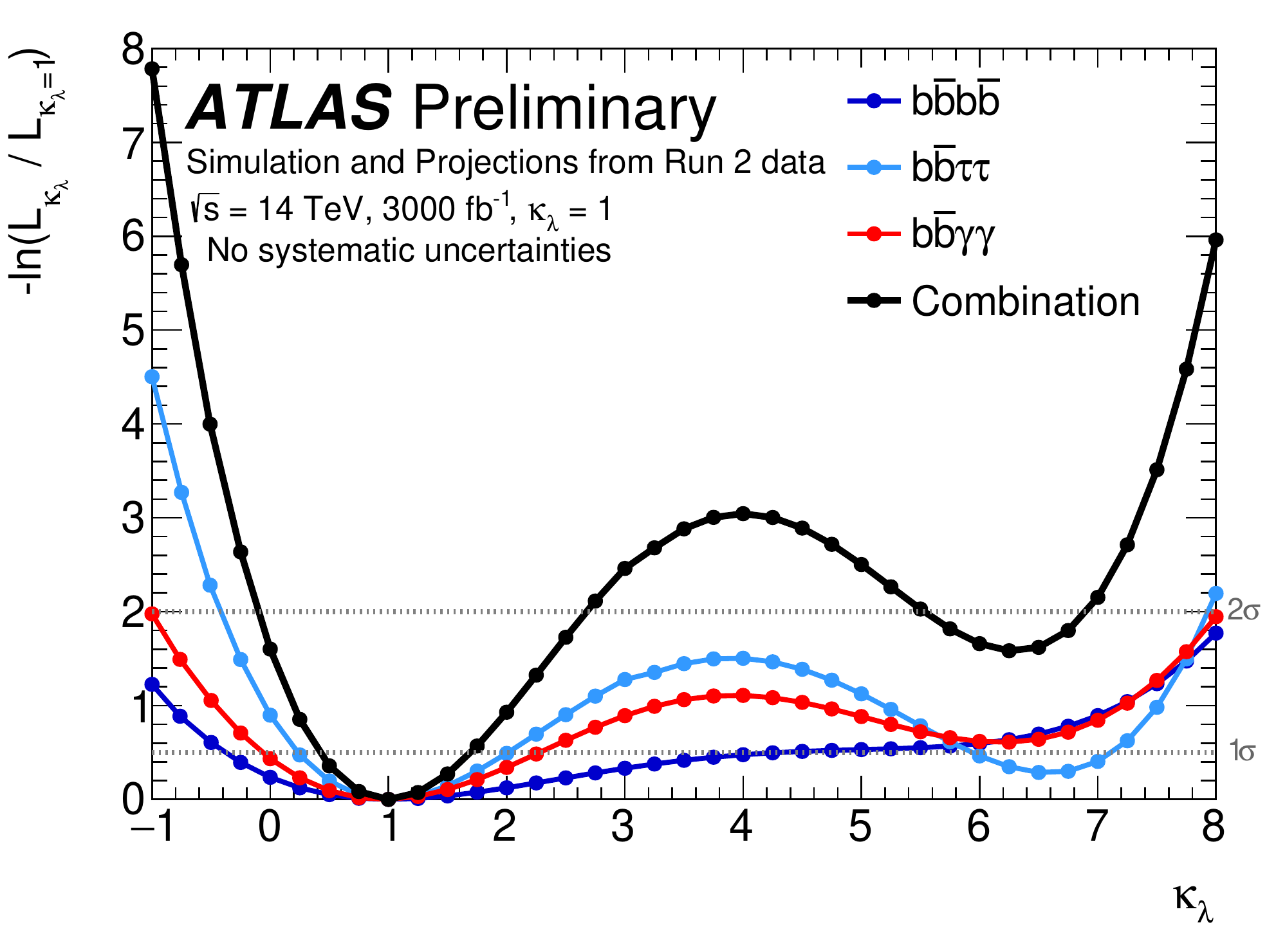}} 
\subfloat[Statistical + systematic uncertainties]{\label{fig:ATLAS_HH_comb_b}\includegraphics[width=0.5\textwidth]{\main/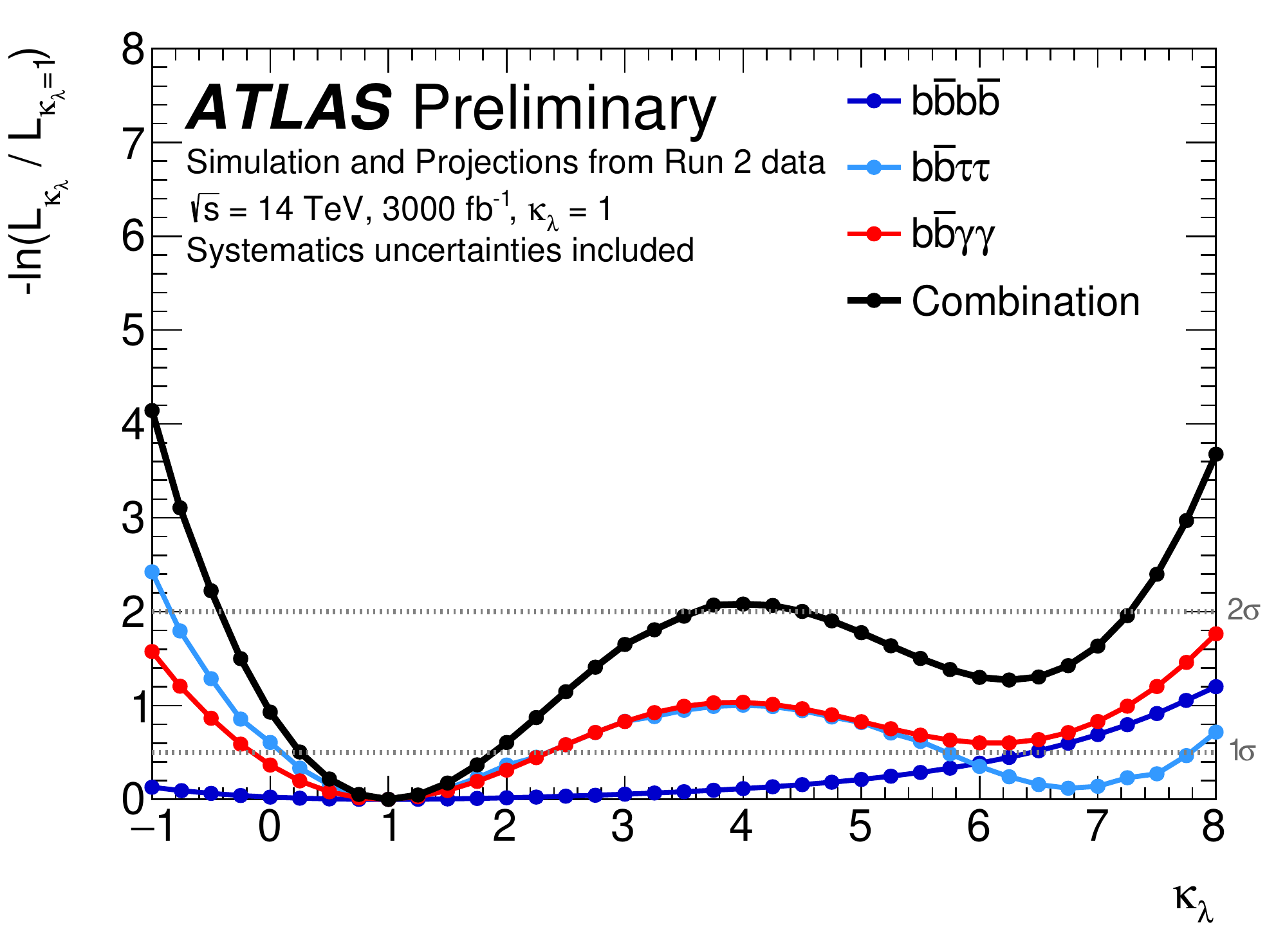}} 
\caption{Maximum likelihood for $\kappa_{\lambda}$ divided by the maximum likelihood for $\kappa_{\lambda}$ = 1 for (a) the fits with only statistical uncertainties and (b) the fits with all systematic uncertainties as nuisance parameters. The black circles show the results for the combination, while the coloured markers show the values coming from the individual channels. The dashed lines at $-\ln\left(L_{\kappa_{\lambda}}/L_{\kappa_{\lambda}=1}\right) = 0.5$ and 2.0 indicate the values corresponding to the $1\sigma$ and $2\sigma$ Confidence Intervals (CI), respectively (assuming an asymptotic $\chi^2$ distribution of the test statistic).} 
\label{fig:ATLAS_HH_comb} 
\end{figure}

The 68\% Confidence Intervals for $\kappa_{\lambda}$, from the likelihood ratio test performed on the Asimov dataset created from the backgrounds and the SM HH signal are $0.4 \leq \kl\ \leq 1.7$ and $0.25 \leq \kl\ \leq 1.9$ with and without systematic uncertainties respectively. The Confidence Intervals per channel are summarised in Table~\ref{tab:ATLAS_HH_kappalambda}. 
The Higgs boson self-coupling is constrained at 95\% confidence level (CL) to $-0.4\leq \kappa_{\lambda} \leq7.3$ ($-0.1\leq \kappa_{\lambda} \leq2.7\cup5.5\leq \kappa_{\lambda} \leq6.9$), with (without) systematic uncertainties.

\begin{table}[htb!]
\begin{center}
\begin{tabular}{lcc} \toprule
 & \textbf{Statistical-only} & \textbf{Statistical + Systematic}\\
\hline
$HH \rightarrow b\bar{b}b\bar{b}$ & $-0.4 \leq \kappa_{\lambda} \leq 4.3$ & $-2.3 \leq \kappa_{\lambda} \leq 6.4$ \\
$HH \rightarrow b\bar{b}\tau\tau$ &  $0.2 \leq \kappa_{\lambda} \leq 2.0 \cup 5.9 \leq \kappa_{\lambda} \leq 7.2$ & $0.1 \leq \kappa_{\lambda} \leq 2.3 \cup 5.7 \leq \kappa_{\lambda} \leq 7.8$ \\
$HH \rightarrow b\bar{b}\gamma\gamma$ &  $-0.1 \leq \kappa_{\lambda} \leq 2.4$ & $-0.2 \leq \kappa_{\lambda} \leq 2.5$ \\
combined &   $0.4 \leq \kappa_{\lambda} \leq 1.7$ & $0.25 \leq \kappa_{\lambda} \leq 1.9$ \\
\bottomrule
\end{tabular}
\end{center}
\caption{68\% Confidence Intervals for $\kappa_{\lambda}$, estimated for an Asimov dataset containing the backgrounds plus SM signal.}
\label{tab:ATLAS_HH_kappalambda}
\end{table}

Assuming the SM HH signal the expected exclusion significance for the \kl\ = 0 hypothesis, i.e. no Higgs self-coupling, is 1.4 and 1.8 standard deviations with and without systematic uncertainties respectively.


\asubsubsection[A. Benaglia, M. Bengala, O. Bondu, L. Borgonovi, S. Braibant, L. Cadamuro, A. Carvalho, C. Delaere, M. Delcourt, N. de Filippis, E. Fontanesi, M. Gallinaro, M. Gouzevitch, J. R. Komaragiri, D. Majumder, K. Mazumdar, F. Monti, G. Ortona, L. Panwar, N. Sahoo, R. Santo, G. Strong, M. Vidal, S. Wertz]{Measurements with the CMS experiment}
\label{sec:HH_CMS}

The work described in this section studies the prospects for \HH production at the HL-LHC with the CMS experiment.
The five decay channels \bbbb, \bbtt, \bbWW ($\PW\PW\to\ell\nu\ell'\nu'$ with $\ell,\ell' = \Pe,\Pgm$), \bbgg, and \bbZZ ($\PZ\PZ\to\ell\ell\ell'\ell'$ with $\ell,\ell' = \Pe,\Pgm$) are explored.
The corresponding branching fractions and the total number of \HH events expected to be produced at the HL-LHC assuming $\sqrt{s} = 14\TeV$ and an integrated luminosity of $3000\fbi$ are reported in Table~\ref{sec3:CMSHH:tab:br_nevent}.

A short description of the analysis strategy and of the results is given here, and further details can be found in Ref.~\cite{CMS-PAS-FTR-18-019}.

\begin{table}[h]
  \begin{center}
    \caption{Branching fraction of the five decay channels considered in the CMS \HH prospects, and corresponding number of events produced at the end of HL-LHC operations assuming $\sqrt{s} = 14\TeV$ and an integrated luminosity of $3000\fbi$. The symbol $\ell$ denotes either a muon or an electron. In the \bbWW decay channel, $\ell$ from the intermediate production of a $\tau$ lepton are also considered in the branching fraction.}
    \label{sec3:CMSHH:tab:br_nevent}
    \begin{tabular}{l  ccccc}
        \hline
        Channel            & $\bbbb$ & $\bbtt$ & $\bbWW(\ell\nu\ell\nu)$ & $\bbgg$ & $\bbZZ(\ell\ell\ell\ell)$ \\
        $\mathcal{B}$ [\%] & 33.9    & 7.3     & 1.7                    & 0.26    & 0.015\\
        Number of events   & 37000   & 8000    & 1830                   & 290     & 17\\
        \hline      
    \end{tabular}
  \end{center}
\end{table}

A parametric simulation based on the \delphes~\cite{deFavereau:2013fsa} software is used to model the CMS detector response in the HL-LHC conditions.
The \delphes simulation accounts for the effects of multiple  hadron interactions (``pileup'') by overlaying simulated minimum-bias events with on average 200 interactions per bunch crossing.
The performance of reconstruction and identification algorithms for electrons, muons, tau decays to hadrons (\tauh) and a neutrino, photons, jets (including the identification of those containing heavy flavour particles), and the missing transverse momentum vector \ptvecmiss is parametrised based on the  results obtained with a full simulation of the CMS detector and dedicated reconstruction algorithms.

\paragraph{The $\HH\to\bbbb$ channel}

While characterised by the largest branching fraction among the $\HH$ final states, the \bbbb decay channel suffers from a large contamination from the multi-jet background that makes it experimentally challenging.
Two complementary strategies are explored here to identify the signal contribution.
For those events where the four jets from the $\HH\to\bbbb$ decay can all be reconstructed separately, also referred to as the ``resolved'' topology, the usage of multivariate methods is explored to efficiently identify the signal contribution in the overwhelming background.
In cases where the invariant mass of the $\HH$ system, $m_{\HH}$, is large, the high Lorentz boost of both Higgs bosons may results in a so-called ``boosted'' event topology where the two jets from a $\Hbb$ decay overlap and are reconstructed as a single, large-area jet.
Resolved topologies correspond to the large majority of SM \HH events, giving the largest sensitivity on this signal.
Boosted topologies help to suppress the multi-jet background and provide sensitivity to BSM scenarios where the differential \HH production cross section is enhanced at high $m_{\HH}$ by the presence of $\Pg\Pg\HH$ and $\PQt\PQt\HH$ effective contact interactions.

In the resolved topology, events are pre-selected by requiring four jets with $\pt > 45~\GeV$ and $|\eta| <$ 3.5 that satisfy the medium b-tagging working point, corresponding to a b jet identification efficiency of approximately 70\% for a light  flavour and gluon jet mis-identification rate of  1\%.
Triggers are assumed to be fully efficient in the phase space defined by this selection.
In scenarios where the minimal jet trigger \pt threshold is increased the loss in sensitivity to the SM signal amounts to approximately 10\% and 25\% for a 10 and 30\GeV increase, respectively.

The four selected b tagged jets are combined into the two Higgs boson candidates $\PH_1$ and $\PH_2$, choosing the pairs of jets with the minimal invariant mass difference.
The invariant mass of the two Higgs candidates is required to satisfy the relation:
\begin{equation}
\sqrt{ \left( \text{m}_{\PH_1} - 120\GeV\right)^2 + \left( \text{m}_{\PH_2} - 120\GeV\right)^2 } < 40\GeV
\end{equation}
i.e. a circular selection where the centre and radius are chosen based on the expected response and resolution of the CMS detector, accounting for the energy loss from undetected neutrinos from B hadron decays.

Because of the very large QCD multi-jet background, a multivariate discriminant, in the form of a boosted decision tree, is trained to identify the \HH signal contribution and used as the discriminant variable.
Other background processes considered are \ttbar and single Higgs boson production.
The output of the BDT discriminant is shown in Fig.~\ref{sec3:CMSHH:fig:bbbb_BDT}.

\begin{figure}[!htb]
\centering 
\includegraphics[width=0.5\textwidth]{\main/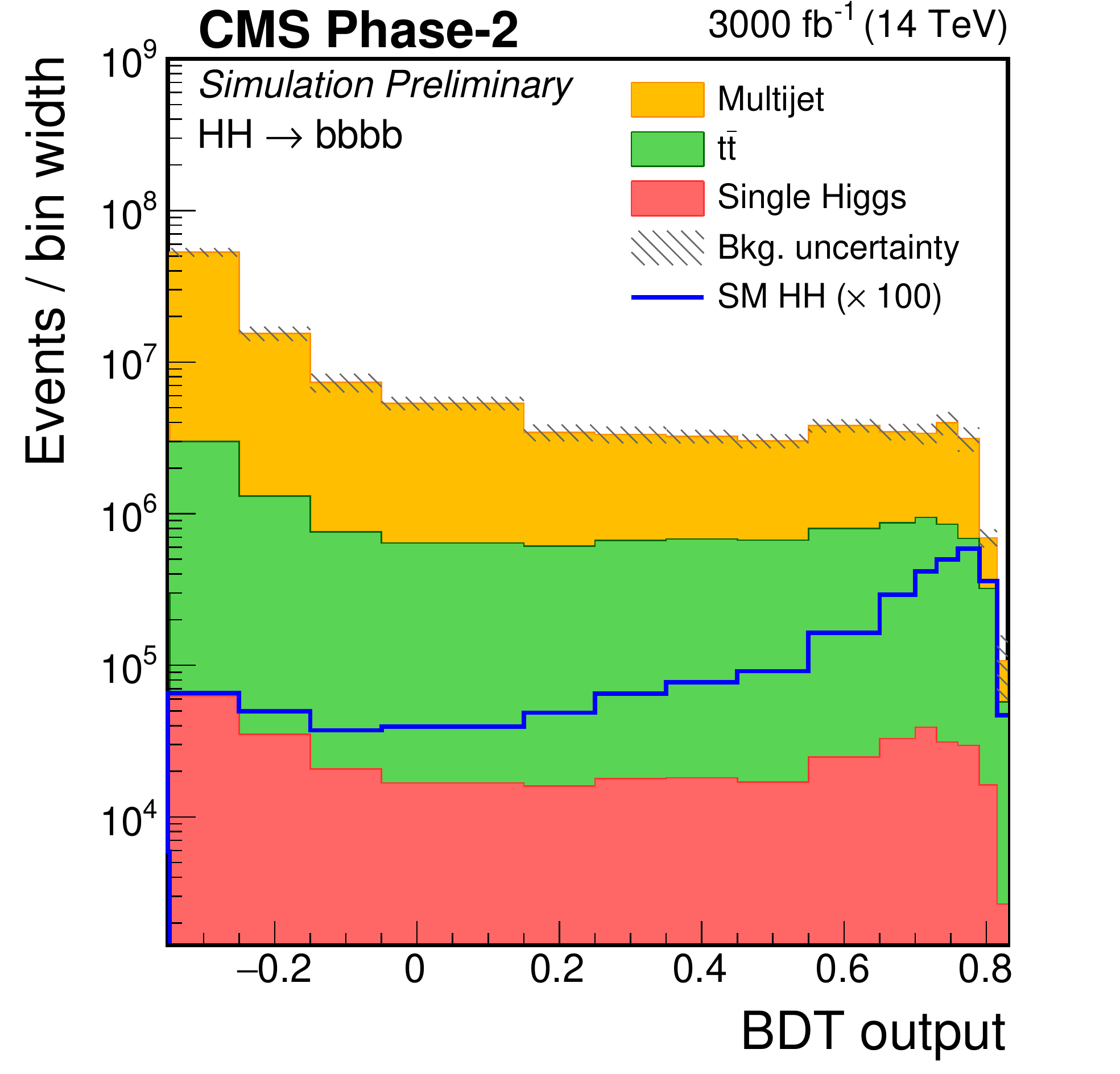}
\caption{BDT output distribution for the signal and background processes considered in the \bbbb resolved search.} 
\label{sec3:CMSHH:fig:bbbb_BDT} 
\end{figure}

The boosted topology offers a good handle to investigate effective Higgs boson contact interactions predicted in BSM scenarios that enhance the \HH production cross section at high $m_{\HH}$ values.
For that reason, the prospects in this channels focus on anomalous couplings and make use of the shape benchmarks signals described in Ref.~\cite{Carvalho2016}.
Large radius jets, clustered with the anti-$k_\text{T}$ algorithm with a cone radius of 0.8 (AK8 jets), are used to identify the  overlapping b jets.
The event is required to contain at least two AK8 jets with $\pt > 300\GeV$ and $|\eta| < 3$.
The two highest $\pt$ jets are chosen in case multiple candidates satisfy such requirements.
The soft drop~\cite{Dasgupta:2013ihk,Larkoski:2014wba} jet grooming algorithm is used to remove soft and collinear components of the jet and retain the two sub-jets associated with the showering and hadronisation of the two b quarks from the $\PH\to\PQb\PAQb$ decay.
A selection is applied on the N-sub-jettiness variable~\cite{Thaler:2011gf} to reduce the background contamination, mostly represented by di-jet production from QCD interactions.
Algorithms for the b jet identification are applied on the sub-jets with a working point corresponding to an efficiency of about 49\% for genuine b jets for a mis-identification rate of light flavour and gluon jets of about 1\%.
Events are divided in two categories if they contain exactly three (3b category) or exactly four (4b category) b-tagged sub-jets.

The invariant mass of the two selected AK8 , $M_{JJ}$, is used to look for the presence of a signal. Its distribution is shown in Fig.~\ref{sec3:CMSHH:fig:bbbb_boosted} for the two event categories.

\begin{figure}[!htb]
\centering 
    \includegraphics[width=0.495\textwidth]{\main/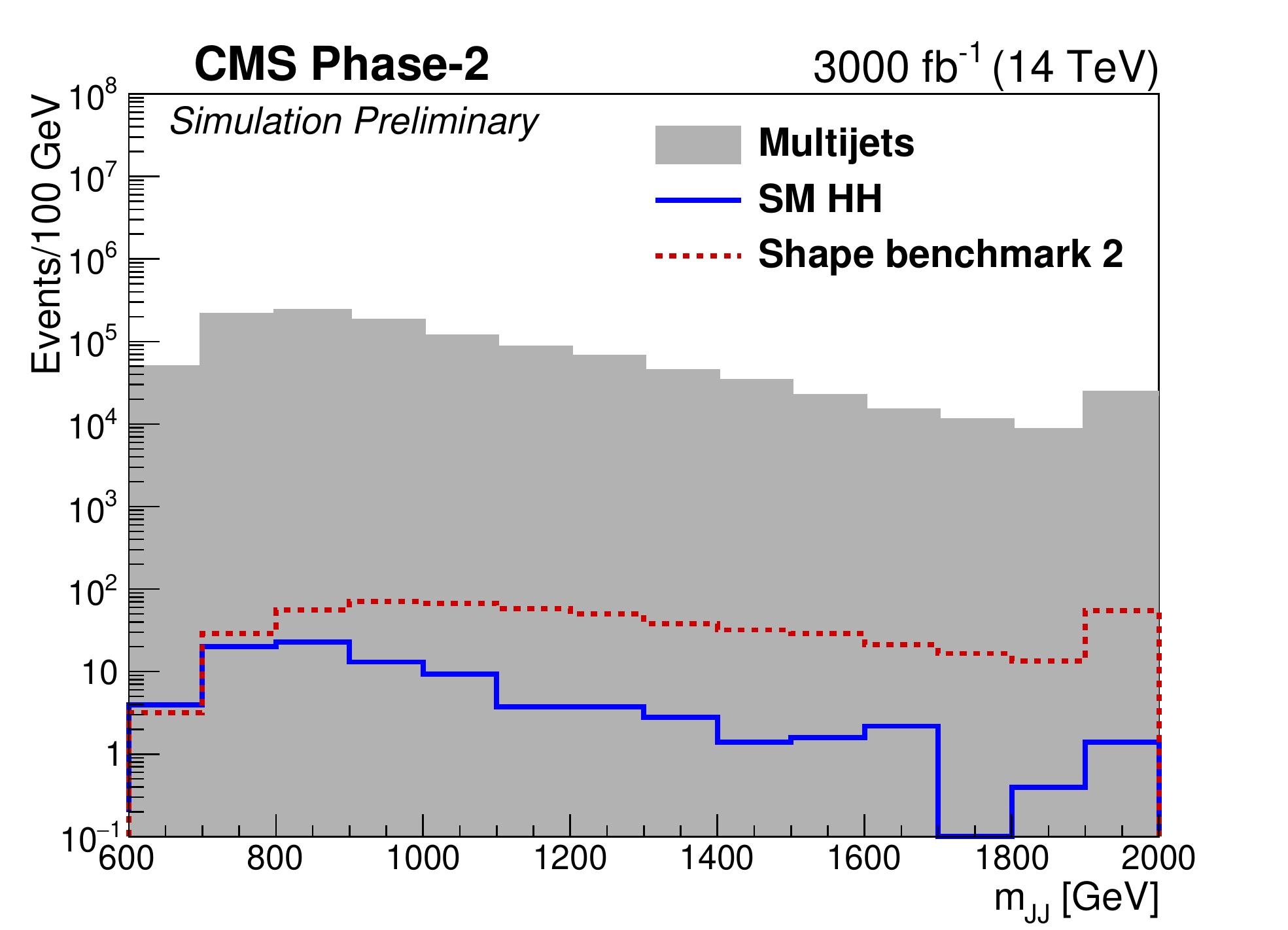}
    \includegraphics[width=0.495\textwidth]{\main/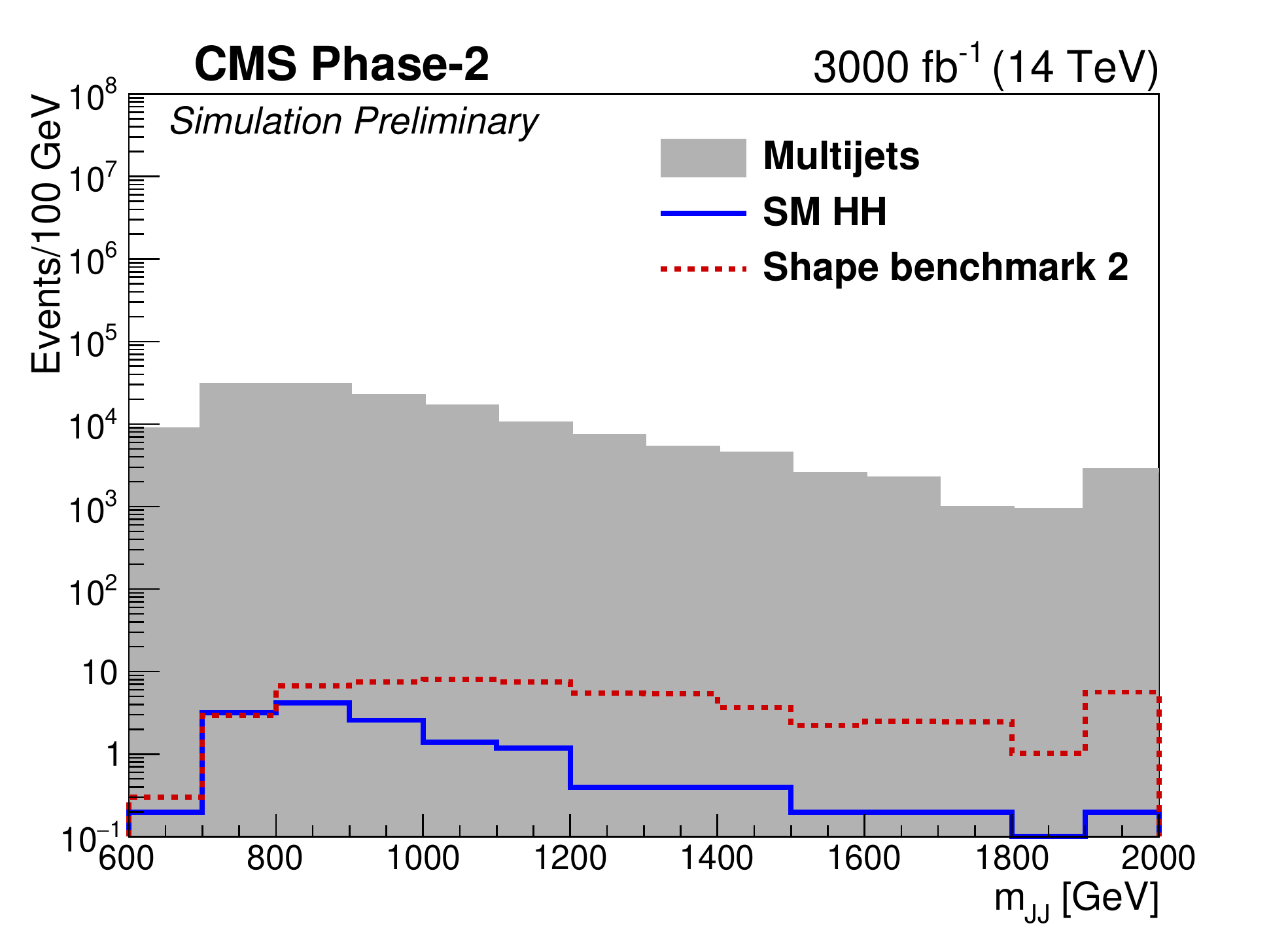}
\caption{Invariant mass of the two selected AK8 jets in the boosted \bbbb \HH search for the multi-jet background and the SM (blue) and shape benchmark 2 (red) signals.
    The distributions on the left are for the $3\PQb$ and those on the right are for the $4\PQb$ sub-jet $\PQb$-tagged categories.
    Both signals are normalised to the SM $\PH\PH$ production cross section for visualisation.} 
\label{sec3:CMSHH:fig:bbbb_boosted} 
\end{figure}

\paragraph{The $HH \rightarrow \bbtt$ channel}

The $\bbtt$ decay channel is experimentally favourable thanks to its sizeable branching fraction of 7.3\% and the moderate background contamination.
Out of the six possible decay channels of the $\tau\tau$ system, the $\mu\tauh$, $\Pe\tauh$, and $\tauh\tauh$ final states are considered here, corresponding together to about 88\% of the total branching ratio.
Events in the three channels are selected requiring the presence of a \tauh candidate in association to an isolated muon, electron, or another \tauh\ depending on the final state considered.
Events in all the three categories above are then required to contain at least two b-tagged jets with $\pt > 30\GeV$ and $|\eta| < 2.4$. 

The main backgrounds are \ttbar and Drell-Yan production of $\tau$ pairs.
Their separation is experimentally challenging because of the incomplete reconstruction of the event due to the presence of neutrinos from $\tau$ decays that escape detection.

A multivariate analysis method is thus used to identify the signal contribution and separate if from the large background.
The usage of state-of-the-art machine learning techniques is studied in this work.
The discriminant consists of a pair of ensembles of ten fully connected deep neural networks (DNN), each with three hidden layers of 100 neurons, trained to separate the \HH signal from the background processes using a wide set of kinematic variables, a few of which are shown for illustration in Fig.~\ref{sec3:CMSHH:fig:bbttinputs}.
Each network is trained using events from all three $\tau\tau$ decay channels, and advanced optimisation techniques are explored and applied to maximise the expected sensitivity.

\begin{figure}[!htb]
\centering 
    \includegraphics[width=0.495\textwidth]{\main/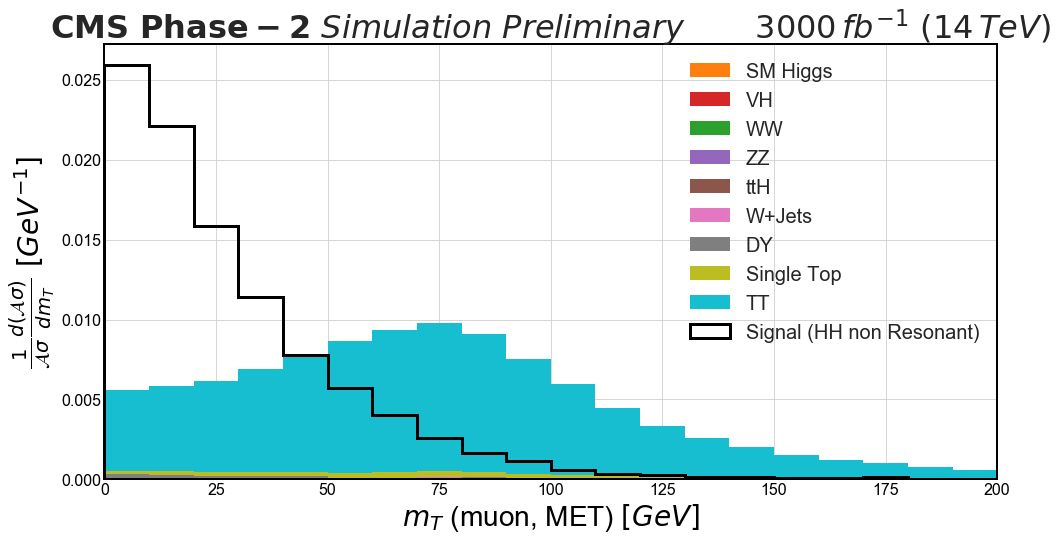}
    \includegraphics[width=0.495\textwidth]{\main/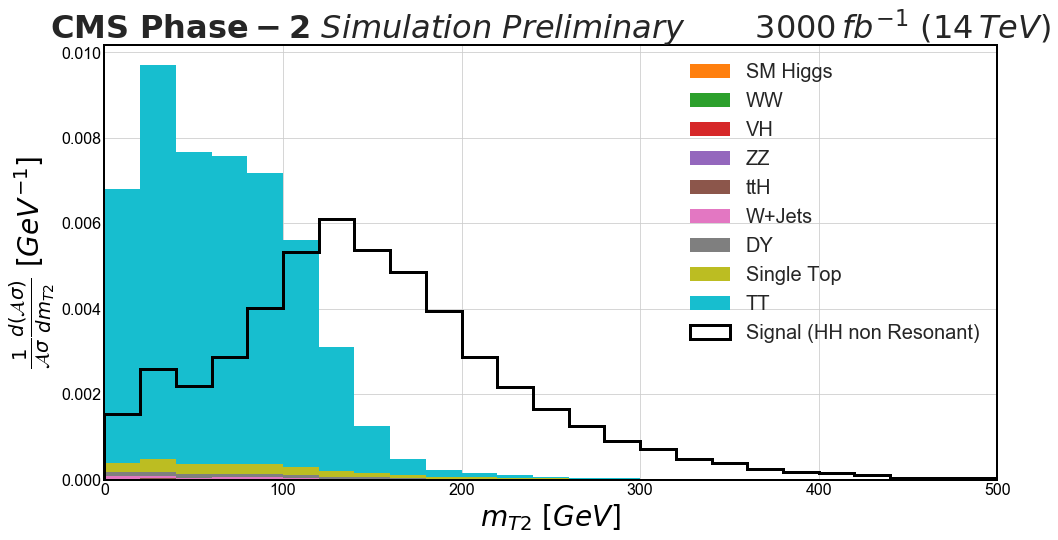}\\
    \includegraphics[width=0.495\textwidth]{\main/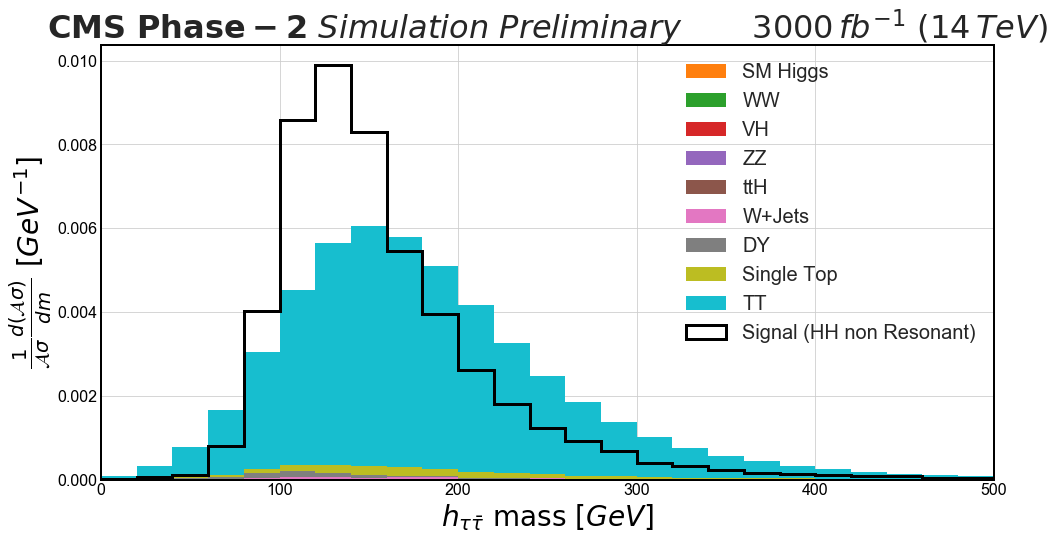}
    \includegraphics[width=0.495\textwidth]{\main/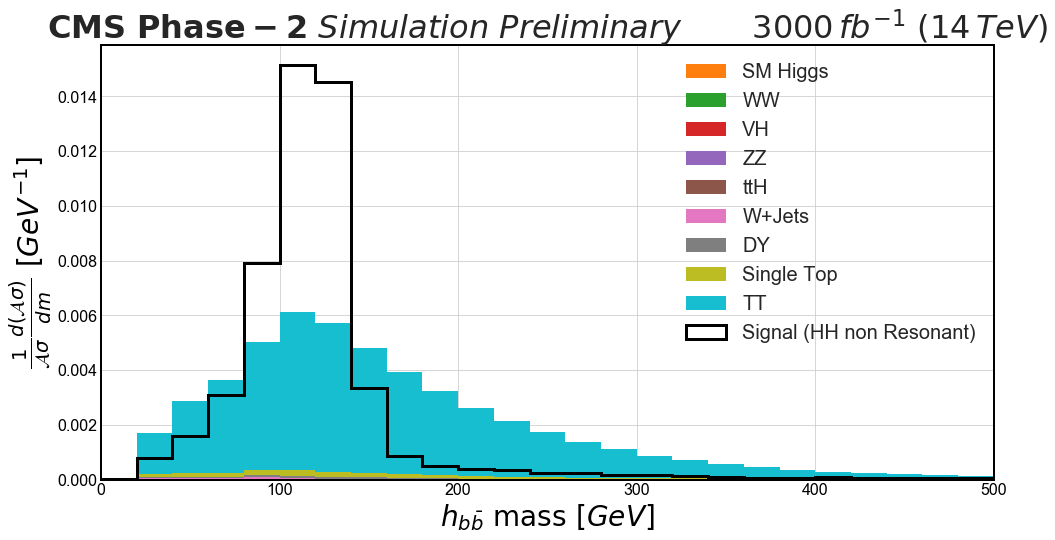}\\    
\caption{Example distributions for some of the discriminant variables used as input of the \bbtt deep neural network: muon transverse mass (top left), system transverse mass $m_\text{T2}$ (top right), and invariant mass of the $\tau\tau$ (bottom left) and bb (bottom right) systems.} 
\label{sec3:CMSHH:fig:bbttinputs} 
\end{figure}

\paragraph{The $HH \rightarrow b\bar{b}\gamma\gamma$ channel}

Despite its low branching fraction, the \bbgg decay channel is one of the most sensitive to \HH production.
It benefits of an excellent di-photon invariant mass ($m_{\gamma\gamma}$) resolution and on the possibility to fully reconstruct all final state objects.
The analysis strategy combines these two aspects and uses a multivariate kinematic discriminant to suppress the background contributions, and the $m_{\gamma\gamma}$ signature to look for the presence of a signal.

The $\PH\to\gamma\gamma$ candidate is built from two photons in the collision event that satisfy identification, isolation, and quality criteria.
Only events where the two photons satisfy $|\eta| < 2.5$ and $100 < m_{\gamma\gamma} < 150$ are considered.
The $\PH\to\text{bb}$ candidate is built from the two jets with the highest b tag discriminant value that satisfy $\pt > 25\GeV$ and $|\eta| < 2.5$.
The background from light flavour jets is suppressed by requiring both jets to satisfy a loose working point of the  b tagging algorithm, corresponding to a 90\% efficiency for a genuine b-jet and 10\% mis-identification rate.
The di-jet invariant mass is required to be between 80 and 190\GeV. 

The backgrounds mainly consist of non-resonant $\gamma\gamma$ production in association with heavy flavour jets, with a smaller contribution from $\gamma\gamma$ plus light flavour jets, and single Higgs boson production in association with top quark ($\Ptop\Ptop\PH$, with $\PH\to\gamma\gamma$).

A multivariate discriminant  in the form of a BDT is used to suppress the $\Ptop\Ptop\PH$ background.
The BDT is trained to identify the presence of decay products from $\PW$ bosons originating from top quark decays, and combines the information on the presence and properties of leptons, additional jets, and helicity angles of the $\HH$ system and its decay products.
A selection on the discriminant is applied, rejecting approximately 75\% of the $\Ptop\Ptop\PH$ events for a 90\% signal efficiency.

A second BDT classifier is trained to separate the $\HH$ signal from the non-resonant di-photon background.
Several variables related to the kinematic properties of the event and to the quality of the selected objects are combined, and background-like events with a low BDT scores are rejected.

Events thus selected are simultaneously classified based on the value of the BDT discriminant described above and on the reduced mass of the four objects selected, defined as:
\begin{equation}
    M_\text{X} = m_{\gamma\gamma\text{jj}} - m_{\gamma\gamma} - m_\text{jj} + 250\GeV,
\end{equation}
where $m_{\gamma\gamma\text{jj}}$, $m_{\gamma\gamma}$, and $m_\text{jj}$ refer respectively to the four body, di-photon, and di-jet invariant masses.
The definition of $M_\text{X}$ mitigates resolution effects by using the expected Higgs boson mass.
Two intervals of the BDT scores are used to define medium and high purity categories (MP and HP), and events in each category are further divided in a low, medium, and high mass category if $250 < M_\text{X} < 350\GeV$, $350 < M_\text{X} < 380\GeV$, or $480 < M_\text{X}~\GeV$, respectively.
While the high mass category is the most sensitive to SM \HH production, low mass categories are important to constrain anomalous values of the Higgs boson self-coupling, that enhance the cross section at the $m_{\HH}$ threshold.

The signal is extracted from a simultaneous fit in each of the $3 \times 2$ categories defined above.
A parametric maximum likelihood fit of the signal and background in the $(m_{\gamma\gamma}, m_\text{jj})$ is used.
An example of the expected event distributions in the high mass and high purity category for the two variables is shown in Fig.~\ref{sec3:CMSHH:fig:bbgg_events}.

\begin{figure}[!htb]
\centering 
    \includegraphics[width=0.495\textwidth]{\main/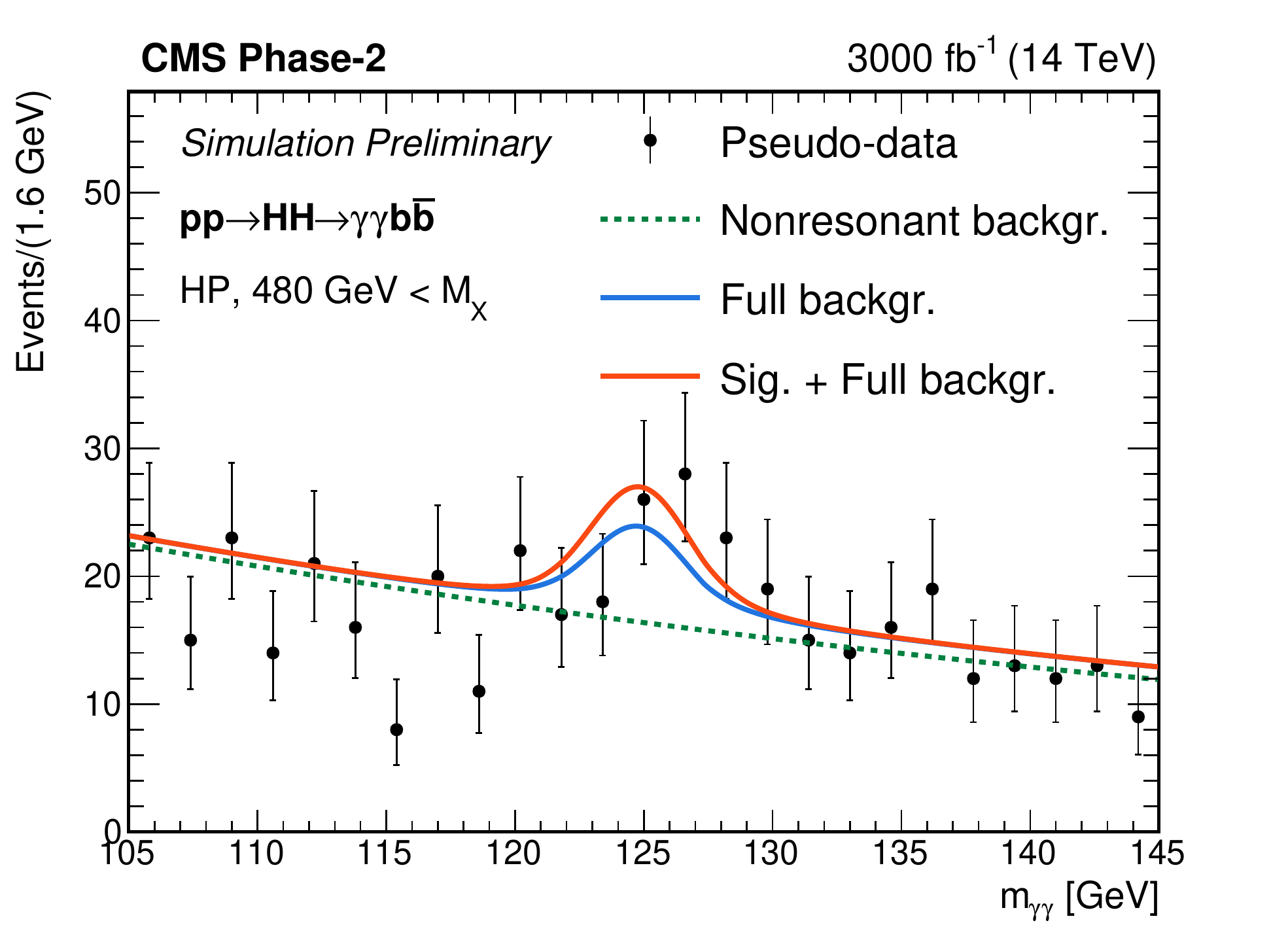}
    \includegraphics[width=0.495\textwidth]{\main/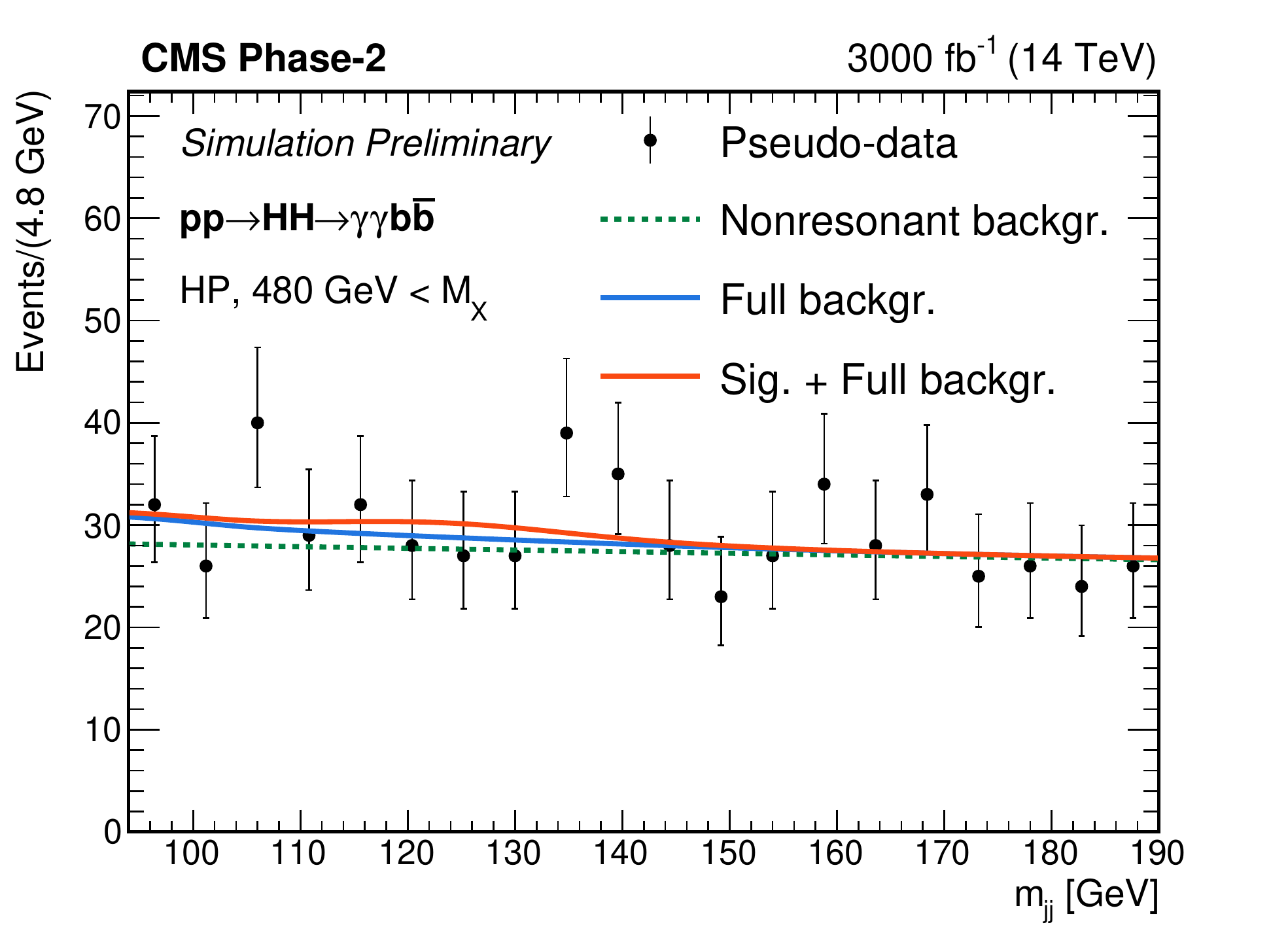}
\caption{Expected distribution of events in the photon (left column) and jet (right column) pair invariant mass for the high mass and high purity event category. The full circles denote pseudo-data obtained from the expected events yields for the sum of the signal and background processes for $3000\fbinv$.} 
\label{sec3:CMSHH:fig:bbgg_events} 
\end{figure}

\paragraph{The $HH \rightarrow \bbWW \to \PQb\PQb\ell\nu\ell\nu$ channel}

We consider here \HH final states containing two \PQb jets  and two neutrinos and two leptons, either electrons or muons.
The decay channels involved are thus $\PH\to\bb$ in association with either a $\PH\to\PZ(\ell\ell)\PZ(\nu\nu)$ or a $\PH \to \PW(\ell\nu)\PW(\ell\nu)$ decay.
While the analysis described in the following is optimised for $\HH\to\bbWW$ decays, that provide the largest branching fraction, the contribution of Higgs boson decays to both $\PW\PW$ and $\PZ\PZ$, globally denoted as $\text{VV}$, is considered.
Decays of the $\text{VV}$ system to tau leptons subsequently decaying to electrons or muons with the associated neutrinos are also considered in the simulated signal samples.
The corresponding branching fraction for the $\text{VV}\to\ell\nu_\ell\ell\nu_\ell$ decay is 1.73~\%.

The dominant and sub-dominant background processes are
the $\ttbar$ production in its fully leptonic decay mode, and
Drell-Yan production of lepton pairs in association with jets.
As both are irreducible background processes, \ie they result in the same final state as the signal, the kinematic properties of the signal and background events are used and combined in an artificial Neural Network (NN) discriminant to enhance the sensitivity.

Events are required to contain two isolated leptons of opposite electric charge, with an invariant mass $m_{\ell\ell} > 12
\UGeV$ to suppress leptonia resonances and $m_{\PZ} - m_{\ell\ell} > 15\GeV$ to suppress Drell-Yan lepton pair production.
The $\PH\to\PQb\PQb$ decay is reconstructed by requiring the presence of two b-tagged jets in the event with $\pt > 20$~\UGeV and $| \eta | < 2.8$, separated from the selected leptons by a distance of $\Delta \text{R} = \sqrt{\Delta \phi^2 + \Delta \eta^2} > 0.3$.



The NN discriminant utilises information related
to object kinematics.
The variables provided as input to the NN exploit the presence of two Higgs  bosons decaying into two b-jets on the one hand, and two leptons and two neutrinos on the other hand, 
which results in different kinematics for the di-lepton and di-jet systems between signal and 
background processes.
The set of variables used as input is:$ m_{\ell\ell}$, $m_\text{jj}$,
$\Delta R_{\ell\ell}$, $\Delta R_{\text{j}\text{j}}$, $\Delta \phi_{\ell\ell, \text{j}\text{j}}$, defined as 
the $\Delta \phi$ between the di-jet and the di-lepton systems, $\pt^{\ell\ell}$, $\pt^{\text{j}\text{j}}$,
min$\left(\Delta R_{\text{j}, \ell}\right)$, and $\mathrm{M}_\mathrm{T}$, defined as
$\mathrm{M}_\mathrm{T} = \sqrt{2 \pt^{\ell\ell} \ptmiss (1 - \cos(\Delta \phi(\ell\ell, \ptmiss)))}$.

The output of the NN is used as the discriminant variable in the three decay channels studied, and its distribution is reported in Fig.~\ref{sec3:CMSHH:fig:bbWW_events}.

\begin{figure}[!htb]
  \begin{center}
    \includegraphics[width=0.32\textwidth]{\main/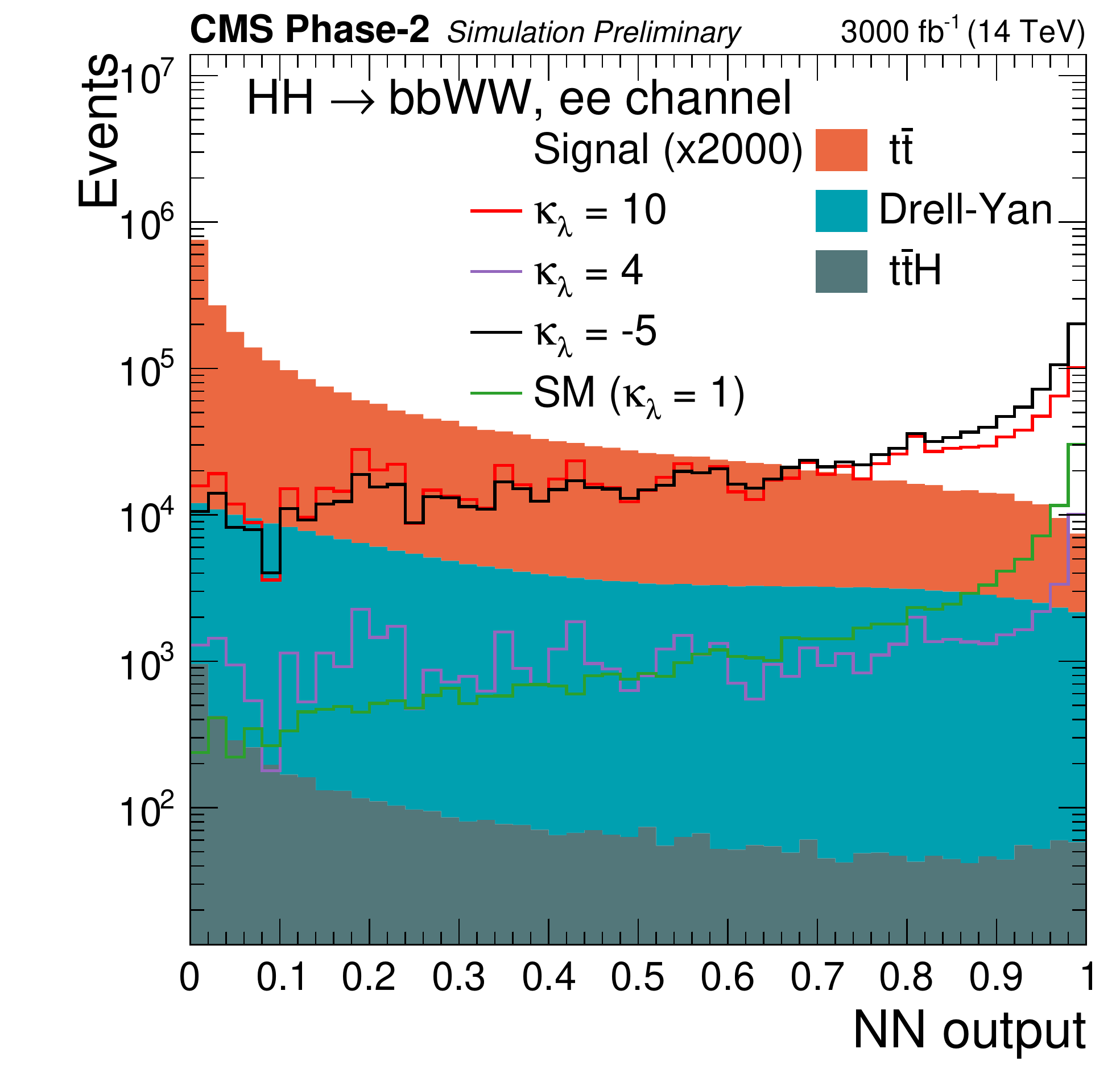}
    \includegraphics[width=0.32\textwidth]{\main/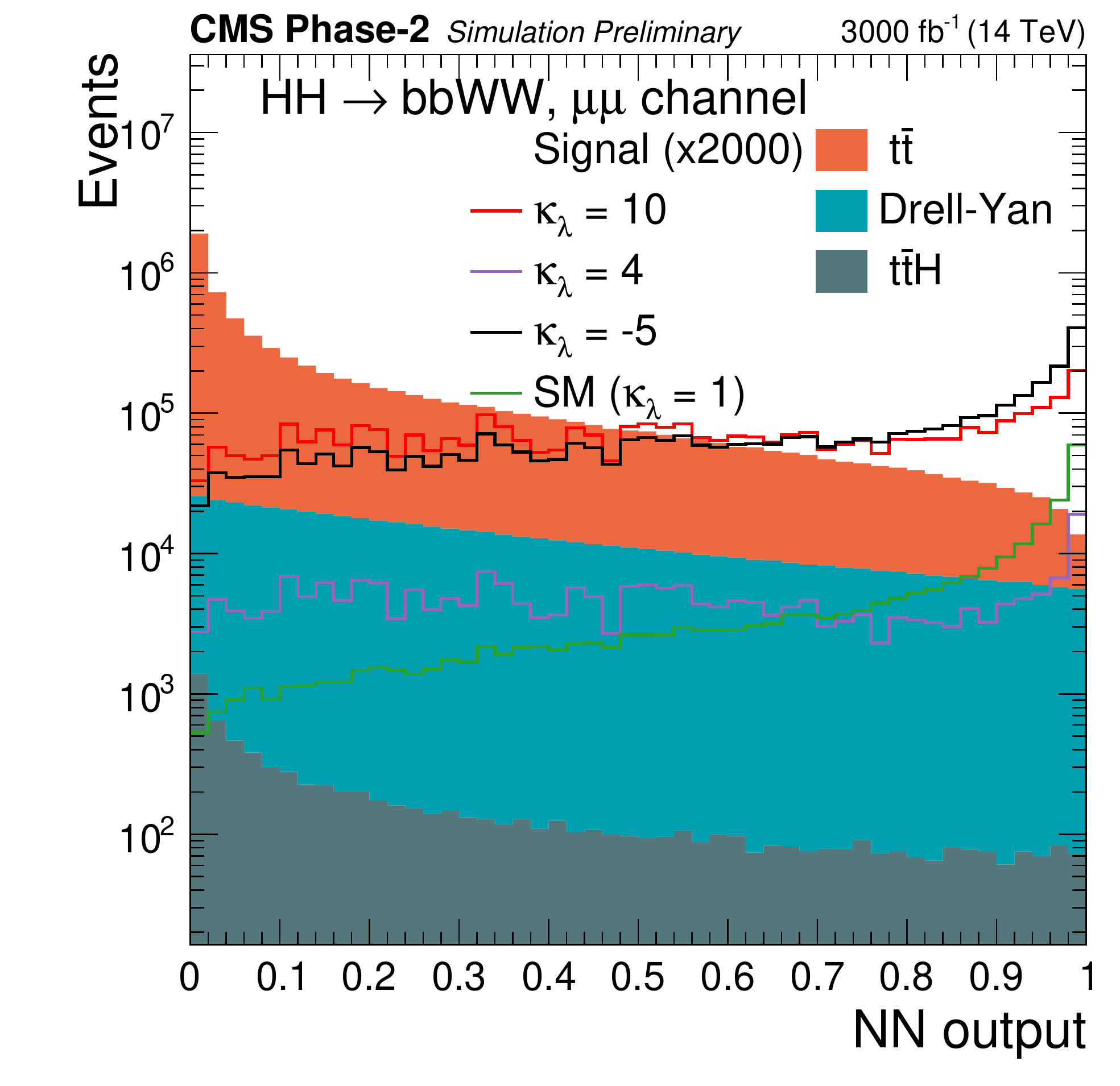}
    \includegraphics[width=0.32\textwidth]{\main/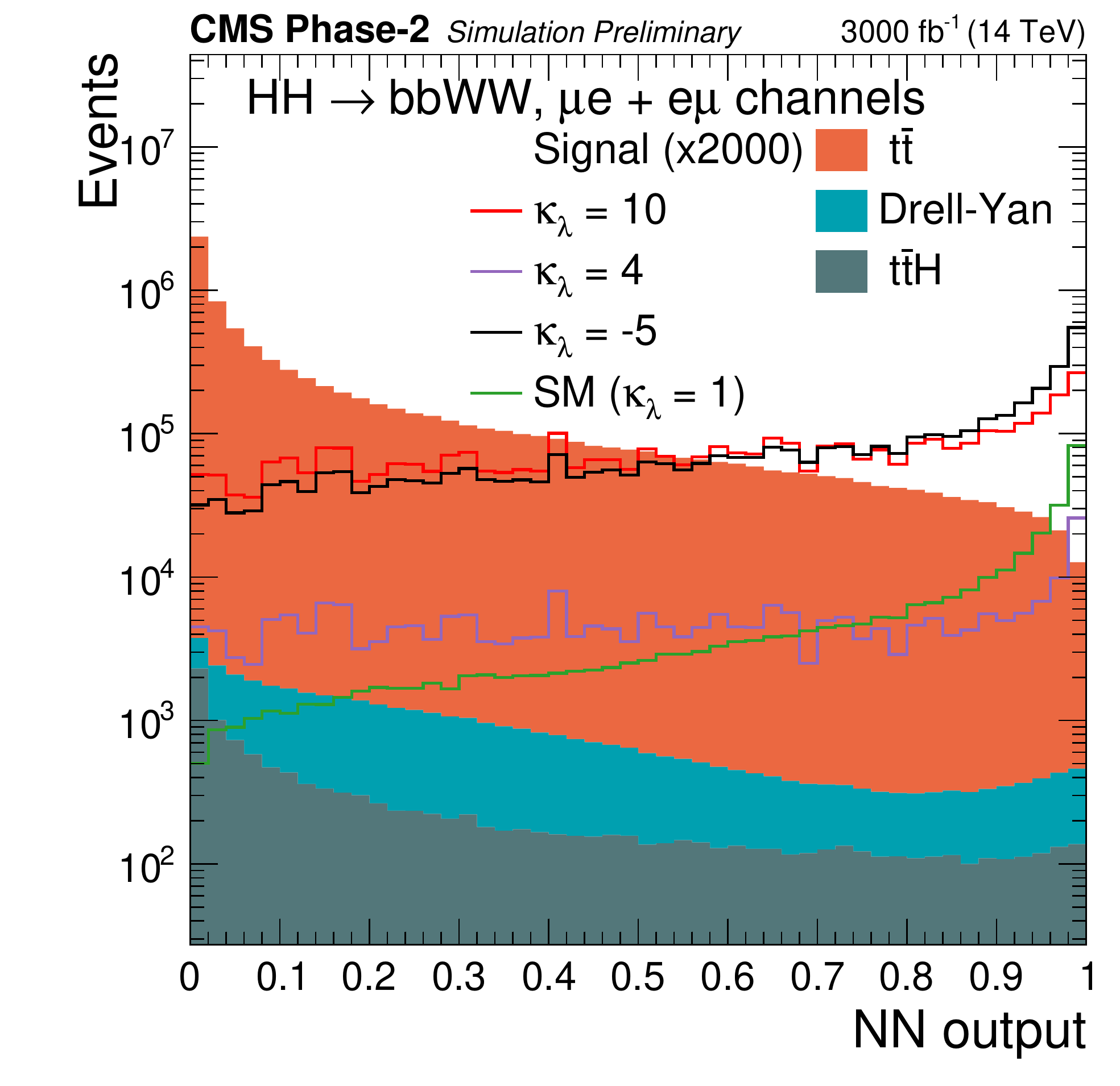}
    \caption{
      The output of the $\bbWW$ NN after the selections, evaluated in the
      $\Pe^{+}\Pe^{-}$~(left) , $\mu^{+}\mu^{-}$~(middle),
      $\Pe^{\pm}\mu^{\mp}$~(right) channels.
    }
    \label{sec3:CMSHH:fig:bbWW_events}
  \end{center}
\end{figure}

\paragraph{The $\HH\to\bbZZ\to \PQb\PQb 4\ell$ channel}

The \HH searches at the LHC have so far focused on final states with a sizeable branching ratio because of the small cross section of this process.
The HL-LHC will open the possibility to study rare but clean decay channels thanks to the large dataset available.
The $\bbZZ(4\ell)$ channel, that is investigated in this work, benefits from the clean four lepton signature to clearly identify signal events in the busy pileup environment of the HL-LHC.

Events are required to have at least four identified and isolated (isolation < 0.7) muons (electrons) with $\pt > 5(7)\GeV$ and $|\eta|< 2.8$. 
The two $\PZ$ boson candidates are formed from pairs of opposite-charge leptons
The $\PZ$ candidate with the invariant mass closest to the nominal Z mass is denoted as $\PZ_1$, while the other one is labelled as $\PZ_2$.
$\PZ$ candidates are required to have an invariant mass in the range [40, 120] \UGeV ($\PZ_1$) and [12, 120] \UGeV ($\PZ_2$), respectively.
The four leptons invariant mass is requested to be in the range [120,130] \UGeV.
At least two (but not more than three) b-tagged jets are also required to be  present  and have an invariant mass.
The jet pair is required to have an invariant mass in the range [80, 160] \UGeV and an angular distance between the 2 jets between 0.5 and 2.3.
The number of events thus selected are used to look for the presence of a signal on top of the background processes, mostly constituted of single Higgs boson production in the $4\ell$ final state.
The distribution of the four lepton invariant mass is shown in  Fig,~\ref{sec3:CMSHH:fig:bbZZ_events}.

\begin{figure}[!htb]
  \begin{center}
    \includegraphics[width=0.49\textwidth]{\main/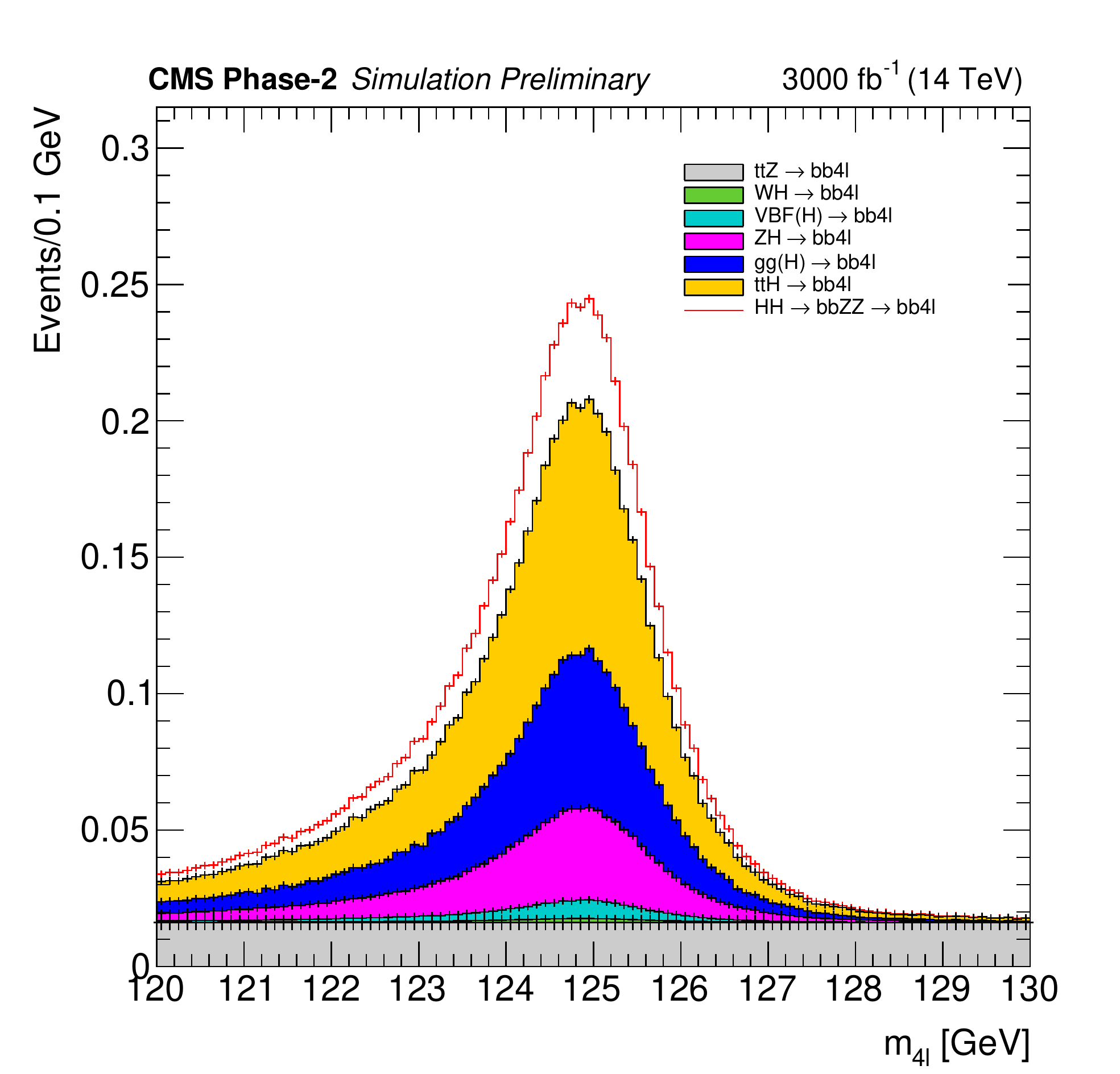}
    \caption{Invariant mass distribution of the four leptons selected at the end of the analysis for the $\PQb\PQb 4\ell$ final state.}
    \label{sec3:CMSHH:fig:bbZZ_events}
  \end{center}
\end{figure}

\paragraph{Combined results}

The five decay channels are combined statistically assuming the SM Higgs boson branching fractions.
Assuming the presence of a signal with the properties predicted by the SM, its total expected  significance is $2.6\sigma$.
If instead the background only hypothesis is assumed, an expected upper limit on the SM \HH signal cross section can be set to 0.77 times the SM prediction.
The contributions from the five decay channels and the combined expected sensitivities are reported in Tab.~\ref{sec3:cMSHH:tab:comb}.

\begin{table}[htp]
\centering
\caption{Significance, upper limit at the 95\% confidence level, and uncertainty on the signal strength $\mu$ of the SM \HH signal for the five channels studied and their combination. Numbers are reported both considering statistical and systematic uncertainties (Stat. + syst.), and neglecting the latter (Stat. only).}
\label{sec3:cMSHH:tab:comb}
\begin{tabular}{l c c c c l@{}l l@{}l}
    \toprule
    \multirow{2}{*}{Channel} & \multicolumn{2}{c}{Significance} & \multicolumn{2}{c}{95\% CL limit on $\sigma_{\HH}/\sigma_{\HH}^\text{SM}$} & \multicolumn{4}{c}{Measured signal strength $\mu$} \\
                             & Stat. + syst.  & Stat. only &  Stat. + syst. & Stat. only & \multicolumn{2}{l}{Stat. + syst.}  & \multicolumn{2}{l}{Stat. only}\\
    \cmidrule(lr){2-3}
    \cmidrule(lr){4-5}
    \cmidrule(lr){6-9}
    \bbbb                     & 0.95 & 1.2  & 2.1  & 1.6 & 1.00 & ${}_{-1.04}^{+1.08}$ & 1.00 & ${}_{-0.83}^{+0.84}$ \\[0.4em]
    \bbtt                     & 1.4  & 1.6  & 1.4  & 1.3 & 1.00 & ${}_{-0.71}^{+0.73}$ & 1.00 & ${}_{-0.64}^{+0.65}$ \\[0.4em]
    $\bbWW(\ell\nu\ell\nu)$   & 0.56 & 0.59 & 3.5  & 3.3 & 1.00 & ${}_{-1.8}^{+1.8}$   & 1.00 & ${}_{-1.7}^{+1.7}$ \\[0.4em]
    \bbgg                     & 1.8  & 1.8  & 1.1  & 1.1 & 1.00 & ${}_{-0.56}^{+0.61}$ & 1.00 & ${}_{-0.53}^{+0.56}$ \\[0.4em]
    $\bbZZ(\ell\ell\ell\ell)$ & 0.37 & 0.37 & 6.6  & 6.5 & 1.0 & ${}_{-2.5}^{+3.2}$    & 1.0 & ${}_{-2.5}^{+3.2}$ \\[0.4em]
    \midrule
    Combination               & 2.6  & 2.8  & 0.77 & 0.71 & 1.00 & ${}_{-0.39}^{+0.41}$ & 1.00 & ${}_{-0.36}^{+0.36}$\\
    \bottomrule
\end{tabular}
\end{table} 



Prospects for the measurement of the Higgs boson self coupling are also studied.
Under the assumption that no \HH signal exists, 95\% CL upper limits on the SM \HH production cross section are derived as a function $\kappa_\lambda = \lambdahhh/\lambdahhh^\text{SM}$, where $\lambdahhh^\text{SM}$ denotes the SM prediction. 
The result is illustrated in Fig.~\ref{sec3:CMSHH:fig:comb_plots}.
A variation of the excluded cross section, directly related to changes in the \HH kinematic properties, can be observed as a function of $\kappa_\lambda$.


Assuming instead that a \HH signal exists with the properties predicted by the SM, prospects for the measurement of the \lambdahhh are derived.
The scan of the likelihood as a function of the \kapl coupling is shown in Fig.~\ref{sec3:CMSHH:fig:comb_plots}.
The projected confidence interval on this coupling corresponds to $[0.35, 1.9]$ at the 68\% CL and to $[-0.18, 3.6]$ at the 95\% CL.
The peculiar likelihood function structure, characterised by two local minimums, is related to the dependence of the total cross section and \HH kinematic properties on $\kapl$, while the relative height of the two minimums depends to the capability of the analyses to access differential $m_{\HH}$ information.
The degeneracy of the second minimum is largely removed thanks to the $\bbgg$ analysis and its $m_{\HH}$ categorisation.


\begin{figure}[htb]
  \centering
   \includegraphics[width=0.48\textwidth]{\main/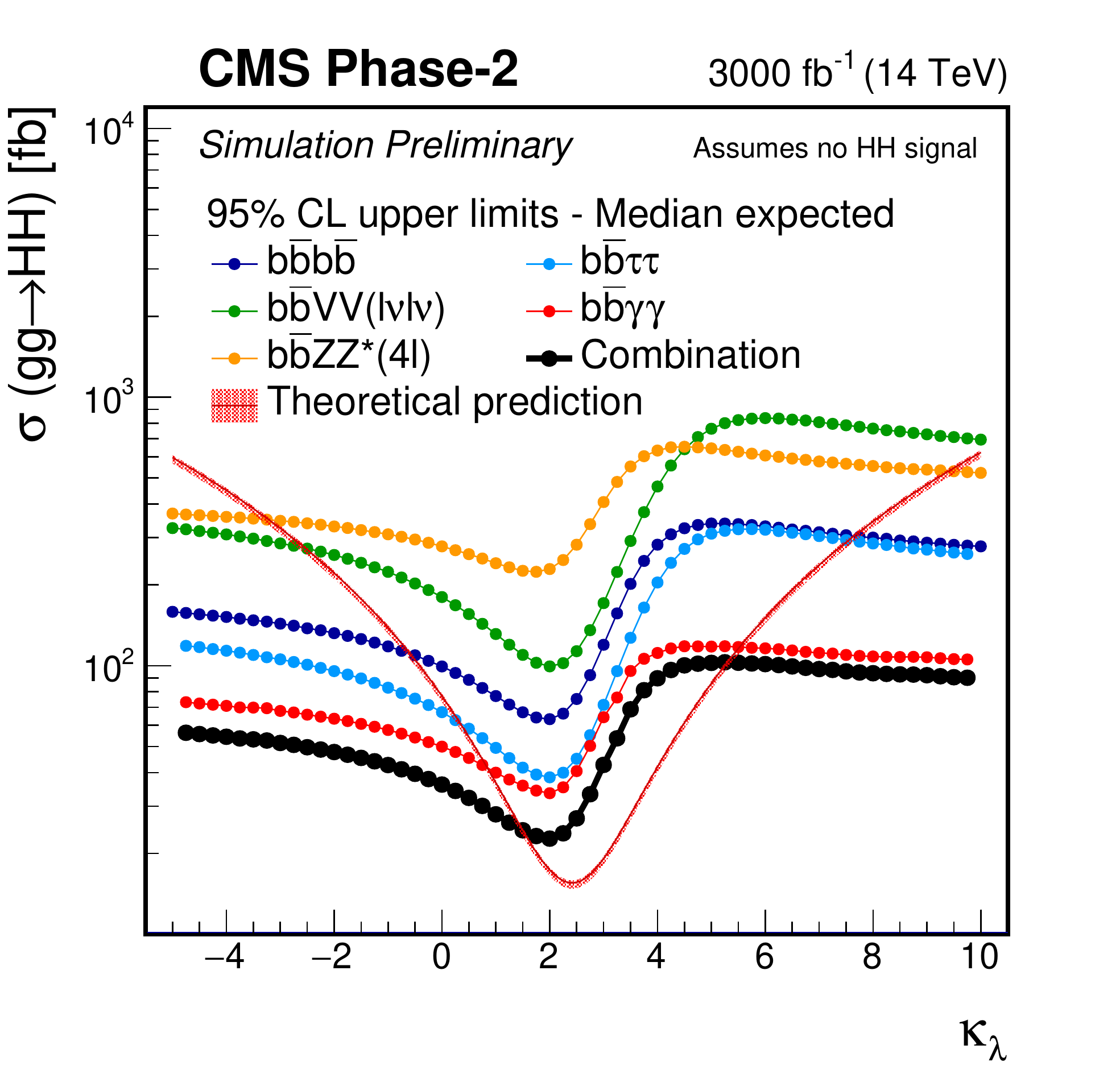}
    \includegraphics[width=0.48\textwidth]{\main/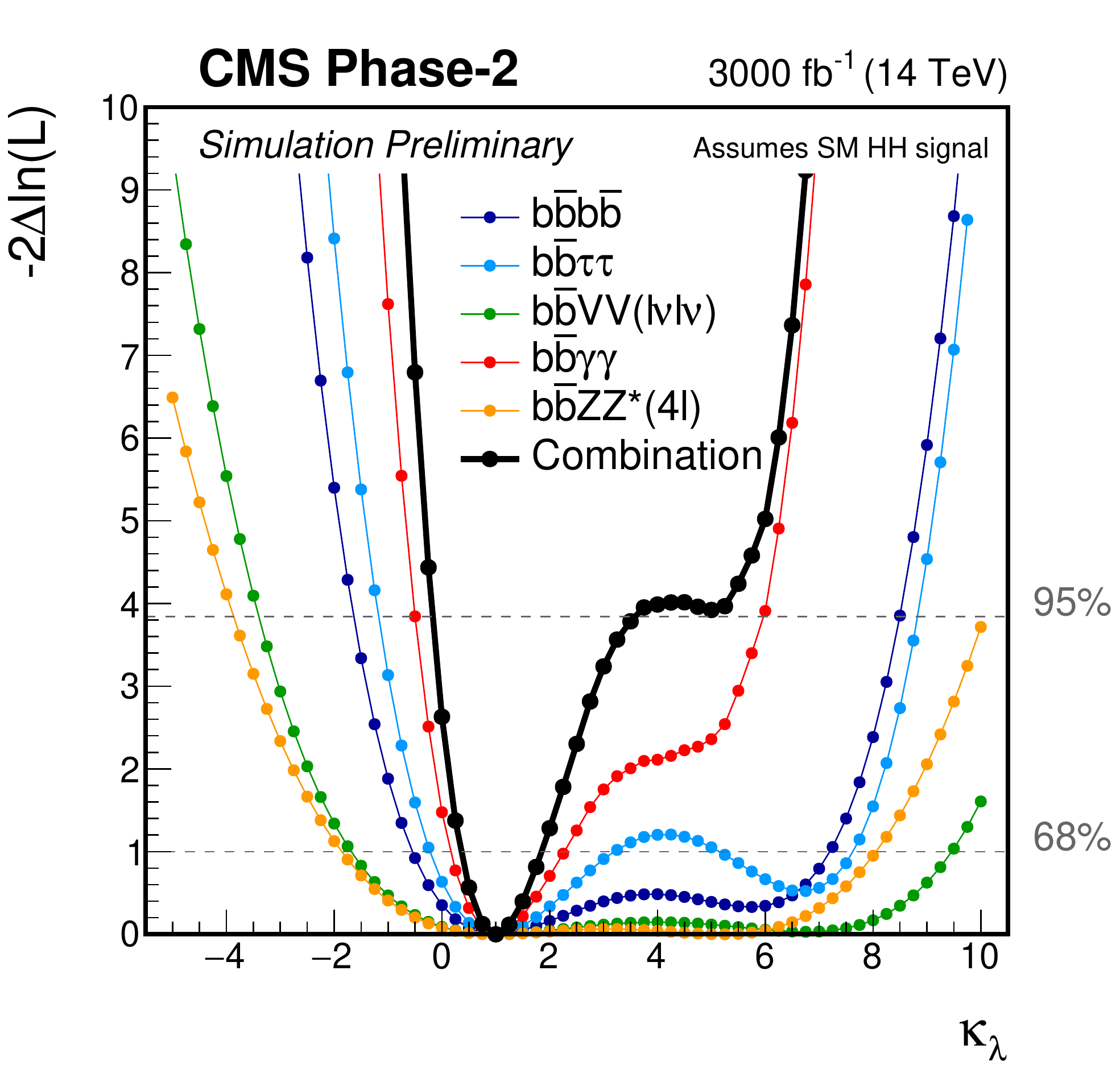}
    \caption{Left: upper limit at the 95\% CL on the \HH production cross section as a function of $\kapl = \lambdahhh/\lambdahhh^\text{SM}$. The red band indicated the theoretical production cross section. Right: expected likelihood scan as a function of $\kappa_\lambda = \lambdahhh/\lambdahhh^\text{SM}$. In both figures the results are shown separately for the five decay channels studied and for their combination.}
    \label{sec3:CMSHH:fig:comb_plots}
\end{figure}

\asubsubsection[L. Cadamuro, E. Petit, D. Wardrope]{Combination of measurements}
\label{sec:HH_Combination}

A simple combination is performed of the measurements from the ATLAS and CMS collaborations described in Sections~\ref{sec:HH_ATLAS} and \ref{sec:HH_CMS}.
The channels are treated as uncorrelated, in particular because the systematic uncertainties that we could expect to be correlated between the experiments, such as the theory uncertainties and the luminosity uncertainty, have little impact on the individual results. Since the measurements in the $HH \rightarrow b\bar{b}VV(ll\nu\nu)$ and $HH \rightarrow b\bar{b}ZZ(4l)$ are only performed by the CMS experiment, the likelihoods for those two channels are scaled to 6000$\ifb$ in the combination.
The significances are added in quadrature and the negative-log-likelihood are simply added together. A summary of the different expected significances, as well as the combination, are shown in Table~\ref{tab:comb_significance}. A combined significance of 4 standard deviation can be achieved with all systematic uncertainties included.

\begin{table}[htb!]
\begin{center}
\begin{tabular}{lcccc} \toprule
 & \multicolumn{2}{c}{\textbf{Statistical-only}} & \multicolumn{2}{c}{\textbf{Statistical + Systematic}}\\
 & ATLAS & CMS & ATLAS & CMS \\
\hline
$HH \rightarrow b\bar{b}b\bar{b}$ & 1.4 & 1.2 & 0.61 & 0.95 \\
$HH \rightarrow b\bar{b}\tau\tau$ & 2.5 & 1.6 & 2.1 & 1.4 \\
$HH \rightarrow b\bar{b}\gamma\gamma$ & 2.1 & 1.8 & 2.0 & 1.8 \\
$HH \rightarrow b\bar{b}VV(ll\nu\nu)$ & - & 0.59 & - & 0.56 \\
$HH \rightarrow b\bar{b}ZZ(4l)$ & - & 0.37 & - & 0.37 \\
combined &  3.5 & 2.8 & 3.0 & 2.6 \\
\hline
 &  \multicolumn{2}{c}{Combined} & \multicolumn{2}{c}{Combined}\\
 &  \multicolumn{2}{c}{4.5} & \multicolumn{2}{c}{4.0}\\
\bottomrule
\end{tabular}
\end{center}
\caption{Significance in standard deviations of the individual channels as well as their combination.}
\label{tab:comb_significance}
\end{table}

Comparisons of the minimum negative-log-likelihoods for ATLAS and CMS are shown in Figure~\ref{fig:comb_HH_experiment}. In those plots the likelihoods for the $HH \rightarrow b\bar{b}VV(ll\nu\nu)$ and $HH \rightarrow b\bar{b}ZZ(4l)$ channels are not scaled to 6000$\ifb$.
A difference of shape between the two experiments can be seen around the second minimum. This difference comes mainly from the $HH \rightarrow b\bar{b}\gamma\gamma$ channel as illustrated in Figure~\ref{fig:comb_HH_experiment_b}. In this channel both experiment use categories of the $m_{HH}$ distributions. But for ATLAS the analysis was optimised to increase the significance of the SM signal so the low values of the $m_{HH}$ distribution are cut by the selection cuts, while for CMS a category of events with low values of $m_{HH}$ is very powerful to remove the second minimum, while having no effect on the SM signal.
The lower precision on $\kappa_{\lambda}$ is slightly better for CMS thanks to the contribution of the $HH \rightarrow b\bar{b}b\bar{b}$ channel, as well as the $HH \rightarrow b\bar{b}VV(ll\nu\nu)$ and $HH \rightarrow b\bar{b}ZZ(4l)$ ones, while the higher precision on $\kappa_{\lambda}$ is similar between the two experiments.

\begin{figure}[!htb]
\centering 
\subfloat[]{\label{fig:comb_HH_experiment_a}\includegraphics[width=0.5\textwidth]{\main/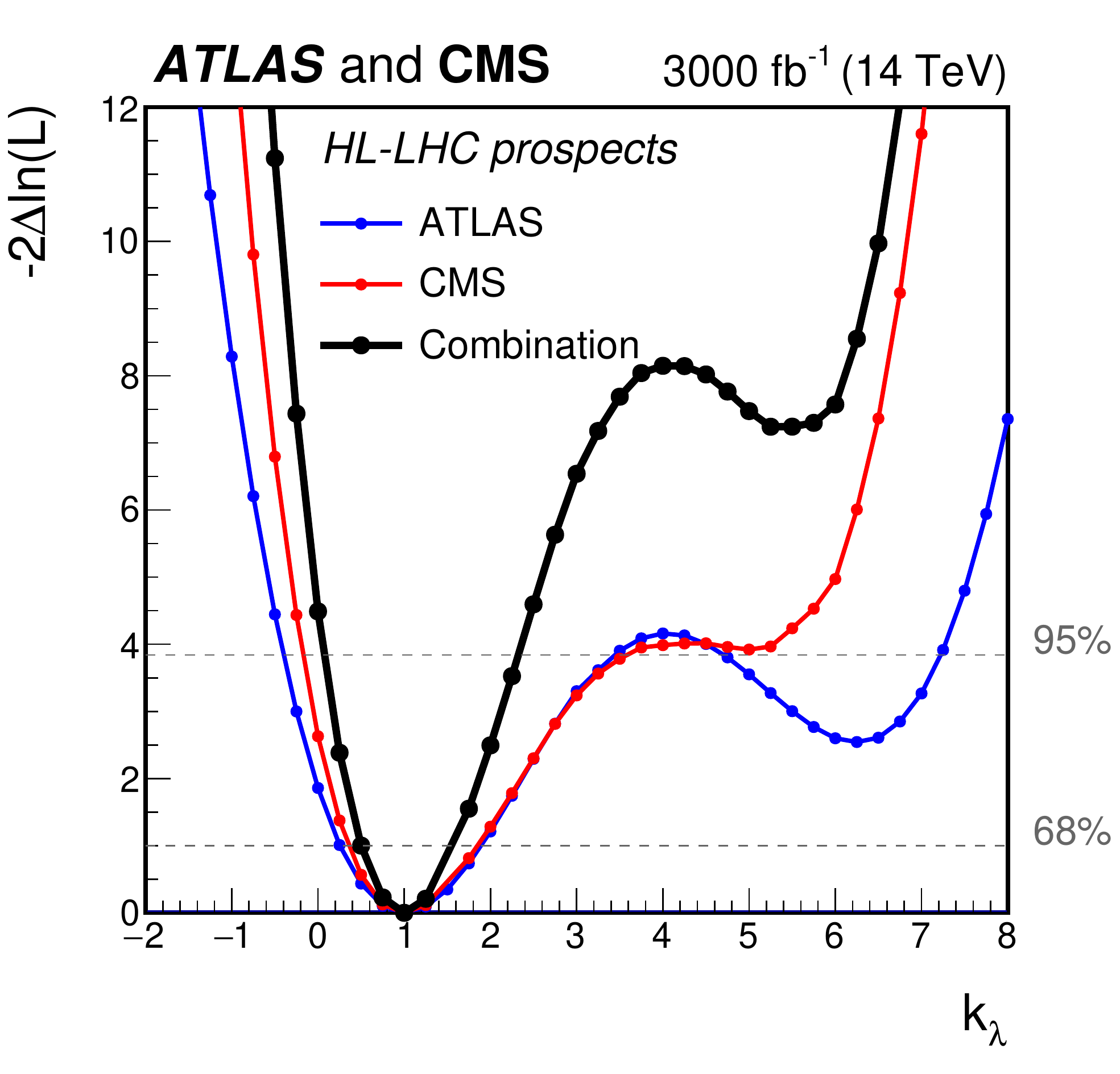}} 
\subfloat[]{\label{fig:comb_HH_experiment_b}\includegraphics[width=0.5\textwidth]{\main/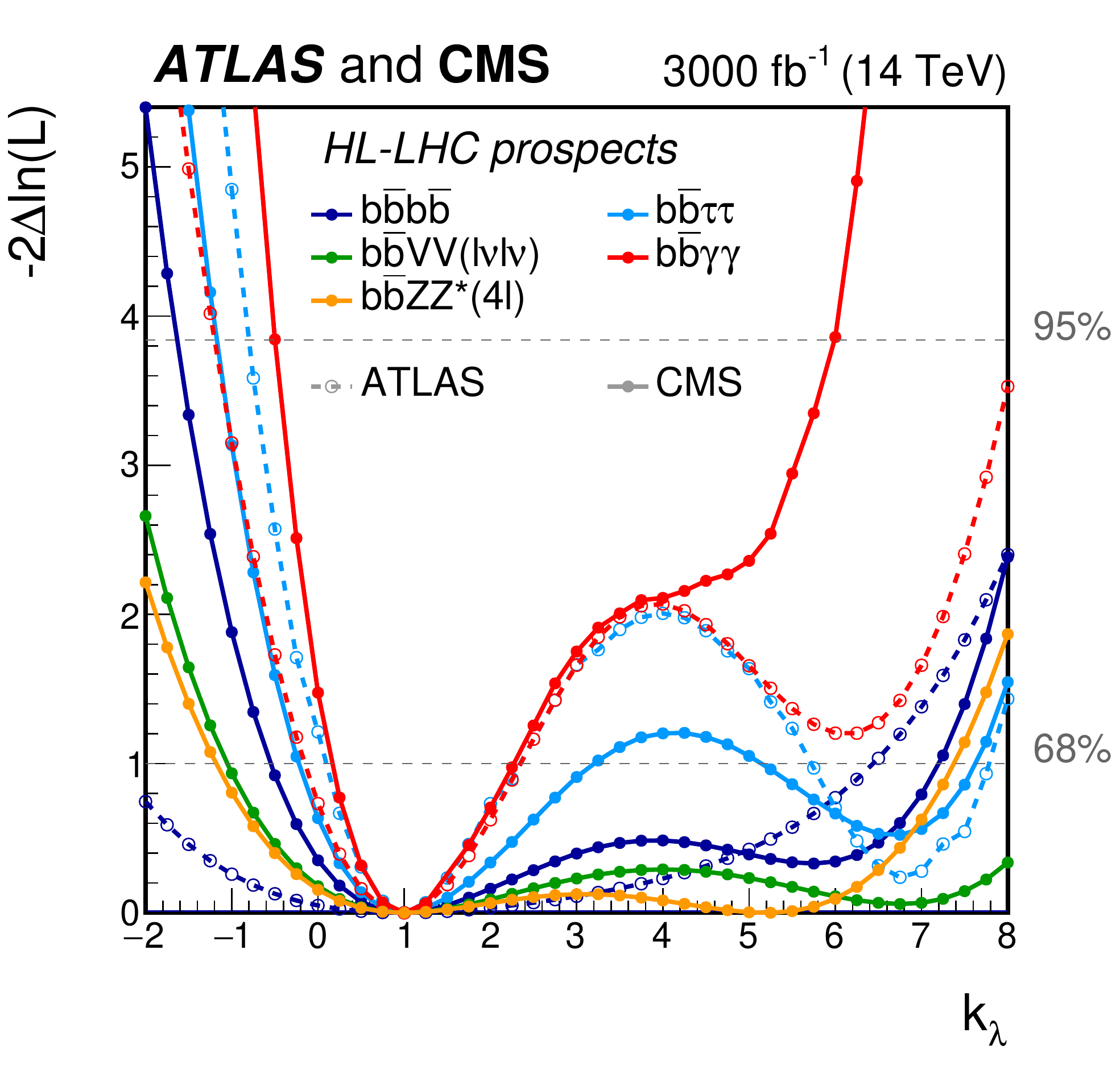}} 
\caption{(a) Minimum negative-log-likelihood as a function of \kl, calculated by performing a conditional signal+background fit to the background and SM signal. (a) The black line corresponds to the combined ATLAS and CMS results, while the blue and red lines correspond to the ATLAS and CMS standalone results respectively. (b) The different colours correspond to the different channels, the plain lines correspond to the CMS results while the dashed lines correspond to the ATLAS results.} 
\label{fig:comb_HH_experiment} 
\end{figure}

The combined minimum negative-log-likelihoods are shown in Figure~\ref{fig:comb_HH}. 
The 68\% Confidence Intervals for $\kappa_{\lambda}$ are $0.52 \leq \kl\ \leq 1.5$ and $0.57 \leq \kl\ \leq 1.5$ with and without systematic uncertainties respectively. The second minimum of the likelihood is excluded at 99.4\% CL. A summary of the 68\% CI for each channel in each experiment, as well as the combination are shown in Figure~\ref{fig:comb_HH_b}.


\begin{figure}[!htb]
\centering 
\subfloat[]{\label{fig:comb_HH_a}\includegraphics[width=0.5\textwidth]{\main/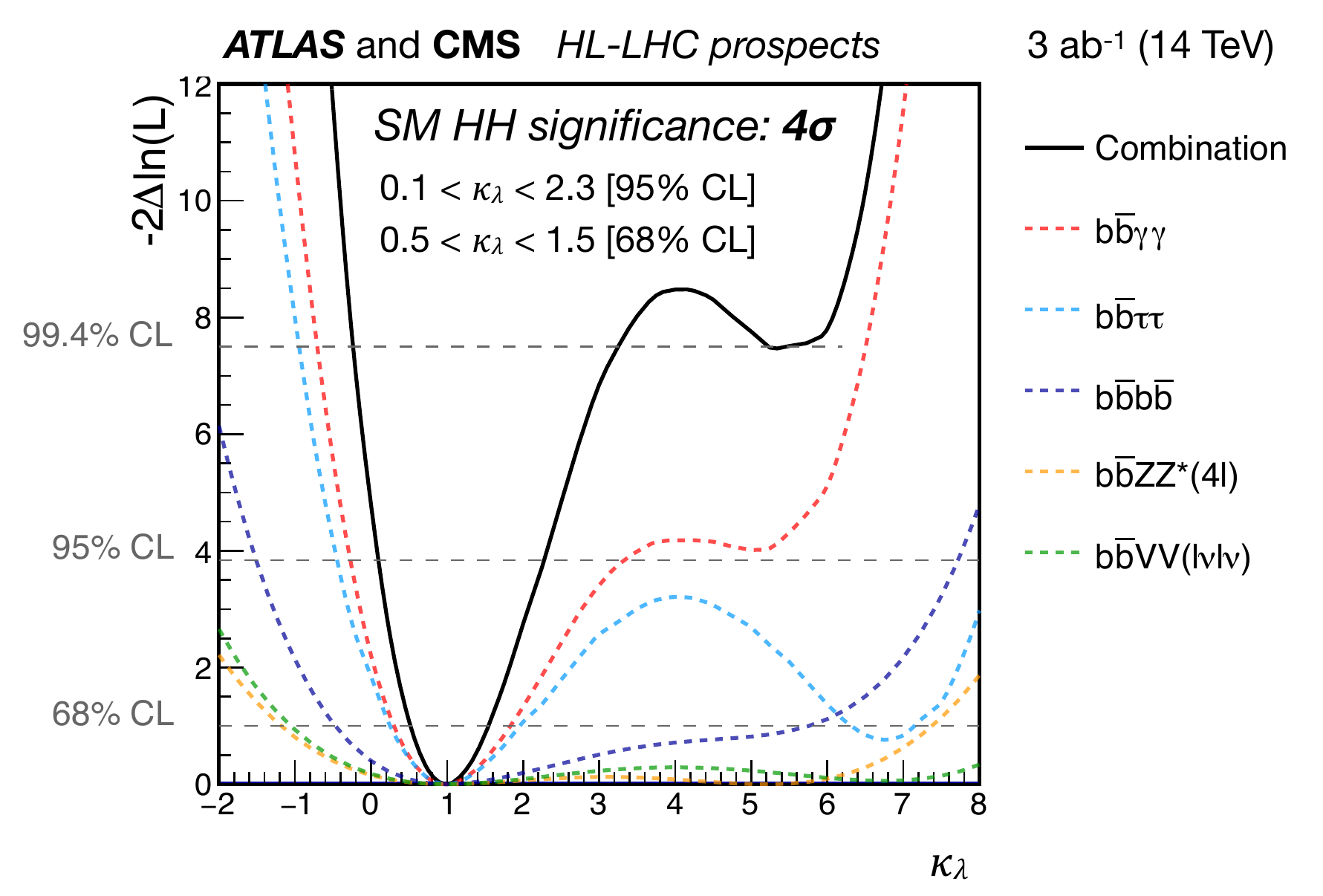}} 
\subfloat[]{\label{fig:comb_HH_b}\includegraphics[width=0.5\textwidth]{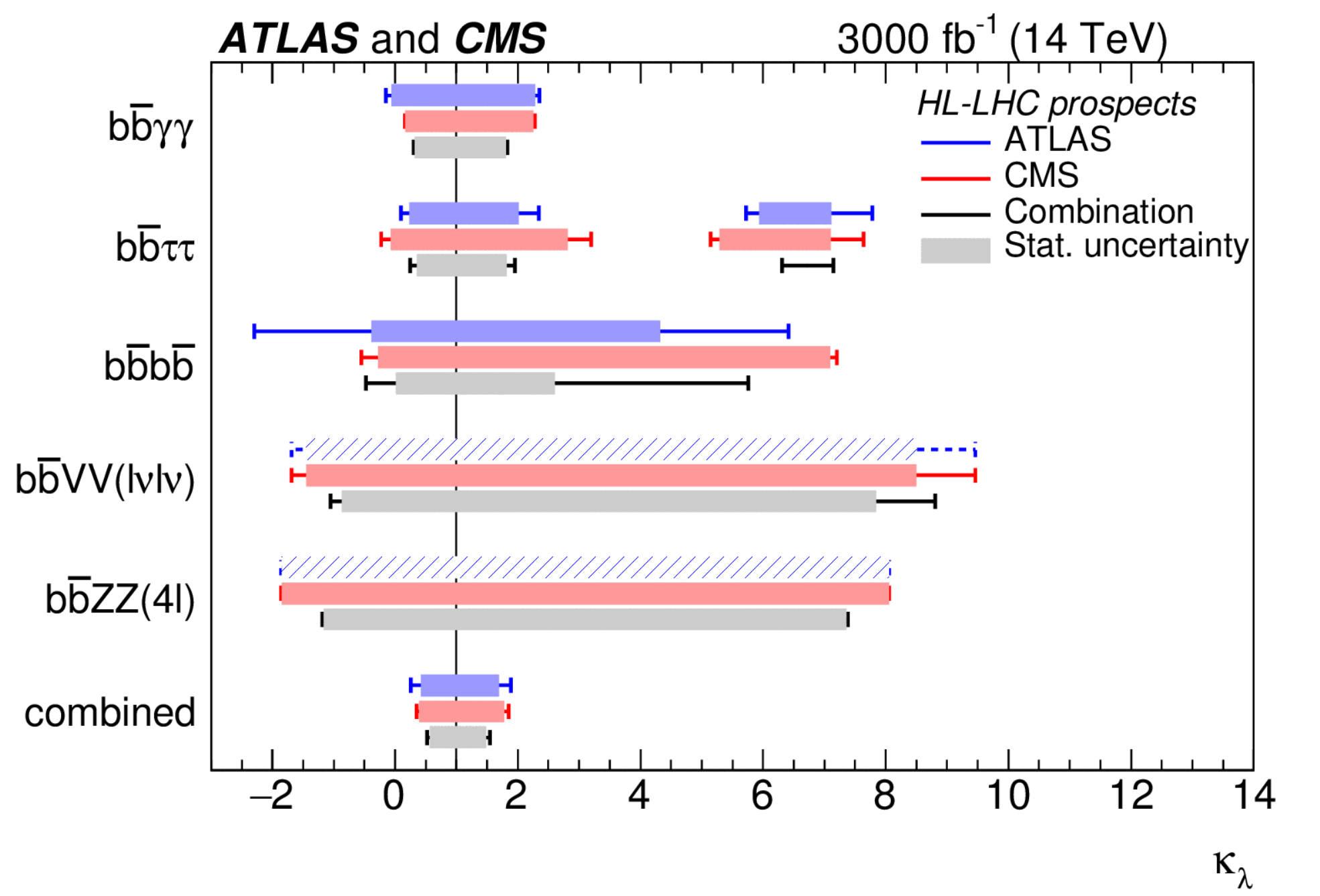}} 
\caption{(a) Minimum negative-log-likelihood as a function of \kl, calculated by performing a conditional signal+background fit to the background and SM signal. The coloured dashed lines correspond to the combined ATLAS and CMS results by channel, and the black line to their combination. The likelihoods for the $HH \rightarrow b\bar{b}VV(ll\nu\nu)$ and $HH \rightarrow b\bar{b}ZZ(4l)$ channels are scaled to 6000\fbinv.(b) Expected measured values of \kl\ for the different channels for the ATLAS in blue and the CMS experiment in red, as well as the combined measurement. The lines with error bars show the total uncertainty on each measurement while the boxes correspond to the statistical uncertainties. In the cases where the extrapolation is performed only by one experiment, same performances are assumed for the other experiment and this is indicated by a hatched bar.} 
\label{fig:comb_HH} 
\end{figure}

\asubsection{Double Higgs measurements and trilinear coupling: alternative methods}
\label{sec:HH_meas_th}

\asubsubsection[J. Han Kim, M. Kim, K. Kong, K.T. Matchev, M. Park]{Prospects for $hh \to (b \bar b)(WW^*) \to (b \bar b)( \ell^+\ell^- \nu_\ell \bar\nu_\ell)$}
In this section, we discuss the discovery prospects for double Higgs production in the $hh \to (b\bar b) (W W^*)$ channel. In order to increase sensitivity in the di-lepton channel \cite{CMS:2015nat,CMS:2017cwx,Adhikary:2017jtu}, we propose a novel kinematic method, which relies on two new kinematic functions, {\it Topness} and {\it Higgsness} \cite{Kim:2018cxf}. They characterise features of the major ($t\bar t$) background and of $hh$ events, respectively. The method also utilises two less commonly used variables, the subsystem $M_{T2}$ (or subsystem $M_2$) \cite{Lester:1999tx,Burns:2008va,Barr:2011xt} for $t\bar t$ and the subsystem $\sqrt{\hat {s}}_{min}$ (or subsystem $M_1$) \cite{Konar:2008ei,Konar:2010ma,Barr:2011xt} for $hh$ production.
For any given event, Topness \cite{Graesser:2012qy,Kim:2018cxf} quantifies the degree of consistency to di-lepton $t\bar t$ production, where there are 6 unknowns (the three-momenta of the two neutrinos, $\vec p_{\nu}$ and $\vec p_{\bar\nu}$) and four on-shell constraints, for
$m_t$, $m_{\bar t}$, $m_{W^+}$ and $m_{W^-}$, respectively. The neutrino momenta can be fixed by minimising the quantity 
\begin{eqnarray}
\chi^2_{ij} \equiv \min_{\tiny \mptvec = \vec p_{\nu T} + \vec p_{ \bar\nu T}}  \left [ 
\frac{\left ( m^2_{b_i \ell^+ \nu} - m^2_t \right )^2}{\sigma_t^4}    +
\frac{\left ( m^2_{\ell^+ \nu} - m^2_W \right )^2}{\sigma_W^4}   \label{eq:tt}  
 + \frac{\left ( m^2_{b_j \ell^- \bar \nu} - m^2_t \right )^2}{\sigma_t^4}  +
\frac{\left ( m^2_{\ell^- \bar\nu} - m^2_W \right )^2}{\sigma_W^4}   \right ]  , 
\end{eqnarray}
subject to the missing transverse momentum constraint, $ \mptvec = \vec p_{\nu T} + \vec p_{ \bar\nu T}$. 
Since there is a twofold ambiguity in the paring of a $b$-quark and a lepton, we define {\it Topness} as the smaller of the two $\chi^2$s,
\begin{eqnarray}
T &\equiv&  { \min} \left ( \chi^2_{12} \, , \, \chi^2_{21} \right ) \, .
\end{eqnarray}

In double Higgs production, the two $b$-quarks arise from a Higgs decay ($h \to b \bar b$), and therefore their invariant mass $m_{bb}$
can be used as a first cut to enhance the signal sensitivity. 
For the decay of the other Higgs boson, $h \to W W^* \to \ell^+ \ell^- \nu \bar\nu$, we define {\it Higgsness} \cite{Kim:2018cxf} as follows: 
\begin{eqnarray}
H &\equiv&    {\rm min} \left [
 \frac{\left ( m^2_{\ell^+\ell^-\nu \bar\nu} - m^2_h \right )^2}{\sigma_{h_\ell}^4}    \right. 
 + \frac{ \left ( m_{\nu  \bar\nu}^2 -  m_{\nu\bar\nu, peak}^2 \right )^2}{ \sigma^4_{\nu}}
 \label{eq:hww}   \\
 && \hspace*{-1cm}  \left. + {\min} \left ( 
\frac{\left ( m^2_{\ell^+ \nu } - m^2_W \right )^2}{\sigma_W^4} + 
\frac{\left ( m^2_{\ell^- \bar \nu} - m^2_{W^*, peak} \right )^2}{\sigma_{W^*}^4}  \, , 
\frac{\left ( m^2_{\ell^- \bar \nu} - m^2_W \right )^2}{\sigma_W^4} + 
\frac{\left ( m^2_{\ell^+ \nu} - m^2_{W^*, peak} \right )^2}{\sigma_{W^*}^4}  
\right )  \right ]   \, , \nonumber
\end{eqnarray}
where $m_{W^*}$ is the invariant mass of the lepton-neutrino pair which resulted from the off-shell $W$. 
It satisfies $0 \leq m_{W^*} \leq m_h- m_W$ and $m_{W^*}^{peak} = \frac{1}{\sqrt{3}} \sqrt{ 2 \left ( m_h^2 + m_W^2 \right ) - \sqrt{m_h^4 + 14 m_h^2 m_W^2 + m_W^4}}$ is the peak in the $m_{W^*}$ distribution.
$m_{\nu\bar\nu}^{peak} = m_{\ell\ell}^{peak} \approx 30$ \UGeV is the location of the peak in the $\frac{\textrm{d}\sigma}{\textrm{d} m_{\nu\bar\nu}}$ or $\frac{\textrm{d}\sigma}{\textrm{d} m_{\ell\ell}}$ distribution \cite{Kim:2018cxf,Cho:2012er}.
%
%
%
%
%
%

The $\sigma$ values in Eqs.\,(\ref{eq:tt}) and (\ref{eq:hww}) result from the experimental uncertainties and intrinsic particle widths. In principle, they can be treated as free parameters and tuned using a neutral network (NN), a boosted decision tree (BDT), etc. In our numerical study, we use $\sigma_t=5$ \UGeV, $\sigma_W=5$ \UGeV, $\sigma_{W^*}=5$ \UGeV, $\sigma_{h_\ell}=2$ \UGeV, and $\sigma_\nu = 10 $ \UGeV. 
The main contribution in Eq.\,(\ref{eq:hww}) comes from the on-shell conditions for the Higgs and the $W$, while the effects of the invariant mass of the two neutrinos and the off-shell $W$ are minor. 

Along with Higgsness and Topness, we adopt the subsystem $\hat{s}_{min}^{(\ell\ell)}$ for $h \to W^\pm W^{*\mp}\to\ell^+\ell^- \nu \bar \nu$ \cite{Konar:2008ei,Konar:2010ma} and the subsystem $M_{T2}$ for the $b \bar b$ system ($M_{T2}^{(b)}$) and the lepton system ($M_{T2}^{(\ell)}$) \cite{Burns:2008va}. 
The variable $ \hat{s}_{min}^{({\rm v})} $ is defined as $ \hat{s}_{min}^{({\rm v})} = m_{{\rm v}}^2 + 2 \left ( \sqrt{ |\vec P_{T}^{\rm v} |^2 + m_{\rm v}^2 }  |\mptvec| - \vec P_{T}^{\rm v} \cdot \mptvec \right )$ \cite{Konar:2008ei,Konar:2010ma,Barr:2011xt}, where $({\rm v})$ represents a set of visible particles under consideration, while $m_{\rm v}$ and $\vec P_{T}^{\rm v}$ are their invariant mass and transverse momentum, respectively. It provides the minimum value of the Mandelstam invariant mass $\hat s$ which is consistent with the observed visible 4-momentum vector.
The $M_{T2}$ is defined as 
$M_{T2} (\tilde m) \equiv \min\left\{\max\left[M_{TP_1}(\vec{p}_{\nu T},\tilde m),\;M_{TP_2} (\vec{p}_{\bar\nu T},\tilde m)\right] \right\}$ where $\tilde m$  is the test mass for the daughter particle and the minimisation over the transverse masses of the parent particles $M_{TP_i}$ ($i=1, 2$) is performed over the transverse neutrino momenta $\vec{p}_{\nu T}$ and $\vec{p}_{\bar\nu T}$ subject to the 
$\mptvec $ constraint  \cite{Lester:1999tx,Burns:2008va,Barr:2011xt,Debnath:2017ktz,Konar:2009wn,Konar:2009qr,Cho:2007qv}.

\begin{figure*}[t]
\centering
\includegraphics[width=5.77cm]{\main/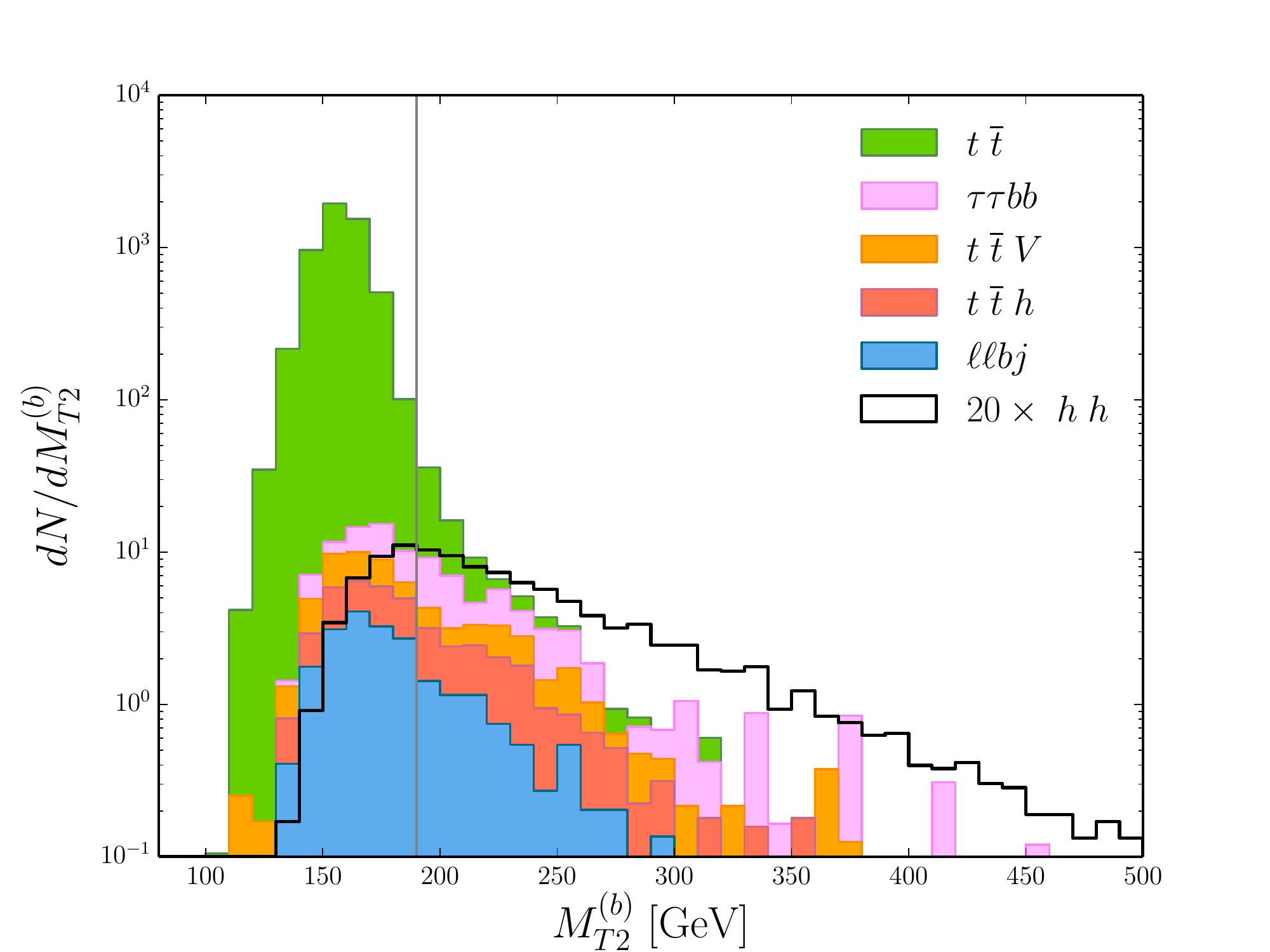} \hspace*{-0.65cm}
\includegraphics[width=5.77cm]{\main/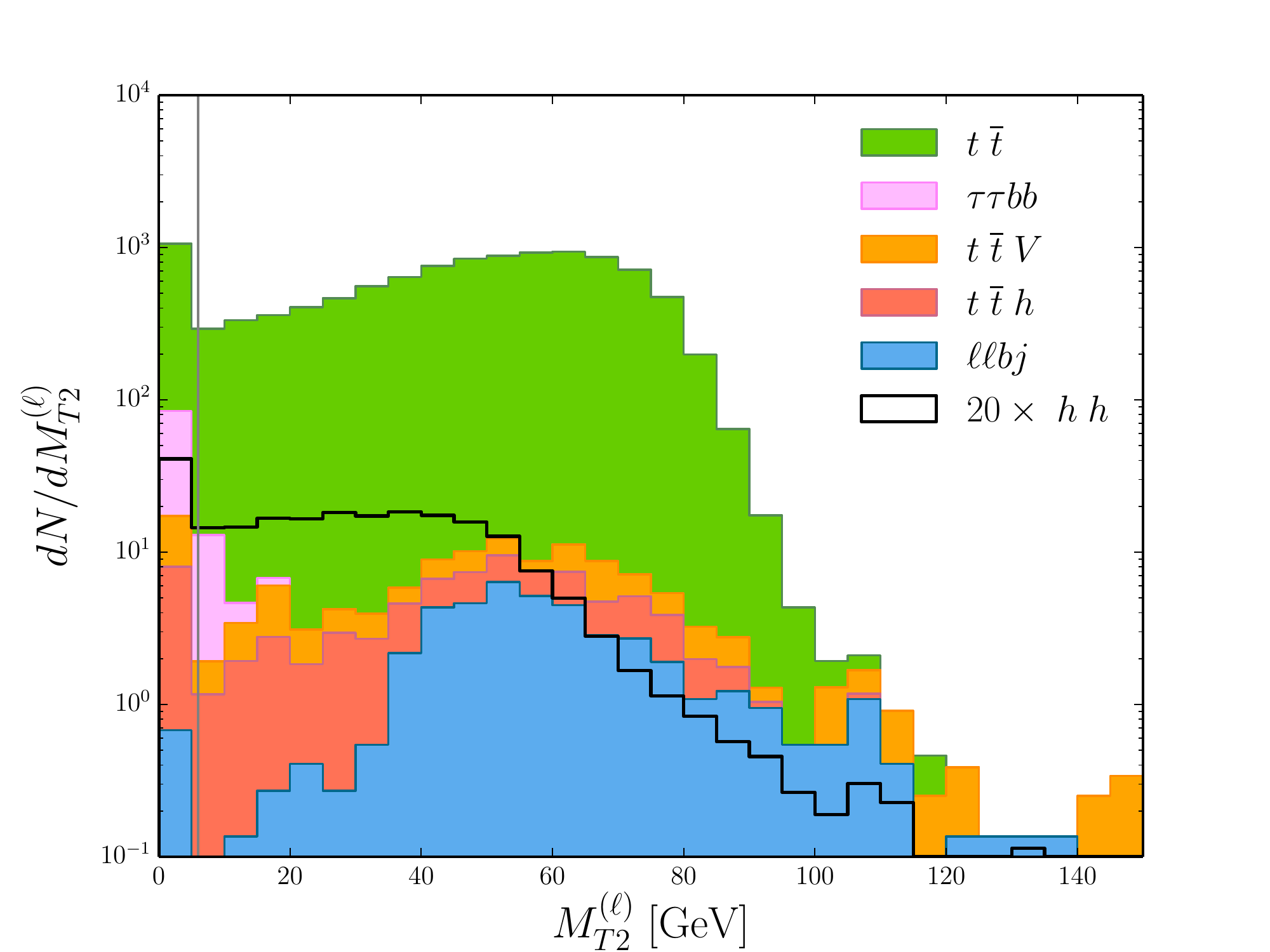}   \hspace*{-0.65cm}
\includegraphics[width=5.4cm]{\main/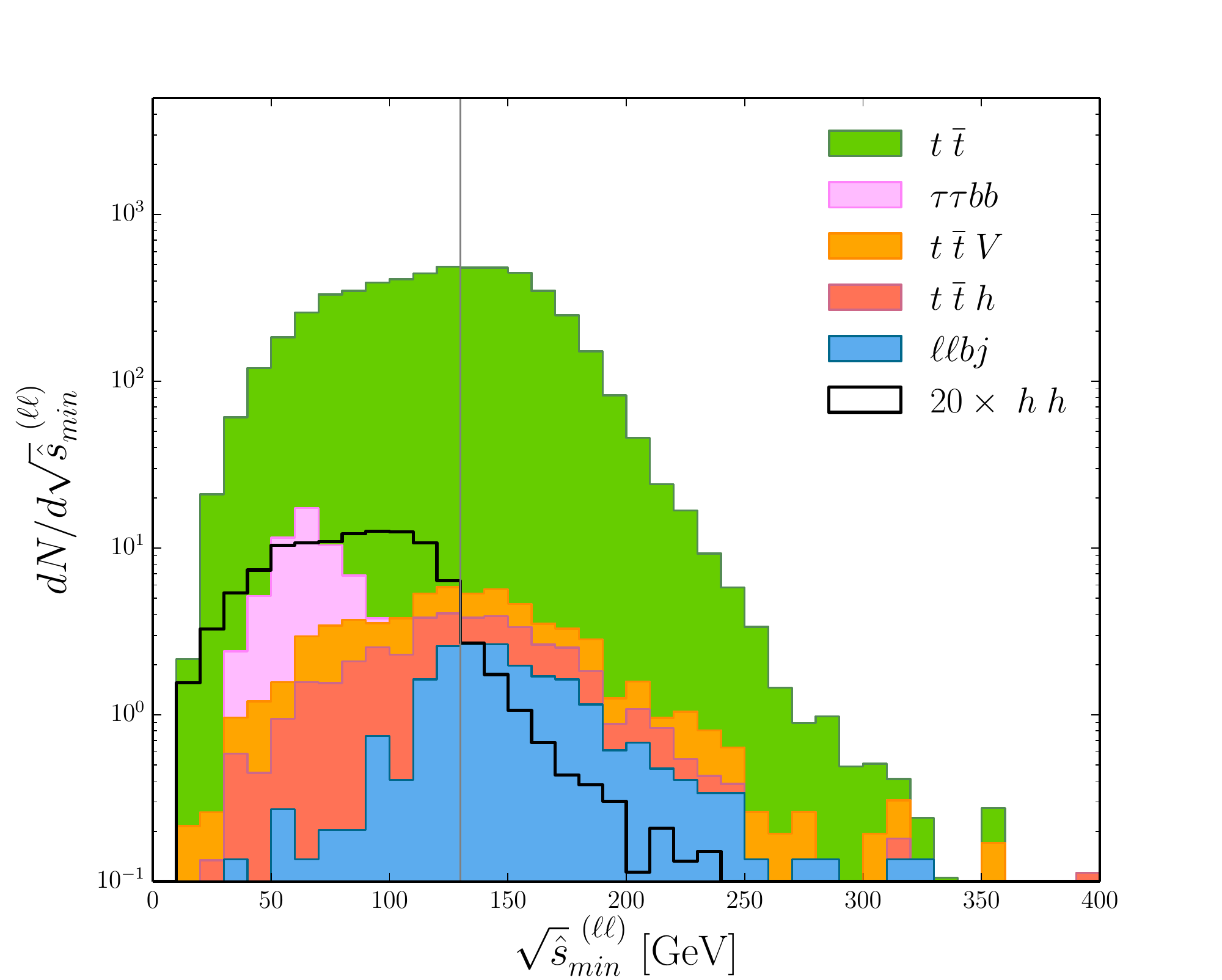} 
\caption{\label{fig:newcuts} 
Distributions for signal ($hh$) and all backgrounds ($t \bar t$, $t\bar t h$, $t \bar t V$, $\ell\ell b j$ and $\tau\tau b b$) for $M_{T2}^{(b)}$, $M_{T2}^{(\ell)}$ and $\sqrt{\hat{s}}_{min}^{(\ell\ell)}$ after loose baseline selection cuts defined in Ref. \cite{Kim:2018cxf}. 
The vertical lines at $M_{T2}^{(b)} = 190$ \UGeV, $M_{T2}^{(\ell)}= 6$ \UGeV and $\sqrt{\hat{s}}_{min}^{(\ell\ell)}=130$ \UGeV mark the optimised cuts.}
\end{figure*}
\begin{figure}[t]
\centering
\includegraphics[width=5.4cm]{\main/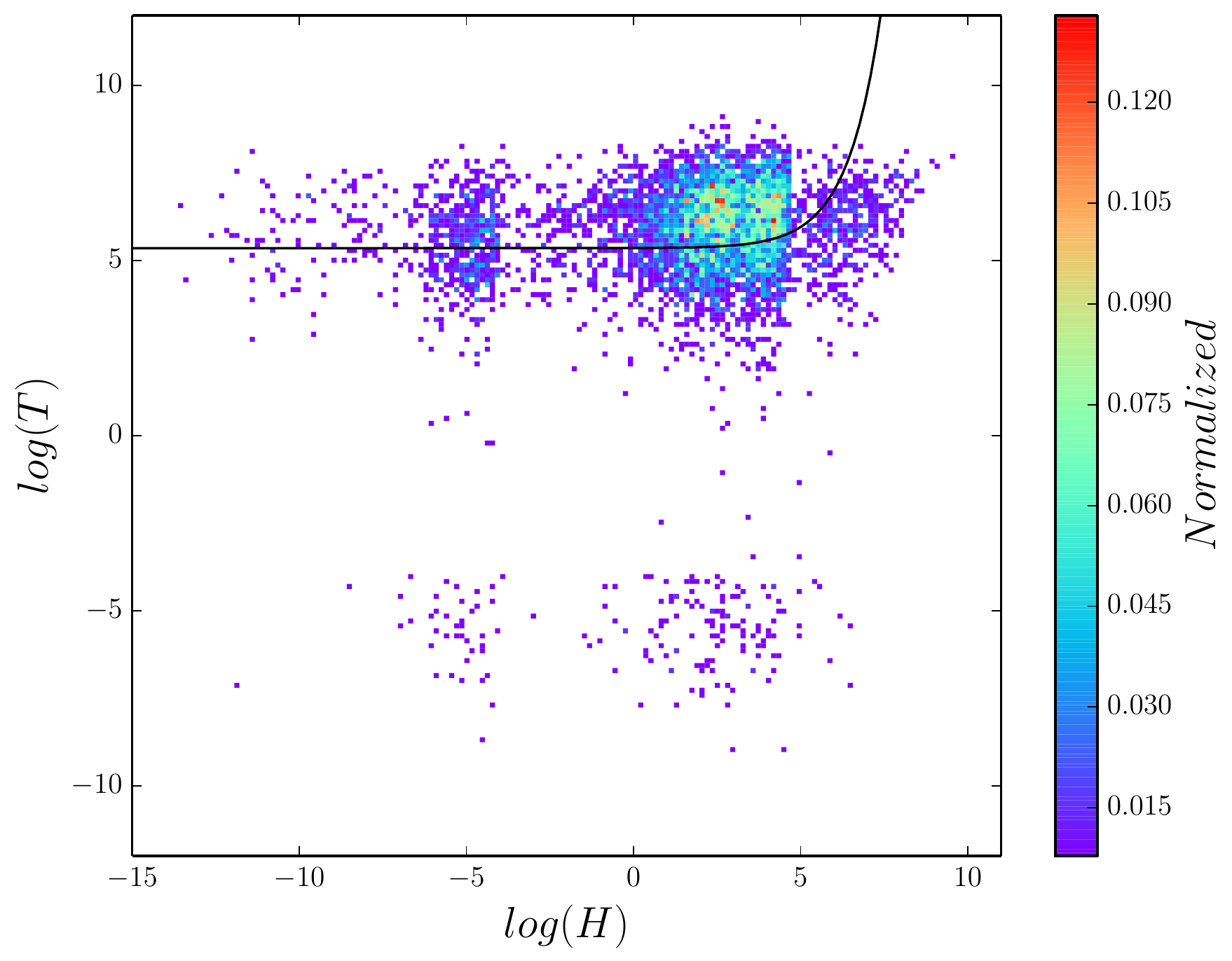}   \hspace*{-0.525cm}
\includegraphics[width=5.2cm]{\main/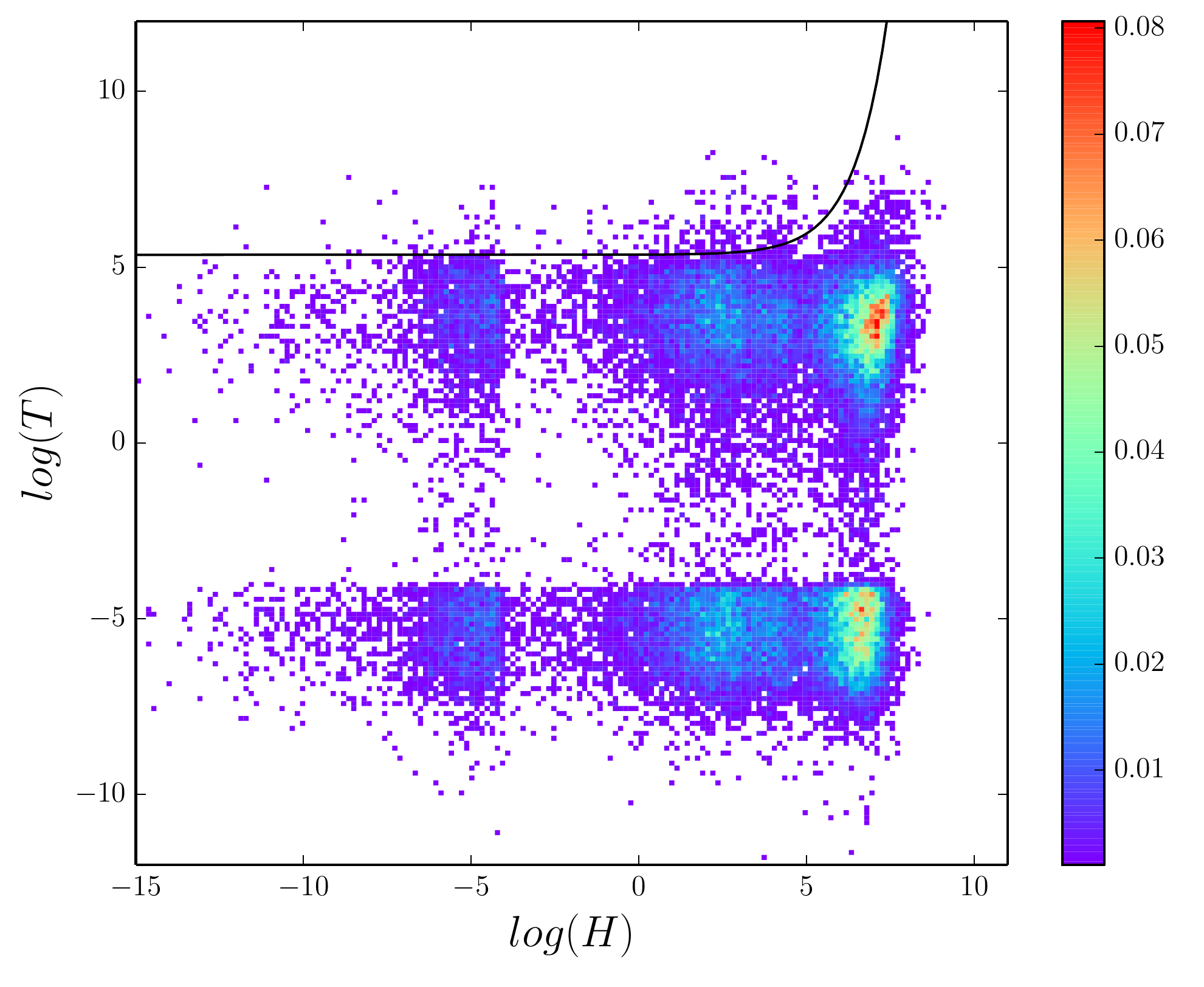} \hspace*{-0.1cm}
\includegraphics[width=5.25cm]{\main/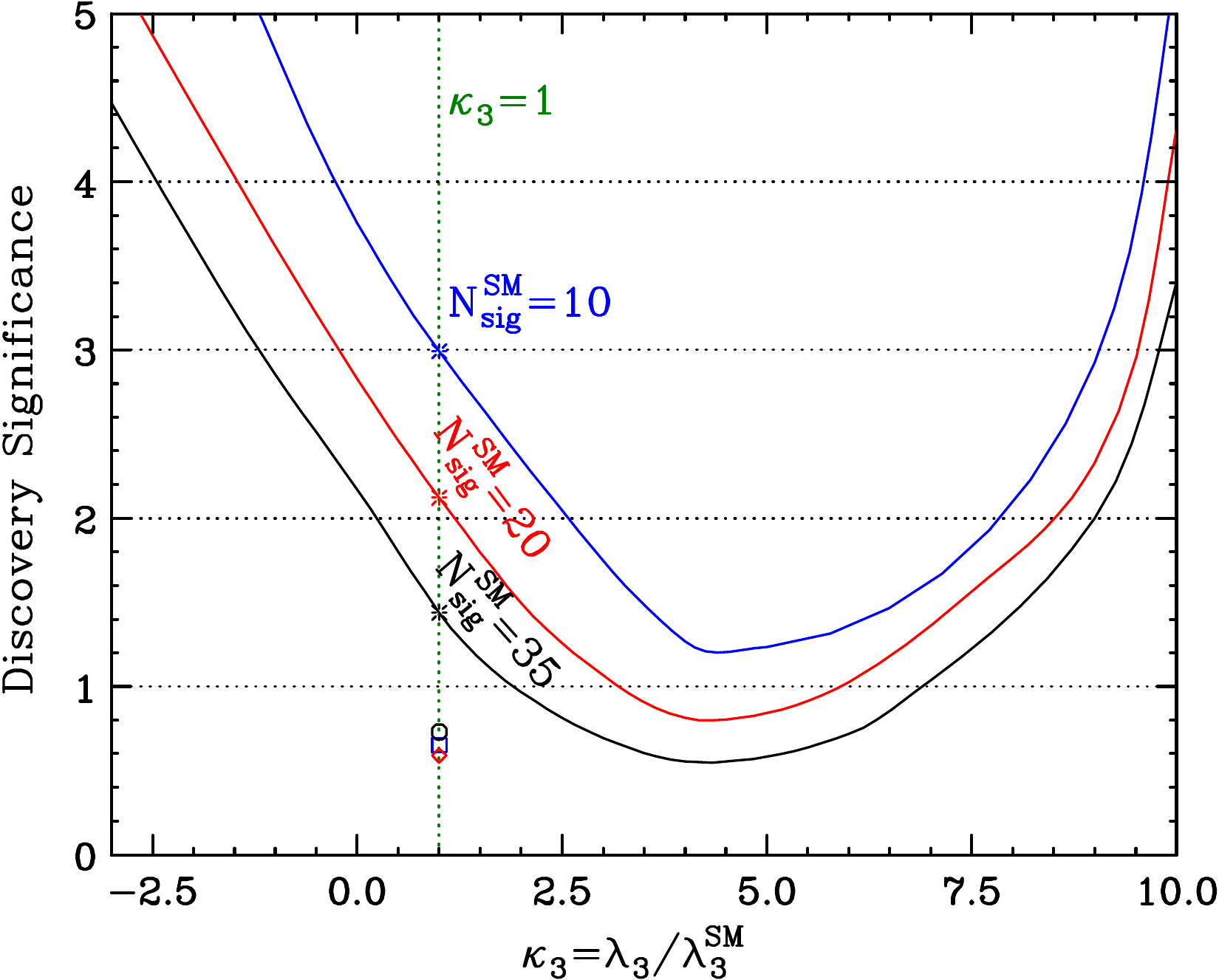} 
\caption{\label{fig:scatter} 
Scatter distribution of ($\log H$, $\log T$) for signal ($hh$ in the left) and backgrounds ($t \bar t$, $t\bar t h$, $t \bar t V$, $\ell\ell b j$ and $\tau\tau b b$ in the middle) after loose baseline selection cuts.
The right panel shows the expected discovery significance at the 14 \UTeV LHC with 3 ab$^{-1}$ as a function of the triple Higgs coupling $\kappa_3$.
We obtain each curve by applying the same set of cuts optimised for the SM point ($\kappa_3=1$) to non-SM points ($\kappa_3 \neq 1$) for ${\rm N_{sig}^{SM}}=35$ in black, ${\rm N_{sig}^{SM}}=20$ in red and ${\rm N_{sig}^{SM}}=10$ in blue.
The curves in the left and middle panel are the optimised cuts for the ${\rm N_{sig}^{SM}}=20$ case.
The three symbols {\color{red}$\Diamond$}, {\color{black}$\bigcirc$} and {\color{blue}$\square$} display the signal significance 
using CMS-NN \cite{CMS:2015nat}, CMS-BDT \cite{CMS:2017cwx} and BDT \cite{Adhikary:2017jtu}, respectively.
}
\end{figure}

Events for the signal and all relevant background processes were simulated as described in Ref.~\cite{Kim:2018cxf}. After basic selection cuts, we use the kinematic information  discussed above for further background suppression. Distributions of $M_{T2}^{(b)}$, $M_{T2}^{(\ell)}$ and $\sqrt{\hat{s}}_{min}^{(\ell\ell)}$ are shown in Fig.~\ref{fig:newcuts}, while scatter distributions of Topness and Higgsness are displayed in Fig.~\ref{fig:scatter}. 
The right panel in Fig.~\ref{fig:scatter} shows the expected signal significance at the HL-LHC as a function of the triple Higgs coupling $\kappa_3$. We obtain each curve by applying the same set of cuts optimised for the SM point ($\kappa_3=1$) to non-SM points ($\kappa_3 \neq 1$) for ${\rm N_{sig}^{SM}}=35$ in black, ${\rm N_{sig}^{SM}}=20$ in red and ${\rm N_{sig}^{SM}}=10$ in blue. The three symbols {\color{red}$\Diamond$}, {\color{black}$\bigcirc$} and {\color{blue}$\square$} show the signal significance using CMS-NN \cite{CMS:2015nat}, CMS-BDT \cite{CMS:2017cwx} and BDT \cite{Adhikary:2017jtu}, respectively.

\begin{figure}[t]
\centering
\includegraphics[width=5cm]{\main/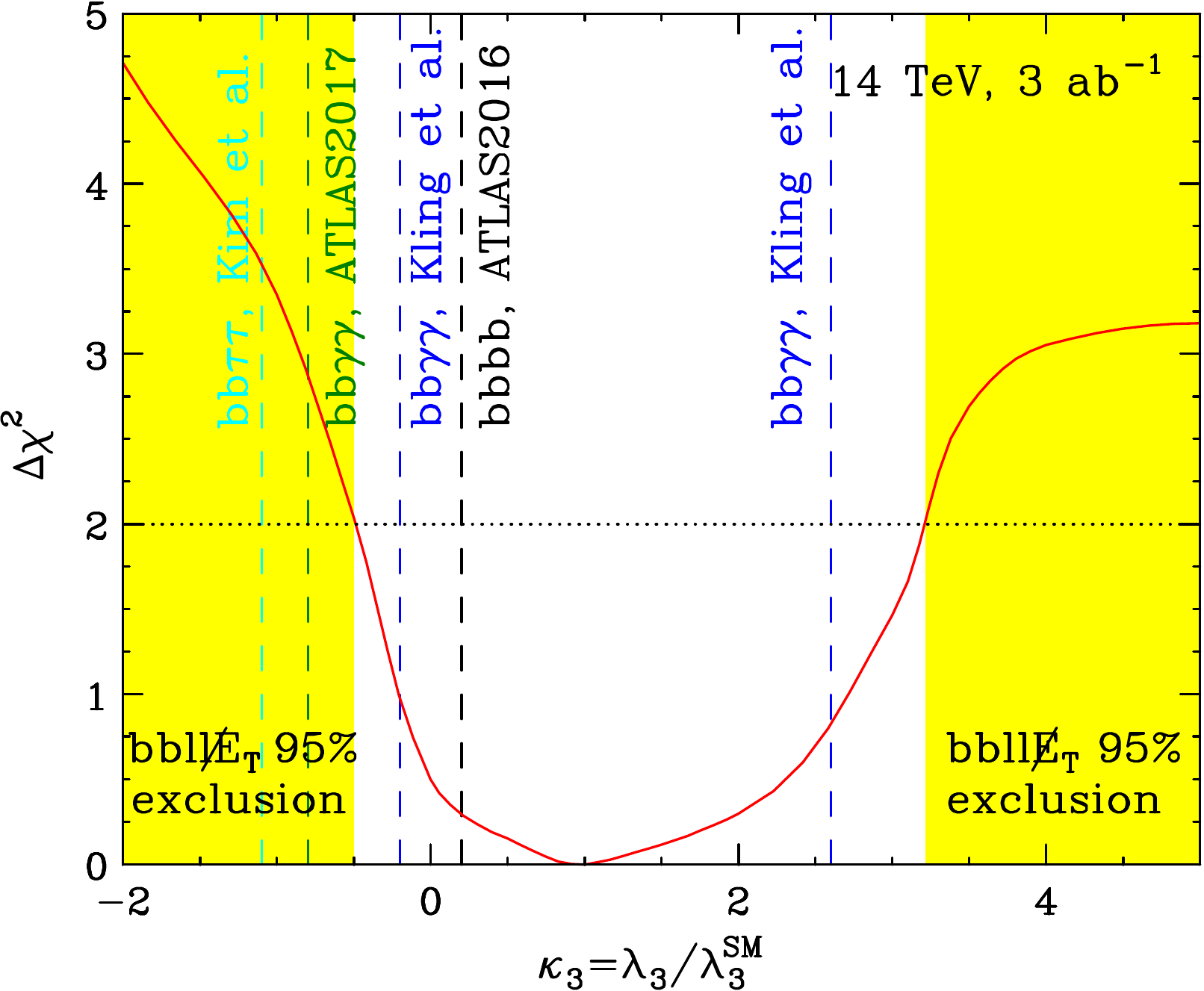}   \hspace*{0.cm}
\includegraphics[width=5.03cm]{\main/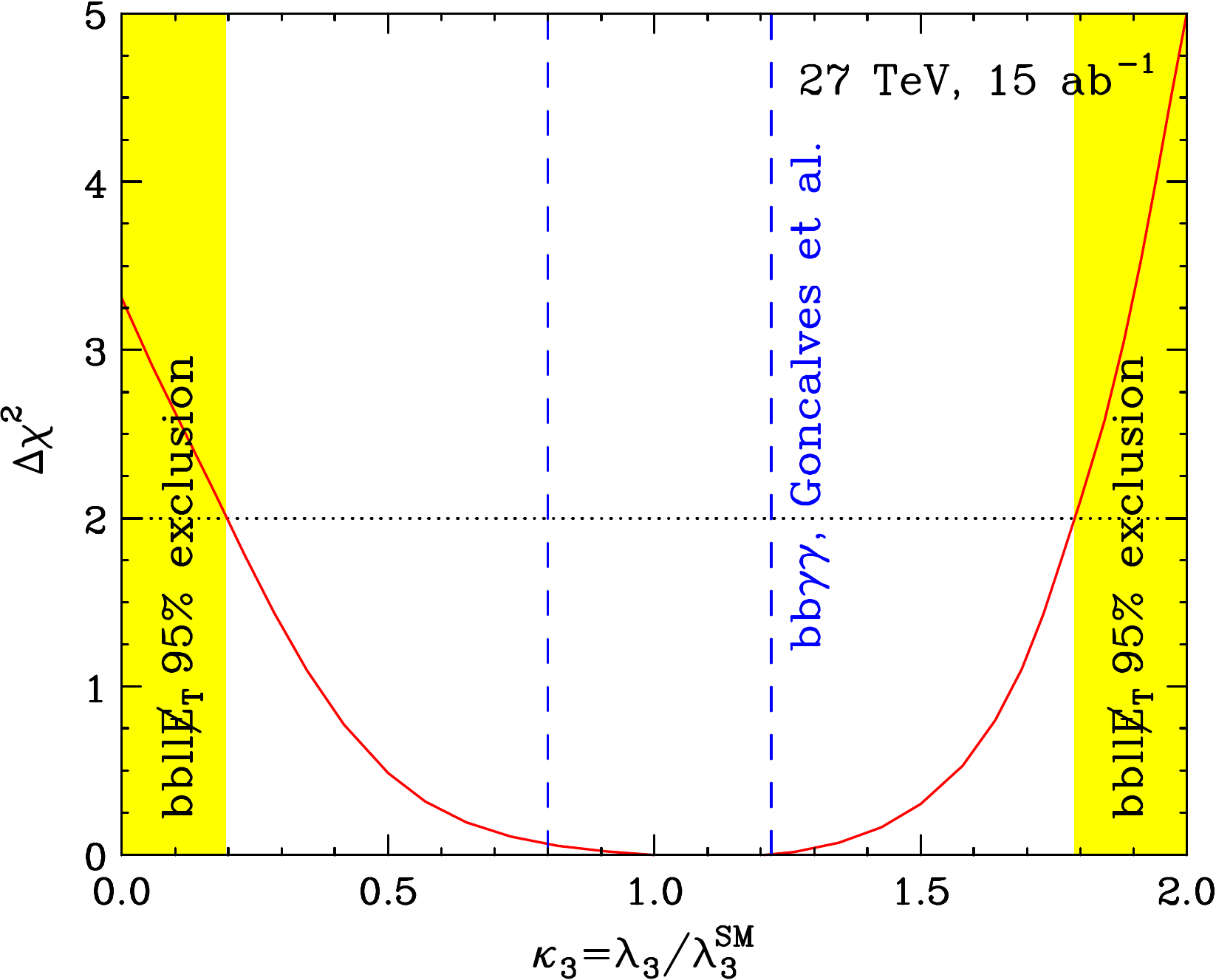}   \hspace*{-0.cm}
\includegraphics[width=5.2cm,height=4.cm]{\main/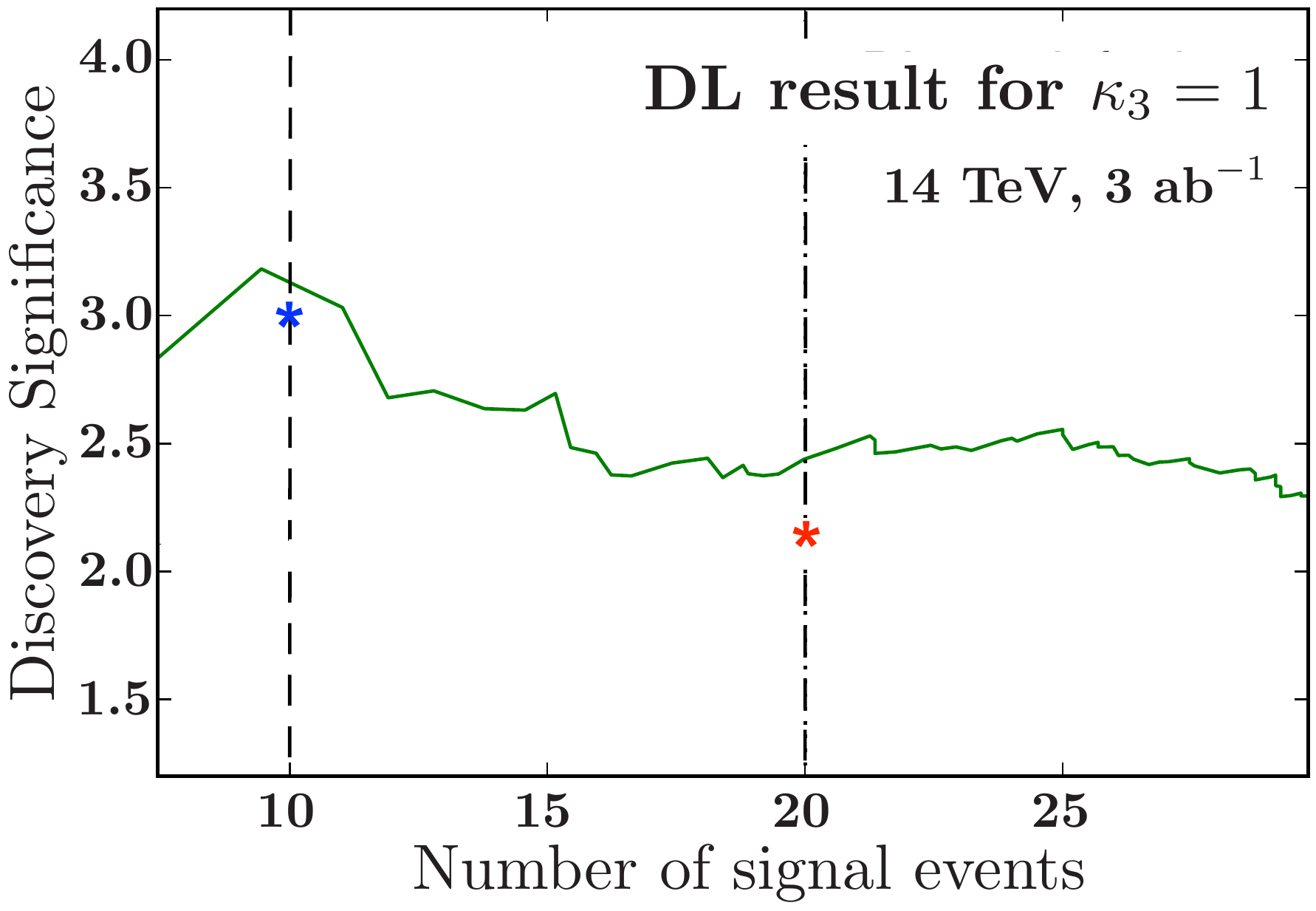}   
\caption{\label{fig:k3precision} 
Significance for observing an anomalous Higgs self-coupling at the 14 \UTeV LHC with an integrated luminosity of 3 ab$^{-1}$ (left) and at 27 \UTeV with 15 ab$^{-1}$ (middle).
Right: the effect of using a Deep Learning algorithm to improve the discovery significance for $\kappa_3=1$ shown in 
the right panel of Fig.~\ref{fig:scatter}.
}
\end{figure}

Finally Fig.~\ref{fig:k3precision} shows the significance for observing an anomalous Higgs self-coupling at the 14 \UTeV LHC with an integrated luminosity of 3 ab$^{-1}$ and at 27 \UTeV with 15 ab$^{-1}$, respectively. 
For the HL-LHC, we follow the analysis presented in Ref.~\cite{Kim:2018cxf}.
The red solid curves are obtained with nominal efficiencies for $b$ (mis-)tagging ($\epsilon_{b \to b} = 0.7$, $\epsilon_{c \to b} = 0.2$ and $\epsilon_{j \to b} = 0.01$) \cite{Sirunyan:2017ezt}. 
The HL-LHC will rule out the Higgs self-coupling outside the range $(-0.5, 3.2)$. The four vertical dashed lines in the left panel represent the expected 95\% CL exclusion of $\kappa_3$ in the $bbbb$ channel (black, from Ref. \cite{ATL-PHYS-PUB-2016-024}), in the $bb\gamma\gamma$ channel (blue, from Ref. \cite{Kling:2016lay} and green from Ref.~\cite{ATL-PHYS-PUB-2017-001}) and in the $bb\tau\tau$ channel (cyan, from Ref. \cite{Kim:2018uty}). We notice that the sensitivity in the $bbWW^*$ channel is comparable to the sensitivity in those other channels.  
For the 27 \UTeV study, we normalise our signal cross section to $139.9$ fb \cite{Grazzini:2018bsd}, and use $K$ factors of $K=1.56$ for $t\bar t$ production \cite{27TeV:ttbar}, $K=1.28$ for $t\bar  t h$ \cite{Demartin:2014fia}, $K=1.54$ for $t \bar t V$ and a conservative $K=2$ for $\ell\ell b \bar b$ and $\tau\tau b\bar b$ \cite{Kim:2018cxf}. 
Our result shows that the 27 \UTeV collider could observe double Higgs production at 5$\sigma$ for a wide range of values for $\kappa_3$ and would be able to exclude $\kappa_3$ outside the range $(0.2, 1.8)$ (for a comparative study in the $bb\gamma\gamma$ channel, see Ref.~\cite{Goncalves:2018qas} (vertical, dashed lines in the middle panel)).

In summary, we obtained a significant increase in the signal sensitivity for $hh$ production in the di-lepton channel compared to previous analyses \cite{CMS:2015nat,CMS:2017cwx,Adhikary:2017jtu}. 
The method can be easily incorporated into more advanced algorithms for further improvement. 
For example, using deep learning (convolutionary neutral network) slightly improves the discovery significance, see the right panel of Fig.~\ref{fig:k3precision}.
The discussed method is very general and can be easily applied to other processes such as the semi-leptonic final state, resonant $hh$ production, non-resonant production with more than one Higgs boson, etc. It is straightforward to generalise the idea to different topologies in searches for other BSM particles as well.


\asubsubsection[A. Alves, T. Ghosh, and K. Sinha]{Prospects for $bb\gamma\gamma $: Bayesian optimisation and BDT}

Searches for double Higgs pair production in the $b\bar{b}\gamma\gamma$ channel are an important target for the future. In this section, we study this problem at the 14 \UTeV LHC in two steps, following \cite{Alves:2017ued}: 

$(i)$ We first propose a Bayesian optimisation approach to select cuts on kinematic variables and study its performance compared to manual and random cuts, taking into account systematic uncertainties. We demonstrate our results with the Python algorithm \texttt{Hyperopt }.

$(ii)$ We next perform a joint optimisation of kinematic cuts and boosted decision trees (BDT) hyper-parameters to further discriminate signal and background events.  For our calculations, we use the \texttt{XGBoost }  implementation of BDTs for Python. 

\asubsubsubsection{Signal and Backgrounds}

For the simulation of the signal, we use {\tt MadGraph5\_aMC@NLO\_v2.3.3}~\cite{Alwall:2011uj}, to generate $p p \rightarrow h h$ process exclusively at the leading order (LO). The simulation of our signal
include both the triangle and box diagrams. We scale our LO cross-section by the partial NNLO K-factor of 2.27~\cite{deFlorian:2013uza}, calculated in the large quark mass limit and use the resulting production cross section of 36.8 fb.

The following backgrounds were taken into account in our study: $(i)$ $b\bar{b}\gamma\gamma$; $(ii)$ $Zh$ with $Z \rightarrow b\bar{b}$ and $h \rightarrow \gamma\gamma$; $(iii)$ $b\bar{b}h$ with $h\to\gamma\gamma$; $(iv)$
$t\bar{t}h \rightarrow b\bar{b}+\gamma\gamma+X$; $(v)$ $jj\gamma\gamma$ where the light-jets $jj$ are mistaken for a $b$-jet pair in the detector; $(vi)$ $b\bar{b}jj$,  where the light-jets $jj$ are mistaken for a photon pair; $(vii)$ $c\bar{c}\gamma\gamma$, where a $c$-jet is mis-tagged as a $b$-jet; $(viii)$ $b\bar{b}\gamma j$, where one light-jet is mistaken for a photon; $(ix)$ $c\bar{c}\gamma j$ where the $c$-jets are mis-tagged as bottom jets and the light-jet as a photon. We note that the $b\bar{b}\gamma j$, $c\bar{c}\gamma\gamma$, and $c\bar{c}\gamma j$ backgrounds were neglected in several early studies. 
 
The cross section normalisations for the backgrounds from $(i)$ - $(v)$ are taken from ref.~\cite{Azatov:2015oxa}, which we consider reliable. In order to obtain the distributions of the kinematic variables of interest, we pass our simulated events to {\tt PYTHIA\_v6.4}~\cite{pythia} for showering, hadronisation and underlying event and finally to {\tt DELPHES\_v3.3}~\cite{deFavereau:2013fsa} for detector simulation. For all further details of our signal and background simulation, we refer to our paper \cite{Alves:2017ued}. 

The following basic cuts were applied on both signal and background:
\begin{eqnarray}
& & p_T(j) > 20\;\hbox{\UGeV},\; p_T(\gamma) > 20\;\hbox{\UGeV},\; |\eta(j)|<2.5,\; |\eta(\gamma)|<2.5 \nonumber \\
& & 100\;\hbox{\UGeV} < |M_{jj}| < 150\;\hbox{\UGeV},\; 100\;\hbox{\UGeV} < |M_{\gamma\gamma}| < 150\;\hbox{\UGeV}\; .
\label{cuts:basicKS} 
\end{eqnarray}
The number of backgrounds events after imposing the basic cuts for 3 ab$^{-1}$ of integrated luminosity is shown in Table~\ref{table:nevKS}.
\begin{table}[!h]
\centering
\begin{tabular}{c|c|c|c|c|c|c|c|c|c}
\hline\hline
signal & $b\bar{b}\gamma\gamma$ & $c\bar{c}\gamma\gamma$ & $jj\gamma\gamma$ & $b\bar{b}\gamma j$ & $t\bar{t}h$ & $c\bar{c}\gamma j$ &  $b\bar{b} h$ & $Zh$ & total backgrounds \\ 
\hline
42.6   & 1594.5  & 447.7   &  160.3 &   137 & 101.1  & 38.2  &  2.4    & 1.8 & 2483 \\
\hline\hline
\end{tabular}
\caption{The number of signal and the various types of backgrounds considered in this work after imposing the basic cuts of eq.~(\ref{cuts:basicKS}) for 3 ab$^{-1}$ of data. We found $b\bar{b} jj$ negligible after cuts and after estimating the probability of the jet pair faking a photon pair.}
\label{table:nevKS}
\end{table}

\asubsubsubsection{Bayesian Optimisation}

The $b\bar{b}\gamma\gamma$ channel has been studied by several groups using cut and count strategies. Once signal and background cross sections are normalised to the proper values, one finds that the analysis of any particular group does not radically outperform that of any other. For a detailed comparison, we refer to Table 2 of \cite{Alves:2017ued}. 

Bayesian optimisation  offers a systematic way to obtain the most optimal cuts on a set of kinematic variables. The algorithm we utilise is implemented in the Python library \texttt{HyperOpt}~, based on the so-called sequential model-based optimisation (SMBO) technique \cite{Bergstra1, Bergstra2, hyperopt}.

The  kinematic variables used in our  study are:
$(i)$ transverse momentum of $b$-jets and photons: $p_T(b)$ and $p_T(\gamma)$; 
$(ii)$ $b\bar{b}$ and $\gamma\gamma$ invariant masses: $M_{bb}$ and $M_{\gamma\gamma}$, where signal events exhibit resonance peaks at $m_h$;
$(iii)$ transverse momentum of $b\bar{b}$ and $\gamma\gamma$: $p_T(bb)$ and $p_T(\gamma\gamma)$; 
$(iv)$ invariant mass of two $b$-jets and two photons: $M_{bb\gamma\gamma}$;
$(v)$ distance between pairs of $b$-jets and photons: $\Delta R(bb)$, $\Delta R(\gamma\gamma)$ and $\Delta R(b\gamma)$, where
$\Delta R=\sqrt{(\Delta\eta)^2+(\Delta\phi)^2}$ in the pseudo-rapidity and azimuthal angle plane $(\eta,\phi)$;
$(vi)$ the fraction $E_T/M_{\gamma\gamma}$ for the two hardest photons in the event; these are variables used in experimental searches as in Ref.~\cite{Aad:2014yja, CMS}.

In Figure~\ref{fig:BayesianKS}, we display the results obtained from the Bayesian optimisation of cuts on the above kinematic variables. We see that after 100-200 trials, the signal significance does not change much and the optimised cuts achieved a significance of 2.81$\sigma$ against 2.1$\sigma$ of the manual search of ref.~\cite{Azatov:2015oxa}, a 34\% improvement. If $b\bar{b}\gamma j$, $c\bar{c}\gamma\gamma$, and $c\bar{c}\gamma j$ backgrounds are incorporated, the Bayesian search reached 2.48$\sigma$ against 1.85$\sigma$ of the cuts of ref.~\cite{Azatov:2015oxa}, again roughly the same improvement. 
The performance of the Bayesian algorithm is also displayed in Figure~\ref{fig:BayesianKS}.

\begin{figure}[!t]
\centering
\includegraphics[scale=0.35]{\main/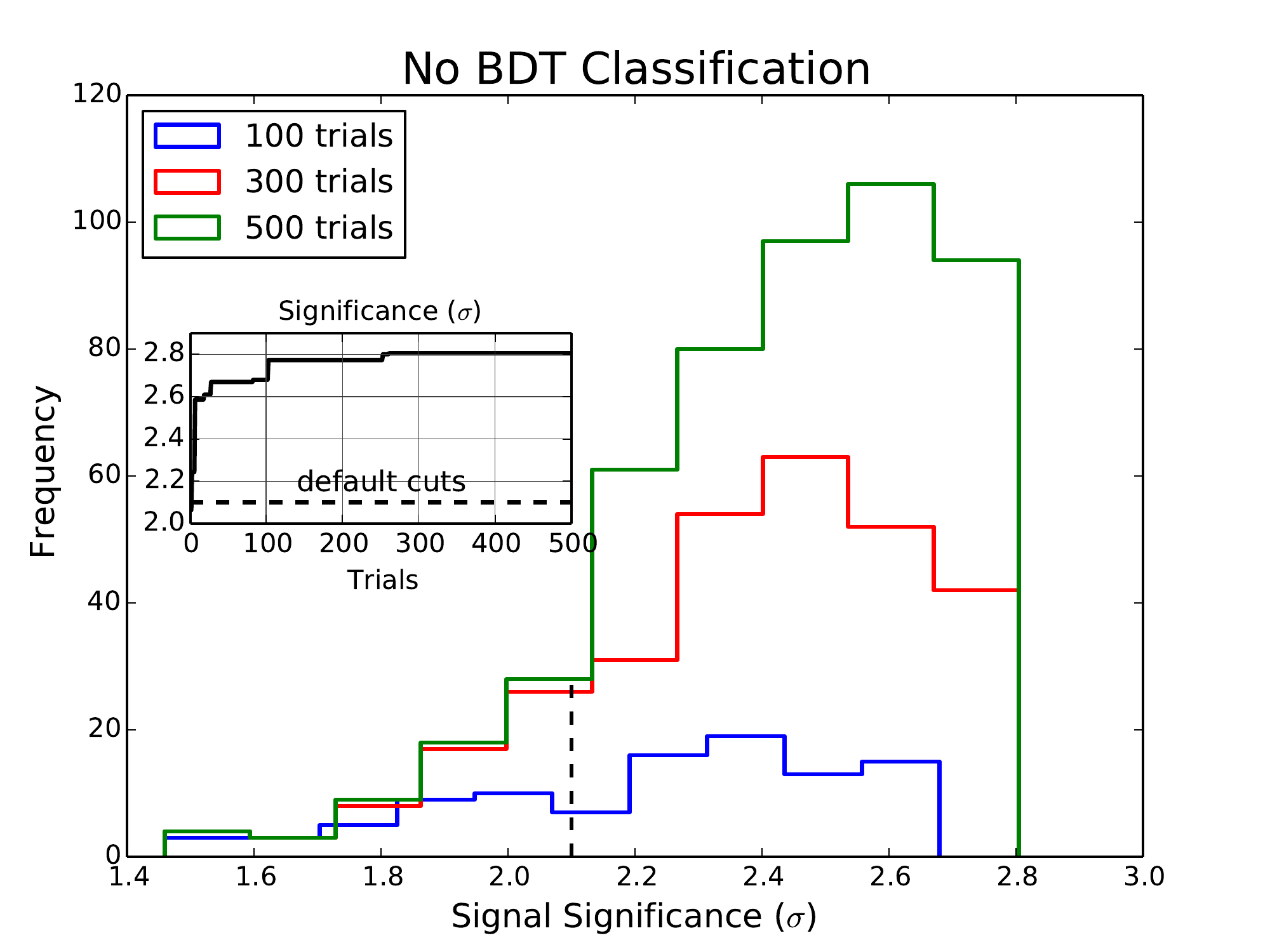}
\includegraphics[scale=0.35]{\main/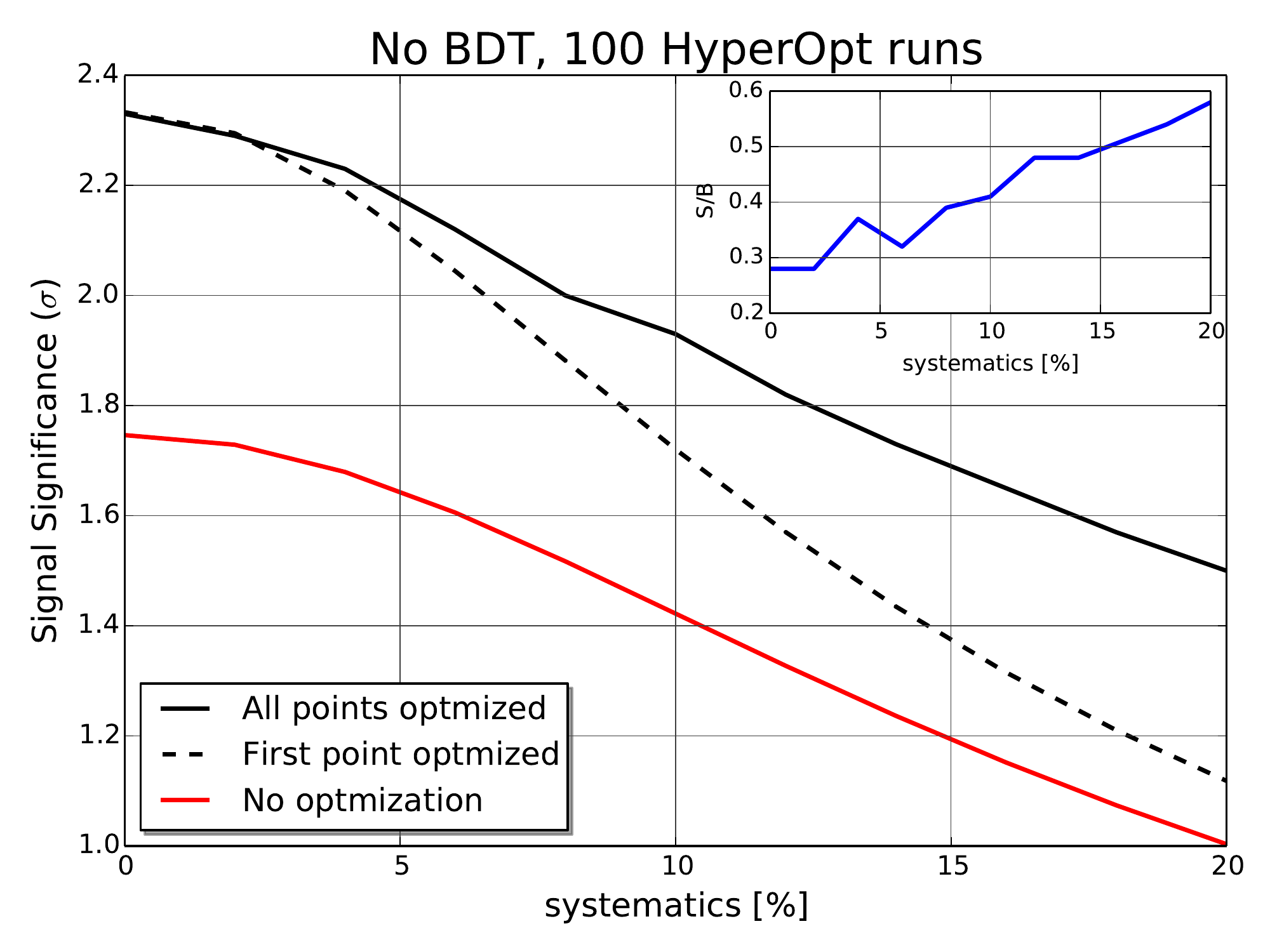}
\caption{\textbf{Left panel:} The left panel shows the optimised search with the TPE algorithm in \texttt{HyperOpt} with no systematic errors. The inset frame in the left plot shows the significance as a function of the number of trials. $S/\sqrt{B}$ is used to compute the signal significance. The black dashed line represents the results obtained with the cuts of Azatov \emph{et. al.}, ref.~\cite{Azatov:2015oxa}. \textbf{Right panel:} The $S/\sqrt{B+(\varepsilon_B B)^2}$ significance metric as a  function of $\varepsilon_B$, the systematic uncertainty in the total background rate. The red line represents the default cuts of Azatov \emph{et. al.}, the black dashed assumes an optimised strategy just for the 0\% systematics point, while for the solid upper line, the algorithm was solicited to learn the best cuts for each systematics level from 0 to 20\%. In the inner plot we show the $S/B$ ratio for the point-to-point optimisation case. }
\label{fig:BayesianKS}
\end{figure}

\asubsubsubsection{BDT Analysis}

We now turn to a discussion of the BDT analysis, for which we utilise the \texttt{XGBoost}  implementation of BDTs for Python. \texttt{XGBoost} is chosen for its good discrimination performance, speed and capacity of parallelisation.  For our  analysis we simulated $\sim 880000$; depending on the cuts, however, the total number of events usually drops to around $100000$--$300000$ events which also turned out to be a sufficient number of samples to keep over-fitting under control.

Using \texttt{HyperOpt}, we perform a joint optimisation of the kinematic variables introduced previously in conjunction with the following BDT hyper-parameters: the \texttt{number of boosted trees}, the \texttt{learning rate}, the \texttt{maximum depth of the trees}, and the minimum sum of instance weight needed in a child to continue the splitting process of the tress, \texttt{min\_child\_weight}. All the BDT results were obtained from a 5-fold cross validation by randomly splitting training and testing samples at the proportion of 2/3 and 1/3 of the total sample, respectively. We allowed for 300 trials in \texttt{HyperOpt}. 

Hyper-parameters like the \texttt{number of boosted trees}, \texttt{maximum depth of the trees} and the \texttt{min\_child\_weight} are directly related to the complexity of the algorithm by controlling the number, size and configuration of the trees. The \texttt{learning rate}, also known as \texttt{shrinkage} in this context, is a parameter that controls the weight new trees have to further model the data. A large value permits a larger effect from new added trees and might lead to more severe over-fitting. There are other parameters which can be eventually used to prevent over-fitting and loss of generalisation power, but we found that tuning these parameters was sufficient to achieve a good performance.

A comparative result of a simple cut and count analysis and a sequential optimisation of cuts and BDT hyper-parameters are presented  Table~\ref{table:resultsbdtKS}. We note that BDT outperforms simple cut and count, even when cutting is performed using Bayesian optimisation. This is due to the better discrimination between the signal and background classes achieved by the machine learning algorithms as they find more profound correlations among the kinematic features and those classes. These correlations cannot be fully explored in simple/manual rectangular cut-and-count analyses. 

\begin{table}[h]
\centering
\begin{tabular}{c|c|c}
\hline
systematics (\%) & Cut-and-count & BDT \\
\hline\hline 
0 & 2.34[1.76] & 3.88 \\
\hline
10 & {\bf 1.93}[1.43] & {\bf 3.57}  \\
\hline
20 & 1.51[1.0] & 3.10  \\
\hline\hline
\end{tabular}
\caption{Signal significances for cut-and-count and BDT for 0, 10 and 20\% systematics. We took all backgrounds into account for the computation of the AMS with optimised cuts and an integrated luminosity of 3 ab$^{-1}$ at the 14 \UTeV LHC. The bold-face numbers represent the significances expected with the level of systematics anticipated by the experimental collaborations in refs.~\cite{ATL-PHYS-PUB-2017-001, CMS}. The numbers inside brackets are the significances computed with the default cuts of Azatov \emph{et. al.}, ref.~\cite{Azatov:2015oxa}, which we took as baseline results.}
\label{table:resultsbdtKS}
\end{table}
\begin{figure}[!t]
\centering
\includegraphics[scale=0.35]{\main/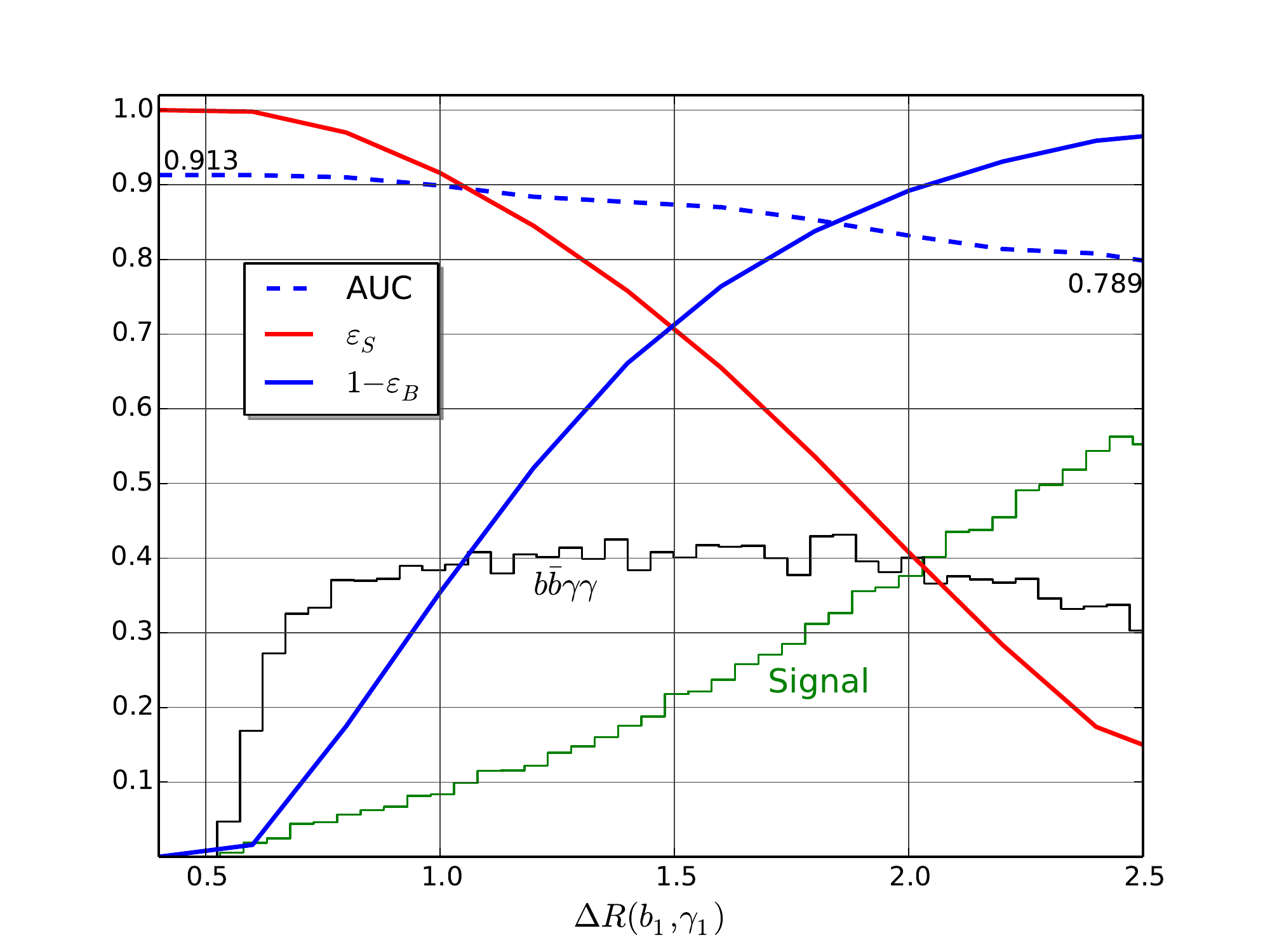}
\includegraphics[scale=0.35]{\main/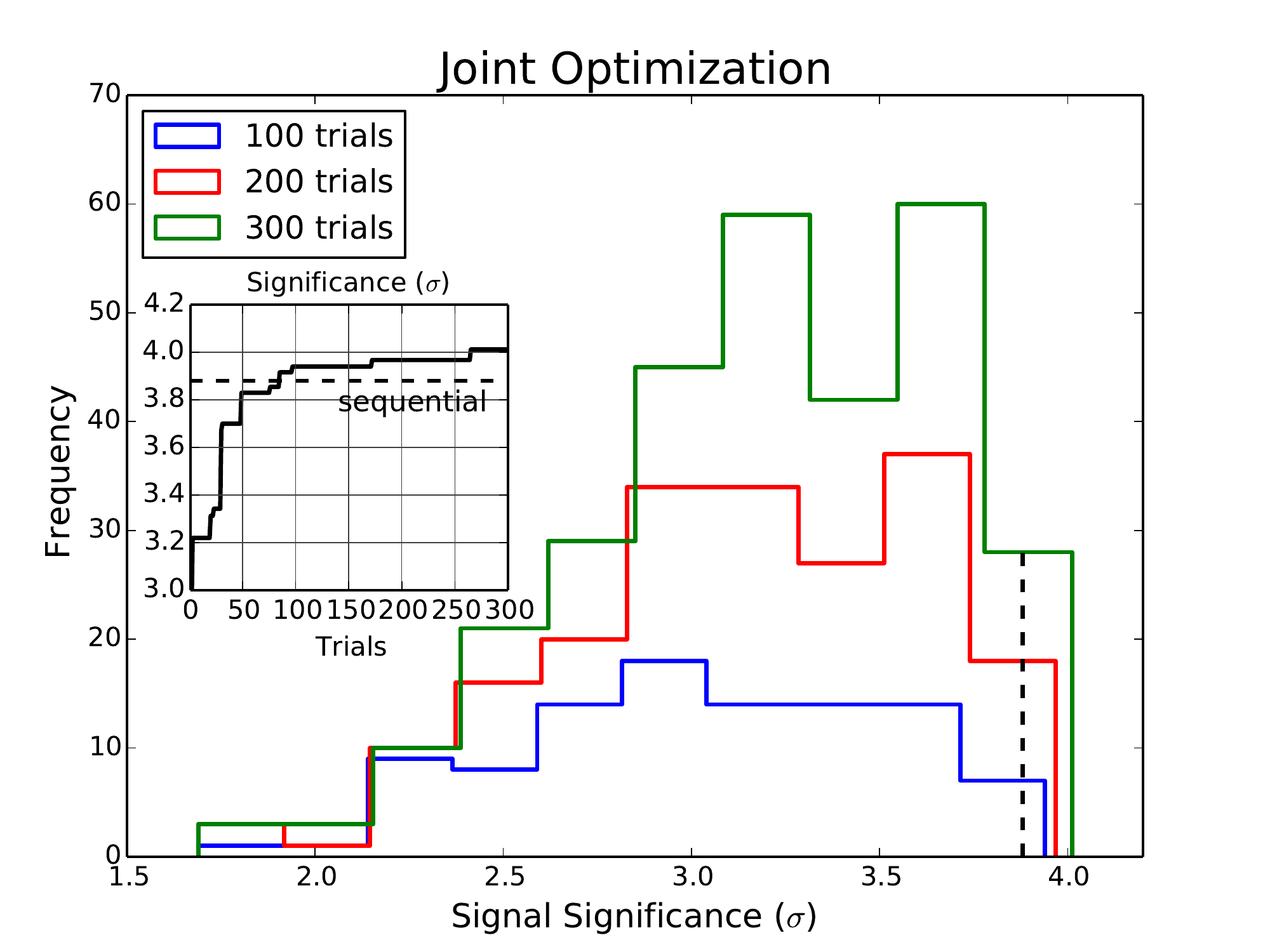}
\caption{\textbf{Left panel:}  We show the results of the effects of imposing hard cuts on $\Delta R_{b_1\gamma_1}$ for the BDT performance, see~\cite{Alves:2017ued} for further details. \textbf{Right panel:} The histogram of number of cut strategies producing a given significance interval in a BDT-aided classification analysis. The inset plot shows the significance as a function of the number of \texttt{HyperOpt} trials. No systematics are assumed, the backgrounds are those of ref.~\cite{Azatov:2015oxa} and the $S/\sqrt{B}$ used to compute the signal significances. The black dashed line represents the results obtained with the default cuts of Azatov \emph{et. al.}, ref.~\cite{Azatov:2015oxa}.}
\label{fig:8resultsKS}
\end{figure}
However, there is a trade-off between the efficiency of the cuts and the ML performance which is usually neglected in phenomenological works where these tools are employed. The reasoning is simple: cutting harder cleans up more backgrounds but weakens the correlations between the kinematic variables and the event classes, thereby decreasing the ML performance. On the other hand, relaxing the cuts makes the correlations stronger helping to boost ML but the discrimination power gained might not be enough to get a good significance with a large number of surviving back- ground events. Hence, a joint optimisation of cuts and BDT hyper-parameters improve the performance of our analysis further. 

The maximum AMS significance is 4.0$\sigma$ for a joint optimisation analysis of cuts and BDT hyper-parameters. The final selections of the kinematic variables and BDT hyper-parameters are the following $ p_T(1) >72\; \hbox{\UGeV},\; p_T(2) >20\; \hbox{\UGeV}; \Delta R_{ij}>0.15,\; \Delta R_{ii}<3.6; M_{b\bar{b}\gamma\gamma}>370\; \hbox{\UGeV},\; p_{T_{ii}}>145\; \hbox{\UGeV},\; M_{b_1\gamma_1}>100\; \hbox{\UGeV}; |M_{bb}-m_h|<27\; \hbox{\UGeV},\; |M_{\gamma\gamma}-m_h|<11\; \hbox{\UGeV}; \hbox{\texttt{number of trees}}=157; \hbox{\texttt{learning rate}}=0.101; \hbox{\texttt{maximum tree depth}}=14; \hbox{\texttt{min\_child\_weight}}=5$. We have denoted $p_T(1)$ as the leading $b$-jet or photon, and $p_T(2)$ as the next-to-leading $b$-jet or photon.

The results are shown in Figure~\ref{fig:8resultsKS}. The left panel 
shows the normalised $\Delta R_{b_1\gamma_1}$ histograms for the signal and the $b\bar{b} \gamma \gamma$ continuum background, the signal efficiency (background rejection) is the red (blue) line, and the area under the Receiver-Operator curve (ROC), AUC, is the dashed line. The bigger the AUC, the better the performance of a cut-and-count analysis based on that distribution. 
On the right panel, we show the histogram of number of cut strategies producing a given significance interval in a BDT-aided joint optimisation analysis. Finding this optimal performance from the competition between hard cuts and an ML algorithm is the core of the method presented in the section.




\asubsection{HE-LHC prospects}
\label{sec:HH_HE}

This section shows prospective results that could be obtained with 15~\abinv\ of data at a centre-of-mass energy of 27~\TeV\ at the HE-LHC.

\asubsubsection[D. Gon\c{c}alves, T. Han, F. Kling, T. Plehn, M. Takeuchi]{Theoretical prospects: from kinematics to dynamics}
\label{sec:THanetal}
sHiggs pair production $pp\rightarrow hh$ offers a direct path to pin
down the Higgs self-coupling $\lambda$ at a hadron collider~\cite{Eboli:1987dy, Dicus:1987ic, Glover:183945, Plehn:1996wb, Djouadi:1999rca, Li:2013rra}.  
Theoretical studies as well as current analyses
point to the $b\bar{b}\gamma\gamma$ decay as the most promising
signature at the LHC~\cite{ATLAS-CONF-2016-004, Aad:2015xja}. 
For the high-luminosity LHC (HL-LHC),
ATLAS and CMS projections indicate a very modest sensitivity to the
Higgs self-coupling~\cite{ATL-PHYS-PUB-2017-001, CMS-PAS-FTR-16-002}. 
In the optimistic scenario that we can neglect systematic
uncertainties, those studies indicate that the LHC will probe the
coupling at 95\% confidence level $-0.8 < \kappa_\lambda < 7.7$, where the SM value is $\kappa_\lambda = \lambda/\lambda_{SM}=1$, 
falling short in precision in comparison to other Higgs 
property measurements at the LHC, and far from satisfactory in probing the
Higgs potential. For example, ${\cal O}(1)$ determination of $\kappa_\lambda$ would be required to 
test some of the EW Baryogenesis models~\cite{Kobakhidze:2015xlz, Chen:2017qcz, Gan:2017mcv, Cao:2017oez, Jain:2017sqm, deVries:2017ncy, Reichert:2017puo, Carena:2018vpt}.

Because of the rapidly growing gluon luminosity at higher energies,
the $hh$ production cross section increases by about a factor of
4~(40) at 27~(100)~\UTeV.  This means that at the HE-LHC with the
anticipated integrated luminosity of $15~\iab$ the number of events in
the $b\bar{b} \; \gamma \gamma$ channel increases by a factor $4
\times 5 = 20$ to around 5k events.  A 100~\UTeV hadron collider with
a projected integrated luminosity of $30~\iab$ features another
increase by a factor $10 \times 2=20$, to around 100k expected Higgs
pair events in the Standard Model.
This estimate shows how the combination of increased energy and
increased luminosity slowly turns Higgs pair production into a valid
channel for precision measurements~\cite{Goncalves:2018yva}. 

\asubsubsubsection{Information in Distributions}

Previous studies have shown that multivariate analysis, taking into 
account kinematic distributions, gives a substantially better reach 
on the Higgs self coupling over the purely rate-based 
analysis~\cite{Goncalves:2018yva,Kling:2016lay,Barger:2014qva,Bauer:2017cov}. In the following, we therefore 
summarise which kinematic features include information about the 
Higgs self-coupling. 

At leading order, Higgs pair production receives contributions both 
from a triangular loop diagram sensitive to the Higgs-self coupling 
and from a box or continuum diagram. The box contribution completely 
dominates the total rate over most of the phase space, making the Higgs 
coupling measurements a challenge. While we can define a number of
kinematic observables describing the continuum backgrounds, the
measurement of the Higgs self-coupling relies on a simple $2 \to 2$
process with two independent kinematic variables.

Three distinct phase space regions provide valuable information on a
modified Higgs self-coupling, all from a large destructive interference 
between the triangle and box contributions. First, there is the 
threshold~\cite{Plehn:1996wb, Djouadi:1999rca, Li:2013rra, Baur:2002rb, Baur:2002qd,Li:2015yia} in the partonic centre of mass energy 
$ m_{hh} \approx 2 m_h$. Based on the effective
Higgs-gluon Lagrangian~\cite{Shifman:1979eb,  Kniehl:1995tn, Spira:2016zna} we can therefore write the
corresponding amplitude for Higgs pair production as
\begin{equation}
\frac{\alpha_s}{12 \pi v}
\left( \frac{\kappa_\lambda \lambda_\text{SM}}{s-m_h^2} - \frac{1}{v} \right) 
\to
\frac{\alpha_s}{12 \pi v^2}
\left( \kappa_\lambda -1 \right) \stackrel{\text{SM}}{=} 0 \; .
\label{eq:higgs_pair}
\end{equation}
While the heavy-top approximation is known to give a poor description
of the signal kinematics as a whole, it does describe the threshold
dependence correctly~\cite{Baur:2002rb, Baur:2002qd,Li:2015yia}. 
This indicates that we can search
for a deviation of the Higgs self-coupling by looking for an
enhancement of the rate at threshold. 
Second, an enhanced sensitivity to the self-coupling appears as top
mass effect. For large positive values of $\lambda$ absorptive imaginary 
parts lead to a significant dip in the combined rate at the threshold 
$m_{hh} \approx 2 m_t $~\cite{Dolan:2012rv, Barr:2013tda, Kling:2016lay}. 
The sharpest interference dip takes place near $\kappa_\lambda\approx 2$
while for negative values of $\kappa_\lambda$ the interference 
becomes constructive.
Finally, the triangular and box amplitudes have a generally different
scaling in the limit $m_{hh} \gg m_h, m_t$
~\cite{Plehn:1996wb, Djouadi:1999rca, Li:2013rra,Dolan:2012rv, Barr:2013tda}. While the triangle amplitude features an 
explicit suppression of either $m_h^2/m_{hh}^2$ or 
$m_t^2/m_{hh}^2$ at high invariant mass, the box diagrams drops 
more slowly towards the high-energy regime. This explains why a 
rate based analysis focusing on the high di-Higgs mass region
only has limited sensitivity. The impact of all three kinematic features can be quantified
statistically and  indicate that essentially the full
information on the Higgs self-coupling can be extracted through a
shape analysis of the $m_{hh}$ distribution~\cite{Bauer:2017cov}. 

In Fig.~\ref{fig:madmax_diff} we present the signal and background 
distributions for three relevant kinematic variables: $m_{hh}$, $p_{T,h}$
and $\Delta R_{\gamma\gamma}$. Using the \textsc{MadMax} 
approach~\cite{Cranmer:2006zs, Plehn:2013paa}, based on the Neyman Pearson Lemma we 
also estimate the maximum significance with which any multi-variate 
analysis will be able to extract an anomalous self-coupling 
$\kappa_\lambda \neq1$. The corresponding differential distribution of 
maximum significance are shown as solid lines in Fig.~\ref{fig:madmax_diff}. 
In addition to the signal features, the significance is limited by the rapidly dropping 
backgrounds, covering both of the above-mentioned regions with an enhanced dependence 
on the triangle diagram. In the absence of background, the significance 
indeed peaks between the production threshold and the top-mass
threshold~\cite{Kling:2016lay}.  The drop towards large values of $m_{hh}$
is a combination of the dominance of the box diagram in the signal and
the limited number of expected signal events.  The significance with
which we can extract modified self-couplings either smaller
($\kappa_\lambda = 0$) or larger ($\kappa_\lambda = 2$) than in the
Standard Model shows a similar phase space dependence. The only
difference is a slightly harder significance distributions for
$\kappa_\lambda = 2$, an effect of the dip at
$m_{hh}\approx 2m_t$.

\begin{figure*}[t]
  \includegraphics[width=0.32\textwidth]{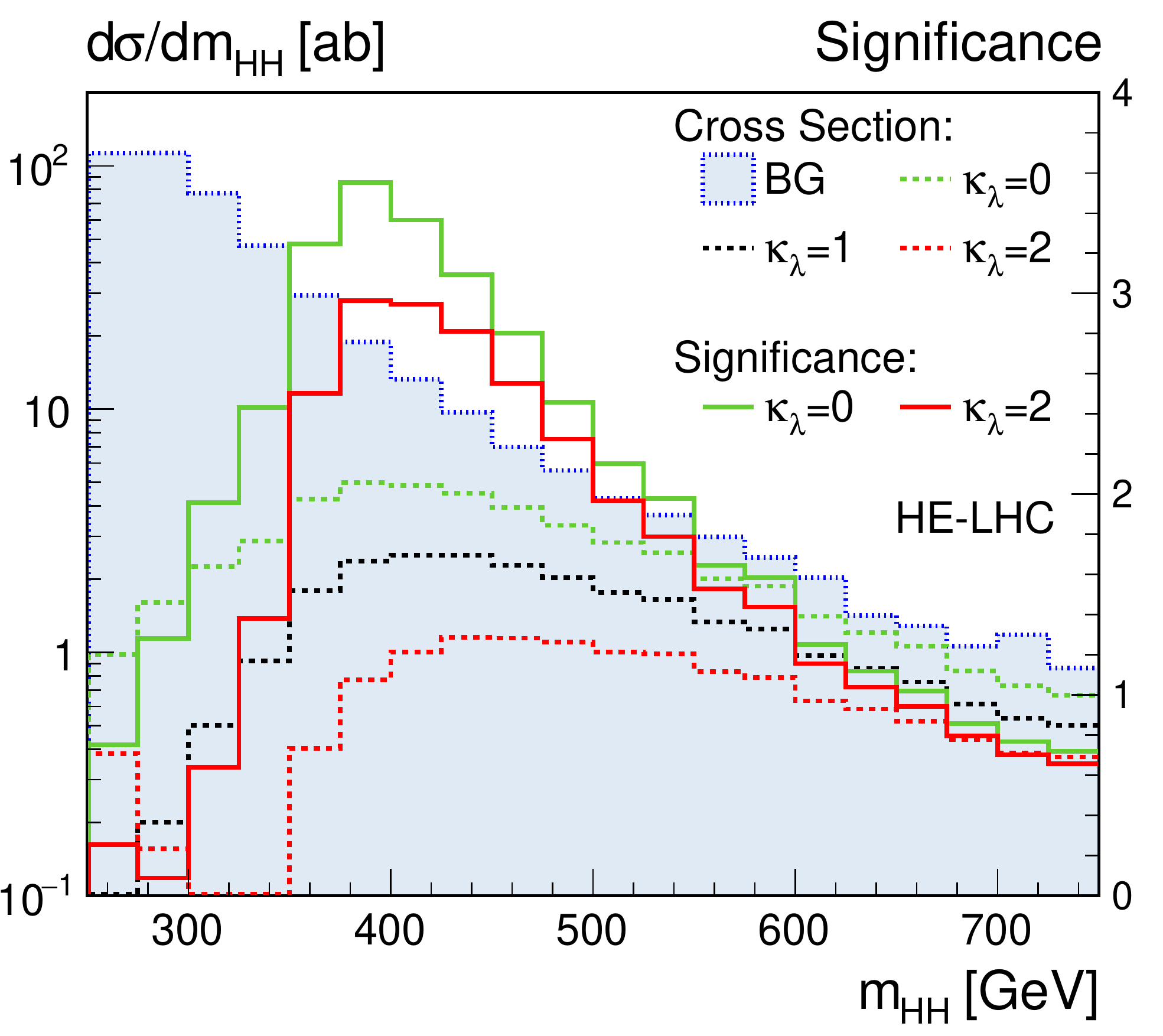}
  \includegraphics[width=0.32\textwidth]{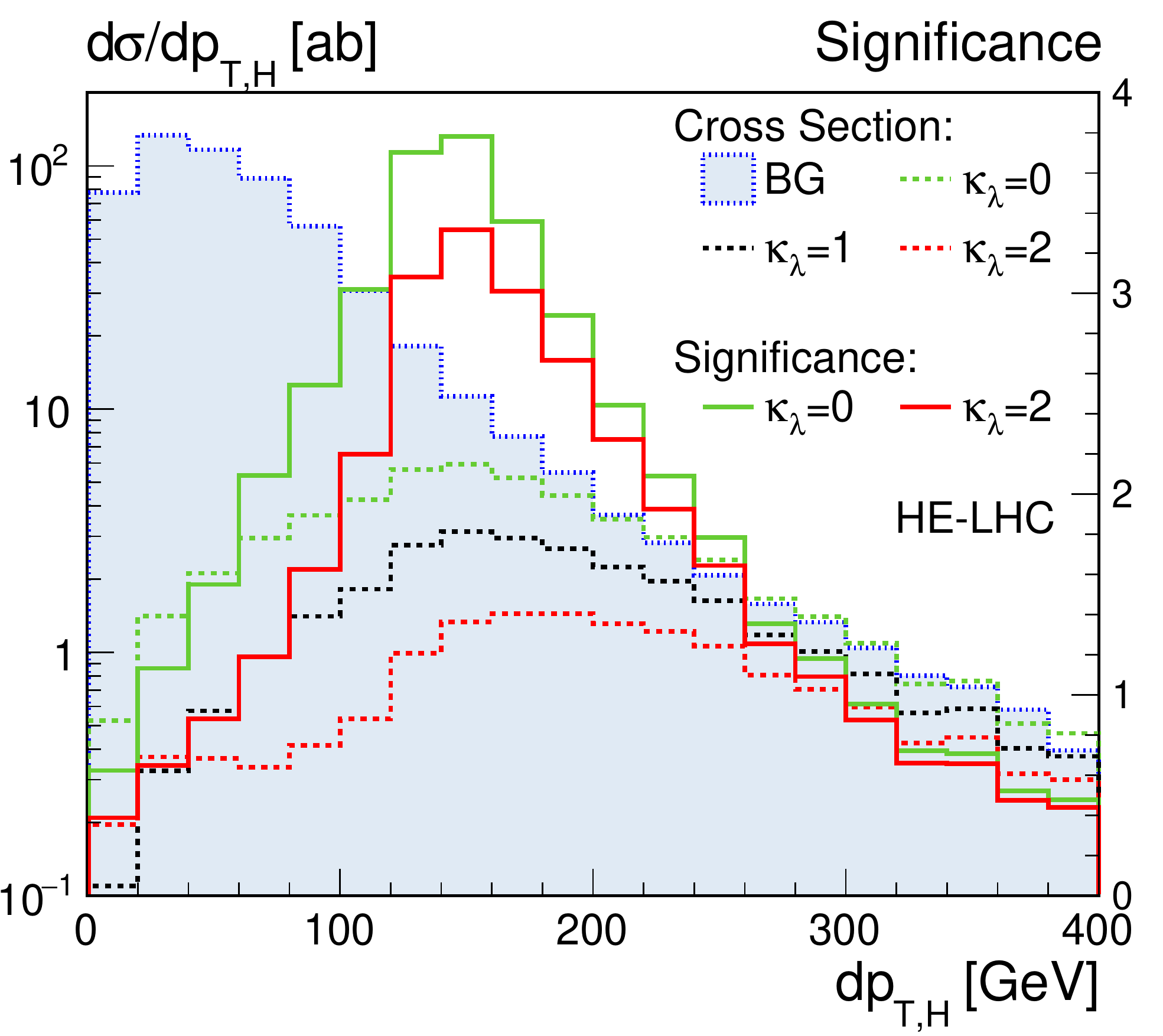}
  \includegraphics[width=0.32\textwidth]{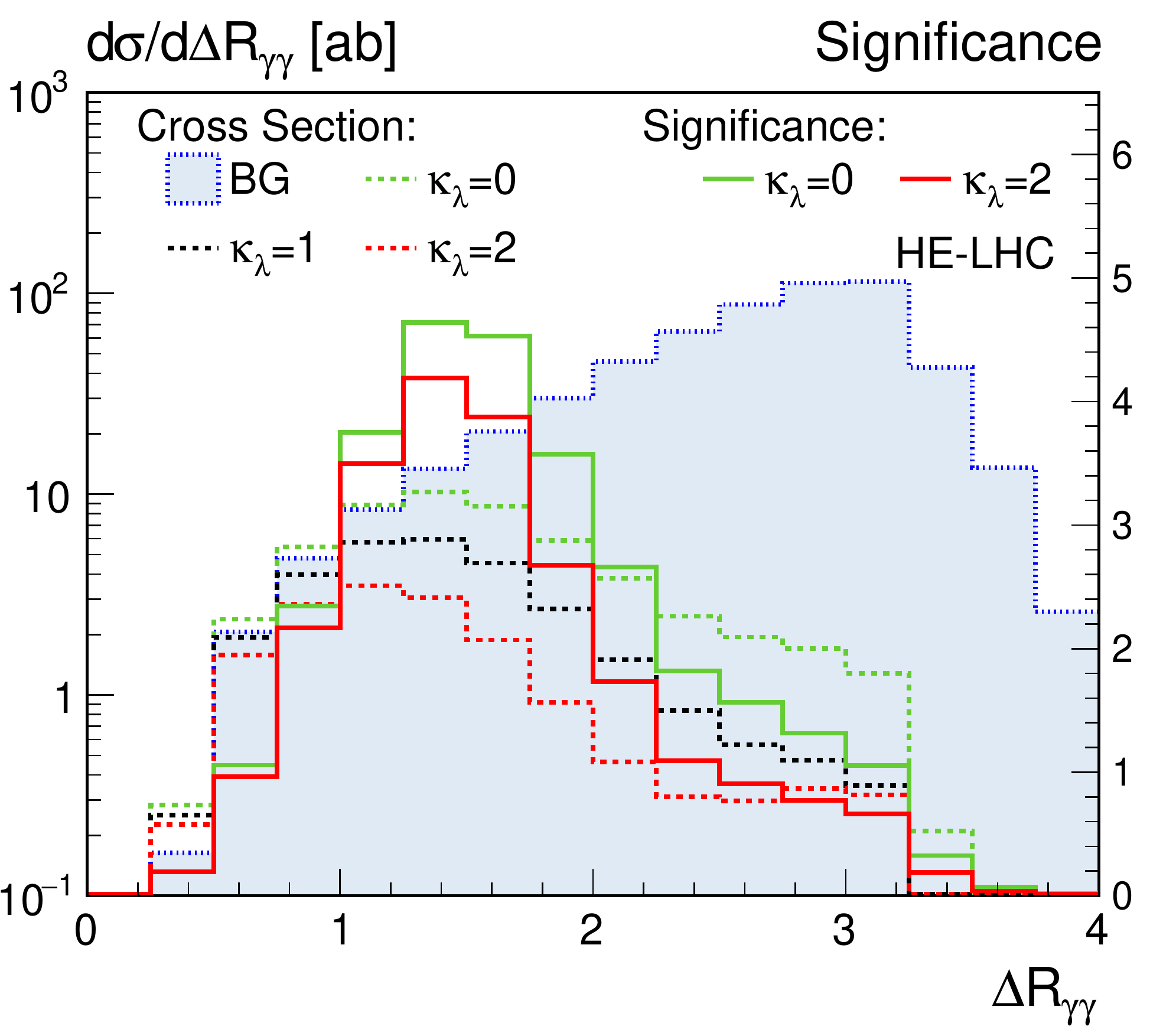}
  \caption{Kinematic distributions (dashed lines with left vertical
    axes) and significance distribution (solid lines with right
    vertical axes) assuming a Higgs self-coupling with
    $\kappa_\lambda=0,1,2$ for the HE-LHC. The significance describes the
    discrimination of an anomalous self-coupling $\kappa_\lambda \neq
    1$ from the SM hypothesis $\kappa_\lambda = 1$.}
\label{fig:madmax_diff}
\end{figure*}

\asubsubsubsection{Detector-Level Analysis}

\begin{figure*}[b!]
\centering 
 \includegraphics[width=.4\textwidth]{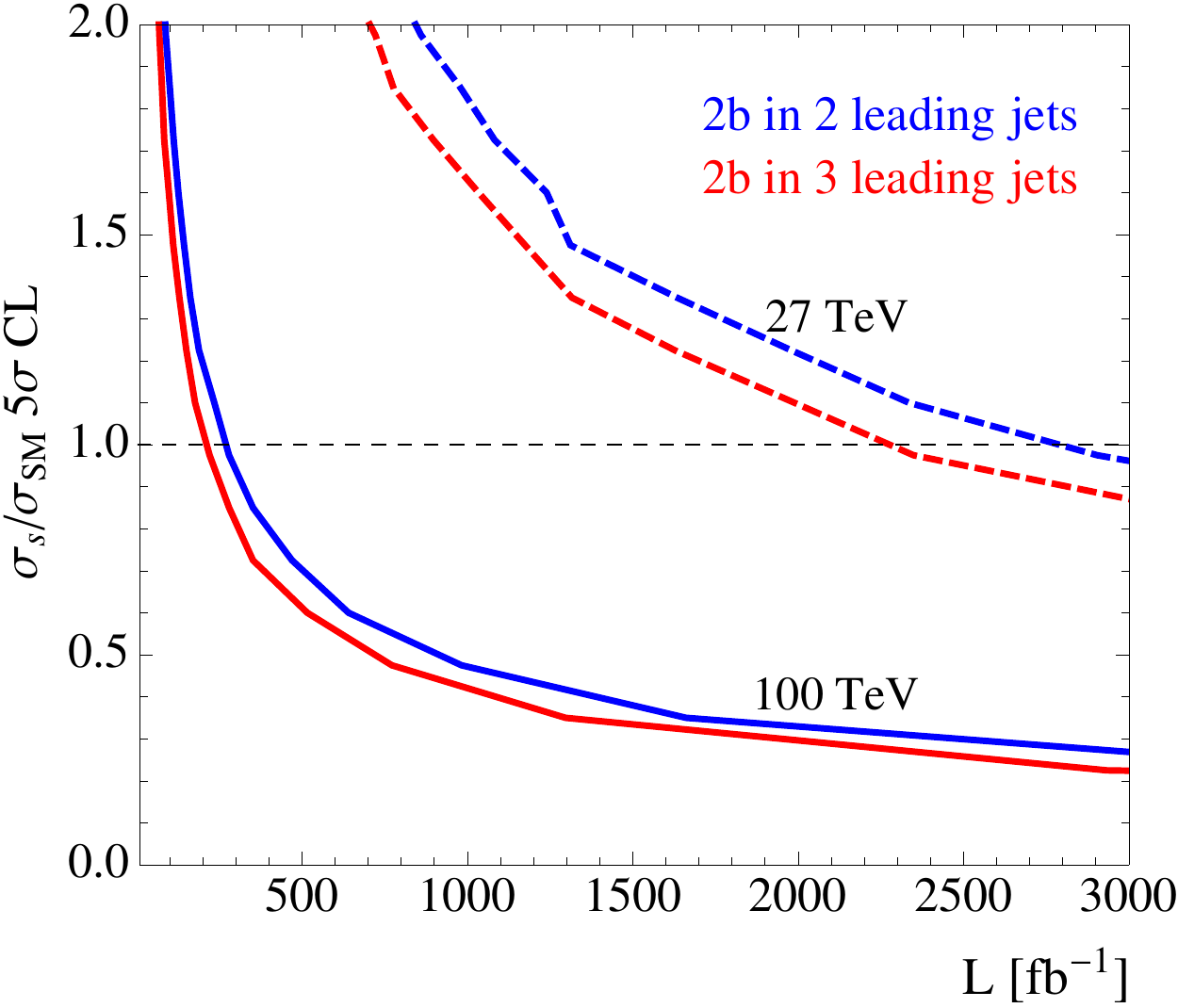} 
 \includegraphics[width=.4\textwidth]{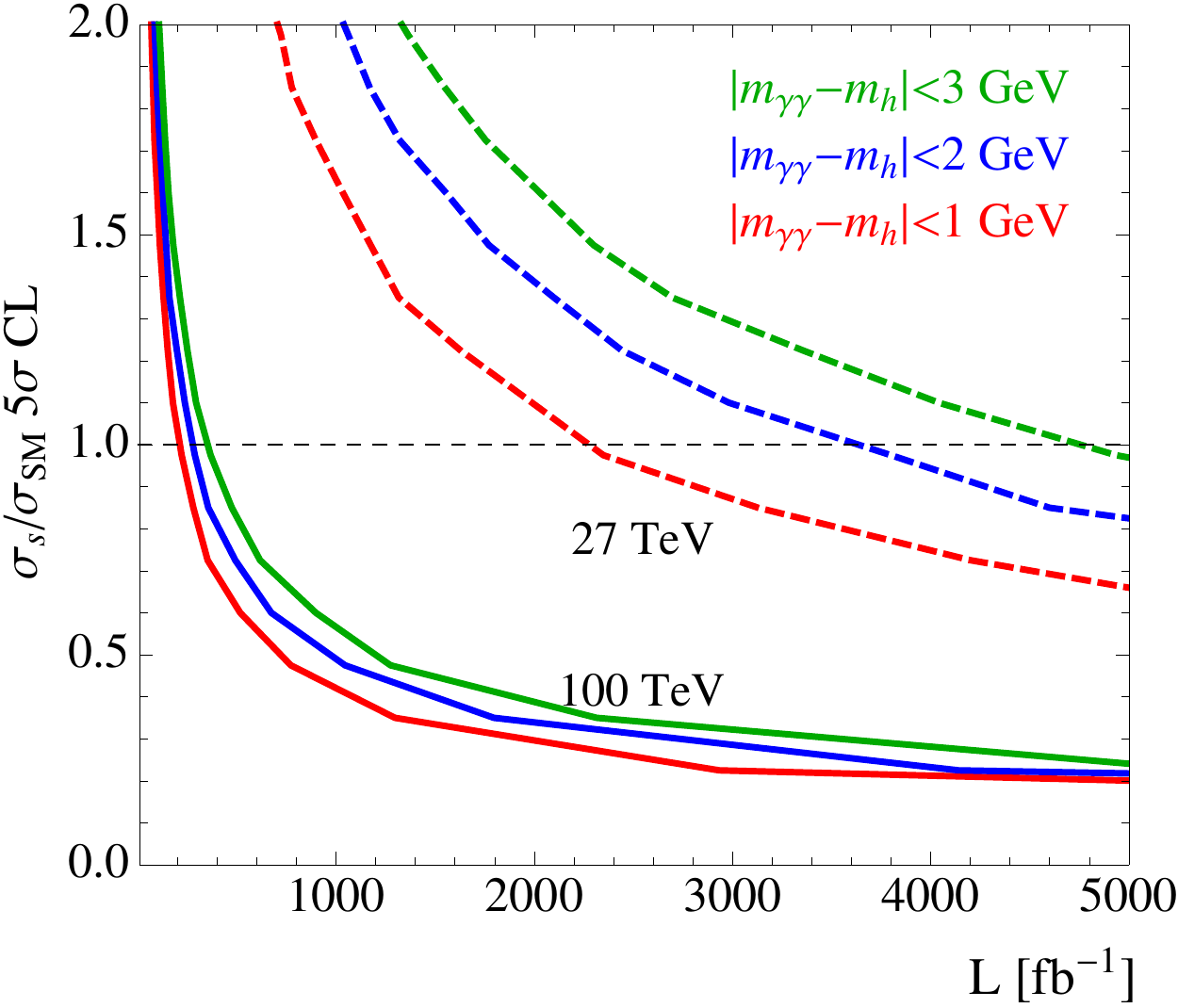}
   \caption{Luminosity required for a $5\sigma$ discover of Higgs pair
     production for the HE-LHC (dashed) and a 100~\UTeV collider (full).
     Left: sensitivity in terms of the total rate, demanding two
     $b$-tags among the two or three leading jets and assuming
     $|m_{\gamma\gamma}-m_h|<1$~\UGeV.  Right: sensitivity for three
     mass windows $|m_{\gamma\gamma}-m_h|<1,2,3$~\UGeV.  We assume the
     SM hypothesis with $\kappa_\lambda=1$ and use a binned
     log-likelihood analysis of the $m_{hh}$ distribution.}
 \label{fig:bound1}
\end{figure*}

Based on our findings above, we now design a detailed analysis strategy to extract the Higgs
self-coupling with a focus on the shape of the $m_{hh}$
distribution~\cite{Goncalves:2018yva}. Our signal is $pp \to hh + X \to b\bar{b} \; \gamma \gamma + X$.
The signal and background samples are generated with
\textsc{MadGraph5}+\textsc{Pythia8}~\cite{Alwall:2011uj, Alwall:2014hca,Sjostrand:2007gs}, including one
extra jet using the \textsc{Mlm} scheme~\cite{Mangano2002}.\medskip

In the final state we demand two $b$-tagged jets and
two isolated photons with the minimal acceptance and trigger cuts
\begin{equation}
p_{T,j}>30~\GeV , \quad  
|\eta_j |<2.5, \quad 
p_{T,\gamma}>30~\GeV, \quad  
|\eta_\gamma| <2.5 , \quad 
\Delta R_{\gamma \gamma, \gamma j, jj} >0.4 \; .
\label{eq:base_selections}
\end{equation}
The background to our $b\bar{b} \; \gamma \gamma$ signal consists of
other Higgs production modes ($t\bar{t}h, Zh$) with $h \to \gamma
\gamma$, continuum $b\bar{b}\gamma\gamma$ production, and of multi-jet
events with light-flavor jets faking either photons or $b$-jets
($jj\gamma\gamma, b\bar{b}\gamma j$)~\cite{Baur:2003gp}.  

The proper simulation of efficiencies and fake rates are a key
ingredient for a realistic background estimate in this analysis.  For
the HE-LHC and the future 100~\UTeV collider we follow the ATLAS
projections~\cite{ATL-PHYS-PUB-2016-026}. The efficiency for a tight photon 
identification can be well parametrised by
\begin{align}
\epsilon_{\gamma\to\gamma} = 0.863 - 1.07 \cdot e^{-p_{T,\gamma}/34.8~\GeV}\;,
\end{align}
and a jet-to-photon mis-identification rate by
\begin{align}
\epsilon_{j\to\gamma} = 
\begin{cases} 
5.30\cdot 10^{-4}  \ \exp\big( -6.5 \left( p_{T,j}/(60.4~\GeV)- 1 \right)^2 \big) 
\quad &\text{ for } p_{T,j} <65~\GeV \;, \\
0.88 \cdot 10^{-4} \  \left[ \exp \big( -(p_{T,j}/(943~\GeV) \big) +248~\GeV/p_{T,j}\right]
\quad &\text{ for } p_{T,j} >65~\GeV \;.
\end{cases}
\end{align}
This leads to a photon efficiency of about 40\% at $p_{T,\gamma}=30$~\UGeV,
saturating around 85\% for $p_{T,\gamma}>150$~\UGeV. Note that the 
Higgs decay products tend to be soft, $p_{T,\gamma}\sim m_h/2$. 
For $b$-tagging, we adopt an efficiency with $\epsilon_b =0.7$ associated 
with mis-tag rates of 15\% for charm quarks and 0.3\% for
light flavors. These flat rates present a conservative estimate from
the two dimensional distribution on $(p_{Tj},\eta_j)$ shown in the
HL-LHC projections~\cite{Kling:2016lay}. Encouragingly, the small light
flavor fake rate projections result in a strong suppression for the
initially dominant $jj\gamma\gamma$ background.

To control the continuum backgrounds, we require two Higgs mass windows,
\begin{align}
 |m_{bb}-m_h|<25~\GeV, \quad 
 |m_{\gamma\gamma}-m_h|<1~\GeV  .
 \label{eq:jreslov}
\end{align}
An obvious way to enhance the Higgs pair signal is to improve the
resolution on the reconstructed photons and $b$-jets from the Higgs
decays.  We adopt the rather conservative resolution for $m_{bb}$ as
in Eq.~\eqref{eq:jreslov}. Any improvement on it in experiments would
be greatly helpful for the signal identification and background
separation.  
\medskip

To take the information in the the differential distribution
$m_{hh}$ into account, we employ a binned log-likelihood analysis based on the 
CL$_{s}$ method, using the full $m_{hh}$ distribution to extract 
$\kappa_{\lambda}$~\cite{Read:2002hq}. As a starting point, we show the $5\sigma$ 
determination on the Higgs pair signal strength for the SM hypothesis $\kappa_{\lambda}=1$ as a 
function of the luminosity in the left panel of Fig.~\ref{fig:bound1}. Here 
we require two $b$-tagged jets among the two or three leading jets. We 
decompose the latter case in two  sub-samples $(bb,bbj)$ and $(jbb,bjb)$. 
We see how exploring the extra-jet emission significantly improves the 
significance as compared to the standard procedure adopted in the 
literature. The $5\sigma$ measurement for HE-LHC is pushed from 
$2.8~\iab$ to below $2.3~\iab$. 

In the right panel of Fig.~\ref{fig:bound1} we show the discovery reach for 
the Higgs pair signal at HE-LHC and a 100~\UTeV collider for three di-photon 
invariant mass resolutions described by a Gaussian width of 
$0.75,1.5,2.25$~GeV and corresponding Higgs mass windows 
$|m_{\gamma\gamma}-m_h|<1,2,3~\GeV$. As resolution of $1.5$~\UGeV 
has already been achieved at the LHC~\cite{CMS:2016zjv}. Higgs pair 
production will be discovered at the HE-LHC with approximately 
$2.5~...~5~\iab$ and at the 100~\UTeV collider with $0.2~...~0.3~\iab$ 
of data, in both cases well below the design luminosity.

\begin{figure}[t!]
\centering 
  \includegraphics[width=.4\textwidth]{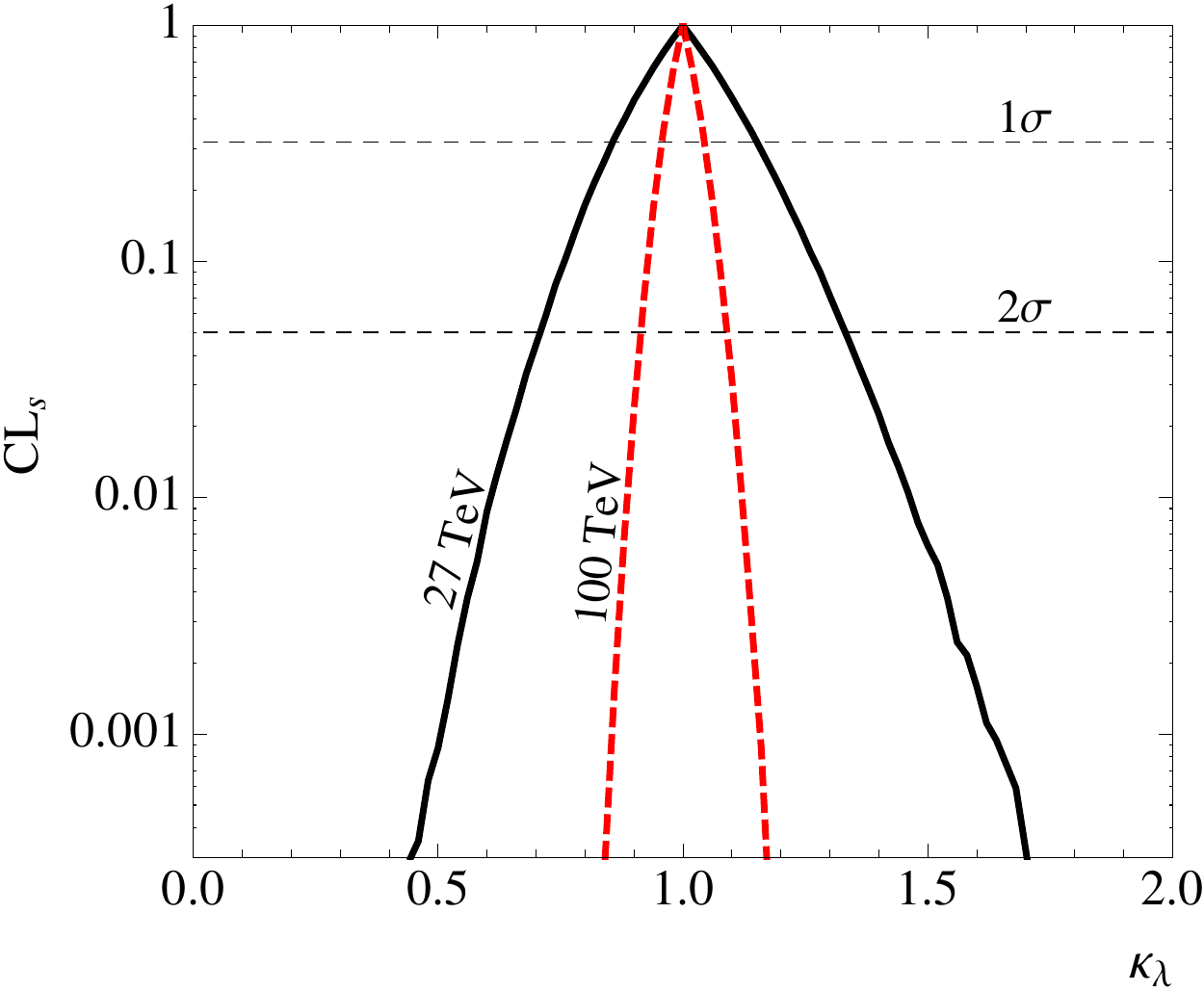}
  \caption{Confidence level for separating an anomalous Higgs
    self-coupling hypothesis from the Standard Model
    $\kappa_{\lambda}=1$.}
 \label{fig:bound2}
\end{figure}

As commented in the introduction, there exist physics scenarios 
in which the Higgs self-coupling could be modified at the level of order 
one deviation from the SM value. The accurate measurement of the 
Higgs self-coupling via Higgs pair production at future colliders has the 
best promise to uncover the new physics associated with the Higgs sector.
In Fig.~\ref{fig:bound2}, we show the accuracy on this measurement. 
We find that the Higgs self-coupling can be measured with a precision
\begin{align}
\kappa_{\lambda} &\approx 1 \pm 15\% \text{    at 68\% CL}\quad \text{and} \quad
\kappa_{\lambda} \approx 1 \pm 30\%   \text{    at 95\% CL}\quad\quad\quad
\text{(HE-LHC, 27~\UTeV, $15~\iab$)}  , \notag \\
\kappa_{\lambda} &\approx 1 \pm 5\% \;\; \text{    at 68\% CL}\quad \text{and}\quad
\kappa_{\lambda} \approx 1 \pm 10\% \text{    at 95\% CL} \quad\quad\quad
\text{(100~\UTeV, $30~\iab$).} 
\end{align}

While our conclusions on the determination of Higgs-self-interaction
at future hadron colliders are robust and important, there is still
room to improve. Although the final state $b\bar b\ \gamma\gamma$ is
believed to be the most sensitive channel because of the background
suppression and signal reconstruction, there exist complementary
channels such as ${gg\to hh \to b\bar b\ \tau^+\tau^-}$, $b\bar
b\ W^+W^-$, $b\bar b\ b\bar b$, etc. The kinematics-based measurement
and the all features related to QCD radiation at higher energies
should be equally applicable to all of them.

\asubsubsection[S. Homiller, P. Meade]{Theoretical prospects: importance of the gluon fusion single Higgs background.}
\label{sec:Homilleretal}

The Higgs self-coupling plays a central role in the spontaneous breaking of electroweak symmetry, and governs a pure elementary scalar interaction -- one that has never been observed in nature. Unfortunately, due to the small rate of $hh$ production, measuring the Higgs self-coupling at a $14\, {\rm \UTeV}$ appears exceedingly difficult unless it deviates substantially from the Standard Model value \cite{ATL-PHYS-PUB-2014-019, ATL-PHYS-PUB-2017-001}. A precision measurement of the Higgs self-coupling is thus one of the primary goals of any higher energy collider.  In this section we use the convention
\begin{equation}\label{eq:v_int}
  V_{\text{int}} = \lambda_3\frac{m_h^2}{2v}h^3 + \lambda_4\frac{m_h^2}{8v^2}h^4
\end{equation}
such that in the SM $\lambda_3=1$.

While the prospects of a $100\, {\rm \UTeV}$ collider in measuring the self-coupling have been well studied \cite{Contino:2016spe}, relatively less attention has been paid to intermediate energy colliders such as HE-LHC. Previous studies indicate that the $hh\rightarrow b\bar{b}\gamma\gamma$ channel has the most promising signature at hadron colliders, and this is expected to be true at $27\, {\rm \UTeV}$ as well. However, the $b\bar{b}\gamma\gamma$ channel still suffers from significant backgrounds from particle mis-identification in the detector, making a dedicated detector study including these effects essential. Finally, as discussed below, single-Higgs production -- including through gluon-fusion -- is a significant background that must be properly understood to accurately project the capabilities of HE-LHC. In what follows, we present a projection of the capabilities of a HE-LHC to measure the self-coupling with these intricacies carefully considered.

\asubsubsubsection{Signal and Background Simulations}

The signal and background samples generated for this study are summarised in Table \ref{t.backgrounds}. We also show the cross sections of $14\, {\rm \UTeV}$ samples generated for validation with previous projections. 

The details of the signal and background simulations mimic those in Ref. \cite{Homiller:2018dgu}. 
The $pp \rightarrow hh \rightarrow b\bar{b}\gamma\gamma$ signal is simulated at leading order using {\sc\small MadGraph5\_aMC@NLO}\ \cite{Alwall:2014hca, Hirschi:2015iia} using the NNPDF2.3LO PDF set \cite{Ball:2014uwa} including all finite top mass effects. The {\sc\small MadSpin}\ package \cite{Artoisenet:2012st} was used for the Higgs boson decays and {\sc\small Pythia~8}\ \cite{Sjostrand:2014zea} for the showering and hadronisation of events. The LO signal is normalised to match the state of the art NNLO/NNLL calculation with finite top mass effects included at NLO in QCD~\cite{Grazzini:2018bsd}. Additional samples with the self-coupling modified to values between $-1$ and $10$ times the SM value were also generated. Representative kinematic distributions of the signal at parton level are shown in Fig. \ref{fig:dihiggs_kinematics}.

\begin{figure}[h]
\centering
	\includegraphics[width=0.32\linewidth]{\main/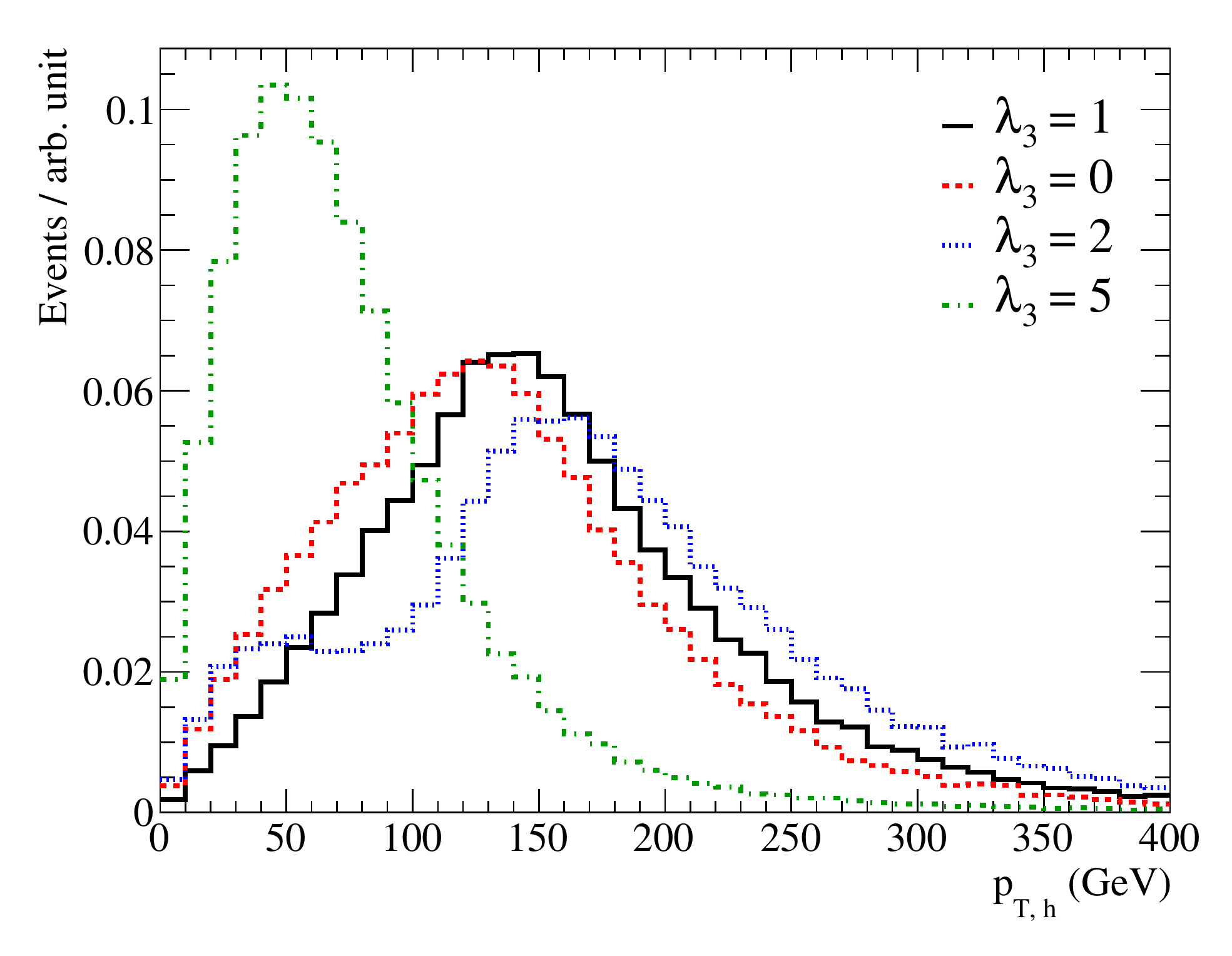}
	\hfill
	\includegraphics[width=0.32\linewidth]{\main/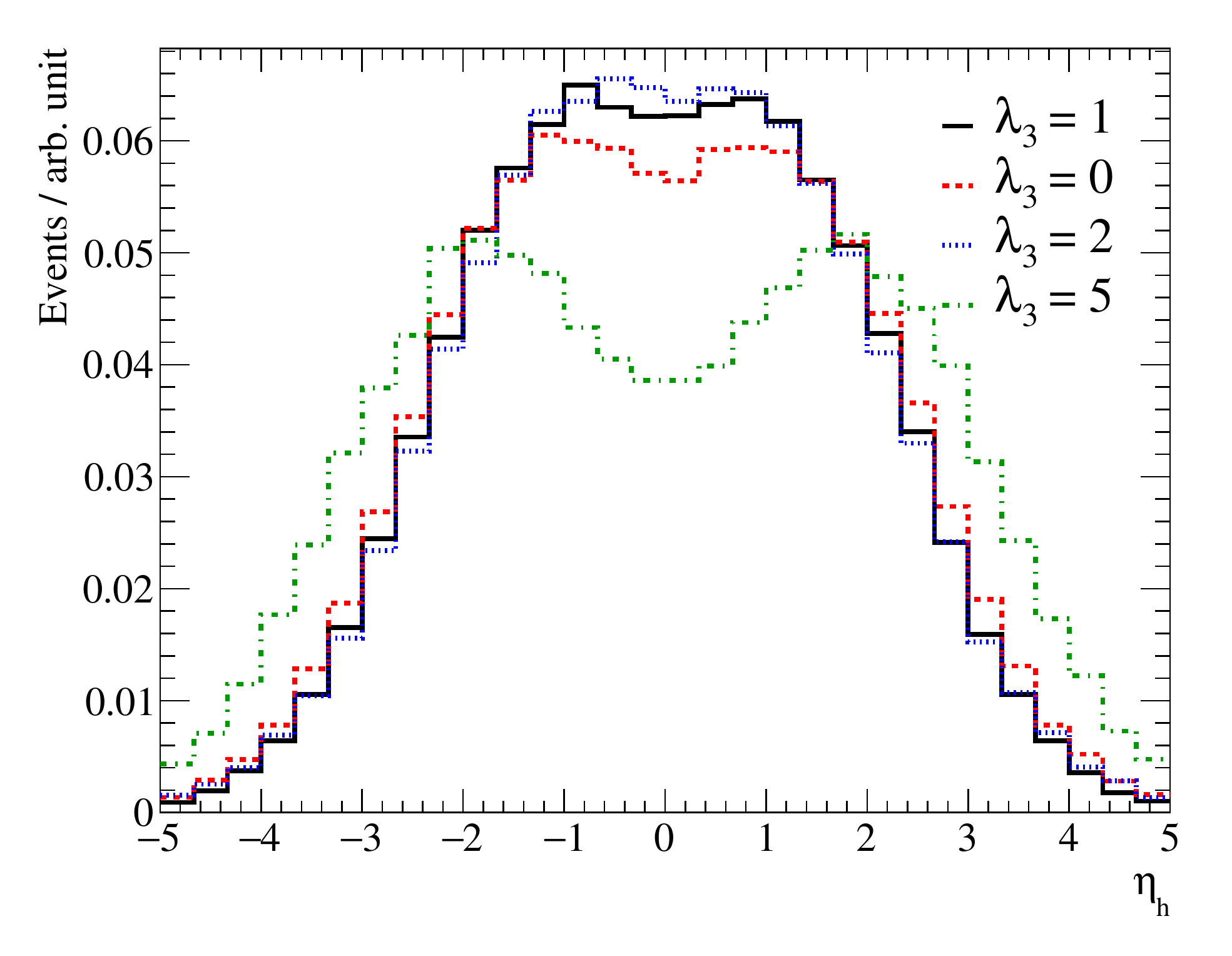}
	\hfill
	\includegraphics[width=0.32\linewidth]{\main/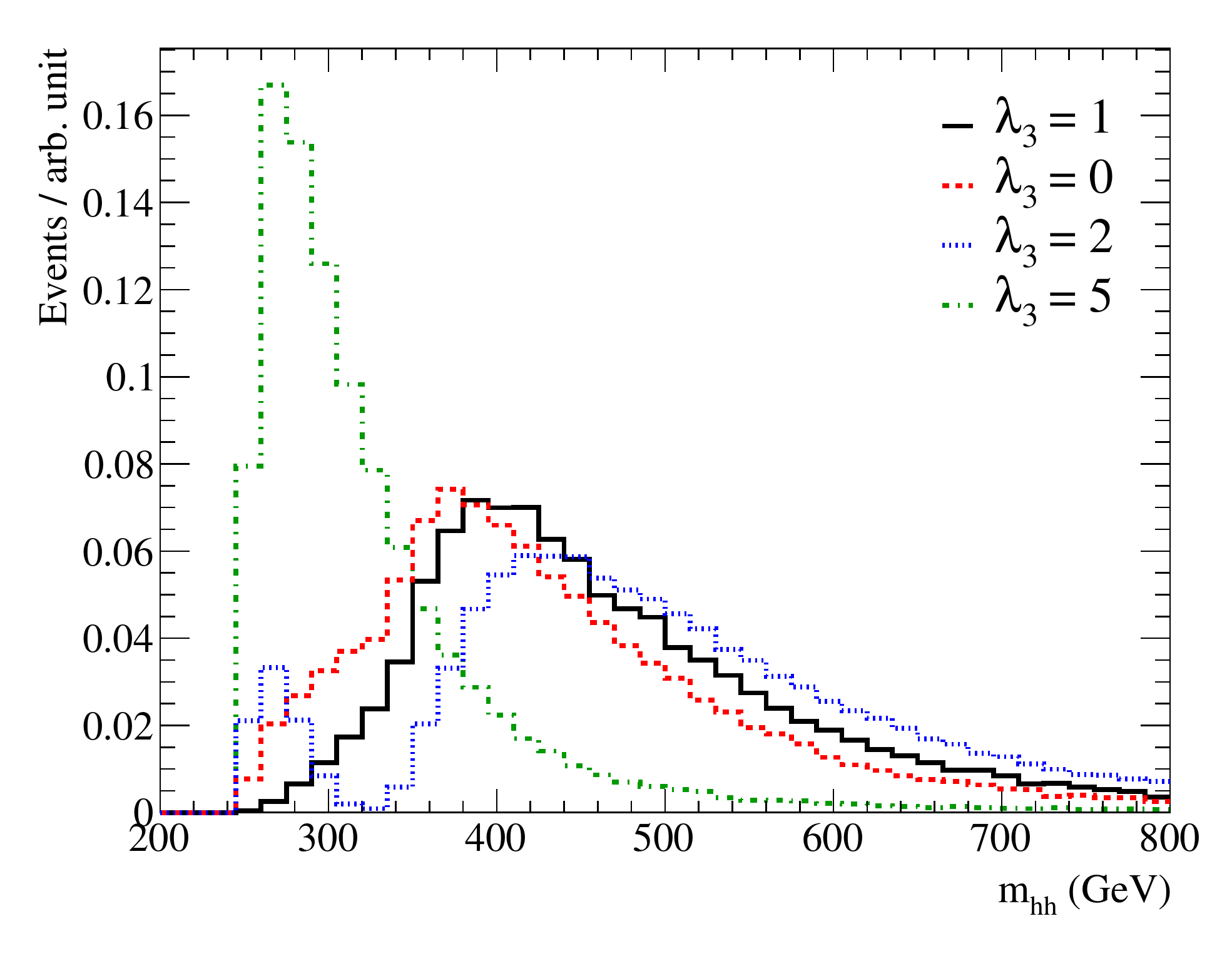}
\caption{(Left:) The transverse-momentum distribution of the true Higgs bosons generated in our 27 \UTeV samples, prior to showering and detector smearing, for several different values of $\lambda_3$. (Centre:) The same, but for the Higgs pseudorapidity. Right: The same, but for the distribution of the true Higgs pair invariant mass.}\label{fig:dihiggs_kinematics}
\end{figure}

Backgrounds to the $b\bar{b}\gamma\gamma$ decay channel include single Higgs production modes, non-resonant QCD backgrounds, as well as $Z(b\bar{b})\gamma\gamma$ and $t\bar{t}(+\gamma)$ production. We include all backgrounds where up to two additional photons or $b$-jets can arise from either misidentified light jets or electrons misidentified as photons.

The background from single Higgs production via gluon fusion ($ggF(\gamma\gamma)$) was generated in {\sc\small MadGraph}\ with up to two extra partons allowed in the matrix element, with no additional $k$-factor due to the already present real emissions. Events from other single Higgs production modes were generated directly in {\sc\small Pythia~8}\ at LO and normalised based on the recommendations in Ref. \cite{deFlorian:2016spz}. The remaining backgrounds were generated in {\sc\small MadGraph}\ interfaced with {\sc\small Pythia~8}\ for showering and hadronisation, with one additional jet allowed in the matrix element with MLM matching \cite{Mangano:2006rw, Alwall:2007fs} to the parton shower.

\begingroup
\renewcommand*{\arraystretch}{.7}
\begin{table}
\centering
\begin{tabular}{|c|c|cc|c|rcl|}
\hline
\multirow{2}{*}{Process}     & \multirow{2}{*}{Generator} & \multicolumn{2}{c|}{$\sigma \cdot BR$ {[}fb{]}} & \multirow{2}{*}{Order QCD} & \multicolumn{3}{c|}{Expected Events} \\
                              			&                   							& 14 \UTeV  			& 27 \UTeV  			&		& \multicolumn{3}{c|}{($27\, {\rm \UTeV}, 15~\text{ab}^{-1}$)} \\ \hline
$h(b\bar{b})h(\gamma\gamma)$  	& {\sc\small MadGraph}/{\sc\small Pythia~8} 	& 0.11      			& 0.41      			& NNLO/NNLL								& $209.6$		& $\pm$ & $0.2$ \\ \hline
$t\bar{t}h(\gamma\gamma)$     		& {\sc\small Pythia~8}           			& 1.40      			& 6.54      			& NLO  										& $286.8$		& $\pm$ & $1.6$\\
$Zh(\gamma\gamma)$            		& {\sc\small Pythia~8}           			& 2.24      			& 5.58      			& NLO 											& $67.1$ 		& $\pm$ & $0.7$ \\
$ggF(\gamma\gamma)$           		& {\sc\small MadGraph}/{\sc\small Pythia~8} 	& 83.2      			& 335.1     		& $\text{N}^{3}\text{LO}$ 				& $349.7$ 	& $\pm$ & $9.5$ \\ \hline
$b\bar{b}\gamma\gamma$        	& {\sc\small MadGraph}/{\sc\small Pythia~8} 	& $3.4\times 10^{2}$     & $9.5\times 10^{2}$     & LO 					& $414.6$ 	& $\pm$ & $10.3$ \\
$c\bar{c}\gamma\gamma$        		& {\sc\small MadGraph}/{\sc\small Pythia~8} 	& $4.4\times 10^{2}$     & $1.5\times 10^{3}$     & LO 					& $185.7$ 	& $\pm$ & $4.2$ \\
$jj\gamma\gamma$              			& {\sc\small MadGraph}/{\sc\small Pythia~8} 	& $5.9\times 10^{3}$     & $1.4\times 10^{4}$     & LO 					& $63.3$ 		& $\pm$ & $3.8$ \\
$b\bar{b}j\gamma$             			& {\sc\small MadGraph}/{\sc\small Pythia~8} 	& $1.1\times 10^{6}$     & $3.4\times 10^{6}$     & LO 					& $199.6$ 	& $\pm$ & $9.4$ \\
$c\bar{c}j\gamma$             			& {\sc\small MadGraph}/{\sc\small Pythia~8} 	& $4.8\times 10^{5}$     	& $1.6\times 10^{6}$     & LO 				& $25.3$ 		& $\pm$ & $3.0$ \\
$b\bar{b}jj$                  					& {\sc\small MadGraph}/{\sc\small Pythia~8} 	& $3.7\times 10^{8}$     	& $1.5\times 10^{9}$     & LO 				& $155.4$ 	& $\pm$ & $8.2$ \\
$Z(b\bar{b})\gamma\gamma$   		& {\sc\small MadGraph}/{\sc\small Pythia~8} 	& 2.61     			& 5.23      			& LO 											& $21.5$ 		& $\pm$ & $0.4$ \\ \hline
$t\bar{t}$                    					& {\sc\small MadGraph}/{\sc\small Pythia~8} 	& $6.7\times 10^{5}$     	& $2.9\times 10^{6}$     	& NNLO 		& $11.6$ 		& $\pm$ & $3.3$ \\
$t\bar{t}\gamma$              				& {\sc\small MadGraph}/{\sc\small Pythia~8} 	& $1.7\times 10^{3}$     	& $7.9\times 10^{3}$     & NLO 				& $145.0$ 	& $\pm$ & $10.3$ \\ \hline
\multicolumn{5}{|c|}{Total Background}					& $1925.8$ & $\pm$ & $22.7$ \\ \hline
\multicolumn{5}{|c|}{Significance ($S/\sqrt{B}$)}		& $4.77$ & $\pm$ & $0.14$ \\ \hline
\end{tabular}
\caption{List of signal and background processes, the event generator used to simulate the matrix element and parton shower, and the cross section of each process along with the corresponding order in QCD at which the cross section is normalised. In the right-most column we show the expected number of events after all the event selection criteria have been applied.}
\label{t.backgrounds}
\end{table}
\endgroup

\asubsubsubsection{Detector Simulation}

To approximate the effects of detector resolution and reconstruction efficiencies, we use {\sc\small Delphes~3}\ with a dedicated card developed to approximate the performance of ATLAS and CMS at HL-LHC. We take this as a reasonable benchmark for the expected performance after the HE-LHC upgrade.

With respect to the {\sc\small Delphes}\ setup used in \cite{Homiller:2018dgu}, the card here has an improved E-Cal resolution and assumes a higher photon identification efficiency, but a somewhat degraded di-jet mass resolution. Aside from resolution and efficiency effects, particle mis-identification in the detector is also an important source of backgrounds to $hh\rightarrow b\bar{b}\gamma\gamma$. To avoid issues with MC statistics, we implement $b$-tagging and jet mis-tagging rates at analysis level using a reweighting scheme, with probabilities taken as functions of the jet $p_T$ as in Ref. \cite{Homiller:2018dgu}. These probabilities correspond to roughly $p_{b\rightarrow b} \approx 70\%$, $p_{c\rightarrow b} \approx 20\%$ and $p_{j\rightarrow b} \lesssim 1\%$. The probability for a light jet to fake a photon in the detector is also included via reweighting at analysis level as a function of $p_T$ (see \cite{Homiller:2018dgu}) which peaks at $5\times 10^{-4}$ for $p_{T,j} \sim 60\, {\rm \UGeV}$ before falling exponentially to $\sim 1\times 10^{-4}$.

\asubsubsubsection{Results and Limits on the Self-Coupling}

To isolate the $hh \rightarrow b\bar{b}\gamma\gamma$ signal, we implement selection cuts as follows:
\begin{itemize}
\setlength\itemsep{0.25em}
  \item At least 2 isolated photons and b-tagged jets with leading $p_T > 60\, {\rm \UGeV}$ and sub-leading $p_T > 35\, {\rm \UGeV}$, all with $|\eta_{\gamma,b}| < 2.5$.
  
\begin{minipage}{0.49\textwidth}
  \item $p_{T,\gamma\gamma}, p_{T,b\bar{b}} > 125\, {\rm \UGeV}$.
  \item $\Delta R_{b\bar{b}}, \Delta R_{\gamma\gamma} < 3.5.$
  \item $|m_{\gamma\gamma} - 125.0\, {\rm \UGeV}| < 4.0\, {\rm \UGeV}$.
  \item $|m_{b\bar{b}} - 125.0\, {\rm \UGeV}| < 25\, {\rm \UGeV}$.
\end{minipage}
\begin{minipage}{0.49\textwidth}
  \item $n_{\text{jets}} < 6$ for jets with $p_T > 30\, {\rm \UGeV}$, $|\eta| < 2.5$.
  \item No isolated leptons with $p_T > 25\, {\rm \UGeV}$.
  \item $|\cos\theta_{hh}| < 0.8$.
\end{minipage}
\end{itemize}

where $\cos\theta_{hh}$ is the decay angle of the Higgs boson pair evaluated in the lab frame (see Fig. \ref{fig:kinematics}).

\begin{figure}[ht]
\centering
	\includegraphics[width=0.49\linewidth]{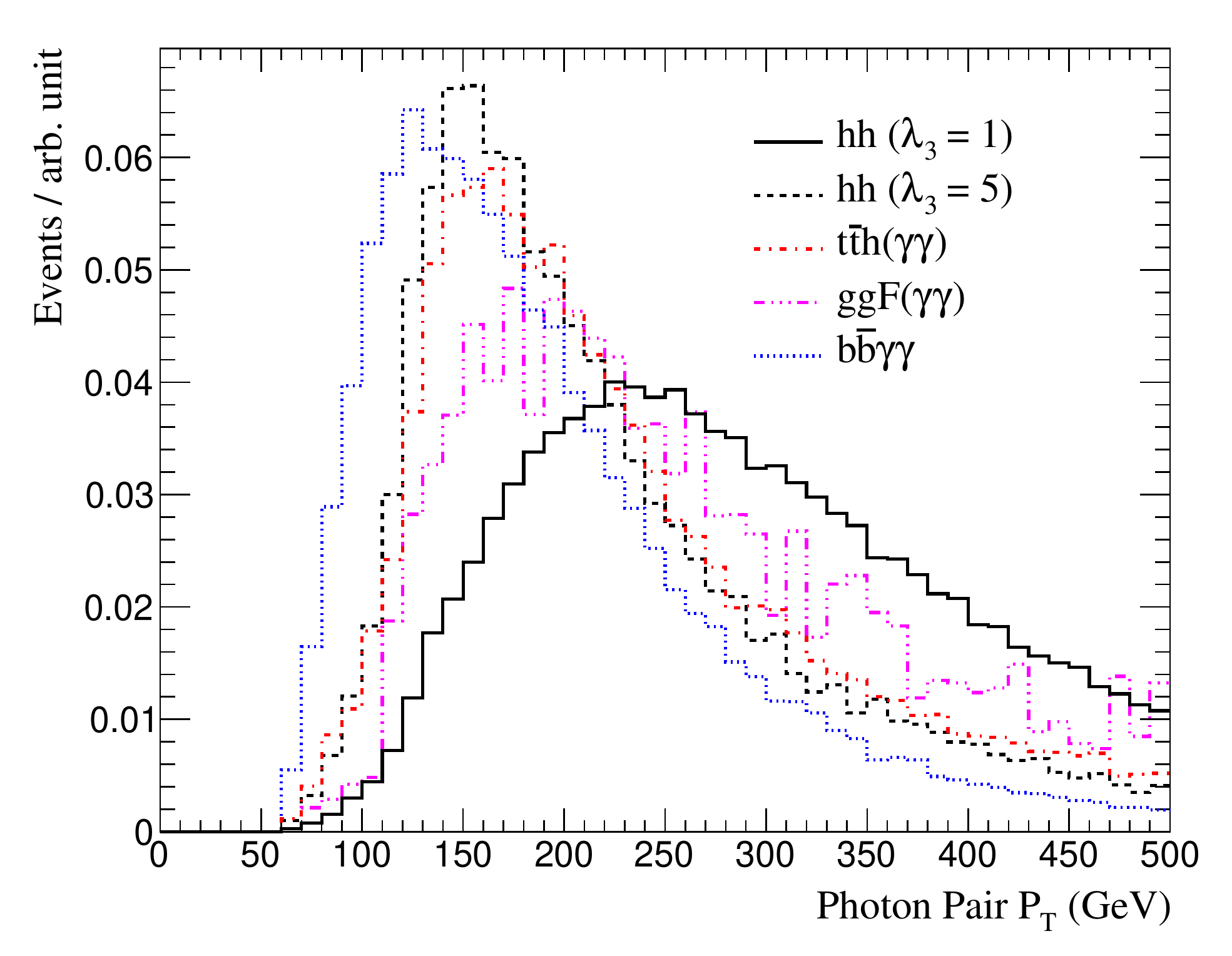}
\hfill
	\includegraphics[width=0.49\linewidth]{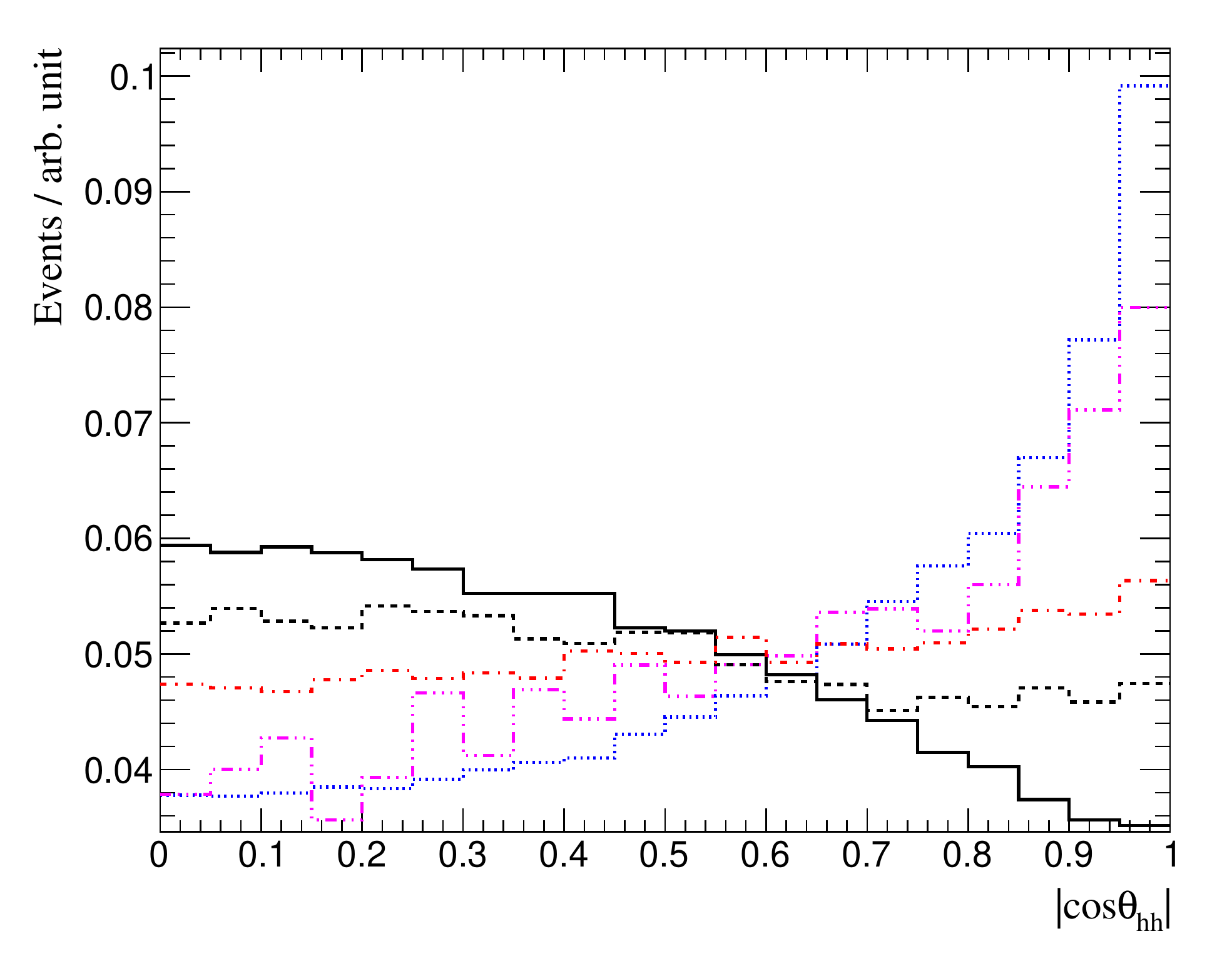}
\caption{Normalised distributions of (Left:) the $p_T$ of the reconstructed $h\rightarrow\gamma\gamma$ and (Right:) the magnitude of $\cos\theta_{hh}$, the Higgs decay angle defined in the text. We show the distributions for the signal with $\lambda_{3} = 1$ and $5$ as well as several representative backgrounds.}
\label{fig:kinematics}
\end{figure}

Note that cuts on the $p_T$ and $\Delta_R$ of the $\gamma\gamma$ and $b\bar{b}$ pair are tightly correlated with the invariant mass of the $hh$ system. As seen in Fig. \ref{fig:kinematics} the photon pair $p_T$ has strong discriminating power for the SM $hh$ signal, but for non-SM values of $\lambda_{3}$, the signal and background become more degenerate.

The final selection efficiency is $3.4 \%$, and the expected number of events from each signal/background channel after applying all the cuts and detector effects is given in Table \ref{t.backgrounds} assuming $15~\text{ab}^{-1}$ integrated luminosity at HE-LHC. The uncertainty for each sample is estimated by partitioning the full event sample in to sub-samples and computing the standard deviation of the results from each sub-sample.

The largest backgrounds are from continuum $bb\gamma\gamma$ and single Higgs production and decay to $\gamma\gamma$. Particularly, we see that the $ggF$ induced mode contributes an $\mathcal{O}(1)$ background, despite being neglected in previous studies. The accurate modelling of the extra jets that arise in the hadron collision is a necessity for properly understanding this contribution.
Other large backgrounds arise from processes where a jet is reconstructed as a photon -- even when two fake photons are needed. Finally, we see that $t\bar{t}$ and $t\bar{t}\gamma$ are not insignificant backgrounds with the set of cuts we've applied. Several of these backgrounds might be mitigated by exploring the additional kinematic information in events with multiple jets, but the single-Higgs production backgrounds are difficult to reduce in light of the true $h\rightarrow \gamma\gamma$ present.

\begin{figure}
	\centering
	\includegraphics[width=8cm]{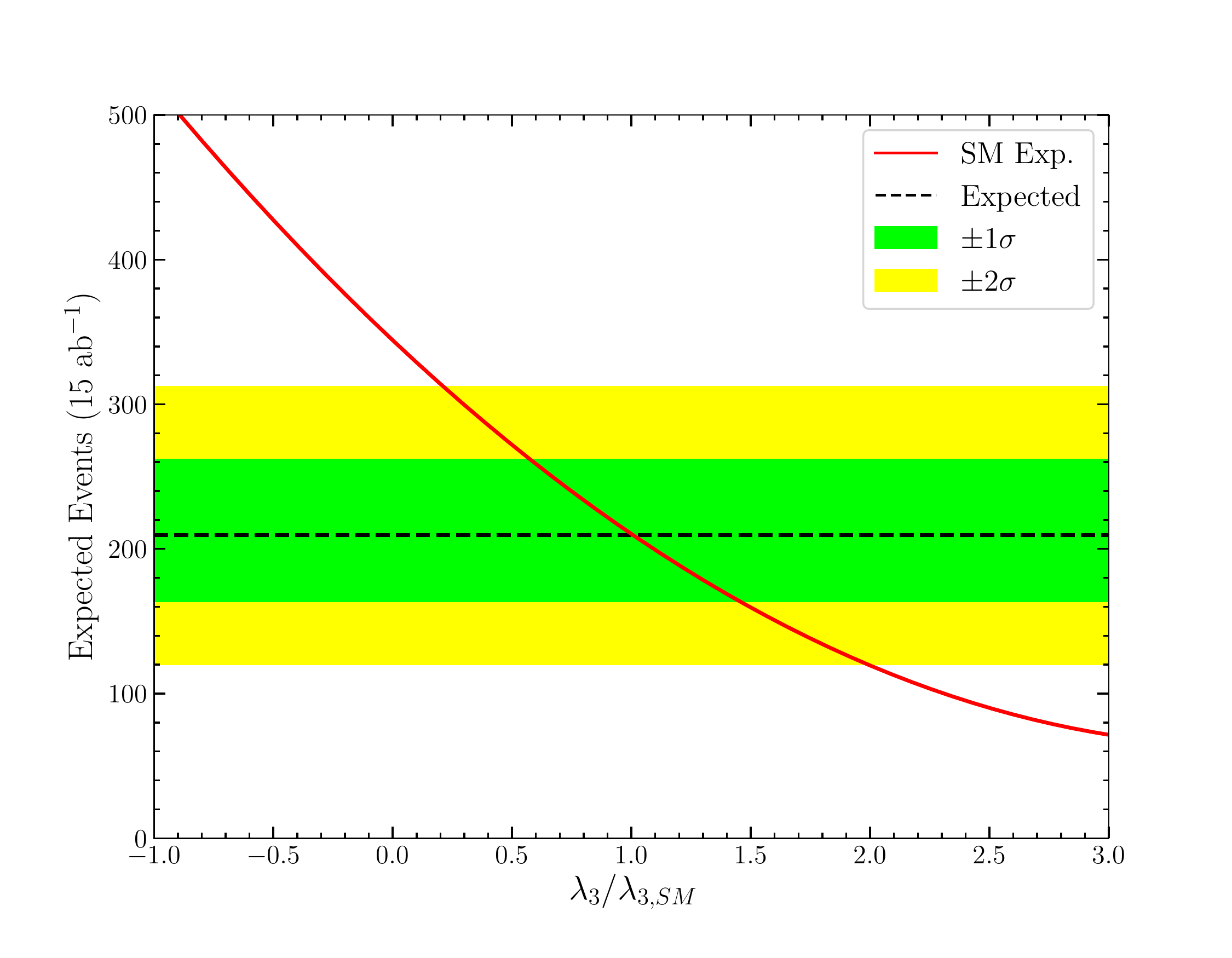}
	\caption{The expected number of signal events in a hypothetical experiment assuming the signal and background rates computed in Table \ref{t.backgrounds} at $L = 15~\text{ab}^{-1}$ for HE-LHC with the regular detector performance assumption. The black dashed line indicates the expected number of events from signal while the green (yellow) regions show the $1\sigma~(2\sigma)$ uncertainty regions arising from a likelihood scan with the statistical and MC uncertainties on the signal and background counts. The red curve shows the expected number of events from signal in a background free measurement as a function of $\lambda_3$, accounting for the changes in the signal acceptance due to kinematic differences at different $\lambda_3$}\label{fig:lam3_limit}
\end{figure}

To understand the attainable precision on $\lambda_{3}$, we assume a hypothetical observation of $S+B$ events after all selection cuts with $S$ and $B$ as in Table \ref{t.backgrounds}. This allows us to derive $68$ and $95\%$ confidence intervals on the expected number of signal events using a likelihood scan, including only the MC and statistical uncertainties. The expected number of signal events with $15~\text{ab}^{-1}$ integrated luminosity is plotted in Fig. \ref{fig:lam3_limit} along with the $1\sigma~(2\sigma)$ regions in green (yellow).

We can also compute the expected number of events at $15~\text{ab}^{-1}$ as a function of $\lambda_{3}$, taking into account both the varying $\sigma_{hh}$ cross section and the modified acceptance due to changes in the signal kinematics. The resulting curve is shown in red in Fig. \ref{fig:lam3_limit}. The intersection of this curve with the $1$ and $2\sigma$ regions indicate the expected precision on $\lambda_3$ in the absence of systematic uncertainties. We find

\begin{equation}
\lambda_{3} \in \left[ 0.58 ,  1.45 \right] \quad \text{at}~ 68\%~\text{C.L.}
\end{equation}

Note that, as a result of the destructive interference between the triangle and box diagrams leading to $hh$ production, there is a degeneracy in the expected number of events around $\lambda_3 \sim 5$. However, the kinematic structure of the $hh$ signal is very different at large values of $\lambda_3$, and such values could be easily rejected using differential measurements (e.g, with $m_{hh} = m_{b\bar{b}\gamma\gamma}$ or $p_{T,hh}$), so the degeneracy can be safely ignored for the purposes of this work.

In conclusion, we find that with a full account of the detector effects and backgrounds to $hh\rightarrow b\bar{b}\gamma\gamma$, a cut based analysis leads to an expected significance of $4.77 \pm 0.14 \sigma$, corresponding to a $45 \%$ measurement of the Higgs self-coupling at $27\, {\rm \UTeV}$ with $15~\text{ab}^{-1}$. Future improvements can be made both by considering other decay channels (e.g., $hh\rightarrow b\bar{b}b\bar{b}, b\bar{b}\tau\tau$, and $b\bar{b}WW$) and by exploiting the additional information present in the $hh$ invariant mass distribution, as discussed elsewhere in this report.

\asubsubsection[P. Bokan, E. Petit, N. Readioff, M. Wielers]{Experimental prospects with the ATLAS detector}
\label{sec:HH_HE_atlas}
The results presented in Section~\ref{sec:HH_ATLAS} were extended to provide estimates of the prospects at the HE-LHC, assuming a centre of mass collision energy of 27~\TeV\ and 15~\abinv\ of data.

The assumption is made that the detector performance will be the one of the HL-LHC ATLAS detector. Comparisons between simulation at centre of mass energy of 14 and 27~\TeV\ show that the kinematic of the Higgs boson decay particles, as well as the $m_{HH}$ distribution are similar. However the pseudorapidity of the particle tends to point more frequently in the forward region, which would decrease the acceptance by around 10\%. This effect is not taken into account and the impact is expected to be small.

The event yields for the various background processes are scaled by the luminosity increase and the cross-section ratio between the two centre of mass energies. For the signal the cross-section of 139.9~fb is used, as described in Section~\ref{sec:HH_NNLO}.

Without including systematic uncertainties a significance of 7.1 and 10.7 standard deviations is expected for the $b\bar{b}\gamma\gamma$ and $b\bar{b}\tau\tau$ channels respectively.
The hypothesis of no Higgs self-coupling can be excluded with a significance of 2.3 and 5.8 standard deviations respectively. Finally the $\kappa_{\lambda}$ parameter is expected to be measured with a 68\% CI precision of 40\% and 20\% for the two channels respectively.
With the $b\bar{b}\gamma\gamma$ channel, if the HL-LHC systematic uncertainties were considered this precision would be 50\%, dominated by the uncertainty on the photon energy resolution. If this uncertainty were divided by a factor 2 then the precision would be 45\%.

\asubsubsection{Comparison of results}

The results presented in Sections~\ref{sec:THanetal} and \ref{sec:HH_HE_atlas} appear to be quite different, with the $\kappa_{\lambda}$ parameter being measured with a precision of 15\% and 40\% respectively at 68\% CL.
Thorough studies were performed to understand the difference. The result from the ATLAS experiment is an extrapolation of the HL-LHC results (which consider a mean pile-up rate of 200) which were optimised to increase the sensitivity to the SM signal, but could be improved for a precise measurement of $\kappa_{\lambda}$. In particular low values of the di-Higgs invariant mass below 400~\GeV\ are suppressed where most of the sensitivity lies.
When a selection similar to the one in Section~\ref{sec:THanetal} is applied to ATLAS simulated samples, 40\% more background events are found. Half of it comes from missing background processes, while the other half comes from differences in the selection because of the effect of pile-up in the ATLAS simulation.
There is also a categorisation based on the \pT-ordering of the jets and b-jets which improves the results as shown in Figure~\ref{fig:bound1} but is hard to reproduce when large pile-up is considered.

In order to get an estimate of the best sensitivity achievable with HE-LHC data, a simple combination of the results of the $b\bar{b}\gamma\gamma$ channel presented in Section~\ref{sec:THanetal} and the results presented with the $b\bar{b}\tau\tau$ presented in Section~\ref{sec:HH_HE_atlas} is performed. No correlations are taken into account in this combination, and no systematic uncertainties are considered. A precision of around 10\% could be then achieved. The $b\bar{b}\tau\tau$ measurement alone is used as an upper value of this precision, so at this point we can consider that the $\kappa_{\lambda}$ parameter could be measured with a precision of 10 to 20\%, as illustrated in Figure~\ref{fig:HH_HELHC_comb}. It should also be noted that the second minimum of the likelihood would be unambiguously excluded at the HE-LHC.

It should be emphasised that these results rely on assumptions of experimental performance in very high pile up environment O(800-100) that would require further validation with more detailed studies, and that no systematic uncertainties are considered at this point. On the other hand these studies do not include the additional decay channels that have already been studied for HL-LHC, and of others that could become relevant at the HE-LHC. Exclusive production modes are also very interesting to take into consideration for this measurement. The potential improvements from these have not yet been assessed yet.

\begin{figure}[!htb]
\centering 
\includegraphics[width=0.7\textwidth]{\main/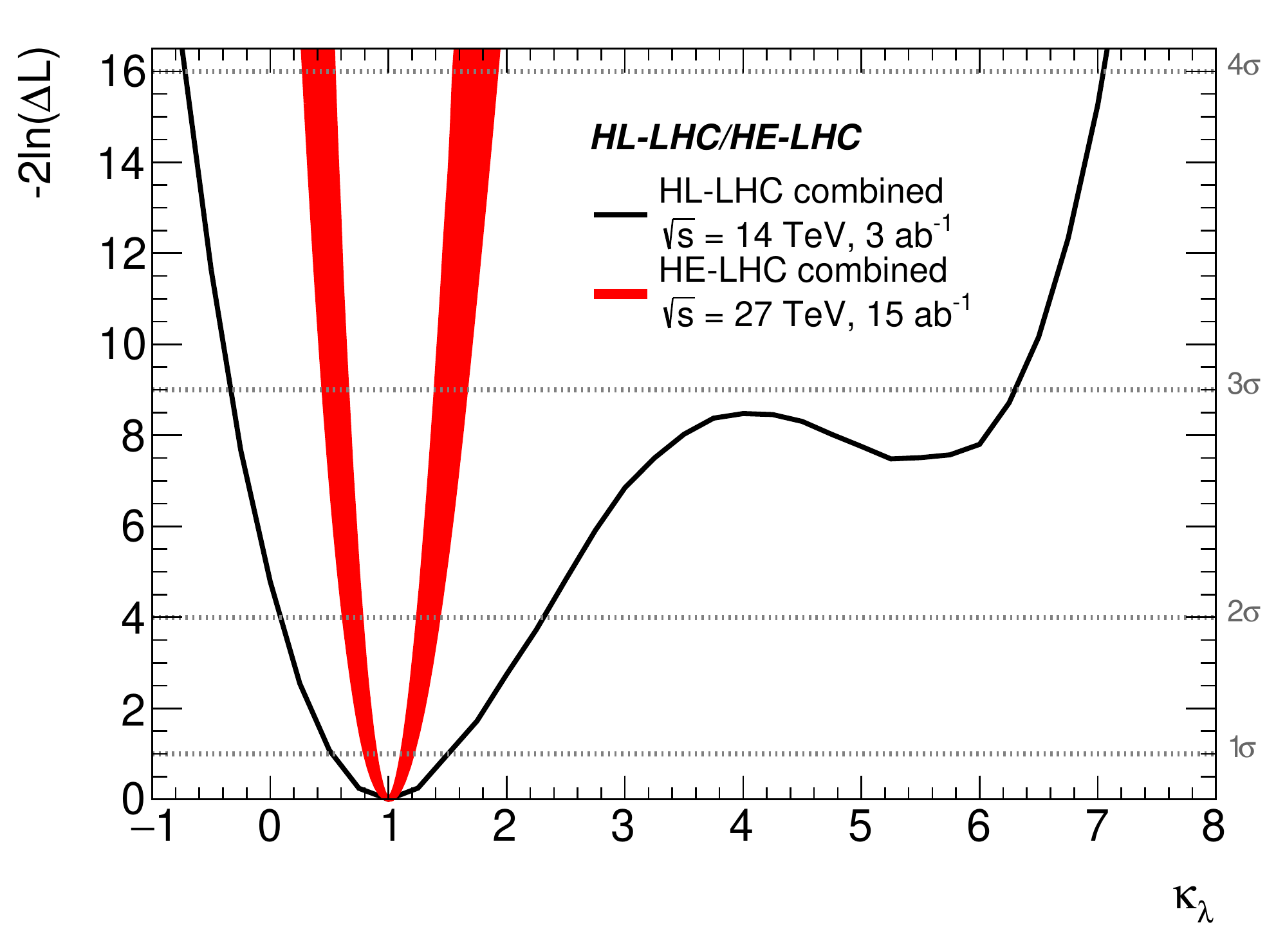}
\caption{Expected sensitivity for the measurement of the Higgs trilinear coupling through the measurement of direct HH production at HE-LHC. The black line corresponds to the combination of ATLAS and CMS measurements with HL-LHC data presented in Section~\ref{sec:HH_Combination}, with systematic uncertainties considered. The red band corresponds to an estimate of the sensitivity using a combination of the $b\bar{b}\gamma\gamma$ and $b\bar{b}\tau\tau$ channels, without systematic uncertainties considered.} 
\label{fig:HH_HELHC_comb} 
\end{figure}


\asubsection{Indirect probes}
\label{sec:HH_indirect}

In this section we discuss the possibility of indirectly extract information on the trilinear self interaction of the Higgs boson via precise measurements of single-Higgs production~\cite{McCullough:2013rea,Gorbahn:2016uoy,Degrassi:2016wml,Bizon:2016wgr,DiVita:2017eyz,Barklow:2017awn,Maltoni:2017ims,DiVita:2017vrr,Maltoni:2018ttu} at the HL-LHC and HE-LHC. This strategy is complementary to the direct measurement via double-Higgs production, which already at leading order,~i.e.~at one loop in the case of $gg \to HH$, depends on the trilinear Higgs self interaction. In the case of single-Higgs production, on the contrary, the Higgs self interactions enter only via one-loop corrections, i.e., at the two-loop level for the gluon-fusion ($ggF$) production mode. The effects of modified Higgs self interactions are therefore generically much smaller, but for single-Higgs production processes the precision of the experimental measurements is and will be much better than for double-Higgs production. This, and the fact that for single-Higgs production many different final states and both inclusive as well as  differential measurements are possible will lead to competitive indirect determinations of the trilinear Higgs self coupling. In \cite{Degrassi:2017ucl,Kribs:2017znd} also electroweak precision observables have been considered to this purpose.

\asubsubsection[W. Bizon, M. Gorbahn, U. Haisch, F. Maltoni, D. Pagani, A. Shivaji, G. Zanderighi, X. Zhao]{Indirect probes through single Higgs boson production}
In the following subsection, we will briefly recall the calculation framework introduced in~\cite{Gorbahn:2016uoy,Degrassi:2016wml}. We also provide  numerical results for the effects due to a modified  trilinear Higgs coupling in the most important inclusive and differential single-Higgs production cross sections as well as the Higgs branching ratios. Based on these results, we will analyse the sensitivity of the HL-LHC and HE-LHC in constraining the trilinear Higgs self interactions. 

\asubsubsubsection{Theoretical framework}
\label{tril-single:theo}

The effects of anomalous Higgs interactions can be extracted from experimental data via the signal strength parameters $\muif$, which
are defined for any specific combination of production and decay channel $i \to H \to f$ as follows 
\be \label{signalstre}
\muif \equiv \mu_i \times \mu^f = \frac{\sigma(i)}{\sigma^{\rm SM} (i)} \times \frac{\br(f)}{\br^{\rm SM}(f)} \, .
\ee 
Here the quantities $\mu_i$ and $\mu^f$ are   the production cross sections $\sigma(i)$ ($i=$ $\ggF$, {\rm VBF}, $WH$, $ZH$, $t \bar tH$, $tHj$) and the branching ratios $\br(f)$ $(f= \gamma\gamma, ZZ, WW, b\bar{b}, \tau\tau, \mu\mu)$ normalised to their SM values, respectively. Assuming on-shell production, the product $\mu_i \times \mu^f$  therefore corresponds to the rate for the $i \to H \to f$ process  normalised to the corresponding SM prediction.

The quantities $\mu_i$ and $\mu^f$ that enter the definition of  $\muif$ in~(\ref{signalstre})  can be expressed as
\begin{equation} \label{eq:muf}
\mu_i = 1 + \dsigmah(i) \,,  \qquad 
\mu^f = 1 + \dBR(f) \,,
\end{equation}
where $\dsigmah(i)$ and $\dBR(f)$ are the deviations induced by an anomalous trilinear Higgs self interaction to the production cross sections and branching ratios, respectively.
This definition can be straightforwardly extended to the differential level and one has $\muif=\mu_i=\mu^f=1$ in the~SM.

In single-Higgs production, the trilinear Higgs self interactions start to enter only at the one-loop level in the case of VBF, $WH$, $ZH$, $t \bar tH$, $tHj$ production, while in the case of $\ggF$ production and the decays $H\to gg, \gamma \gamma$ one has to calculate two-loop EW corrections. The appearance of the quadrilinear Higgs self coupling in single-Higgs processes is further delayed by one loop order.

For the strategy discussed here, the anomalous trilinear Higgs self interactions can be equivalently parametrised either via an anomalous trilinear coupling  
\begin{equation}
\tril \equiv \ktre \trilsm
\end{equation}
where $\lambda_3^{\rm SM} = m_H^2/(2 v^2)$ with $v = (\sqrt{2} G_F)^{-1/2} \simeq 246 \, {\rm \UGeV}$ the EW vacuum expectation value, or via the corresponding dimension-six operator 
\begin{equation} \label{O6}
{\cal O}_6 = - \frac{\lambda_3^{\rm SM} c_6}{v^2} \, |\Phi|^6  \,,
\end{equation}
with $\Phi$~denoting the usual SM Higgs doublet. In the normalisation adopted in (\ref{O6}), the simple relation 
\begin{equation}
\ktre = 1 + c_6 \,,
\end{equation}
is obtained and allows to translate constraints on the coupling modifier $\ktre$ into bounds on the Wilson coefficient $c_6$ and vice versa. 

\begin{figure}
	\centering
	\includegraphics[width=0.4\linewidth]{\main/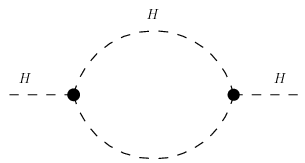}\\
	\includegraphics[width=0.4\linewidth]{\main/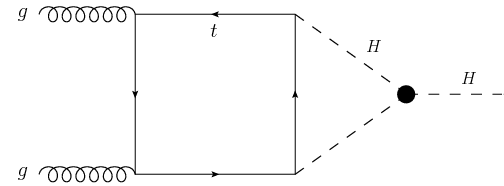}
	\includegraphics[width=0.4\linewidth]{\main/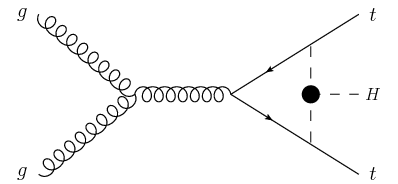}
	\caption{Examples of NLO contribution of the Higgs self-coupling to single Higgs observables. Top: Contribution to the Higgs self-coupling, which generates a global correction to all amplitudes. Bottom: Examples of diagrams contributing to the $ggF$ (left) and $t\bar{t}H$ (right) production modes.}
	\label{fig:singlehdiagrams}
\end{figure}

In the presence of modified trilinear Higgs self interactions, all single-Higgs production and decay channels receive two types of contributions~\cite{Gorbahn:2016uoy,Degrassi:2016wml}, as shown in Fig~\ref{fig:singlehdiagrams}: firstly, a process and kinematic dependent one, denoted as $C_1$ hereafter, which is linear in $c_6$ or $\ktre$ and second, a universal one proportional to the Higgs wave function renormalisation constant $Z_H$, which is proportional to $\ktre^2$ and therefore contains both a linear and quadratic piece  in $c_6$. The quantity~$\dsigmah(i)$ introduced in~\eqref{eq:muf} as well as any differential distribution related to it can thus be written as\footnote{This equation is in reality a linearised version of the complete formula that is used for extracting the results in Section ... and involves the Higgs wave function resummation \cite{Degrassi:2016wml,Maltoni:2017ims}. Also \eqref{BRgeneral} is a linear expansion. } 
\begin{equation} \label{dsigmakt}
\dsigmah(i)  = \left ( \kappa_3 - 1\right ) C_1^{\sigma}  + \left ( \kappa_3^2 - 1\right ) \delta Z_H = c_6 C_1^{\sigma} +  \left ( 2 c_6 + c_6^2 \right )  \delta Z_H \,, 
\end{equation}
where $\delta Z_H$ denotes the one-loop correction to the Higgs wave function renormalisation constant associated to modifications of the trilinear Higgs self coupling. In the case of the decays, the effects due to  Higgs wave function renormalisation cancel in the branching ratios, and as a result the quantities $\dBR (f)$ defined in~\eqref{eq:muf} take the following form 
\begin{equation}
\dBR(f)  =   \left ( \kappa_3 - 1\right )  \, \big (C_1^{\Gamma}-C_1^{\Gamma_{\rm tot}} \big ) = c_6 \, \big (C_1^{\Gamma}-C_1^{\Gamma_{\rm tot}} \big )  \,.
\label{BRgeneral} 
\end{equation}
Here $C_1^{\Gamma_{\rm tot}}$ is an effective term that describes the process dependent  corrections to the total decay width of the Higgs boson. 

In the following we provide the values of the $C_1$ coefficients that are used in the numerical analyses presented in section~\ref{sec4}. The given values correspond to the input 
\begin{equation}
\begin{split}
& \hspace{0.5cm}  G_F = 1.1663787\times  10^{-5}~{\rm \UGeV}^{-2} \,, \qquad  m_W = 80.385 \, {\rm \UGeV} \, , \\ 
& m_Z = 91.1876 \, {\rm \UGeV}\,, \qquad m_H = 125 \, {\rm \UGeV}\, , \qquad m_t = 172.5 \, {\rm \UGeV} \,.
\end{split}
\end{equation}
For these parameters one finds numerically~\cite{Degrassi:2016wml} 
\begin{equation}
\delta Z_H = -1.536\times 10^{-3} \,, \qquad C_1^{\Gamma_{\rm tot}} = 2.3 \times 10^{-3} \,.
\end{equation}
In the calculations of production cross sections and distributions, the renormalisation and factorisation scales are taken to be $ \mu_R = \mu_F = \frac{1}{2} \sum_f m_f$ with $m_f$ the masses of the particles in the final state and   {\tt PDF4LHC2015}~\cite{Butterworth:2015oua}  parton distribution functions are used. On the other hand, the dependence of the $C_1$ coefficients on $ \mu_R$,  $\mu_F$ and the PDF set is negligible.

\begin{table}[t!]
\begin{center}
\begin{tabular}{|c |c| c| c| c| c| c|}
\hline
$C_1^{\sigma}~[\%]$	& ${\ggF}$ &
${\text{VBF}} $ &  ${WH}$ & ${ZH}$ &
${\tth}$ &  ${tHj}$ \\ 
\hline \hline
$13 \, {\rm \UTeV}$ & $0.66$ & $0.64$  & $1.03$ & $1.19$ & $3.51$ & $0.91$ \\ \hline
$14 \, {\rm \UTeV}$ & $0.66$ & $0.64$  & $1.03$ & $1.18$ & $3.47$ & $0.89$ \\ \hline
$27 \, {\rm \UTeV}$ & $0.66$ & $0.62$  & $1.01$ & $1.16$ & $3.20$ & $0.79$ \\
 \hline
\end{tabular}
\end{center}
\caption{$C_1^{\sigma}$ coefficients for inclusive single-Higgs production cross sections at different CM energies.}
\label{c1s}
\end{table}
 
\begin{table}[t!]
\begin{center}
\begin{tabular}{|c |c| c| c| c| c| c|}
\hline
   $p_T(H)$~[\UGeV] & $[0, 25]$  & $[25, 50]$ & $[50, 100]$ & $[100, 200]$ &  $[200, 500]$ &  $>500$ \\ 
\hline
\hline
VBF & $0.97$ & $0.88$ & $0.73$ & $0.58$ & $0.45$ & $\phantom{-}0.29$ \\
\hline
$ZH$ & $2.00$ &$1.75$ & $1.21$ & $0.51$ & $0.01$ & $-0.10$ \\
\hline 
$WH$ & $1.70$ & $1.49$ & $1.04$ & $0.44$ & $0.01$ & $-0.09$ \\
\hline
$t\bar tH$ & $5.31$ & $5.07$ & $4.38$ & $3.00$ & $1.27$ & $\phantom{-}0.17$ \\
\hline
${tHj}$ & $1.23$ & $1.18$ & $1.02$ & $0.74$ & $0.33$ & $-0.06$ \\
\hline
\end{tabular}
 \caption{$C_1^{\sigma}$ coefficients for single-Higgs production processes at $13 \, {\rm \UTeV}$ in different $p_T(H)$ bins.}
\label{tab:c1-xs_13}
\end{center}
\end{table}

 \begin{table}[t!]
\begin{center}
\begin{tabular}{|c |c| c| c| c| c| c|}
\hline
    $p_T(H)$~[\UGeV] & $[0, 25]$  & $[25, 50]$ & $[50, 100]$ & $[100, 200]$ &  $[200, 500]$ &  $>500$ \\ 
\hline
\hline
VBF & $0.65$ & $0.65$ & $0.65$ & $0.62$  & $0.52$ & $\phantom{-}0.29$ \\
\hline
$ZH$ & $2.00$ & $1.74$ & $1.21$ & $0.50$ & $0.00$ & $-0.10$ \\
\hline 
$WH$ & $1.70$ & $1.49$ & $1.04$ & $0.44$ & $0.01$ & $-0.09$ \\
\hline
$t\bar tH$ & $5.00$ & $4.78$ & $4.14$ & $2.86$ & $1.23$ & $\phantom{-}0.22$ \\
\hline
${tHj}$ & $1.06$ & $1.03$ & $0.91$ & $0.69$ & $0.33$ & $\phantom{-}0.02$ \\
\hline
\end{tabular}
 \caption{Same as table~\ref{tab:c1-xs_13} but for a CM energy of $27 \, {\rm \UTeV}$.}
\label{tab:c1-xs_27}
\end{center}
\end{table}

\begin{table}[t!]
\begin{center}
\begin{tabular}{|l|l|l|l|l|l}
\hline 
$C_1^{\Gamma}~[\%]$	& ${\gamma\gamma}$ &${ZZ}$& ${WW}$ &${gg}$\\
\hline \hline 
on-shell $H$  &0.49 & 0.83& 0.73 & 0.66 \\
\hline 
\end{tabular}
\end{center}
\caption{ $C_1^{\Gamma}$ coefficients for the phenomenologically relevant decay modes of the Higgs boson.}
\label{c1g}
\end{table}

In table~\ref{c1s} we list the values of $C_1^{\sigma}$ for the various production modes at different centre of mass~(CM) energies. One first notices that ${WH}$, ${ZH}$ and especially ${\tth}$ production depend stronger on the anomalous trilinear Higgs self coupling than the $\ggF$, the ${\rm VBF}$ and the $t H j$ channel.  Furthermore, in the case of ${WH}$, ${ZH}$  and ${\tth}$ production the loop corrections contributing to $C_1^{\sigma}$ feature a Sommerfeld enhancement, which results in an increased sensitivity to anomalous trilinear Higgs self interactions at low energies~\cite{Degrassi:2016wml,Bizon:2016wgr,Maltoni:2017ims}. This feature is illustrated in tables \ref{tab:c1-xs_13} and \ref{tab:c1-xs_27} where we give  the values of $C_1^{\sigma}$ in bins of the Higgs transverse momentum $p_T(H)$  for $pp$ collisions at $13 \, {\rm \UTeV}$ and $27 \, {\rm \UTeV}$, respectively.\footnote{Results for a different binning or different observables can be easily obtained with the code presented in~\cite{Maltoni:2017ims}.} Table~\ref{c1g} finally provides the values of  the $C_1^{\Gamma}$ coefficients for the decay modes of the Higgs boson that are relevant in our numerical study. 

Notice that all the formulas and numbers presented in this subsection take into account only  effects associated to an anomalous trilinear Higgs self coupling. The extension to more general and physically motivated scenarios that include also other new-physics effects is simple and has been worked out in~\cite{DiVita:2017eyz,Maltoni:2017ims}. It consists in adding to  (\ref{dsigmakt}) and (\ref{BRgeneral}) the effects of other anomalous  interactions such as a modified top Yukawa coupling or altered/new gauge-Higgs vertices. In the next subsection, we perform a global analyses of the constraints on $\lambda_3$ that the HL-LHC and the HE-LHC should be able to set.    We thereby follow the  lines of the study~\cite{DiVita:2017eyz}, using the results for the coefficients $C_1$ provided above.


 As discussed in refs.~\cite{Maltoni:2017ims, DiVita:2017eyz}, the constraints  that can be set on $c_6$ critically depend on the interplay between the following aspects:
\begin{itemize}
\item The number of additional parameters related other anomalous interactions.
\item The number of independent measurements considered in the analysis.
\item The inclusion of differential information.
\item The assumptions on the theoretical and experimental (statistical and systematic) errors.
\end{itemize}
   
In the section~\ref{sec:HH_Global_fit} we explore this interplay for the cases of the HL- and HE-LHC following the lines of the study presented in refs.~\cite{DiVita:2017eyz} augmented with the new results provided in this section. Independent analyses performed by the ATLAS and CMS collaborations with a full-fledged treatment of all the correlations among experimental uncertainties are desirable, and the first steps towards this are being presented in the next section~\ref{sec:CMS_HH_indirect}. It is worth noting that, when other anomalous interactions are also considered, the effects of  $Z_H^{\rm BSM}$ are degenerate with those in general affecting the Higgs wave-function normalisation, typically parametrised via the Wilson coefficient $\mathcal{C}_{H}$. Thus, the coefficients $C_1^{\sigma}$ and therefore the differential distributions have a primary role in the extraction of the information on $\kappa_3$ from measurements of  single Higgs production.

We also recall that limits on $\ktre$ or equivalently $c_6$ obtained with this strategy are sensible only when $|\ktre|<20$; as discussed in refs.~\cite{Degrassi:2016wml} this limit guarantees that the perturbative loop expansion is converging and that the leading missing higher orders depending on $\ktre-1=c_6$ are below 10\% level. On the contrary, as discussed in refs.~\cite{DiLuzio:2017tfn, Maltoni:2018ttu}, when the information from double Higgs production is considered a more cautious limit $|\ktre|<6$ should be adopted in order to achieve both perturbative unitarity and the convergence of the loop expansion.


\asubsubsection[N. Wardle, J. Langford]{Indirect probes of the trilinear coupling through differential distributions measurements with the CMS detector}
\label{sec:CMS_HH_indirect}

As detailed in the previous section, an alternative approach to probing the Higgs boson self-coupling is to measure deviations of the inclusive and differential Higgs boson production rates. Contributions to single Higgs boson production from the Higgs boson self-coupling are sizeable for production in association with a pair of top quarks ($\ttH$) or a single top-quark ($\tH$). The contributions are  greatest in these production modes due to the large mass of the top quark. Differential cross section measurements, in particular as a function of the Higgs boson 
transverse momentum $\pT^{H}$, allow one to disentangle the effects of modified Higgs boson self-coupling values from 
other effects such as the presence of anomalous top--Higgs couplings.


The differential cross-section measurements, described in section~\ref{sec:ttHdiffxs}, 
are used to extract a 
constraint on the Higgs boson self-coupling ($\lambda_{3}$), by parametrising deviations from SM predictions as described in the previous section. The kinematic dependence of these deviations are determined by reweighting signal events, on an event by event basis, using the tool described in Ref.~\cite{EWreweightingtool}, which calculates $\lambda_3$-dependent corrections to the tree level cross-sections as a function of the kinematic properties of the event, and is encapsulated as a 
varying $C_{1}$ coefficient. The value of $C_{1}$ depends on both the Higgs boson production mode and the kinematic properties of the event. Table~\ref{tab:ttHdiff_CMS_c1_values} shows the values of $C_1$ calculated in the fiducial region for $\ttH$ and $\tH$ production, in each bin of $\pTH$.

\begin{table}[t!]
\begin{center}
\begin{tabular}{|c |c| c| c| c| c| c|}
\hline
   $p_T(H)$~[\UGeV] & $[0, 45]$  & $[45, 80]$ & $[80, 120]$ & $[120, 200]$ &  $[200, 350]$ &  $>350$ \\ 
\hline
\hline
ttH & $5.31$ & $4.73$ & $3.92$ & $2.79$ & $1.42$ & $0.42$ \\
\hline
tH & $1.32$ &$1.19$ & $1.00$ & $0.75$ & $0.40$ & $0.06$ \\
\hline 
VH & $1.66$ & $1.23$ & $0.77$ & $0.35$ & $0.02$ & $-0.09$ \\
\hline
\end{tabular}
 \caption{Process dependant $C_{1}$ values for each bin of $\pTH$.}
\label{tab:ttHdiff_CMS_c1_values}
\end{center}
\end{table}

In addition, the contribution from $\VH$ production is included by similarly calculating the $C_{1}$ values for $\VH,~\hgg$ events. For the contribution of $\ggH$ and to account for modifications of the $\hgg$ decay width, the parametrisations which have been calculated for inclusive events provided in Ref.~\cite{Degrassi:2016wml} are used directly.

Figure~\ref{fig:ttHdiff_CMS_klambda_scan} shows a scan of the profile log-likelihood as a function of $\kappa_{\lambda}$. In the scan, all other Higgs boson couplings are assumed to attain their SM values. For the purposes of constraining $\kappa_{\lambda}$, theoretical uncertainties in the differential $\ttH$~+~$\tH$ cross section, as described in section~\ref{sec:ttHdiffxs}, are included in the signal model. The results when only including the hadronic or leptonic categories are shown in addition to the result obtained from their combination. 

\begin{figure}[htb!]
        \centering
        \includegraphics[width=0.6\textwidth]{\main/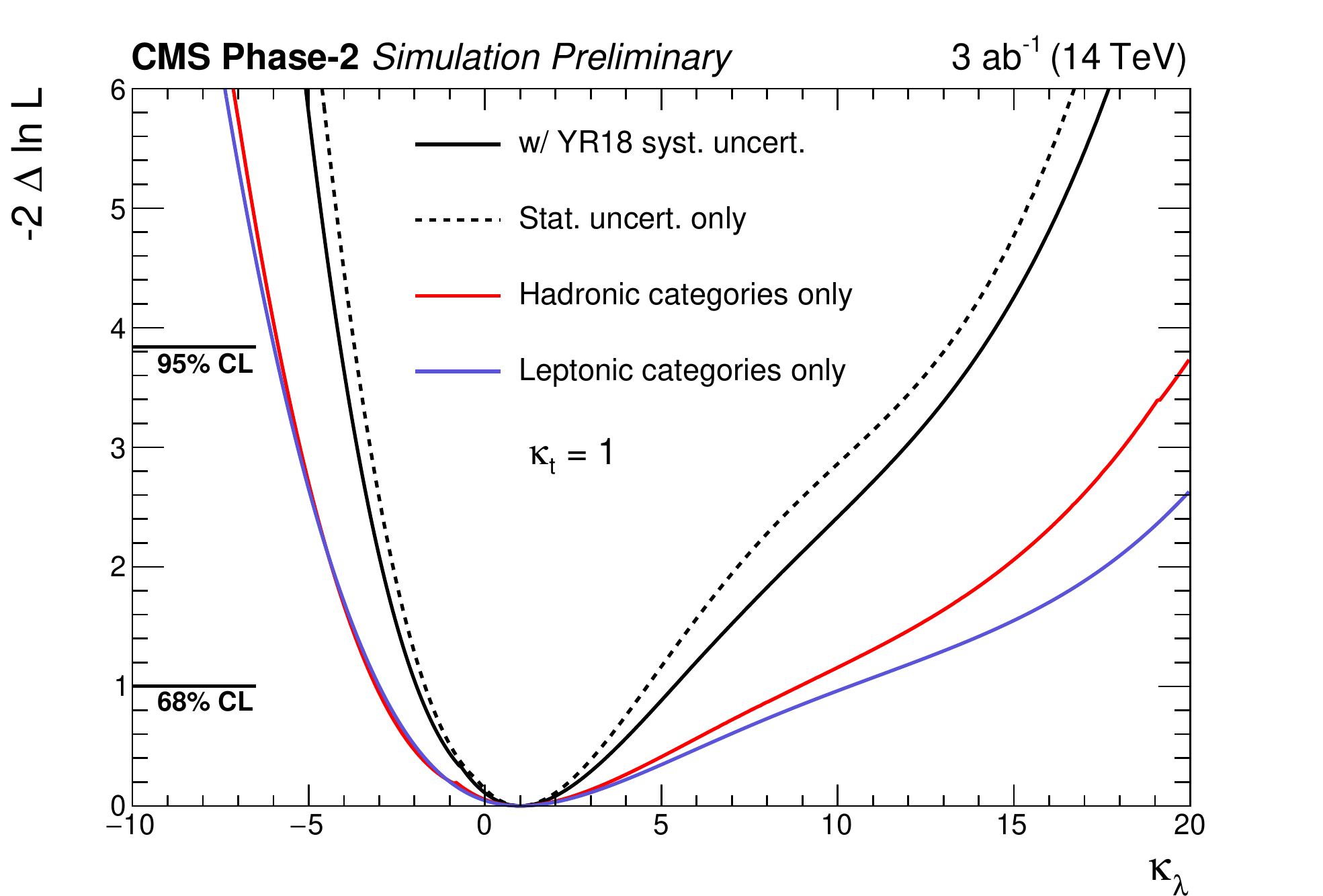}
        \caption{Profile log-likelihood scan as a function of $\kappa_\lambda$. The individual contributions of the statistical and systematic uncertainties are separated by performing a likelihood scan with all systematics removed. Additionally, the contributions from the hadronic and leptonic channels have been separated, shown in red and purple, respectively.}
        \label{fig:ttHdiff_CMS_klambda_scan}
\end{figure}

The profiled log-likelihood in the region around 
$5<\kappa_\lambda<15$ results from the behaviour of the parametrisations which modify the predicted cross sections. For the $\ttH$ production mode, the derivative of the predicted cross section with respect to $\kappa_\lambda$ changes sign in this region, such that the predicted cross section is relatively stable for different values of $\kappa_\lambda$. This degeneracy is however somewhat resolved by the  other production modes for which the change in sign occurs at different values of $\kappa_\lambda$. With $3\,\text{ab}^{-1}$ of data collected by CMS at the HL-LHC, this result shows that a constraint of $-4.1 < \kappa_\lambda < 14.1$ at the 95\% confidence level (CL) is achievable from the differential cross-section measurement of a single Higgs boson decay channel produced in association with tops, using data from only one of the two general purpose detectors at the HL-LHC.

The $\ttH$~+~$\tH$ differential cross section measurements are also sensitive to other potential BSM effects, such as those which give rise to anomalous top--Higgs couplings. A two-dimensional profile log-likelihood scan is shown in  Fig.~\ref{fig:ttHdiff_CMS_klambda_2Dscan} as a function of $\kappa_\lambda$ and $\mu_{H}$. The parameter $\mu_{H}$ is a multiplicative scaling factor which is common to all Higgs boson production modes and all $\pTH$ bins. Even with this additional parameter, constraints on $\kappa_\lambda$ are still achievable, owing to the information retained in the shape of the $\pTH$ distribution. The constraint on $\kappa_\lambda$ is $-7.1 < \kappa_\lambda < 14.1$ at the 95\% CL, when the log-likelihood is also profiled with respect to $\mu_{H}$. 

\begin{figure}[htb!]
        \centering
        \includegraphics[width=0.6\textwidth]{\main/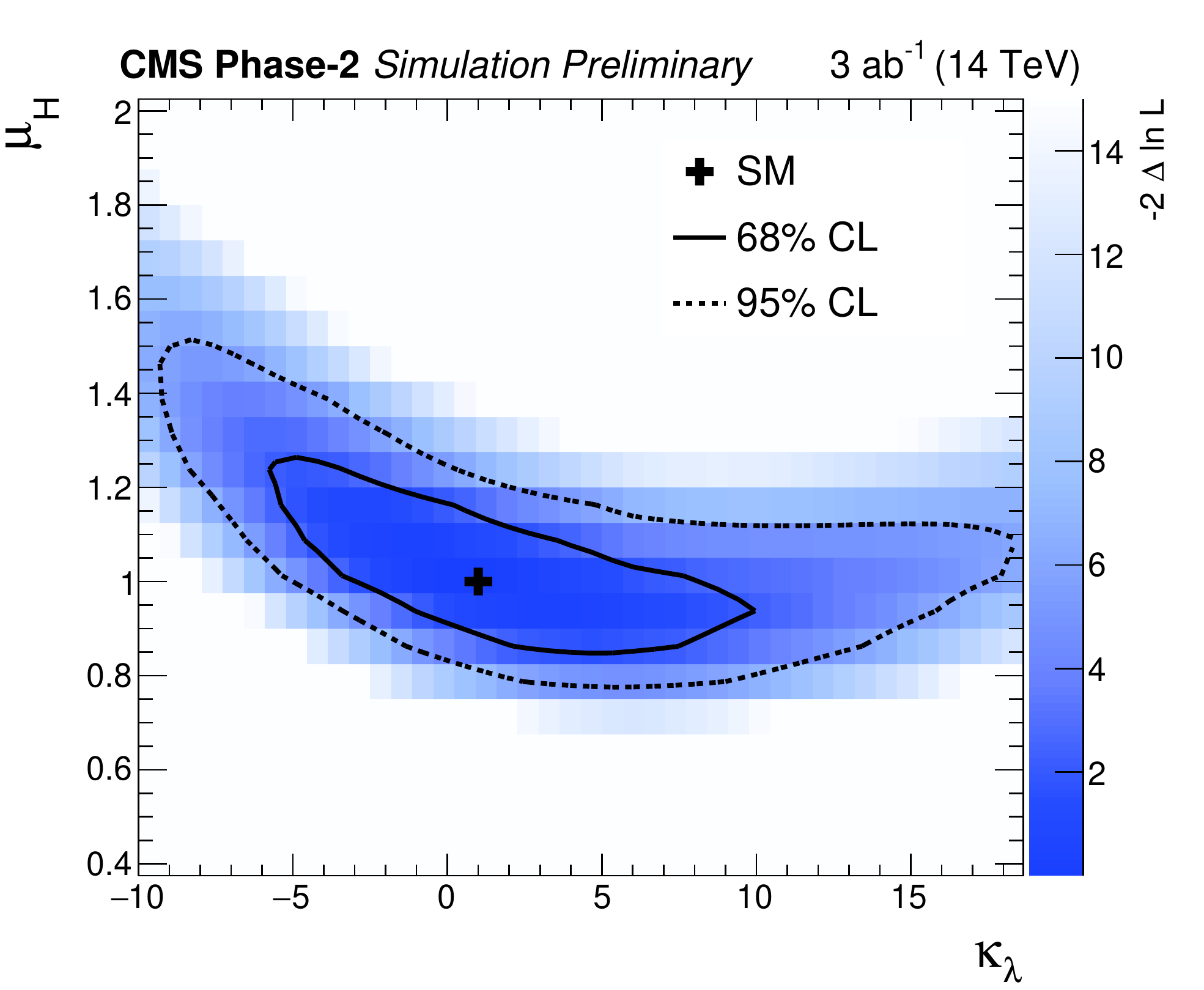}
        \caption{Results of the two-dimensional likelihood scan in $\kappa_\lambda$-vs-$\mu_{H}$, where $\mu_{H}$ allows all Higgs boson production modes to scale relative to the SM prediction. The 68\% and 95\% confidence level contours are shown by the solid and dashed lines respectively. The SM expectation is shown by the black cross.}
        \label{fig:ttHdiff_CMS_klambda_2Dscan}
\end{figure}

\asubsubsection[S. Di Vita, G. Durieux, C. Grojean, J. Gu, Z. Liu, G. Panico, M. Riembau, T. Vantalon]{Global fit}
\label{sec:HH_Global_fit}
Assuming that the trilinear coupling is the only coupling deviating from its SM value, single Higgs observables can give competitive bounds with double Higgs production, see Refs.~\cite{Gorbahn:2016uoy,Degrassi:2016wml,Bizon:2016wgr,Degrassi:2017ucl,Maltoni:2017ims} \footnote{Electroweak processes where the Higgs trilinear coupling enter at the two loop level have also been studied in \cite{Kribs:2017znd}.}. Nevertheless, departures of the Higgs self-coupling from its SM prediction signal the existence of new dynamics that, in general, would leave an imprint on other Higgs couplings as well which have a strong impact on the bound as shown by Ref.~\cite{DiVita:2017eyz}. The importance of a global fit is therefore two-fold, namely to assess the robustness of the studies that take into account deformations exclusively in the Higgs trilinear coupling, and to single out the sensitivity on the single-Higgs couplings that is required to minimise the impact of the possible correlations.
\medskip

\begin{figure}
	\centering
	\includegraphics[width=0.45\linewidth]{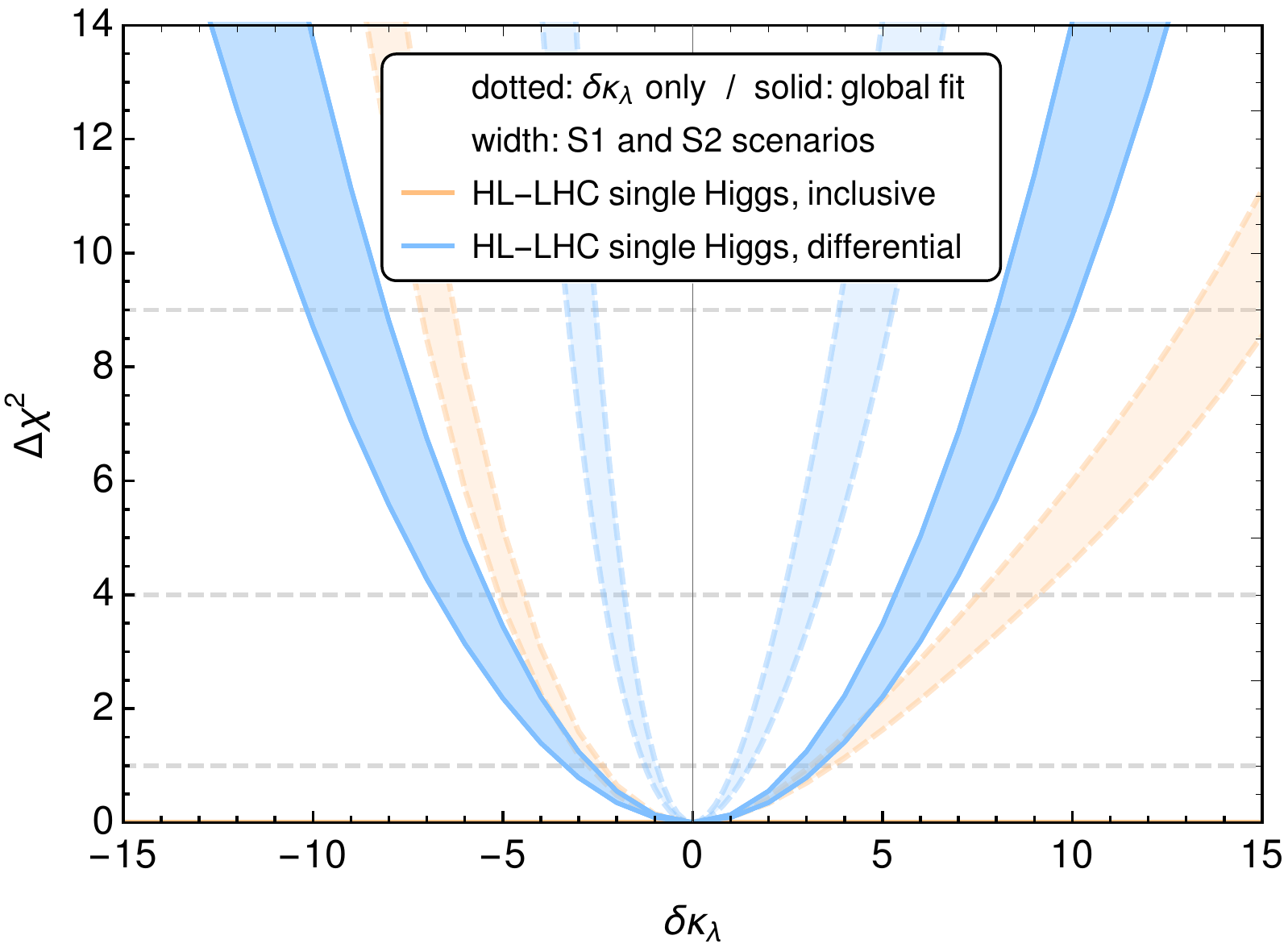}\hfill
	\includegraphics[width=0.45\linewidth]{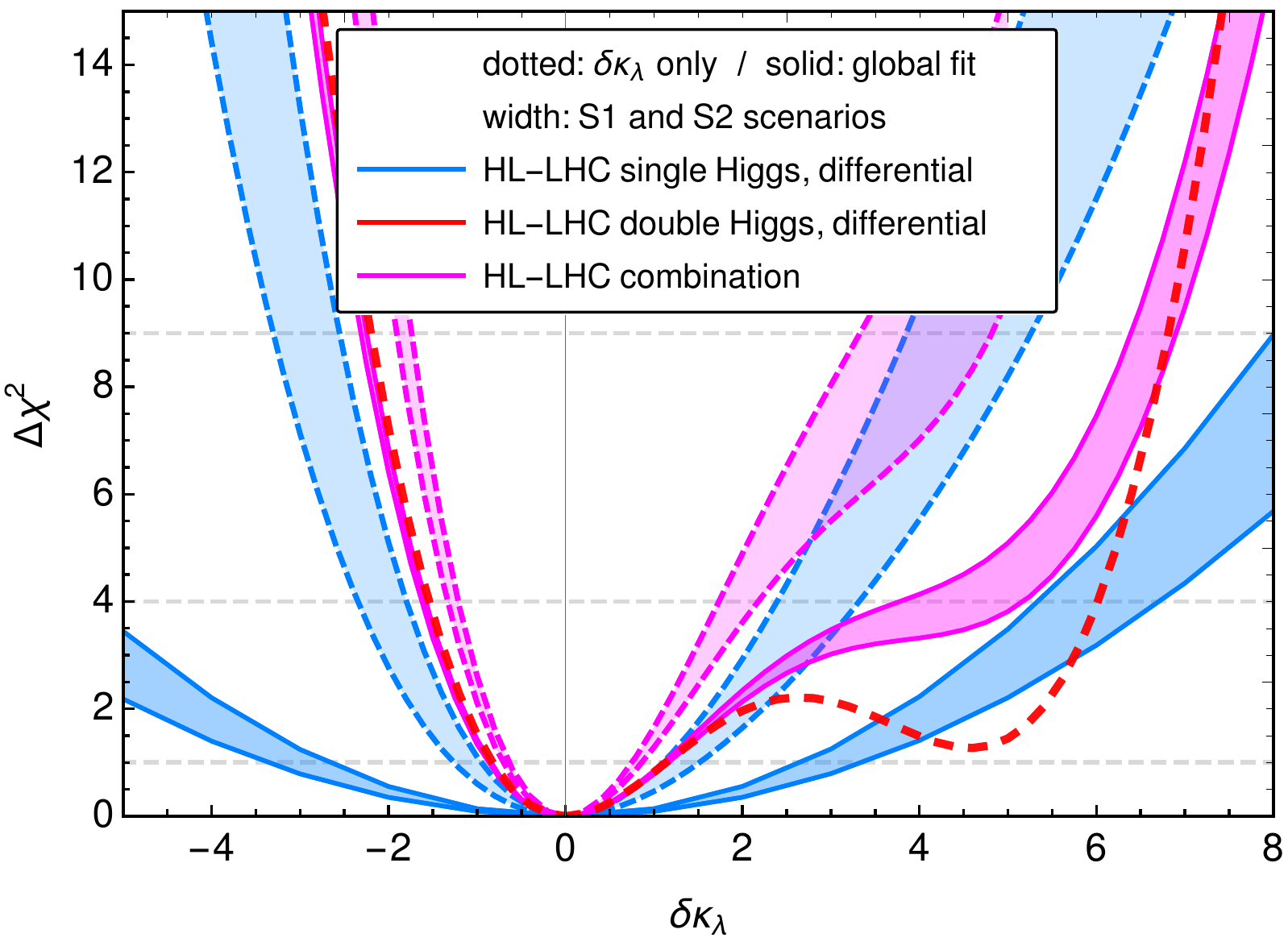}
	\caption{$\chi^2$ analysis of the Higgs self-coupling $\delta \kappa_\lambda$ using single- and double-Higgs processes for the HL-LHC at 13\,\UTeV and 3\,ab$^{-1}$. The widths of the lines correspond to the differences between the scenarios S1 and S2. \textbf{Left:} Comparison of the constraints obtained using inclusive single-Higgs processes (orange), with the ones using differential observables (blue). Dashed is an exclusive fit while solid is the result of a global fit. \textbf{Right:} Comparison of the constraints from differential single Higgs (blue), with those from differential double-Higgs data (dashed red) and its combination (pink).}
	\label{fig:hllhcchi2}
\end{figure}

To include the effect of the different deformations away from the SM, we use the EFT framework described in Ref.~\cite{DiVita:2017eyz}, where 9 parameters describe the deviations of the single-Higgs couplings. In particular, we consider three\footnote{If other fermionic decay channels can be observed, further parameters can be included, with no effect on the number of degrees of freedom.} parameters for the Yukawa interactions ($\delta y_t,\,\delta y_b,\,\delta y_\tau,$), two for the contact interactions with gluons and photons ($c_{gg}\,,c_{\gamma\gamma}$), rescalings of the SM $hZZ$ and $hWW$ interactions (parametrised by one coefficient, $\delta c_z$, if custodial symmetry is unbroken), and three coefficients ($c_{zz},c_{z\square},c_{z\gamma}$) parametrising interactions of the Higgs with the electroweak bosons that have non-SM tensor structures. Note that two combinations of the last three parameters are constrained by di-boson data, showing an interesting interplay between the gauge and the Higgs sectors. A global fit on the Higgs self-coupling, parametrised by $\delta\kappa_\lambda$ (which is zero in the SM) using only inclusive single Higgs observables, and taking into account the additional 9 EFT deviations described above, suffers from a flat direction. To lift it, it is necessary to include data from differential measurements of those processes, since the single-Higgs deformations and $\delta\kappa_\lambda$ tend to affect the distributions in complementary ways.
\medskip

As input for the uncertainties we consider the S1 and S2 scenarios, corresponding to the projected uncertainties on the inclusive signal strengths of the different production and branching ratio modes of the Higgs, recommended by ATLAS and CMS. The projections of the differential uncertainties are estimated by first rescaling the statistical uncertainty on each bin. This gives an overestimation of the actual reach since it assumes a flat background distribution while it tends to peak at lower invariant masses. We use therefore the CMS analysis on $h\to \gamma \gamma$ in $t\bar{t} h$ production as a template for the tilt of the background. With this we get a good agreement with the CMS fit on the trilinear coupling using this channel only, and we use it as a simple guess for the rest of the uncertainties.
\medskip

The global fit for the HL-LHC is summarised in Fig.~\ref{fig:hllhcchi2}. In the left plot, we show in green the $\Delta\chi^2$ including only single-Higgs data, both in an exclusive study (dotted, pale colour), and after profiling over all the other parameters (solid, strong colour). The width of the lines correspond to the difference between the scenarios S1 and S2. The fact that the lines are not very separated means that the constraints are mostly statistics dominated. In orange, we consider only inclusive measurements, while in blue we include the differential information. We can see that, in a global fit, the constraint on the trilinear coupling is worsened due to correlations (mainly with the top Yukawa $\delta y_t$ and the contact interaction with gluons $c_{gg}$, and, to a lesser degree, between $\delta y_b$ and $\delta c_z$. The inclusion of the differential information allows to partially remove the flat directions. In the right plot we compare the constraints using differential observables with the ones obtained from double Higgs production, taken from the study in Ref.~\cite{Azatov:2015oxa}, in dashed red. We include the combination in pink. While double-Higgs is clearly driving the bound, differential single-Higgs data is nonetheless relevant as it can help lift the degenerate minima around $\delta \kappa_\lambda\sim 5$.
\medskip

\begin{figure}
	\centering
	\includegraphics[width=0.45\linewidth]{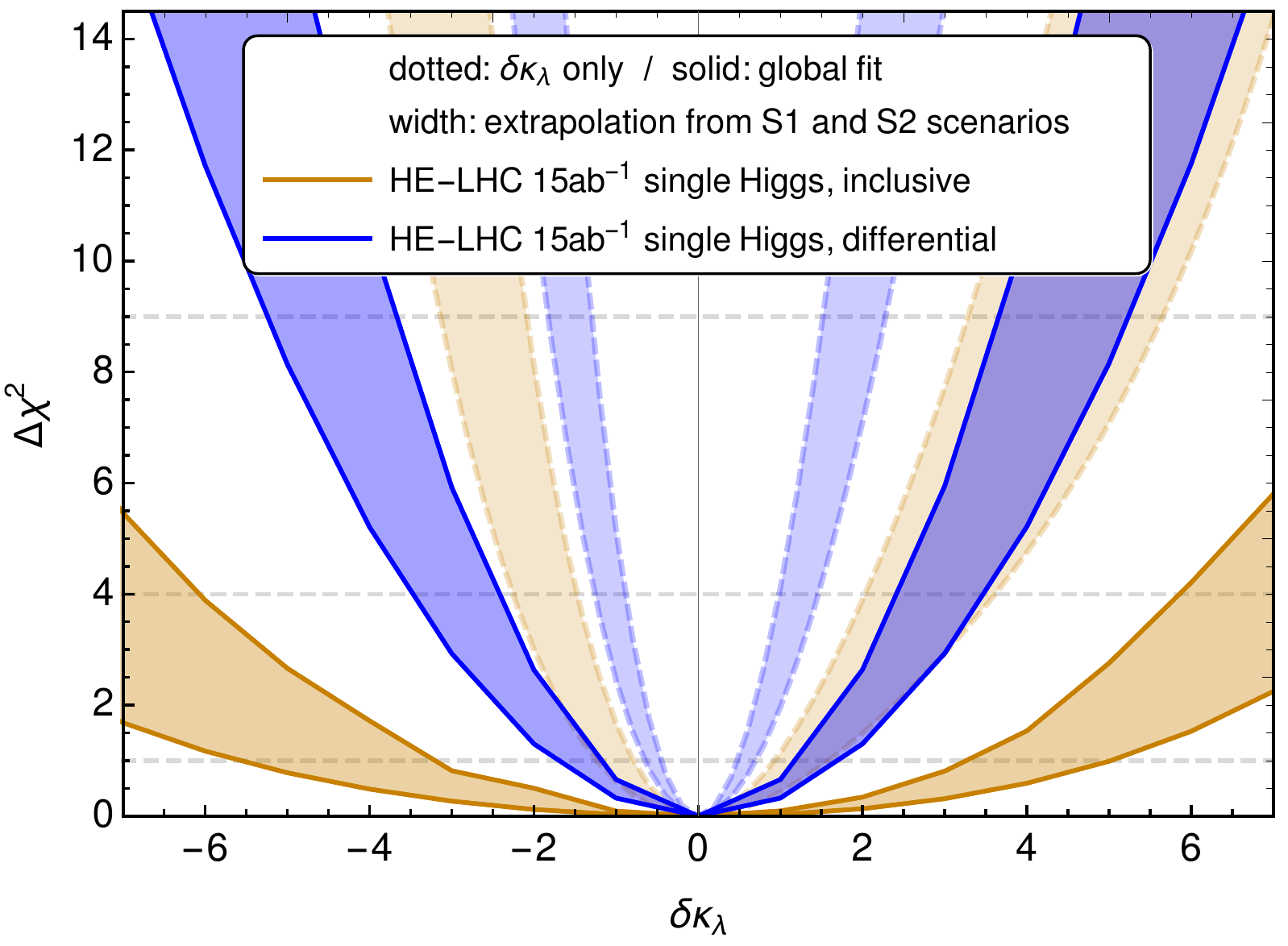}\hfill
	\includegraphics[width=0.45\linewidth]{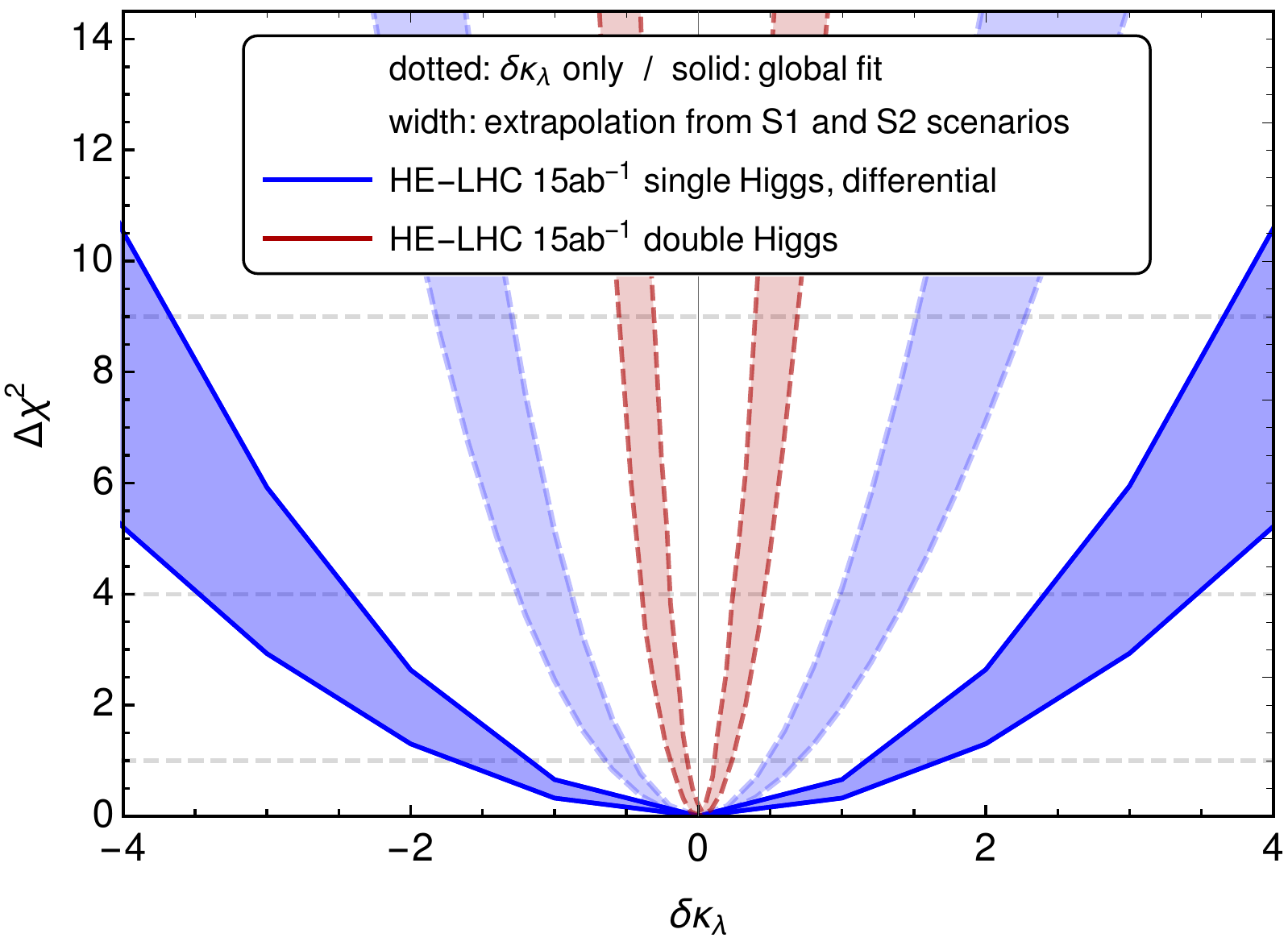}
	\caption{$\chi^2$ analysis of the Higgs self-coupling $\delta \kappa_\lambda$ using single- and double-Higgs processes for the HE-LHC at 27\,\UTeV and 15\,ab$^{-1}$. The widths of the lines correspond to the differences between the projected uncertainties from the scenarios S1 and S2. \textbf{Left:} Comparison of the constraints obtained using inclusive single-Higgs processes (orange), with the ones using differential observables (blue). Dashed is an exclusive fit while solid is the result of a global fit. \textbf{Right:} Comparison of the constraints from differential single Higgs (blue), with those from differential double-Higgs data (dashed red).}
	\label{fig:helhcchi2}
\end{figure}

We now discuss projections for the HE-LHC at 27\,\UTeV with 15\,ab$^{-1}$ of integrated luminosity. For the uncertainties we perform a simple extrapolation where the theory and systematic uncertainties are kept the same as in the HL-LHC projections, while the statistical uncertainty is rescaled accordingly \cite{Goncalves:2018qas}. We show the results in Fig.~\ref{fig:helhcchi2}. In the left plot, in brown, we present the $\chi^2$ analysis using the projections for the single-Higgs channels at HE-LHC at the inclusive level. Inclusive measurements are able to lift the flat direction due to the measurement of the $th+j$ production and the $z\gamma$ decay. In blue we present the results using differential observables. We note that the with of the line indicates that, contrary to the HL-LHC case, the constraint is limited by systematics, as expected. In the right plot we compare the constraints with the projections coming from double Higgs production measurement.
\medskip

\asubsection{Implications of the HH measurements}
\label{sec:HH_implications}

\asubsubsection[M. Bauer, M. Carena, A. Carmona]{Implications for flavor models}
In the Two-Higgs-Doublet Model (2HDM), the term $H_1 H_2\equiv H_1^T (i\sigma_2) H_2 $ is a SM singlet which can however be charged under an additional $U(1)$ flavor symmetry. This is an interesting possibility that allows to generate the different fermion masses with a Froggatt-Nielsen (FN) mechanism where the flavon is replaced by the $H_1 H_2$ operator. In this way, the new physics scale $\Lambda$ where the higher dimensional FN operators are generated is tied to the electroweak scale, leading to much stronger phenomenological consequences. Let us assume for concreteness a type-I like 2DHM with the following Yukawa Lagrangian 
\begin{align}
\label{eq:yuk1}
\mathcal{L}_Y&\supset  y^u_{ij} \left(\frac{H_1 H_2}{\Lambda^2}\right)^{n_{u_{ij}}}\bar{q}_L^{i}H_1 u_{R}^j+y^d_{ij} \left(\frac{H_1^{\dagger} H_2^{\dagger}}{\Lambda^2}\right)^{n_{d_{ij}}}\bar{q}_L^{i}\tilde{H}_1 d_{R}^j+y_{ij}^{\ell}\left(\frac{H_1^{\dagger}H_2^{\dagger}}{\Lambda^2}\right)^{n_{e_{ij}}}\bar{\ell}_{L}^i\tilde{H}_1 e_R^j
+\mathrm{h.c.}\,,
\end{align}
where $\tilde{H}_1\equiv i\sigma_2 H^{\ast}_1$ as usual and the charges $n_{u,d,e}$ are a combination of the $U(1)$ charges of $H_1$, $(H_1H_2)$ and the different SM fermion fields. For simplicity, we set the flavor charges of $(H_1 H_2)$ and $H_2$ to $0$ and $1$, respectively, such that  
\begin{align}
n_{u_{ij}}=a_{q_i}-a_{u_j},\quad n_{d_{ij}}=-a_{q_i}+a_{d_j},\quad n_{e_{ij}}=-a_{\ell_i}+a_{e_j},
\end{align}
if we denote by $a_{q_i},a_{u_i}, \ldots,$ the $U(1)$ charges of the SM fermions. In general, the fermion masses are given by
\begin{align}\label{eq:epsilon}
m_\psi=y_{\psi} \varepsilon^{n_\psi} \frac{v}{\sqrt{2}}, \quad \varepsilon = \frac{v_1 v_2}{2\Lambda^2}=\frac{t_\beta}{1+t_\beta^2}\frac{v^2}{2\Lambda^2},
\end{align}
with the vacuum expectation values $\langle H_{1, 2}\rangle=v_{1, 2}$ and $t_\beta \equiv v_1/v_2$. Besides being able to accommodate the observed hierarchy of SM fermion masses and mixing angles for the right assignment of flavor charges \cite{Bauer:2015fxa, Bauer:2015kzy}, this framework can lead to enhanced diagonal Yukawa couplings between the Higgs and the SM fermions while having suppressed flavour changing neutral currents (FCNCs). If we denote by $h$ and $H$ the two neutral scalar mass eigen-states, with $h$ being the observed $125$ \UGeV Higgs, the couplings between the scalars $\varphi=h,H$ and SM fermions $\psi_{L_i, R_i}= P_{L,R} \psi_i$ in the mass eigen-basis read 
\begin{equation}\label{eq:newlagrangian}
\mathcal{L}= g_{\varphi \psi_{L_i} \psi_{R_j} }\, \varphi \,\bar \psi_{L_i} \psi_{R_j}+\mathrm{h.c.}
\end{equation}
with $i$, such that $u_i=u, c,t$,\,  $d_i=d,s,b$ and $e_i=e,\mu,\tau$. This induces 
flavor-diagonal couplings 
\begin{align}\label{eq:diagocoup}
g_{\varphi \psi_{L_i}\psi_{R_i}}= \kappa^\varphi_{\psi_i}\, \frac{m_{\psi_i}}{v} =\left(g^{\varphi}_{\psi_i}(\alpha,\beta)+n_{\psi_i}\, f^\varphi(\alpha, \beta)\right)\frac{m_{\psi_i}}{v},
\end{align}
as well as flavor off-diagonal couplings
 \begin{align}\label{eq:foffc}
g_{\varphi \psi_{L_i}\psi_{R_j}}&=  f^\varphi(\alpha, \beta)\left(\mathcal{A}_{ij}\frac{m_{\psi_j}}{v}-\frac{m_{\psi_i}}{v}\mathcal{B}_{ij} \right)\,.
\end{align}
The flavor universal functions in \eqref{eq:diagocoup} and \eqref{eq:foffc} read
\begin{align}
g^h_{\psi_i}=\frac{c_{\beta-\alpha}}{t_\beta}+s_{\beta-\alpha}\,,\qquad g^H_{\psi_i}=c_{\beta-\alpha}-\frac{s_{\beta-\alpha}}{t_\beta}\,,
\end{align}
and
\begin{align}\label{eq:fFs}
f^h(\alpha,\beta)&=c_{\beta-\alpha}\Big(\frac{1}{t_\beta}-t_\beta\Big)+2s_{\beta-\alpha}\,,\quad f^H(\alpha,\beta)=-s_{\beta-\alpha}\Big(\frac{1}{t_\beta}-t_\beta\Big)+2c_{\beta-\alpha}\,,
\end{align}
where $c_{x}\equiv \cos x$, $s_x\equiv \sin x$. One can see that, unless all flavor charges for a given type of fermions are equal, the off-diagonal elements in matrices $\mathcal{A}$ and $\mathcal{B}$ lead to FCNCs which are chirally suppressed by powers of the ratio~$\varepsilon$, see \cite{Bauer:2017cov} for more details.

The scalar couplings to the different gauge bosons are the same as in a normal type-I 2HDM while the scalar coupling between the heavy Higgs $H$ and two SM Higgs scalars $h$, as well as the triple Higgs coupling can be expressed as \cite{Boudjema:2001ii, Gunion:2002zf}
\begin{align}
\label{eq:mainn1}
g_{Hhh}&=\frac{c_{\beta-\alpha}}{v}\!\left[\big(1\!-\!f^h(\alpha,\beta)s_{\beta-\alpha}\big)\big(3M_A^2\!-\!2m_h^2\!-\!M_H^2\big)\!-\!M_A^2\right],\\
g_{hhh}&= -\frac{3}{v}\!\left[f^h(\alpha,\beta)c_{\beta-\alpha}^2(m_h^2-M_A^2)+m_h^2s_{\beta-\alpha}\right],\label{eq:mainn2}
\end{align}
where $M_A$ is the pseudoscalar mass. The $U(1)$ flavor symmetry restricts the number of allowed terms in the scalar potential forbidding e.g. terms proportional to $H_1 H_2$. The interesting feature is that one can rewrite such self scalar interactions with the help of the function $f^h(\alpha,\beta)$, since it is somehow related to the combination $H_1 H_2^{\dagger}$ appearing in both the scalar potential and the higher dimensional operators generating the different Yukawa couplings. Therefore, the parameter space for which $f^h(\alpha,\beta)\gg 1$ and  $c_{\beta-\alpha}\neq 0$ leads to maximally enhanced diagonal couplings of the SM Higgs to fermions \eqref{eq:diagocoup} as well as to an enhancement of the trilinear couplings \eqref{eq:mainn1} and \eqref{eq:mainn2}. For maximally enhanced Yukawa couplings, the mass of the heavy Higgs $H$ cannot be taken arbitrarily large and resonant Higgs pair production has to be present. This correlation between the enhancement of the Higgs Yukawa couplings $\kappa^h_{\psi}$ and $\text{Br}(H \to hh)$ is illustrated for $M_H=M_A=M_{H^\pm}=500$ \UGeV in Fig. \ref{fig:BRvskappa} where we plot the dependence of $\text{Br}(H \to hh)$ on $c_{\beta-\alpha} $ and $ t_\beta$.  The dashed contours correspond to constant values of $|\kappa_{\psi}^h|$ for $n_{\psi}=1$. This correlation does not dependent of the factor $n_\psi$, although  $n_\psi > 1$ leads to a larger enhancement. The two exceptions for which this correlation breaks down are the limits $c_{\beta-\alpha}\approx 0$ and $c_{\beta-\alpha}\approx \pm 1$. Whereas the second case is strongly disfavoured by SM Higgs couplings strength measurements, the first one (which corresponds to the decoupling limit) is at odds with the flavor model, for it requires large values of the spurion $\mu_3\propto M_A$ which softly breaks the $U(1)$ flavor symmetry.

 \begin{figure}
 \begin{center}
 \includegraphics[width=.8\textwidth]{\main/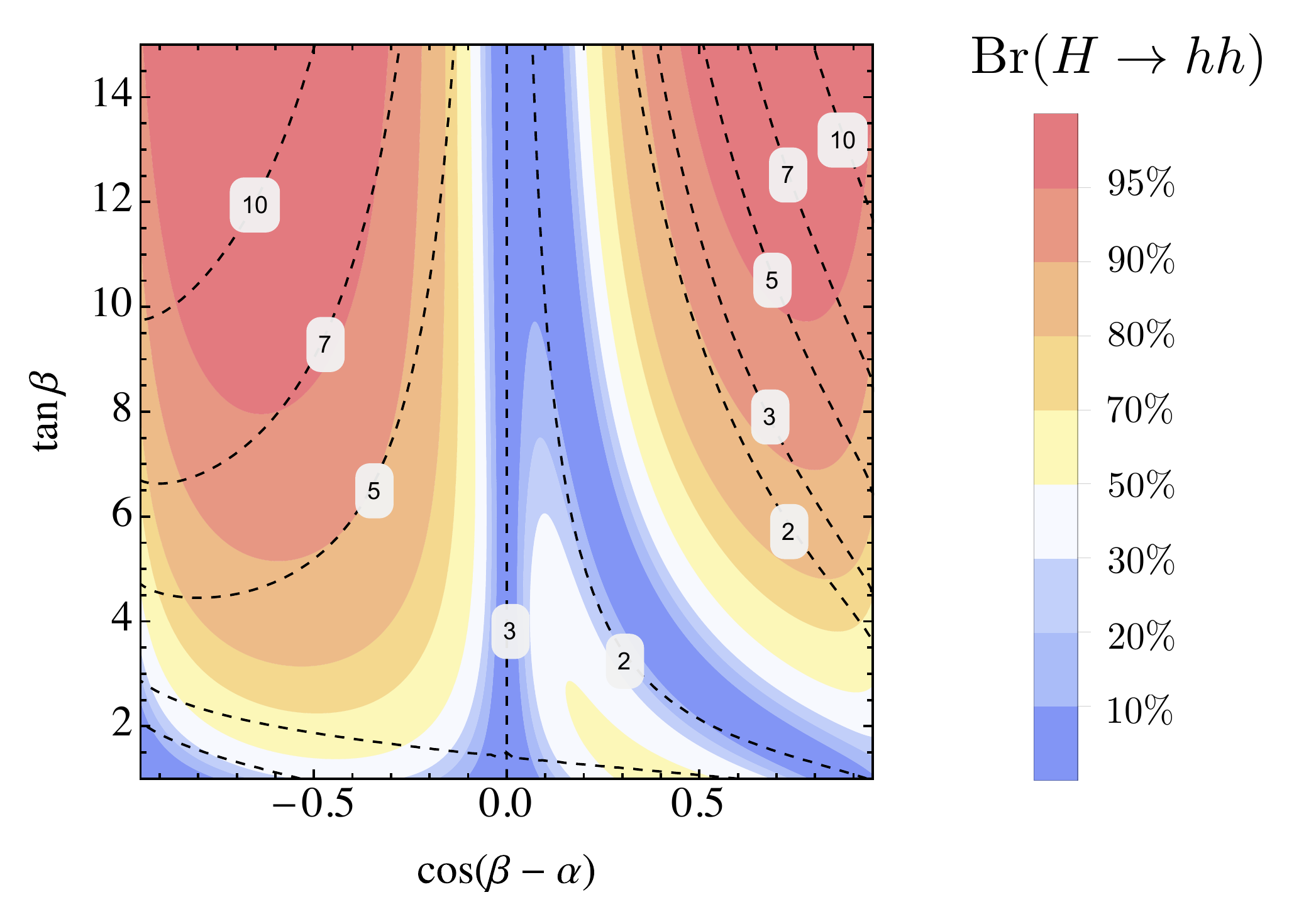}
 \caption{\label{fig:BRvskappa} $\text{Br}(H \to hh)$ as a function of $\cos(\beta-\alpha)$ and $ \tan\beta$ for $M_H=M_{H^\pm}=550$ \UGeV and $M_A=450$ \UGeV. The dashed contours correspond to constant values $|\kappa_{\psi}^h|$ for $n_{\psi}=1$.}
 \end{center}
  \vspace{-.6cm}
 \end{figure}
\begin{figure*}
\includegraphics[width=.465\textwidth]{\main/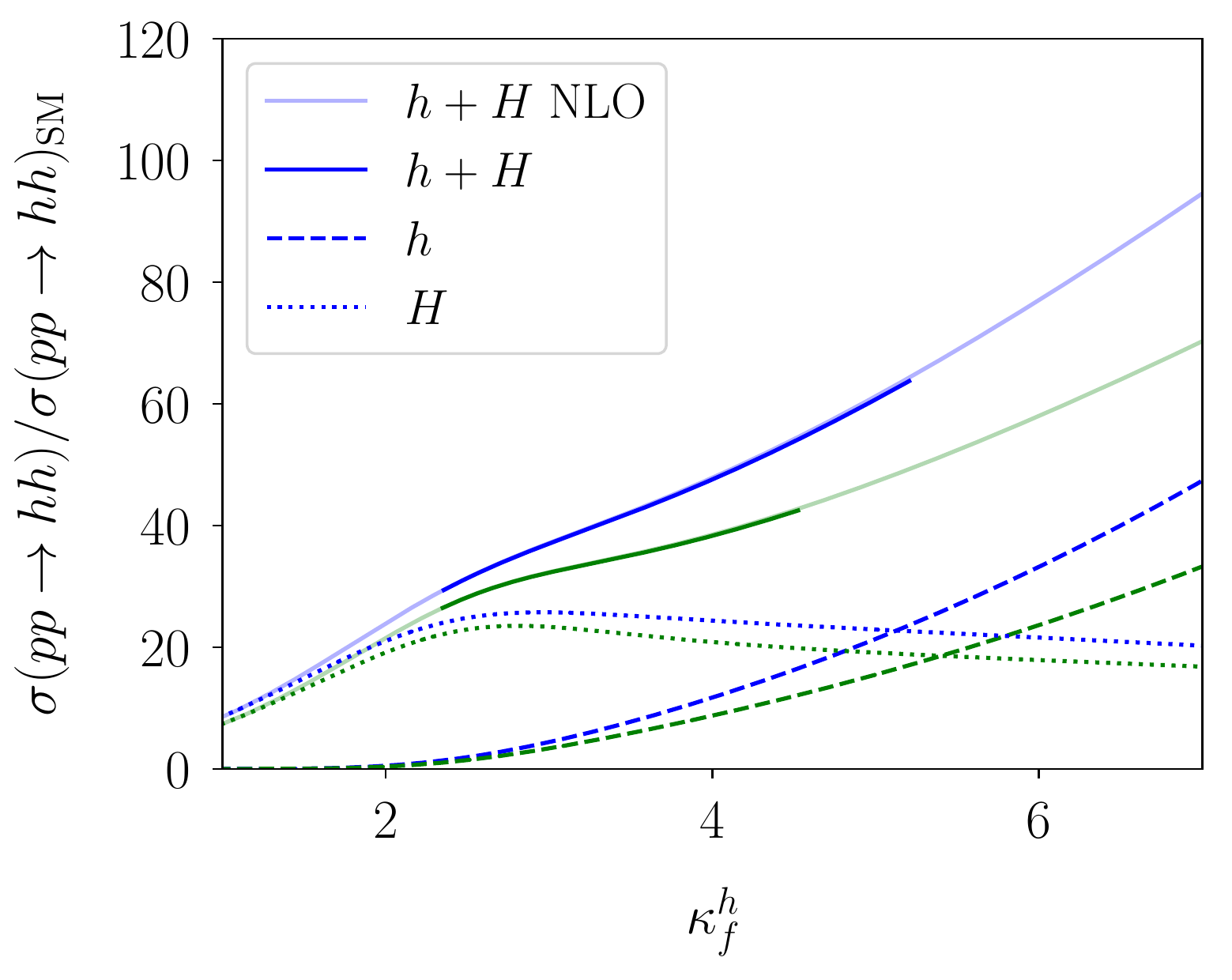}\includegraphics[width=.49\textwidth]{\main/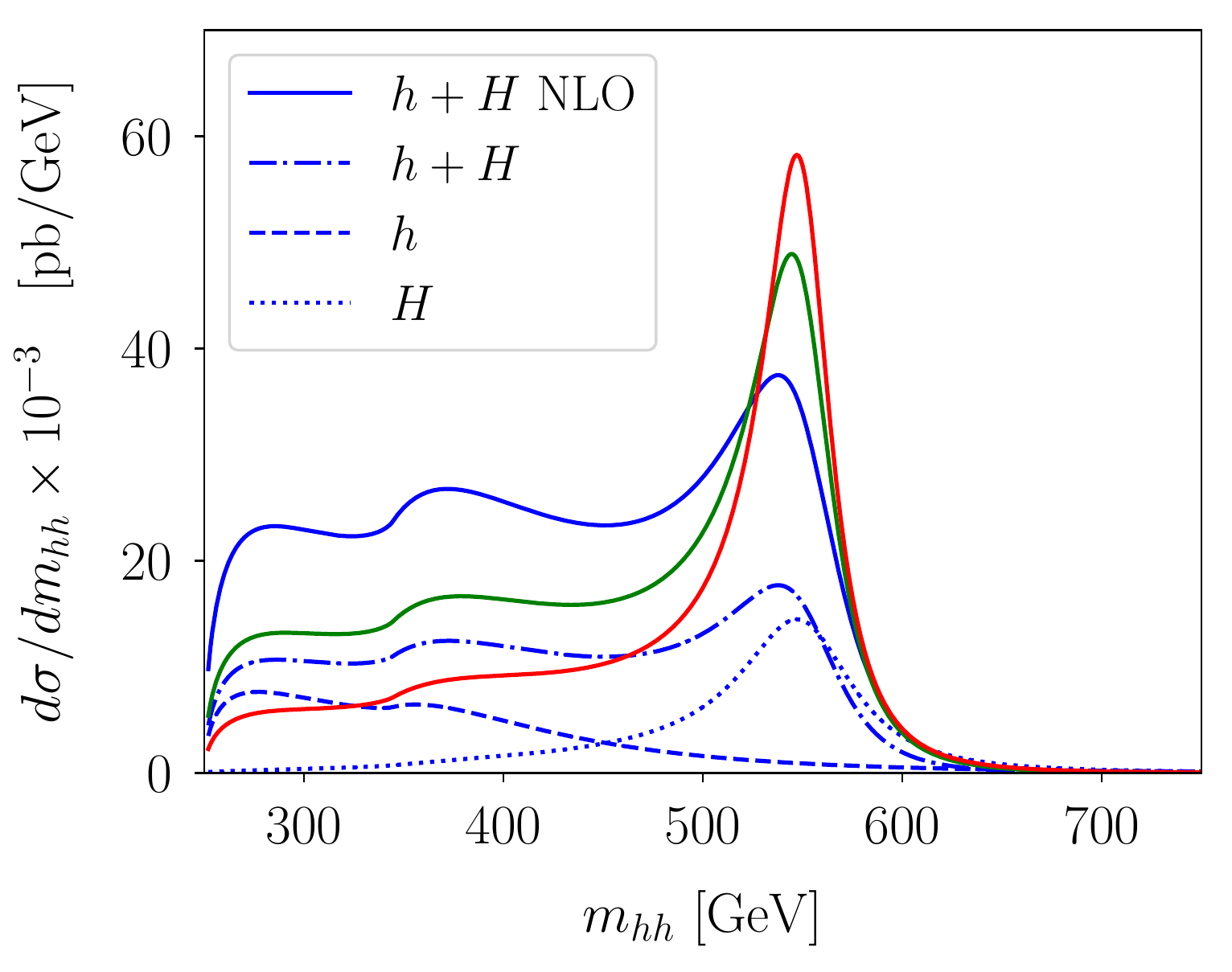}
	\caption{\label{fig:xsecc} Left: Cross section for Higgs pair production in units of the SM prediction as a function of $\kappa_{\psi}^h$ for $c_{\beta-\alpha}=-0.45~(-0.4)$ and $M_A=450$ \UGeV , $M_H=M_{H^\pm}=550$ \UGeV in blue (green) at $\sqrt{s}=27 $ \UTeV. Right: Invariant mass distribution for the different contributions to the signal with $c_{\beta-\alpha}=-0.45$ and $\kappa^h_{\psi}=5$ (blue),  $\kappa_{\psi}^h=4$ (green) and $\kappa_{\psi}^h=3$ (red) at $\sqrt{s}=27 $ \UTeV, respectively.} 
	\label{fig:hhflavourxsanddistr}
\end{figure*}

The enhancement in $\text{Br}(H\to hh)$ shown in Figure~\ref{fig:BRvskappa} is partially cancelled in the production cross section $\sigma(gg\to H)$ for large values of $t_{\beta}$ due to the fact that $\sigma(gg\to H)\propto 1+1/t_\beta^2-(\kappa_t^h)^2$, with $\kappa_t^h \approx 1$. However, the cross-section $\sigma(gg\to h\to hh)$ is not suppressed for such values of $t_{\beta}$ and the combination of both contributions leads to a continuous enhancement in the di-Higgs cross-section. There is therefore a non-trivial interplay between resonant and non-resonant contributions, which we illustrate in the left panel of Fig.~\ref{fig:xsecc}, where we plot both contributions assuming as a function of $\kappa_{\psi}^h$ for fixed values of $c_{\beta-\alpha}$ (which is a monotonic function of $t_{\beta}$). We assume a centre-of-mass energy of $\sqrt{s}=27$ \UTeV and set $M_A=450$ \UGeV and $M_{H}=M_{H^{\pm}}=550$ \UGeV, while choosing two different values of $c_{\beta-\alpha}=-0.45$ and $-0.4$. Dashed (dotted) lines correspond to the non-resonant (resonant) contributions, whereas the solid lines represent the full $\sigma(gg\to hh)$ in the 2HDM in units of the SM prediction, both at LO and NLO.   Solid lines show the NLO results, while the solid shaded lines mark the values of $\kappa_{\psi}$ excluded by perturbativity and unitarity constraints~\cite{Eriksson:2009ws}.  More details about the calculation of the signal and plots for $\sqrt{s}=13$ \UTeV can be found in Ref.~\cite{Bauer:2017cov}.  The values of $\kappa_{\psi}^h$ in Fig.~\ref{fig:xsecc} correspond to $n_{\psi}=1$ but values of $\mathcal{O}(10)$ and larger are obtained for $n_{\psi}>1$. We also show in the right panel of Fig.~\ref{fig:xsecc} the invariant mass distribution for the different contributions to the di-Higgs signal for $c_{\beta-\alpha}=-0.45$ and three different values of $\kappa_{\psi}^{h}=3, 4$ and $5$. The interesting feature is that, when the enhancement in the Higgs Yukawa couplings is large enough, the interference between both non-resonant and resonant contributions turns the broad peak into a shoulder in the $d\sigma/ dm_{hh}$ distribution for the total cross section, as shown for the case $\kappa_{\psi}^h=5$ by the blue line in the right panel of Fig.~\ref{fig:xsecc}. Resolving such shape in the  invariant mass distribution can be quite challenging. We encourage a dedicated analysis considering the corresponding $d\sigma/dm_{hh}$ templates to maximise the sensitivity to features in the di-Higgs invariant mass distribution from the simultaneous enhancement of  $g_{hhh}$, $g_{Hhh}$ and $\kappa^h_{\psi}$.\\

\asubsubsection[J. Kozaczuk, A.J. Long, J.M. No, M.J. Ramsey-Musolf]{Implications for theories of electroweak phase transition}
\noindent{\it Introduction.} Explaining the origin of the cosmic matter-antimatter asymmetry is a key open problem at the interface of high energy physics and cosmology. A number of scenarios have been proposed, ranging in energy scales from $\sim 10^{12}$ and above to below the electroweak scale and corresponding to different eras in cosmic history. One of the most compelling -- electroweak baryogenesis --  ties the generation of the asymmetry to electroweak symmetry breaking (for a review, see Ref.~\cite{Morrissey:2012db}). In this scenario, the universe must have undergone a first order phase from the electroweak symmetric to broken phase at a temperature $T_\mathrm{EW} \sim 100$ \UGeV. If such a transition occurred, then there must have also existed sufficiently active CP-violating interactions to produce the observed asymmetry. Neither requirement is satisfied by the Standard Model. The symmetry-breaking transition for a 125 \UGeV Higgs boson is known to be of a crossover type, and the CP-violating interactions encoded in the Cabibbo-Kobayashi-Maskawa matrix are too feeble to have produced the observed asymmetry. Thus, viable electroweak baryogenesis (EWBG) requires physics beyond the standard model that couples to the Higgs boson.

The HE LHC would provide new opportunities to search for this BSM physics. Studies of di-Higgs production, measurements of the Higgs triple self-coupling, and precision tests of other Higgs couplings are particularly interesting as probes of the new interactions needed for a first order electroweak phase transition (EWPT). For the first order phase transition to be sufficiently strong, so as to provide the needed conditions for EWBG, the new interactions must be mediated by particles with masses below roughly one \UTeV, making them accessible to $pp$ collisions at $\sqrt{s} = 27$ \UTeV. While a definitive program of searching for these interactions would likely require higher centre of mass energy, the HE LHC would significantly extend the discovery reach over what is accessible at the HL LHC. Below, we provide a few key examples that illustrate this possibility.

\noindent{\it Higgs Potential at Finite Temperature.} The nature of the EWSB transition is governed by the temperature-dependent Higgs potential, $V_\mathrm{EFF}(\varphi, T)$. In the regime where $T\gg M_W$, this potential takes the 
simple form
\begin{equation}
V_\mathrm{EFF}(\varphi,T) = D(T^2-T_0^2)\varphi^2 -(ET+e)\varphi^3 + {\bar\lambda}\varphi^4+\cdots ~~.
\label{eq:HiggspotT}
\end{equation}
In the SM one has $e=0$, while $D$, $T_0$, $E$ and ${\bar\lambda}$ are all non-vanishing functions of the zero temperature parameters of the theory (e.g., gauge, Yukawa, and Higgs self-couplings). At any temperature, the minimum of energy is obtained when $\varphi$ equals its vacuum expectation value $v(T)$, with $v(0) = 246$ \UGeV. The Higgs boson field is just the difference $h=\varphi-v(0)$. 

At sufficiently high temperatures, the minimum of the potential resides at the origin, {\it i.e.}, $v(T) = 0$. As the universe cools, however, the minimum eventually moves away from the origin, corresponding to the onset of EWSB. The details of this evolution, and the nature of the transition (first order, second order, or crossover) depends on the values of the couplings in Eq.~(\ref{eq:HiggspotT}). Since the latter are determined by the $T=0$ interactions, measurements of Higgs boson properties allow one to infer the thermal history of electroweak symmetry breaking. 

Assuming the SM form of the $T=0$ Higgs potential and Higgs couplings to other SM particles, lattice studies imply that for a 125 \UGeV Higgs boson, the symmetry-breaking transition is of a cross over type\cite{Rummukainen:1998as,Csikor:1998eu,Laine:1998jb,Gurtler:1997hr}. Thus, one of the three ``Sakharov conditions'' for successful baryogenesis\cite{Sakharov:1967dj} -- out of equilibrium dynamics -- would not have been satisfied, thereby precluding EWBG. However, the presence of additional bosons that interact with the Higgs boson could yield a first order EWPT even for a 125 \UGeV Higgs boson (see e.g., \cite{Morrissey:2012db,Assamagan:2016azc}). A sufficiently strong first order EWPT may arise if these interactions induce changes in the zero-temperature vacuum structure of the scalar potential and/or generate finite-temperature quantum corrections that modify the parameters in Eq.~(\eqref{eq:HiggspotT}). In addition, the presence new neutral scalars that may also obtain vacuum expectation values may allow for a richer thermal history than in the SM universe, including the presence of new symmetry-breaking phases that preceded the presence of the ``Higgs phase''\cite{Patel:2012pi,Patel:2013zla,Blinov:2015sna,Ramsey-Musolf:2017tgh}. 

\noindent{\it Collider Probes.} Existing searches for new scalars at the LHC, together with present measurements of Higgs boson properties, generally rule out a strong first order transition if the new scalars are charged under S(3$)_C$\cite{Katz:2014bha,Katz:2015uja}. In contrast, interactions involving scalars that carry only EW quantum numbers (EW multiplets) or no SM quantum numbers at all (singlets) are considerably less constrained. Cross sections for directly producing these scalars can be as small as a few fb when model parameters are consistent with a strong first order EWPT. If one of these scenarios is realised in nature, then one may or may not be able to discover it at the HL LHC. The higher energy and integrated luminosity of the HE LHC would significantly expand this discovery potential. 

Perhaps, the simplest illustration of this potential is the extension of the SM scalar sector with a single real singlet scalar\cite{Espinosa:1993bs,Choi:1993cv,Ham:2004cf,Profumo:2007wc,Cline:2012hg,Espinosa:2011ax,No:2013wsa,Curtin:2014jma,Kotwal:2016tex,Brauner:2016fla,Huang:2016cjm,Chen:2017qcz,Huang:2017jws}, the ``xSM'' \cite{Barger:2007im} (for analogous studies with a complex singlet, see \cite{Jiang:2015cwa,Chiang:2017nmu}). 
The xSM contains two Higgs-like scalars, $\rm h_{1}$ and $\rm h_{2}$ that are admixtures of the neutral component of the SM Higgs doublet and the singlet. For a wide range or model parameters, the interactions in the xSM scalar potential can  lead to a strong first order EWPT when the SM-like state $\rm h_1$ has a mass of 125 \UGeV. The associated collider signatures direct and indirect effects:  direct production of scalar pairs; include modifications of the Higgs self-coupling, which may be as as large as $\mathcal{O}(1)$ or small as a few percent; and a shift in the associated production ($\rm Zh_1$) cross section. 

We consider first scalar pair production. In pp collisions, a pair of SM-like scalars $\rm h_1$ can be produced through an on-shell $\rm h_2$, corresponding to the so-called ``resonant di-Higgs production''.  Each $\rm h_1$ then decays to the conventional Higgs boson decay products, yielding various combinations. The possibilities for discovery through the ``resonant di-Higgs production'' process are illustrated in Fig.~\ref{fig:ewpt_resdihiggs}, where the results are obtained by combining the 4\texttau\ and b\={b}\textgamma\textgamma\ final states\cite{Kotwal:2016tex} (for early studies of resonant di-Higgs production, see, {\it e.g.}. Ref.~\cite{Baur:2003gp}). Each coloured band gives the projected significance $N_\sigma$ of observation as a function of the $\rm h_2$ mass, with the $N_\sigma$ range obtained by varying over all other model parameters consistent with a strong first order EWPT, constraints from EW precision observables, and present LHC Higgs signal strength determinations.  The maximum $\rm h_2$ mass consistent with a strong first order EWPT is just below 900~\UGeV.  Results are shown for different prospective centre of mass energies.

At the time this work was completed, no analysis had been performed for $\sqrt{s} = 27 $ \UTeV and 15 ab$^{-1}$ integrated luminosity. Consequently, we show in the left panel the reach for the LHC and a 100 \UTeV $pp$ collider and in the right panel the corresponding reach for $\sqrt{s} = 50$, 100, and 200 \UTeV with 30 ab$^{-1}$. As one can see, the HL-LHC discovery potential is limited to a relatively modest portion of the light $\rm h_2$ parameter space, whereas the FCC-hh with 30~ab$^{-1}$ would enable discovery over the entire first order EWPT-viable parameter space in this model. Interpolating by eye, one can anticipate that the reach for the HE LHC will lie somewhere between that of the LHC and the 50 \UTeV band in the right panel. 

\begin{figure}[hbtp]
  \begin{center}
    \includegraphics[width=0.49\linewidth]{\main/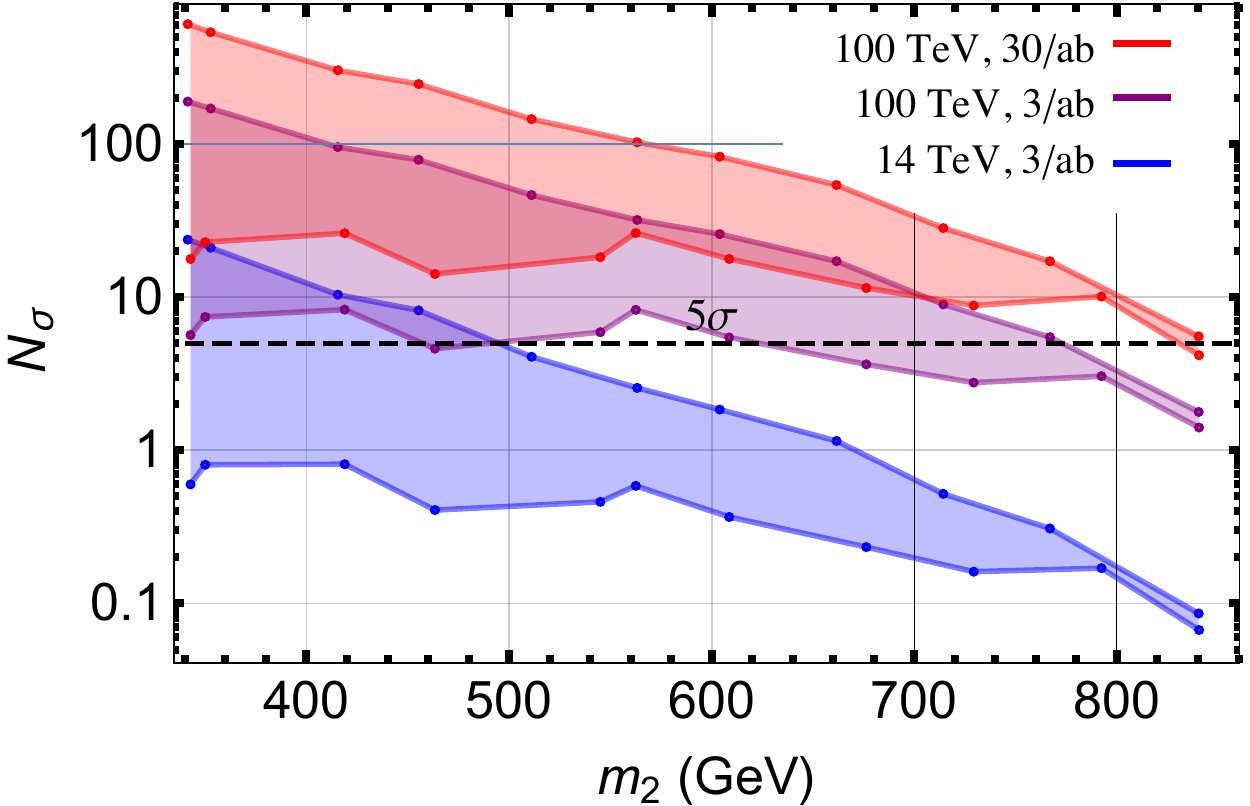}
    \includegraphics[width=0.49\linewidth]{\main/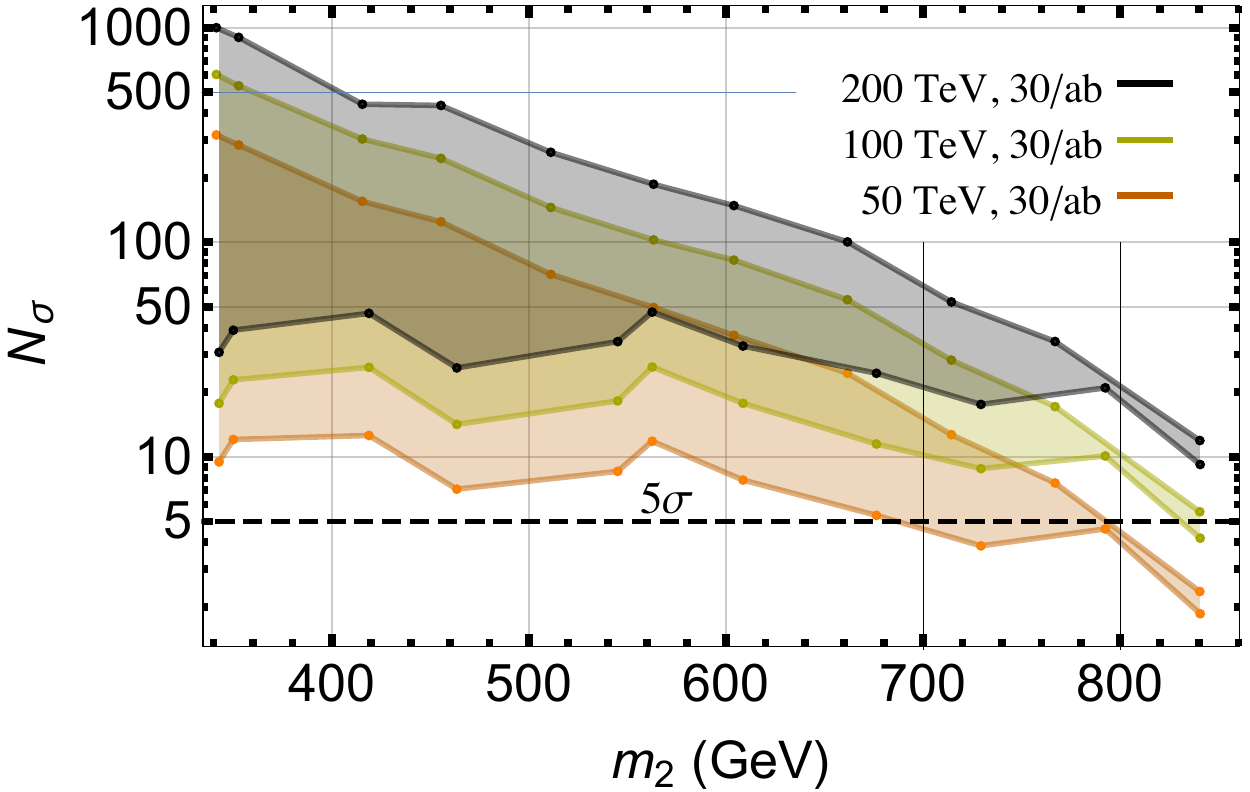}
    \caption{
    Discovery potential for the singlet-induced strong first order EWPT using resonant di-Higgs production combining 4\texttau\ and b\={b}\textgamma\textgamma\ final states\cite{Kotwal:2016tex}. Vertical axis gives significance as a function of the singlet-like scalar mass $m_2$. Left panel gives comparison of the reach for the HL-LHC (blue band) and the FCC-hh with 3 ab$^{-1}$ and 30 ab$^{-1}$ (purple and red bands, respectively). Right panel shows the prospective reach for different centre of mass energies, assuming 30 ab$^{-1}$.    
        }
    \label{fig:ewpt_resdihiggs}
  \end{center}
\end{figure}

It is worth noting that the foregoing analyses are based on the assumption that the di-Higgs production process is dominated by the resonant amplitude. As discussed in Section 9.6.2, inclusion of interference with non-resonant amplitudes may lead to an enhanced sensitivity, particularly at higher values of the singlet-like mass $m_2$. The corresponding gain in going to the HE-LHC may be as much as a 40-50\% increase in mass reach compared to that of the HL-LHC, depending on the choice of other model parameters.

Another class of signatures providing important information about the couplings in the Higgs potential in singlet-extended Higgs sectors involves pair production of the new scalar itself. These processes can complement resonant di-Higgs production in their coverage of the parameter space. For example, the process $p p\rightarrow h_2 h_2 \rightarrow 3\ell 3\nu j j$ was analysed in Ref.~\cite{Chen:2017qcz} and shown to provide good sensitivity to the first-order EWPT-compatible parameter space at both the high luminosity LHC and a future 100 \UTeV collider for masses below the di-Higgs threshold. While the analogous study has not been performed for the 27 \UTeV HE-LHC, $h_2 h_2$ production should still provide sensitivity to the couplings in the potential responsible for strengthening the EWPT, improving over the reach of the HL-LHC. In models in which a new $\mathbb{Z}_2$ symmetry is imposed on the singlet scalar, the VBF-like topology $p p \rightarrow j j h_2 h_2$ can be used to access the relevant Higgs portal coupling. In this case, $h_2$ is stable and escapes the detector as missing energy. Ref.~\cite{Curtin:2014jma} showed that this process at 100 \UTeV can probe first-order EWPTs for relatively low scalar masses. The analogous studies for the 14 \UTeV HL-LHC and 27 \UTeV HE-LHC remain to be done. 

Beyond direct production, the HE LHC will provide opportunities to observe indirect signatures of a strong first order EWPT through modifications of Higgs couplings. Considering first the xSM, the mixing between the doublet and singlet states will lead to modifications of the Higgs triple self coupling. This possibility is indicated in Fig.~\ref{fig:ewpt_self}, where we show he correlation between the critical temperature for the first order EWPT and the triple self coupling. The vertical axis gives the ratio of the xSM triple self-coupling of Higgs-like state $h_1$ to its Standard Model value, corresponding to the quantity $\kappa_\lambda$  introduced earlier in this chapter.  According to the analysis presented in the first part of Section~\ref{sec:THanetal}, a 15\% determination of $\kappa_\lambda$ may be possible using the $bb\gamma\gamma$ channel (however, see a parallel analysis later in that section for a less optimistic projection). This sensitivity corresponds roughly to the width of the green band in Fig.~\ref{fig:ewpt_self}. One can see that there exists a wide range of xSM parameter choices that would lead to an observable deviation of $\kappa_\lambda$ with the HE LHC.

\begin{figure}[hbtp]
  \begin{center}
    \includegraphics[width=0.49\linewidth]{\main/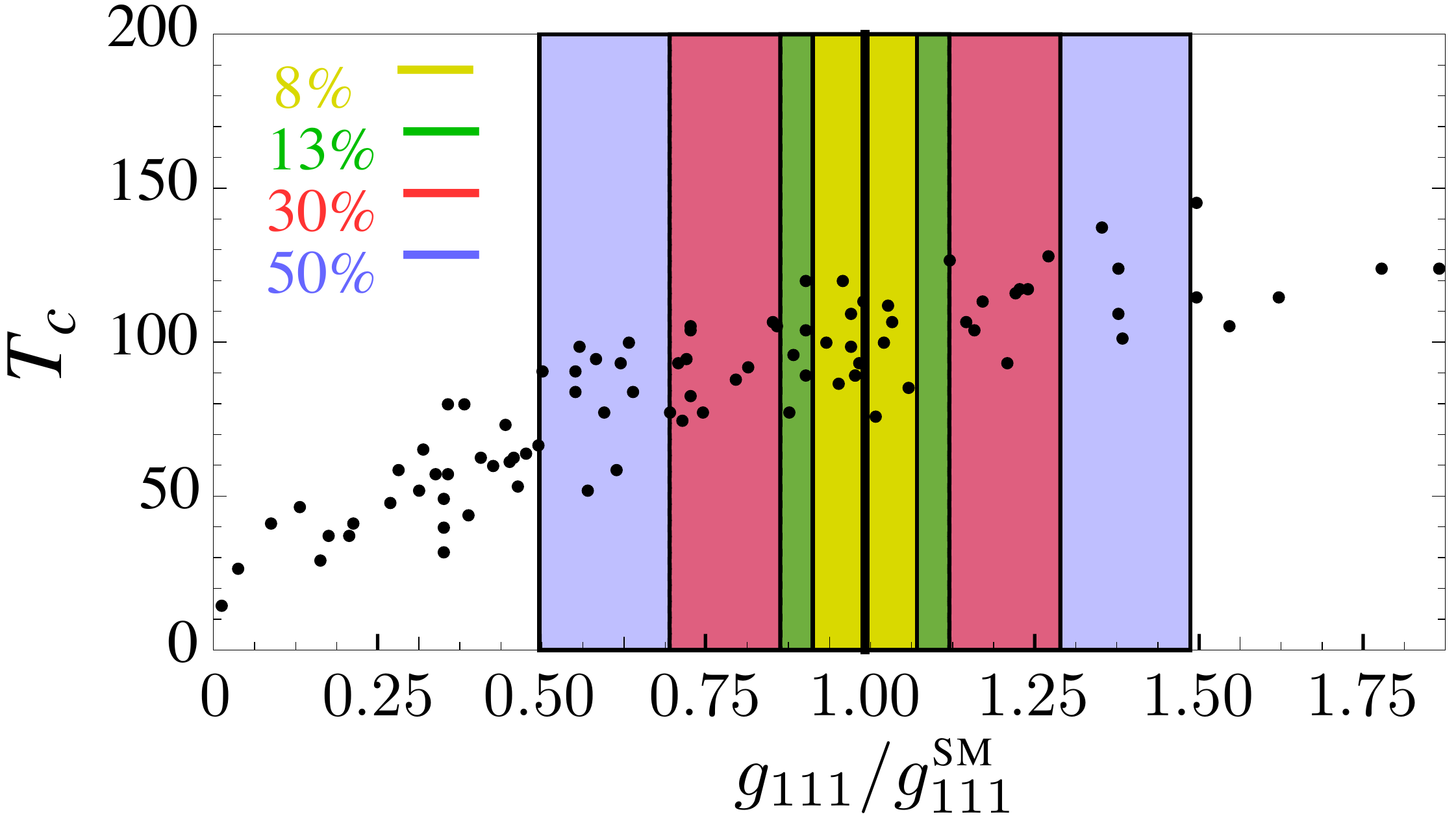}
    \caption{
    Parameter space scan for a singlet model extension of the Standard Model. The points indicate a first order phase transition. These points lead to signals observable at future colliders. Shown is the correlation between critical temperature $T_c$ (vertical axis) and the triple Higgs ($h_1 h_1 h_1$) coupling scaled to its SM value (horizontal axis). SM prediction for the latter is indicated as $g_{111}/g_{111}^\mathrm{SM}=1$. Adapted from Ref.~\cite{Profumo:2014opa}   
        }
    \label{fig:ewpt_self}
  \end{center}
\end{figure}

\begin{figure}[hbtp]
  \begin{center}
  \includegraphics[width=0.49\linewidth]{\main/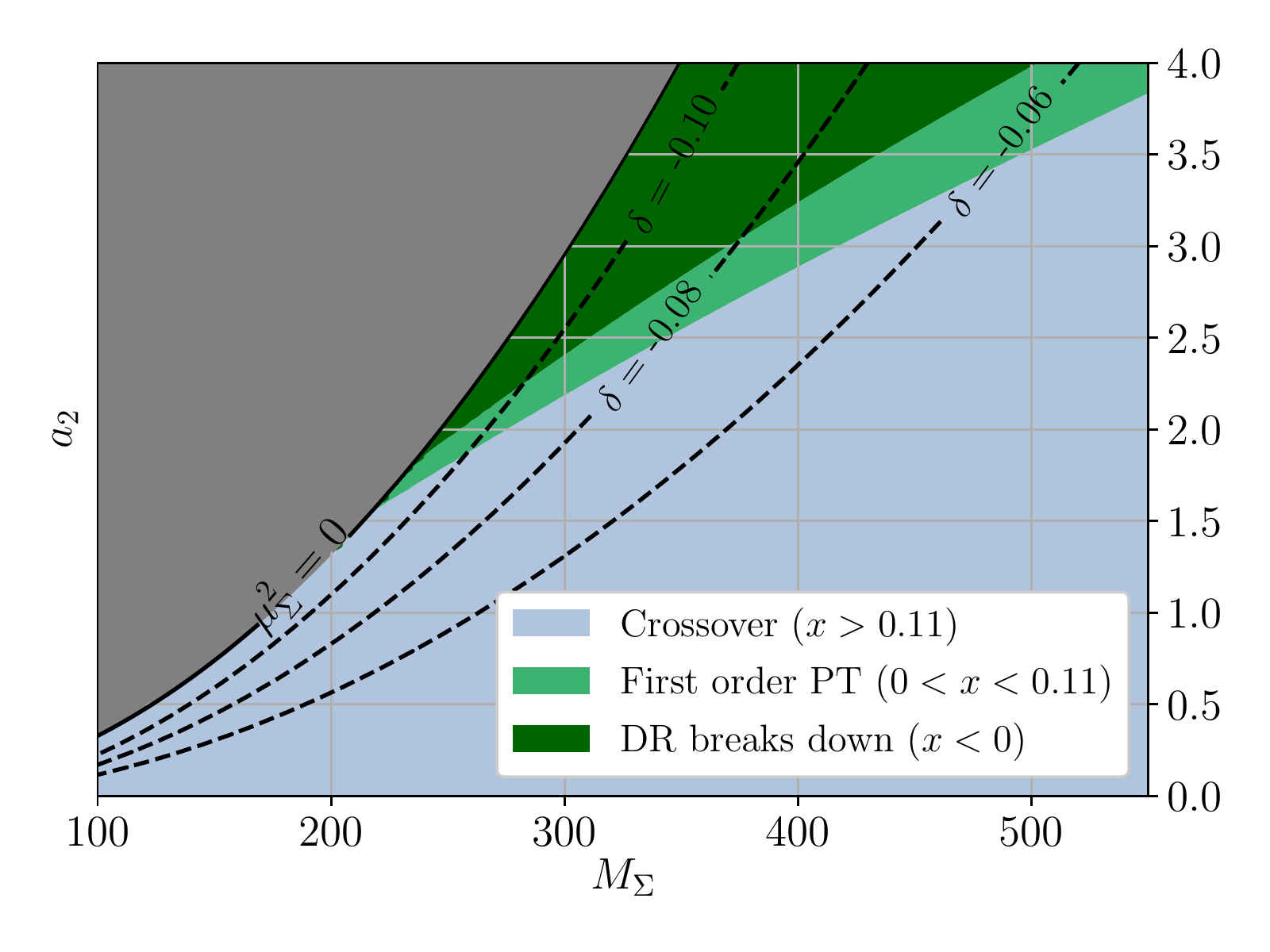}
    \caption{
    EW phase diagram for the real triplet extension of the SM scalar sector. Horizontal axis gives the triplet mass and vertical axis indicates the triplet-Higgs coupling. Light blue and green regions correspond to cross-over and first order transitions. Dark green (``DR breaks down'') and grey regions indicate parameter choices for which the present non-perturbative computations are not applicable. Dashed lines indicate relative shift $\delta$ in the Higgs di-photon decay rate compared to its Standard Model value. From Ref.~\cite{Niemi:2018asa}.    
        }
    \label{fig:ewpt_triplet}
  \end{center}
\end{figure}

Going beyond the SM, one may also anticipate a strong first order EWPT in scalar sector extensions carrying electroweak charge. Among the most widely studied ones, such scenario is the 2HDM. The authors of Refs.~~\cite{Dorsch:2013wja,Dorsch:2014qja} have shown that the strong phase transition would be correlated with the presence of the $\rm A^0\to Z H^0$ decay and that a nearly definitive probe of this possibility could be achieved with the LHC. An interesting alternative is a scalar EW triplet with vanishing hyper-charge.Interactions between the latter and the Standard Model doublet could lead to breaking of electroweak symmetry through either a single transition to the Higgs phase or through a succession of transitions\cite{Patel:2012pi}. Recently, the latter possibility has been explored in Ref.~\cite{Niemi:2018asa} using non-perturbative methods. In this work, it is shown how a precise measurement of the Higgs di-photon decay rate could probe the nature of the transition in this scenario. Fig.~\ref{fig:ewpt_triplet} illustrates this possibility. The horizontal and vertical axes give the triplet mass and coupling to the Higgs boson, respectively. The light blue and green regions correspond to a cross-over transition and first order phase transition, respectively. The dashed lines indicate the relative reduction in the Higgs di-photon decay rate relative to the prediction for the Standard Model. When combined with knowledge of the triplet mass, a precise measurement of the di-photon decay rate would indicate whether the transition is first order or crossover. As shown in Fig. 30, one expects to achieve a 1.8\% ($1\sigma$) determination of the Higgs-di-photon coupling parameter $\kappa_\gamma$ with the HL-LHC.

\label{sec:HH_EWPT}


\asubsection{Summary}

A measurement of the Higgs self-coupling is not only one of the last corners of the SM to be experimentally tested, but also a particularly interesting one due to its important implications on our knowledge of the Higgs potential, and the direct implications on the nature of electroweak symmetry breaking, the stability of our universe's vacuum and the matter-antimatter asymmetry. In this chapter we presented a study on the capabilities of the HL-LHC and HE-LHC programs to elucidate on these fundamental questions.

The Higgs self-coupling appears at tree level in the production of Higgs boson pairs. The SM cross section for $pp\to \HH$ computed at full NLO precision is $32.88^{13.5\%}_{-12.5\%}\,\text{fb}$ at $\sqrt{s}=14\,\text{\UTeV}$ and $127.7^{11.5\%}_{-10.4\%}\,\text{fb}$ at $\sqrt{s}=27\,\text{\UTeV}$, a factor 4 increase between the two centre of mass energies. The full NLO dependence on the trilinear coupling has been computed and illustrated in Table~\ref{tab:sigmatot} for selected coupling strength values. The NLO cross section has been computed for a set of benchmark points in the nonlinear EFT formalism as reported in Table~\ref{sigmatot}.

The ATLAS projections at HL-LHC for the $b\bar{b}b\bar{b}$, $b\bar{b}\tau\bar{\tau}$ and $b\bar{b}\gamma\gamma$ decay modes are summarised in Fig.~\ref{fig:ATLAS_HH_comb}. Considering the systematic uncertainties, the $b\bar{b}b\bar{b}$ channel gives the constraints $-2.3\leq \kappa_\lambda \leq 6.4$ at 68\% CL; the  $b\bar{b}\tau\bar{\tau}$ allows to resolve the region between the two minima giving $0.1\leq \kappa_\lambda\leq 2.3 \cup 5.7\leq \kappa_\lambda \leq 7.8$; and the $b\bar{b}\gamma\gamma$ gives the best precision with the interval $-0.2 \leq \kappa_\lambda \leq 2.5$.

The CMS projections include, on top of the channels studied by ATLAS, the decays $b\bar{b}\ell\nu\ell\nu$ and $b\bar{b}\ell\ell\ell\ell$, with projected constraints $-1.7\leq \kappa_\lambda \leq 9.6$ and $-1.8\leq \kappa_\lambda \leq 8.1$ respectively. For the other channels, the projected reach is $-0.6\leq \kappa_\lambda \leq 7.2$ for $b\bar{b}b\bar{b}$, $-0.2\leq \kappa_\lambda \leq 3.2 \cup 5.2\leq \kappa_\lambda \leq 7.7$ for  $b\bar{b}\tau\bar{\tau}$ and $0.3\leq \kappa_\lambda \leq 2.3$ for $b\bar{b}\gamma\gamma$.

Experiments provide a combination of the projections, summarised in Fig~\ref{fig:comb_HH_experiment} and Fig.~\ref{fig:comb_HH}. A combined significance of $4\sigma$ for the SM \HH signal is predicted.
The 68\% CL intervals are $0.52\leq \kappa_\lambda \leq 1.5$ and $0.57\leq \kappa_\lambda \leq 1.5$ with and without systematic uncertainties respectively. The second minimum is excluded at 99.4\% CL, and the hypothesis corresponding to the absence of self-coupling  ($\kappa_\lambda=0$) is excluded at the 95\% CL.

Further improvements in the analysis are proposed. The Topness and Higgsness variables can further increase the signal sensitivity for $\HH$ production, and improved multivariate methods are proposed to improve the background discrimination in the $b\bar{b}\gamma\gamma$ channel.

For the future HE-LHC upgrade, ATLAS presented projections for the $b\bar{b}b\bar{b}$ and $b\bar{b}\tau\bar{\tau}$ channels, with an expected sensitivity of around 10\%-20\% on $\kappa_\lambda$.

A complementary strategy to extract the trilinear from the LHC data consists in considering NLO corrections to single Higgs observables that depend on the self-coupling. This dependence has been computed and is presented in Tables~\ref{c1s},\ref{tab:c1-xs_13},\ref{tab:c1-xs_27},\ref{c1g}. The CMS experiment has provided a first analysis based on this strategy using $t\bar{t}\text{H}$ production with the decay $\text{H}\to \gamma\gamma$, with a resulting constraint of $-2\leq \kappa_\lambda\leq 5.5$, see Fig.~\ref{fig:ttHdiff_CMS_klambda_scan}. Considering that other parameters may affect single Higgs process, this has been studied under the perspective of a global fit, presented in Fig.~\ref{fig:hllhcchi2} estimating a $-3\leq \kappa_\lambda \leq 3$ reach at 68\% CL, complementing the constraints from double Higgs boson production. At HE-LHC, the projections are presented in Fig.~\ref{fig:helhcchi2}.

Implications of the $\HH$ measurements involve a wide variety of models. We present the interpretation within a flavour model which implies large Yukawa couplings for the light quarks, modifying the di-Higgs invariant mass distribution non-trivially, see Fig.~\ref{fig:hhflavourxsanddistr}. Also, Higgs boson pair production plays a crucial role on understanding the nature of the electroweak symmetry breaking phase transition, which might imply potentially observable effects if there are new states modifying the potential so that the matter-antimatter asymmetry can be explained via electroweak baryogenesis.

\begin{figure}[h]
  \centering
    \includegraphics[width=0.9\textwidth]{\main/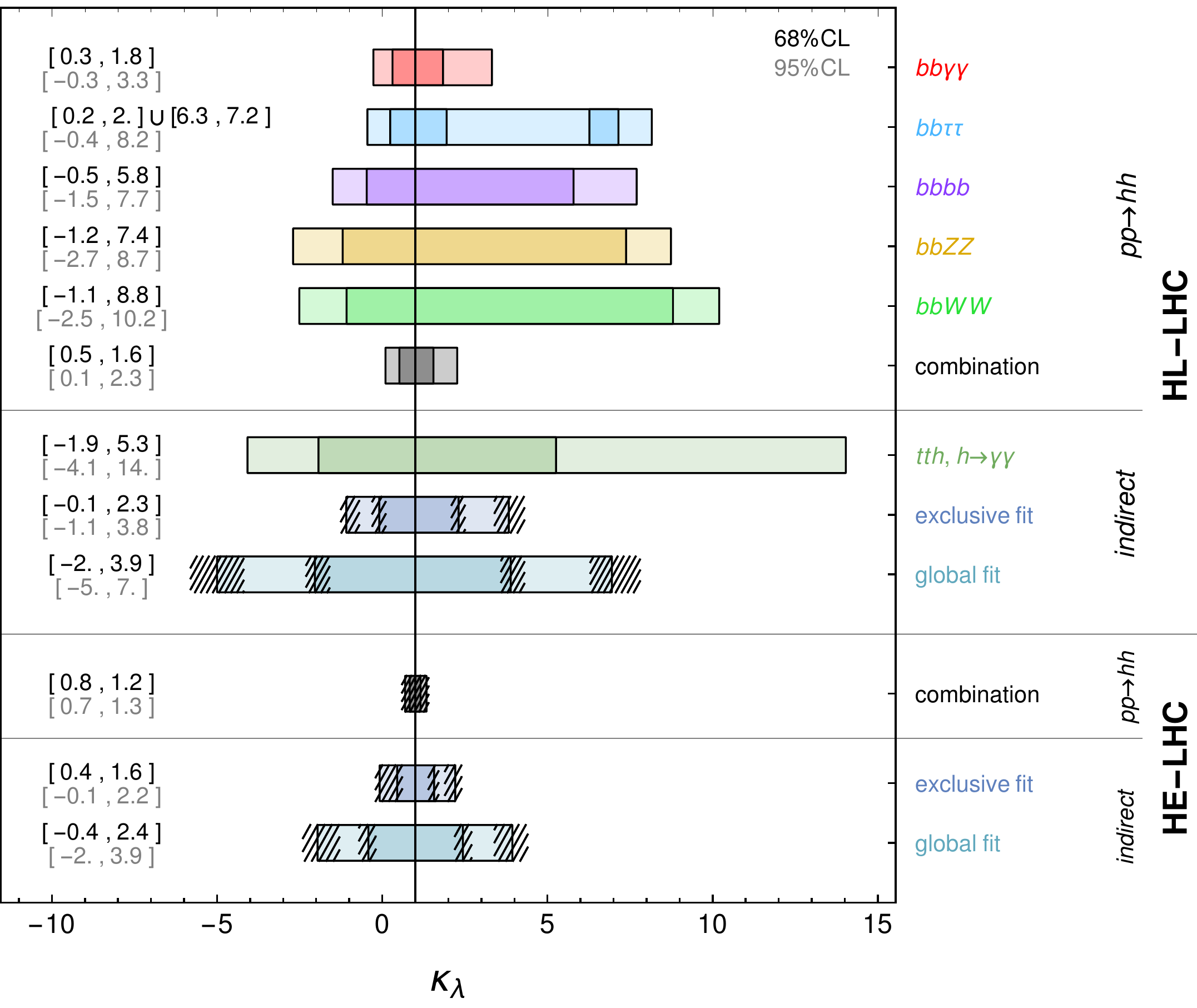}
\caption{Summary plot for the different expected constraints on the Higgs boson self-coupling $\kappa_\lambda$ at HL-LHC and at HE-LHC. The dashed lines correspond to uncertainties on the values, when any.}
\label{fig:tril_summaryplot}
\end{figure}

\newpage

\clearpage
\asection[F. Riva]{High Energy Probes}
\label{sec4}
An important aspect of both the HL and HE programs, is their enhanced sensitivity to the high-energy tails of kinematic distributions.  
These constitute genuinely \emph{new} observables with which we can realistically  conceive \emph{high-energy precision tests} that have been impossible at previous experimental facilities.
 There are two reasons that make this precision program appealing. The first is that we can \emph{define} the high-energy region as that in which statistical uncertainty becomes comparable to systematic uncertainty: in this way, high-energy precision probes are, by construction, expected to deliver the fastest relative improvement with accumulated luminosity, contrary to other observables which are, sooner or later, saturated by systematics.

The second reason why these tests are particular interesting is that the very framework in which precision tests are conceived, that of Effective Field Theories (EFTs) -   see section~\ref{sec:eftintro} -  implies effects that grow with the energy $E$. In particular, at the dimension-6 level, the expected growth is $\propto E^2$, implying a quadratic enhancement of the signal. 
As an order of magnitude estimate, an LHC $O(10\%)$ measurement at 1 \UTeV, is equivalent in precision to a $O(0.1\%)$ measurement at LEP (where at the $Z$-pole $E\approx 100$ \UGeV). For this reason, the HL-LHC (and even more so its HE upgrade) constitute an important continuation of the precision program. 

In this chapter, we provide a perspective on the importance of these high-energy probes, and collect a number of contributions that target energy-growing effects in the EFT framework.
Our focus is of course Higgs-physics; yet, in the  high-energy  limit $E\gg m_{W}$, the longitudinal polarisations of gauge bosons are also associated with the scalar sector, as can be understood by the Equivalence Theorem~\cite{Chanowitz:1985hj,Wulzer:2013mza}, where external longitudinally-polarised vector states are represented in Feynman diagrams as the corresponding
scalar Goldstone bosons, up to corrections of order $m_W/E$ from diagrams with gauge external lines.
This brings us to study, in wider generality, processes involving gauge  and Higgs bosons. 

Sections~\ref{WZlong} and \ref{sec:ZHWZeft} discuss the reach to modified Higgs sectors from $WZ$, $WH$ and $ZH$ high-$p_T$
distributions, while section~\ref{sec:WZtrans} focuses on an additional class of effects, associated with new physics in the transverse polarisations of vectors, that also modifies $WZ$ processes. 
Drell-Yann processes also constitute a very clean and powerful probe for new physics coupling to the electroweak bosons, as we discuss in section~\ref{DY}. In sections \ref{sect-illus} and \ref{sec:VVHHcont} we motivate and study modifications of the $hhWW$ coupling, that can be tested in VBF processes
and in section~\ref{sec4:hwh} we provide a generic perspective on the effects associated with  Higgs couplings modifications at high-energy.
Finally, a more complete EFT discussion of the VBF topology appears in section~\ref{sec:VBFdim6eft} and in section \ref{sect-ssWW} for the same-sign case.

\asubsection[R. Franceschini, G. Panico, A. Pomarol, F. Riva,
  A. Wulzer]{Electroweak Precision Tests in High-Energy Di-boson
  Processes} \label{WZlong}

In this section we classify the leading new-physics effects that can be tested
in di-boson channels, showing that they can be encapsulated in four real ``high-energy primary'' (HEP) parameters~\cite{Franceschini:2017ab} .
We also assess the reach on these parameters at the HL-LHC and at future hadronic colliders, focusing in particular
on the fully leptonic $WZ$ channel that appears particularly promising.

We are interested in processes which fulfil two conditions. First, their amplitudes must receive BSM contributions that grow with $E^2$ at the leading order (i.e., $d=6$) in the EFT operator expansion. Second, the  SM amplitudes must be constant and sizeable
at high energy, in such a way that, at the linear order in the EFT Wilson coefficient,  the $E^2$-growth of the BSM amplitudes
results into a $E^2$-growth of the differential cross-sections thanks to the SM-BSM interference. 
As explained in detail in Ref.~\cite{Franceschini:2017ab}, only $pp \to V_LV_L $ and $pp \to V_L h$ (see section~\ref{sec:ZHWZeft}) production
processes  enjoy quadratic energy growth at the interference level; we thus focus on these in the rest of the
section.\footnote{Notice however that promising strategies to circumvent the non-interference problem have been recently proposed \cite{Panico:2017frx,Azatov:2017kzw}, which allow for instance to ``resurrect'' interference effects in transverse vector bosons production, see also section~\ref{sec:WZtrans}.}

In order to assess the leading energy behaviour, it is sufficient to study the amplitude in the unbroken phase,
where the EW bosons are massless and the $G_{\rm{SM}}={\textrm{SU}}(2)_L\times{\textrm{U}}(1)_Y$ symmetry is exact.
Given that the Goldstone bosons live in the Higgs doublet $H$, together with the Higgs particle, $G_{\rm{SM}}$ implies
that the high-energy behaviour of the former ones are connected with the latter. This is the reason why $V_LV_L$
and $V_Lh$ production processes, collectively denoted as $\Phi\Phi'$ in what follows, should be considered together.

Focusing our interest to the production of $\Phi\Phi'$ out of a quark $q'$ with helicity $\lambda'$ and an anti-quark ${\overline{q}}$ with helicity $\lambda$ we can restrict the form of the BSM amplitudes that interfere with SM one.
At order  $E^2/M^2$ in the EFT expansion the relevant BSM effects can be parametrised as corrections to the $J=1$ partial wave amplitudes~\cite{Franceschini:2017ab}, namely
\begin{equation}\label{amp0}
\delta{\mathcal{A}}\left(q^\prime_{\pm}{\overline{q}}_{\mp}\rightarrow\Phi\Phi'\right)= f^{\Phi\Phi'}_{q^\prime_{\pm}{\overline{q}}_{\mp}}(s)\sin\theta=  \frac{1}{4} A^{\Phi\Phi'}_{q^\prime_{\pm}{\overline{q}}_{\mp}}\, E^2 \sin\theta^*\,,
\end{equation}
where $\theta^*$ is the scattering angle in the $\Phi\Phi'$ centre of mass, and $E=\sqrt{s}$ is the centre of mass energy.

Eq.~(\ref{amp0}) shows that at the leading order in the SM EFT expansion each di-boson process is sensitive at high energy to a single constant new-physics parameter $A^{\Phi\Phi'}_{q^\prime_{\pm}{\overline{q}}_{\mp}}$ for every combination of initial or final states. This can be taken real since its imaginary part does not interfere with the SM. In addition, the SM symmetry group, which is restored in the high-energy limit, as previously explained, implies  several relations among these parameters~\cite{Franceschini:2017ab}. 
As a consequence, only $4$ HEP parameters are enough to parametrise the BSM effects we are interested in. This is very non-trivial from an EFT perspective, since a total of $6$ anomalous couplings coming from $d=6$ effective operators contribute to longitudinal di-boson processes. These couplings can be identified as
${\delta g^Z_{uL}}$, ${\delta g^Z_{uR}}$, ${\delta g^Z_{dL}}$, ${\delta g^Z_{dR}}$, ${\delta g_1^Z}$ and ${\delta \kappa_{\gamma}}$ in the notation of Ref.~\cite{Gupta:2014rxa}.
The relations between the HEP parameters and the $4$ combinations of the low-energy primaries that produce growing-with-energy effects are reported in the third column of table~\ref{Wilsons}, while
\begin{table}[t]
\begin{center}
\begin{tabular}{c|c|c}
Amplitude& High-energy primaries& Low-energy primaries  \\\hline
\rule[-1.4em]{0pt}{3.2em}$\bar u_L d_L\to W_LZ_L,W_Lh$ & $\sqrt{2}a_q^{(3)}$ & $ \displaystyle\sqrt{2}\frac{g^2}{m_W^2}\left[c_{\theta_W}({\delta g^Z_{uL}}-{\delta g^Z_{dL}})/g-c_{\theta_W}^2{\delta g_1^Z} \right]$ \\
\hline
\rule[-.6em]{0pt}{1.7em}$\bar u_L u_L\to W_LW_L$& \multirow{ 2}{*}{$a_q^{(1)}+a_q^{(3)}$}& \multirow{ 2}{*}{$\displaystyle-\frac{2g^2}{m_W^2}\left[Y_L t^2_{\theta_W}{\delta\kappa_\gamma}+T_Z^{u_L}{\delta g_1^Z}+c_{\theta_W}{\delta g^Z_{dL}}
 /g\right]$}\\
\rule[-.55em]{0pt}{1.45em}$\bar d_L d_L\to Z_Lh$& &\\
\hline
\rule[-.6em]{0pt}{1.7em}$\bar d_L d_L\to W_LW_L$& \multirow{ 2}{*}{$a_q^{(1)}-a_q^{(3)}$}& \multirow{ 2}{*}{$\displaystyle-\frac{2g^2}{m_W^2}\left[Y_L t^2_{\theta_W}{\delta\kappa_\gamma}+T_Z^{d_L}{\delta g_1^Z}+c_{\theta_W}{\delta g^Z_{uL}}
/g\right]$}\\
\rule[-.55em]{0pt}{1.45em}$\bar u_L u_L\to Z_Lh$& & \\
\hline
\rule[-1.2em]{0pt}{3.em}$\bar f_R f_R\to W_LW_L,Z_Lh$& $a_{f}$& $\displaystyle-\frac{2g^2}{m_W^2}\left[Y_{f_R} t^2_{\theta_W}{\delta\kappa_\gamma}+T_Z^{f_R}{\delta g_1^Z}+c_{\theta_W}{\delta g^Z_{fR}}/g\right]$
 \end{tabular}
  \caption{Parameter combinations (in the high- and in the low-energy primary bases) that control $E^2$-enhanced effects in each polarised longitudinal di-boson production process. Here, $T_Z^f=T_3^f-Q_fs^2_{\theta_W}$ and $Y_{L,f_R}$ is the hyper-charge of the left-handed and right-handed quark (e.g., $Y_L=1/6$).}
\label{Wilsons}
\end{center}
\end{table}
the relations between the HEP and the Wilson coefficients in the SILH basis~\cite{Giudice:2007fh}, see table~\ref{tab:dim6ops}, are given by
\be
 a_q^{(3)}=\frac{g^2}{M^2}(c_W+c_{HW}-c_{2W})\ ,\ \  \ \
  a_q^{(1)}=\frac{g'^2}{3M^2}(c_B+c_{HB}-c_{2B})\,, \label{a2c}
\ee
and
\be
a_u=-2a_d=4 a_q^{(1)}\,.
\label{unifersal}
\ee
These relations can also be written using  the $\hat S$, $\hat T$, $W$ and $Y$ parameters (we follow the notation of Ref.~\cite{Barbieri:2004qk}) in addition to the two anomalous  triple gauge couplings (aTGC), $\delta g_1^Z$ and  $\delta \kappa_\gamma$. We have
\be
 a_q^{(3)}=-\frac{g^2}{m_W^2}\left(c^2_{\theta_W}\delta g_1^Z+W\right)\ ,\ \  \ \
  a_q^{(1)}=\frac{g'^2}{3m_W^2}\left(\hat S-\delta\kappa_\gamma+c^2_{\theta_W} \delta g_1^Z-Y\right)\,,
\label{heptosilh}  
\ee
which  can  be useful in order to  compare HEP analyses  from  LHC  with  other experiments, such as LEP.

In the Warsaw basis \cite{Grzadkowski:2010es}, the HEP are transparently identified with contact interactions between quarks and scalars \footnote{These relations, as well as those in  \eq{a2c}, are obtained by computing the di-boson helicity amplitudes in the presence of the EFT operators, and matching with the results of the low energy primaries. See Ref.~\cite{Franceschini:2017ab} for details.}, see Table~\ref{tab:dim6ops}
\be
a_u=4\frac{c_R^u}{M^2}\ ,\  \ a_d=4\frac{c_R^d}{M^2}\ ,\ \  a^{(1)}_q=4\frac{c^{(1)}_L}{M^2}\ ,\ \  a^{(3)}_q=4\frac{c^{(3)}_L}{M^2}\,.
\ee

To illustrate the HL/HE-LHC reach on the high-energy primaries we focus on $WZ$ production. This channel gives access to the
$a_q^{(3)}$ primary and has a very high sensitivity to new physics~\cite{Franceschini:2017ab}. We consider the fully leptonic final state
$$
pp\to W^{\pm}Z + {\rm{jets}}\to \ell\nu\ell^{\prime}\bar{\ell}^{\prime} + {\rm{jets}} \,, \;\;\; {\rm{with}}\;\;l,l'=e,\mu\,,
$$ 
which is likely to be measured with good accuracy and can benefit from a straightforward reconstruction of the final-state
leptons and a very low reducible background~\cite{Aad:2016ett}. At the experimental level
the situation might not be too much different from the neutral Drell-Yan process, in which a measurement with $2\%$
relative systematic uncertainty of the differential cross-section was performed, with run-$1$ LHC data,
up to \UTeV energies~\cite{Aad:2016zzw}.
A systematic uncertainty of $5\%$ might be considered as a realistic goal for the differential cross-section measurement in the leptonic $WZ$ channel. 

The main obstacle to obtain sensitivity to new physics is the potentially large contribution of the other polarisations, which for our purposes constitute a background, since they are insensitive to the new physics parameter $a_{q}^{(3)}$. 
Due to the symmetry structure, the emission of transversely polarised $W$ and $Z$ bosons in the central rapidity region is disfavoured~\cite{Franceschini:2017ab}. No suppression is instead expected for longitudinally polarised gauge bosons, therefore it is advantageous to concentrate our analysis on central scattering region, $|\cos \theta^*| \sim 0$, or, equivalently, at large $\ptv$ ($\ptv > 1\ {\rm \UTeV}$). We stress that other di-boson processes, e.g. $pp \to WW$, do not enjoy this suppression of transverse vector boson emission, therefore are expected to be less sensitive probes of the high energy primaries. 

We now estimate the reach on $a_q^{(3)}$ based on a full NLO simulation of the $pp \to 3\ell\nu$ process, see Ref.~\cite{Franceschini:2017ab}. 
We consider generation-level leptons momenta, but we include an overall detector efficiency for reconstructing the three leptons that we estimate around 50\%~\cite{ATLAS:2016iqc}. We furthermore apply standard acceptance cuts on the leptons (see Table~\ref{tab:cuts}).
The same-flavor and opposite-charge lepton pair with invariant mass closer to the $Z$ boson mass is taken as the $Z$ candidate and the remaining lepton  is taken to be the decay product of the $W$ boson. The missing transverse energy vector of the event ($\cancel{\vec{E}}_T$ ) is estimated from the generation-level transverse neutrino momentum, to which we apply a Gaussian smearing with standard deviation
$ \sigma_{{\met}_{i}}^2 = (0.5)^2 \cdot \sum_f |p_i| \cdot {\rm{\UGeV}}\,$.

\begin{table}
\centering
\begin{tabular}{c|c}
acceptance cuts &
$p_{T,\ell}>30 \UGeV \ ,\quad |\eta_{\ell}|<2.4$\\
\hline
\multirow{2}{*}{analysis cuts} &
$\deltapt/p_{T,V}<
0.5$\\
& $|\cos \thetastar| \leq 
0.5$
\end{tabular}
\caption{List of acceptance and analysis cuts.}\label{tab:cuts}
\end{table}

In order to highlight the production of longitudinally polarised vector bosons in the central rapidity region, it is useful to eliminate events with hard real radiation, which tend to be more abundant for our background of transverse polarised gauge bosons. To tame real radiation events in a controlled way we employ a selection on the transverse momentum of the $WZ$ system, denoted by $\deltapt=|{\vec{p}}_{T,W}+{\vec{p}}_{T,Z}|\,$.\footnote{Alternatively, a jet veto might be considered, which however could lead to lower accuracy because of the experimental and theoretical uncertainties in jets reconstruction.
See also Ref.~\cite{Campanario:2014lza} for a different approach.} We require $\deltapt$ to be smaller than $50\%$ of the transverse momentum of the gauge bosons in the event, $p_{T,V} = \min(p_{T,W}, p_{T,Z})\,$. We also impose a cut on the scattering angle in the $WZ$ centre of mass frame
$|\cos \thetastar| \leq  0.5\,$. The cuts are summarised in Table~\ref{tab:cuts}.

The kinematic variables described so far allow us to determine $p_{T,Z}$ and $p_{T,W}$, and in turn $p_{T,V}$ and $p_{T,VV}$, used to construct the binned distribution and for the selection cuts. In order to extract $|\cos \thetastar|$ the neutrino rapidity is reconstructed by the standard technique of imposing the invariant mass of the neutrino plus lepton system to be as close as possible to the physical $W$ boson mass.
A twofold ambiguity in the reconstruction is resolved by imposing the $|\cos \thetastar|$ cut on both solutions, i.e.~by retaining for the analysis only events such that both the possible neutrino configurations satisfy the selection criteria.

We study the $3$ collider energy options that correspond to the LHC ($14$~\UTeV), to the High-Energy LHC (HE-LHC, $27$~\UTeV) and to the FCC-hh ($100$~\UTeV). In each case we consider suitably designed $p_{T,V}$ bins, namely 
\begin{eqnarray}\label{finalbins}
{\textrm{LHC\hspace{-10pt}}}&{\textrm{:}}& p_{T,V}\in\{100,150,220,300,500,750,1200 \}\,,\\
{\textrm{HE-LHC\hspace{-10pt}}}&{\textrm{:}}& p_{T,V}\in\{150,220,300,500,750,1200, 1800 \}\,,\nonumber\\
{\textrm{FCC\hspace{-10pt}}}&{\textrm{:}}& p_{T,V}\in\{220,300,500,750,1200, 1800, 2400 \}\,.\nonumber
\end{eqnarray}
The binning is chosen such as to cover the kinematic regime that is accessible at each collider and it is taken as fine as possible in order to maximise the BSM sensitivity. On the other hand, a minimum bins size $\Delta p_{T,V}/ p_{T,V}\gtrsim30\%$ is required in order to avoid a degradation of the accuracy due to the $p_{T,V}$ resolution.

The predicted cross-sections are used to construct the $\chi^2$, under the assumption that observations  agree with the SM, and are eventually used to derive $95\%$~CL bounds on $a_q^{(3)}$. The uncertainties in each bin are the sum in quadrature of the statistical error, obtained from the SM expected events yield, and of a systematical component (uncorrelated across bins) which we take as a fixed fraction ($\delta_{\textrm{syst}}$) of the SM expectations. With this procedure we obtain, for different collider energies and luminosities and for $\delta_{\textrm{syst}}=5\%$
\begin{eqnarray}
{\textrm{HL-LHC, }} 3\,{\textrm{ab}}^{-1}\hspace{-10pt} & {\textrm{:}} & a^{(3)}_q  \in  [-4.9, 3.9] \,10^{-2}\,\UTeV^{-2}\nonumber\\
{\textrm{HE-LHC, }} 10\,{\textrm{ab}}^{-1}\hspace{-10pt} & {\textrm{:}} & a^{(3)}_q  \in  [-1.6, 1.3] \,10^{-2}\,\UTeV^{-2}\nonumber\\
{\textrm{FCC-hh, }} 20\,{\textrm{ab}}^{-1}\hspace{-10pt} & {\textrm{:}} & a^{(3)}_q  \in  [-7.3, 5.7] \,10^{-3}\,\UTeV^{-2}
\label{boundsf}
\end{eqnarray}
We see that the HE-LHC will improve the HL-LHC reach by a factor of $3$, while a gain of nearly one order of magnitude would be possible with the FCC-hh collider. The FCC-hh reach is comparable with the one of CLIC, as extracted from the analysis in Ref.~\cite{Ellis:2017kfi}.
 
 
The results of \eq{boundsf}  rely on BSM cross-section predictions obtained by integrating up to very high centre of mass energies, formally up to the collider threshold. Therefore these limits assume that the description of the underlying BSM model offered by the EFT is trustable in the whole relevant kinematic regime, i.e.~that the cutoff $M$ of the BSM EFT is high enough. We quantify how large $M$ concretely needs to be for our results to hold by studying~\cite{Racco:2015dxa,Pobbe:2017wrj,Biekoetter:2014jwa} how the limit deteriorates if only events with low $WZ$ invariant mass, $m_{\textsc{wz}}<m_{\textsc{wz}}^{\rm{max}}$ are employed. This obviously ensures that the limit is consistently set within the range of validity of the EFT provided the EFT cutoff $M$ is below $m_{\textsc{wz}}^{\rm{max}}$. The results are reported in figure~\ref{fig:bounds_future}. Since the $95\%$~CL interval is nearly symmetric around the origin only the upper limit is reported in the figure for shortness.

\begin{figure}[t]
\centering
\includegraphics[width=0.7\textwidth]{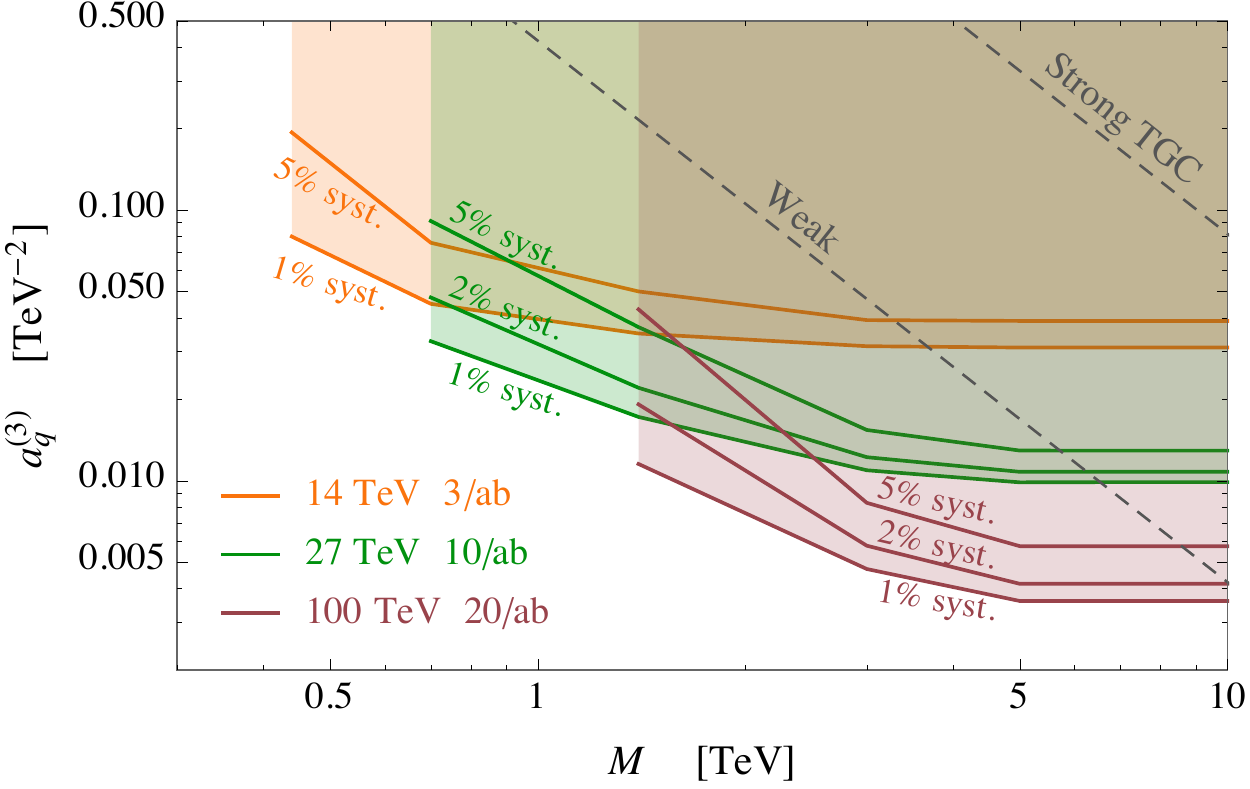}
\caption{Expected $95\%$ CL bounds from fully leptonic $WZ$ on the high-energy primary parameter $a^{(3)}_q$ as a function of the new physics scale $M$. The plots reports the results for the HL-LHC (orange lines), HE-LHC (green lines) and FCC-hh (brown lines) for different
values of the systematic uncertainties.} 
\label{fig:bounds_future}
\end{figure}

Several conclusions can be drawn from figure~\ref{fig:bounds_future}. First of all we see that the reach saturates for $m_{\textsc{wz}}^{\rm{max}}$ below around $1.5$~\UTeV at the HL-LHC if the systematic uncertainties are low, meaning that the limits obtained without $m_{\textsc{wz}}$ cut apply to theories with cutoff $M$ above that threshold.  The threshold grows to around $3$ and $4$~\UTeV at the HE-LHC and at the FCC-hh, respectively. The figures show that $\delta_{\textrm{syst}}=5\%$ is sufficient to probe ``Weak'' theories in all cases, but it also shows that the impact of larger or smaller uncertainties on the reach can be significant. Systematic errors at the $\delta_{\textrm{syst}}=5\%$ level already make an appreciable difference with respect to $\delta_{\textrm{syst}}=1\%$. This is due to the fact that the low-$p_{T,V}$ bins have small statistical error and the reach in those bins benefits from lower systematics. The effect is even more pronounced at the HE-LHC and at the FCC-hh, where even with $\delta_{\textrm{syst}}=2\%$ the reach deteriorates significantly with respect the ideal case $\delta_{\textrm{syst}}=1\%$. The fact that more accurate measurements would improve the reach of future colliders is an element that should be taken into account in the design of the corresponding detectors.

\asubsection[S. Banerjee, R.S. Gupta, C. Englert, M. Spannowsky]{WH/ZH at high energy/luminosity}\label{sec:ZHWZeft}

In this section we perform a collider study of the Higgs-strahlung process, $pp \to Z(\ell^+ \ell^-)h(b\bar 
b)$ in the Standard Model Effective Field Theory (SMEFT) framework. We will see that the leading high energy contribution to the $pp \to Zh$ process comes from the four 
contact interactions  $hZ_\mu \bar{u}_{L,R} \gamma^\mu u_{L,R}$ and $hZ_\mu \bar{d}_{L,R} \gamma^\mu 
d_{L,R}$ that appear in the dimension-6 Lagrangian. These are the same four EFT directions, the so called ``high 
energy primaries'' that control  high energy $Wh,~WW$ and $WZ$ production (see Ref.~\cite{Franceschini:2017xkh}). The (pseudo-)observables involved in these di-boson processes (anomalous TGCs and $Z$-pole observables)  have already been constrained at LEP.  We show in this note that    because of the higher  energies accessible at the LHC one can obtain bounds on these observables that are at least an order of magnitude stronger than those obtained at LEP.

The vertices in the dimension 6 Lagrangian that contribute 
to the $ff \to Vh$ (where $V=W,Z$) process in unitary gauge are as follows, 
\begin{eqnarray}
&&\Delta {\cal L}_6\supset \sum_f {\delta g^Z_{f}} Z_\mu \bar{f} \gamma^\mu f +\delta g^W_{ud} (W^+_\mu \bar{u}_L \gamma^\mu d_L+h.c.)
+ {g^h_{VV}}\,  h \left[W^{+\, \mu} W^-_\mu+\frac{1}{2c_{\theta_W}^2}Z^\mu Z_\mu\right]\nonumber \\
&&+  \delta g^h_{ZZ}\, h \frac{Z^\mu Z_\mu}{2c_{\theta_W}^2} + \sum_f g^h_{Zf}\,\frac{h}{v}Z_\mu \bar{f} \gamma^\mu f+g^h_{Wud}\,\frac{h}{v}(W^+_\mu \bar{u}_L \gamma^\mu d_L+h.c.)
+{\kappa_{Z \gamma }}\frac{h}{v} A^{\mu\nu} Z_{\mu\nu} \nonumber \\ 
&&+\kappa_{WW}\,\frac{h}{v}
 W^{+\, \mu\nu}W^-_{\mu\nu}
+\kappa_{ZZ}\,\frac{h}{2v} Z^{\mu\nu}Z_{\mu\nu}\,.
\label{lagr}
\end{eqnarray}
Here we have used the Lagrangian presented in Ref.~\cite{Gupta:2014rxa, Pomarol:2014dya}, where $\alpha_{em}$, $m_Z$ and $m_W$   have been used as input parameters and any corrections 
to the SM vector boson propagators  have 
been traded in favour of the vertex corrections. After summing over all $V$-polarisations, the leading 
piece in the high energy cross-section deviation for $ff \to Vh$ , is proportional to the four  contact interactions:  $g^h_{Zf}$, with $f=u_L, u_R,d_L$ and $d_R$.\footnote{There exists a basis independent constraint at the dimension-6 level, $\sqrt{2}\;g^h_{Wud}= (g^h_{Zd_L}-g^h_{Zu_L})$. } Table~\ref{dirn}, shows the linear combinations of Wilson coefficients contributing to the four 
$g^h_{Zf}$ couplings in different EFT bases.  The aforementioned directions are shown in the BSM Primary basis of Ref.~\cite{Gupta:2014rxa}, 
where the Wilson coefficients are already constrained pseudo-observables. In this basis we see that these can be written in terms of already constrained LEP (pseudo)observables. 

Given the inability to control the polarisation of the initial state partons in a hadron collider, 
the process, in reality, only probes two of the above four directions. Taking only the interference 
term, we find these directions to be
\begin{equation}
g^Z_{\textbf{u}}= g^h_{Zu_L}+\frac{g^Z_{u_R}}{g^Z_{u_L}} g^h_{Zu_R},\; 
g^Z_{\textbf{d}}= g^h_{Zd_L}+\frac{g^Z_{d_R}}{g^Z_{d_L}} g^h_{Zd_R}\,.
\end{equation}
 At a given energy, a linear combination of the up-type and down-type coupling 
 deviations, enters the interference term for the 
 $pp \to Zh$ process, $g^Z_{\textbf{p}}= g^Z_{\textbf{u}}+ \frac{{\cal L}_d(\hat{s})}{{\cal L}_u(\hat{s})}g^Z_{\textbf{d}},$
where ${\cal L}_{u,d}$ is the $u \bar{u}$, $d \bar{d}$ luminosity at a given partonic centre of mass 
energy. The luminosity ratio changes very little with energy: between 0.65 and 0.59 as 
$\sqrt{\hat{s}}$ is varied from 1 to  2 \UTeV. Thus, to a good approximation, $pp \to Zh$ probes 
the single direction, 
\begin{equation}
g^Z_{\textbf{p}}=g^h_{Zu_L} -0.76~g^h_{Zd_L}   - 0.45~g^h_{Zu_R} + 0.14~g^h_{Zd_R}  \,.
\label{compdir}
\end{equation}
using ${\hat{s}}=(1.5~\text{\UTeV})^2$. Using Tab.~\ref{dirn}, one can now write this in terms of the LEP-constrained pseudo-observables,
\begin{eqnarray}
g^h_{Z\textbf{p}}&=&2~\delta g^h_{Zu_L} -1.52~\delta g^h_{Zd_L}   - 0.90~\delta g^h_{Zu_R} + 0.28~\delta g^h_{Zd_R}\nonumber\\
&&-0.14~\delta \kappa_\gamma   -0.89~\delta g^Z_1 \nonumber\\
g^h_{Z\textbf{p}}&=&-0.14~(\delta \kappa_\gamma-\hat{S}+Y) -0.89~\delta g^Z_1  -1.3~W
\label{diru}
\end{eqnarray}
where the first  and second lines apply respectively to the general and universal case (third and fourth row of Table~\ref{dirn}). 
 
\begin{table*}[!t]
\begin{center}
\small
\begin{tabular}{|c|c|}
\hline
&EFT directions probed by high energy $ff \to Vh$ production \\
\hline
\hline
 Warsaw Basis~\cite{Grzadkowski:2010es} & $- \frac{2 g}{c_{\theta_W}}\frac{v^2}{\Lambda^2}(|T_3^f|c^{1}_L-T_3^f c^{3}_L+(1/2-|T_3^f|)c_f)$\\
BSM Primaries~\cite{Gupta:2014rxa} &$   \frac{2 g}{c_{\theta_W}}Y_f t_{\theta_W}^2 \delta \kappa_\gamma+2 \delta g^Z_{f}- \frac{2 g}{c_{\theta_W}}(T^f_3 c_{\theta_W}^2 + Y_f s_{\theta_W}^2)\delta g_1^Z $\\
 SILH Lagrangian~\cite{Giudice:2007fh} &$  \frac{g}{c_{\theta_W}}\frac{m_W^2}{\Lambda^2}(2 |T_3^f|\hat{c}_W-2t_{\theta_W}^2 Y_f \hat{c}_B)$\\
 Universal observables &$ \frac{2 g}{c_{\theta_W}}Y_f t_{\theta_W}^2  (\delta \kappa_\gamma-\hat{S}+Y)- \frac{2 g}{c_{\theta_W}}(T^f_3 c_{\theta_W}^2 + Y_f s_{\theta_W}^2)\delta g_1^Z- \frac{2 g}{c_{\theta_W}}T^f_3 W$\\
 High Energy Primaries~\cite{Franceschini:2017xkh} &$ -\frac{2 m_W^2}{g c_{\theta_W} }(|T_3^f| a_q^{(1)}-T_3^f a_q^{(3)}+(1/2-|T_3^f|)a_f)$\\
\hline
 \end{tabular}
  \caption{ The linear combinations of Wilson coefficients contributing to the contact interaction couplings $g^h_{Zf}$where  $f=u_L, d_L, u_R, d_R$. the direction for a given $f$  can  be read off from this table by substituting the corresponding  value of the $SU(2)_L$ and $U(1)_Y$ quantum numbers  $T_3^f$ and $Y_f$. Here $\hat{c}_W=c_W+c_{HW}-c_{2W}$ and $\hat{c}_B= c_B+c_{HB}-c_{2B}$. For the nomenclature of the operators, their corresponding Wilson coefficients and observables see for eg. Ref.~\cite{Franceschini:2017xkh}.}
  \label{dirn}
\end{center}
\end{table*}


To estimate  the cut-off for 
our EFT, note that the ${g}^h_{Vf}$ couplings arise from current-current operators that can be generated, for instance, 
by integrating out at tree-level a heavy $SU(2)_L$ triplet (singlet) vector $W'^a$ ($Z'$) that couples 
to SM fermion currents, $\bar{f} \sigma^a \gamma_\mu f$ ($\bar{f} \gamma_\mu f)$ with a coupling 
$g_f$ and to the Higgs current $i{H}^\dagger \sigma^a \lra{D}_\mu H$ ($i{H}^\dagger \lra{D}_\mu H$) 
with a coupling $g_H$. This gives ${g}^h_{Zf}\sim g_H g g_f v^2/\Lambda^2$,
where $\Lambda$ is the mass of the massive vector and thus the cut-off for our EFT description. A universal coupling to the SM fermions can arise  
via kinetic mixing of the heavy vector with the SM gauge bosons; this would give $g_f=g/2$ ($g_f=g' Y$), such that,
\begin{equation}
  \label{cutoff}
{g}^h_{Zu_L,d_L} \sim \frac{g_H g^2  v^2}{ 2\Lambda^2},\;
{g}^h_{Zu_R,d_R} \sim \frac{g_H g g' Y_{u_R,d_R} v^2}{\Lambda^2}.
\end{equation}
 For a given  set of couplings   $\{{g}^h_{Zu_L} ,{g}^h_{Zd_L} ,{g}^h_{Zu_R} ,{g}^h_{Zd_R} \}$, the cut-off is  evaluated using 
 Eq.~\ref{cutoff} with $g_H=1$ (note that this is somewhat larger than the value corresponding to the SM $hZZ$ coupling) and taking the smallest of the four values. \\ \\

For our collider analysis, we consider  $Z(\ell^+\ell^-)h$ production from a pair of quarks as well as from a 
pair of gluons.  For the decay of the Higgs boson,  we find that at an integrated luminosity of  300 fb$^{-1}$,  the di-photon mode is not feasible as it yields less than 5 events at high energies ($p_{T,Z}>150$ \UGeV). We thus focus on the decay $h\to b\bar b$ to obtain large statistics. The dominant backgrounds are then  
$Zb\bar{b}$ and the irreducible $Zh$ production in SM. Reducible contributions also arise from $Z+$ jets production (where we include $c$-quarks but do not require that they are explicitly tagged). We employ the BDRS 
approach~\cite{Butterworth:2008iy} and demand a fat jet with a cone radius of $R=1.2$. More details of the Monte-Carlo analysis, the QCD corrections, the detailed cut-based 
and multivariate analyses (MVA) can be found in Ref.~\cite{Banerjee:2018bio}. Finally, we find a cut-based (MVA) SM $Zh$ to $Zb\bar{b}$ ratio of $\sim 0.26 \; 
(0.50)$.

\begin{figure}[!t]\centering
\includegraphics[width=0.45\textwidth]{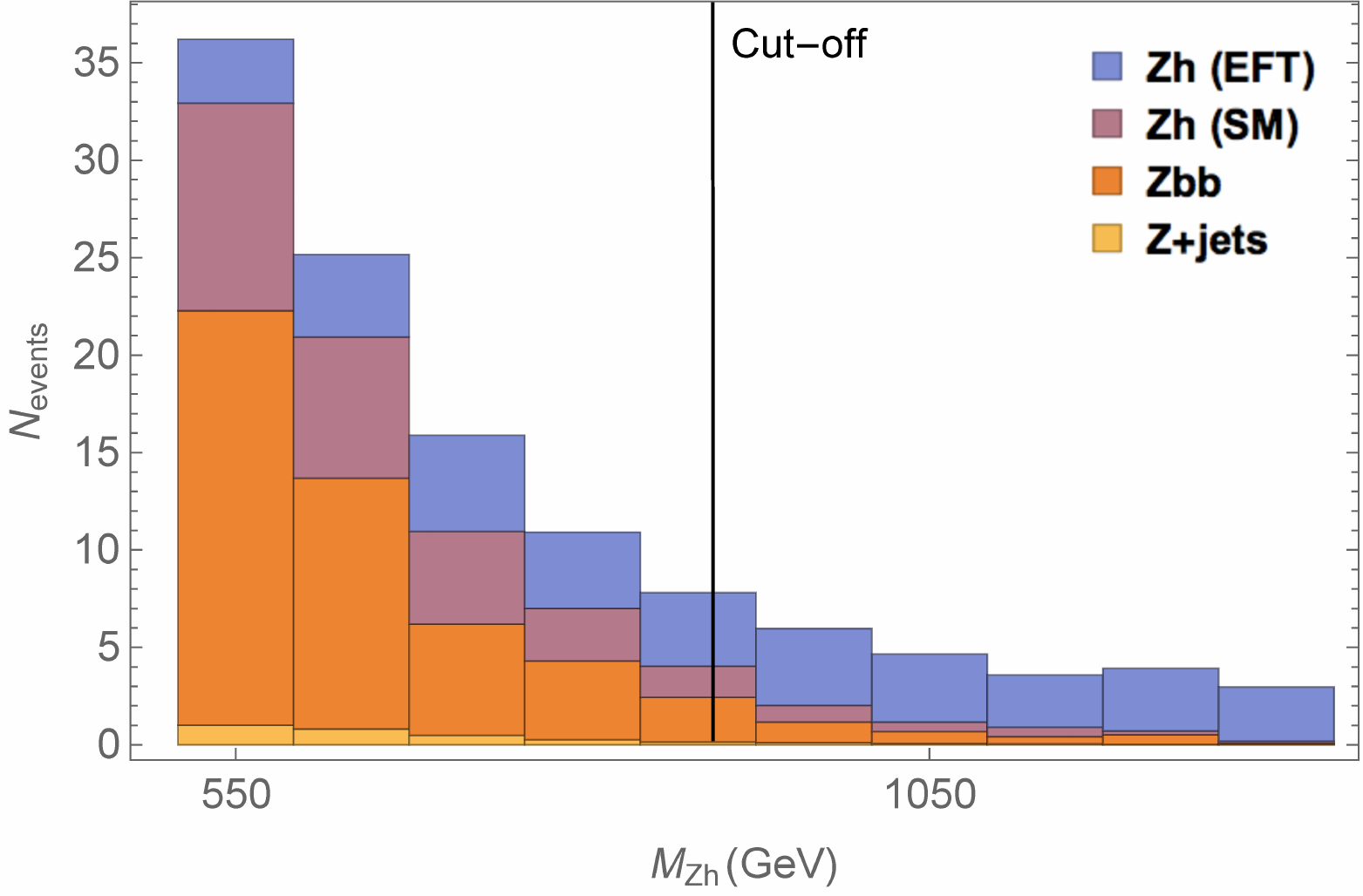}
\caption{The differential distribution of events at an integrated luminosity of 300 fb$^{-1}$ with respect to $M_{Zh}$ for the EFT signal as well as the different backgrounds. The EFT signal corresponds to the point $\{{g}^h_{Zu_L} ,{g}^h_{Zd_L} ,{g}^h_{Zu_R} ,{g}^h_{Zd_R} \}=\{-0.005, 0.0001 , -0.010, 0.005 \}$ which is allowed by  LEP bounds. The vertical line shows the cut-off evaluated using Eq.~\ref{cutoff}.}
\label{bars}
\end{figure}

To discriminate between the EFT signal and the irreducible SM $Zh(b\bar b)$ background we study  the 
growth of the EFT cross-section at high energies. This can be seen in Fig.~\ref{bars}  where we show the differential distribution with respect to $M_{Zh}$, the invariant mass of the leptons and the fat jet, for the EFT signal as well as for the different backgrounds. The EFT signal corresponds to a point that can be excluded in our analysis but is allowed by the LEP constraint.  To fully utilise the shape deviation of the EFT signal with respect to the background, 
we perform a binned log likelihood analysis assuming a 5$\%$ systematic error taking only events below the cut-off (evaluated as explained below Eq.~\ref{cutoff}). To obtain the   95$\%$ CL exclusion curve, we assume that 
the observed number of events would agree with the SM. \\ \\

Taking into account only the SM-BSM interference term, we find the following per-mille level bounds for 300 (3000) fb$^{-1}$, 
\begin{equation}
g^h_{Z\textbf{p}} \in  \left[-0.004,0.004\right] \; (\left[-0.001,0.001\right])
\label{obo}
\end{equation}
The above bounds    translate to a lower bound on the scale of new physics given by 2.4 \UTeV (4.4 \UTeV) at 300 fb$^{-1}$ (3000 fb$^{-1}$) using Eq.~\ref{cutoff}.  To compare the above projections with existing LEP bounds, one can now extract bounds on the LEP observables contributing to $g^h_{Z\textbf{p}}$ in Eq.~\ref{diru} by turning them  on one by one. We show the results  in Tab.~\ref{lepb}. For the TGCs $\delta g^Z_1$ and $\delta \kappa_\gamma$, our projections are much stronger than the LEP bounds and  in the case of the $Z$-pole observables $\delta g^Z_f$, that parametrise the deviations of the $Z$ coupling to quarks, they are comparable. 

\begin{figure}[!t]\centering
\includegraphics[width=0.45\textwidth]{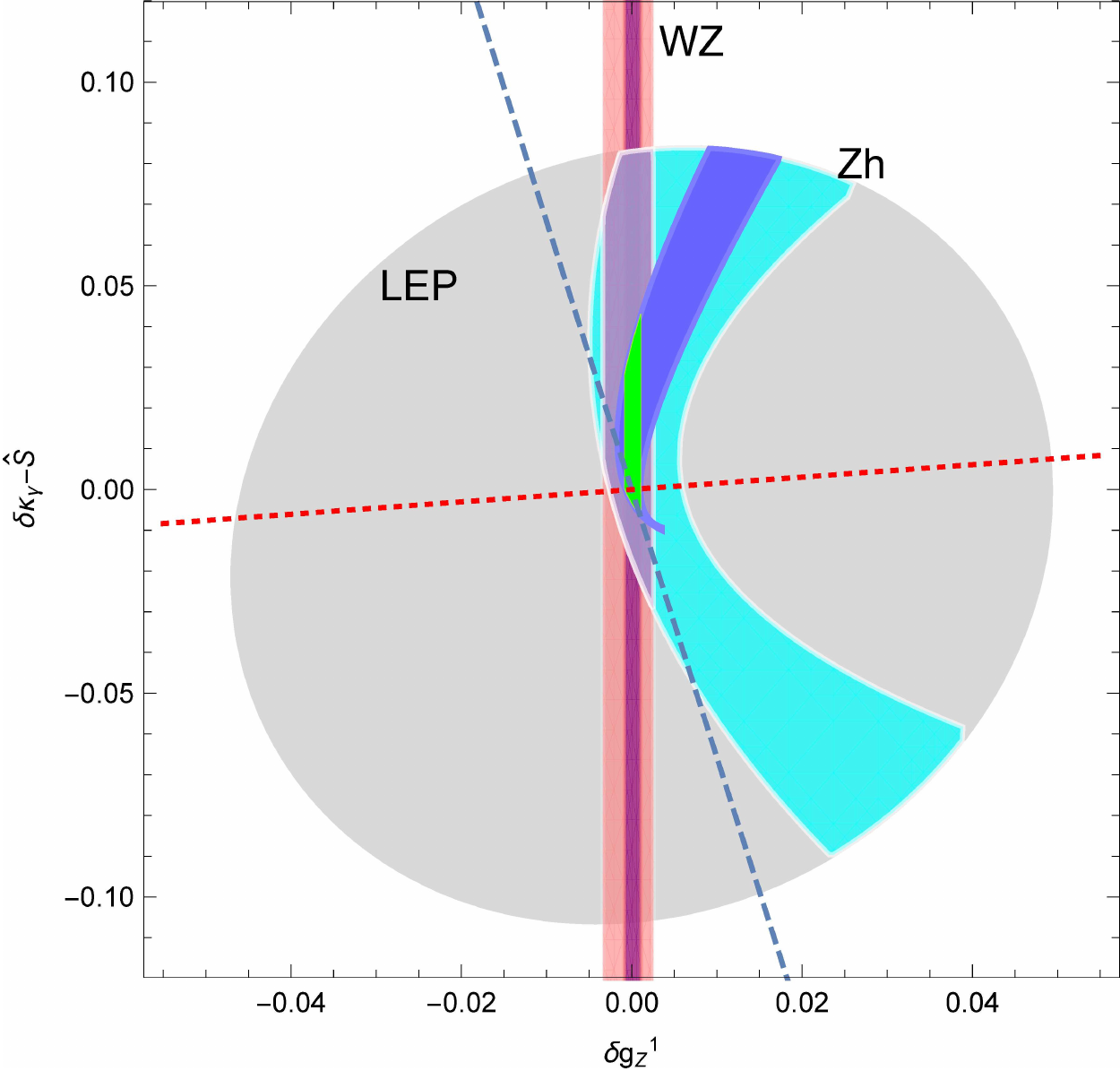}
\caption{The light blue (dark blue) region above shows the projection for the allowed region with 300 fb$^{-1}$ (3 ab$^{-1}$) data from the 
$pp \to Zh$ process  in the $\delta \kappa_\gamma-\hat{S}$ vs $\delta g^Z_1$ plane for universal models. We show in grey the allowed region after LEP bounds (taking the TGC $\lambda_\gamma=0$, a conservative choice) are imposed.  In pink (dark pink)  we show the region that corresponds to the projection from the $WZ$ process  with 300 fb$^{-1}$ (3 ab$^{-1}$) data derived in Ref.~\cite{Franceschini:2017xkh} and the purple (green) region shows the region that survives from a combination of the $Zh$ and $WZ$ projections with 300 fb$^{-1}$ (3 ab$^{-1}$) data.}\label{bounds}
\end{figure} 

For the universal case, we perform a more detailed analysis. The results are shown in the 
$\delta \kappa_\gamma-\hat{S}$ vs. $\delta g^Z_1$ plane  in Fig.~\ref{bounds} for the interesting class of models where $W=Y=0$~\cite{Franceschini:2017xkh}. The  
direction related to the $pp \to Zh$ interference term, \textit{i.e.}, $g^h_{Z\textbf{p}}=0$ (see Eq.~\ref{compdir} and the second line of Eq.~\ref{diru}) is shown by the dashed blue line, whereas the direction orthogonal to it is shown by the dotted red line.  Once the LEP II bounds~\cite{LEP2} from the $e^+e^- \to W^+W^-$ 
process are imposed, the allowed region that remains is shown by the grey shaded area. We show the results of this work  in blue (light (dark) blue for results at
300 (3000) fb$^{-1}$).  The shape of the allowed region arises due to the fact that the interference term vanishes along the dashed blue line and the squared term increases in magnitude as we move away from the origin. This curves the allowed region away from the dashed line as we move away from the origin.  The accidental cancellation of the interference term means that our bounds are susceptible to dimension-8 effects along this direction. On the other hand our bounds are more robust and not susceptible to such effects in the orthogonal direction shown by the red dotted line.

\begin{table}[t]
\begin{center}

\begin{tabular}{c|c|c}
&Our Projection &LEP Bound\\\hline
$\delta g^Z_{u_L}$         & $\pm0.002~(\pm0.0007)$ & $-0.0026\pm 0.0016$\\
$\delta g^Z_{d_L}$         & $\pm0.003~(\pm0.001)$  & $0.0023\pm 0.001$\\
$\delta g^Z_{u_R}$         & $\pm0.005~(\pm0.001)$  & $-0.0036\pm 0.0035$\\
$\delta g^Z_{d_R}$         & $\pm0.016~(\pm0.005)$  & $0.0016\pm 0.0052$\\
$\delta g^Z_1$             & $\pm0.005~(\pm0.001)$  & $0.009^{+0.043}_{-0.042}$\\
$\delta \kappa_\gamma$     & $\pm0.032~(\pm0.009)$  & $0.016^{+0.085}_{-0.096}$\\
$\hat{S}$                  & $\pm0.032~(\pm0.009)$ & $0.0004 \pm 0.0007$\\
$W$                        & $\pm0.003~(\pm0.001)$  & $0.0000 \pm 0.0006$\\
$Y$                        & $\pm0.032~(\pm0.009)$  & $0.0003\pm 0.0006$
 \end{tabular}
  \caption{Comparison of the bounds obtained in this work with existing LEP bounds obtained by turning on the LEP observables in Eq.~\ref{diru} one by one and using Eq.~\ref{obo}. The LEP bounds on the $Z$ coupling to quarks has been obtained from Ref.~\cite{Falkowski:2014tna}, the bound on the TGCs from Ref.~\cite{LEP2}, the bound on $\hat{S}$ from Ref.~\cite{Baak:2012kk} and  finally the bounds on $W,Y$ have been obtained from Ref.~\cite{Barbieri:2004qk}. Except for the case of the bounds on $\delta g^Z_f$, all of the bounds in the last column were derived by turning on only the given parameter and putting all other parameters  to zero. The numbers outside (inside) brackets, 
  in the second column, denote our bounds with $\mathcal{L} = 300 \; (3000)$ fb$^{-1}$.  \label{lepb}}
\end{center}
\end{table}

As $VV$ production constrains the same set of operators as the $Vh$ 
production in Fig.~\ref{bounds}, we also show the projected bound from the $WZ$ process at 300 
fb$^{-1}$ obtained in Ref.~\cite{Franceschini:2017xkh} (see section~\ref{WZlong}). Only the 
purple region remains when both these bounds are combined  at 300 fb$^{-1}$. This shrinks further to the green region  at 3000 fb$^{-1}$. A drastic reduction in the allowed LEP region is   thus possible by considering the $pp \to Zh$ at high energies.

\asubsection[A. Azatov, D. Barducci, J. Elias-Mir\'o,
  E. Venturini]{Novel measurements of anomalous triple gauge
  couplings} \label{sec:WZtrans} 

In this work we are interested in the measurement of the Standard Model (SM) Triple Gauge Couplings (TGCs). This is a classic test of the SM and a possible measurement of deviations from its expectations would signify an invaluable piece of information for the theory beyond the SM. A consistent way to parametrise such possible deviations is through the SM Effective Field Theory (EFT) approach.
We are going to consider the SM EFT as defined in \cite{Azatov:2017kzw,Azatov:2019xxn}, in particular we are going to focus on the measurement of the EFT operator $\mathcal{O}_{3W}$, see Table~\ref{tab:dim6ops}, which is associated to the the anomalous triple gauge coupling (aTGC) $\lambda_z$.

A precise determination of the TGC stems from the measurement of  the $2\to 2$  cross section $\sigma ( q\bar q \rightarrow VV )$~\cite{Sirunyan:2017bey,Aad:2016ett}. Naive dimensional analysis and standard EFT reasoning predicts that the energy scaling of such cross-section is given by
\bea
\label{eq:sigtt}
\begin{split}
\sigma(q\bar q \to V V)\sim \frac{g_{\text{SM}}^4}{E^2}\bigg[ 
 1 & +\overbrace{c_i\frac{E^2}{\Lambda^2}}^\text{BSM$_6\times\,$SM}
       +\overbrace{c_i^2 \frac{E^4}{\Lambda^4}}^\text{BSM$_6$$^2$}+ \dots \bigg]\, , 
\end{split} \label{genxsec}
\eea
where the first factor $g_\text{SM}^4/E^2$ accounts for the energy flux of the initial quarks, $c_i$ are the  relevant Wilson coefficients, and we have omitted numerical factors.
In (\ref{genxsec}) we have explicitly indicated dimension six squared  ($\text{BSM$_6$$^2$}$) and SM-dimension six interference terms ($\text{BSM$_6\times\,$SM}$).\footnote{Note that operators of dimension 7 necessarily violate either baryon or lepton number. We assume the scale of such symmetry violation to be very large and therefore irrelevant for di-boson physics at the LHC.}  The ellipses in (\ref{genxsec}) are due to corrections from operators of dimension $\geq 8$, which we will neglect. The leading  such term is an interference term of the type $\text{BSM$_8\times\,$SM}$ and it is of order $O(E^4/\Lambda^4)$.

A closer inspection however reveals that the  $2\to 2$ di-boson production through the dimension six operator $\mathcal{O}_{3W}$  has an interference piece with a suppressed energy scaling. 
Indeed, the energy scaling of such process is  
\begin{equation}
\sigma ( q\bar q \rightarrow VV ) \sim \frac{g_\text{SM}^4}{E^2}\bigg[  1 +  C_{3W}\frac{m_V^2}{\Lambda^2}  +C_{3W}^2 \frac{E^4}{\Lambda^4}  + O(E^4/\Lambda^4) \bigg]. \label{sup}
\end{equation} 
This is a consequence of the  helicity selection rules, see    
\cite{Dixon:1993xd,Azatov:2016sqh,Azatov:2017kzw,Panico:2017frx}.
The suppressed energy scaling can be problematic for the correct EFT  
interpretation of the $\sigma(q\bar q \to V V)$ measurement. 
Namely, in view of (\ref{sup}), the sensitivity on $C_{3W}$ is largely 
dominated by the quadratic piece $\text{BSM}_6^2$, which is 
$O(E^4/\Lambda^4)$. 
 Furthermore, in this case, the measurements become insensitive to the sign of the Wilson coefficient.
The main objective of the present work is to 
improve the sensitivity to the linear piece $\text{BSM$_6\times\,$SM}$.
We will present two classes of solutions to achieve this goal. Firstly, in section \ref{angs} we  will show that the differential angular cross-section 
 of the process $q\bar q \rightarrow VV\rightarrow 4\psi$ has a large sensitivity on $\text{BSM$_6\times\,$SM}$ compared to the inclusive cross-section.   Secondly, in section \ref{jetsol} we will show that accounting for extra radiation $q\bar q \rightarrow VV+j$ also results in an improved sensitivity on the leading piece $\text{BSM$_6\times\,$SM}$.
These measurements are specially interesting in a HL/HE phase of the LHC, for which we show the prospects below.


\begin{figure}[t]
\begin{center}
\includegraphics[scale=0.4]{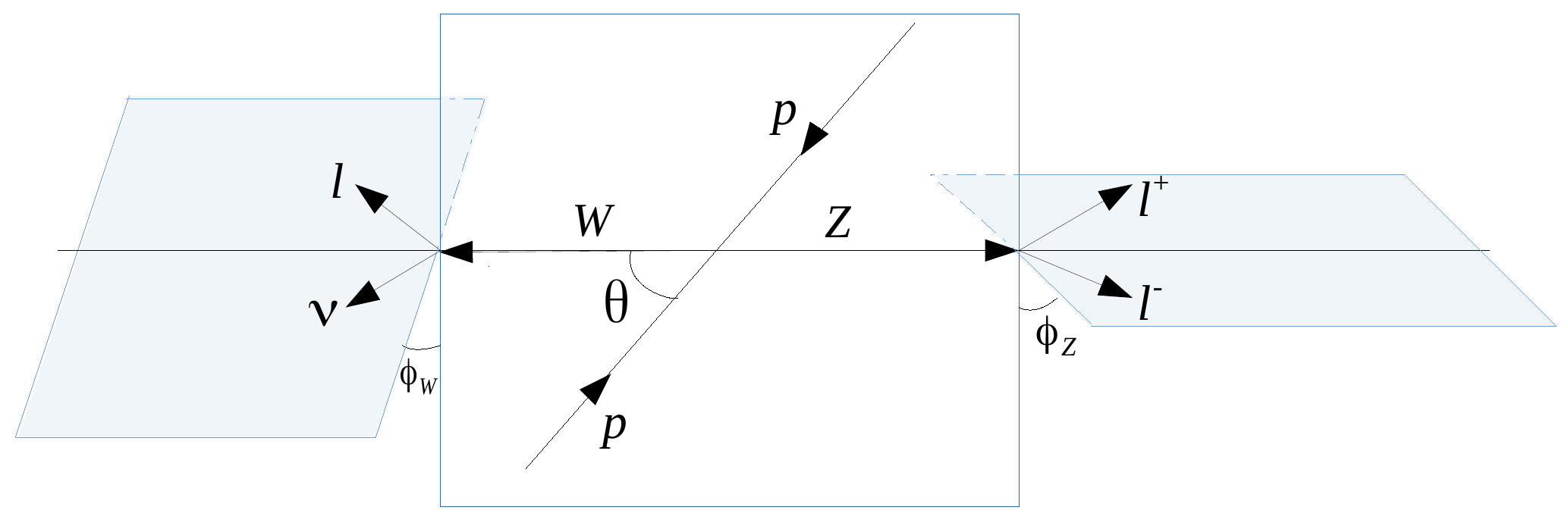}
\end{center}
\vspace{-.3cm}
\caption{Angles for $2\to 4$ scattering. \label{fig:ang}}
\end{figure}

Next we will present two ways to  improve the sensitivity to the aTGC $\lambda_z$ by restoring the energy growth
$g_\text{SM}^4/E^2\left[1+c_i E^2/\Lambda^2+  \dots\right]$ of the interference piece $\text{BSM$_6\times\,$SM}$ of the  $\mathcal{O}_{3W}$ operator.\\

\noindent
{\bf Interference resurrection via angular distributions}\\
\label{angs}
The first way of enhancing the interference term is by noting that in a collider experiment instead of the  $2\to 2$ process we actually measure a $2\to 4$ scattering, i.e. vector bosons decay into fermions $q\bar q\to V_1V_2\to 4 \psi$.  
Let us start by considering  the differential cross section for the production of the polarised particles $W_{T+} Z_T \rightarrow W_{T+}  l_+ \bar l_-$\footnote{ We ignore the  longitudinal Z polarisation which is sub-dominant at the LHC \cite{Baur:1994ia}.}
\be
\frac{d\sigma(q\bar q \rightarrow W_{T_+} l_- \bar l_+)}{d\text{LIPS} } =
 \frac{1}{2 s}  \frac{ \left|\sum_i  
(  {\cal M}^\text{SM}_{q\bar q \rightarrow W_{T_+}Z_{i}}+
 {\cal M}^\text{BSM}_{q\bar q \rightarrow W_{T_+}Z_{i}}
)
  {\cal M}_{Z_{i}\to  l_- \bar l_+}   \right|^2}{(k_Z^2-m_Z^2)^2+m_Z^2\Gamma_Z^2} 
\,  , \label{xsecan}
\ee
where   sum runs over intermediate Z polarisations and $d\text{LIPS}\equiv (2\pi)^4\delta^4(\sum p_i -p_f) \prod_i {d^3 p_i}/\left(2 E_i(2\pi)^3\right)$ is the Lorentz Invariant differential 
Phase Space.  
Then in the narrow width approximation the leading contribution to the interference, i.e. the cross term $\text{SM}\times\text{BSM}_6$ in \ref{xsecan},   is given by
$
d\sigma_\text{int}(q\bar q \rightarrow W_{T_+} l_- \bar l_+)/ d\phi_Z \propto E^2/\Lambda^2 \cos(2\phi_Z) 
$, 
where  $\phi_Z$  is the azimuthal angle between the plane defined by the decaying leptons and the plane defined by the collision and $WZ$ momenta, see Fig.~\ref{fig:ang}.  Note that $d\sigma_\text{int}(q\bar q \rightarrow W_{T_+} l_- \bar l_+)/ d\phi_Z $ has the energy growth expected from naive dimensional analysis, see  Eq.~\ref{genxsec}.
An analogous derivation goes through if we also consider the decay of the W gauge boson. The differential interference term   for the process $q\bar q\to $WZ$\to 4 \psi$ is  unsuppressed and modulated as
\bea
 \frac{d\sigma_\text{int}(q\bar q \rightarrow W Z\rightarrow  4\psi)}{d\phi_Z\, d\phi_W}  \propto \frac{E^2}{\Lambda^2}\left( \cos(2\phi_Z)+\cos(2\phi_W)\right) , \label{simp}
\eea
where $\phi_{W,Z}$ are the corresponding azimuthal angles. 
Integrating  \ref{simp} over the fermion phase space the   interference term vanishes as expected from the discussion above. 
Since the dependence on the two azimuthal angles is additive, integrating  over $\phi_W$ leads to a differential cross-section that is modulated by $\cos(2\phi_Z)$ and that features $E^2/\Lambda^2$ energy growth. 
We will use the result in Eq.~\ref{simp}  to prove the aTGC $\lambda_Z$, with an increased  overall sensitivity  to both the magnitude and sign of the Wilson coefficient.

 Following Ref.~\cite{Panico:2017frx}, we make a few remarks on the experimental measurement  of $\phi_{Z,W}$ in Eq.~\ref{simp}.
 The angle $\phi_Z$  can be determined up to an ambiguity  $\phi_Z\leftrightarrow \phi_Z \pm \pi$,
 since experimentally we can only measure the charges but not the helicities of the  leptons from $Z$  decay.
 The reconstruction of the $W$  azimuthal  angle $\phi_W$ in the $l\nu$ final state  suffers from an ambiguity  $\phi_W \leftrightarrow \pi-\phi_W$ due to the twofold ambiguity in the determination of the neutrino momentum.
Interestingly, none of these ambiguities  affects  Eq.~\ref{simp}.\\

\noindent
{\bf Interference resurrection via jet emission}\\
\label{jetsol}
A second way to resurrect the expected energy growth of the 
interference term is based on the observation that the helicity 
selection rule holds only at leading-order \cite{Azatov:2017kzw}. 
So the next-to-leading-order (NLO) effects will necessarily lead to the 
enhancement of the interference.
Virtual effects are expected to be suppressed by a factor ${\cal{O}}( 
\alpha_s/4\pi)$ with respect to the contributions coming from azimuthal modulation discussed in the previous section. A complete study at NLO accuracy for the operator ${\cal O}_{3W}$ together with its CP-odd counterpart can be found in~\cite{Azatov:2019xxn}.
Alternatively we will consider processes with an extra hard jet emission, which will improve on the  signal over the  background ratio. 
 In this case,  since   we are dealing with the hard
  $2\to 3$ process,
 the same polarisation configuration $q\bar{q}\to V_{\pm}V_{\pm}g_{\mp}$ is allowed both in SM and in the BSM five point amplitude with the $\mathcal{O}_{3W}$ insertion. Therefore the interference is not suppressed and the leading quadratic energy scaling is restored by requiring an extra (hard) QCD radiation.\\

\noindent
{\bf Results}\\
\emph{HL-LHC.} In order to test the sensitivity of the High-Luminosity (HL) phase of the LHC on the $\mathcal{O}_{3W}$ with the proposed solution to the non-interference behaviour we proceed in the following way. We generate with {\tt MadGraph5 aMC@NLO}~\cite{Alwall:2014hca} parton level events for $pp \to W^{\pm} Z $ decaying into a four leptons (electron and muon) final state together with events for the same process where we allow for a jet emission in the initial state. We perform two different analyses (see~\cite{Azatov:2017kzw} for more details): an inclusive one where we restrict to events up to $p_T^j<100\;$ \UGeV and do not bin on the $\phi_Z$ variable and an exclusive one where we bin both on the jet transverse momentum and on $\phi_Z$, where for the latter we define two bins with the threshold $|\cos(\phi_Z)|=1/\sqrt{2}$. All together the results for the bound on the $C_{3W}$ Wilson coefficient are reported in Fig.~\ref{fig:LHC13} as a function of the maximum transverse mass of the $WZ$ system, which allows to have an estimate of the validity of the EFT computation, see again~\cite{Azatov:2017kzw} for a detailed discussion~\footnote{These results are obtained by keeping both the linear and the quadratic terms in the cross section determination.}.

One might wonder if a simulation beyond the parton level accuracy might spoil these results. To this end we have performed a more detailed simulation by showering the events through {\tt PYTHIA 8}~\cite{Alwall:2014hca} and simulating the detector response via {\tt Delphes 3}~\cite{deFavereau:2013fsa}. By analysing the density of events in the two azimuthal bins we found that with respect to the parton level case the relative difference is of at most a few \%, thus making our parton level analysis solid.

  \begin{figure}[ht]
\begin{center}
 \includegraphics[width=0.39\textwidth]{\main/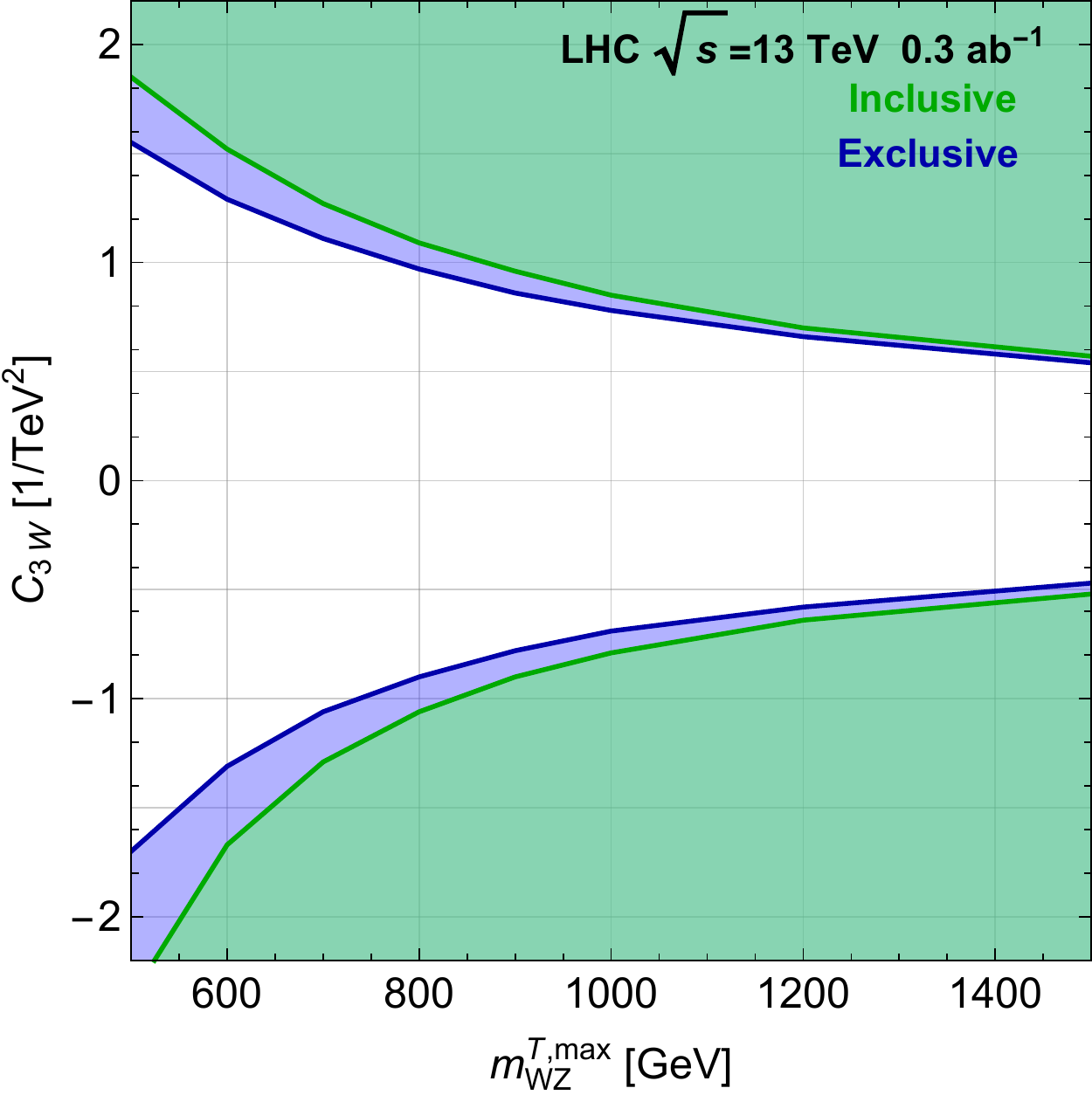}{}\hspace{2cm}
 \includegraphics[width=0.39\textwidth]{\main/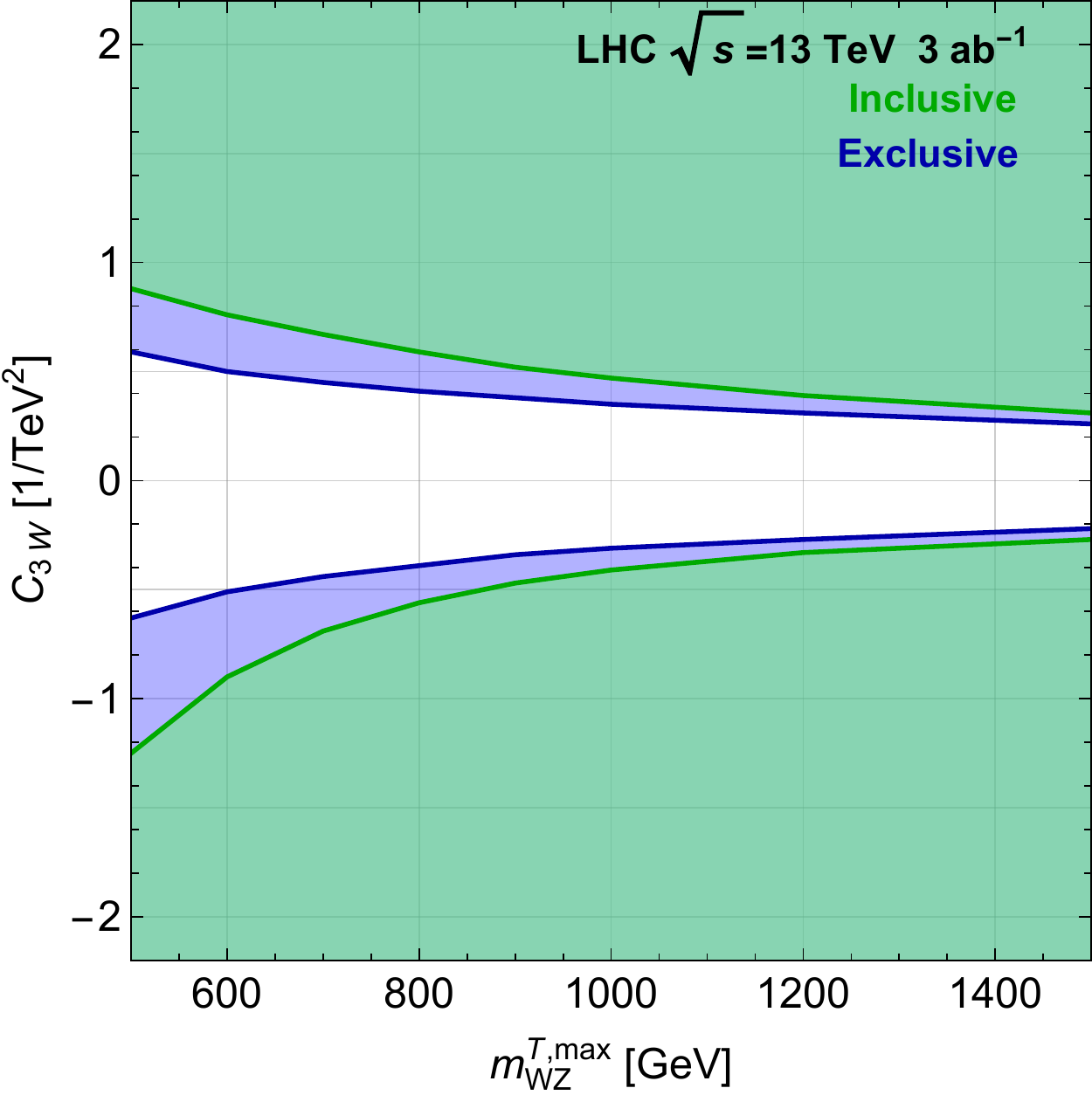}{}
\end{center}
\caption{Bounds on the $C_{3W}$ Wilson coefficient for the inclusive and exclusive categories at the LHC 13 for 300 fb$^{-1}$ (left) and 3000 fb$^{-1}$ (right) of integrated luminosity.}
\label{fig:LHC13}
\end{figure}

\noindent
\emph{HE-LHC.}
We now estimate the reach of a future HE phase of the LHC with
$\sqrt{s}=27\;$\UTeV. For these preliminary results we adopt the same binning, both in $\phi_Z$ and in jet transverse momentum, of the previous section. We show the results in Fig.~\ref{fig:LHC27}. We found a slight increase of order $~30\%$ on the reach on $C_{3W}$. We expect that a dedicated HE analysis will lead to a further improvement of these bounds; this can be done by exploiting in a more efficient way the high energy tails of the differential distributions. In~\cite{Azatov:2019xxn} we also present a complete NLO study for the HE-LHC stage.

  \begin{figure}[ht]
\begin{center}
 \includegraphics[width=0.39\textwidth]{\main/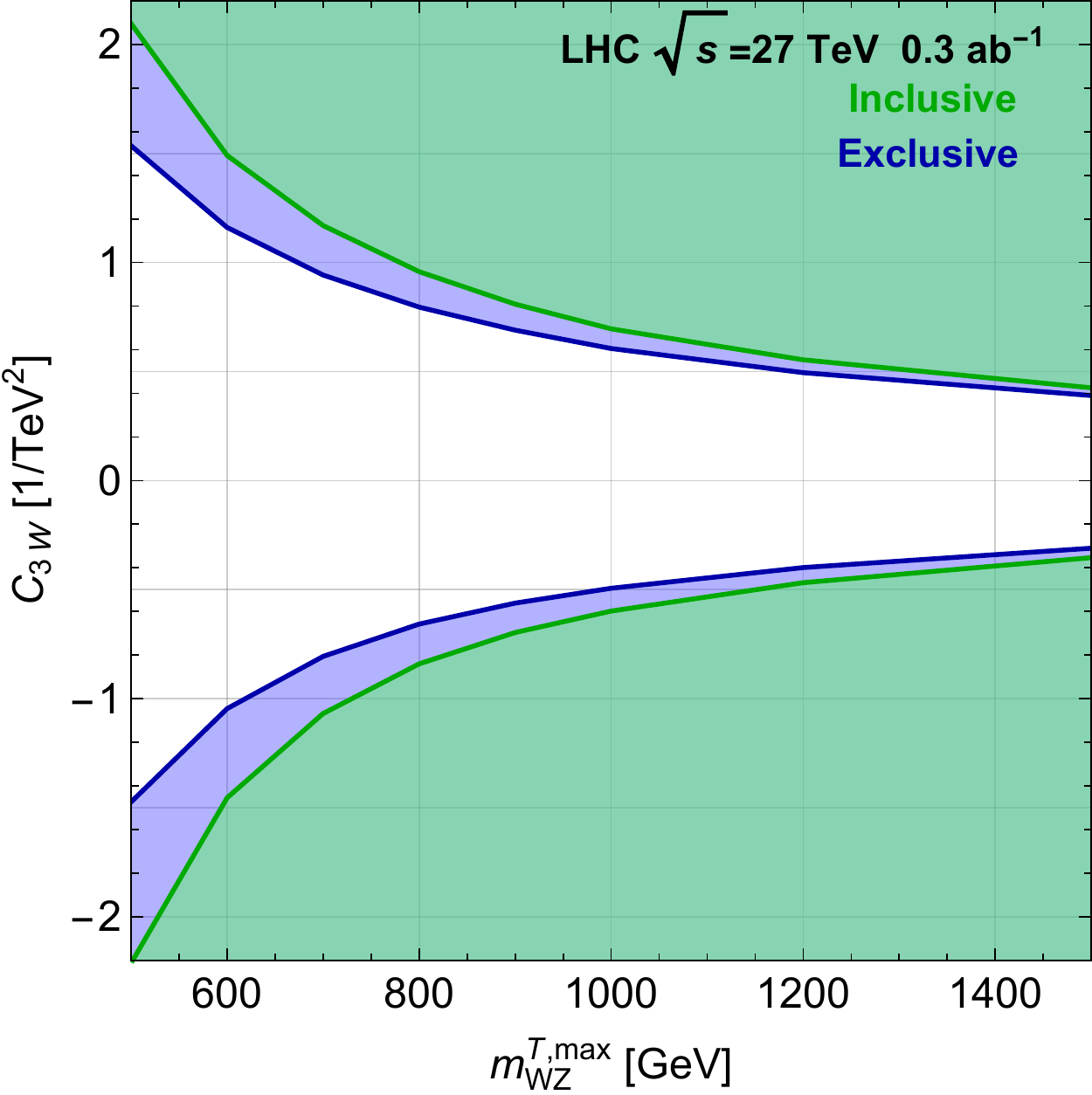}{}\hspace{2cm}
 \includegraphics[width=0.39\textwidth]{\main/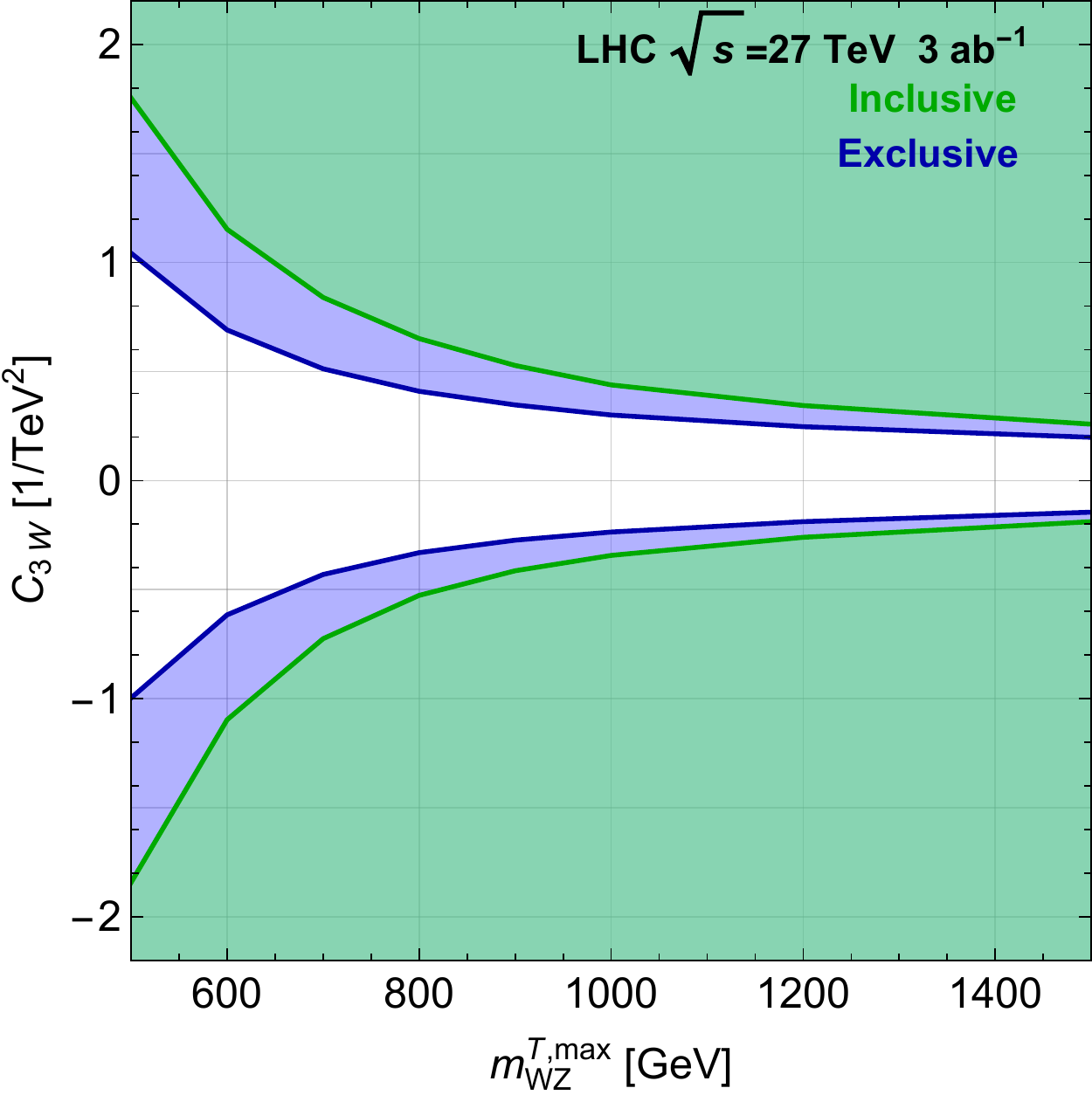}{}
\end{center}
\caption{Bounds on the $C_{3W}$ Wilson coefficient for the inclusive and exclusive categories at the LHC 27 for 300 fb$^{-1}$ (left) and 3000 fb$^{-1}$ (right) of integrated luminosity.}
\label{fig:LHC27}
\end{figure}

\asubsection[S. Alioli, M. Farina, G. Panico, D. Pappadopulo,
  J.T. Ruderman, R. Torre, A. Wulzer]{Electroweak Precision Tests in
  High-Energy Drell-Yan Processes} \label{DY}
The simplest process allowing to set constraints on EFT operators is Drell-Yan di-lepton and lepton-neutrino production at high invariant mass, for concreteness we focus on oblique corrections only, generalisations being rather obvious. These corrections can be parametrised in the electroweak sector by the four oblique parameters $S$, $T$, $W$ and $Y$. These correspond to four operators that modify the propagators of the $W$ and $Z$ bosons both on the pole ($S$ and $T$) and off the pole, i.e.~on the tails ($W$ and $Y$). Hadron colliders can hardly compete with lepton colliders for pole observable. However, due to the enhancement of the kinematic distributions with respect to the corresponding SM ones at high energy, hadron colliders are particularly suited to study off-pole observables like $W$ and $Y$. Deviations from the SM proportional to $W$ and $Y$ can be parametrised through the two operators from table~\ref{tab:dim6ops},
\begin{equation}
 -\frac{W}{2m_W^2}\mathcal{O}_{2W}   \,, \quad  -\frac{Y}{2m_W^2}\mathcal{O}_{2W}   
 \end{equation}
They modify the neutral and charged gauge boson propagators as
\begin{equation}\label{eq:propagators}
\begin{array}{ll}
& P_{\hspace{-1pt}{N}}\hspace{-1pt} = \hspace{-2.5pt}\left[\hspace{-2pt}
\begin{array}{cc}
 \frac{1}{q^2}-\frac{t^{2}{\rm{\sc{W}}}+{\rm{\sc{Y}}}}{m_Z^2} & \frac{t \left(\left({\rm{\sc{Y}}}+{\rm\hat{T}}\right) c^2+s^2 {\rm{\sc{W}}}-{\rm\hat{S}}\right)}{\left(c^2-s^2\right) \left(q^2-m_Z^2\right)}+\frac{t ({\rm{\sc{Y}}}-{\rm{\sc{W}}})}{m_Z^2} \\
 \star & \frac{1+ {\rm\hat{T}}-{\rm{\sc{W}}}-t^{2}{\rm{\sc{Y}}}}{q^2-m_Z^2}-\frac{t^{2}{\rm{\sc{Y}}}+{\rm{\sc{W}}}}{m_Z^2} \\
\end{array}
\hspace{-2pt}
\right] \\
& P_{\hspace{-1pt}{C}}\hspace{-1pt} =
\begin{array}{c}
\frac{1+\left( \left({\rm\hat{T}}-{\rm{\sc{W}}}-t^2 {\rm{\sc{Y}}}\right)-2 t^2 \left({\rm\hat{S}}-{\rm{\sc{W}}}-{\rm{\sc{Y}}}\right)\right)/(1-t^2)}{(q^{2}-m_W^2)}-\frac{{\rm{\sc{W}}}}{m_W^2}\,,
\end{array}
\end{array}
\end{equation}
Studying the tails of the invariant mass distribution of two leptons and of the transverse mass of lepton-neutrino, one can set constraints on these observables. For details on the procedure see ref.~\cite{Farina:2016rws}, also extended to di-jet and multi-jet analyses in ref.s~\cite{Alioli:2017nzr,Alioli:2017jdo}. The prospect results for the HL-LHC and HE-LHC are shown in fig.~\ref{fig:DY-WY}
\begin{figure}[ht]
\centering
\includegraphics[width=0.49 \textwidth ]{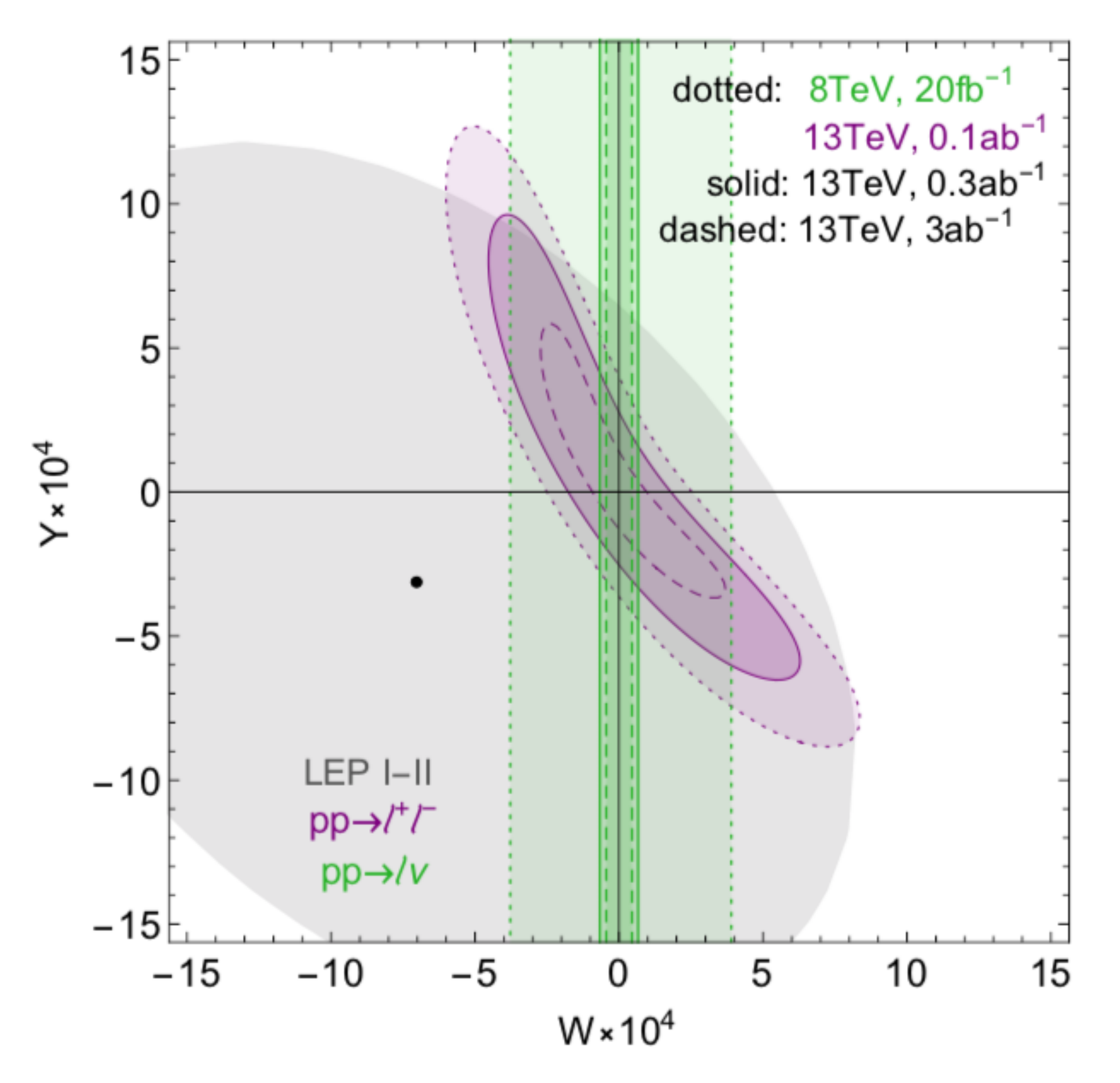}
\includegraphics[width=0.468 \textwidth ]{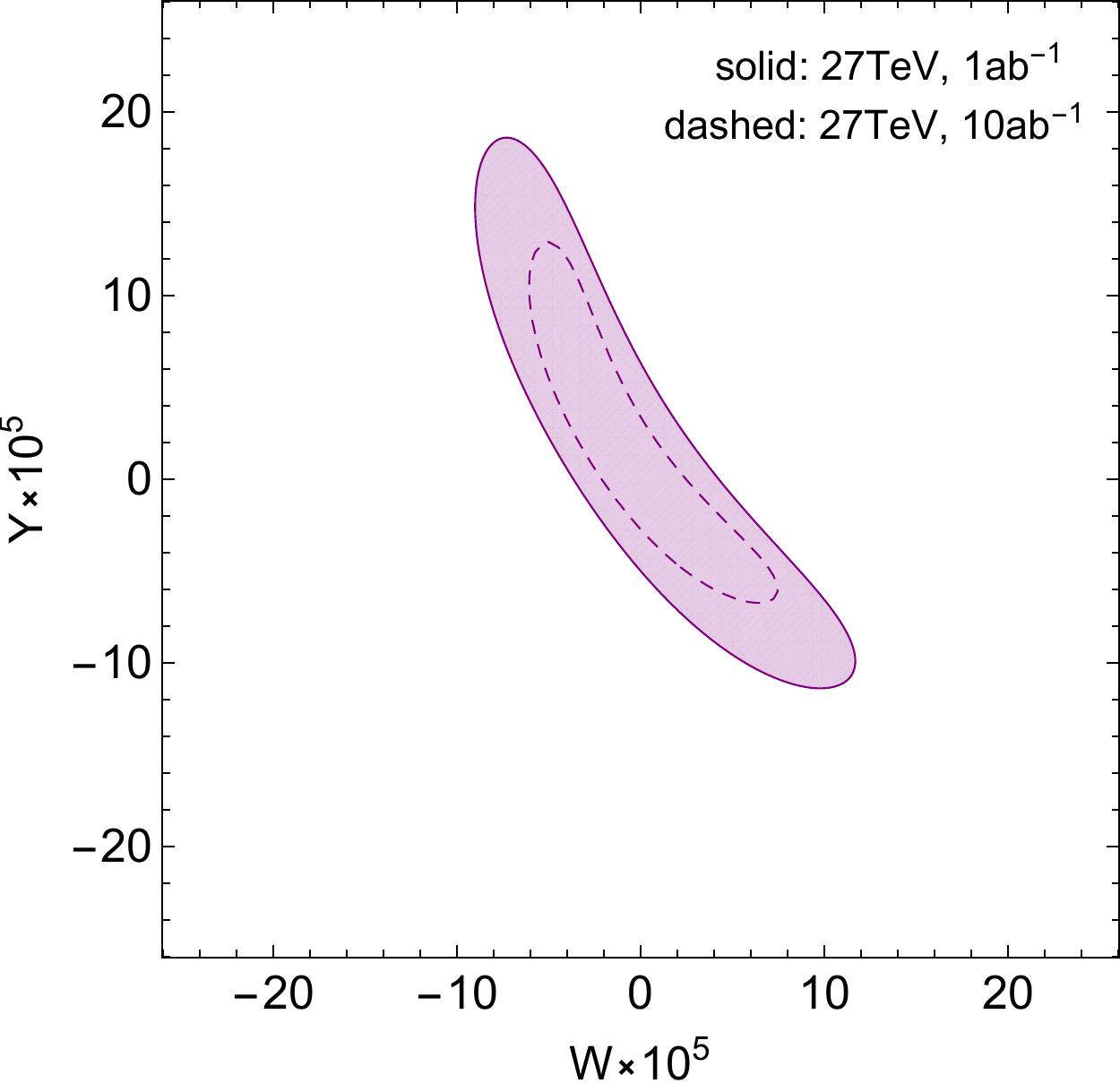}
\caption{Left: LHC and HL-LHC. Right: HE-LHC}
\label{fig:DY-WY}
\end{figure}

\asubsection[D. Liu, I. Low, Z. Yin]{Testing the universal Higgs non-linearity} 
\label{sect-illus}

In this section we motivate precision measurements on the tensor structures of one Higgs couplings with two electroweak gauge bosons (HVV) and two Higgses couplings with two electroweak gauge bosons (HHVV) in HE/HL LHC. There exist special relations between HVV and HHVV couplings in composite Higgs models that are universal, independent of the  symmetry breaking pattern invoked in a particular model. These "universal relations" are controlled by a single input parameter, the decay constant $f$ of the pseudo-Nambu-Goldstone Higgs boson. Testing the universal relations requires measuring the tensor structures of HVV and HHVV couplings to high precision.  In particular,  HHVV interactions remains as one of the few untested predictions of the Standard Model Higgs boson, which can be probed through the double Higgs production in the vector boson fusion (VBF) channel at the LHC. Below we summarise the main results. The phenomenological details and theoretical foundation can be found in Refs.~\cite{Low:2014nga,Low:2014oga,Liu:2018vel,Liu:2018qtb}


At the leading two-derivative order, the HVV and HHVV couplings in composite Higgs models in the unitary gauge is given by the following simple expression:
\begin{equation}
\label{eq:twod}
{\cal L}^{(2)} = \frac12 \partial_\mu h\partial^\mu h+\frac{g^2f^2}{4}\sin^2\left(\theta+h/f\right)\left(W_\mu^+W^{-\, \mu}+\frac1{2\cos^2\theta_W}Z_\mu Z^\mu\right)\ ,
\end{equation}
where  $v=246$ \UGeV, $f$ is the decay constant of the composite Higgs boson and $\sin\theta=v/f$. This result is independent of the symmetry breaking pattern of the strong composite sector in the UV, apart from the overall normalisation of $f$, which does depend on the UV model.

At the four-derivative level, we parametrise the HVV and HHVV couplings  as follows:
\begin{equation}
\label{eq:Cis}
{\cal L}^{(4)} =  \sum_i\frac{m_W^2}{m_\rho^2} \left(C^h_i {\cal I}^h_i  +C^{2h}_i {\cal I}^{2h}_i\right) ,
\end{equation}
where the definition of the  operators ${\cal I}^h_i$ and ${\cal I}^{2h}_i$ are presented in Table~\ref{tab:oneh} and Table~\ref{tab:twoh}. On the other hand, $C^h_i$ and $C^{2h}_i$ are Wilson coefficients which depend on six unknowns ($\theta, c_3, c_4^\pm, c_5^\pm$) in composite Higgs models and on four unknowns ($c_W, c_B, c_{HW}, c_{HB}$)  in the Standard Model Effective Field Theory (SMEFT). In the above $m_\rho=g_\rho f$ is the typical mass scale of the new composite resonances. The different Lorentz structures lead to different angular distributions in the decay products and, therefore, can be measured accordingly. At the LHC Run 1,  testing the tensor structure of HVV couplings was among the top priorities and gave confidence to the Higgs nature of the 125 \UGeV resonance. (See, for example, Ref.~\cite{Sirunyan:2017tqd}.) A similar program for HHVV coupling is currently lacking and should be pursued at HE/HL LHC.

\begin{table}[!t]
\begin{center}
\caption{Single Higgs coupling coefficients $C_i^h$ for the non-linearity case (NL) and the  purely dimension-6 contributions (D6) in SMEFT. Here  $c_w, t_w$ and $c_\theta$ denote $\cos\theta_W, \tan\theta_W$ and $\cos\theta$, respectively, where $\theta_W$ is the weak mixing angle. ${\cal D}^{\mu\nu}$ denotes $\partial^\mu \partial^\nu - \eta^{\mu\nu} \partial^2$. Hermitian conjugate of an operator is implied when necessary.}\label{tab:oneh}
\begin{tabular}{lc c c c ccc}\hline\hline
${\cal I}^{h}_i$ &   $  C^h_i$ (NL)& $  C^h_i$ (D6)\\
 \hline
(1) $  \frac{h}{v} Z_{\mu} {\mathcal{D}}^{\mu \nu} Z_{\nu}$ & $ \frac{4 c_{ 2 w} }{c^2_{w}}  (-2 c_3  + c_4^- )  
+\frac{4}{c^2_{w}}   c_4^+ c_\theta $ &$ 2 (c_W + c_{HW}) +2 t^2 _{w} (c_B + c_{HB})$\\
(2)  $ \frac{h}{v}  Z_{\mu\nu} Z^{\mu\nu}$ & $  -\frac{2  c_{2 w}}{c^2_{w}}  ( c_4^-  + 2 c_5^- ) -\frac{2 }{c^2_{w}}   ( c_4^+  - 2 c_5^+ ) c_\theta    $&$-( c_{HW} + t^2_{w} c_{HB})$\\
(3)  $ \frac{h}{v}  Z_{\mu} {\mathcal{D}}^{\mu \nu} A_{\nu}$ & $8  ( - 2 c_3  +   c_4^- )  t_{ w} $& $ 2 t_w (c_W + c_{HW}-c_B - c_{HB})  $ \\
(4) $ \frac{h}{v}  Z_{\mu\nu} A^{\mu\nu}$ & $-4  (  c_4^-   +2 c_5^- )  t_{ w} $&  $ -  t_w(c_{HW}- c_{HB})$\\
(5)  $ \frac{h}{v}  W^+_{\mu} {\mathcal{D}}^{\mu \nu} W^-_{\nu}$ & $4 (-2 c_3  + c_4^- ) 
+  4 c_4^+ c_\theta $ & $2 (c_W + c_{HW})$\\
(6) $ \frac{h}{v}  W^+_{\mu\nu} W^{-\mu\nu} $ & $ -4( c_4^- + 2 c_5^-)   -4 ( c_4^+  - 2 c_5^+) c_\theta $&   $-2 c_{HW} $ \\
\hline\hline
\end{tabular}
\end{center}
\end{table}

\begin{table}[!h]
\begin{center}
\caption{The coupling coefficients $C_i^{2h}$ involve two Higgs bosons  for universal  non-linearity case (NL) and the  dimension-six case in SMEFT (D6). A cross ($\times$) means there is no contribution at the order we considered. Notice $C_i^{2h}=C_i^h/2$ for  SMEFT at the dimension-6 level. $c_{2\theta}$ and $s_\theta$ denote $\cos 2\theta$ and $\sin\theta$, respectively.}\label{tab:twoh}
\begin{tabular}{c c c c c}
\hline\hline
${\cal I}^{ 2 h}_i$ &  $ C^{ 2h}_i$ (NL) &  $   C^{2h}_i$ (D6) \\
 \hline
(1) $  \frac{h^2}{v^2} Z_{\mu} {\mathcal{D}}^{\mu \nu} Z_{\nu}$ & $ \frac{2c_{ 2 w} }{c^2_{w}} ( - 2 c_3  +  c_4^-  ) c_\theta   +   \frac{2 }{c^2_{w}} c_4^+ c_{2 \theta}  $ & $\frac12 C_1^h$ 
 \\
(2) $  \frac{h^2}{v^2} Z_{\mu\nu} Z^{\mu\nu}$ & $ - \frac{c_{2 w}}{c^2_{w}}  (c_4^-  + 2 c_5^-  ) c_\theta   - \frac{1}{c^2_{w}}  (c_4^+ - 2 c_5^+ ) c_{2 \theta}   $ &$\frac12 C_2^h$ \\
(3) $  \frac{h^2}{v^2} Z_{\mu} {\mathcal{D}}^{\mu \nu} A_{\nu}$ & $ 4  t_{ w}  ( -2c_3   + c_4^- ) c_\theta $ & $\frac12 C_3^h$ \\
(4) $  \frac{h^2}{v^2} Z_{\mu\nu} A^{\mu\nu}$ & $ -2  t_{ w}   (   c_4^-  + 2 c_5^-  )  c_\theta $ &   $\frac12 C_4^h$  \\
(5)  $ \frac{h^2}{v^2} W^+_{\mu} {\mathcal{D}}^{\mu \nu} W^-_{\nu}$ & $ 2  (- 2c_3 + c_4^- ) c_\theta  + 2 c_4^+ c_{2 \theta}   $ & $\frac12 C_5^h$ \\
(6) $ \frac{h^2}{v^2} W^+_{\mu\nu} W^{-\mu\nu}$ & $  -2 ( c_4^- + 2 c_5^- )c_\theta   -2  (c_4^+  - 2 c_5^+ ) c_{2\theta}  $ &  $\frac12 C_6^h$\\
(7) $      \frac{(\partial_\nu h)^2} {v^2}Z_\mu Z^{ \mu}  $& $\frac{8 }{c^2_{w}}  c_1 s^2_\theta$ &  $\times$\\
(8)  $   \frac{\partial_\mu h \partial_\nu h}{v^2} \, Z^\mu Z^\nu  $& $\frac{8 }{c^2_{w}}  c_2  s^2_\theta$&  $\times$\\
(9) $      \frac{(\partial_\nu h)^2}{v^2} W^+_\mu W^{- \mu}  $& $16 c_1 s^2_\theta$ & $\times$ \\
(10) $  \frac{ \partial^\mu h \partial^\nu h }{v^2}\, W^+_{\mu} W^{-}_\nu  $& $16  c_2  s^2_\theta$ &  $\times$\\
\hline\hline
\end{tabular}
\end{center}
\end{table}

In general, we have two different Lorentz structure in the HVV couplings:
\begin{equation}
\label{eq:lstruc1}
\frac{h}{v} V_{1\,\mu} \mathcal{D}^{\mu\nu} V_{2\,\nu} \ ,\qquad \frac{h}{v} V_{1\,\mu\nu}  V_2^{\mu\nu} \ ,
\end{equation}
where  ${\mathcal{D}}^{\mu\nu} = \partial^\mu\partial^\nu-\eta^{\mu\nu}\partial^2$ and $V_{1,2} \in\{W, Z,\gamma\}$ with electric charge conservation implicitly indicated. For HHVV couplings we have: 
\begin{equation}
\frac{h^2}{v^2} V_{1\,\mu} {\mathcal{D}}^{\mu\nu} V_{2\,\nu}\ , \quad  \frac{h^2}{v^2} V_{1\,\mu\nu}  V_2^{ \mu\nu} \ , \qquad \frac{\partial_\mu h \partial_\nu h}{v^2} V_1^\mu V_2^{\nu}, \qquad \frac{\partial_\mu h \partial^\mu h}{v^2} V_1^\mu V_{2\mu} \ .
\label{eq:lstruc2}
\end{equation}
The ultimate goal then will be to measure these different tensor structures at CLIC.

From Table~\ref{tab:oneh} and Table~\ref{tab:twoh}, we can extract relations among $C_i^{h}$ and $C_i^{2h}$ that only depend on the $\theta$. We call them "universal relations" as they represent universal predictions of a composite Higgs boson, whose nonlinear interactions are dictated by the underlying shift symmetries acting on the four components of the Higgs doublet \cite{Low:2014nga,Low:2014oga,Liu:2018vel,Liu:2018qtb}.  Some examples of universal relations involving both HVV and HHVV couplings are:
\begin{eqnarray}
\label{eq:ident3}
&& \frac{C^{2h}_3}{C^h_3} = \frac{C^{2h}_4}{C^h_4} =\frac12 \cos\theta  \ ,\\
\label{eq:ident4}
&& \frac{C^{2h}_5 - C^{2h}_3 /2t_w} {C^h_5 - C^{h}_3/2t_{w}} =  \frac{C^{2h}_6 -C^{2h}_4/t_{w} }{C^h_6 - C^{h}_4/t_{w}} = \frac{\cos2\theta}{2\cos\theta} \approx \frac12\left(1 - \frac32 \xi \right)\, ,\\
&&\frac{s_{2w} \,C^{2h}_1 - c_{2w}\,   C^{2h}_3} {s_{2w}\, C^h_1-c_{2w} \,  C^{h}_3} =\frac{s_{2w}\, C^{2h}_2 - c_{2w}\,   C^{2h}_4} {s_{2w}\, C^h_2-c_{2w}\,   C^{h}_4}  = \frac{\cos2\theta}{2\cos\theta}\approx \frac12\left(1 - \frac32 \xi \right)\ .
\end{eqnarray}
These relations depend on one single parameter $\theta$ or, equivalently, $\xi=v^2/f^2$. In other words, they can be used to over-constrain the parameter $f$. If the 125 \UGeV Higgs boson indeed arises as a pseudo-Nambu-Goldstone boson, the decay constant $f$ as measured from the different universal relations must be consistent with one another.


In order to test the universal relations, it is necessary to measure the tensor structures of HHVV couplings. This is where the HE/HL LHC could have an advantage over circular lepton colliders. At a hadron collider, $C^h_i$ can be measured from single Higgs decays into four leptons in a fashion similar to the analysis performed in Ref.~\cite{Sirunyan:2017tqd}, while measurements on $C^{2h}_i$ would have to rely on double Higgs production in the VBF  channel and the associated production with a $Z$ boson. The production topology is displayed in Fig.~\ref{fig:testhh}.

\begin{figure}[!t]
\centering
    \subfloat[Double Higgs production through vector boson fusion.]{\includegraphics[height=3.3cm]{\main/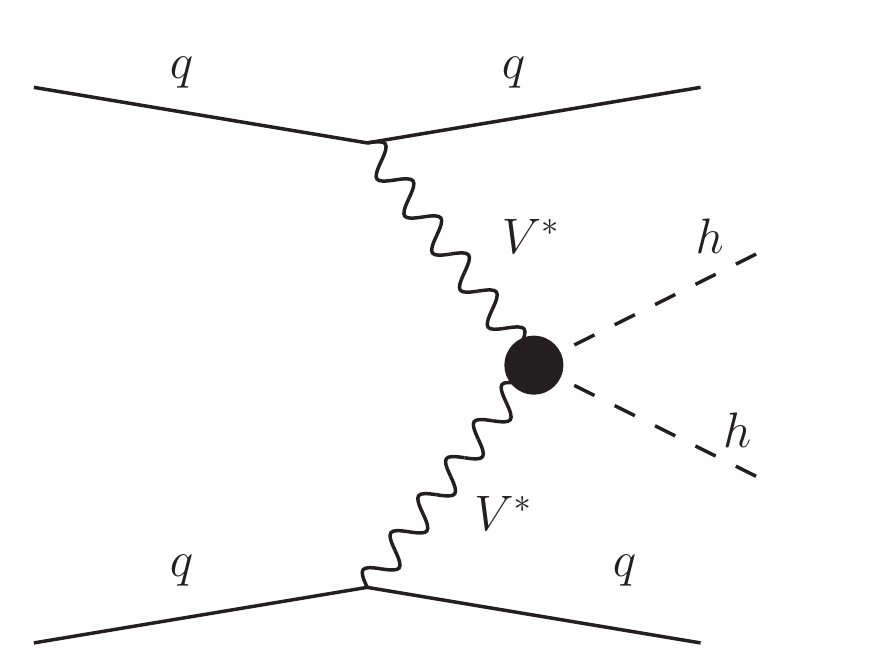}}
  \quad \ \ 
  \subfloat[Double Higgs production in association with a vector boson.]{\includegraphics[height=2.8cm]{\main/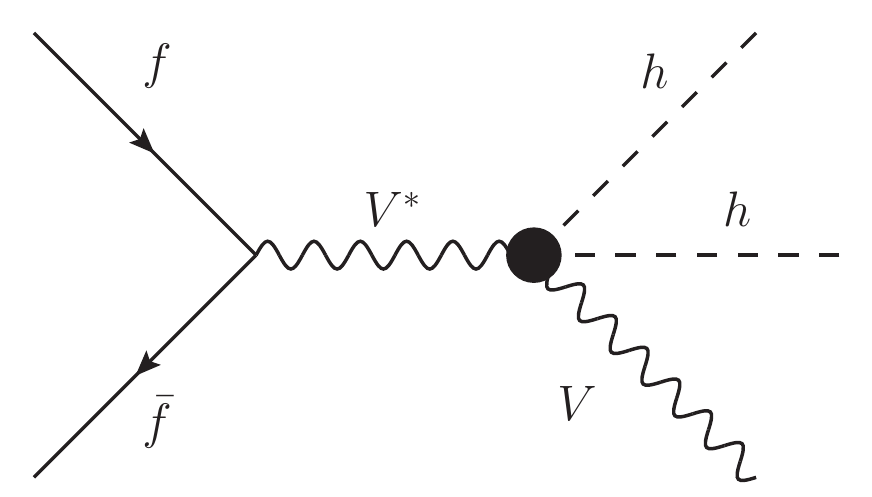}}
  \caption{Production and decay topology of venues to test the HHVV couplings at the LHC. A black dot represents contributions from various Feynman diagrams. \label{fig:testhh}}
\end{figure}

In Fig.~\ref{fig:xshh}  we show the double Higgs production rate in the VBF channel and the associated production channel in a hadron collider as a function of the centre-of-mass energy $\sqrt{S}$, adopting the computation done in Ref.~\cite{Frederix:2014hta}. The VBF rate at 13 \UTeV is less than 1 fb, while at 27 \UTeV the cross-section is about 3 fb, which offers the best chance to probe the HHVV couplings at the LHC. 

\begin{figure}[!t]
\centering
   \includegraphics[width=9cm]{\main/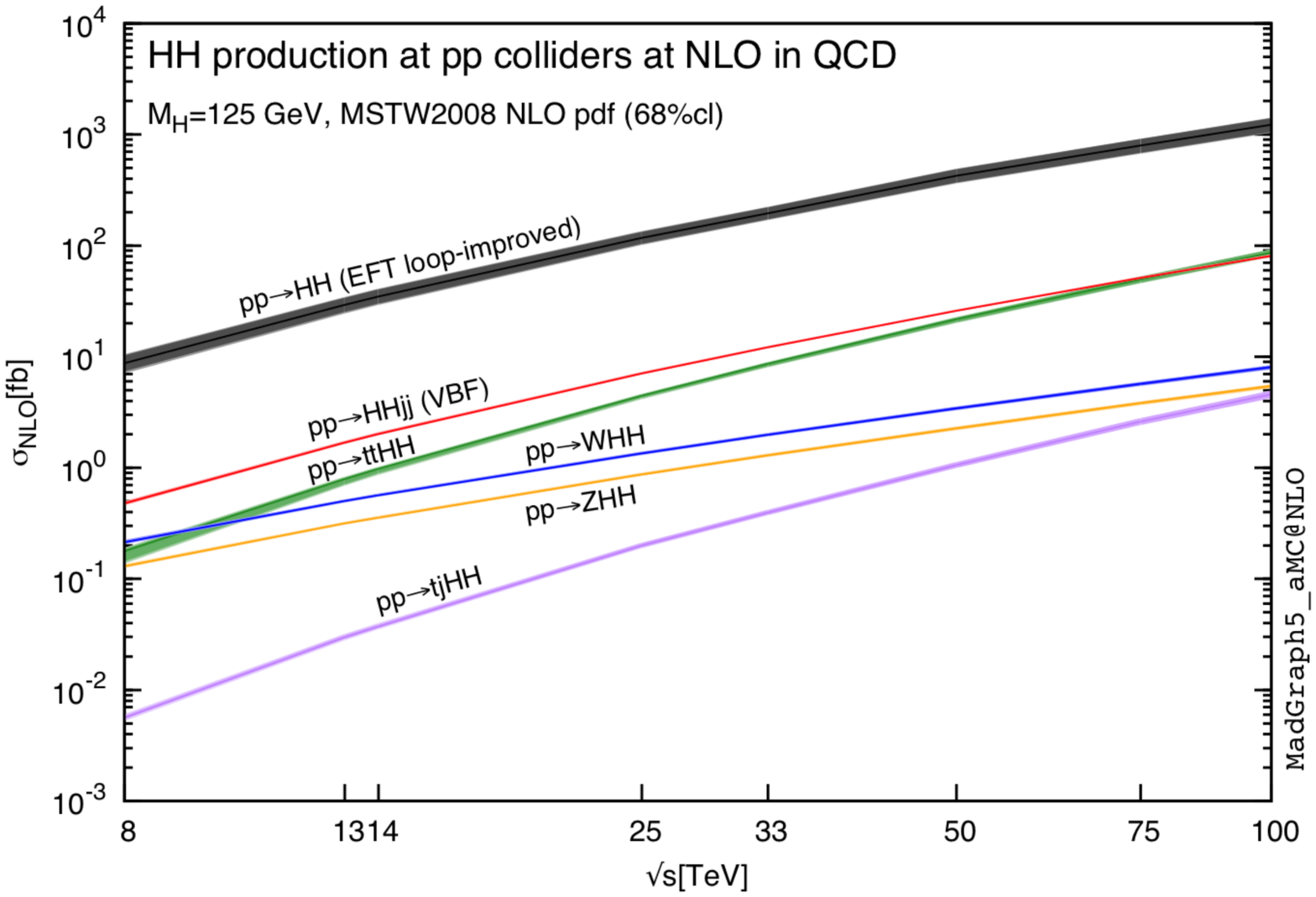}
  \caption{Double Higgs production rates, including the VBF  and associate production channels, at a hadron collider. This figure is adopted from Ref.~\cite{Frederix:2014hta}. \label{fig:xshh}}
\end{figure}

The analysis of the HHVV coupling is further discussed in the next section \ref{sec:VVHHcont}.

\asubsection[F. Bishara, R. Contino, J. Rojo]{Higgs pair production in vector-boson fusion at the HL-LHC}
\label{sec:VVHHcont}

While the dominant production channel of Higgs boson pairs at hadron colliders is the gluon-fusion mechanism, other channels are also of phenomenological relevance. In particular, Higgs pair production in weak vector-boson fusion~\cite{Bishara:2016kjn} is interesting since it probes the
strength of the  Higgs non-linear interactions with vector bosons at high
energies. This process can therefore provide unique information
to test the nature of the Higgs boson,
whether it is a composite or elementary state, and whether or not it emerges as
a Nambu-Goldstone boson (NGB) of some new dynamics at the \UTeV
scale~\cite{Giudice:2007fh,Contino:2010mh,Contino:2013gna}.

The production of Higgs pairs in the  VBF
channel~\cite{Giudice:2007fh,Contino:2010mh,Dolan:2013rja,Brooijmans:2014eja,Liu-Sheng:2014gxa,Dolan:2015zja} proceeds via the soft emission of two vector bosons from the incoming protons  followed by 
the hard $VV \to hh$ scattering, with $V=W,Z$.
In the SM, the VBF inclusive cross section
at 14 \UTeV is around 2 fb (see Fig.~\ref{fig:xshh}), more than one
order of magnitude smaller than in gluon fusion.
Higher order QCD corrections are moderate ($\sim 10\%$) as
expected for an electroweak process.
Despite the small rate, Higgs pair production via VBF is relevant since
even small modifications of the SM couplings induce a striking increase of the
cross section as a function of the di-Higgs invariant mass, for instance in models 
where the Higgs is a composite pseudo-NGB (pNGB) of new strong dynamics at the \UTeV scale~\cite{Kaplan:1983fs}.
In these theories, the Higgs anomalous couplings imply a growth of the $VV\to hh$ cross section with the
partonic centre-of-mass energy, $\hat{\sigma} \propto \hat s/f^4$, where $f$ is the pNGB decay constant~\cite{Giudice:2007fh}.
This enhanced sensitivity to the underlying strength of the Higgs interactions makes 
double Higgs production via VBF a key process to test the nature of the
electroweak symmetry breaking dynamics
and to constrain the $hhVV$ quartic coupling in a model-independent way.

Here we review the feasibility of measuring and interpreting
the VBF Higgs pair production at the HL-LHC in
the $hh\to b\bar{b}b\bar{b}$ final state.
While QCD multi-jet backgrounds are huge, this final state turns out
to be within the reach of the HL-LHC thanks to the unique VBF topology,
characterised by two forward jets well separated in rapidity and with a large
invariant mass and a reduced hadronic activity in the central region. 
In addition, the di-Higgs system will acquire a substantial boost
in the presence of BSM dynamics, and jet
substructure techniques~\cite{Salam:2009jx,Gouzevitch:2013qca,Behr:2015oqq} 
make it possible to fully exploit the
high-energy limit and optimise the signal significance.

To describe the deviations of the Higgs couplings with respect to their SM values we follow~\cite{Contino:2010mh} where a general parametrisation of the	couplings of a light Higgs-like scalar $h$ to the SM vector bosons and fermions was introduced.
In this formalism, assuming that the couplings of the Higgs boson to SM fermions scale with their masses and
do not violate flavor, the resulting effective Lagrangian is given by
\begin{equation}
\label{eq:lagrangian}
\begin{split}
{\cal L}  \supset
& \, \frac{1}{2}(\partial_\mu h)^2 - V(h) +\frac{v^2}{4}{\rm Tr}\!\left (D_\mu \Sigma^\dagger D^\mu \Sigma\right )
\left [ 1+2c_V\, \frac{h}{v}+c_{2V}\,\frac{h^2}{v^2}+\dots\right ] \\
&-m_i\,\bar \psi_{Li}\, \Sigma\left (1+c_{\psi}\,\frac{h}{v} + \dots \right)\psi_{Ri}\,+\,{\rm h.c.}\, ,
\end{split}
\end{equation}
where $V(h)$ denotes the Higgs potential, 
\begin{equation}
V(h) = \frac{1}{2} m_h^2 h^2 + c_3\, \frac{1}{6} \left( \frac{3m_h^2}{v} \right) h^3 + c_4\, \frac{1}{24} \left( \frac{3m_h^2}{v^2} \right) h^4 + \dots
\end{equation}
The parameters $\cv$, $\cvv$, $c_{\psi}$, $\ccc$, and $c_4$ are in general
arbitrary coefficients, normalised so that they equal 1 in the SM.
In this contribution we focus on the determination of $\cvv$ by means
of di-Higgs VBF production in the $b\bar{b}b\bar{b}$ final state.

\noindent
{\bf Analysis strategy}\\
Signal and background events are simulated at leading-order (LO) by means of
matrix-element  generators and then processed through a parton shower (PS).
The dominant background is given by QCD multi-jet production,
while other backgrounds, such as top-quark pair production and 
Higgs pair production via
gluon-fusion, turn out to much smaller.
After the parton shower, events are clustered with 
{\sc\small FastJet} v3.0.1~\cite{Cacciari:2011ma} using the
anti-$k_t$ algorithm~\cite{Cacciari:2008gp} with a jet radius $R=0.4$.
The resulting jets are then processed through a $b$-tagging algorithm,
where a jet is tagged as $b$-jet with probability $\eps(b\text{-tag})=0.75$ 
if it contains a $b$-quark with $p_T^b > 15\,$\UGeV.
In order to account for $b$-jet mis-identification (fakes),
jets which do not meet this requirement
are also tagged as $b$-jets with probability $\eps(c\text{-mistag})=0.1$ or $\eps(q,g\text{-mistag})=0.01$
depending on whether they contain a $c$-quark or not.
Only events with four or more jets, of which at least two
must be $b$-tagged, are retained at this stage.

Subsequently to $b$-tagging, events are
classified through a scale-invariant tagging procedure~\cite{Gouzevitch:2013qca,Behr:2015oqq}.
This step is crucial to efficiently reconstruct the Higgs boson candidates and
suppress the otherwise overwhelming QCD backgrounds while at the same time
taking into account all the relevant final-state topologies.
The basic idea of this method is to robustly merge three event
topologies -- { boosted, intermediate} and { resolved} --
into a common analysis.
This is
particularly relevant for our study given that
the degree of boost of the di-Higgs
system strongly depends
on the deviations of $\cvv$ from its SM value.

Acceptance cuts to match detector coverage are applied to signal and background events.
We require the $p_T$ of both the light and $b$-tagged jets
to be larger than 25 \UGeV, while 
the pseudo-rapidities of light
and $b$-tagged jets, $\eta_j$
and $\eta_b$, are limited
by the coverage of the forward calorimeters and
by the tracking region where $b$-tagging can be applied
respectively.
We also impose a set of selection cuts tailored
to the VBF topology which is characterised by two forward and very energetic
jets with little hadronic activity between them. In particular, we cut on the
rapidity separation $\Delta y_{jj}\equiv|y_j^{\rm lead}-y_j^{\rm sub-lead}|>5$ and
the invariant mass $m_{jj}>700$ \UGeV of the two VBF-tagging jets, and impose a central
jet veto (CJV) on  the hardest non-VBF light jet in the central region. The VBF
tagging jets are defined as the pair of light jets satisfying the acceptance
cuts with the largest invariant mass $m_{jj}$.
Moreover, a CJV cut  is imposed  in VBF analyses to veto light
jets with pseudo-rapidity $\eta_{j_3}$ lying between those of the VBF-tagging
jets, $\eta_j^{\max}>\eta_{j_3}>\eta_j^{\min}$, above a $\pT{}$ threshold of 45 \UGeV.

Figure~\ref{fig:mhhdist} (right) shows the $m_{hh}$ distribution
after all analysis cuts for both
for the signal (SM and $c_{2V}=0.8$) and the total background.
For $c_{2V}=0.8$, the crossover between the resolved and
boosted categories takes place
at $m_{hh}\simeq 1.5$ \UTeV,
although this specific value depends on the choice of the jet
radius $R$~\cite{Gouzevitch:2013qca}.
Unsurprisingly, we find that
background events are always dominated by the resolved topology.
The decomposition of the total background in terms of individual processes as a
function of $m_{hh}$ is shown in Fig.~\ref{fig:mhhdist} (left),
where each component is
stacked on top of each other. We see how the $4b$ background dominates
for large $m_{hh}$ while the $2b2j$ one is instead the most important for small
$m_{hh}$. 

\begin{figure}[h!]
	\centering
	\begin{minipage}{0.49\textwidth}\centering
		\includegraphics[width=\textwidth]{\main/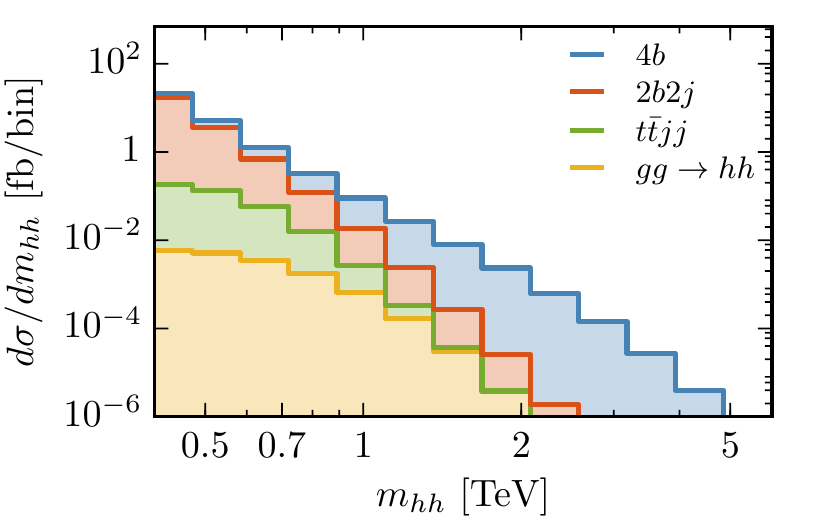}
	\end{minipage}
	\begin{minipage}{0.49\textwidth}\centering
		\includegraphics[width=\textwidth]{\main/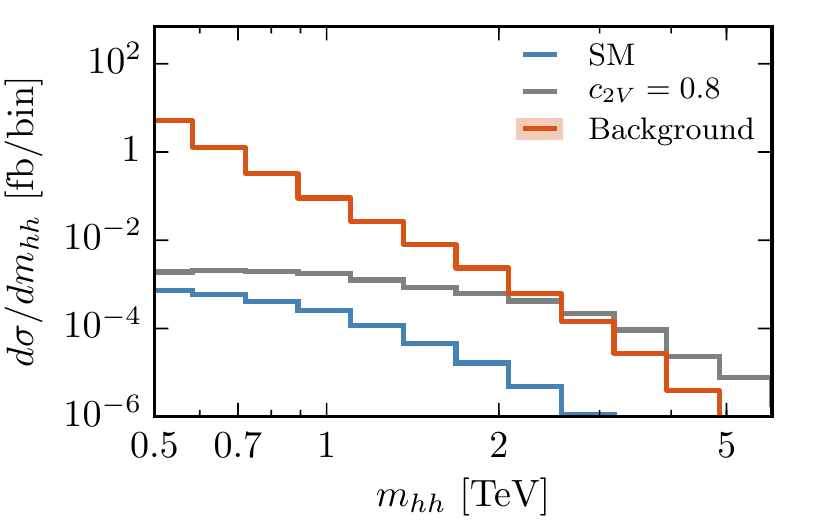}
	\end{minipage}
	\caption{\small Left: Decomposition of the total
		background into individual processes as a function of $m_{hh}$
		after all analysis cuts have been imposed.
		Right: the di-Higgs invariant mass distribution after all analysis cuts
		for the signal (SM and $c_{2V}=0.8$) and the total background.
	}
	\label{fig:mhhdist} 
\end{figure}

\begin{table}[h]\centering
	\small
	\renewcommand{\arraystretch}{1.5}
	\begin{tabular}{llcccc}
		\toprule[0.1em]
		& & \multicolumn{4}{c}{Cross-sections (fb) } \\
		&  &  Acceptance & VBF  & Higgs reco. & $m_{hh}>500$ \UGeV  \\\midrule 
		\multirow{3}{*}{	{\large $14\,$\UTeV}} 
		& Signal SM &  0.011 & 0.0061 & 0.0039 & 0.0020 \\
		& Signal $c_{2V}=0.8$ & 0.035 & 0.020 & 0.017 & 0.011 \\
		& Bkgd (total)   & $1.3\xtento{5}$ & $4.9\xtento{3}$ & 569 & 47   \\
		\bottomrule[0.1em]
	\end{tabular} 
	\caption{\small Cross sections, in fb, after the
		successive application of the acceptance, VBF cuts, and Higgs reconstruction cuts
		for signal events (SM and  $\cvv=0.8$) and for
		the total background. \label{tab:xsecs}
	}
\end{table}

The cross-sections after the
successive application of the acceptance, VBF cuts, and Higgs reconstruction cuts
for signal events (SM and  $\cvv=0.8$) and for
the total background is shown in Tab.~\ref{tab:xsecs}. We find that the VBF di-Higgs signal in the SM is rather small
already after the basic acceptance cuts. However, the signal event
yield is substantially increased for $c_{2V}\ne 1$ as illustrated by the
benchmark value of $\cvv = 0.8$ leading to more than a factor~3\,(5) enhancement
compared to the SM after the acceptance (all analysis) cuts. The fact that this
cross-section enhancement for the $c_{2V} = 0.8$ scenario is more marked at the
end of the analysis is not a coincidence: our selection cuts have been designed
so as to improve the sensitivity to $c_{2V}$ by increasing the signal
significance in the large-$m_{hh}$ region.
Note however that even after
all analysis cuts the background is still much larger than the signal (either SM
or $c_{2V}=0.8$) at the level of inclusive rates. It is only by exploiting  the
large-$m_{hh}$ region that the former can be made small enough to achieve high
signal significances.

\noindent
{\bf Projections for the HL-LHC}\\
Following the analysis strategy outlined in the previous section,
we can now estimate the expected precision on the 
determination of the $\cvv$ coupling at the HL-LHC.
In the left panel of Fig.~\ref{fig:post-pdf} we show the
posterior probabilities for $\cvv$ at 14 \UTeV,
from where we can assess the expected precision  
its measurement at the HL-LHC assuming SM couplings.
The corresponding 68\% probability intervals for the determination of $\cvv$ at the HL-LHC 
are are listed in Table~\ref{tab:resultsdcvv} for two different 
scenarios for the background cross section.
\begin{table}[h!]
	\begin{center}
		\begin{tabular}{r|@{\hskip 0.15in}c @{\hskip 0.2in}c}
			\toprule[1pt]
			&\multicolumn{2}{c}{68\% probability interval on $\dcvv$}\\[0.1cm]
			& $1\times\sigma_\text{bkg}$&
			$3\times\sigma_\text{bkg}$ \\[0.1cm]
			\hline
			& & \\[-0.3cm]
			LHC$_{14}$ & [$-0.37$,\,$0.45$] & [$-0.43$,\,0.48] \\[0.2cm]
			HL-LHC & [$-0.15$,\,0.19] & [$-0.18$,\,0.20]\\[0.2cm]
			\bottomrule[1pt]
		\end{tabular}
	\end{center}
	\vspace{-0.3cm}
	\caption{\small Expected precision (at 68\% probability level) for
		the measurement of $\dcvv$ at the HL-LHC for
		SM values of the Higgs couplings, for two scenarios for the background cross section.
		\label{tab:resultsdcvv}
	}
\end{table}

\begin{figure}[h!]
	\centering
	\begin{minipage}{0.49\textwidth}\centering
		\includegraphics[width=\textwidth]{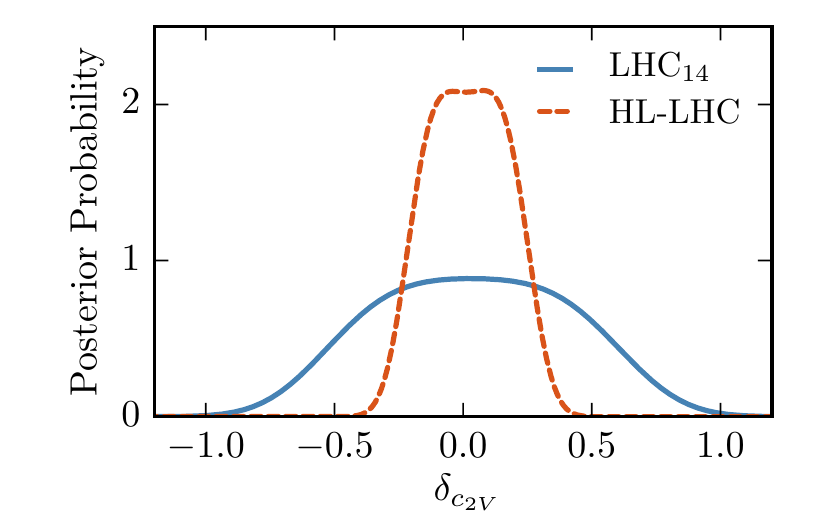}
	\end{minipage}
	\begin{minipage}{0.49\textwidth}\centering
		\includegraphics[width=\textwidth]{\main/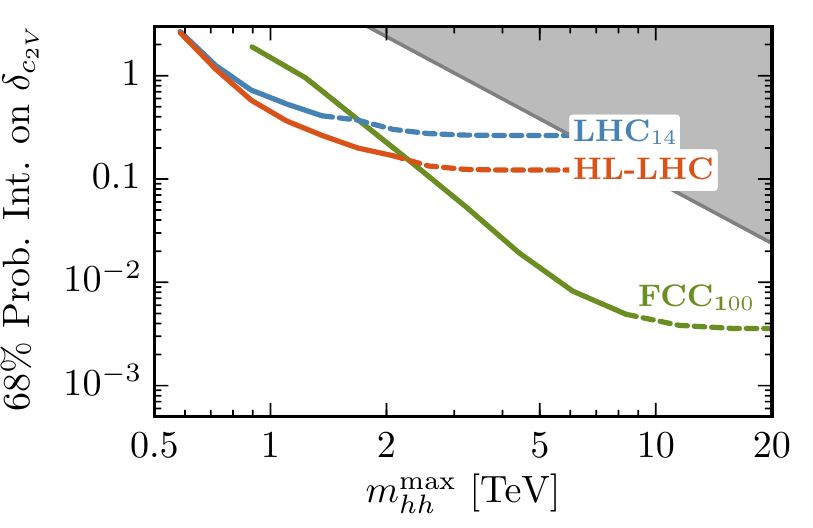}
	\end{minipage}
	\caption{\small Left: posterior probability densities for $\dcvv$ at the HL-LHC.
		Right: expected
		precision for a measurement of $\dcvv$ at the 68\% CL
		as a function of $m_{hh}^{\max}$, where
		the grey area indicates the region where $\dcvv > \dcvv^\text{max}$.
		The transition between solid and dashed curves occurs at the last bin wit hat least one event.
	}
	\label{fig:post-pdf} 
\end{figure}

From Table~\ref{tab:resultsdcvv}, we find that the $c_{2V}$ coupling, for which
there are currently no direct experimental constraints, can be measured 
with a precision of
around  $_{-15\%}^{+19\%}$ at the HL-LHC. 
It is interesting to compare these results with the experimental precision
expected on the fiducial VBF di-Higgs cross section after all
analysis cuts, expressed in terms of $\mu$, the signal strength parameter normalised to the SM result 
We find that the 95\% CL upper limits on $\mu$ for the nominal background cross section
is $\mu\le 109$ with 300 fb$^{-1}$, and $\mu\le 49$ at the HL-LHC.
This result highlights  that the high precision expected on
$c_{2V}$ can obtained despite the loose constraints expected
on the VBF di-Higgs cross section itself.

The results of Table~\ref{tab:resultsdcvv}
have been obtained by making full use of the information
contained on the di-Higgs invariant mass distribution $m_{hh}$.
However, the EFT expansion might break down at large enough values of
$m_{hh}$, corresponding to large partonic centre-of-mass energies,
and some assessment on the validity of our procedure is thus required.
In particular, results can be consistently derived within the EFT
framework only if the new physics scale $\Lambda$ is larger than the largest value of $m_{hh}$ included in the analysis.
Constraining $\Lambda$
requires making assumptions on the structure of the UV dynamics extending the SM~\cite{Contino:2016jqw}.
For example, for the case where the new physics is characterised by a single 
coupling strength~$g_*$ and mass scale $\Lambda$~\cite{Giudice:2007fh}, one expects $
\dcvv \approx g_*^2 v^2/\Lambda^2$, so that
for maximally strongly-coupled UV completions (with $g_* \simeq 4\pi$)
it is possible to derive the upper limit
$\dcvv<16\pi^2 v^2/\Lambda^2$
which connects $\dcvv$ with the new physics scale~$\Lambda$.
The validity of the EFT can thus be monitored
by introducing a restriction $m_{hh} \leq  m_{hh}^\text{max}$, and
then determining how the sensitivity on $\dcvv$ varies as a function of 
$m_{hh}^\text{max}$~\cite{Contino:2016jqw}.
The precision on ${\dcvv}$
is shown in Fig.~\ref{fig:post-pdf}~(right) as a function of  $m_{hh}^{\max}$,
where the grey area indicates the region where $\dcvv > \dcvv^\text{max}=16\pi^2v^2/m_{hh}^\text{max}$.
As expected, increasing $m_{hh}^\text{max}$
leads to stronger constraints.
We therefore find that in the kinematic region accessible at the HL-LHC the
EFT description of the di-Higgs VBF process should be valid.

\asubsection[B. Henning, D. Lombardo, M. Riembau, F. Riva]{Higgs Couplings in High-Energy Multi-boson Processes} \label{HwH}
\label{sec4:hwh}

In this section, based on Ref.~\cite{Henning:2018kys}, we present a novel program to test the Higgs couplings \emph{off-shell} and at high-energy, via their contributions to the physics of longitudinally polarised gauge bosons. We will show that this program is potentially competitive with \emph{on-shell} measurements, but it also offers endless opportunities of refinements and improvements.

Our leitmotiv is that \emph{any} observable modification of a SM coupling will produce in \emph{some} process a growth with energy. In some sense, this is obvious: since the SM is the only theory that can be extrapolated to arbitrarily high-energy, any departure from it can have only a finite range of validity, a fact that is made manifest by a disproportionate growth in some scattering amplitude. Theories with a finite range of validity are, by definition, EFTs; for this reason the best vehicle to communicate our message is  the EFT language where deviations on Higgs couplings come from the operators $\op_{BB},\op_{WW},\op_{y_t},\op_{6}$, etc. We stress nevertheless that at, tree level, the very same conclusions can be reached in the $\kappa$ framework \cite{Heinemeyer:2013tqa} or in the unitary-gauge framework of Ref.~\cite{deFlorian:2016spz,Gupta:2014rxa}.

The operators of that we will be interested on have the form $|H|^2\times \op^{\mathrm SM}$, with $\op^{\mathrm SM}$ a dimension-4  SM operator (i.e. kinetic terms, Higgs potential, and Yukawas) times
\begin{equation}
|H|^2=\frac{1}{2}\left(v^2+ 2 hv + h^2+2\phi^+\phi^-+(\phi^0)^2\right)
\end{equation}
where $h$ is the physical Higgs boson and $\phi^{\pm,0}$ are the would-be longitudinal polarisations of $W$- and $Z$- bosons.
In this contribution we focus on the last two terms, and study processes with longitudinal gauge bosons instead of processes with an on-shell Higgs; we dub this search strategy ``Higgs without Higgs''  - HwH in short \cite{Henning:2018kys}.  
For each modification of a Higgs coupling we identify a process where couplings different from the SM ones induce a high-energy growth in the amplitude with respect the SM,
\begin{eqnarray}
&&\kappa_t: p p \to j t+ V_LV^\prime_L\label{prockt}\\
&&\kappa_\lambda:  p p \to  j j h + V_LV^\prime_L\,,\hspace{.3cm}p p \to jj+4V_L, \label{prockh36}\\
 &&\kappa_{\gamma\gamma,Z\gamma}:  p p \to j j +V^\prime V,\label{prockga} \\ 
&&\kappa_V:  p p \to jj+V_LV^\prime_L,\label{prockv}\\
&&\kappa_g:  p p \to W_L^+W_L^-, Z_LZ_L ,\ \label{prockg}
\end{eqnarray} 
where $V_LV^\prime_L\equiv\{W_L^\pm W_L^\pm,W_L^\pm W_L^\mp,W_L^\pm Z_L,Z_LZ_L\}$ (similarly $4V_L$ a generic longitudinally polarised final state) and $V^{(\prime)}$ any (longitudinal or transverse) vector, including photons.
In the following paragraphs we explore these processes in turn and provide a first estimate of the potential HwH reach at the HL-LHC in comparison with the reach from Higgs couplings measurements. 
Our results are based on leading order (LO) MadGraph simulations \cite{Alwall:2014hca}, where the Higgs couplings have been modified using FeynRules \cite{Christensen:2008py} and checked against the model of Ref. \cite{Falkowski:2015wza}.

\begin{figure}[t]
\begin{center}
\includegraphics[width=0.32\textwidth]{\main/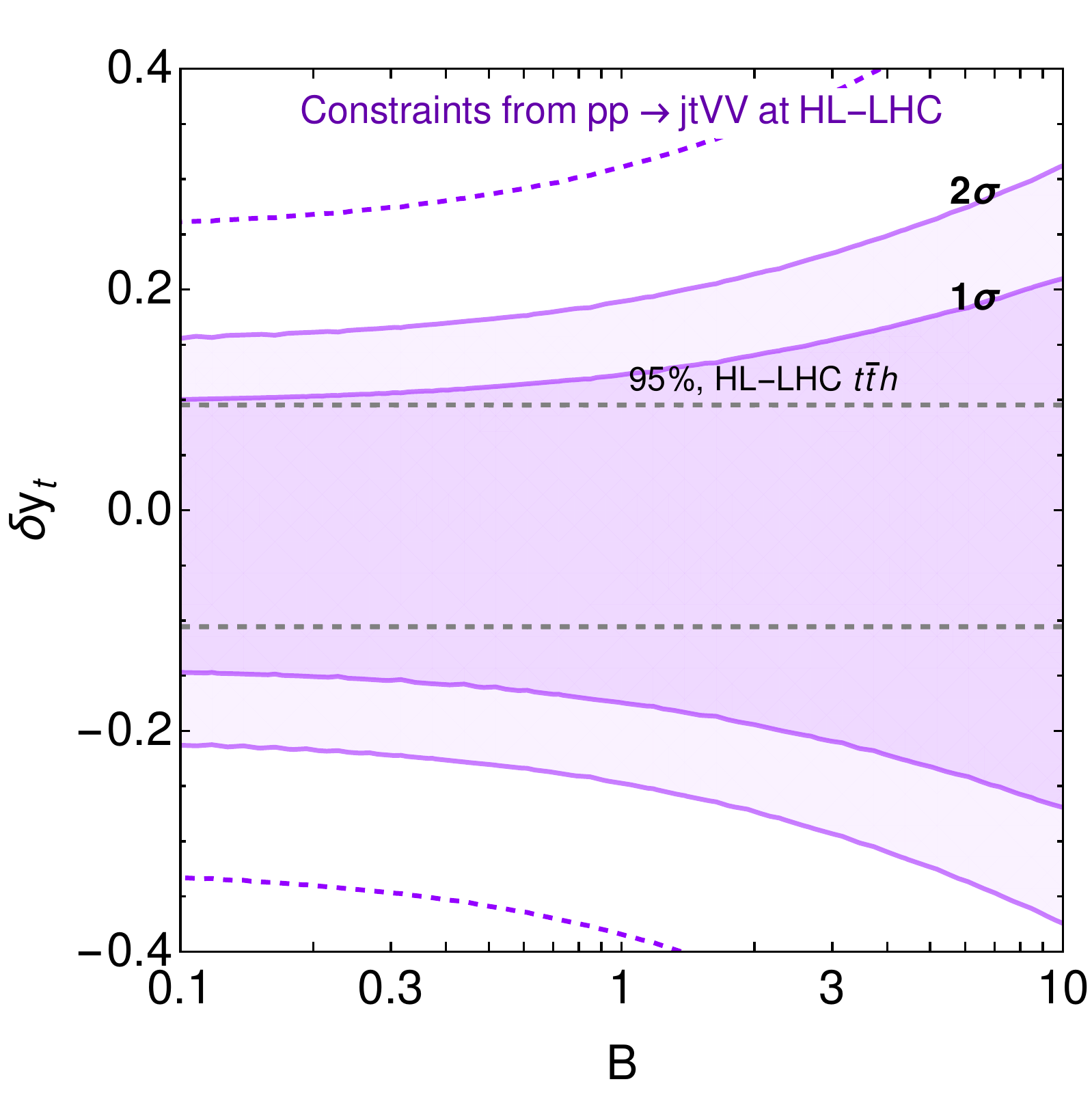}
\includegraphics[width=0.32\textwidth]{\main/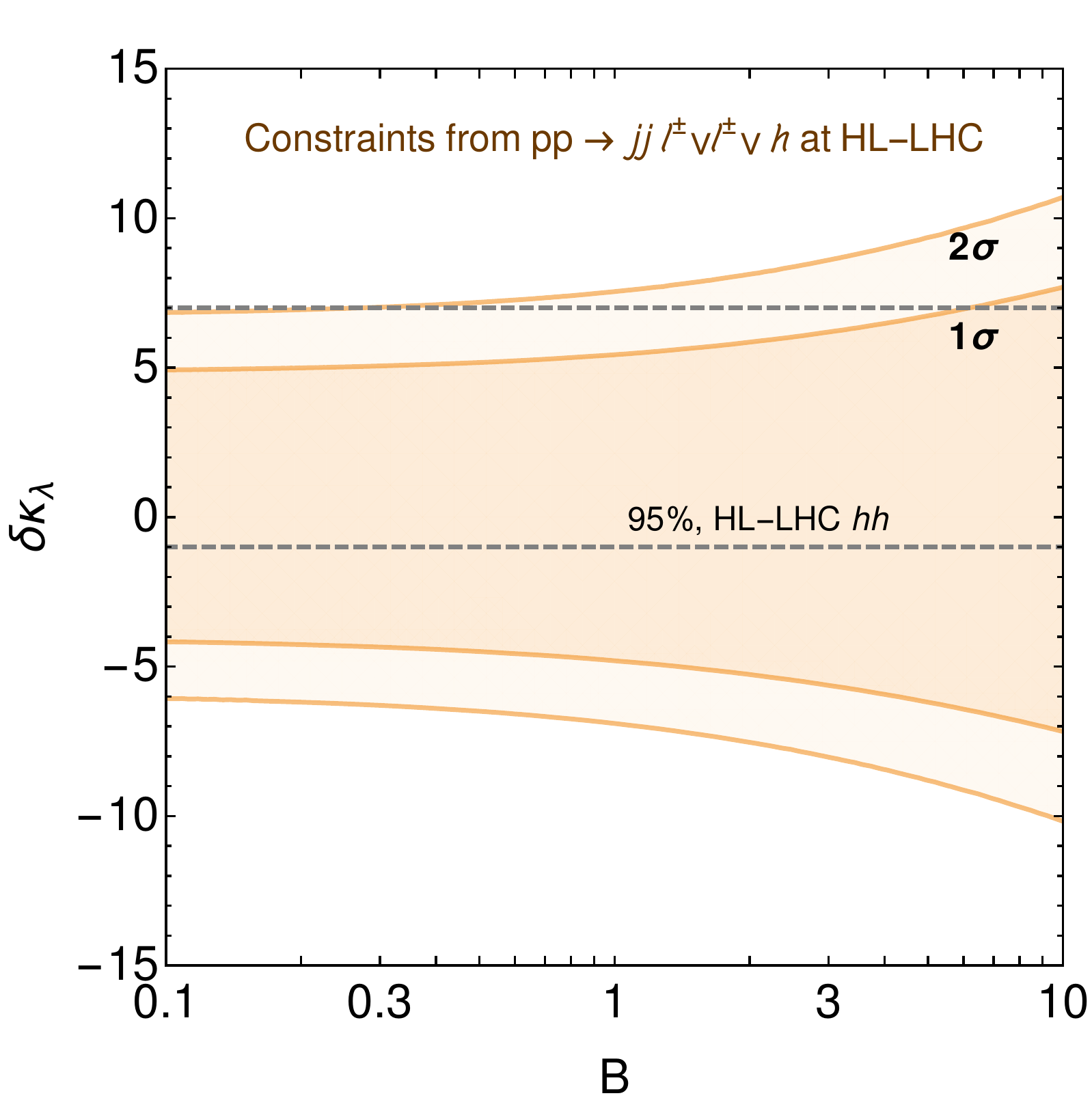}
\includegraphics[width=0.30\textwidth]{\main/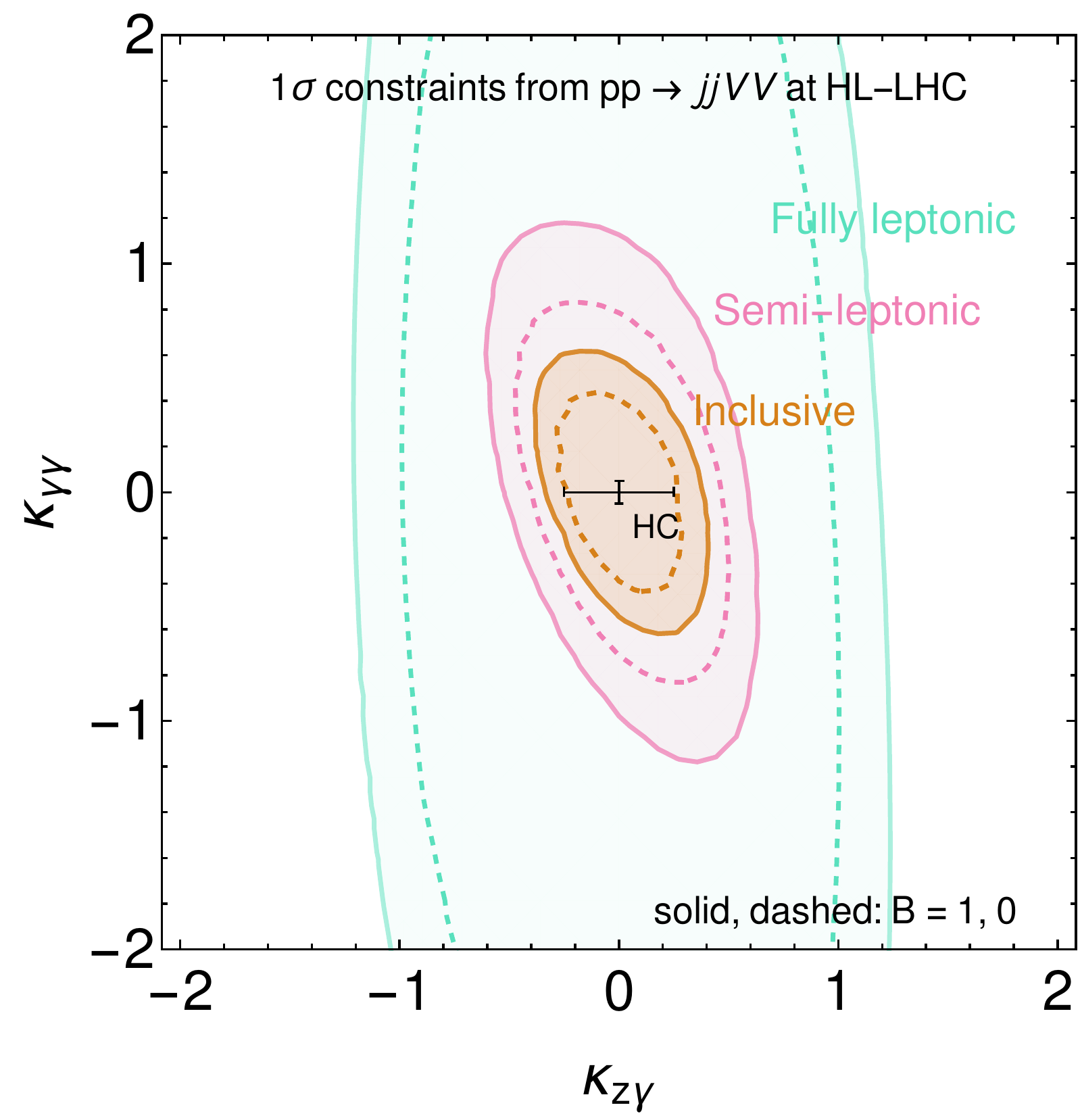}
\end{center}
     \caption{\footnotesize \emph{LEFT: HL-LHC (3000 fb$^{-1}$) sensitivity on modifications of the top quark Yukawa $\delta y_t$ from the process in \eq{prockt} (shaded bands), and from measurements of Higgs couplings (95\%C.L., dashed grey lines); $B$ controls additional backgrounds (for $B=1$ the analysis includes a number of background events equal to the SM signal); $1\sigma$ results without the $0\ell$ and $1\ell$ categories correspond to the dashed purple line. CENTRE: same but for modifications of the Higgs trilinear $\delta\kappa_\lambda$. RIGHT: $1\sigma$ reach for modification of the Higgs-$\gamma\gamma$ and $Z\gamma$ rates, using high-$E$ measurements (green,pink,brown bands correspond to leptonic,semi-leptonic, and also hadronic final states)  or Higgs couplings (black error bars). 
}}\label{fig:reach}
\end{figure} 

\vspace{5mm}
\noindent
{\bf The top Yukawa.}
Modifications of the Yukawa coupling of the Higgs boson to top quarks is reputedly difficult to measure on the $h$ resonance~\cite{Aaboud:2018urx}; however, an anomalous top quark Yukawa induces a quadratic energy growth in the five point amplitude involving a bottom quark, a top, and three longitudinal bosons $W_LV_LV_L$. This amplitude leads to a process with a final state consisting of a top quark, a forward jet and two longitudinally polarised vector bosons in the final state.\footnote{See also Ref.~\cite{Degrande:2018fog} that  studies $tHj$ final states which exhibits linear $E$-growth with modifications of the top-Yukawa.}

 The top carries a large transverse momentum $p_T^t$ due to the hardness of the process, which makes it a good discriminator. We consider two categories, for $p_T^t>250(500)$~\UGeV. A forward jet with $|\eta_j|>2.5$, $p_T^j>30$~\UGeV and $E_j>300$ \UGeV is required. The signal is classified by  counting the number of extra leptons reconstructed in the event. The following table shows the number of signal events  at the 14 \UTeV HL-LHC  with 3000 fb$^{-1}$, for $p_T^t>250$~\UGeV / $p_T^t>500$~\UGeV,

\begin{center}
\begin{tabular}{|c|c|c|c|c|c|}
\hline
Process& 0$\ell$ & 1$\ell$ & $\ell^\pm\ell^\mp$ & $\ell^\pm\ell^\pm$ & $3\ell(4\ell)$\\
 \hline
 $W^\pm W^\mp$    &  3449/567 & 1724/283 & 216/35 & - & -  \\
 $W^\pm W^\pm$    &  2850/398 & 1425/199 & - & 178/25 & -  \\
 $W^\pm Z$   &  3860/632& 965/158 & 273/45 & - & 68/11 \\
 $Z Z$    &  2484/364& - & 351/49 & - &  (12/2) \\\hline
\end{tabular}
 \end{center}
 
The categories with two or more leptons have small background.
The largest source of background for the hadronic modes comes from $\bar t t jj\to tWbjj$ where a $bj$ pair is taken to reconstruct a $W/Z$-boson.
The initial $\bar{t}tjj$ cross section is approximately six orders of magnitude bigger than the ones we are interested in, but we have verified that simple cuts on the invariant mass of the $bj$ pair, on the rapidity of the forward jet, on $p_T^t$, and on the separation between the $W$ and the $b$, as well as vector boson tagging techniques \cite{ATLAS-CONF-2018-016}, can reduce this background to a level that is comparable with the signal.

We broadly parametrise this and other backgrounds by a uniform rescaling $B$ of the SM signal expectation in each bin (so that for $B=1$ we add an irreducible background equal to the SM signal in each channel), and show the estimated reach in the left panel of Fig.~\ref{fig:reach}. The dashed grey lines compare our results with those from Higgs Couplings  measurements (see section \ref{sec2}. For illustration we also show  results that focus on channels with at least 2 leptons with a dashed purple line: here the backgrounds are much smaller. The large number of events left in the zero and one lepton categories makes it possible to extend the analysis to higher energies, where not only the effects of the energy growth will be enhanced, but also the background reduced.

\vspace{5mm}
\noindent
{\bf The Higgs self coupling.}
Measurements of the Higgs self-coupling have received enormous attention in collider studies.  In the di-Higgs channel at HL-LHC precision can reach $\delta\kappa_\lambda\in[-1.8,6.7]$ at 95\%C.L.~\cite{ATL-PHYS-PUB-2017-001} using the $b\bar{b}\gamma\gamma$ final state. 
Here we propose the processes of Eq.~(\ref{prockh36}) with VBS scattering topology and a  multitude of longitudinally polarised vector bosons.
The modified coupling $\delta \kappa_\lambda$, or the operator $\op_6$, induces a linear growth  with energy w.r.t. the SM in processes with $jj h V_LV_L $ final state, and a quadratic growth in processes with  $jj V_LV_LV_LV_L$. For the former, the same-sign $W^\pm W^\pm h jj$  with leptonic $(e,\mu)$ decays is particularly favourable for its low background: two same-sign leptons (2ssl) and VBS topology offers a good discriminator against background, allowing for $h\to\bar b b$ decays. For illustration we focus on this channel in which the SM gives $N_{\textrm{SM}}\simeq 50$ events. {Backgrounds from $ t \bar t jj$ enter with a mis-identified lepton, but it can be shown that they can be kept under control with the efficiencies reported in \cite{Khachatryan:2015hwa} and with VBS cuts on the forward jets. A potentially larger background is expected to come from fake leptons, but the precise estimation of it is left for future work.}

The results -shown in the centre panel of Fig.~\ref{fig:reach}- are very encouraging: this simple analysis can match the precision of the by-now very elaborate di-Higgs studies.
There are many directions in which this approach can be further refined: \emph{i)} including the  many other final states, both for the vector decays and for the Higgs decay \emph{ii)} including the  $E^2$-growing $jj V_LV_LV_LV_L$ topologies, \emph{iii)} taking into account differential information.
Moreover, the process studied grows only linearly with energy w.r.t. the SM amplitude with transverse vectors in the final state, but it grows quadratically w.r.t. the SM final states, so \emph{iv)} measurements of the polarisation fraction can improve this measurement.

\vspace{5mm}
\noindent
{\bf Higgs to $\gamma\gamma,Z\gamma$.}
These decay rates are loop-level and small in the SM: their measurement implies therefore  tight constraints on possible large (tree-level) BSM effects, which in the EFT language are captured by the operators $\op_{WW,BB}$.\footnote{The same operators also affect the  $h$ couplings to $Z_TZ_T$ and $W_TW_T$. The same qualitative analysis can be performed with focus on the $hA_{\mu\nu}A^{\mu\nu}$ and  $hA_{\mu\nu}Z^{\mu\nu}$ vertices, but we prefer to work here with the gauge invariant $\op_{WW,BB}$ operators.}
These also enter in high-energy VBS \eq{prockga}, and they represent a beautiful additional motivation (together with $\kappa_V$, see below) to study these processes, which at present are often interpreted in the context of anomalous quartic gauge couplings (QGC)~\cite{Eboli:2006wa}, corresponding to dimension-8 operators.

We perform a simple analysis of vector boson scattering (VBS) with $W^\pm W^\pm, ZZ, WZ,Z\gamma$ final states. For the first three we use the usual cuts on the forward jets: $|\delta_{jj}|>2.5$, $p_T^j>30$~\UGeV and $m_{jj}>500$~\UGeV \cite{Aad:2014zda}. A kinematic variable that captures the hardness of the $2\to 2$ process is the scalar sum of the $p_T^V$ of the vector bosons, and therefore we bin the distribution in bins of 250 \UGeV up to 2 \UTeV. For the $Z\gamma$ final state, we follow the analysis for aQGC of \cite{Aaboud:2017pds}.

The combined results are displayed in the right panel of Fig.~\ref{fig:reach}, for fully leptonic, semi-leptonic and fully hadronic decays, for backgrounds  $B=0,1$ where, as explained above, $B=1$ corresponds to an additional background of the same order as the SM. Note that we translated the constraints on $c_{BB},c_{WW}$ to the $\kappa_{\gamma\gamma},\kappa_{z\gamma}$. We find that the $ZZ,Z\gamma$ final states provide the best reach.
For comparison, the individual reach from HL-LHC measurements of HC (section \ref{sec2}) is shown by the black error bars. These clearly offer an unbeatable sensitivity in the $h\gamma\gamma$ direction; the $hZ\gamma$ direction is however less tested, and our simple analysis of high-energy probes shows promising results.

\vspace{5mm}
\noindent
{\bf Higgs to $W^+W^-,ZZ$.}
It is  known that modifications of the tree-level $hZZ$ and $hW^+W^-$ SM couplings (assumed here to be controlled by a unique parameter, corresponding for instance to $\op_H$ in the SILH basis \cite{Giudice:2007fh}) imply a quadratic $E$-growth in longitudinal VBS. This is discussed in detail in Ref.~\cite{Contino:2010mh} (and \cite{Contino:2013gna} for linear colliders), where it is pointed out that, in the SM, the longitudinal component is suppressed by an accidental factor $\sim 2000$, which is equivalent to a very large irreducible background. This motivated studies of VBS $hh$ pair production instead, see \cite{Bishara:2016kjn}, finding at $1\sigma$, $\delta\kappa_V\lesssim 8\%$, comparable to $\delta\kappa_V\lesssim 5\%$ from HC.\footnote{The authors of \cite{Bishara:2016kjn} assume separate couplings of the vector bosons to $h$ or $h^2$; when the Higgs is part of a doublet, these are proportional. Moreover, the numbers we report here are indicative: both HC measurements and the di-Higgs analysis have  optimistic and pessimistic scenarios in which these numbers might differ.}

\vspace{5mm}
\noindent
{\bf Higgs to $gg$.}
This coupling modifies the main production mode at hadron colliders and is, therefore, very well measured. 
The most interesting high-energy process that can be associated with this coupling is $gg\to ZZ$, which has  been discussed in Refs.~\cite{Azatov:2014jga,Cacciapaglia:2014rla,Azatov:2016xik}. 
Using the results from Ref.~\cite{Azatov:2014jga} we estimate HwH versus HC reach at the end of the HL-LHC, in particular we have considered a scenario with and one without the background and three different decay channels . We find that 
\begin{eqnarray}
&\textrm{HC:}\,\,\,\, & |\kappa_g|\lesssim 0.025\nn\\
&\textrm{HwH:}&|\kappa_g| \lesssim 0.24\, /\, 0.06 \, /\, 0.01\\
&\textrm{HwH}&(\textrm{no }\bar qq\to Z_TZ_T):\,\, |\kappa_g| \lesssim 0.09  \, /\, 0.02  \, /\, 0.005\nn
\end{eqnarray}
where the numbers stand for the fully leptonic / semi-leptonic / fully hadronic channels.

The partonic $\bar qq\to Z_TZ_T$ process represents here the main irreducible background, as it does not interfere with our $gg\to Z_LZ_L$  amplitude with longitudinal polarisation. Its reduction would constitute an important aspect of HwH analyses. Notice that, unfortunately, in the SM  the $gg\to Z_LZ_L$ process is extremely suppressed at high-$E$, to the benefit of the transverse $TT$ one, see Ref.~\cite{Glover:1988rg}.  This implies that the $SM-BSM$ interference is also suppressed.

Despite these difficulties, which might be overcome in more refined analyses (along the lines of \cite{Panico:2017frx,Azatov:2017kzw}), the high-$E$ results remain competitive  in the semi-leptonic and fully hadronic channels, assuming that the background from $\bar qq\to Z_TZ_T$ can be efficiently suppressed. 

\vspace{1cm}

In summary, the preliminary results are very positive,
 especially given the potential of improvements that we foresee.  
Simple cut-and-count analyses were shown, in some cases, to match the precision of sophisticated Higgs Coupling measurements.
For instance, the $jjW^\pm W^\pm h$  channel with leptonic decays, allows a precision comparable to di-Higgs production in measuring the Higgs self-coupling. Similarly, modifications of the top Yukawa can be measured in  the many $j t+ V_LV^\prime_L$ final states to a precision in the ballpark of Higgs coupling measurements.
VBS processes and $ZZ$ at high-energy offer further alternative possibilities to test the Higgs coupling to electroweak gauge bosons and to gluons, respectively.

\asubsection[R. Covarelli, R. Gomez-Ambrosio]{Dimension-6 EFT effects on Vector Boson Scattering at high energies}\label{sec:VBFdim6eft}

In this note we assess the sensitivity of vector boson scattering (VBS) processes to different dimension-6 ($\mathrm{dim=6}$) operators. We focus here on the $ZZ$ final state, decaying to 4 charged leptons. This experimental channel, currently statistically limited at the LHC \cite{Sirunyan:2017fvv}, will become more interesting at the HL-LHC because of the attainable selection purity. The full reconstruction of the final states also gives access to cleaner observables with respect to final states involving $W$ bosons, where neutrino 4-momenta must be inferred using approximated methods.
This analysis can nevertheless be repeated analogously to other VBS final states. 

In \cite{Gomez-Ambrosio:2018pnl} we studied the purely electroweak component of the $p p \to Z Z j j$ process, referred to as VBS(ZZ). Sensitivity to several $\mathrm{dim=6}$ operators has been demonstrated, as well as the impact of such EFT contribution on the VBS cross-section and triple and quartic gauge couplings (TGCs and QGCs). 

Here we update predictions for the HL-LHC setup and show the kinematic distributions for a handful of relevant operators. For the $\mathrm{dim=6}$ parametrisations we use the \emph{Warsaw basis} from \cite{Grzadkowski:2010es}, following the notation and classification from \cite{Jenkins:2013zja}. Other technical details can be found in the original publication \cite{Gomez-Ambrosio:2018pnl}.

\begin{figure}[h!]
\includegraphics[scale=0.3]{\main/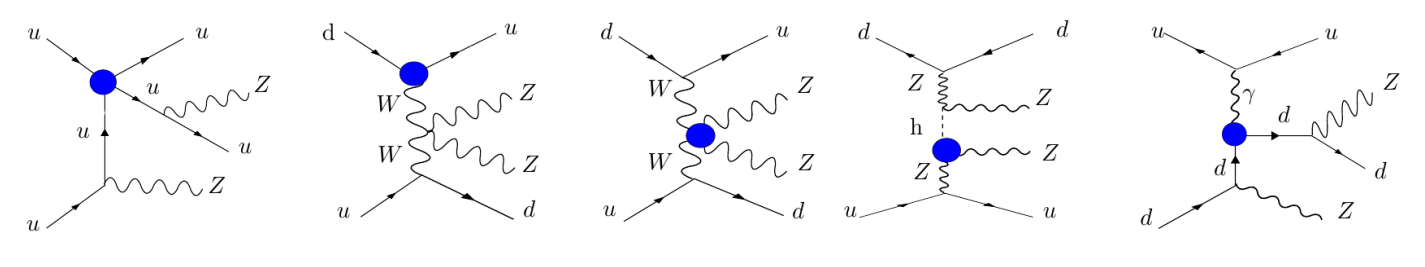}
\caption{Examples of some EFT diagrams for the VBS(ZZ) signal. The blobs represent $\mathrm{dim=6}$ insertions.}
\end{figure}

\noindent {\bf Effective Field Theory parametrisation}\\
We consider the standard SMEFT parametrisation of eq.~(\ref{EFTLAG}).\footnote{In particular, we assume CP symmetry, neglecting the CP-odd operators since their impact on VBS cross-sections and differential distributions is negligible. However it is well known that certain variables of these processes (namely spin correlations and polarisations) can be sensible to CP-violation.} Further, the SMEFT amplitudes and cross sections can be parametrised as
\begin{equation}
\mathcal{A}_{EFT} = \mathcal{A}_{SM} + \frac{g'}{\Lambda^2} \mathcal{A}_{6} +   \frac{g'^2}{\Lambda^4} \mathcal{A}_{8} + \dots 
\end{equation}
\begin{equation}
 \sigma_{EFT} \sim \vert \mathcal{A}_{SM} \vert^2 + {\color{red} 2\frac{g'}{\Lambda^2} \mathcal{A}_{SM} \mathcal{A}_{6} } + {\color{blue} \frac{g'2}{\Lambda^4} \left(  2 \mathcal{A}_{SM} \mathcal{A}_{8} + \vert \mathcal{A}_{6} \vert^2 \right)} + \dots
\end{equation}
Here, we assume the linear contribution (red) of the EFT effects to be leading. Analysis of the $\mathrm{dim=6}$ quadratic terms and the $\mathrm{dim=8}$ interference terms (both in blue) will be subject of further studies.  In particular, $\mathrm{dim=8}$ are commonly associated with quartic gauge couplings and such contribution, albeit sub-leading, would represent some added value to the linear $\mathrm{dim=6}$ prediction. \\
 
\noindent {\bf Definition of the fiducial region}\\
The VBS(ZZ) process has a very peculiar experimental signature, with two energetic forward jets and 4 identifiable charged leptons ($\ell, \ell' = \mu$ or $e$).
The electroweak component of the process $p p \to Z Z j j \to \ell \bar{\ell} \, \ell' \bar{\ell'} j j$ is defined and isolated through some experimental cuts. The ones used in the CMS analysis (in the measurement of the fiducial cross-section) can be found in \cite{Sirunyan:2017fvv}. Here we define a similarly VBS-enriched region, with a relaxed $m_{jj}$ selection: 
%
\begin{eqnarray}
p_T (j) > 30 \textrm{\UGeV} \quad \Delta \eta (j_1 j_2) > 2.4\quad m_{jj} > 100 \textrm{\UGeV}\quad \textrm{\emph{on-shell}}Z_1,Z_2
\end{eqnarray}
%


\noindent {\bf EFT analysis}\\ 
In tables~\ref{tab:signal} and \ref{tab:background} we show the sensitivities to different $\mathrm{dim=6}$ operators of the VBS(ZZ) process, as well as of its main background at LHC: the di-boson production channel from quark-antiquark annihilation associated to gluon radiation (studied in depth by CMS for LHC runs I and II in \cite{Sirunyan:2018vkx}, QCD(ZZ)). 

Further, in figure \ref{fig:plots} we show differential distributions for a subset of operators. In particular we chose the three operators that directly affect triple and quartic gauge couplings, with the following notation, which differs from that of table~\ref{tab:dim6ops},
\begin{equation}
\mathcal{O}_W = \frac{3!}{g}\mathcal{O}_{3W}\quad
\mathcal{O}_{HW} = \frac{1}{g^2}\mathcal{O}_{WW}\quad
\mathcal{O}_{HWB} = \frac{1}{gg^\prime}\mathcal{O}_{WB}
\end{equation}
However, as reported in tables~\ref{tab:signal} and \ref{tab:background}, there are other relevant operators for the VBS process, 
for example $\mathcal{O}_{\ell \ell}$, the 4-lepton operator that affects $G_F$, or $\mathcal{O}_{HB}$ that enters the $Z$ boson propagator. More details can be found for example in \cite{Ghezzi:2015vva}.
 
Figure \ref{fig:plots} should be interpreted as follows: we select one paradigmatic operator (for example $\mathcal{O}_W$), and see how much does its interference term affect the VBS and di-boson signals ($2.5\%$ in this case). As the VBS(ZZ) cross section is still mostly unconstrained experimentally, while the QCD(ZZ) has a $21\%$ uncertainty in the 2-jet bin \cite{Sirunyan:2018vkx}, we know the bounds within which we can vary this coefficient. If we assume for example a $10 \%$ positive interference with the total cross-section, we observe that such a small contribution to the total cross-section can represent a large modification in certain bins of the differential distributions. This advantage is twofold: with this procedure we can select the optimal bin(s) for the study and fit of each EFT operator; and, by applying unitarity considerations, we can constrain the values of the Wilson coefficients further. In our example, a contribution of $10 \%$ in $\mathcal{O}_W$, still allowed for the total rate, has a large impact on the high energy bins of the $p_T (Z_1)$ distribution. 
\\
 
\noindent {\bf Conclusions}\\
The VBS(ZZ) and QCD(ZZ) final states, still largely unexplored at the LHC, will be an important source of constraints on $\mathrm{dim=6}$ EFT operators at the HL-LHC. We have shown the impact that values of Wilson coefficients still experimentally allowed have on differential distributions that are easily accessible experimentally in this channel. 

\begin{figure}[htbp]
  \begin{minipage}[b]{0.5\textwidth}
    \includegraphics[width=1.02\textwidth]{\main/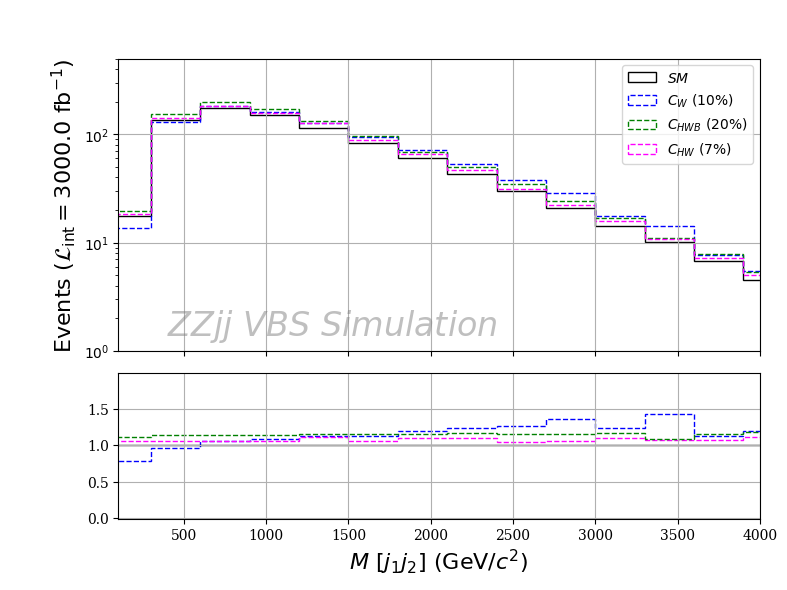}
  \end{minipage}
  \begin{minipage}[b]{0.5\textwidth}
    \includegraphics[width=1.02\textwidth]{\main/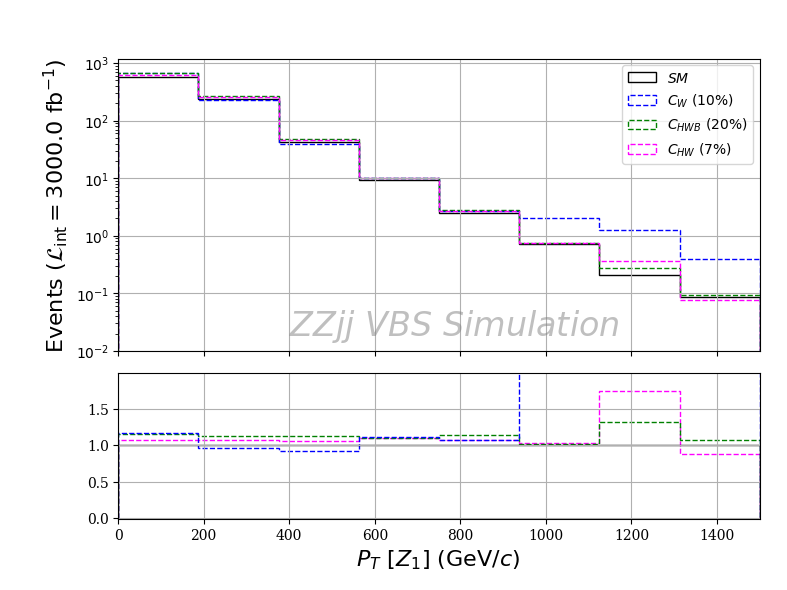}
  \end{minipage}
  \caption{Two generic simulations showing the EFT effects on key differential distributions: invariant mass of the di-jet system (left) and transverse momentum of the leading Z boson (right). We selected arbitrary values for the Wilson coefficients $\lbrace c_W, c_{HW}, c_{HWB} \rbrace$. Notice that the notation differs from table \ref{tab:dim6ops}.}
  \label{fig:plots}
\end{figure}


\begin{table*}[]
\begin{center}
\small
\begin{tabular}{|c|c|}
\hline
VBS Signal & Signal strengths (Linear EFT) \\
\hline
 Class 1:  &  $\mathcal{O}_W = c_W \cdot 2.5 \%$ \\
 Class 3: & $\mathcal{O}_{HD} = c_{HD} \cdot 6.0 \%$ \\
 Class 4: & $ \mathcal{O}_{HW} = c_{HW} \cdot 5 \% $, $ \mathcal{O}_{CHB} = c_{HB} \cdot 0.2 \% $,  $ \mathcal{O}_{HWB} = c_{HWB} \cdot 14 \% $\\
 Class 7: & $ \mathcal{O}_{Hl^{(3)}} = c_{Hl^{(3)}} \cdot 48 \% $,
 $ \mathcal{O}_{Hq^{(1)}} = c_{Hq^{(1)}} \cdot 2 \% $, \\ 
 & $ \mathcal{O}_{Hq^{(3)}} = c_{Hq^{(3)}} \cdot 46 \% $,  $ \mathcal{O}_{Hu} = c_{Hu} \cdot 0.8 \% $ 
 \\
 Class 8a: $(L \bar{L})(L \bar{L})$  &  ($G_F \to$) $ \mathcal{O}_{\ell \ell} = c_{\ell \ell} \cdot 24 \% $ , 
 $ \mathcal{O}_{qq^{(1)}} = c_{qq^{(1)}} \cdot 12 \% $, \\ &
$ \mathcal{O}_{qq^{(11)}} = c_{qq^{(11)}} \cdot 14 \% $, 
$ \mathcal{O}_{qq^{(33)}} = c_{qq^{(33)}} \cdot 100 \% $, 
$ \mathcal{O}_{qq^{(31)}} = c_{qq^{(31)}} \cdot 75 \% $
 \\
\hline 
 \end{tabular}
  \caption{  Different sensitivities to each of the Warsaw basis operators. The operators that are not listed do not intervene in the process, or do it in a negligible way. Each sensitivity $\epsilon_i$ is calculated as $\epsilon_i  = \vert \frac{\sigma_{EFT} - \sigma_{SM}}{\sigma_{SM}} \vert$, and they include a standard EFT pre-factor $\frac{v^2}{\Lambda^2}\vert_{\Lambda=1\UTeV}$ which needs to be taken into account if substituting values for the $c_i$ in the table. NB: we quote the absolute value for the sensitivities $\epsilon$. Notice that the notation differs from table \ref{tab:dim6ops}.}
  \label{tab:signal}
\end{center}
\end{table*}

\begin{table*}[]
\begin{center}
\small
\begin{tabular}{|c|c|}
\hline
ZZ Di-boson & Sensitivities (Linear EFT) \\
\hline
 Class 1:  &  $\mathcal{O}_G = 2.5 \%$, \quad $\mathcal{O}_W = 2.5 \%$ \\
Class 3: & $\mathcal{O}_{HD} = 6.0 \%$ \\
Class 4: & $ \mathcal{O}_{CHW} = 0.2 \% $, $ \mathcal{O}_{CHG} =  8 \% $, $ \mathcal{O}_{CHB} =  0 \% $, $ \mathcal{O}_{CHWB} =  12 \% $ \\
Class 7: & $ \mathcal{O}_{Hl^{(3)}} = c_{Hl^{(3)}} \cdot 25 \% $ ,
$ \mathcal{O}_{Hq^{(1)}} = c_{Hq^{(1)}} \cdot 3 \% $, \\ & 
$ \mathcal{O}_{Hq^{(3)}} = c_{Hq^{(3)}} \cdot 31 \% $,
$ \mathcal{O}_{Hu} = c_{Hu} \cdot 1.1 \% $
\\
Class 8a: $(L \bar{L})(L \bar{L})$  &  ($G_F \to$) $ \mathcal{O}_{\ell \ell} = c_{\ell \ell} \cdot 12 \% $ , $ \mathcal{O}_{qq^{(1)}} = c_{qq^{(1)}} \cdot 1.0 \% $, \\ &
$ \mathcal{O}_{qq^{(11)}} = c_{qq^{(11)}} \cdot 1.3 \% $, 
$ \mathcal{O}_{qq^{(33)}} = c_{qq^{(33)}} \cdot 8.4 \% $, 
$ \mathcal{O}_{qq^{(31)}} = c_{qq^{(31)}} \cdot 8.0 \% $
\\
\hline
 \end{tabular}
  \caption{Sensitivities to the different $\mathrm{dim=6}$ operators in the di-boson production channel, main background for the VBS(ZZ) at LHC. A large sensitivity does not necessarily mean that a large EFT effect is expected, since the corresponding Wilson coefficient might as well be very small. Notice that the notation differs from table \ref{tab:dim6ops}. 
  }
  \label{tab:background}
\end{center}
\end{table*}

\newpage

\asubsection[G. Chaudhary, J. Kalinowski, M. Kaur, P. Koz{\'o}w, S. Pokorski, J. Rosiek, K. Sandeep, M. Szleper, S. Tkaczyk]{Same-sign WW
  scattering and EFT applicability}\label{sect-ssWW}

Although any statistically significant deviation in data from the  Standard Model(SM) predictions would be a manifestation of a BSM physics, the question is what we can learn about its scale and its strength before discovering new particles. The appropriate tool for answering this question is the Effective Field Theory(EFT) approach: the information about the  scale $\Lambda$ and the strength $C$ of new physics is encoded in  the Wilson coefficients of the higher dimension operators, $f_i=C^m/\Lambda^n$.
The usefulness of any EFT analysis of a given process relies on the assumption that only a few higher-dimension terms 
in the expansion ${\cal L}={\cal L}_{SM}+\sum_i f_i^{(6)}{\cal O}_i^{(6)}+ \sum_i f_i^{(8)}{\cal O}_i^{(6)}+\ldots$ 
provide adequate approximation to an unknown UV completion.   
This assumption  introduces a strong model-dependent aspect and therefore it is convenient to introduce the concept of EFT ``models" defined by the choice of operators  and the values of their Wilson coefficients $({\cal O}_i^{(d)},f_i^{(d)})$.  Our focus is on the proper use of the EFT "models" in their range of validity for the WW scattering in purely leptonic W decay channels where the $WW$ invariant mass cannot be determined experimentally. A full explanation of the concept is to be found in  \cite{Kalinowski:2018oxd} and here we summarise the main points..

Following a common practice we take one operator at a time setting others to 
zero,  which effectively defines the EFT ``model", 
and consider   the process 
$pp\rightarrow 2jW^+ W^+ \rightarrow 2j l^+\nu l'^+\nu^\prime$. 
The EFT ``model" can be maximally valid up to the invariant mass $M$ of the 
$W^+W^+$ system 
$M<\Lambda\leq M^U$, where $M^U=M^U(f)$ is the perturbative partial wave unitarity bound  in the chosen EFT "model".  
If  the kinematic range $M_{max}$  at the LHC is greater than $\Lambda$, there is necessarily a 
contribution to observables from the region $\Lambda < M < M_{max}$.  
Two questions arise:1) what is the discovery region in the space $(\Lambda, f)$ for the chosen EFT "model", 2) if a deviation from  the SM predictions is indeed observed,
how to verify the chosen EFT  ``model"  by fitting it to a set of experimental distributions $D$ 
and in what range of $\Lambda, f_i$ such a fit is really meaningful?
\begin{figure}
\includegraphics[width=0.8\linewidth]{\main/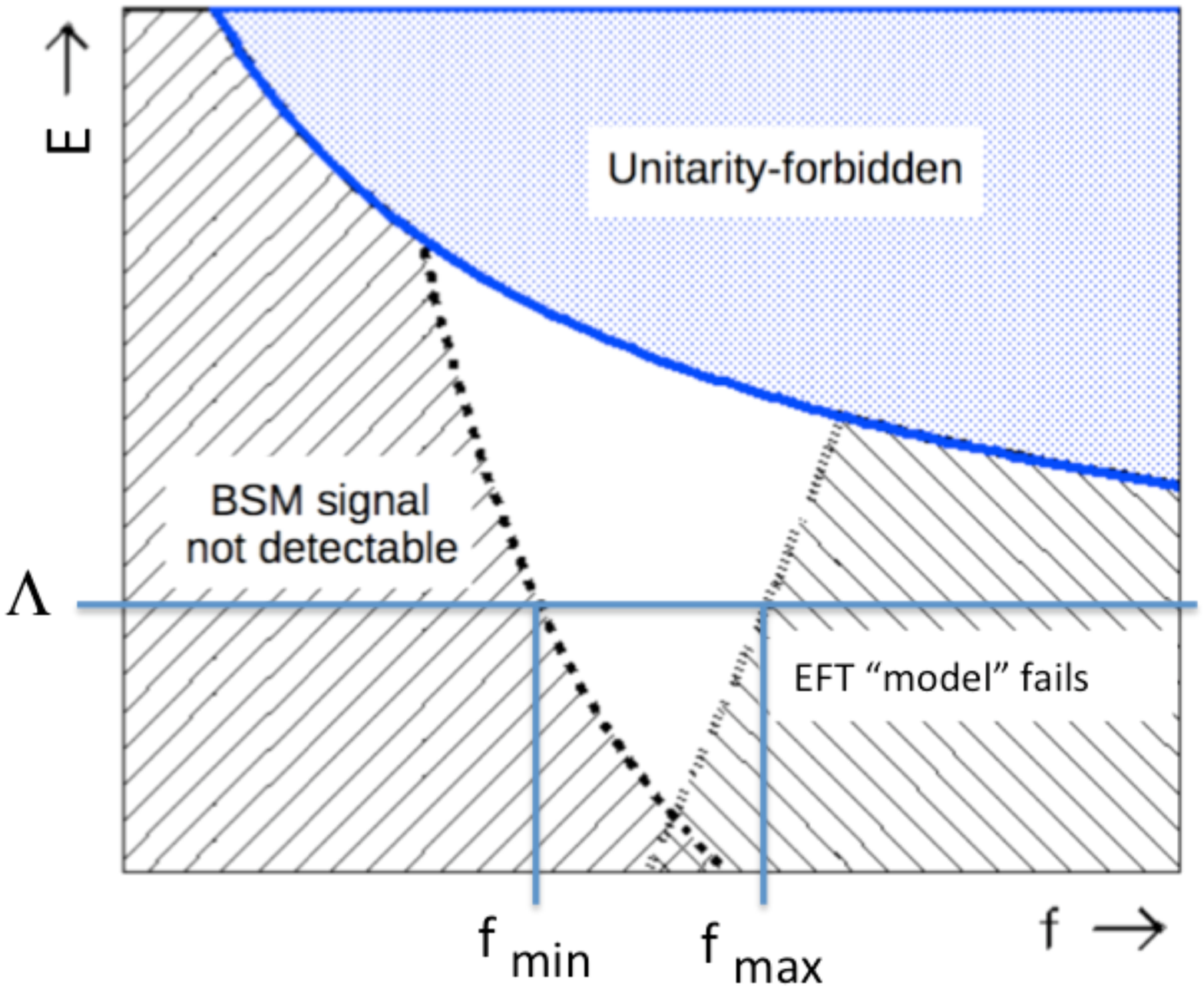}
\caption{
Cartoon plot showing the regions in $f_i$ and $\Lambda$ in terms of BSM signal observability
and applicability of the EFT ``model"  for  the same-sign WW process with purely 
leptonic decays.  The white triangle shows the region
where the BSM physics can be studied within the chosen EFT "model".}
\label{fig:cartoonplot}
\end{figure}

For a given EFT ``model" the unitarity bound is very different for  different helicity amplitudes.
As  $M^U$ we take  the $lowest$ value from T-matrix diagonalisation of the $W^+W^+$ and $W^+W^-$, universally for all helicity 
amplitudes. 
The BSM signal  $S$ of the EFT ``model" $({\cal O}_i^{(d)},\,f_i^{(d)})$ 
can be defined as the deviation from SM predictions observed in the distribution of some observable $D$,  
$S=D^{model}-D^{SM}$.
A quantitative estimate of the signal can be written as
\small
\begin{equation} 
D^{model}=\int^{\Lambda}_{2M_W}\frac{d\sigma}{dM} \Bigl|_{model} dM +\int_{\Lambda}^{M_{max}}\frac{d\sigma}{dM}\Bigr|_{SM} dM\,,
\label{dsigma}
\end{equation}
\normalsize
which comes uniquely from the operator that defines the ``model" in its range
of validity and assumes only the SM contribution in the region $M>\Lambda$.
BSM contribution from the region above $\Lambda$   may enhance the signal, but it may also 
preclude proper description of the data in the EFT  "model",
which makes sense {\it if and only if} this additional contribution is small enough  
compared to the contribution from the validity region.  
For a  quantitative estimate of this 
contribution we define a second estimate in which  all the helicity amplitudes above 
$\Lambda$ are assumed to remain 
constant at their respective values they reach at $\Lambda$
\small
\begin{equation}
D^{model}=\int^{\Lambda}_{2M_W}\frac{d\sigma}{dM}\Bigl|_{model} dM +\int_{\Lambda}^{M_{max}}\frac{d\sigma}{dM}\Bigr|_{A=const} dM.
\label{unitarized}
\end{equation}
\normalsize
For $\Lambda = \Lambda_{max}$ this prescription regularises the helicity amplitudes that violate unitarity at $M^U$. 
We adopt the criterion that the EFT ``model" is tested for values of $(\Lambda\leq M^U, f_i)$ 
when the signals computed from Eq.(\ref{dsigma}) are statistically consistent 
within 2$\sigma$ with the signals computed with Eq.(\ref{unitarized}).

The observability of  the EFT "model" predictions imposes some minimum value of $f_{min}$, while the description within the EFT "model" imposes some 
maximum value of $f_{max}$ such that signal
estimates computed from Eqs.(\ref{dsigma}) and (\ref{unitarized}) remain 
statistically consistent.  
For $\Lambda=M^U$ a finite interval of $f_i$ values is possible, while 
for $\Lambda <M^U$ the  respective limits on $f_i$  depend on the actual value of $\Lambda$. It is illustrated in 
a cartoon plot in Fig.~\ref{fig:cartoonplot}, where the white "triangle" is bounded from 
above by the unitarity bound $M^U(f_i)$, from the left by the signal significance criterion 
and from the right by the consistency criterion.
The EFT "model" could be the right framework to describe the  BSM signal as long as the ``triangle"
shown in our cartoon plot is not empty.

Our preferred strategy for data analysis is as follows:\\
a)  Measure  distributions $D$ that
offer the highest sensitivity to the studied EFT "model",\\
b) if deviations from the SM are observed, fit the values of 
$(\Lambda\leq M^U, f_i)$  according to 
Eq.(\ref{unitarized}),\\
c) using the fitted values of $f_i$ and $\Lambda$ recalculate 
$D$  templates 
according to 
Eq.(\ref{dsigma}),\\
d) check statistical consistency between estimates based on Eqs.(\ref{dsigma}) and (\ref{unitarized}). \\
 Physics conclusions from the obtained $(\Lambda, f_i)$ values can only be drawn
if such a consistency is found.  
%
%
Stability of the result against alternative 
regularisation methods  would provide a measure of uncertainty of the procedure - too much
sensitivity to the region above $\Lambda$ means the procedure is destined to fail
and  that data cannot be described within the chosen EFT "model".

%

To demonstrate our strategy we considered EFT ``models" defined by 
one-at-a-time dimension-8 operator that affects $WWWW$ couplings. Details of the simulation of events for  the process 
$pp \to jj\mu^+\mu^+\nu\nu$ (at 14 \UTeV with 3/ab integrated luminosity) and their processing according to our strategy can be found in \cite{Kalinowski:2018oxd}. 
Assuming $\Lambda$ equal to the respective unitarity bounds,
the lower and upper limits for the values of $f$ for each dimension-8 operator, for positive
and negative $f$ values, as well as the applicability "triangles" in the $(\Lambda,f_i)$ plane for each operator have been calculated.
  These limits define the (continuous) sets of testable 
EFT ``models" based on the choice of single dimension-8 operators.

Following the above strategy we have calculated the expected reach for the dim-8 operator $O_{M7}$ at the HE-LHC and compared it with the obtained reach for the HL-LHC (14 \UTeV) from Ref.~\cite{Kalinowski:2018oxd}, assuming in each case an integrated luminosity of 3/ab.  Fig.\ref{fig:fM7} shows the respective ``EFT triangles".  It is evident that increasing the proton energy allows to explore much lower values of the Wilson coefficients, with lower limits for a 5$\sigma$ BSM discovery being shifted by as much as almost an order of magnitude.  On the other hand, the upper limit on consistent EFT description shifts likewise by a similar amount. This is due to the fact that by increasing the collision energy more and more events come from the region, where $M>\Lambda$ and therefore shrinking the range of Wilson coefficients that satisfy our consistency criterion.  Overall, the area of the actual ``EFT triangle" does not get significantly larger for 27 \UTeV compared to 14 \UTeV, even when viewed in a log scale.  

To summarise:  we have analysed the physics potential of "EFT models" defined by the choice of single dimension-8 operators in the same-sign WW scattering process in the purely leptonic decay modes.  
We argue that usage of EFT ``models" in the analysis of purely
leptonic $W$ decay channels requires bounding the possible contribution from
the region $M_{WW} > \Lambda$, no longer described by the ``model",
and ensuring it does not significantly distort the measured distributions 
compared to what they would have looked from the region of EFT validity alone and 
propose a data analysis strategy to satisfy the above requirements.  
We find that the "triangles"  turned to be rather narrow, even when going from 14 to 27 \UTeV of $pp$ beam energy . This result reinforces our former conclusion that study of BSM effects by means of varying single Wilson coefficients has little physics potential and future data analysis should be rather focused on simultaneous fits of many operators to the combined data from all VBS processes.  We find this conclusion to hold equally regardless of the actual beam energy.

\begin{figure}
\centering
\includegraphics[width=0.7\linewidth]{\main/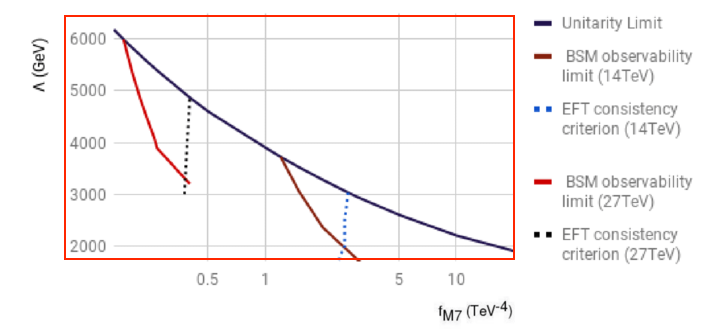}
\caption{
Regions in $f_{M7}$ and $\Lambda$ in terms of BSM signal observability
for the EFT ``model"  based on ${\cal O}_{M7}$ operator at the HL- and HE-LHC.}
\label{fig:fM7}
\end{figure}

%
Acknowledgements: 
Work partially supported by the National Science Centre (Poland) grants
DEC-2015/18/M/ST2/00054,  DEC-2016/23/G/ST2/04301 (SP),
DEC-2015/19/B/ST2/02848 (JR) and 
HARMONIA project 
UMO-2015/18/M/ST2/00518  (JK) as well as the COST Action CA16108. 
ST is supported by Fermi Research Alliance, LLC under Contract No.~De-AC02-07CH11359 with the US DoE. The work of PK  has been supported by the Spanish MINECO project FPA2016-78220-C3-1- P (Fondos FEDER).

\newpage

\asection[Z. Liu, M. Xiao]{Higgs boson mass and width}\label{sec5}

\asubsection[F. Caola, R. R\"ontsch]{Theory review}
The total decay width is an important property of the Higgs boson,
as it contains information about the interactions of the Higgs with all other
fundamental particles, and is predictable both in the Standard Model and its extensions.
Therefore, measuring this property is an important part of Higgs studies. 
Direct measurements of the Higgs width are very challenging at hadron colliders,
as these require a scan of the invariant mass profile of the Higgs decay
products.
This is limited by detector resolution to roughly $\sim 1~{\rm \UGeV}$,
which is  three orders of magnitude larger than the SM prediction of  $\gHsV\sim 4~{\rm \UMeV}$.
Current LHC measurements have already attained this level of precision. Although no explicit projection has been made, it is expected that the direct
method will only be able to constrain the Higgs width to $\mathcal O(100)$
times the SM value.\footnote{Lower bounds
on the Higgs width can be obtained
from lifetime measurements.}

Given this situation, there has been considerable interest in devising indirect probes of the
Higgs width. In general, a standard Higgs analysis in the $H\to X$ decay channel 
measures the production cross section times branching ratio,
$\sigma \sim \sigma_{\rm prod}\times \gHsVtoX/\gHsV,$
and is thus  only sensitive to a combination of the coupling and the width. Schematically,
\begin{equation}
\sigma \sim \frac{g_{\rm prod}^2 \times g_{\rm dec}^2}{\gHsV},
\label{eq:width_standard}
\end{equation}
where $g_{\rm prod}$ and $g_{\rm dec}$ are the couplings that enter the Higgs production
and decay channels, respectively. An independent measurement of the couplings and the decay width is 
therefore not possible from such analyses. The idea
behind all indirect determinations of $\gHsV$ is to find an observable whose dependence on
$g_i$ and $\gHsV$ is different from Eq.~(\ref{eq:width_standard}), which allows one
to lift the coupling/width degeneracy. Indirect determinations can be broadly separated in two
classes: \emph{on-shell} methods, which rely on interference effects on the Higgs resonant
peak, and \emph{off-shell} methods, which combine on-peak and off-peak information. In the
following, we provide a quick overview of these methods, emphasising their strengths and weaknesses. 

The starting point of the \underline{on-shell
methods}~\cite{Dixon:2003yb,Martin:2012xc,Dixon:2013haa,Campbell:2017rke}
is the observation that measurements in the $H\to X$ decay channel
receive a contribution both from the signal $p p \to H \to X$ process
and from the continuum background $p p \to X$, and the two
interfere. Schematically, the amplitude for the process can be written
as
\begin{equation}
\mathcal A_{p p \to X} = \frac{S \mhsV^2}{s-\mhsV^2 + i \mhsV\gHsV} + B,
\end{equation}
where $S\propto g_{\rm prod}\times g_{\rm dec}$ is the signal part and $B$ is the background 
contribution. This leads to
\begin{equation}
|\mathcal A_{p p \to X}|^2 = \frac{\mhsV^4}{(s-\mhsV^2)^2 + \mhsV^2\gHsV^2}
\times\left[
|S|^2 + \frac{(s-\mhsV^2)}{\mhsV^2} 2{\rm Re}(S B^*) + \frac{\gHsV}{\mhsV} 2{\rm Im}(S B^*)
\right] + |B|^2.
\label{eq:width_gammagamma}
\end{equation}
Here, $|S|^2\propto g_{\rm prop}^2 \times g_{\rm dec}^2$, but 
$SB^*\propto g_{\rm prop}\times g_{\rm dec}$, so a combined determination of the signal $|S|^2$ and
interference $SB^*$ contributions can lift the coupling/width degeneracy of 
Eq.~(\ref{eq:width_standard}), thus giving access to $\gHsV$. 
For this method to be effective, one needs to consider channels where
the interference is large. The best candidate is the $gg\to H\to \gamma\gamma$ channel: indeed,
in this case both the $gg\to H$ production and the $H\to\gamma\gamma$ are loop induced, as is
the continuum contribution $gg\to\gamma\gamma$. This implies that at least naively there is a loop
enhancement factor in the interference w.r.t. the pure signal, thus making the former noticeable. 

The \emph{real part of the interference} in
Eq.~(\ref{eq:width_gammagamma}) is anti-symmetric around the Higgs peak,
so it does not affect the total rate. However, it leads to a
distortion in the shape of the $m_{\gamma\gamma}$ distribution around
the Higgs peak, which in turns translates into a slight shift in the
reconstructed Higgs mass~\cite{Martin:2012xc}.  The size of this mass
shift is proportional to the interference contribution, whose
dependence on couplings and width is different from
Eq.~(\ref{eq:width_standard}). A measurement of the mass shift then
allows for a determination of $\gHsV$. This can be done for example by
comparing the mass extracted in the $\gamma\gamma$ channel with that
determined in the $4l$ channel, where these interference effects are negligible. However,
even if the $4l$ channels lead to a very good mass determination
once high enough statistics  have been accumulated, extracting the
mass shift from a $\gamma\gamma$ vs $4l$ comparison introduces
additional systematics. One way to circumvent this issue is to 
consider only the  $\gamma\gamma$ decay mode and to compare different kinematic regions, although detailed systematic studies within this approach have not yet been done.
This is possible since the interference
is strongly dependent on the transverse momentum of the Higgs~\cite{Dixon:2013haa}. 
In particular, hard radiation tends to lessen this effect somewhat.
Another candidate for a reference mass could be obtained from studying Higgs production
in association with two hard jets. Indeed, in this case there are
cancellations between the $ggF$ and $VBF$ contributions and the net result
for the interference is very small~\cite{Coradeschi:2015tna}. Theoretical predictions for
the mass shift are under good control, with the interference being
known to NLO in QCD~\cite{Dixon:2013haa,deFlorian:2013psa,Martin:2013ula} 
and matched to parton shower~\cite{Gleisberg:2008ta,Hoche:2015sya}. It turns
out that radiative corrections deplete the interference contribution somewhat.
Although it is well known that higher order corrections are important for 
Higgs physics, for this analysis the main limitation comes from experimental
systematics, namely the detector response, which must be properly modelled
to   extract the interference contribution from the measured mass shift. 
In the SM, the mass shift at the LHC is rather small, 
$\Delta m_{\gamma\gamma}\sim\mathcal O (30)~{\rm \UMeV}$. This implies that
it will be extremely difficult for this method to access the region  
$\gHsV \lesssim  10 \times \gHsVSM$. Detailed projections at the HL-LHC can be found in Sec.~\ref{sec:5_interference_imag}.

The \emph{imaginary part of the interference}~\cite{Dixon:2013haa,Campbell:2017rke} 
in Eq.~(\ref{eq:width_gammagamma})
is symmetric around the Higgs peak, so it leads to a change in the rate. Unfortunately,
because of helicity conservation this imaginary part is highly suppressed at LO.
Higher order corrections provide a new mechanism to generate an imaginary part, lifting this suppression~\cite{Dixon:2013haa}.
However, because the bulk of the interference effectively enters at NLO,
the anticipated loop enhancement factor in the interference relative to the pure signal (mentioned above)
is not present, and the actual size of the effect is quite small. In the SM, 
it reduces the total rate by
about $2\%$, 
which makes it challenging to observe, and the effect is further diluted 
by additional radiation~\cite{Campbell:2017rke}. Thus this technique requires very good
control on the total rate, both experimental and theoretical. To reduce the former, it is 
profitable to consider cross-section ratios; for example, the $\gamma\gamma$ to $4l$ ratio
is projected to be measured at the few percent level.
However, this introduces additional experimental and theoretical systematics, including 
theoretical model dependence since one would need to make assumptions about the structure of Higgs 
couplings.
For this reason, it may be advantageous to perform the interference effect extraction in 
the $\gamma\gamma$ channel alone,
by considering different kinematic regions.
As with the real part of the interference,  this effect is also quite sensitive to the transverse momentum of the Higgs, with the bulk of the interference effect confined to the small $p_t$ region, as shown in an NLO analysis in Ref.~\cite{Campbell:2017rke}. However, since the interference is essentially an NLO effect, as discussed above,  the residual theoretical uncertainty at this order is still quite sizeable. Moreover, a fine-grained comparison of the low and high 
Higgs $p_t$ regions requires very good theoretical control. For the former, this is notoriously
complicated as several different effects are at play, see e.g.~\cite{Caola:2018zye} and references 
therein for a recent discussion of this point. Because of this, 
assuming a few percent experimental accuracy, the width extraction from this method would be
limited by theoretical uncertainties. Although computing higher order corrections for this
effect is well beyond our current ability, it is reasonable to assume that the situation
will improve on the HL/HE-LHC timescale, along the lines described in Section~\ref{sec2_HXSWG1}.
Currently, it is expected that this technique will lead to bounds of the order
$\gHsV \sim \mathcal O(10)\times \gHsVSM$, see section~\ref{sec:5_interference_real} for details. 

The main advantage of the on-shell width determinations discussed above is that they require minimal theoretical assumptions on potential
BSM effects. This is because couplings are extracted at the same energy scale, ideally
from the same process. However, since interference effects scale like 
$g_{\rm prod}\times g_{\rm dec}$ at the first power, the constraints on the width are relatively mild.
Indeed, if one assumes that the on-shell rates are kept fixed, a linear dependence on the coupling 
translates into a square root dependence on the width. 

Another option to constrain the width is 
\underline{off-shell methods}~\cite{Kauer:2012hd,Caola:2013yja,Campbell:2013una,Campbell:2013wga}, 
which are based on the following observation.
Schematically, the cross section can be written as
\begin{equation}
\sigma \sim \frac{g_{\rm prod}^2 \times g_{\rm dec}^2}{(s-\mhsV^2)^2 + \mhsV^2 \gHsV^2}.
\label{eq:width_gensigma}
\end{equation}
On the resonant peak, this leads to the usual relation in Eq.~(\ref{eq:width_standard}). Typically,
most of the cross section is concentrated there. In the $VV$ decay channel though there is a sizeable
contribution from the off-shell $s \gg \mhsV^2$ region~\cite{Kauer:2012hd}: 
indeed, Higgs decay to vector bosons is 
strongly enhanced at high energy. In the far
off-shell region, Eq.~(\ref{eq:width_gensigma}) reduces to 
$\sigma\sim (g_{\rm prod}^2\times g_{\rm dec}^2)/s^4$. Assuming that the on-peak rates are kept
fixed, this quadratic dependence on the couplings translates into a linear dependence on $\gHsV$, allowing this quantity
to be constrained by a comparison of on- and off-shell rates. 

However, it is important to stress that to extract $\gHsV$ from off-shell
measurements one has to assume that on-shell and off-shell couplings are the same. Since the two
are evaluated at very different energy scales, this introduces a theoretical model dependence that is not present in the on-shell methods. 
Indeed, there are several new physics scenarios where BSM effects decorrelate on- and off- shell 
couplings, see e.g.~\cite{Azatov:2014jga,Englert:2014aca,Logan:2014ppa,Englert:2014ffa} and sections~\ref{sec4:hwh},~\ref{sec:VBFdim6eft} where some of these effects are discussed explicitly. 
These include, for example, new light degrees of freedom coupled to the Higgs,
additional Higgs states, or anomalous $HVV$ couplings. Therefore, to constrain the width using an off-shell analysis,
 it is important to perform complementary measurements to control potential BSM effects. 
This was studied in detail for the case of $HVV$ anomalous couplings in~\cite{Anderson:2013afp}. 
 Projections at the HL-LHC will be presented in section~\ref{sec:5_offshell}. 
In general, off-shell measurements offer the opportunity to investigate Higgs interactions at 
high energy scale, thus leading to interesting information that is not limited to the width
extraction. For example, in combination with measurements of boosted Higgs, $HH$ and $t\bar tH$, 
an off-shell analysis can help lifting the degeneracy between $ggH$ and $t\bar tH$ 
couplings~\cite{Azatov:2014jga}. The off-shell program will clearly benefit from the 
increased statistics and energy of the HL/HE upgrade. For example, this would allow for 
off-shell studies in the VBF production mode~\cite{Campbell:2015vwa}. 
Although the rate here is very small,
by looking at same-sign vector boson final states one can significantly reduce backgrounds. 
Although it is estimated that HL-LHC measurements in this channel would lead to constraints
at the same level of current ones in the $ggF$ channel~\cite{Campbell:2015vwa}, 
the completely different 
production mechanism
makes them complementary to the $ggF$ constraints, thus allowing for a less model dependent
interpretation.
Aside from these considerations, it is interesting to study the potential of 
future LHC upgrades to constrain $\gHsV$ under the assumption that 
no large decorrelations between on- and off-shell couplings occur. Because of the linear dependence on
the width discussed above, such constraints are rather powerful. Indeed, assuming a reasonable
reduction in the theoretical uncertainty in the HL-LHC timescale, 
it will be possible to probe values close to the SM value
$\gHsV\sim 4~{\rm \UMeV}$. Projections under different assumptions for the theoretical uncertainty
are reported in section~\ref{sec:5_offshell}. 

A reliable theoretical description of the off-shell region is non trivial. First, there is a large
$q\bar q \to VV$ background, which needs to be properly subtracted to access the signal yield. 
More important, there is an irreducible $gg\to VV $ continuum background that interferes with the
signal process $gg\to H\to VV$. The interference effect is sizeable and destructive, which is a 
consequence of the Higgs mechanism ensuring unitarity in the SM. Because of the large interference,
it is necessary to have
good theoretical control not only on the signal process but also on the continuum background 
amplitude. This is non trivial, since the $gg\to VV$ process is loop induced, so higher order
corrections -- expected to be large given the $gg$ initial state -- involve multi-loop amplitudes.
Moreover, at large invariant masses, the contribution of virtual top quarks to the amplitude becomes
dominant. Its proper description would then require multi-loop amplitudes involving internal massive
states, which are extremely challenging to compute. For this reason, exact predictions for the
background amplitude are only known to LO in the off-shell region. NLO corrections are known below
the top threshold, and only in an approximate form 
above~\cite{Caola:2016trd,Campbell:2016ivq,Campbell:2014gua,Caola:2015psa,Caola:2015rqy,Bonvini:2013jha,Alioli:2016xab}. 
Nevertheless, recent developments in 
numerical techniques~\cite{Borowka:2016ehy} 
make NLO computations for the background feasible in the near
future. One subtle point in this discussion is the role of quark-initiated reactions. On the one
hand, they appear naturally in the computation of NLO corrections to $gg\to VV$ from initial
state splitting. On the other hand this kind of contribution -- although separately finite and 
gauge invariant -- only forms a small subset of the whole $qg \to VV q$ process at 
$\mathcal O(\alpha_s^3)$, which are part of the genuine N$^3$LO corrections to the 
quark-initiated $q\bar q\to VV$ process. Therefore, only including  
the contribution coming from initial
state splitting in the $gg\to VV$ process, although formally possible, may not entirely
capture the correct physics. In general, this problem is not particularly relevant because
the gluon channel provides the bulk of the contribution. This is however no longer the case if
strong requirements on extra jet activities (typical e.g. for $WW$ analysis) are imposed. 
Understanding this issue is an interesting theoretical problem, and the high statistics
available at the HL/HE-LHC motivates its detailed investigation. Another issue that should be 
investigated is the impact of electroweak corrections, which can be sizeable at high energy. Once again, 
although they are currently unknown, it is natural to expect progress in this direction within the HL-LHC timescale. 

The modelling of the $gg\to H\to VV$ process is under better control than the background one. 
Still, since
in the far off-shell region the top loop cannot be approximated by a contact interaction,
computations are still much harder than in the on-shell region, where such an approximation is justified. As a consequence, exact results are
only known to NLO. A full computation of NNLO corrections would require significant advances on 
current technology, which are however likely to occur in the HL-LHC timescale. It is reasonable
to expect~\cite{deFlorian:2016spz} 
that the $K$-factor for the exact theory is rather similar to that obtained from calculations in which
the top loop is integrated out. In the absence of an exact calculation, one can use this 
approximation to estimate rates at the HL/HE LHC. 

The HL/HE-LHC upgrade will improve off-shell analysis in several ways. On the one hand,
the larger statistics will allow for a better discrimination of the $q\bar q\to VV$ vs
$gg\to VV$ background and -- crucially -- interference. Currently, this is done by using
the different kinematic behaviour of these contributions. Clearly, a higher statistical
sample would allow for more powerful discrimination. Furthermore, increasing the collider
energy would lead to a larger fraction of gluon initiated events w.r.t. quark initiated
events. For example, the $(gg\to H\to VV)/(q\bar q\to VV)$ ratio increases by a factor of roughly
1.5 in the off-shell region when the centre-of-mass energy is increased from 14 \UTeV to 27 \UTeV. Furthermore, the increase in the total rate at the HE-LHC
will lead to a significant number of off-shell events in the few-\UTeV region. This would
allow for precise investigations of the Higgs sector in the high-energy region, which could 
shed light on the unitarity structure of the SM.

\asubsection[G. Barone, A. Gabrielli]{Measurement of the Higgs boson mass}
The measurement of the Higgs boson mass by the ATLAS and CMS
experiments at the LHC is~\cite{Tanabashi:2018oca}: $$m_H = 125.18 \pm
0.16 \; \rm \UGeV$$

This precision is reached with the two high resolution Higgs boson
measurement channels the $H\rightarrow ZZ^* \rightarrow 4\ell$ and
$H\rightarrow \gamma \gamma$. At the LHC Run 2, the precision in the
latter channel is already limited by the systematic uncertainty on the
photon energy scale. The photon energy response is calibrated using
both electrons from Z decays (which requires to be extrapolated from
electrons to photons) and radiative Z events reconstructed with two
charged leptons and a photon, which is limited by statistics in the
transverse momentum range of interest. The most precise measurement is
obtained in the $4\mu$ and the $2e2\mu$ where the on-shell Z mass
constraint can be applied to the $2e$ system.

Detailed studies of the calibration of the muons, electrons and
photons with the very large HL-LHC sample have not been done yet,
however it is plausible that the mass of the Higgs boson will be
measured with a precision of 10-20 \UMeV, assuming that with the higher
statistics the analysis will be further optimised to gain in
statistical precision and that systematic uncertainties on the muon
transverse momentum scale will significantly improve with the higher
statistics.

\label{sec:5_HiggsMass}



\asubsection[M. Xiao, U. Sarica]{Constraints from off-shell Higgs boson production}

Extracting the Higgs width from \offshell measurements are performed in ATLAS~\cite{Aaboud:2018puo} and CMS~\cite{CMS:2018bwq} experiments using LHC Run 1 and Run2 data. The latest constraints on the Higgs width are $< 14.4$~\UMeV and $<9.2$~\UMeV from ATLAS and CMS, respectively. Theoretical basis of the measurements is that on-shell and off-shell couplings are the same. Developed on the experimental analyses, the expected precision on the Higgs width at the luminosity of 3000~$\fbi$ is given in this section.  

The CMS projection adopted the same analysis strategy as defined in the Run~2 analysis~\cite{CMS-PAS-HIG-18-002}, where the $4\ell$ final state is used. Events are selected and put into different categories that are sensitive to ggF, VBF and VH production modes. The invariant mass of the four leptons, matrix-element based discriminant separating the major signal and background events and discriminants sensitive to the width are used in each category. The ratio between \offshell and \onshell event yields and the shape of the observables are sensitive to the Higgs width. To extrapolate to 3000~$\fbi$, event yields are scaled with luminosity. Assumptions on the uncertainties are made, and two scenarios are considered~\cite{CMS-PAS-FTR-18-011}:
\begin{itemize}
\item {\bf ``Run 2 systematic uncertainties'' scenario (S1):} All systematic uncertainties are kept constant with
  integrated luminosity. The performance of the CMS detector is
  assumed to be unchanged with respect to the reference analysis;

\item {\bf ``YR18 systematic uncertainties'' scenario (S2):} Theoretical uncertainties are scaled down by a factor of two,
  while experimental systematic uncertainties are scaled down with the
  square root of the integrated luminosity until they reach a defined
  minimum value based on estimates of the achievable accuracy with the
  upgraded detector.
\end{itemize} 

The projections are shown in Fig.~\ref{fig:GH-projections}. Limits on \GH are given for an approximate S2 in which the experimental systematics are not reduced, while the theoretical systematics are halved with respect to S1. The $10\%$ additional uncertainty applied on the QCD NNLO K factor on the \glufu background process is kept the same in this approximated S2 in order to remain conservative on the understanding of these corrections on this background component. It is also noted that the uncertainties on the signal and background QCD NNLO K factors are smaller in the Run~2 analysis~\cite{CMS:2018bwq} than in previous projections using Run~1 data~\cite{ATL-PHYS-PUB-2015-024}. The expected \GH precision in S2 is $4.1 ^{+1.0}_{-1.1}$ \UMeV. 

\begin{figure*}[!htbp]
\centering
\includegraphics[width=0.45\textwidth]{\main/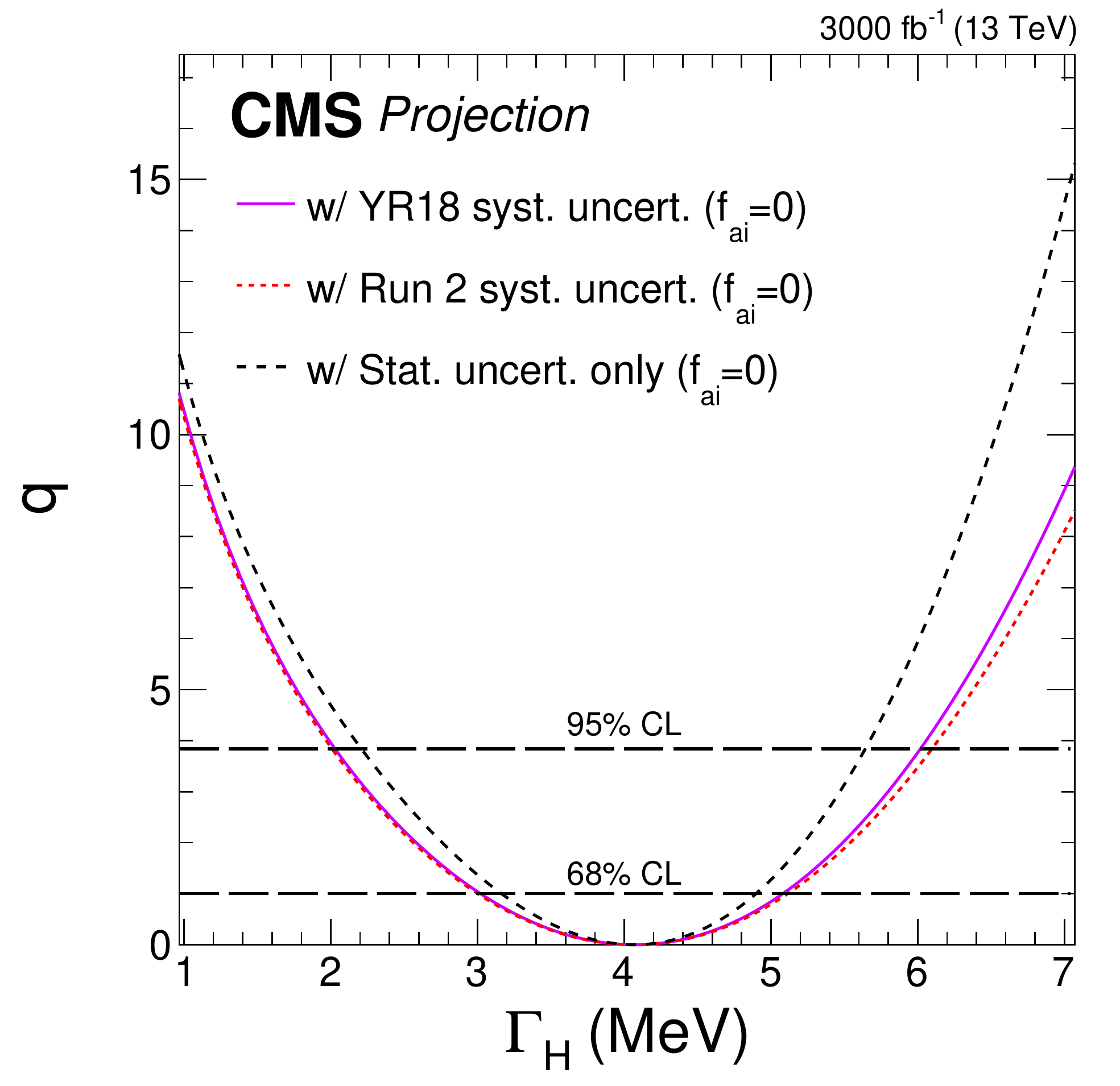}
\caption
{
Likelihood scans for projections on \GH at $3000~\fbi$~\cite{CMS-PAS-FTR-18-011}. Scenarios S2 (solid magenta) and S1 (dotted red) are compared to the case where all systematics (dashed black) are removed. The dashed horizontal lines indicate the 68\% and 95\% CLs. 
}
\label{fig:GH-projections}
\end{figure*}

The ATLAS projection~\cite{ATL-PHYS-PUB-2015-024} is based on the ATLAS Run1 analysis~\cite{PhysRevD.91.012006}. $\PH\to\PZ\PZ\to4\ell$ final state is used. Events are selected and put in ggF, VBF and VH categories. The invariant mass of the four leptons and a matrix-element based discriminant sensitive to both signal background separation and width variation are used. In the extrapolation to 3000 $\fbi$, event yields are scaled with luminosity and the change in the centre mass of energy. Only theoretical uncertainties are taken into account, as the experimental ones have a negligible impact. The treatment of theoretical uncertainty is close to Run1 analysis, with more conservative ones below: 
\begin{itemize}
	\item {The k-factor uncertainty for the gg initial state signal, background and their interference is taken as 30\%. Based on the latest theory papers, this uncertainty is considered to be conservative, and is 10\% is the CMS projection result.}
	\item {The background to signal k-factor ratio $R^{B}_{H}(\rm{mZZ})$ uncertainty, two benchmarks are considered: 10\% and 30\%. }
\end{itemize}
The expected precision on $\GH$ at 3000 $\fbi$ is $4.2^{+1.5}_{-2.1}$ \UMeV as shown in Fig.~\ref{fig:ATLAS_projections}. It is more conservative than the CMS result, and the cause of it was discussed above. 
\begin{figure*}[!htbp]
\centering
\includegraphics[width=0.45\textwidth]{\main/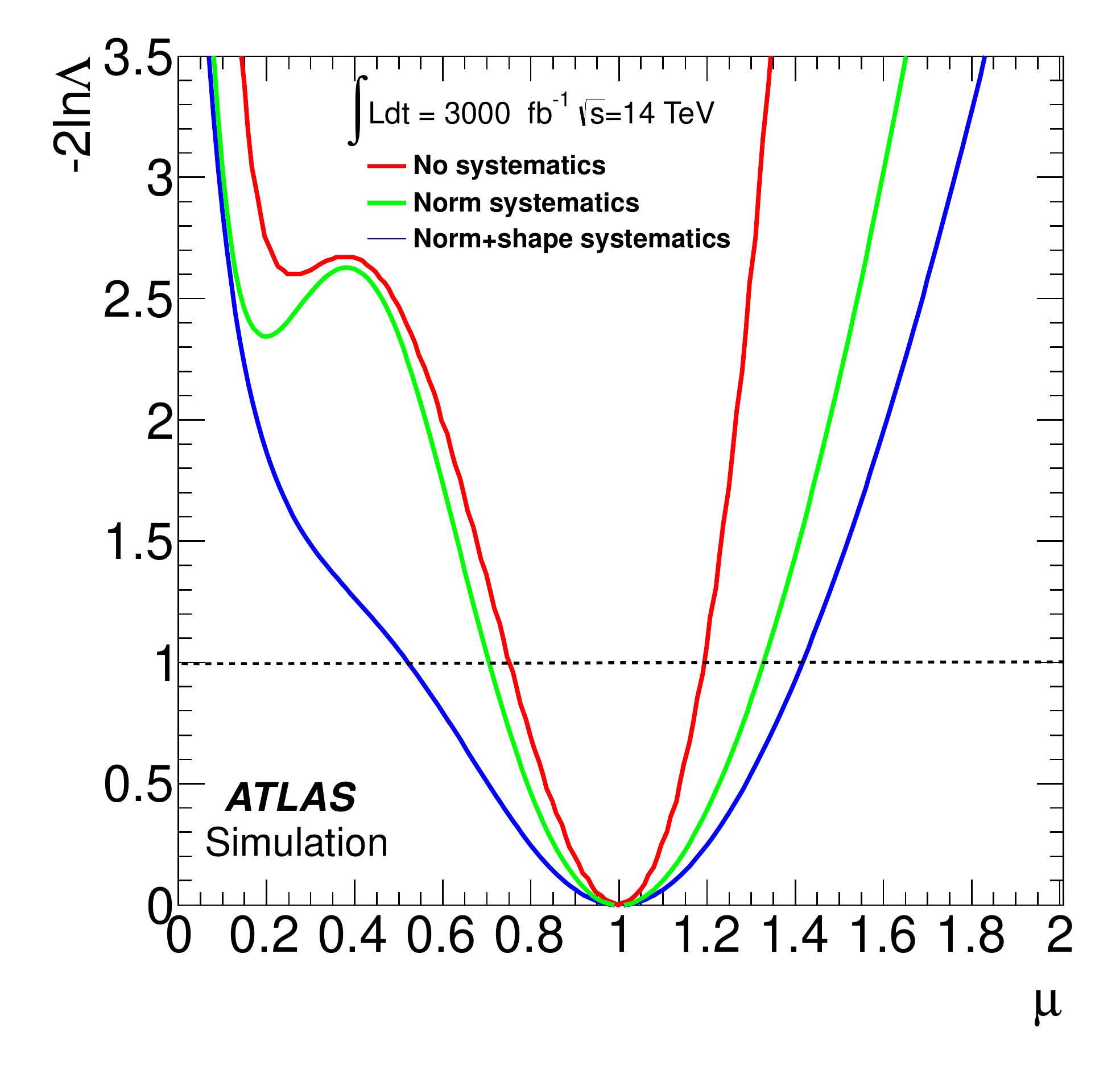}
\caption{Likelihood scans on $\mu_{\rm{off-shell}}$ with and without systematic uncertainties. The error on $\mu$ is computed at the $1\sigma$ level and the uncertainty on $R^{B}_{H}(\rm{mZZ})$ is set to 30\%.}
\label{fig:ATLAS_projections}
\end{figure*}

In conclusion, it is reasonable to believe the realistic precision from ATLAS at 3000~$\fbi$ will be better than the number above. Using the CMS numbers, we con estimate that with CMS and ATLAS measurements combined, the precision on the width can reach $4.1 ^{+0.7}_{-0.8}$ \UMeV.

\label{sec:5_offshell}

\asubsection[J. Campbell, M. Carena, R. Harnik, Z. Liu. This manuscript has been authored by Fermi Research Alliance, LLC under Contract No. DE-AC02-07CH11359 with the U.S. Department of Energy, Office of Science, Office of High Energy Physics.]{Width from the di-photon interference rate}
\label{sec:5_interference_real}


The SM Higgs total decay width can be constrained from the change in on-shell Higgs rates due to interference effects between the Higgs signal and the QCD background~\cite{Campbell:2017rke}. This change in rates requires the existence of a so-called strong phase in the amplitudes, that can be present both  in the Higgs signal and in the continuum background, as is the case in the SM. We shall demonstrate that,
the different scaling behaviour between   the strong phase induced interference  and the Breit-Wigner parts of the on-shell Higgs rate may allow the placement of bounds on, or even measurements of, the Higgs boson total width.
Both theoretical and experimental uncertainties are the leading  limiting factors in this program. On the other hand, without  the strong phase induced interference effects, fits to on-shell Higgs rates can only place bounds on the total width by making definite theoretical assumptions~\cite{Duhrssen:2004cv,LHCHiggsCrossSectionWorkingGroup:2012nn,Dobrescu:2012td}. 

It is useful to write the amplitude for $gg \to h \to \gamma\gamma$ in a form which explicitly factors out the loop-induced couplings to gluons
($F_{gg}$) and photons ($F_{\gamma\gamma}$),
\begin{equation}
A_h \equiv A_{gg\to h\to\gamma\gamma}  \propto \frac {\hat s} {\hat s-m_h^2+i\Gamma_h m_h} F_{gg} F_{\gamma\gamma} \,.
\label{eq:Ah}
\end{equation}
Both the Higgs couplings $F_{gg}$ and $F_{\gamma\gamma}$ as well as the background amplitude $A_\mathrm{bkg}$
receive absorptive contributions that arise from loops of particles that are sufficiently light to be on shell. 
The resulting induced phases are usually dubbed `strong phases' in the flavor literature and we will adopt this terminology here.\footnote{Strong phases, which are CP even, get their name because they often arise in flavor physics from QCD dynamics. This is in contrast with CP odd weak phases, e.g., the relative size of the Higgs couplings to $F\tilde F$ versus $FF$. 
}
In the presence of a strong phase we can write the interference term as
\begin{eqnarray}
|\mathcal{M}_h|^2_\mathrm{int} &&\equiv 2 \Re[A_h A_\mathrm{bkg}^*]= 
\frac {2 |A_\mathrm{bkg}||F_{gg}||F_{\gamma\gamma}|} {(\hat s - m_h^2)^2+\Gamma_h^2 m_h^2} \label{eq:phase}\\
&&\!\!
\times \!\left[ (\hat s-m_h^2) \cos (\delta_\mathrm{bkg}-\delta_h) \!+ \!m_h\Gamma_h \sin (\delta_\mathrm{bkg}-\delta_h) \right]\!\!, \nonumber
\end{eqnarray}
where we have taken $\delta_h=\mathrm{arg}[ F_{gg}]+\mathrm{arg}[F_{\gamma\gamma}]$ and $\delta_\mathrm{bkg}=\mathrm{arg}[ A_\mathrm{bkg}]$ as the signal and background strong phases, respectively.
The first term in the square bracket is the contribution to the interference term that does not modify the overall rate upon integration over $\hat s$. The second term is the subject of this work and leads to a modified rate in the presence of a strong phase. For convenience, we define $|\mathcal{M}_h|^2_\mathrm{int}=\mathcal{R}_h^\mathrm{int}+\mathcal{I}_h^\mathrm{int}$ and $\delta_s=\delta_\mathrm{bkg}-\delta_h$ such that
\bea
\mathcal{R}_h^\mathrm{int}&\equiv& \frac {2|A_\mathrm{bkg}||F_{gg}||F_{\gamma\gamma}|} {(\hat s - m_h^2)^2+\Gamma_h^2 m_h^2} (\hat s-m_h^2) \cos \delta_s \nonumber \\
\mathcal{I}_h^\mathrm{int}&\equiv& \frac {2|A_\mathrm{bkg}||F_{gg}||F_{\gamma\gamma}|} {(\hat s - m_h^2)^2+\Gamma_h^2 m_h^2} m_h\Gamma_h \sin \delta_s.
\eea

\begin{figure}[htbp]
\begin{center}
\includegraphics[scale=0.4,clip]{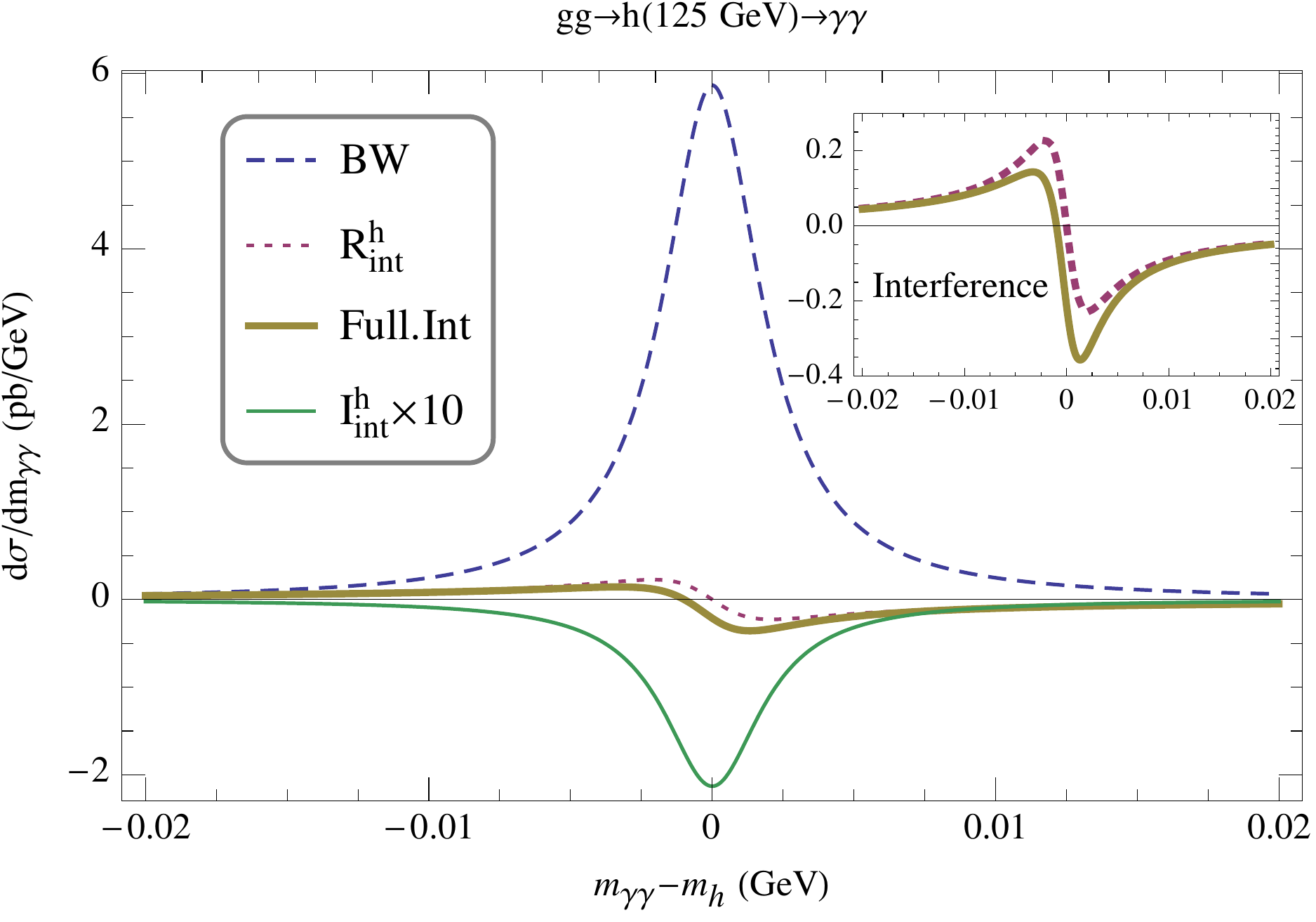}
\caption{
The line-shape induced by various contributions to the cross-section for $gg \to h \to \gamma\gamma$ in the SM.
The Breit-Wigner line-shape, with no interference, is shown in blue (dashed) while the effect of $\mathcal{R}_h^\mathrm{int}$ and
$\mathcal{I}_h^\mathrm{int}$ (multiplied by a factor of 10) are shown in red (dotted) and green (solid), respectively.  The overall
effect of the interference in the full NLO calculation is given by the brown (solid) line. The insert in the top right is a magnification of the corresponding interference line-shapes.
}
\label{fig:sig_shape}
\end{center}
\vspace*{-0.6cm}
\end{figure}

In the SM the dominant contribution to $\mathcal{I}_h^\mathrm{int}$ comes from the phase of the background amplitude at two loops~\cite{Dicus:1987fk, Dixon:2003yb}. The signal amplitude also contains a
strong phase, mainly due to bottom quark loops. 
We have performed a calculation of the interference effect that accounts for absorptive effects from both signal and background.
In Fig.~\ref{fig:sig_shape} we illustrate the features of the interference effects.
The line shape, the differential cross-section as a function of~$\hat s$, is shown
for the pure Breit-Wigner (only $|A_h|^2$), and for the interference contributions $\mathcal{I}_h^\mathrm{int}$ and $\mathcal{R}_h^\mathrm{int}$ as well as for the sum of both. 
For visualisation, the interference contribution $\mathcal{I}_h^\mathrm{int}$ has been magnified by a factor of 10. 
In this figure we show the line-shapes obtained including NLO effects with virtual corrections only. After summing over different interfering helicity amplitudes, we obtain averaged strong phases $\delta_h= (\pi+0.036)$ and $\delta_\mathrm{bkg}=-0.205$ for the signal and background, respectively.

As a concrete example that demonstrates the potential of this novel effect, without loss of generality
we can consider excursions in the flat direction corresponding to,
\be
\frac{|F_{gg}|^2|F_{\gamma\gamma}|^2}{|F^{\rm SM}_{gg}|^2|F^{\rm SM}_{\gamma\gamma}|^2}
 = \frac{\Gamma_h}{\Gamma^{\rm SM}_h} \,.
\label{eq:flatdir}
\ee
The total Higgs cross section can then be written as,
\be
\sigma = \sigma_\mathrm{BW}^\mathrm{SM}\!\! \left(\!\!1\!+\! \frac {\sigma_\mathrm{int}^\mathrm{SM}} {\sigma_\mathrm{BW}^\mathrm{SM}}\sqrt{\frac {\Gamma_h} {\Gamma_h^\mathrm{SM}}}\right)
\simeq \sigma_\mathrm{BW}^\mathrm{SM} \!\!\left(\!\!1\!-\!2\%\sqrt{\frac {\Gamma_h} {\Gamma_h^\mathrm{SM}}}\right)\!\!.\label{eq:deltasigma}
\ee
%
The result of a full NLO calculation of the interference effect are presented in Fig.~\ref{fig:width}, that shows the relative size of the interference effect as a function of the total width, normalised to its SM value, for parameter excursions defined by Eq.~(\ref{eq:flatdir}). 
\footnote{For details of the 
NLO calculation
, see the supplemental material 
with Refs~\cite{Bern:1991aq,Bern:2001df,Bern:1995db,Bern:1993mq,Bern:2002jx,deFlorian:2013psa,Catani:1996vz,Campbell:2011bn,Dulat:2015mca,Djouadi:1997rj,Degrassi:2005mc,Passarino:2007fp,Dittmaier:2011ti}.}
The variation of the interference effect with the total width is shown imposing a 20~\UGeV $p_T^{h}$-veto, with and without LHC cuts on the final state photons. 
Since the interference effect is largest at small scattering angles, the photon cuts reduce the expected interference.  
This small consideration in the SM leads to much
bigger differences for $\Gamma_h \gg \Gamma_h^{\rm SM}$.
Observe that in the SM the interference contribution is destructive. However,  if the sign of $F_{gg} F_{\gamma\gamma}$ were flipped, ($\delta_s\to \pi+\delta_s$), the interference effect would lead to an enhancement of the di-photon rate rather than a suppression. 
The theoretical scale uncertainty is shown in the bottom panel of Fig.~\ref{fig:width} and amounts to about 
$^{+50\%}_{-30\%}$. For example, the interference effect is $-(2.20^{+1.06}_{-0.55})\%$ without photon cuts for SM Higgs.
\begin{figure}[htbp]
\begin{center}
\includegraphics[scale=0.4,clip]{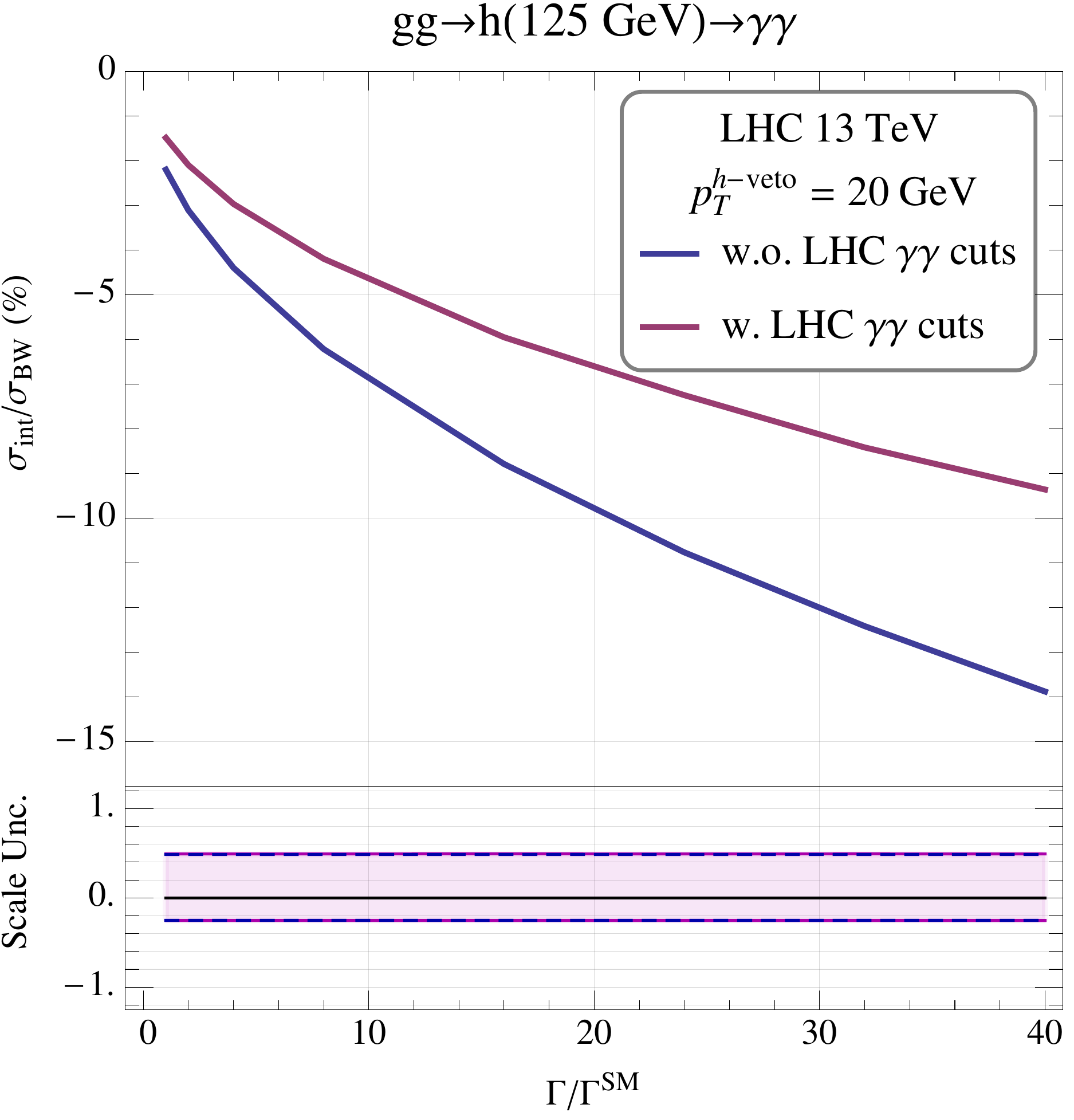}
\caption{
The total signal rate change due to the interference effect as a function of the Higgs total width normalised to its SM value, while keeping
the Breit-Wigner cross section identical to that of the SM Higgs. The magenta and blue (solid) lines represent the cases with and without LHC cuts
on the final state photons, respectively. The lower panel shows the scale variation uncertainties for these interference terms  as bands delimited by the blue (dashed)
 and magenta (solid) lines. The curves are obtained with a veto on the Higgs boson $p_T$  at 20 \UGeV. 
}
\label{fig:width}
\end{center}

\end{figure}
Although a measurement at the 2\% level may be challenging at the LHC, 
this shows that a precise measurement of the $gg\to h \to \gamma\gamma$ rate can place a limit on the width of the Higgs boson. 
In this respect a measurement of the ratio of the $\gamma\gamma$ rate to the $4\ell$ rate is a promising route to reduce many of the systematic and theoretical, e.g. PDF and other parametric, uncertainties.

The best measured channels at the LHC, $gg\to h\to \gamma\gamma$ and $gg\to h \to 4\ell$, provide the most accurate cross section ratio, projected to 
 be measurable at the 4\% level~\cite{ATL-PHYS-PUB-2014-016}.   In contrast to single cross section measurements, the precision on this ratio is statistically limited.
Keeping the current theoretical uncertainty band in mind, the projected sensitivity of 4\% on the ratio of $\gamma\gamma$ to $4\ell$ yields
can be translated into an upper  limit of 22, 14, and 8 on $\Gamma_h/\Gamma_h^\mathrm{SM}$ at 1-$\sigma$ level, for low, central and high theoretical expectations
on this interference effect, respectively.\footnote{This limit is worse by one order of magnitude than the off-shell Higgs measurement that constrains the Higgs total width~\cite{Kauer:2012hd,Caola:2013yja,Campbell:2013una}. However, unlike the off-shell Higgs measurement, our effect is independent from the assumptions on the high-energy behaviour of the Higgs boson and the absence of new physics contribution in the off-shell region. 
For more detailed discussion, see e.g., chapter I.8 of the Higgs Yellow Report~\cite{deFlorian:2016spz} and Refs.~\cite{Englert:2014aca,Logan:2014ppa}.} 
This assumes that the couplings to photons and $Z$ bosons maintain their SM ratio and the photon and gluon couplings respect Eq.~(\ref{eq:flatdir}).
The Higgs cross section precisions are anticipated to improve by a factor of three or so from statistical improvement at the HE-LHC with 27 \UTeV centre of mass energy and 15 $\fbi$ of integrated luminosity. 
This can be naively translated into  lower and upper limits on the Higgs total width of $\Gamma_h/\Gamma_h^\mathrm{SM}< 5$ at 1-$\sigma$ level using the central value from our NLO theory calculation. 

In summary, we discuss the change in the $gg\to h\to \gamma\gamma$ on-shell rate, due to interference between the Higgs signal and the QCD background amplitudes, as a way to provide a novel handle to constrain - or even measure~-  the Higgs boson total width. 
We perform a full NLO calculation at order $\alpha_s^3$ of the interference effect and find that in the standard model it leads to a reduction of the on-shell rate by $\sim 2\%$. 
The proposed  method for gaining sensitivity to the Higgs boson width is complementary to other methods that have been discussed in the literature. 
Altogether our study aims at motivating  a more thorough examination of Higgs precision physics  taking into account the  strong phase induced interference effect in different Higgs boson observables.

\asubsection[C. Becot]{Mass shift from the di-photon interference}
\label{sec:5_interference_imag}
A detailed study of the impact of the phenomenon of interference between the signal process $gg \rightarrow H \rightarrow \gamma\gamma$ and its irreducible background process $gg \rightarrow \gamma\gamma$ has been made in \cite{ATL-PHYS-PUB-2016-009}. It will be summarised below.

All the results from \cite{ATL-PHYS-PUB-2016-009} have been made with simulated data for the $\sqrt{s} = 8$ \UTeV dataset. A full-fledged extrapolation to higher centre of mass energies remains to be done. However, the two processes that interfere together, are induced by the same initial state and are at a similar mass scale. All the results are based on a difference between results made with simulated data where the interference have been implemented in the simulation, and simulated data where it is not. In the following, we will therefore consider that any increase of the cross-section with respect to $\sqrt{s}$ will cancel out in the difference.

The main goal of \cite{ATL-PHYS-PUB-2016-009} was a robust estimate of the mass shift induced by the interference between $gg \rightarrow H \rightarrow \gamma\gamma$ and its irreducible background process $gg \rightarrow \gamma\gamma$ within the Standard Model (SM). This was achieved by using the Monte-Carlo generator Sherpa 2.0, that provides a specific plug-in allowing to generate datasets of weighted events, corresponding either to the the signal term ($gg \rightarrow H \rightarrow \gamma\gamma$ amplitude), the irreducible gluon-induced background term ($gg \rightarrow \gamma\gamma$ amplitude), their interference term or a any combination of these processes. The Feynman diagram describing the different processes involved are given in fig. \ref{fig:intef_diags}. 

\begin{figure}
    \centering
    \includegraphics[width=0.85\textwidth]{\main/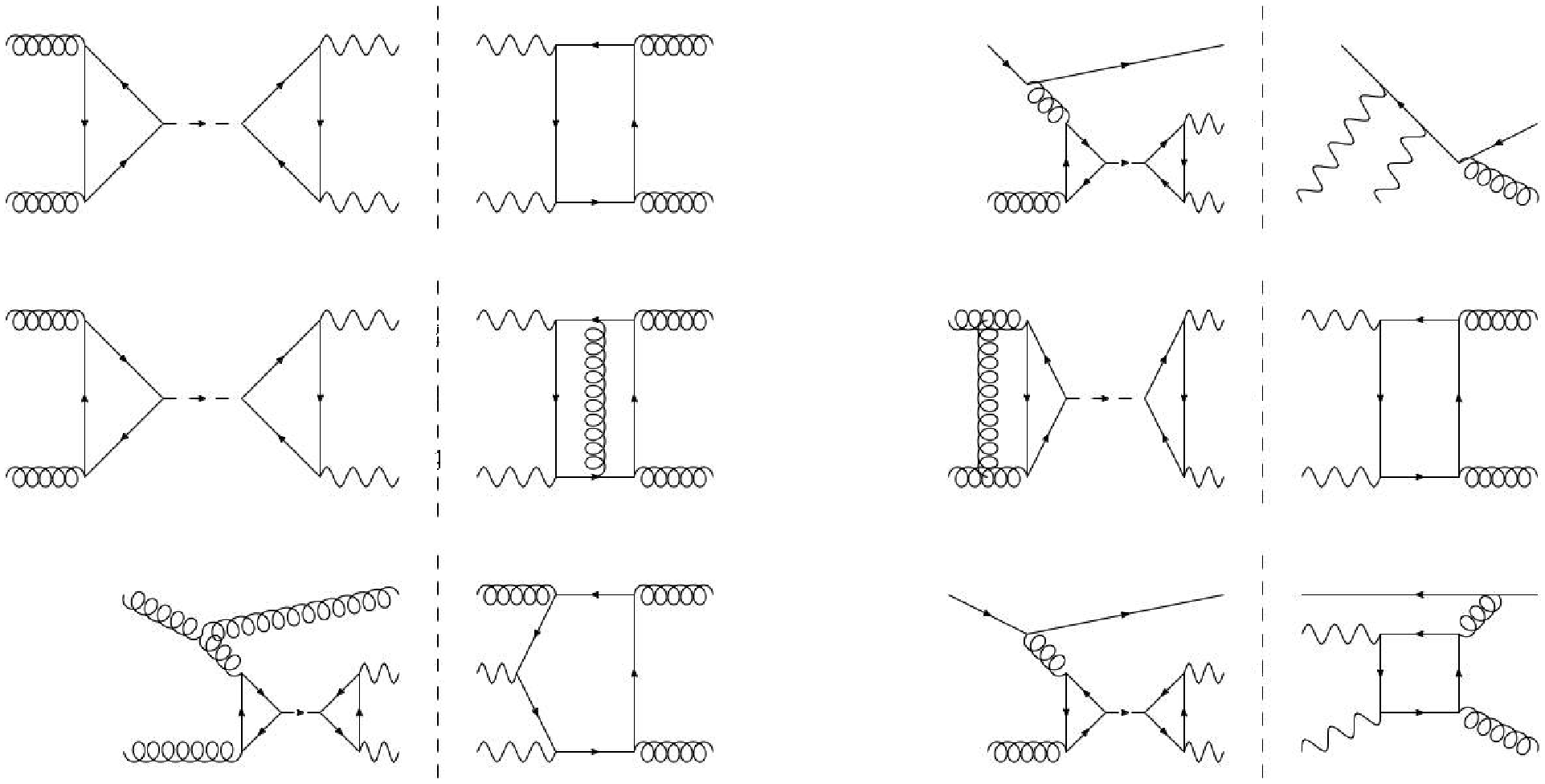}
    \caption{Feynman diagrams describing the various processes involved in the phenomenon of interferences between $H\rightarrow\gamma\gamma$ and its background,}
    \label{fig:intef_diags}
\end{figure}

It was the first time that Sherpa 2.0 was used in an analysis, and in particular the only time its interference module has been used. The distribution of the di-photon pair transverse momentum ($P_T^{\gamma\gamma}$) has been studied in detailed, as it is a preliminary requirement to be able to recast the mass analysis described in \cite{Aad:2014aba}. To get the best match of the $P_T^{\gamma\gamma}$ distribution between this simulation and the state of the art estimates, Sherpa 2.0 had to be tuned. This was done by varying the parameter $CSS\_IS\_AS\_FAC$ that controls the scale at which the parameter $\alpha_S$ is evaluated during parton shower evolution for the initial state. Simulated data samples were generated at several values of this parameter and compared to prediction for the Higgs boson transverse momentum from HRes2.0. The value of $CSS\_IS\_AS\_FAC$ for which both predictions were agreeing the best has been kept for the simulations used for the final result. The distributions of $P_T^{\gamma\gamma}$ obtained for a simulation of the background has been compared between ResBos and Sherpa2.0 for the best value of $CSS\_IS\_AS\_FAC$. They were found to be in reasonable agreement.

Then large samples of weighted signal (S) and interference (I) events were simulated. The background component was determined using a data-driven method (B). The background functions described in \cite{Aad:2014aba} were re-used, and they were fitted to the $\sqrt{s} = 8$ \UTeV dataset. This allowed to construct two varieties of simulated $m_{\gamma\gamma}$ templates, one that was made of $S+B$ and the other of $S+B+I$. These samples were then folded by the energy resolution and photon efficiencies used in \cite{Aad:2014aba} to mimic the detector simulation. The di-photon mass distributions induced by these different terms can be found in fig. \ref{fig:interfer_lineshape}. The Higgs boson mass ($m_H$) is measured separately on both templates, giving the values of $m_H$ including the impact of interference ($m_H^{S+B+I}$), and without ($m_H^{S+B}$). Then the impact of the interference term itself on $m_H$ is determined as \begin{equation}
    \Delta m_H = m_H^{S+B+I}-m_H^{S+B}.
\end{equation}

\begin{figure}
    \centering
    \includegraphics[width=0.6\textwidth]{\main/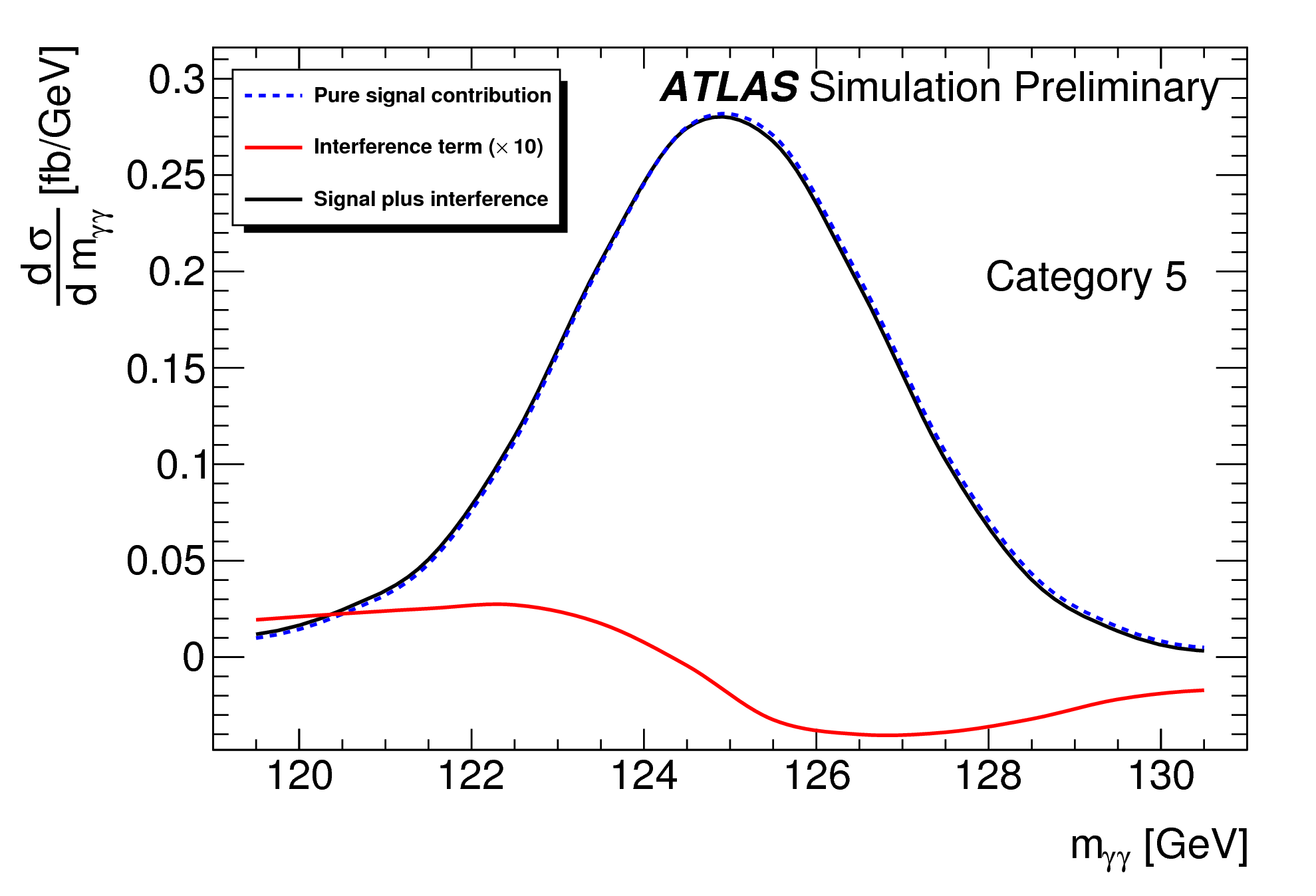}
    \caption{Di-photon invariant mass distribution for signal and interference terms.}
    \label{fig:interfer_lineshape}
\end{figure}

Several uncertainties have been considered. First a non-closure of 3 \UMeV has been observed while measuring $m_H^{S+B}$ from the S+B simulated sample, and is propagated as an experimental uncertainty. The only other experimental uncertainty stems from the choice of background function, and has been estimated at 3 \UMeV by trying different background functions.

For theoretical uncertainties we considered both scale variations as well as K-factor variations, as we will describe now.
The renormalisation  and factorisation scales were varied by a factor of 2, between $\frac{1}{2}~m_{\gamma\gamma}$ and $2~m_{\gamma\gamma}$ (the nominal value is $m_{\gamma\gamma}$). The resummation scale itself was varied between $\frac{1}{4}~m_{\gamma\gamma}$ and $2~m_{\gamma\gamma}$. Samples were re-generated at these different values of the scales for S and I, and the value of $\Delta m_H$ was re-evaluated. This gave a small uncertainty of $\pm 4$ \UMeV, which can be explained by the migration of events between transverse momentum categories that leads to a cancellation of these variations in the difference $\Delta m_H$.

The signal K-factor ($k_S = \dfrac{NNLO}{NLO}$) has been varied by 0.1, between $k_S = 1.35$ and $k_S = 1.55$. The background K-factor $k_B$ is not known, and it was decided to vary it between 1 and $k_S$. The recipe gave an uncertainty of $\pm 7$ \UMeV, which is the dominant uncertainty. Constraining $k_B$ could lead to a huge improvement, but it requires to separate the component $gg \rightarrow \gamma\gamma$ from the inclusive $pp \rightarrow \gamma\gamma$ production. This is expected to be complicated but might be achieved by the study of angular distributions in $pp \rightarrow \gamma\gamma+jets$.

This study conducted to the following estimate of the Higgs boson mass shift induced by the interference process within the Standard Model :

\begin{equation}
    \Delta m_H = -35 \pm 8 {\rm{(theory)}} \pm 4 {\rm{(experimental) \UMeV}}.
\end{equation}

The mass shift was also determined for larger widths, in the specific scenario where the width is modified but the expected number of events around the di-photon peak (S+I), is not. The mass shift was determined to be $\Delta m_H = -313\pm 72$ \UMeV for $\Gamma_H = 300$ \UMeV and $\Delta m_H = -453 \pm 106$ \UMeV for $\Gamma_H = 600$ \UMeV. By the end of HL-LHC such deviations could be probed by comparing the mass measured in the high-precision $H\rightarrow 4l$ channel and the one measured in $H\rightarrow\gamma\gamma$. The following difference :
\begin{equation}
    \delta m_H^{4l-\gamma\gamma} = m_H^{4l} - m_H^{\gamma\gamma}
\end{equation}
will be largely dominated by the systematic uncertainty on $m_H^{\gamma\gamma}$. Assuming it is at the same level than the one of Run 2 \cite{Aaboud:2018wps}, it would lead to an uncertainty on the measured $\delta m_H^{4l-\gamma\gamma}$ of 290 \UMeV, hence a shift of 580 \UMeV could be excluded at 95\% C.L. Also assuming that $\delta m_H^{4l-\gamma\gamma}$ scales linearly with the Higgs boson width, it would allow in this naive model, to set upper limits on the Higgs boson width at $\Gamma_H < 1$ \UGeV at 95\% C.L.

This is only using the difference between the Higgs boson mass measured in its 4 leptons decay channel and the one measured in its di-photon decay channel. Now, it is also known that the interference term will have a bigger impact in some parts of the phase space where the signal to background ratio is smaller. This is for instance the case at low $p_T^{\gamma\gamma}$, and it could be used to carry out the same inference internally in the $H\rightarrow\gamma\gamma$ channel, comparing the masses measured in two (or more) $p_T^{\gamma\gamma}$ bins. A full-fledged prospect study of this analysis remains to be done, as the only attempt carried out so far \cite{ATL-PHYS-PUB-2013-014} suffered from large mis-modellings of the kinematic distributions and of the cross-sections. The success of such an analysis will rely on very precise predictions for the kinematic distributions, which is not yet available. This will require the development of new higher-order calculations and resummed predictions, and could be helped by the development of new Monte-Carlo tools using better showering algorithms.



\newpage

\asection[A. Magnan, B. Nachman, T. Robens and T. Stefaniak]{Invisible decays of the Higgs boson}\label{Sec:6Invisible}

Invisible decays of the $125~\mathrm{\UGeV}$ Higgs boson are a generic prediction of new physics models that feature a light dark matter (DM) particle which couples directly or indirectly to the Standard Model Higgs field. The invisible branching ratio in the SM is very small (0.1\%) so any observable rate would be evidence for physics beyond the SM.  LHC searches for this signature require the Higgs boson to be produced in association with a taggable object, most importantly, a $Z$-boson, extra forward jets (as appearing in the vector boson fusion process), or a single high-$p_T$ jet. Furthermore, the invisible Higgs decay to the DM particles inevitably suppresses the branching fractions for the Higgs decays to SM particles. This --- along with possible model-dependent alterations of the Higgs couplings --- leads to a modification of the LHC Higgs signal rates of channels with SM final states with respect to their SM expectation, which can be probed with precision Higgs measurements. 

In so-called \emph{Higgs portal} models the SM Higgs field acts as a mediator between the visible SM sector and a hidden DM sector. {Commonly,} an additional symmetry {is introduced that} prohibits interactions of single hidden sector fields with SM fields, thus allowing only pair production of hidden sector particles and rendering the lightest hidden sector particle a stable DM candidate. 
The \emph{Higgs portal} and its generalisation to other non-SM Higgs bosons are found in many BSM scenarios (see e.g.~Refs.~\cite{McDonald:2008up,McDonald:2008ua,ArkaniHamed:2006mb,Profumo:2017ntc} and~\cite{Barger:2008jx,Cohen:2011ec,Englert:2011yb,Goudelis:2013uca,Bai:2012nv,Berlin:2015wwa} for models with and without supersymmetry). 

The invisible decay of the Higgs is experimentally challenging because the missing transverse energy spectrum is relatively soft, where resolution and pileup effects are non-negligible.  The issues associated with pileup, both from pileup jets and from pileup-induced resolution degradation, will only become more severe beyond Run 3.  Significant recent advances in constituent-based pileup mitigation techniques will likely play a key role in maintaining and possibly improving the MET performance~\cite{Cacciari:2014gra,Bertolini:2014bba,Berta:2014eza,Komiske:2017ubm,Martinez:2018fwc}.  Furthermore, lepton identification and pileup jet rejection will both improve with the increased tracking acceptance planned by both ATLAS and CMS~\cite{Klein:2017nke,Collaboration:2283187,Collaboration:2283189,Collaboration:2293646,Collaboration:2017mtb,Collaboration:2285585}. Current analyses with $\sim 30$ fb$^{-1}$ place limits on the invisible branching ratio of the Higgs boson at about 20- 25\%~\cite{Khachatryan:2016whc,Aad:2015pla,Aaboud:2017bja,Aad:2015txa,Sirunyan:2018owy,Aaboud:2018sfi}.  The systematic uncertainty is about the same size as the statistical uncertainty; this means that the factor of 100 increase in statistics will not necessarily translate into an improvement by a factor of 10.  Early projections from ATLAS and CMS~\cite{CMS-PAS-FTR-16-002,ATL-PHYS-PUB-2013-014} predict limits that are a factor of 3-5 below the current result. The main limiting systematic uncertainty is from using $W\rightarrow l\nu$ to estimate the $Z\rightarrow \nu\nu$ in the dominant VBF channel. Advances in this theory input over the next decade could significantly improve the achievable precision. Already, CMS has shown that optimistic projections with reduced systematic uncertainties are realistic - the 2016 analysis~\cite{Sirunyan:2018owy} follows optimistic (reduced systematic uncertainties) scaling from the 2015 projection~\cite{CMS-PAS-FTR-16-002}.

Currently, the VBF production dominates the branching ratio limit.  This is because the VBF mode has a large cross-section (about 10\% of the total) and the main background $Z\rightarrow\nu\nu$ is qualitatively different (QCD production) from the same background in the $VH$ mode (EW production).  However, it is not clear which mode will dominate after Run 3, since there will be a non-trivial change in experimental conditions that will make both triggering and background rejection more difficult for both the VBF and $VH$ modes.  At the same time, there are many interesting opportunities to improve both channels from new detector capabilities (extended trackers and timing detectors) as well as new analysis techniques (e.g. quark/gluon tagging).

In this report we assess the prospects of probing \emph{Higgs portal} models directly with future searches for invisible Higgs decays, as well as indirectly with precision Higgs rate measurements, at the LHC in the high luminosity phase with $3~\mathrm{ab}^{-1}$. Furthermore, we shall highlight the complementarity between these two probes, as well as with other constraints, e.g.~with current and future limits from DM direct detection experiments and limits from LEP Higgs searches. Searches for invisible decays of the Higgs boson also have important implications for ``nearly invisible'' decays into neutral long-lived particles that are predicted by many models~\cite{Curtin:2018mvb}. Dedicated searches set much stronger limits~\cite{Aad:2015uaa,ATLAS:2016jza,Aaij:2016xmb,CMS:2014hka}, but the Higgs to invisible search is a model-independent constraint on all possibilities.  Projections for dedicated searches as well as proposals for dedicated detectors at the LHC complex~\cite{Chou:2016lxi,Gligorov:2017nwh,Gligorov:2018vkc} are not discussed in this section.

This contribution is organised as follows. We briefly review in Section~\ref{sec6:exp} the experimental input for the HL-LHC that we use in our study. In Section~\ref{sec6:effC} we first employ an effective description of the generic phenomenological Higgs features that appear in this class of models. We then focus on two specific realisations of the Higgs portal: in Section~\ref{sec6:minimalHP} we discuss the \emph{minimal Higgs portal}, where the SM Higgs field directly couples to an additional DM field through a quartic interaction; in Section~\ref{sec6:singletHP} we show results for the \emph{scalar singlet portal}, where an additional scalar singlet is acting as a mediator between the visible and hidden sector. We conclude in Section~\ref{sec6:conclusions}.

\asubsection{Main channels for direct searches}\label{sec:expinp}

Given the VBF Higgs (VBFH) production presents the best sensitivity, this channel is chosen to investigate the sensitivity of the search with the HL-LHC~\cite{CMS-PAS-FTR-18-016}. The CMS phase-2 detector is simulated using Delphes~\cite{deFavereau:2013fsa} (fast parametrisation), with on average 200 interactions per bunch crossing.  A
cut-and-count approach similar to the one described in the analysis from Ref.~\cite{Sirunyan:2018owy} is used.

The VBF\PH signal samples are produced using \POWHEG
v2.0~\cite{Alioli:2010xd,Nason:2009ai} at next-to-leading order
in perturbative QCD, assuming 100\% branching ratio \BHinv of the Higgs boson
to invisible final states, and normalised using the SM inclusive Higgs
boson production cross sections provided in Ref.~\cite{deFlorian:2016spz}.  Full-simulation samples produced at 13\,\UTeV are used to derive the gluon-fusion contribution, applied as a fraction of the Delphes expected VBF\PH yields.

The main backgrounds are processes involving vector bosons (W,Z)
produced in association with jets, either through QCD or electroweak
(EWK) vertices.  Monte Carlo samples for these backgrounds are
generated at leading order using \MGvATNLO v2.2.2~\cite{Alwall:2014hca} interfaced with \PYTHIA v8.205~\cite{Sjostrand:2014zea} or higher. SM processes involving top quarks also contribute to the
background, and are simulated using a combination of the \POWHEG and \MGvATNLO generators. Backgrounds arising from QCD multi-jet events are simulated using \MGvATNLO interfaced with \PYTHIA, imposing a minimum threshold of 1000\,\UGeV on the di-jet mass at parton level.

The objects studied are as defined for the analysis in Ref.~\cite{Sirunyan:2018owy}, with extended coverage in pseudorapidity $\eta$. Electrons passing loose
identification criteria, with a transverse momentum p$_T>10$\,\UGeV, and $|\eta|<2.8$ are vetoed. Similarly, muons passing loose
identification criteria with p$_T>20$\,\UGeV and $|\eta|<3.0$ are
vetoed. Taus passing loose identification criteria with p$_T>20$\,\UGeV
and $|\eta|<2.8$ are vetoed. Jets are reconstructed using the
anti-k$_T$ algorithm~\cite{Cacciari:2008gp,Cacciari:2011ma} with a parameter size of
0.4.  The jets are required to have p$_T>30$\,\UGeV and $|\eta|<5.0$,
and are corrected for pileup effects using the ``Puppi''
algorithm~\cite{Bertolini:2014bba}.

 A b-tagging algorithm is used to tag jets that originate from decays
of B hadrons (b jets).  The algorithm uses a combination of vertexing
and timing information, and a working point with an efficiency of
around 60\% and a mis-tagging rate below 1\% is defined to identify b
jets.  Events containing any identified b jets are vetoed.

The leading and sub-leading jets in the event are required to have
p$_{T}>80$ and $40$\,\UGeV, respectively, and be in opposite hemispheres of
the detector. These two jets form the VBF di-jet pair, and further
requirements are applied on the invariant mass M$_{\text{jj}}$, and
their separations in pseudorapidity $|\Delta\eta_{\text{jj}}|>4.0$ and
azimuthal angle $|\Delta\phi_{\text{jj}}|<1.8$.

To reject the QCD multi-jet background, for which the transverse
missing energy arises from jet mis-measurements, the \METvec vector is
required to not be aligned with a jet using min$[\Delta\phi$(jet
\ptvec$>30$\,\UGeV, \METvec)]$>0.5$. The magnitude of the vector sum of
the p$_{T}$ of all jets with p$_{T}>30$\,\UGeV is defined as \MHT.

The analysis uses five non-overlapping event regions: the signal
region (SR) where events containing charged leptons ($\ell$, where
$\ell=\Pe$ or $\Pgm$) are vetoed, and four control regions (CR) with
exactly one electron or muon ($\PW\rightarrow\Pe\nu$ CR and
$\PW\rightarrow\Pgm\nu$ CR) or exactly two electrons or two muons
($\PZ\rightarrow\Pe\Pe$ CR and $\PZ\rightarrow\Pgm\Pgm$ CR).  In the
$\PW\rightarrow\Pe\nu$ and $\PW\rightarrow\Pgm\nu$ CRs, to further
reject QCD multi-jet backgrounds, the transverse mass, defined as
$\sqrt{2\pt^{\ell}\MET\left[1
- \cos{\Delta\phi(\ell,\METvec)}\right]}$, where p$_{T}^{\ell}$ is the
transverse momentum of the lepton and $\Delta\phi(\ell,\METvec)$ is
the azimuthal angle between the lepton momentum and $\METvec$ vectors,
is required to be less than 160\,\UGeV. In the W$\rightarrow$e$\nu$ CR a
selection on \MET$>60$\,\UGeV is also applied due to the higher QCD
multi-jet contamination than in the muon channel. In the
$\PZ\rightarrow\Pe\Pe$ and $\PZ\rightarrow\Pgm\Pgm$ CRs, the di-lepton
mass is required to be between 60 and 120\,\UGeV. To account for the
higher single-electron trigger thresholds that will be required at the
HL-LHC , the leading electron p$_{T}$ is required to be above 40\,\UGeV,
for both the $\PW\rightarrow\Pe\nu$ and $\PZ\rightarrow\Pe\Pe$ CRs.

The lower threshold on the \MET is varied from 130 to 400\,\UGeV in 10 to
50\,\UGeV steps. Likewise, the lower threshold on M$_{\text{jj}}$ is
varied from 1000 to 4000\,\UGeV in 100\,\UGeV steps. The statistical
uncertainty on the MC is considered to be negligible, assuming the
available MC samples will have at least 10 times the integrated
luminosity available in the data.  For each (\MET, M$_{\text{jj}}$)
selection, the yields are extracted in the four control regions and in
the signal region, and a likelihood is constructed as the product of
five Poisson terms, one per region.  Upper limits on the Higgs boson
production cross section times
\BHinv are placed at the 95\% CL using the CLs
criterion~\cite{Read:2002hq,Junk:1999kv,Dittmaier:2012vm}, with a profiled
likelihood ratio as the test statistic in which systematic
uncertainties are incorporated via nuisance
parameters~\cite{Chatrchyan:2013lba,CMS-NOTE-2011-005}. Asymptotic
formula are used to determine the distribution of the test statistic
under signal and background hypotheses~\cite{Cowan:2010js}.

The scenario considered for the systematic uncertainties is described
in table~\ref{tab:systs}, together with the systematic uncertainties that were 
considered in Ref.~\cite{Sirunyan:2018owy}, for comparison.

\begin{table}[ht]
  \centering
  {
  \begin{tabular}{l|c|c}
    \hline
    Systematic & From Ref.~\cite{Sirunyan:2018owy} & This analysis \\
    \hline
    e-ID & 1\%(reco)$\oplus$1\%(idiso) & 1\% \\
    $\mu$-ID & 1\%(reco)$\oplus$1\%(id)$\oplus$0.5\%(iso) & 0.5\% \\
    \hline
    e-veto & 0.6\%(reco)$\oplus$1.5\%(idiso) & 1\% \\
    $\mu$-veto on QCD V+jets & 5\%(reco)$\oplus$5\%(id)$\oplus$2\%(iso) & 2\% \\
    $\mu$-veto on EWK V+2jets & 10\%(reco)$\oplus$10\%(id)$\oplus$6\%(iso) & 6\% \\
    $\tau$-veto & 1--1.5\% for QCD--EWK  & 0.5--0.75\% \\
    b-tag-veto & 0.1\% (sig) 2\% (top) & 0.05\% (sig) 1\% (top) \\
    \hline
    JES & 14\%(sig) 2\%(W/W) 1\%(Z/Z) & 4.5\%(sig) 0.5\%(W/W) 0.2\%(Z/Z) \\
    Integrated luminosity & 2.5\% & 1\% \\
    QCD multi-jet & 1.5\%  & 1.5\% \\
    \hline
    Theory on W/Z ratio & 12.5\% & 7\% \\
    ggH normalisation & 24\% & 20\% \\
    \hline
\end{tabular}
    } \caption{Impact on the signal and background yields from the
    different sources of systematic uncertainty considered in
    Ref.~\cite{Sirunyan:2018owy} and for the HL-LHC setup considered
    in this analysis.}  \label{tab:systs}
\end{table}

The analysis is expected to be systematics dominated, with the
dominant systematic uncertainties due to the muon and electron
efficiencies (e-ID and $\mu$-ID in table~\ref{tab:systs}), both in the control and signal regions, and the jet
energy scale (JES) evaluated for the signal (sig) or on the ratios of W and Z yields in signal and control regions (W/W and Z/Z in table~\ref{tab:systs}). In Ref.~\cite{Sirunyan:2018owy}, due to the limited size of the di-lepton
samples, the knowledge of the ratio of the cross sections of the W to
Z boson production was used as a constraint between the two
backgrounds, leading to an increased sensitivity. The theoretical
uncertainty on this ratio is set at 12.5\% from studies of missing
higher order QCD and EWK corrections~\cite{Sirunyan:2018owy}, for both
QCD and EWK production. Once 300\fbinv of data will be available, this
constraint will play a smaller role. It is expected that improvements
in theoretical calculations of the ratio will lead to half the current
theoretical uncertainty, namely 7\%.  This uncertainty is expected to
have an impact of at most 3--5\% for the selection with the largest
expected significance and is therefore neglected in the results
presented herein. However, the uncertainty will be relevant when
considering very tight selection criteria on \MET and M$_{\text{jj}}$,
i.e. when the statistical uncertainty in the CRs becomes dominant.

The most stringent upper limits are achieved in the regions with lower
thresholds on M$_{\text{jj}}$ and \MET of 2500\,\UGeV and 190\,\UGeV,
respectively, for the 3000\fbinv scenario. The minimum is rather flat
between M$_{\text{jj}}$ values of 2300 and 3000\,\UGeV, and between
\MET values of 170 and 220\,\UGeV, indicating limited impact from the size of the MC
samples. The upper limits degrade steeply as the \MET threshold
increases above 250\,\UGeV. The behaviour is similar for the 300 and
1000\fbinv scenarios, with best thresholds found at lower values
of \MET (170\,\UGeV) and M$_{\text{jj}}$ (1500 and 1800\,\UGeV respectively)
due to the interplay between the size of the control regions and the
systematic uncertainties.

Distributions in M$_{jj}$ for
the leading jet pair and \MET in the signal region are shown in
figure~\ref{fig:plotsdijetmet}, for an integrated luminosity of 3000 fb$^{-1}$. The
corresponding expected yields are shown in
table~\ref{tab:yieldsYR18}. The uncertainties shown represent the statistical 
uncertainties due to the limited size of the Delphes samples and are not used in the calculations
of the final limits.

 \begin{table}[h!]
     \centering

\begin{tabular}{l|c|c|c|c|c}
\hline
Process & SR          &     $\PW\rightarrow\Pe\nu$ CR      &   $\PW\rightarrow \Pgm\nu$ CR      &   $\PZ\rightarrow\Pe\Pe$ CR      & $\PZ\rightarrow \Pgm\Pgm$ CR  \\
\hline
VBF\PH & 47812 $\pm$ 584       & - & - & - & - \\
ggH & 972  & - & - & - & - \\
\hline
Z$\rightarrow \ell\ell$ (EWK) & 103 $\pm$ 8   & 398 $\pm$ 16 & 641 $\pm$ 20 & 1342 $\pm$ 30 & 1889 $\pm$ 35 \\
Z$\rightarrow \ell\ell$ (QCD) & 451 $\pm$ 90   & 944 $\pm$ 126 & 1048 $\pm$ 116 & 1347 $\pm$ 118 & 2297 $\pm$ 158 \\
\hline
Z$\rightarrow\nu\nu$ (EWK) & 15275 $\pm$ 358       & - & - & - & - \\
Z$\rightarrow\nu\nu$ (QCD) & 20968 $\pm$ 599       & - & - & - & - \\
\hline
W$\rightarrow\mathrm{e}\nu$ (EWK) & 3358 $\pm$ 62      & 18986 $\pm$ 146 & 72 $\pm$ 9 & 33 $\pm$ 6 & - \\
W$\rightarrow \mu\nu$ (EWK) & 3426 $\pm$ 62     & 7 $\pm$ 3 & 29360 $\pm$ 181 & - & 17 $\pm$ 4 \\
W$\rightarrow \tau\nu$ (EWK) & 3595  $\pm$ 64   & 55 $\pm$ 8 & 87 $\pm$ 10 & - & - \\
\hline
W$\rightarrow\mathrm{e}\nu$ (QCD) & 3994  $\pm$ 999   & 13376 $\pm$ 1656 & 170 $\pm$ 168 &  - & - \\
W$\rightarrow \mu\nu$ (QCD) & 6891 $\pm$ 1388  & - & 23322 $\pm$ 2096 & - & - \\
W$\rightarrow \tau\nu$ (QCD) & 4308 $\pm$ 938  & -       & -       & -       & - \\
\hline
Top & 2050 $\pm$  132   &  2171 $\pm$ 143 & 3735 $\pm$ 188 & 107 $\pm$ 36 & 130 $\pm$ 39 \\
QCD & -    & -       & -       & -       & -\\
\hline
\end{tabular}
\caption{Number of events expected after the final selection, M$_{\text{jj}}>2500$\,\UGeV and \MET$>190$\,\UGeV, with an integrated luminosity of 3000\fbinv. The uncertainties are the statistical uncertainties from the Delphes samples.}
\label{tab:yieldsYR18}
  \end{table}

\begin{figure}[htbp]
  \centering
    \includegraphics[width=0.49\textwidth]{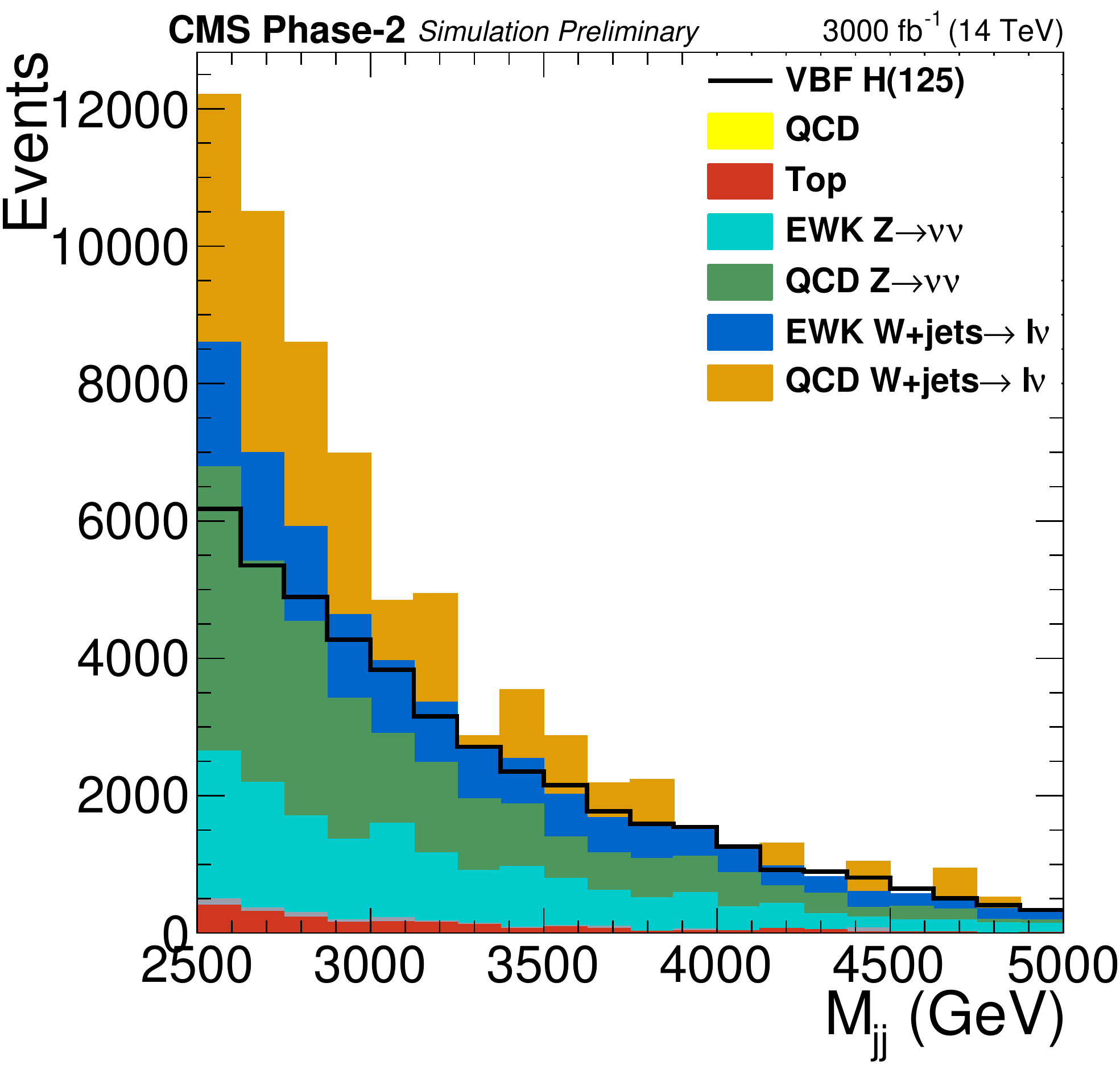}
    \includegraphics[width=0.49\textwidth]{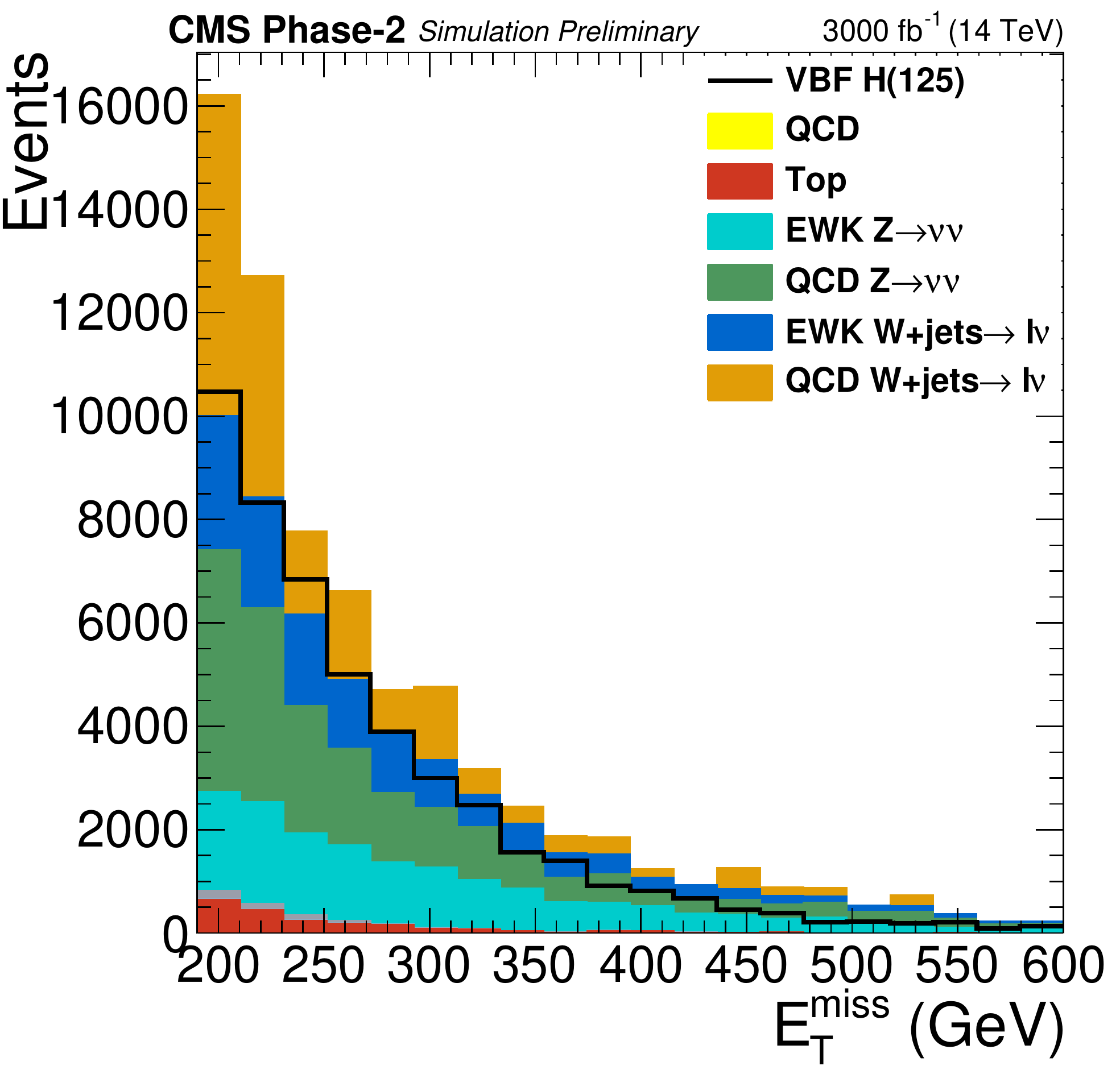}
\caption{Distributions of M$_{jj}$ (left) and \MET (right) in the signal region for the final selection, M$_{jj}>2500$\,\UGeV and \MET$>190$\,\UGeV~\cite{CMS-PAS-FTR-18-016}.}
  \label{fig:plotsdijetmet}
\end{figure}

The 95\% CL upper limits for an integrated luminosity of 3000\fbinv
are shown in figure~\ref{fig:1Dlimits}, left, as a function of the
thresholds applied on \MET assuming the MC statistical uncertainties
are negligible, for the final selections described above\footnote{A previous phenomenological study \cite{Biekotter:2017gyu} focusing on jet tagging has found a similar behaviour for $~\MET\,\gtrsim\,250\GeV$.} In the best
case, the lowest 95\% CL limit on \BHinv, assuming Standard Model
production, is expected to be at 3.8\%, for thresholds values of
2500\,\UGeV (190\,\UGeV) on the di-jet mass (\MET). If the \MET resolution
was to be a factor of 2 worse, the re-optimisation of the selection
leads to minimum thresholds of 1800\,\UGeV (250\,\UGeV) on the di-jet mass
(\MET), but a similar 95\% CL limit. The limits are shown for
different integrated luminosities in figure~\ref{fig:1Dlimits}, right.

\begin{figure}[htbp]
  \centering
    \includegraphics[width=0.49\textwidth]{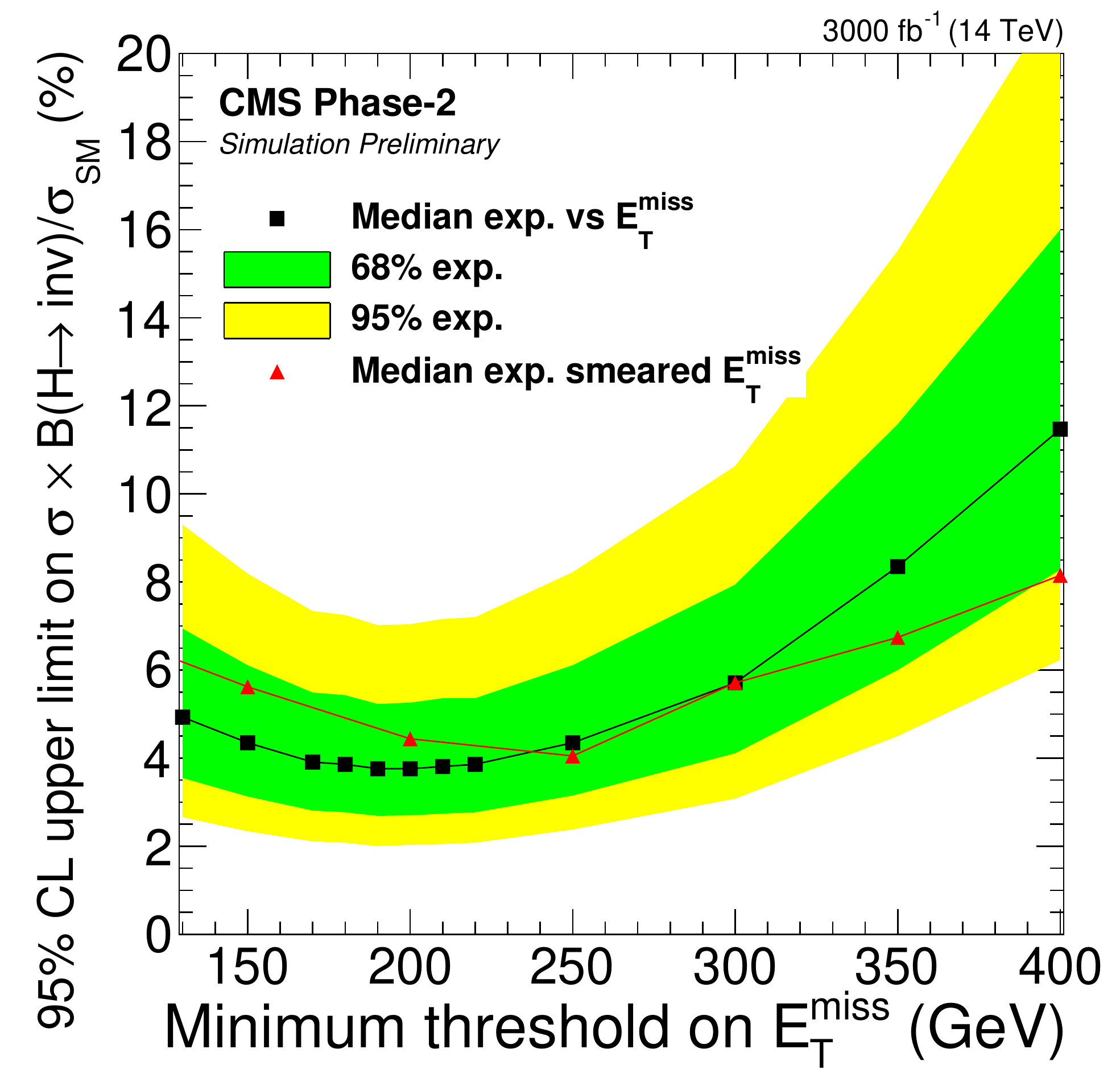}
    \hfill
    \includegraphics[width=0.49\textwidth]{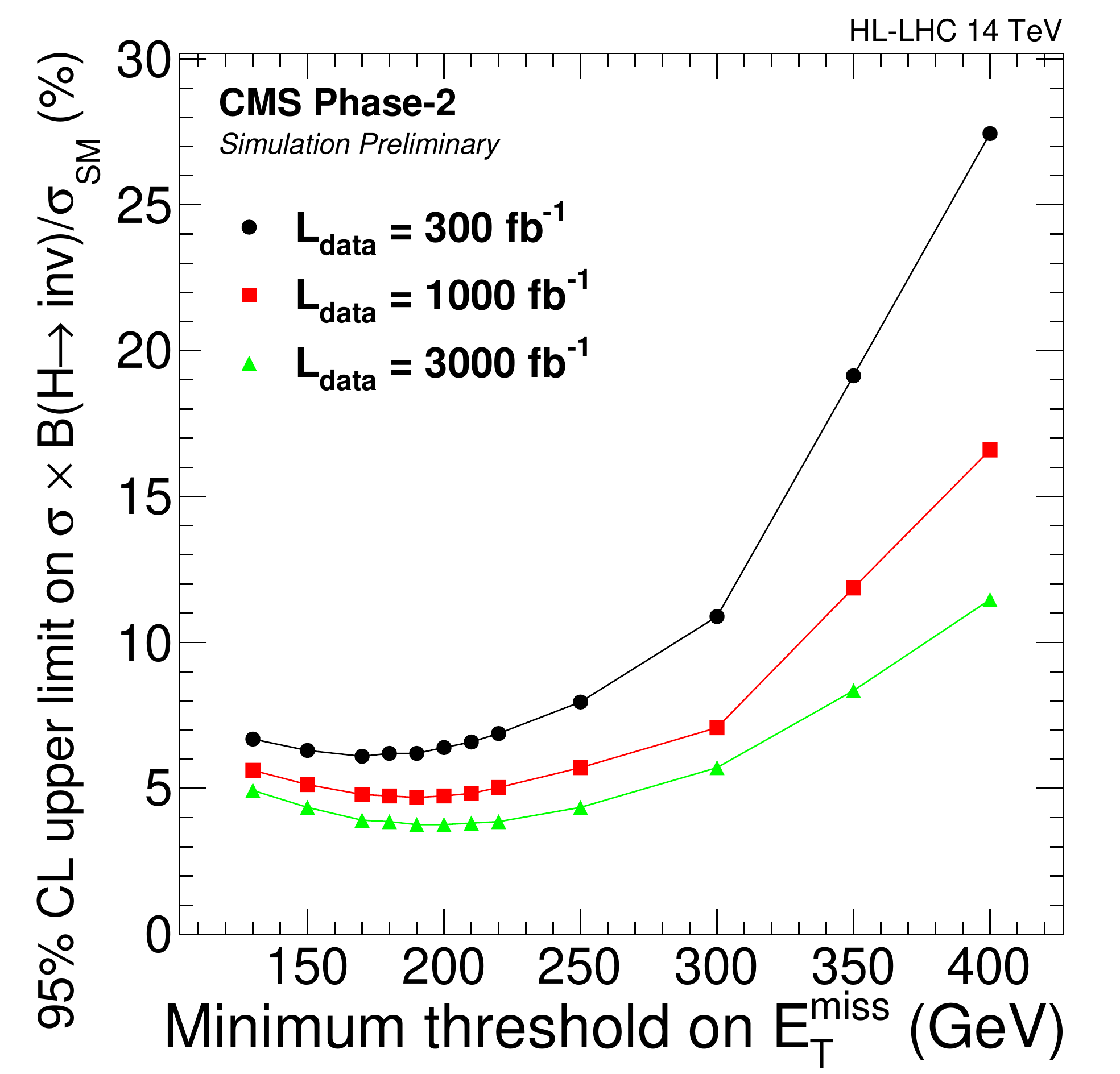}
\caption{Left: 95\% CL limits on \BHinv as a function of the minimum threshold on \MET, for M$_{\text{jj}}>2500$\,\UGeV and an integrated luminosity of 3000\fbinv ~\cite{CMS-PAS-FTR-18-016}. Right: 95\% CL limits for scenarios with different integrated luminosities.}
  \label{fig:1Dlimits}
\end{figure}

The performance of pileup mitigation techniques will have a significant impact on the projected sensitivity for the final VBF result. ATLAS has conducted a study to show the impact of pileup jets on the invisible Higgs branching ratio limit in the VBF channel~\cite{ATL-PHYS-PUB-2018-038} using full detector simulations based on Geant4\cite{Agostinelli:2002hh,Allison:2006ve} and the complete detector simulation and event reconstruction. The branching ratio limit can vary by a factor of 4 when no explicit pileup jet mitigation is used to the case when truth information is used to remove all pileup jets. Therefore, the development of improved pileup jet mitigation will be an important development to empower the invisible Higgs decay analyses in the future.

\asubsection{Interpretation and combination with precision Higgs boson measurements}
\asubsubsection{Experimental input}
\label{sec6:exp}
{For the VBF production channel, the projected HL-LHC limit on the invisible Higgs decay rate from the CMS experiment amounts to $4\%$, see Section~\ref{sec:expinp}. For the $VH$ production channel ATLAS projected a limit of around $8\%$ in 2013}~\cite{ATL-PHYS-PUB-2013-014}.
Assuming ATLAS (CMS) performs equally well as CMS (ATLAS) in the $\text{VBF}$ ($VH$) channel, and neglecting possible correlations of experimental and theoretical uncertainties~\cite{CMS-PAS-JME-15-001}, a combination of these limits results in
\begin{align}\label{eq:brinvcombi}
\left(\mu_{\text{VBF},VH} \cdot \BRHinv \right)^\text{HL-LHC} \le 2.5\%,
\end{align}
where $\mu_{\text{VBF},VH}$ is a common signal strength modifier of the VBF and $VH$ production cross sections. In our theory interpretations below, we take Eq.~\eqref{eq:brinvcombi} as a benchmark value for the prospective ATLAS and CMS combined limit on $\BRHinv$.

We implemented the ATLAS and CMS HL-LHC projections for Higgs signal strength measurements for the individual production times decay modes (see Section~\ref{sec2:exp_combination}) into the code \texttt{Higgs\-Signals}~\cite{Bechtle:2013xfa,Bechtle:2014ewa}, including the corresponding correlation matrices.
We consider the projections for both future scenarios S1 (with Run~2 systematic uncertainties) and S2 (with YR18 systematic uncertainties)~\cite{HLHELHCCommonSystematics}, see Sec.~\ref{sec:expcomb_prodtimesdecay}, Tab.~\ref{tab:summary_A1_5PD}.
Note that correlations of theoretical rate uncertainties between the future ATLAS and CMS measurements are taken into account in our fit via \texttt{HiggsSignals}. 

We furthermore study the impact of a future electron-proton collider option (LHeC) at CERN~\cite{AbelleiraFernandez:2012cc,uta,uta2,Zimmermann:2651305}, assuming a $60~\mathrm{\UGeV}$ electron beam, a $7~\mathrm{\UTeV}$ proton beam and an integrated luminosity of $1~\mathrm{ab}^{-1}$. We implemented the prospective signal strength measurements at the LHeC presented in Ref.~\cite{uta} into \texttt{HiggsSignals}.\footnote{In addition to the experimental precision quoted in Ref.~\cite{uta} we assume a theoretical uncertainty of $1\%$ ($1.5\%$) on the charged (neutral) current production cross section, as well as a $1\%$ luminosity uncertainty.} The projected limit on the invisible Higgs decay rate is around $5\%$~\cite{uta,Tang:2015uha,kuzeLHeC,bc1,bc2,uta2} \footnote{Optimisation of the signal selection, advanced background estimation techniques and details of the detector design may improve this limit down to about $(3-4)\%$~\cite{Utacommunications}.}.
In combination with the above CMS and ATLAS projections, we obtain
\begin{equation*}
\left(\mu_{\text{VBF},VH,\text{NC}} \cdot \text{BR}_\text{inv}\right)^{\text{HL-LHC}\oplus\,\text{LHeC}}\,\leq\,2.25\%
\end{equation*}
as upper limit on the branching ratio of an invisible Higgs decay mode. {Here, {we assume} the common signal strength modifier $\mu$ also applies to the neutral current (NC) Higgs production cross section at the LHeC.}

\asubsubsection{Effective description of Higgs portal models}
\label{sec6:effC}

In this section we discuss the HL-LHC prospects in the context of an effective parametrisation of Higgs rate modifications that are commonly predicted by Higgs portal models, using the coupling scale factor ($\kappa$) framework~\cite{Heinemeyer:2013tqa} (see also Section \ref{sec2:exp_kappa}).  Herein, the scale factors $\kappa_X$ ($X = W, Z, g, \gamma, b, \tau, \dots$) are introduced for every relevant Higgs coupling to SM particle $X$. The partial widths and cross sections associated with these Higgs couplings are then rescaled by $\kappa^2_X$ (see Refs.~\cite{Heinemeyer:2013tqa,Bechtle:2014ewa} for more details). In addition, we treat the branching fraction for invisible Higgs decays, $\BRHinv$, as free parameter.

In particular, we investigate two scenarios for the Higgs coupling modifications:
\begin{enumerate}
\item[(\emph{i})] a universal scale factor for all Higgs couplings to SM particles, $\kappa \equiv \kappa_X$ ($X = W, Z, g, \gamma, b, \tau, \dots$);
\item[(\emph{ii})] additional free parameters $\kappa_g$ and $\kappa_\gamma$ that rescale the loop-induced Higgs couplings to gluons and photons, respectively. The remaining (tree-level) Higgs couplings to SM particles are again rescaled universally with $\kappa \equiv  \kappa_X$ ($X = W, Z, b, \tau, \dots$).
\end{enumerate}

\begin{figure}
\centering
\includegraphics[width=0.49\textwidth]{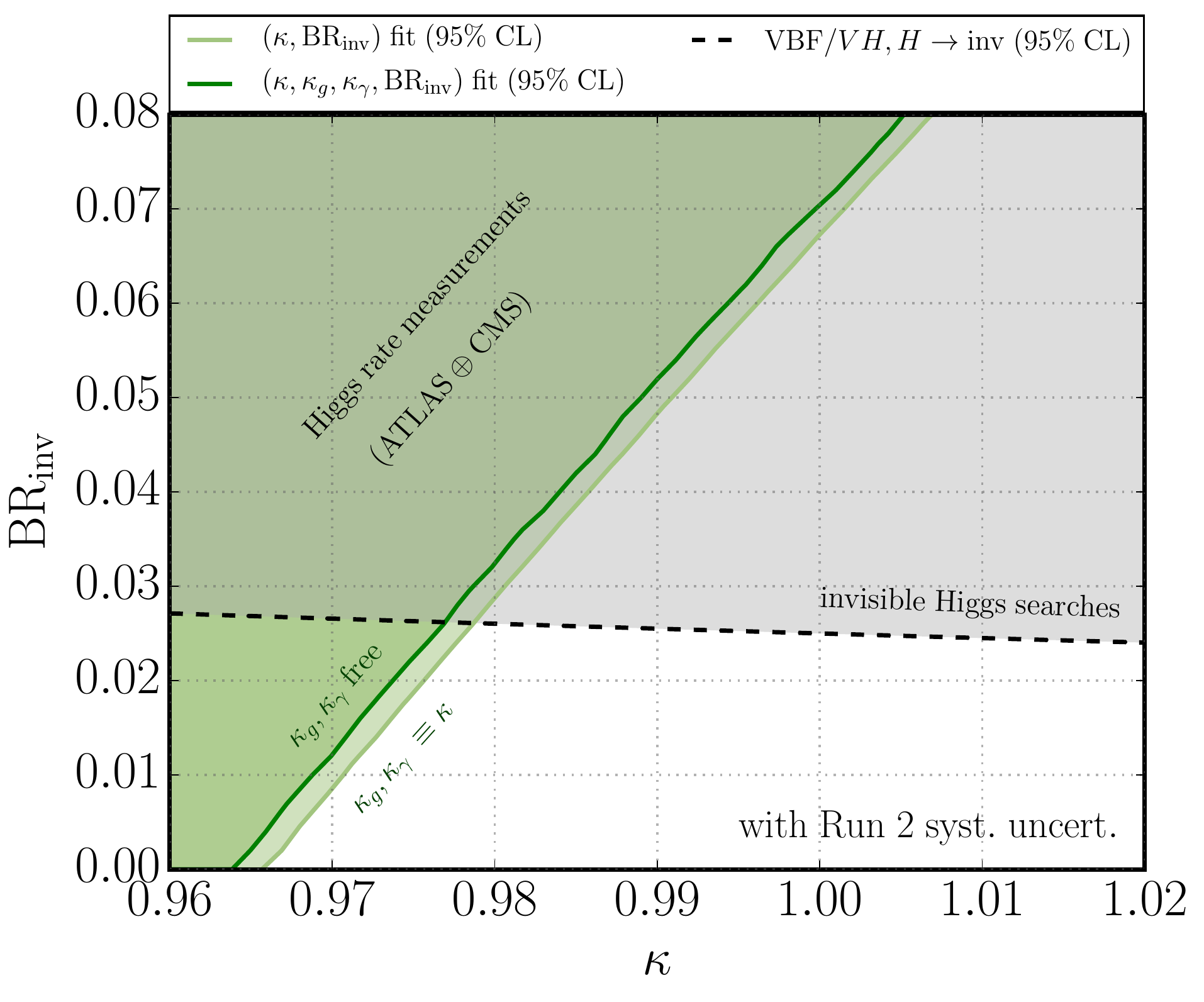}
\hfill
\includegraphics[width=0.49\textwidth]{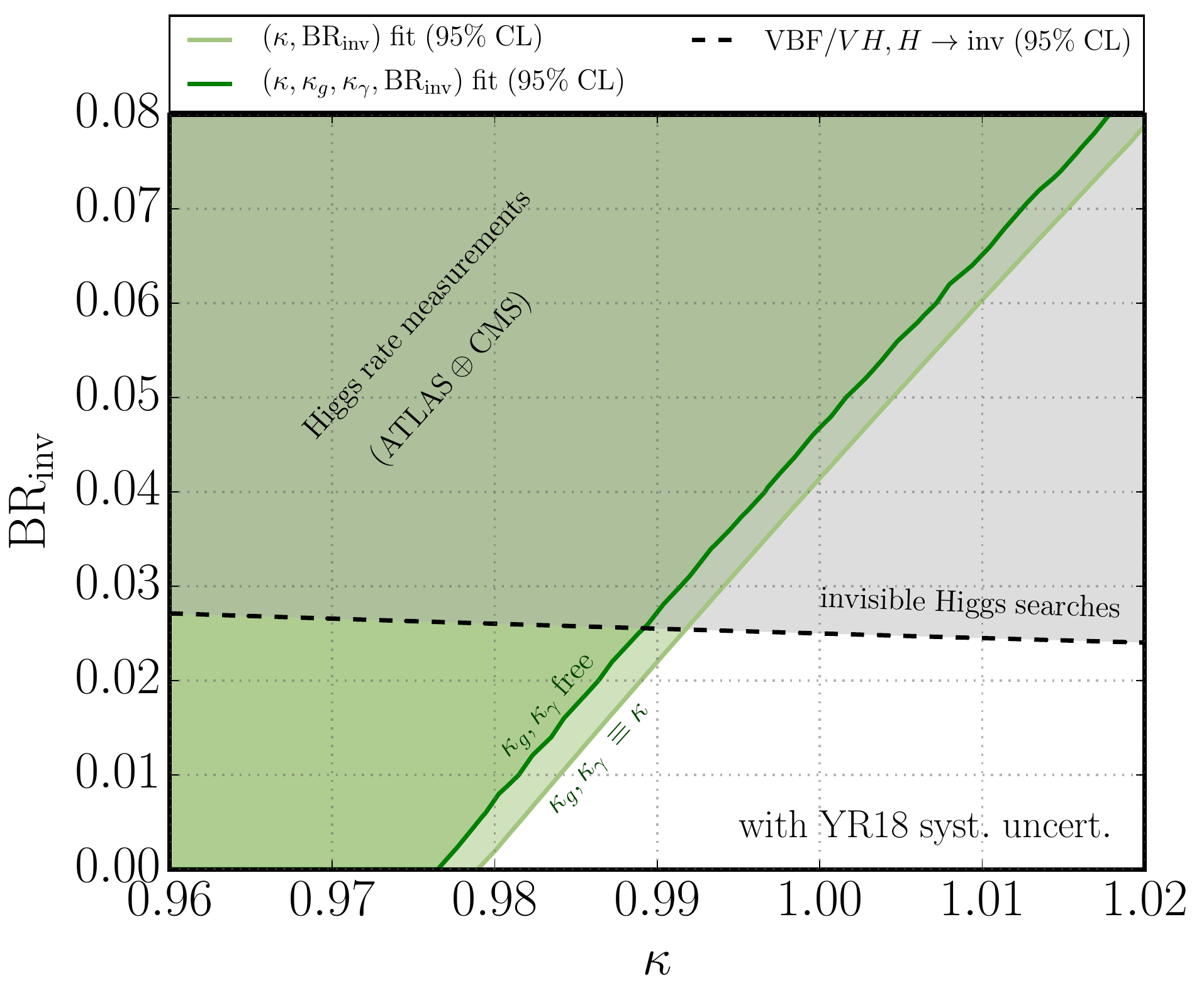}\\
\includegraphics[width=0.49\textwidth]{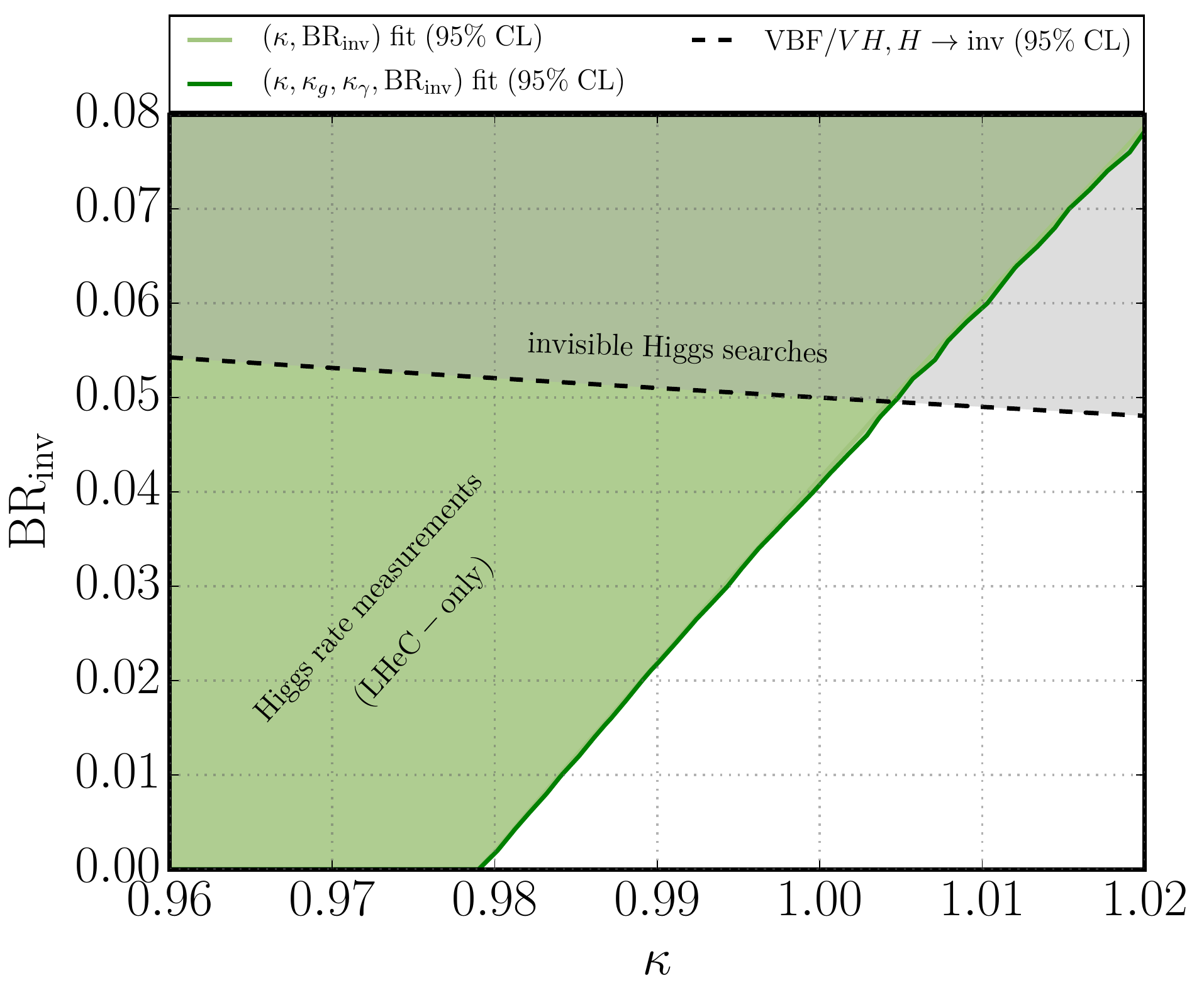}
\hfill
\includegraphics[width=0.49\textwidth]{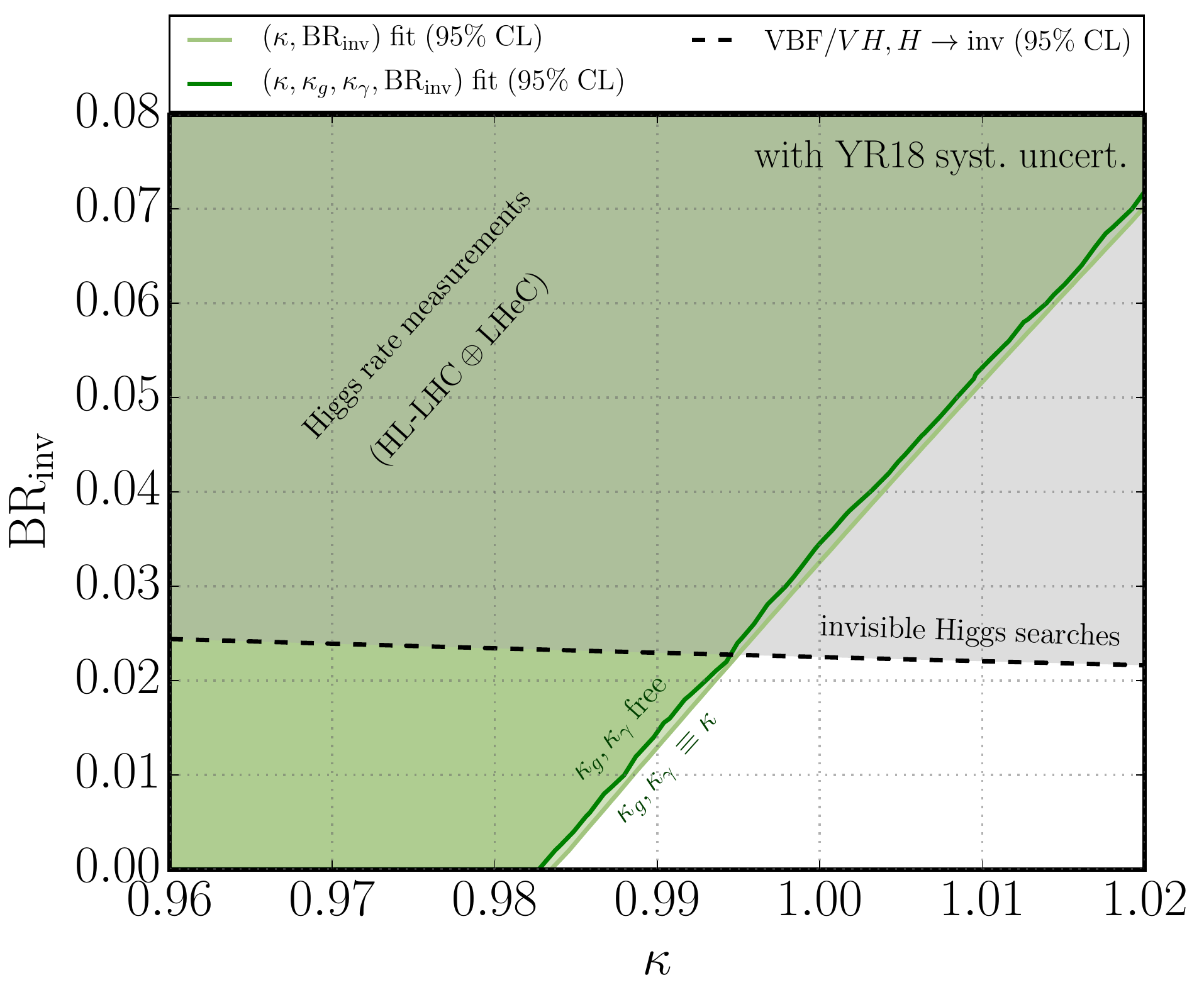}
\caption{Projected $95\%~\mathrm{C.L.}$ limit in the $(\kappa, \BRHinv)$ plane inferred from Higgs rate measurements (\emph{green regions}) and direct invisible Higgs searches (\emph{black dashed line}) at the HL-LHC and LHeC. We show results for the two future HL-LHC scenarios S1 [with Run~2 systematic uncertainties] {\sl (top left)} and S2 [with YR18 systematic uncertainties] {\sl (top right)} (see text for details), as well as for the LHeC {\sl (bottom left)} and the combination of LHeC and HL-LHC~[S2] {\sl (bottom right)}. The light green area shows the limit from Higgs rates obtained by assuming no new physics contributions to the loop-induced Higgs couplings to gluons and photons, $\kappa = \kappa_g = \kappa_\gamma$, whereas for the dark green area $\kappa_g$ and $\kappa_\gamma$ are marginalised free parameters.}
\label{fig:effC}
\end{figure}

We employ the program \texttt{HiggsSignals}~\cite{Bechtle:2013xfa,Bechtle:2014ewa} to perform a $\chi^2$ fit to the projected HL-LHC {and/or LHeC} Higgs rate measurements (see Section~\ref{sec6:exp})  in each scenario. The resulting future $95\%~\mathrm{C.L.}$ limit is shown in Fig.~\ref{fig:effC} as a light and dark green area for scenario (\emph{i}) and (\emph{ii}), respectively. {The \emph{top panels} display the HL-LHC projections for future scenarios S1 [with Run~2 systematic uncertainties] (\emph{left}) and S2 [with YR18 systematic uncertainties] (\emph{right}), while the \emph{bottom panels} show the projections for LHeC (\emph{left}) and the combination of LHeC with HL-LHC S2 measurements (\emph{right}).} {In Tab.~\ref{tab:effC_limits} we summarise the lower limits on the Higgs signal strength of channels with SM final states, $\kappa^2 (1-\BRHinv)$, as well as the upper limits on the invisible Higgs decay rate, $\mathrm{BR}_\text{inv}$, assuming SM Higgs coupling strengths ($\kappa \equiv 1$), for the four future collider scenarios and for the two global fit scenarios.} Note that these results do not strictly require the additional Higgs decay mode to yield an invisible final state.

\begin{table}
\centering
\begin{tabular}{ l |  l | cccc}
\hline
fit setup &  quantity   & HL-LHC S1   & HL-LHC S2   & LHeC	  & LHeC $\oplus$ HL-LHC S2 \\
\hline
\multirow{2}{*}{$(\kappa, \mathrm{BR}_\text{inv})$}                         &  $\kappa^2 (1-\BRHinv)$ &  $\ge 0.933$ & $\ge 0.958$ & $\ge 0.959$  & $\ge 0.967$ \\
                                                                            &  $\BRHinv$ ($\kappa\equiv 1$) & $\le 6.7\%$ & $\le 4.2\% $ & $\le 4.1\%$ & $\le 3.3\%$\\
\hline									  
\multirow{2}{*}{$(\kappa, \kappa_g, \kappa_\gamma,  \mathrm{BR}_\text{inv})$}   &  $\kappa^2 (1-\BRHinv)$ & $\ge0.930$ &  $\ge0.954$ & $\ge0.959$ & $\ge0.966$\\
                                                                                &   $\BRHinv$ ($\kappa\equiv 1$) & $\le7.0\%$ & $\le4.6\%$ & $\le4.1\%$ & $\le3.4\%$\\

\hline
\end{tabular}
\caption{{Comparison of prospective $95\%~\mathrm{C.L.}$ limits on the Higgs signal strength for SM final states, $\kappa^2(1-\mathrm{BR}_\text{inv})$, and the invisible Higgs decay rate, $\mathrm{BR}_\text{inv}$ (assuming SM Higgs couplings, $\kappa =1$), for HL-LHC scenarios S1 and S2, LHeC, and the combination of LHeC and HL-LHC (assuming scenario S2). First (second) row shows the results obtained in the fit parametrisation (\emph{i}) [(\emph{ii})].}}
\label{tab:effC_limits}
\end{table}

These results are compared in Fig.~\ref{fig:effC} with the prospective future limits from direct searches for invisible Higgs decays (see Section~\ref{sec6:exp}). At the HL-LHC, assuming scenario S1 (S2), direct invisible Higgs searches are more sensitive than Higgs rates if deviations from the SM Higgs couplings are small, $\Delta \kappa \equiv 1- \kappa \lesssim 2~(1)\%$. For larger suppression of the Higgs couplings the Higgs rates will provide the strongest constraint. In contrast, if we allow for an enhancement of the Higgs couplings, $\kappa > 1$, the invisible Higgs searches will provide the strongest constraint (besides other bounds on the Higgs total decay width, see Sec.~\ref{sec5}).

At the LHeC the prospective indirect Higgs rate constraints are comparable to the HL-LHC S2 prospects, reaching a precision of $\Delta \kappa \lesssim (2.1-2.3)\%$ independently of the invisible Higgs decay rate, in both fit parametrisations considered here.\footnote{The complementarity of LHeC and HL-LHC Higgs rate measurements is much stronger in more general coupling fit setups, e.g., when independent scale factors for the Higgs-$W$-$W$ and Higgs-$Z$-$Z$ couplings are considered~\cite{uta}.} On the other hand, the direct invisible Higgs searches at the LHeC are weaker than at the HL-LHC. In combination with the HL-LHC (assuming future scenario S2), the bounds from the Higgs rates can further be improved to coupling deviations of $\Delta \kappa \lesssim1.7\%$. 

{Compared with the sensitivity of Higgs rate measurements during Run~1 of the LHC~\cite{Khachatryan:2016vau} to the invisible decay rate, $\BRHinv \lesssim \mathcal{O}({20\%})$} (at $95\%~\mathrm{C.L.}$), we find that the sensitivity improves by roughly a factor of $3$--$5$  at the HL-LHC (depending on the evolution of systematic uncertainties). In combination with LHeC results we expect the indirect limit to improve by a factor of up to $6$.

\asubsection{Higgs portal interpretations}

\asubsubsection{Minimal Higgs Portal}
\label{sec6:minimalHP}

In the minimal Higgs portal model, we impose a quartic interaction of the SM Higgs doublet field $H$ with the DM field, which could be either a scalar ($S$)~\cite{Silveira:1985rk}, a vector ($V^\mu$)~\cite{Lebedev:2011iq} or a fermion ($\chi$)~\cite{Kim:2006af} (see Refs.~\cite{Kanemura:2010sh,Djouadi:2011aa} for a comprehensive overview):
\begin{align}
\mathcal{L} &\supset -\tfrac{1}{4} \lambda_{hSS} H^\dagger H S^2 \qquad (\mbox{scalar~DM}) \quad \mbox{or} \label{eq:scalarDM}\\
\mathcal{L} &\supset +\tfrac{1}{4}  \lambda_{hVV} H^\dagger H V_\mu V^\mu \qquad (\mbox{vector~DM}) \quad \mbox{or} \label{eq:vectorDM}\\
\mathcal{L} &\supset -\tfrac{1}{4}  \tfrac{\lambda_{h\chi\chi}}{\Lambda} H^\dagger H \bar{\chi} \chi \qquad (\mbox{fermion~DM}), \label{eq:fermionDM}
\end{align}
respectively. Besides these operators the Lagrangian contains an explicit mass term of the DM field, allowing us to use the mass of the DM particle, $M_\text{DM}$, as a free model parameter. In addition, the Lagrangian $\mathcal{L}$ contains DM self-interaction operators, however, these are irrelevant to our study.

If DM is light, $M_\text{DM} < M_H /2 \simeq 62.5\UGeV$, the above interactions lead to the invisible Higgs decay into two DM particles. An upper limit on $\BRHinv$ can therefore be translated into an upper limit on the portal coupling $\lambda$ of above operators, Eqs.~\eqref{eq:scalarDM}-\eqref{eq:fermionDM}, depending on $M_\text{DM}$.  At the same time, the portal coupling $\lambda$ governs the DM phenomenology. For DM masses $M_\text{DM} \lesssim M_H/2$ the relic abundance of the DM particles is driven by the $s$-channel annihilation through the exchange of the Higgs boson.\footnote{Assuming a standard cosmological history and thermal freeze-out dark matter, the minimal Higgs portal scenario with light DM is tightly constrained, with only a narrow mass range around $M_\text{DM} \simeq M_H/2$ being  allowed. However, this can be relaxed in alternative cosmological scenarios and DM production mechanisms, see e.g.~Refs.~\cite{Hardy:2018bph,Bernal:2018ins,Bernal:2018kcw}.} As the DM--nucleon elastic scattering amplitudes are directly proportional to the portal coupling~\cite{Kanemura:2010sh}, it can be additionally constrained by DM direct detection experiments.  These are sensitive to the elastic scattering of the DM particles with nuclei, mediated by the Higgs boson. Hence, in turn, the upper limit on $\lambda$ can be translated into an upper limit on the (spin-independent) DM-nucleon scattering cross section, $\sigma_{\text{DM}-\text{nucleon}}$ (see Refs.~\cite{Kanemura:2010sh,Djouadi:2011aa}).

In Fig.~\ref{fig:minimalHP} we show the current [\emph{left panel}] and prospective [\emph{right panel}] upper limits on $\sigma_{\text{DM}-\text{nucleon}}$ inferred from a current and HL-LHC prospective upper limit on $\BRHinv$ of $20\%$ and $2.5\%$, respectively.\footnote{For current  ATLAS and CMS results for the minimal Higgs portal interpretation see Refs.~\cite{Aad:2015pla,Khachatryan:2016whc,Aaboud:2018sfi,Sirunyan:2018owy,ATLAS-CONF-2018-054}.} These are shown for scalar [\emph{blue curve}], fermion [\emph{red curve}]  and vector [\emph{green curve}] DM. The uncertainty bands on these curves correspond to the uncertainty in the Higgs-nucleon coupling form factor, where we use the recent result from Ref.~\cite{Hoferichter:2017olk}. For comparison we include in Fig.~\ref{fig:minimalHP} current limits from DM direct detection experiments \textsc{Xenon10}~\cite{Angle:2011th}, \textsc{Xenon100}~\cite{Aprile:2012nq} and \textsc{Xenon1T}~\cite{Aprile:2018dbl}, prospective limits from \textsc{XenonNT}~\cite{Aprile:2015uzo} and \textsc{SuperCDMS} at SNOLAB~\cite{Agnese:2016cpb}. For completeness, we also show the favoured parameter regions from excesses seen in the \textsc{DAMA/LIBRA}~\cite{Savage:2008er}, \textsc{Cresst}~\cite{Angloher:2011uu}, \textsc{CDMS~II}~\cite{Agnese:2013rvf} and \textsc{CoGeNT}~\cite{Aalseth:2012if} experiments.\footnote{Note that the limits and favoured regions from DM direct detection experiments assume for the incoming flux of DM particles that the observed relic density in the Universe is fully saturated by this one DM particle species.} We furthermore indicate by the grey area in Fig.~\ref{fig:minimalHP} the \emph{neutrino floor}, i.e.~the parameter region that is inaccessible to DM direct detection experiments due to the irreducible neutrino flux background~\cite{Billard:2013qya}\footnote{Note that a complete study of these minimal portals would need to include further theoretical and experimental constraints on the models parameter space; see e.g. \cite{Athron:2017kgt,Athron:2018hpc} for recent discussions.}.

Currently, the inferred limit from invisible Higgs searches yields the most sensitive constraint in the low mass region, $M_\text{DM} \lesssim 6, 10~\mbox{and}~30\UGeV$ for scalar, fermion and vector DM, respectively, while at larger DM masses the \textsc{Xenon1T} limit is more constraining. In particular, in the fermion and vector DM case, the $\BRHinv$ limit probes deep into the parameter region that is inaccessible to direct detection experiments. A future limit on the invisible Higgs decay rate from the HL-LHC will improve the limits on $\sigma_{\text{DM}-\text{nucleon}}$ by almost one order of magnitude, which pushes the limit for light scalar DM close to the neutrino floor. For fermion (vector) DM in the mass range $10\UGeV \lesssim M_{\text{DM}} \lesssim 20~(60)\UGeV$, in case of a future excess seen in the \textsc{XenonNT} data, complementary measurements of an invisible Higgs decay at the HL-LHC may be possible. 

\begin{figure}
\centering
\includegraphics[width=0.49\textwidth]{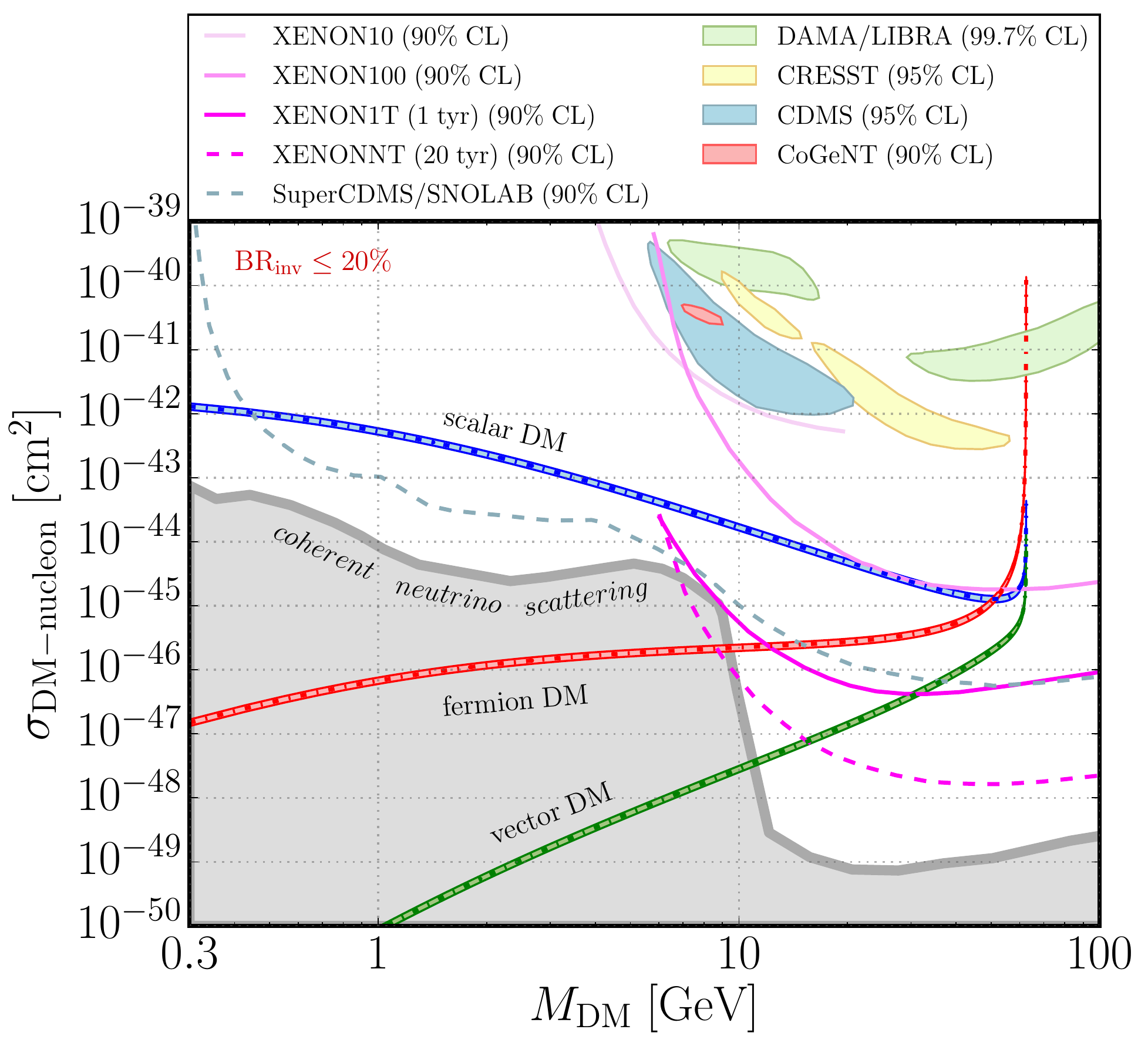}
\hfill
\includegraphics[width=0.49\textwidth]{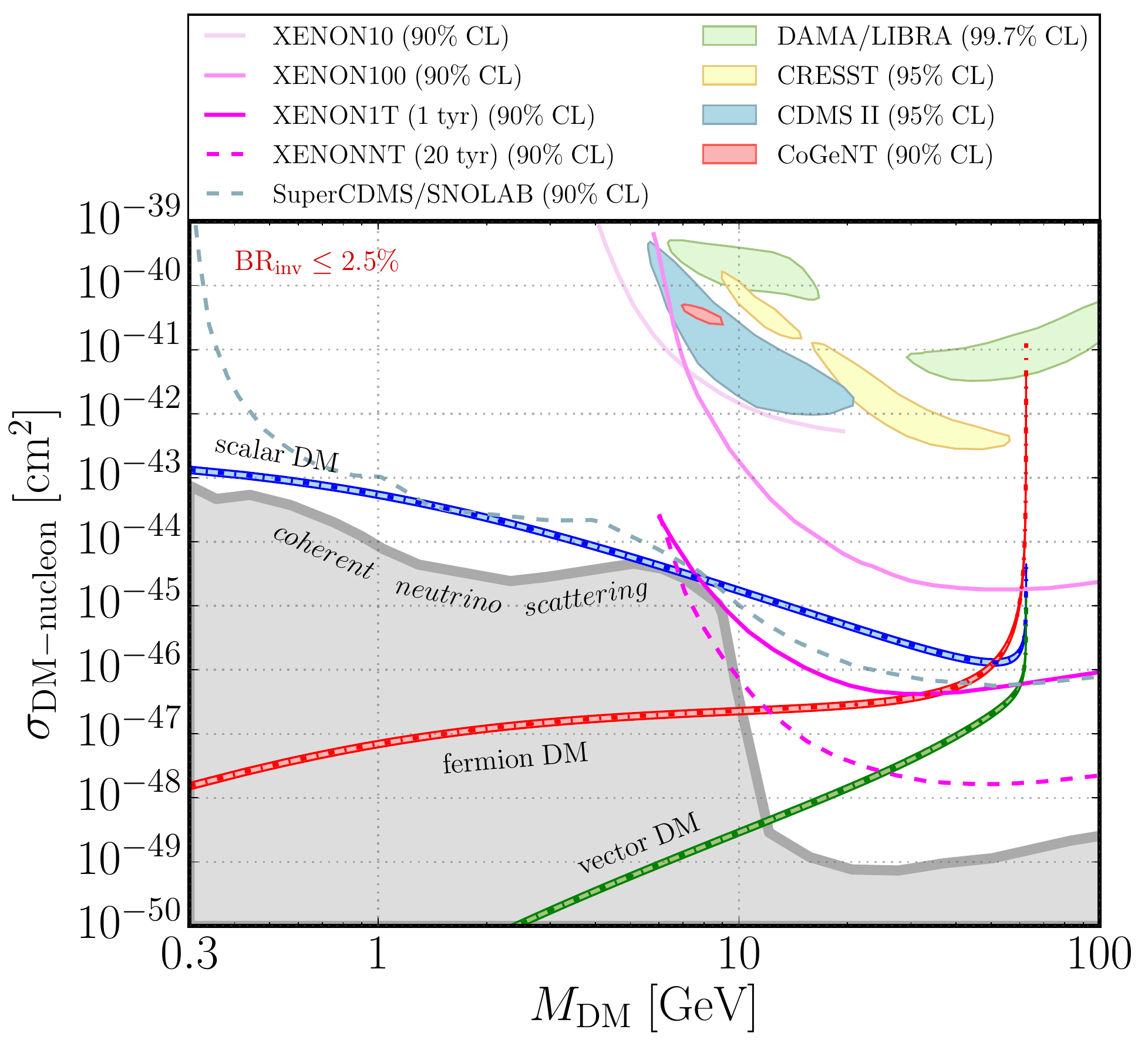}
\caption{\label{fig:mini} Implications for the minimal Higgs portal model: Comparison of current (\emph{left figure}) and future HL-LHC (\emph{right figure}) limits from invisible Higgs searches  with limits from DM direct detection experiments on the spin-independent DM-nucleon scattering cross section, $\sigma_{\mathrm{DM}-\mathrm{nucleon}}$, as a function of the DM mass, $\MDM$. The inferred limits from invisible Higgs searches are shown for scalar DM (\emph{blue curve}), fermion DM (\emph{red  curve}) and for vector DM (\emph{green  curve}). In addition we show present limits (\emph{solid lines}), favoured regions (\emph{filled areas}) and future sensitivity (\emph{dashed lines}) of the DM direct detection experiments \textsc{Xenon10}~\cite{Angle:2011th}, \textsc{Xenon100}~\cite{Aprile:2012nq}, \textsc{Xenon1T}~\cite{Aprile:2018dbl}, \textsc{XenonNT}~\cite{Aprile:2015uzo}, \textsc{SuperCDMS} at SNOLAB~\cite{Agnese:2016cpb}, \textsc{DAMA/LIBRA}~\cite{Savage:2008er}, \textsc{Cresst}~\cite{Angloher:2011uu}, \textsc{CDMS~II}~\cite{Agnese:2013rvf} and \textsc{CoGeNT}~\cite{Aalseth:2012if} (\emph{see legend}). The grey area indicates regions inaccessible to DM direct detection experiments due to the irreducible neutrino flux background~\cite{Billard:2013qya}.}
\label{fig:minimalHP}
\end{figure}

\asubsubsection{Scalar singlet portal}
\label{sec6:singletHP}
We now turn our discussion to a model that features an additional scalar singlet in the visible sector, which provides the portal interaction to the hidden DM sector. In contrast to the minimal Higgs portal discussed in Section~\ref{sec6:minimalHP}, this model allows for a modification of the $125\UGeV$ Higgs couplings, and thus for a non-trivial interplay between direct invisible Higgs searches and Higgs rate measurements at the HL-LHC. For illustration, we focus here on the case of scalar DM, the other cases (fermion and vector DM) can be treated analogously. The model is inspired by Refs.~\cite{Englert:2011yb,Robens:2015gla}.

The SM Higgs sector is extended by two real scalar singlet fields, $S$ and $X$. Imposing a $\mathbb{Z}_2$ symmetry described by the transformation $S\to -S$, $X\to -X$, the model is characterised by the scalar potential $\mathcal{V} = \mathcal{V}_\mathrm{visible}  + \mathcal{V}_\mathrm{hidden}$, where
\begin{align}
\label{Eq:Vvis}\mathcal{V}_\mathrm{visible} &=   \mu_{\Phi}^2 \Phi^\dagger \Phi + \lambda_\Phi (\Phi^\dagger \Phi)^2 + \mu_S^2 S^2 + \lambda_S S^4 + \lambda_{\Phi S} \Phi^\dagger \Phi S^2,\\
\label{Eq:Vhid} \mathcal{V}_\mathrm{hidden} &= \frac{1}{2}\left[\mu^2_{X} X^2 + \lambda_X X^4 + \lambda_{SX} S^2 X^2 + \lambda_{\Phi X} \Phi^\dagger \Phi X^2\right]. 
\end{align}
For the sake of simplicity we assume that the quartic interaction between the scalar doublet $\Phi$ and the DM scalar $X$ can be neglected, $\lambda_{\Phi X} \approx 0$. After electroweak symmetry breaking the scalar $\mathrm{SU}(2)_L$ doublet field $\Phi$ is given by 
$\Phi \equiv \left(0 \quad \phi + v\right)^T/\sqrt{2}$, with the vacuum expectation value (VEV) $v \approx 246\UGeV$. 
We assume the scalar field $S$ to acquire a non-zero VEV, $v_S$, which softly breaks the $\mathbb{Z}_2$ symmetry, such that the singlet field $S$ is given by $S \equiv (s + v_S)/\sqrt{2}$. Through the last term in Eq.~\eqref{Eq:Vvis} the non-zero VEVs induce a mixing of the physical degrees of freedom of these two fields, $\phi$ and $s$,
\begin{align}
\begin{pmatrix} h \\ H \end{pmatrix} = \begin{pmatrix} \cos \alpha & -\sin\alpha \\ \sin\alpha & \cos\alpha \end{pmatrix} \begin{pmatrix} \phi \\ s \end{pmatrix},
\end{align}
with the masses of the physical states $h$ and $H$ given by
\begin{align}
M_{h/H}^2 = \lambda_\Phi v^2 + \lambda_S v_S^2 \mp \sqrt{\left(\lambda_\Phi v^2 - \lambda_S v_S^2\right)^2 + \left(\lambda_{\Phi S} v v_S\right)^2},
\end{align}
and the mixing angle $\alpha \in [-\tfrac{\pi}{2}, \tfrac{\pi}{2}]$ given by
\begin{align}
\tan 2\alpha = \frac{\lambda_{\Phi S} v v_s}{\lambda_S v_S^2 - \lambda_\Phi v^2}.
\end{align}
In contrast, $X$ does not acquire a VEV. As a result $X$ is stable and thus a possible DM candidate, with a mass given by
$M_X^2  = \mu_X^2  + \lambda_{SX} v_S^2/2$.

In this analysis, we assume $M_H = 125.09\UGeV$, and $M_h < M_H$. Furthermore, we discard the quartic interaction term $\propto \lambda_X$ in Eq.~\eqref{Eq:Vhid} as this operator is irrelevant for our study. With this, the model can then be parametrised in terms of the following input quantities:
\begin{align}
M_h, \cos\alpha, v_S, M_X, \lambda_{SX}.
\end{align}
The couplings of the Higgs bosons $h$ and $H$ to SM gauge bosons and fermions are universally suppressed by the mixing,
\begin{align}
g_h/g_{h,\mathrm{SM}} = \cos \alpha , \qquad g_H/g_{H,\mathrm{SM}} = \sin\alpha.
\end{align}
If the DM scalar $X$ is light enough the portal coupling $\lambda_{SX}$ gives rise to decays of the Higgs bosons $h$ and $H$ to the invisible $XX$ final state. The partial decay widths are given by
\begin{align}
\label{eq:invdecaywidth}
\begin{array}{l} \Gamma(h\to XX) = \sin^2\alpha \cdot \Gamma_{XX} (M_h), \\ \Gamma(H\to XX) = \cos^2\alpha \cdot \Gamma_{XX} (M_H), \end{array} \quad \mbox{with} \quad \Gamma_{XX}(M) = \frac{\lambda_{SX}^2 v_S^2 }{32\pi M} \sqrt{1 - \frac{4M_X^2}{M^2}}.
\end{align}
Furthermore, if $M_h \le M_H/2$, the heavier Higgs boson $H$ can decay into $hh$, with the partial width given by
\begin{align}
\Gamma(H\to hh) = \frac{\lambda_{Hhh}^2}{32\pi M_H} \sqrt{1 - \frac{4M_h^2}{M_H^2}},
\end{align}
and the effective $Hhh$ coupling\footnote{Note that the relative sign between the two terms in Eq.~\eqref{Eq:lamHhh} differs with respect to Eq.~(13) in Ref.~\cite{Englert:2011yb}.}
\begin{align}
\label{Eq:lamHhh}
\lambda_{Hhh} = & - 3\sin2\alpha \left[ \lambda_S v_S \sin\alpha  + \lambda_{\Phi} v \cos\alpha \right] \nonumber\\
& - \tan2\alpha \left( \lambda_S v_S^2 - \lambda_{\Phi} v^2 \right) \left[ (1-3\sin^2\alpha) \frac{\cos\alpha}{v} + (1-3 \cos^2\alpha) \frac{\sin\alpha}{v_S} \right].
\end{align}

Through the successive decay of the lighter Higgs boson $h$ into either final states with SM particles (denoted as `SM') or the invisible $XX$ final state, this gives rise to the following signatures\footnote{{Note that LHC searches for the semi-invisible and visible final states are highly complementary to invisible Higgs searches in this model.}}
\begin{align}
H \to hh \to \left\{ \begin{array}{l} (\mathrm{SM}) (\mathrm{SM}) ,\quad \mbox{(visible)}, \\ (\mathrm{SM}) (XX) ,\quad \mbox{(semi-invisible)}, \\
(XX) (XX) ,\quad \mbox{(invisible)}.
\end{array} \right.
\end{align}

The branching ratio of the invisible decay of the SM-like Higgs boson $H$ is given by
\begin{align}
\BRHinv = \mathrm{BR}(H\to XX) + \mathrm{BR}(H\to hh) \cdot \mathrm{BR}(h\to XX)^2.
\end{align}
 
\begin{figure}[t]
\centering
\includegraphics[width=0.5\textwidth]{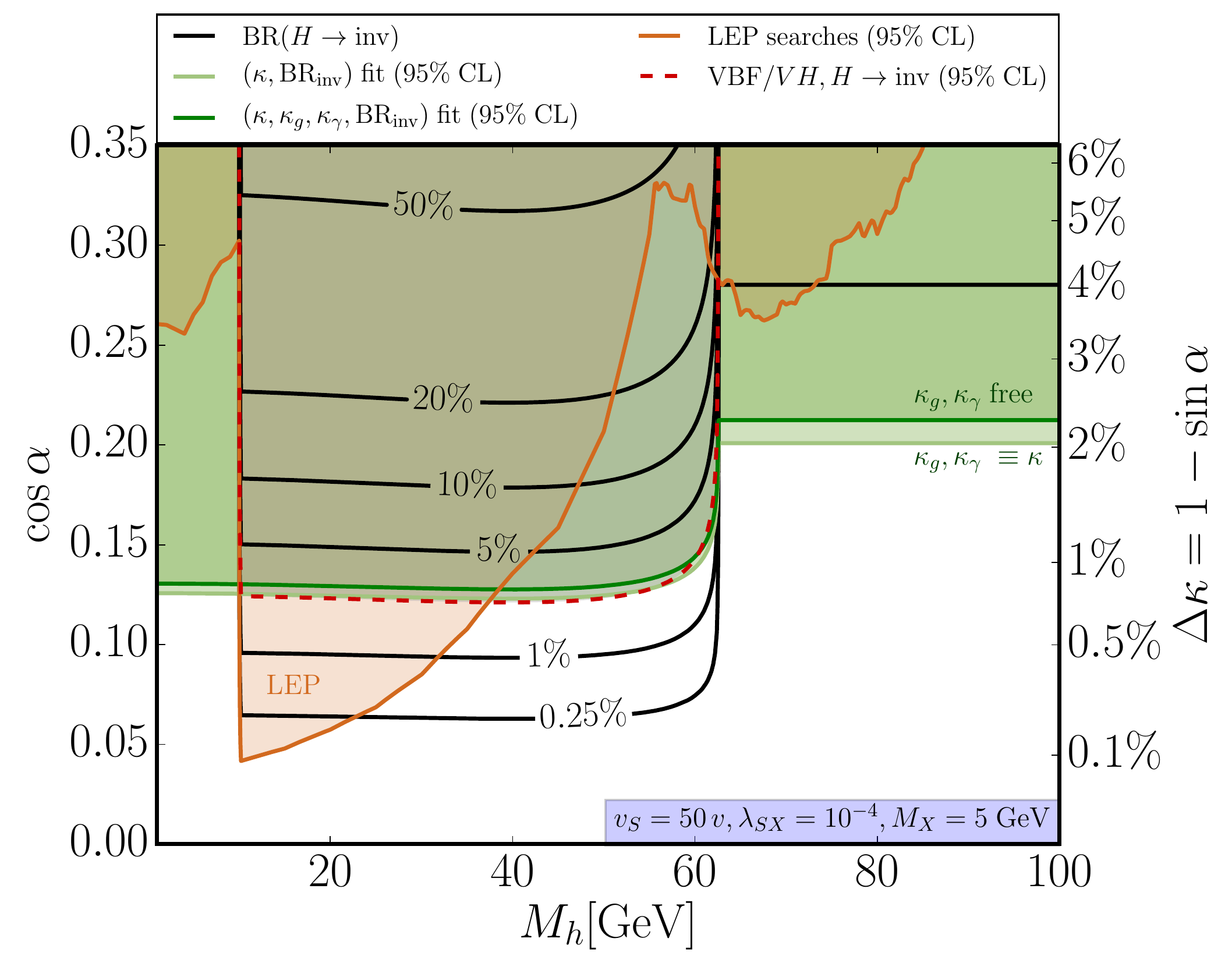}\hfill
\includegraphics[width=0.5\textwidth]{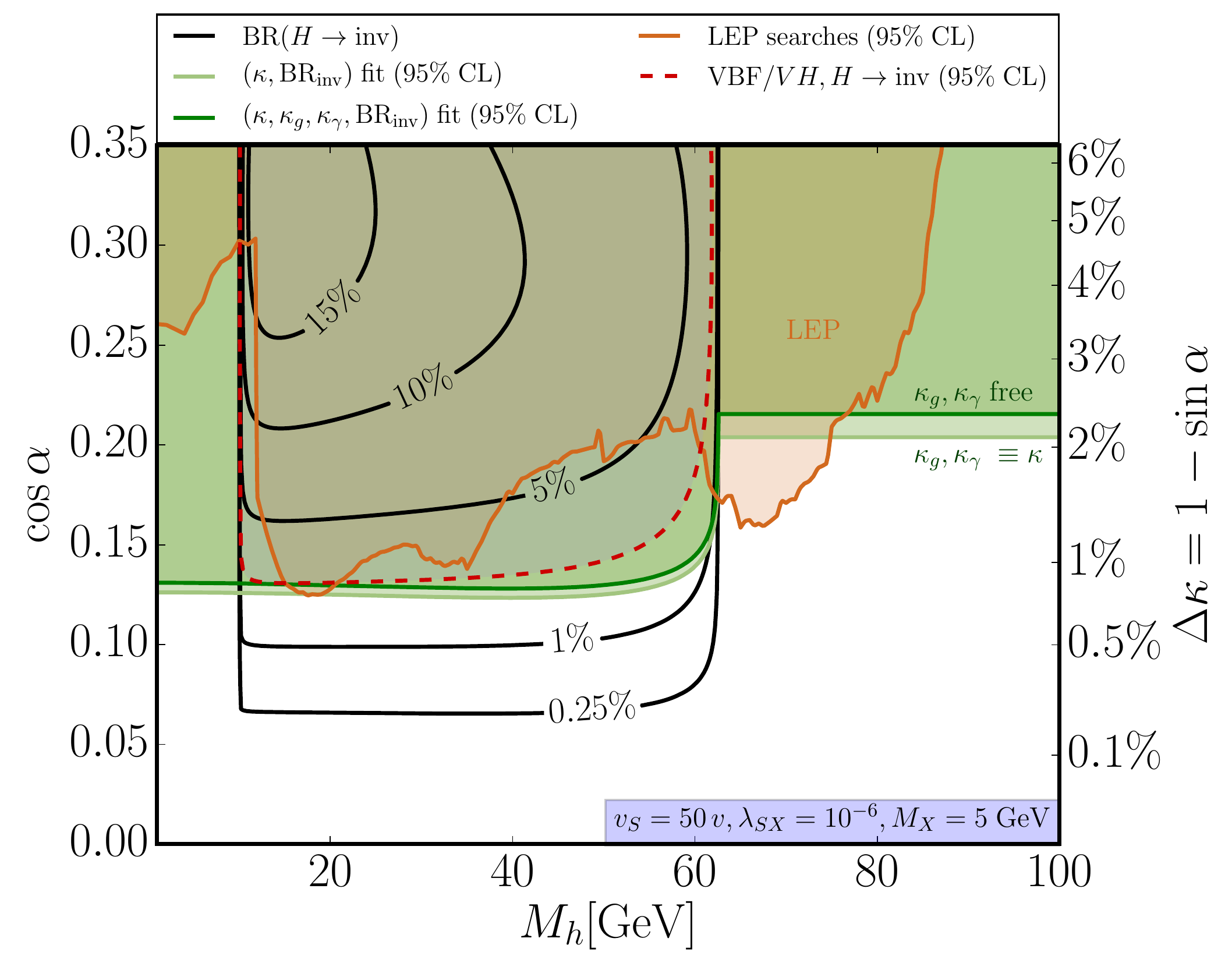}\\
\includegraphics[width=0.5\textwidth]{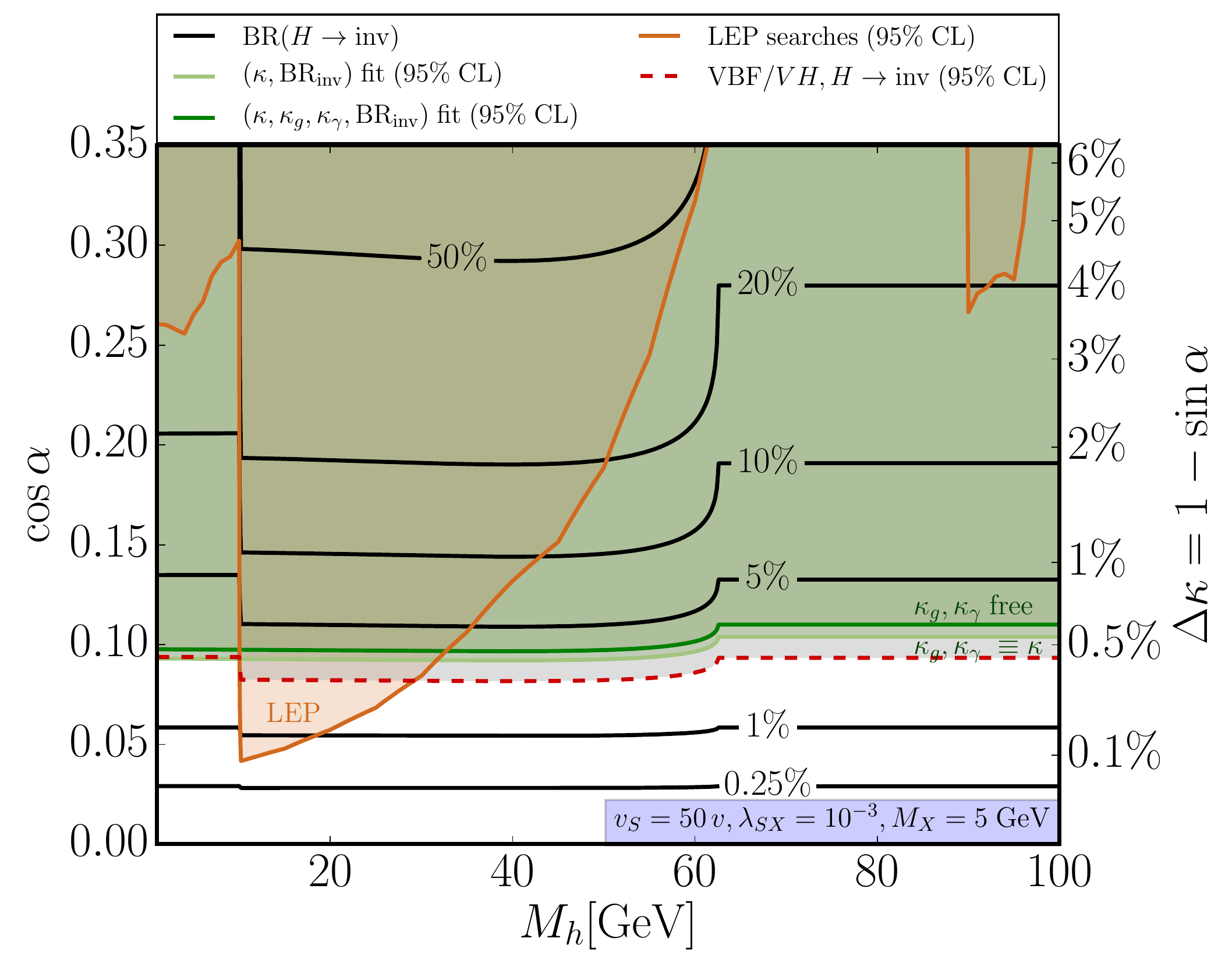}\hfill
\includegraphics[width=0.5\textwidth]{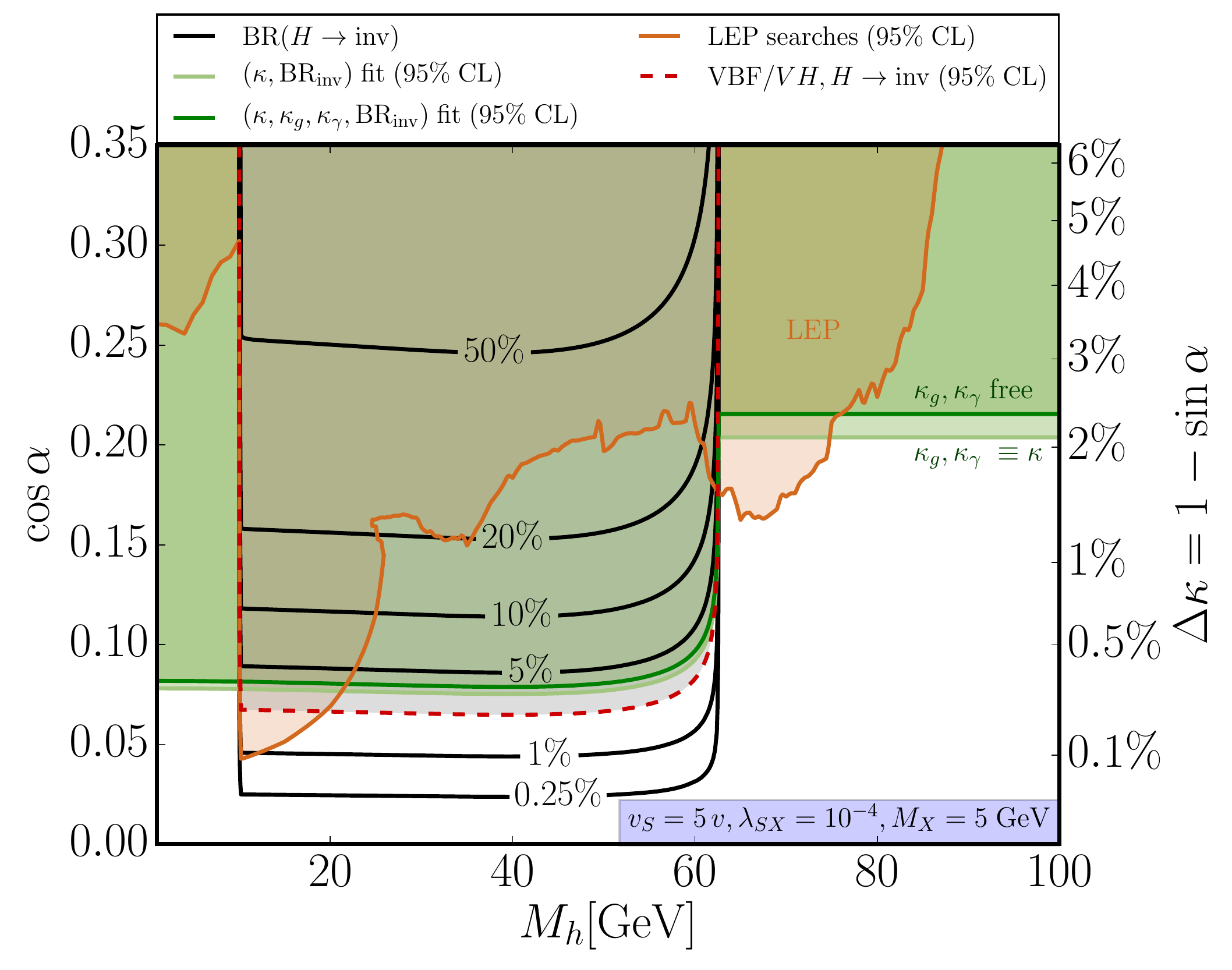}
\caption{Implications for the scalar singlet portal model, shown in the $(M_h, \cos\alpha)$ parameter plane for a DM mass of $M_X = 5\UGeV$ and $(v_S, \lambda_{SX}) = (50\,v, 10^{-4})$ [\emph{top left}], $(50\,v, 10^{-6})$ [\emph{top right}], $(50\,v, 10^{-3})$ [\emph{bottom left}] and $(5\,v, 10^{-4})$ [\emph{bottom right}]. Black solid contours show the invisible Higgs decay rate, $\mathrm{BR}(H\to \mathrm{inv})$, the red dashed contour/grey area indicates the expected HL-LHC limit from invisible Higgs searches, pale and bright green contours/areas indicate the indirect constraints from HL-LHC Higgs rate measurements (using the two parametrisations, see Section~\ref{sec6:effC}), and the orange contour/area marks the excluded region from LEP searches. See text for more details.}
\label{Fig:singletportal}
\end{figure}
 
 In Fig.~\ref{Fig:singletportal} we show the invisible Higgs decay rate $\BRHinv$ in the $(M_h, \cos\alpha)$ plane, for fixed DM mass $M_X = 5\UGeV$, and four choices $(v_S, \lambda_{SX}) = (50\,v, 10^{-4})$ [\emph{top left}], $(50\,v, 10^{-6})$ [\emph{top right}], $(50\,v, 10^{-3})$ [\emph{bottom left}] and $(5\,v, 10^{-4})$ [\emph{bottom right}]. For better illustration, the secondary y-axis shows the deviation from the SM coupling strength, $\Delta \kappa = 1 -\sin\alpha$, of the heavier Higgs boson $H$. The $\BRHinv$ prediction is given by the black solid contours. Various constraints (at $95\%\CL$) are included in the figures: future direct invisible Higgs searches (\emph{red dashed contour}/\emph{grey area}), future indirect limits from Higgs rate measurements {at the HL-LHC (assuming S2)} using the two parametrisations of Section~\ref{sec6:effC} [cf.~Fig.~\ref{fig:effC}] (\emph{solid pale/bright green contour and area}), and LEP searches for the lighter Higgs boson $h$ (\emph{orange contour and area}), obtained via \texttt{HiggsBounds}~\cite{Bechtle:2008jh,Bechtle:2011sb,Bechtle:2013wla}. For the latter, the relevant experimental analyses are searches for $e^+e^-\to Zh$ production with $h$ either decaying to invisible particles~\cite{Searches:2001ab,Abdallah:2003ry,Achard:2004cf,Abbiendi:2007ac} or to SM particles (in particular, $b\bar{b}$)~\cite{Schael:2006cr}, as well as the decay mode independent analysis by OPAL~\cite{Abbiendi:2002qp}.

In all four panels of Fig.~\ref{Fig:singletportal} we can identify two kinematic thresholds for the invisible $H$ decay: at $M_h = M_H/2 \simeq 62.5\UGeV$, where the cascade decay $H\to hh\to (XX)(XX)$ becomes available for decreasing $M_h$, and at $M_h = 2 M_X = 10\UGeV$, where the decay $h\to XX$ kinematically closes for smaller $M_h$, and thus the $H\to hh$ decay cannot further lead to an invisible final state. Above the first threshold, $M_h > M_H/2$, and below the second threshold, $M_h < 2M_X$, the invisible $H$ decay is solely given by {the direct decay} $H\to XX$. 

For the parameter choice $(v_s, \lambda_{SX}) = (50\,v, 10^{-4})$ (\emph{top left panel}), the direct invisible Higgs searches at the HL-LHC {will} provide similar bounds as the indirect constraints from the Higgs rates for the mass range $M_h \in [2 M_X, M_H/2]$. However, in the mass range $M_{h}\sim (10-40)\UGeV$, the LEP searches for invisible $h$ decays will still yield the strongest exclusion. Note that the Higgs rate measurements are always constraining the sum $\BRNP = \mathrm{BR}(H\to XX) + \mathrm{BR}(H\to hh)$, regardless of whether the decay $H\to hh$ leads to an invisible final state. Hence, they remain to be sensitive in the low mass region $M_h < 2 M_X$.

For a larger Higgs-portal interaction, $\lambda_{SX} = 10^{-3}$ (\emph{bottom left panel}), the direct decay $H\to XX$ becomes more prominent, leading to sizeable $\BRHinv$ even at smaller $\cos\alpha$. Here, direct invisible Higgs searches at HL-LHC will be most constraining and will supersede the LEP limits except in the mass range $M_h\sim (10-33)\UGeV$. {In contrast,} for very small Higgs-portal interaction, $\lambda_{SX} = 10^{-6}$ (\emph{top right panel}) {the invisible Higgs decay rates are much smaller. Nevertheless,} future indirect constraints from Higgs rate measurements will supersede the LEP limits {in almost the entire mass range except for $M_h$ values between $62$ to $75\UGeV$.}
Note that the LEP exclusion arises from $e^+e^- \to Zh \to Z(b\bar{b})$ searches~\cite{Schael:2006cr}.

If we decrease the VEV of the singlet field, $v_s = 5\,v$ (\emph{bottom right panel}), the effective $Hhh$ coupling becomes larger, leading to a more pronounced $H\to hh$ decay if kinematically accessible. Hence, in the region $M_{h} < M_H/2$, the HL-LHC constraints both from direct invisible Higgs searches and Higgs rate measurements are very strong and supersede the LEP constraints in almost the entire mass range up to $M_h \lesssim M_H/2$. {In this case, the direct invisible Higgs searches} are slightly more {sensitive} than the {Higgs rate measurements}.

In summary, the example parameter choices made in Fig.~\ref{Fig:singletportal} illustrate an interesting interplay between past LEP searches for a light Higgs boson $h$, future HL-LHC searches for an invisibly decaying SM-like Higgs boson $H$, and future HL-LHC precision Higgs rate measurements. Depending on the parameter choice, each experimental probe can be the most sensitive/constraining one, which highlights their complementarity and strongly motivates a corresponding experimental program at the HL-LHC.

\asubsection{Conclusions}
\label{sec6:conclusions}

Higgs portal models are intriguing and simple new physics scenarios that contain a dark matter candidate which can be tested at collider as well as astrophysical experiments. The HL-LHC will be able to constrain the Higgs boson--dark matter coupling constant and probe the parameter regime down to an invisible Higgs decay rate of 2$\%$. For low dark matter masses, $M_\text{DM}\,\lesssim\,30\UGeV$, these bounds are typically more constraining than limits obtained from dark matter direct detection experiments. For a specific model with two visible scalar states and a scalar dark matter candidate, we presented scenarios for which future HL-LHC {searches} will supersede complementary constraints from LEP searches for a light scalar boson. In summary, the future HL-LHC measurements of the Higgs signal strength, as well as {direct searches} for the invisible decay of the observed Higgs boson, promise to provide important insight within the framework of Higgs portal models. {The sensitivity can further be improved by the future electron-proton collider LHeC. In particular, the indirect constraints from Higgs rate measurements will improve substantially if HL-LHC and LHeC results are combined.}

\newpage

\asection[K. Nikopoulos, A. Schmidt, L. Sestini, Y. Soreq]{Higgs flavor and rare decays}\label{sec7}

\asubsection{Introduction}

In this section we cover the current status and future prospects for measuring the different Higgs couplings to fermions, these go under the generic name of ``Higgs and Flavor". 
The Higgs mechanism of the SM predict that the Yukawa couplings are proportional to the fermion mass and CP conserving, or more precisely 
\begin{align}
    y^{\rm SM}_f = \sqrt{2} m_f / v \, ,
\end{align}
where the tree-level flavor changing couplings are zero.  Currently,
only the third generation Yukawa couplings were directly measured and
found to be in agreement with the SM prediction, see
Refs.~\cite{Aad:2015vsa,Aaboud:2018pen,Sirunyan:2018kst,Aaboud:2018zhk,
  Aaboud:2017jvq,Aaboud:2018urx,Sirunyan:2018shy} for recent results
on $h\tau\bar{\tau}$, $h b\bar{b}$ and $h \bar{t} t$.  However, for
the Higgs coupling to first and second generations there are only
upper bounds
~\cite{Aaboud:2018fhh,Aaboud:2017ojs,Perez:2015aoa,Altmannshofer:2015qra,
  Kagan:2014ila}.

Below, we adapt the generalised $\kappa$ framework to describe deviations of the Higgs couplings from their SM values due to new physics~(NP).
In particular, we define 
\begin{equation}
\label{eq:L_eff}
\begin{split}
	\mathcal{L}_{\rm eff} &= -\kappa_{f_i} \frac{m_{f_i}}{v} h \bar f_i f_i + i \tilde \kappa_{f_i} \frac{m_{f_i}}{v} h \bar f_i \gamma_5 f_i  
	- \left[\left( \kappa_{f_i f_j} + i \tilde \kappa_{f_i f_j} \right) h \bar f_L^i f_R^j +{\rm h.c.}\right]_{i\neq j}, 
\end{split}
\end{equation}
where a sum over fermion type $f=u,d,\ell$ and generations $i,j=1,2,3$ is implied.
The first two terms are flavour-diagonal with the first term CP-conserving and
the second CP-violating.  The terms in square brackets are flavour
violating. The real (imaginary) part of the coefficient is CP
conserving (violating). In the SM, we have $\kappa_{f_i}=1$ while $\tilde
\kappa_{f_i}=\kappa_{{f_i}{f_j}}=\tilde \kappa_{{f_i}{f_j}}=0$.

The different Higgs Yukawa couplings can be probed by direct and
indirect methods.  Direct methods include $t\bar{t}h$ (for
top~\cite{Aaboud:2017jvq,Aaboud:2018urx,Sirunyan:2018shy}), $Vh, h\to
b\bar{b}, c\bar{c} $~(for
bottom~\cite{Aaboud:2018zhk,Sirunyan:2018kst} and
charm~\cite{Aaboud:2018fhh}), $h\to \ell^+ \ell^-$ (for
leptons~\cite{Aaboud:2018pen,Aaboud:2017ojs,Aad:2015vsa}) and
exclusive decays for photon and vector
meson~\cite{Aad:2015sda,Khachatryan:2015lga,Aaboud:2018txb,Aaboud:2017xnb}
(for light quarks).  In addition, the upper bound on the Higgs total
width from $t\to ZZ^*$ and $h\to\gamma\gamma$ signal shapes is an
unavoidable constraint on the rates to any light
particles~\cite{Perez:2015aoa}.  In principle, one can use the
off-shell Higgs width measurement~\cite{Englert:2014aca,Englert:2014ffa},
but it involves assumptions about the ratio between off-shell and
on-shell Higgs productions.  In addition, there are several indirect
probes of the different Higgs Yukawa couplings, such as kinematic
distributions~\cite{Soreq:2016rae,Bishara:2016jga}.  A global fit of
the Higgs data also provides a bound on the different Yukawa via the
bound on the non SM decays of the Higgs (up to small effects on the
Higgs production,
see~\cite{Delaunay:2013pja,Kagan:2014ila,Perez:2015aoa}), however,
this bound is subject to different assumptions.

The Higgs production and decay signal strengths from the CMS collaboration~\cite{Sirunyan:2018koj} and from ATLAS for $h\to c\bar{c}$~\cite{Aaboud:2018fhh} (most recent at the time of writing) from a global fit which includes the direct observation of $t\bar t h$ production are
\begin{align}
    \mu^{ttH} &= 1.18^{+0.30}_{-0.27}\, ,\qquad \mu^{bb}=1.12^{+0.29}_{-0.29}\, , \qquad
    \mu^{cc} < 105 \, , \nonumber\\ \mu^{\tau\tau} &= 1.20^{+0.26}_{-0.24}\, ,\qquad   \mu^{\mu\mu}=0.68^{+1.25}_{-1.24}\,.
\end{align}
In terms of modifications of the flavor-diagonal and CP-conserving Yukawas,  the best fit values are
\begin{align}
\label{eq:kappalimits}
    \kappa_t &= 1.11^{+0.12}_{-0.10}, & \kappa_b &= -1.10^{+0.33}_{-0.23},\nonumber\\
    \kappa_\tau &= 1.01^{+0.16}_{-0.20}, & \kappa_\mu &= 0.79^{+0.58}_{-0.79}.
\end{align}
See also \cite{CMS-PAS-HIG-15-002,CMS-PAS-HIG-17-031}.
The light quarks $u$, $d$, and $s$ and the charm Yukawa can be constrained from a global fit of Higgs data and precision EW measurements at LEP. Floating all couplings in the fit results in the following upper bounds~ \cite{Perez:2015aoa,Kagan:2014ila},
\begin{align}
\kappa_u&<3.4 \cdot 10^{3},  &\kappa_d&<1.7 \cdot 10^3, &\kappa_s&< 42, & \kappa_c \lesssim 6.2\nonumber.
\end{align}
While for the electron Yukawa, the upper bound on $\mbox{BR}(h\to e^+e^-)$ at the LHC translates to an upper bound, $|\kappa_e|<611$~\cite{Khachatryan:2014aep,Altmannshofer:2015qra}. And for future prospects, see~\cite{Perez:2015lra,Brivio:2015fxa,Koenig:2015pha,Aad:2015sda,Bodwin:2014bpa,Bodwin:2013gca}.

The upper bounds on $\kappa_{c,s,d,u}$ roughly
correspond to the size of the SM bottom Yukawa coupling and are thus much bigger than the corresponding SM Yukawa couplings. The upper bounds can be saturated only if one allows for large cancellations between the
contribution to fermion masses from the Higgs VEV and an equally large
but opposite in sign contribution from NP. We will show that in models
of NP motivated by the hierarchy problem, the effects of NP are
generically well below these bounds. A summary of the projected limits on $\kappa_{c,s,d,u}$ is given in Fig.~\ref{fig:sec7:summary} using the methods outlined in this section: exclusive decays of the Higgs, fits of differential cross-sections, constraints from the total Higgs width assuming a value of $200~\mathrm{\UMeV}$, a global fit of Higgs production cross-sections, and direct searches for a $c\bar{c}$ final state.
\begin{figure}
    \centering
    \includegraphics[width=0.7\textwidth]{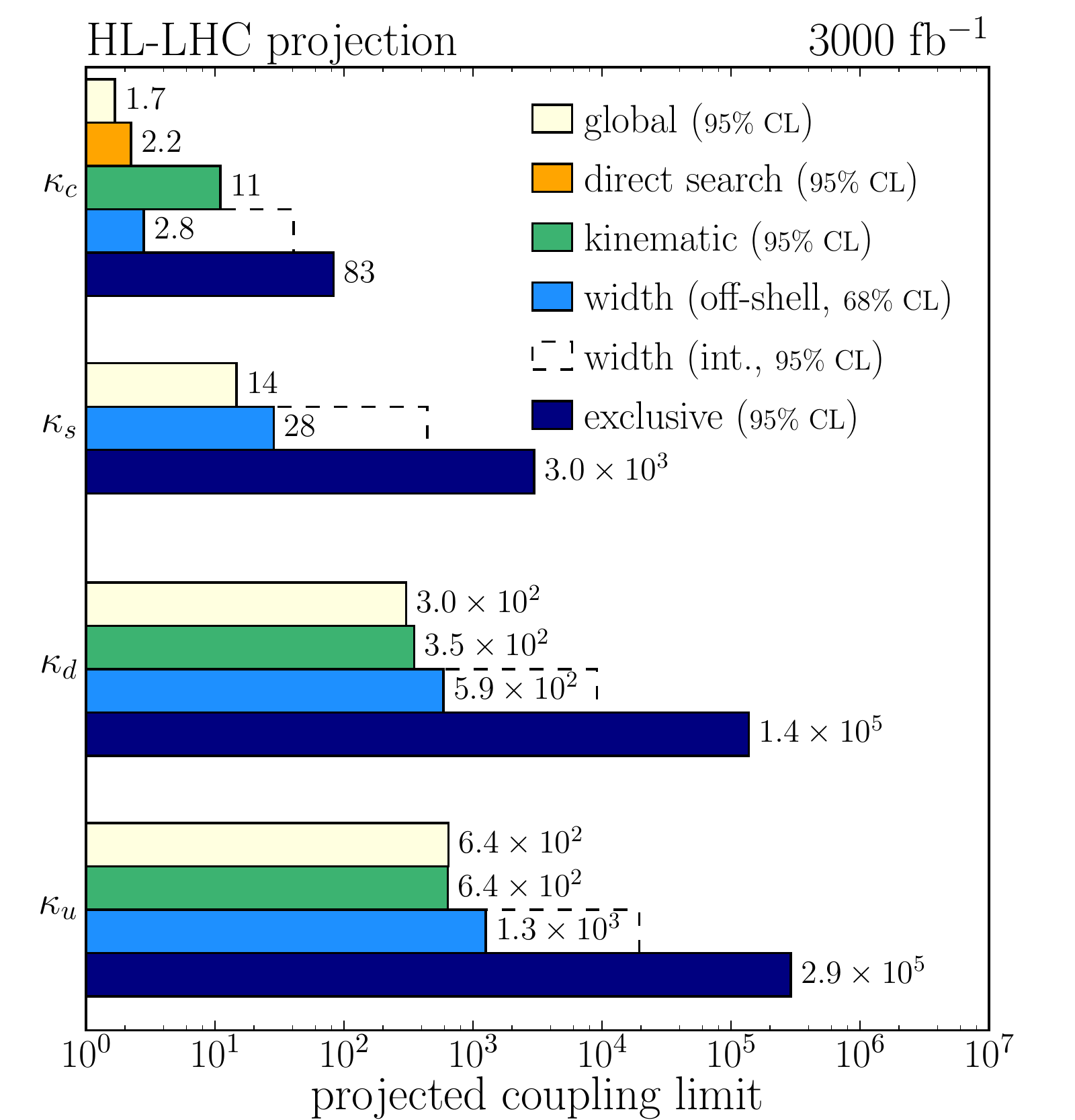}
    \caption{Summary of the projected HL-LHC limits on the light quark Yukawa couplings, including charm.}
    \label{fig:sec7:summary}
\end{figure}

The CP-violating flavour-diagonal Yukawa couplings, $\tilde \kappa_{f_i}$,
are well constrained from bounds on the electric dipole moments~(EDMs)
\cite{Brod:2013cka,Chien:2015xha,Altmannshofer:2015qra,Egana-Ugrinovic:2018fpy,Brod:2018pli} under the assumption of no 
cancellation with other contributions to EDMs beyond the Higgs contributions.
For the electron Yukawa, the latest ACME measurement~\cite{Baron:2013eja,Andreev:2018ayy} results into an upper bound of $\tilde\kappa_e<1.9\times 10^{-3}$~\cite{Altmannshofer:2015qra}. Whereas for the bottom and charm Yukawas, the strongest limits come from the neutron EDM~\cite{Brod:2018pli}. Using the NLO QCD theoretical prediction, this translates into the upper bounds $\tilde\kappa_b<5$ and $\tilde\kappa_c<21$ when theory errors are taken into account.
For the light quark CPV Yukawas, measurement of the Mercury EDM places a strong bound on the up and down Yukawas of $\tilde\kappa_u<0.1$ and $\tilde\kappa_d<0.05$~\cite{Brod:2018xyz} (no theory errors) while the neutron EDM measurement gives a weaker constraint on the strange quark Yukawa of $\tilde\kappa_s<3.1$~\cite{Brod:2018xyz} (no theory errors).

The flavour violating Yukawa couplings are well constrained by the low-energy
flavour-changing neutral current measurements
\cite{Harnik:2012pb,Blankenburg:2012ex,Gorbahn:2014sha}. A notable
exception are the flavour-violating couplings involving a tau lepton. The
strongest constraints on $\kappa_{\tau\mu}, \kappa_{\mu\tau},
\kappa_{\tau e}, \kappa_{e \tau}$ are thus from direct searches of flavour-violating Higgs decays at
the LHC \cite{Sirunyan:2017xzt,Aad:2016blu}.
Finally, the LHC can also set bounds on rare FCNC top decays involving a Higgs~\cite{Aaboud:2017mfd,Khachatryan:2016atv,Aad:2015pja,Aad:2014dya}. The strongest current bound, for example, is $\sqrt{|\kappa_{ct}|^2+|\kappa_{tc}|^2}<0.06$ at 95\%CL where the latest ATLAS bound was converted to a bound on the Yukawa modifier at leading order.

\asubsection[F. Bishara]{New Physics benchmarks for modified Higgs couplings}

Here we review the expected sizes of $\kappa_{f_i}$ in popular models of weak scale NP models, some of them motivated by the hierarchy problem. 
Tables~\ref{tab:upyukawa}, ~\ref{tab:downyukawa},
and~\ref{tab:leptyukawa}, adapted from
\cite{Bishara:2015cha,Dery:2014kxa,Dery:2013aba,Dery:2013rta,Bauer:2015kzy},
summarise the predictions for the effective Yukawa couplings,
$\kappa_f$, in the Standard Model, multi-Higgs-doublet models
(MHDM) with natural flavour conservation (NFC)~\cite{Glashow:1976nt,
  Paschos:1976ay}, a ``flavourful'' two-Higgs-doublet model beyond NFC
(F2HDM)~\cite{Altmannshofer:2015esa, Altmannshofer:2016zrn, 
Altmannshofer:2017uvs, Altmannshofer:2018bch} the Minimal Supersymmetric Standard Model (MSSM) at tree level, 
a single Higgs doublet with
a Froggat-Nielsen mechanism (FN)~\cite{Froggatt:1978nt}, the
Giudice-Lebedev model of quark masses modified to 2HDM
(GL2)~\cite{Giudice:2008uua}, NP models with minimal flavour violation
(MFV)~\cite{D'Ambrosio:2002ex}, Randall-Sundrum models
(RS)~\cite{Randall:1999ee}, and models with a composite Higgs where
Higgs is a pseudo-Nambu-Goldstone boson (pNGB)~\cite{Dugan:1984hq,
  Georgi:1984ef, Kaplan:1983sm, Kaplan:1983fs}. The flavour-violating
couplings in the above set of NP models are collected in Tables
\ref{tab:upFVyukawa} and \ref{tab:downFVyukawa}. Next, we briefly
discuss each of the above models, and show that the effects are either
suppressed by $1/\Lambda^2$, where $\Lambda$ is the NP scale, or are
proportional to the mixing angles with the extra scalars.

\begin{table}[t]
\begin{center}
\begin{tabular}{l  c  c  c c  }
\toprule[0.1em]
Model	& $\kappa_t$ & $\kappa_{c (u)}/\kappa_t$  & $\tilde \kappa_t/\kappa_t$ & $\tilde \kappa_{c (u)}/\kappa_t$ \\ \midrule[0.05em]
SM	& 1	& 1 & 0 & 0 \\
MFV &$1+\frac{\Re(a_uv^2+2b_u m_t^2)}{\Lambda^2}$
&$1-\frac{2\Re(b_u)m_t^2}{\Lambda^2}$
&$\frac{\Im(a_uv^2+2b_u m_t^2)}{\Lambda^2}$ & $\frac{\Im(a_u
  v^2)}{\Lambda^2} $ \\
NFC & $V_{hu}\,v/v_u$	& 1 &  0 &0 \\
F2HDM & $\cos\alpha/\sin\beta$	& $-\tan\alpha/\tan\beta$ & $\mcO\left(\frac{m_c}{m_t}\frac{\cos(\beta-\alpha)}{\cos\alpha\cos\beta}\right)$ & $\mcO\left(\frac{m_{c(u)}^2}{m_t^2} \frac{\cos(\beta-\alpha)}{\cos\alpha\cos\beta}\right)$ \\
MSSM	& $\cos\alpha/\sin\beta$	&1 &0  &0\\
FN & $1+\mcO\left(\frac{v^2}{\Lambda^2}\right)$ &
	$1+\mcO\left(\frac{v^2}{\Lambda^2}\right)$ &
	$\mcO\left(\frac{v^2}{\Lambda^2}\right)$ &
	$\mcO\left(\frac{v^2}{\Lambda^2}\right)$ \\
GL2 	& $\cos\alpha/\sin\beta$& $\simeq 3(7)$ & 0 & 0 \\
RS &$1-{\mathcal O}\Big(\frac{ v^2}{m_{KK}^2}\bar Y^2\Big)$&$1+{\mathcal O}\Big(\frac{ v^2}{m_{KK}^2}\bar Y^2\Big)$ &${\mathcal O}\Big(\frac{ v^2}{m_{KK}^2}\bar Y^2\Big)$ &${\mathcal O}\Big(\frac{ v^2}{m_{KK}^2}\bar Y^2\Big)$ \\
pNGB & $1+{\mathcal O}\Big(\frac{ v^2}{f^2}\Big)+{\mathcal O}\Big(y_*^2 \lambda^2 \frac{ v^2}{M_*^2}\Big)$ & $1+{\mathcal O}\Big(y_*^2 \lambda^2 \frac{ v^2}{M_*^2}\Big)$ & ${\mathcal O}\Big(y_*^2 \lambda^2 \frac{ v^2}{M_*^2}\Big)$ & ${\mathcal O}\Big(y_*^2 \lambda^2 \frac{ v^2}{M_*^2}\Big)$ \\
\bottomrule[0.1em]
\end{tabular}
\caption{Predictions for the flavour-diagonal up-type Yukawa couplings
  in a sample of NP models (see text for details).
}
\label{tab:upyukawa}
\end{center}
\end{table}

\begin{table}[h]
\begin{center}
\begin{tabular}{ l   c  c c c}
\toprule[0.1em]
Model	& $\kappa_b$ & $\kappa_{s(d)}/\kappa_b$ & $\tilde \kappa_b/\kappa_b$ & $\tilde \kappa_{s(d)}/\kappa_b$ \\ \midrule[0.05em]
SM	& 1 & 1 &0 &0\\
MFV & $1+\frac{\Re(a_d v^2 +2 c_d m_t^2)}{\Lambda^2}$&$1-\frac{2\Re(c_d)m_t^2}{\Lambda^2}$&$ \frac{\Im(a_d v^2+2 c_d m_t^2)}{\Lambda^2}$&$ \frac{\Im(a_d v^2+2 c_d |V_{ts(td)}|^2 m_t^2)}{\Lambda^2}$ \\
NFC & $V_{hd}\,v/v_d$	& 1 &0 &0\\
F2HDM & $\cos\alpha/\sin\beta$	& $-\tan\alpha/\tan\beta$ & $\mcO\left(\frac{m_s}{m_b}\frac{\cos(\beta-\alpha)}{\cos\alpha\cos\beta}\right)$ & $\mcO\left(\frac{m_{s(d)}^2}{m_b^2} \frac{\cos(\beta-\alpha)}{\cos\alpha\cos\beta}\right)$ \\
MSSM	 & $-\sin\alpha/\cos\beta$	&1 &0 &0\\
FN & $1+\mcO\left(\frac{v^2}{\Lambda^2}\right)$ &
	$1+\mcO\left(\frac{v^2}{\Lambda^2}\right)$ &
	$\mcO\left(\frac{v^2}{\Lambda^2}\right)$ &
	$\mcO\left(\frac{v^2}{\Lambda^2}\right)$ \\
GL2	& $-\sin\alpha/\cos\beta$	& $\simeq 3(5)$ & 0 & 0 \\
RS &$1-{\mathcal O}\Big(\frac{ v^2}{m_{KK}^2}\bar Y^2\Big)$&$1+{\mathcal O}\Big(\frac{ v^2}{m_{KK}^2}\bar Y^2\Big)$ &${\mathcal O}\Big(\frac{ v^2}{m_{KK}^2}\bar Y^2\Big)$ &${\mathcal O}\Big(\frac{ v^2}{m_{KK}^2}\bar Y^2\Big)$ \\
pNGB & $1+{\mathcal O}\Big(\frac{ v^2}{f^2}\Big)+{\mathcal O}\Big(y_*^2 \lambda^2 \frac{ v^2}{M_*^2}\Big)$ & $1+{\mathcal O}\Big(y_*^2 \lambda^2 \frac{ v^2}{M_*^2}\Big)$ & ${\mathcal O}\Big(y_*^2 \lambda^2 \frac{ v^2}{M_*^2}\Big)$ & ${\mathcal O}\Big(y_*^2 \lambda^2 \frac{ v^2}{M_*^2}\Big)$\\
\bottomrule[0.1em]
\end{tabular}
\caption{Same as Table \ref{tab:upyukawa} but for down-type Yukawa
  couplings. 
}

\label{tab:downyukawa}
\end{center}
\end{table}

\begin{table}[t]\centering
\begin{tabular}{ l   c  c c c}\toprule[0.1em]
Model	& $\kappa_\tau$ & $\kappa_{\mu(e)}/\kappa_\tau$ & $\tilde \kappa_\tau/\kappa_\tau$ & $\tilde \kappa_{\mu(e)}/\kappa_\tau$ \\ \midrule[0.05em]
SM & 1 & 1 & 0 & 0\\
MFV & $1+\frac{\Re\left(a_\ell\right) v^2}{\Lambda^2}$ &
	$1-\frac{2\Re{\left(b_\ell\right)}m_\tau^2}{\Lambda^2}$ &
	$\frac{\Im\left(a_\ell\right) v^2}{\Lambda^2}$ &
	$\frac{\Im\left(a_\ell\right) v^2}{\Lambda^2}$\\
NFC & $V_{h\ell}\,v/v_\ell$	& 1 &0 &0\\
F2HDM & $\cos\alpha/\sin\beta$	& $-\tan\alpha/\tan\beta$ & $\mcO\left(\frac{m_\mu}{m_\tau}\frac{\cos(\beta-\alpha)}{\cos\alpha\cos\beta}\right)$ & $\mcO\left(\frac{m_{\mu(e)}^2}{m_\tau^2} \frac{\cos(\beta-\alpha)}{\cos\alpha\cos\beta}\right)$ \\
MSSM & $-\sin\alpha/\cos\beta$ & 1 & 0 & 0 \\
FN & $1+\mathcal{O}\left(\frac{v^2}{\Lambda^2}\right)$ & $1+\mathcal{O}\left(\frac{v^2}{\Lambda^2}\right)$ & $\mathcal{O}\left(\frac{v^2}{\Lambda^2}\right)$  & $\mathcal{O}\left(\frac{v^2}{\Lambda^2}\right)$ \\
GL2 &  $-\sin\alpha/\cos\beta$ & $\simeq 3(5)$ & 0 & 0 \\
RS & $1+\mcO\left(\yvmkk\right)$ &
	$1+\mcO\left(\yvmkk\right)$ &
	$\mcO\left(\yvmkk\right)$ &
	$\mcO\left(\yvmkk\right)$ \\
\bottomrule[0.1em]
\end{tabular}
\caption{Same as Table \ref{tab:upyukawa} but for lepton Yukawa
  couplings. NP effects in the pNGB model are negligible and therefore we do not report them here.
}
\label{tab:leptyukawa}
\end{table}

\begin{table}[t]
\begin{center}
\begin{tabular}{l  c  c  c }
\toprule[0.1em]
Model	& $\kappa_{ct (tc)}/\kappa_t$ & $\kappa_{ut (tu)}/\kappa_t$  & $\kappa_{uc (cu)}/\kappa_t$ \\ \midrule[0.05em]\vspace{0.15cm}
MFV &$ \frac{\Re\big( c_u m_b^2 V_{cb}^{(*)}\big)}{\Lambda^2}\frac{\sqrt2 m_{t(c)}}{v} $~&~ $ \frac{\Re\big( c_u m_b^2 V_{ub}^{(*)}\big)}{\Lambda^2} \frac{\sqrt2 m_{t(u)}}{v}$~&~ $ \frac{\Re\big( c_u m_b^2 V_{ub(cb)}V_{cb(ub)}^{*}\big)}{\Lambda^2} \frac{\sqrt2 m_{c(u)}}{v}$\\\vspace{0.15cm}
F2HDM & $\mcO\left(\frac{m_c}{m_t}\frac{\cos(\beta-\alpha)}{\cos\alpha\cos\beta}\right)$ & $\mcO\left(\frac{m_u}{m_t}\frac{\cos(\beta-\alpha)}{\cos\alpha\cos\beta}\right)$ & $\mcO\left(\frac{m_c m_u}{m_t^2}\frac{\cos(\beta-\alpha)}{\cos\alpha\cos\beta}\right)$ \\\vspace{0.15cm}
FN &  $\mathcal{O}\left(\frac{v m_{t(c)}}{\Lambda^2} |V_{cb}|^{\pm 1}\right)$ &
	$\mathcal{O}\left(\frac{v m_{t(u)}}{\Lambda^2} |V_{ub}|^{\pm 1}\right)$ &
	$\mathcal{O}\left(\frac{v m_{c(u)}}{\Lambda^2} |V_{us}|^{\pm 1}\right)$\\\vspace{0.15cm}
GL2	& $\epsilon (\epsilon^2)$ & $\epsilon (\epsilon^2)$ & $\epsilon^3$ \\\vspace{0.15cm}
RS & $\sim \lambda^{(-)2} \frac{m_{t(c)}}{v} \bar Y^2\frac{v^2}{m_{KK}^2} $&$\sim \lambda^{(-)3} \frac{m_{t(u)}}{v} \bar Y^2\frac{v^2}{m_{KK}^2} $&$\sim \lambda^{(-)1} \frac{m_{c(u)}}{v} \bar Y^2\frac{v^2}{m_{KK}^2} $ \\\vspace{0.15cm}
pNGB & ${\mathcal O}(y_*^2 \frac{m_t}{v}\frac{\lambda_{L (R),2} \lambda_{L(R),3}m_W^2}{M_*^2})$ & ${\mathcal O}(y_*^2 \frac{m_t}{v}\frac{\lambda_{L (R),1} \lambda_{L(R),3}m_W^2}{M_*^2})$  & ${\mathcal O}(y_*^2 \frac{m_c}{v}\frac{\lambda_{L (R),1} \lambda_{L(R),2}m_W^2}{M_*^2})$ \\
\bottomrule[0.1em]
\end{tabular}
\caption{Same as Table \ref{tab:upyukawa} but for flavour-violating up-type Yukawa couplings. In the SM,
  NFC and the tree-level MSSM the Higgs Yukawa couplings are flavour
  diagonal. The CP-violating $\tilde \kappa_{ff'}$ are obtained by replacing the real part, ${\Re}$, with the imaginary part, ${\Im}$. All the other models predict a zero contribution to these flavour changing couplings.
}
\label{tab:upFVyukawa}
\end{center}
\end{table}
\begin{table}[t]
\begin{center}
\begin{tabular}{l  c  c  c }
\toprule[0.1em]
Model	&   $\kappa_{bs (sb)}/\kappa_b$ & $\kappa_{bd (db)}/\kappa_b$  & $\kappa_{sd (ds)}/\kappa_b$ \\ \midrule[0.05em]\vspace{0.15cm}
MFV &$\frac{\Re\big(c_d m_t^2 V_{ts}^{(*)}\big)}{\Lambda^2} \frac{\sqrt2m_{s(b)}}{v}$~&~$\frac{\Re\big(c_d m_t^2 V_{td}^{(*)}\big)}{\Lambda^2} \frac{\sqrt2 m_{d(b)}}{v}$~&~$\frac{\Re\big(c_d m_t^2 V_{ts(td)}^*V_{td(ts)}\big)}{\Lambda^2} \frac{\sqrt2 m_{s(d)}}{v}$ \\\vspace{0.15cm}
F2HDM & $\mcO\left(\frac{m_s}{m_b}\frac{\cos(\beta-\alpha)}{\cos\alpha\cos\beta}\right)$ & $\mcO\left(\frac{m_d}{m_b}\frac{\cos(\beta-\alpha)}{\cos\alpha\cos\beta}\right)$ & $\mcO\left(\frac{m_s m_d}{m_b^2}\frac{\cos(\beta-\alpha)}{\cos\alpha\cos\beta}\right)$ \\\vspace{0.15cm}
FN &  $\mathcal{O}\left(\frac{v m_{b(s)}}{\Lambda^2} |V_{cb}|^{\pm 1}\right)$ &
	$\mathcal{O}\left(\frac{v m_{b(d)}}{\Lambda^2} |V_{ub}|^{\pm 1}\right)$ &
	$\mathcal{O}\left(\frac{v m_{s(d)}}{\Lambda^2} |V_{us}|^{\pm 1}\right)$\\\vspace{0.15cm}
GL2	&$\epsilon^2 (\epsilon)$ & $\epsilon$ & $\epsilon^2(\epsilon^3)$\\\vspace{0.15cm}
RS & $\sim \lambda^{(-)2} \frac{m_{b(s)}}{v} \bar Y^2\frac{v^2}{m_{KK}^2} $&$\sim \lambda^{(-)3} \frac{m_{b(d)}}{v} \bar Y^2\frac{v^2}{m_{KK}^2} $&$\sim \lambda^{(-)1} \frac{m_{s(d)}}{v} \bar Y^2\frac{v^2}{m_{KK}^2} $ \\\vspace{0.15cm}
pNGB & ${\mathcal O}(y_*^2 \frac{m_b}{v}\frac{\lambda_{L (R),2} \lambda_{L(R),3}m_W^2}{M_*^2})$ & ${\mathcal O}(y_*^2 \frac{m_b}{v}\frac{\lambda_{L (R),1} \lambda_{L(R),3}m_W^2}{M_*^2})$  & ${\mathcal O}(y_*^2 \frac{m_s}{v}\frac{\lambda_{L (R),1} \lambda_{L(R),2}m_W^2}{M_*^2})$\\
\bottomrule[0.1em]
\end{tabular}
\caption{Same as Table \ref{tab:upFVyukawa} but for flavour-violating down-type Yukawa couplings. 
}
\label{tab:downFVyukawa}
\end{center}
\end{table}

\begin{table}[t]
\begin{center}
\begin{tabular}{l  c  c  c }
	\toprule[0.1em]
	Model	&   $\kappa_{\tau\mu (\mu\tau)}/\kappa_\tau$ & $\kappa_{\tau e (e\tau)}/\kappa_\tau$  & $\kappa_{\mu e (e\mu)}/\kappa_\tau$ \\ \midrule[0.05em]\vspace{0.15cm}
	F2HDM & $\mcO\left(\frac{m_\mu}{m_\tau}\frac{\cos(\beta-\alpha)}{\cos\alpha\cos\beta}\right)$ & $\mcO\left(\frac{m_e}{m_\tau}\frac{\cos(\beta-\alpha)}{\cos\alpha\cos\beta}\right)$ & $\mcO\left(\frac{m_\mu m_e}{m_\tau^2}\frac{\cos(\beta-\alpha)}{\cos\alpha\cos\beta}\right)$ \\\vspace{0.15cm}
		FN & $\mathcal{O}\left(\frac{v m_{\mu(\tau)}}{\Lambda^2} |U_{23}|^{\mp1}\right)$ & $\mathcal{O}\left(\frac{v m_{e(\tau)}}{\Lambda^2}|U_{13}|^{\mp1}\right)$ & $\mathcal{O}\left(\frac{v m_{e(\mu)}}{\Lambda^2}|U_{12}|^{\mp1}\right)$ \\\vspace{0.15cm}
GL2	&$\epsilon^2 (\epsilon)$ & $\epsilon$ & $\epsilon^2(\epsilon^3)$\\
	RS & $\sim\sqrt{\frac{m_{\mu(\tau)}}{m_{\tau(\mu)}}}\,\yvmkk$ &
		$\sim\sqrt{\frac{m_{e(\tau)}}{m_{\tau(e)}}}\,\yvmkk$&
		$\sim\sqrt{\frac{m_{e(\mu)}}{m_{\mu(e)}}}\,\yvmkk$\\
	\bottomrule[0.1em]
\end{tabular}
\caption{Same as Table  \ref{tab:upFVyukawa} but for flavour-violating lepton Yukawa couplings. 
}
\label{tab:leptFVyukawa}
\end{center}
\end{table}
\underline{\it Dimension-Six Operators with Minimal Flavor Violation (MFV).}
We first assume that there is a mass gap between the SM and NP. Integrating out the NP states leads to dimension six operators (after absorbing the modifications of kinetic terms using equations of motion~\cite{AguilarSaavedra:2009mx}), 
\begin{equation}
\begin{split}\label{eq:EFT:MFV}
	\mathcal{L}_{\rm EFT} &=  \frac{Y_u^\prime}{\Lambda^2}\bar{Q}_L H^c u_R
        (H^\dagger H)+ \frac{Y_d^\prime}{\Lambda^2} \bar{Q}_L H d_R
        (H^\dagger H)+\frac{Y_\ell^\prime}{\Lambda^2} \bar{L}_L H \ell_R
        (H^\dagger H)+\text{h.c.}\,, 
\end{split}
\end{equation}
which correct the SM Yukawa interactions. 
        Here $\Lambda$ is the NP scale and $H^c =
i\sigma_2H^\ast$. The fermion mass matrices and
Yukawa couplings after EWSB are 
\begin{equation}
M_f=\frac{v}{\sqrt2}\Big(Y_f +Y_f' \frac{v^2}{2 \Lambda^2}\Big)\,,
\qquad y_f=Y_f +3 Y_f' \frac{v^2}{2 \Lambda^2}\,, \qquad \quad f=u,d, \ell\,, 
\end{equation}
Because $Y_f$ and $Y_{f}'$ appear in two different combinations in
$M_f$ and in the physical Higgs Yukawa couplings, $y_f$, the two, in general, cannot be made diagonal in the
same basis and will lead to flavour-violating Higgs couplings.

In Tables \ref{tab:upyukawa}-\ref{tab:leptFVyukawa} we show the resulting $\kappa_f$ assuming MFV, i.e.,  that the flavour breaking in the NP sector is only due to the SM
Yukawas \cite{D'Ambrosio:2002ex, Chivukula:1987py,
  Gabrielli:1994ff, Ali:1999we, Buras:2000dm, Buras:2003jf,
  Kagan:2009bn}. This gives  $Y_u^{\prime}= a_uY_u +
        b^{\phantom\dagger}_uY^{\phantom\dagger}_uY_u^\dagger
        Y^{\phantom\dagger}_u + c^{\phantom\dagger}_u
        Y^{\phantom\dagger}_dY_d^\dagger
        Y^{\phantom\dagger}_u+\cdots\,,$ and similarly for $Y_d'$ with $u\leftrightarrow d$, while
        $a_q, b_q, c_q\sim {\mathcal O}(1)$ and are in general complex. For leptons we follow \cite{Dery:2013rta} and assume that the SM $Y_{\ell}$ is the only flavour-breaking spurion even for the neutrino mass matrix (see also \cite{Cirigliano:2005ck}). Then $Y_\ell'$ and $Y_{\ell}$ are diagonal in the same basis and there are no flavour-violating couplings. The flavour-diagonal $\kappa_\ell$ are given in Table \ref{tab:leptyukawa}.

\underline{\it Multi-Higgs-doublet model with natural flavour conservation~(NFC).}
Natural flavour conservation in multi-Higgs-doublet models is an assumption that only one doublet, $H_u$, couples to the up-type quarks, only one Higgs doublet, $H_d$, couples to the down-type quarks, and only one doublet, $H_\ell$ couples to leptons (it is possible that any of these coincide, as in the SM where $H=H_u=H_d=H_\ell$)~\cite{Glashow:1976nt, Paschos:1976ay}. The neutral scalar components of $H_i$ are $(v_i+h_i)/\sqrt2$, where
$v^2=\sum_i v_i^2$. The dynamical fields $h_{i}$ are a linear combination of the neutral
Higgs mass eigen-states (and include $h_u$ and $h_d$). We thus have
$h_i=V_{hi} h + \ldots$, where $V_{hi}$ are elements of the unitary
matrix $V$ that diagonalises the neutral-Higgs mass terms and we only
write down the contribution of the lightest Higgs, $h$. NFC means that there are no tree-level Flavor Changing Neutral Currents (FCNCs) and no $CP$
violation in the Yukawa interactions $\kappa_{qq'}=\tilde \kappa_{qq'}=0\,, \tilde \kappa_q=0$. 

There is a universal shift in all up-quark Yukawa couplings, $\kappa_u = \kappa_c = \kappa_t = V_{hu}{v}/{v_u}$. Similarly, there is a (different) universal shift in all down-quark Yukawa couplings and in all lepton Yukawa couplings, see Tables \ref{tab:upyukawa} - \ref{tab:leptyukawa}.

\underline{\it Higgs sector of the MSSM at tree level.}
The MSSM tree-level Higgs potential and the couplings to quarks are
the same as in the type-II two-Higgs-doublet model, see, e.g.,
\cite{Haber:1984rc}. This is an example of a 2HDM with natural flavour
conservation in which $v_u=\sin\beta\, v$, $v_d=\cos\beta\,
v$. The mixing of $h_{u,d}$ into the Higgs mass-eigen-states $h$ and
$H$ is given by $h_u=\cos \alpha h+\sin\alpha H$, $h_d = -\sin\alpha h +\cos\alpha H$,
where $h$ is the observed SM-like Higgs. The up-quark Yukawa couplings are rescaled universally, 
$	\kappa_u = \kappa_c = \kappa_t=\cos\alpha/\sin\beta$, and similarly the down-quark Yukawas, 
	$\kappa_d = \kappa_s = \kappa_b = -\sin\alpha/\cos\beta$.
The flavour-violating and CP-violating Yukawas are zero\footnote{Note that beyond the tree level, in fine-tuned regions of parameter space the loops of sfermions and gauginos can lead to substantial corrections to these expressions \cite{Aloni:2015wvn}.}. In Tables \ref{tab:upyukawa}-\ref{tab:leptyukawa} we limit ourselves to the tree-level expectations, which are a good approximation for a large part of the MSSM parameter space.

In the alignment limit, $\beta-\alpha=\pi/2$ \cite{Gunion:2002zf,Carena:2013ooa,Dev:2014yca,Carena:2014nza,Dev:2015bta,Haber:2015pua,Carena:2015moc}, the Yukawa couplings tend toward their
SM value, $\kappa_i=1$.  The global fits to Higgs data in type-II 2HDM
already constrain $\beta-\alpha$ to be not to far from $\pi/2$~\cite{Carmi:2012in,
  Falkowski:2013dza, Grinstein:2013npa} so that the couplings of the light Higgs are also constrained to be close to their SM values. 
  Note that the decoupling limit of the 2HDM, where the heavy Higgs bosons become much heavier than the SM Higgs, implies the alignment limit while the reverse is not necessarily true \cite{Carena:2013ooa}.

\underline{\it Flavorful two-Higgs-doublet model.}
In~\cite{Altmannshofer:2015esa} a 2HDM setup was introduced in which one Higgs doublet couples only to top, bottom and tau, and a second Higgs doublet couples to the remaining fermions (see also~\cite{Botella:2016krk,Ghosh:2015gpa,Das:1995df,Blechman:2010cs}).
Such a 2HDM goes beyond NFC and therefore introduces FCNCs at tree level. However, the Yukawa couplings
of the first Higgs doublet to the third generation fermions preserve a $U(2)^5$ flavour symmetry, only
broken by the small couplings of the second Higgs doublet. This approximate $U(2)^5$ symmetry
leads to a strong suppression of the most sensitive flavour violating transitions between the second and first generation.

The non-standard flavour structure
of this ``flavourful'' 2HDM scenario leads to flavour non-universal modifications of all Higgs couplings. To be more precise $\kappa_t \neq \kappa_c = \kappa_u$, $\kappa_b \neq \kappa_s = \kappa_d$, and $\kappa_\tau \neq \kappa_\mu = \kappa_e$. CP violation in Higgs couplings can arise but is strongly suppressed by small fermion masses, see Tables \ref{tab:upyukawa} - \ref{tab:leptyukawa}. Also potentially sizeable flavour violating Higgs couplings involving the third generation fermions arise, see Tables \ref{tab:upFVyukawa} - \ref{tab:leptFVyukawa}.
As in all 2HDMs, the Higgs couplings approach their SM values in the alignment limit, $\beta-\alpha=\pi/2$.

\underline{\it A single Higgs doublet with Froggatt-Nielsen mechanism (FN).}
The Froggatt-Nielsen~\cite{Froggatt:1978nt} mechanism provides a simple explanation of the size and hierarchy of the SM Yukawa couplings. In the simplest realisation this is achieved by a $U(1)_H$ horizontal symmetry under which different generations of fermions carry different charges. The $U(1)_H$ is broken by a spurion, $\epsilon_H$.
The entries of the SM Yukawa matrix are then parametrically suppressed by powers of $\epsilon_H$ as, for example, in the lepton sector
\begin{equation}
\big(Y_\ell\big)_{ij}\sim \epsilon_H^{H(L_i)-H(e_j)},
\end{equation}
where $H(e,L)$ are the FN charges of the right- and left-handed charged lepton, respectively. The dimension 6 operators in \eqref{eq:EFT:MFV} due to electroweak NP have similar flavour suppression, $\big(Y_\ell'\big)_{ij}\sim \epsilon_H^{H(e_j)-H(L_i)} v^2/\Lambda^2$ \cite{Dery:2013rta,Dery:2014kxa}. After rotating to the mass eigen-basis, the lepton masses and mixing angles are then given by~\cite{Leurer:1993gy,Grossman:1995hk}
\begin{equation}
m_{\ell_i}/v \sim \epsilon_H^{|H(L_i)-H(e_i)|},\quad
	|U_{ij}|\sim \epsilon_H^{|H(L_i)-H(L_j)|},
\end{equation}
giving the Higgs Yukawa couplings in Tables~\ref{tab:leptyukawa}~and~\ref{tab:leptFVyukawa} in the row labelled `FN' \cite{Dery:2014kxa}.
Similarly for the quarks, after rotating to the mass eigen-basis, the masses and the mixings are given by~\cite{Leurer:1993gy} 
\begin{equation}
m_{u_i(d_i)}/v \sim \epsilon_H^{|H(Q_i)-H(u_i(d_i))|},\quad
|V_{ij}|\sim \epsilon_H^{|H(Q_i)-H(Q_j)|},
\end{equation}
where $V$ is the Cabibbo-Kobayashi-Maskawa (CKM) mixing matrix and $H(u,d,Q)$ are the FN charges of the right-handed up and down and the left-handed quark fields, respectively.

\underline{\it Higgs-dependent Yukawa couplings (GL2)} In the model of
quark masses introduced by Giudice and Lebedev~\cite{Giudice:2008uua},
the quark masses, apart from the top mass, are small because they
arise from higher dimensional operators. The original GL proposal is
ruled out by data, while the straightforward modification to a 2HDM
(GL2) is
\begin{equation}\label{eq:Higgs-dep}
\begin{split}
{\cal L}_{f}&=c_{ij}^{u} \bigg( \frac{H_1^\dagger H_1}{M^2}
  \bigg)^{n_{ij}^{u}} \, \bar Q_{L,i} u_{R,j} H_1 + c_{ij}^{d} \bigg( \frac{H_1^\dagger H_1}{M^2}
  \bigg)^{n_{ij}^{d}} \, \bar Q_{L,i} d_{R,j} H_2 + \phantom{c}\\
  &\qquad c_{ij}^\ell \bigg( \frac{H_1^\dagger H_1}{M^2}
  \bigg)^{n_{ij}^{\ell}} \, \bar L_{L,i} e_{R,j} H_2 +\text{h.c.}\,,
\end{split}
\end{equation}
where $M$ is the mass scale of the mediators.  In the original GL
model $H_2$ is identified with the SM Higgs, $H_2=H$, while
$H_1=H^c$. Taking $c_{ij}^{u,d}\sim {\mathcal
  O}(1)$, the ansatz $n_{ij}^{u,d}=a_i +b_j^{u,d}$ with $a=(1,1,0)$,
$b^d=(2,1,1)$, and $b^u=(2,0,0)$ then reproduces the hierarchies of
the observed quark masses and mixing angles for $\epsilon \equiv
v^2/M^2 \approx 1/60$. The Yukawa couplings are of the form $
y_{ij}^{u,d} = (2n_{ij}^{u,d} + 1) (y_{ij}^{u,d})_\text{SM}$. The SM
Yukawas are diagonal in the same basis as the quark masses,
while the $y_{ij}^{u,d}$ are not.  Because the bottom Yukawa is
largely enhanced, $\kappa_b \simeq 3$, this simplest version of the GL
model is already excluded by the Higgs data. Its modification, GL2, is
still viable, though \cite{Bishara:2015cha}. For
$v_1/v_2=\tan\beta\sim 1/\epsilon$ one can use the same ansatz for
$n_{ij}^{u,d}$ as before, modifying only $b^d$, so that $b^d=(1,0,0)$,
with the results shown in Tables
\ref{tab:upyukawa}-\ref{tab:leptFVyukawa}. For leptons we use the same
scalings as for right-handed quarks. Note that the $H_1^\dagger H_1$
is both a gauge singlet and a flavour singlet. From symmetry point of
view it is easier to build flavour models, if $H_1 H_2$ acts as a
spurion in \eqref{eq:Higgs-dep}, instead of $H_1^\dagger H_1$. This
possibility is severely constrained phenomenologically, though
\cite{Bauer:2015fxa,Bauer:2015kzy}.

\underline{\it Randall-Sundrum models (RS).}
The Randall-Sundrum warped extra-dimensional model has been proposed to address
the hierarchy problem and simultaneously explain the hierarchy of the SM fermion
masses ~\cite{Randall:1999ee, Gherghetta:2000qt, Grossman:1999ra, Huber:2000ie,
	Huber:2003tu}. Integrating out the Kaluza-Klein (KK) modes of mass $m_{KK}$, and
working in the limit of a brane-localised Higgs, keeping only terms of leading
order in $v^2/m_{KK}^2$, the SM quark mass matrices are given
by~\cite{Azatov:2009na} (see also~\cite{Casagrande:2008hr, Bauer:2009cf,
	Malm:2013jia, Archer:2014jca, Blanke:2008zb, Blanke:2008yr, Albrecht:2009xr,
	Agashe:2006wa, Agashe:2014jca}, and Ref.~\cite{Dillon:2014zea} for a bulk Higgs
scenario) \begin{equation}\label{eq:RS:Mdu} M^{d(u)}_{ij}=\big[F_q
	Y_{1(2)}^{5D}F_{d(u)}\big]_{ij} v\,. \end{equation} The $F_{q,u,d}$ are $3\times
3$ matrices of fermion wave-function overlaps with the Higgs and are diagonal
and hierarchical. Assuming flavour anarchy, the 5D Yukawa matrices,
$Y_{1,2}^{5D}$, are general $3\times 3$ complex matrices with $\bar Y\sim
{\mathcal O}(1)$ entries, but usually $\bar Y \lesssim 4$, see, e.g.,
\cite{Archer:2014jca}. At leading order in $v^2/m_{KK}^2$ the Higgs Yukawas are
aligned with the quark masses, i.e., $M_{u,d}=y_{u,d} v/{\sqrt2}+{\mathcal
	O}(v^2/m_{KK}^2)$. The mis-alignments are generated by tree-level KK quark
exchanges, giving \begin{equation}\label{eq:misalignment}
	\big[y_{u(d)}\big]_{ij}-\frac{\sqrt2}{v}\big[M_{u,d}\big]_{ij}\sim -
	\frac{1}{3}F_{q_i} \bar Y^3 F_{u_j(d_j)}\frac{v^2}{m_{KK}^2}\,. \end{equation}

For the charged leptons, there are two choices for generating the hierarchy in
the masses~\cite{Azatov:2009na}. If left- and right-handed fermion profiles are
both hierarchical (and taken to be similar) then the misalignment between the
masses and Yukawas is $\sim \sqrt{{m_i m_j}/{v^2}} \times \mcO\big(\bar Y^2
v^2/m_{KK}^2\big)$. If only the right-handed profiles are hierarchical the
misalignment is given by (see also
Tables~\ref{tab:leptyukawa}~and~\ref{tab:leptFVyukawa})
\begin{equation}\label{eq:lept-misalignment}
	\big[y_\ell\big]_{ij}-\frac{\sqrt2}{v}\big[M_\ell\big]_{ij}\sim -
	\frac{1}{3}\yvmkk\,\frac{m_j^\ell}{v}\,. \end{equation} The Higgs mediated FCNCs
are suppressed by the same zero-mode wave-function overlaps that also suppress
the quark masses, \eqref{eq:RS:Mdu}, giving rise to the RS GIM
mechanism~\cite{Cacciapaglia:2007fw, Agashe:2004cp, Agashe:2004ay}. Using the
fact that the CKM matrix elements are given by $V_{ij}\sim F_{q_i}/F_{q_j}$ for
$i<j$, Eq.~\eqref{eq:misalignment}, one can rewrite the $\kappa_i$ as in
Tables~\ref{tab:upyukawa}-\ref{tab:downFVyukawa}. The numerical analysis of
Ref.~\cite{Azatov:2009na} found that for diagonal Yukawas typically
$\kappa_i<1$, with deviations in $\kappa_{t(b)}$ up to $30\%(15\%)$, and in
$\kappa_{s,c (u,d)}$ up to $\sim 5\%(1\%)$.
For the charged leptons one obtains deviations in $\kappa_{\tau\mu(\mu\tau)}\sim 1(5)\times 10^{-5}$~\cite{Azatov:2009na}.
These estimates were obtained fixing the mass of the first KK gluon excitation to
$3.7$~\UTeV, above the present ATLAS bound \cite{ATLAS-CONF-2015-009}.

\underline{\it Composite pseudo-Goldstone Higgs (pNGB).}
Finally, we assume that the Higgs is a
pseudo-Goldstone boson arising from the spontaneous breaking of a
global symmetry in a strongly coupled sector, and  couples to the composite
sector with a typical coupling $y_*$ \cite{Dugan:1984hq,
  Georgi:1984ef, Kaplan:1983sm, Kaplan:1983fs} (for a review, see~\cite{Panico:2015jxa}).
Assuming partial compositeness, the SM fermions couple linearly to composite operators
$O_{L,R}$,
$\lambda_{L,i}^q \bar Q_{L,i} O_R^i+\lambda_{R,j}^u \bar u_{R,j}
O_L^j+h.c. \,,$
where $i,j$ are flavour indices~\cite{Kaplan:1991dc}. This is the 4D dual of fermion
mass generation in 5D RS models.  The SM masses and Yukawa
couplings arise from expanding the two-point functions of the
$O_{L,R}$ operators in powers of the Higgs field~\cite{Agashe:2009di}.

The new ingredient compared to the EFT analysis in \eqref{eq:EFT:MFV}
is that the shift symmetry due to the pNGB nature of the Higgs
dictates the form of the higher-dimensional operators. The flavour
structure and the composite Higgs coset structure completely factorise
if the SM fields couple to only one composite operator. The general
decomposition of Higgs couplings then becomes \cite{Agashe:2009di}
(see also \cite{Gillioz:2012se,Delaunay:2013iia,Azatov:2014lha})
\begin{equation}\label{eq:comp}
	Y_u \bar{Q}_L H u_R + Y_u^\prime\bar{Q}_L H u_R
        \frac{(H^\dagger H)}{\Lambda^2}+\ldots \quad \to \quad
        c_{ij}^u \, P(h/f) \, \bar{Q}_L^i H u_R^j \,,
\end{equation}
and similarly for the down quarks. Here $f\gtrsim v$ is the equivalent
of the pion decay constant, while $P(h/f)=a_0+a_2 (H^\dagger H/f^2)+\ldots
$ is an analytic function whose form is fixed by the  pattern of the
spontaneous breaking and the embedding of the SM fields in the global
symmetry of the strongly coupled sector. In \eqref{eq:comp} the flavour
structure of $Y_u$ and $Y_u'$ 
is the same. The resulting corrections to the quark Yukawa couplings 
are therefore strictly diagonal, 
\begin{equation}\label{eq:kappaq:estimate}
\kappa_q\sim 1+{\mathcal O}\big({v^2}/{f^2}\big).
\end{equation}
For example, for the models based on the breaking of $SO(5)$ to
$SO(4)$, the diagonal Yukawa couplings can be written as
$\kappa_q=(1+2m-(1+2m+n)(v/f)^2)/{\sqrt{1-(v/f)^2}}$, where $n,m$ are
positive integers~\cite{Pomarol:2012qf}. The Minimal Composite Higgs
Model 4 (MCHM4) corresponds to $m=n=0$, while MCHM5 is given by
$m=0,n=1$.

The flavour-violating contributions to the quark Yukawa couplings
arise only from corrections to the quark kinetic terms~\cite{Agashe:2009di},
\begin{equation}
\bar q_L i \slashed q_L \frac{H^\dagger H}{\Lambda^2}, \,\,
\bar u_R i \slashed u_R \frac{H^\dagger H}{\Lambda^2},
\dots\,,
\end{equation}
due to the exchanges of composite vector resonances
with typical mass $M_* \sim \Lambda$. After using the equations of
motion 
these give (neglecting relative ${\mathcal O}(1)$ contributions
in the sum)~\cite{Agashe:2009di, Azatov:2014lha, Delaunay:2013pja},
\begin{equation}
\kappa_{ij}^u\sim 2 y_*^2 \frac{v^2}{M_*^2}
\Big(\lambda_{L,i}^q\lambda_{L,j}^q \frac{m_{u_j}}{v}
+\lambda_{R,i}^u\lambda_{R,j}^u \frac{m_{u_i}}{v}\Big)\,,
\end{equation}
and similarly for the down quarks. If the strong sector is CP
violating, then $\tilde \kappa_{ij}^{u,d}\sim \kappa_{ij}^{u,d}$.

The exchange of composite vector resonances also contributes to the
flavour-diagonal Yukawa couplings, shifting the
estimate~\eqref{eq:kappaq:estimate} by 
$\Delta \kappa_{q_i}\sim 2 y_*^2 \frac{v^2}{M_*^2}
\Big[\big(\lambda_{L,i}^q\big)^2+ \big(\lambda_{R,i}^u\big)^2\Big] \,$.
This shift can be large for the quarks with a large composite
component if the Higgs is strongly coupled to the vector resonances,
$y_*\sim 4\pi$, and these resonances are relatively light, $M_*\sim
4\pi v\sim 3$ \UTeV. The left-handed top and bottom, as well as the
right-handed top, are expected to be composite, explaining the large
top mass (i.e., $\lambda_{L,3}^q\sim \lambda_{R,3}^u\sim 1$). In the
anarchic flavour scenario, one expects the remaining quarks to be
mostly elementary (so the remaining $\lambda_i\ll 1$).  
If there is some underlying flavour alignment, it is also possible that
the light quarks are composite. This is most easily achieved in the
right-handed sector~\cite{Redi:2011zi, Redi:2012uj,
  Delaunay:2013iia}.

In the case of the lepton sector, if we assume that there are no
hierarchies in the composite sector~\cite{Redi:2013pga} (see also~\cite{Csaki:2008qq,delAguila:2010vg,Hagedorn:2011un,Hagedorn:2011pw}), then the NP
effects in the flavour diagonal and off-diagonal Yukawas are
negligible. For this reason, we do not report them in Tables \ref{tab:leptyukawa} and \ref{tab:leptFVyukawa}.

\asubsection[O.A. De Aguiar Francisco, M. Schlaffer, L. Sestini, E. Stamou]{Inclusive Search with Flavor tagging (charm and strange)}

\asubsubsection{Charm quark tagging}

In the SM, the coupling of the Higgs to bottom quarks is small, i.e.,
$y_b^{\text{SM}}\simeq 0.016$ at $\mu=m_H$, and 
its coupling to charm quarks even smaller by roughly four times,
i.e., $y_c^{\text{SM}}\simeq 0.0036$ at $\mu=m_H$.
Nevertheless, due to phase-space the process $H\to b\bar b$ is the dominant 
decay mode of the Higgs in the SM.
This situation has not only made a roughly $30\%$ precise measurement of such 
a small coupling possible at Run I of the LHC, but has also created 
opportunities to measure possible order one deviations in the 
coupling of the Higgs to charm quarks.

An important difference between the charm- and to some extent also the strange-quark (see section~\ref{sec:strange-tagging}) with respect to up- and 
down-quarks is that it is possible to pursue an inclusive approach in
identifying the flavour of the final state particles by $c$-tagging jets.
The underlying geometrical/kinematic input necessary for $c$-tagging
is similar to $b$-tagging with the most relevant one being the identification
of displaced vertices due to the lifetime of $c$-hadrons.
$c$-tagging has been used early on in Run I of the LHC by ATLAS and CMS in searches for supersymmetry, e.g., Refs.~\cite{Aad:2014nra,Aad:2015gna}.
Its usefulness in relations to Higgs physics was first discussed in Ref.~\cite{Delaunay:2013pja} and subsequently
used in Ref.~\cite{Perez:2015aoa} to recast ATLAS's and CMS's Run~I analyses for $h\to b\bar b$ to provide the first direct LHC 
constraint on the charm Yukawa. 

The inclusive method of probing the charm-quark Yukawa is in many ways complementary 
to searches for exclusive decays (see discussion of section~\ref{sec:exclusiveHiggs}) or 
searches for deviations in Higgs distributions (see section \ref{sec:HiggsDist}).
For example, in the inclusive approach an underlying assumption is 
that the Higgs coupling to $WW$ and $ZZ$ ---entering Higgs production--- is SM-like,
while the interpretation of Higgs distributions assumes no additional new physics contribution that affects them in a significant way.
An important difference between the inclusive and the exclusive approach is that the latter relies on interference with the SM $H\to \gamma\gamma$ amplitude while the former does not.
Therefore, in principle the exclusive approach may be sensitive to the sign and 
$CP$ properties of the coupling to which the inclusive approach is insensitive to.
At the same time, measurements of exclusive decays of the Higgs are challenging due 
to the small probability of fragmenting into the specific final state and 
large QCD backgrounds, which is why the inclusive approach appears to be the most promising one to probe deviations in the magnitude of the Higgs to charm coupling.

\begin{figure}[t]
	\centering
	\includegraphics[width=\textwidth]{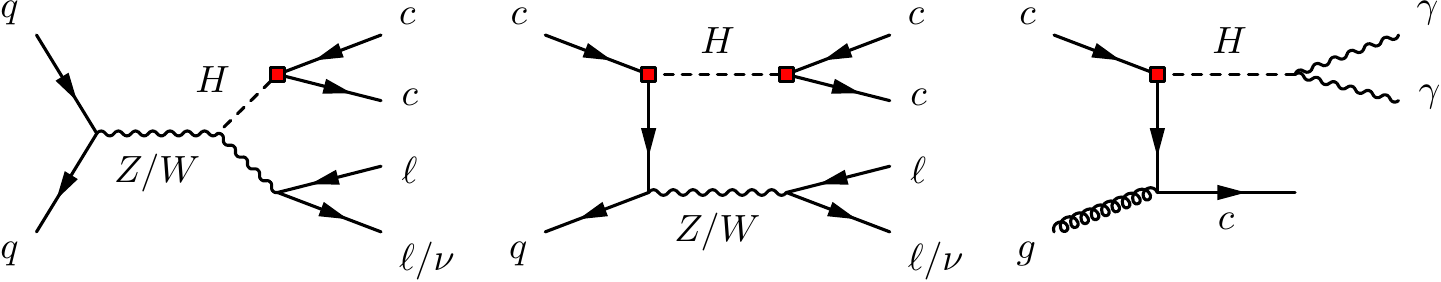}
	\caption{Left panel, leading-order production of Higgs in association with a heavy gauge boson ($Z/W$) and 
		subsequent decays. Central panel, additional production channel of Higgs in association 
		with a heavy gauge boson that becomes relevant for large $y_c$ \cite{Perez:2015aoa}.
		Right panel, leading-order diagram to search for non-SM $y_c$ in Higgs production in 
		association with a charm-quark \cite{Brivio:2015fxa}.
	\label{fig:higgsproductionHcc}
	}
\end{figure}

The most straight-forward way of inclusively probing the charm-quark Yukawa is by 
expanding the search for $H\to b\bar b$ to search for 
$pp \to (Z/W\to\ell\ell/\nu) (H\to c\bar c)$ \cite{Perez:2015aoa}
(left and central panel in Fig.~\ref{fig:higgsproductionHcc}).
Another possibility discussed in Ref.~\cite{Brivio:2015fxa} is to search for deviations in Higgs production 
in association with a charm quark in which the Higgs is produced from a charm-quark in the proton 
parton-distribution functions (right panel in Fig.~\ref{fig:higgsproductionHcc}).
We focus here on the measurement from $pp\to VH$ events proposed in Ref.~\cite{Perez:2015aoa}
and recently performed on a $36.1$~fb$^{-1}$ sample of $ZH$ data by ATLAS \cite{Aaboud:2018fhh}
at $\sqrt{s}=13$~\UTeV.
The following two key elements for this measurement are discussed below:
\begin{itemize}
	\item[i)] The experimental sensitivity in discriminating between $c$-jets from
		background $b$- and light-jets.
	\item[ii)] Disentangling the charm-quark coupling from the 
		bottom-quark Yukawa (breaking the degeneracy).
\end{itemize}

\begin{figure}[h]
	\centering
	\includegraphics[width=0.7\textwidth]{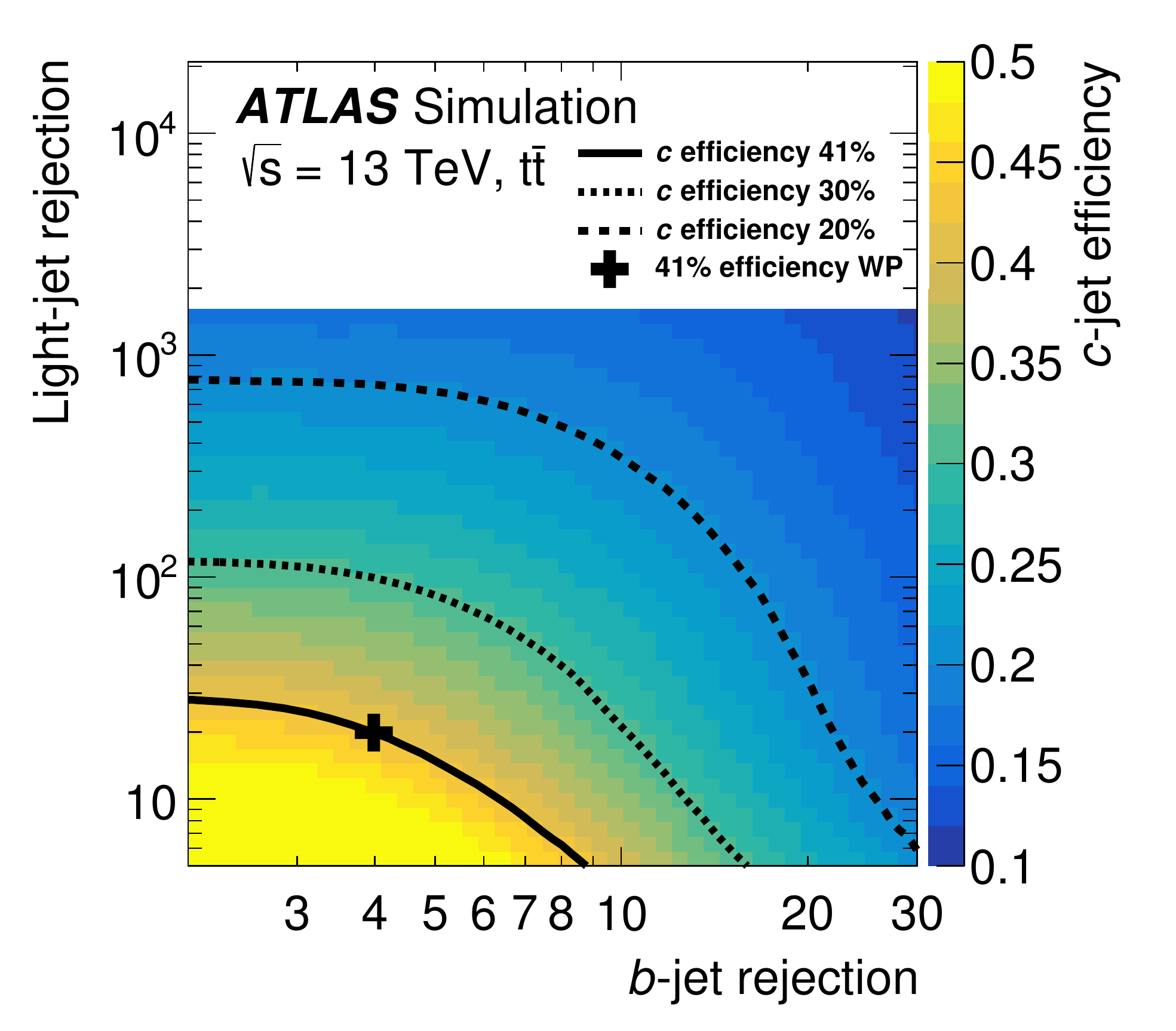}
	\caption{Correlation of $c$-tagging efficiency with $b$- and light-quark-jet rejection
	in ATLAS's $c$-tagger employed in the analysis of Ref.~\cite{Aaboud:2018fhh}.
\label{fig:ATLASctagefficiency}}
\end{figure}

Jet flavour tagging algorithms rely on Monte-Carlo simulations  to assign a probability 
for a given jet to be produced from a specific quark-flavour. 
Therefore, the efficiency / confidence in associating a jet to a specific quark is 
correlated with the confidence to reject other hypotheses, e.g., production from light-quarks.
The $c$-tagging tagging working point chosen in the ATLAS analysis \cite{Aaboud:2018fhh}  has an  efficiency of approximately $41\%$
to tag $c$-jets and rejection factors of roughly $4$ and $20$ for $b$- and light-quark-jets, respectively.
In Figure~\ref{fig:ATLASctagefficiency} 
the correlation between $c$-tagging efficiency and rejection factors is shown.
The observed limit is $\sigma(pp\to ZH)\text{BR}(H\to c\bar c)<2.7$\,pb at $95\%$ CL.
To translate this cross-section bound to a non-trivial constraint on $y_c$ it is essential 
to include the additional production channel from large charm Yukawa 
(central panel in Fig.~\ref{fig:higgsproductionHcc}) as demonstrated in Refs.~\cite{Perez:2015aoa}.
The additional production channel is affected by the kinematics, e.g., $p_T$ of the 
$Z$ and thus depends on the details of the analysis. 
This ``unfolding'' / reinterpretation of the analysis is thus best performed by 
the analysis itself and cannot be avoided to obtain non-trivial constraints on the Yukawa itself.
Note that at the moment the systematic uncertainties are approximately a factor of two
larger than the statistical uncertainties of the $36.1$\,fb$^{-1}$ sample used in the analysis;
the largest systematic uncertainty is associated to flavour-tagging and the tagging of $c$-jets in particular.

Given the rather similar lifetime of $b$ and $c$ hadrons, there is always
a non-negligible ``contamination'' of the $c$-jet sample 
from jets originating from $b$ quarks~\cite{Perez:2015aoa}.
An inclusive $H\to c\bar c$ analysis probing $y_c$ must thus either assume a SM value 
for the bottom Yukawa (as done in Ref.~\cite{Aaboud:2018fhh}) or allow the 
simultaneous variation of $y_b$ and $y_c$ to break the degeneracy.  
One possibility to achieve this is discussed  in Refs.~\cite{Perez:2015aoa,Perez:2015lra} where more than one tagging working point with different ratios of $c$-tagging to $b$-tagging efficiency are applied.

\begin{figure}[]
	\centering
	\includegraphics[width=0.45\textwidth]{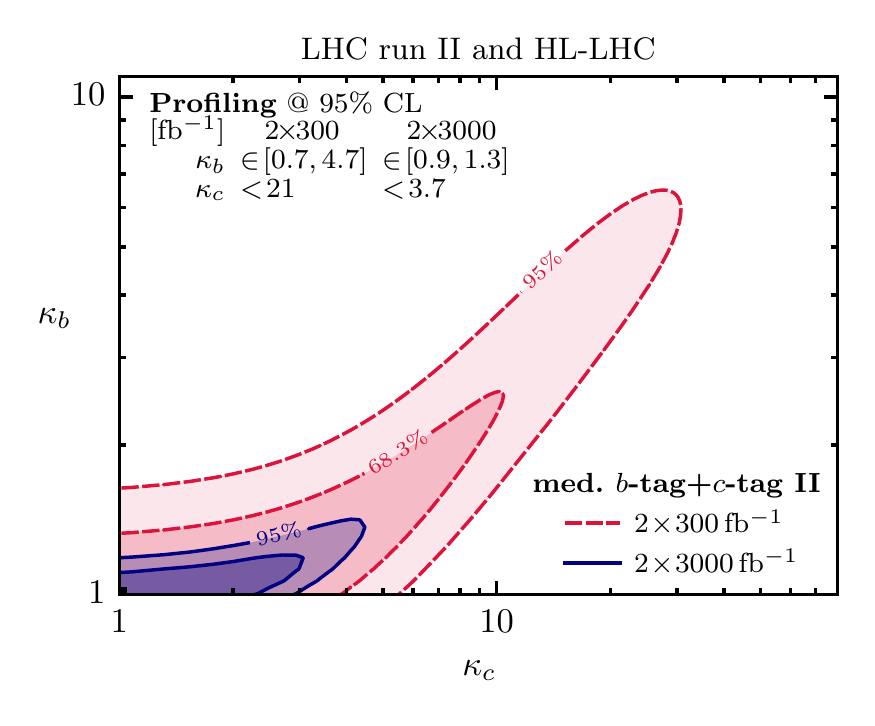}
	\includegraphics[width=0.45\textwidth]{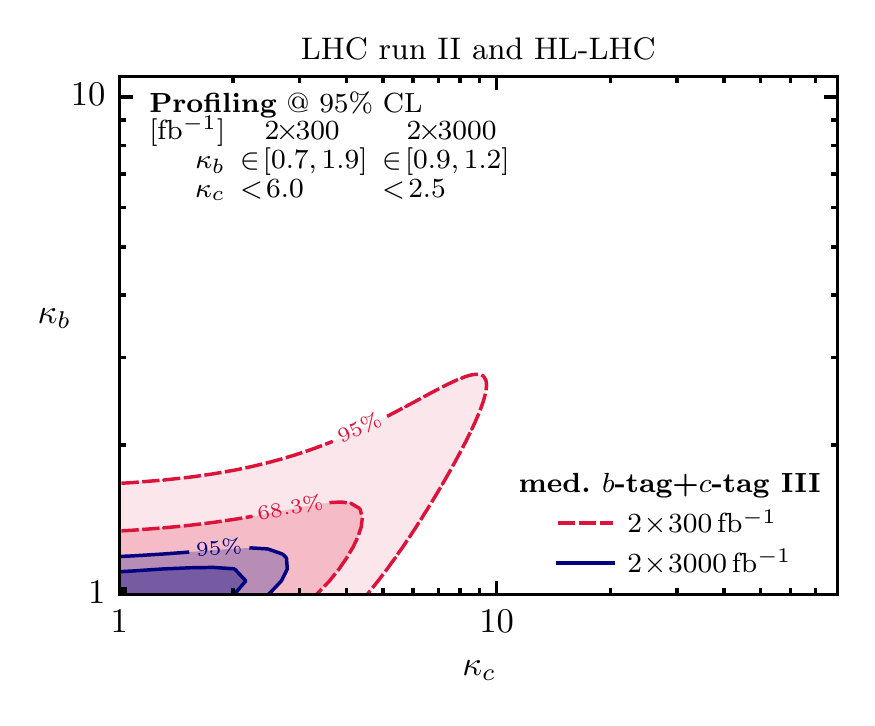}
	\caption{Projections for measuring charm Yukawa modifications from an inclusive 
		$H\to c\bar c$ search at $\sqrt{s}=14$\,\UTeV using two different 
		$c$-taggers (left and right panel) \cite{Perez:2015lra}.
		In red the $95\%$ CL region employing an integrated luminosity 
		of $2\times 300$~fb$^{-1}$ and in blue the region 
		employing $2\times 3000$~fb$^{-1}$.
	\label{fig:inclusiveforecast}}
\end{figure}

The prospects of measuring 
the rate of $pp\to ZH(\to c\bar c)$ at the HL-LHC are published in  \cite{ATL-PHYS-PUB-2018-016}.
The study uses the Run II analysis \cite{Aaboud:2018fhh} and rescales the results 
to an integrated luminosity of $3000$~fb$^{-1}$.
Possibilities to reduce the systematic uncertainties are discussed as well.
The analysis finds that, if there is no significant NP contribution, 
an upper bound  on the signal strength 
 of $\mu_{ZH(c\bar c)} <6.3$ at $95\%$ CL can be set.
This result is to be compared with  Ref.~\cite{Perez:2015lra}
in which  the prospects for measuring $H\to b\bar b$ at $\sqrt{s}=14$\,\UTeV 
\cite{ATLAS-collaboration:2012iza} are recast 
 to obtain an inclusive measurement of $H\to c \bar c$.
In the left panel of Figure~\ref{fig:inclusiveforecast} a $c$-tagging efficiency of $30\%$ ($c$-tag I) is used while $50\%$ ($c$-tag II)  is used in the right panel. 
In both cases the $b$-jet rejection was chosen to be $5$ and the light-jet rejection
$200$.
These two tagging working points cover the currently employed tagging working point in which
the $c$-tagging efficiency is approximately $41\%$.
In the analysis both the charm and the bottom quark are treated as free variables;
the bottom-Yukawa direction is profiled away to project the sensitivity to the 
charm-quark Yukawa.
It was found that with $2\times 3000$\,fb$^{-1}$ at $\sqrt{s}=14$\,\UTeV the 
high-luminosity stage of the LHC probes values of $y_c/y_c^{\text{SM}}\simeq 21 (6)$ with
$c$-tag I ($c$-tag II)  at $95\%$ CL, indicated by the blue regions in Figure~\ref{fig:inclusiveforecast}.

Even though the LHCb experiment operates at lower luminosity  compared to ATLAS and CMS, it has unique capabilities for discrimination between $b$- and $c$-jets thanks to its excellent vertex reconstruction system~\cite{Aaij:2015yqa}. With the secondary vertex tagging (SV-tagging) LHCb achieved an identification efficiency of 60$\%$ on $b$-jets, of 25$\%$ on $c$-jets and a light jets (light quarks or gluons) mis-identification probability of less than 0.2$\%$. Further discrimination between light and heavy jets and between $b$- and $c$-jets is achieved by exploiting the secondary vertex kinematic properties, using Boosted Decision Tree techniques (BDTs): for instance an additional cut on the BDT that separates $b$- from $c$-jets removes 90$\%$ of $H\rightarrow b \bar{b}$ while retaining 62$\%$ of $H\rightarrow c \bar{c}$ events  ~\cite{LHCb:2016yxg}. In the $H\rightarrow c \bar{c}$ search it is crucial to remove the $H\rightarrow b \bar{b}$ contribution since it represents an irreducible background source.

The LHCb acceptance covers $\sim 5$\% of the associated production of $W/Z+H$ at 13 \UTeV.  Figure~\ref{fig:hbbetas} shows the coverage of  LHCb for the $b\bar{b}$ pair produced in the decay of the  Higgs boson in association with a vector boson. When the two $b$-jets are within the acceptance, the lepton from $W/Z$ tends to be in acceptance as well ($\sim 60$\% of times). Due to the forward geometry, Lorentz-boosted Higgs bosons are  likely to be properly reconstructed.

\begin{figure}[ht]
	\centering
	\includegraphics[width=0.5\linewidth]{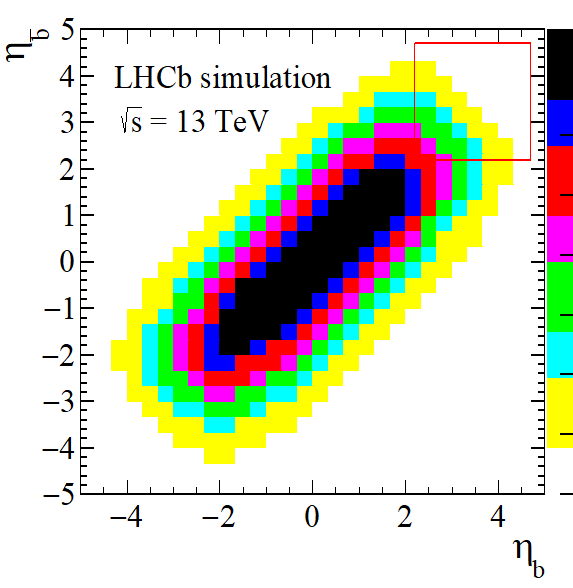}
	\caption{2D histogram showing the coverage of the LHCb acceptance for the $b\bar{b}$ pair produced by the Higgs decay in associated production with a W or a Z boson.}
	\label{fig:hbbetas}
\end{figure}

LHCb set  upper limits on the $V+H(\rightarrow b \bar{b})$ and $V+H(\rightarrow c \bar{c})$ production~\cite{LHCb:2016yxg} with data from LHC Run I.
Without any improvements in the analysis or detector, the extrapolation of this  to 300fb$^{-1}$ at 14\,\UTeV leads to a sensitivity of $\mu^{cc}\lesssim 50\,$. 

Detector improvements are expected in future upgrades, in particular in impact parameter resolution which directly affects the $c$-tagging performance. 
If the detector improvement is taken into account, the $c$-jet tagging efficiency with the SV-tagging is expected to  improve as shown in the Figure~\ref{fig:c-tag-eff}.  
A further improvement is expected from the electron reconstruction due to upgraded versions of the electromagnetic calorimeter. Electrons are used in the identification of the vector bosons associated with the Higgs.
Therefore, with these improvements, the expected limit can be pushed down to $\mu^{cc}\lesssim 5-10$ which corresponds to a limit of 2-3 times the Standard Model prediction on the charm Yukawa coupling.
This extrapolation does not include improvements in analysis techniques: for instance  Deep Learning methods can be applied to exploit correlations in  jets substructure properties to reduce the backgrounds.

\begin{figure}[ht]
	\centering
	\includegraphics[width=0.99\linewidth]{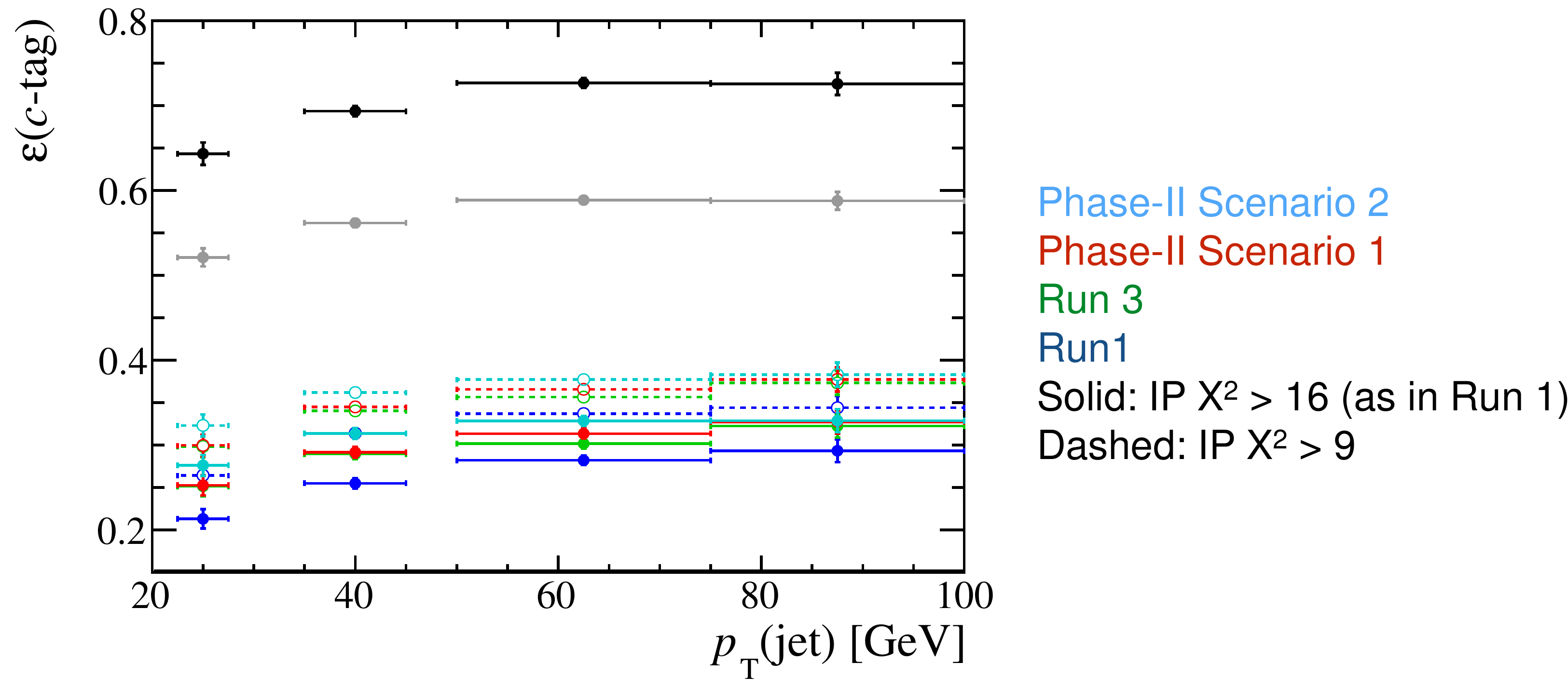}
	\caption{LHCb $c$-jet SV-tagging efficiency for different scenarios in the HL-LHC conditions.}
	\label{fig:c-tag-eff}
\end{figure}

\FloatBarrier

\subsubsection{Strange quark tagging}
\label{sec:strange-tagging}

Tagging strange jets from Higgs decays provides an alternative method to exclusive Higgs decays~\cite{Kagan:2014ila, Perez:2015lra, Koenig:2015pha, Aaboud:2016rug, Aaboud:2017xnb,Alte:2016yuw} for constraining the Yukawa coupling of the strange quark. See \Bref{Soreq:2016rae, Yu:2016rvv, Bishara:2016jga, Gao:2016jcm} for approaches using event shape and kinematic
observables. The main idea behind the strange tagger described in~\Bref{Duarte-Campderros:2018ouv} is that
strange quarks---more than other partons---hadronise to prompt kaons that carry a large fraction of the jet momentum. Based on this idea a tagger is constructed to allow for an estimate of the
capabilities in measurements involving strange quarks. Although the current focus at LHC is on
mainly on charm and bottom tagging, recognising strange jets has been attempted before at
DELPHI~\cite{Boudinov:1998fao} and SLD~\cite{Kalelkar:2000ig}, albeit in $Z$ decays.

The shown results are based on an analysis of event samples of Higgs and $W$ events generated with
\texttt{PYTHIA} 8.219~\cite{Sjostrand:2006za, Sjostrand:2014zea}. In each of the two hemispheres of the resonance decay, the charged pions and kaons stemming from the resonance are selected with an
assumed efficiency of 95\%. Similarly, $K_s$ are identified with an efficiency of 85\% if they decay
within \Unit{80}{cm} of the interaction point into a $\pi^+\pi^-$ pair that allows to reconstruct
the decaying neutral kaon. Among the two lists of Kaon candidates---one per hemisphere---one Kaon of
each list is chosen for further analysis such that the scalar sum of their momenta is maximised
while rejecting charged same-sign pairs. The events are separated into the categories
charged-charged~(CC), charged-neutral~(CN) and neutral-neutral~(NN) with a relative abundance of about CC:CN:NN$\approx 9:6:1$ from isospin considerations and branching ratios by the charges of the selected Kaon candidates.

All selected candidates are required to carry a large momentum $p_{||}$ along the hemisphere
axis. This cut allows to reduce the background from gluon jets as gluons radiate more than quarks
and therefore tend to spread their energy among more final state particles. In addition, charged Kaons need to be produced promptly, in order to reject heavy flavor jets. This latter requirement is implemented by a cut on the impact parameter $d_0$ after the truth value has been smeared by the detector resolution.
\begin{figure}[t]
  \centering
  \includegraphics[width=0.6\textwidth]{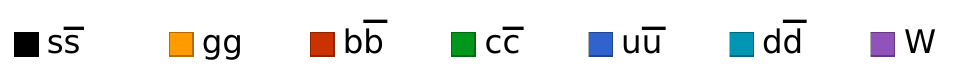}\\
  \includegraphics[width=0.49\textwidth]{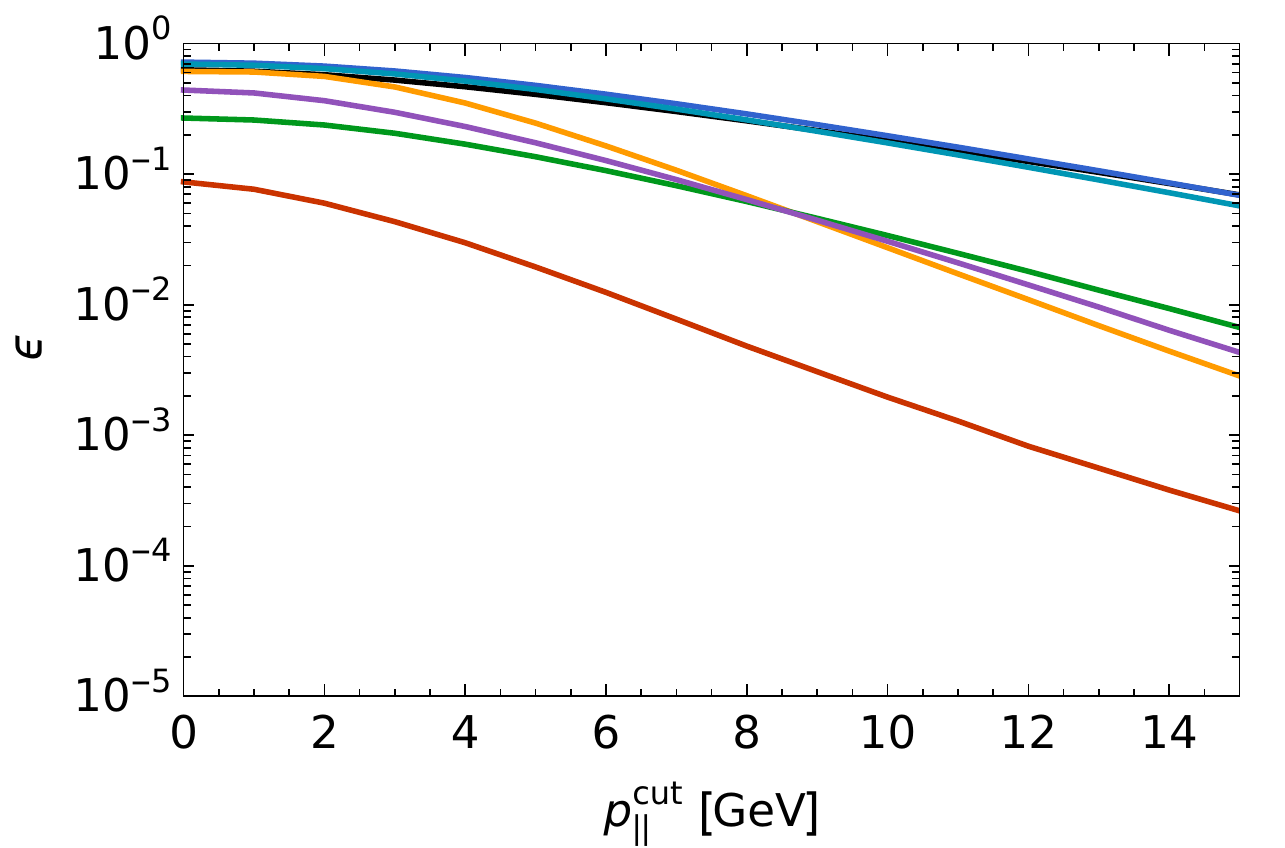}
  \includegraphics[width=0.49\textwidth]{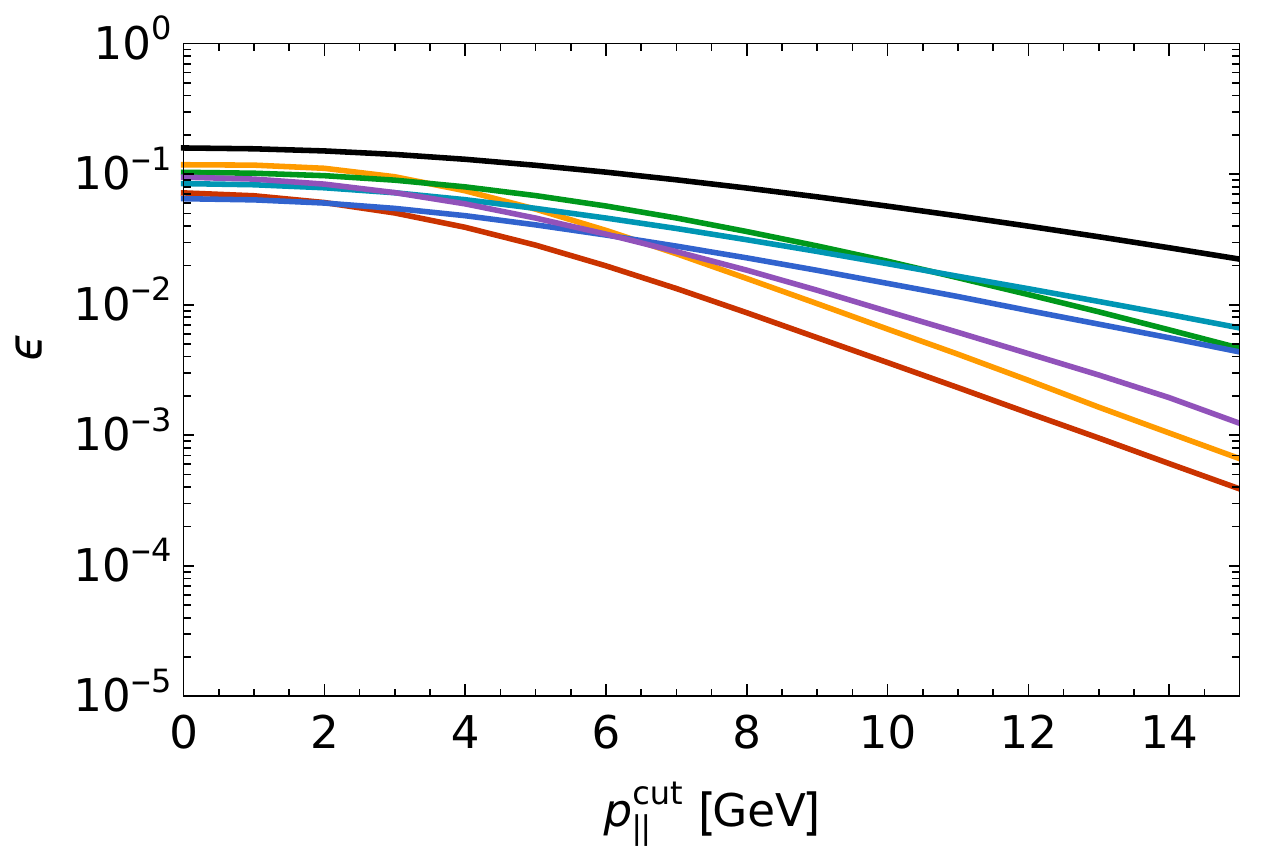}
  \caption{Efficiencies as function of the cut on $p_{||}$ and for $d_0<$\Unit{14}{$\mu$m} to
    reconstruct the different Higgs decay channels and $W$ decays as $s\bar s$ event by the
    described tagger. The left plot shows the CC channel, the right the CN channel.}
  \label{fig:stagger-efficiencies}
\end{figure}

The efficiencies obtained in the CC and CN channel for a cut of $d_0<$\Unit{14}{$\mu$m} are shown in
Fig.~\ref{fig:stagger-efficiencies}. While there is clearly still ample room for improvement, this
simple tagger shows already a good suppression by orders of magnitude of the bottom, charm and gluon
background. Due to missing particle identification, the efficiencies for first-generation jets and
strange jets are degenerate in the CC channel. However, in the CN channel, due to the required
$K_s$, a suppression of pions is achieved that breaks this degeneracy. This is particularly
interesting in light of the HL-LHC, where a large background from first generation jets is expected.

\asubsection[Y. Soreq]{Exclusive Higgs decays}
\label{sec:exclusiveHiggs}

Exclusive Higgs decays to a vector meson ($V$) and a photon, $h\to V\gamma$, directly probe the Higgs bottom, charm~\cite{Bodwin:2013gca,Bodwin:2014bpa} strange, down and up~\cite{Kagan:2014ila} quark Yukawas, as well as to the flavor violating couplings. 
For improved theory predictions see~\cite{Koenig:2015pha}.
Within the LHC, the Higgs exclusive decays are the only direct probe of the $u$ and $d$ Yukawa couplings. If $s$-tagging will be implemented at the LHC, than the strange Yukawa will be probed both inclusive and exclusive as charm and bottom.  
On the experimental side, both ATLAS and CMS report first upper bounds on $h\to J/\psi\gamma$~\cite{Aad:2015sda,Khachatryan:2015lga}, $h\to\phi\gamma$ and $h\to\rho\gamma$~\cite{Aaboud:2016rug,Aaboud:2017xnb}. 
The $h \to VZ, ZW$ modes as a probe of the Higgs electroweak coupling are discussed  in~\cite{Isidori:2013cla}. 
Finally, $Z$ exclusive decays are considered in~\cite{Grossmann:2015lea,Alte:2015dpo} and can be served as a test of QCD factorisation. 

The Higgs exclusive decays which involve $V=\rho,\omega,\phi,J/\psi,\Upsilon$ are sensitive to the diagonal Yukawa couplings. 
These receive contributions from two amplitudes which are denoted as direct and indirect, see Fig.~\ref{fig:hexlusive}. 
The direct amplitude, first analysed in~\cite{Keung:1983ac}, involves a hard $h\to q\bar{q}\gamma$ vertex and sensitive to the $q$-quark Yukawa. 
The indirect process is mediated by $h\gamma\gamma$ vertex which is followed by a $\gamma^*\to V$ fragmentation. 
Since the indirect contribution is larger than the direct, the largest sensitivity to the Higgs $q$-quark coupling is via the interference between the two diagrams. 
\begin{figure}[t]
\begin{center}
\includegraphics[width=0.65\textwidth]{\main/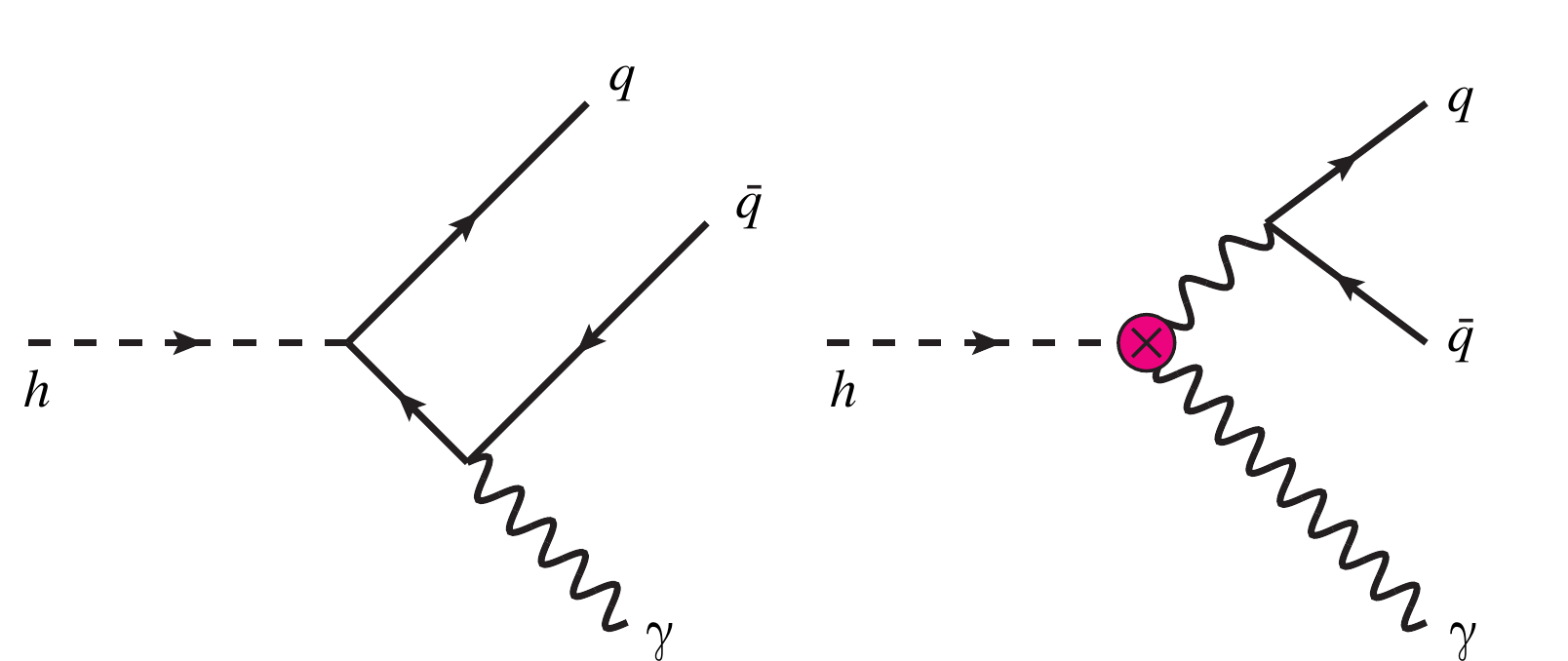}
\caption{The two contributions to $h \to V\gamma$ with $V=\rho,\omega,\phi,J/\psi,\Upsilon$. Left: the direct amplitude, proportional to  the $q$-quark Yukawa; Right: indirect amplitude involve the $h\gamma\gamma$ vertex.}
\label{fig:hexlusive}
\end{center}
\end{figure}

It is beneficial to consider the ratio between $h\to V\gamma$ and $h\to\gamma\gamma$ or $h\to ZZ^*\to4\ell$ as various of theoretical uncertainties and the dependence of the Higgs total width are cancelled~\cite{Perez:2015aoa,Koenig:2015pha}. Moreover, since the Higgs production is inclusive for all of these modes, it cancelled in the ratio to large extension. Thus, we can write 
\begin{align}
	\cR_{V\gamma,f} 
=	\frac{\mu_{V\gamma} }{\mu_{f}}  \frac{ \BR^{\rm SM}_{h\to V\gamma}}{ \BR^{\rm SM}_{h\to f}}  
	\simeq  
	\frac{\Gamma_{h\to V\gamma}}{ \Gamma_{h\to f}}  \, ,
\end{align}
where $f=ZZ^*, \gamma\gamma\,$, $\mu_X = \sigma_h \BR_{X}/\sigma^{\rm SM}_h \BR^{\rm SM}_{X}$, the superscript ``SM" denotes the SM values and we assume a perfect cancellation of the production mechanism. 
For simplicity, we assume CP even Higgs coupling and find 
\begin{align}
	\cR_{V\gamma,f} 
=& 	\alpha_{V,f} \abs{1 - \left(\Delta^R_V+ i \Delta^I_V \right)\frac{\bar{\kappa}_V}{\kappa^{\rm eff}_{\gamma\gamma}} + \Delta^U_V}^2  \, , 	
\end{align}
with
\begin{align}	
	\alpha_{V,\gamma\gamma}
=&	6\frac{\Gamma_{V\to e^+e^-}}{\alpha \, m_V} \left( 1 - \frac{m^2_V}{m^2_h} \right)^2 \, , \quad\quad
	\alpha_{V,ZZ^*}
= \abs{\frac{\kappa^{\rm eff}_{\gamma\gamma}}{\kappa_{Z}}}^2  \frac{\Gamma^{\rm SM}_{h\to\gamma\gamma}}{\Gamma^{\rm SM}_{h\to ZZ^*\to 4\ell}}\alpha_{V,\gamma\gamma} \, ,
\end{align}
where $\kappa_X$ is the normalised coupling with respect to its SM value. Below, we adopted the numerical values of $\Delta^X_V$ from Ref.~\cite{Koenig:2015pha}.
The advantage of use $h\to\gamma\gamma$ for the normalisation is that there are only two unknown - the Higgs coupling to di-photon and the quark Yukawa. 
However, since $h\to ZZ^*$ is a very clean channel is serve as a good channel to use for the normalisation. 
Moreover, by combing the Higgs data with the electroweak precision measurements, the Higgs coupling to $ZZ$ is known to a few percent level~\cite{Falkowski:2013dza,deBlas:2016ojx}, thus, there is no additional large uncertainty.  
We note that with the current data the bounds evaluating by using $\cR_{V\gamma,ZZ^*}$ are slightly stronger than the ones from $\cR_{V\gamma,\gamma\gamma}$. 

For the interpretation of the experimental results in term of bounds on the different Yukawa coupling we follow Refs.~\cite{Perez:2015aoa,Perez:2015lra}. 
Denoting the 95\,\%\,CL bound on the ratio $\cR_{V\gamma,f}$ as $\cR_{V\gamma,f}^{95}$ we can write 
\begin{align}
	\frac{\Delta^R_V - \sqrt{  \frac{(\Delta^R_V)^2 + (\Delta^I_V)^2    }{\alpha_{V\gamma,f}} \cR_{V\gamma,f}^{95}   - (\Delta^I_V)^2 } }{ (\Delta^R_V)^2 + (\Delta^I_V)^2 } 
	< \frac{\bar{\kappa}_V}{\kappa^{\rm eff}_{\gamma\gamma}} < 
	\frac{\Delta^R_V + \sqrt{  \frac{(\Delta^R_V)^2 + (\Delta^I_V)^2    }{\alpha_{V\gamma,f}} \cR_{V\gamma,f}^{95}   - (\Delta^I_V)^2 } }{ (\Delta^R_V)^2 + (\Delta^I_V)^2 }  \, ,
\end{align}
where we neglect $\Delta^U_V$ as it is a small correction. 
Moreover, neglecting $\Delta^I_V$ we get simplified formula, which hold to good accuracy, 
\begin{align}
	\label{eq:Rsimp}
	\frac{1 - \sqrt{  \cR_{V\gamma,f}^{95}/\alpha_{V\gamma,f} } }{ \Delta^R_V} 
	< \frac{\bar{\kappa}_V}{\kappa^{\rm eff}_{\gamma\gamma}} < 
	\frac{1 + \sqrt{  \cR_{V\gamma,f}^{95}/\alpha_{V\gamma,f} } }{ \Delta^R_V}  \, .
\end{align}
Table~\ref{tab:hexclusive} summarises the current experimental status along with the theory interpretation in terms of light quarks Yukawa.  
\begin{table}[t]
\begin{center}
\begin{tabular}{|c|c|c|c|}
\hline
mode & $\BR_{h\to V\gamma}<$ & $\cR_{V\gamma,ZZ^*}<$ &Yukawa range  \\
\hline\hline 
$J/\psi\,\gamma$  &
$1.5\times 10^{-3}$\, 8\,\UTeV~\cite{Aad:2015sda,Khachatryan:2015lga} & 
$9.3$ &
$-295\kappa_Z + 16\kappa^{\rm eff}_{\gamma\gamma} < \kappa_c < 295\kappa_Z + 16\kappa^{\rm eff}_{\gamma\gamma} $\\
\hline
$\phi\,\gamma$ &
$4.8 \times 10^{-4}$\, 13\,\UTeV~\cite{Aaboud:2016rug,Aaboud:2017xnb} & 
$3.2$ &
$-140\kappa_Z + 10\kappa^{\rm eff}_{\gamma\gamma} < \bar{\kappa}_s < 140\kappa_Z + 10\kappa^{\rm eff}_{\gamma\gamma} $\\
\hline
$\rho\,\gamma$ &
$8.8 \times 10^{-4}$\, 13\,\UTeV~\cite{Aaboud:2017xnb} & 
$5.8$ &
$-285\kappa_Z + 42\kappa^{\rm eff}_{\gamma\gamma} < 2\bar{\kappa}_u + \bar{\kappa}_d < 285\kappa_Z + 42\kappa^{\rm eff}_{\gamma\gamma} $\\
\hline\hline
\end{tabular}
\end{center}
\caption{The current upper bounds, assuming SM Higgs production, on the different exclusive Higgs decays and the interpretation in terms of the Higgs Yukawa couplings. Note that $\bar{\kappa}_q = y_q/y^{\rm SM}_b\,$. The quoted bounds are at 95\,CL. }
\label{tab:hexclusive}
\end{table}%

The prospects for probing light quark Yukawa within future LHC runs and for future colliders are estimated in Ref.~\cite{Perez:2015lra}, which we follow here. 
One of the important implications of the first upper bounds on the different exclusive modes is that the measurement is background dominated. Thus, even for future runs, without significant improvement of the analysis, we expect only upper bounds.  
Given an upper bound on $\cR^{95}_{V\gamma,f}(E_1,\cL_1)$, where $E_1\,(\cL_1)$ stands for the collider energy\,(integrated luminosity), the estimated bound with $E_2$ and $\cL_2$ is
\begin{align}
	\label{eq:Rscale}
	\cR^{95}_{V\gamma,f}(E_2,\cL_2)
=	\cR^{95}_{V\gamma,f}(E_1,\cL_1) \sqrt{ \frac{1}{R_E} \frac{\sigma^{\rm SM}_{h,E_1} \cL_1  }{\sigma^{\rm SM}_{h,E_2} \cL_2} } \, ,
\end{align}
where $\sigma^{\rm SM}_{h,E_{1,2}}$ is the SM Higgs production cross section, $R_{E} = (S_{E_1}^{\rm SM}/B_{E_1})/(S_{E_2}^{\rm SM}/B_{E_2})$ with $S(B)$ the number of signal\,(background) events, which encoded the difference in the analysis details and assumed to be 1 here. 
In Table~\ref{tab:exclusiveproj},we combine Eqs.~\eqref{eq:Rsimp} and \eqref{eq:Rscale} along with the current bounds to estimate the future projections of probing the different light quark Yukawa. 
We note that the estimation in Table~\ref{tab:exclusiveproj} is in agreement with the ATLAS projection of $h\to J/\psi\gamma$~\cite{ATL-PHYS-PUB-2015-043}, which quote $\cR_{J/\psi\gamma,ZZ^*}<0.34^{+0.14}_{-0.1}$
 
\begin{table}[t]
\begin{center}
\begin{tabular}{|c|c|c|c|}
\hline
mode & collider energy & $\cR_{V\gamma,ZZ^*}<$ &Yukawa range ($\kappa_V=\kappa^{\rm eff}_{\gamma\gamma}=1$)   \\
\hline\hline 
\multirow{ 3}{*}{$J/\psi\,\gamma$} &
14\,\UTeV & 
$0.47\sqrt{L_3}$ &
$16 - 67 L_3^{1/4}  < \kappa_c < 16 + 67 L_3^{1/4} $\\
&
27\,\UTeV & 
$0.28\sqrt{L_3}$ &
$16 - 52 L_3^{1/4}  < \kappa_c < 16 + 52 L_3^{1/4} $\\
&
100\,\UTeV & 
$0.12\sqrt{L_3}$ &
$16 - 33 L_3^{1/4}  < \kappa_c < 16 + 33 L_3^{1/4} $\\
\hline
\multirow{ 3}{*}{$\phi\,\gamma$} &
14\,\UTeV & 
$0.33\sqrt{L_3}$ &
$11 - 46 L_3^{1/4} < \bar{\kappa}_s < 11 + 46 L_3^{1/4} $\\
&
27\,\UTeV & 
$0.20\sqrt{L_3}$ &
$11 - 35 L_3^{1/4} < \bar{\kappa}_s < 11 + 35 L_3^{1/4} $\\
&
100\,\UTeV & 
$0.083\sqrt{L_3}$ &
$11 - 23 L_3^{1/4} < \bar{\kappa}_s < 11 + 23 L_3^{1/4} $\\
\hline
\multirow{ 3}{*}{$\rho\,\gamma$} &
14\,\UTeV & 
$0.60\sqrt{L_3}$ &
$44 - 93 L_3^{1/4} < 2\bar{\kappa}_u + \bar{\kappa}_d < 44 + 93 L_3^{1/4} $\\
&
27\,\UTeV & 
$0.36\sqrt{L_3}$ &
$44 - 72 L_3^{1/4} < 2\bar{\kappa}_u + \bar{\kappa}_d < 44 + 72 L_3^{1/4} $\\
&
100\,\UTeV & 
$0.15\sqrt{L_3}$ &
$44 - 47 L_3^{1/4} < 2\bar{\kappa}_u + \bar{\kappa}_d < 44 + 47 L_3^{1/4} $\\
\hline\hline
\end{tabular}
\end{center}
\caption{The projection for Yukawa range for future $pp$ colliders with centre of mass energy of $14$, $27$ and $100\,$\UTeV. In the above table we define $L_3\equiv (3/{\rm ab})/\cL\,$. }
\label{tab:exclusiveproj}
\end{table}%

In addition to the Higgs diagonal Yukawa, in principle, Higgs exclusive decays can probe off-diagonal couplings by measuring modes such as $h\to B^*_s \gamma$~\cite{Kagan:2014ila}. 
These processes receive contribution only from the direct amplitude and there is not enhancement from interference with the relative large indirect amplitude. 
Moreover, the Higgs flavor violating couplings are strongly constrained by meson mixing~\cite{Blankenburg:2012ex,Harnik:2012pb}. Thus, the expected rates are too small to be observe. 
For a detailed discussion on the $h \to VZ,VW$ channels see~\cite{Alte:2016yuw}.

\asubsection[Y. Soreq]{Lepton flavor violating decays of the Higgs}

The flavour violating Yukawa couplings are well constrained by the low-energy
flavour-changing neutral current measurements
\cite{Harnik:2012pb,Blankenburg:2012ex,Gorbahn:2014sha}. A notable
exception are the flavour-violating couplings involving a tau lepton. The
strongest constraints on $\kappa_{\tau\mu}, \kappa_{\mu\tau},
\kappa_{\tau e}, \kappa_{e \tau}$ are thus from direct searches of flavour-violating Higgs decays at
the LHC \cite{Sirunyan:2017xzt,Aad:2016blu}.
Currently, the CMS 13\,\UTeV with 35.9\,fb$^{-1}$~\cite{Sirunyan:2017xzt} is the strongest constrain 
\begin{align}
    \sqrt{y_{\mu\tau}^2 + y^2_{\tau\mu}} < 1.43 \times 10^{-3}\, , \qquad
    \sqrt{y_{e\tau}^2 + y^2_{\tau e}} < 2.26 \times 10^{-3}\, ,
\end{align}
which corresponds to upper 95\,\%~CL on the branching ratio of $0.25\,$\% and $0.61\,$\% for $\mu\tau$ and $e\tau$, respectively. In addition, we note that once can directly measure the difference between the branching ratios of $h\to\tau e$ and $h\to\tau\mu$ as proposed in~\cite{Bressler:2014jta}.
Naively, assuming that both systematics and statistical error scale with square root of the luminosity, one can expect that the sensitivity of the HL-LHC with $3000\,$fb$^{-1}$ will be  around the half per-mil level for the branching ratio of $h\to e\tau$ or $\to \mu\tau$.

The LHC can also set bounds on rare FCNC top decays involving a Higgs~\cite{Aaboud:2017mfd,Khachatryan:2016atv,Aad:2015pja,Aad:2014dya}. The strongest current bound, for example, is $\sqrt{|\kappa_{ct}|^2+|\kappa_{tc}|^2}<0.06$ at 95\,\%~CL.

\asubsection[F. Yu, A. Schmidt, T. Klijnsma, Y. Soreq]{Yukawa constraints from Higgs distributions}
\label{sec:HiggsDist}

\asubsubsection{Determinations of Higgs boson coupling modifiers using differential distributions}

The distribution of the transverse momentum $p_T$ of the Higgs boson has been considered before as a probe of high scale new physics running in the $ggh$ loop \cite{Arnesen:2008fb,Biekotter:2016ecg,Brehmer:2015rna,Dawson:2015gka,Schlaffer:2014osa,Grojean:2013nya,Langenegger:2015lra,Bramante:2014hua,Buschmann:2014twa,Azatov:2013xha,Banfi:2013yoa,Buschmann:2014sia}. 
In addition, the soft spectrum is an indirect probe of the Higgs coupling to light quarks~\cite{Soreq:2016rae,Bishara:2016jga}. 
Higgs production modes due to quark fusion, which are negligible in the SM, have two effects on the distributions of kinematic variables. 
First, the Sudakov peak will be at lower $p_T$ around 5\,\UGeV vs 10\,\UGeV for gluon fusion, see~\cite{Collins:1984kg}. This is because the effective radiation strength of gluons is several times larger than that of quarks, $\alpha_s N_c$ vs. $\alpha_s (N^2_c-1)/(2N_c)$, with $N_c=3$. This leads to harder $p_T$ spectra for gluon fusion compared to quark scattering. 
Therefore, the $u\bar{u}$ or $d\bar{d}$ scattering leads to a much sharper peak at lower $p_T$ compared to  $gg$ scattering~\cite{Soreq:2016rae}.
Second, in the SM, the Higgs production is dominated by  gluon fusion, where the two gluons carry similar partonic $x$. This leads to a peak at zero Higgs rapidity. However, for $u\bar u$ or $d\bar d$ fusion, the valance quark will carry larger partonic $x$ than the sea anti-quark. This leads to a peak in the forward direction. 
In case of enhanced $s$ or $c$ Yukawa couplings, the dominant effect is the one loop of the quarks in the $gg\to hj$ process, which has double logarithms behaviour and peaks towards lower $p_T$ of the Higgs boson~\cite{Baur:1989cm}. This will also result in a softer Higgs $p_T$ spectrum~\cite{Bishara:2016jga}, which can be used to constrain the charm and strange Yukawa. 
The impact of effect on various kinematic distributions is shown in Fig.~\ref{fig:HiggsDist}. 
Many theoretical and experimental uncertainties are cancelled in  the normalised kinematic distributions, $(1/\sigma)d\sigma/dX$ with $X=p_T, y_h$, see for example~\cite{Soreq:2016rae}.
Thus, the use of them will result in a better sensitivity for probing the light quark Yukawa. 
\begin{figure}[t]
\begin{center}
\includegraphics[width=0.35\textwidth]{\main/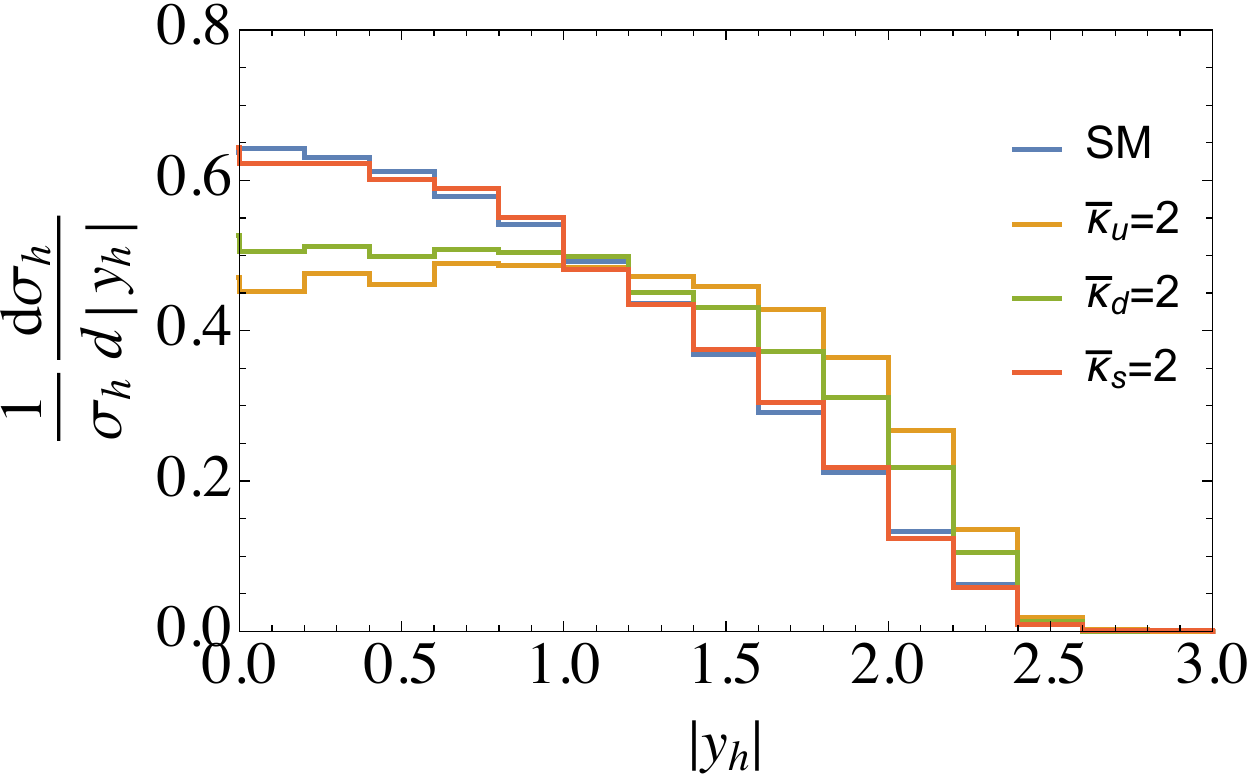}~~~~~~~~
\includegraphics[width=0.35\textwidth]{\main/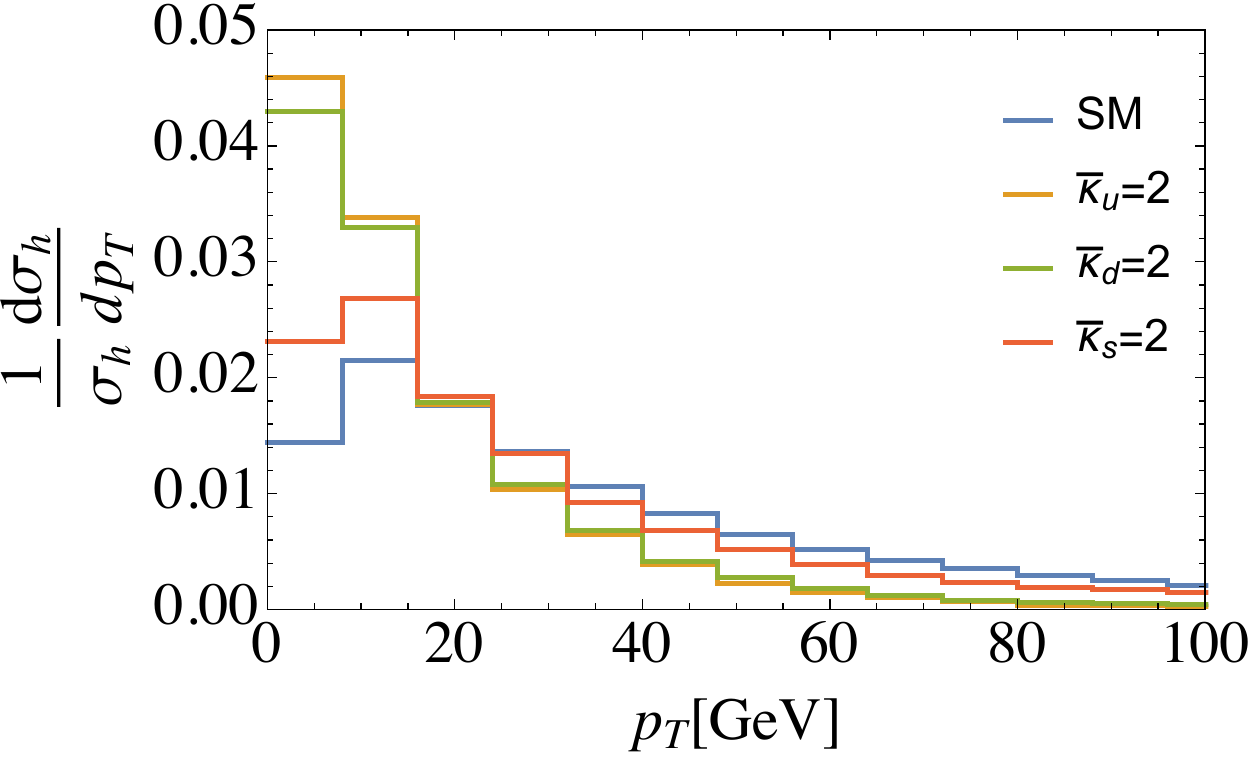}\\
\includegraphics[width=0.35\textwidth]{\main/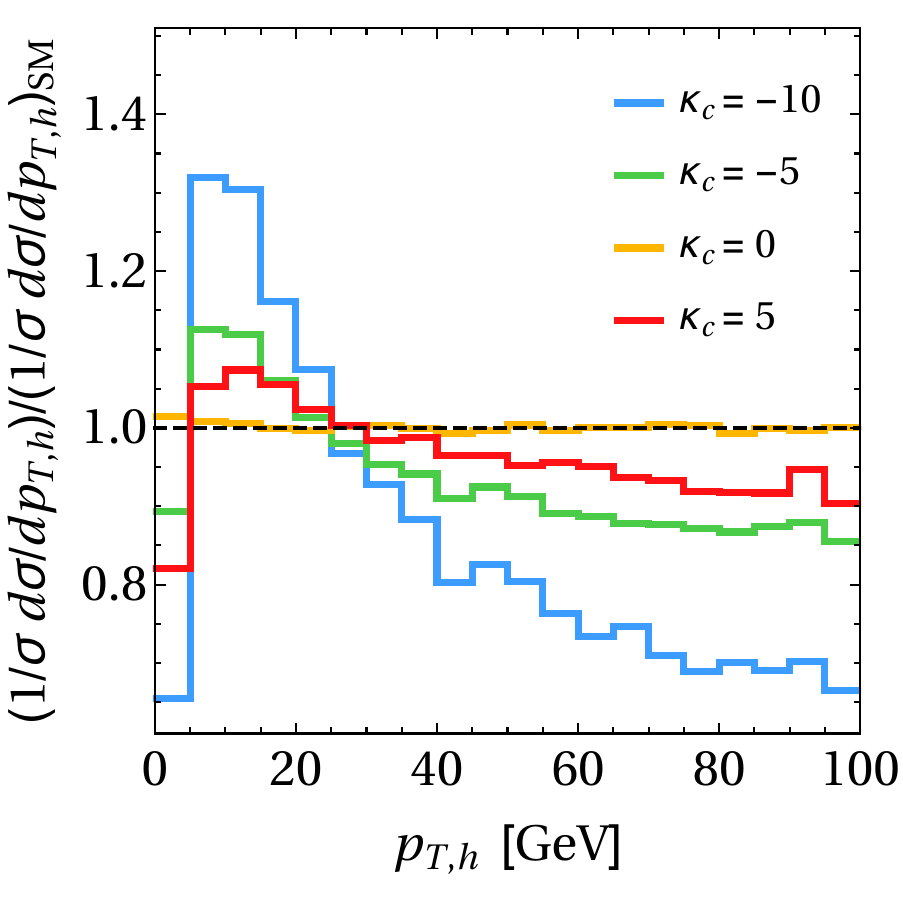}
\caption{Normalised distributions of kinematic variables of the Higgs boson. 
Left-top\,(Right-top) : $y_h\,(p_T)$ distribution for enhanced $u$, $d$ and $s$ Yukawa compared to the SM~\cite{Soreq:2016rae};
Bottom: $p_T$ distribution for enhanced $c$ Yukawa~\cite{Bishara:2016jga}. 
}
\label{fig:HiggsDist}
\end{center}
\end{figure}

In Ref.~\cite{Soreq:2016rae}, the 8\,\UTeV ATLAS results~\cite{Aad:2015lha} have been used to evaluate a first bound on the $u$ and $d$ Yukawa from kinematic distributions. The resulting $95\,\%$ CL regions obtained from the  $p_T$ distribution are
\begin{align}
	\bar{\kappa}_u = y_u/y^{\rm SM}_b < 0.46 \, , \qquad\qquad
	\bar{\kappa}_d = y_d/y^{\rm SM}_b< 0.54 \, ,
\end{align}
which are stronger than the fits to the inclusive Higgs production cross sections. 
These upper bounds are found to be stronger than the expected due to an under-fluctuation of the data in the first $p_T$ bin. 
The bounds from the  rapidity distribution are found to be weaker. 
The sensitivities expected for Run II  are shown in Fig.~\ref{fig:HiggsDistFuture}. 
\begin{figure}[t]
\begin{center}
\includegraphics[width=0.35\textwidth]{\main/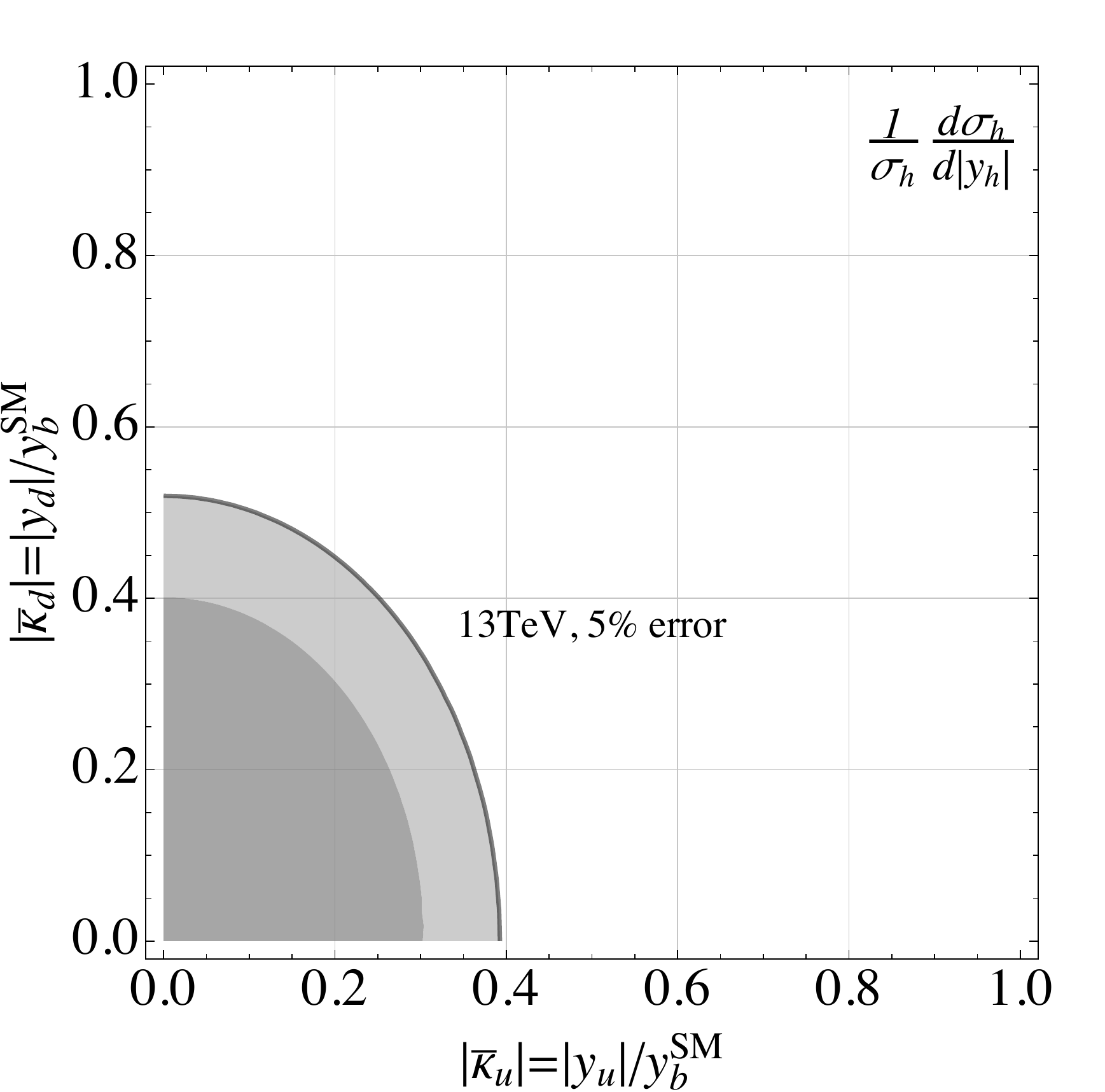}~~~~~~
\includegraphics[width=0.35\textwidth]{\main/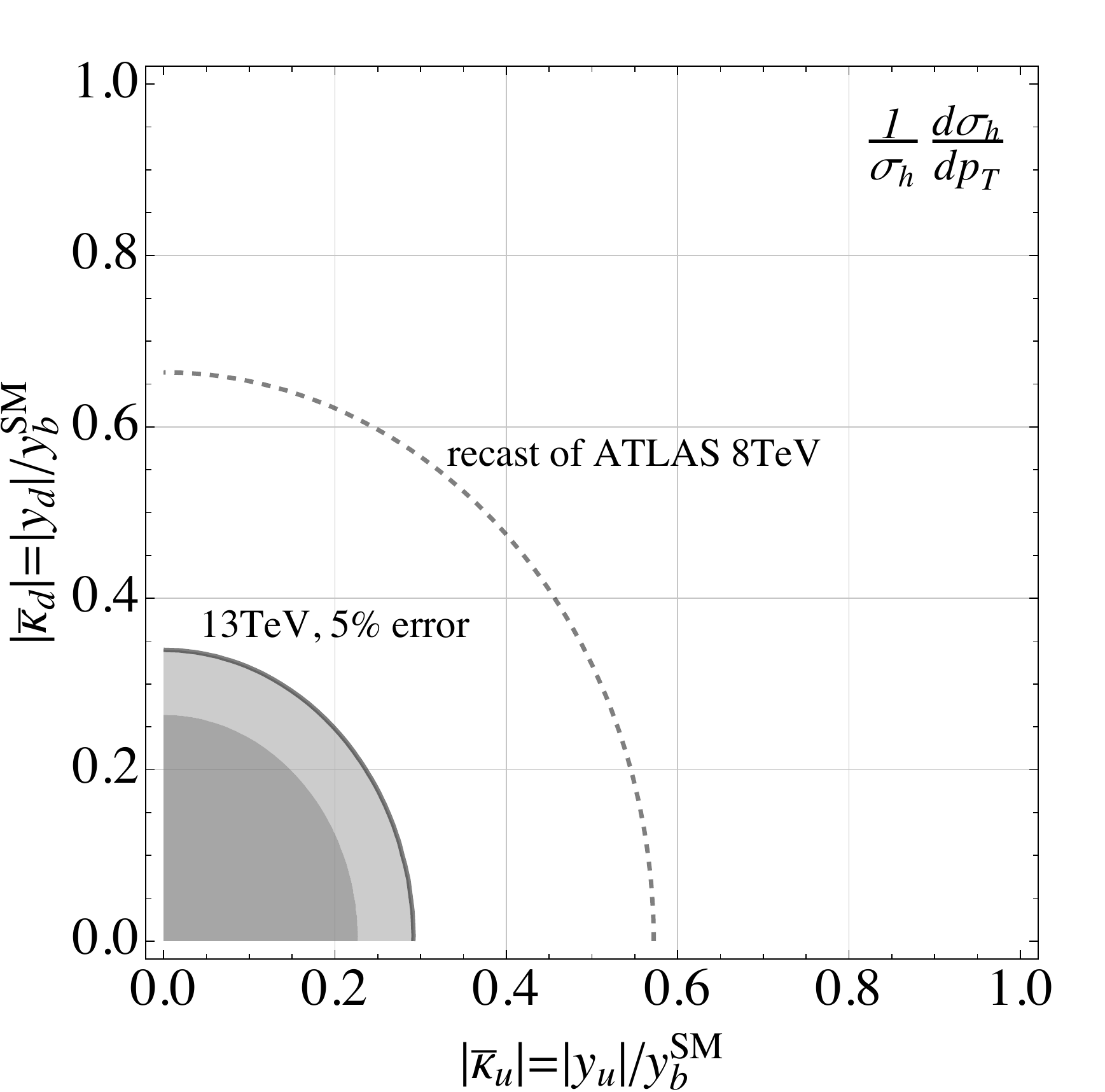}\\
\includegraphics[width=0.35\textwidth]{\main/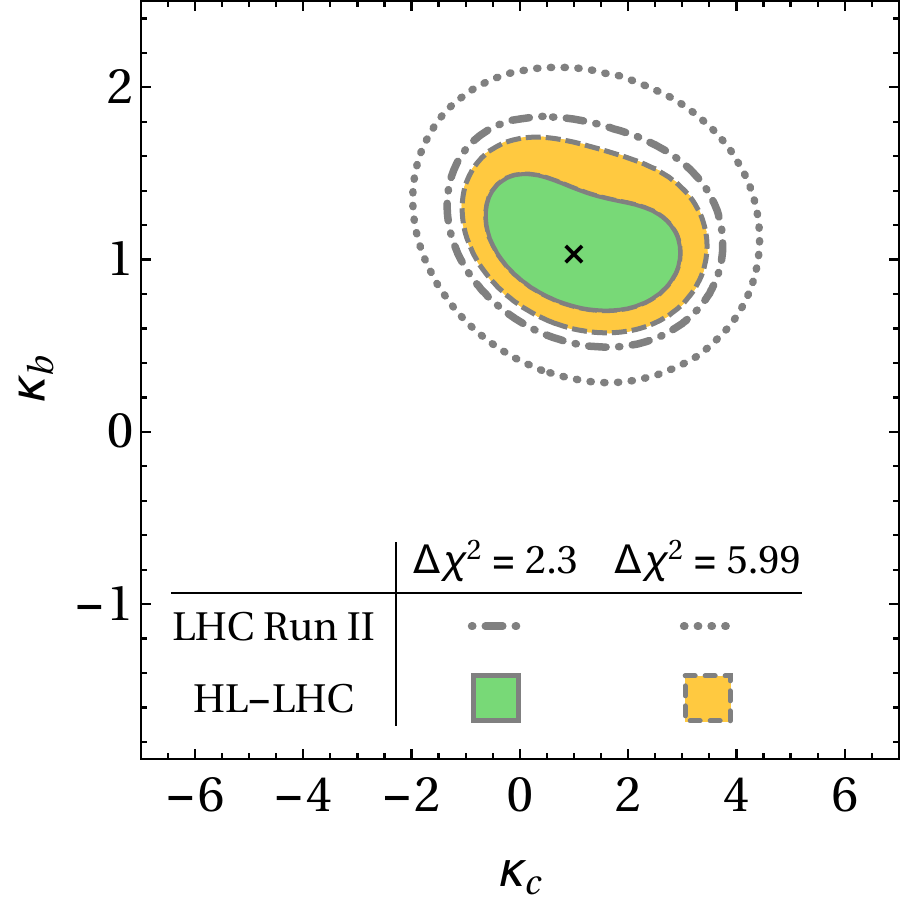}
\caption{The sensitivity of  different angular distributions probing the light quark Yukawa couplings at 13\,\UTeV.  
Left-top: $u$ and $d$ quark Yukawa from $y_h$ distribution~\cite{Soreq:2016rae};
Right-top: $u$ and $d$ quark Yukawa from $p_T$ distribution~\cite{Soreq:2016rae}; 
Bottom: $b$ and $c$ quark Yukawa from $p_T$ distribution~\cite{Bishara:2016jga}. 
}
\label{fig:HiggsDistFuture}
\end{center}
\end{figure}

CMS interpreted the 13\,\UTeV Higgs $p_T$ spectrum with luminosity of $35.9\,$fb$^{-1}$ to obtain bounds on the $c$ and $b$ Yukawa couplings~\cite{Sirunyan:2018sgc}. 
The resulting 95\,\% CL intervals are
\begin{align}
    -4.9 < \kappa_c <4.8 \, ,  \qquad
    -1.1 < \kappa_b < 1.1 \, ,
\end{align}
if the branching fractions  depend on $\kappa_b$ and $\kappa_c$. In case the branching fractions are allowed to float freely the results are
\begin{align}
    -33 < \kappa_c < 38 \, , \qquad
    -8.5 < \kappa_b < 18 \, .
\end{align}
The former bound on the $c$ Yukawa is stronger than the bound from the global fit of the 8\,\UTeV Higgs data along with the electroweak precision data, allowing all Higgs couplings to float~\cite{Perez:2015aoa}. However, it relies on strong assumptions that the Higgs couplings (besides $c$ or $b$) are SM like, and it is mostly sensitive to the cross section and not to the angular shape as the latter bound.

These bounds on the $c$ Yukawa are weaker\,(stronger) then the bounds from the global fit of the 8\,\UTeV Higgs data along with the electroweak precision data allowing all Higgs coupling to float~\cite{Perez:2015aoa}  

In the following these constraints on Higgs boson couplings obtained in Ref.~\cite{Sirunyan:2018sgc} are projected to an integrated luminosity of $3000$fb$^{-1}$,
using the expected differential distributions at $3000$fb$^{1}$ 
presented in Sec.~\ref{sec:diffxs} and detailed in Ref.~\cite{CMS-PAS-FTR-18-011}.

The Higgs boson coupling fits are based on a combination of $p_T$
distributions from the $\hgg$~\cite{Sirunyan:2018kta} decay channels
obtained at $\sqrt{s}=13\,$\UTeV.
Furthermore, a search for the Higgs boson produced with large \pt and decaying to a bottom quark-antiquark ($\bb$) pair, which enhances the sensitivity at high $\pth$, is included in the $\kappat/\cg$ fit.
The Higgs boson coupling fits are performed using an simultaneous extended maximum likelihood fit to the di-photon mass, four-lepton mass, and soft-drop mass $\msd$~\cite{Dasgupta:2013ihk,Larkoski:2014wba} spectra in all the analysis categories of the $\hgg$, $\hzz$, and $\hbb$ channels, respectively.
For more details on the treatment of the input measurements, see Ref.~\cite{Sirunyan:2018kta}.

The treatment of the decay of the Higgs boson affects the Higgs boson coupling fits.
Assuming full knowledge of how the Higgs decays, \ie assuming no beyond-the-SM contributions, the inclusive Higgs production cross section adds a strong constraint on the Higgs boson couplings in the fit.
This result is obtained by parametrising the branching fractions as functions of the Higgs boson couplings.
Likewise, the constraints on the Higgs boson couplings excluding the information from the inclusive cross section are of interest in order to evaluate the discriminating power of the differential distributions.
This result is implemented by letting the branching fractions be determined in the fit without any prior constraint.

The expected one and two standard deviation contours of the $\kappac/\kappab$ fit with the branching fractions \hyphenation{parametrized} as functions of the Higgs boson couplings at a projected integrated luminosity of $3000$\fbinv is shown in Fig.~\ref{fig:kbkc_couplingdependentBRs}, for both scenarios of systematic uncertainty.
For the $\hgg$ channel the systematic uncertainties dominate if kept at the current level (\ie in Scenario 1), but when scaled down according to the Scenario 2 prescription the systematic uncertainties are within the same order of magnitude as the statistical ones.

The same fits, but now with the branching fractions implemented as nuisance parameters with no prior constraint, are shown in \ref{fig:kbkc_floatingBRs}.
As this fit is dominated by statistical uncertainties even at very high integrated luminosities, the smaller systematic uncertainties in Scenario 2 have only a minor impact.

\begin{figure}[hbtp]
  \begin{center}
    \includegraphics[width=0.49\linewidth]{\main/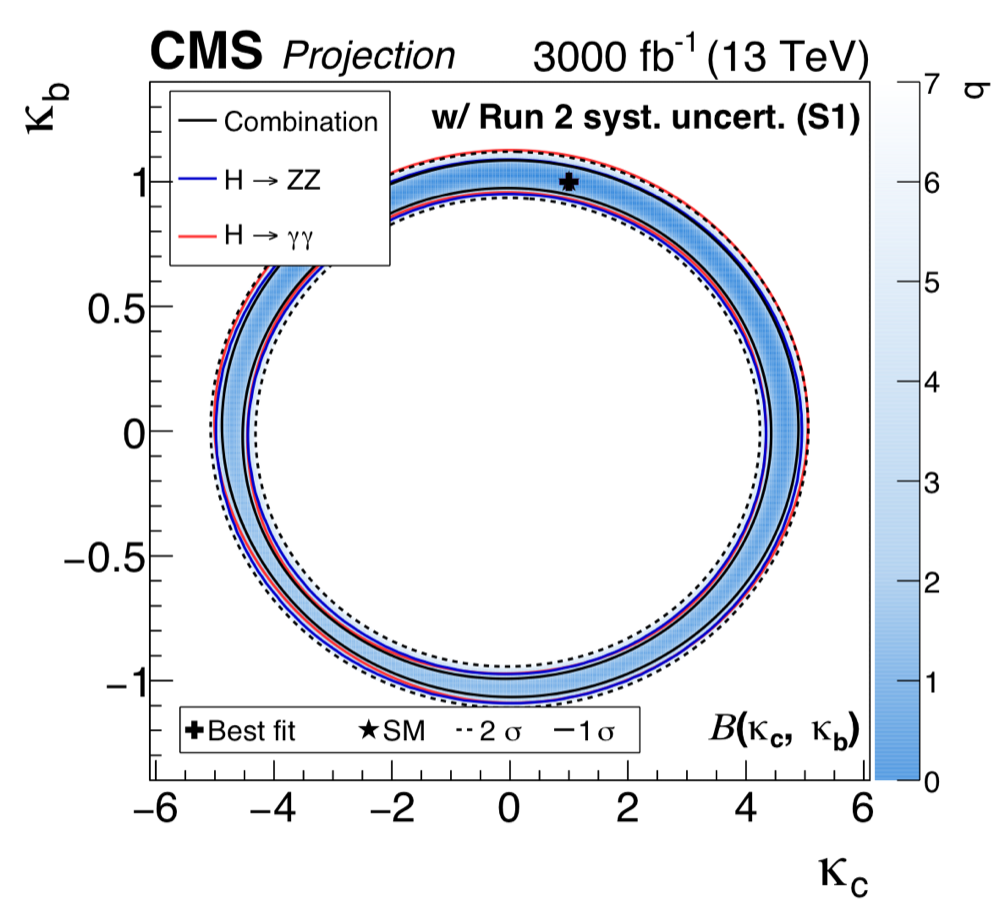}
    \includegraphics[width=0.49\linewidth]{\main/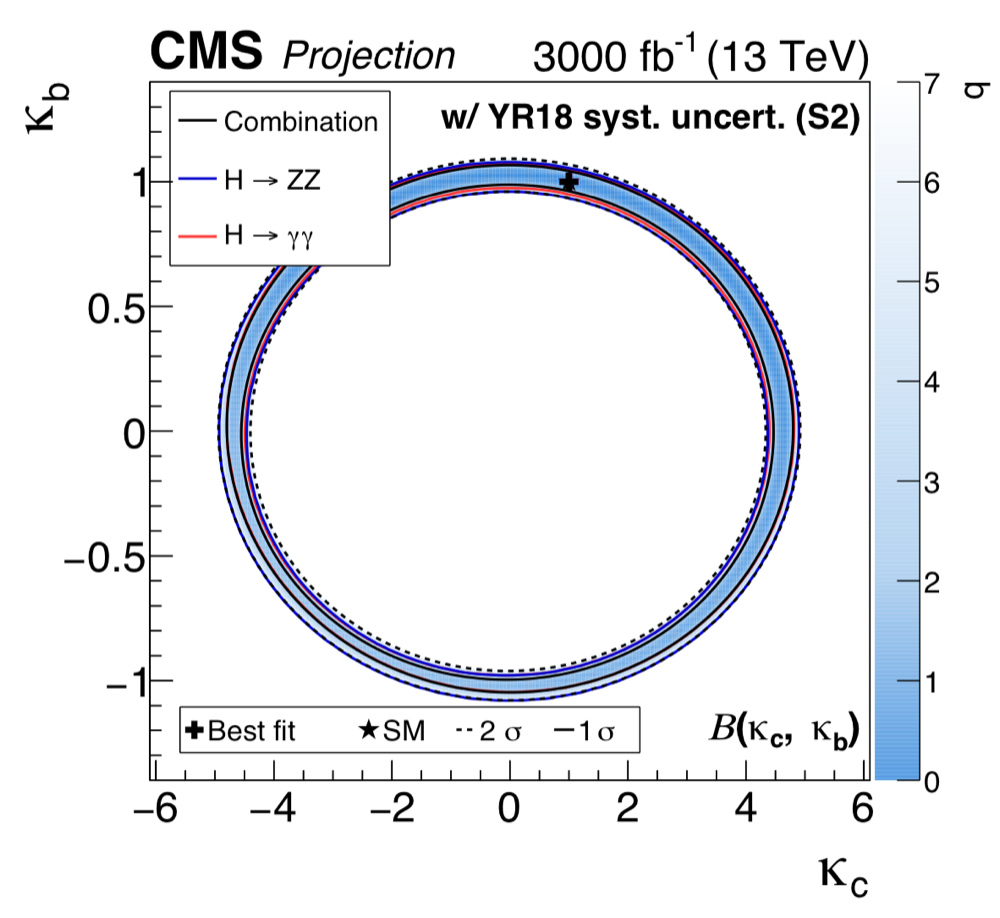}
    \caption{
        Simultaneous fit to data for $\kappab$ and $\kappac$, assuming a coupling dependence of the branching fractions for Scenario 1 (\UcmsLeft) and Scenario 2 (\UcmsRight).
        The one standard deviation contour is drawn for the combination ($\hgg$ and $\hzz$), the $\hgg$ channel, and the $\hzz$ channel in black, red, and blue, respectively.
        For the combination the two standard deviation contour is drawn as a black dashed line, and the negative log-likelihood value on the coloured axis.
        }
    \label{fig:kbkc_couplingdependentBRs}
  \end{center}
\end{figure}

\begin{figure}[hbtp]
  \begin{center}
    \includegraphics[width=0.49\linewidth]{\main/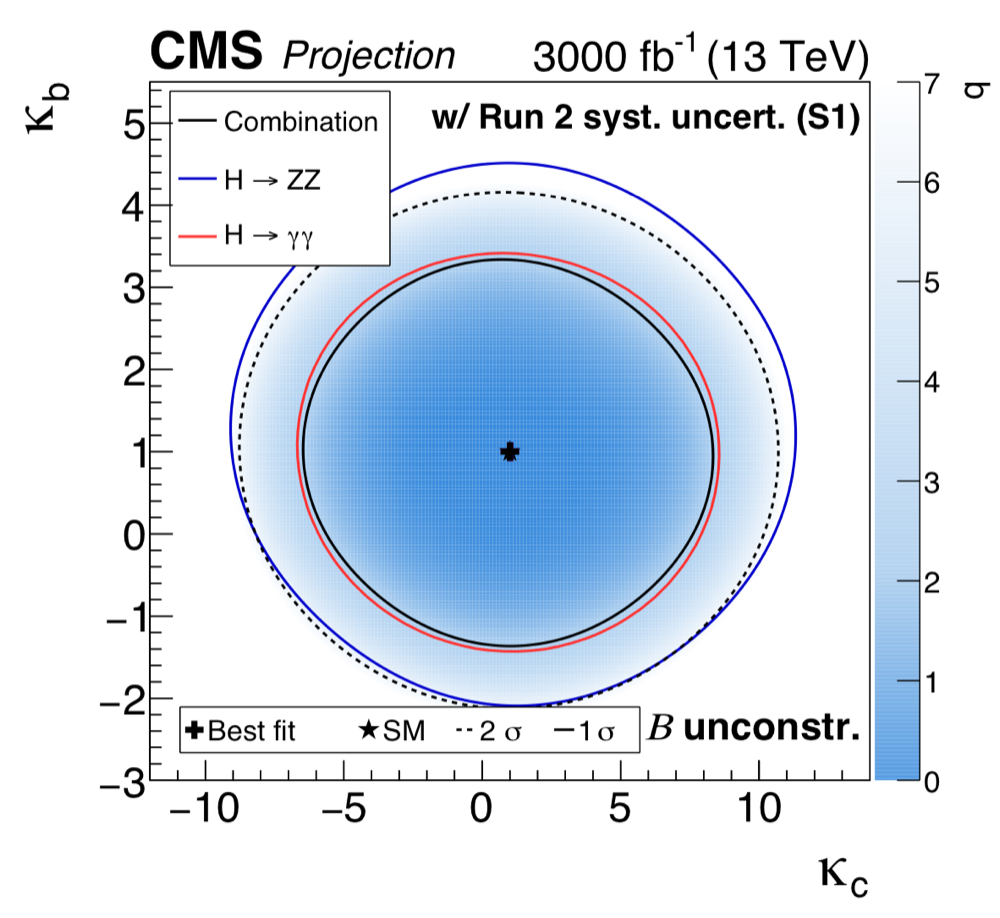}
    \includegraphics[width=0.49\linewidth]{\main/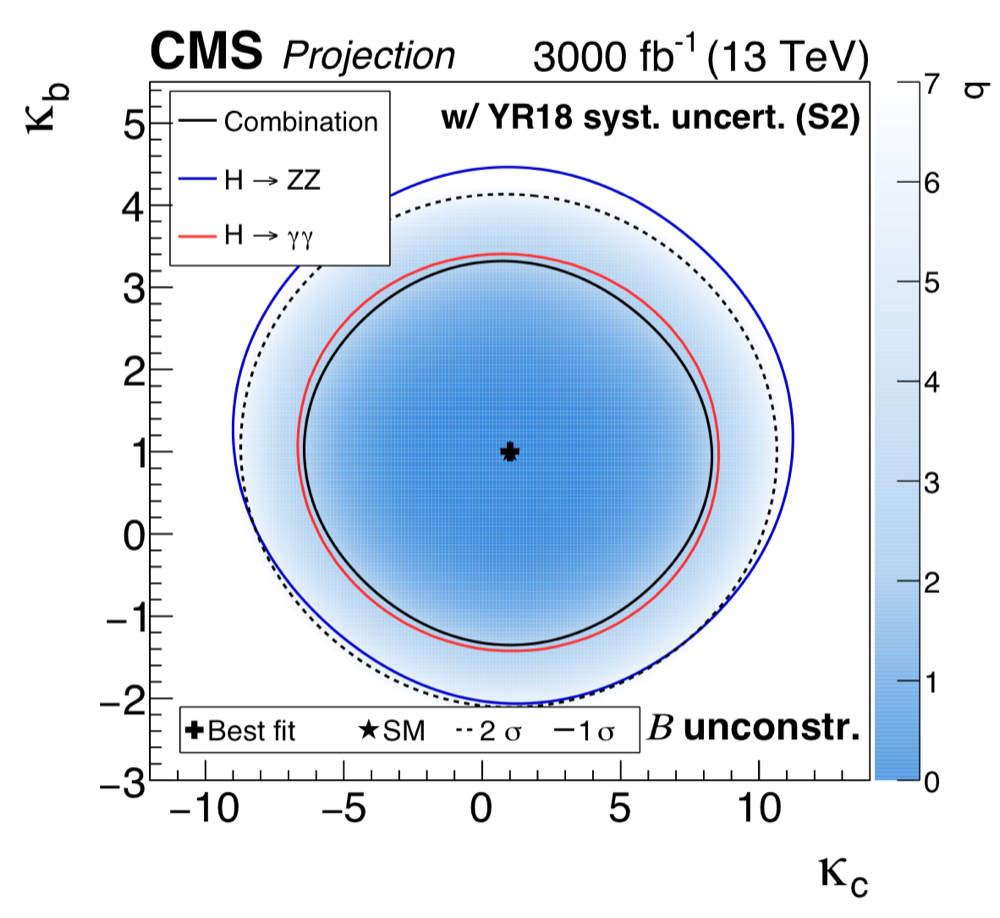}
    \caption{
        As Fig.~\ref{fig:kbkc_couplingdependentBRs}, but with the branching fractions implemented as nuisance parameters with no prior constraint.
        %
        }
    \label{fig:kbkc_floatingBRs}
  \end{center}
\end{figure}

\asubsubsection{$W^\pm h$ charge asymmetry}

The $W^\pm h$ charge asymmetry, introduced in~\cite{Yu:2016rvv}, is a
new, production-based probe for constraining the light quark Yukawa
couplings.  In contrast to decay-based probes, which rely on rare or
sub-dominant Higgs decay modes, production-based probes can take
advantage of the dominant Higgs decays with high signal-to-background
ratios.  

The main observable is the charge asymmetry between $W^+ h$ and $W^-$
production,
\begin{align}
	A 
= 	\frac{ \sigma (W^+ h) - \sigma (W^- h)}
	{\sigma (W^+ h) + \sigma (W^- h) } \, ,
\end{align}
In the SM, the inclusive HE-LHC charge asymmetry is expected to be $17.3\%$, while the HL-LHC charge asymmetry is expected to be $21.6\%$.  In either
case, the charge asymmetry is driven by the proton PDFs and the fact
that the dominant $W^\pm h$ production mode stems from Higgs bosons
radiating from $W^\pm$ intermediate lines, where the Yukawa-mediated
diagrams are negligible.  If the quark Yukawas are not SM-like,
however, the charge asymmetry can either increase or decrease,
depending on the overall weight of the relevant PDFs.  In particular,
the charge asymmetry will increase if the down or up quark Yukawa
couplings are large, reflecting the increased asymmetry of $u \bar{d}$
vs.~$\bar{u} d$ PDFs; the charge asymmetry will decrease if the
strange or charm Yukawa couplings are large, reflecting the symmetric
nature of $c \bar{s}$ vs.~$\bar{c} s$ PDFs.  The sub-leading correction
from the Cabibbo angle-suppressed PDF contributions determines the
asymptotic behaviour for extremely large Yukawa enhancements.

\begin{figure}[tb!]
  \begin{center}
 \includegraphics[width=0.9\textwidth, angle=0]{\main/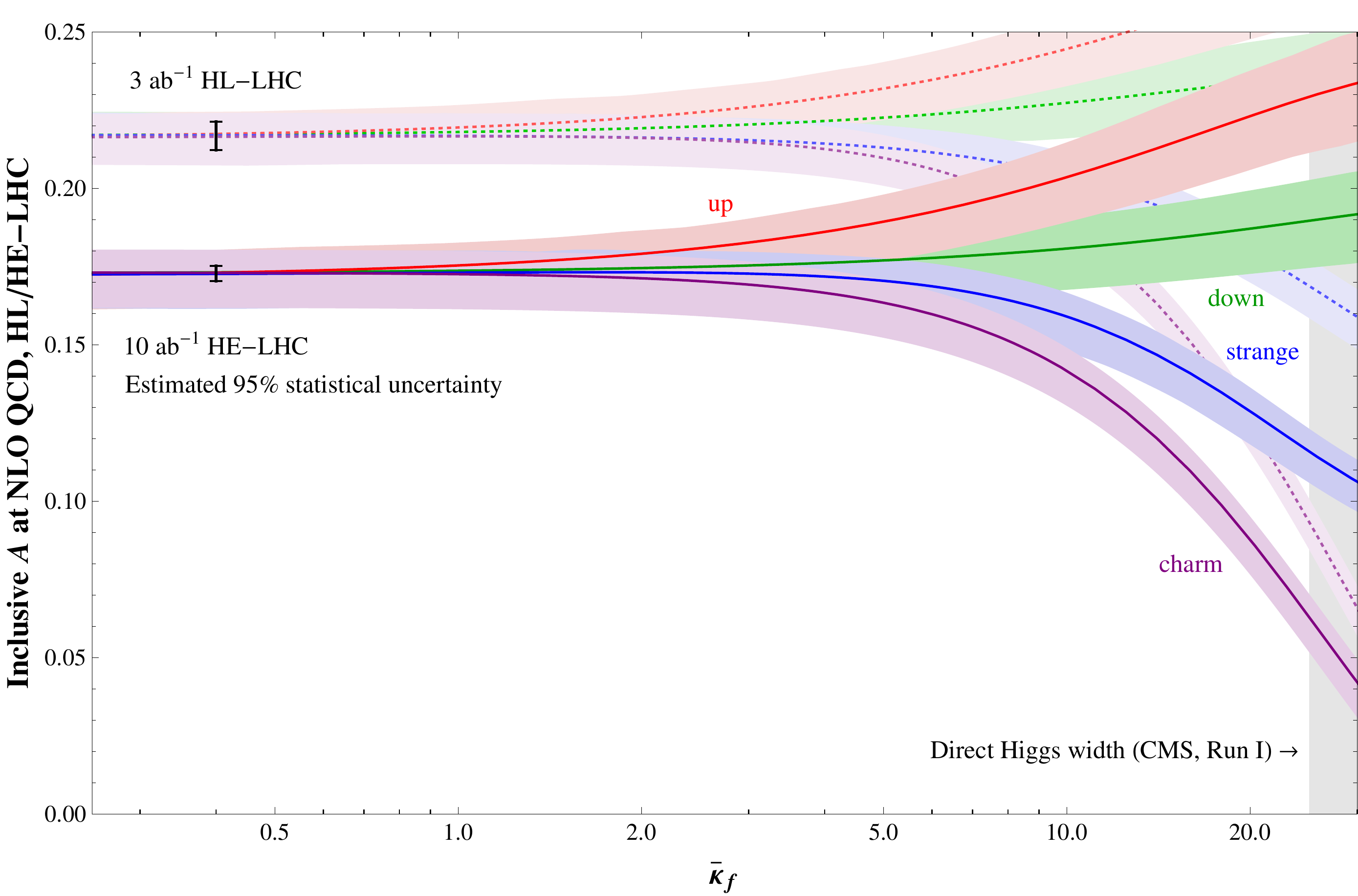}
 \caption{Inclusive charge asymmetry for $W^\pm h$ production at the
   27 \UTeV HE-LHC (solid coloured bands), and 14 \UTeV HL-LHC (dotted
   coloured bands), calculated at NLO QCD from MadGraph\_aMC@NLO using
   NNPDF 2.3 as a function of individual Yukawa rescaling factors
   $\bar{\kappa}_f$ for $f = u$ (red), $d$ (green), $s$ (blue), and
   $c$ (purple).  Shaded bands correspond to scale uncertainties at
   $1\sigma$ from individual $\sigma(W^+ h)$ and $\sigma(W^- h)$
   production, which are conservatively taken to be fully
   uncorrelated.  The expected statistical errors from this
   measurement using 10 ab$^{-1}$ of HE-LHC data and 3 ab$^{-1}$ of
   HL-LHC data are also shown.}
  \label{fig:asymmetry}
  \end{center}
\end{figure}

The effect of individual $d$, $u$, $s$, or $c$ quark Yukawa
enhancements on the inclusive charge asymmetry is shown in
Figure~\ref{fig:asymmetry}, in units of $\bar{\kappa}_f = y_f /
y_{\text{SM, b}}$, evaluated at the Higgs mass scale.  Since $W^\pm h$
production probes lower Bjorken-$x$ at the HE-LHC compared to the
HL-LHC, the expected SM charge asymmetry is lower at the higher energy
collider.  In Figure~\ref{fig:asymmetry}, we also display the expected
$0.45\%$ statistical sensitivity to the charge asymmetry coming from
an HL-LHC simulation study~\cite{Yu:2016rvv} in the $W^\pm h \to
\ell^\pm \ell^\pm jj \nu \nu$ final state.  To estimate the HE-LHC
sensitivity, we simply rescale by the appropriate luminosity ratio,
giving $0.25\%$, since we expect that the increase in both signal and
background electroweak rates to largely cancel.  We also indicate the
constraint from the direct Higgs width constraint using Run I data
from CMS~\cite{Yu:2016rvv}.  The bands denote the change in the charge
asymmetry from the varying the renormalisation  and factorisation
scales within a factor of 2.

We see that the expected statistical sensitivity supersedes the
combined theoretical uncertainty in the PDF evaluation.  Hence, in
addition to being an important consistency check of the SM regarding
enhanced light quark Yukawa couplings, the charge asymmetry
measurement in different Higgs channels can be used to help determine
PDFs at the HE-LHC, assuming light quark Yukawa couplings are SM-like.
Separately, enhanced light quark Yukawa couplings would also generally
be expected to decrease the Higgs signal strengths, necessitating the
introduction of other new physics to be consistent with current Higgs
measurements~\cite{Yu:2016rvv}.  If the signal strengths are fixed to
SM expectation and the central prediction is used, the HE-LHC charge
asymmetry measurement could constrain $\bar{\kappa}_f \lesssim 2-3$
for up and charm quarks, and $\bar{\kappa}_f \lesssim 7$ for down or
strange quarks.


\asubsection[E. Gouevia, R. Harnik, B. Le, L. Lechner, Y. Li, A. Martin, A. Onofre, R. Schoefbeck, D. Spitzbart, E. Vryonidou, D. Zanzi]{CP Violation}


The CP-violating flavor conserving Yukawa couplings, $\tilde \kappa_{f_i}$, can be directly probed at the HL-LHC. Here we focus on the $\tau$ and top phases, assuming that the low energy constrained can be avoided. 
At low energy, the different flavor diagonal CP violating coupling are bounded by EDMs~\cite{Brod:2013cka,Chien:2015xha,Altmannshofer:2015qra,Egana-Ugrinovic:2018fpy,Brod:2018pli}. 
For the electron Yukawa, the latest ACME measurement~\cite{Baron:2013eja,Andreev:2018ayy} results into an upper bound of $\tilde\kappa_e<1.9\times 10^{-3}$~\cite{Altmannshofer:2015qra}. Whereas for the bottom and charm Yukawas, the strongest limits come from the neutron EDM~\cite{Brod:2018pli}. Using the NLO QCD theoretical prediction, this translates into the upper bounds $\tilde\kappa_b<5$ and $\tilde\kappa_c<21$ when theory errors are taken into account.
For the light quark CPV Yukawas, measurement of the Mercury EDM places a strong bound on the up and down Yukawas of $\tilde\kappa_u<0.1$ and $\tilde\kappa_d<0.05$~\cite{Brod:2018xyz} (no theory errors) while the neutron EDM measurement gives a weaker constraint on the strange quark Yukawa of $\tilde\kappa_s<3.1$~\cite{Brod:2018xyz} (no theory errors).

\asubsubsection{$t\bar{t} h$}

CP violation in the top quark-Higgs coupling is strongly constrained by EDM measurements and Higgs rate measurements~\cite{Brod:2013cka}. However, these constraints assume that the light quark Yukawa couplings and $hWW$ couplings have their SM values. If this is not the case, the constraints the phase of the top Yukawa coupling relax.
    
Assuming the EDM and Higgs rate  constraints can be avoided, the CP structure of the top quark Yukawa can be probed directly in $pp \to t\bar t h$. Many simple observables, such as $m_{t\bar t h}$ and $p_{T,h}$ are sensitive to the CP structure, but require reconstructing the top quarks and Higgs.

Some $t\bar t h$ observables have been proposed recently that access the CP structure without requiring full event reconstruction. These include the azimuthal angle between the two leptons in a fully leptonic $t\bar{t}$ decay with the additional requirement that the $p_{T,h} > 200\, \text{\UGeV}$~\cite{Buckley:2015vsa}, and the angle between the leptons (again in a fully leptonic $t/\bar t$ system) projected onto the plane perpendicular to the $h$ momentum~\cite{Boudjema:2015nda}. These observables only require that the Higgs is reconstructed and are inspired by the sensitivity of $\Delta \phi_{\ell^+\ell^-}$ to top/anti-top spin correlations in $pp \to t\bar t$~\cite{Mahlon:1995zn}. The sensitivity of both of these observables improves at higher Higgs boost (and therefore higher energy), making them promising targets for the HE-LHC, though no dedicated studies have been carried out to date.

\begin{figure}[h]
\centering
\begin{minipage}{0.45\textwidth}
 \includegraphics[width=\textwidth,clip]{\main/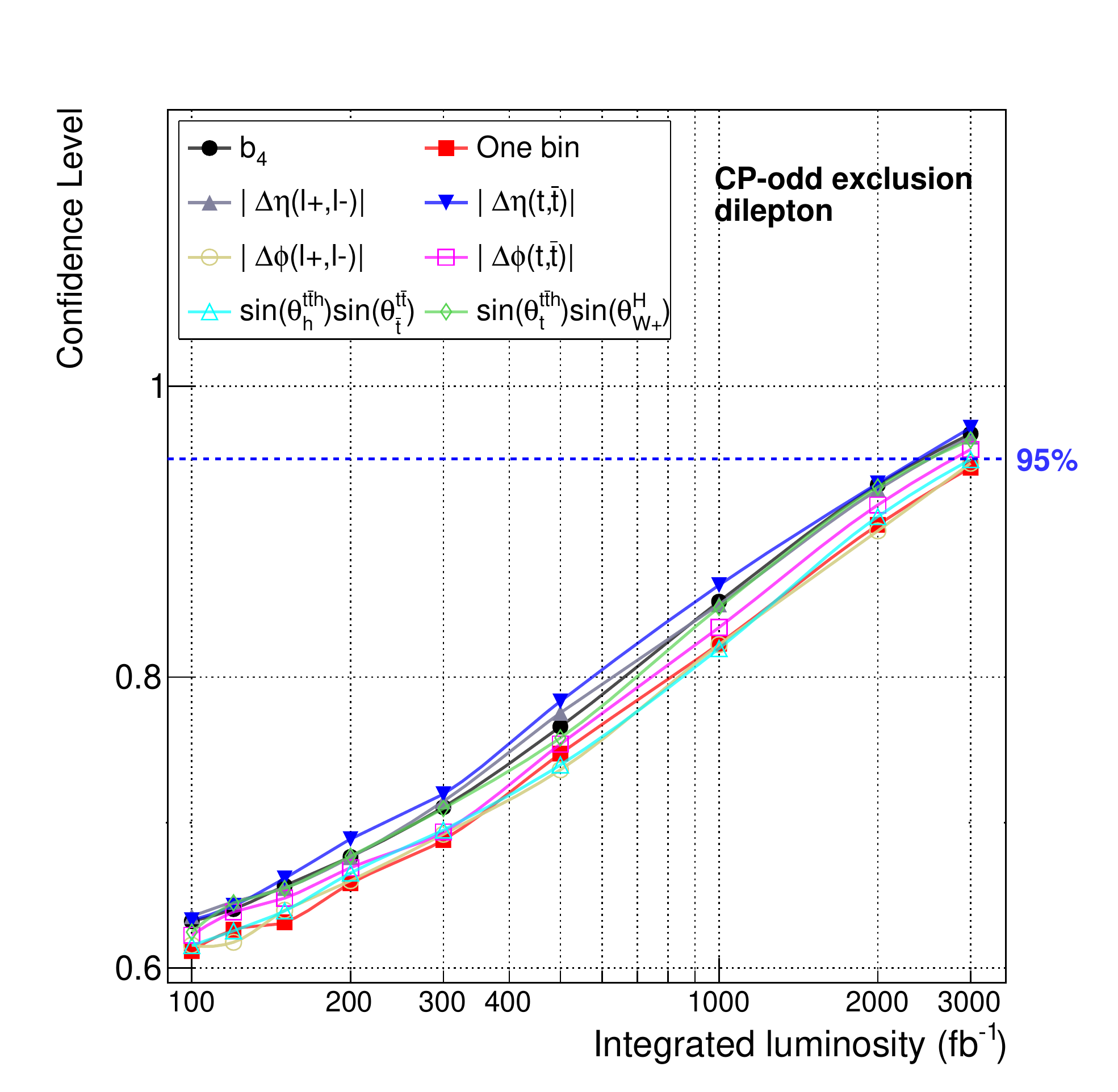}
 \caption{Expected CL, assuming the SM, for exclusion of the pure CP-odd scenario, as a function of the integrated luminosity, using the $t\bar{t}h$ ($h\rightarrow b\bar{b}$) di-leptonic analysis only. A likelihood ratio computed from the binned distribution of the corresponding discriminant observable was used as test statistic.}
 \label{dilep}
\end{minipage}
\hfill
\begin{minipage}{0.45\textwidth}
 \includegraphics[width=\textwidth,clip]{\main/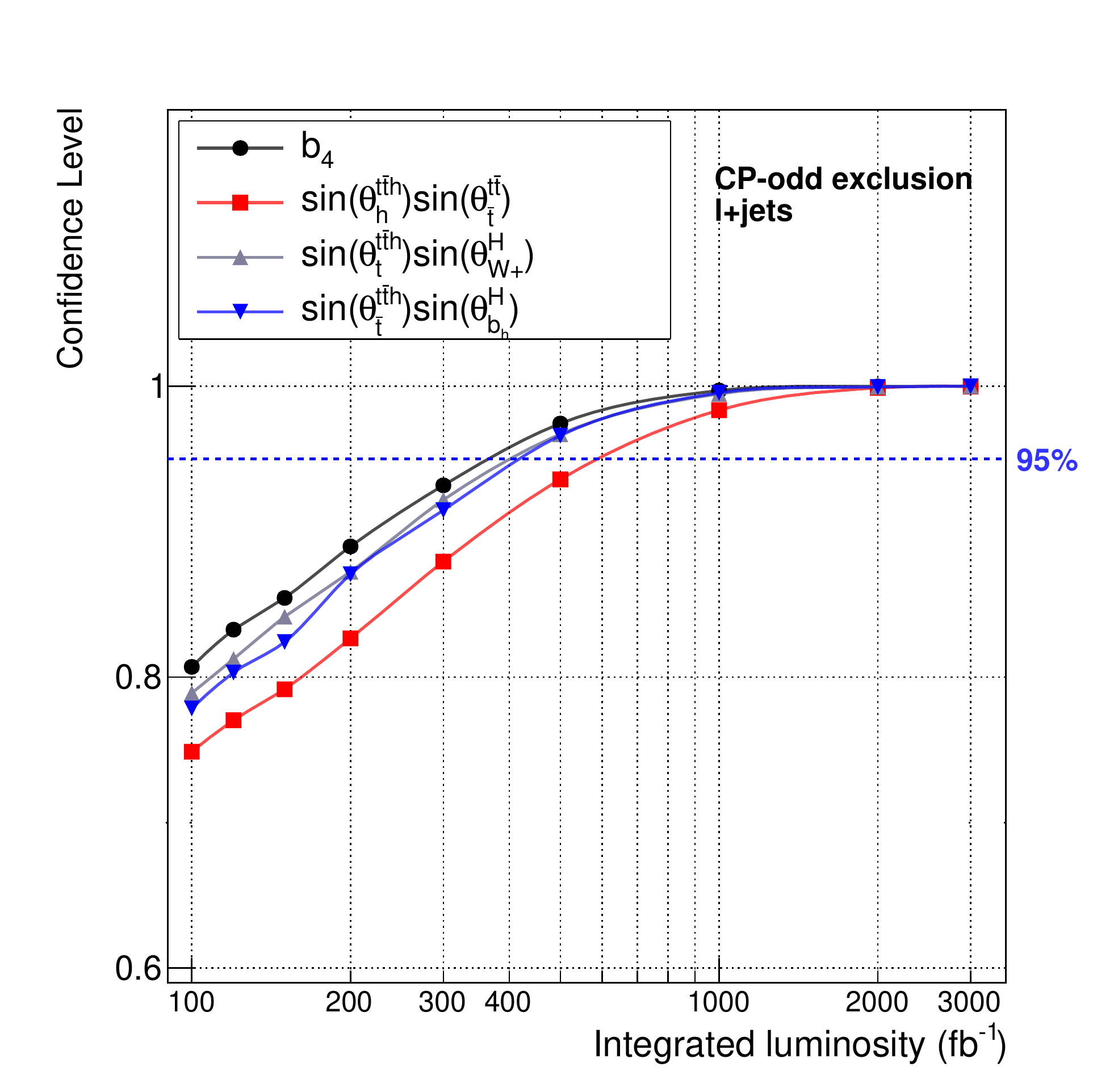}
 \caption{Expected CL, assuming the SM, for exclusion of the pure CP-odd scenario, as a function of the integrated luminosity, using the $t\bar{t}h$ ($h\rightarrow b\bar{b}$)  semi-leptonic analysis only. A likelihood ratio computed from the binned distribution of the corresponding discriminant observable was used as test statistic.}
 \label{semilep}
\end{minipage}
\begin{minipage}{0.45\textwidth}
 \includegraphics[width=\textwidth,clip]{\main/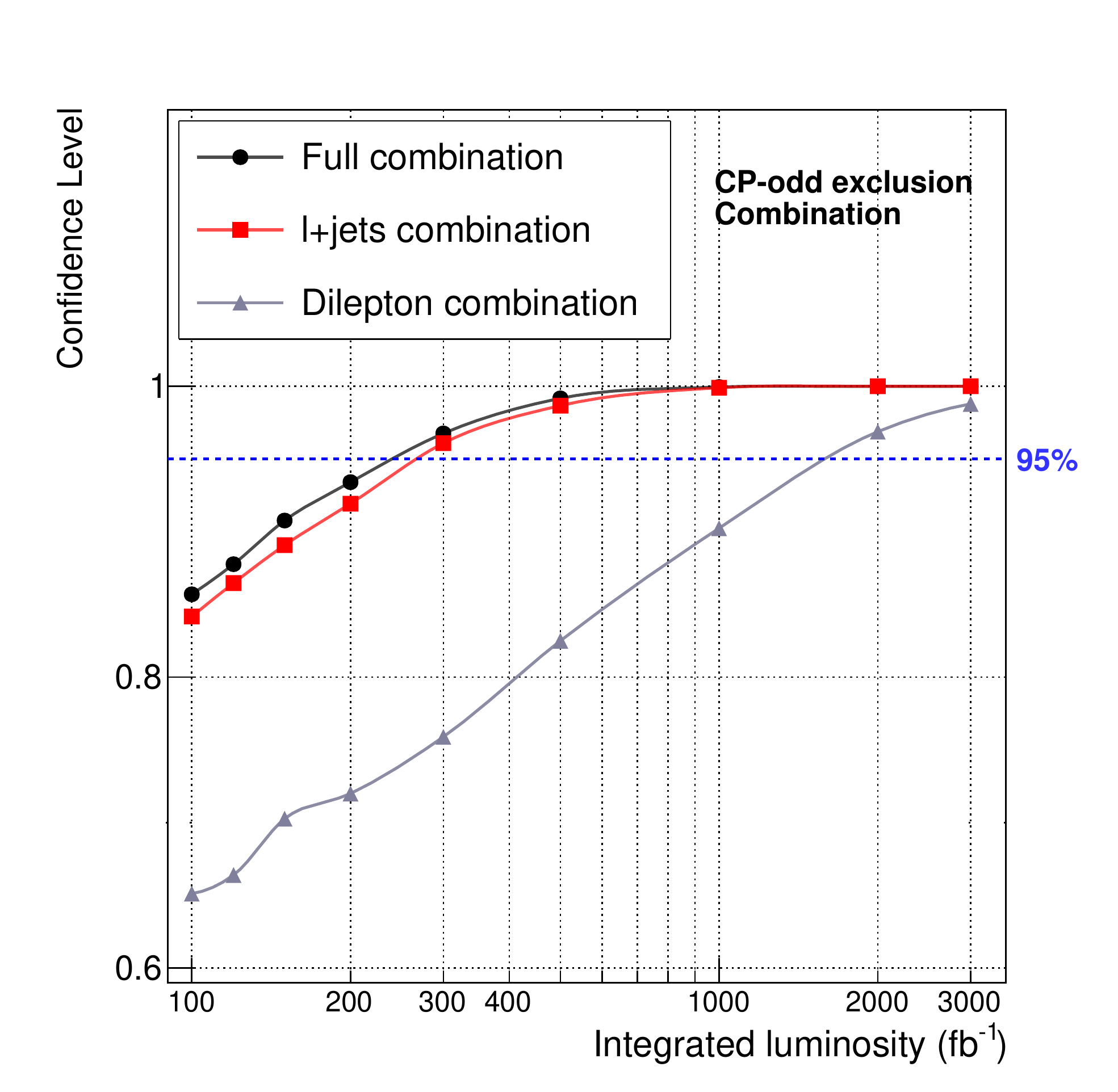}
 \caption{Expected CL, assuming the SM, for exclusion of the pure CP-odd scenario, as a function of the integrated luminosity, combining observables in each individual channel and combining both channels. The observables were treated as uncorrelated.}
 \label{combination}
\end{minipage}
\hfill
\begin{minipage}{0.45\textwidth}
 \includegraphics[width=\textwidth,clip]{\main/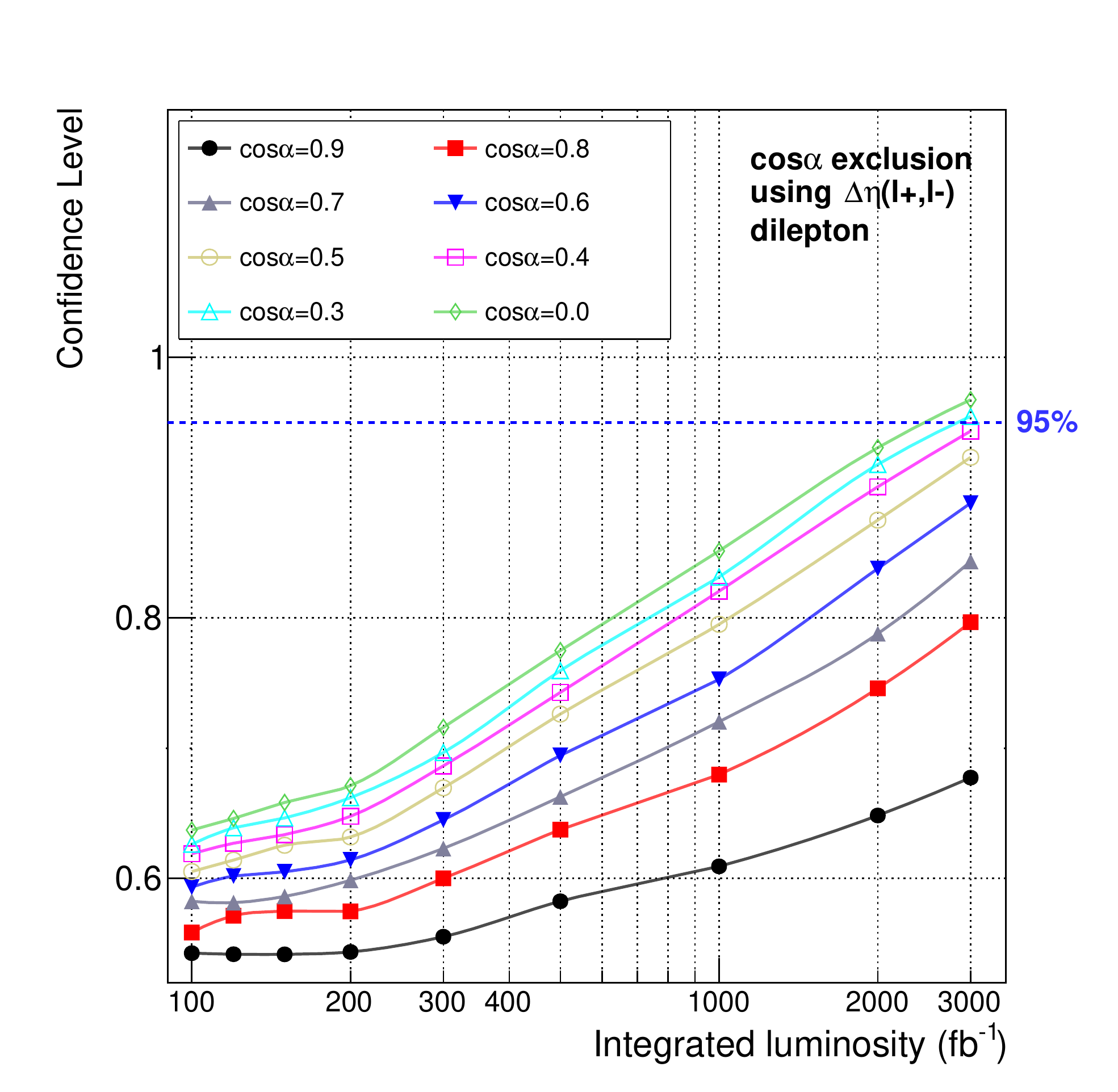}
 \caption{Expected CL, assuming the SM, in the $t\bar{t}h$ ($h\rightarrow b\bar{b}$) di-leptonic analysis alone, for the exclusion of various $\cos\alpha$ values between 0 and 1, as a function of the integrated luminosity. The results shown used the $\Delta\eta(\ell^+,\ell^-)$ as discriminant variable.}
 \label{cosaScan}
\end{minipage}
\end{figure}

Departures from the SM top quark Yukawa interactions with the Higgs boson can be considered by including a $CP$-odd component in the effective Lagrangian, i.e., ${\cal L}_{tth}=y_t\bar t (\cos\alpha+i\gamma_5\sin\alpha)t h$. The pure CP-even (CP-odd) coupling can be recovered by setting $\cos\alpha=1$ ($\cos\alpha=0$). Samples of $t\bar t h$($h\rightarrow b\bar{b}$) events were generated at the LHC for $\sqrt{s}=13$~\UTeV, with {\sc MadGraph5\_aMC@NLO}~\cite{Alwall:2014hca}, for several mixing angles, using the \texttt{HC\_NLO\_X0} model~\cite{Artoisenet:2013puc}. All relevant SM background processes were also generated using {\sc MadGraph5\_aMC@NLO}. The analyses of the $t\bar t h$ ($h\rightarrow b\bar b$) events were carried out in the di-leptonic and semi-leptonic decay channels of the $t\bar t$ system. Delphes~\cite{deFavereau:2013fsa} was used for a parametrised detector simulation and both analyses used kinematic fits to fully reconstruct the $t\bar t h$ system. The results were extrapolated, as a function of luminosity, up to the full luminosity expected at the HL-LHC (3000~fb$^{-1}$).

Figure \ref{dilep} (\ref{semilep}) shows the expected CL, assuming the SM, for exclusion of the pure CP-odd scenario, as a function of the integrated luminosity, using the di-leptonic (semi-leptonic) analysis only. 
The CL were obtained using a di-leptonic (semi-leptonic) signal-enriched region containing events with at least 4 jets and 3 $b$-tagged jets (with 6 to 8 jets and 3 or 4 $b$-tagged jets) in which a likelihood ratio was computed from binned distributions of various discriminant observables~\cite{AmorDosSantos:2017ayi,Azevedo:2017qiz,Demartin:2014fia}. 
Only statistical uncertainties were considered. Figure \ref{combination} shows CL obtained from the combination of different observables in each channel i.e., $\Delta\eta(\ell^+,\ell^-)$, $\Delta\phi(t,\bar t)$ and $\sin(\theta^{t\bar tH}_{t})\sin(\theta^{H}_{W^+})$ in the di-leptonic channel and, $b_4$ and $\sin(\theta^{t\bar tH}_{\bar t})\sin(\theta^{H}_{b_H})$ in the semi-leptonic channel. The combination of the two channels is also shown for comparison. The observables were treated as uncorrelated. Figure \ref{cosaScan} shows a comparison between the CL obtained in the di-leptonic analysis alone, for the exclusion of several values of $\cos\alpha$ (between 0 and 1), taking $\Delta\eta(\ell^+,\ell^-)$ as discriminant variable.

The main conclusions of these studies can be summarised in what follows: i) many angular observables (Fig.~\ref{dilep} and Fig.~\ref{semilep}) are available with the potential of discriminating between different mixing-angles ($\cos\alpha$) in the top quark Yukawa coupling; ii) the sensitivity of the semi-leptonic final state of $t\bar{t}h$~($h\rightarrow b\bar{b}$) is higher than the di-leptonic channel alone, which requires roughly five times more luminosity for the same confidence level (Fig.~\ref{combination}); iii) the combination of the semi-leptonic and di-leptonic channels improves visibly the sensitivity with respect to the di-leptonic channel, providing a powerful test of the top quark-Higgs interactions in the fermionic sector.

\asubsubsection{$\tau\bar{\tau} h$}

The most promising direct probe of CP violation in fermionic Higgs
decays is the $\tau^+ \tau^-$ decay channel, which benefits from a
relatively large $\tau$ Yukawa giving a SM branching fraction of
$6.3\%$. Measuring the CP violating phase in the tau Yukawa requires a measurement of the linear polarisations of both $\tau$ leptons and and the azimuthal angle between them. This can be done by analysing tau substructure, namely the angular distribution of the various components of the tau decay products.

The main $\tau$ decay modes studied include $\tau^\pm \to
\rho^\pm (770) \nu$, $\rho^\pm \to \pi^\pm \pi^0$~\cite{Bower:2002zx,
  Desch:2003mw, Desch:2003rw, Harnik:2013aja, Askew:2015mda,
  Jozefowicz:2016kvz} and $\tau^\pm \to \pi^\pm
\nu$~\cite{Berge:2008wi, Berge:2008dr, Berge:2011ij}.  Assuming CPT
symmetry, collider observables for CP violation must be built from
differential distributions based on triple products of three-vectors.
In the first case, $h \to \pi^\pm \pi^0 \pi^\mp \pi^0 \nu \nu$,
angular distributions built only from the outgoing charged and neutral
pions are used to determine the CP properties of the initial $\tau$
Yukawa coupling.  In the second case, $h \to \pi^\pm \pi^\mp \nu \nu$,
there are not enough reconstructible independent momenta to construct an observable sensitive to CP violation, requiring additional kinematic information such as the $\tau$ decay impact parameter.

In the kinematic limit when each outgoing neutrino is taken to be
collinear with its corresponding reconstructed $\rho^\pm$ meson, the
acoplanarity angle, denoted $\Phi$, between the two decay planes
spanned by the $\rho^\pm \to \pi^\pm \pi^0$ decay products is exactly
analogous to the familiar acoplanarity angle from $h \to 4 \ell$
CP-property studies.  Hence, by measuring the $\tau$ decay products in
the single-prong final state, suppressing the irreudicible $Z \to
\tau^+ \tau^-$ and reducible QCD backgrounds, and reconstructing the
acoplanarity angle of $\rho^+$ vs.~$\rho^-$, the differential
distribution in $\Phi$ gives a sinusoidal shape whose maxima and
minima correspond to the CP-phase in the $\tau$ Yukawa coupling.  

An optimal observable using the collinear approximation was derived in~\cite{Harnik:2013aja}. Assuming 70\% efficiency for tagging hadronic $\tau$ final states, and
neglecting detector effects, the estimated sensitivity for the
CP-violating phase of the $\tau$ Yukawa coupling using 3 ab$^{-1}$ at
the HL-LHC is 8.0$^\circ$.  A more sophisticated
analysis~\cite{Askew:2015mda} found that detector resolution effects
on the missing transverse energy distribution degrade the expected
sensitivity considerably, and as such, about 1 ab$^{-1}$ is required
to distinguish a pure scalar coupling (CP phase is zero) from a pure
pseudoscalar coupling (CP phase is $\pi/2$).

A study on the prospect for the measurement of the $\mathcal{CP}$ state of the Higgs boson in its couplings to $\tau$ leptons has been conducted considering 3000~fb$^{-1}$ of $pp$ collision data at $\sqrt{s}=14$ \UTeV collected with the ATLAS detector at the HL-LHC~\cite{ATL-PHYS-PUB-2019-008}. This study investigates the sensitivity for such measurement utilising the $H\to\tau\tau$ decays where both $\tau$ leptons decay via $\tau^{\pm}\to\rho^{\pm}\nu\tau\to\pi^{\pm}\pi^0\nu_\tau$. These decays have a large branching ratio (25\%) and offer a simple way to construct an observable that is sensitive to the $\mathcal{CP}$-violating phase $\phi_\tau$. Such observable is the acoplanarity angle $\varphi ^*_{\mathcal{CP}}$~\cite{Desch:2003rw}, that is the angle between the $\tau$ decay planes that are spanned by each pion pair in the frame where the vectorial sum of the pion momenta is vanishing. The distribution of $\varphi ^*_{\mathcal{CP}}$ is expected to have sinusoidal shape with a phase that varies linearly with $\phi_\tau$. However, in order to observe such modulation, events need to be categorised based on the sign of the product of the asymmetries $y_{\pm} = (E_{\pm}-E_0)/(E_{\pm}+E_0)$ of the energies of the pions from each $\tau$ decay. In fact, events with opposite $y_{+}y_{-}$ have modulations shifted by $\pi$ in phase.

This study is based on the measurement of the $H\tau\tau$ coupling with 36.1~$\mathrm{fb}^{-1}$ of $\sqrt{s}=13$ \UTeV data \cite{Aaboud:2018pen} and on an extrapolation of this measurement to the HL-LHC scenario~\cite{ATL-PHYS-PUB-2018-054}. The same selections on the hadronically decaying $\tau$ leptons and on the di-$\tau$ events are applied as in the 13~\UTeV measurement. This assumes that at the HL-LHC the online and offline selections on the hadronically decaying $\tau$ leptons will be similar to those applied during the LHC Run-2~\cite{Collaboration:2285584}. Optimisations of the event selection for the higher centre-of-mass energy, the exclusive $\tau$ decay of interest, and the detector effects impacting on the resolution of the observable $\varphi ^*_{\mathcal{CP}}$ have not been investigated, but are expected to improve the sensitivity. 

The performance of the upgraded ATLAS detector~\cite{Collaboration:2285585} has been evaluated with Gaussian smearings of the particle momenta simulated at particle level.
The precision in measuring the direction of the $\pi^{0}$ four-vector is taken to be the same as in Ref.~\cite{Aad:2015unr}. HL-LHC simulation studies show that the $\pi^{0}$ $p_{T}$ and directional resolutions in correctly reconstructed decays, do not degrade in pileup of $<\mu>\sim200$ by more than a few percent compared to Run-1. However, scenarios with worse resolutions are also considered, since this resolution is expected to have a leading effect on the precision of the $\varphi ^*_{\mathcal{CP}}$ reconstruction.
The uncertainties include only the statistical uncertainties of the expected data sample.

This study shows that even with $\pi^{0}$ resolutions 1.5 times as large as in the LHC Run-2, the pseudoscalar hypothesis could be excluded at $2\sigma$ analysing only the $H\to\tau\tau$ decays where both $\tau$ leptons decay via $\tau^{\pm}\to\rho^{\pm}\nu\tau\to\pi^{\pm}\pi^0\nu_\tau$ (about $6\%$ of the $H\to\tau\tau$ events). The $\mathcal{CP}$-violating phase could be measured at 68.3\% confidence level within $\pm 18^{\circ}$ and $\pm 33^{\circ}$ assuming the nominal or a twice as large $\pi^{0}$ resolution, respectively. Higher sensitivities are expected when more $\tau$ decays are included.

At the HE-LHC, the increased signal cross section for Higgs production
is counterbalanced by the increased background rates, and so the main
expectation is that improvements in sensitivity will be driven by the
increased luminosity and more optimised experimental methodology.
Rescaling with the appropriate luminosity factors, the optimistic
sensitivity to the $\tau$ Yukawa phase from acoplanarity studies is
4-5$^\circ$, while the more conservative estimate is roughly an order
of magnitude worse.

\newpage

\asection{Global effective field theory fits}\label{sec8:globalfit}
\label{sec8}

The absence to date of conclusive signals of physics beyond the Standard Model (SM) at the Large Hadron Collider (LHC) suggests
that there might be a separation of scales between the SM, and whatever may lie beyond it at some higher energy.
This motivates using the Standard Model Effective Field Theory (SMEFT) as a tool to search indirectly for new physics in LHC data,
given its (near) model-independence, capacity for systematic improvement, and ability to exploit simultaneously multiple datasets. For more motivation and details, see section~\ref{sec:eftintro}.
Beside the high-energy effects discussed in section \ref{sec4}, the HL and HE-LHC have a great potential in this context via the global fit to electroweak precision (EWPO) and Higgs
data, thanks to the higher precision they will reach both in the
measurement of some of the crucial input parameters of global EW fits
(e.g. $M_W$, $m_t$, $M_H$, and
$\sin^2\theta_{\mathrm{eff}}^{\mathrm{lept}}$) and the measurement
of Higgs-boson total rates.

This section focuses  on new physics effects, parametrised by extending
the SM Lagrangian via gauge-invariant dimension-six operators in \eq{EFTLAG} and estimates the reach on the Wilson coefficients provided by a global analysis, that is, an analysis with multiple observables and coefficients.
A global analysis of constraints on the Wilson coefficients of the SMEFT is of critical importance when more than a few coefficients are allowed to be non-zero,
since many SMEFT operators contribute to multiple observables, so that different classes of measurement should not be analysed in isolation.
The importance of this feature increases as
measurements from the LHC compete in precision with previous generation precision experiments.

Section~\ref{sec8:smeft} proposes a global fit  based on the ATLAS and CMS signal strength extrapolations from section~\ref{sec2:exp_combination} and on the use of Simplified Template Cross Sections (STXS) and estimates of the $WW$ production rate. Section~\ref{sec8:universalewpo} focuses on universal theories and exploits, instead of STXS, the information obtained in the analysis of high-energy observables in section~\ref{sec4}; moreover it includes projections of electroweak precision observables in the context of HL and HE LHC. Finally section~\ref{sec8:cubicetc} puts emphasis on the impact of a global fit on measurements of the Higgs cubic self-coupling at the HE LHC.

\label{sec8:intro}

\asubsection[J. Ellis, C.W. Murphy, V. Sanz, T. You]{Prospective SMEFT Constraints from HL- and HE-LHC Data}
\label{sec8:smeft}

In this note, after reviewing the SMEFT framework and our previous results~\cite{Ellis:2018gqa,Ellis:2014dva, Ellis:2014jta, Murphy:2017omb}, we present projections for the
prospective sensitivities of measurements with the approved High-Luminosity LHC (HL-LHC) project and the proposed High-Energy LHC (HE-LHC) project
to Wilson coefficients. 
Our projections are based on ATLAS and CMS estimates of the accuracies with which they
could measure Higgs production rates together with our estimates of the possible accuracies of STXS
measurements, assuming plausible future reductions in theoretical and systematic errors.

We focus on dimension-6 operators, and work to linear order in the Warsaw basis~\cite{Grzadkowski:2010es},
so as to make a consistent EFT expansion to order ${\cal O}(\Lambda^{-2})$. 
We choose $\alpha$, $G_F$, and $M_Z$ as input parameters for our computations, though we note that
the choice of input scheme does not have much impact on the results of a fit to Wilson coefficients if a sufficiently global analysis is performed~\cite{Brivio:2017bnu}.
There are 2499 baryon-number-preserving dimension-6 Wilson coefficients in the SMEFT~\cite{Alonso:2013hga}.
Here we assume a $U(3)^5$ flavor symmetry between the operator coefficients for the five lighter SM fermion fields,
which reduces the number of (real) coefficients to 76.
However only 20 of those parameters are relevant for the di-boson, electroweak precision and Higgs observables that we consider here.

In the Warsaw basis, the 11 operators from table \ref{tab:dim6ops} relevant for di-boson measurements and electroweak precision observables, 
whether through direct contributions or shifts in input parameters, can be written as~\footnote{The operator definitions and normalisations used here differ slightly from those used in the Warsaw basis in Ref.~\cite{Ellis:2018gqa}. For convenience we list here the relations between the Wilson coefficients in the two notations:
\begin{align*}
&\bar{C}_H = \frac{v^2}{\Lambda^2}C_6  \, , \,  \bar{C}_{H\ell}^{(3)} = \frac{v^2}{\Lambda^2} C_{HL}^{(3)}  \, , \, \bar{C}_{H\ell}^{(1)} = \frac{v^2}{\Lambda^2} C_{HL}  \, , \,  \bar{C}_{\ell\ell} = \frac{v^2}{\Lambda^2} C_{LL}  \, , \,  \bar{C}_{HD} = \frac{v^2}{\Lambda^2} C_{HD}  \, , \, \\ 
& \bar{C}_{HWB} = \frac{v^2}{\Lambda^2} g g^\prime C_{WB}  \, , \,  \bar{C}_{He,Hu,Hd} = \frac{v^2}{\Lambda^2} C_{He,Hu,Hd}  \, , \,  \bar{C}_{Hq}^{(3)} = \frac{v^2}{\Lambda^2} C_{HQ}^{(3)} \, , \, \bar{C}_{Hq}^{(1)} = \frac{v^2}{\Lambda^2} C_{HQ}  \, , \,  \\
& \bar{C}_W = \frac{v^2}{\Lambda^2} \frac{g}{3!} C_{3W} \, , \,  \bar{C}_{eH, dH, uH} = \frac{v^2}{\Lambda^2} C_{ye, yd, yu}  \, , \, \bar{C}_{H\square} = \frac{v^2}{\Lambda^2}  \frac{1}{2} C_H \, , \,  \bar{C}_{HW} = \frac{v^2}{\Lambda^2} g^2 C_{WW}  \, , \, \\
& \bar{C}_{HB} = \frac{v^2}{\Lambda^2} {g^\prime}^2 C_{BB}  \, , \, \bar{C}_{HG} = \frac{v^2}{\Lambda^2} g_s^2 C_{GG}  \, , \,  \bar{C}_G = \frac{v^2}{\Lambda^2} \frac{g_s}{3!} C_{3G}  \, .
\end{align*}
}  
{\small
\begin{align}
\label{eq8:L11}
\mathcal{L}_\text{SMEFT}^\text{Warsaw} &\supset \frac{C_{HL}^{(3)}}{\Lambda^2} (H^\dag i\overleftrightarrow{D}^I_\mu H)(\bar\ell \tau^I \gamma^\mu \ell) + \frac{C_{HL}}{\Lambda^2}(H^\dag i\overleftrightarrow{D}_\mu H)(\bar \ell \gamma^\mu \ell) +\frac{C_{LL}}{\Lambda^2}(\bar \ell \gamma_\mu \ell)(\bar \ell \gamma^\mu \ell) \nonumber \\
&+ \frac{C_{HD}}{\Lambda^2}\left|H^\dag D_\mu H\right|^2 + \frac{C_{WB}}{\Lambda^2} g g^\prime H^\dag \tau^I H\, W^I_{\mu\nu} B^{\mu\nu}  \nonumber  \\ 
&+ \frac{C_{He}}{\Lambda^2} (H^\dag i\overleftrightarrow{D}_\mu H)(\bar e \gamma^\mu e) \nonumber + \frac{C_{Hu}}{\Lambda^2} (H^\dag i\overleftrightarrow{D}_\mu H)(\bar u \gamma^\mu u) + \frac{C_{Hd}}{\Lambda^2} (H^\dag i\overleftrightarrow{D}_\mu H)(\bar d \gamma^\mu d) \nonumber \\
&+ \frac{C_{HQ}^{(3)}}{\Lambda^2} (H^\dag i\overleftrightarrow{D}^I_\mu H)(\bar q \tau^I \gamma^\mu q) + \frac{C_{HQ}}{\Lambda^2} (H^\dag i\overleftrightarrow{D}_\mu H)(\bar q \gamma^\mu q) + \frac{C_{3W}}{\Lambda^2} \frac{g}{3!} \epsilon^{IJK} W_\mu^{I\nu} W_\nu^{J\rho} W_\rho^{K\mu} \, ,
\end{align}
}
where Hermitian conjugate operators are implicit.
The flavor indices are trivial, except for the four-lepton operator, $C_{LL} = C_{\substack{LL \\ e\mu\mu e}} = C_{\substack{LL \\ \mu ee\mu}}$~\cite{Cirigliano:2009wk}, 
and are also left implicit.
There are an additional nine operators that contribute to Higgs measurements, 
{\small
\begin{align}
\label{eq8:L9}
\mathcal{L}_\text{SMEFT}^\text{Warsaw} &\supset  \frac{C_{y_e}}{\Lambda^2} y_e (H^\dag H)(\bar \ell e H) + \frac{C_{y_d}}{\Lambda^2} y_d (H^\dag H)(\bar q d H) + \frac{C_{y_u}}{\Lambda^2} y_u (H^\dag H)(\bar q u \widetilde H) \nonumber \\
&+ \frac{C_{3G}}{\Lambda^2} \frac{g_s}{3!} f^{ABC} G_\mu^{A\nu} G_\nu^{B\rho} G_\rho^{C\mu}  + \frac{C_{H}}{\Lambda^2} \frac{1}{2}\left(\partial^\mu |H|^2\right)^2  + \frac{C_{uG}}{\Lambda^2} y_u (\bar q \sigma^{\mu\nu} T^A u) \widetilde H \, G_{\mu\nu}^A   \nonumber \\
&+ \frac{C_{WW}}{\Lambda^2} g^2 H^\dag H\, W^I_{\mu\nu} W^{I\mu\nu} + \frac{C_{BB}}{\Lambda^2} {g^\prime}^2 H^\dag H\, B_{\mu\nu} B^{\mu\nu} + \frac{C_{GG}}{\Lambda^2} g_s^2 H^\dag H\, G^A_{\mu\nu} G^{A\mu\nu} \, .
\end{align}}
The explicit appearance of the Yukawa matrices in Eq.~\eqref{eq8:L9} is necessary to preserve formally the $U(3)^5$ flavor symmetry.
We neglect here $\mathcal{O}_6$ which is discussed in detail in sections \ref{sec3} and \ref{sec8:cubicetc}.
All of the operators in Eq.~\eqref{eq8:L11} except $\mathcal{O}_{3W}$ affect Higgs measurements at leading order, and
$\mathcal{O}_{3W}$ contributes to Higgs processes at next-to-leading order~\cite{Alonso:2013hga, Hartmann:2015oia, Hartmann:2015aia, Dedes:2018seb, Dawson:2018liq}.
We note also that Higgs production in association with a top-quark pair probes additional terms in the 
SMEFT~\cite{Maltoni:2016yxb, AguilarSaavedra:2018nen, Ellis:2018gqa} that do not contribute to the other observables we consider.
The only one we consider explicitly is $C_{uG}$, which makes the largest contribution to $t \bar{t} h$ production~\cite{Ellis:2018gqa}~\footnote{For 
an alternative analysis including all the operators, see~\cite{Murphy:2017omb}.}.

In our global fit to the current data we have used the predictions for electroweak precision observables and $WW$ scattering at LEP~2 in the 
Warsaw basis from Refs.~\cite{Berthier:2016tkq, Brivio:2017vri}, and predictions for LHC observables are made using $\mathtt{SMEFTsim}$~\cite{Brivio:2017btx}.
The following data are used in our global fit:

\begin{itemize}
\item \textit{Pre-LHC data}:  
We use 11 $Z$-pole observables from LEP 1 and one from SLC, as given in Ref.~\cite{ALEPH:2005ab}, as well as the $W$ mass measurement from the 
Tevatron~\cite{Aaltonen:2013iut}.
In addition, we use all the LEP~2 data for the processes $e^+ e^- \to W^+ W^- \to 4f$,
as compiled in Ref.~\cite{Berthier:2016tkq}, the original experimental papers being Refs.~\cite{Heister:2004wr, Achard:2004zw, Abbiendi:2007rs, Schael:2013ita}. 
These measurements probe eleven directions in the SMEFT, which can be mapped to the operators in Eq.~\eqref{eq8:L11}.

\item \textit{LHC Run 1 data}: 
We use all the 20 Higgs signal strengths from Table 8 of Ref.~\cite{Khachatryan:2016vau}.
We also use the ATLAS and CMS combination for the $h \to \mu^+ \mu^-$ signal strength~\cite{Khachatryan:2016vau}, and the ATLAS $h \to Z \gamma$ signal strength~\cite{Aad:2015gba}. 
We also include the $W$ mass measurement from ATLAS~\cite{Aaboud:2017svj}.

\item \textit{LHC Run 2 data}: 
We use 25 measurements from CMS~\cite{Sirunyan:2017dgc, Sirunyan:2017elk, Sirunyan:2018mvw, Sirunyan:2018shy, CMS-PAS-HIG-16-042, Sirunyan:2018ouh, Sirunyan:2017exp, Sirunyan:2017khh}, and 23 measurements from ATLAS~\cite{Aaboud:2017ojs, Aaboud:2017xsd, Aaboud:2017rss, Aaboud:2017jvq, ATLAS-CONF-2018-004, ATLAS-CONF-2017-047, ATLAS-CONF-2016-112}, including experimental correlations whenever possible. 
In addition, we include one ATLAS measurement at 13 \UTeV of the differential cross section for $p p \to W^+ W^- \to e^{\pm} \nu \mu^{\mp} \nu$
that requires $p_T > 120$~\UGeV for the leading lepton~\cite{Aaboud:2017qkn}.
\end{itemize}

\vspace{5mm}
\noindent
{\bf Selection of Current Results.} 
\label{sec8:curr}
In this Section we summarise the main result of the global fit to current data in Ref.~\cite{Ellis:2018gqa}. Fig.~\ref{fig8:HEPfitvsEMSY} summarises the sensitivities to operator coefficients in the Warsaw basis.
It shows the 95\% CL bounds in \UTeV on the Wilson coefficients, as obtained in~\cite{Ellis:2018gqa} from marginalised (orange) and individual (green) fits to the 20 dimension-6 operators 
entering in electroweak precision tests, di-boson and Higgs measurements at LEP, SLC, Tevatron, and LHC Runs 1 and 2.
Also shown, in blue, are the analogous results of the individual fits of the HEPfit Collaboration~\cite{Ciuchini:2013pca, deBlas:2016ojx}~\footnote{See 
also~\cite{luca:talk, otto:talk}. For a recent fit in another basis see Ref.~\cite{Alves:2018nof}, and
for a recent fit in the nonlinear Effective Theory see Ref.~\cite{deBlas:2018tjm}.}. 
We note that the degree of agreement in the individual operator fits is generally good.

\begin{figure}
  \centering
\includegraphics[width=0.8\textwidth]{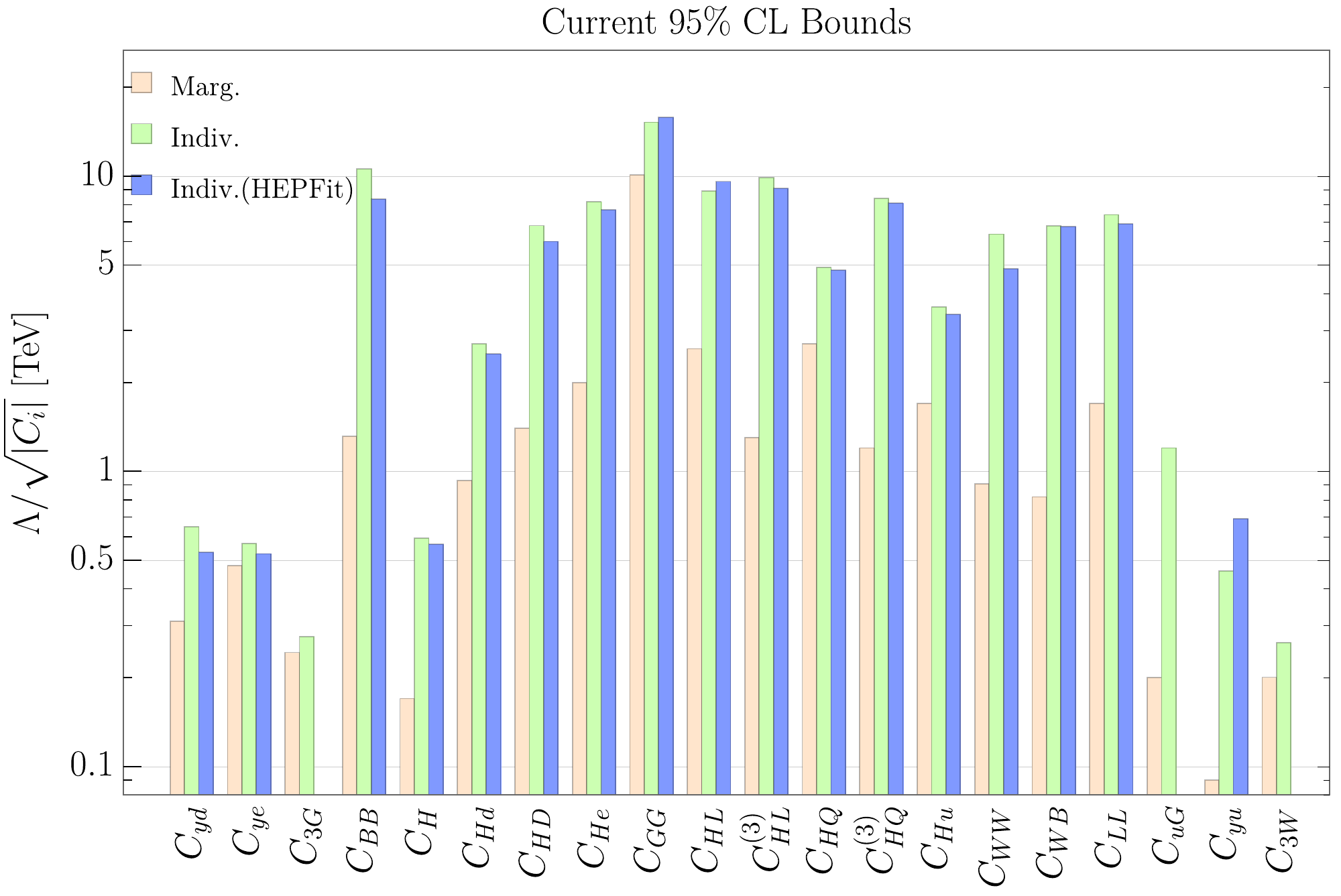}
\vspace{-0.2cm}
 \caption{\it Current constraints on dimension-6 operators. The orange and green bars are the marginalised and individual limits from Ref.~\cite{Ellis:2018gqa}, respectively. 
 The analogous results of the HEPfit Collaboration~\cite{Ciuchini:2013pca, deBlas:2016ojx} for the case where only one operator is switched on at a time are shown in blue.}
   \label{fig8:HEPfitvsEMSY}
\end{figure} 

Table~\ref{tab8:FIpM} gives the Fisher information contained in a given dataset per number of measurements in that dataset,
for each coefficient in the case where one operator is switched on at a time.
The datasets are categorised as in Ref.~\cite{Ellis:2018gqa}, where a
cross indicates no (current) sensitivity.
This Table is the per-measurement analogue of Table 5 of Ref.~\cite{Ellis:2018gqa}, and
shows that LHC di-boson measurements are more powerful than LEP 2 di-boson measurements for constraining triple gauge couplings (TGCs).
In the Warsaw basis the three TGCs correspond to $C_{WB}$, $C_{3W}$, and a linear combination of $C_{HD}$, $C_{HL}^{(3)}$, $C_{WB}$, and $C_{LL}$.
We note, however, that $Z$-pole and Higgs-pole measurements are more constraining for all but $C_{3W}$. 

\begin{table}
\centering
 \begin{tabular}{|c|c|c|c|c|c|c|} \hline
Coefficient & $Z$-pole + $m_W$ & $WW$ at LEP2 & Higgs Run1 & Higgs Run2 & LHC $WW$ high-$p_T$  \\  \hline \hline
 $C_{yd}$ & $\times$ & $\times$ & 10 & 8.1 & $\times$  \\
 $C_{ye}$ & $\times$ & $\times$ & 2.9 & 1.3 & $\times$ \\
 $C_{3G}$ & $\times$ & $\times$ & 0.5 & 9.1 & $\times$  \\
 $C_{BB}$ & $\times$ & $\times$ & $9.9 \cdot 10^5$ & $2.0 \cdot 10^6$ & $\times$  \\
 $C_{H}$ & $\times$ & $\times$ & 8.1 & 15 & 0.1  \\
$C_{Hd}$ & $7.4 \cdot 10^3$ & $\times$ & 2.0 & 1.5 & 9.8  \\
 $C_{HD}$ & $4.3 \cdot 10^5$ & 51 & 4.6 & 4.5 & $5.5 \cdot 10^2$  \\
$C_{He}$ & $6.5 \cdot 10^5$ & 14 & $1.1 \cdot 10^{-2}$ & $3.7 \cdot 10^{-2}$ & $\times$  \\
 $C_{GG}$ & $\times$ & $\times$ & $9.8 \cdot 10^5$ & $8.6 \cdot 10^5$ & $1.5 \cdot 10^4$  \\
$C_{HL}$ & $1.1 \cdot 10^6$ & 51 & $1.1 \cdot 10^{-2}$ & $3.6 \cdot 10^{-2}$ & $4.6 \cdot 10^{-3}$  \\
 $C_{HL}^{(3)}$ & $1.7 \cdot 10^6$ & $1.3 \cdot 10^3$ & 51 & 49 & $3.5 \cdot 10^3$  \\
 $C_{HQ}$ & $6.4 \cdot 10^4$ & $\times$ & 2.3 & 1.0 & 37  \\
$C_{HQ}^{(3)}$ & $4.9 \cdot 10^5$ & $9.1 \cdot 10^2$ & $5.9 \cdot 10^2$ & $3.3 \cdot 10^2$ & $5.0 \cdot 10^3$  \\
 $C_{Hu}$ & $1.4 \cdot 10^4$ & $\times$ & 18 & 12 & 83  \\
$C_{WW}$ & $\times$ & $\times$ & $9.1 \cdot 10^4$ & $1.8 \cdot 10^5$ & $7.0 \cdot 10^{-3}$ \\
$C_{WB}$ & $3.3 \cdot 10^6$ & $1.9 \cdot 10^2$ & $3.0 \cdot 10^5$ & $5.7 \cdot 10^5$ & $2.2 \cdot 10^3$ \\
 $C_{LL}$ & $5.5 \cdot 10^5$ & $3.3 \cdot 10^2$ & 16 & 21 & $6.0 \cdot 10^2$ \\
$C_{uG}$ & $\times$ & $\times$ & 18 & 97 & $\times$  \\
 $C_{yu}$ & $\times$ & $\times$ & 0.4 & 1.8 & $\times$ \\
 $C_{3W}$ & $\times$ & 6.7 & $\times$ & $\times$ & 19 \\ \hline
\end{tabular}
\caption{\it Impacts of different sets of measurements on the fit to individual Wilson coefficients in the Warsaw basis as measured by the Fisher information contained in a given dataset per number of measurements in that dataset for each coefficient. A cross indicates no (current) sensitivity.}
\label{tab8:FIpM}
\end{table}
\vspace{5mm}

\noindent
{\bf Future Projections.}\label{sec8:proj}
We use the same framework as described in Ref.~\cite{Ellis:2018gqa} to project how the sensitivities to the Wilson coefficients of the SMEFT will change at HL- and HE-LHC. Our projection strategy is as follows:
we leave all pre-LHC, and LHC Run-1 measurements unchanged; we adopt the CMS and ATLAS WG2 recommendations for the HL- and HE-LHC sensitivities, using the HL-LHC S2 scenario for the experimental projections on the signal strength uncertainties and their associated correlation matrix (provided separately by ATLAS and CMS);
and the remaining statistical uncertainties at HL- and HE-LHC are assumed to scale naively with the integrated luminosities and cross sections:
\begin{equation}
\frac{\delta\mathcal{O}_{\text{HL}, i}}{\delta\mathcal{O}_{\text{today}, i}} = \sqrt{\frac{L_{\text{today}, i}}{L_{\text{HL}}}} \; , \;
\frac{\delta\mathcal{O}_{\text{HE}, i}}{\delta\mathcal{O}_{\text{today}, i}} = \sqrt{\frac{\sigma_{13, i}}{\sigma_{27, i}} \frac{L_{\text{today}, i}}{L_{\text{HE}}}}  . \nonumber
\end{equation}
For almost all the measurements by ATLAS(CMS) $L_{\text{today}, i} = 36.1(35.9)~\text{fb}^{-1}$,
and we use the benchmark luminosities $L_{\text{HL}} = 3~\text{ab}^{-1}$ and $L_{\text{HE}} = 15~\text{ab}^{-1}$ for all the measurements in the respective HL- and HE-LHC extrapolations.
The cross sections $\sigma_{13, i}$ and $\sigma_{27, i}$ refer to the SM cross sections in the signal region for a given measurement $i$ at 13 and 27 \UTeV, respectively. At HE-LHC the S2 systematics are reduced by half. For the STXS and $WW$ measurements used in Ref.~\cite{Ellis:2018gqa}, since no official projections have been made yet, we extrapolate the statistical part as described above and treat the systematics as unchanged for HL-LHC and halve them for HE-LHC. The correlations between experimental measurements are assumed to be unchanged.

We stress that both our projection scenarios are pessimistic in the sense that they do not take into account the additional channels~\cite{gilbert:talk}, finer binning~\cite{deFlorian:2016spz, Hays:2290628, ATLAS-CONF-2017-047} and extension of the STXS method to larger kinematic regions that will become available as more data are collected.
Furthermore, as suggested by Table~\ref{tab8:FIpM}, our projections under-utilise LHC di-boson scattering measurements~\cite{Azatov:2017kzw, Franceschini:2017xkh, Dawson:2018dcd}. 

\begin{figure}
  \centering
 \includegraphics[width=0.8\textwidth]{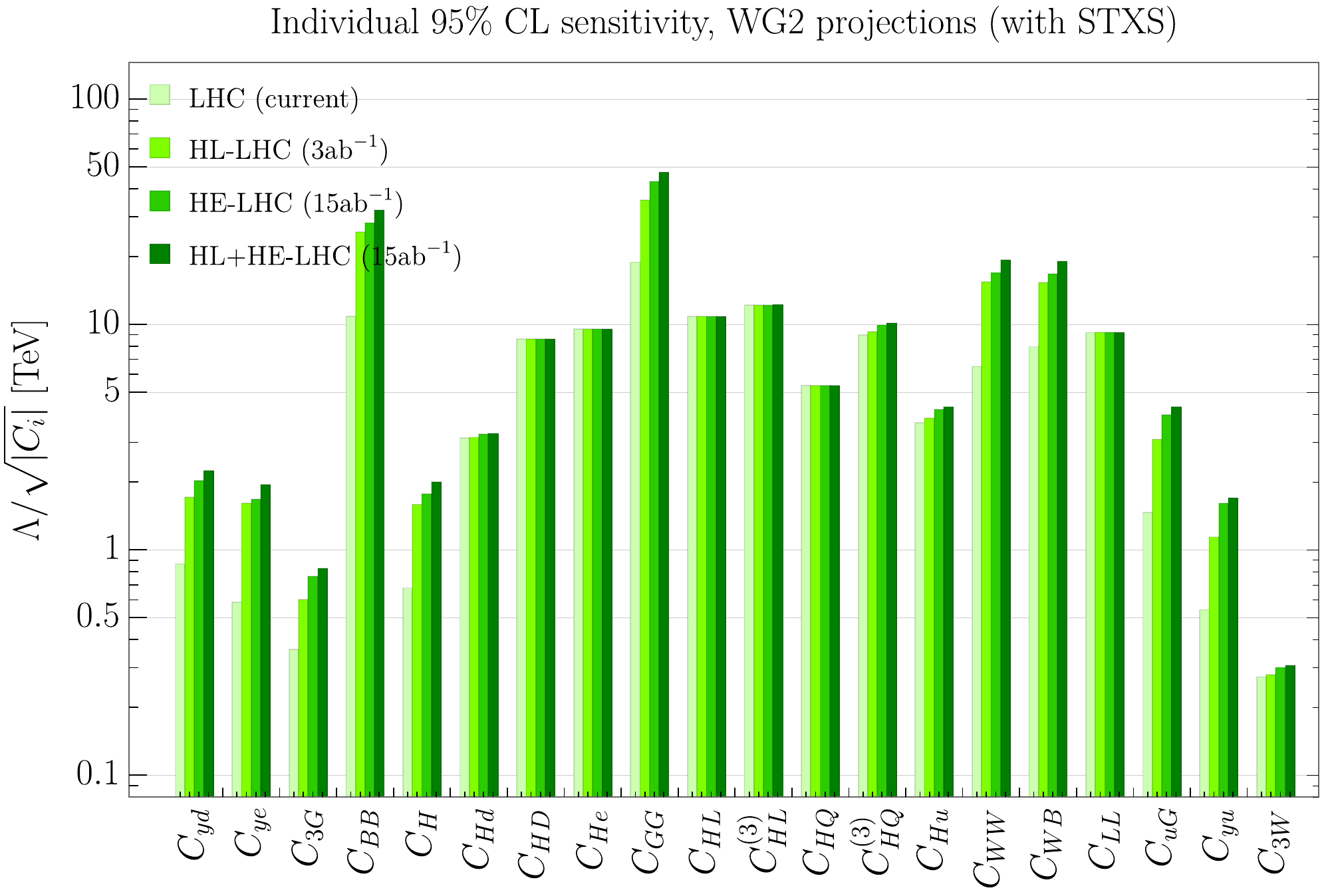}
 \vspace{-0.2cm}
 \caption{\it Individual 95\% CL projected sensitivities for LHC, HL-LHC, HE-LHC, and combined HL/HE-LHC in increasingly darker shades of green. The vertical axis gives the reach to the scale of new physics divided by the dimensionless Wilson coefficient, in units of \UTeV. }
   \label{fig:indiv}
\end{figure} 

\begin{figure}
  \centering
 \includegraphics[width=0.8\textwidth]{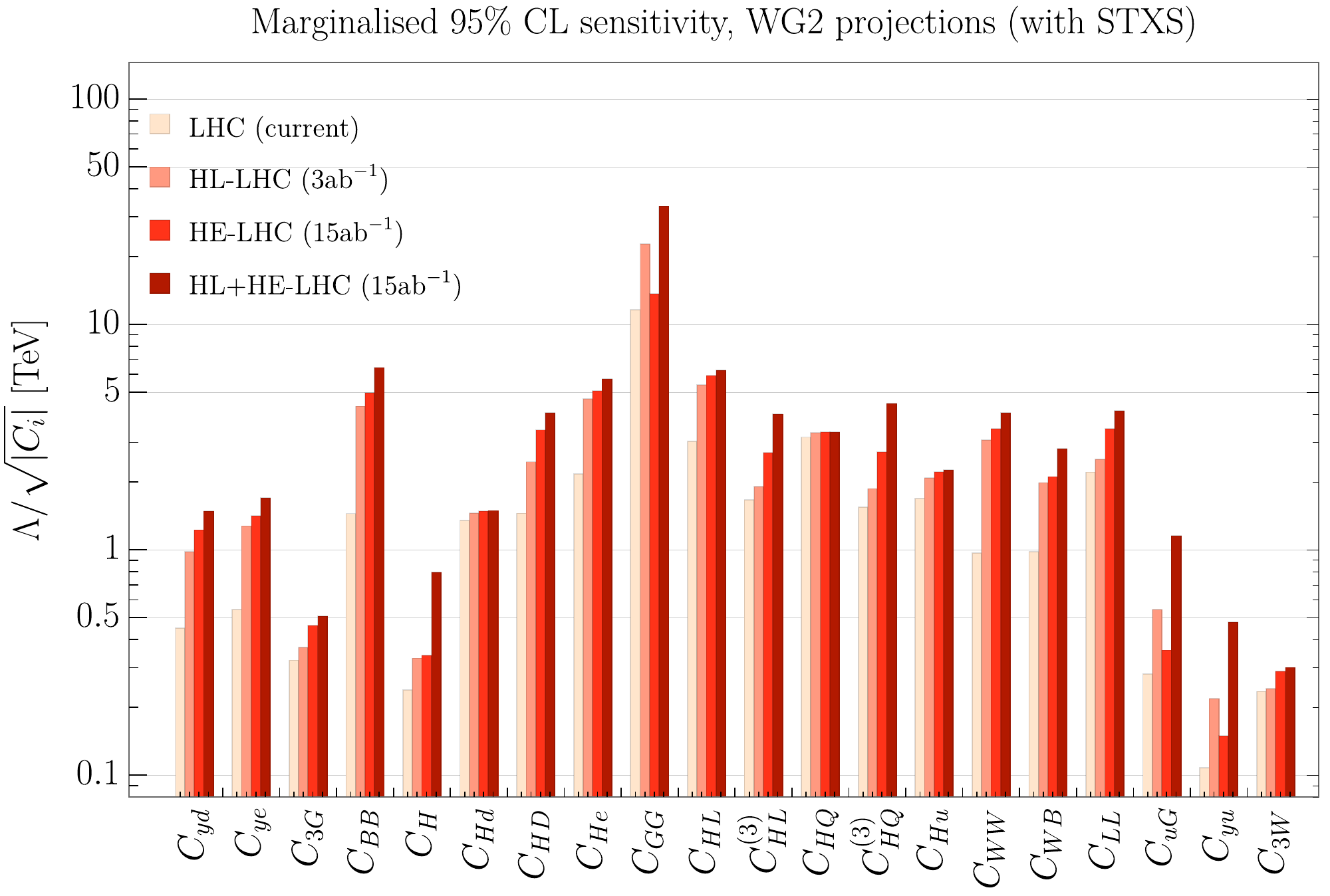}
 \vspace{-0.2cm}
 \caption{\it  Marginalised 95\% CL projected sensitivities for LHC, HL-LHC, HE-LHC, and combined HL/HE-LHC in increasingly darker shades of red. The vertical axis gives the reach to the scale of new physics divided by the dimensionless Wilson coefficient, in units of \UTeV.}
   \label{fig:marg}
\end{figure} 

The results of our projections are shown in Figures~\ref{fig:indiv} and~\ref{fig:marg} for individual and marginalised 95\% CL sensitivities respectively. The vertical axis is the operator scale in units of \UTeV divided by the square root of the dimensionless Wilson coefficient. Increasingly darker shades for the four bars represent the sensitivities of the LHC up to now, HL-LHC with 3 ab$^{-1}$, HE-LHC with 15 ab$^{-1}$, and combining HL- and HE-LHC results (neglecting any correlations between the two). We see that in the individual bounds the sensitivity for each operator increases correspondingly, except for those that are only constrained by electroweak precision tests, which remain unchanged from their current LEP limits. In the marginalised case even the latter operators benefit from an improvement in the limits on other operators. We note that in going from HL- to HE-LHC there is a decrease in the marginalised sensitivity for $C_{GG}, C_{uG}$, and $C_{yu}$, despite improvements in their individual bounds. This is because $C_{uG}$ and $C_{yu}$ enter only in $tth$ production, and the relative increase in their contribution with respect to $C_{GG}$ in going from 13 to 27 \UTeV opens up a relatively flat direction in the parameter space, reducing the sensitivity to all three coefficients. This degeneracy can be broken by measurements at different energies, as shown in the combined HL/HE-LHC fit, or by including measurements involving top quarks.

In general, we see that data from HL- and/or HE-LHC would extend the sensitivity to new physics into the multi-\UTeV range
for most operator coefficients, extending to tens of \UTeV for $C_{GG}$.




\newcommand{\HEPfit}{{\tt HEPfit}}


\asubsection[J. de Blas, M. Ciuchini, E. Franco, S. Mishima, M. Pierini, L. Reina, L. Silvestrini]{Global constraints on universal new physics at the HL/HE-LHC}
\label{sec8:universalewpo}


\label{sec:general-framework}

To systematically study the effects of new physics on EWPO and
Higgs-boson observables we consider a SM effective field theory
 that adds to the Lagrangian of the SM new effective
interactions of the SM fields in the form of higher-dimension ($d>4$)
local operators that preserve the SM gauge symmetry, \eq{EFTLAG}.
While using, e.g., the
complete basis of dimension-six interactions presented in
Ref. \cite{Grzadkowski:2010es} one can test new physics effects in 
a more general way (provided one has enough experimental inputs),
for the purpose of the fit presented in this Section we are interested only in those
new physics effects that arise in the context of the so-called {\it universal theories}~\cite{Barbieri:2004qk,Wells:2015uba}.
In the EFT framework universal theories can be defined such that, via field re-definitions, all new physics effects
can be captured by operators involving SM bosons only. Note that this includes not only theories where the
new particles couple to the SM bosonic sector, but also scenarios where the interactions 
occur via the SM fermionic currents. Therefore, this class of theories automatically 
satisfy minimal flavour violation, so the results of the global fit to Higgs and electroweak data 
are not affected by the strong constraints set by flavour measurements. Furthermore,
we will assume only CP-preserving interactions. 
In particular, we will focus on the following non-redundant set of operators, among those defined in Table~\ref{tab:dim6ops}\footnote{In principle, the physics at hadron colliders also allows to test the universal interactions ${\cal O}_{2G}$ and ${\cal O}_{3G}$. Due to the absence of HL-LHC projections for the relevant processes that can be used to constrain such operators, we do not include them in the global fits presented here.}: 
\begin{equation}
\label{eq:universal}
\{\mathcal{O}_{H}, \mathcal{O}_{H D} ,
\mathcal{O}_{6} ,
\mathcal{O}_{G G} ,
\mathcal{O}_{B B} ,
\mathcal{O}_{W W} ,
\mathcal{O}_{W B} ,
\mathcal{O}_{HB} ,
\mathcal{O}_{HW} ,
\mathcal{O}_{2B} ,
\mathcal{O}_{2W} ,
\mathcal{O}_{3W} ,
\mathcal{O}_{y} \}.
\end{equation}
Of course, we note that the HL-LHC data allows to constrain EFT effects beyond the context of this class of universal new physics, 
e.g. constraining independently Higgs couplings to different types of fermions, or operators modifying the EW interactions
in a non-universal way. Therefore, the results presented in this section are to be understood not as an exhaustive exploration 
of the HL/HE-LHC capabilities, but as the interpretation within a particularly broad and well-motivated class
of scenarios of physic beyond the SM.

The global fit of EWPO and Higgs data is performed using the
\HEPfit~package~\cite{hepfitsite}, a general tool to combine direct
and indirect constraints on the SM and its extensions in any
statistical framework.  The default fit procedure, which we use here,
follows a Bayesian statistical approach and uses BAT (Bayesian
Analysis Toolkit)~\cite{Caldwell:2008fw}. We use flat priors for all
input parameters, and build the likelihood assuming Gaussian
distributions for all experimental measurements. The output of the fit
is therefore given as the posterior distributions for each input
parameter and observable, calculated using a Markov Chain Monte
Carlo method.

For the results in this section we use the SMEFT class in \HEPfit~for 
the calculation of the dimension-6 effects in EWPO and Higgs signal strengths.
The EFT expressions for these physical observables are truncated
consistently with the dimension-6 expansion of the SMEFT Lagrangian, 
retaining only terms of order
$1/\Lambda^2$, i.e.
\begin{equation}
O=O_{\mathrm{SM}} + \sum_i a_i \frac{C_i}{\Lambda^2}\,.
\end{equation}
For the SM prediction of
all EWPO we include all available higher-order corrections, including the latest
theoretical developments in the calculation of radiative corrections
to the EWPO of \cite{Dubovyk:2016aqv,Dubovyk:2018rlg}.
On the other hand,  for the SM predictions of
Higgs production cross sections and decay rates we use the results
quoted in \cite{deFlorian:2016spz} and in the current report. 
The new physics corrections to most
Higgs production cross sections are obtained using {\tt Madgraph},
with our own implementation of the dimension-6 SMEFT Lagrangian in a
{\tt FeynRules} UFO model, except for the corrections to the
gluon-gluon fusion production cross section that is computed
analytically. The corrections to Higgs decay rates are
also computed using {\tt Madgraph}, or analytically following
the calculations presented in the {\tt eHdecay} code.

One of the advantages of \HEPfit~is its modularity, allowing for an
easy implementation of new physics models or additional observables.
Taking advantage of this, we have extended the fits to EWPO plus Higgs 
signal strengths to include several of the studies presented in this report, 
and in particular those presented in the di-Higgs or the High Energy probes sections.
We provide details of the observables in the fits for the HL-LHC or HE-LHC 
scenarios in what follows, before presenting our results.


\subsubsection*{HL-LHC inputs for the fit}

We include both the projected improvements 
in the Higgs signal strength measurements from ATLAS and CMS in the HL-LHC scenario,
as well as also the corresponding projections for the measurements of EWPO.
For the details of the EWPO analysis we refer to the corresponding HL-LHC study 
presented in the report of the activities of the WG1 of this workshop~\cite{Azzi:2650160}.
The HL-LHC Higgs signal strength projections are implemented directly from the values
provided by ATLAS and CMS in the two scenarios for systematic and theory uncertainties
denoted as S1 --- which assumes the same uncertainties as in current data--- and S2 ---where systematics
are improved with the luminosity and theory uncertainties are reduced.
The correlations between the ATLAS and CMS sets of inputs were not available at the time
these fits were performed, and therefore we combined them in an uncorrelated manner. One must take into 
account that such correlations, especially between theory uncertainties, can be sizeable. Therefore,
the results of this uncorrelated combination can be somewhat optimistic.
Finally, estimates for the $H\to Z\gamma$ channel are not available within the set of CMS projections, so we 
assume measurements with the same precision as ATLAS.

These results have been further combined with several of the studies presented in this report. 
  For the HL-LHC studies we include:
  \begin{enumerate}
  {\item The differential distribution in $M_{HH}$ in the $HH\to b\bar{b}\gamma\gamma$ channel presented in Section~\ref{sec:HH_Global_fit}. This was available only for the HL-LHC scenario. No difference in term of the assumption for systematics is applied between the S1 and S2 scenarios.}
  {\item The results from the study of the invariant mass distribution, $M_{ZH}$ in the $ZH, H\to b\bar{b}$ channel in the boosted regime from Section~\ref{sec:ZHWZeft}. Results for S1 scenario assume 5\% systematics while a systematic uncertainty of only 1\% is applied in the S2 case. 
  }
  \end{enumerate}
 Furthermore, we combine these results with those obtained from the high-energy measurements in the di-boson channel presented in:
 \begin{enumerate}
 \setcounter{enumi}{2}
 {\item Section~\ref{sec:WZtrans}: We include the bounds on the operator ${\cal O}_{3W}$ from the exclusive analysis. As in the $ZH$ case, we assume a systematic uncertainty of 5\% (1\%) for the S1 (S2) scenario.}
 {\item The invariant $p_{T,V}$ distributions in $pp\to WZ$ production from the analysis in Section~\ref{WZlong}. The systematics are applied as in the previous observables.}  
 \end{enumerate}
Finally, as we assume in our fit that the new physics only affects ``low-energy'' observables in a universal way, we also include:
\begin{enumerate}
\setcounter{enumi}{4}
{\item The sensitivity study to the $W$ and $Y$ parameters ---which one can map into $C_{2W}$ and $C_{2B}$, respectively--- in Drell-Yan
production from Section~\ref{DY}. Systematics uncertainties are fixed in this case, with no difference between the S1 and S2 fits.}
\end{enumerate}
%


\subsubsection*{HE-LHC inputs for the fit}

The same observables included in the HL-LHC study are also used in the HE-LHC scenario. 
For the case of the Higgs signal strengths, we follow the agreed ATLAS and CMS guidelines for the calculation of
the HE-LHC projections.
Starting from the precisions given in the HL-LHC S2 signal strengths, we scale the statistical uncertainties of the projections
according to the cross section and luminosities of the HE-LHC scenario, i.e.
\begin{equation}
\delta_{\mathrm{stat}} \mu_{\mbox{\tiny HE-LHC}} = \sqrt{\frac{\sigma_{pp\to H}^{\mathrm{14 \UTeV}}\times 3~\!\mathrm{ab}^{-1}}{\sigma_{pp\to H}^{\mathrm{27 \UTeV}}\times 15~\! \mathrm{ab}^{-1}}} \delta_{\mathrm{stat}} \mu_{\mbox{\tiny HL-LHC}} .
\end{equation}
We consider this as the {\it ``Base''} HE-LHC scenario. A more optimistic scenario was also suggested, reducing the 
systematics and theory uncertainties by a factor of 2. While we also include this scenario in our fits ---we denote it by {\it ``Opt.''}---
we must note that it does not come from a thoughtful extrapolation of the possible reduction of uncertainties but, instead, 
should be consider only as an hypothesis.

The study of EWPO observables is kept as in the HL-LHC scenario. Similarly, the studies of the $M_{HH}$ and $M_{ZH}$ differential distributions presented in the corresponding sections of this report where available only for $\sqrt{s}=14$ \UTeV. For the HE-LHC fit we use, instead of the $M_{HH}$ distribution, the results on the Higgs self-coupling from Section~\ref{sec:THanetal}. For  the $M_{ZH}$ distribution we include the same result as for HL-LHC. All the other analyses are available for the HE-LHC scenario.\footnote{The analysis of the Drell-Yan constraints on $WY$ only contains projections at 27 \UTeV for the neutral channel $pp\to \ell^+ \ell^-$. For the charged current $pp\to \ell \nu$ we use same constraints on the $W$ parameter given at 14 \UTeV.} Whenever possible, the sensitivities have been scaled to the
expected luminosity of 15 ab$^{-1}$.


\subsubsection*{The Global EFT fit for Universal new physics}
\label{sec:results}

The main results of the fits to universal new physics in the HL-LHC and HE-LHC are illustrated in Figures~\ref{fig:dim6U_HLLHC} and \ref{fig:dim6U_HELHC}, respectively.
The results are shown as the 95\% probability limits on the {\it new physics interaction scale}, $\Lambda/\sqrt{\vert C_i\vert }$, associated to each operator ${\cal O}_i$. (We also show, in the right axis, the value translated into the sensitivities to the ratios $\vert C_i\vert /\Lambda^2$, which give the linear new physics correction to each observable.)
The limits are compared with those obtained from current data from LEP/SLD, the LHC Runs 1 and 2~\cite{deBlas:2016ojx,deBlas:2016nqo,deBlas:2017wmn,hepfit:2019}, and the LHC sensitivity for $W$ and $Y$ from Ref.~\cite{Farina:2016rws}. We also indicate, with dashed lines, the {\it exclusive} bounds obtained assuming that the new physics only generates one operator at a time. The difference between the global and exclusive limits indicates the presence of large correlations between the dimension-six interactions in the global fit.

As it is apparent, there is a significant improvement in the sensitivity to new physics for many types of interactions. This is particularly evident for those operators entering in the high-energy probes. For instance, the sensitivity to ${\cal O}_{3W}$ effects is largely improved with respect to the bounds obtained from LEP 2 $e^+ e^- \to W^+ W^-$ data. The operators ${\cal O}_{HW,HB}$ also induce effects that grow with the energy in $pp\to ZH$ or $pp\to WZ$ (for ${\cal O}_{HW}$), thus leading to more stringent bounds compared to the fit using current data, which does not include the effect of high-energy observables in those channels. The operators ${\cal O}_{2W,2B}$ are mainly constrained from their effects in Drell-Yan. The improvement in the limits on these operators is more visible in the exclusive bounds, because of the absence of HE-LHC estimates for the Drell-Yan analysis in the charged-current channel (See Section~\ref{DY}).

Finally, it is worth noticing the improvements on some of the operators whose effects in the observables we consider are constant with the energy.  Such improvement is therefore the result of a pure increase in the experimental precision in the measurements of the corresponding observables. This is the case for the interactions ${\cal O}_{GG,WW,BB}$. Their main effect is to generate tree-level contributions to loop-induced Higgs boson observables, e.g. $gg\to H$ or $H\to \gamma\gamma, Z\gamma$, whose precision, especially for the rare decay channels, will be largely increased with more luminosity.

\begin{figure}[h]
\centering
  \hspace{-0.1cm}\includegraphics[width=.8\textwidth]{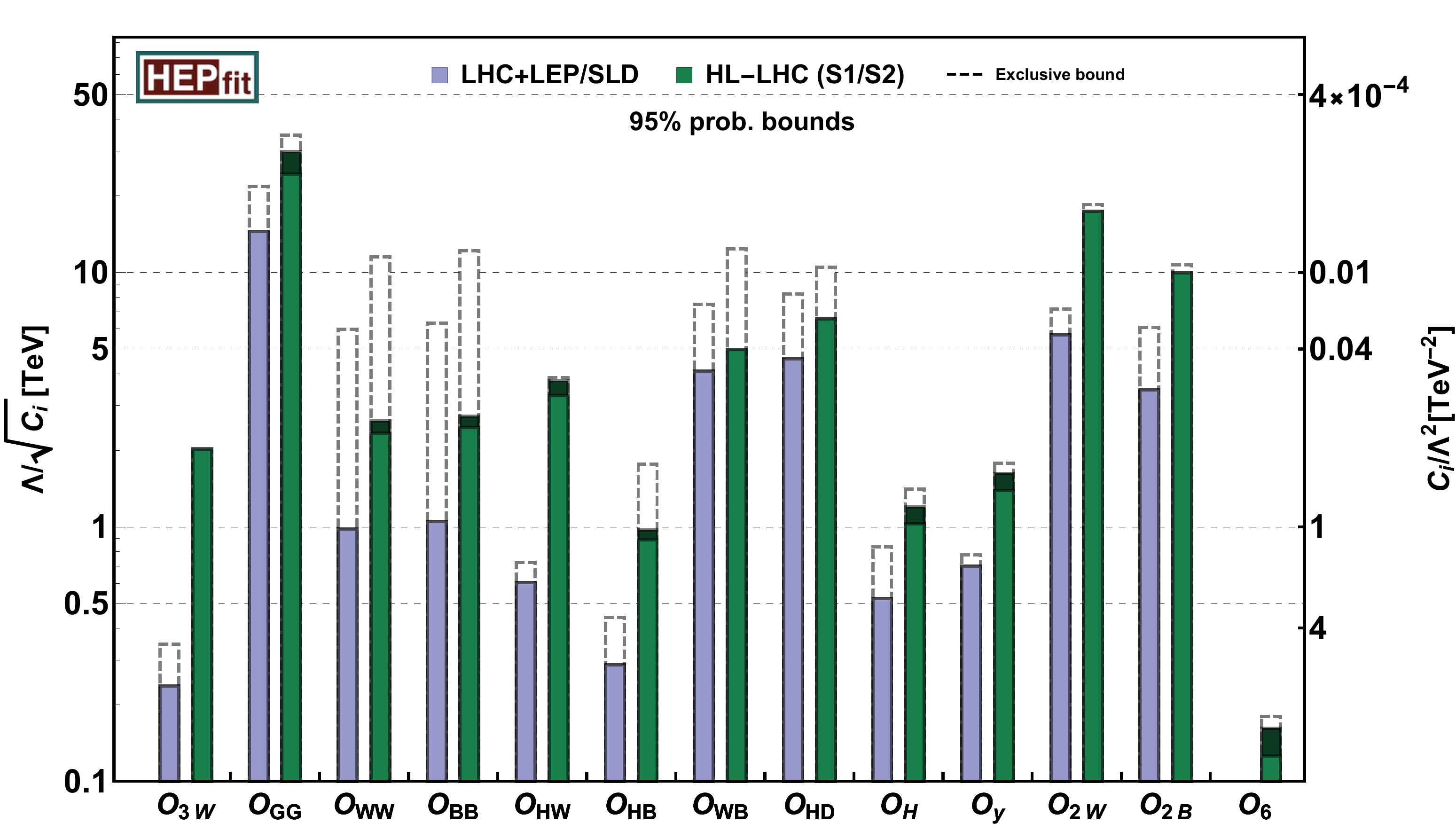}
  \vspace{-0.2cm}
  \caption{$95\%$ probability limits on the new physics interaction scale $\Lambda/\sqrt{\vert C_i\vert}$ [\UTeV] (left axis) and coefficients $\vert C_i\vert /\Lambda^2$ [\UTeV$^{-2}$] (right axis) associated to each dimension-six operator from the global fit to universal new physics at the HL-LHC (green bars, light and dark shades indicate the S1 and S2 assumptions for systematics, respectively). The limits are compared with the ones from current data (in blue), as well as those obtained assuming only one operator at a time (dashed lines).~\label{fig:dim6U_HLLHC}}
\end{figure}

\begin{figure}[h]
\centering
  \hspace{-0.1cm}\includegraphics[width=.8\textwidth]{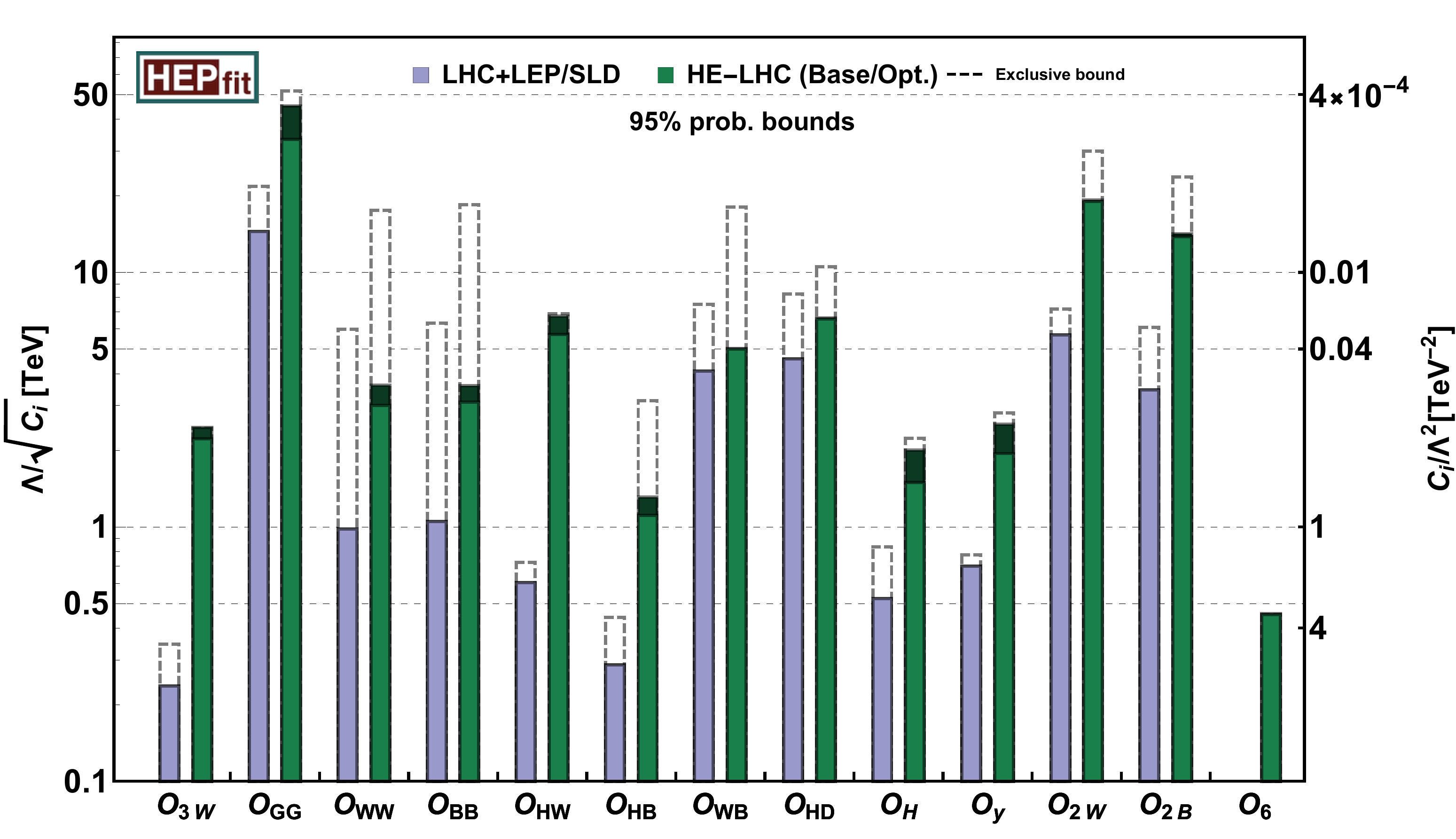}
  \vspace{-0.2cm}
   \caption{$95\%$ probability limits on the new physics interaction scale $\Lambda/\sqrt{\vert C_i\vert }$ [\UTeV] (left axis) and coefficients $\vert C_i\vert/\Lambda^2$ [\UTeV$^{-2}$] (right axis) associated to each dimension-six operator from the global fit to universal new physics at the HE-LHC (green bars, light and dark shades indicate the {\it Base} and {\it optimistic} assumptions for systematics, respectively). The limits are compared with the ones from current data (in blue), as well as those obtained assuming only one operator at a time (dashed lines).~\label{fig:dim6U_HELHC}}
\end{figure}


\FloatBarrier

\asubsection[A. Biek\"otter, D. Gon\c{c}alves, T. Plehn, M. Takeuchi, D. Zerwas]{Global analysis including the Higgs self-coupling}
\label{sec8:cubicetc}



In record time Higgs physics has moved from a spectacular discovery of
a new particle to a systematic and comprehensive study of its
properties~\cite{Dawson:2018dcd}. 
One of the main goals of the physics program of a 27~\UTeV hadron 
collider 
is the measurement of the Higgs self-coupling in Higgs pair
production~\cite{Baur:2003gp,Kling:2016lay,Goncalves:2018yva,Homiller:2018dgu,Barger:2013jfa,Barr:2014sga},
testing for example a possible first-order electroweak
phase transition as an ingredient to
baryogenesis. 
We show that a 27~\UTeV hadron collider will for the first time 
deliver a meaningful measurement of this fundamental physics 
parameter~\cite{Biekotter:2018jzu}. \medskip

We perform a global study of Higgs physics at a 27~\UTeV hadron collider
using the effective dimension-6 Lagrangian~\cite{Brivio:2017vri} 
\begin{align}
\mathcal{L}_\text{eff} 
= & - \frac{\alpha_s }{8 \pi} \frac{f_{GG}}{\Lambda^2} \mathcal{O}_{GG}  
    + \frac{f_{BB}}{\Lambda^2} \mathcal{O}_{BB} 
    + \frac{f_{WW}}{\Lambda^2} \mathcal{O}_{WW} 
    + \frac{f_B}{\Lambda^2} \mathcal{O}_B 
    + \frac{f_W}{\Lambda^2} \mathcal{O}_W 
    + \frac{f_{WWW}}{\Lambda^2} \mathcal{O}_{WWW} \notag \\ 
  &  + \frac{f_{\phi 2}}{\Lambda^2} \mathcal{O}_{\phi 2} 
    + \frac{f_{\phi 3}}{\Lambda^2} \mathcal{O}_{\phi 3} 
    + \frac{f_\tau m_\tau}{v \Lambda^2} \mathcal{O}_{e\phi,33} 
    + \frac{f_b m_b}{v \Lambda^2} \mathcal{O}_{d\phi,33} 
    + \frac{f_t m_t}{v \Lambda^2} \mathcal{O}_{u\phi,33} \notag \\
  & + \text{invisible decays} \; .
\label{eq:ourleff}
\end{align}
with the operators defined in Ref.~\cite{Butter:2016cvz}, 
and specifically
\begin{align}
\mathcal{O}_{\phi 2} = \frac{1}{2} \, \partial^\mu (\phi^\dagger \phi) \partial_\mu (\phi^\dagger \phi)  
\qquad \qquad
\mathcal{O}_{\phi 3} = -\frac{1}{3} \, (\phi^\dagger \phi)^3 \; ,
\label{eq:ope23}
\end{align}
describing a modified Higgs potential. In addition, we include invisible 
Higgs decays as an additional contribution to the total width or, equivalently, 
the invisible branching ratio.
\begin{figure}
\centering
\includegraphics[width=0.65\textwidth]{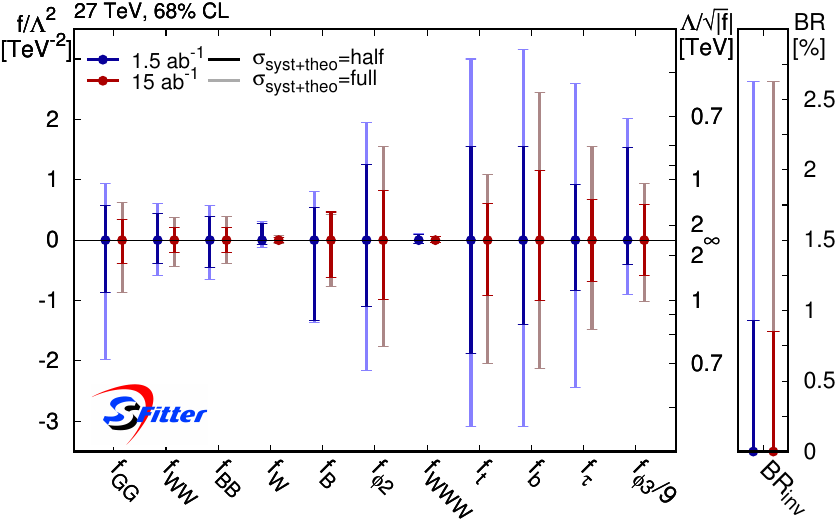}
\caption{Result from the global Higgs analysis in terms of dimension-6
  operators. All limits are shown as profiled over all other Wilson
  coefficients. Figure from Ref.~\cite{Biekotter:2018jzu}.}
\label{fig:d6}
\end{figure}

The self-coupling with its unique relation to the Higgs potential is
not yet included in most global analyses of SM-like Higgs 
couplings~\cite{Ellis:2018gqa,Alves:2018nof,Biekotter:2018rhp}
because of the modest reach of the LHC Run~II.  
However, for a 27~\UTeV
collider with an integrated luminosity of $15~\iab$ we quote the
expected reach~\cite{Goncalves:2018yva}
\begin{align}
\frac{\lambda_{3H}}{\lambda_{3H}^\text{(SM)}}
=\begin{cases} 
1 \pm 15\% &  68\%~\text{C.L.} \\
1 \pm 30\% &  95\%~\text{C.L.} 
\end{cases}
\label{eq:reach_lam}
\end{align}
We can translate this range into the conventions of
Eq.\eqref{eq:ourleff} assuming that the underlying new physics
only affects $\mathcal{O}_{\phi 3}$,
\begin{align}
V = \mu^2 \; \frac{(v+H)^2}{2} 
  + \lambda \; \frac{(v+H)^4}{4} 
  + \frac{f_{\phi 3}}{3 \Lambda^2} \; \frac{(v+H)^6}{8} \; .
\end{align}
The reach of the dedicated one-parameter self-coupling
analysis becomes
\begin{align}
\lambda_{3H} 
= \lambda_{3H}^\text{(SM)}
  \left( 1 + \frac{2 v^2}{3 m_H^2} \; \frac{f_{\phi 3} v^2}{\Lambda^2} \right) 
\qquad \text{and} \qquad
\left| \frac{\Lambda}{\sqrt{f_{\phi 3}}} \right| \gtrsim
\begin{cases}
1~\text{\UTeV} &  68\%~\text{C.L.} \\
700~\text{\UGeV} &  95\%~\text{C.L.} 
\end{cases}
\label{eq:reach_d6}
\end{align}

Our global analysis containing all operators in Eq.\eqref{eq:ourleff}
is based on a re-scaling of the number of signal and background events 
in the 8~\UTeV analysis~\cite{Corbett:2015ksa} to 27~\UTeV, assuming two
experiments. For the invisible Higgs searches we use an in-house extrapolation of
the WBF analysis~\cite{Eboli:2000ze} from Ref.~\cite{Biekotter:2017gyu} to 27~\UTeV.  
Because the effective Lagrangian of Eq.\eqref{eq:ourleff} includes new
Lorentz structures, especially valuable information comes from the
kinematic distributions~\cite{Englert:2015hrx,Corbett:2015ksa} listed in Tab.~\ref{tab:distr}.
They are particularly relevant for our analysis of the Higgs self-coupling,
where the full kinematic information from Higgs pair production encoded in
the $m_{HH}$ distribution allows us to separate the effects of
$\mathcal{O}_{\phi 2}$ and $\mathcal{O}_{\phi  3}$~\cite{Kling:2016lay,Goncalves:2018yva}.
Following Refs.~\cite{Bizon:2016wgr,Maltoni:2017ims,DiVita:2017eyz} we neglect the loop effects
of $\mathcal{O}_{\phi 3}$ on single Higgs production, because they will
hardly affect a global Higgs analysis.  
\medskip

\begin{table}[t!]
\centering
\begin{tabular}{lrrr}
\toprule
channel  & observable & \# bins & range [\UGeV] \\
\midrule
$WW \rightarrow (\ell \nu)(\ell \nu)$          & $m_{\ell\ell'}$ & $10$ & $0-4500$\\
$WW \rightarrow (\ell \nu)(\ell \nu)$          & $p_T^{\ell_1}$ & $8$ & $0-1750$\\
$WZ \rightarrow (\ell \nu)(\ell \ell)$          & $m_T^{WZ}$ & $11$ & $0-5000$\\
$WZ \rightarrow (\ell \nu)(\ell \ell)$          & $p_T^{\ell\ell}$ ($p_T^Z$) & $9$ & $0-2400$\\
WBF, $H\rightarrow \gamma\gamma$          & $p_T^{\ell_1}$ & $9$ & $0-2400$\\
$VH \rightarrow (0\ell) (b \bar{b})$ & $p_T^V$ & $7$ & $150-750$\\
$VH \rightarrow (1\ell) (b \bar{b})$ & $p_T^V$ & $7$ & $150-750$\\
$VH \rightarrow (2\ell) (b \bar{b})$ & $p_T^V$ & $7$ & $150-750$\\
$HH \rightarrow (b \bar{b}) (\gamma \gamma)$, $2j$ & $m_{HH}$ & $9$ & $200-1000$ \\
$HH \rightarrow (b \bar{b}) (\gamma \gamma)$, $3j$ & $m_{HH}$ & $9$ & $200-1000$ \\
\bottomrule
\end{tabular}
\caption{Distributions included in the analysis. The number of bins
includes an overflow bin for all channels. Table from Ref.~\cite{Biekotter:2018jzu}.}
\label{tab:distr}
\end{table}

In Fig.~\ref{fig:d6} we show the expected reach for the global 
Higgs analysis in terms of dimension-6 operators for a 27~\UTeV LHC upgrade. 
For all measurements we assume the SM predictions,
which means that our best-fit points will always be the SM values.
To illustrate the importance of precision predictions we
will show results with the current theory uncertainties as well as an
assumed improvement of theory and systematics by a factor two.
Asymmetric
uncertainty bands arise because of correlations, but also reflect
numerical uncertainties.  Different colours correspond to assumed
integrated luminosities of $1.5~\iab$ and $15~\iab$.  
Most of the effective operators
benefit from an increased statistics, because larger luminosity
extends the reach of kinematic distributions, which in their tails are
always statistically limited. In contrast, the Yukawa couplings
$f_{b,\tau}$, which do not change the Lorentz structure, are mostly
limited by the assumed systematic and theory uncertainties.
Consequently, the reach for operators which modify the Lorentz
structure of some Higgs interaction exceeds the reach for the
Yukawa-like operators or the reach for the operator $\mathcal{O}_{\phi 2}$,
which introduces a wave function renormalisation  for the Higgs field
and only changes the kinematics of Higgs pair production.

\begin{figure}[t!]
\centering
\includegraphics[width=0.5\textwidth]{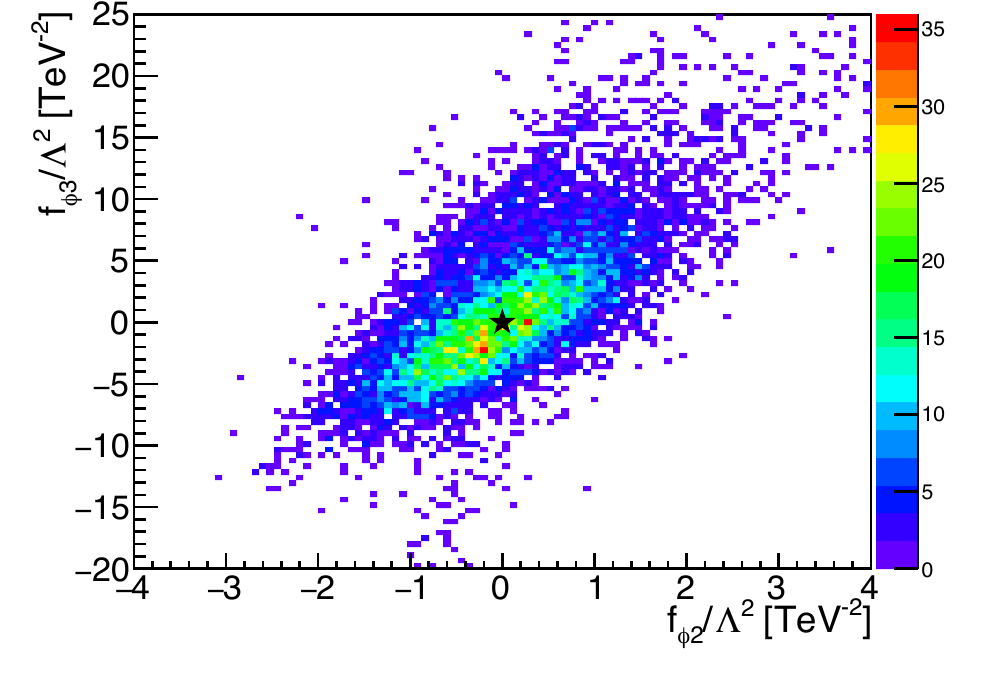}
\caption{Correlations between the leading operators describing Higgs
  pair production. Figure from Ref.~\cite{Biekotter:2018jzu}.}
\label{fig:corr}
\end{figure}
For the operator $\mathcal{O}_{\phi 3}$, which modifies the Higgs potential
as part of a global analysis, we find the limits
\begin{alignat}{9}
\frac{\Lambda}{\sqrt{|f_{\phi 3}|}} &> 430~\text{\UGeV}
&\qquad \text{68\% C.L.} \notag \\
\frac{\Lambda}{\sqrt{|f_{\phi 3}|}} &> 245~\text{\UGeV}
\quad (f_{\phi 3} >0)
\quad \text{and} \quad 
\frac{\Lambda}{\sqrt{|f_{\phi 3}|}} > 300~\text{\UGeV}
\quad (f_{\phi 3} <0) 
&\qquad \text{95\% C.L.} 
\label{eq:reach_d6_3}
\end{alignat}
These limits are diluted from the one-parameter analysis quoted in
Eq.\eqref{eq:reach_d6}, largely because of the combination with
$\mathcal{O}_{\phi 2}$. We can directly compare the
effects from $\mathcal{O}_{\phi 2}$ and $\mathcal{O}_{\phi 3}$ for similar values of
$f/\Lambda^2$ as a function of the momentum flowing through the
triple-Higgs vertex or $m_{HH}$. In that case we find that the
momentum dependence in $\mathcal{O}_{\phi 2}$ matches the effects from
$\mathcal{O}_{\phi 3}$ for $m_{HH} \gtrsim 1$~\UTeV, with either relative
sign. 
In Fig.~\ref{fig:corr} we show 
the correlation between $\mathcal{O}_{\phi 2}$ and $\mathcal{O}_{\phi 3}$
which is not
accounted for in the usual Higgs pair analyses.
The global analysis obviously
reduces the reach for the Higgs self-coupling modification compared to
a one-parameter analysis, but still indicates that a 27~\UTeV hadron
collider will for the first time deliver a meaningful measurement of
this fundamental physics parameter.

\newpage

\asection[M. Borsato, M. Flechl, S. Gori, L. Zhang]{Searches for beyond the standard model Higgs physics}\label{sec9}
HL and HE-LHC will have an unprecedented opportunity to explore not only the Higgs sector of the SM, but also extended Higgs sectors, as well as rare BSM processes involving the 125 \UGeV Higgs boson. 

In this section, we first discuss the prospects for detecting 125 \UGeV Higgs exotic decays, i.e. decays to light NP particles that eventually decay back to SM particles (Sec. \ref{Sec:9.1Exo}). Higgs exotic decays to collider stable particles (Higgs invisible decays) are already discussed in Sec. \ref{Sec:6Invisible}. Here we report the study of a large array of (semi-) visible decays as $h\to \phi\phi$, where $\phi$ is either a long lived scalar (Secs. \ref{Sec:9.1.1}, \ref{Sec:9.1.2}) or it promptly decays back to two SM fermions (Secs. \ref{Sec:9.1.3}-\ref{Sec:9.1.8}). So far, the LHC has produced $\mathcal O(10)$ million Higgs bosons at its Run 1 and Run 2. This is only $\mathcal O(5\%)$ ($\mathcal O(5\permille)$) the amount of Higgs bosons that will be produced by the HL (HE)-LHC. The huge Higgs statistics that will be collected at the HL/HE-LHC will be a key ingredient for the success of a Higgs exotic decay program. 

In Secs. \ref{sec:Hff}, \ref{sec:XZZ}, we then present the ATLAS and CMS prospects for searching for new heavy Higgs bosons, either in fermionic final states ($H\to\tau\tau$, as particularly motivated in Supersymmetric theories), or in bosonic final states ($H\to ZZ$, as motivated in theories that deviate from the alignment limit). These two sections are followed by a phenomenological section (Sec. \ref{Sec:9.4}) that discusses the reach of possible additional searches for neutral and charged Higgs bosons and the role of interference effects in heavy Higgs searches. The interplay of the 125 Higgs coupling measurements and searches for new degrees of freedom is discussed in Secs. \ref{Sec:9.5}-\ref{Sec:9.7} for several BSM models (the Minimal Supersymmetric SM, Twin Higgs models and composite Higgs models). 

Several well motivated BSM models can also predict new Higgs bosons with a mass below 125 \UGeV. The prospects to probe these light Higgs bosons and the corresponding models are discussed in Sec. \ref{Sec:9.8}. Particularly, light Higgs bosons could be uniquely probed at the HL-stage of the LHCb experiment.


\asubsection{Exotic decays of the 125 \UGeV Higgs boson}\label{Sec:9.1Exo}
\asubsubsection[A. Gandrakota, Y. Gershtein, R. Glein, B. Greenberg, F. Paucar-Velasquez, R. Patel, K. Ulmer]{First Level Track Jet Trigger for Displaced Jets at High Luminosity LHC}\label{Sec:9.1.1}

The high luminosity LHC program offers many exciting opportunities to search for rare processes. It is expected that the
LHC will accumulate 3\abinv of proton-proton collisions at 14\UTeV.
The CMS detector will undergo major upgrades to all subsystems, including the tracker~\cite{cmstdr-014},
the barrel~\cite{cmstdr-barrel} and end-cap~\cite{cmstdr-ec} calorimeters, the muon system~\cite{cmstdr-mu},
and the trigger~\cite{cmstdr-017}. 

The bandwidth limitations of the first level (L1) trigger
are one of the main problems facing current searches 
for exotic Higgs boson decays, as well as many other signals beyond the standard model.
The process where the Higgs boson decays to two new light scalars that in turn decay to jets, \Hphiphi, is an important example. If the scalar $\phi$ has a
macroscopic decay length, the offline analysis has no background from SM processes, but the majority of the signal events do not get recorded because they fail to be selected by the L1 trigger.
The main obstacle is the high rate for low transverse momentum jets, which is made worse by additional extraneous \pp collisions in the
high luminosity environment.

In this note~\cite{CMS-PAS-FTR-18-018}, 
we investigate the capabilities of L1 track finding~\cite{cmstdr-014} to increase the L1 trigger efficiency for such signals.
We focus on small or moderate decay lengths of the new particles, 1--50\Umm, and assume, as is demonstrated by
many analyses~\cite{ll1, ll2, ll3}, that the offline selection can remove all SM backgrounds with only a moderate loss of efficiency.

The investigation has two major thrusts. First, we propose a jet clustering algorithm that uses the L1 tracks found with a primary vertex constraint.
Second, we consider the extension of the L1
track finder to off-pointing tracks, and develop a jet lifetime tag for tracks with $|\eta| < 1.0$. 
Future work will include: expanding the off-pointing track finding at L1 to the full acceptance of the outer tracker;
matching the track jets with high transverse energy (\ET) deposits in the electromagnetic calorimeter; and finding new ways to evaluate
track quality to suppress ``fake'' tracks that result from finding the wrong combination of track hits. 

While in this study we focus on the specific Higgs boson decay to light scalars (see Ref.~\cite{bsmh} for extensive review of physics motivations for such decays),
the results and the proposed triggers are relevant for a broad spectrum of new physics searches, with or without macroscopic decay lengths.

\paragraph{Signal and background simulation}

In these studies, the Phase-2 CMS detector is simulated using \GEANTfour~\cite{geant}.
Event samples corresponding to 200 collisions per bunch crossing (pileup) \cite{cmstdr-017} are used for the evaluation of trigger rates.

The following signal samples are considered:
\begin{enumerate}
\item Displaced single muons, generated with a uniform distribution of transverse momentum (\pt) between 2 and 8\UGeV, uniform
in $\eta$ between -1 and 1, and with impact parameter \dtrans distributed as a Gaussian with width $\sigma=2\Ucm$.
\item The decay of the SM Higgs boson $\PH(125)\to \phi\phi\to \bbbar\bbbar$, with $\phi$ masses of 15, 30, and 60\UGeV, and \ctau of 0, 1, and 5\Ucm.
The production of the Higgs boson via gluon fusion is simulated by \POWHEG v2.0 \cite{powheg}, while the hadronisation and decay is performed by \PYTHIA v8.205 \cite{pythia}.
\item The decay of a heavy SM-like Higgs boson with mass 250\UGeV, $\PH(250)\to \phi\phi\to \bbbar\bbbar$, with $\phi$ masses of 15, 30, and 60\UGeV, and \ctau of 0, 1, and 5\Ucm.
The production of the heavy SM-like Higgs boson via gluon fusion, its decay, and its hadronisation are all simulated with \PYTHIA 8 \cite{pythia}.
\end{enumerate}

\paragraph{Track jets}
\label{sec:trackjets}

The tracker is the most granular detector participating in the L1 decision, and therefore the most resilient to pileup. 
Track finding at L1 relies on selection at the front end of tracker hits that originated from high transverse momentum particles.
This is achieved through use of the so-called \pt-modules consisting of two sensors separated by a few mm \cite{cmstdr-014}. A particle crossing a tracker module
produces a pair of hits in the two sensors. Such pairs form a ``stub'' if the azimuthal difference between the hits in the two sensors of a module is consistent
with a prompt track with $\pt \gtrsim 2\UGeV$.

In this section, we describe a simple jet clustering algorithm implementable in firmware, and compare it with anti-\kt jets~\cite{Cacciari:2008gp} with a size parameter of $R = 0.3$,
as produced by \FASTJET~\cite{fastjet}.

A simplified algorithm for L1 track jets is used to facilitate the firmware implementation for the L1 trigger applications. 
L1 track jets are found by grouping tracks in bins of \zo, the point of closest approach to the $z$-axis, for the tracks. The bins are overlapping, staggered by
half a bin, so that each track ends up in two bins, eliminating inefficiencies at bin edges. In each \zo bin, the \pt of the tracks are summed in bins of
$\eta$ and azimuthal angle $\phi$ with bin size $0.2 \times 0.23$. A simplified nearest-neighbour clustering is performed, and the total $\HT=\sum \pt^{\text{trk}}$ in the \zo bin is calculated.
The \zo bin with the highest \HT is chosen. Jets obtained through this algorithm are referred to as ``TwoLayer Jets.'' 
For the studies below, \zo bins with size 6 cm are used. Jets with $\ET>50\,(100)\UGeV$ are required to have at least two (three) tracks.

The track purity depends on the number of stubs in the track and the $\chi^2$ of the track fit. High-\pt tracks are much less pure than low-\pt tracks,
with fake tracks distributed approximately uniformly in $1/\pt$ while real tracks are mostly low-\pt.
To mitigate the effect of high-\pt fake tracks, any track with a reconstructed \pt above 200\UGeV is assigned a \pt of 200\UGeV. The track quality selection used in this analysis
is summarised in Table~\ref{tab:tracksel}. 

\begin{table*}[htb]
\centering
\caption{Track selection for jet finding. The $\chi^2$ selections are per degree of freedom for a 4-parameter track fit. \label{tab:tracksel}}
\begin{tabular}{r|ccc}
\hline
track \pt & 4 stubs  & 5 stubs & 6 stubs \\
\hline
2--10\UGeV   & $\chi^2<15$ & $\chi^2<15$ & accept \\
10--50\UGeV & reject      & $\chi^2<10$ & accept \\
${>}50\UGeV$   & reject      & $\chi^2<5$ & $\chi^2<5$ \\
\hline
\end{tabular}
\end{table*}

We have verified that the TwoLayer trigger algorithm gives similar performance to a full jet clustering using the anti-\kt algorithm with a size parameter $R = 0.3$, as implemented in \FASTJET.
Figure~\ref{fig:jeteff} shows the efficiency to reconstruct a track jet as function of the generator-level jet \pt.
Figure~\ref{fig:rates} shows the calculated L1 trigger rates for an \HT trigger (scalar sum of \pt of all jets above threshold)
and a quad-jet trigger (at least four jets above threshold) as a function of the threshold.
\HT is computed from track jets with $\pt>5\UGeV$.

The rates are computed based on a fixed number of colliding bunches. The trigger rate is computed as
\begin{equation*}
  \text{Rate} = \effLT \Nbunches \fLHC,
\end{equation*}
where $\Nbunches=2750$ bunches for 25\Uns bunch spacing operation, $\fLHC = 11246\UHz$, and \effLT is the efficiency to pass a given L1 threshold as determined
in simulation. For both the L1 trigger efficiency and rate, the performance of the TwoLayer hardware algorithm is compatible with the performance from the more sophisticated algorithm
from \FASTJET.

\begin{figure*}[hbtp]\centering
 \includegraphics[width=0.45\textwidth]{\main/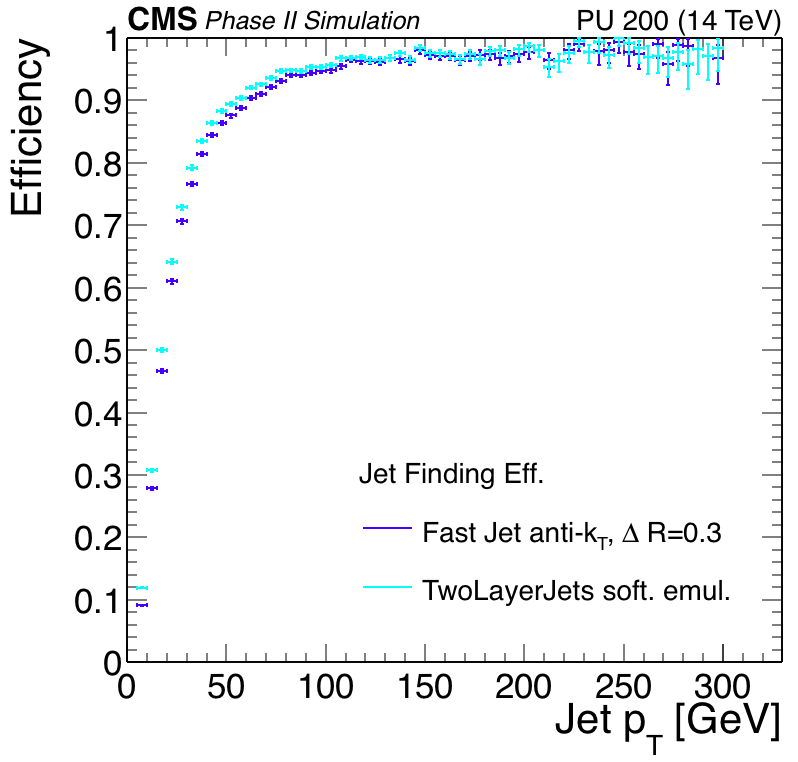}
 \caption{The efficiency for a jet to give rise to a L1 track jet as a function of the generator-level \pt of the jet.
 The light and dark blue lines correspond to the trigger clustering (TwoLayer Jets) and anti-\kt with $R =0.3$ (\FASTJET), respectively.}
  \label{fig:jeteff}
\end{figure*}%

\begin{figure*}[hbtp]\centering
 \includegraphics[width=0.45\textwidth]{\main/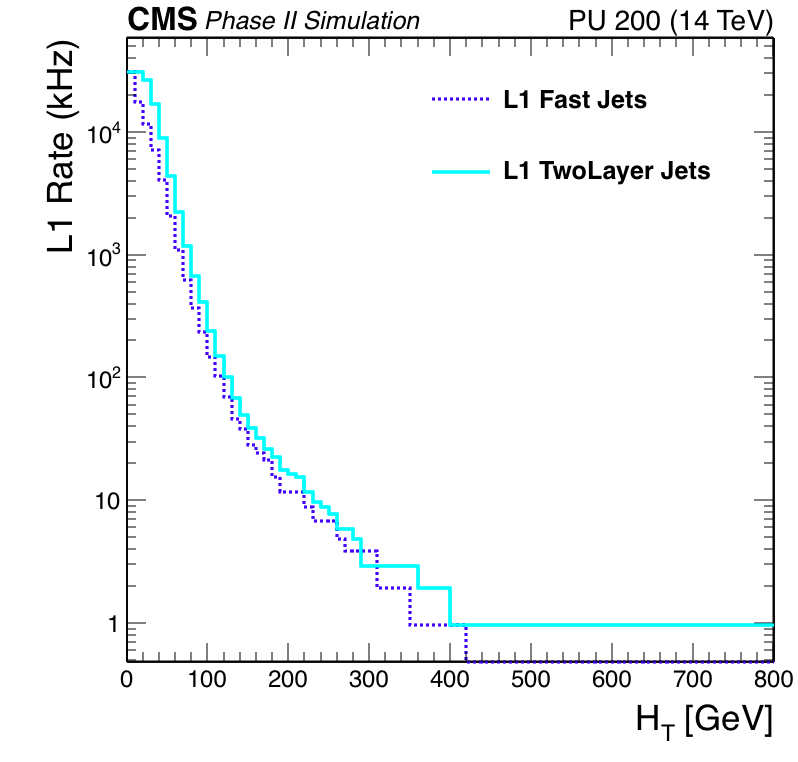}
 \includegraphics[width=0.45\textwidth]{\main/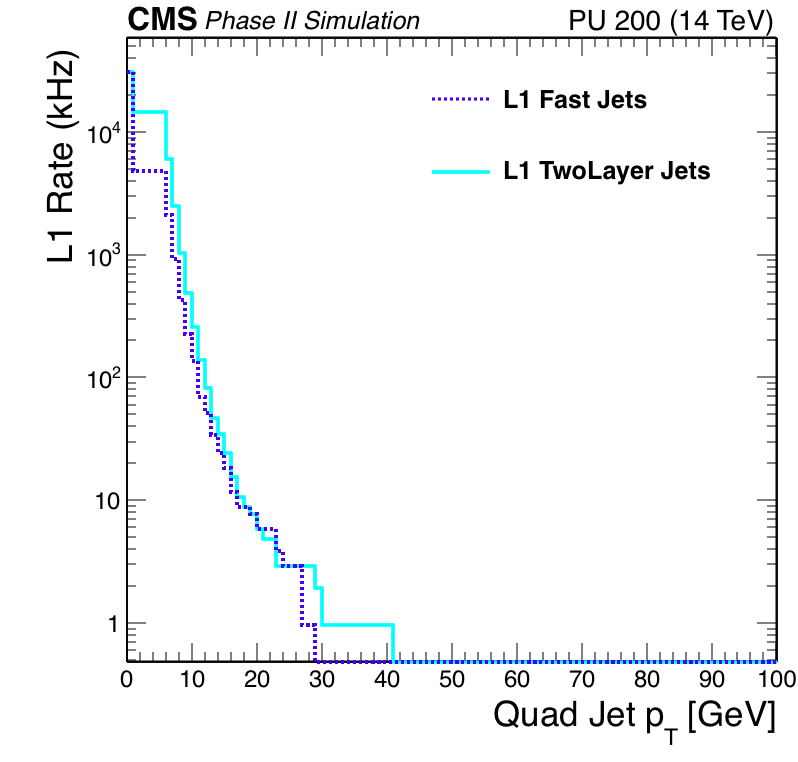}
 \caption{Calculated L1 trigger rates for track jet based \HT (left) and quad-jet (right) triggers.
 The light and dark blue lines correspond to the trigger clustering (TwoLayer Jets) and anti-\kt with $R =0.3$ (\FASTJET), respectively.}
  \label{fig:rates}
\end{figure*}%

\paragraph{Displaced track finding}

In this section, we briefly describe the performance of an algorithm for reconstruction of tracks with non-zero impact parameter. This approach extends the baseline L1 Track Trigger design to 
handle tracks with non-zero impact parameter and to include the impact parameter in the track fit. This enhanced design is feasible without greatly altering the track finding approach, but will 
require more computational power than the current proposal, which considers only prompt tracks. Tracks passing the selection are clustered using the same algorithm as described in Section~\ref{sec:trackjets}, 
and clusters containing tracks with high impact parameters are flagged as displaced jets. Though the baseline design of the L1 Track Trigger currently is optimised to find prompt tracks, these 
studies show that an enhanced L1 Track Trigger can extend the L1 trigger acceptance to include new BSM physics signals.

\begin{figure*}[hbtp]\centering
 \includegraphics[width=0.49\textwidth]{\main/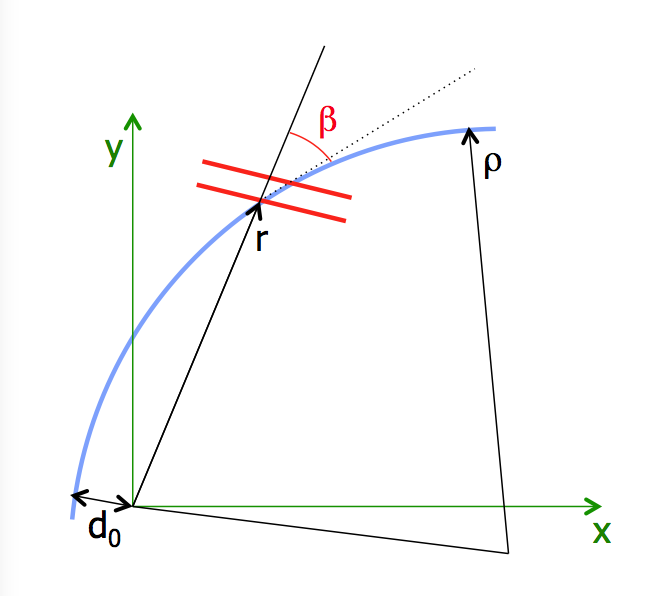}
 \caption{A sketch of a track crossing a \pt-module.}
  \label{fig:bend}
\end{figure*}%

A track with a sufficiently small impact parameter can produce a stub.
For tracks with large \pt (i.e. large curvature radius $\rho$) and small \dtrans,
the bending angle $\beta$ between the track and the prompt infinite momentum track, as shown in Fig. \ref{fig:bend}, is
\begin{equation*}
  \beta \approx \frac{r}{2\rho} - \frac{\dtrans}{r}.
\end{equation*}

Therefore, for a given \dtrans, one expects the stubs to be formed more efficiently as the radius of the module $r$ increases.
Fig. \ref{fig:stubeff} shows the efficiency for a displaced muon to produce a stub as a function of the signed transverse momentum and the impact parameter
of the muon, as measured in the full \GEANTfour-based simulation of the Phase-2 detector. 

\begin{figure*}[hbtp]\centering
 \includegraphics[width=0.9\textwidth]{\main/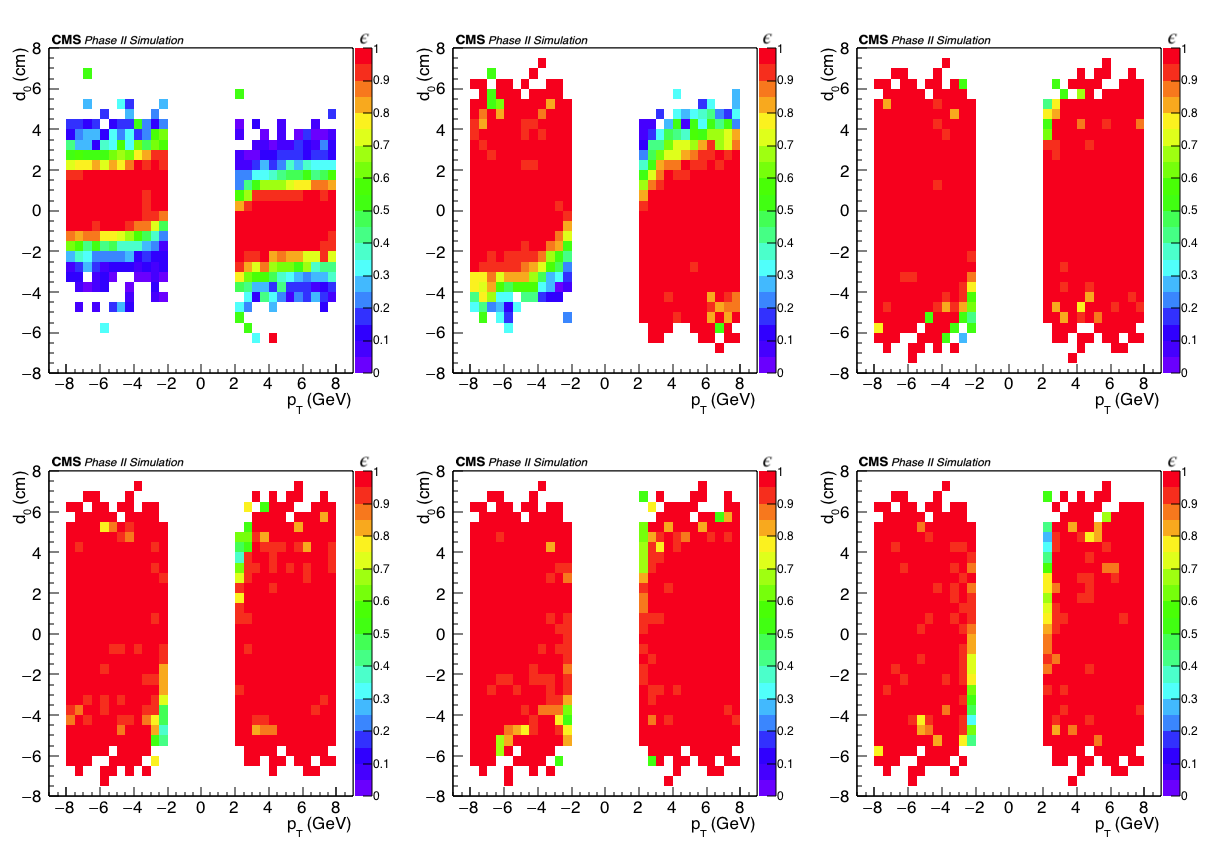}
 \caption{The efficiency for a displaced muon to form stubs in the six barrel layers of the Phase-2 tracker, as a function of the signed muon \pt and impact parameter.
 The top row shows, from left to right, layers 1, 2, and 3;
 the bottom row shows layers 4, 5, and 6.
 The sample is comprised of 2000 muons generated with uniformly distributed transverse momentum between 2 and 8\UGeV and pseudorapidity $|\eta|<1$, and with the impact parameter \dtrans
 distributed as a Gaussian with width of 2\Ucm.}
  \label{fig:stubeff}
\end{figure*}

A special version of the tracklet algorithm \cite{cmstdr-014} has been developed that is capable of reconstructing tracks with impact parameters of a few cm.
For now, the reconstruction is limited to the barrel region ($|\eta|<1.0$). Preliminary feasibility studies show
that the algorithm will have similar performance in the entire outer tracker coverage. 

Fig. \ref{fig:trackeff} shows the track reconstruction efficiency requiring at least four and at least five stubs on the track. As expected, allowing
only four stubs on a track gives a higher efficiency for high impact parameter tracks.

\begin{figure*}[hbtp]\centering
 \includegraphics[width=0.9\textwidth]{\main/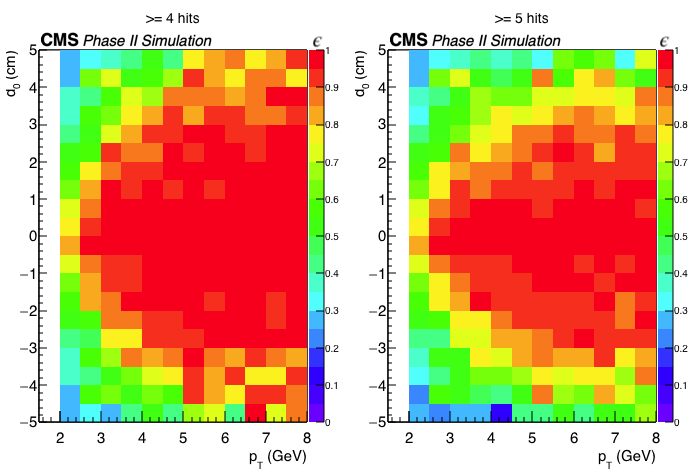}
 \caption{The efficiency for a displaced muon to be reconstructed as a track with at least four stubs (left) and at least five stubs (right).}
  \label{fig:trackeff}
\end{figure*}%

For the extended track finding algorithm, two track fits are performed:
a 3-parameter $r\phi$ fit yielding $1/\rho$, $\phi_0$, and \dtrans, and a 2-parameter $rz$ fit
yielding $t$ and \zo. The bend consistency variable is defined as
\begin{equation*}
\text{consistency} = \frac{1}{\Nstubs} \sum_{i=1}^{\Nstubs}\left(\frac{\beta_i - \beta_i^{\text{exp}}}{\sigma_{i}}\right)^2,
\end{equation*}
where \Nstubs is the total number of stubs comprising the track, $\beta_i$ and $\beta_i^{\text{exp}}$ are the measured and expected bend angles for stub $i$,
and $\sigma_i$ is the expected bend angle resolution.

Two track categories are defined, loose and tight.
The selection is summarised in Table \ref{tab:dtracksel}.

\begin{table*}[htb]
\centering
\caption{Track selection criteria for jet finding with extended L1 track finding. \label{tab:dtracksel}}
\begin{tabular}{c|ccc|ccc}
\hline
           & \multicolumn{3}{c|}{Loose}                        & \multicolumn{3}{c}{Tight} \\
\Nstubs & $\chi^2_{r\phi}$ & $\chi^2_{rz}$ & consistency & $\chi^2_{r\phi}$ & $\chi^2_{rz}$ & consistency \\
\hline
$4$      &  ${<}0.5$ & ${<}0.5$ & ${<}1.25$ & \multicolumn{3}{c}{reject} \\
$\geq 5$ &  ${<}5.0$ & ${<}2.5$ & ${<}5.0$  & ${<}3.5$ & ${<}2.0$ & ${<}4.0$ \\
\hline
\end{tabular}
\end{table*}

A jet is required to have at least two tracks passing the tight selection. If two or more tight tracks
in a jet have $|\dtrans|>0.1\Ucm$, the jet is tagged as a displaced jet.

\paragraph{Results}

Figure \ref{fig:reff} shows the rate of the track jet \HT trigger as a function of the efficiency
of the heavy SM-like Higgs boson signal. While for prompt $\phi$ decays one can realistically achieve 20\%
efficiency at an L1 rate of 25\UkHz, the efficiency quickly drops with the decay length, since the displaced tracks are not reconstructed for \dtrans values above a few mm.

\begin{figure*}[hbtp]\centering

 \includegraphics[width=0.49\textwidth]{\main/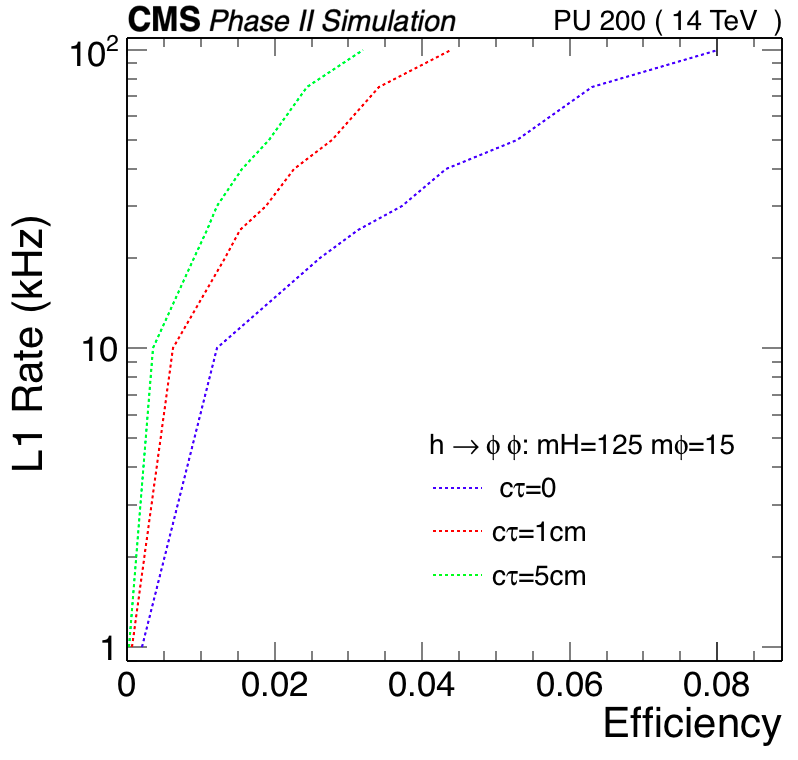}
 \includegraphics[width=0.49\textwidth]{\main/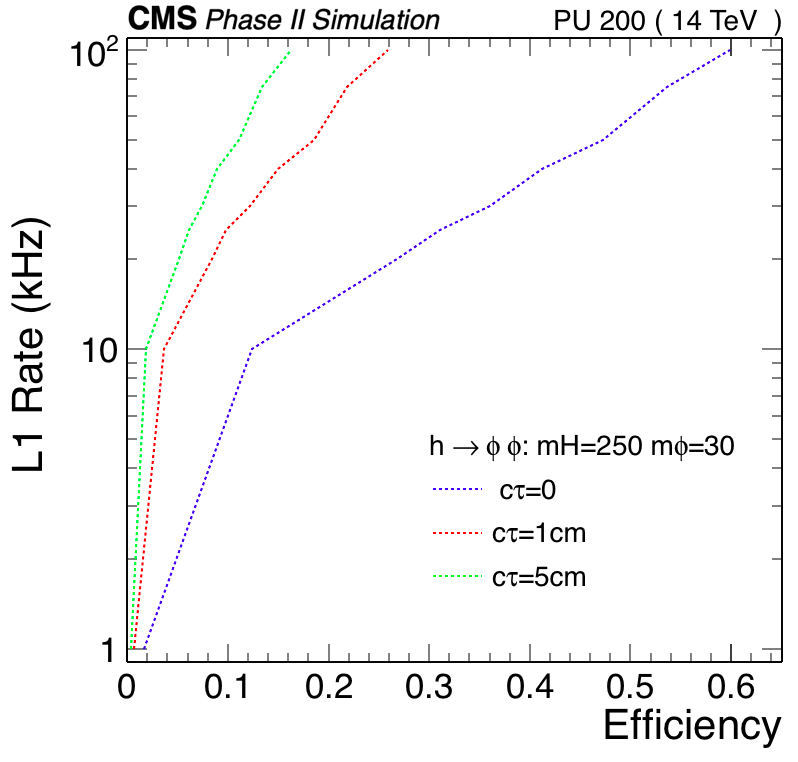} \\
 \includegraphics[width=0.49\textwidth]{\main/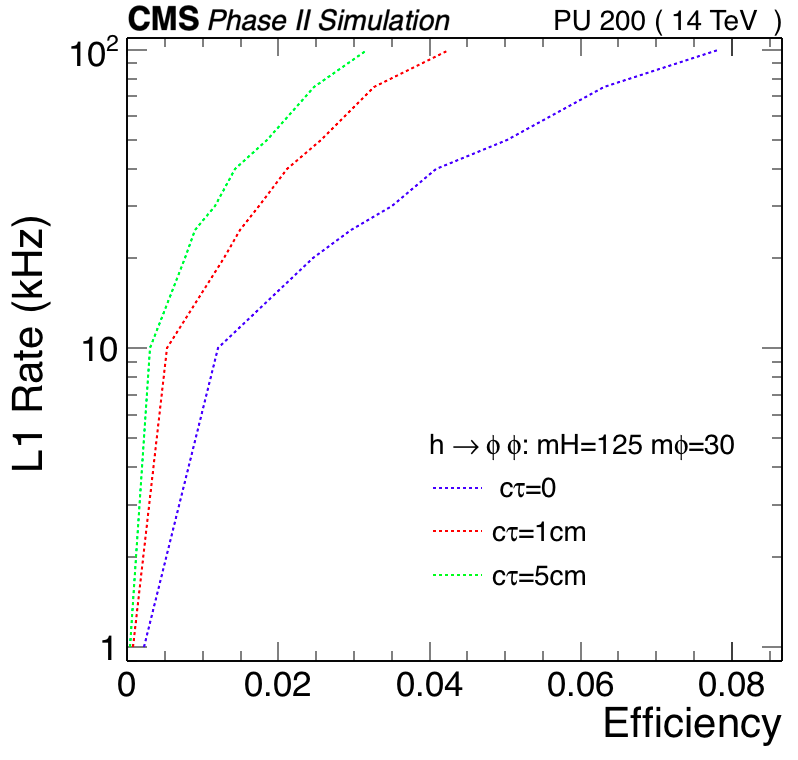}
\includegraphics[width=0.49\textwidth]{\main/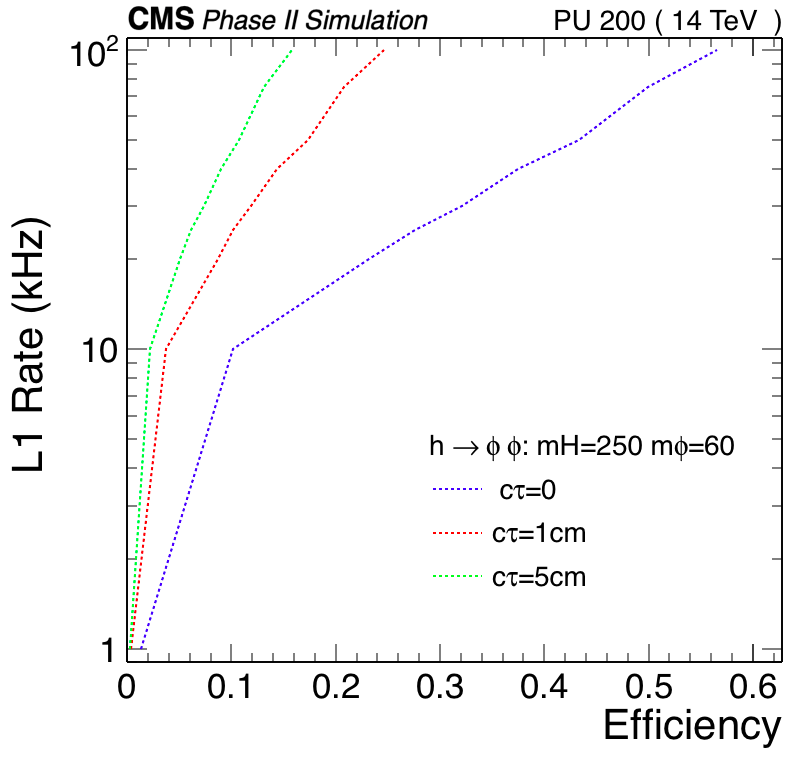}
 \caption{The rate of the track jet \HT trigger as a function of signal efficiency for the SM Higgs boson (left) and the heavy SM-like Higgs boson (right) using prompt track finding.}
  \label{fig:reff}
\end{figure*}%

The rate for the \HT trigger using the extended track finding is shown in Fig. \ref{fig:rate_disp}, with and without 
a requirement of at least one jet with a displaced tag. The displaced tag requirement suppresses the rate by more than
an order of magnitude. The displaced tracking and the trigger that requires a jet with a displaced tag make the signals with low \HT accessible for displaced jets.

\begin{figure*}[hbtp]\centering
 \includegraphics[width=0.49\textwidth]{\main/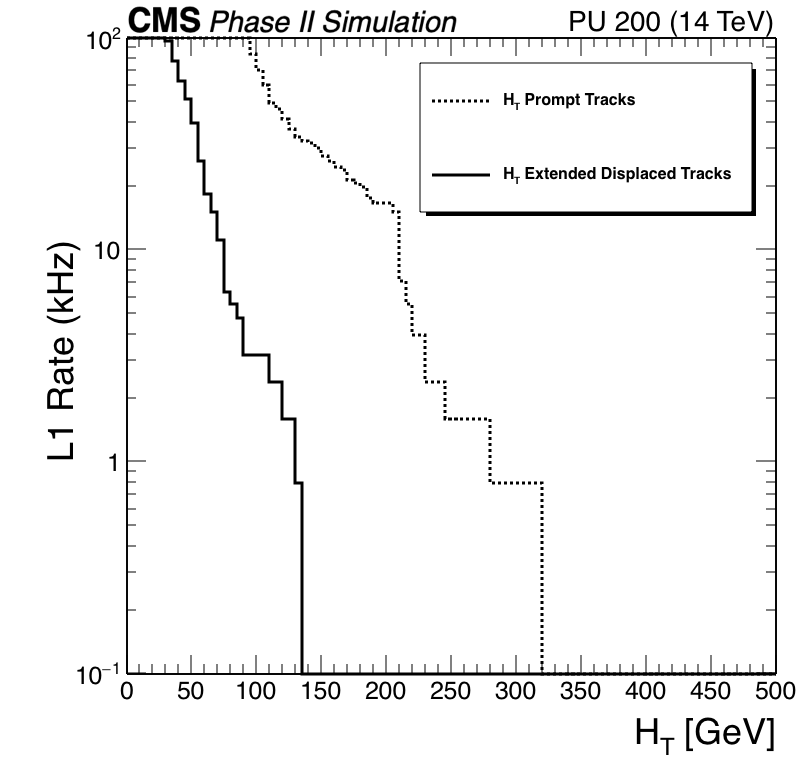}
 \caption{The rate of the track jet \HT trigger using extended track finding with (solid line) and without (dashed line) a
 requirement of at least one jet with a displaced tag.}
  \label{fig:rate_disp}
\end{figure*}

In order to compare the results with prompt and extended track reconstruction, one needs to make a correction for
the rapidity coverage: prompt tracks are found in $|\eta|<2.4$, while the extended track algorithm currently only
reconstructs tracks in $|\eta|<1.0$.  For the feasible thresholds, the rate for $|\eta|<0.8$ and $|\eta|<2.4$ differ by a factor of five.  To scale the efficiency for finding track jets to the full
$|\eta|<2.4$ range, we derive a scale factor (SF) based on efficiency in the full $\eta$ range and the central $\eta$ range. The signal efficiency SFs range from 4--6, 
which is comparable to the increase in the L1 rate.
We have confirmed that such extrapolation works for the track jets clustered with prompt tracks. Figure \ref{fig:reff_dispextra} shows the expected trigger rate as a function of efficiency for the SM and the heavy SM-like Higgs bosons.

\begin{figure}[hbtp]
  \centering
    \includegraphics[width=0.49\textwidth]{\main/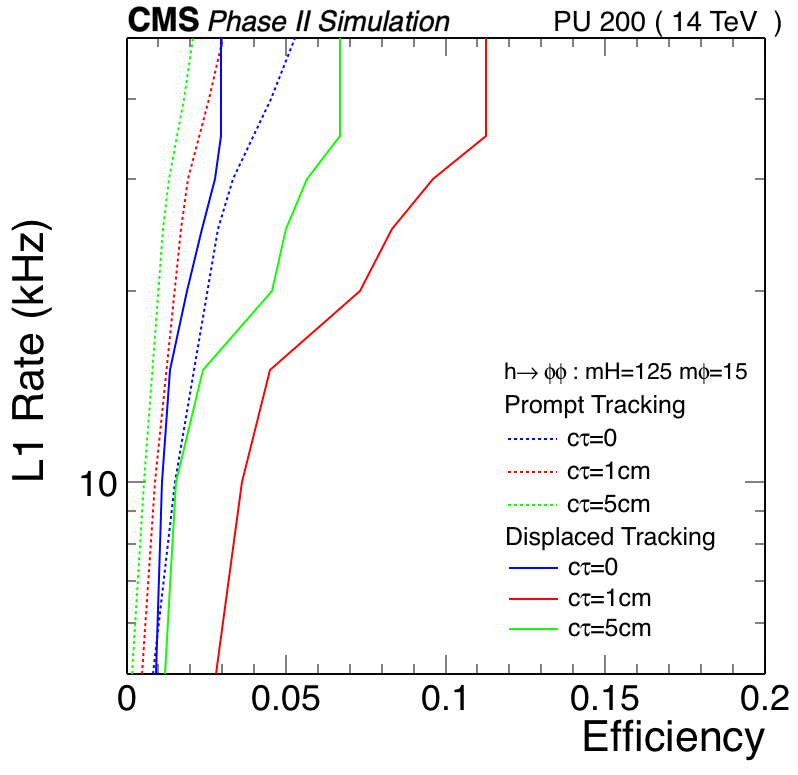}
    \includegraphics[width=0.49\textwidth]{\main/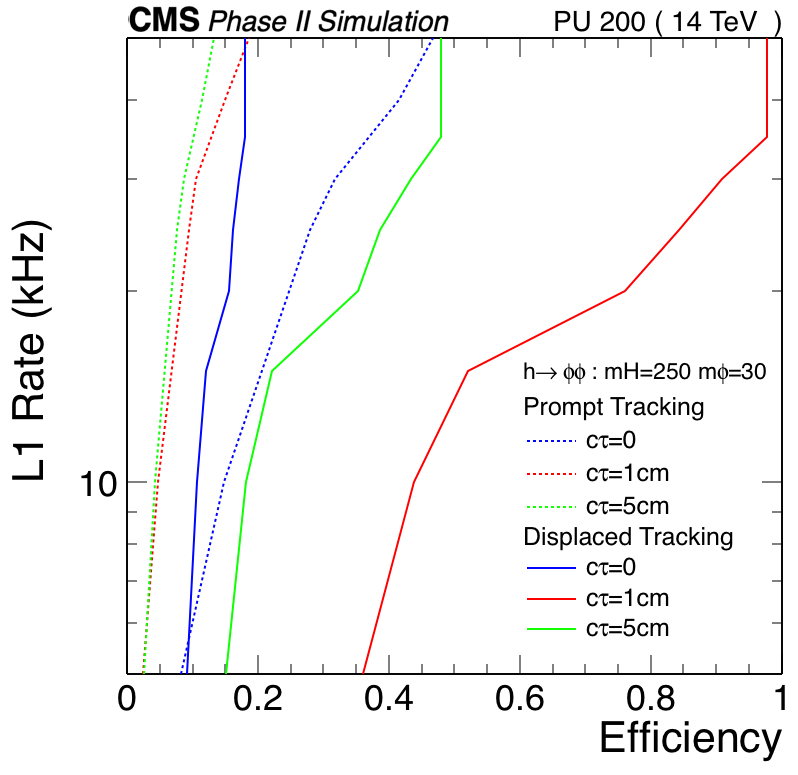}\\
    \includegraphics[width=0.49\textwidth]{\main/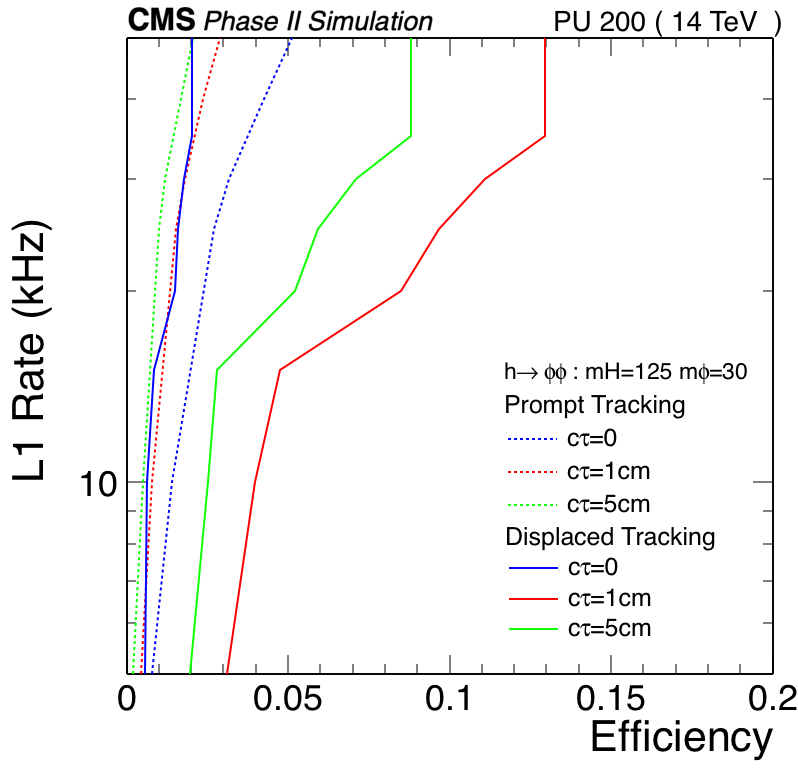}
    \includegraphics[width=0.49\textwidth]{\main/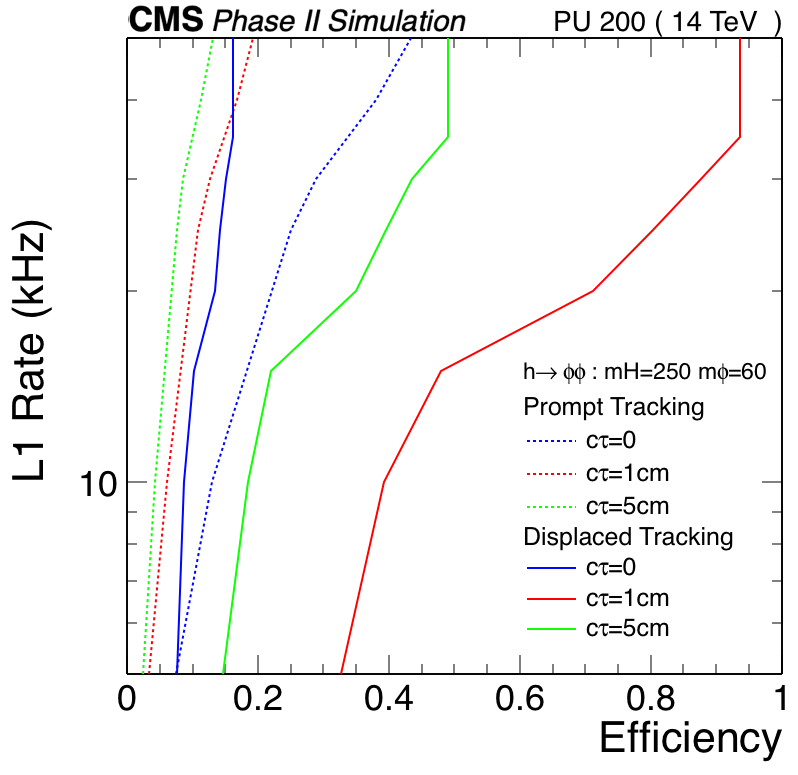}
    \caption{The rate of the track jet \HT trigger as a function of signal efficiency using extended track finding for the SM Higgs (left)
    and the heavy SM-like Higgs (right). The extended track finding performance is extrapolated to the full outer tracker acceptance as described in text.}
    \label{fig:reff_dispextra}
\end{figure}

The available bandwidth for the triggers described above, if implemented, will be decided as a part of the full trigger menu optimisation.
Here, we consider two cases, 5 and 25\UkHz. The expected event yield for triggers using extended and prompt tracking are shown in Fig. \ref{fig:money},
assuming branching fraction $\mathcal{B}[\Hphiphi] = 10^{-5}$ for the SM Higgs boson.
For the heavy Higgs boson, the expected number of produced signal events is set to be the same as for the SM Higgs by requiring
$\sigma_{\pp\to \PH(250)} \mathcal{B}[\BSMHphiphi] = 10^{-5} \sigma_{\pp\to \PH(125)}$.

\begin{figure}[hbtp]
  \centering
    \includegraphics[width=0.49\textwidth]{\main/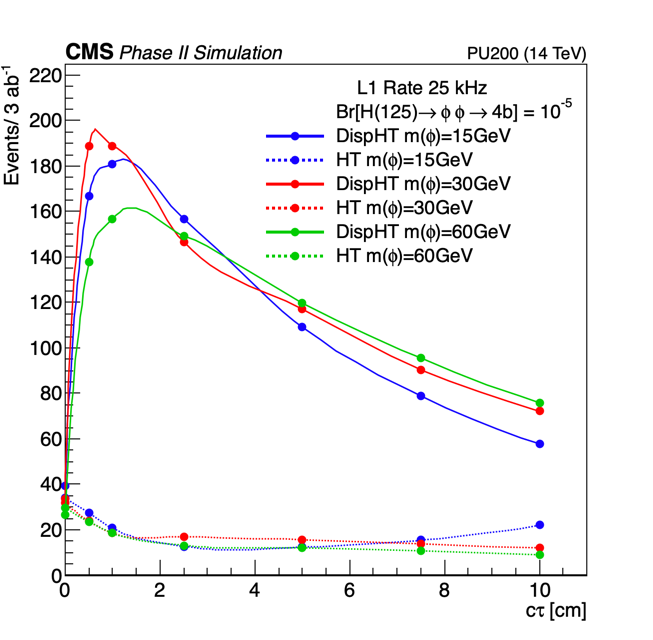}
    \includegraphics[width=0.49\textwidth]{\main/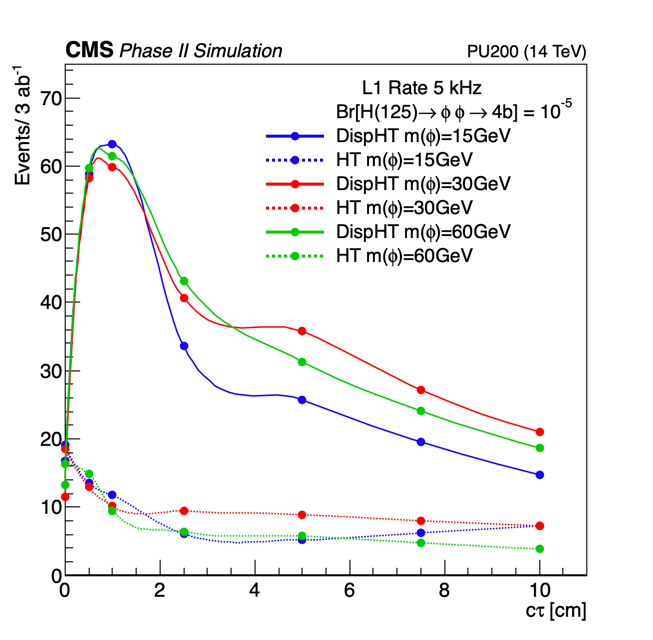}\\
    \includegraphics[width=0.49\textwidth]{\main/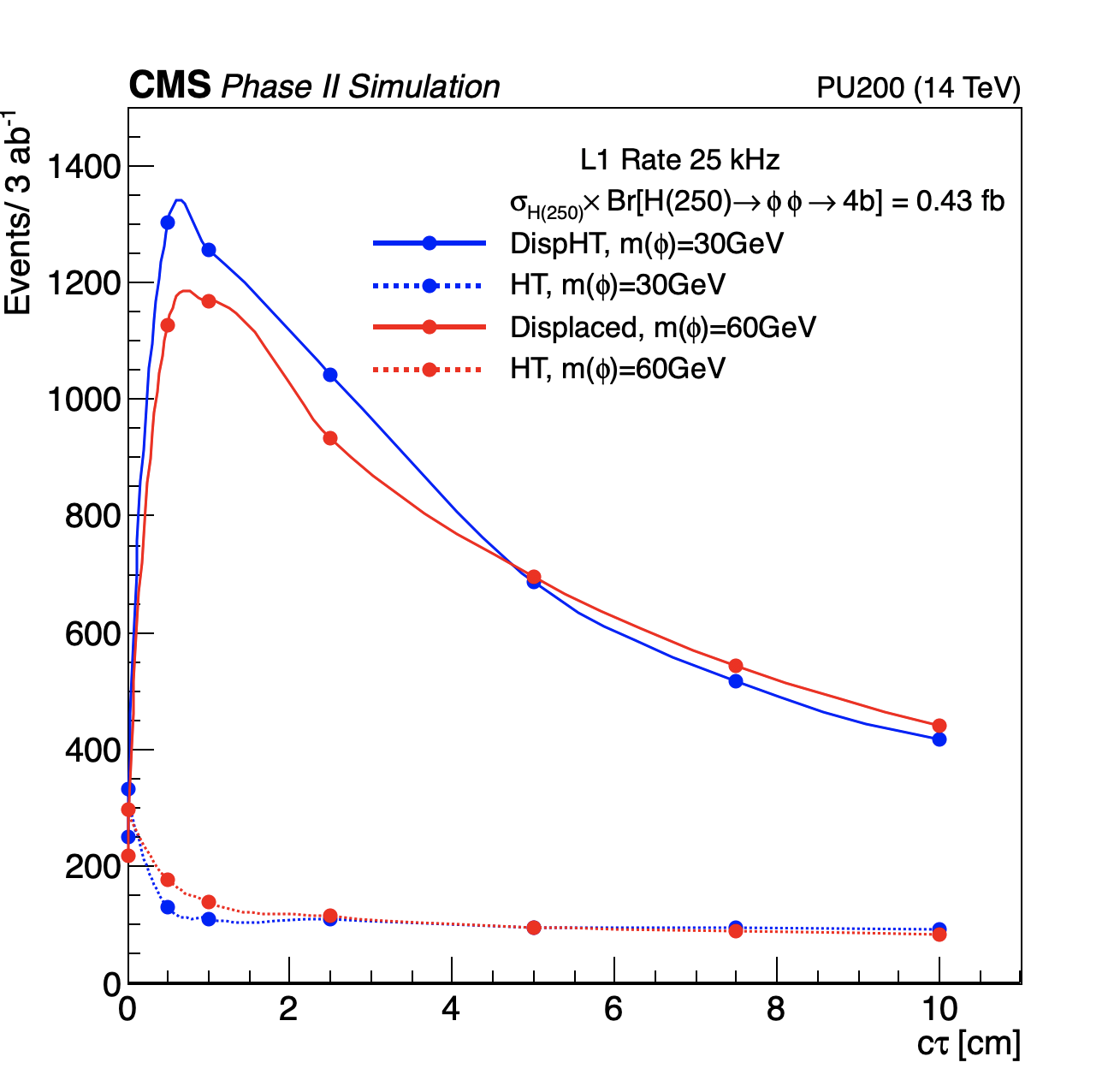}
    \includegraphics[width=0.49\textwidth]{\main/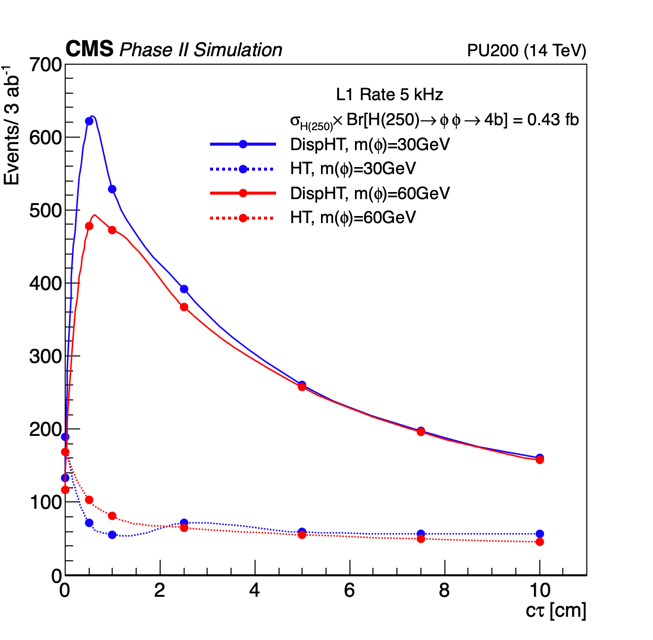}
    \caption{This plot shows the number of triggered Higgs events (assuming ${\mathcal{B}[\Hphiphi] = 10^{-5}}$, corresponding to 
1700 events) as a function of \ctau for two choices for the trigger rates: 25\UkHz (left), 5\UkHz (right).
Two triggers are compared: one based on prompt track finding (dotted lines) and another that is based on extended track finding 
with a displaced jet tag (solid lines). }
    \label{fig:money}
\end{figure}

\paragraph{Conclusion}

We have studied the upgraded CMS detector's ability to trigger on events with long lived particles decaying into jets.
Currently, such events pass the L1 trigger only if the total transverse energy in the event is above a few hundred \UGeV.
This is an important blind spot for searches, especially for the rare exotic Higgs boson decays like
 \Hphiphi.

In this note, a new L1 trigger strategy based on the Phase-2 CMS detector's ability to find tracks at L1 is explored.
Using L1 tracks for jet reconstruction significantly suppresses pile-up and allows to accept events with lower \HT.
For the exotic Higgs decays considered, given the total Phase-2 dataset of 3\abinv and branching fraction of $10^{-5}$, 
CMS would collect $\mathcal{O}(10)$ events, which should be sufficient for discovery.
We also considered a plausible extension of the L1 track finder to consider tracks with impact parameters of a few cm.
That approach improves the yield by more than an order of magnitude. The gains for the extended L1 track finding
are even larger for the events with larger \HT, as demonstrated by the simulations of heavy Higgs boson decays.

\title{
Higgs exotic decays into long-lived particles
}

\asubsubsection[J. Liu, Z. Liu, L.T. Wang]{Higgs exotic decays into long-lived particles}\label{Sec:9.1.2}

The Higgs boson is a well-motivated portal to new physics sectors, introducing a vast variety of exotic decays~\cite{Curtin:2013fra}.
The presence of long-lived particles (LLP) can be a striking feature of many new physics models~\cite{Liu:2018wte,Barbier:2004ez,Giudice:1998bp,Meade:2010ji,Arvanitaki:2012ps,ArkaniHamed:2012gw,Liu:2015bma,Chacko:2005pe, Burdman:2006tz,Kang:2008ea,Craig:2015pha, Davoli:2017swj}. 
At the same time, vast swaths of the possible parameter space of the LLP remain unexplored by LHC searches.
LHC general purpose detectors, ATLAS and CMS, provide full angular coverage and sizeable volume, making them ideal for LLP searches.
However, searches for LLPs that decay within a few centimetre of the interaction point suffer from large SM backgrounds. 
LLPs produced at the LHC generically travel slower than the SM background and decay at macroscopic distances away from the interaction point. Hence, they arrive at outer particle detectors with a sizeable time delay. 

Recently, precision timing upgrades with a timing resolution of 30 picoseconds have been proposed to reduce pile-up for the upcoming 
runs with higher luminosity, including MIP Timing Detector (MTD)~\cite{Collaboration:2296612}  by the CMS collaboration for 
the barrel and end-cap region in front of the electromagnetic calorimeter, the High Granularity Timing Detector~\cite{Allaire:2018bof} 
by the ATLAS collaboration in end-cap and forward region,  and similarly multiple precision timing upgrades~\cite{Bediaga:2018lhg} by the 
LHCb collaboration.
As a strategy applicable to a broad range of models, we propose the use of a generic Initial State Radiation (ISR) jet to time-stamp the hard collision and require only a single LLP decay inside the detector with significant time delay. Such a strategy can greatly suppress the SM background and 
reach a sensitivity two orders of magnitude or more better than traditional searches in a very larger parameter space~\cite{Aad:2015uaa, CMS:2014wda,Coccaro:2016lnz, Liu:2015bma}. 
With a general triggering and search strategy that can capture most LLP decays, 
we show a striking improvement in sensitivity and coverage for LLPs. In addition to the MTD at CMS, we also consider a hypothetical timing layer on the outside of the ATLAS Muon Spectrometer (MS) as an estimate of the best achievable reach of our proposal for LLPs with long lifetimes.  \\

Higgs decaying to glueballs with subsequent decays into SM jet pairs is our benchmark model here. This occurs in model~\cite{Craig:2015pha} where the Higgs is the portal to a dark QCD sector whose lightest states are the long-lived glueballs. 
Typical energy of the glueball is set by the Higgs mass, and the time delay depends on glueball mass. 
Time-stamping the hard collision is achieved by using an ISR jet: 
\bea
pp\rightarrow h+j&,\ h\rightarrow X+X,\ X\rightarrow {\rm SM},   
\eea
where $X$ represents the LLP.

While particle identification and kinematic reconstruction are highly developed,  
usage of timing information has so far been limited since prompt signatures are often assumed. 
Such an assumption could miss a crucial potential signature of an LLP, a significant time delay. 
Here we outline a general BSM signal search strategy that uses the timing information
and the corresponding background consideration.
A typical signal event of LLP is shown in the left panel of Fig.~\ref{fig:ctaulimitHiggs}. 
The LLP, $X$, travels a distance $\ell_X$ into a detector volume and decays into two light SM particles 
$a$ and $b$, which then reach timing detector at a transverse distance $L_{T_2}$ away from the beam axis. Typically, the SM particles travel at velocities close to the speed of light.
For simplicity, we consider neutral LLP signals where background from charged particles can be vetoed using 
particle identification and isolation.
The decay products of $X$ arrive at the timing layer with a time delay
$
\Delta t^i_{\rm delay} = \frac{\ell_X}{\beta_X} + \frac{\ell_i}{\beta_i} - \frac{\ell_{\rm SM}}{\beta_{\rm SM}}
$
for $i$th decay products from $X$ and $\beta_i \simeq \beta_{\rm SM} \simeq 1 $. 
It is necessary to have prompt particles from production or decay, or ISR, which arrives at timing layer with 
the speed of light, to derive the time of the hard collision at the primary vertex (to ``time-stamp'' the hard collision).\\

\begin{figure}[t]
    \centering
    \includegraphics[width=0.45\columnwidth]{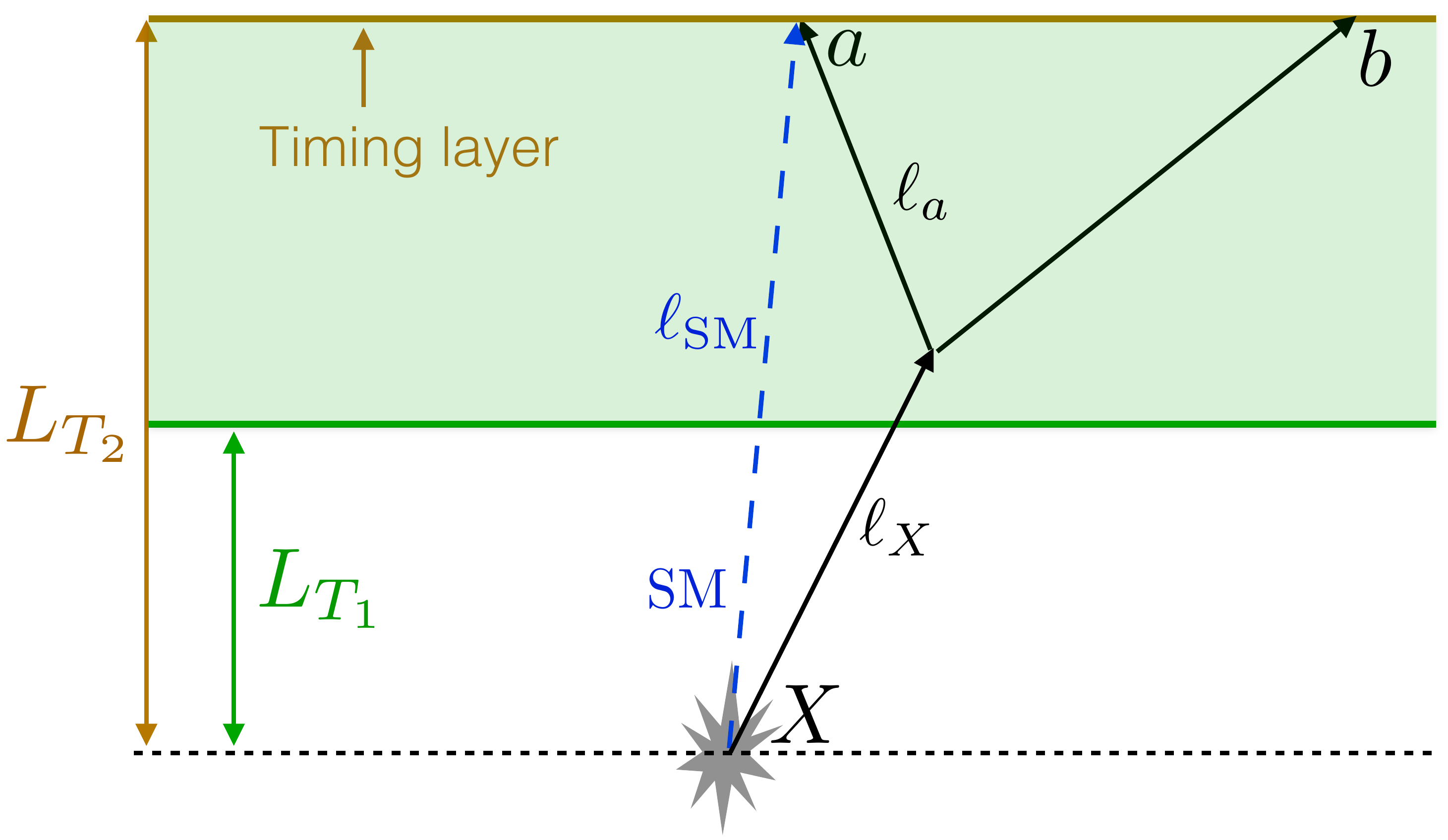} 
    \includegraphics[width=0.45\columnwidth]{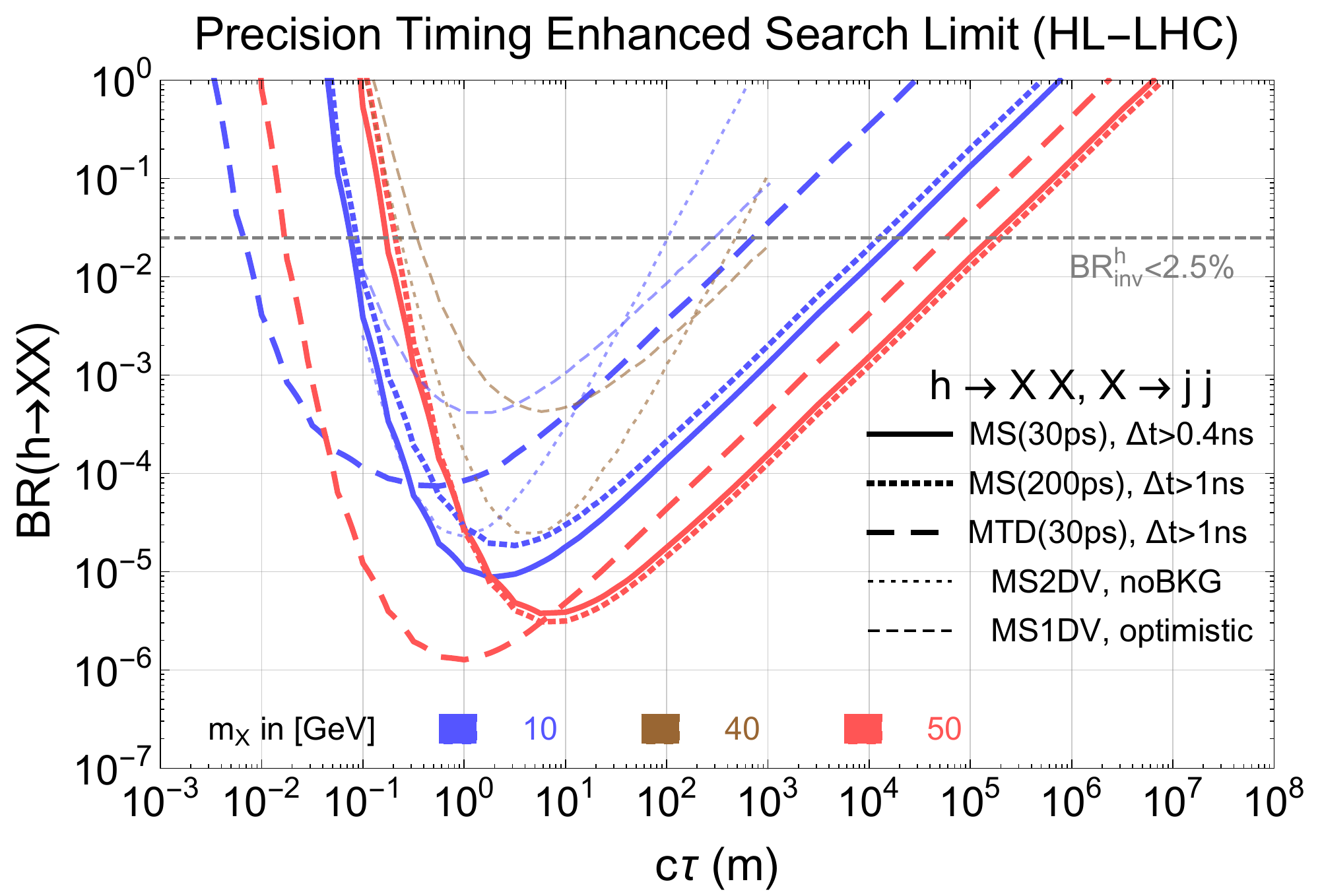} 
    \caption{ Left Panel: An event topology with an LLP $X$ decaying into two light SM particles $a$ and $b$. A timing 
    layer, at a transverse distance $L_{T_2}$ away from the beam axis (horizontal grey dotted line), is placed at the end of the detector volume 
    (shaded region). The trajectory of a reference SM background particle is also shown (blue dashed line).
    The grey polygon indicates the primary vertex.
    Right Panel: the $95\%$ C.L. limit on $\text{BR}(h \to XX)$ for signal process $pp \to j h$ with subsequent decay 
    $h\to X X$ and $X \to j j$. Different colours indicate different masses of the particle $X$. 
    The thick solid and dotted (thick long-dashed) lines indicate MS (MTD) searches with different timing cuts. The numbers in parentheses are 
    the assumed timing resolutions. Other 13 \UTeV LHC projections \cite{Coccaro:2016lnz, Bernaciak:2014pna} are plotted in thin lines. Figures are taken from Ref.~\cite{Liu:2018wte}.
    }
    \label{fig:ctaulimitHiggs}
\end{figure}

We consider events with at least one ISR jet to time-stamp the primary vertex (PV) and one delayed SM object coming from the LLP decay. We propose two searches using the time delay information:
\begin{center}
	\scalebox{0.95}{
    \begin{tabular}{|c|c|c|c|c|c|c|c|}
        \hline
        & $L_{T_2}$ & $L_{T_1}$ & Trigger & $\epsilon_{\rm trig}$ & $\epsilon_{\rm sig}$ & $\epsilon_{\rm fake}^{j}$ & Ref. \\ \hline
        MTD & 1.17 m & 0.2 m & DelayJet & 0.5 & 0.5 & $10^{-3}$ & \cite{Collaboration:2296612} 
        \\ \hline
        MS & 10.6 m & 4.2 m & MS RoI & 0.25, 0.5 & 0.25 & $5.2\times 10^{-9}$ & \cite{Aad:2015uaa}
        \\
        \hline
    \end{tabular}}
\end{center}
The size of the detector volume is described by transverse distance to the beam pipe from $L_{T_1}$ to $L_{T_2}$, where the timing layer
is located. For both searches, we assume a similar timing resolution of 30~ps.
For the MS search, because of the larger time delay and much less background due to ``shielding'' by inner detectors, 
a time resolution of 0.2 - 2 ns could achieve a similar physics reach. The $\epsilon_{\rm trig}$, 
$\epsilon_{\rm sig}$ and $\epsilon_{\rm fake}^j$ are the efficiencies for trigger, signal selection and a QCD jet faking the delayed jet signal with $p_T>30$~\UGeV in MTD and MS searches, respectively. 

For the MTD search, we assume a new trigger strategy of a delayed jet using the CMS MTD.
This can be realised by putting a minimal time delay cut when comparing the prompt time-stamping jet with $p_T > 30$~\UGeV with the arrival time of another jet at the timing layer. 
The MTD signal, after requiring minimal decay transverse distance of 0.2 m ($L_{T_1}$), will not 
have good tracks associated with it. Hence, the major SM background is from trackless jets.
The jet fake rate of $\epsilon_{\rm fake}^{\rm j, MTD} = 10^{-3}$ is estimated
using {\tt Pythia}~\cite{Sjostrand:2007gs} by simulating the jets with minimal $p_T$ of 30~\UGeV and studying the anti-kt jets with $R=0.4$, where all charged constituent hadrons are too soft ($p_T<1~\GeV$). 
For the MS search, we use the MS Region 
of Interest (MS RoI) trigger from a very similar search~\cite{Aaboud:2017iio} as a reference, with an efficiency 
of $\epsilon_{\rm trig}=0.25$ and 0.5 for the two benchmark BSM signals, and a signal selection efficiency of 
$\epsilon_{\rm sig}=0.25$. The backgrounds are mainly from the punch-through jets, and the  fake efficiency can be inferred 
to be $\epsilon_{\rm fake}^{\rm j, MS} = 5.2 \times 10^{-9}$, normalised to 1300 fake MS barrel events at 8 \UTeV~\cite{Aaboud:2017iio}.
For detailed discussion on the background estimation, see Ref.~\cite{Liu:2018wte}.
\\

To emphasise the power of timing, we rely mostly on the timing information to suppress background and make only 
minimal cuts. We only require one low $p_T$ ISR jet, with $p_T^j > 30~\text{\UGeV}$ and $|\eta_j| < 2.5$.
In both signal benchmarks, we require that at least one LLP decays inside the detector. We generate signal events using {\tt MadGraph5}~\cite{Alwall:2014hca} at parton level. 
After detailed simulations of the delayed arrival time, we derive the projected sensitivity using the 
cross-sections obtained in Ref.~\cite{Greiner:2015jha}. The $95\%$ C.L. sensitivity is shown in Fig.~\ref{fig:ctaulimitHiggs}. 
We assume $X$ decays to SM jet pairs with 100\% branching fraction.
The MTD and MS searches, with $30$ ps timing resolution, are plotted in thick dashed and solid lines, respectively. 
For MS, the best reach of $\text{BR}(h\to XX)$ is about a few times $10^{-6}$ for $c \tau \sim 10$~m. 
The reach is relatively insensitive to the mass of $X$ when $m_X> 10$~\UGeV because $X$ are moving slowly enough
to pass the timing cut. 
For the MS search, a less precise timing resolution 
($200$ ps) has also been considered with cut $\Delta t > 1$ ns. After the cut, the backgrounds from same-vertex hard collision (SV) and pile-up (PU) for MS search are $0.11$ and $7.0 \times 10^{-3}$ respectively, and the SV background dominates.  
The reach for heavy $X$ is almost not affected, 
while reduced by a factor of around two for light X.

In the right panel of Fig.~\ref{fig:ctaulimitHiggs}, we compare MTD and MS (thick lines) with 13 \UTeV  HL-LHC (with $3 ~\text{ab}^{-1}$ integrated luminosity) projections, two displaced vertex (DV) at 
MS using zero background assumption (thin dotted) and one DV at MS using a data-driven method with optimistic background estimation (thin dashed) from \cite{Coccaro:2016lnz}. 
The projected limits from invisible Higgs decay at the 14 \UTeV HL-LHC (see Sec. \ref{sec6:exp} of this report) is also shown in the right panel of Fig.~\ref{fig:ctaulimitHiggs}. \\

Exploiting timing information can significantly enhance 
the sensitivities of LLP searches at the HL-LHC. 
To emphasise the advantage of timing, we made minimal requirements on the signal, with one ISR jet
and a delayed signal. 
The temporal behaviour of the SM and detector background are not yet well understood. This novel investigation~\cite{Liu:2018wte} requires further studies on the background behaviour at the HL- and HE-LHC to further realise the proposed trigger and analysis.
Further optimisation can be developed for more dedicated searches. 
The time-stamping ISR jet can be replaced by other objects, such as leptons or photons. Depending on the underlying signal 
and model parameters, one can also use prompt objects from signal production and decay. In addition, for specific searches, 
one could also optimise the selection of the signal based on the decay products of LLPs. 
Finally, we emphasise that the current LLP searches are complimentary to the timing based searches discussed here. Once combined,
the current searches should in general gain better sensitivity for heavy LLP. These future perspectives can be further extended and realised at the HE-LHC with more advanced phenomenological studies with detector, trigger and analysis, as well as higher statistics on the Higgs bosons.

\asubsubsection[C. Caillol]{Projection of CMS search for exotic $\PH\to aa \to 2b 2\tau$}\label{Sec:9.1.3}
This analysis looks for decays of the Higgs boson to pairs of pseudoscalar bosons in the final state of two
$\tau$ leptons and two $b$ quarks~\cite{Sirunyan:2018pzn}. The $\tau\tau$ pair is reconstructed as $e\mu$, $\mu\tau_h$, or $e\tau_h$, depending on the decay modes of the $\tau$ leptons. The symbol $\tau_h$ denotes a $\tau$ lepton decaying hadronically.
Only one $b$ jet with $p_T>20$ \UGeV is required to be reconstructed and tagged as originating from a $b$ quark
because the $b$ jets originating from the pseudoscalar boson are typically soft.
An improved signal sensitivity is obtained by dividing the events in four different
categories depending on the visible invariant mass of the $\PQb$ jet and the $\tau$ candidates, denoted $m_{b\tau\tau}$.
 The thresholds that define the categories depend on the final state.
The categories with low $m_{b\tau\tau}$ are enriched in signal events, while the categories with large $m_{b\tau\tau}$ help to
constrain the backgrounds. The results are extracted with a maximum likelihood fit of the visible $\Pgt\Pgt$ mass spectrum.
The dominant backgrounds at low $\ma$ are $\ttbar$ production as well as events with jets misidentified as $\Pgt$
candidates, whereas the Drell--Yan background starts to contribute for $m_a>45$ \UGeV values. This analysis
is only sensitive to pseudoscalar masses above 15 \UGeV. The sensitivity of the analysis mostly comes from the low $m_{b\tau\tau}$
category, which is statistically limited, and the statistical uncertainty strongly dominates the results.

The extrapolations summarised in Ref.~\cite{CMS-PAS-FTR-18-035},  and presented in this section and the next two assume that the CMS experiment will have a similar level of detector and triggering performance during the HL-LHC operation as it provided during the LHC Run~2
period~\cite{CMSCollaboration:2015zni, Klein:2017nke, Collaboration:2283187, Collaboration:2293646, Collaboration:2283189}.  
The results of extrapolations, hereafter named projections, are presented for different assumptions on the
size of systematic uncertainties that will be achievable by the time of HL-LHC:
\begin{itemize}
\item {\bf ''Run 2 systematic uncertainties'' scenario:} This scenario assumes that performance of the experimental methods at the HL-LHC will be unchanged with respect to the LHC Run 2 period, and there will be no significant improvement in the theoretical descriptions of relevant physics effects. All experimental and
theoretical systematic uncertainties are assumed to be unchanged with respect to the ones in
the reference Run~2 analyses, and kept constant with integrated luminosity.
\item {\bf ''YR18 systematics uncertainties'' scenario:} This scenario assumes that there will be further
advances in both experimental methods and theoretical descriptions of relevant physics effects.
Theoretical uncertainties are assumed to be reduced by a factor two with respect to the ones in the reference
Run~2 analyses. For experimental systematic uncertainties, it is assumed that those will be reduced by the square root
of the integrated luminosity until they reach a defined lower limit based on estimates of
the achievable accuracy with the upgraded detector.
\end{itemize}

In these scenarios, all the uncertainties related to the limited number of simulated events are neglected,
 under the assumption that sufficiently large simulation samples will be available by the time the HL-LHC becomes operational.

For all scenarios, the intrinsic statistical uncertainty in the measurement is reduced by a
factor $1/\sqrt{\text{R}_\text{L}}$, where $\text{R}_\text{L}$ is the projection integrated luminosity divided by that of the
reference Run~2 analysis.

Table~\ref{tab:systematics} summarises the Run~2 uncertainties for which a lower limit value is set in
the ''YR18 systematics uncertainties'' scenario. Systematic uncertainties in the identification and isolation efficiencies
for electrons and muons are expected to be reduced to around $0.5\%$. The hadronic $\tau$ lepton ($\tau_{\mathrm{h}}$) performance is assumed to remain similar to the current level and therefore the associated
uncertainties are not reduced in this scenario. The uncertainty in the overall jet energy scale (JES) is expected to
reach around 1\% precision for jets with $\pt > 30 \GeV$, driven primarily by improvements for the
absolute scale and jet flavour calibrations. The missing transverse momentum uncertainty is obtained by propagating the JES uncertainties in its computation, yielding a reduction by up to a half of the Run~2 uncertainty. For the identification of b-tagged jets, the uncertainty in the selection efficiency of b (c) quarks, and in misidentifying a light jet is expected to remain similar to the current level, with only the statistical component reducing with increasing integrated luminosity. The uncertainty in the integrated luminosity of the data sample could be reduced down to 1\% by a better understanding of the calibration and fit models employed in its determination, and making use of the finer granularity and improved electronics of the upgraded detectors.

\begin{table}[hbtp]
\footnotesize
\centering
\caption{The sources of systematic uncertainty for which limiting values are applied in ''YR18 systematics uncertainties'' scenario.
}

\begin{tabular}{@{} l p{3cm} c c @{}}
Source       &   Component         & Run~2 unc.             & Projection minimum unc. \\
\hline
Muon ID   &           & $1$--$2\%$            & $0.5\%$        \\
Electron ID &           & $1$--$2\%$            & $0.5\%$        \\
Photon ID   &           & $0.5$--$2\%$            & $0.25$--$1\%$    \\
Hadronic $\Pgt$ ID &         & $6\%$               & Same as Run~2    \\
Jet energy scale &  Absolute    & $0.5\%$               & $0.1$--$0.2\%$   \\
         &  Relative    & $0.1$--$3\%$            & $0.1$--$0.5\%$   \\
         &  Pileup      & $0$--$2\%$            & Same as Run~2  \\
         &  Method and sample & $0.5$--$5\%$          & No limit     \\
         &  Jet flavour       & $1.5\%$             & $0.75\%$     \\
         &  Time stability      & $0.2\%$             & No limit     \\
Jet energy res.   &     & Varies with $\pt$ and $\eta$      & Half of Run~2   \\
$\ptvecmiss$ scale       &     & Varies with analysis selection        & Half of Run~2   \\
b-Tagging   & b-/c-jets (syst.)       & Varies with $\pt$ and $\eta$    & Same as Run~2\\
          & light mis-tag (syst.)   & Varies with $\pt$ and $\eta$       & Same as Run~2      \\
           & b-/c-jets (stat.)    & Varies with $\pt$ and $\eta$       & No limit        \\
          & light mis-tag (stat.)   & Varies with $\pt$ and $\eta$       & No limit           \\
Integrated lumi.    &         & $2.5\%$                         & $1\%$          \\
\end{tabular}
\label{tab:floors}
\vspace{0.5cm}
\label{tab:systematics}
\end{table}

Upper limits at 95\% CL on $(\sigma(h)/\sigma_{\textrm{SM}}) \mathcal{B}(h \to aa \to 2b2\tau)$ are shown in Fig.~\ref{fig:bbtt_proj} for different integrated luminosities and systematic uncertainty scenarios. In this expression, $\sigma_\textrm{SM}$ denotes the SM production cross section of the Higgs boson, whereas $\sigma(\Ph)$ is the $\Ph$ production cross section.
The limits improve proportionally to the square root of the integrated luminosity,
as the analysis is statistically limited. For an integrated luminosity of 3000\fbinv, the difference between the limits in the  systematic scenarios of Run 2 and YR18 is of the order of 5\%, and the limits become another 5\% better if all systematic uncertainties are neglected.

\begin{figure*}[hbpt]
\centering
        \includegraphics[width=0.49\textwidth]{\main/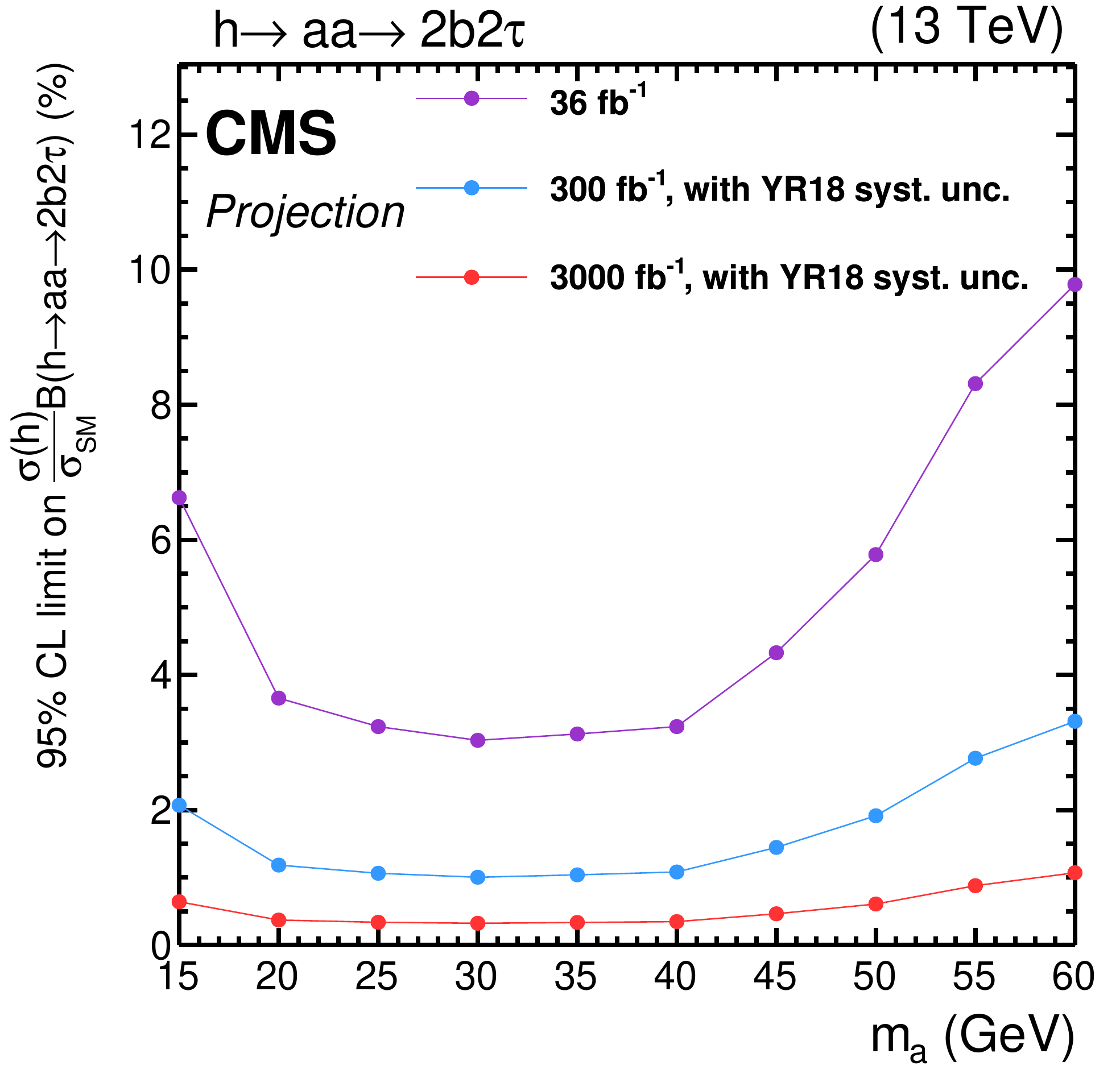}
        \includegraphics[width=0.49\textwidth]{\main/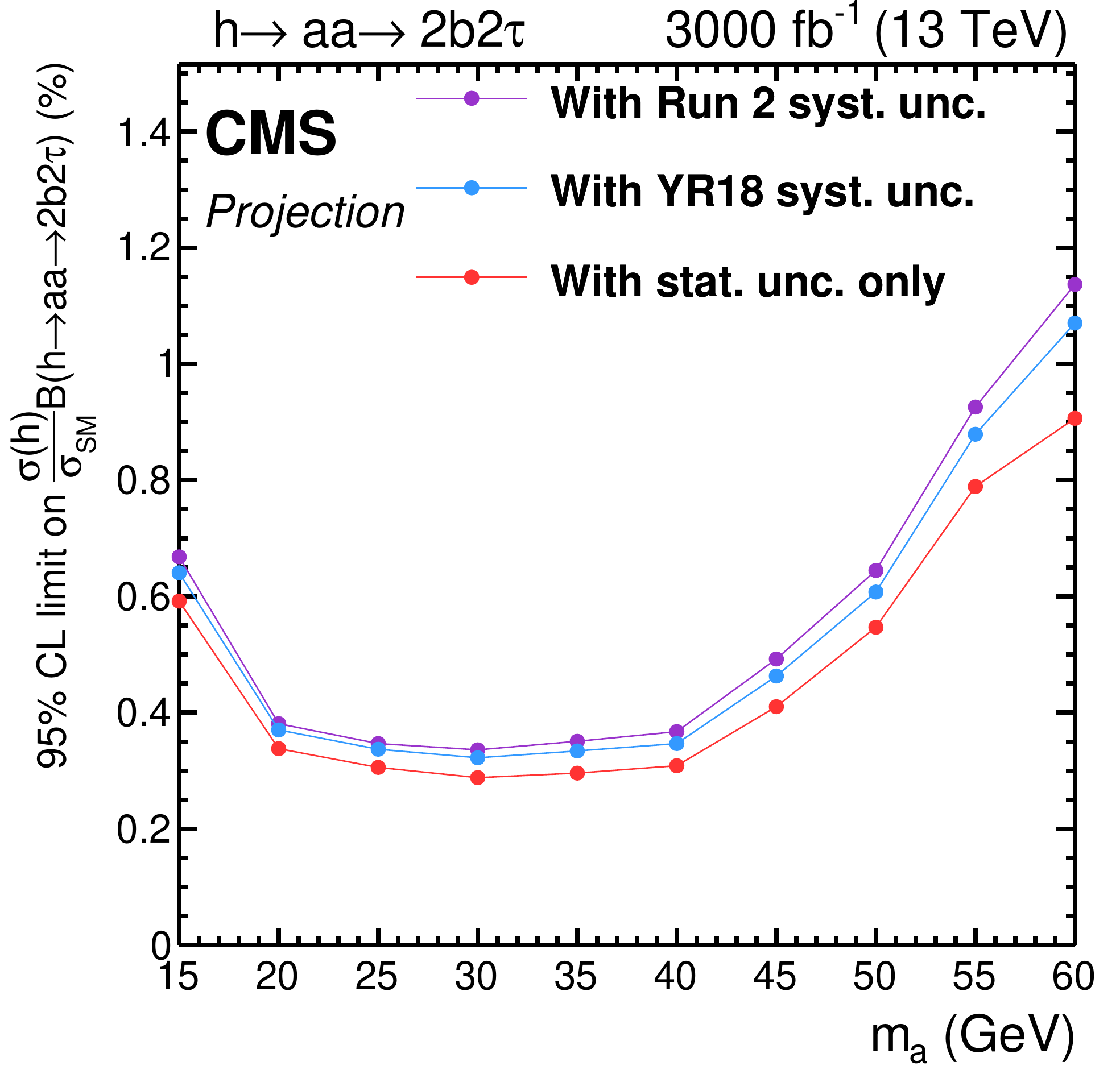}\\
    \caption{Left: Projected expected limits on $(\sigma(h)/\sigma_{\textrm{SM}})$ times the branching fraction for $h\to aa\to 2b 2\tau$, for 36, 300, and 3000\fbinv. Right: Projected expected limits $(\sigma(h)/\sigma_{\textrm{SM}}) \mathcal{B}(h\to aa \to 2b2\tau)$, comparing different scenarios for systematic uncertainties for an integrated luminosity of 3000\fbinv.}
    \label{fig:bbtt_proj}
\end{figure*}

The limits of the $h\to aa$ analyses can be converted to limits on $\mathcal{B}(h\to aa)$
in two-Higgs-doublet models extended with
a scalar singlet (2HDM+S)~\cite{PhysRevD.90.075004}, for a given type of model, $m_a$, and $\tan\beta$. The limits in the four types of 2HDM+S are shown
in Fig.~\ref{fig:summary_bbtt}, assuming 3000\fbinv of data with YR18 systematic uncertainties. The colour scale
indicates the upper limits on $(\sigma(h)/\sigma_{\textrm{SM}}) \mathcal{B}(h\to aa)$ that can be set assuming some values for $m_a$ and $\tan\beta$.

\begin{figure*}[hbpt]
\centering
        \includegraphics[width=0.49\textwidth]{\main/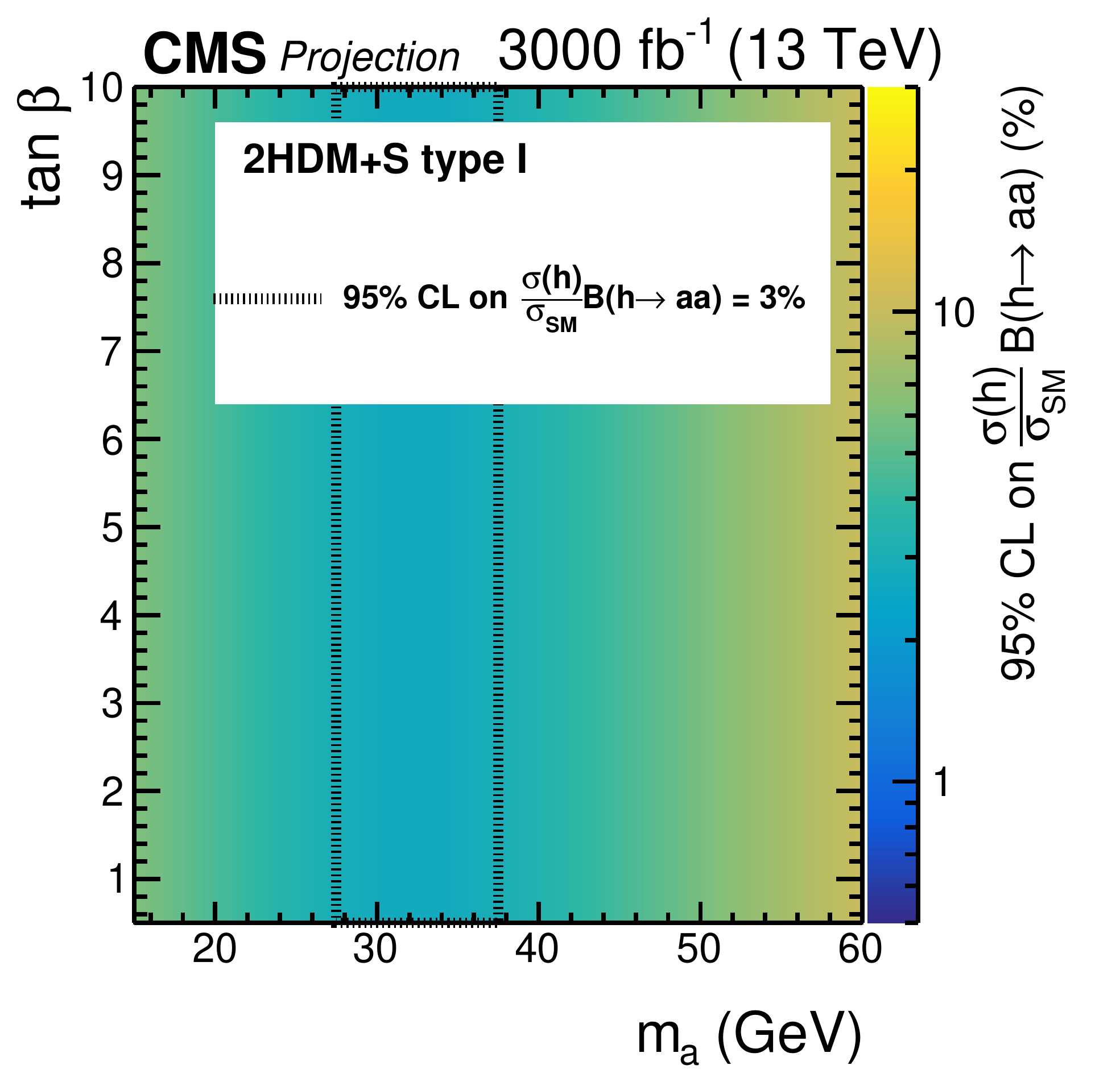}
        \includegraphics[width=0.49\textwidth]{\main/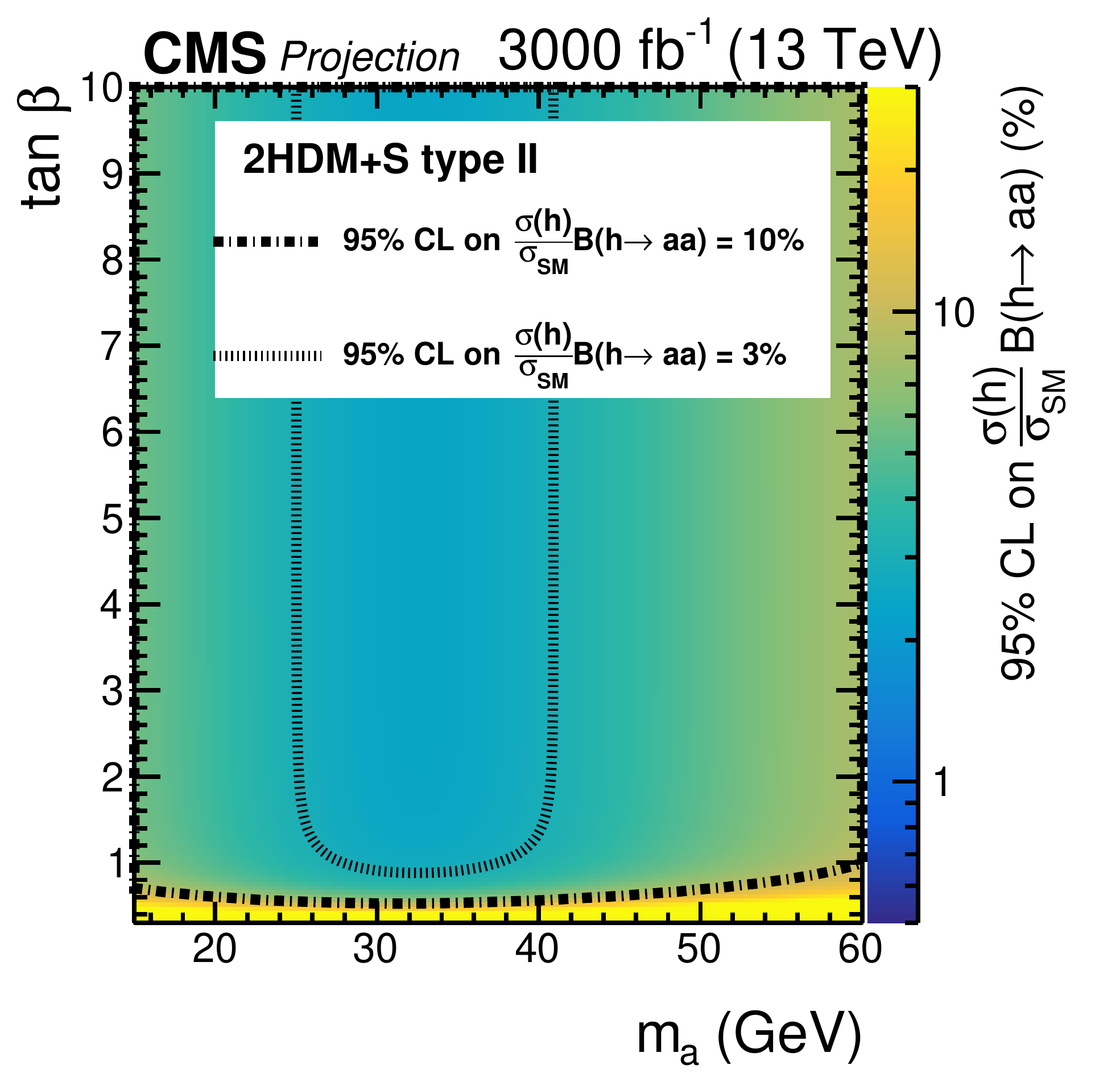} \\
        \includegraphics[width=0.49\textwidth]{\main/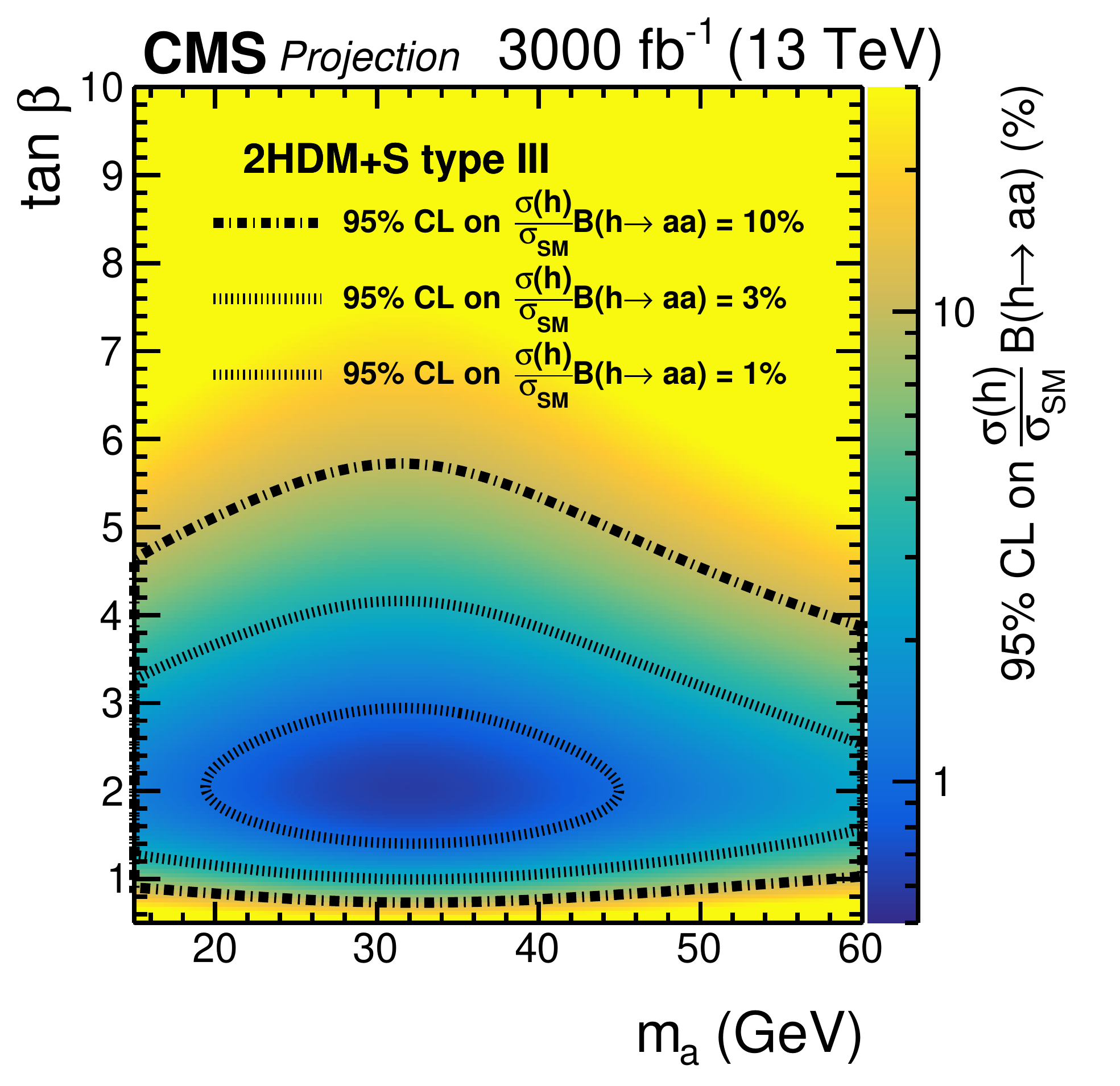}
        \includegraphics[width=0.49\textwidth]{\main/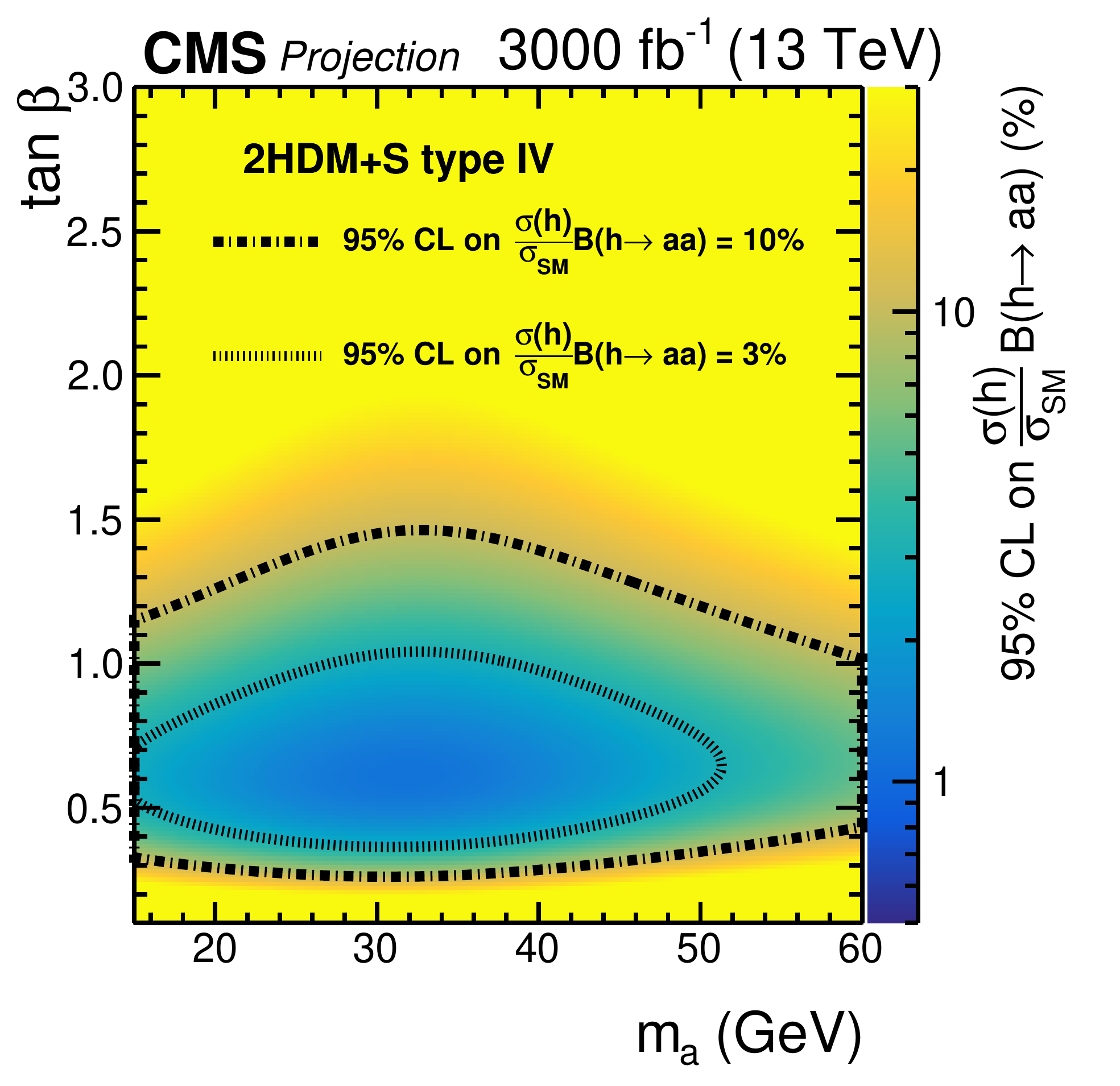}
    \caption{Expected upper limits on $(\sigma(h)/\sigma_{\textrm{SM}}) \mathcal{B}(h\to aa)$ for 3000 fb$^{-1}$ of data with YR18 systematic uncertainties for the $2b 2\tau$ final state in 2HDM+S type-1 (top left), type-2 (top right), type-3 (bottom left), and type-4 (bottom right).}
    \label{fig:summary_bbtt}
\end{figure*}

\asubsubsection[C. Caillol]{Projection of CMS search for exotic $\PH\to aa \to 2\mu 2\tau$}
This analysis searches for the exotic decay of the Higgs boson to a pair of light pseudoscalars, in the
final state of two muons and two $\tau$ leptons~\cite{Sirunyan:2018mbx}. Pseudoscalar
masses between 15 and 62\GeV are investigated; in this mass range the decay products from the pseudoscalars are
not collimated.
Several $\tau\tau$ pair possibilities are studied in this analysis: $e\mu$, $e\tau_h$,
$\mu\tau_h$, and $\tau_h\tau_h$. In the case where there are 3 muons, the highest-$p_t$ one is paired with
the opposite-sign muon that has the highest-$\pt$ among the other two, while the last muon is considered as
originating from a $\tau$ lepton decay.
To reduce the backgrounds from \PZ\PZ, Z+jets, and \PW\PZ+jets productions, the invariant mass of the muon pair
is required to be above the visible invariant mass of the $\tau\tau$ pair, and the
visible invariant mass of the four objects is required to be less than 110--130 \UGeV depending on the final state.
The limits are extracted with an unbinned maximum likelihood fit of the di-muon mass spectrum.
The backgrounds are characterised by a rather flat di-muon mass spectrum, while the signal $h\to aa \to 2\mu2\tau$
forms a narrow peak in the di-muon mass spectrum.
The number of expected background events below the signal peak is almost zero,
especially at low di-muon mass, and the analysis is strongly statistically dominated.

In the ''YR18 systematics uncertainties'' scenario, in addition to the limiting values detailed in Table \ref{tab:floors}, the uncertainty in the normalisation of the reducible background is not allowed to go lower than 20\% of the value used in Run-2. The corresponding limits for the $h \to aa \to 2\mu2\tau$ search are shown in Fig.~\ref{fig:mmtt_proj}.
They scale approximately inversely with the integrated luminosity at low $m_a$ because the
analysis is close to background-free, while they tend to scale inversely with the square root of the integrated luminosity at higher $m_a$, where the background is more important. This leads to the large improvement at low $m_a$ for 3000\fbinv of collected data shown in Fig.~\ref{fig:mmtt_proj}. The analysis is statistically limited, even with 3000\fbinv of data. The difference between the Run 2 and YR18 systematic uncertainties in terms of upper limits
is up to 5\%, and is the largest at high $m_a$.

\begin{figure*}[hbpt]
\centering
        \includegraphics[width=0.49\textwidth]{\main/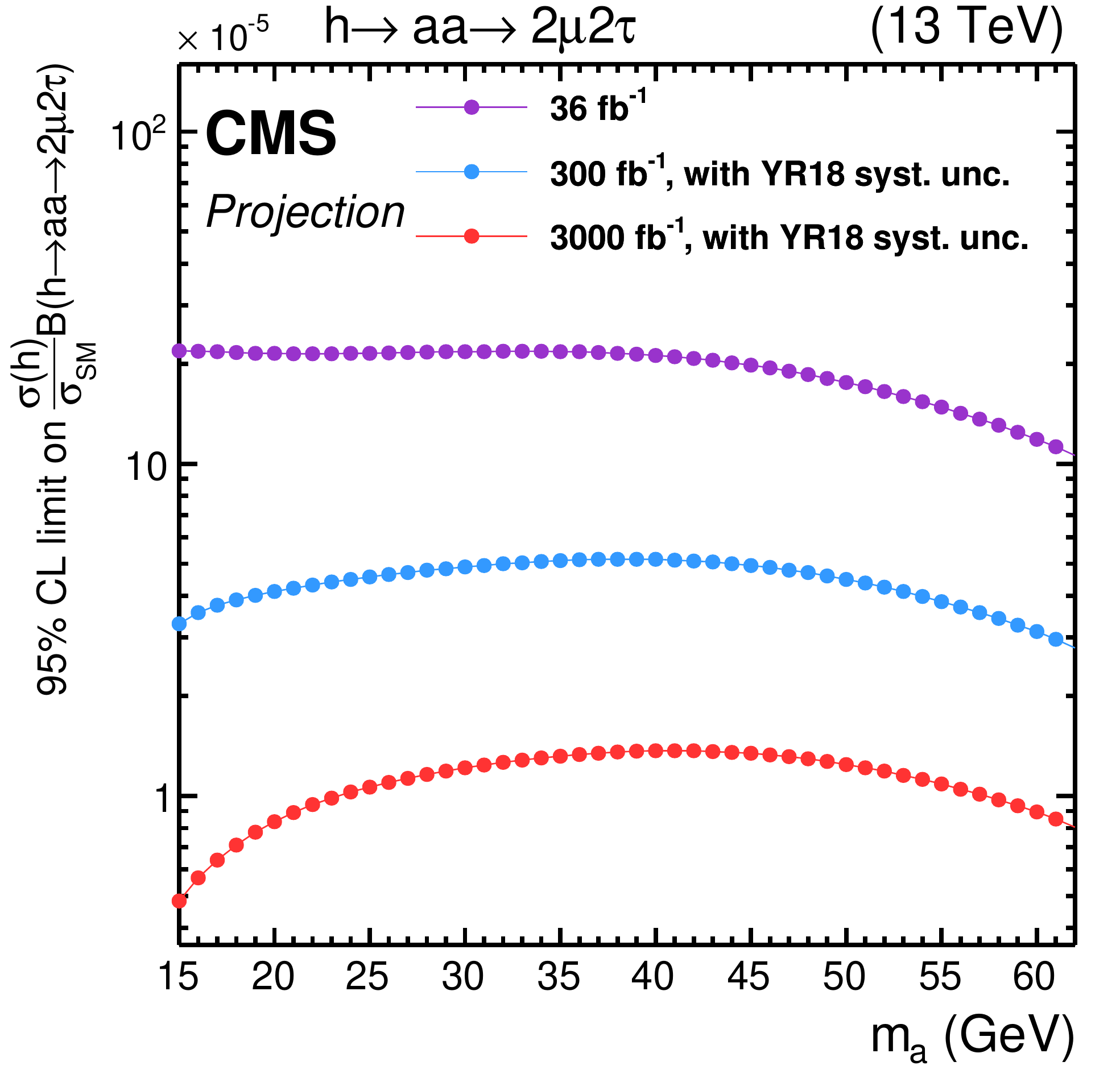}
        \includegraphics[width=0.49\textwidth]{\main/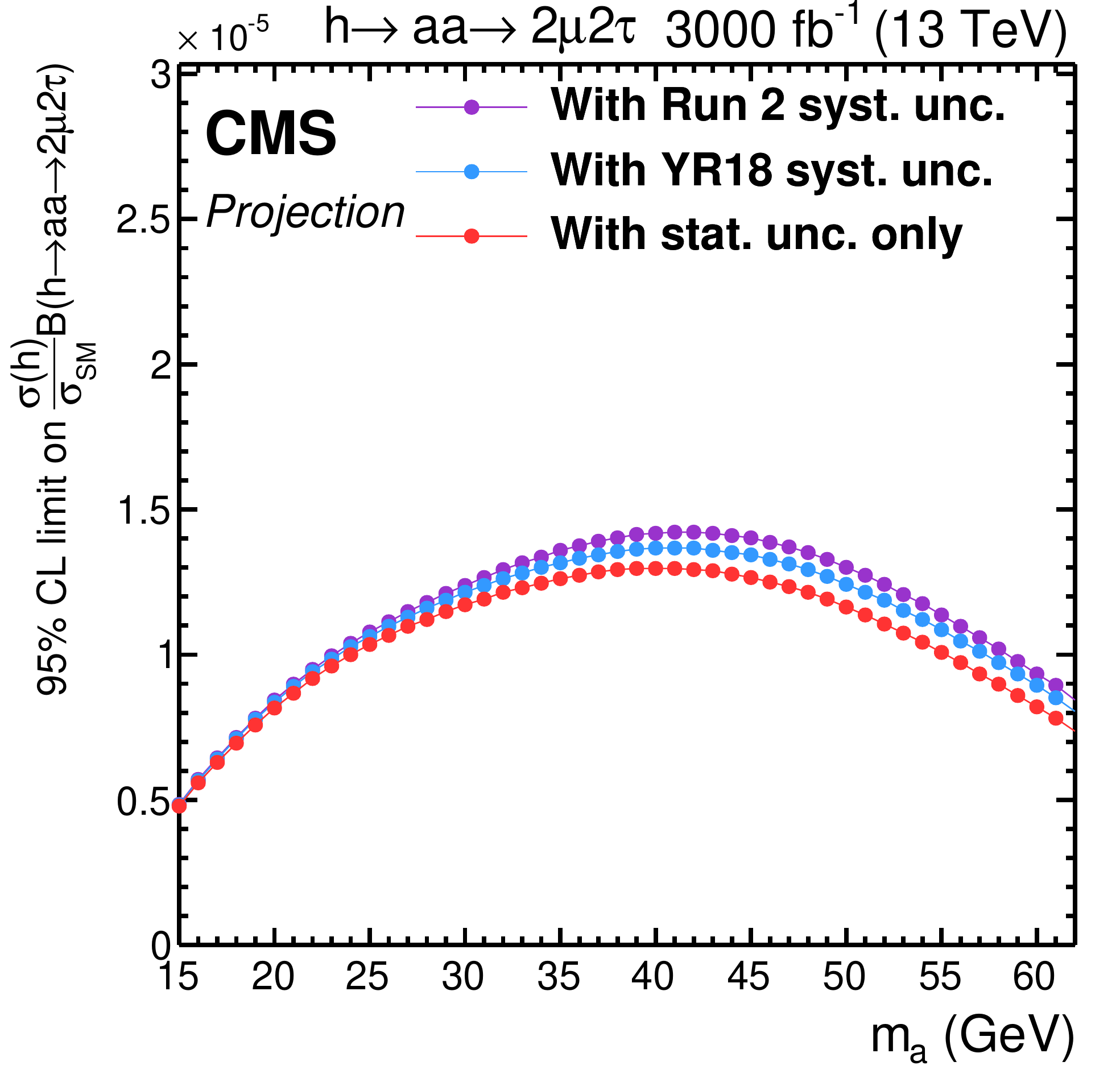}\\
    \caption{Left: Projected expected limits on $(\sigma(h)/\sigma_{\textrm{SM}}) \mathcal{B}(h \to aa \to 2\mu 2\tau)$, for 36, 300, and 3000 fb$^{-1}$. Right: Projected expected limits on $(\sigma(h)/\sigma_{\textrm{SM}}) \mathcal{B}(h\to aa \to 2\mu2\tau)$, comparing different scenarios for systematic uncertainties for an integrated luminosity of 3000\fbinv.}
    \label{fig:mmtt_proj}
\end{figure*}

The limits in the four types of 2HDM+S are shown
in Fig~\ref{fig:summary_mmtt} for the $2\mu 2\tau$ analysis, assuming 3000\fbinv of data in the ''YR18 systematics uncertainties'' scenario.

\begin{figure*}[hbpt]
\centering
        \includegraphics[width=0.49\textwidth]{\main/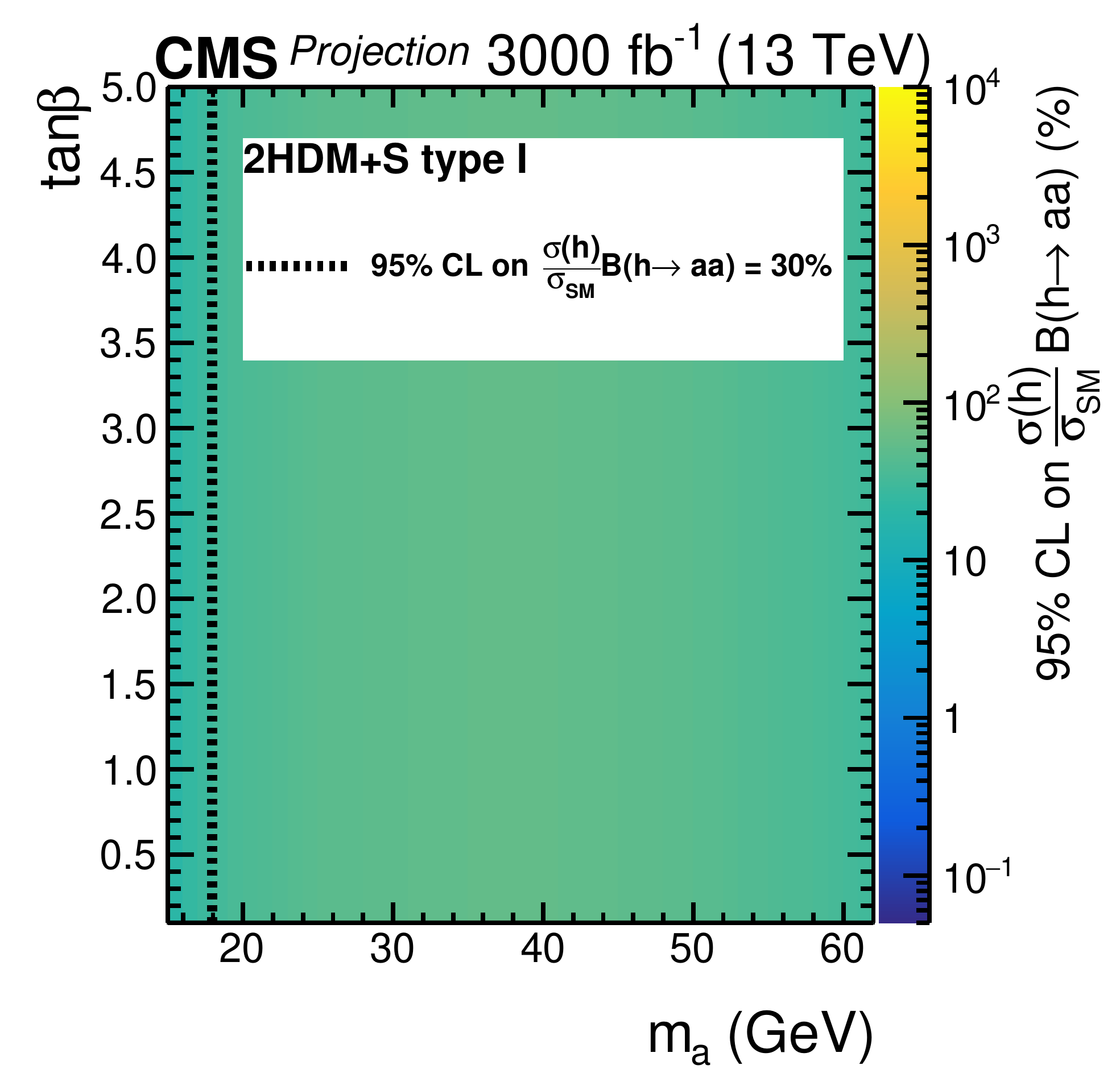}
        \includegraphics[width=0.49\textwidth]{\main/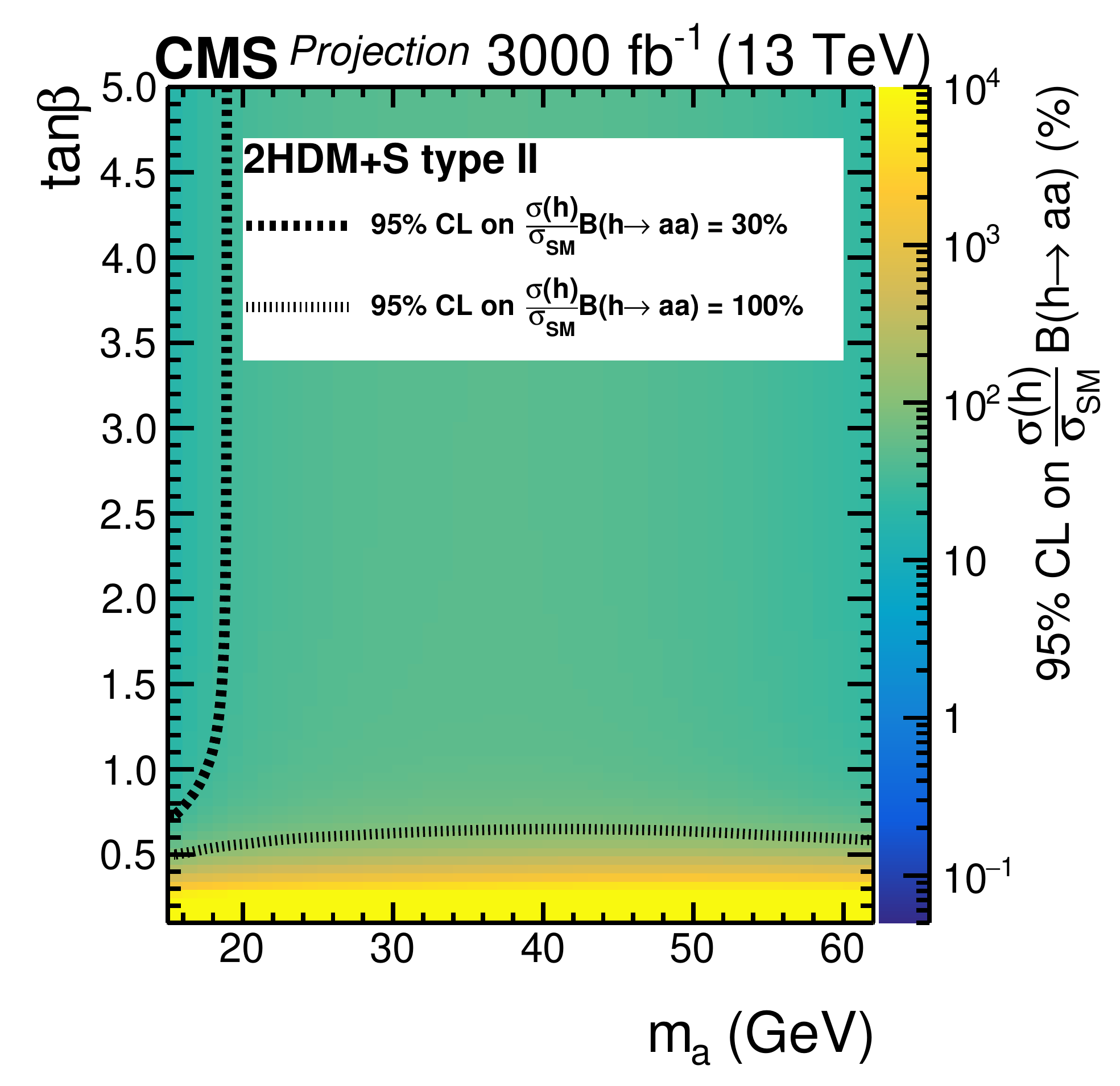} \\
        \includegraphics[width=0.49\textwidth]{\main/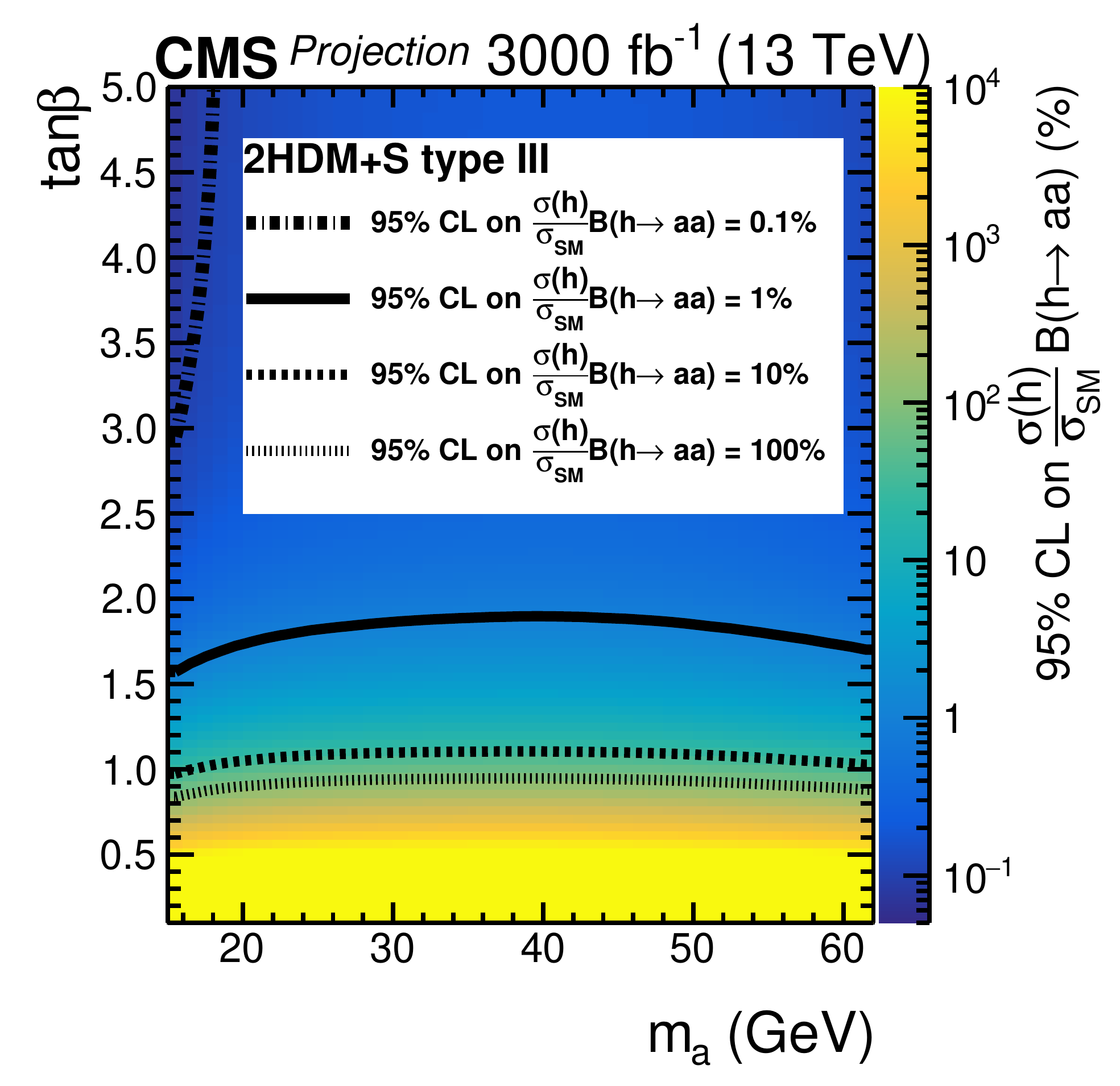}
        \includegraphics[width=0.49\textwidth]{\main/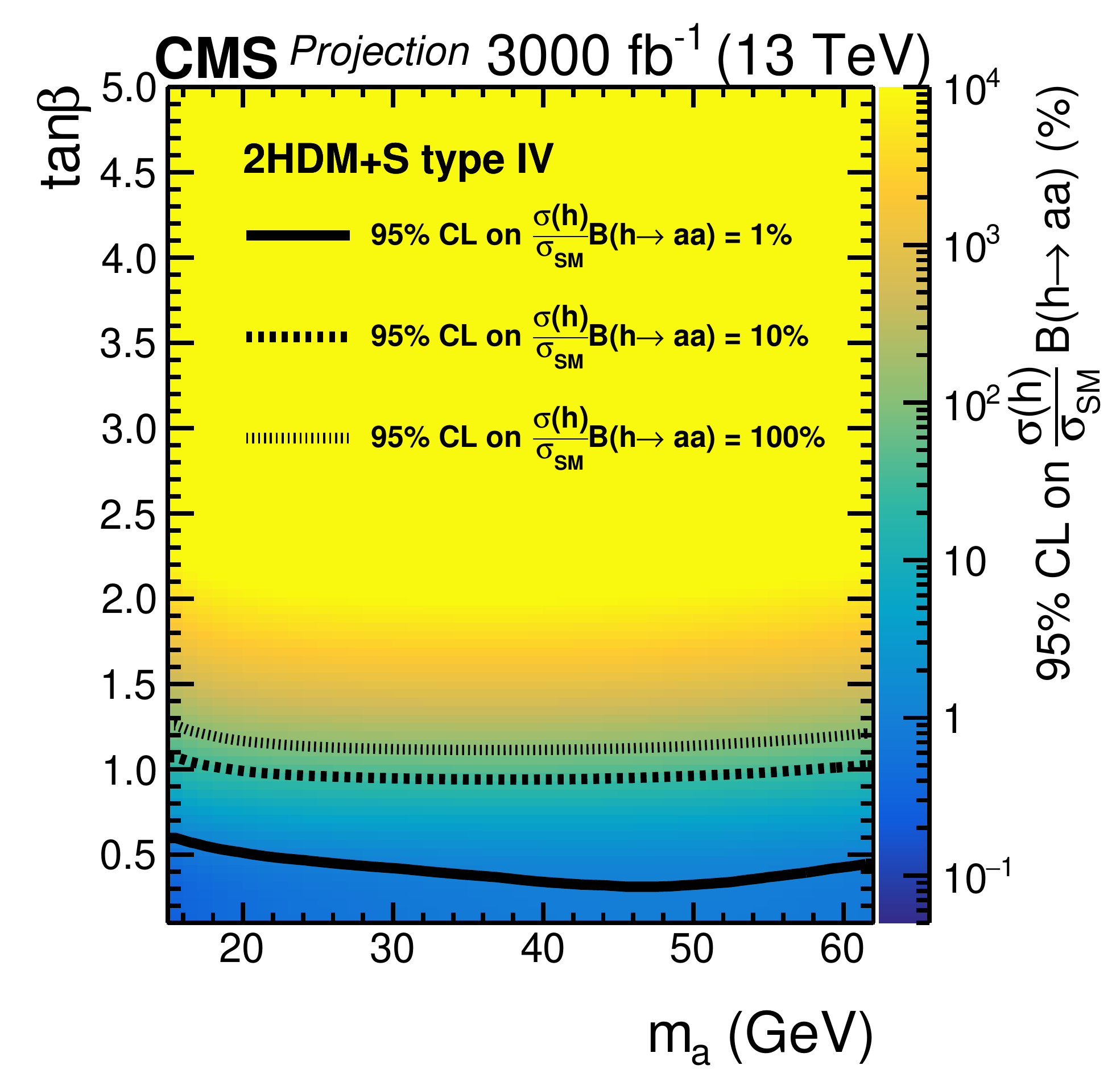}
    \caption{Expected upper limits on $(\sigma(h)/\sigma_{\textrm{SM}})\mathcal{B}(h\to aa)$ for 3000\fbinv of data with YR18 systematic uncertainties for the $2\mu 2\tau$ final state in 2HDM+S type-1 (top left), type-2 (top right), type-3 (bottom left), and type-4 (bottom right).}
    \label{fig:summary_mmtt}
\end{figure*}

\asubsubsection[Y.M. Zhong]{Exotic  decays of the Higgs to $2b2\mu$}\label{Sec:2b2muExo}
We assess the potential of an exotic Higgs decay search for $h \to 2X \to b\bar{b}\mu^+ \mu^-$ to constrain theories with light CP-even ($X = s$) and CP-odd ($X = a$) singlet scalars at the HL-LHC. This contribution is based on~\cite{Curtin:2014pda}. The decay channel $h \to 2X \to b\bar{b}\mu^+ \mu^-$ may represent the best discovery avenue for many models, such as the 2HDM model with an additional complex scalar singlet. It has competitive reach, and is less reliant on low-$p_{\rm T}$ $b-$ and $\tau-$reconstruction compared to other channels like $4b$, $4\tau$, and $2\tau2\mu$.

To estimate the reach of $h \to 2X \to b\bar{b}\mu^+ \mu^-$ search at the 14 \UTeV LHC, we take $X =a$ for simplicity. (The results for a scalar, $X=s$, will be similar.) The dominant backgrounds are Drell-Yan (DY) production with associated jets, \ie $Z/\gamma^*+2b/2c/2j$, where $Z/\gamma^*$ produces a muon pair. A secondary background arises from $t\bar t$ production. Backgrounds from di-boson production ($\PZ\PZ, WW, \PW\PZ$) have small enough cross sections so that we can neglect them. It is also possible for QCD multi-jet events, with two jets being mis-identified as muons, to contribute to the background. We find this can be neglect for analysis with $b$-tags. Signal, as well as DY and $t\bar t$ backgrounds, are simulated at LO by \texttt{Sherpa 2.1.1}~\cite{Gleisberg:2008ta} for $\sqrt{s}=14$ \UTeV with the CT10 PDF, and matched up to three jets. The Higgs production cross section for the signal is normalised to the cross section presented in Sec.~\ref{sec2_HXSWG1} of this report, $\sigma_{ggF} \simeq 54.72$ pb.

Two types of analyses have been included. A conventional analysis that uses standard anti-$k_t$ jets with a jet radius of $R\sim 0.4$. Two $b$-tags at 70\% $b$-tagging efficiency working point~\cite{ATLAS:2014cal}  are imposed to the final states. A missing transverse energy cut of $\slashed{E}_T < 30 \UGeV$ suppresses the $t \bar t$ background. In addition we make use of the double-resonance structure of the signal by imposing invariant mass cuts 
\begin{align}
|m_{b_1 b_2 \mu_1 \mu_2}-m_h|< 15\UGeV,\quad |m_{b_1 b_2}-m_a|< 15\UGeV,\quad|m_{\mu_1\mu_2}-m_a| < 1\UGeV,
\end{align}
separately for each $m_a$. After imposing the above cuts, we then perform a simple counting experiment to estimate the reach. The expected bounds are approximately independent of scalar mass for $m_a \geq 30 \UGeV$. For $m_a < 20 \UGeV$, the signal efficiency drops dramatically because the two $b$'s from the $a$-decay become collimated. The second analysis is based on the mass drop tagger (MDT)~\cite{Butterworth:2008sd}, a jet substructure technique, to improve the search sensitivity for the low -$m_a$ region. After clustering a $b$-tagged Cam- bridge/Aachen (C/A) jet with a jet radius of $R=0.8$, we resolve its hardest sub-jets that satisfy the MDT criteria ($\mu < 0.67$, $y>0.09$) by undoing the last step of the C/A clustering. We then apply the same missing energy and invariant mass cuts to the sub-jets as the conventional analysis.

The results of the combined substructure and conventional analyses are shown in Fig.~\ref{bbmumu}. The figure shows a fairly flat sensitivity of  $\mathrm{Br}(h \to 2a \to 2b2\mu) \lesssim \text{few}\times10^{-4}$ for 14 \UTeV LHC with 30 ${\rm fb}^{-1}$ data in the range $15\UGeV \leq m_a \leq 60 \UGeV$. With either 300 or 3000 ${\rm fb}^{-1}$ of data, the projected sensitivity increases to  $10^{-4}$, and $\text{few}\times 10^{-5}$, respectively. For HE-LHC (27 \UTeV with $15\abinv$), we expected the number of signal and DY background events to be respectively increased by a factor of $\sim15$ and $\sim12$ in comparison with those of the HL-LHC. This leads to a HE-LHC reach at around $\lesssim 10^{-5}$, i.e., a factor of $15/\sqrt{12}\approx 4$ better than those of the HL-LHC. In the figure, we also show the 95\% CL bounds from the 13 \UTeV ATLAS analysis with 36.1 ${\rm fb}^{-1}$ data~\cite{Aaboud:2018esj} for this Higgs decay mode (grey shaded region). For a range of $m_a$ values, the ATLAS bounds are better than our projections  by a factor of $\sim 2$. This may due to more dedicated analysis techniques such as kinematic-likelihood fit~\cite{Aaboud:2018esj}, which improve the invariant mass resolutions. Based on the above comparison, we expect the real HL-LHC and HE-LHC reach could be better than our conservative projections.

\begin{figure}[h]
\begin{center}
\includegraphics[width=0.65\textwidth]{\main/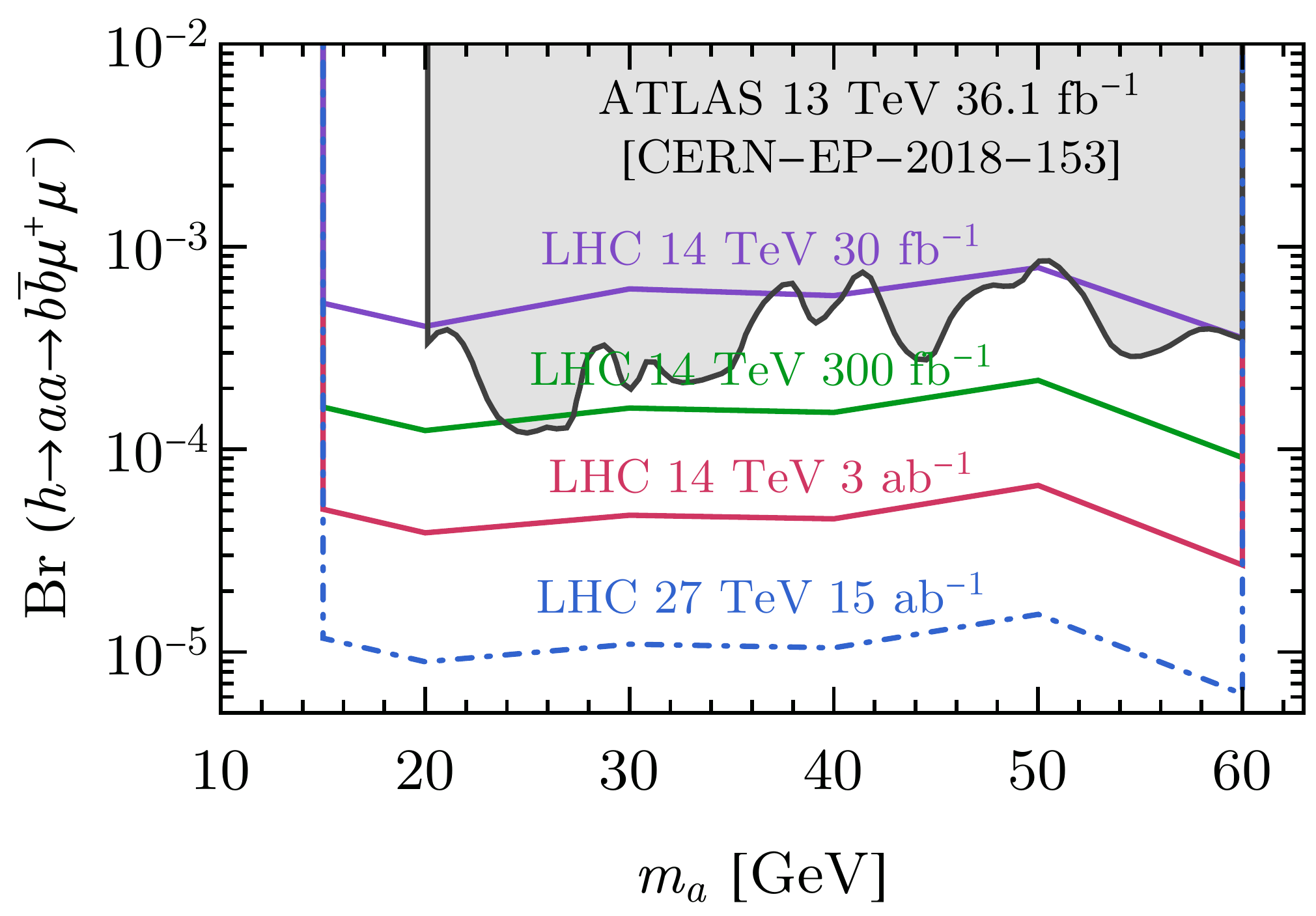}
\caption{\small Combined $95\%$ CL projected reaches on $\Br(h\to aa \to b\bar{b}\mu^+ \mu^-)$ for 30 (purple), 300 (green), and 3000 (red) ${\rm fb}^{-1}$ at 14 \UTeV~\cite{Curtin:2014pda} and 15 $\abinv$ at 27 \UTeV (dash-dotted blue). The 95\% CL upper limits from the 13 \UTeV ATLAS search with 36.1 ${\rm fb}^{-1}$ data~\cite{Aaboud:2018esj} is shown as the grey shaded region.}
\label{bbmumu}
\end{center}
\end{figure}

\asubsubsection[D. Curtin]{Exotic Higgs Decays to dark photons}\label{Sec:4lExo}

Dark photons, or simply a broken or unbroken Abelian gauge interaction, are natural ingredients of hidden sectors. (See e.g.~\cite{Jaeckel:2010ni,Hewett:2012ns,Essig:2013lka,Alexander:2016aln,Battaglieri:2017aum} for recent reviews.) Its ubiquity in such theories is particularly important because it can connect the hidden sector to the SM via two portals: the \emph{photon portal} (strictly speaking hyper-charge portal) and the \emph{Higgs portal}. The former refers to a renormalisable kinetic mixing between the dark photon~\cite{Holdom:1985ag,Galison:1983pa,Dienes:1996zr} and the SM hyper-charge gauge boson, while the latter refers to the mixing between the SM Higgs and a ``dark Higgs'', $S$, that may be responsible for generating a nonzero dark photon mass. 
The most general minimal Abelian dark photon model, with no other hidden sector matter but with a dark Higgs, was studied in detail in~\cite{Curtin:2014cca}. It was found that exotic Higgs decays are an important probe of such scenarios, and a Madgraph \cite{Alwall:2011uj} model, the Hidden Abelian Higgs Model, was supplied to conduct the necessary Monte Carlo studies. In this section, we briefly summarise the main results, include constraints from recent searches, and obtain new sensitivity projections for the HE-LHC.

There are two relevant groups of terms in the model Lagrangian. One is responsible for kinetic mixing between SM hyper-charge, $U(1)_Y$, and the broken dark Abelian gauge symmetry, $U(1)_D$:
\begin{equation}\label{eq:KM}
\mathcal{L} \subset -\frac{1}{4} \,\hat B_{\mu\nu}\, \hat B^{\mu\nu} - \frac{1}{4} \,\hat Z_{D\mu\nu}\, \hat Z_D^{\mu\nu}  + \frac{1}{2}\,\frac{\epsilon}{\cos\theta} \,\hat Z_ {D\mu\nu}\,\hat B^{\mu\nu} + \frac{1}{2}\, m_{D,0}^2\, \hat Z_D^\mu \, \hat Z_{D\mu}\, .
\end{equation}
The hatted fields indicate the original fields with non-canonical
kinetic terms, before any field re-definitions. 
The $U(1)_Y$ and
$U(1)_D$ field strengths are respectively $\hat B_{\mu\nu}
=\partial_\mu \hat B_{\nu} - \partial_\nu \hat B_{\mu}$ and $\hat
Z_{D\mu\nu} =\partial_\mu \hat Z_{D\nu} - \partial_\nu \hat Z_{D\mu}$,
$\theta$ is the Weinberg mixing angle, and $\epsilon$ is the kinetic
mixing parameter.
The most general renormalisable potential for the SM and dark Higgs fields is 
\begin{equation}
\label{eq:HM}
V_0 (H,S) =   -\mu^2|H|^2 +  \lambda |H|^4 -\mu_S^2 |S|^2 + \lambda_S |S|^4 +
   \kappa  |S|^2|H|^2\, .
\end{equation}
Here $H$ is the SM Higgs doublet, while $S$ is the SM-singlet `dark
Higgs' with $U(1)_D$ charge $q_S$.  The Higgs portal coupling, $\kappa$, which links the dark and SM Higgs fields is again a renormalisable parameter that controls the mixing between the SM Higgs boson, $h$, and the uneaten component of the dark Higgs, $s$.

\begin{figure}
\begin{center}
\begin{tabular}{ccc}
\includegraphics[width=0.4 \textwidth]{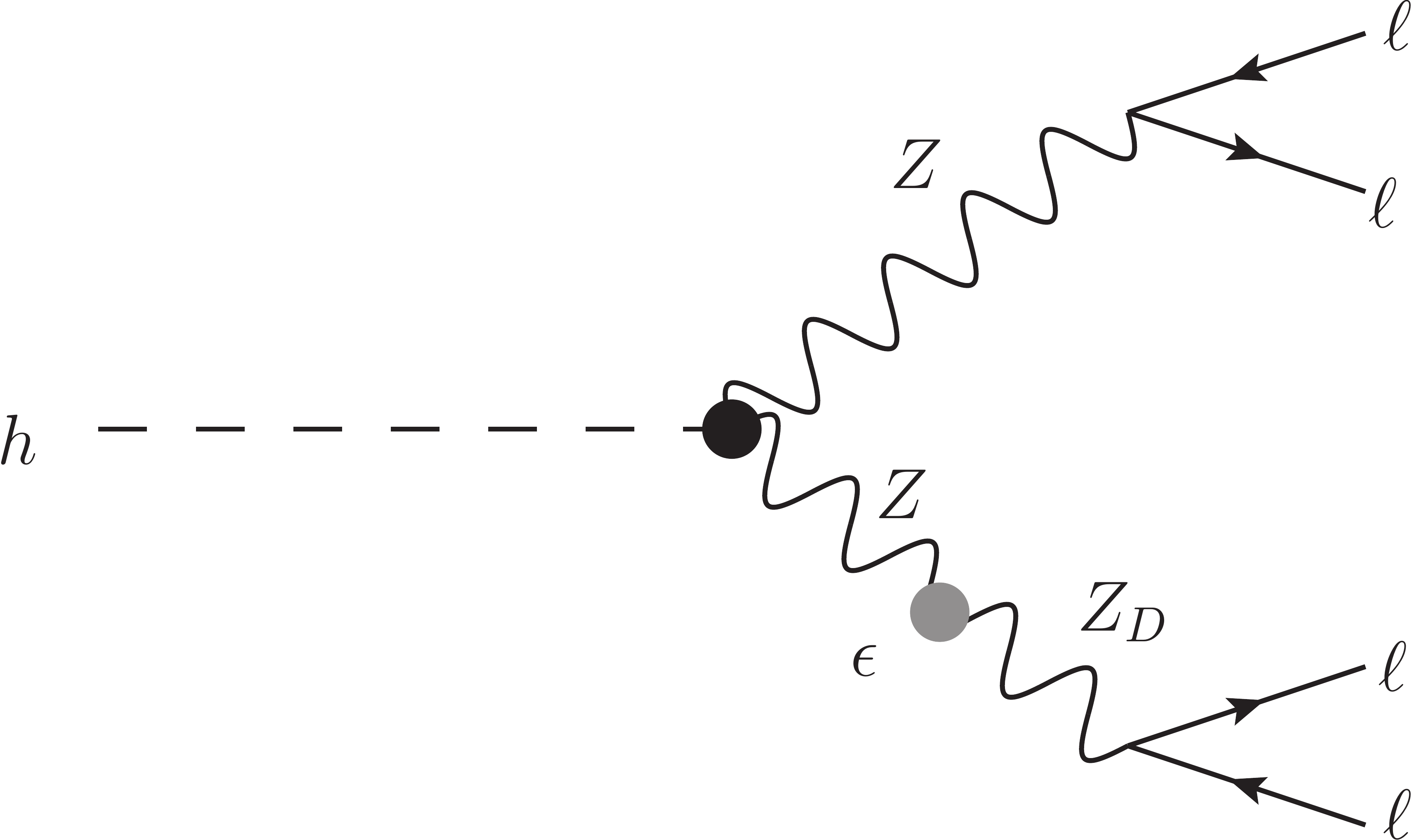}
& \hspace{4mm} &
\includegraphics[width=0.4\textwidth]{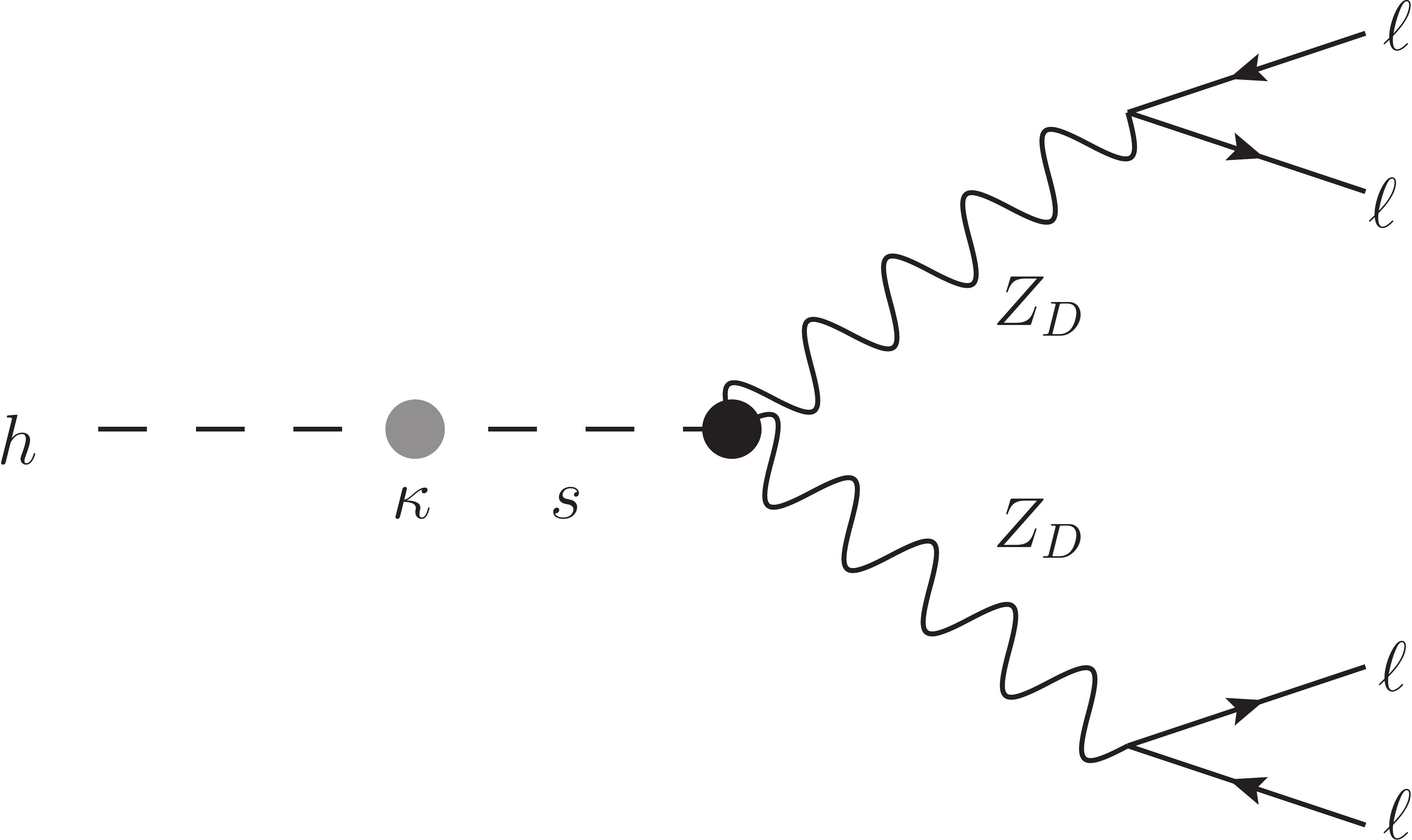}
\end{tabular}
\end{center}
\caption{ Exotic Higgs decays to four leptons induced by intermediate
  dark photons in the Higgsed dark $U(1)_D$ model. \emph{Left:} $h\to Z_D
  Z^{(*)} \to 4\ell$ via the photon portal. \emph{Right:} $h \to Z_D Z_D
  \to 4\ell$ via the Higgs portal. }
\label{f.ZDfeynman}
\end{figure}


This simplified model gives rise to two kinds of exotic Higgs decays, shown in Fig.~\ref{f.ZDfeynman}. The first is the decay through the \emph{photon portal}: kinetic mixing between $Z$ and $Z_D$ allows for $h\to Z_D Z^{(\star)}$, with $\mathrm{Br} \propto \epsilon^2$. The second is the decay through the \emph{Higgs portal}: mixing between $h$ and $s$ allows for $h \to Z_D Z_D$ with $\mathrm{Br} \propto \kappa^2$. We discuss these decays in more detail below, but we note that dark photon models can give rise to other signals as well. 
Kinetic mixing gives rise to DY-like production of dark photons and a resulting di-lepton resonance via $p p \to Z_D \to \ell^+ \ell^-$. This probes the same coupling as $h \to Z_D Z^{(\star)}$ and, as we discuss below, tends to have greater sensitivity. 
If the dark Higgs and dark photon masses are in a suitable range, the so-called ``platinum channel'' becomes available~\cite{Izaguirre:2018atq}, where $h \to 2s \to 4 Z_D \to 8 \ell$ with $\mathrm{Br} \propto \kappa^2$. We do not discuss this channel in detail here, but this final state, if kinematically available, is extremely conspicuous, and the corresponding low-background search could have significantly greater sensitivity to exotic Higgs decay branching ratios than the example of $h \to Z_D Z_D$ we study in this section. 
The mass spectrum could also allow for exotic $Z$-decays~\cite{Blinov:2017dtk} via an intermediate dark Higgs, $Z \to Z_D s \to Z_D Z_D Z_D$. 
Finally, all these signatures could be dressed up or augmented by signatures of a non-minimal hidden sector, where the dark photon/Higgs could decay into invisible stable particles and/or LLPs (see e.g. \cite{Alexander:2016aln, Curtin:2018mvb}).
%
The space of possible signatures is clearly very rich. Even so, the simple benchmark decays we examine here give a feeling for the physics reach of the HL- and HE-LHC in probing these kinds of theories. 

\bigskip

\paragraph{Decays through the photon portal}

\begin{figure}
\begin{tabular}{m{0.5 \textwidth} m{0.4\textwidth}}
\includegraphics[width=0.5\textwidth]{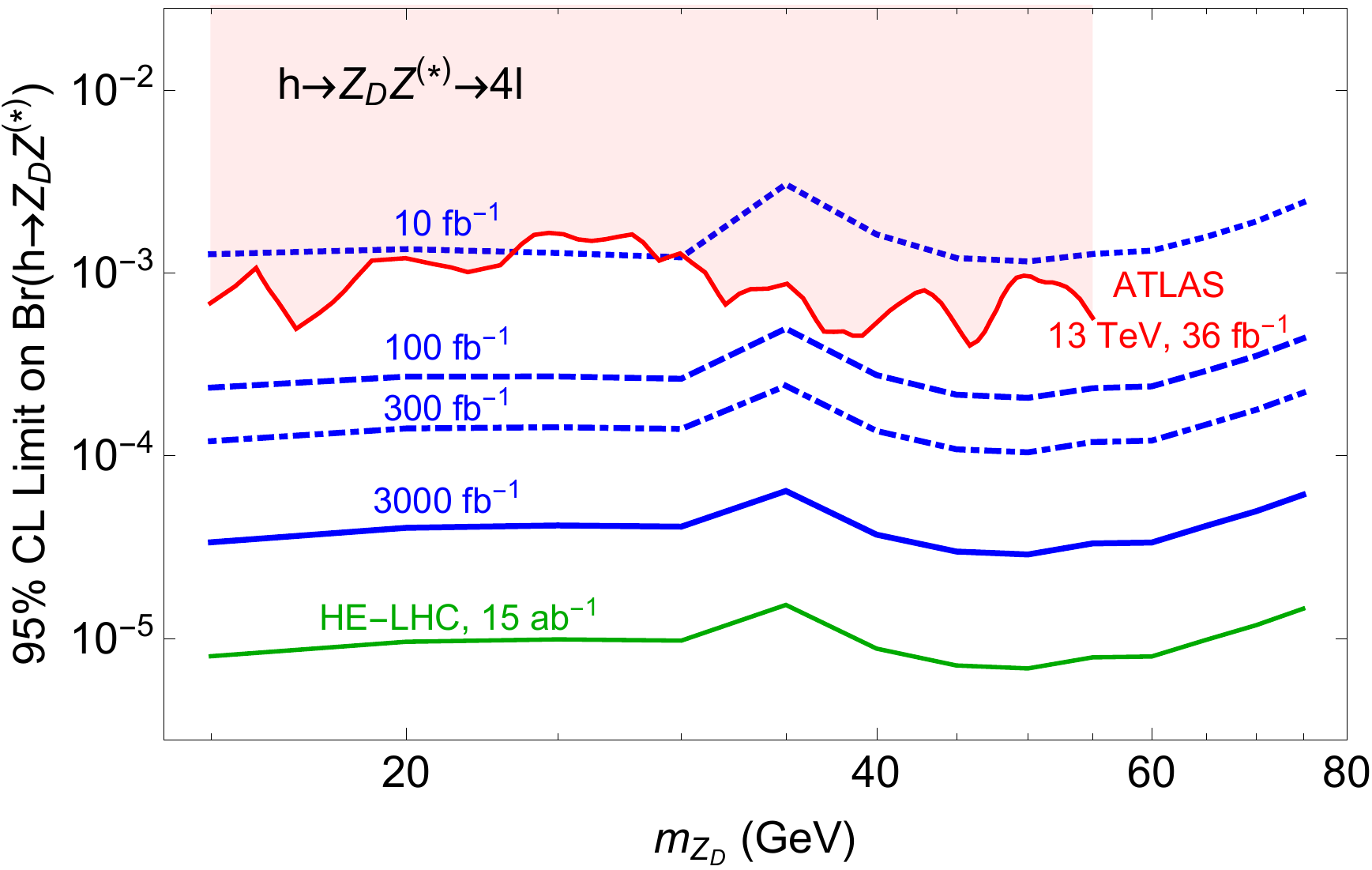}
&
\includegraphics[width=0.4\textwidth]{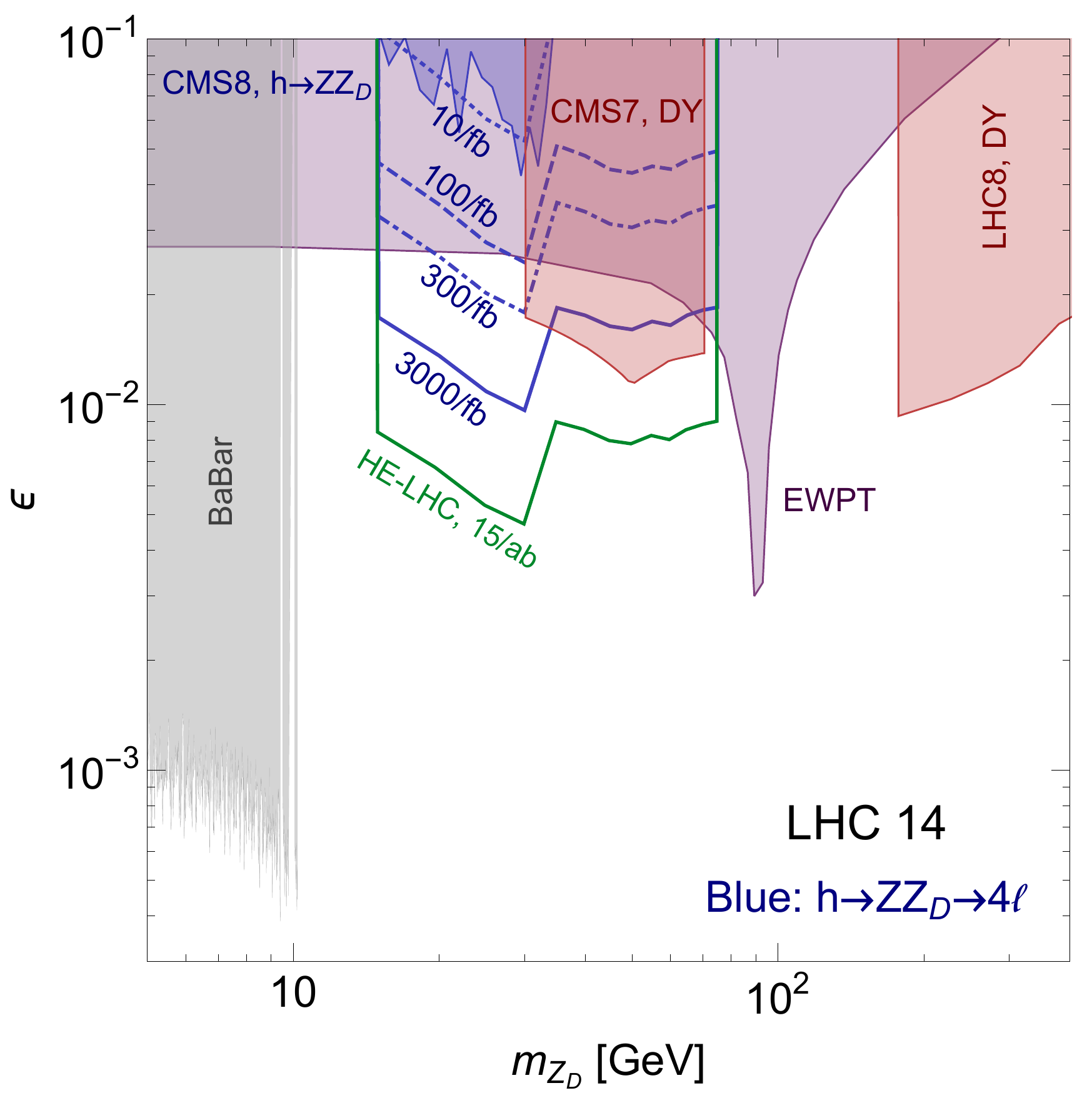}
\end{tabular}
\caption{
Sensitivity of the (HL-)LHC and HE-LHC to  $h \to Z_D Z^{\star} \to 4 \ell$ decays, as a function of dark photon mass and exotic Higgs decay branching ratio (left) or dark photon kinetic mixing parameter $\epsilon$ (right). Figures taken from~\cite{Curtin:2014cca} with the addition of the HE-LHC projection and the recent experimental limit from~\cite{Aaboud:2018fvk}.
 Blue contours, taken from~\cite{Curtin:2014cca}, correspond to the reach of $14 \TeV$ $pp$ collisions with the HL-LHC sensitivity indicated by a solid blue curve. The green contour corresponds to the HE-LHC at $\sqrt{s} = 27 \TeV$ with 15 $\mathrm{ab}^{-1}$ of luminosity, and is derived by rescaling the 14 \UTeV projections for 27 \UTeV signal and background cross sections, see text for details. The red shaded region on the right shows the exclusions from the recent ATLAS search at 13 \UTeV with 36 fb$^{-1}$~\cite{Aaboud:2018fvk}.
}
\label{f.darkphotonZZD}
\end{figure}

Kinetic mixing of the dark photon can allow the Higgs to undergo the decay $h \to Z_D Z^{\star}$ shown in the left panel of Fig.~\ref{f.ZDfeynman}. A search for the four-lepton ($e$ or $\mu$) final state has the best sensitivity, making use of the known invariant mass of the Higgs and the assumed mass peak in the invariant mass of one of the lepton pairs. The HL-LHC sensitivity of such a search was estimated in~\cite{Curtin:2014cca} and is shown in Fig.~\ref{f.darkphotonZZD}. 
Exotic Higgs branching ratios of few$\times 10^{-5}$ can be probed at the HL-LHC. 
The projected limits of~\cite{Curtin:2014cca} can be approximately rescaled for the HE-LHC if the increase in signal and background cross section (as a function of $m_{Z_D}$) are known. 
The signal increases simply in accordance with the greater Higgs production cross section at 27 \UTeV compared to 14 \UTeV.
The increase in background generally depends on $m_{Z_D}$, since this determines the applied invariant mass cuts. To estimate this background increase, we simulate the two main backgrounds to the four-lepton final state, di-$Z/\gamma$ and $h \to Z Z^*$ production, in Madgraph at parton level for 14 and 27 \UTeV and apply the analysis cuts of~\cite{Curtin:2014cca}.
The resulting increase in background rate is quite $m_{Z_D}$-independent, since the background is dominated by SM Higgs decays. We therefore adopt a uniform factor of 4.2  for the HE-LHC branching ratio sensitivity increase compared to the HL-LHC, and the resulting projection is shown as the green contour in Fig.~\ref{f.darkphotonZZD}.
We also show the recent exclusions obtained by the ATLAS 13 \UTeV search for this decay with 36 fb$^{-1}$~\cite{Aaboud:2018fvk}, which agrees roughly with our projections for LHC reach.

The reach in exotic Higgs branching ratio is impressive, below the $10^{-5}$ level at the HE-LHC.
This allows exotic Higgs decay sensitivities on the kinetic mixing parameter better than $10^{-2}$, surpassing, therefore, the model independent bound from electroweak precision measurements, 
see Fig.~\ref{f.darkphotonZZD} (right). 
Even so, exotic Higgs decays do not lead to the most stringent probe of $\epsilon$ in this model. Instead, simple DY production of $Z_D$ and search for the resulting di-lepton resonance on top of the $Z^*$ background still has the greatest reach in $\epsilon$, see Fig.~\ref{f.darkphotonDY} from~\cite{Curtin:2014cca}. For this figure, HE-LHC constraints are also derived from the 14 \UTeV projections by rescaling the signal and background cross sections as a function of $m_{\ell \ell} \approx m_{Z_D}$. Here, the HE-LHC could reach sensitivities better than $\epsilon \sim 10^{-3}$. 

It is important to point out that while exotic Higgs decays may not be the most sensitive probe of kinetic mixing in this scenario, they nevertheless serve an important function in diagnosing the dark sector. Discovery of a resonance in the DY spectrum could indicate a conventionally coupled $Z'$ or a kinetically mixed dark photon. However, a  discovery of the  corresponding exotic Higgs decay would strongly suggest the latter scenario.

\begin{figure}
\begin{center}
\includegraphics[width=0.5\textwidth]{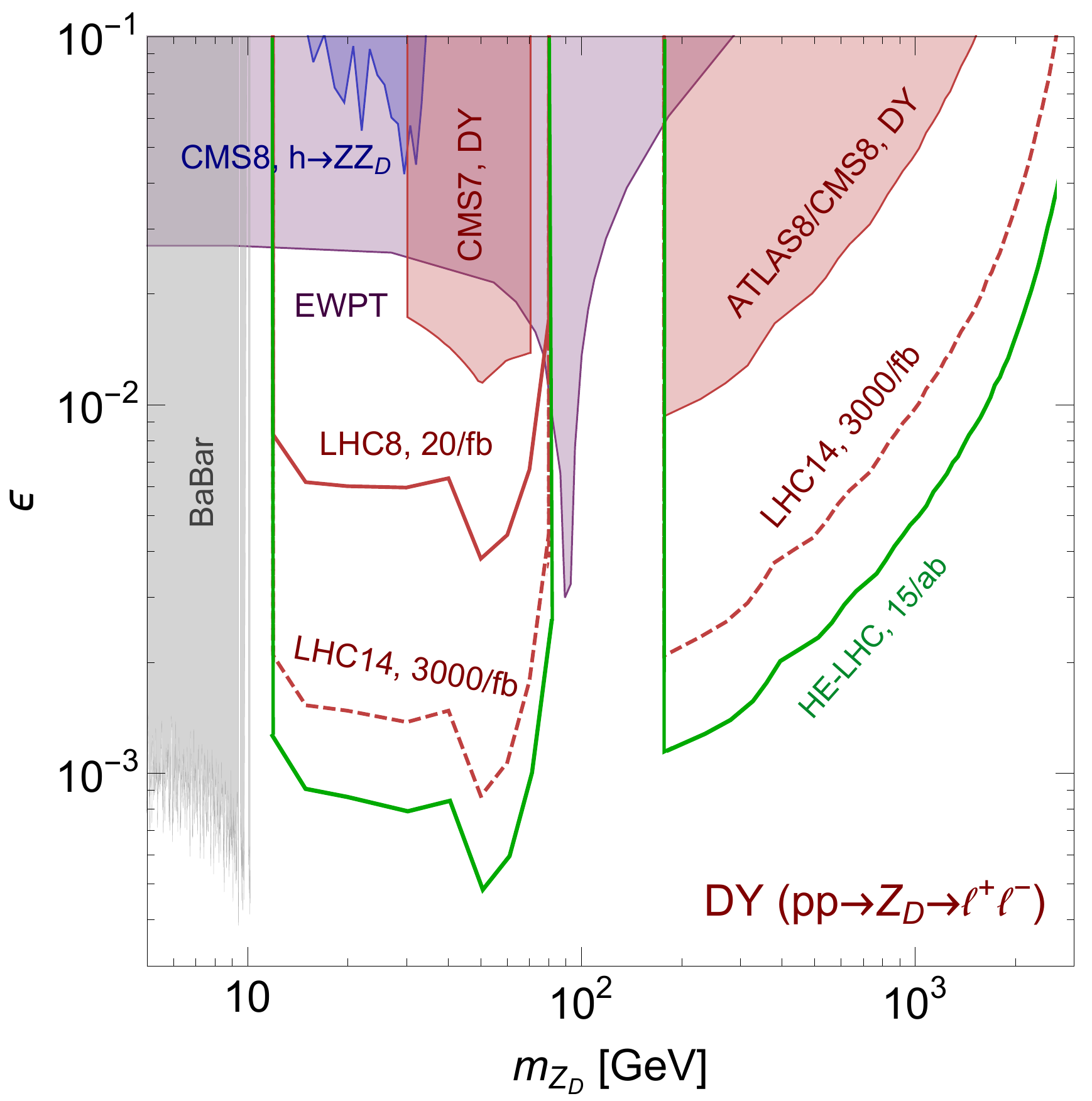}
\end{center}
\caption{
Sensitivity of the (HL-)LHC and HE-LHC to DY production of $Z_D$ and decay to two leptons,  as a function of dark photon mass and kinetic mixing parameter $\epsilon$. Figure taken from~\cite{Curtin:2014cca}, indicating the sensitivity of searches at 14 \UTeV (red contours) with the HL-LHC sensitivity indicated by the red dashed curve. The added green contour corresponds to the HE-LHC at $\sqrt{s} = 27 \TeV$ with 15 $\mathrm{ab}^{-1}$ of luminosity, and is derived by rescaling the 14 \UTeV projections for 27 \UTeV signal and background cross sections, see text for details. 
}
\label{f.darkphotonDY}
\end{figure}

\bigskip

\paragraph{Decays through the Higgs portal}

If the dark photon obtains its mass from a dark Higgs mechanism, one would generally expect there to be nonzero mixing with the SM Higgs. This can lead to exotic Higgs decays to dark photons, as shown in the right panel of Fig.~\ref{f.ZDfeynman}.

The signal is independent of $\epsilon$ as long as $\epsilon$ is large enough for $Z_D$ to decay promptly. 
Again, the four-lepton final state is the best search target, and the requirement of two di-lepton invariant masses coincident at $m_{Z_D}$ and $m_{4 \ell} \approx m_h$ is a very stringent signal requirement that greatly suppresses backgrounds. 
The sensitivity projections for the HL-LHC are shown in Fig.~\ref{f.darkphotonZDZDprompt}, with exotic Higgs branching ratios as small as $\times 10^{-6}$ being observable. 
We rescale these limits for the HE-LHC in an identical manner to the previous two analyses, with the resulting projection shown as the green contour. The low background of the search means sensitivity increases almost proportional to the increased signal rate at higher energy and luminosity, allowing branching ratios for $h\to Z_D Z_D$ below $10^{-7}$ (and, therefore, branching ratios for $h\to Z_D Z_D\to 4\ell$ below $10^{-8}$) to be probed. This corresponds to tiny Higgs portal couplings of $\kappa \sim 10^{-5}$ (see right panel of the figure). 


\begin{figure}
\begin{tabular}{m{0.45 \textwidth} m{0.45\textwidth}}
\includegraphics[width=0.45\textwidth]{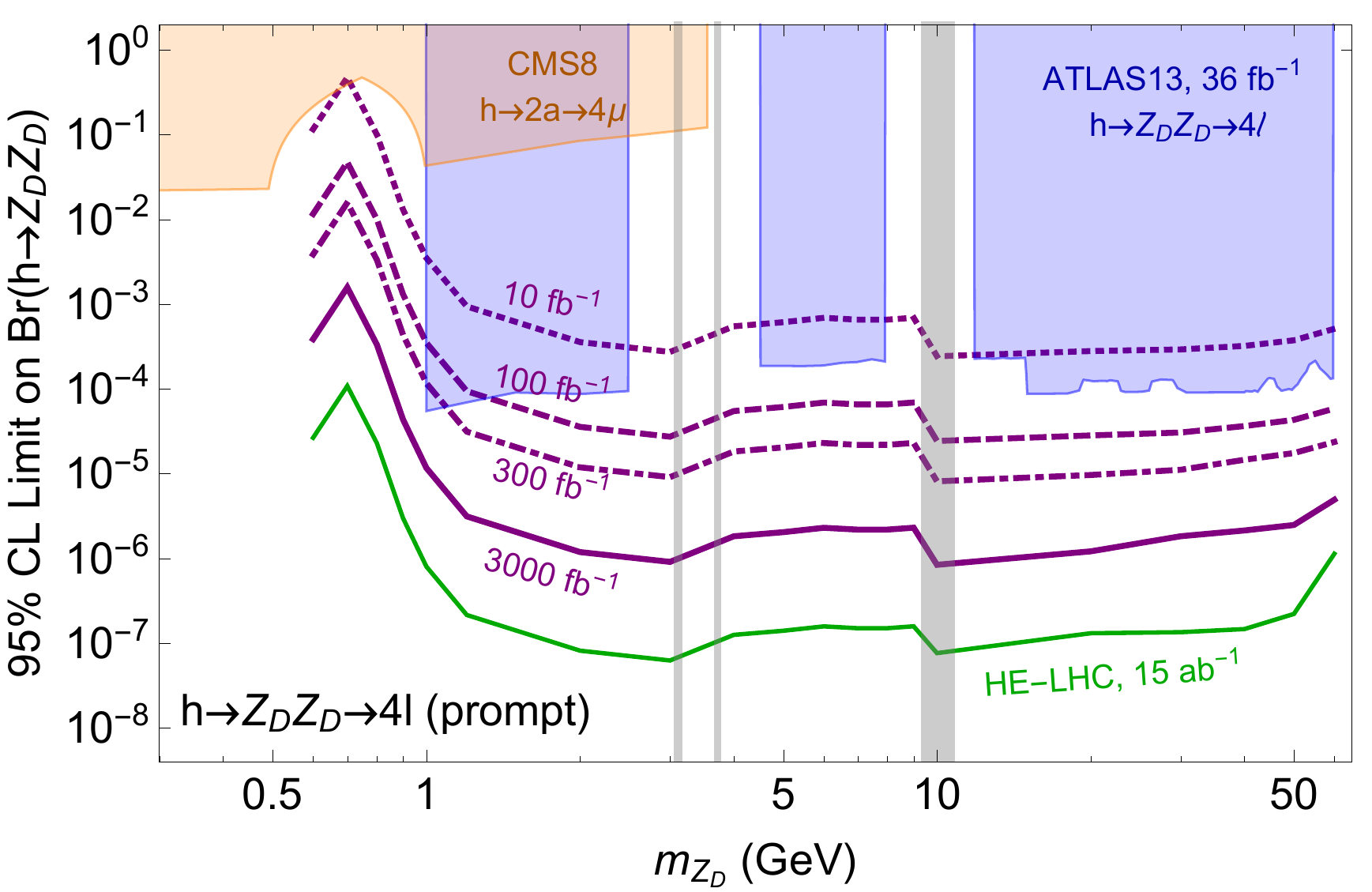}
&
\includegraphics[width=0.45\textwidth]{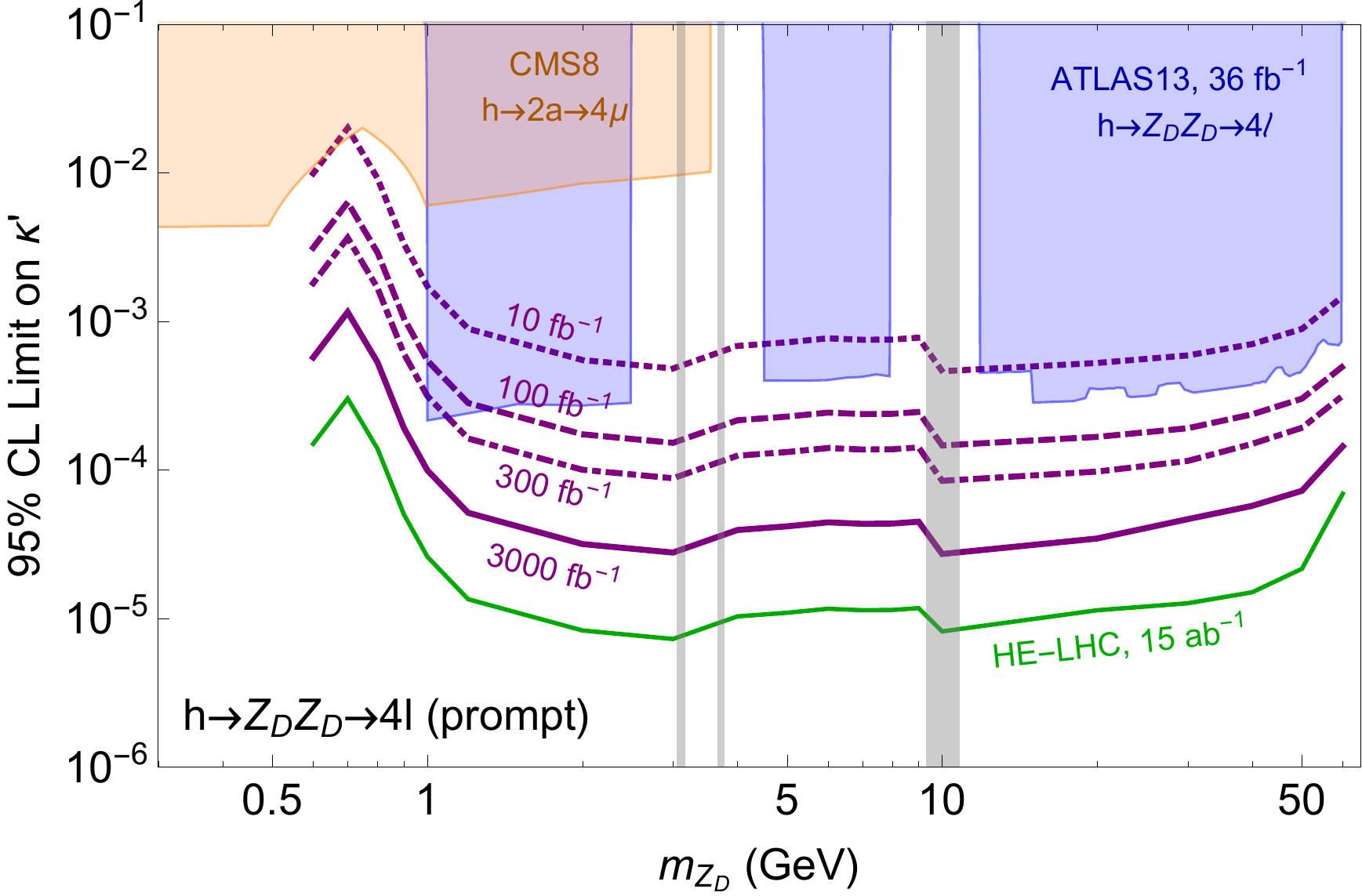}
\end{tabular}
\caption{
Sensitivity of the (HL-)LHC and HE-LHC to  $h \to Z_D Z_D$ decays, where each $Z_D$ decays to $ee$ or $\mu \mu$ promptly. Limits are shown as a function of dark photon mass and exotic Higgs decay branching ratio (left) or Higgs mixing parameter $\kappa$ (right). 
Figures taken from~\cite{Curtin:2014cca} with the addition of the HE-LHC projection and the recent experimental limit from~\cite{Aaboud:2018fvk}.
 Purple contours, taken from~\cite{Curtin:2014cca}, correspond to the reach of $14 \TeV$ $pp$ collisions with the HL-LHC sensitivity indicated by a solid purple curve. The green contour corresponds to the HE-LHC at $\sqrt{s} = 27 \TeV$ with 15 $\mathrm{ab}^{-1}$ of luminosity, and is derived by rescaling the 14 \UTeV projections for 27 \UTeV signal and background cross sections, see text for details. The blue shaded regions show the exclusions from the recent ATLAS search at 13 \UTeV with 36 fb$^{-1}$~\cite{Aaboud:2018fvk}.
}
\label{f.darkphotonZDZDprompt}
\end{figure}


For very small $\epsilon$, the $Z_D$ decay is displaced, becoming a long-lived particle. In this case, exotic Higgs decays play a uniquely important role in probing the dark sector, since the photon portal could be far too small to serve as a production mechanism, while the Higgs portal could be wide open. 
This was analysed in~\cite{Curtin:2014cca,Curtin:2018mvb} and can lead to $\epsilon$ as small as few$\times 10^{-9}$ to be probed in Higgs exotic decays at the HL-LHC.





\asubsubsection[M. Bauer, M. Neubert, A. Thamm]{Exotic Higgs decays to axion-like particles: $h \to Za$ and $h \to aa$}\label{Sec:9.1.8}
In this section, we discuss the exotic Higgs decays $h\to aa$ and $h\to Za$, where $a$ is a light pseudoscalar particle often called an axion-like particle (ALP). Its interactions with SM particles are described by dimension-5 operators or higher when assuming that the ALP respects a shift symmetry apart from a soft breaking through an explicit mass term \cite{Georgi:1986df} 
\begin{equation}\label{Leff}
\begin{aligned}
   {\cal L}_{\rm eff}^{D\le 5}
   &= \frac12 \left( \partial_\mu a\right)\!\left( \partial^\mu a\right)
    - \frac{m_{a,0}^2}{2}\,a^2 
    + \sum_f \frac{c_{ff}}{2}\,\frac{\partial^\mu a}{\Lambda}\,\bar f\gamma_\mu\gamma_5 f  + g_s^2\,C_{GG}\,\frac{a}{\Lambda}\,G_{\mu\nu}^A\,\tilde G^{\mu\nu,A} \\[-1mm]
   &\quad\mbox{}+ e^2\,C_{\gamma\gamma}\,\frac{a}{\Lambda}\,F_{\mu\nu}\,\tilde F^{\mu\nu}
    + \frac{2e^2}{s_w c_w}\,C_{\gamma Z}\,\frac{a}{\Lambda}\,F_{\mu\nu}\,\tilde Z^{\mu\nu}
    + \frac{e^2}{s_w^2 c_w^2}\,C_{ZZ}\,\frac{a}{\Lambda}\,Z_{\mu\nu}\,\tilde Z^{\mu\nu} \,,
\end{aligned}
\end{equation}
where $m_{a,0}$ is the explicit symmetry breaking mass term, $s_w$ and $c_w$ are the sine and cosine of the weak mixing angle, respectively, and $\Lambda$ sets the new physics scale and is related to the ALP decay constant by $\Lambda/|C_{GG}| = 32 \pi^2 f_a$. Note that an exotic $Z$-decay $Z\to\gamma a$ proceeds through the $C_{\gamma Z}$ operator.  Interactions with the Higgs boson, $\phi$, are described by the dimension-6 and 7 operators
\begin{equation}\label{LeffD>5}
   {\cal L}_{\rm eff}^{D\ge 6}
   = \frac{C_{ah}}{\Lambda^2} \left( \partial_\mu a\right)\!\left( \partial^\mu a\right) \phi^\dagger\phi
    + \frac{C_{Zh}}{\Lambda^3} \left( \partial^\mu a\right) 
    \left( \phi^\dagger\,iD_\mu\,\phi + \mbox{h.c.} \right) \phi^\dagger\phi + \dots  \,,
\end{equation}
where the first operator mediates the decay $h\to aa$, while the second one is responsible for $h\to Za$. Note that a possible dimension-5 operator coupling the ALP to the Higgs current is redundant unless it is introduced by integrating out a heavy new particle which acquires most of its mass through electroweak symmetry breaking \cite{Bauer:2016ydr,Bauer:2016zfj,Bauer:2017nlg,Bauer:2017ris}. The exotic Higgs decay rates into ALPs are given by  
\begin{align}
\Gamma(h \to Za)&=\frac{m_h^3}{16\pi\,\Lambda^2}|C_{Zh}^\text{eff}|^2\lambda^{3/2}\Big(\frac{m_Z^2}{m_h^2},\frac{m_a^2}{m_h^2}\Big)\,,\\
\Gamma(h \to aa)&=\frac{m_h^3\,v^2}{32\pi\,\Lambda^4}|C_{ah}^{\rm eff}|^2\left(1-\frac{2m_a^2}{m_h^2}\right)^2\sqrt{1-\frac{4m_a^2}{m_h^2}}\,,
\end{align}
where $\lambda(x,y)=(1-x-y)^2-4xy$ and we define $C_{Zh}^{\rm eff} = C_{Zh}^{(5)} + C_{Zh} v^2/2 \Lambda^2$ to take into account possible contributions from a dimension-5 operator which originates from integrating out chiral heavy new physics. The relevant partial widths for this study are the decay of the ALP into photons and leptons. For the derivation and one-loop contributions we refer the reader to \cite{Bauer:2017ris}
\begin{align}
 \Gamma(a\to\gamma\gamma)  &= \frac{4\pi\alpha^2 m_a^3}{\Lambda^2}\,\big| C_{\gamma\gamma}^\text{eff} \big|^2 \,, \\
 \Gamma(a\to \ell^+ \ell^-)&=\frac{m_a m_\ell^2}{8\pi\Lambda^2} \left| c_{\ell\ell}^\text{eff}\right|^2 \sqrt{1-\frac{4m_\ell^2}{m_a^2}}\,.
\end{align}

Future hadron colliders can significantly surpass the reach of the LHC in searches for ALPs. In particular, searches for ALPs produced in exotic Higgs and $Z$ decays profit from the higher centre-of-mass energies and luminosities of the proposed high-energy LHC (HE-LHC), planned to replace the LHC in the LEP tunnel with $\sqrt{s}=27 \,$\UTeV, and the ambitious plans for a new generation of hadron colliders with $\sqrt{s}=100\,$\UTeV at CERN (FCC-hh) and in China (SPPC). As benchmark scenarios we assume integrated luminosities of $3$\,ab$^{-1}$ at the LHC, $15$\,ab$^{-1}$ at the HE-LHC and $20$\,ab$^{-1}$ at the FCC-hh.
At hadron colliders, ALP production in association with electroweak bosons suffers from large backgrounds. Previous studies of these processes have therefore focused on invisibly decaying (or stable) ALPs, taking advantage of the missing-energy signature \cite{Mimasu:2014nea,Brivio:2017ije}. In contrast, here we focus on ALPs produced in the decays of a Higgs boson, $h\to Za$ and $h \to a a$ (for more details see \cite{Bauer:2018uxu}).

%
\begin{figure}
\begin{center}
\includegraphics[width=1.3\textwidth]{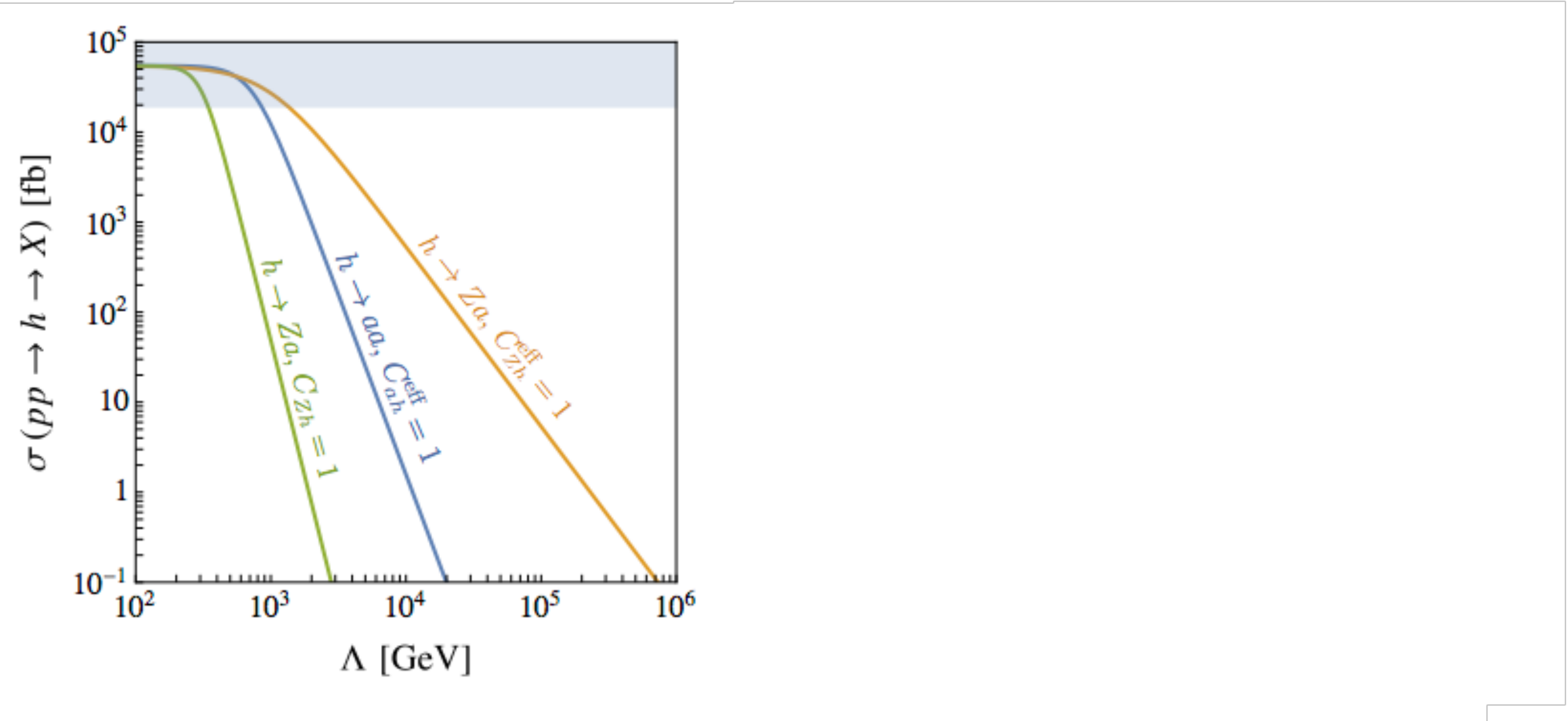}
    \end{center}
\caption{\label{fig:ALPpsec} Production cross sections of ALPs produced in Higgs decays at the LHC ($\sqrt{s} = 14\,$\UTeV) versus the new-physics scale $\Lambda$. We set $m_a=0$ and fix the relevant Wilson coefficients to 1. For the green contour we fix $C_{Zh}^{(5)}=0$ and only consider the dimension-7 coupling in \eqref{LeffD>5}.
The grey regions are excluded by Higgs coupling measurements (BR$(h\to{\rm{BSM}}<0.34)$) \cite{Khachatryan:2016vau}.}
\end{figure}
%

Exotic decays are particularly interesting, because even small couplings can lead to appreciable branching ratios and be as large as several percent \cite{Bauer:2017nlg, Bauer:2017ris}. 
This allows us to probe large new-physics scales $\Lambda$, as illustrated in Figure \ref{fig:ALPpsec}, where we show the cross sections of the processes $pp \to h \to Z a$ and $pp \to h \to aa$  at the LHC with $\sqrt{s} = 14\,$\UTeV. The figure nicely reflects the different scalings of the dimension-5, 6, and 7 operators in the effective ALP Lagrangian. The 
shaded region is excluded by the present Higgs coupling measurements constraining general beyond the SM decays of the Higgs boson, $\text{Br}(h\to\text{BSM})<0.34$ 
\cite{Khachatryan:2016vau}. This leads to constraints on the coefficients $|C_{Zh}^{\textrm{eff}}| < 0.72\,(\Lambda/\textrm{\UTeV})$ and $|C_{ah}^{\rm eff}| < 1.34\,(\Lambda/\textrm{\UTeV})^2$.

%
\begin{figure}[t]
\begin{center}
\includegraphics[width=0.85\textwidth]{\main/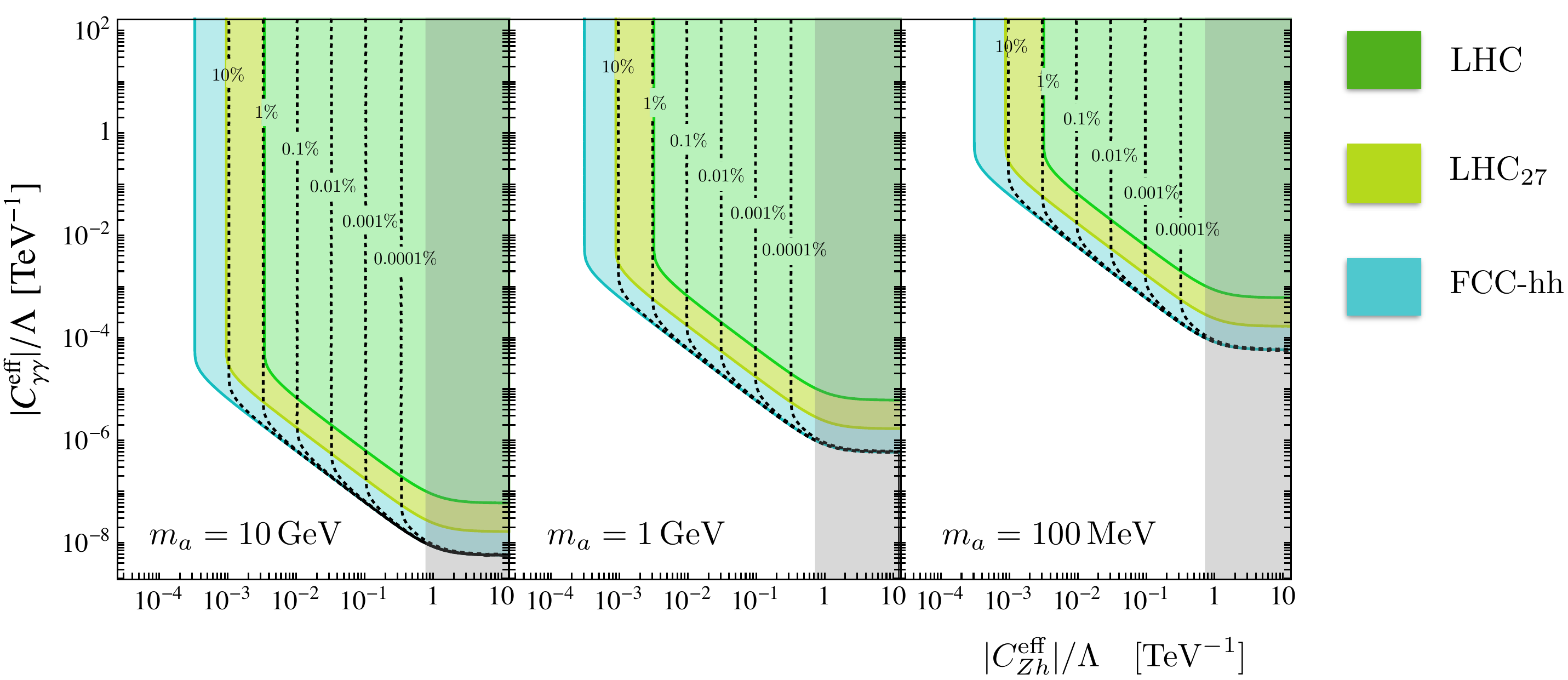}
\includegraphics[width=0.85\textwidth]{\main/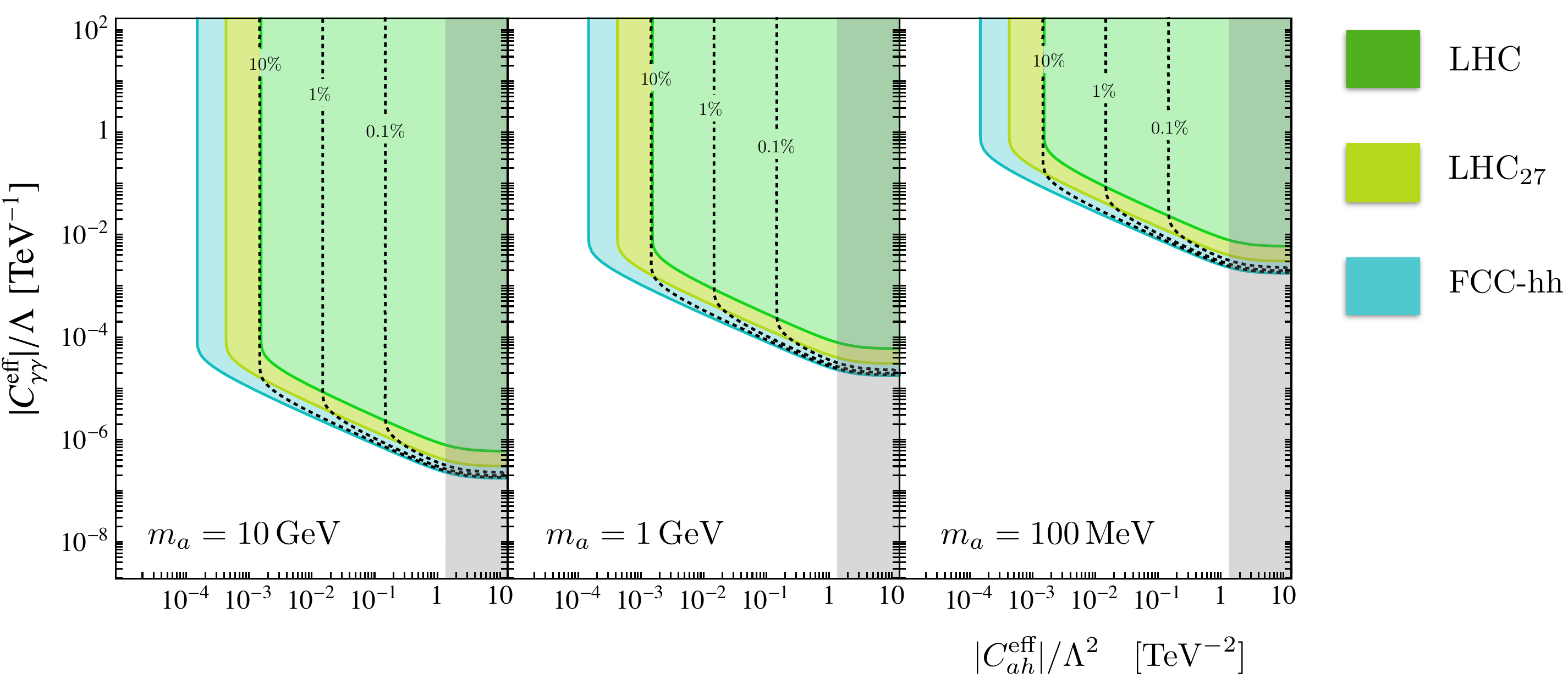}
\end{center}
\vspace{-0mm}
\caption{\label{fig:pphZa} Projected reach in searches for $h \to Za \to \ell^+\ell^-+2\gamma $ and $h \to aa \to 4\gamma $ decays with the LHC with $3$\,ab$^{-1}$
(green), HE-LHC with $15$\,ab$^{-1}$ (light green) and a $100\,$\UTeV collider with $20$\,ab$^{-1}$ (blue). The parameter region with the solid contours correspond to a branching ratio of $\text{Br}(a\to 
\gamma\gamma)=1$, and the contours showing the reach for smaller branching ratios are dotted. The grey areas indicate the regions excluded by the present upper bound on the BSM Higgs width coming from Higgs coupling fits (BR$(h\to{\rm{BSM}}<0.34)$) \cite{Khachatryan:2016vau}.}
\end{figure}
%

Light or weakly coupled ALPs can be long-lived, and thus only a fraction of them decay inside the detector and can be reconstructed. The average ALP decay length perpendicular to the beam axis is given by
\begin{equation}\label{eq:Lperp}
   L_a^\perp(\theta) = \frac{\sqrt{\gamma_a^2 - 1}}{\Gamma_a}\,\sin\theta \,, 
\end{equation}
where $\Gamma_a$ denotes the total width of the ALP, $\theta$ is the scattering angle (in the centre-of-mass frame) and $\gamma_a$ specifies the relativistic boost factor. Using the fact that most Higgs bosons are produced in the forward direction at the LHC and approximating the ATLAS and CMS detectors (as well as future detectors) by infinitely long cylindrical tubes, we first perform a Lorentz boost to the rest frame of the decaying boson. In this frame the relevant boost factors for the Higgs decay into ALPs are given by
\begin{align}
\gamma_a=\begin{cases}\displaystyle{ \frac{m_h^2-m_Z^2+m_a^2}{2m_am_h}}\,,& \text{for}\,\, h \to Z a\,,\\[14pt]
\displaystyle \frac{m_h}{2m_a}\,,&\text{for}\,\, h \to aa\,.\\
\end{cases}
\end{align}
We can compute the fraction of ALPs decaying before they have travelled a certain distance $L_{\rm det}$ from the beam axis, finding 
\begin{equation}\label{eq:ffactors}
\begin{aligned}
   f_{\rm dec}^{a} &= \int_0^{\pi/2}\!d\theta \,  \sin\theta
    \left( 1 - e^{-L_{\rm det}/L_a^\perp(\theta)} \right) , \\
   f_{\rm dec}^{aa} &= \int_0^{\pi/2}\!d\theta \,  \sin\theta
    \left( 1 - e^{-L_{\rm det}/L_a^\perp(\theta)} \right)^2,
\end{aligned}
\end{equation} 
where $f^a_\text{dec}$ is relevant for $h \to Z a$ decays and $f^{aa}_\text{dec}$ applies to $h\to aa$ decays. 

For prompt ALP decays, we demand all final state particles to be detected in order to reconstruct the decaying SM particle. For the decay into photons we 
require the ALP to decay before the electromagnetic calorimeter which, at ATLAS and CMS, is situated approximately $1.5\,$m from the interaction point, and we thus take $L_{\rm det} = 
1.5\,$m. Analogously, the ALP should decay before the inner tracker, $L_{\rm det} = 2\,$cm, for an $e^+e^-$ final state to be detected. We also require $L_\text{det}
=2\,$cm for muon and tau final states in order to take full advantage of the tracker information in reconstructing these events. We define the effective branching ratios
\begin{align}
\text{Br}(h\to Za\to Y\bar{Y}+X\bar{X})\big\vert_\text{eff}&=\text{Br}(h\to Za)\,\text{Br}(a\to X\bar{X})f_\text{dec}^a\,\text{Br}(Z\to Y\bar{Y})\,,\label{eq:LHChZa}\\
\text{Br}(h\to aa\to X\bar{X}+X\bar{X})\big\vert_\text{eff}&=\text{Br}(h\to aa)\,\text{Br}(a\to X\bar{X})^2f_\text{dec}^{aa}\,,\label{eq:LHChaa}
\end{align}
where $X=\gamma, e, \mu, \tau$ and $Y=\ell, {\rm hadrons}$. Multiplying the effective branching ratios by the appropriate Higgs production cross section and luminosity allows us to derive results for a specific collider. The Higgs production cross section at $14\,$\UTeV is given by $\sigma(pp\to h)= 54.72\,$pb (see Sec.~\ref{sec2_HXSWG1} of this report). We use the reference cross section $\sigma(gg 
\to h) = 146.65\,$pb at $\sqrt{s} = 27\,$\UTeV. At $\sqrt{s} = 100\,$\UTeV, the relevant cross section is $\sigma(gg \to h) = 802\,$pb ~\cite{Mangano:2016jyj}. We require $100$ signal events, since this is what is typically needed to suppress backgrounds in new-physics searches with prompt Higgs decays \cite{Khachatryan:2016vau, Chatrchyan:2013vaa,Aad:2015bua} (see also \cite{Bauer:2017ris} for 
further discussion).
We do not take advantage of the additional background reduction obtained by cutting on a secondary vertex in the case where the ALP lifetime becomes appreciable. A dedicated analysis by the experimental collaborations including detailed simulations of the backgrounds is required to improve on our projections.

%
\begin{figure}[t]
\begin{center}
\includegraphics[width=0.85\textwidth]{\main/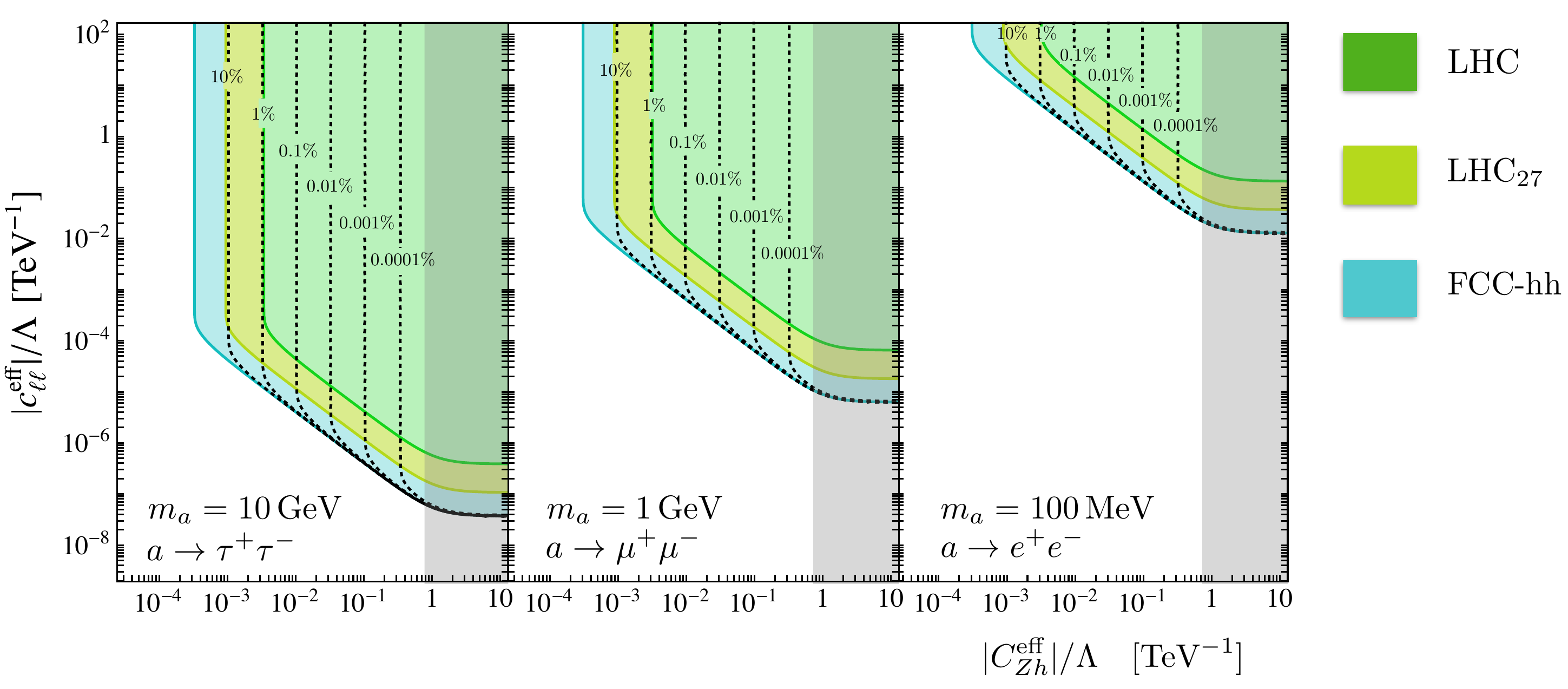}
\includegraphics[width=0.85\textwidth]{\main/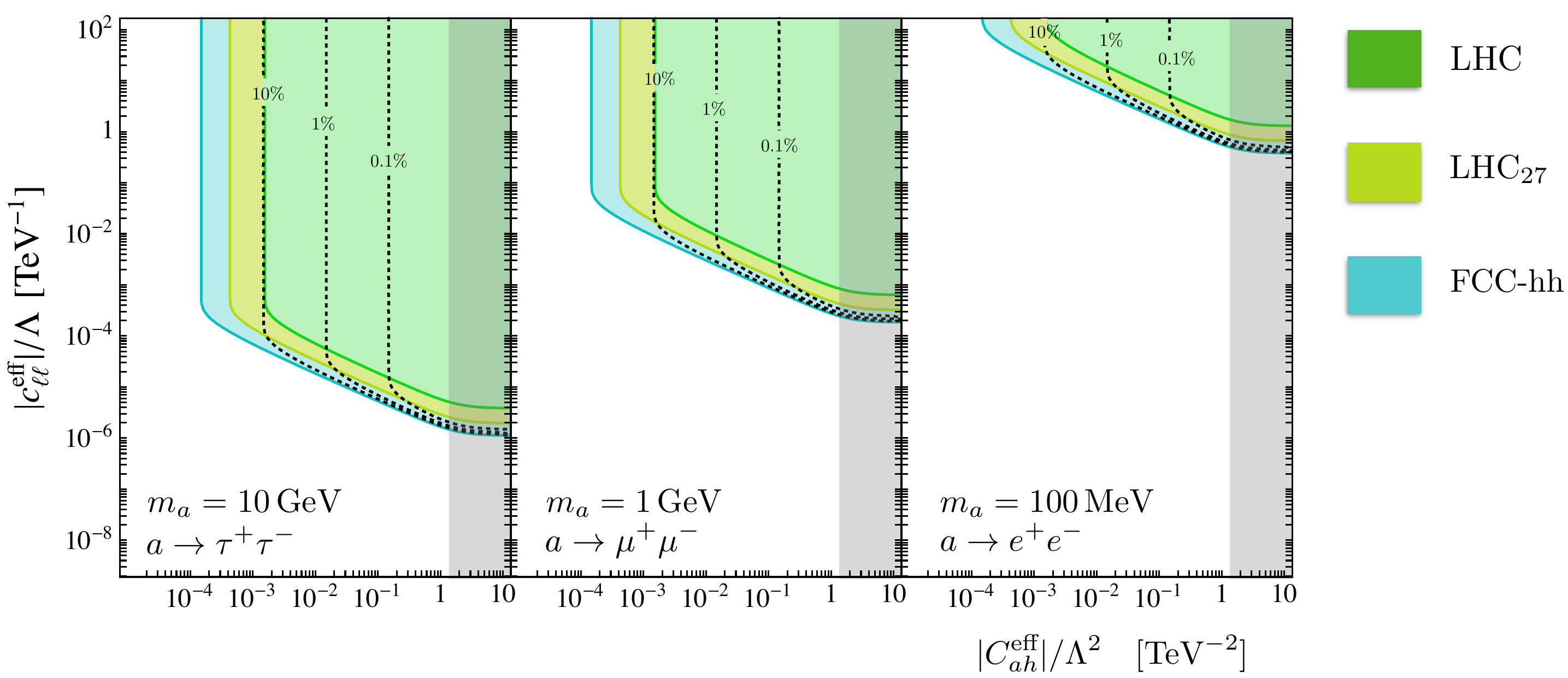}
\end{center}
\vspace{-3mm}
\caption{\label{fig:pphZalep} Projected reach in searches for $h \to Za \to \ell^+\ell^-+\ell^+\ell^- $ and $h \to aa \to 4\ell $ decays with the LHC with $3$\,ab$^{-1}$
(green), HE-LHC with $15$\,ab$^{-1}$ (light green) and a $100\,$\UTeV collider with $20$\,ab$^{-1}$ (blue). The parameter region with the solid contours correspond to a branching ratio of $\text{Br}(a\to 
\ell^+\ell^-)=1$, and the contours showing the reach for smaller branching ratios are dotted. The grey areas indicate the regions excluded by the present upper bound on the BSM Higgs width coming from Higgs coupling fits.}
\end{figure}
%

In Figure~\ref{fig:pphZa}, we display the reach for observing 100 events at the HL-LHC, HE-LHC and FCC-hh (in green, light green and blue respectively) in searches for $pp\to h \to Za\to \ell^+\ell^-\gamma\gamma$ (upper panels) and  $pp\to h \to aa\to 4 \gamma$ (lower panels) for $m_a= 10\,$\UGeV, $1\,$\UGeV and $100\,$\UMeV (left, middle, right panel) in the $|C_{Zh}^{\rm eff}|/\Lambda - |C_{\gamma\gamma}^{\rm eff}|/\Lambda$ and $|C_{ah}^{\rm eff}|/\Lambda^2 - |C_{\gamma\gamma}^{\rm eff}|/\Lambda$ planes respectively. We assume $\text{Br}(a\to \gamma\gamma)=1$ and indicate the reach of the FCC-hh obtained in the case that $\text{Br}(a\to \gamma\gamma)<1$ by the black dotted lines. 
In $h \to Za$ searches, the HL-LHC can reach values of $|C_{Zh}^{\rm eff}|/\Lambda$ down to $3 \times 10^{-3} (\Lambda/\textrm{\UTeV})$ for all ALP masses. The HE-LHC improves this reach by a factor of $3$ to $1 \times 10^{-3} (\Lambda/\textrm{\UTeV})$, while the FCC-hh increases the reach by an order of magnitude to values as small as $3 \times 10^{-4} (\Lambda/\textrm{\UTeV})$. In $|C_{\gamma\gamma}^{\rm eff}|/\Lambda$, the HL-LHC is sensitive to values larger than $10^{-7}, 10^{-5}$ and $10^{-3} (\Lambda/\textrm{\UTeV})$ for $m_a = 10\,$\UGeV, $1\,$\UGeV and $100\,$\UMeV, respectively, and the largest allowed value of $|C_{Zh}^{\rm eff}|/\Lambda = 0.72\,(\Lambda/\textrm{\UTeV})$. The sensitivity of the HE-LHC (FCC-hh) increases by a factor $3$ ($10$). 

The process $h \to aa$ can access $|C_{ah}^{\rm eff}|/\Lambda^2 = 1.5, 0.4, 0.15 \times 10^{-3} (\Lambda/\textrm{\UTeV})^2$ at the HL-LHC, HE-LHC and FCC-hh, respectively. In $|C_{\gamma\gamma}^{\rm eff}|/\Lambda$, the HL-LHC is sensitive to values larger than $8 \times 10^{-7}, 8 \times 10^{-5}$ and $8 \times 10^{-3} (\Lambda/\textrm{\UTeV})$ for $m_a = 10\,$\UGeV, $1\,$\UGeV and $100\,$\UMeV, respectively, for the largest allowed value of $|C_{ah}^{\rm eff}|/\Lambda$. Both the HE-LHC as well as the FCC-hh improve this reach by a factor $2$ each. For all considered ALP masses, the $h\to Z a$ decay could be observed at a $100\,$\UTeV collider for $\text{Br}(a\to \gamma\gamma)\gtrsim 10^{-6}$ and the $h\to a a$ decay could be fully reconstructed for $\text{Br}(a\to \gamma\gamma)\gtrsim 0.01$.

The results are similar for leptonic ALP decays. In Figure~\ref{fig:pphZalep} we show the reach in the $|c_{\ell\ell}^\text{eff}|/\Lambda - |C_{Zh}^\text{eff}|/\Lambda$ plane (upper row) and  $|c_{\ell\ell}^\text{eff}|/\Lambda - |C_{ah}^\text{eff}|/\Lambda^2$ plane (lower row) for ALP decays into taus (left), muons (middle) and electrons (right). In $|C_{Zh}^\text{eff}|/\Lambda$ the reach coincides with the one of the same process with ALP decays into photons. For $h \to Za$ the HL-LHC can probe values $|c_{\ell\ell}^\text{eff}|/\Lambda = 6 \times 10^{-7}, 10^{-4}, 2 \times 10^{-1}\,(\Lambda/\textrm{\UTeV})$ for $m_a = 10\,$\UGeV, $1\,$\UGeV and $100\,$\UMeV, respectively. The HE-LHC (FCC-hh) increases this reach by a factor 3 (10). Similarly for $h \to aa$ the HL-LHC is sensitive to values $|c_{\ell\ell}^\text{eff}|/\Lambda = 6 \times 10^{-6}, 10^{-3}, 2 \,(\Lambda/\textrm{\UTeV})$ which the HE-LHC and FCC-hh can increase by a factor $2$ each.

\asubsection{LHC searches for additional heavy neutral Higgs bosons in fermionic final states}\label{sec:Hff}
\asubsubsection[L. Zhang, Y. Liu]{Projection of Run-2 ATLAS searches for MSSM heavy neutral Higgs bosons}
The studies presented in this section have also been published in ~\cite{ATL-PHYS-PUB-2018-050}.

\paragraph{Introduction}

The discovery of a Standard Model-like Higgs boson ~\cite{ATLASHiggsJuly2012, CMSHiggsJuly2012}
at the Large Hadron Collider~\cite{LHC} has provided important insight into the mechanism of
electroweak symmetry breaking. However, it remains possible that the discovered particle is part
of an extended scalar sector, a scenario that is favoured by a number of theoretical arguments~\cite{Djouadi:2005gj,Branco:2011iw}.
Searching for additional Higgs bosons is among the main goals of
the High-Luminosity LHC programme~\cite{ecfa15}. The Minimal Supersymmetric Standard
Model~\cite{Djouadi:2005gj,Fayet:1976et,Fayet:1977yc} is one of the well motivated extensions
of the SM\@. Besides the SM-like Higgs boson, the MSSM requires two additional neutral Higgs bosons:
one CP-odd ($A$) and one CP-even ($H$), which in the following are generically called $\phi$.
At tree level, the MSSM Higgs sector depends on only two non-SM parameters, which can be chosen
to be the mass of the CP-odd Higgs boson, $m_A$, and the ratio of the vacuum expectation values
of the two Higgs doublets, $\tan\beta$. Beyond tree level, a number of additional parameters
affect the Higgs sector, the choice of which defines various MSSM benchmark scenarios, such as \mhmodp and hMSSM.
The couplings of the additional MSSM Higgs bosons to down-type fermions are enhanced with respect to
the SM Higgs boson for large $\tan\beta$ values, resulting in increased branching fractions to
$\tau$-leptons and $b$-quarks, as well as a higher cross section for Higgs boson production
in association with $b$-quarks.

The projections presented in this section are extrapolations of the recent results obtained by ATLAS using
the $36.1~\ifb$ \RunTwo dataset~\cite{ATLASRun2Ditau}.  The MSSM Higgs boson with masses of
0.2--2.25\TeV and $\tan\beta$ of 1--58 is searched for in the \lephad and \hadhad decay modes,
where \taulep represents the leptonic decay of a $\tau$-lepton, whereas \tauhad represents the hadronic decay.
The main production modes are gluon--gluon fusion and in association with $b$-quarks.
To exploit the different production modes, events containing at least one $b$-tagged jet enter the $b$-tag
category, while events containing no $b$-tagged jets enter the $b$-veto category. The total transverse
mass ($\mTtot$), as defined in Ref.~\cite{ATLASRun2Ditau}, is used as the final discriminant between
the signal and the background.

In making these extrapolations, the assumption is made that the planned upgrades to the ATLAS detector and
improvements to reconstruction algorithms will mitigate the effects of the higher pileup which can reach up
to 200 in-time pileup interactions, leading to the overall reconstruction performance matching that of the current detector.
Furthermore, the assumption is made that the analysis will be unchanged in terms of selection and
statistical analysis technique, though the current analysis has not been re-optimised for the HL-LHC datasets.

\paragraph{Extrapolation method}
\label{sec:extrapolation method}
To account for the integrated luminosity increase at HL-LHC, signal and background distributions are scaled by
a factor of $3000/36.1$. Furthermore, to account for the increase in collision energy from $13~\TeV$
to $14~\TeV$, the background distributions are further scaled by a factor 1.18 which assumes the same
parton-luminosity increase for quarks as that for gluons. The cross section of signals in various scenarios
at $14~\TeV$ are given in Ref.~\cite{deFlorian:2016spz}. Possible effects on the kinematics and the $\mTtot$
shape due to the collision energy increase are neglected for this study. The scaled $\mTtot$ distributions
for the four signal categories and one for the top control region are shown in
Figures~\ref{fig:mTtotDistributionsSR} and~\ref{fig:mTtotDistributionsCR}. These distributions are used in
the statistical analysis.

\begin{figure}[!ht]
    \centering
        \subfloat[\lephad $b$-veto category]{\includegraphics[width=0.45\columnwidth]{\main/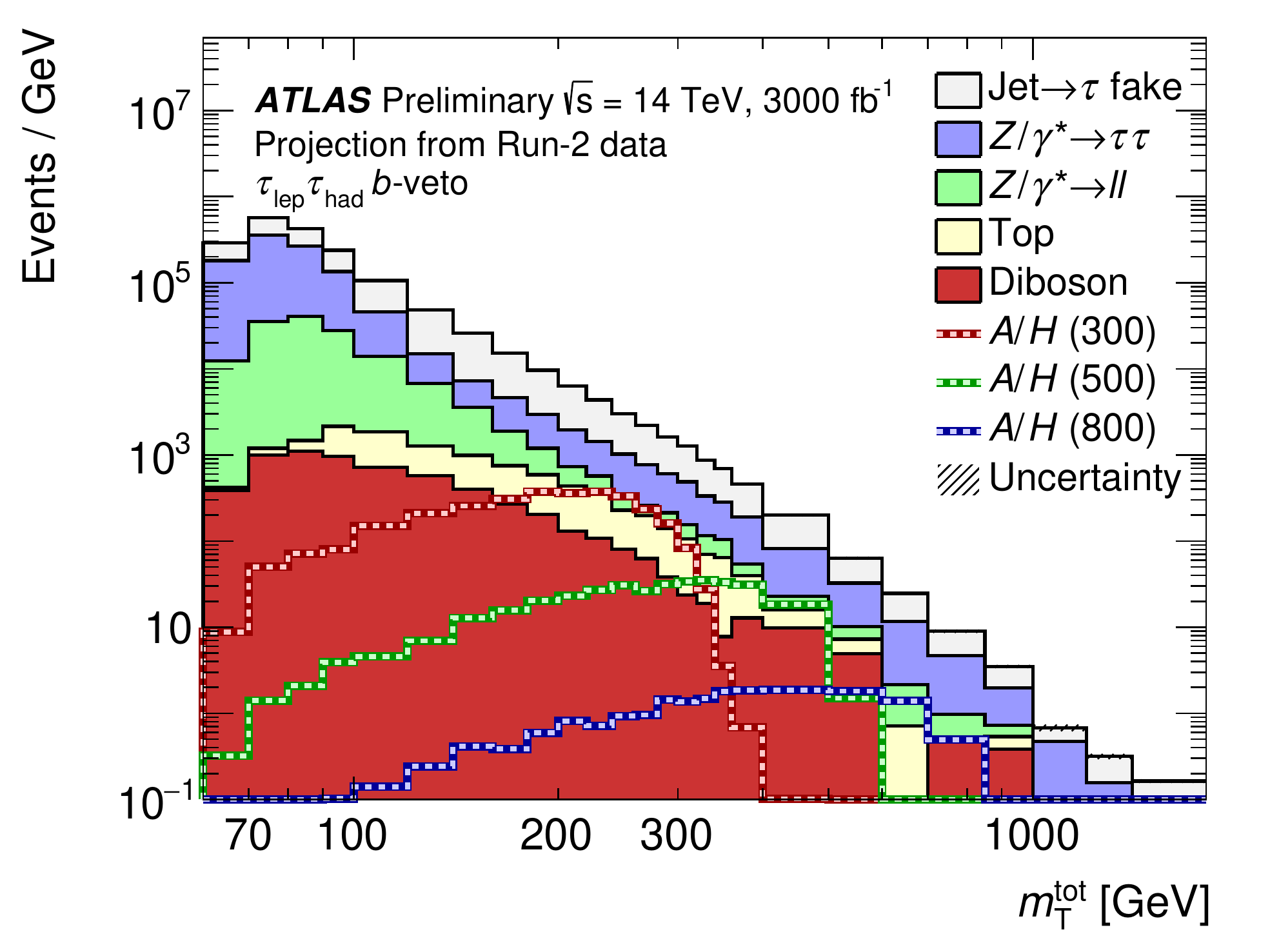}}
        \qquad
        \subfloat[\lephad $b$-tag category]{\includegraphics[width=0.45\columnwidth]{\main/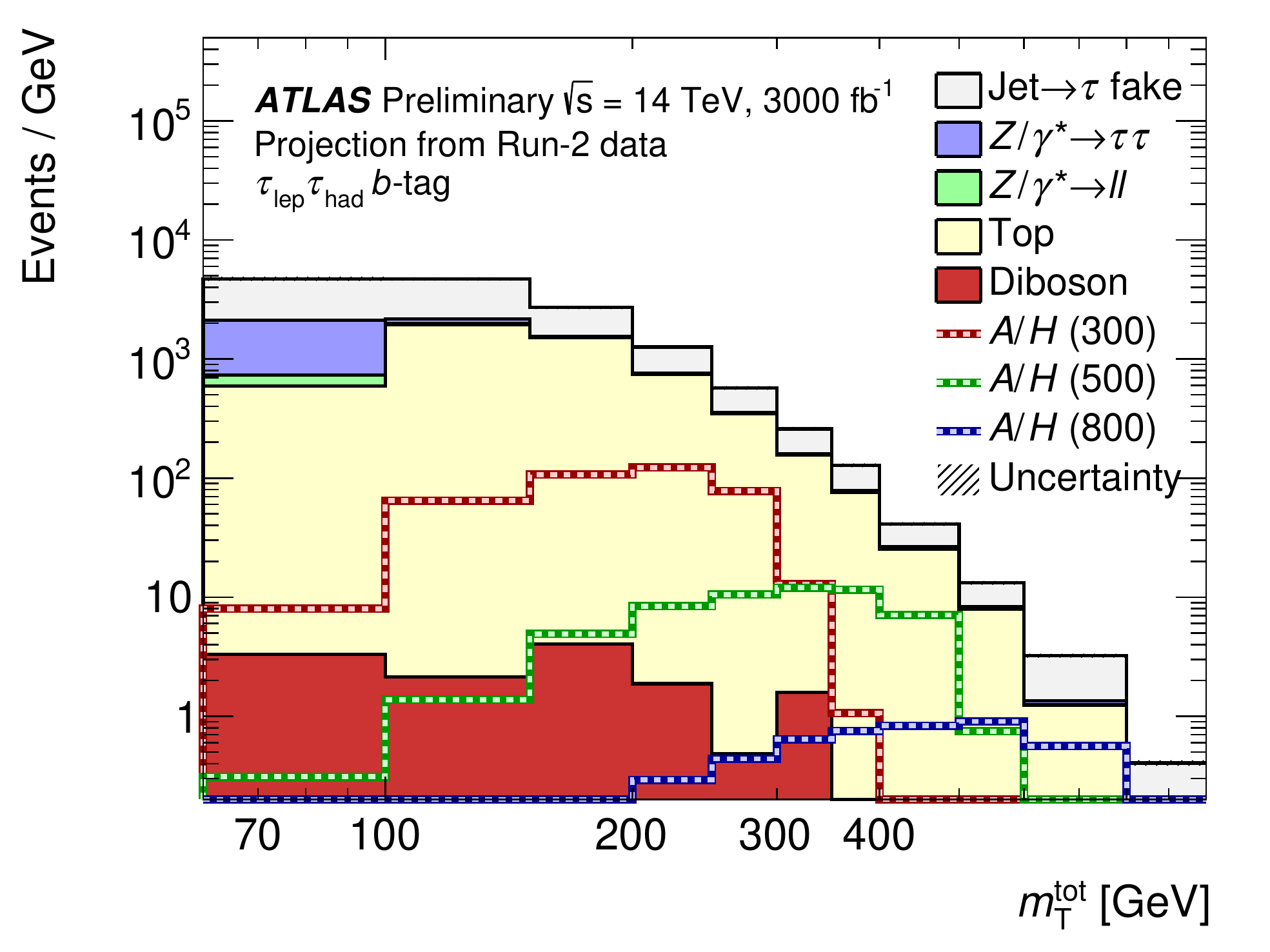}}
        \qquad
        \subfloat[\hadhad $b$-veto category]{\includegraphics[width=0.45\columnwidth]{\main/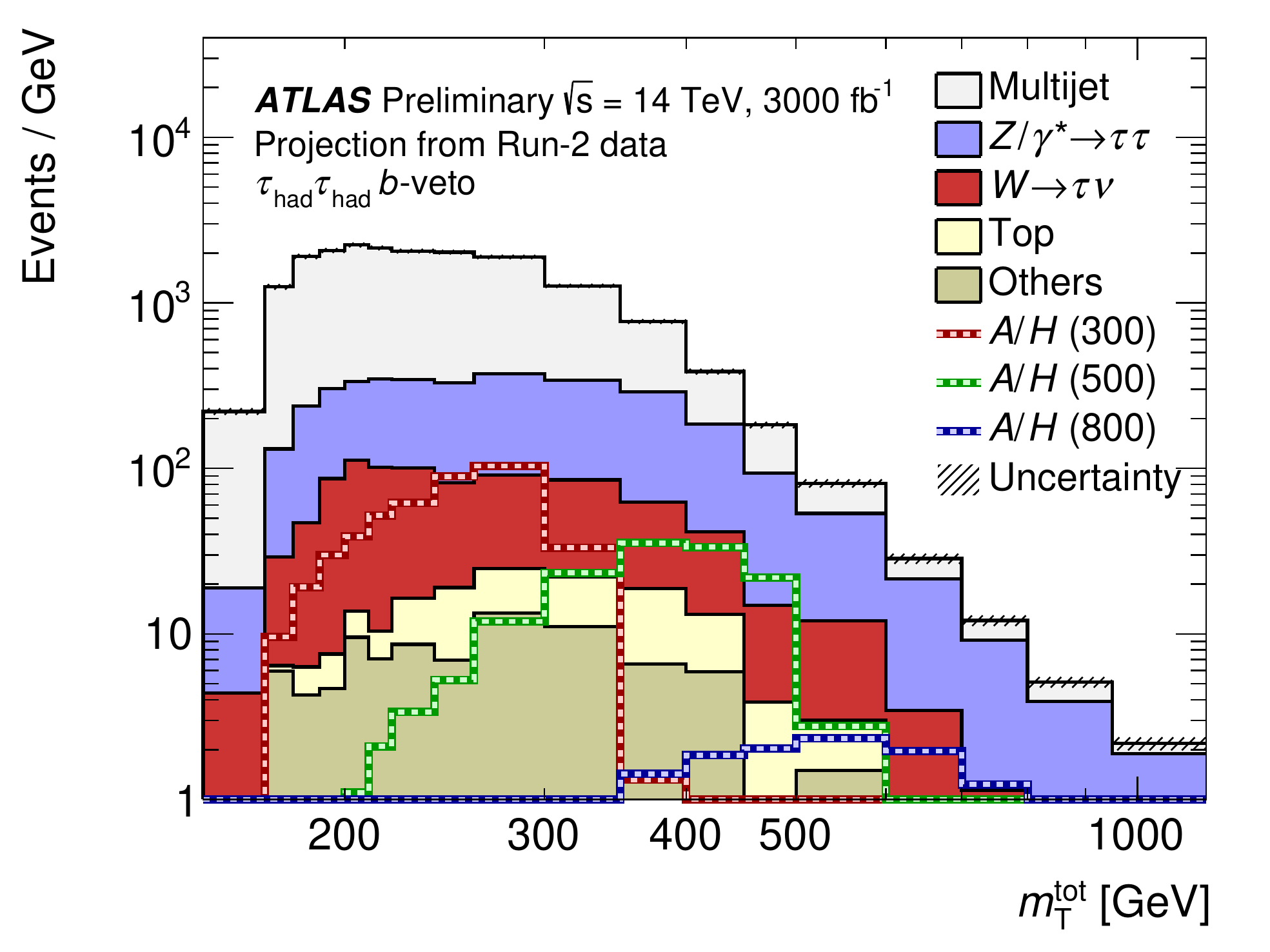}}
        \qquad
        \subfloat[\hadhad $b$-tag category]{\includegraphics[width=0.45\columnwidth]{\main/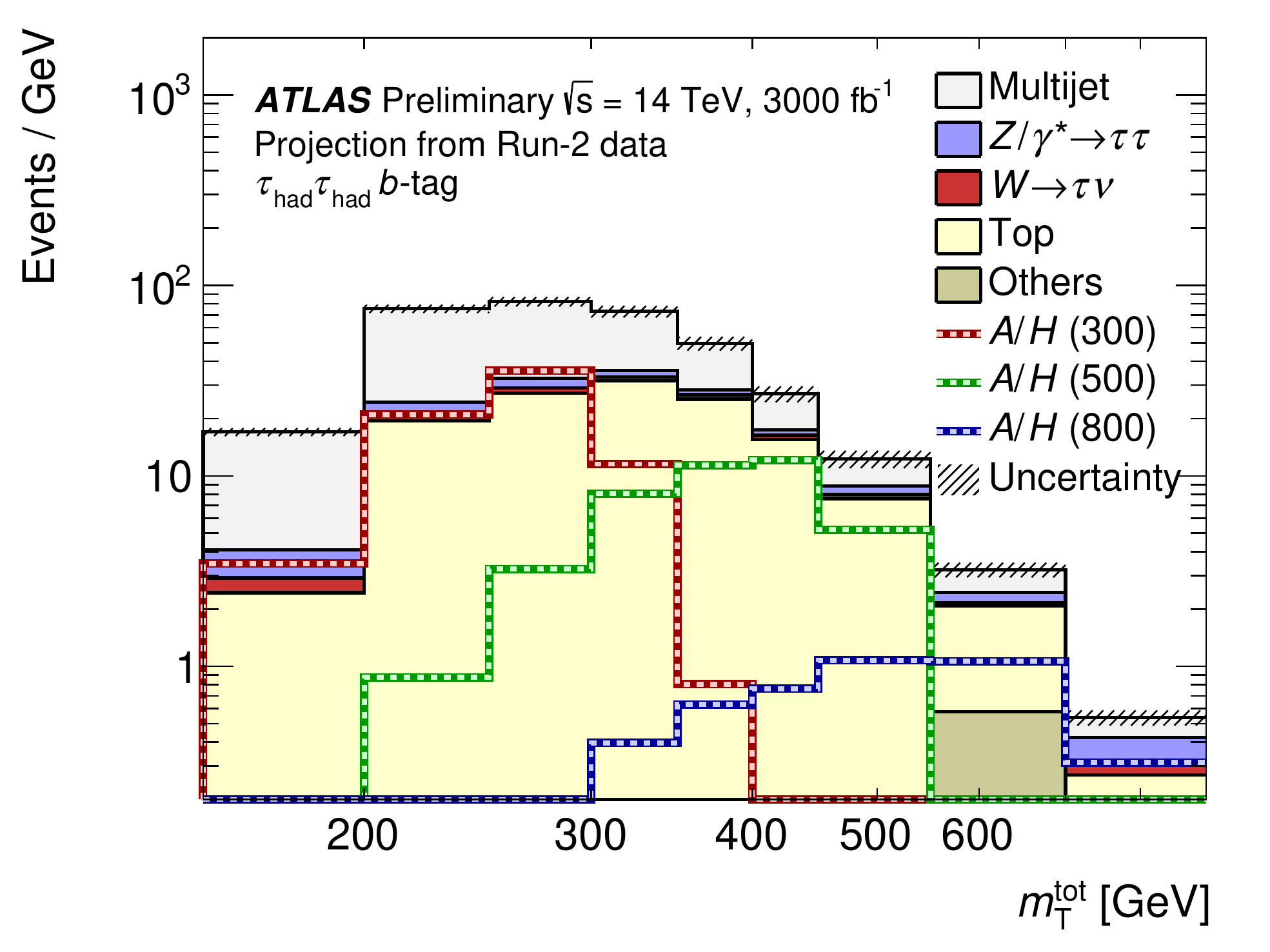}}
        \caption{Distributions of $\mTtot$ for each signal category. The predictions and uncertainties (including both statistical and systematic components) for the background processes are obtained from the fit under the hypothesis of no signal. The combined prediction for $A$ and $H$ bosons with masses of 300, 500 and 800~\GeV and $\tan\beta= 10$ in the hMSSM scenario are superimposed. }

    \label{fig:mTtotDistributionsSR}
\end{figure}

\begin{figure}[!ht]
    \centering
        \qquad
        \includegraphics[width=0.5\columnwidth]{\main/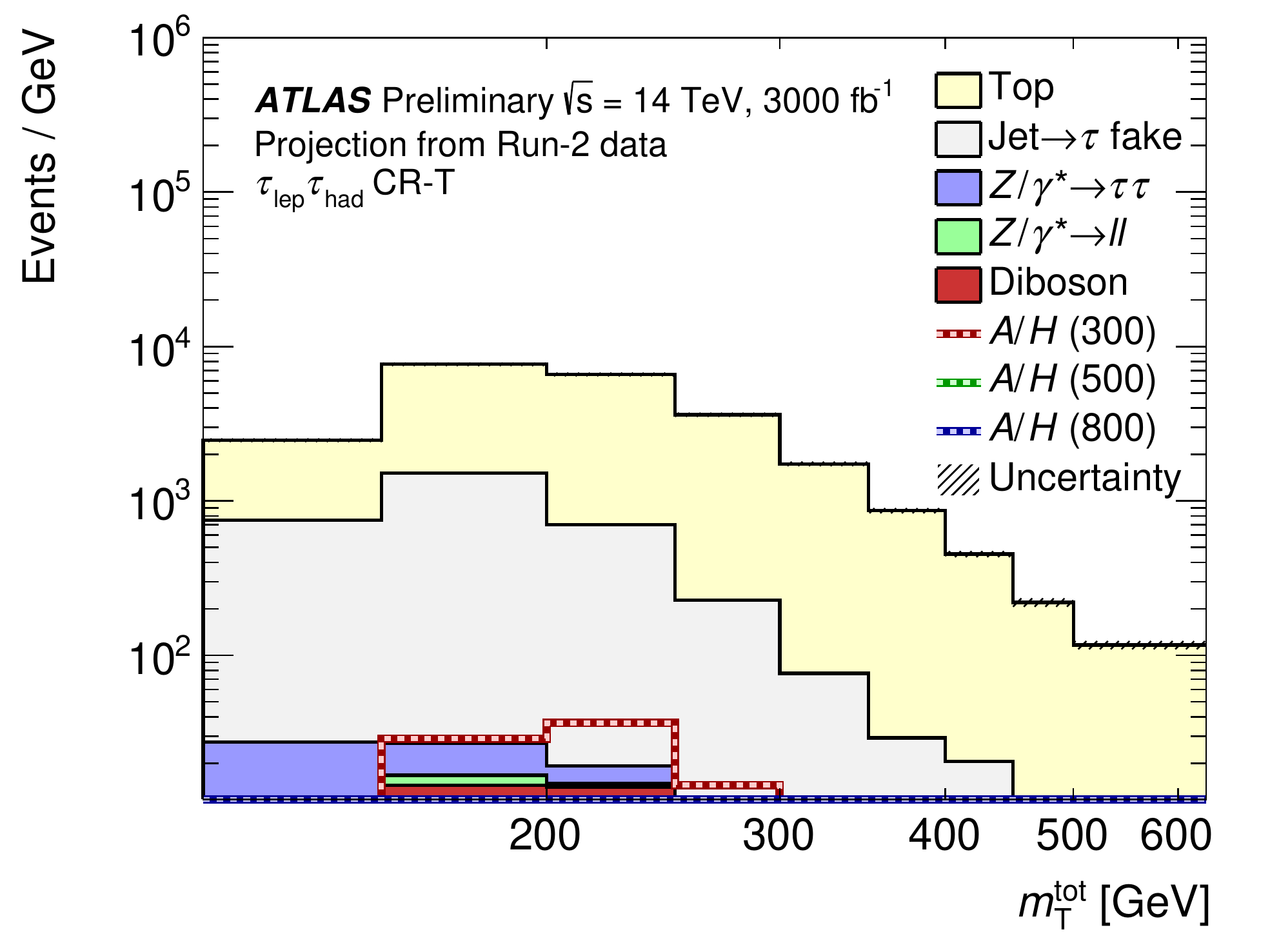}
        \caption{Distribution of $\mTtot$ distributions in in the top quark enriched control region of the \lephad channel.}
    \label{fig:mTtotDistributionsCR}
\end{figure}

The larger dataset at HL-LHC will give the opportunity to reduce the systematic uncertainties.
The ``Baseline'' scenario for the systematic uncertainty reduction compared to current Run 2 values follows
the recommendation of Ref.~\cite{ATLAS_PERF_Note}, according to which the systematic uncertainties
associated with $b$-tagging, $\tauhad$ (hadronic $\tau$ decay) and theoretical uncertainties
due to the missing higher order,
the PDF uncertainty, etc., are reduced. The systematic uncertainties associated with the reconstruction
and identification of the high-$\pt$ $\tauhad$ is reduced by a factor of 2 and becomes the leading systematic
uncertainty for a heavy Higgs boson with mass $m_{\phi} > 1~\TeV$. The systematic uncertainty associated
with the modelling of the jet to $\tauhad$ fake background is assumed to be the same as in the current analysis.
For the jet to $\tauhad$ fake background from multi-jet in \hadhad channel, the modelling uncertainty is mainly
due to the limited data size in the control region and is reduced by a factor of 2. The statistical uncertainties
on the predicted signal and background distributions, defined as the ``template stat.\@ uncertainty'', is determined
by the size of the MC samples and of the data sample in the control region where the \tauhad~ fake factor is applied.
The impact of the template stat.\@ uncertainty is negligible in the Run 2 analysis. Assuming large enough MC
samples will be generated for HL-LHC and sufficient data will be collected at HL-LHC, the uncertainties due to
the sample size is ignored in this extrapolation study.
To quantify the importance of the reduction of systematic uncertainties compared to current Run 2 values,
results (labelled as ``Unreduced'') will also be given with current Run 2 values
except for ignoring the template stat.\@ uncertainty.

\paragraph{Results}
\label{sec:result}
The $\mTtot$ distributions from the \lephad(separately in the electron and muon channels) and \hadhad signal regions, as well as the top control region, are used in the final combined fit to extract the signal. The statistical framework used to produce the \RunTwo results is documented in Ref.~\cite{ATLASRun2Ditau} and is adapted for this HL-LHC projection study. The results are given in terms of exclusion limits~\cite{CLs_2002}, as well as the 5 $\sigma$ discovery reach for gluon--gluon fusion and $b$-quarks association production modes.

\paragraph{Impact of systematic uncertainties}
The impact of systematic uncertainties on the upper limit of the cross section times branching ratio ($\sigma\times BR(\phi\to\tau\tau)$) in the Baseline scenario are calculated by comparing the expected 95\% CL upper limit in case of no systematic uncertainties, $\mu^{95}_{\text{stat}}$, with a limit calculated by introducing a group of systematic uncertainties, $\mu^{95}_i$, as described in Ref.~\cite{ATLASRun2Ditau}. The systematic uncertainty impacts are shown in Figure~\ref{fig:sysimpact, ggH} for gluon--gluon fusion production and Figure~\ref{fig:sysimpact, bbH} for $b$-quarks association production as a function of the scalar boson mass. The major uncertainties are grouped according to their origin, while minor ones are collected as ``Others'' as detailed in Ref.~\cite{ATLASRun2Ditau}.

The impact of systematic uncertainties is significant, as they degrade the expected limits by about 10--150 percent. In the low mass range, the leading uncertainties arise from the estimation of the dominant jet to $\tauhad$ fake background. At high masses, the leading uncertainty is from the reconstruction and identification of high-$\pt$ $\tauhad$. Because the $\mu^{95}_{\text{stat}}$ is mainly determined by the data statistical uncertainty, regions with low statistics are dominated by this uncertainty. In Figure~\ref{fig:sysimpact, ggH} the impact of the $\tauhad$ related systematic uncertainties decreases after $1~\TeV$ due to the fact that the results at the higher mass regime are more limited by the data statistical uncertainty, while in Figure~\ref{fig:sysimpact, bbH} the data statistical uncertainty in the \PQb-tag category dominates in the high mass regime which leads the high-$\pt$ $\tauhad$ systematic uncertainty less outstanding.

\begin{figure}[!ht]
    \centering
        \subfloat[gluon--gluon fusion production\label{fig:sysimpact, ggH}]{\includegraphics[width=0.4\columnwidth]{\main/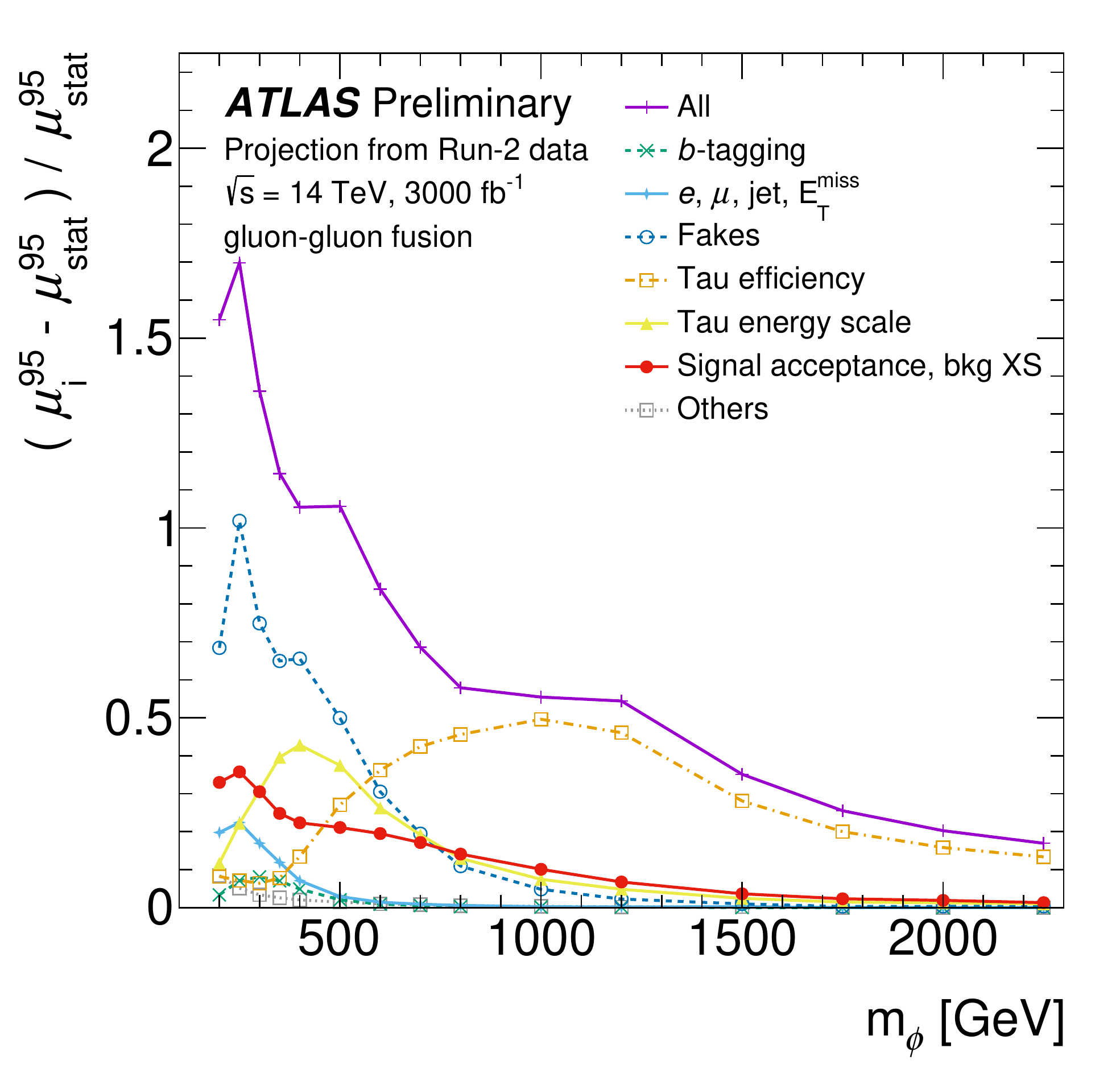}}
        \qquad
        \subfloat[$b$-associated production\label{fig:sysimpact, bbH}]{\includegraphics[width=0.4\columnwidth]{\main/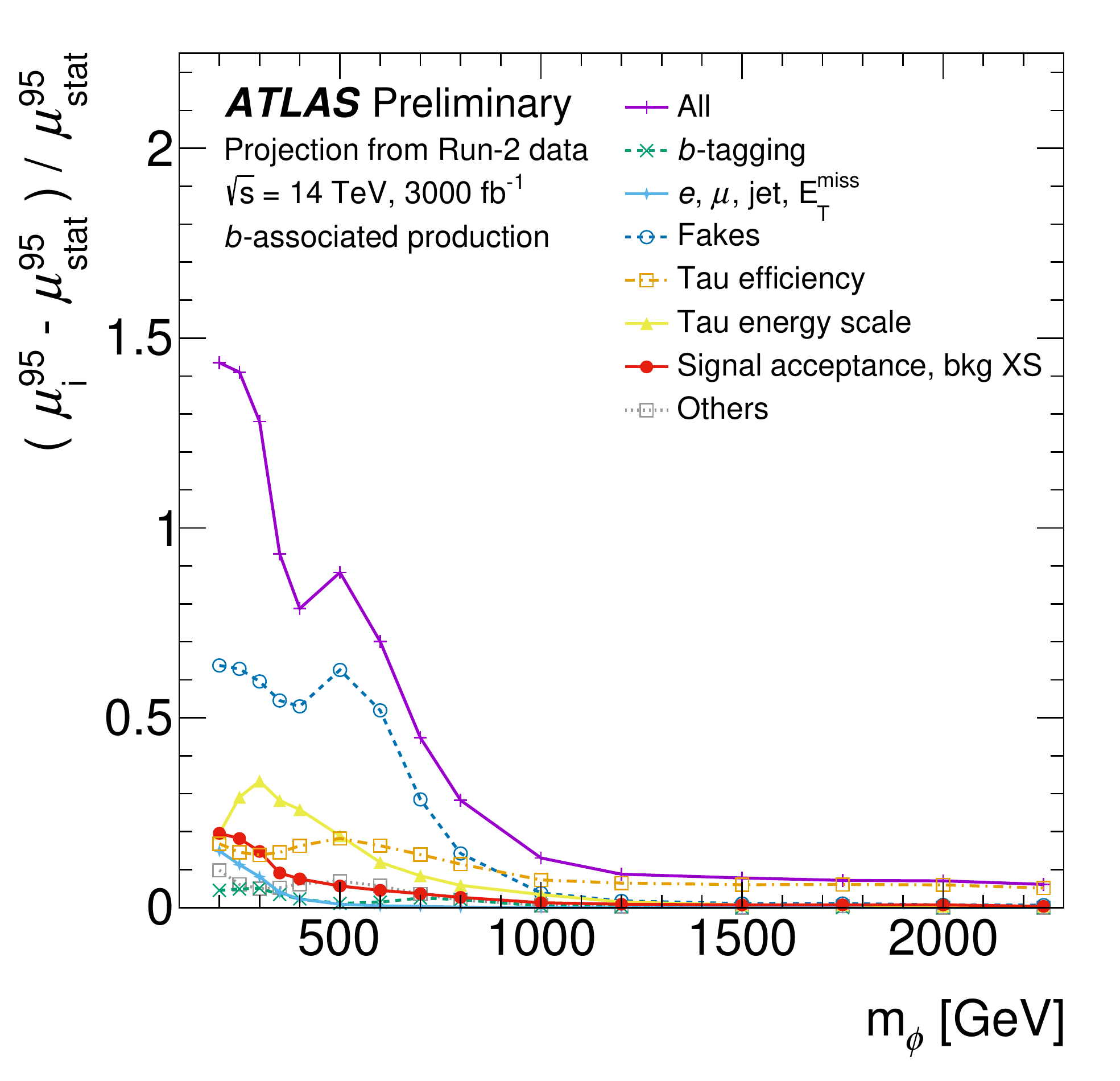}}
        \caption{Impact of major groups of systematic uncertainties (Baseline) on the $\phi\to\tau\tau$ 95\% CL cross section upper limits as a function of the scalar boson mass, separately for the (a) gluon--gluon fusion and (b) $b$-associated production mechanisms.}
    \label{fig:sysimpact}
\end{figure}

\paragraph{Cross section limits and discovery reach}
Figure \ref{fig:modelInde} shows the upper limits on the gluon--gluon fusion and $b$-quark associated production cross section times the branching fraction for $\phi\rightarrow\tau\tau$. To demonstrate the impact of systematic uncertainties, the expected exclusion limits with different systematic uncertainty scenarios are shown, as well as the Run 2 expected results~\cite{ATLASRun2Ditau}. The peaking structure around $m_\phi = 1~\TeV$ in figure \ref{fig:modelInde, ggH} is due to the impact of the high-$\pt$ $\tauhad$ systematic uncertainty. The 5 $\sigma$ sensitivity line in the same figure illustrates the smallest values of the cross section times the branching fraction for which discovery level can be reached at HL-LHC: as clearly shown, the region where discovery is expected at HL-LHC extends significantly below the currently expected Run 2 exclusion region.

\begin{figure}[!ht]
    \centering
        \subfloat[gluon--gluon fusion production\label{fig:modelInde, ggH}]{\includegraphics[width=0.4\columnwidth]{\main/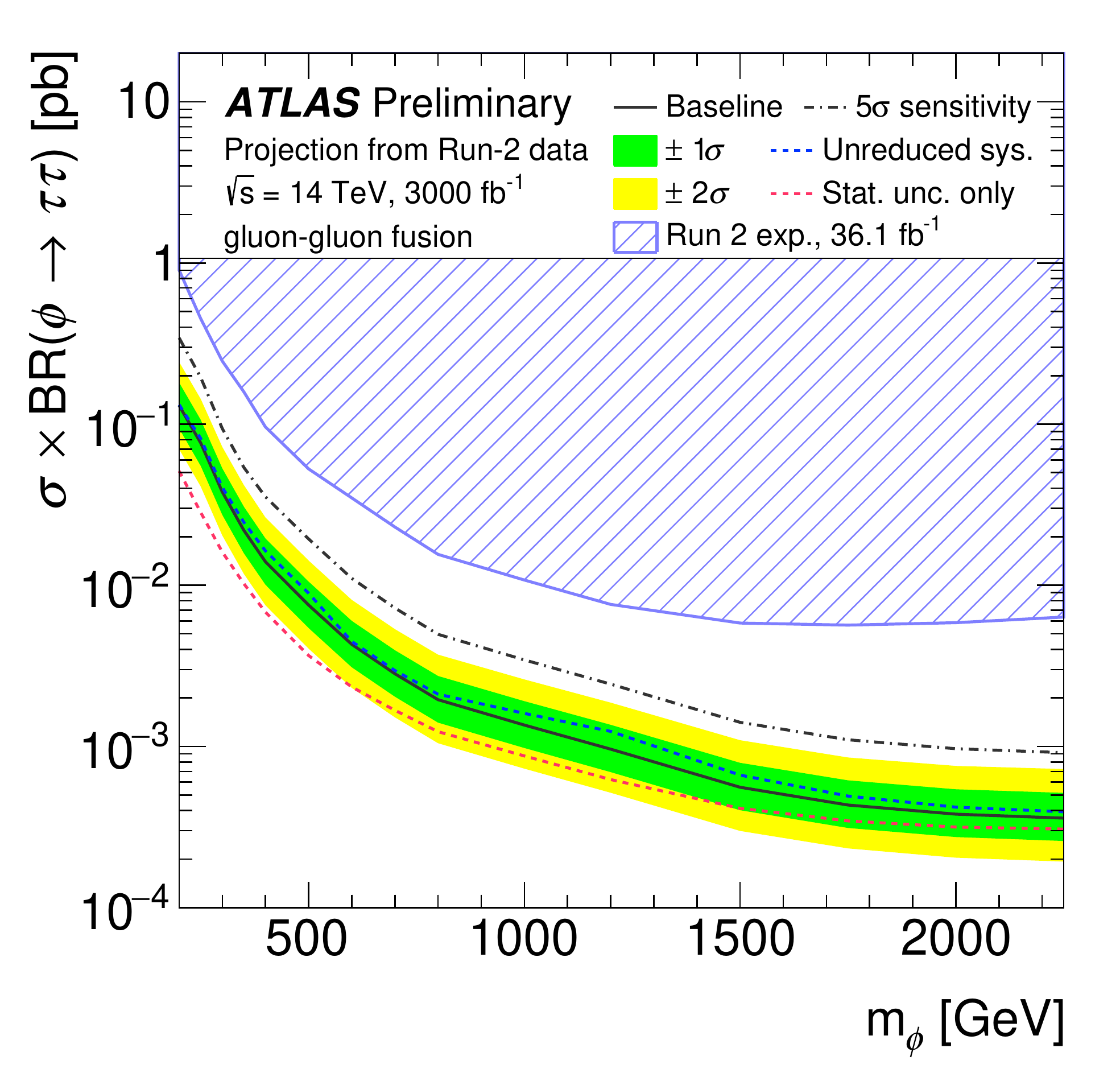}}
        \qquad
        \subfloat[$b$-associated production\label{fig:modelInde, bbH}]{\includegraphics[width=0.4\columnwidth]{\main/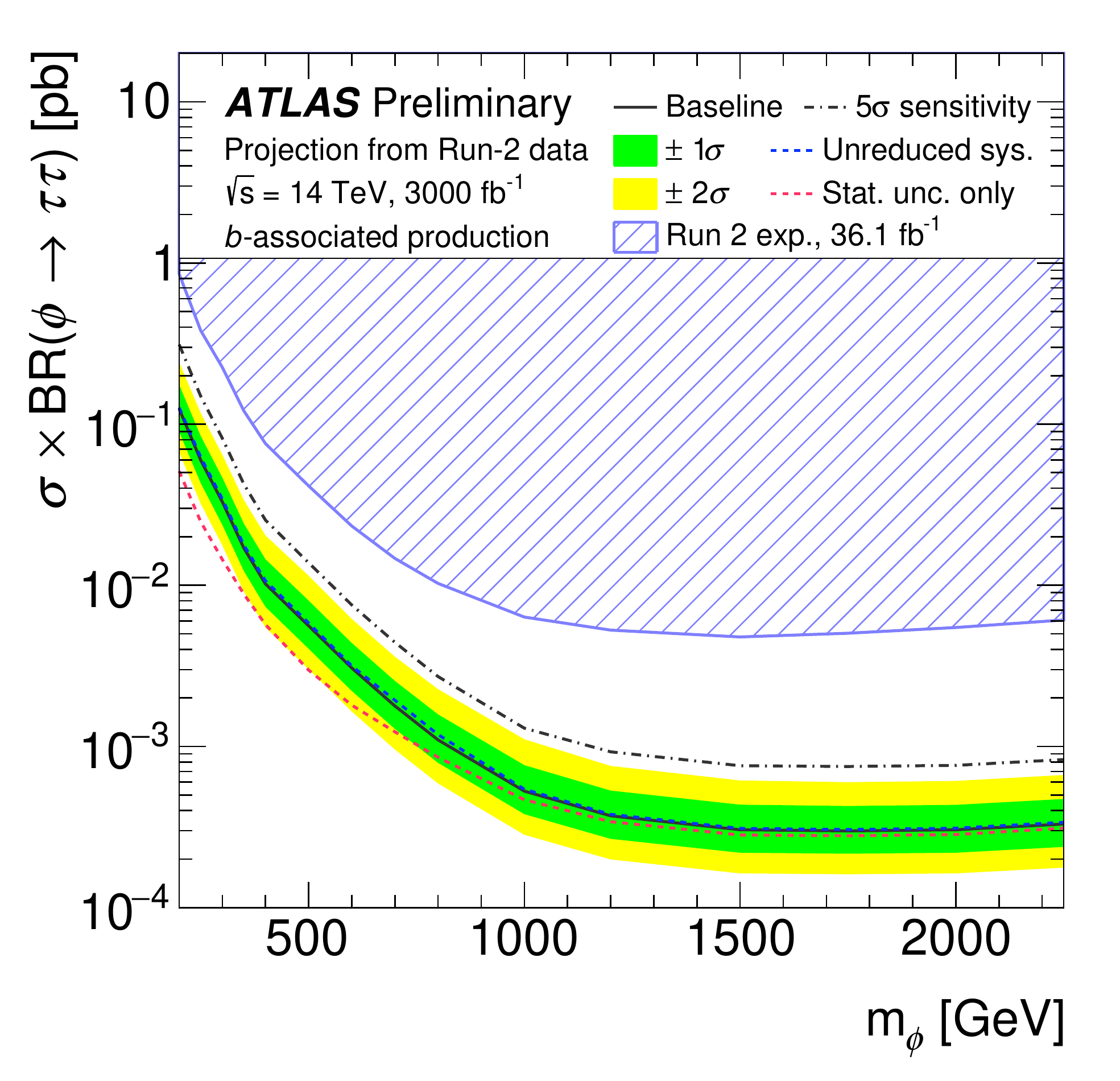}}
        \caption{
         Projected 95\% CL upper limits on the production cross section times the $\phi\rightarrow\tau\tau$ branching fraction for a scalar boson $\phi$ produced via (a) gluon--gluon fusion and (b) $b$-associated production, as a function of scalar boson mass. The limits are calculated from a statistical combination of the \ehad, \muhad and \hadhad channels. ``Baseline'' uses the reduced systematic uncertainties scenario described in the text. ``Unreduced sys.\@'' uses the same systematic uncertainties as the Run 2 analysis while ignoring the template stat.\@ uncertainty. ``Stat.\@ unc.\@ only'' represents the expected limit without considering any systematic uncertainty. ``5 $\sigma$ sensitivity'' shows the region with the potential of 5 $\sigma$ significance in the Baseline scenario.
        }
    \label{fig:modelInde}
\end{figure}


\paragraph{MSSM interpretation}
Results are interpreted in terms of the MSSM\@. The cross section calculations follow the exact procedure used in Ref.~\cite{ATLASRun2Ditau}, apart from the centre of mass energy is switched to 14~\TeV. Figure~\ref{fig:model1} shows regions in the $m_A$--$\tan\beta$ plane excluded at 95\% CL or discovered with 5 $\sigma$ significance in the hMSSM and \mhmodp scenarios. In the hMSSM scenario, $\tan\beta > 1.0$ for $250~{\rm{\UGeV}}< m_A < 350$~\GeV and $\tan\beta > 10$ for $m_A = 1.5~\TeV$ could be excluded at 95\% CL\@. When $m_A$ is above the $A/H\to \ttbar$ threshold, this additional decay mode reduces the sensitivity of the $A/H\to\tau\tau$ search for low $\tan\beta$. In the MSSM \mhmodp scenario, the expected 95\% CL upper limits exclude $\tan\beta >2$ for $250~{\rm{\UGeV}}< m_A< 350$~\GeV and $\tan\beta > 20$ for $m_A =1.5~\TeV$.

\begin{figure}[!ht]
    \centering
        \subfloat[hMSSM scenario]{\includegraphics[width=0.4\columnwidth]{\main/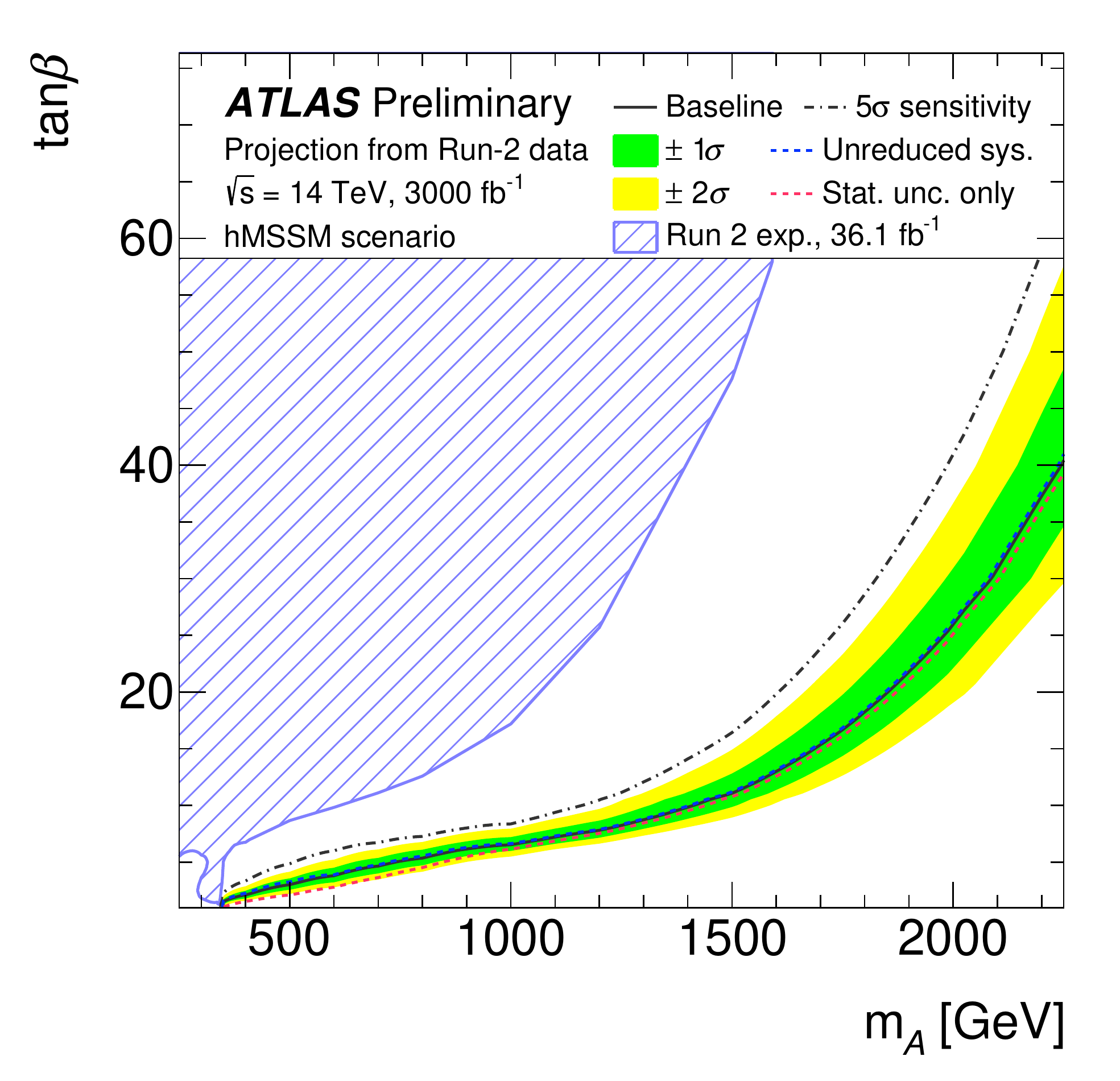}}
        \qquad
        \subfloat[\mhmodp scenario]{\includegraphics[width=0.4\columnwidth]{\main/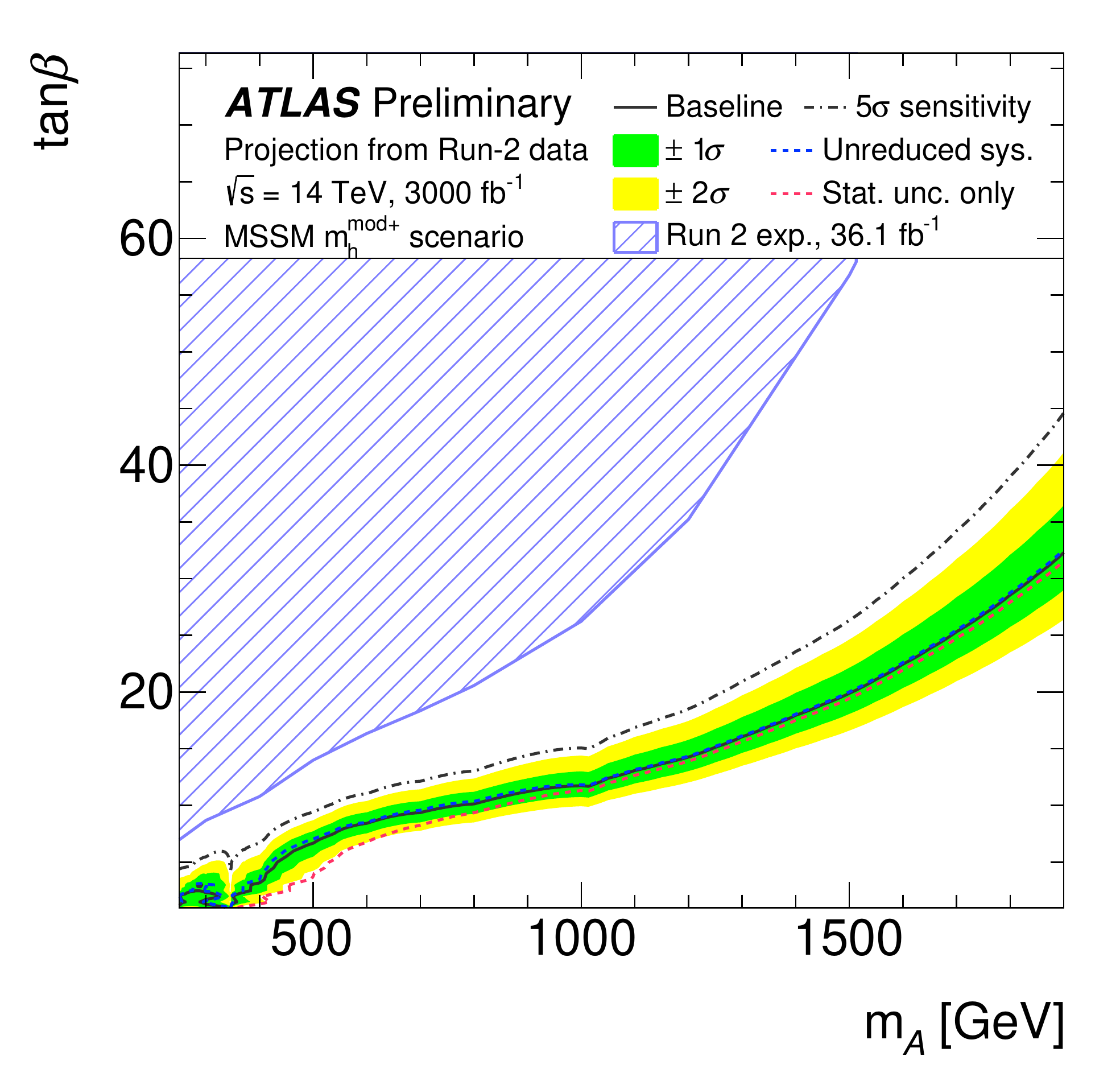}}
        \caption{Projected 95\% CL limits on $\tan\beta$ as a function of $m_\phi$ in the MSSM (a) hMSSM and (b) \mhmodp scenarios. The limits are calculated from a statistical combination of the \ehad, \muhad and \hadhad channels. ``Baseline'' uses the reduced systematic uncertainties scenario described in the text. ``Unreduced sys.\@'' uses the same systematic uncertainties as the Run 2 analysis while ignoring the template stat.\@ uncertainty. ``Stat.\@ unc.\@ only'' represents the expected limit without considering any systematic uncertainty. ``5 $\sigma$ sensitivity'' shows the region with the potential of 5 $\sigma$ significance in the Baseline scenario.
        }
    \label{fig:model1}
\end{figure}

\FloatBarrier

\paragraph{Conclusion}
\label{sec:conclusion}
The $H/A \to \tau\tau$ analysis documented in \cite{ATLASRun2Ditau} has been extrapolated to estimate the sensitivity with $3000~\ifb$ of the HL-LHC dataset. The expected upper limits at 95\% CL or, in alternative, the 5 $\sigma$ discovery reach in terms of cross section for the production of scalar bosons times the branching fraction to di-tau final states have been estimated. The region with 5 $\sigma$ discovery potential at HL-LHC extends significantly below the currently expected Run 2 exclusion region. The expected limits are in the range 130--0.4~$\fb$ (130--0.3~$\fb$) for gluon--gluon fusion ($b$-associated) production of scalar bosons with masses of 0.2--2.25~$\TeV$. A factor of 6 to 18 increase in the sensitivity compared to the searches with the 36.1~$\ifb$ Run 2 data~\cite{ATLASRun2Ditau} is projected. In the context of the hMSSM scenario, in the absence of a signal, the most stringent limits expected for the combined search exclude $\tan\beta > 1.0$ for $250~{\rm{\UGeV}} < m_A < 350$~\GeV and $\tan\beta > 10$ for $m_A = 1.5~\TeV$ at 95\% CL\@. The systematic uncertainties degrade the exclusion limit on $\sigma\times BR(\phi\to\tau\tau)$ by more than a factor of 2 for $m_\phi<500~\GeV$ and about 10\%--20\% for $m_\phi= 2~\TeV$. While the uncertainty on the estimate of fake $\tauhad$ dominates at low $m_{\phi}$, the uncertainty on high-$\pt$ $\tauhad$ reconstruction and identification is the leading systematic uncertainty at $m_\phi > 1.0~\TeV$.

\asubsubsection[M. Flechl]{Projection of Run-2 CMS searches for MSSM heavy neutral Higgs bosons}\label{sec:CMShtt}
Searches for Minimal Supersymmetric Standard Model  Higgs bosons have been performed by CMS using
the 2016 data from the LHC Run~2~\cite{HIG-18-014,HIG-16-018,HIG-17-020}.
So far,
no significant evidence for physics beyond the SM has been found.
However, the LHC to date has delivered only a small fraction of the
integrated luminosity expected over its lifetime.
Searches that are currently limited by statistical precision
will see significant extensions in their reach as larger data sets are collected.
Among the searches that will benefit are those for MSSM Higgs bosons.

In this section,
projections are presented for the reach that can be expected at higher luminosities
in searches for heavy neutral Higgs bosons that decay to a pair of tau leptons~\cite{CMS-PAS-FTR-18-017}.
The projections are based on the most recent CMS publication
for this search~\cite{HIG-17-020},
performed using 35.9\fbinv of data collected during 2016
at a centre-of-mass energy of 13 \UTeV.
All the details of the analysis,
including the simulated event samples, background estimation methods,
systematic uncertainties, and different interpretations are described in Ref.~\cite{HIG-17-020}.
Only details of direct relevance to the projection are presented here.

The analysis is a direct search for a neutral resonance
decaying to two tau leptons.
The following tau lepton decay mode combinations are considered: $\mutau$, $\etau$,
$\tautau$, and $\emu$, where $\tauhad$ denotes a hadronically decaying tau lepton.
In all these channels, events are separated into those that contain
at least one b-tagged jet and those that do not contain any b-tagged jet. 
The goal of this categorisation is to increase sensitivity to the dominant MSSM production modes: 
gluon fusion ($ggF$) and production in association with b quarks ($bbH$).
The final discriminant is the total 
transverse mass, defined in Ref.~\cite{HIG-17-020}.
The signal hypotheses considered consist of additional Higgs
bosons in the mass range from 90 \UGeV to 3.2 \UTeV.
The projection of the limits 
is performed by scaling all the signal
and background processes to integrated luminosities
of 300 and 3000\fbinv,
where the latter integrated luminosity corresponds to the total that is
expected for the High-Luminosity LHC. 

A previous CMS projection of the sensitivity for
MSSM Higgs boson decays to a pair of tau leptons at the HL-LHC
is reported in Ref.~\cite{FTR-16-002}.
The results are presented in terms of model independent
limits on a heavy resonance (either \PH\ or {\PA},
generically referred to as \PH\ below)
decaying to two tau leptons,
and are also interpreted in the context of MSSM benchmark scenarios.
\paragraph{Projection methodology}
\label{sec:method}
Three scenarios are considered for the projection of the size of
systematic uncertainties to the HL-LHC:
\begin{itemize}
\item
statistical uncertainties only: all systematic uncertainties are neglected;
\item
  Run 2 systematic uncertainties: all systematic uncertainties are held
  constant with respect to luminosity, i.e., they are assumed to be
  the same as for the 2016 analysis;
\item
  YR18 systematic uncertainties: systematic uncertainties are assumed to
  decrease with integrated luminosity
  following a set of assumptions described below.
\end{itemize}

In the YR18 scenario,
selected systematic uncertainties decrease
as a function of luminosity until they reach a certain minimum value. 
Specifically, all pre-fit uncertainties of an experimental nature
(including statistical uncertainties in control regions and
in simulated event samples) 
are scaled proportionally to the square root of the integrated luminosity.
The following minimum values are assumed:
\vspace{-0.2cm}
\begin{itemize}
\item
muon efficiency: 25\% of the 2016 value, corresponding to an average absolute uncertainty of about 0.5\%; 
\item
electron, $\tauhad$, and b-tagging efficiencies: 50\% of the 2016 values, 
corresponding to average absolute uncertainties
of about 0.5\%, 2.5\%, and 1.0\%, respectively;
\item
jet energy scale: 1\% precision for jets with $\pt > 30$ \UGeV; 
\item
  estimate of the background due to jets mis-reconstruction
  as $\tauhad$~\cite{HIG-15-007},
  for the components that are not statistical in nature:
  50\% of the 2016 values;
\item
luminosity uncertainty: 1\%;
\item
  theory uncertainties: 50\% of the 2016 values,
  independent of the luminosity for all projections.
\end{itemize}
Note that for limits in which the Higgs boson mass is larger than about 1 \UTeV,
the statistical uncertainties dominate and
the difference between the systematic uncertainties found from
the different methods has a negligible impact on the results.

The lightest Higgs boson, \Ph, is excluded from the SM versus MSSM hypothesis test for the following reason:
With increasing luminosity, the search will become sensitive to this boson. However, the current benchmark scenarios do not
incorporate the properties of the \Ph boson with the accuracy required at the time of the HL-LHC.
Certainly the benchmark scenarios will evolve with time in this respect. Therefore the signal hypothesis includes
only the heavy $\PA$ and \PH bosons, to demonstrate the search potential only for these.
\paragraph{Model-independent limits}
\label{sec:model_indep}
The model independent 
95\% \CL upper
limit on the cross sections for the ggH and bbH 
production modes, 
with the subsequent decays $\PH\to\tau\tau$,
are shown in Figs.~\ref{fig:model_indep} and~\ref{fig:model_indep2} for 
integrated luminosities of 300, 3000 and 6000\fbinv. 
For the limit on one process, e.g., gluon fusion, the normalisation for the other process, e.g. b-associated production, is treated as a freely varying parameter in the fit performed prior to the limit calculation.
The 6000\fbinv limit is an approximation of the sensitivity with the complete HL-LHC 
dataset to be collected by the ATLAS and CMS experiments, corresponding to an integrated 
luminosity of 3000\fbinv each. The approximation assumes that the results of the 
two experiments are uncorrelated and that their sensitivity is similar. The first 
assumption is fulfilled to a high degree because the results are statistically limited; 
the validity of the second assumption is evident by comparing previous limits and 
projections. 
\begin{figure}[htbp]
\begin{center}
\subfloat[ggH]{\includegraphics[width=0.45\textwidth]{\main/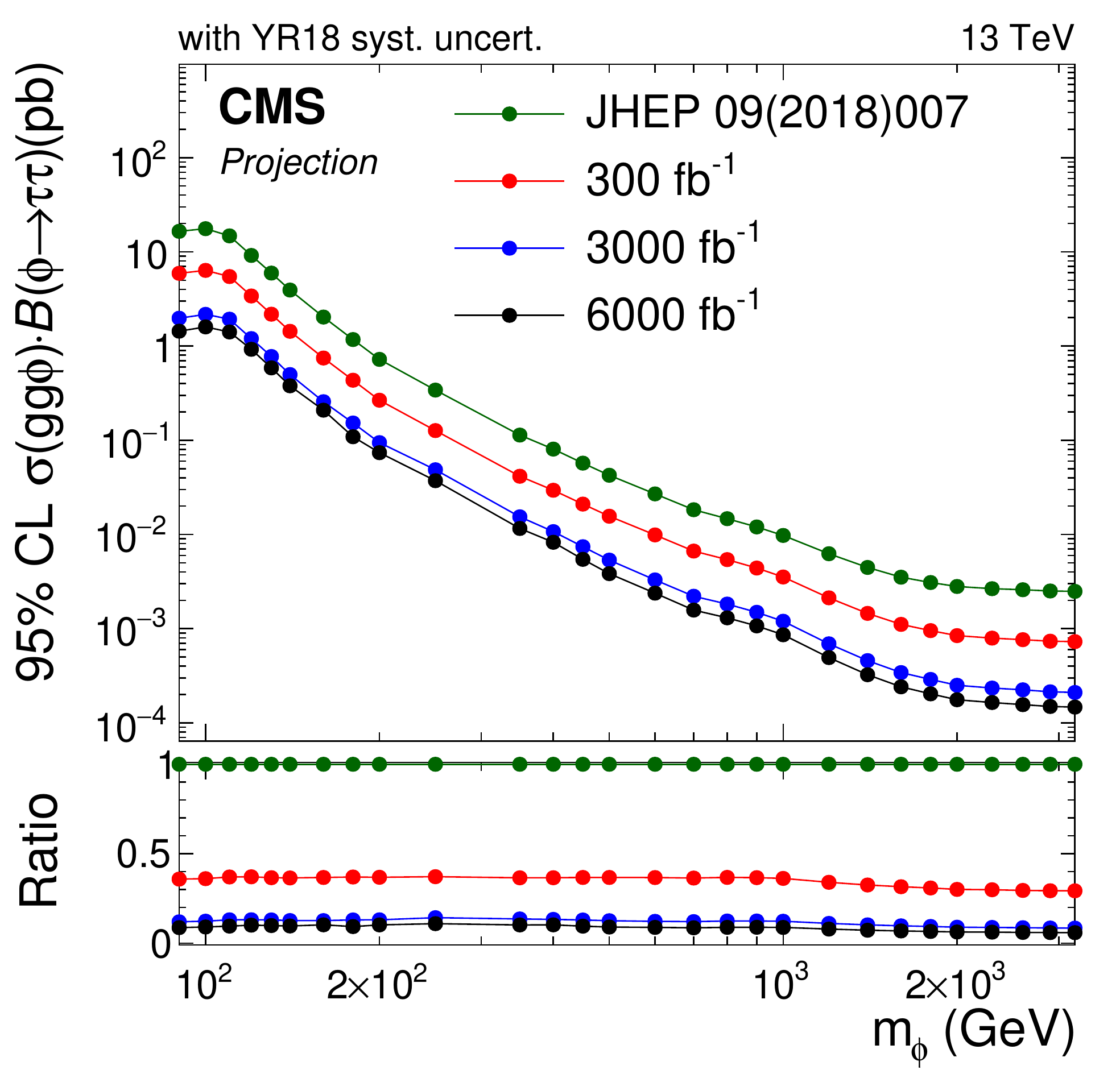}}
\subfloat[bbH]{\includegraphics[width=0.45\textwidth]{\main/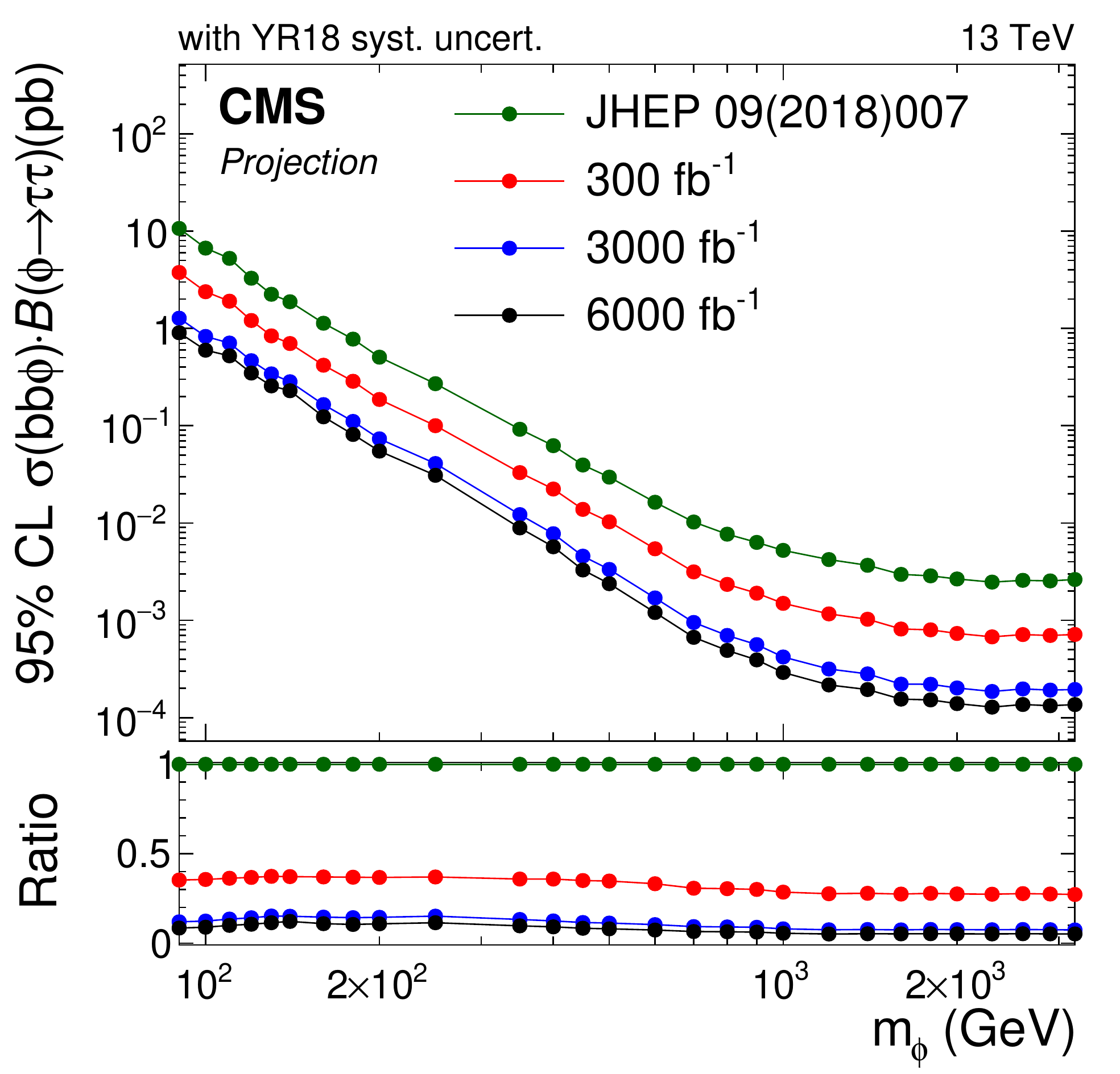}}
\end{center}
\caption{Projection of expected model independent
  95\% CL upper limits based on 2016 CMS data~\cite{HIG-17-020} for 
$ggH$ and $bbH$ production with subsequent \htt decays, with YR18 systematic 
uncertainties~\cite{CMS-PAS-FTR-18-017}. 
The limit shown for 6000\fbinv is an approximation of the 
sensitivity with the complete HL-LHC dataset to be collected by the ATLAS and 
CMS experiments, corresponding to an integrated luminosity of 3000\fbinv each.
The limits are compared to the CMS result using 2016 data~\cite{HIG-17-020}.}
\label{fig:model_indep}
\end{figure}

For both production modes, 
the improvement in the limits at high mass values
scales similarly to the square root of the integrated luminosity,
as expected from the increase in statistical precision.
The improvement at very low mass is almost entirely a consequence of reduced systematic uncertainties
and not the additional data in the signal region. 
The difference between the Run 2 and YR18 scenarios results mostly from of the treatment 
of two kinds of systematic uncertainty of a statistical nature: 
the uncertainty related to the number of simulated events 
and that related to the number 
of events in the data control regions.
\begin{figure}[htbp]
\begin{center}
\subfloat[$ggH$]{\includegraphics[width=0.45\textwidth]{\main/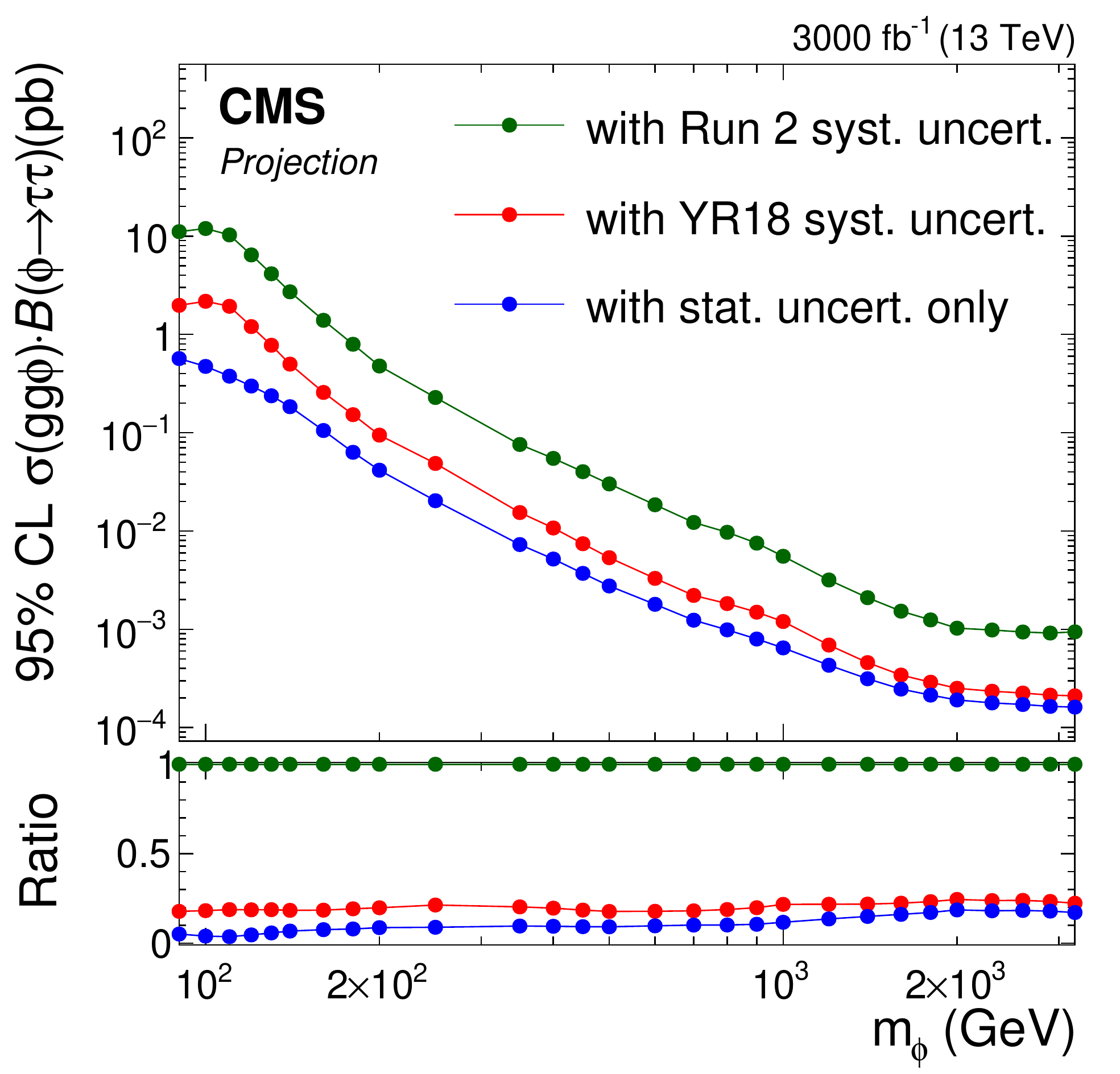}}
\subfloat[$bbH$]{\includegraphics[width=0.45\textwidth]{\main/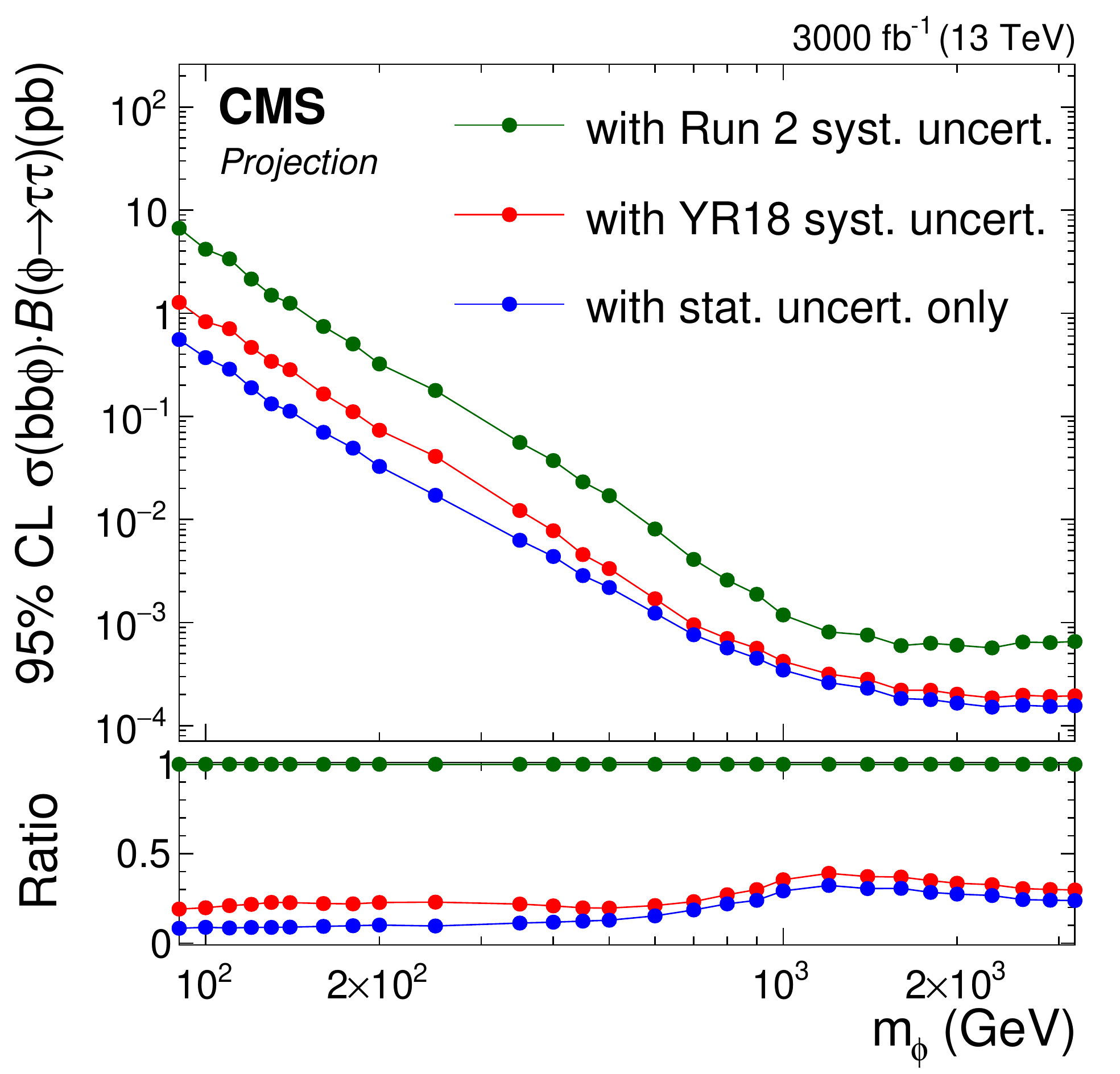}}
\end{center}
\caption{Projection of expected model-independent limits based on 2016 CMS data~\cite{HIG-17-020} 
for $ggH$ and $bbH$ production with subsequent \htt decays, comparing different 
scenarios for systematic uncertainties for an integrated luminosity of 3000\fbinv.}
\label{fig:model_indep2}
\end{figure}

Figure~\ref{fig:model_indep2d}  shows the exclusion contours corresponding to 68\% and 95\% CL for a scan of the likelihood as a function of both the gluon fusion and the b-associated cross section, for a few representative mass points.
\begin{figure}[htbp]
\begin{center}
\subfloat[$ggH$]{\includegraphics[width=0.45\textwidth]{\main/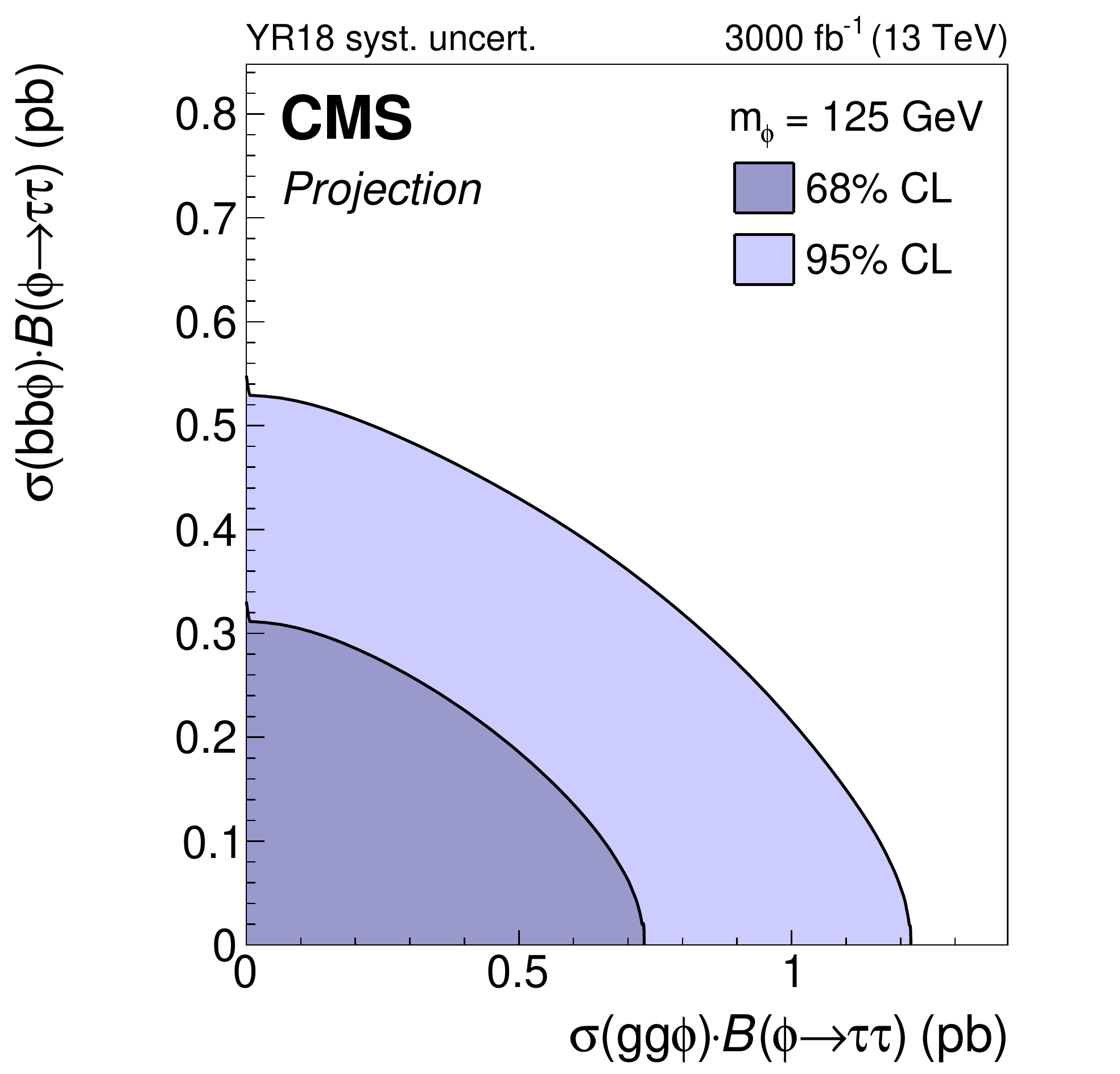}}
\subfloat[$ggH$]{\includegraphics[width=0.45\textwidth]{\main/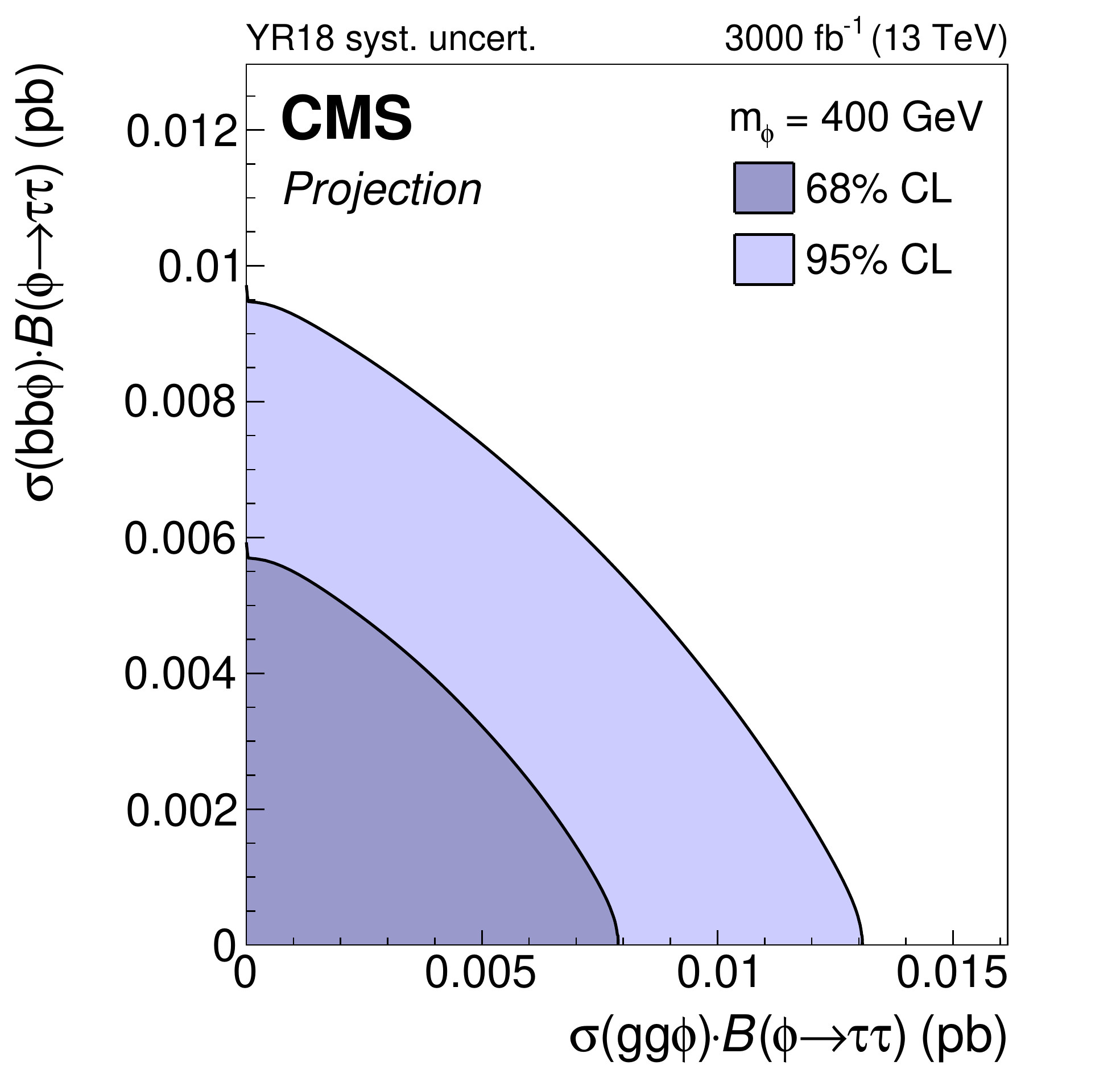}}\\
\subfloat[$ggH$]{\includegraphics[width=0.45\textwidth]{\main/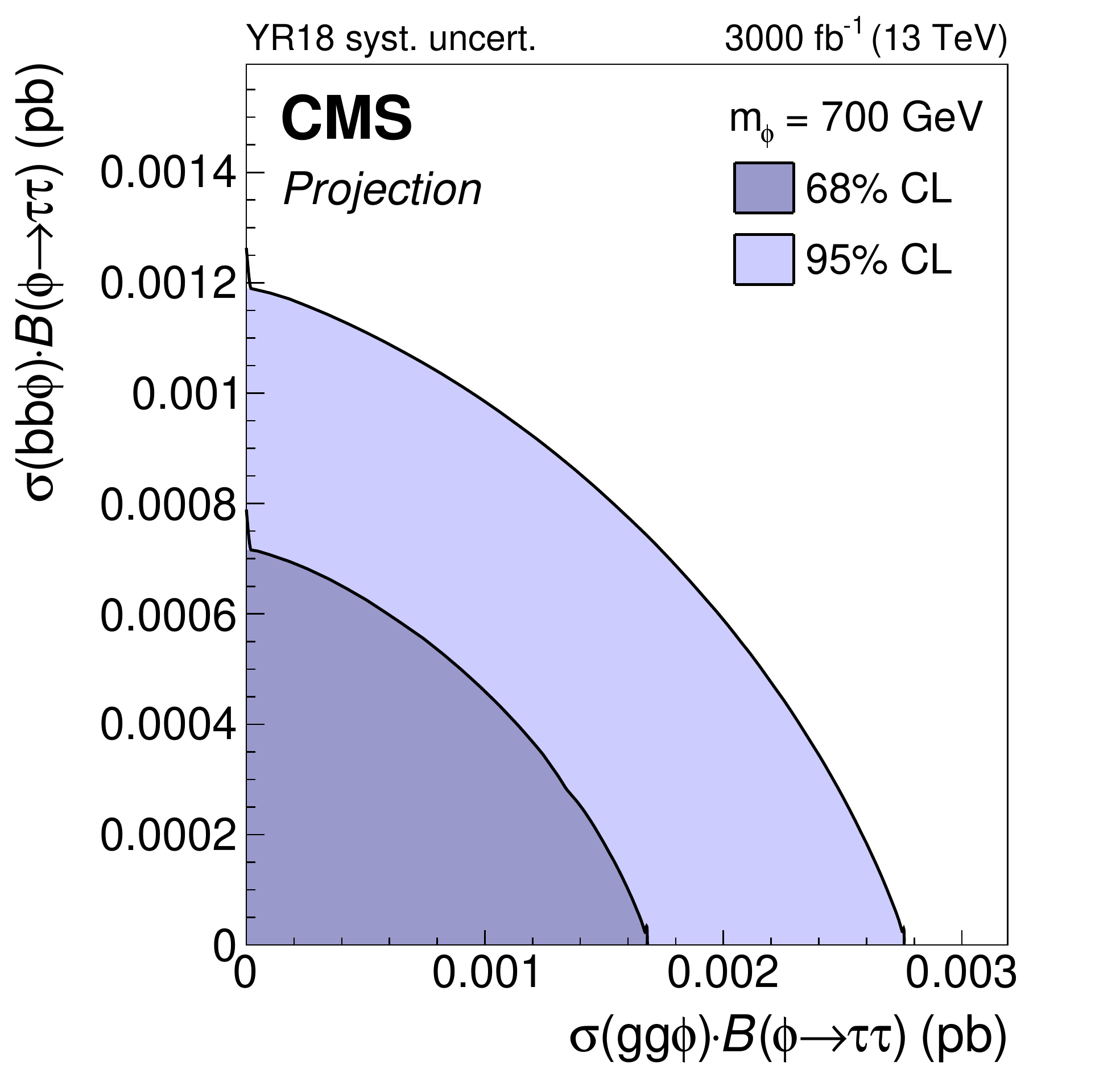}}
\subfloat[$ggH$]{\includegraphics[width=0.45\textwidth]{\main/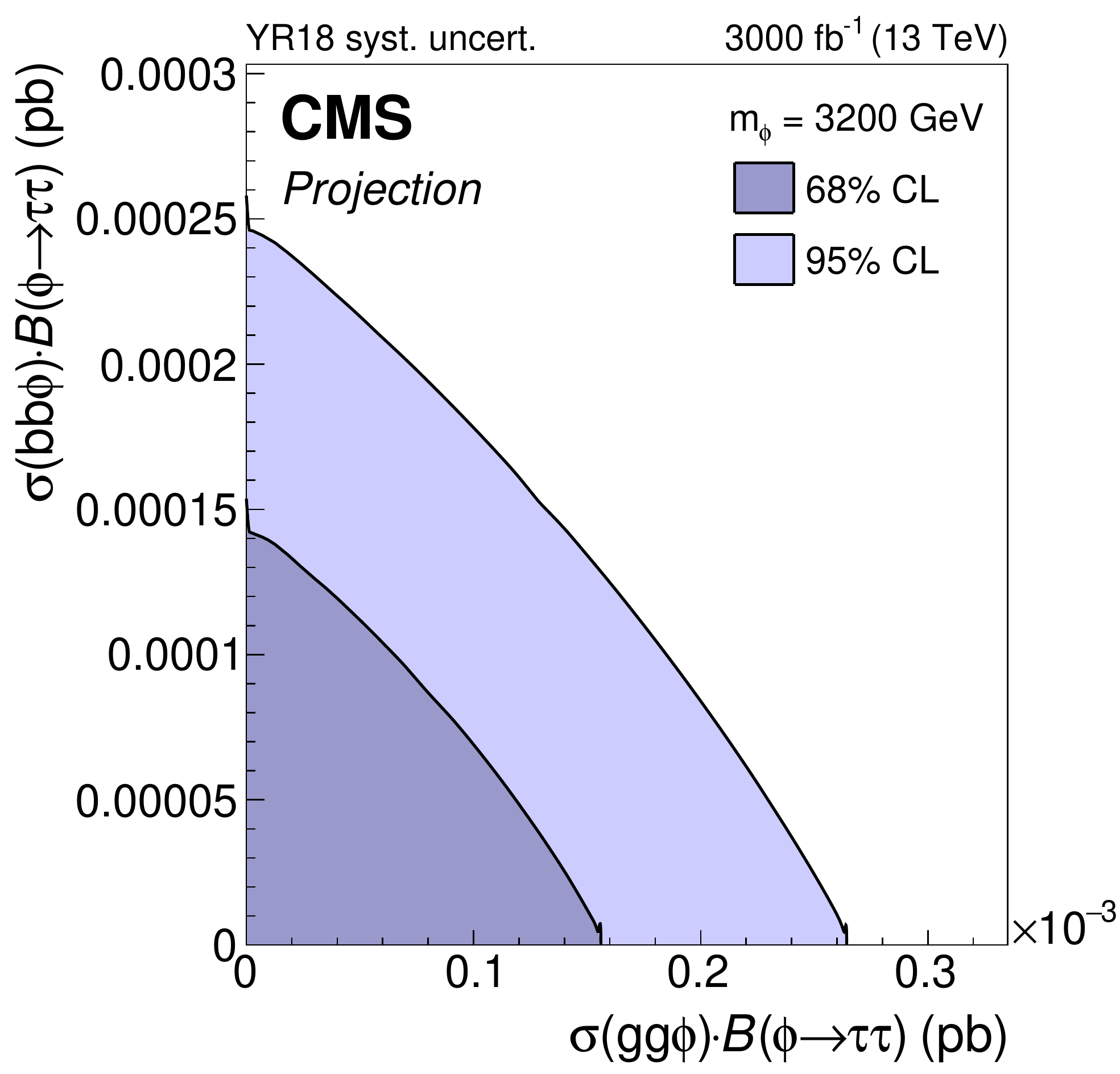}}
\end{center}
\caption{Projection of expected model-independent limits based on 2016 CMS data~\cite{HIG-17-020} for a scan of the likelihood for the $ggH$ and $bbH$ production cross sections with subsequent \htt decays, 
for an integrated luminosity of 3000\fbinv and with YR18 systematic uncertainties.}
\label{fig:model_indep2d}
\end{figure}

\paragraph{Model-dependent limits}
\label{sec:model_dep}
At the tree level, the Higgs sector of the MSSM can be specified by suitable choices for two variables, 
often chosen to be the mass $m_{\PA}$ of the pseudoscalar Higgs boson
and $\tan\beta$, the ratio of the \mbox{vacuum} expectation
values of the two Higgs doublets.
The typically large radiative corrections are 
fixed based on experimentally and phenomenologically sensible choices for
the supersymmetric parameters, each choice defining a particular benchmark scenario.
Generally, MSSM scenarios assume that the 125 \UGeV Higgs boson is the lighter scalar \Ph, 
an assumption that is compatible with the current experimental constraints 
for at least a significant portion of the $m_{\PA}$--$\tan\beta$ parameter space.
The di-tau lepton final state provides the most sensitive direct search for additional
Higgs bosons predicted by the MSSM for intermediate and high values of tan$\beta$, 
because of the enhanced
coupling to down-type fermions.
\begin{figure}[htbp]
\begin{center}
\subfloat[$\mhmodp$]{\includegraphics[width=0.47\textwidth]{\main/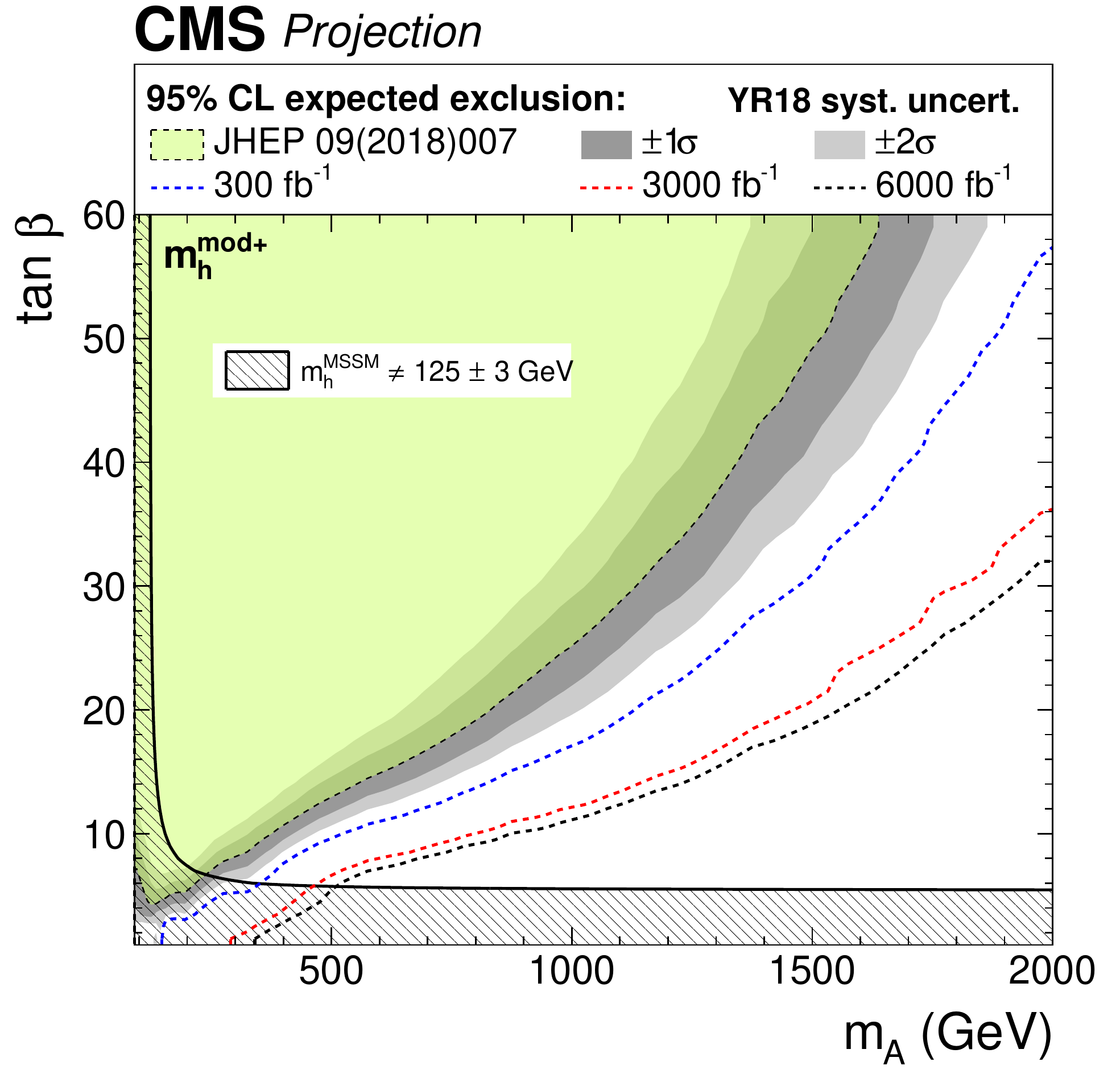}}\\
\subfloat[hMSSM]{\includegraphics[width=0.47\textwidth]{\main/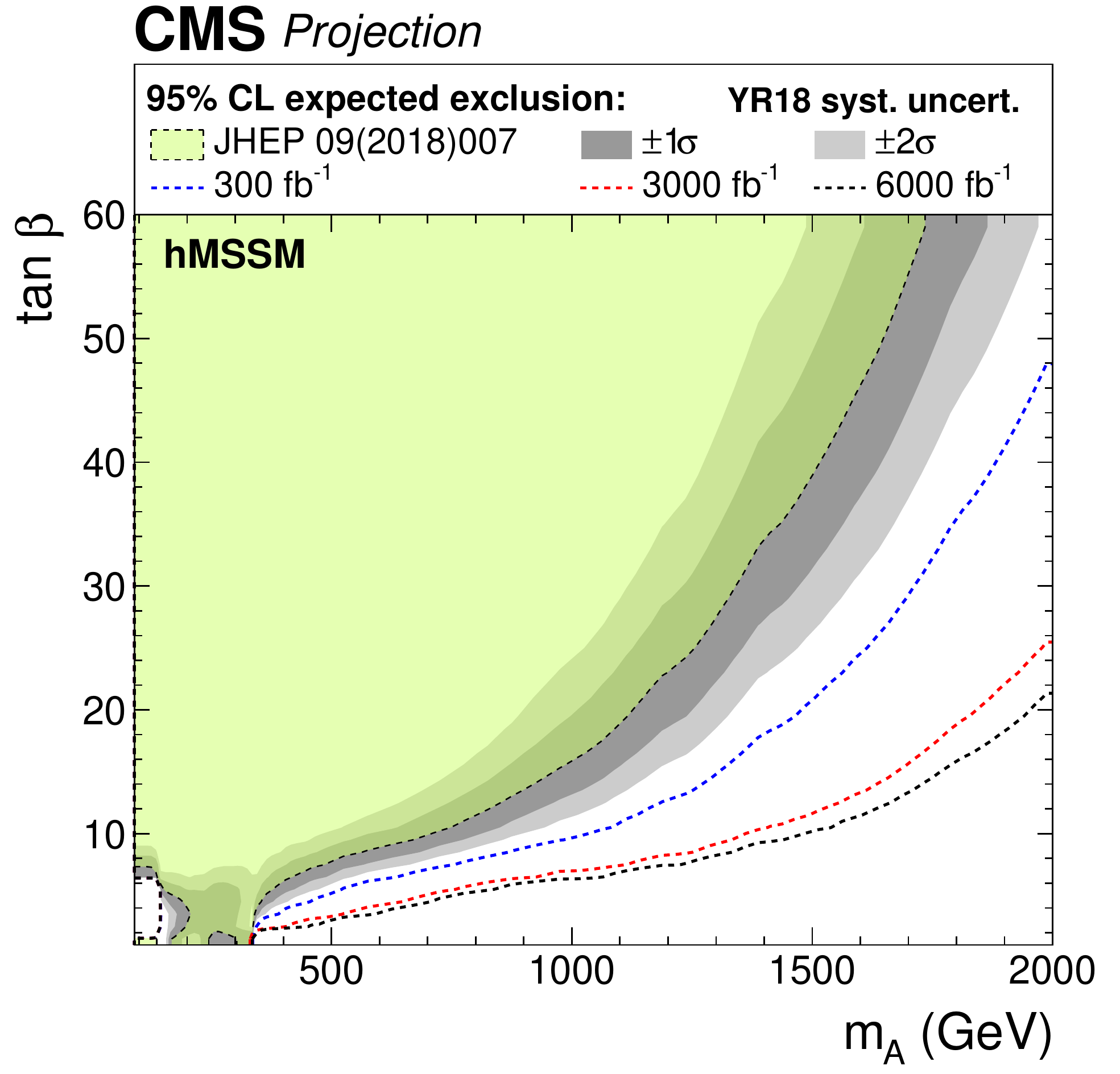}}
\subfloat[tau-phobic]{\includegraphics[width=0.47\textwidth]{\main/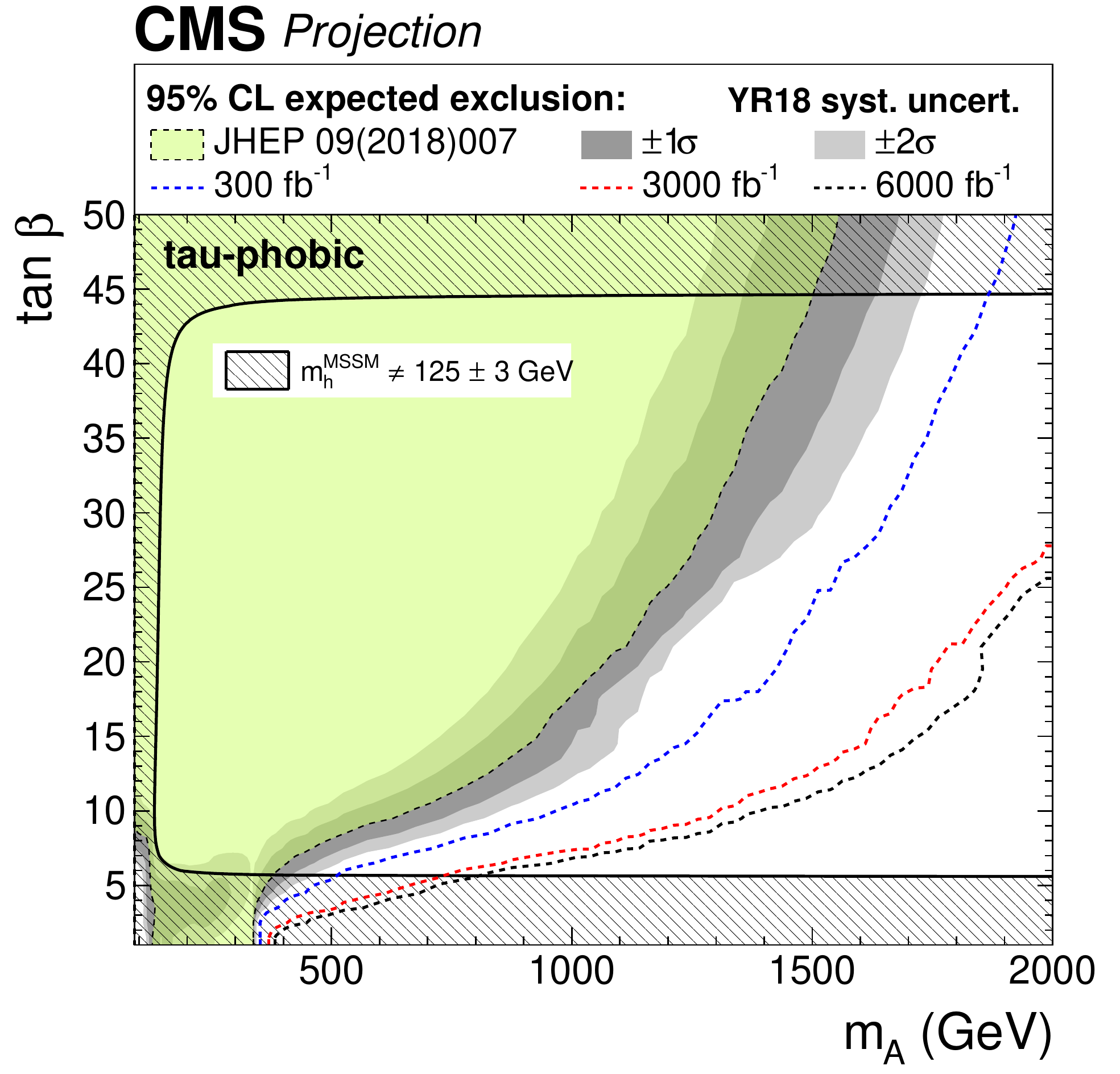}}
\end{center}
\caption{Projection of expected MSSM \htt 95\% CL upper limits based on 2016 data~\cite{HIG-17-020} for different benchmark 
scenarios, with YR18 systematic uncertainties~\cite{CMS-PAS-FTR-18-017}. The limit shown for 6000\fbinv is an approximation of the sensitivity with 
the complete HL-LHC dataset to be collected by the ATLAS and CMS experiments, corresponding to an integrated luminosity of 3000\fbinv each. 
The limits are compared to the CMS result using 2016 data~\cite{HIG-17-020}; for the tau-phobic scenario, 
it is a new interpretation of the information given in this reference. 
}
\label{fig:model_mssm1}
\end{figure}

The analysis results are interpreted in terms of these benchmark scenarios based on the profile likelihood ratio of the 
background-only and the tested signal-plus-background hypotheses. 
For this purpose, the predictions from both production modes and both heavy neutral Higgs bosons are combined.
Figure~\ref{fig:model_mssm1} shows the results~\cite{HIG-17-020} 
for three different benchmark scenarios:
the $\mhmodp$ and tau-phobic scenarios~\cite{Carena:2013ytb} and the hMSSM~\cite{Djouadi:2013uqa,Bagnaschi:2015hka}.
The sensitivity reaches up to Higgs boson masses
of 2 \UTeV for values of $\tan \beta$ of 36, 26, and 28
for the $\mhmodp$, the hMSSM, and the tau-phobic scenarios,
respectively.
Even at low mass, improvements are expected but in this case they are mostly 
a consequence of reduced systematic uncertainties and not 
the additional data in the signal region.
\paragraph{Conclusions}
\label{sec:conclusions}
The HL-LHC projections of the most recent results on searches for neutral 
MSSM Higgs bosons decaying to $\tau$ leptons have been shown,
based on a data set of proton-proton collisions at $\sqrt{s}=13\TeV$
collected in 2016, 
corresponding to a total integrated luminosity of 35.9\fbinv.
The assumed integrated luminosity for the HL-LHC is 3000\fbinv.
In terms of cross section, 
an order-of-magnitude improvement in sensitivity
is expected for neutral Higgs boson masses above 1 \UTeV 
since here the current analysis is statistically limited by the 
available integrated luminosity. For lower masses, an improvement of 
approximately a factor of five is expected for realistic
assumptions on the evolution of the systematic uncertainties.
For the MSSM benchmarks,
the sensitivity will reach up to Higgs boson masses
of 2 \UTeV for values of $\tan \beta$ of 36, 26, and 28  
for the $\mhmodp$, the hMSSM, and the tau-phobic scenarios,
respectively.

\asubsection{LHC searches for additional heavy neutral Higgs bosons in bosonic final states}\label{sec:XZZ}
\asubsubsection[M. Xiao]{Projection of Run-2 CMS searches for a new scalar resonance decaying to a pair of Z bosons}
\asubsubsubsection{Introduction}
CMS and ATLAS collaborations have performed searches for a heavy scalar partner of the SM Higgs boson decaying into a pair of Z bosons~\cite{Aaboud:2018bun,Sirunyan:2018qlb}. The CMS search for a heavy scalar partner of the SM Higgs boson using $35.9 \mathrm{fb}^{-1}$ of pp collision data~\cite{Sirunyan:2018qlb} will be referred to as HIG-17-012 throughout this section. In HIG-17-012, the search for a scalar resonance $\PX$ decaying to $\PZ \PZ$ is performed over the mass range $130\UGeV < m_{\PX} < 3$\UTeV, where three final states based on leptonic or hadronic decays of Z boson, $\PX\to \PZ\PZ\to 4\ell$, $2\ell2\Pq$, and $2\ell2\nu$ are combined. Because of the different resolutions, efficiencies, and branching fractions, each final state contributes differently depending on the signal mass hypothesis. The most sensitive final state for the mass range of 130--500\UGeV is $4\ell$ due to its best mass resolution, whereas, for the intermediate region of 500--700\UGeV, $2\ell2\nu$ is most sensitive. For masses above 700\UGeV the $2\ell2\Pq$ provides the best sensitivity. In this paper, we are particularly interested in the sensitivity in the high mass region, thus only $2\ell2\Pq$ is used. 

In the $2\ell2\Pq$ final state, events are selected by combining leptonically
and hadronically decaying \PZ\ candidates. The lepton pairs (electron or muon) of opposite sign and same flavor 
with invariant mass between 60 and 120 \UGeV are constructed.
Hadronically decaying $\PZ$ boson candidates are reconstructed using two distinct
techniques, which are referred to as ``resolved'' and ``merged''.
In the resolved case, the two quarks from the $\PZ$ boson decay form two distinguishable narrow jets, while in the merged case a single wide jet with a large $p_T$ is taken as a hadronically decaying Z candidate.

The two dominant production mechanisms of a scalar boson are gluon fusion ($ggF$) and EW production, the latter dominated by vector boson fusion (VBF) with a small contribution of
production in association with an EW boson $ZH$ or $WH$ ($VH$). We define the parameter $f_{VBF}$ as the fraction of the EW production cross section with respect to the total cross section. The results are given in two scenarios: $f_{VBF}$ floated, and $f_{VBF}=1$. In the expected result, the two scenarios correspond to $ggF$ and VBF production modes, respectively. To increase the sensitivity to the different production modes, events are categorised into VBF and inclusive types. Furthermore, since a large fraction of signal events is enriched with \PQb\ quark jets due to the presence of $\PZ\to\PQb\PAQb$ decays, a dedicated category is defined.

The invariant mass of $\PZ \PZ$ and a dedicated discriminant separating signal and background distributions are compared between observation and expected background to set limits on the production cross section. 

Further details of the HIG-17-012 analysis, including simulation samples, event categorisation, background estimation methods, systematic uncertainties, and different interpretations are described in Ref~\cite{Sirunyan:2018qlb}. Only details of direct relevance to the projection of the HIG-17-012 are documented in Ref.~\cite{CMS-PAS-FTR-18-040} and in the following.  

\asubsubsubsection{Extrapolation procedure}

A projection of this analysis is carried out by scaling all the signal
and background processes to an integrated luminosity of 3000$\mathrm{fb}^{-1}$, expected to be collected at the high-luminosity LHC. This projection assumes that the CMS experiment will have a similar level of detector and triggering performance during the HL-LHC operation as it provided during the LHC Run~2 period. It does not take into account the small cross section change due to the small change in the centre of mass energy from 13 \UTeV to 14 \UTeV. The results of projection are presented for different assumptions based on the size of systematic uncertainties that is estimated for HL-LHC. 

\asubsubsubsection{Results}
\begin{figure}[htbp]
        \centering
                \includegraphics[width=0.48\textwidth]{\main/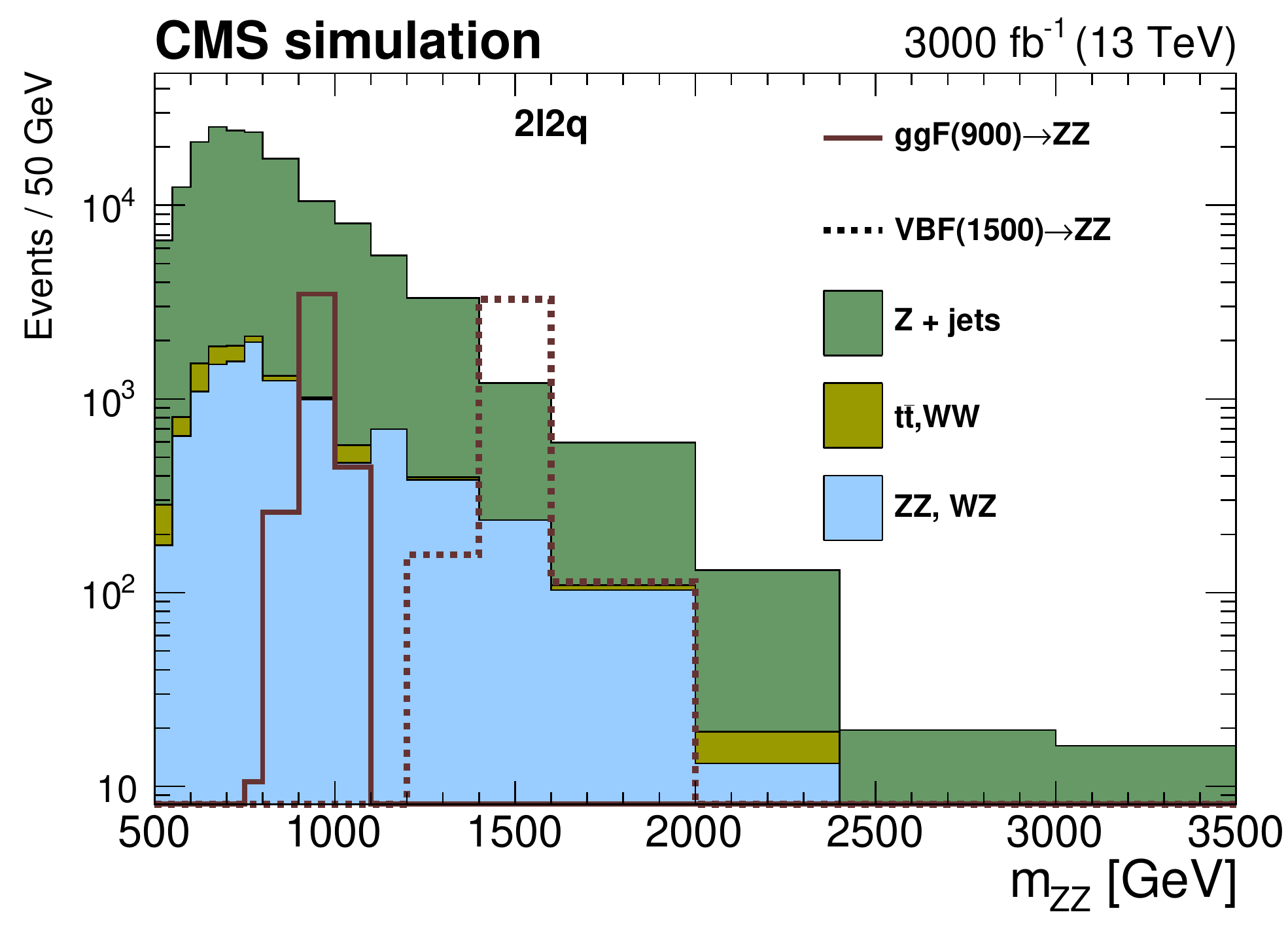}
                \caption{The $m_{\PZ\PZ}$ distribution of the merged category events expected at 3000$\mathrm{fb}^{-1}$. Examples of a 900 \UGeV $ggF$ signal and a 1500 \UGeV VBF signal are given. the cross section corresponds to 10 times the excluded limit. 
        \label{fig:mzz}
        }
            \end{figure}
The $m_{\PZ\PZ}$ distribution of the merged category events expected at 3000 $\mathrm{fb}^{-1}$ is shown in Figure~\ref{fig:mzz}. Figure~\ref{fig:combinedresult} shows upper limits at the 95\% confidence level on the $\Pp\Pp\to\PX\to\PZ\PZ$ cross section
$\sigma_{\PX} \mathcal{B}_{{\PX}\to\PZ\PZ}$ as a function of $m_{\PX}$ for a narrow resonance whose $\Gamma_{\PX}$ is much smaller than experimental resolution.

In the mass range between 550--3000 \UGeV, the excluded cross section of the scalar decaying to a pair of Z bosons is 0.7--5\fb for the VBF production mode and 0.8--9\fb for the $ggF$ production mode. This represents a factor of 10 improvement with respect to the results obtained using Run 2 data. The differences between the two scenarios are minor and mostly present in the low mass region. It is because the search will still be limited by statistical uncertainties. Among all the systematic uncertainties, the theoretical uncertainty from higher order QCD corrections on the $gg\to ZZ$ background and the signal is the most dominant for the $ggF$ search. The next important ones are the shape and yield uncertainties of the Z$+$jets background. They are determined from a data control region and are scaled with $1/\sqrt{L}$ in YR18 scenario. It is expected that at HL-LHC, the Z$+$jets background will have huge statistics, and the understanding of it will be at percent level. The effect of systematics in this search has mild effect, if no $1/\sqrt{L}$ scaling is applied, the difference in the limit is 10\% at low mass and almost none in the high mass region. In the HIG-17-012 analysis, Z$+$jets fake rates are derived from LO MC samples, and differences with respect to NLO samples are assigned as systematic uncertainty. This major source is treated as a theoretical uncertainty and is scaled by 0.5 in the YR18 scenario. 
The results for wide resonances are not given in this note for simplicity. The Run-2 result has shown that the excluded cross section for a 30\% width resonance will be 40\% higher at 1 \UTeV, compared to a narrow resonance assumption. 

\begin{figure}[htbp]
  \centering
  \includegraphics[width=0.48\textwidth]{\main/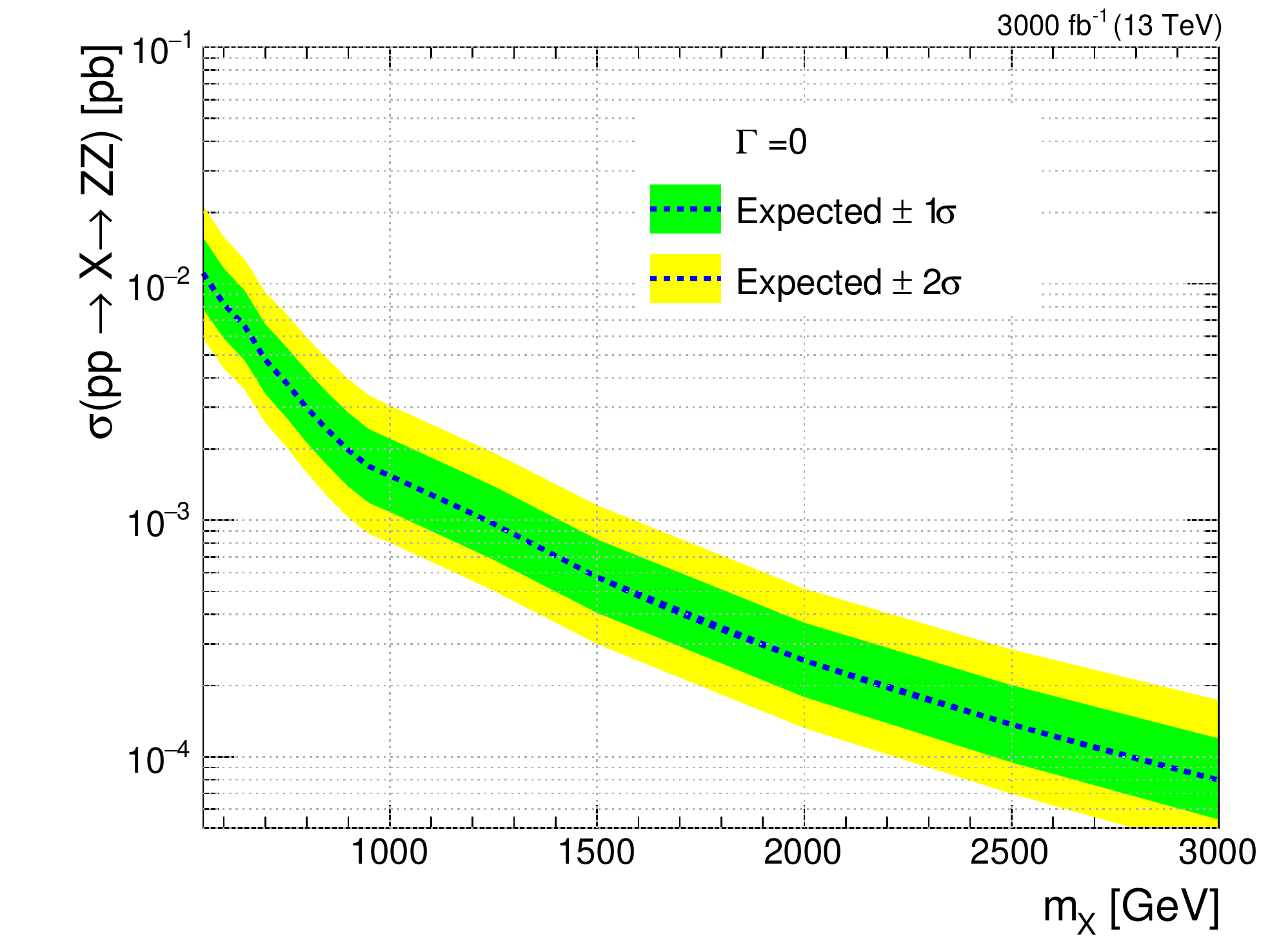}
  \includegraphics[width=0.48\textwidth]{\main/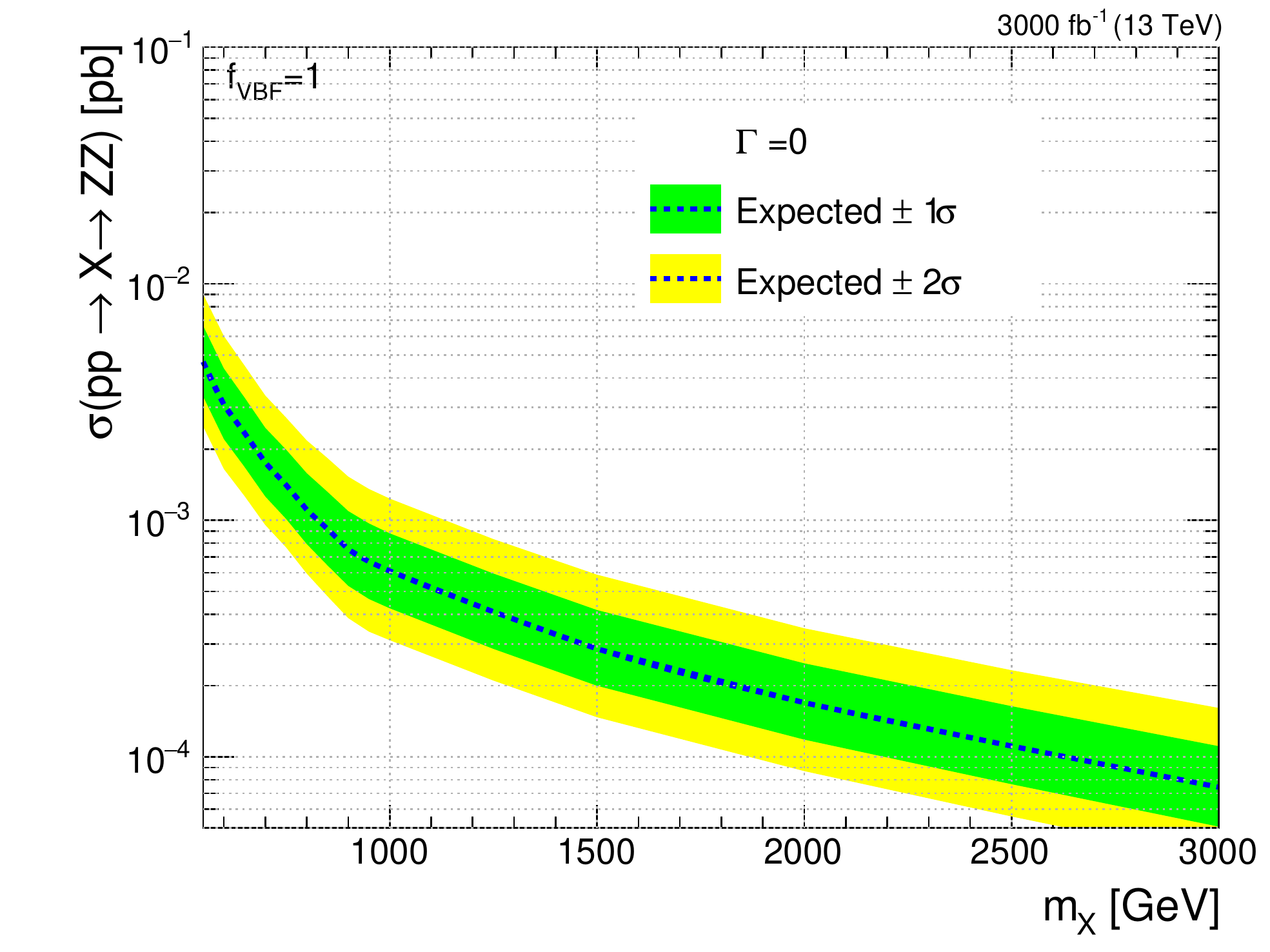}\\
  \includegraphics[width=0.48\textwidth]{\main/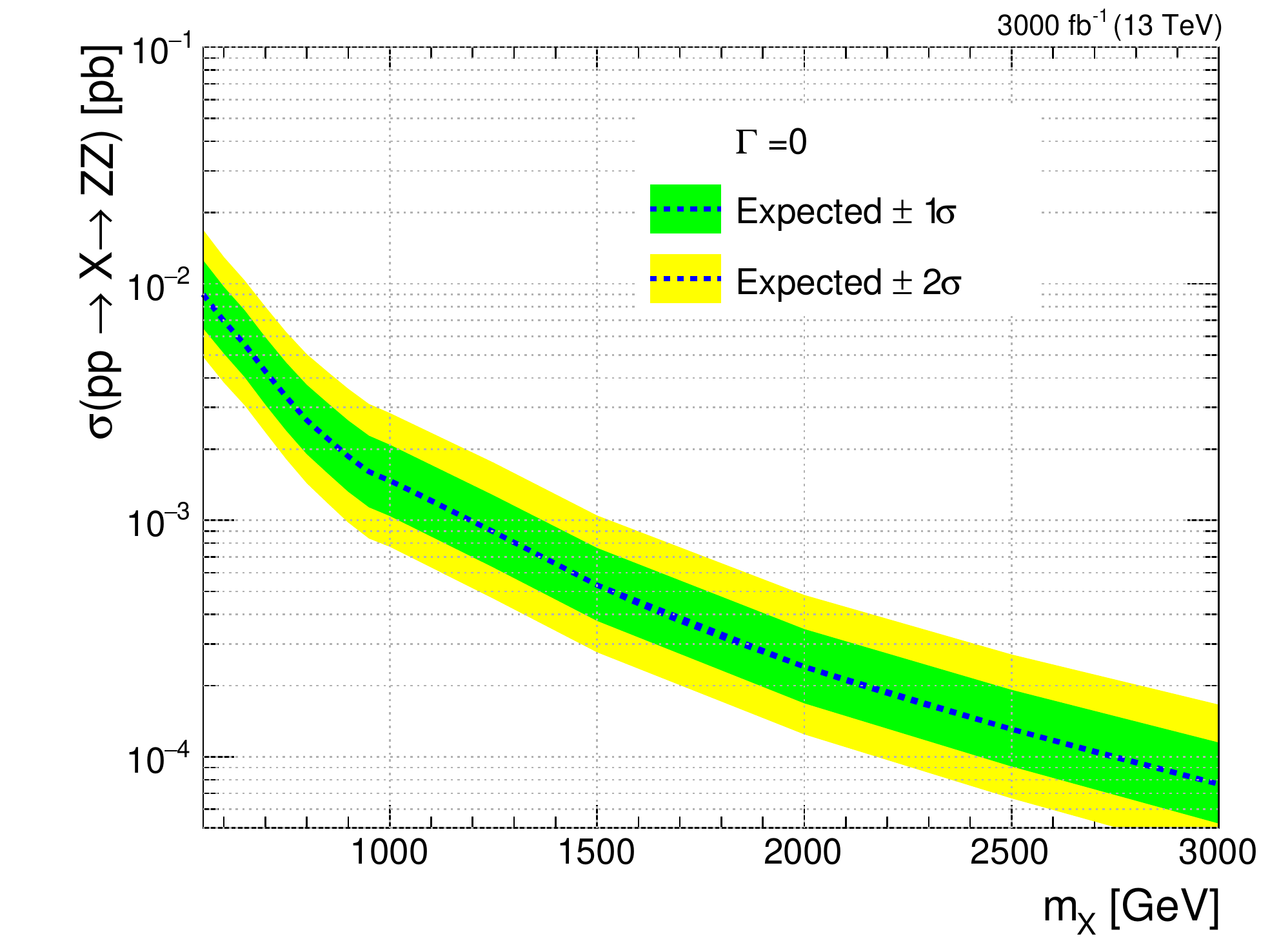}
  \includegraphics[width=0.48\textwidth]{\main/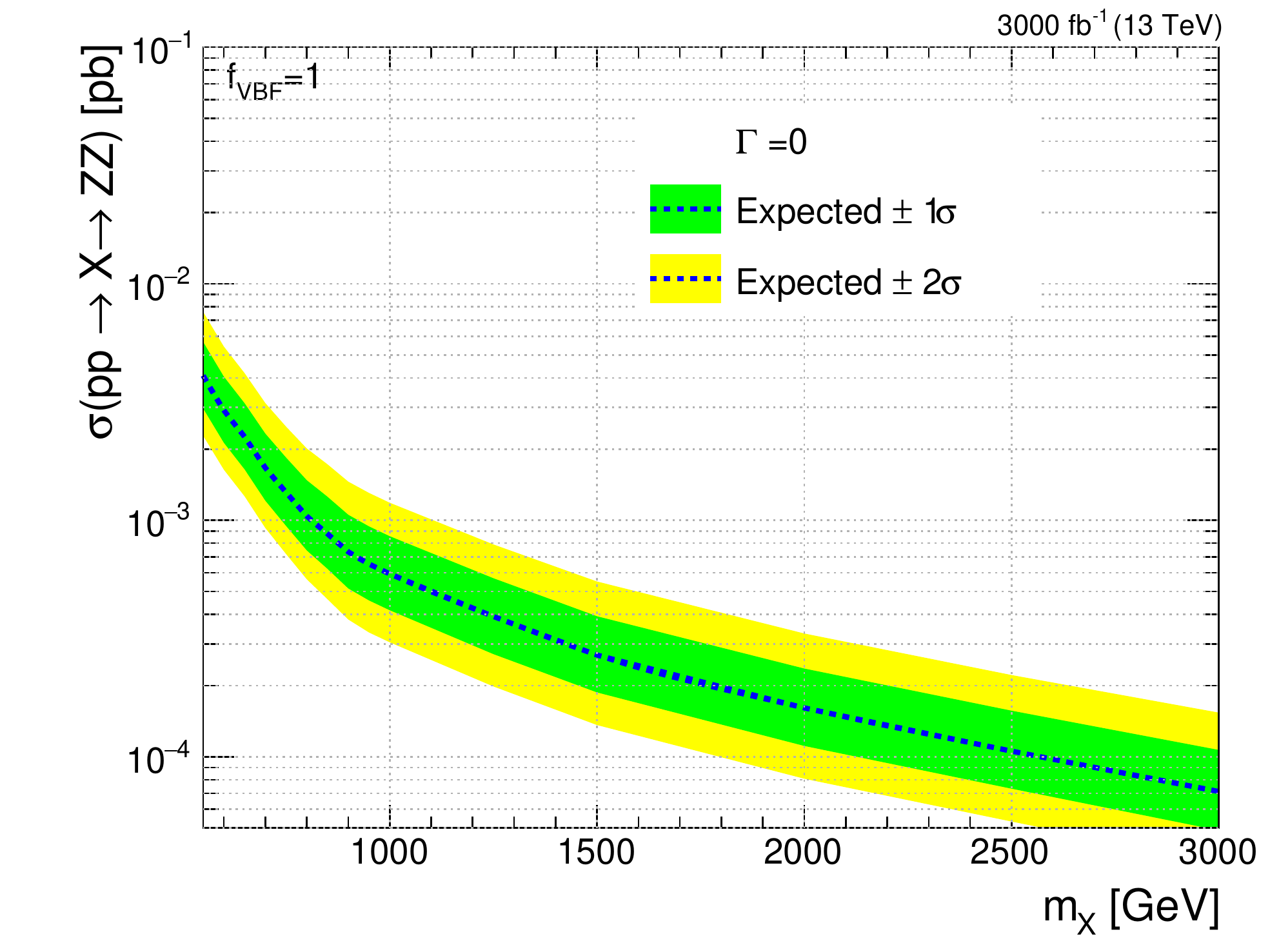}
  \caption{Expected upper limits at the 95\% CL on the
    $\Pp\Pp\to\PX\to\PZ\PZ$ cross section as a function of $m_{\PX}$,
    with $f_{\mathrm{VBF}}$ as a free parameter (left) and fixed to 1
    (right). Scenario 1 (top) and scenario 2 (bottom) are shown. The
    scalar particle X is assumed to have a more narrow decay width
    than the detector resolution. The results are shown for the
    $2\ell2\Pq$ channel.\label{fig:combinedresult}}
\end{figure}

\asubsection{Additional channels for heavy Higgs bosons}\label{Sec:9.4}

\asubsubsection[K. Mimasu, J.M. No]{Sensitivity to heavy Higgs bosons from the 2HDM in "Higgs-to-Higgs" decays}
Searches for heavy scalars are highly complementary to coupling measurements of the 125 \UGeV Higgs $h$ as probes of extended Higgs sectors. Di-boson search channels $H \to WW$, $ZZ$ probe the parameter space for which the 125 \UGeV Higgs is not SM-like, together with Higgs coupling measurements. Both suffer a significant loss in sensitivity to new physics scenarios in the limit of a SM-like 125 \UGeV Higgs, as the couplings $g_{HVV}$ ($V = W^{\pm},\,Z$) vanish in such case. For two-Higgs-doublet-model scenarios, this corresponds to the so-called alignment limit~\cite{Gunion:2002zf},
where searches for heavy scalars through non-standard ``Higgs-to-Higgs"
decay channels~\cite{Coleppa:2014hxa,Dorsch:2014qja,Li:2015lra,Kling:2016opi,Dorsch:2016tab} 
as well as through fermionic decay channels~\cite{Craig:2015jba,Gori:2016zto} become the key avenue to find these new states, and are crucial to cover the parameter space of 2HDMs.

In this section, we focus on HL-LHC and HE-LHC probes of 2HDM neutral scalars via ``Higgs-to-Higgs" decays, and briefly discuss also their interplay with direct searches of such states in fermionic decay channels. Specifically, we consider a general scalar potential for two Higgs doublets 
with a softly broken $\mathbb{Z}_2$ 
symmetry\footnote{In specific BSM scenarios featuring two Higgs doublets, it is possible that $\lambda_6 \left|H_1\right|^2 (H_1^{\dagger}H_2+\mathrm{h.c.})$  and $\lambda_7 \left|H_2\right|^2 (H_1^{\dagger}H_2+\mathrm{h.c.})$ terms, which explicitly break the $\mathbb{Z}_2$ symmetry, get generated radiatively even if absent at tree-level (e.g. in the MSSM). Being suppressed by $1/(4\pi)^2$, their impact on the present analysis should nevertheless be mild.} (and no CP violation), given by 
\begin{eqnarray}	
\label{2HDM_potential}
V(H_1,H_2) &= &\mu^2_1 \left|H_1\right|^2 + \mu^2_2\left|H_2\right|^2 - \mu^2\left[H_1^{\dagger}H_2+\mathrm{h.c.}\right] 
+\frac{\lambda_1}{2}\left|H_1\right|^4 +\frac{\lambda_2}{2}\left|H_2\right|^4 \nonumber \\
&+& \lambda_3 \left|H_1\right|^2\left|H_2\right|^2
+\lambda_4 \left|H_1^{\dagger}H_2\right|^2+ \frac{\lambda_5}{2}\left[\left(H_1^{\dagger}H_2\right)^2+\mathrm{h.c.}\right]\, . 
\end{eqnarray}
Regarding the couplings of the two doublets $H_{1,2}$ to fermions, we consider  
a Type-I and a Type-II 2HDM scenarios, with the parameters $t_{\beta} \equiv \mathrm{tan}\,\beta$ and $c_{\beta -\alpha} \equiv \mathrm{cos}\,(\beta-\alpha)$ controlling the coupling strength of the various 2HDM scalars to fermions and gauge bosons, respectively (see e.g.~\cite{Branco:2011iw} for a review).

Apart from the 125 \UGeV Higgs $h$, the 2HDM scalar sector includes two neutral states 
$H$ and $A$, respectively CP-even and CP-odd, as well as a charged scalar $H^{\pm}$.
Focusing on the neutral scalars, the decay $A \to Z H$ ($H \to Z A$) yields a powerful probe of the parameter space region with a sizeable mass splitting $m_A > m_H + m_Z$ ($m_H > m_A + m_Z$)~\cite{Coleppa:2014hxa,Dorsch:2014qja}. We first obtain the present 13 \UTeV LHC limits on the 2HDM parameter space from the search $A \to Z H$ ($Z\to \ell\ell$, $H \to b\bar{b}$) by ATLAS with $36.1$ fb$^{-1}$~\cite{Aaboud:2018eoy} (see also~\cite{Khachatryan:2016are,CMS:2016qxc} for 
corresponding searches by CMS), considering in particular the alignment limit $c_{\beta -\alpha} = 0$.
Our signal cross sections and branching fractions are obtained respectively with {\sc SusHi}~\cite{Harlander:2012pb} and {\sc 2HDMC}~\cite{Eriksson:2009ws}, and we use the 
publicly available observed 95$\%$ C.L. signal cross section limits in the ($m_A$, $m_H$) plane from~\cite{Aaboud:2018eoy}. In order to derive a sensitivity projection of this search to HL-LHC and HE-LHC with $3000$ fb$^{-1}$ of integrated luminosity, we first perform a luminosity rescaling of the present ATLAS expected sensitivity, assuming that the background uncertainties are statistically dominated ({\it i.e.} we rescale the present expected sensitivity by $\sqrt{\mathcal{L}_2/\mathcal{L}_1} = \sqrt{3000/36.1}$).
We then perform a further rescaling of the sensitivity from $\sqrt{s} = 13$ \UTeV to 
$\sqrt{s} = 14$ \UTeV (HL-LHC) and $\sqrt{s} = 27$ \UTeV (HE-LHC) under the assumption that the
ratio of acceptance times cross section ($\mathcal{A} \times \sigma$) for the SM background for 27 \UTeV and 14 \UTeV w.r.t. 13 \UTeV are the same as the ratio of $A$ production cross section\footnote{That is, we assume signal and background increase by the same amount in going from 13 \UTeV to 14 \UTeV, or 13 \UTeV to 27 \UTeV. This is a conservative assumption particularly for high masses $m_A$.}. 
The present LHC bounds and projected sensitivities for $p p \to A \to Z H$ ($Z\to \ell\ell$, $H \to b\bar{b}$) are shown in Figure~\ref{AZH_HL-LHC} in the ($m_A$, $\mathrm{tan}\beta$) plane for Type II (left) and Type I (right) 2HDM, considering respectively $m_A = m_H + 100$ \UGeV (top) and $m_A = m_H + 200$ \UGeV (bottom).
We note that, since the limits from~\cite{Aaboud:2018eoy} neither extend above 
$m_A = 800$ \UGeV nor go below $m_H = 130$ \UGeV, our corresponding projections based on those limits cannot 
extend beyond those parameter regions either.  

\begin{figure}[h]
\begin{center}
\includegraphics[width=0.48\textwidth]{\main/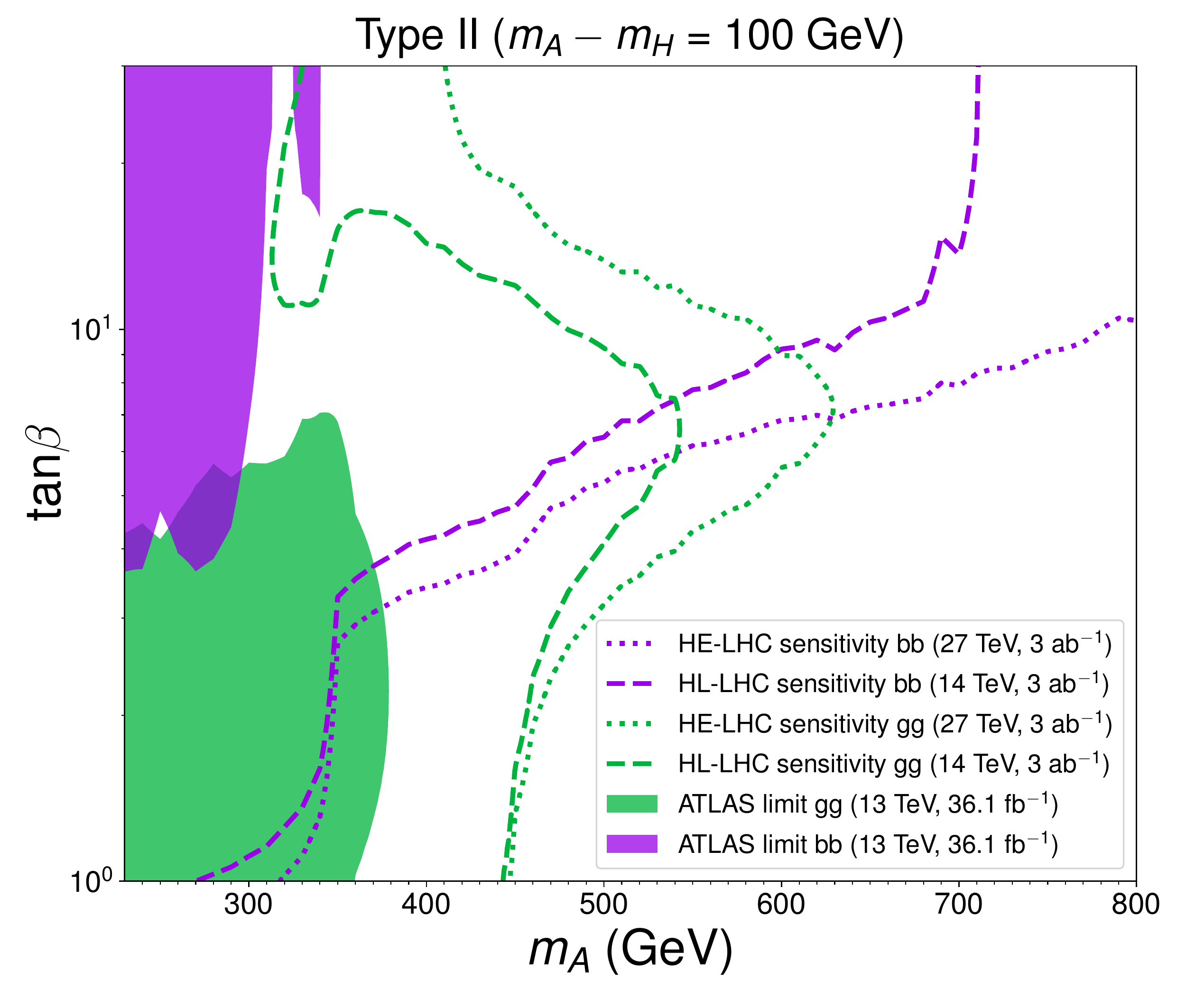}
\hspace{3mm}
\includegraphics[width=0.48\textwidth]{\main/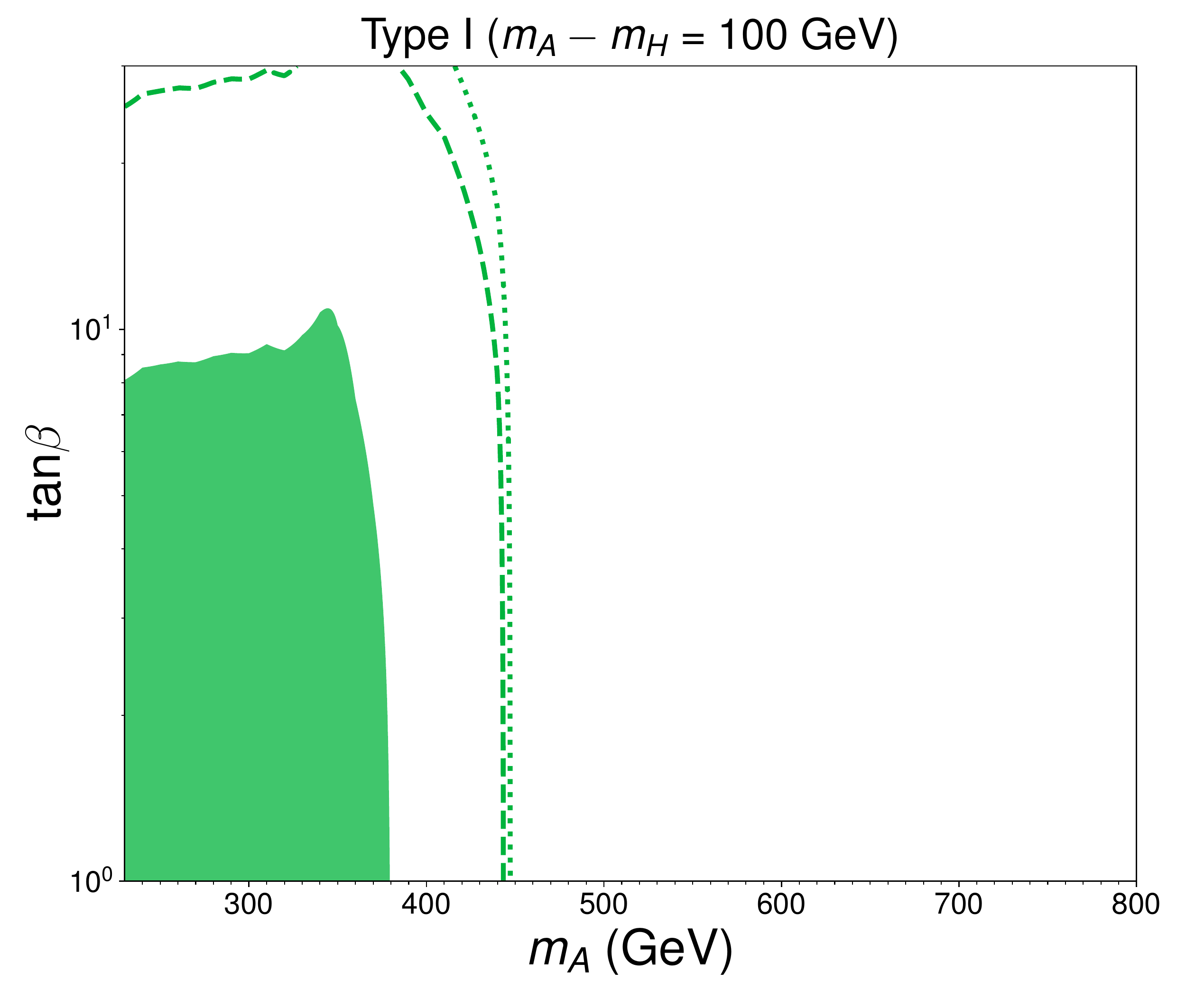}
\includegraphics[width=0.48\textwidth]{\main/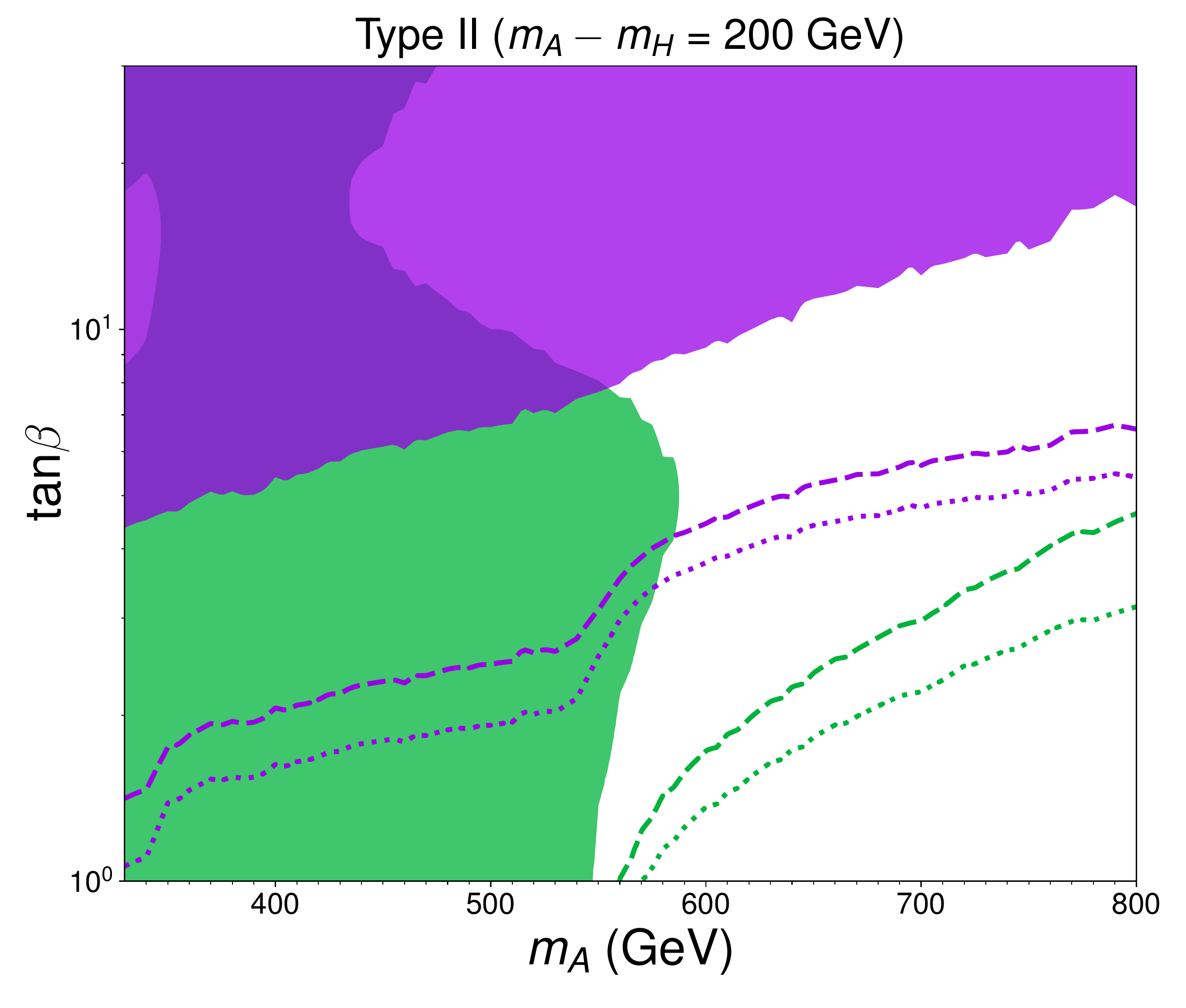}
\hspace{3mm}
\includegraphics[width=0.48\textwidth]{\main/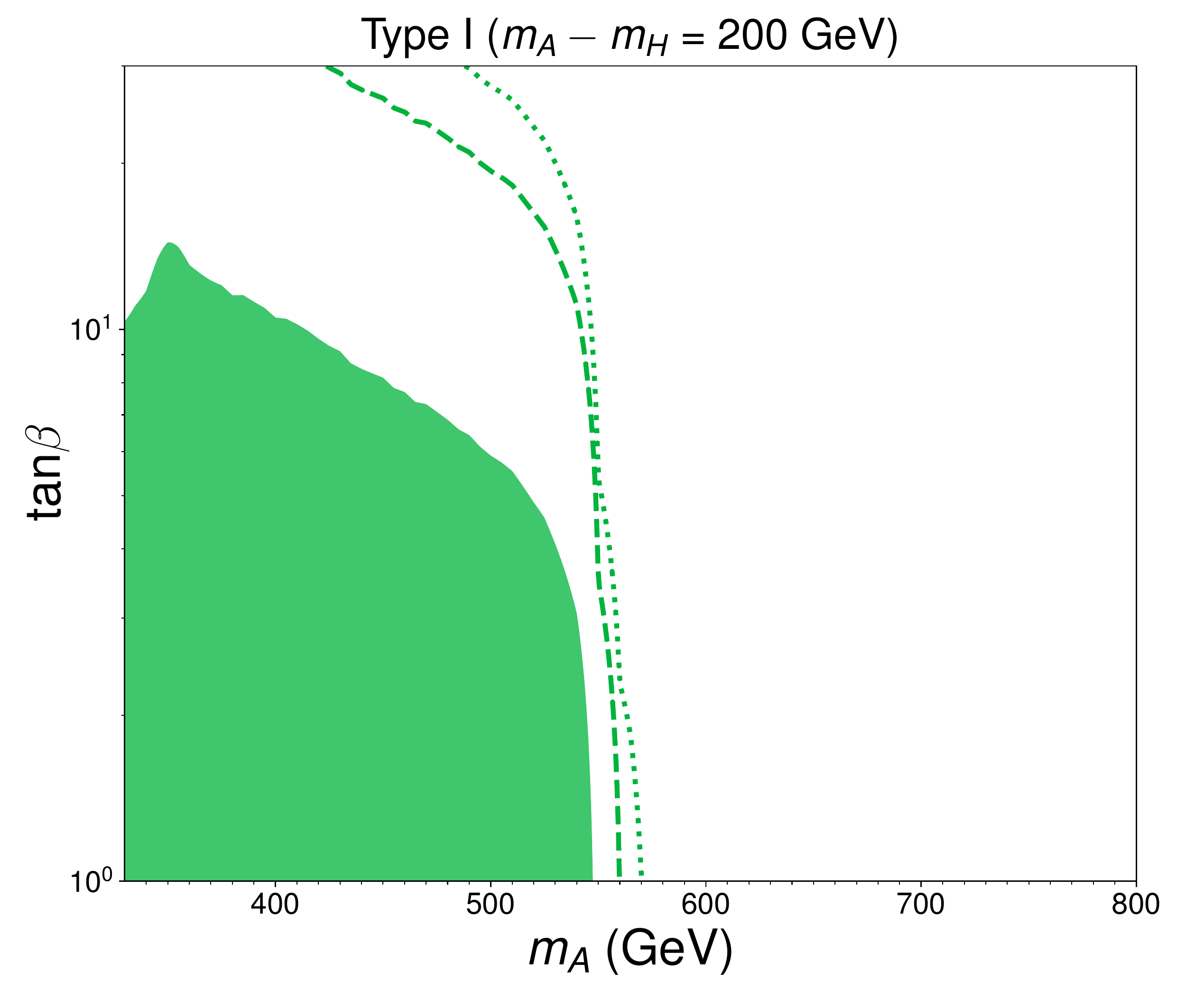}
\caption{\small $95\%$ C.L. exclusion sensitivity for $p p \to A \to Z H \to \ell\ell b \bar{b}$  
in the ($m_{A}$, tan$\beta$) plane for 2HDM Type II (left) and Type I (right), for $m_A = m_H +100$ \UGeV (top) and $m_A = m_H +200$ \UGeV (bottom). The present bounds~\cite{Aaboud:2018eoy} are shown as solid regions, while
$3000$ fb$^{-1}$ projections for 14 \UTeV HL-LHC and 27 \UTeV HE-LHC are respectively shown as dashed and dotted lines. Limits from gluon fusion are shown in green, and limits from $bb$-associated production (for Type II 2HDM) are shown in purple.}
\label{AZH_HL-LHC}
\end{center}
\end{figure}

We then study the interplay of the $p p \to A \to Z H \to \ell\ell b \bar{b}$ search with searches for heavy scalars in fermionic decay modes, e.g. $H/A \to \tau\tau$. For this purpose, we consider 
the above benchmarks $m_A = m_H + 100$ \UGeV and $m_A = m_H + 200$ \UGeV for Type II 2HDM, and translate the present ATLAS $H \to \tau\tau$ limits with $36.1$ fb$^{-1}$~\cite{Aaboud:2017sjh} and the 14 \UTeV HL-LHC sensitivity projections for $H \to \tau\tau$ from Section~\ref{sec:Hff} to the ($m_{A}$, $\mathrm{tan}\beta$) plane using {\sc SusHi} and {\sc 2HDMC}, assuming $\mathrm{cos}(\beta - \alpha) = 0$. The results are shown in Figure~\ref{TauTau_AZH_HL-HE} together with the combined HL-LHC sensitivity of $A \to Z H$ from gluon fusion and $bb$-associated production, highlighting the complementary between ``Higgs-to-Higgs" decays and direct searches in fermionic final states. We also note that a limiting factor of the latter (as currently searched for by the experimental collaborations) for low tan$\beta$ and $m_H > 340$ \UGeV is the small branching fraction $H \to \tau \bar{\tau}$ due to the opening of the $H \to t \bar{t}$ decay. This region of parameter space would therefore be efficiently explored via a search for $p p \to A \to Z H$ ($Z\to \ell\ell$, $H \to t\bar{t}$) (see e.g.~\cite{Dorsch:2016tab,Haisch:2018djm}).

\begin{figure}[h!]
\begin{center}
\includegraphics[width=0.48\textwidth]{\main/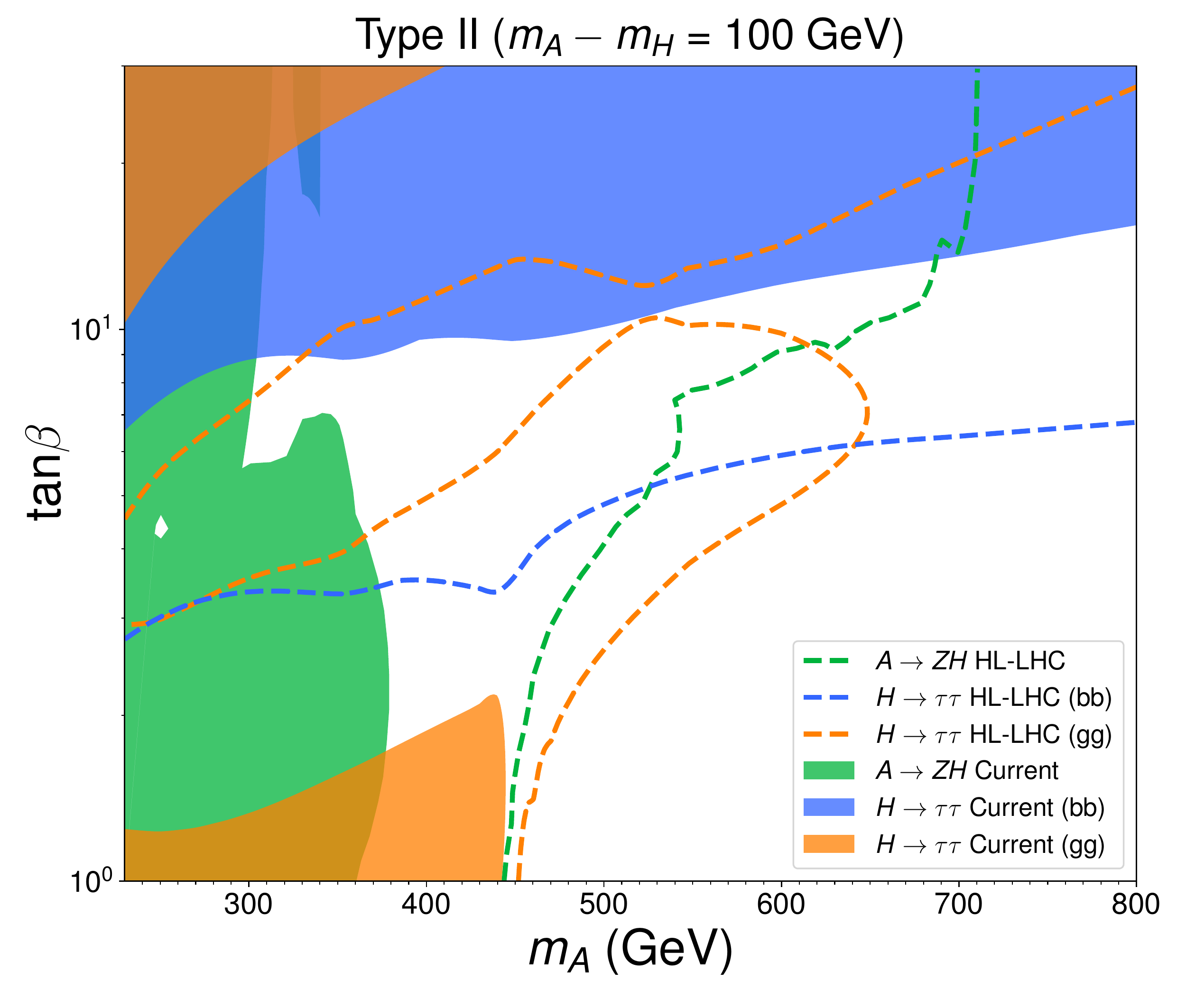}
\hspace{3mm}
\includegraphics[width=0.48\textwidth]{\main/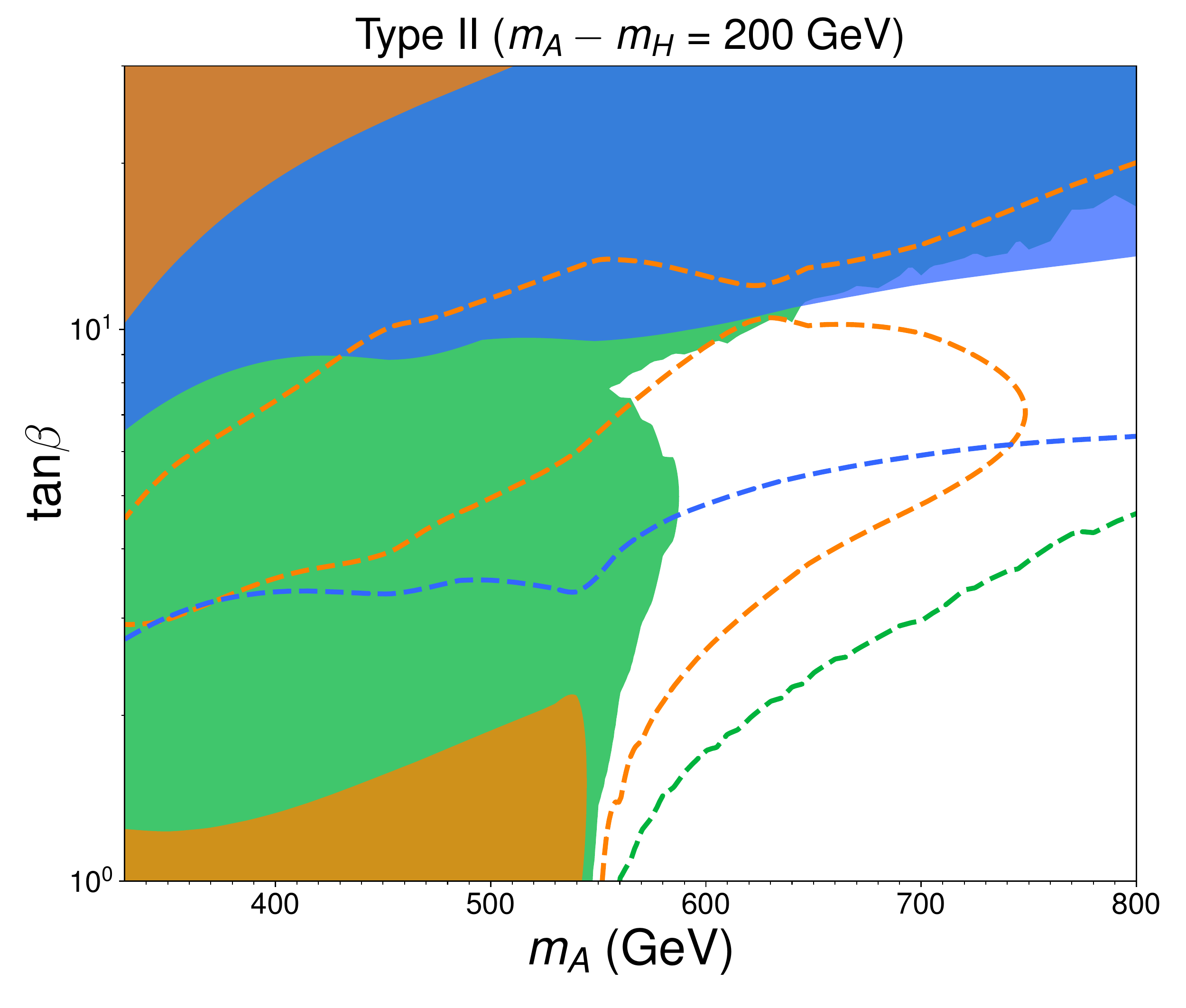}
\caption{\small 2HDM Type II $95\%$ C.L. exclusion sensitivity for $p p \to A \to Z H \to \ell\ell b \bar{b}$  (green)
in the ($m_{A}$, tan$\beta$) plane for $m_A = m_H +100$ \UGeV (left) and $m_A = m_H +200$ \UGeV (right), combining 
gluon fusion and $bb$-associated production (see Figure~\ref{AZH_HL-LHC}). We show for comparison
the $95\%$ C.L. exclusion sensitivity for $p p \to H \to \tau\tau$ in gluon fusion (orange) 
and $bb$-associated production (blue). Present bounds are shown as solid regions, while $3000$ fb$^{-1}$ projections for 14 \UTeV HL-LHC are shown as dashed lines.}
\label{TauTau_AZH_HL-HE}
\end{center}
\end{figure}

In addition, we analyse the prospects for probing ``Higgs-to-Higgs" decay channels within the 2HDM, when one of the scalars involved in the decay is the SM-like 125 \UGeV Higgs. We focus here on the decay $A \to Z h$, which vanishes in the alignment limit $\mathrm{cos}(\beta - \alpha) = 0$, but may yield the dominant decay mode of $A$ even close to alignment. Following a similar procedure to the one discussed above, in Figure~\ref{A_hZ_HL-HE} we show the present 95$\%$ C.L. signal cross section limits in the ($m_{A}$, tan$\beta$) plane for 2HDM Type I, fixing $m_A = m_H = m_{H^{\pm}}$ and $\mathrm{cos}(\beta-\alpha) = 0.1$ (for tan$\beta > 1$, this value of $\mathrm{cos}(\beta-\alpha)$ is barely within the reach of Higgs coupling measurements at HL-LHC, see e.g.~\cite{ATL-PHYS-PUB-2014-017}), from the LHC 13 \UTeV ATLAS search for $p p \to A \to Z h \to \ell\ell b \bar{b}$ with $36.1$ fb$^{-1}$~\cite{Aaboud:2017cxo}. We also show the projected 14 \UTeV HL-LHC 95$\%$ C.L. sensitivity with $3$ ab$^{-1}$, as well as the 27 \UTeV HE-LHC sensitivity by a rescaling of the HL-LHC limits, under the assumption that the ratio of ($\mathcal{A} \times \sigma$) for the SM background from 14 \UTeV to 27 \UTeV is the same as the ratio of signal production cross section.
As Figure~\ref{A_hZ_HL-HE} highlights, the search for $p p \to A \to Z h \to \ell\ell b \bar{b}$ yields a powerful probe of the 2HDM parameter away for the alignment limit, probing up to tan$\beta \sim 60$ at HE-LHC.

\begin{figure}[h!]
\begin{center}
\includegraphics[width=0.6\textwidth]{\main/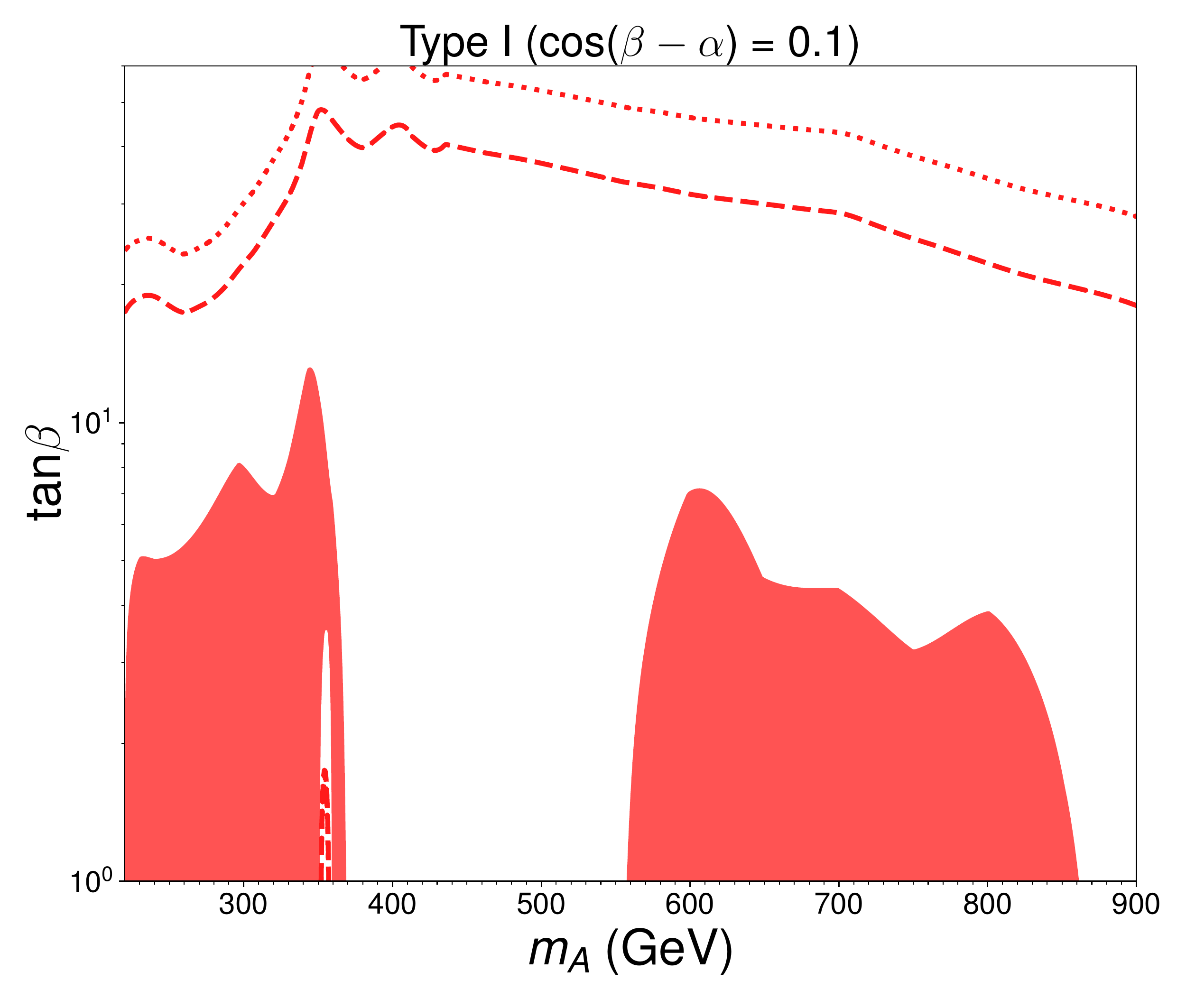}
\caption{\small 2HDM Type I $95\%$ C.L. exclusion sensitivity for $p p \to A \to Z h \to \ell\ell b \bar{b}$ 
in the ($m_{A}$, tan$\beta$) plane, for $\mathrm{cos}(\beta-\alpha) = 0.1$. 
Present bounds are shown as solid regions, while $3000$ fb$^{-1}$ projections for 14 \UTeV HL-LHC and 27 \UTeV HE-LHC
are shown as dashed and dotted lines, respectively.}
\label{A_hZ_HL-HE}
\end{center}
\end{figure}

\asubsubsection[M. Carena, Z. Liu, M. Riembau. This manuscript has been authored by Fermi Research Alliance, LLC under Contract No. DE-AC02-07CH11359 with the U.S. Department of Energy, Office of Science, Office of High Energy Physics. ZL is also supported by the NSF under Grant No. PHY1620074 and by the Maryland Center for Fundamental Physics (MCFP).]{Interference effects in heavy Higgs searches}\label{Sec.9.4.2}
The singlet SM extension  serves as the simplest, yet elusive benchmark to test  a sufficiently strong  first-order phase transition (EWPT) compatible with the Higgs boson mass measurements at the LHC.
The singlet without $Z_2$ protection could mix with the SM Higgs and (in most cases) a promptly decaying scalar particle would provide a rich phenomenology at colliders. The singlet scalar could be produced resonantly  and decay back to pairs of SM particles, dominantly into $WW$, $ZZ$, $HH$ and $t\bar t$. The signal of  a singlet scalar resonance decaying into $HH$ is a smoking-gun  for singlet enhanced EWPT~\cite{Profumo:2007wc,Chen:2014ask,Dawson:2015haa,Kotwal:2016tex,Chen:2017qcz,Huang:2017jws,Robens:2016xkb,Lewis:2017dme,Dawson:2016ugw,Huang:2017nnw,deFlorian:2016spz} (see also the discussion in Section 3.6.2).

Searches for resonant di-Higgs production have received much attention by both the ATLAS and CMS collaborations~\cite{Aaboud:2016xco,Aad:2015xja,Sirunyan:2017djm,CMS-PAS-HIG-17-006,CMS-PAS-HIG-17-008,CMS-PAS-HIG-17-009}. In the case of a singlet resonance,  constraints from SM precision measurements  render these searches more challenging. From one side  precision measurements  imply that  the singlet-doublet mixing parameter is constrained to be small over a large region of parameter space.  On the other side, the singlet only couples to SM particles through mixing with the SM Higgs doublet. This results in a reduced di-Higgs production via singlet resonance decays. In particular, the singlet resonance amplitude  becomes of the same order as the SM  triangle  and box diagram amplitudes. Most important, in this work we shall show that a large relative phase between the SM box diagram and the singlet triangle diagram becomes important. This special on-shell interference  effect  has been  commonly overlooked in the literature and turns out to have important phenomenological implications.
We shall choose the spontaneous $Z_2$ breaking scenario of the SM plus singlet to demonstrate the importance of the novel on-shell interference effect for the resonant singlet scalar searches in the di-Higgs production mode.

\asubsubsubsection{Model framework}
\label{sec:model}

We will consider the simplest extension of the SM that can assist the scalar potential to induce a strongly first-order electroweak phase transition, consisting of an additional real scalar singlet with a $Z_2$ symmetry. The scalar potential of the model can be written as
\be
V(s,\phi)=-\mu^2 \phi^\dagger \phi -\frac 1 2 \mu_s^2 s^2 +  \lambda (\phi^\dagger \phi)^2 + \frac {\lambda_s} {4} s^4 + \frac {\lambda_{s\phi}} 2 s^2 \phi^\dagger\phi,
\label{eq:potential}
\ee
where $\phi$ is the SM doublet
\footnote{ $\phi^T=(G^+,\frac 1 {\sqrt{2}} (h+ i G^0 +v))$, where $G^{\pm,0}$ are the Goldstone modes.}
and $s$ represents the new real singlet field. In the above, we adopt the conventional normalisation for the couplings of the SM doublets and match the other couplings with the singlet with identical normalisation. We allow for spontaneous $Z_2$ breaking with the singlet $s$ acquiring a vacuum expectation value $v_s$, since this case allows for interesting collider phenomenology of interference effects. As we shall show  later, the (on-shell) interference effects commonly exist for  loop-induced processes in BSM phenomenology and it is the focus of this paper. The CP even neutral component $h$ of the Higgs doublet field $\phi$ mixes with the real singlet scalar $s$, defining the new mass eigen-states $H$ and $S$
\bea
\binom{h}{s} =  \begin{pmatrix}
 \cos\theta & \sin\theta \\
 -\sin\theta & \cos\theta 
 \end{pmatrix}
 \binom{H}{S},
\eea
where $\theta$ is the mixing angle between these fields.
The five free parameters in Eq.~(\ref{eq:potential}) can  be traded by the two boundary conditions 
\be
m_{H}= 125~\UGeV,~~v=246~\UGeV
\ee
and the three ``physical'' parameters,
\be
m_S,~~\tan\beta(\equiv \frac  {v_s} v),~{\rm and~}\sin\theta,\label{eq:basis}
\ee
where $\tan\beta$ characterises the ratio between the VEVs of the doublet and the singlet scalar fields, respectively. Detailed relations between the bare parameters and physical parameters can be found in Ref.~\cite{Carena:2018vpt}.

\asubsubsubsection{Enhancing the di-Higgs signal via interference effects}
\label{sec:interference}

The on-shell interference effect may enhance or suppress the conventional Breit-Wigner resonance production. 
Examples in Higgs physics known in the literature, such as $gg\to h\to\gamma\gamma$~\cite{Campbell:2017rke} and $gg\to H\to t\bar t$~\cite{Dicus:1994bm,Carena:2016npr,Gori:2016zto,Craig:2015jba,Jung:2015gta}, are both destructive.
We discuss in detail in this section the on-shell interference effect between the resonant singlet amplitude and the SM di-Higgs box diagram. We shall show that in the singlet extension of the SM considered in this paper, the on-shell interference effect is generically constructive and could be large in magnitude, thus enhances the signal production rate.

The interference effect between two generic amplitudes can be denoted as non-resonant amplitude $A_{nr}$ and resonant amplitude $A_{res}$.
The resonant amplitude $A_{res}$, defined as
\be
A_{res} = a_{res} \frac {\hat s} {\hat s - m^2 + i \Gamma m},
\ee
has a pole in the region of interest and 
we parametrise it as the product of a fast varying piece containing its propagator and a slowly varying piece $a_{res}$ that generically is a product of couplings and loop-functions. The general interference effect can then be parametrised as~\cite{Carena:2016npr,Campbell:2017rke},
\bea
|\mathcal{M}|_{int}^2 &=& 2\Re(A_{res}\times A_{nr}^*)\,=\,2\left(\mathcal{I}_{int} + \mathcal{R}_{int}\right),\nonumber \\
\mathcal{R}_{int} &\equiv& |A_{nr}||a_{res}|\frac {\hat s (\hat s - m^2)} {(\hat s - m^2)^2+\Gamma^2 m^2} \cos(\delta_{res}-\delta_{nr})\nonumber \\
\mathcal{I}_{int} &\equiv& |A_{nr}||a_{res}|\frac {\hat s \Gamma m} {(\hat s - m^2)^2+\Gamma^2 m^2} \sin(\delta_{res}-\delta_{nr}),
\label{eq:decomposition}
\eea
where $\delta_{res}$ and $\delta_{nr}$ denote the complex phases of $a_{res}$ and $A_{nr}$, respectively.

The special interference effect $\mathcal{I}_{int}$ only appears between the singlet resonant diagram and the SM box diagram. This interference effect is proportional to the relative phase between the loop functions $\sin(\delta_\vartriangleright-\delta_\square)$ and the imaginary part of the scalar propagator which is sizeable near the scalar mass pole.

\asubsubsubsection{Differential distribution}

\begin{figure}[t]  
  \centering
  \includegraphics[width=.60\textwidth]{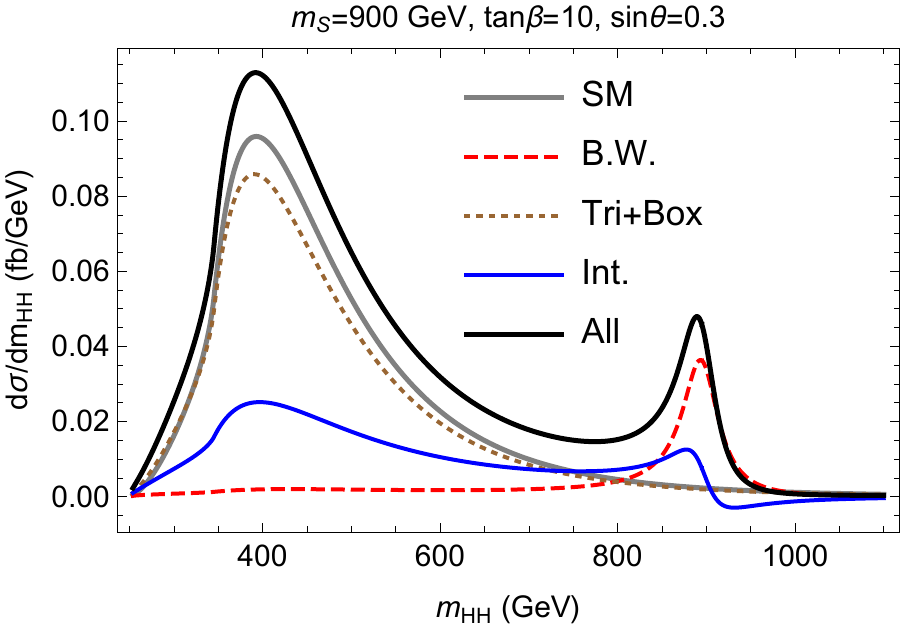}
  \caption{
  The differential di-Higgs distribution for a benchmark point of the singlet extension of the SM shown in linear scale and over a broad range of the di-Higgs invariant mass. The full results for the SM and the singlet SM extension  are shown by the  grey and black curves, respectively. In the singlet extension of the SM, the contributions from the resonant singlet diagram, the non-resonant diagram and the interference between them are shown in red (dashed), brown (dotted) and blue curves, respectively.  
  }
  \label{fig:phenoshape} 
\end{figure}

We present in this section our analysis of the differential distribution of the Higgs pair invariant mass to estimate the relevance of the interference effects discussed in the previous section. We choose one of the best channels, $pp\to HH \to b\bar b \gamma\gamma$, as the benchmark channel to present the details of our analysis. Furthermore, we discuss another phenomenologically relevant piece of interference in the far off-shell region of the singlet scalar. We display the discovery and exclusion reach  for both HL-LHC and HE-LHC  for various values of $\tan\beta$ in the $m_S$-$\sin\theta$ plane. 

In Fig.~\ref{fig:phenoshape} we display the differential cross section as a function of the Higgs pair invariant mass for a benchmark point with a heavy scalar mass of 900 \UGeV, mixing angle $\sin\theta=0.3$ and $\tan\beta=10$. 
The differential cross section is shown in linear scale for a broad range of di-Higgs invariant masses,  including the low invariant mass regime favoured by parton distribution functions at hadron colliders.

We choose this benchmark to show well the separation of the scalar resonance peak and the threshold enhancement peak above the $t\bar t$-threshold. The SM Higgs pair invariant mass distribution is given by the  grey curve while the black curve depicts the di-Higgs invariant mass distribution from the singlet extension of the SM. 
It is informative to present all three pieces that contribute to the full result of the di-Higgs production, namely, the resonance contribution (red, dashed curve), the SM non-resonance contribution (box and triangle diagrams given by the brown, dotted curve), and the interference between them (blue curve). 
Note that the small difference between the ``Tri+Box'' and the ``SM'' line shapes is caused by the doublet-singlet scalar mixing, which leads to a $\cos\theta$ suppression of the SM-like Higgs coupling to top quarks as well as a modified SM-like Higgs trilinear coupling $\lambda_{HHH}$.
We observe that the full results show an important enhancement in the di-Higgs production across a  large range of invariant masses. This behaviour is anticipated from the decomposition analysis in the previous section. There is a clear net effect from  the interference curve shown in blue.  Close to the the scalar mass pole at 900 \UGeV, the on-shell interference effect enhances the Breit-Wigner resonances peak (red, dashed curve) by about 25\%. Off-the resonance peak, and especially at the threshold peak, the interference term (blue curve) enhances the cross section quite sizeably as well. Hence, a combined differential analysis in the Higgs pair invariant mass is crucial in probing the singlet extension of the SM. 

\asubsubsubsection{Discovery and exclusion reach at the HL- and HE-LHC} 

\begin{figure}[t]
  \centering
  \includegraphics[width=1\textwidth]{\main/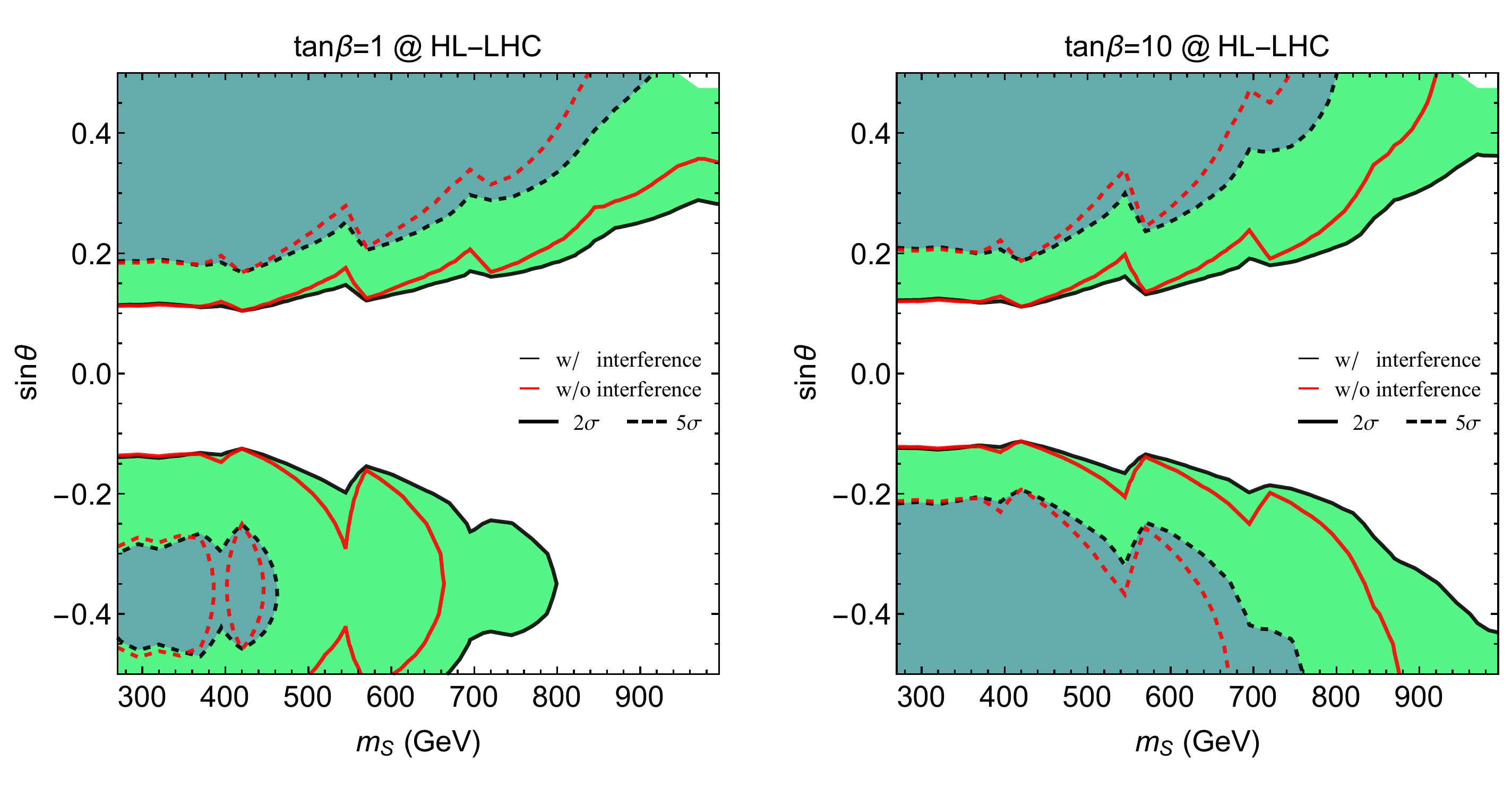}
  \caption{Projected exclusion and discovery limits at HL-LHC in the $m_S$-$\sin\theta$ plane with the line-shape analysis detailed in the text for $\tan\beta=1$ (left panel) and $\tan\beta=10$ (right panel). The shaded regions bounded by dashed/solid curves are within the discovery/exclusion reach of the HL-LHC. The black and red lines represent the projection with and without the inclusion of the interference effects between the singlet resonance diagram and the SM Higgs pair diagram, respectively.
  }
  \label{fig:HLprojection}
\end{figure} 

Using the analysis detailed in Ref.~\cite{Carena:2018vpt} through the $pp\to HH \to \gamma\gamma b\bar b$ channel, we obtain the discovery and exclusion projections for the HL-LHC and HE-LHC.
In Fig.~\ref{fig:HLprojection} we show the projected 2-$\sigma$ exclusion and 5-$\sigma$ discovery reach for the HL-LHC in the $m_S$-$\sin\theta$ plane for $\tan\beta=1$ (left panel) and $\tan\beta=10$ (right panel) in solid and dashed curves, respectively. The shaded regions are within the reach of the HL-LHC for discovery and exclusion projections. To demonstrate the relevance of the interference effects discussed in the previous sections, we show both the results obtained with and without the inclusion of the interference effects in black and red contours, respectively. 

We observe in Fig.~\ref{fig:HLprojection} that the inclusion of the interference effects extend the projections in a relevant way. For example, considering the $\tan\beta=10$ case in the right panel for $\sin\theta\simeq 0.35$ the interference effect increase the exclusion limit on $m_S$ from 850~\UGeV to 1000~\UGeV.
Note that the on-shell interference effect is larger for heavier scalar mass $m_S$. 

\begin{figure}[t]
  \centering
\includegraphics[width=1\textwidth]{\main/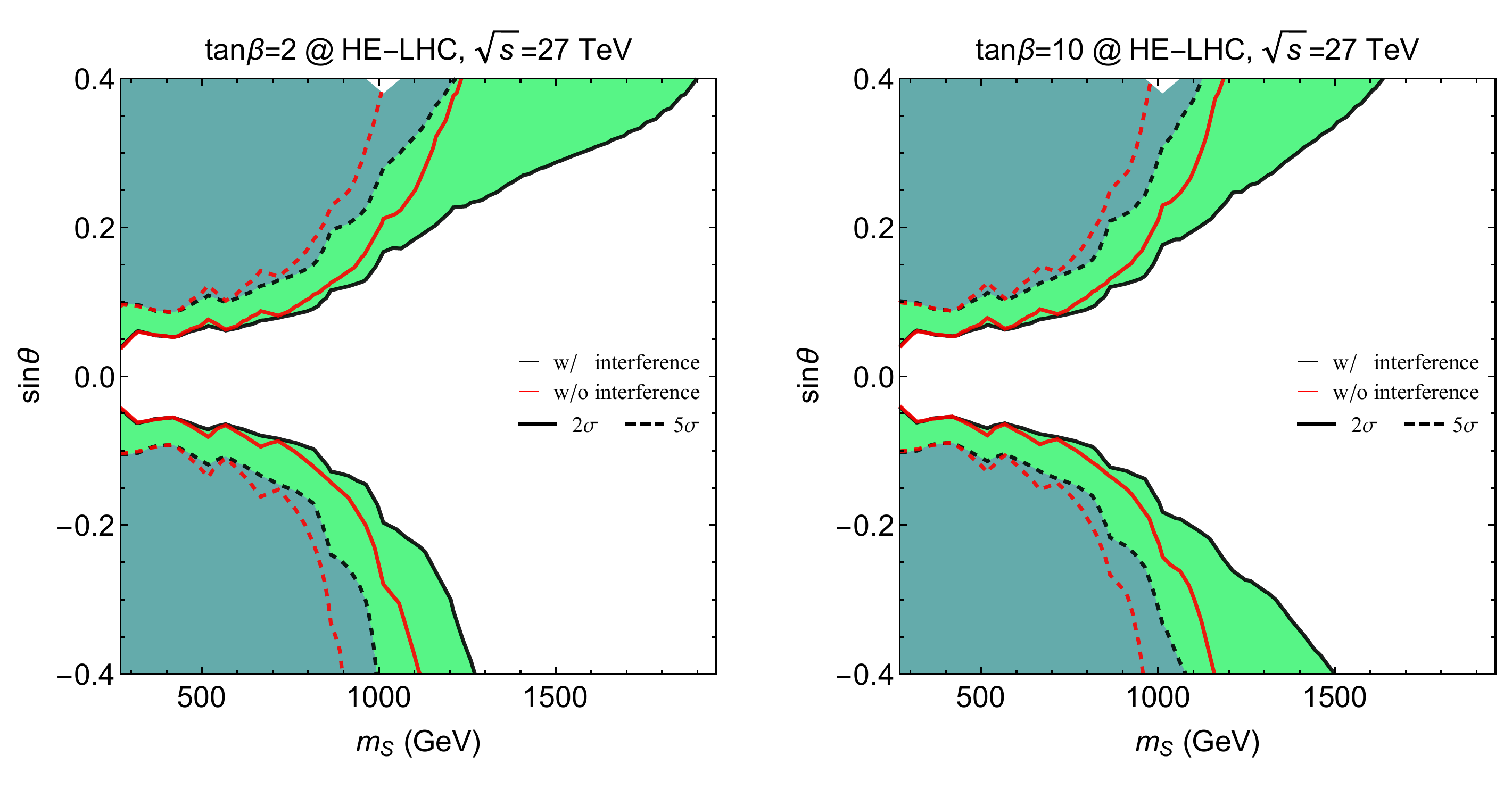}
  \caption{
  Similar to Fig.~\ref{fig:HLprojection}, projected exclusion and discovery limits at HE-LHC with 27 \UTeV centre of mass energy and an integrated luminosity of 10~$\abi$ for $\tan\beta=2$ (left panel) and $\tan\beta=10$ (right panel).
  }
  \label{fig:HEprojection}
\end{figure} 

In Fig.~\ref{fig:HEprojection} we show the projections for the HE-LHC in a analogous fashion as in Fig.~\ref{fig:HLprojection}. The discovery and exclusion reach for heavy scalars can be significantly extended by the HE-LHC operating at 27 \UTeV centre of mass energy with 10 $\abi$ of integrated luminosity. We show the results for $\tan\beta=2$ (left panel) and $\tan\beta=10$ (right panel). 
For example, considering the $\tan\beta=2$ case in the right panel of Fig.~\ref{fig:HEprojection}, for $\sin\theta\simeq 0.35$ the exclusion reach increases from 1200 to 1800~\UGeV, once more showing the importance of including the on-shell interference effects.

\asubsubsubsection{Summary and outlook}
\label{sec:conclude}

In this study, we analyse the interference effects in the $gg\to HH$ process in the presence of a heavy scalar resonance.
We focus on the novel effect of the on-shell interference contribution and discuss it in detail considering the framework of the singlet extension of the SM with spontaneous $Z_2$ breaking. 
The interference pattern between the resonant heavy scalar contribution and the SM non-resonant triangle and box contributions show interesting features. 
We highlight the constructive on-shell interference effect that uniquely arises between the heavy scalar resonance diagram and the SM box diagram, due to a large relative phase between the loop functions involved.
We observe that the on-shell interference effect can be as large as 40\% of the Breit-Wigner resonance contribution and enhances notably the total signal strength, making it necessary taking into account in heavy singlet searches.

To better evaluate the phenomenological implications of the interference effects in the di-Higgs searches, we carried out a line-shape analysis in the $gg\to HH \to \gamma\gamma b\bar b$ channel, taking into account both the on-shell and off-shell interference contributions. We find that both for the HL-LHC and HE-LHC, the proper inclusion of the interference effects increases the discovery and exclusion reach significantly.


\asubsubsection[A. Aboubrahim, P. Nath]{MSSM charged Higgs bosons}
In this section, we discuss the potential of HL-LHC and HE-LHC for discovering a heavy charged Higgs boson~\cite{Aboubrahim:2018tpf} ($m_{H^{\pm}} > m_t$) in a class of high scale models, specifically SUGRA models~\cite{Chamseddine:1982jx,Nath:1983aw,Hall:1983iz} (for a review see~\cite{Nath:2016qzm}) consistent with the experimental constraints on the light Higgs mass at $\sim 125$ \UGeV and dark matter relic density (for a recent related work see~\cite{Aboubrahim:2018bil}).  We will focus on models where the
radiative  electroweak symmetry breaking is likely realised on the hyperbolic branch~\cite{Chan:1997bi}
and where the Higgs mixing parameter $\mu$ can be relatively small. Specifically we consider supergravity models with  non-universalities in  the Higgs sector and in the gaugino sector so that 
the extended parameter space of the models we consider is given by $m_0, ~A_0, ~m_1, ~m_2, ~m_3, ~m^0_{H_u}, ~m^0_{H_d}, ~\tan\beta, ~\text{sgn}(\mu)$.
Here $m_0$ is the universal scalar mass, $A_0$ is the universal trilinear coupling, 
$m_1,  m_2, m_3$ are the masses of the $U(1)$, $SU(2)$, and $SU(3)_C$ gauginos, and $m^0_{H_u}$ and $m^0_{H_d}$ are the 
masses of the up and down Higgs bosons all at the GUT scale. To satisfy the relic density constraint in these models often
one needs  co-annihilation (see, e.g.,~\cite{Aboubrahim:2017aen,Aboubrahim:2017wjl} and the references therein).

The largest production mode of the charged Higgs at hadron colliders is the one that proceeds in association with a top quark (and a low transverse momentum $b$-quark), $pp \rightarrow t[b]H^{\pm}+X$. This production mode can be realised in two schemes, namely, the four and five flavour schemes (4FS and 5FS, respectively), where in the former, the $b$-quark is produced in the final state and  in the latter it is considered as part of the proton's sea of quarks and folded into the parton distribution functions. The cross-sections of the two production modes $q\bar{q}, gg \rightarrow tbH^{\pm} ~\text{(4FS)}$ and $gb \rightarrow tH^{\pm} ~\text{(5FS)}$, are evaluated at next-to-leading order (NLO) in QCD with \code{MadGraph\_aMC@NLO-2.6.3}~\cite{Alwall:2014hca} using \code{FeynRules}~\cite{Alloul:2013bka} UFO files~\cite{Degrande:2011ua,Degrande:2014vpa} for the Type-II 2HDM. The simulation is done at fixed order, i.e., no matching with parton shower. The couplings of the 2HDM are the same as in the MSSM, but when calculating production cross-sections in the MSSM, one should take into account the SUSY-QCD effects. In our case, gluinos and stops are rather heavy and thus their loop contributions to the cross-section are very minimal. In this case, the 2HDM is the decoupling limit of the MSSM and this justifies using the 2HDM code to calculate cross-sections. For the 5FS, the bottom Yukawa coupling is assumed to be non-zero and normalised to the on-shell running $b$-quark mass. In the 5FS, the process is initiated via gluon-$b$-quark fusion while in the 4FS it proceeds through either quark-antiquark annihilation (small contribution) or gluon-gluon fusion. At finite order in perturbation theory, the cross-sections of the two schemes do not match due to the way the perturbative expansion is handled but one expects to get the same results for 4FS and 5FS when taking into account all orders in the perturbation. In order to combine both estimates of the cross-section, we use the Santander matching criterion~\cite{Harlander:2011aa} whereby 
\begin{equation}
\sigma^{\rm matched}= ({\sigma^{4\rm FS}+\alpha\sigma^{5\rm FS}})/{1+\alpha},
\label{matched}
\end{equation}  
with $\alpha=\ln\left(\frac{m_{H^{\pm}}}{\bar{m}_b}\right)-2$. The uncertainties are combined as $\delta\sigma^{\rm matched}=\frac{\delta\sigma^{4\rm FS}+\alpha\delta\sigma^{5\rm FS}}{1+\alpha}$. The results are shown in Table~\ref{tab1:xsec}.

\begin{table}[th]
\begin{center}
\resizebox{\linewidth}{!}{\begin{tabulary}{1.12\textwidth}{l|CC|CC|CC|CC}
\hline\hline\rule{0pt}{3ex}
Model  & \multicolumn{2}{c}{$\sigma^{\rm 4FS}_{\rm NLO}(pp\rightarrow tbH^{\pm})$} & \multicolumn{2}{c}{$\sigma^{\rm 5FS}_{\rm NLO}(pp\rightarrow tH^{\pm})$} & \multicolumn{2}{c}{$\sigma^{\rm matched}_{\rm NLO}$} & $\mu_F=\mu_R$ & $\bar{m}_b$\\
  & 14 \UTeV & 27 \UTeV & 14 \UTeV & 27 \UTeV  & 14 \UTeV & 27 \UTeV & \multicolumn{2}{c}{(\UGeV)} \\ 
\hline\rule{0pt}{3ex} 
\!\!(a) & 49.0$^{+12.6\%}_{-13.1\%}$ & 272.8$^{+9.2\%}_{-10.3\%}$ & 71.8$^{+6.6\%}_{-5.7\%}$ & 397.1$^{+7.0\%}_{-6.6\%}$  & 65.9$^{+8.1\%}_{-7.6\%}$ & 365.4$^{+7.6\%}_{-7.5\%}$ & 183.6 & 2.72  \\
(b)  & 34.5$^{+10.6\%}_{-12.1\%}$ & 204.6$^{+8.1\%}_{-9.6\%}$ & 58.3$^{+7.0\%}_{-5.9\%}$ & 336.1$^{+6.9\%}_{-6.5\%}$ & 52.4$^{+7.9\%}_{-7.4\%}$ & 303.5$^{+7.2\%}_{-7.3\%}$ & 197.9 & 2.70  \\
(c)  & 29.1$^{+11.1\%}_{-12.3\%}$ & 175.9$^{+8.2\%}_{-9.7\%}$ & 48.8$^{+6.7\%}_{-5.7\%}$ & 285.9$^{+6.4\%}_{-6.0\%}$ & 43.9$^{+7.8\%}_{-7.3\%}$ & 259.0$^{+6.8\%}_{-6.9\%}$ & 205.6 & 2.69 \\
(d) & 24.8$^{+10.9\%}_{-12.3\%}$ & 149.9$^{+7.1\%}_{-9.1\%}$ & 42.6$^{+6.3\%}_{-5.3\%}$ & 264.8$^{+6.8\%}_{-6.2\%}$ & 38.3$^{+7.4\%}_{-6.9\%}$ & 237.2$^{+6.8\%}_{-6.9\%}$ & 215.9 & 2.68 \\
(e) & 18.4$^{+11.2\%}_{-12.4\%}$ & 120.1$^{+8.3\%}_{-9.8\%}$ & 32.3$^{+5.9\%}_{-4.9\%}$ & 206.7$^{+6.4\%}_{-6.0\%}$ & 29.0$^{+7.1\%}_{-6.7\%}$ & 186.3$^{+6.8\%}_{-6.9\%}$ & 229.6 & 2.67 \\
(f)  & 13.6$^{+11.3\%}_{-12.5\%}$ & 93.2$^{+7.8\%}_{-9.5\%}$ & 25.1$^{+6.1\%}_{-5.2\%}$ & 169.6$^{+6.7\%}_{-6.0\%}$ & 22.4$^{+7.3\%}_{-6.9\%}$ & 152.1$^{+7.0\%}_{-6.8\%}$ & 248.2 & 2.65  \\
(g) & 13.1$^{+10.5\%}_{-12\%}$ & 95.8$^{+7.6\%}_{-9.5\%}$ & 26.0$^{+6.2\%}_{-5.6\%}$ & 185.1$^{+6.7\%}_{-6.0\%}$ & 23.1$^{+7.2\%}_{-7.0\%}$ & 165.0$^{+6.8\%}_{-6.8\%}$ & 264.6 & 2.64 \\
(h) & 11.2$^{+10.3\%}_{-12.0\%}$ & 85.1$^{+7.5\%}_{-9.4\%}$ & 22.7$^{+6.1\%}_{-5.8\%}$ & 168.3$^{+6.8\%}_{-5.9\%}$ & 20.2$^{+7.0\%}_{-7.2\%}$ & 149.9$^{+6.9\%}_{-6.7\%}$ & 278.2 & 2.63 \\
(i)  & 7.8$^{+11.7\%}_{-12.6\%}$ & 61.1$^{+8.1\%}_{-9.8\%}$ & 15.8$^{+6.0\%}_{-6.0\%}$ & 121.0$^{+6.9\%}_{-6.0\%}$ & 14.0$^{+7.2\%}_{-7.4\%}$ & 107.9$^{+7.2\%}_{-6.8\%}$ & 292.9 & 2.62 \\
(j) & 5.5$^{+12.6\%}_{-13.0\%}$ & 48.9$^{+9.1\%}_{-10.3\%}$ & 11.6$^{+6.7\%}_{-6.5\%}$ & 99.4$^{+6.4\%}_{-5.5\%}$ & 10.3$^{+7.9\%}_{-7.8\%}$ & 88.7$^{+6.9\%}_{-6.5\%}$ & 329.9 & 2.60 \\
\hline
\end{tabulary}}\end{center}
\caption{The NLO production cross-sections, in fb, of the charged Higgs in association with a top (and bottom) quark in the five (and four) flavour schemes along with the matched values at $\sqrt{s}=14$ \UTeV and $\sqrt{s}=27$ \UTeV for the ten benchmark points in~\cite{Aboubrahim:2018tpf}. The running $b$-quark mass, $\bar{m}_b$, is also shown evaluated at the factorisation and normalisation scales, $\mu_F=\mu_R=\frac{1}{3}(m_t+\bar{m}_b+m_{H^{\pm}})$.}
\label{tab1:xsec}
\end{table}

For the parameter points considered, the $H^{\pm}\rightarrow\tau\nu$ channel has the smallest branching ratio but it is of interest since jets can be tau-tagged and the tau has leptonic and hadronic decay signatures. For the considered signal final states (fully hadronic), the SM backgrounds are mainly $t\bar{t}$, $t$+jets, $W/Z/\gamma^*$+jets, di-boson production and QCD multi-jet events which can fake the hadronic tau decays. The simulation of the charged Higgs associated production, $t[b]H^{\pm}$, is done at fixed order in NLO while the SM backgrounds are done at LO (which are then normalised to their NLO values) using \code{MadGraph} interfaced with LHAPDF~\cite{Buckley:2014ana} and \code{PYTHIA8}~\cite{Sjostrand:2014zea} which handles the showering and hadronisation of the samples.  For the SM backgrounds a five-flavor MLM matching~\cite{Mangano:2006rw} is performed on the samples. Detector simulation and event reconstruction is performed by \code{DELPHES-3.4.2}~\cite{deFavereau:2013fsa} using the beta card for HL-LHC and HE-LHC studies.

The selection criteria depends on the flavour scheme under consideration. For the 4FS (5FS) we apply a lepton veto and at least five (four) jets, two (one) of which are (is) b-tagged and one is tau-tagged. To discriminate the signal from background we use gradient boosted decision trees, \code{GradientBoost}, which proves to be more powerful than the conventional cut-based analysis. A large set of variables have been tried in the BDT training and the ones which produced the best results were kept. The kinematic variables entering into the training of the BDTs are: 
\begin{align}
E^{\rm miss}_T, ~~~ E^{\rm miss}_T/\sqrt{H_T}, ~~~m_{T2}^{\rm jets}, ~~~m_T^{\tau}, ~~~p_T^{\tau}, ~~~E^{\rm miss}_T/m_{\rm eff},  \nonumber \\
m_T^{\rm min}(j_{1-2},E^{\rm miss}_T), ~~~\Delta\phi(p^{\tau}_T,E^{\rm miss}_T), ~~~N_{\rm tracks}^{\tau}, ~~~\sum_{\rm tracks} p_T \, .
\end{align}
The training and testing of the samples is carried out using \code{ROOT}'s~\cite{Antcheva:2011zz} own TMVA (Toolkit for Multivariate Analysis) framework~\cite{Speckmayer:2010zz}. After the training and testing phase, the variable ``BDT score" is created. We apply the selection criteria (as given in~\cite{Aboubrahim:2018tpf}) along with a BDT score cut $>0.95$ on the SM background and on each of the 4FS and 5FS signal samples to obtain the remaining cross-sections. The signal cross-sections are combined using Eq.~(\ref{matched}) in order to evaluate the required minimum integrated luminosity for a $\frac{S}{\sqrt{S+B}}$ discovery at the $5\sigma$ level. The results for both the 14 and 27 \UTeV cases are shown in Fig.~\ref{fig1}. 

\begin{figure}[H]
 \centering
   \includegraphics[width=0.49\textwidth]{\main/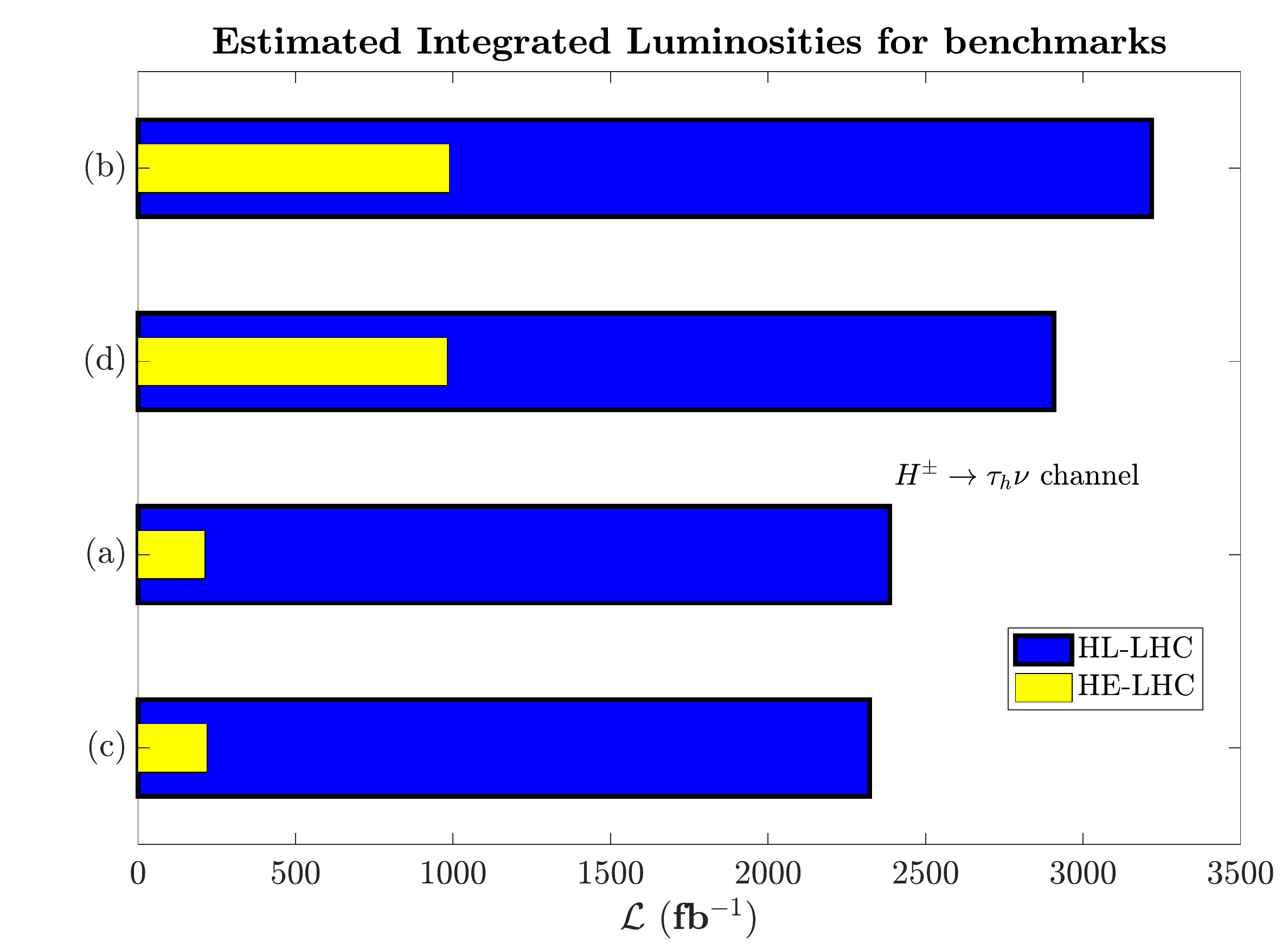} 
      \includegraphics[width=0.49\textwidth]{\main/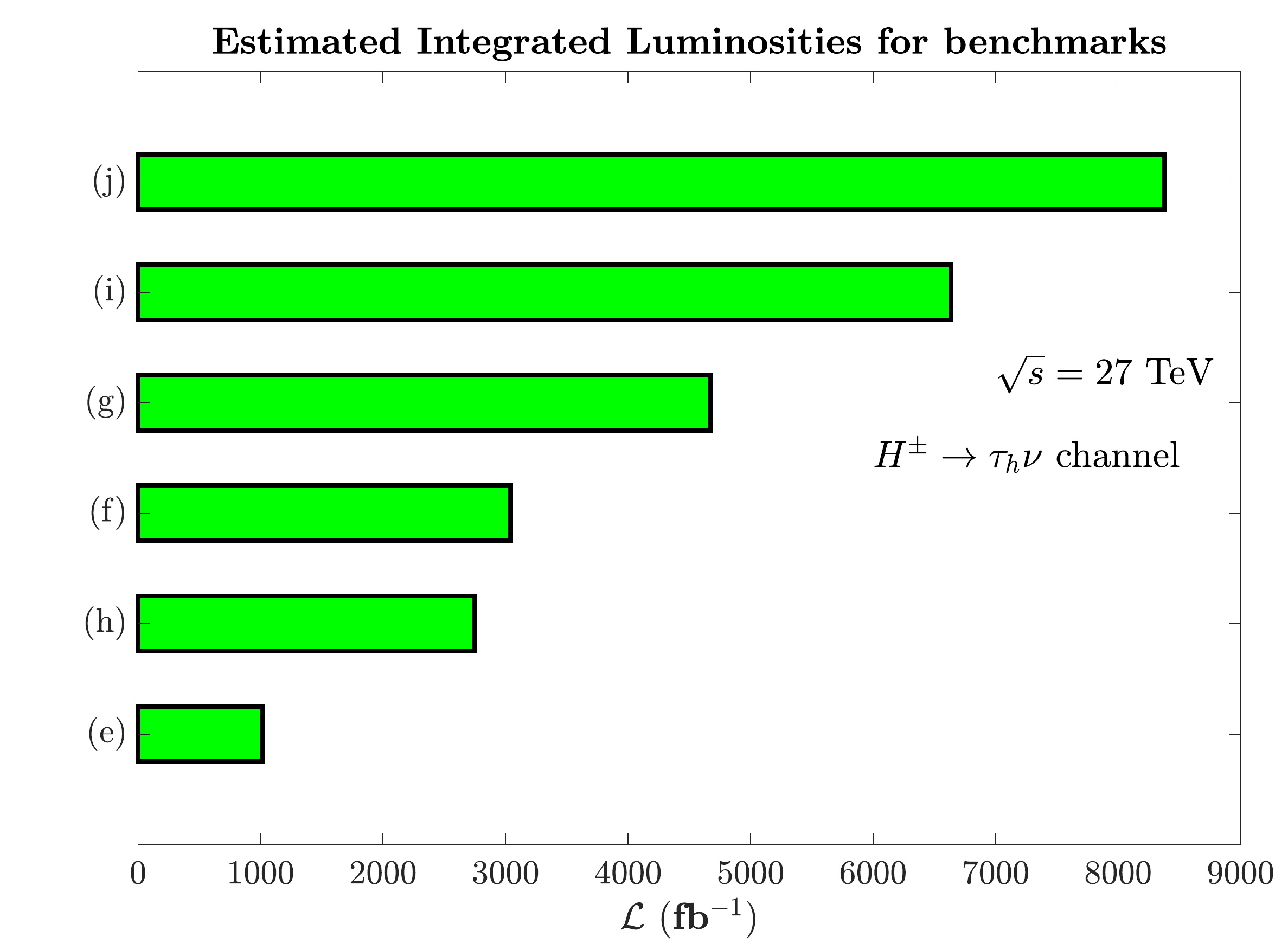}
   \caption{The evaluated integrated luminosities, $\mathcal{L}$ ($\ifb$), for ten benchmark points. Left plot: calculated $\mathcal{L}$ for points discoverable at both HL-LHC and HE-LHC. Right plot: calculated $\mathcal{L}$ for points discoverable only at HE-LHC.}
	\label{fig1}
\end{figure}

One can see from Fig.~\ref{fig1} that four of the ten points may be discoverable at the HL-LHC as it nears the end of its run where a maximum integrated luminosity of 3000$\ifb$ will be collected. Given the rate at which the HL-LHC will be collecting data, points (a)-(d) will require $\sim 7$ years of running time. On the other hand, the results from the 27 \UTeV collider show that all points may be discoverable for integrated luminosities much less than $15\iab$. 

\textbf{Acknowledgements:}
This research was supported in part by the NSF Grant PHY-1620575.


\asubsection[P. Bechtle, S. Heinemeyer, S. Liebler, T. Stefaniak, G. Weiglein]{Direct and indirect sensitivity to heavy Higgs bosons using MSSM benchmark scenarios}\label{Sec:9.5}

The LHC keeps measuring the properties of the discovered Higgs boson with increasing precision. So far the measured properties are, within current experimental and theoretical uncertainties, in agreement with the SM predictions~\cite{Khachatryan:2016vau}. The MSSM~\cite{Nilles:1983ge,Haber:1984rc,Gunion:1984yn} is one of the best studied models with an extended Higgs sector. It predicts two scalar partners for all SM fermions as well as fermionic partners to all SM bosons. Contrary to the case of the SM, the MSSM contains two Higgs doublets.
This results in five physical Higgs bosons instead of the single Higgs boson in the SM. In the absence of $\mathcal{CP}$-violating phases, these are the light and heavy $\mathcal{CP}$-even Higgs bosons,  
$h$ and $H$, the $\mathcal{CP}$-odd Higgs boson, $A$, and the charged Higgs bosons, $H^\pm$.

In order to facilitate collider searches for the additional MSSM Higgs bosons, a set of new benchmark scenarios for MSSM Higgs boson searches at the LHC have been proposed recently~\cite{Bahl:2018zmf}. The scenarios are compatible -- at least over wide portions
of their parameter space -- with the most recent LHC results for the
Higgs-boson properties and the bounds on masses and couplings of new
particles. Each scenario contains one $\mathcal{CP}$-even scalar with mass around $125$\,\UGeV and SM-like couplings. However, the scenarios differ importantly in the phenomenology of the additional, so far undetected Higgs bosons.

The search for the additional Higgs bosons will continue at the LHC
Run~3 and subsequently at the HL-LHC. These benchmark scenarios, due to their distinct phenomenology of the additional Higgs bosons, serve well to assess the reach of current and future colliders. The reach can either be direct, via the search for new Higgs bosons, or indirect, via the precise measurements of the properties of the Higgs boson at $\sim 125$\,\UGeV. 

\subsubsection*{Experimental and theoretical input}

In order to analyse the potential of the HL-LHC in the exploration of the MSSM Higgs sector we evaluate the direct and indirect physics reach in two of the benchmark scenarios proposed in Ref.~\cite{Bahl:2018zmf}. The first scenario is the $M_h^{125}$: it is characterised by relatively heavy superparticles, such that the Higgs phenomenology at the LHC resembles that of a Two-Higgs-Doublet-Model with MSSM-inspired Higgs couplings. The second scenario is the $M_h^{125}(\tilde{\chi})$. It is characterised by light electroweakinos (EWinos), resulting in large decay rates of the heavy Higgs bosons $H$ and $A$ into charginos and neutralinos, thus diminishing the event yield of the $\tau^+\tau^-$~final state signatures that are used to search
for the additional Higgs bosons at the LHC. In addition, the branching ratios of the Higgs boson at $125$\,\UGeV into a pair of photons is enhanced for small values of $\tan\beta$ due to the EWinos present in the loop.

We assess the reach of direct LHC searches in the  $\tau^+\tau^-$
final state by applying the model-independent $95\%~\mathrm{CL}$
limit projections for $6~\mathrm{ab}^{-1}$ from the CMS experiment, see Sec.~\ref{sec:model_indep}, Fig.~\ref{fig:model_indep}.\footnote{We thank Martin Flechl for helpful discussions.} 
We implemented these limits --- presented as one-dimensional (marginalised) cross section limits on either the
gluon fusion or $b\bar{b}$-associated production mode --- in the program
\texttt{HiggsBounds}~\cite{Bechtle:2008jh,Bechtle:2011sb,Bechtle:2013wla,Bechtle:2015pma} to obtain the projected $95\%~\mathrm{CL}$ exclusion in our scenarios.

We estimate the indirect reach through Higgs rate measurements by
using detailed HL-LHC signal strength projections for the individual
Higgs production times decay modes, including the corresponding
correlation matrix, as evaluated by the ATLAS and CMS experiment assuming YR18 systematic uncertainties (S2), see Sec.~\ref{sec:expcomb_prodtimesdecay}, Tab.~\ref{tab:summary_A1_5PD}.  We furthermore take cross-correlations of theoretical
rate uncertainties between future ATLAS and CMS measurements into
account. All this is done with the use of the program
\texttt{HiggsSignals}~\cite{Bechtle:2013xfa}. 

The theory predictions are obtained from
\texttt{FeynHiggs}~\cite{Heinemeyer:1998yj,Heinemeyer:1998np,Degrassi:2002fi,Frank:2006yh,Hahn:2013ria,Bahl:2016brp,Bahl:2017aev,Bahl:2018qog},
as well as from \texttt{SusHi}~\cite{Harlander:2012pb, Harlander:2016hcx,
Harlander:2005rq,Harlander:2002wh,
Harlander:2002vv,Anastasiou:2014lda,Anastasiou:2015yha,
Anastasiou:2016cez,Degrassi:2010eu, Degrassi:2011vq,
Degrassi:2012vt,Actis:2008ug}
for gluon fusion and matched predictions for bottom-quark
annihilation~\cite{Bonvini:2015pxa,Bonvini:2016fgf,Forte:2015hba, Forte:2016sja}. We determine the theoretical uncertainties on the Higgs production cross sections as in Ref.~\cite{Bahl:2018zmf}. 
For the light Higgs rate measurements we use the SM uncertainties following Ref.~\cite{deFlorian:2016spz}.

\subsubsection*{Projected HL-LHC reach}
 
Our projections in the $M_h^{125}$ and the $M_h^{125}(\tilde{\chi})$ scenario in the
($M_A, \tan\beta$) plane are presented in the left and right panel of Fig.~\ref{fig:bench}, respectively. 
We furthermore include the current limit (magenta dotted line) for the indirect reach of the LHC in the two benchmark scenarios, as evaluated in Ref.~\cite{Bahl:2018zmf}, as well as the expected limit from current direct BSM Higgs searches by ATLAS~\cite{Aaboud:2017sjh} (red dashed line) and CMS~\cite{Sirunyan:2018zut} (green dashed line) in the $\tau^+\tau^-$ final state, using $\sim 36~\mathrm{fb}^{-1}$ of data from Run~II at $13~\mathrm{\UTeV}$.

\begin{figure}
\begin{center}
\includegraphics[width=0.5\textwidth]{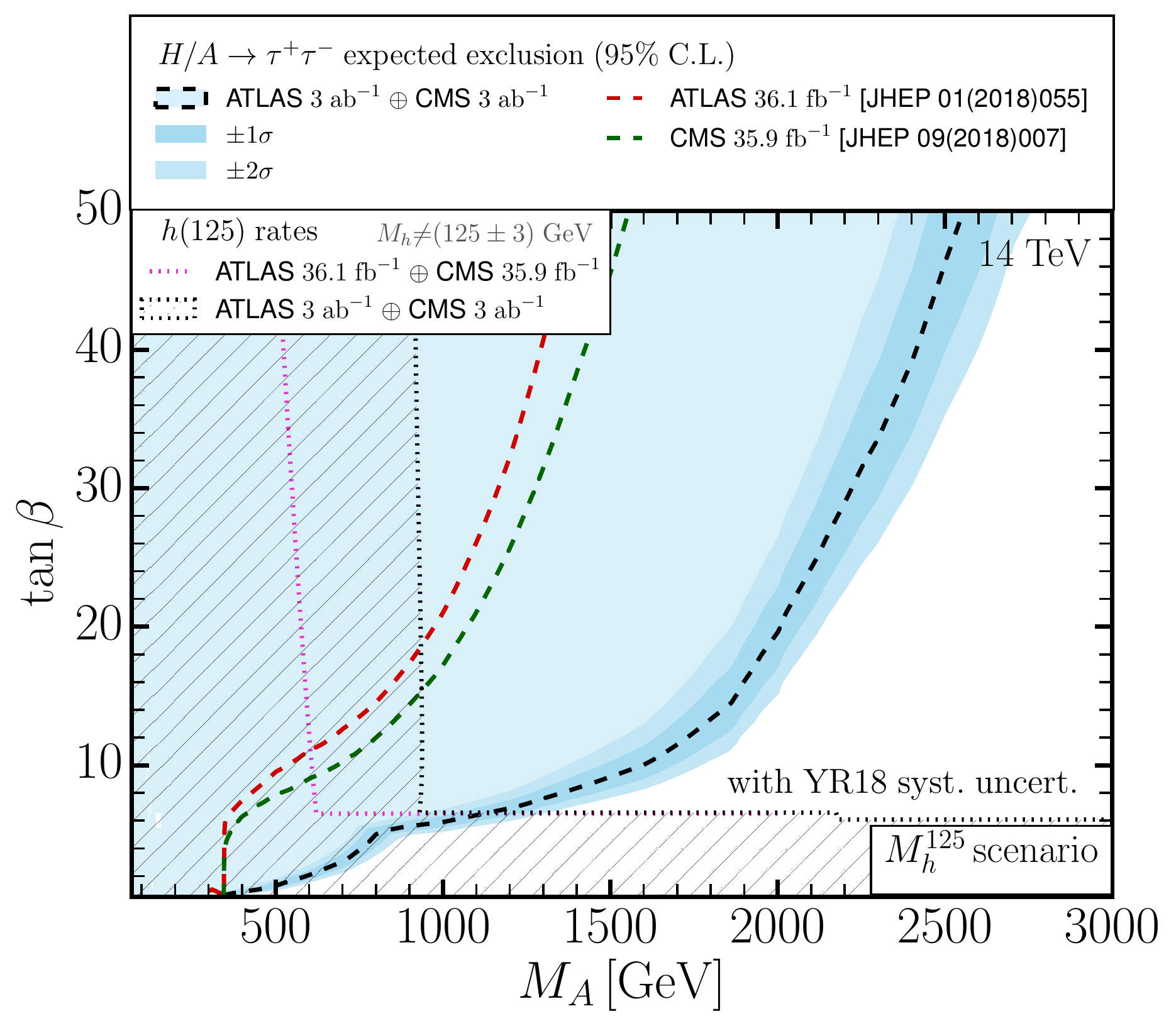}\hfill
\includegraphics[width=0.5\textwidth]{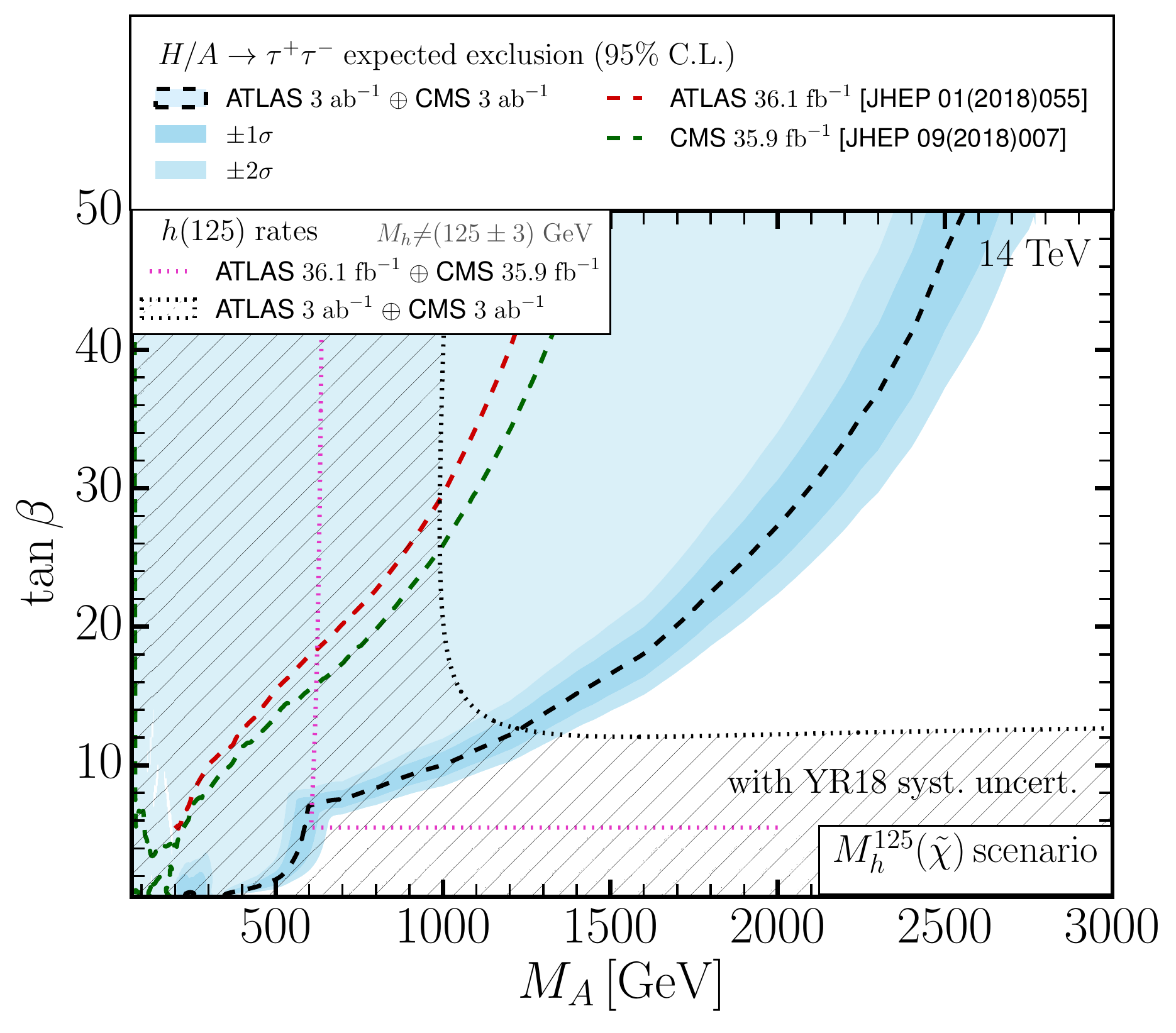}
\end{center}
\caption{HL-LHC projections in the $M_h^{125}$ (\emph{left}) and
  $M_h^{125}(\tilde{\chi})$ (\emph{right}) scenario, assuming YR18 systematic uncertainties (S2). The dashed black curve and blue filled region indicate the HL-LHC reach via direct heavy Higgs searches in the $\tau^+\tau^-$ channel with $6~\mathrm{ab}^{-1}$ of data (with the dark blue regions indicating the $1$ and $2\sigma$ uncertainty), whereas 
  the red and green dashed lines show the expected limit from current searches in this channel by ATLAS~\cite{Aaboud:2017sjh} and CMS~\cite{Sirunyan:2018zut}, respectively. The current and future HL-LHC sensitivity via  combined ATLAS and CMS Higgs rate measurements is shown as magenta and black dotted contours, respectively (the latter being accompanied with a hatching of the prospectively excluded region).}
\label{fig:bench}
\end{figure}

Within the $M_h^{125}$ scenario the reach via measurements of the Higgs signal strengths extends to $M_A$ values of around $900$\,\UGeV. The horizontal contour excluding $\tan\beta$ values less than 6 is due to the light Higgs mass being below $122$\,\UGeV, where the interpretation of the observed Higgs signal in terms of the light $\mathcal{CP}$-even MSSM Higgs boson $h$ becomes invalid. The direct heavy Higgs searches in the $\tau^+\tau^-$ final state will probe the parameter space up to $M_A \le 2550$\,\UGeV for $\tan\beta = 50$, and up to $M_A \le 2000$\,\UGeV at $\tan\beta = 20$. 

The picture is somewhat different in the $M_h^{125}(\tilde{\chi})$ scenario. Here the large branching ratio of the heavy neutral Higgs boson decaying to charginos and neutralinos leads to a strongly reduced direct reach of heavy Higgs to $\tau^+\tau^-$ searches. While at large values of $\tan\beta \sim 50$  the reach is only slightly weaker than in the $M_h^{125}$ scenario, at $\tan\beta = 20$ it is significantly reduced to $M_A \le 1700$\,\UGeV. In order to overcome this, dedicated searches for the decays of $H$ and $A$ to charginos and neutralinos will have to be devised. On the other hand, Higgs rate measurements are an important complementary probe. They exclude
$M_A \le 950$\,\UGeV and $\tan\beta \le 12.5$. While the bound in $M_A$
is induced through Higgs coupling modifications arising from
non-decoupling, values of $\tan\beta\le 12.5$ feature a too-large enhancement of the $h\to \gamma\gamma$ partial width. The combination of direct and indirect bounds yields
a lower limit of $M_A \ge 1250$\,\UGeV in the $M_h^{125}(\tilde{\chi})$ scenario.

In summary, the HL-LHC has the potential, using the combined direct and indirect reach, to probe the MSSM Higgs sector up to $M_A \sim 900$-$1000$\,\UGeV and possibly beyond, depending on the details of the MSSM scenario. Values larger than that, as predicted, e.g., by GUT based models~\cite{Buchmueller:2013rsa,Buchmueller:2014yva,Bechtle:2015nua,Bagnaschi:2016afc,Bagnaschi:2016xfg,Costa:2017gup} or Finite Unified Theories~\cite{Heinemeyer:2013nza,Heinemeyer:2018roq,Heinemeyer:2018zpw}, or allowed by global fits of the phenomenological MSSM~\cite{Bechtle:2016kui,Bagnaschi:2017tru} would remain uncovered. To explore these regions an energy upgrade and/or refined Higgs signal strength measurements (e.g.\ at an $e^+e^-$ collider~\cite{Moortgat-Picka:2015yla}) will be
necessary.

\asubsection[D. Redigolo]{Direct and indirect sensitivity to heavy Twin Higgs bosons}
The existence of additional Higgs bosons is motivated by many
approaches to physics beyond the Standard Model. Here we focus on the simple case of a second Higgs which is a singlet of the SM gauge group. This is motivated in many BSM constructions addressing the naturalness problem of the electroweak scale like Supersymmetry or Compositeness. Independently on naturalness, an extra singlet arises in minimal scenarios to get a first-order EW phase transition which is necessary for EW baryogenesis. 

There is now an extensive suite of LHC searches for additional Higgs bosons decaying promptly into SM
final states. Among these, di-boson searches are particularly promising to hunt for an additional Higgs singlets (see Refs.~\cite{Sirunyan:2018qlb,Aaboud:2018knk,Sirunyan:2017isc,ATLAS:2017spa}). A first important question for HL-LHC is to understand how these direct searches correlate with Higgs coupling deviations. This question has already been addressed for the singlet Higgs at HL-LHC in Ref.~\cite{Buttazzo:2015bka} (see also Sec.~6.1.4 in the WG3 physics report \cite{CidVidal:2018eel} and Sec.~\ref{sec6:singletHP} of this report) and we summarise it here for completeness. In short, one can prove that there is limited room for discovery of the second Higgs singlet in direct production unless deviations in the SM Higgs coupling bigger than $5\%$ will be found at HL-LHC. This is a great motivation for future machines exploring the high energy frontier in SM visible decays of the second Higgs (see for example Ref.~\cite{Buttazzo:2018qqp} for an assessment of the reach of high-energy lepton colliders). 

Here we show that the situation is radically different if the second Higgs singlet has exotic displaced decays following Ref.~\cite{Alipour-fard:2018mre}. We focus on the case of a singlet Higgs decaying into a pair of long-lived particles (LLPs), whose decays within the detector volume set them qualitatively apart from promptly-decaying or detector-stable
particles. These type of displaced decays are often present in extensions of the Higgs sector which entail rich hidden sectors coupled primarily through the Higgs portal to additional Higgs-like scalars (see Ref.~\cite{Curtin:2018mvb} for a recent summary of the theory motivations).  

On the experimental side, the signatures of displaced LLP's pairs produced from the decay of a heavy singlet Higgs are sufficiently distinctive that they may be identified by analyses with little or no Standard Model backgrounds even at HL-LHC, making them a promising channel for discovering additional Higgs bosons. By recasting present LHC searches for a pair of displaced tracks with different displacements \cite{Aad:2015uaa,CMS:2014wda,Aaboud:2018aqj,Sirunyan:2018pwn}, we show that the discovery potential of exotic decays of the second Higgs singlet exceeds the asymptotic reach of SM Higgs coupling deviations and provides a natural avenue for the further development of searches for additional Higgs bosons. This is a promising next step to complete the experimental coverage of extended Higgs sectors at the LHC, especially because analogous decays of the 125 \UGeV Higgs to LLPs may be challenging to discover at the LHC due to trigger thresholds (see Ref.~\cite{Curtin:2013fra} for a summary and Refs.~\cite{Clarke:2015ala,Csaki:2015fba,Curtin:2015fna,Pierce:2017taw} for collider studies of displaced signal from SM Higgs decays).
  
On the theory side many models addressing dark matter, baryogenesis and the hierarchy problem can be mapped to the singlet simplified model we discuss here as long as the interactions of the heavy Higgs are controlled primarily by the Higgs portal. As an example, we present the concrete case of the Twin Higgs (TH) construction, quantifying the asymptotic reach of LHC searches for a Twin Higgs decaying into a pure glue hidden valley as originally proposed in Ref.~\cite{Craig:2015pha}. 

Regarding the physics opportunities of HE-LHC, we refer to \cite{CidVidal:2018eel} for a discussion of the visible decays of the singlet Higgs. A reliable assessment of the reach in exotic displaced decays strongly depends on the details of the trigger opportunities of HE-LHC and it is left for the future.    

\subsubsection{The simplified model with a long lived singlet scalar}

We introduce the effective Lagrangian of a CP-even scalar up to dimension four:
\begin{equation}
\mathcal{L}_{\text{visible}}=\frac{1}{2}(\partial_\mu S)-\frac{1}{2} m_S^2 S^2-a_{HS} S\vert H\vert^2-\frac{\lambda_{HS}}{2} S^2\vert H\vert^2-\frac{a_S}{3} S^3-\frac{\lambda_S}{4} S^4\ .\label{eq:everybody}
\end{equation}
After electroweak symmetry breaking, the singlet mixes with the uneaten CP-even component of the Higgs doublet and we can write the mixing angle $\gamma$ as
\begin{equation}
\gamma\simeq\frac{v(a_{HS}+\lambda_{HS} f)}{m_\phi^2}\quad\ ,\qquad H=\begin{pmatrix}\pi^+\\ \frac{v+h}{\sqrt{2}}\end{pmatrix}\quad\ ,\qquad S=f+\phi\ ,
\end{equation}   
where $m_\phi$ is the mass of the singlet in the mass basis, $v=246\text{ \UGeV}$ is the electroweak vacuum expectation value and $f$ is the VEV of the singlet $S$. The formula shows how the mixing between the singlet and the SM Higgs is controlled by the spontaneous and/or explicit breaking of a discrete $\mathbb{Z}_2$ symmetry under which the singlet is odd ($S\to -S$) and the SM Higgs even $(H\to H)$.  In what follows, we focus mainly on the scenario where the singlet takes a VEV at the minimum of a $\mathbb{Z}_2$-invariant potential. Then the $\mathbb{Z}_2$-breaking is spontaneous and we have
\begin{equation}
\gamma\simeq \frac{\lambda_{HS}}{\lambda_S}\cdot\frac{v}{f}\qquad ,\qquad m_\phi^2\simeq \lambda_S f^2\ . \label{eq:mixing}
\end{equation} 
 This is explicitly realised in Twin Higgs scenarios where $\lambda_S\simeq \lambda_{HS}$ and $\gamma\sim \frac{v}{f}$ (see Refs~\cite{Chacko:2005pe,Barbieri:2005ri}). The phenomenology of the singlet and the SM-like Higgs can be summarised as follows:
\begin{align}
&\frac{g_{hVV,f\bar f}}{g_{hVV,f\bar f}^{SM}}=\cos^2\gamma \label{eq:SMHiggs}\ ,\\
&\sigma_\phi=\sin^2\gamma\cdot \sigma_{h}(m_\phi)\label{eq:Sproduction}\ ,\\
&\text{BR}_{\phi\to f\bar{f},VV}=\text{BR}_{h\to f\bar{f},VV}(1-\text{BR}_{\phi\to hh\ \label{eq:SBR}
})\ ,
\end{align}
where $g_{hVV}/g_{hVV}^{SM}$ and $g_{hf\bar f}/g_{hf\bar f}^{SM}$ refer to the couplings of the SM-like Higgs to SM vectors and fermions, respectively, normalised to the SM prediction. The couplings of the SM-like Higgs in Eq. (\ref{eq:SMHiggs}) are reduced by an overall factor, leading to a reduced production cross section in every channel but unchanged branching ratios. The production cross section of the heavy singlet $\sigma_\phi$ in Eq. (\ref{eq:Sproduction}) is the one of the SM Higgs boson at mass $m_\phi$ rescaled by the mixing angle.  The branching ratios of the singlet into SM gauge bosons in Eq.~\eqref{eq:SBR} are rescaled by a common factor depending on the branching ratio into $hh$. The latter is model dependent but in the limit $m_\phi\gg m_W$ an approximate $SO(4)$ symmetry dictates $\text{BR}_{\phi\to hh }\simeq \text{BR}_{\phi\to ZZ}\simeq \text{BR}_{\phi\to WW} /2$.

\begin{figure}[t]
\includegraphics[width=.8\textwidth]{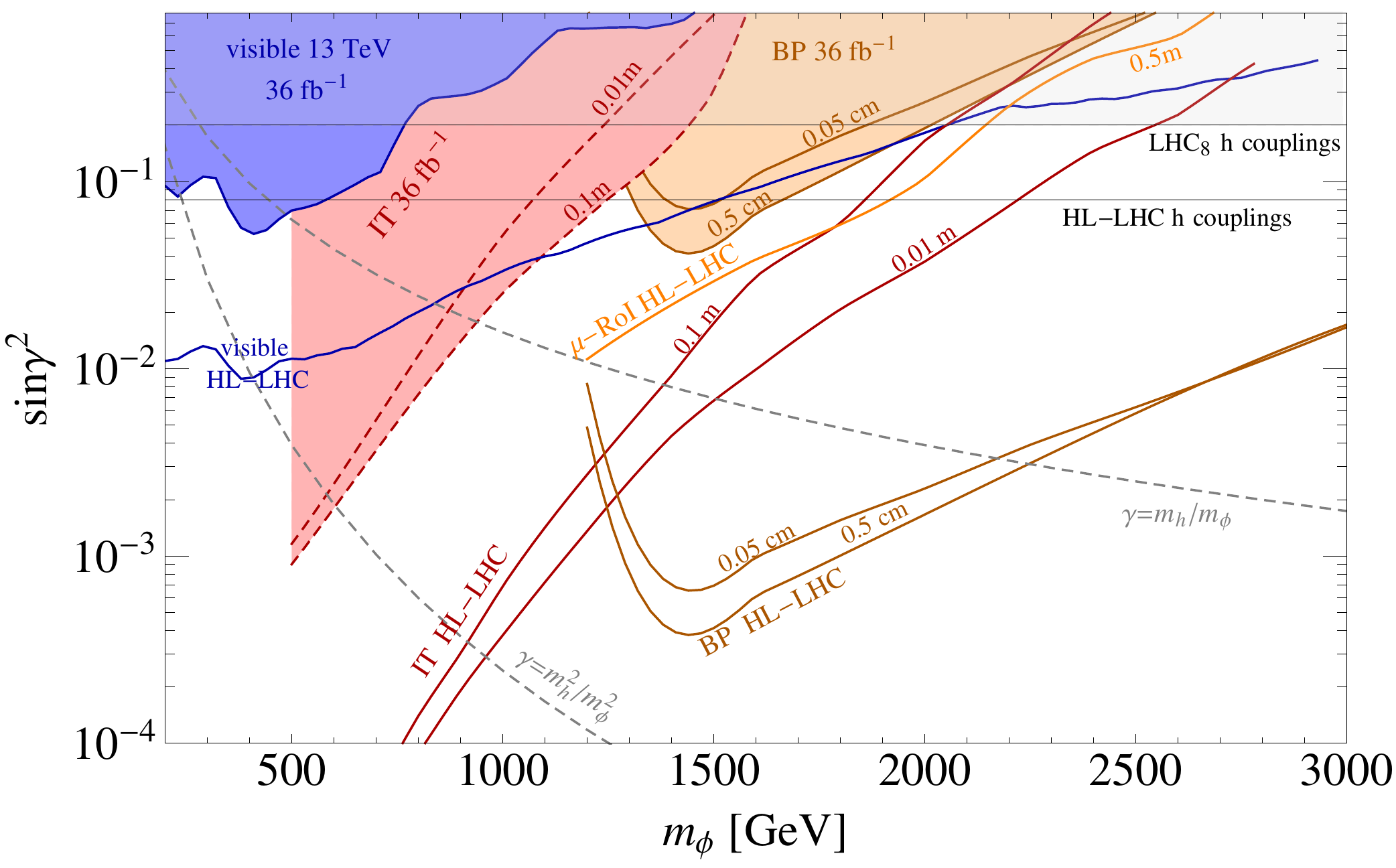}
\centering
\caption{Parameter space of the singlet Higgs as a function of $m_\phi$ and $\sin^2 \gamma$. See text for details.
\label{fig:singlet}
}
\end{figure}
We summarise in Fig.~\ref{fig:singlet} the relative strength of existing and future di-boson and di-Higgs searches at the LHC \cite{Sirunyan:2018qlb,Aaboud:2018knk,Sirunyan:2017isc,ATLAS:2017spa}, as well as constraints coming from the precision measurement of Higgs couplings (taking for definiteness the values in \cite{Dawson:2013bba}). 

We now want to add to the setup in Eq. (\ref{eq:everybody}) the reach of present and future displaced searches.  We consider the singlet $S$ to be a portal to a generic dark sector. In this case the singlet $S$ can decay abundantly to a pair of approximately long lived  dark states without suppressing the signal rate. A simple example motivated by Twin Higgs constructions \cite{Craig:2015pha} and Hidden Valley models~\cite{Strassler:2006ri,Han:2007ae}) is 
\begin{equation}
\mathcal{L}_{\text{displaced}}=-\frac{a_{SX}}{2} S X^2-\frac{b_{SX}}{2} S^2 X-\frac{\lambda_{SX}}{4} S^2 X^2-\frac{\lambda_{SX}}{4} \vert H\vert^2 X^2-\frac{m_X^2}{2} X^2\ ,
\end{equation} 
where  the extra dark singlet scalar daughter $X$ is odd under an approximate $\mathbb{Z}_2$-symmetry like $S$ and $a_{SX}\simeq b_{SX}\simeq0$. For $m_S>2 m_X$ the singlet $S$ will decay into pairs of scalar daughters with a width $\Gamma_{\text{displaced}}=\frac{\lambda_{SX}^2 f^2}{8\pi m_S}$ which is now independent of the mixing in Eq. (\ref{eq:mixing}). The width of $X$ into SM states is proportional to the $\mathbb{Z}_2$-breaking operators and can be arbitrarily suppressed. In next section we show how the Twin Higgs gives an explicit realisation of this simplified model where the singlet $X$ is identified with the lightest glueball. The mass of the glueball is naturally light because of dimensional transmutation in the dark sector, and the decay of the singlet $S$ into dark states unsuppressed because of the rich structure of the hidden sector where heavier states shower down to the lightest glueball. 

We consider the present bound and future projections at HL-LHC of the following searches \footnote{The several $\epsilon$ in the equations below account for the detector acceptance and efficiency for the signal.}:
\begin{itemize}
\item The {\bf muon region of interest trigger ($\mu$-RoI)}  analysis of  ATLAS at 13 \UTeV \cite{Aaboud:2018aqj} is tailored to tag displaced decays with decay length $0.5\text{ m}\lesssim c\tau\lesssim 20\text{ m}$. The 13 \UTeV search is an update of a previous 8 \UTeV analysis \cite{Aad:2015uaa} which remains background-free with trigger performance comparable to the old search. The 95\% C.L. exclusion limit is given by 
\begin{equation}
\sigma_\phi^{13\text{ \UTeV}}\cdot\text{ BR}=  0.083\text{ fb}\cdot \frac{L}{36.1\text{ fb}^{-1}}\cdot \frac{1}{\epsilon(m_\phi,m_X,c\tau_X)}\ .
\end{equation}
\item The { \bf displaced di-jet pairs in the inner tracker (IT)} analysis of CMS at 8 \UTeV \cite{CMS:2014wda} is mostly sensitive to displacement with decay length $\text{5 mm}\lesssim c\tau\lesssim 1 \text{ m}$. The 95\% C.L. exclusion limit is given by 
\begin{equation} 
\sigma_\phi^{8\text{ \UTeV}}\cdot\text{ BR}= 0.23\text{ fb}\cdot \frac{L}{18.5\text{ fb}^{-1}}\cdot \frac{1}{\epsilon(m_\phi,m_X,c\tau_X)} \ .
\end{equation}
provided that $m_X$ is not so much smaller than $m_\phi$ that the average boost of $X$ collimates its decay products.
\item The search based on { \bf two  displaced vertices in the beampipe (BP)} at CMS 13 beam-pipe \cite{Sirunyan:2018pwn} is dedicated to very small displacements $c\tau\lesssim 1\text{ mm}$. The 95\% C.L. exclusion limit is given by 
\begin{equation}
\sigma_\phi^{13\text{ beam-pipe}}\cdot\text{ BR}= 0.078\text{ fb}\cdot \frac{L}{38.5\text{ fb}^{-1}}\cdot \frac{1}{\epsilon(m_\phi,m_X,c\tau_X)}
\end{equation}
This analysis is only effective for $m_\phi \gtrsim 1$ beam-pipe due to the substantial $H_T$ requirement, and is correspondingly only sensitive to larger values of $m_X$.
\end{itemize}
The above searches provide a quite extensive coverage in the $X$ lifetime. We refer to Ref.~\cite{Alipour-fard:2018mre} for a careful explanation and validation of the recasting. For the projection at HL-LHC we follow the procedure of Ref.~\cite{Thamm:2015zwa,Buttazzo:2015bka} for visible searches. As far as displaced searches are concerned we rescale the bounds linearly with the luminosity, assuming their background to remain constant at higher luminosity. This extrapolation is probably optimistic, however the new challenges to control the backgrounds for LLP searches at high luminosity could be compensated  by future hardware and trigger improvements as proposed for example in \cite{Gershtein:2017tsv,Liu:2018wte}.

In principle there are three branching ratios that determine the relative contribution of displaced searches: the branching ratio into prompt or ``visible'' final states, $\text{BR}_{\text{visible}}$; the branching ratio into long-lived or ``displaced'' final states, $\text{BR}_{\text{displaced}}$, and an additional branching ratio into detector-stable or ``invisible'' final states, $\text{BR}_{\text{invisible}}$. In Fig.~\ref{fig:singlet} we fix a representative values of LLP mass $m_X$ and of the proper lifetime $\tau_X$ (indicated in the plot) and we assumed $ \text{BR}_{\text{displaced}} \simeq \text{BR}_{\phi \rightarrow ZZ}$ and $\text{BR}_{\text{invisible}} = 0$. 

From Fig.~\ref{fig:singlet} we see that in the absence of singlet Higgs decays into LLPs, the sensitivity of direct searches at the HL-LHC is unlikely to surpass limits from Higgs coupling measurements for $m_\phi \gtrsim 1.5$ beam-pipe. However, for singlet Higgses decaying partly into LLPs, the potentially considerable reach of searches for displaced decays makes a direct search program competitive with Higgs coupling measurement to much higher values of $m_\phi$. The primary weakness of the displaced searches is at high $m_\phi$, low $m_X$, and large $c \tau$, where the muon RoI search loses sensitivity. Optimal coverage of this region could in principle be provided by MATHUSLA \cite{Curtin:2018mvb}.

\begin{figure}[h!]
\centering
\includegraphics[width=.8\textwidth]{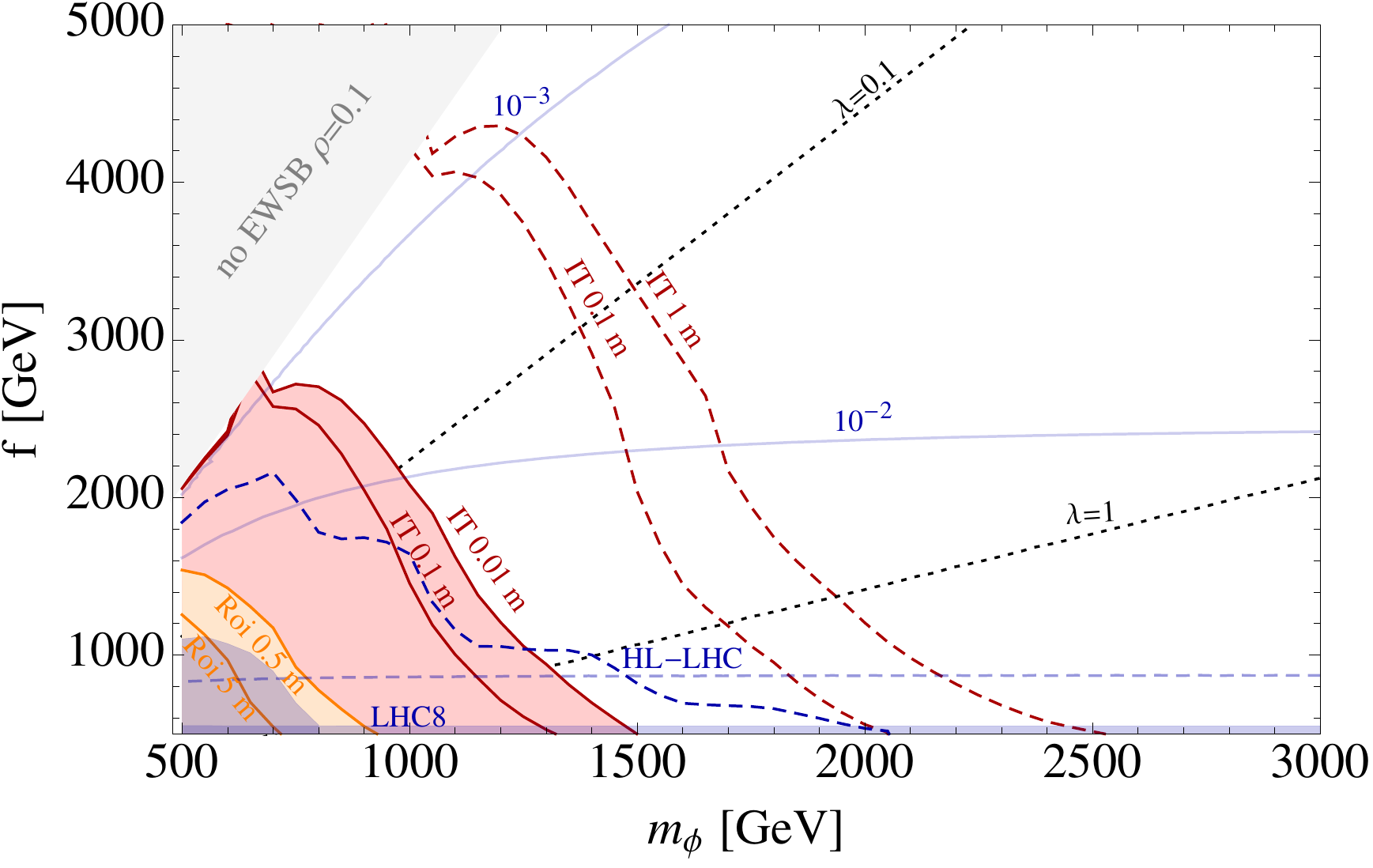}
\caption{Parameter space of the Fraternal Twin Higgs model as a function of the Twin Higgs mass, $m_\phi$, and $f$ overlaid with current and projected constraints from direct searches. See text for details.
\label{fig:twinmoney}
}
\end{figure}

\subsubsection{A specific realisation: the Twin Higgs model}
In all the Twin Higgs models the SM Higgs sector is extended by adding the twin Higgs $H_B$ which is a doublet under a mirror EW gauge group $SU(2)_B$ and a singlet under the SM gauge group. The most general renormalisable potential reads 
\begin{equation}
V=\lambda\left(\vert H_A\vert^2+\vert H_B\vert^2\right)^2-m^2\left(\vert H_A\vert^2+\vert H_B\vert^2\right)+\kappa\left(\vert H_A\vert^4+\vert H_B\vert^4\right)+\tilde{\mu}^2\vert H_A\vert^2+\rho \vert H_A\vert^4,\label{eq:V_TH}
\end{equation}
where $\lambda$ and $m^2 (>0)$ are the $SU(4)$ preserving terms, $\kappa$ preserves the $\mathbb{Z}_2$ mirror symmetry that exchanges $A\leftrightarrow B$, but breaks $SU(4)$, and $\tilde\mu$ and $\rho$ are the $\mathbb{Z}_2$ breaking terms.

The requirement to reproduce the EW scale $v$ and the Higgs mass $m_h$ fixes 2 out of the 5 free parameters in Eq.~\eqref{eq:V_TH}.
We choose the three remaining free parameters as the spontaneous breaking scale $f$, the physical singlet mass $m_\phi=4\lambda f$, and the $Z_2$-breaking quartic $\rho_{\rm}$.

In TH models the fine-tuning is parametrically reduced with respect to the ones of regular SUSY or Composite Higgs scenarios by $\lambda_{\text{SM}}/\lambda$, where $\lambda_{\text{SM}}\simeq 0.13$, see e.g.\ Ref.~\cite{Katz:2016wtw,Contino:2017moj}. In models where the $Z_2$-breaking is mostly achieved by the quartic $\rho_{\rm hard}$ the additional gain in fine-tuning is given by $\lambda_H/\vert\lambda_H-\rho_{\rm}\vert$, which is maximised for $\rho$ as close as possible to the SM quartic. This gain is however limited by the irreducible IR contributions to $\kappa$, as discussed in Ref.~\cite{Katz:2016wtw}. 
In Fig. \ref{fig:twinmoney}, we show the status of a representative slicing of parameter space of the Fraternal Twin Higgs model. We refer the reader to Ref.~\cite{Alipour-fard:2018mre} for details on the calculation of the Twin Higgs rates into visible and displaced final states. As a simplifying assumption the glueball final states are estimated by the LLP pair-production simplified model for the purposes of illustrating the potential reach of LLP searches. For each point in the figure, the mass of the lightest glueball is fixed to 50 \UGeV and a specific values of $c \tau$ is assumed to highlight the sensitivity of the different LLP searches. Very much in the spirit of the Fraternal Twin Higgs \cite{Craig:2015pha} a fixed glueball and $c \tau$ can be obtained by varying the value of the dark QCD coupling and the the one of the dark bottom Yukawa affecting quite mildly the fine-tuning   of the Twin Higgs construction. The quartic $\rho = 0.1$ is chosen here because it leads to a broader parameter space with successful EWSB for a light twin Higgs mass compared to $\rho = 0$. For positive $\rho$ the rate for $\phi \rightarrow hh$ is enhanced compared to the case of $\rho = 0$, so that limits from prompt decays in current data and HL-LHC projections (in blue) are driven by $\phi \rightarrow hh \rightarrow 4b$. 

The 13 beam-pipe ATLAS muon RoI search and our extrapolation to 13 beam-pipe, 36 fb$^{-1}$ of the 8 beam-pipe CMS inner tracker search \cite{CMS:2014wda} suggest that LLP searches with current 13 beam-pipe LHC data have the potential to provide broad coverage of the parameter space for Twin Higgs masses up to $\sim 1.5$ beam-pipe. Suitable searches at the HL-LHC could potentially extend coverage to masses of order $\sim 2.5$ beam-pipe, significantly exceeding the reach of searches for prompt decay products of the Twin Higgs and the sensitivity of Higgs coupling deviations. Of course the coverage of direct and displaced searches is quite sensitive to varying the lightest glueball mass and lifetime and more work is required to map out this parameter space completely.

\asubsection[C. Bautista, L. de Lima, R.D. Matheus, E. Pont\'on, L.A.F. do Prado, A. Savoy-Navarro]{Production of $t\bar{t}h$ and $t\bar{t}h h$ at the LHC in Composite Higgs models}\label{Sec:9.7}
\label{sec9:MCHMtthh}
With the discovery of the Higgs boson~\cite{Aad:2012tfa,
CMSHiggsJuly2012} the question of whether this resonance is a
composite state has gained new prominence.  We consider the effects of
Higgs compositeness~\cite{Kaplan:1983fs, Kaplan:1983sm, Georgi:1984ef, Georgi:1984af, Dugan:1984hq} on the $t\bar{t}h$ and
$t\bar{t}  h h$ processes.  The first process has already been
observed~\cite{Aaboud:2018urx, Sirunyan:2018hoz}, and is consistent
with the SM expectation, although with large uncertainties of order
20\%.  The second process is of particular interest, due to the contribution of charge 2/3 vector-like ``top partners" decaying in the $tH$ channel. Searches focusing on
this channel have been presented in~\cite{Aaboud:2018xuw}, and
combined ones that consider the $bW$, $tZ$ and $tH$ channels
already put strong constraints on such vector-like
resonances~\cite{Aaboud:2018pii, Sirunyan:2018omb}.

In this work, we point out the non-resonant ${t\bar t}hh$ process is of considerable interest, since in light of these strong bounds it will very often account for a large fraction of the total ${t\bar t}hh$ cross-section. Furthermore, it carries
information about the compositeness nature of the Higgs boson that is distinct and complimentary to the effect of the heavy fermion resonances. We also point out that the non-resonant
${t\bar t}hh$ process is closely connected to ${\bar t}th$, but would be expected to display larger deviations from the
SM prediction.  We present here a first step in the analysis of such
processes in the context of the HL-LHC and HE-LHC for the ``Minimal Composite Higgs
Models" (MCHM)~\cite{Agashe:2004rs}, in which the Higgs doublet is identified with the pseudo-Nambu-Goldstone bosons originating from the breaking $SO(5) \to SO(4)$ by new strong dynamics, with $SO(4)$ weakly gauged by the SM gauge group. We refer the
reader to the full review in~\cite{Panico:2015jxa} for complete
details on Composite Higgs.  Further details on our work
 can be found in the companion paper~\cite{MCHMtthh}.


\subsubsection*{Theoretical Framework}
\label{MCHM}
In composite Higgs models, physical states are linear superpositions of the strong sector composite resonances
and the SM-like ``elementary" states with the same quantum numbers, realising the paradigm of partial
compositeness ~\cite{KAPLAN1991259}.
We focus on the top sector,
which is the most relevant to the processes we study. (See Section~\ref{sect-illus} for a complementary study on Higgs coupling to gauge bosons).  Here we present only the essential features of the
 analysis, referring the reader to the companion
work~\cite{MCHMtthh} for further details. Two concrete
realisations of the fermionic sector are adopted. Both share an elementary sector denoted by $q_L$ and
$t_R$, transforming  as $(3,2,1/6)$ and $(3,1,2/3)$ under the SM gauge group.
%

\medskip
\noindent
\textit{\small \bf The MCHM$_5$}
\medskip

In this ``minimal" extension, one considers fermion resonances in a 
${\bf 5}$ of $SO(5)$, which splits into a $SO(4)$ 4-plet,
$\Psi_4$, with mass $M_4$, and a $SO(4)$ singlet, $\Psi_1$, with mass $M_1$.
%
\be
\Psi_4 ~\sim~ (X_{5/3}, X_{2/3}, T, B)~;~   \Psi_1 ~\sim~ \tilde{T}~.
\label{comp5content}
\ee

 The states $(X_{5/3}, X_{2/3})$ transform as a $SU(2)_L$ doublet with $Y = 7/6$, while $(T, B)$ and $\tilde{T}$ transform like $q_L$ and $t_R$, respectively. These are not mass eigen-states due to the mixing with elementary states, described here by ~\footnote{In principle, one can
choose different Yukawa couplings for the terms involving the ${\bf 4}$-plet, $\Psi_4$, and the singlet, $\Psi_1$.
See~\cite{MCHMtthh}.}
%
%
%
\bea
\mathcal{L}^{\mathbf{5}}_{\rm mix} &=&
y_L f \, \bar{q}_L^{\mathbf{5}} U \left[ \Psi_{\mathbf{4}} + \Psi_{\mathbf{1}} \right] 
+ y_R f \, \bar{t}_R^{\mathbf{5}} U \left[ \Psi_{\mathbf{4}} + \Psi_{\mathbf{1}} \right] + \mathrm{h.c.}
\label{Lmix5}
\eea
where $U$ parametrises the Higgs field and $f$ is the ``Higgs decay
constant".  All the features required for our analysis follow from diagonalisation of the charge 2/3 fermion mass matrix, which is given in Ref.~\cite{MCHMtthh}, while the
 remaining resonances have masses $M_{X_{5/3}} = M_4$ and $M_B =
\sqrt{M_4^2 + y_{L}^2 f^2}$. Deviations from a SM Higgs due to compositeness are characterised by the parameter $\xi = v^2/f^2$ (here $v = 246~{\rm \UGeV}$). Consistency with current
Higgs measurements results in $\xi\lesssim 0.1$, or $f \gtrsim 800$
\UGeV~\cite{Falkowski:2013dza, Carena:2014ria, Buchalla:2014eca,
Sanz:2017tco, Liu:2017dsz, Banerjee:2017wmg, deBlas:2018tjm}.  

%
%

\medskip
\noindent
\textit{\small \bf The MCHM$_{14}$}
\medskip

In the second scenario, the composite states span a ${\bf 14}$ of
$SO(5)$~\cite{Pomarol:2012qf, Panico:2012uw, Montull:2013mla, 
Carena:2014ria, Carmona:2014iwa, Kanemura:2016tan, Gavela:2016vte, 
Liu:2017dsz}. Under $SO(4)$, in addition to a 4-plet and a singlet, as in
Eq.~(\ref{comp5content}), we have an additional $SO(4)$ nonet:
\bea
\Psi_{\bf 9} &\sim& (U_{8/3}, U_{5/3}, U_{2/3}, Y_{5/3}, Y_{2/3}, Y_{-1/3}, Z_{2/3}, Z_{-1/3}, Z_{-4/3})~.
\eea
The $U$'s, $Y$'s and $Z$'s transform as $SU(2)_L$ triplets, with
hyper-charges $Y = 5/3$, $2/3$ and $-1/3$, respectively. The 
Lagrangian of the MCHM$_5$ is supplemented by terms
involving $\Psi_9$, whose mass is denoted $M_9$, and which mixes with the elementary states in an analogous manner to $\Psi_1$ and $\Psi_4$.
%
%
%
%
We give the full charge $2/3$ and $-1/3$ mass matrices as well as the complete Lagrangian in Ref.~\cite{MCHMtthh}. The remaining states have masses
$M_{X_{5/3}} = M_4$, $M_{U_{8/3}} = M_{U_{5/3}} = M_{Y_{5/3}} =
M_{Z_{-4/3}} = M_9$.

%
%


An important distinction between the two scenarios is that when the
mixing is dominated by the nonet, the leading order operator coupling
the top quark to the Higgs doublet is the non-renormalisable operator
${\bar q}_L \tilde{H} t_R H^\dagger H$.  In contrast, mixing through a
4-plet or singlet lead to the SM operator ${\bar q}_L \tilde{H} t_R$
(plus corrections that are higher order in $v/f$).  In the former case
the ratio of the top Yukawa coupling to the top mass is three times
larger than in the second case.  Cases where the nonet plays a
comparable role to the 4-plet or singlet can then lead to interesting
enhancements in the top Yukawa coupling, which are not present in the
MCHM$_5$.

The scenarios under consideration can also affect the Higgs decays. Once the light fermion representations are chosen, and assuming their mixing angles are small, one can express the partial widths as
a rescaling of the corresponding SM widths. For further details, we refer to ~\cite{MCHMtthh}.
\subsubsection*{Parameter Space and Results}
\label{analysis}
Taking all parameters to be real for simplicity, the free parameters can be taken to be $f$, $|M_1|$, $|M_4|$, ${\rm sign}(M_1)$, $y_L$ and $y_R$, common for both models, plus $|M_9|$ and ${\rm sign}(M_4)$ for the MCHM$_{14}$.
Out of these, we choose to fix $y_R$ to reproduce the top mass. Running of this mass from the scale of the resonances, typically around $2-3$ \UTeV to the relevant scales for $t\bar{t}hh$ of the order of a
couple hundred \UGeV is taken into account to a first approximation by using a running top mass of $\bar{m}_t = 150$ \UGeV for the diagonalisation of the mass
matrix, and of $m_t = 173$ \UGeV, for the kinematic quantities.   We take the Higgs mass as an independent parameter,
referring the reader to~\cite{MCHMtthh} for further discussion on this point.

We consider the following ranges for the parameters (those common to both models take the same range):
\begin{align*}
|M_1|& \in [800, 3000]~{\rm \UGeV}~, &  |M_4| &\in [1200, 3000]~{\rm \UGeV}~,  &  M_9 &\in [1300, 4000]~{\rm \UGeV}~,  \\
f&\in [800, 2000]~{\rm \UGeV}~,          &  y_L&\in [0.5, 3] ~.
\end{align*}
%
We take $y_L < 3$ and check that $y_R < 4$, in order to remain in the (semi-) perturbative regime.
%
\begin{figure}[t]
\centering
\includegraphics[width=0.475\textwidth]{\main/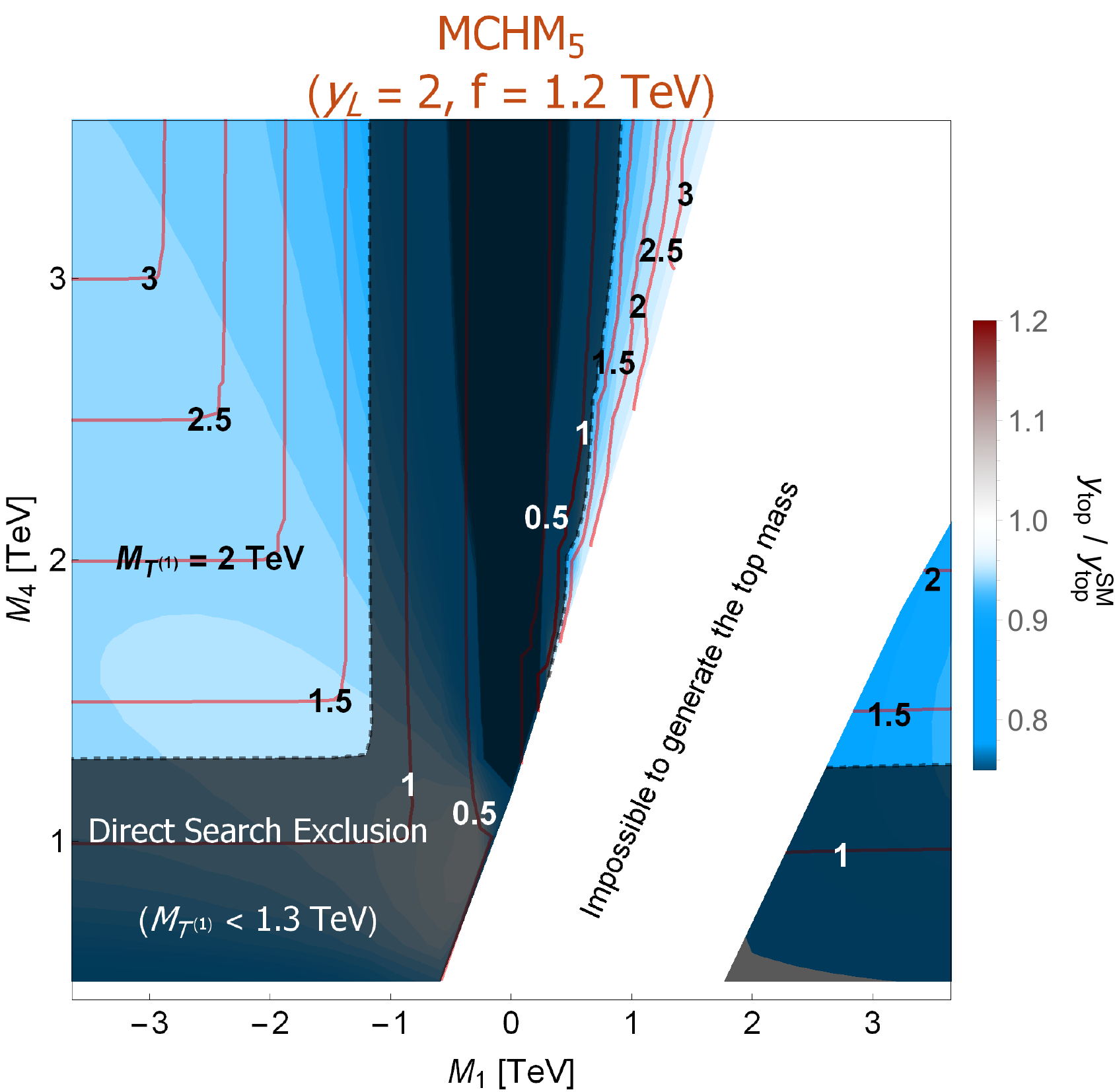}
\hspace{3mm}
\includegraphics[width=0.485\textwidth]{\main/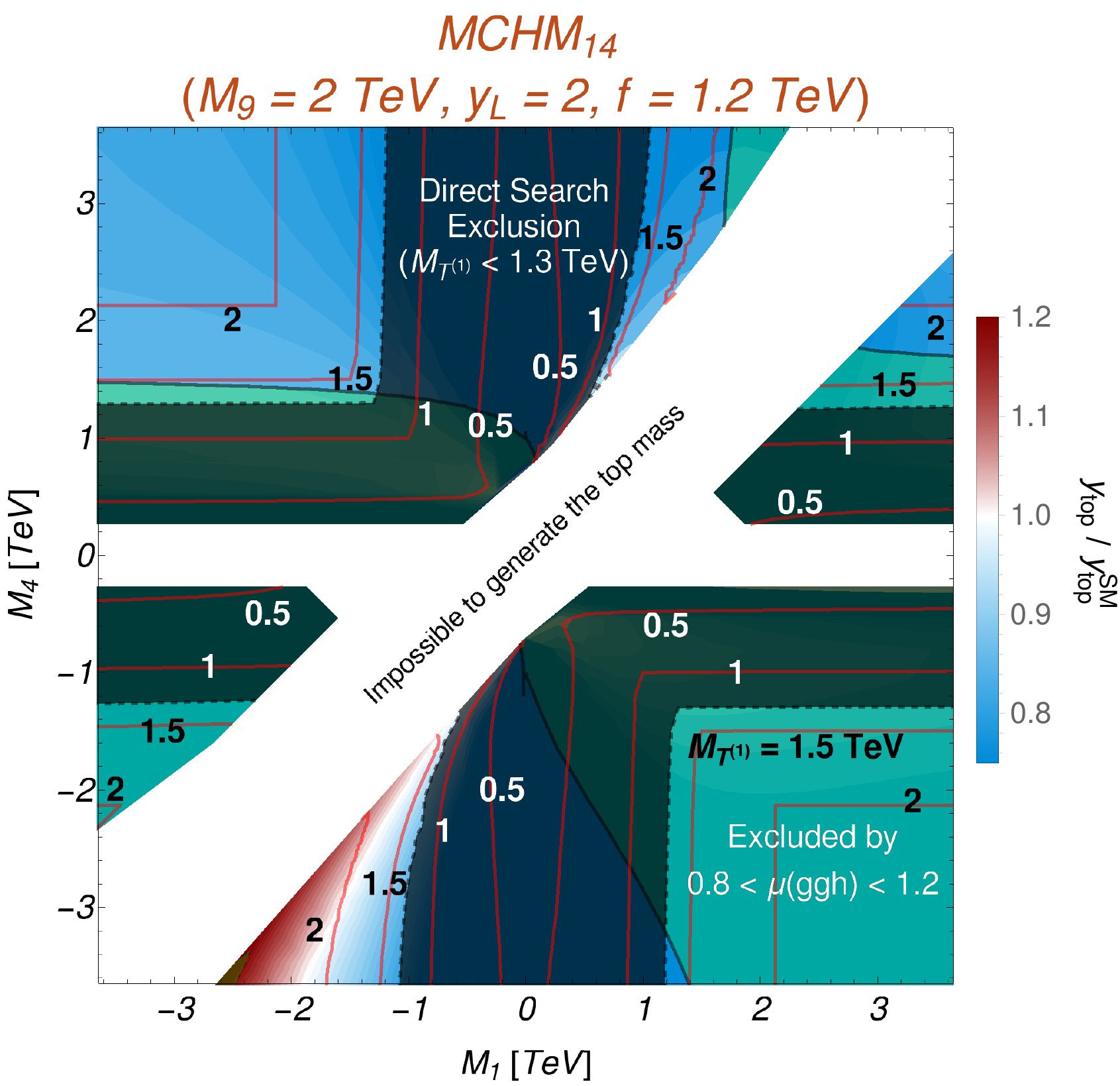}
\caption{Display of the values of the normalised top Yukawa coupling,
$y_{\rm top} / y_{\rm top}^{\rm SM}$, in the $M_1$-$M_4$ plane.  Blue
colours indicate a suppression and red colours an enhancement.  Also
shown the curves of constant $M_{T^{(1)}}$, the mass of the lightest $Q=2/3$
vector-like resonance.  The darker bands indicate the approximate
current direct exclusion of top partner VLQ resonances, assuming decays into
$bW$, $tZ$ and $tH$ ~\cite{Aaboud:2018pii, Sirunyan:2018omb}.}
\label{fig:ytvsM1M4}
\end{figure}
In Fig.~\ref{fig:ytvsM1M4} we show the normalised top Yukawa coupling,
$y_{\rm top} / y_{\rm top}^{\rm SM}$, in the $M_1$-$M_4$ plane for
both the MCHM$_5$ and MCHM$_{14}$ scenarios.  We fix $y_L = 2$ and $f
= 1200$~\UGeV, and $M_9 = 2$~\UTeV for the MCHM$_{14}$.  In the MCHM$_5$,
the scaling with $f$ is, to first approximation, given by the function
$(1-2\xi)/\sqrt{1-\xi}$, while for the MCHM$_{14}$ it is
intertwined with the other parameters in a more complicated way.  We
see that the MCHM$_5$ always displays a suppression of the top Yukawa
coupling compared to the SM limit, while the MCHM$_{14}$ can display
an enhancement in certain regions of parameter space, as pointed out
in~\cite{Liu:2017dsz}.  We also show in the figure, curves of constant
$M_{T^{(1)}}$ (red lines) and the approximate direct exclusion region
(dark bands).  The white area corresponds to the region in parameter
space where it is not possible to reproduce the top quark mass.  We
also show the region where the $ggh$ coupling deviates by more than
20\% from unity, as this region is expected to be in tension with the
current constraints on Higgs couplings~\cite{Khachatryan:2016vau}.

\subsubsubsection*{The $t\bar{t}h$ Process}
\label{tth}
To an excellent approximation, the $t\bar{t} h$ process in the MCHM
is related to the corresponding SM process by a simple rescaling of $\sigma_{\rm MCHM}(t\bar{t}h) ~=~ \left( y_t/y_t^{\rm SM} \right)^2 \, \sigma_{\rm SM}(t\bar{t}h)~.
\label{sigmatth}$
All the modifications due to Higgs compositeness, or mixing with
vector-like fermions, enter only through the top Yukawa coupling.
Therefore, only a modification in the total rate is expected, but not in kinematic distributions.
%
%

\subsubsubsection*{The $t\bar{t}hh$ Process}
\label{tthh}
For the $t\bar{t}hh$ process there are two qualitatively different
contributions:
\begin{enumerate}
\item Resonant processes, involving the production and decay (in the
$th$ channel) of heavy vector-like states of charge 2/3 (top partners).
\item Non-resonant processes: these are defined by the diagrams that
do not involve the production of the vector-like resonances.
\end{enumerate}
\begin{figure}[t]
\centering
\includegraphics[width=0.48\textwidth]{\main/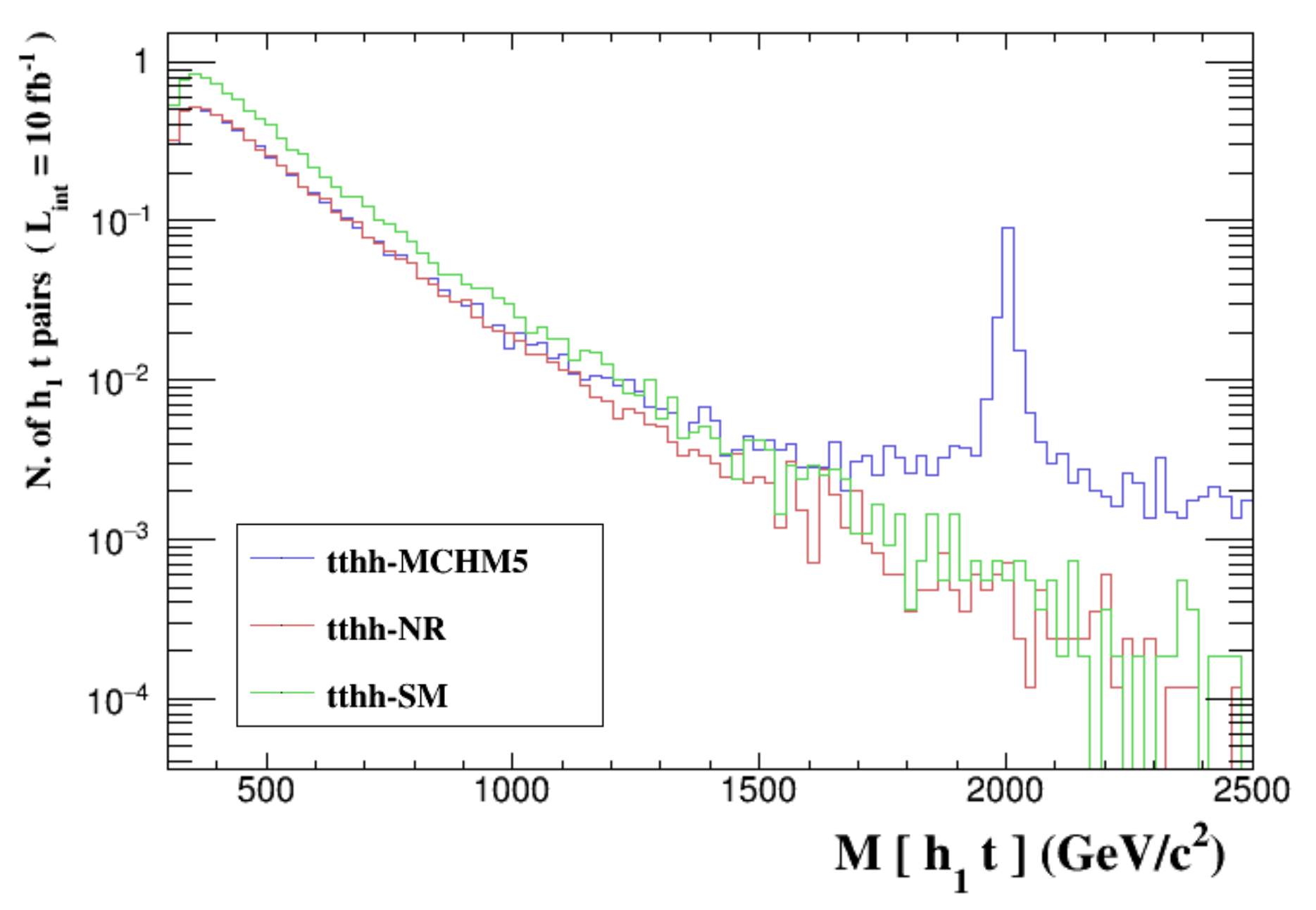}
\put(-110,35){\footnotesize14 \UTeV}
\hspace{3mm}
\includegraphics[width=0.48\textwidth]{\main/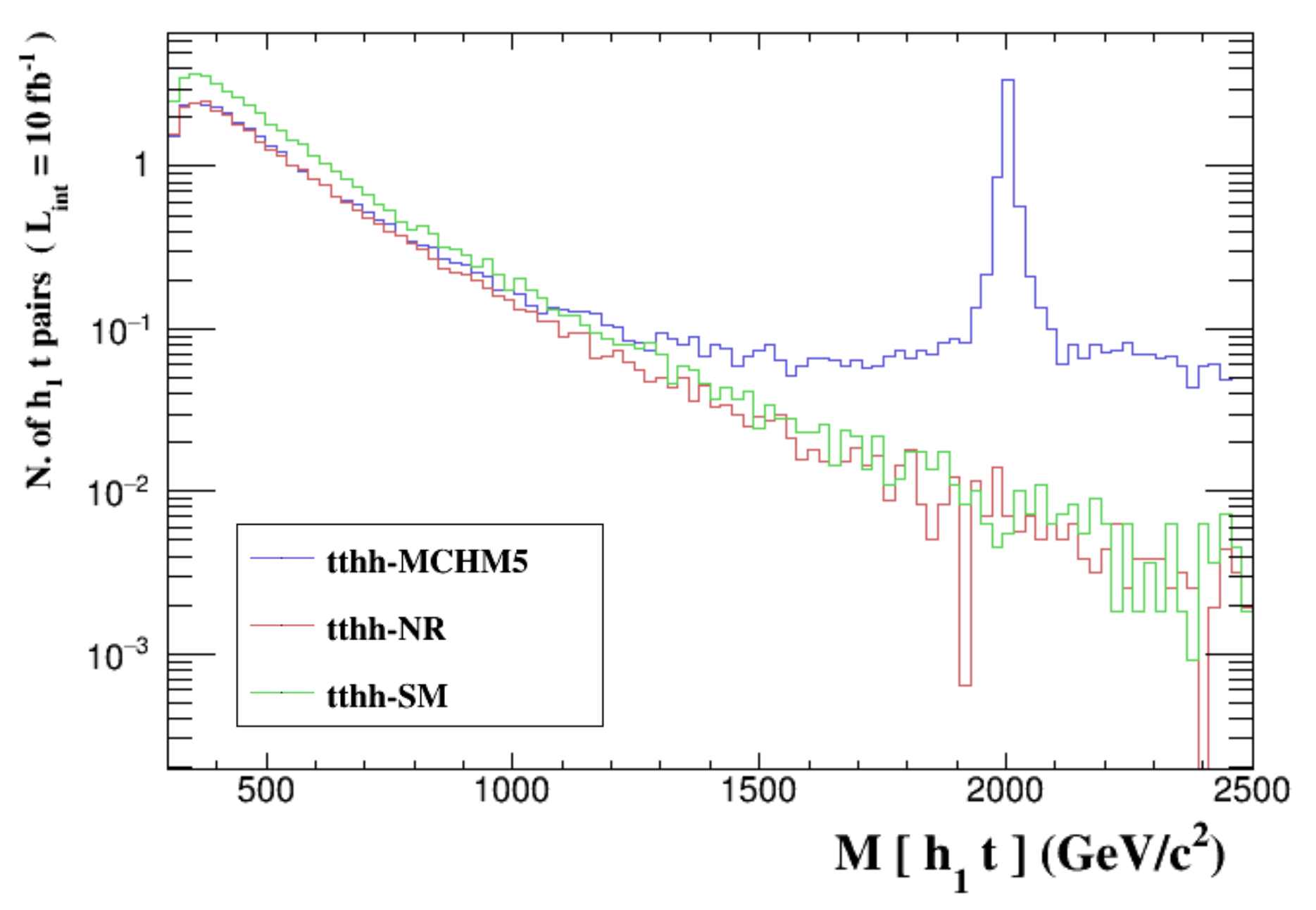}
\put(-110,35){\footnotesize 27 \UTeV} 
\caption{Distribution of the invariant mass of the top quark and the
hardest Higgs boson in the MCHM$_5$ ($M_{1} = -2500$~\UTeV, $M_4 =
2$~\UTeV, $f=1.8$~\UTeV, $y_L=1$).  The blue histogram shows the
distribution of the full $t\bar{t}hh$ process in the MCHM$_5$, while
the NR-$tthh$ cross-section is shown in red.  For comparison, we also
show in green the SM $t\bar{t}hh$ distribution.  Plots generated with
MadAnalysis~5~\cite{Conte:2012fm}.}
\label{fig:htdist}
\end{figure}
These contributions may be used to define corresponding ``non-resonant" (NR-$tthh$), and ``resonant" cross sections. The later can lead to important enhancements depending on the masses, while the former carries distinct information. We find that, to an excellent approximation, the total $t\bar{t}hh$ cross-section is given by the sum of these two cross-sections.

In Fig.~\ref{fig:htdist} we show the $ht$ invariant mass distribution
for the resonant and non-resonant processes for a particular point in
the MCHM$_5$.  For comparison, we also show the SM $t\bar{t}hh$
cross-section.  We see that the NR-$tthh$ follows the SM cross-section,
but displays a suppression.  We also see that the relative importance
of the resonant process w.r.t.~the non-resonant one increases with
larger c.m. energies.  The cross-section for both processes also
increases significantly with the c.m. energy (by a factor of 7 in the
total $t\bar{t}hh$ cross-section when going from 14 to 27~\UTeV, and by
a factor of 5 when restricted to NR-$tthh$). 

\subsubsubsection*{The Non-Resonant $t\bar{t}hh$ Process}
\label{NRtth}
The diagrams in the MCHM scenarios contributing to the NR-$tthh$ process
fall into three categories:
\begin{enumerate}
\item Those that involve only the $ttH$ vertex.
\item Those that involve the trilinear Higgs self-interaction (Section~\ref{sec3}):
$\lambda = \left[ (1 - 2 \xi)/\sqrt{1 - \xi} \right] \lambda_{\rm
SM}$.
\item Those that involve the $ttHH$ vertex (``double Higgs" Yukawa vertex).
\end{enumerate}
The first two categories correspond to sets of \textit{diagrams} that
are identical to those in the SM. The third type involves diagrams
that have no counterpart in the SM~\cite{Contino:2012xk}.  The latter
is closely connected to the Higgs compositeness aspect of the MCHM
scenarios, and it would therefore be extremely interesting if one
could get information about such effects experimentally.

By turning off in turn the double Higgs and the trilinear coupling, we find that the effects of the former are
typically at the couple to few percent level in MCHM$_5$ and MCHM$_{14}$ if the $t\bar{t}h$ signal strength,
$\mu(t\bar{t}h) \equiv \sigma(t\bar{t}h)/{\sigma(t\bar{t}h)}_{SM} < 1$, 
and at most $2\%$ in MCHM$_{14}$ if $\mu(t\bar{t}h) > 1$, with a mild dependence
on the c.m. energy (at~14 and 27~\UTeV) in all cases, while the later contributes around $15\%$ in MCHM$_5$ and MCHM$_{14}$ if $\mu(t\bar{t}h) < 1$, and $10\%$ in MCHM$_{14}$ if $\mu(t\bar{t}h) > 1$ at a c.m. energy of 14~\UTeV, decreasing to a few
percent at higher c.m. energies in all cases.  For comparison, the trilinear coupling in the SM $t\bar{t}hh$ cross-section
contributes about $20\%$, with a very mild c.m. energy dependence.  Thus, the NR-$tthh$ is largely determined
by the top Yukawa, being related to the SM process, to a first approximation, by a scaling factor $(y_t / y_t^ {\rm SM})^4$.
This explains the result seen in Fig.~\ref{fig:htdist}, with the
suppression arising from the suppression of the top Yukawa coupling in
the MCHM$_5$.


The previous observation also leads to a strong correlation between the $t\bar{t}h$ and the NR-$tthh$ processes, as shown in Fig.~\ref{fig:nrtthhvstth}. Due to the different scaling with the top Yukawa coupling, the deviations from the SM in the NR-$tthh$ process are larger than those in $t\bar{t}h$.
\subsubsubsection*{Set of Example Points}
\label{benchmarks}
We show in Table \ref{fig:benchmarkTable1} 
a number of points selected as examples that illustrate, in
more detail, the properties of the MCHM$_5$ and MCHM$_{14}$.  
These properties are reflected in Figs.~\ref{fig:nrtthhvstth}, ~\ref{fig:tthvsf} and~\ref{fig:tthhvsMT4}, where these points are indicated. The MCHM$_5$ points are labelled as P$_i$, i=1 to 5, and MCHM$_{14}$ points as P'$_j$, with j=1 to 4.
\begin{table}[t]
\centering
\includegraphics[width=0.85\textwidth]{\main/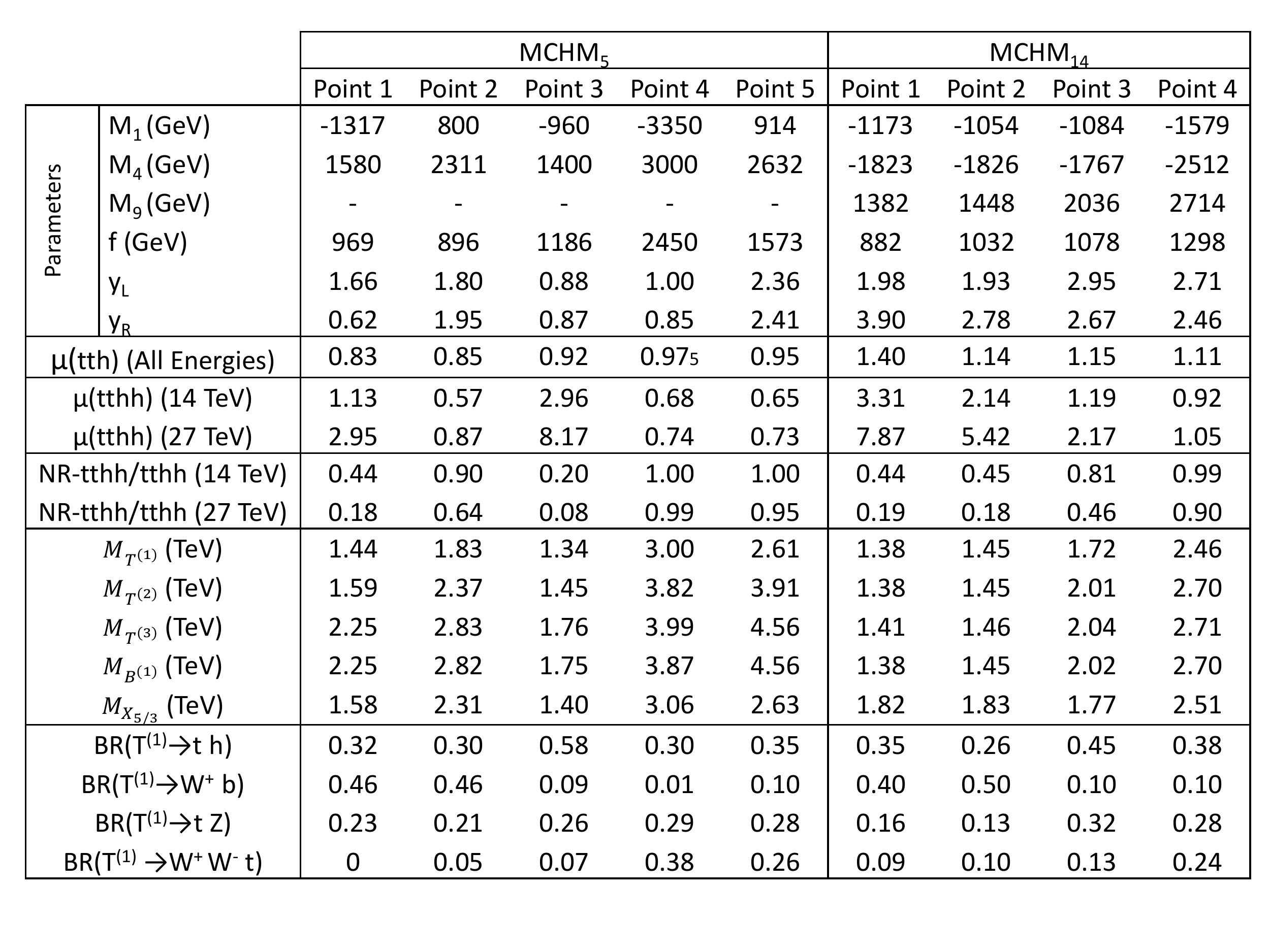}
\caption{Sample points for MCHM$_5$ with M$_1$ M$_4$ same sign and opposite sign and for MCHM$_{14}$ with M$_1$ and M$_4$ both $<0$ and $\mu({\rm ttH})>1$.}
\label{fig:benchmarkTable1}
\end{table}
\begin{figure}[t]
\centering
\includegraphics[width=0.4\textwidth]{\main/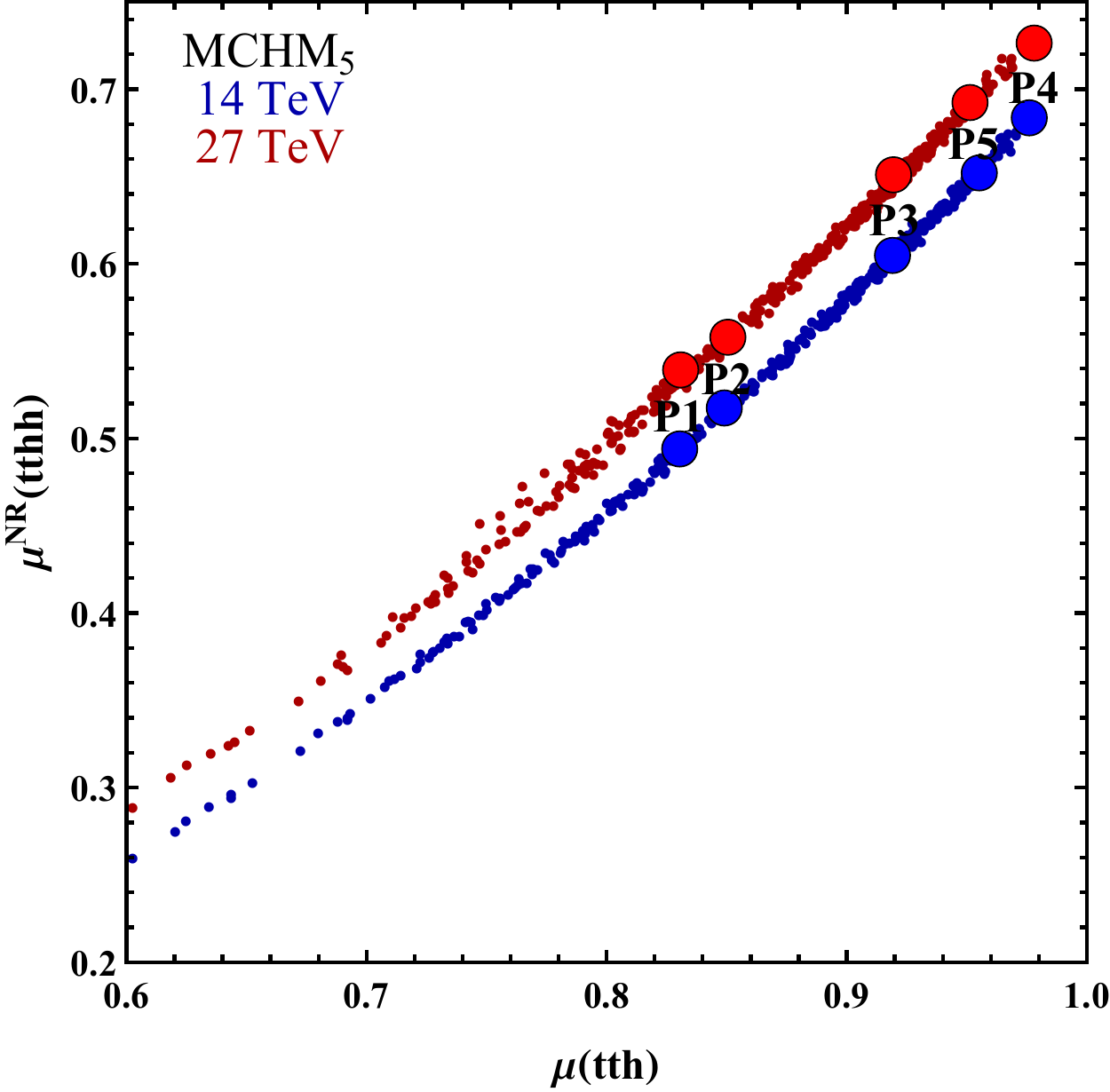}
\hspace{1.5cm}
\includegraphics[width=0.4\textwidth]{\main/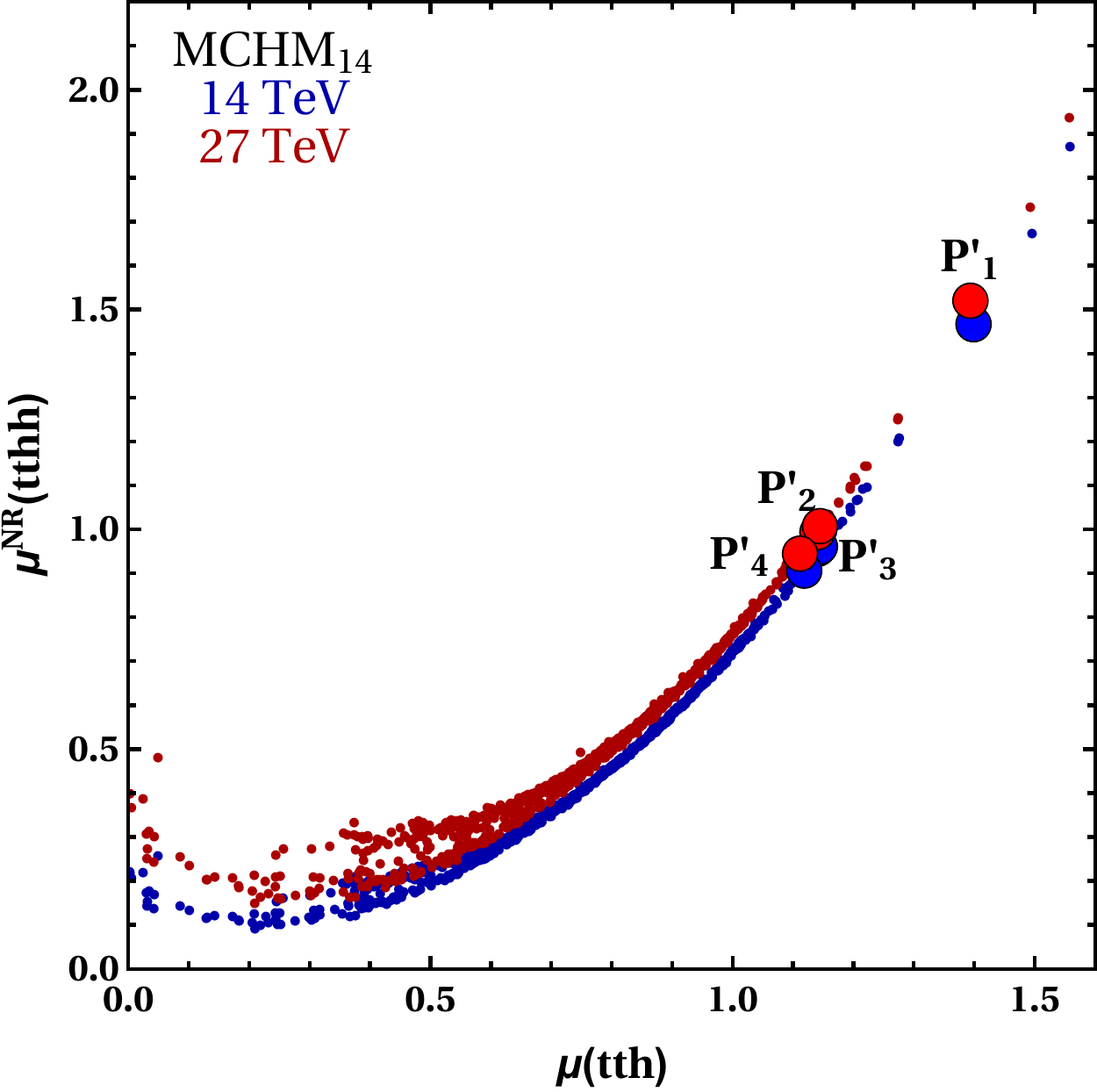}
\caption{Correlation between the $t\bar{t}h$ and
non-resonant $t\bar{t}hh$ signal strengths ($\mu$), for 14 and 27~\UTeV c.m.~energies. The left (right) plots correspond to the MCHM$_5$ (MCHM$_{14}$)}
\label{fig:nrtthhvstth}
\end{figure}
The points for the MCHM$_5$ exhibit a suppression in $\mu(t\bar{t}h)$ that ranges from about 15\% (roughly at the current
95\% C.L.~limit~\cite{Aaboud:2018urx, Sirunyan:2018hoz}) to a few
percent, a sensitivity that might be achievable by the end of the HL
phase of the LHC run, with smaller deviations from the SM for larger values of $f$ (Fig~\ref{fig:tthvsf},a). The Table ~\ref{fig:benchmarkTable1} and Fig~\ref{fig:tthhvsMT4} show that the
$t\bar{t}hh$ process can exhibit an enhancement for light enough
resonances, increasing with higher c.m.~energy, as expected. For points 2, 4 and 5 in the MCHM$_{5}$, the resonant production is not enough to produce an
enhancement in $t\bar{t}hh$ compared to the SM, although these points correspond to two different cases; the resonances for Point 2 are slightly beyond the current direct limit whereas, on the contrary, much beyond that limit for points 4 and 5. In this case, the
$t\bar{t}hh$ process is easily dominated by the NR-$tthh$ process, as
defined above.  

The set of example points for MCHM$_{14}$ in Table~\ref{fig:benchmarkTable1} exhibits an enhancement of the top Yukawa
coupling, due to the effect described above and reflected in Fig~\ref{fig:tthvsf},b.  These
enhancements can easily be of the order of 10-20\%.  Interestingly,
Point 1 shows that the enhancement can be as large as 40\% (while
being consistent with a sufficiently small deviation in the $ggh$
vertex~\cite{MCHMtthh}). The four points display as well, an enhancement
in the $t\bar{t}hh$ process.  While about half of the rate is due to
resonant production in Points 1 and 2, for points 3 and 4 the
enhancement arises dominantly from the non-resonant process,
reflecting the enhancement in the top Yukawa coupling. All the selected points for MCHM$_{14}$ lie in the
$M_1 < 0$, $M_4 < 0$ quadrant of the right panel of
Fig.~\ref{fig:ytvsM1M4}. The properties of the other quadrants are
qualitatively rather similar to those of the MCHM$_5$ (see~\cite{MCHMtthh}).

For completeness, Table~\ref{fig:benchmarkTable1} includes the
spectrum of the 5 resonances in the MCHM$_{5}$, and of the 3 lightest 2/3 resonances, the lightest B resonance and the lightest 5/3 resonance out of the total of 14 resonances of the MCHM$_{14}$, as well as the BRs for the lightest $Q=2/3$ one.  It
decays mostly into the standard $th$, $Wb$ and $tZ$ channels (with BRs
that are model dependent), but in some cases it has non-negligible
non-standard BRs, such as into the $W^+W^- t$ channel.  
\begin{figure}[!htb]
\centering
\includegraphics[width=0.4\textwidth]{\main/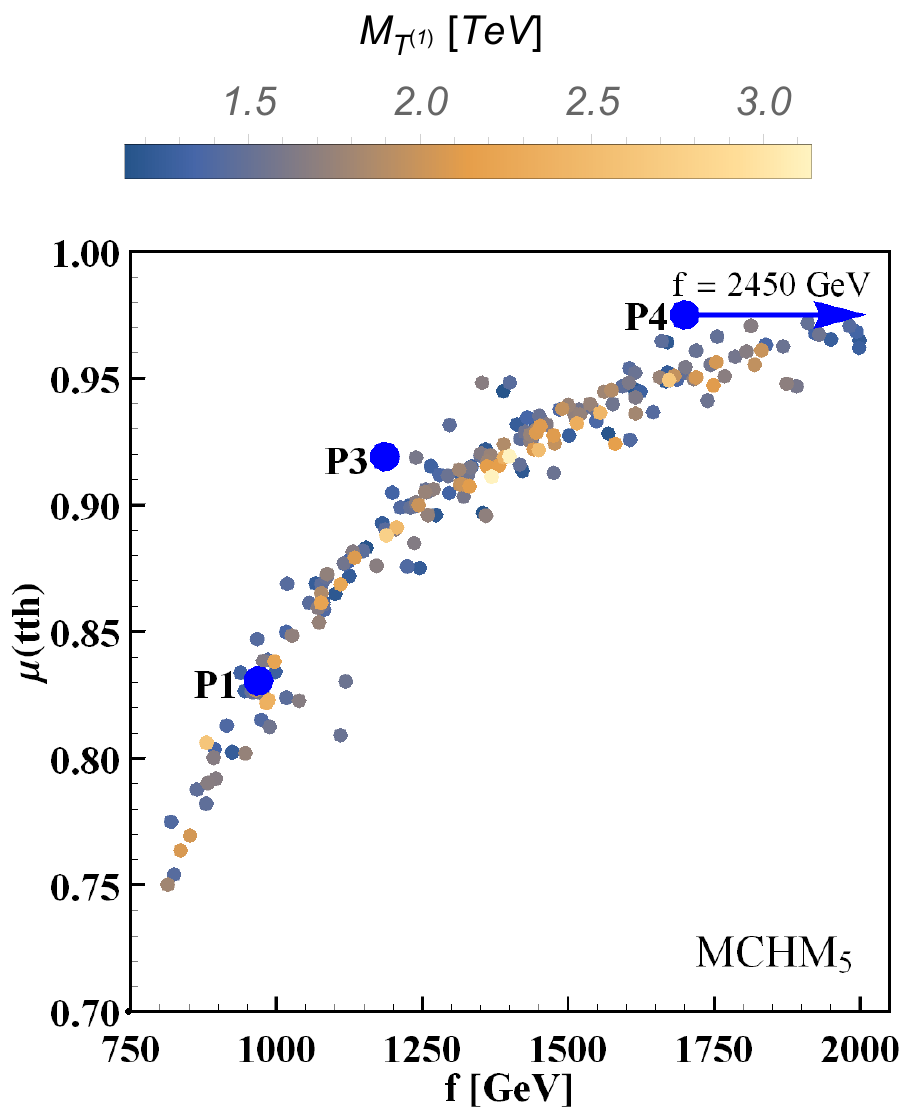}
\put(-35,45){a)}
\hspace{1.5cm}
\includegraphics[width=0.417\textwidth,scale=1.2]{\main/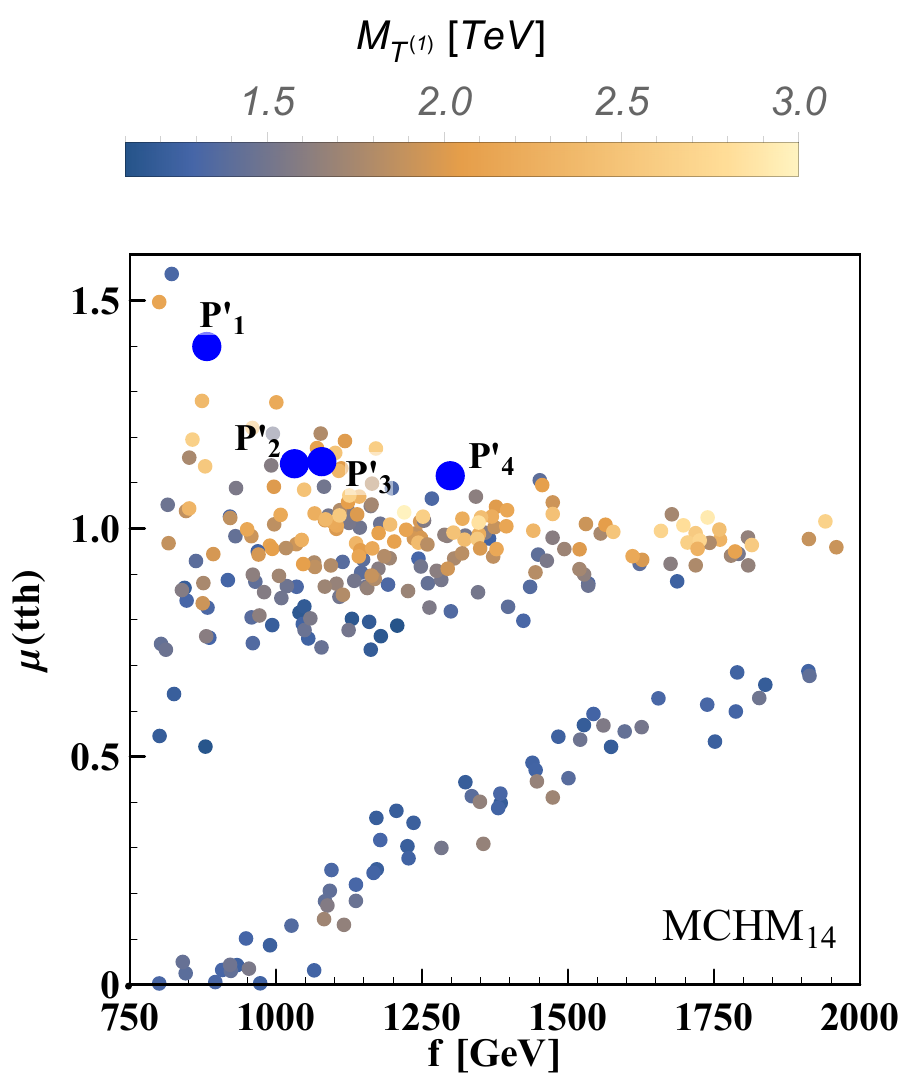}
\put(-45,45){b)}
\caption{The $t\bar{t}h$ signal  strength as a function of the $f$-scale,
for 14 and 27~\UTeV c.m. energies, with colour coded the lightest
vector-like mass.  The left (right) plots correspond to Q2 of MCHM$_5$ (Q3 of MCHM$_{14}$). The blue arrow indicates that the point P4 is outside the horizontal range of the plot with f=2450 \UGeV.}
\label{fig:tthvsf}
\end{figure}
\begin{figure}[!htb]
\centering
\includegraphics[width=0.4\textwidth]{\main/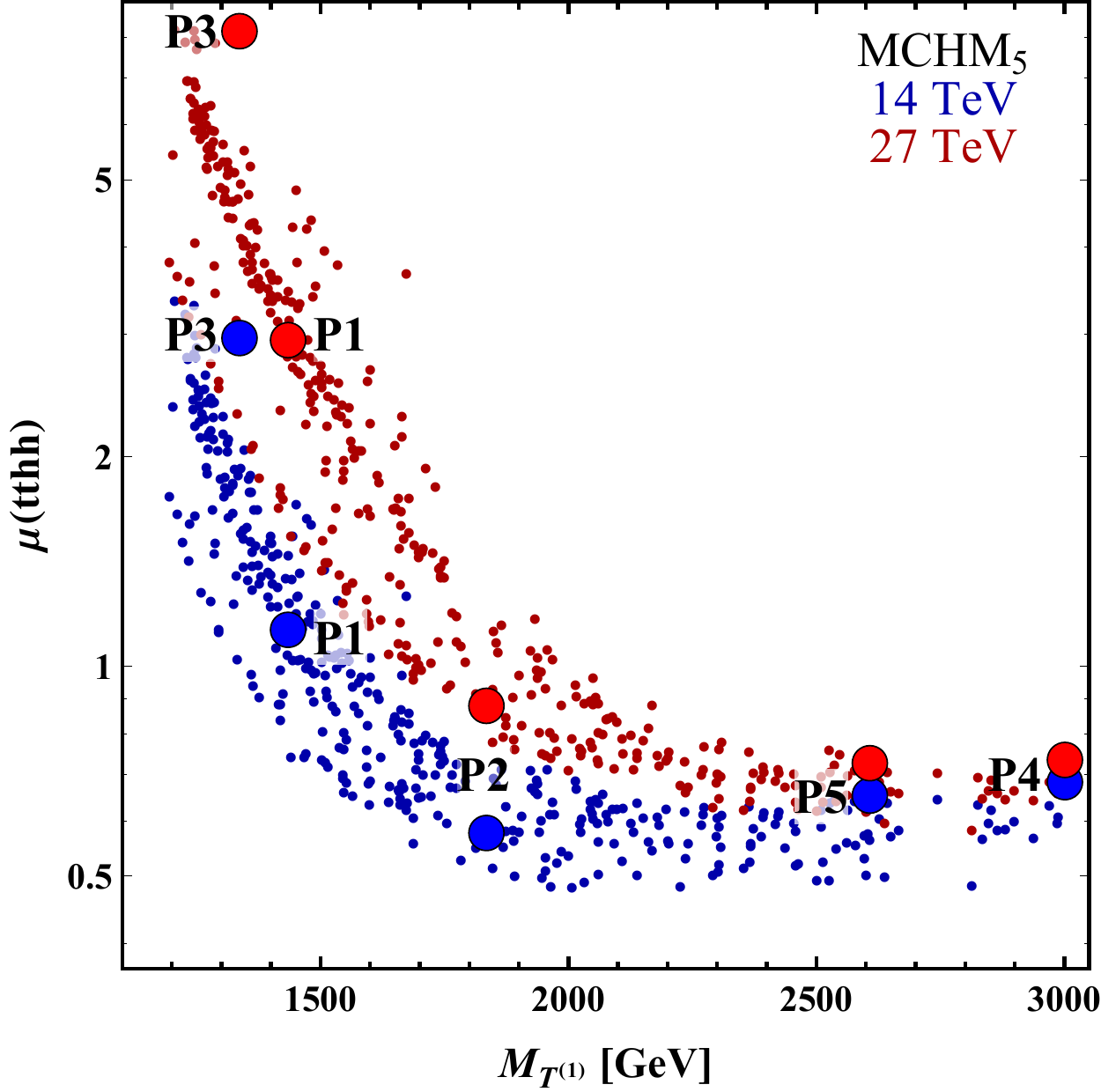}
\put(-60,171){a)}
\hspace{1.5cm}
\includegraphics[width=0.4\textwidth]{\main/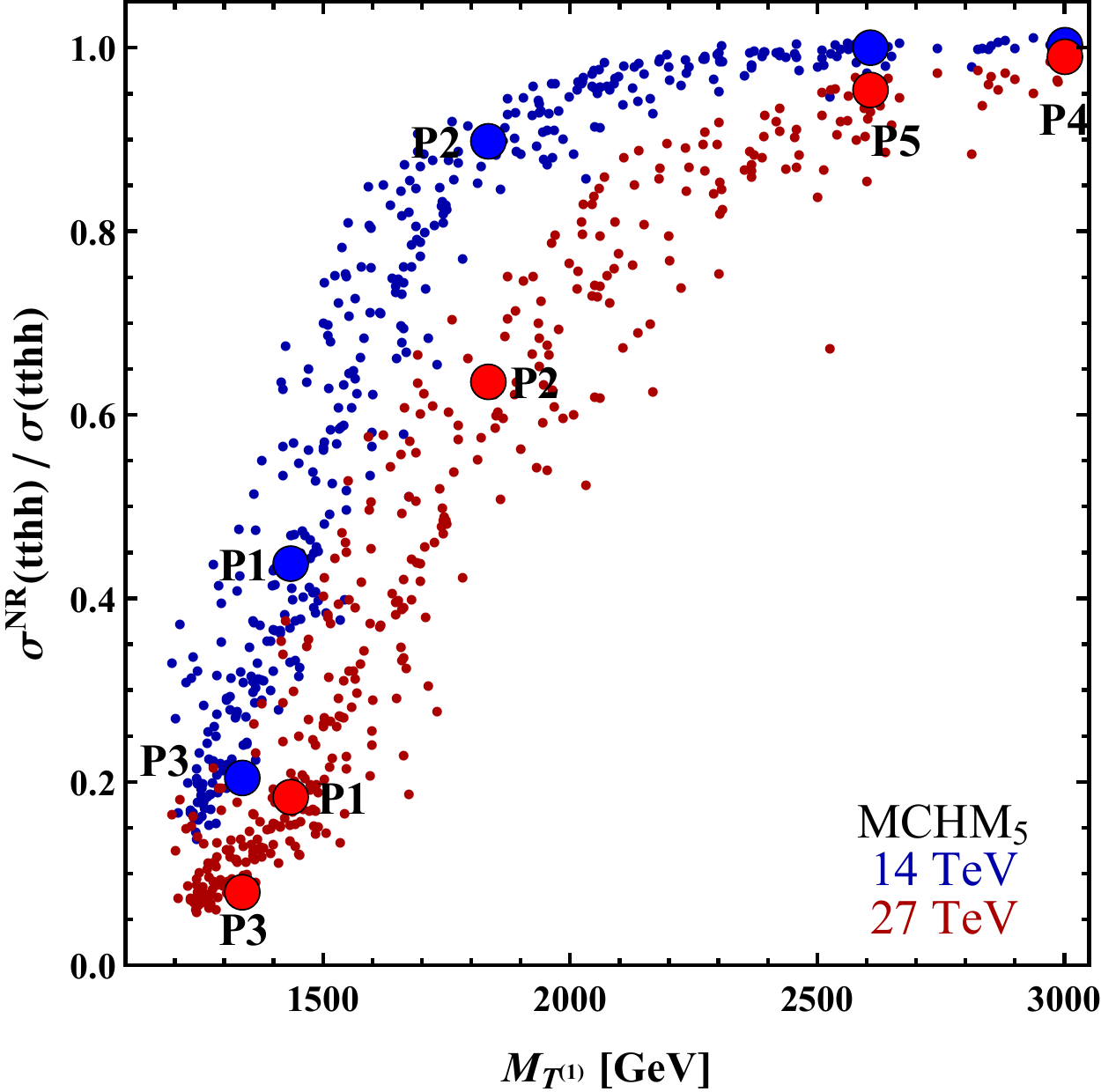}
\put(-60,45){b)}
\caption{The left plot shows the $t\bar{t}hh$ signal  strength as a function of the
lightest $Q = 2/3$ vector-like mass, T$^{(1)}$ for 14 and 27~\UTeV c.m. energies for the MCHM$_5$. The right plot shows the ratio between the non-resonant $t\bar{t}hh$ cross section and the total $t\bar{t}hh$ cross section as a function of T$^{(1)}$
for 14 and 27~\UTeV c.m. energies for the MCHM$_5$.}
\label{fig:tthhvsMT4}
\end{figure}
%
\subsubsection*{Experimental perspectives}
\label{perpectives}
%
A deviation from the SM in the $t{\bar t}h$ production is an essential measurement for MCHM. An increase will reject the MCHM$_5$ scenario and greatly refine the areas of the parameter space where MCHM$_{14}$ would be valid. A deficit instead, would make MCHM$_5$ and MCHM$_{14}$ both possible. The measurement of this observable is expected to be achieved within 5\% accuracy at the HL-LHC (Sections~\ref{sec2:top},\ref{sec2:exp_combination},\ref{sec2:exp_kappa}) and thus with very high accuracy at HE-LHC. The $t{\bar t}hh$ production process plays a major role in MCHM searches. Deviations from the SM expectation (deficit or increase) can be significant in both MCHM scenarios. The $t{\bar t}hh$ production cross-section is around 1 fb (Section~\ref{sec:HH_NLO}) at tree level whereas $t{\bar t}h$ is about 500 times larger (Section~\ref{sec2_HXSWG1}). Therefore the aim at HL-LHC will be to evidence this process and discover if a strong deviation from SM. Higher energy together with higher luminosity (HE-LHC) will further explore MCHM.

\textbf{Acknowledgements:}
This work was supported by the S\~ao Paulo Research Foundation
(FAPESP) under Grants No.~2016/01343-7, No.~2013/01907-0,
No.~2015/26624-6 and No.~2018/11505-0 and by Science Without Borders/CAPES for UNESP-SPRACE under the Grant No.~88887.116917/2016-00.


\asubsection{New Higgs bosons below the 125 \UGeV Higgs mass}\label{Sec:9.8}

\asubsubsection[S. Heinemeyer, J. Santiago, R. Vega-Morales]{Searches for low mass Higgs bosons (below 120 \UGeV)}

\asubsubsubsection{Introduction}

Many extensions of the Standard Model Higgs sector allow for new
charged and neutral Higgs bosons that can be lighter than the Higgs
boson discovered~\cite{Aad:2012tfa,Chatrchyan:2012xdj} at $\approx 125$~\UGeV. 
However, as the observed (heavier) Higgs boson shows itself to be 
increasingly SM-like~\cite{Falkowski:2013dza} in its couplings to $WW$
and $ZZ$ pairs~\cite{Khachatryan:2014kca,Khachatryan:2016vau,Sirunyan:2017exp,Sirunyan:2017tqd,Aaboud:2017oem,Falkowski:2013dza},
as well as to fermions~\cite{Aaboud:2018zhk,CMS:2018abb}, we are in
general pushed into an `alignment without decoupling'
limit~\cite{Gunion:2002zf,Carena:2013ooa}, which has been examined in a
number of recent
studies~\cite{Craig:2013hca,Carena:2014nza,Carena:2015moc,Bernon:2015wef,Profumo:2016zxo,Bechtle:2016kui,Haber:2017erd,Bahl:2018zmf}. In
this limit, the 125~\UGeV Higgs boson has SM like couplings without having
to decouple the other Higgs bosons which might be present allowing them
to be lighter than 125~\UGeV. In what follows we work in the alignment
without decoupling limit focusing on new Higgs bosons in the mass range
$65 - 120$~\UGeV, between the SM-like Higgs mass and its two body decay
threshold. 

In 2HDMs alignment occurs when one of the
neutral CP-even Higgs mass eigen-states is approximately aligned in field
space with the direction of the vacuum expectation value~\cite{Carena:2013ooa,Bernon:2015wef}. For non-doublet
electroweak multiplets (as well as singlets~\cite{Robens:2015gla}), one
obtains an `aligned' SM-like Higgs when the
non-doublet~\cite{Georgi:1985nv,Killick:2013mya} Higgs \emph{VEV} is
small, which typically also suppresses the Higgs mixing
angle~\cite{Haber:1978jt,Hartling:2014zca}. Furthermore, in the singlet
and non-doublet multiplet cases, the new Higgs bosons are (at least
approximately) fermiophobic, making them generically harder to
detect~\cite{Akeroyd:1998ui,Akeroyd:1995hg,Delgado:2016arn,Vega:2018ddp}
either directly or indirectly as we discuss more below.  

In this section we summarise the relevant experimental constraints on
light Higgs bosons in the mass range $65 - 120$~\UGeV. We also discuss
models which can realise light Higgs bosons and highlight promising
search signals at the LHC. This includes searching for deviations in
Higgs couplings since, as emphasised in~\cite{Bernon:2015wef}, even in
the deep alignment regime where one might naively expect everything to
be very SM-like, precise measurements of the 125~\UGeV Higgs boson signal
strengths could uncover the existence of an extended Higgs sector. Some
projections for the HL and HE LHC are also made. 
The aim is to encourage new experimental analysis, targeting
specifically searches for light Higgs bosons at the HL/HE-LHC.

\asubsubsubsection{Experimental constraints on light Higgs bosons}\label{sec:limits}

In the mass range and alignment limit we consider, the most relevant
constraints for the \emph{anti}-aligned neutral Higgs bosons but with
significant couplings to SM fermions, come from CMS $b\bar{b}X$ with $X
\to \tau\bar{\tau}$ searches~\cite{Khachatryan:2015baw} as well as
ATLAS~\cite{Aad:2014vgg} and CMS~\cite{Khachatryan:2014wca} searches for
$X \to \tau\bar{\tau}$ decays in both the $gg \to X$ and $b\bar{b}X$
production modes. Similarly, the searches in the di-photon channel
  place important bounds~\cite{CMS-PAS-HIG-17-013,ATLAS-CONF-2018-025}.%
\footnote{It is interesting to note that the CMS search in the di-photon
  channel~\cite{CMS-PAS-HIG-17-013} shows an excess of events at $\sim
  96$~\UGeV, in the same mass range where the LEP searches in the 
  $b \bar b$ final state observed a $2\,\sigma$ excess~\cite{Schael:2006cr}.}
A recent CMS search~\cite{CMS:2015mba} for new
resonances decaying to a $Z$ boson and a light resonance, followed by 
$Z \to \ell\bar\ell$ and the light resonance decaying to $b\bar{b}$ or
$\tau\bar{\tau}$ pairs, has also been shown to impose severe
constraints~\cite{Bernon:2015wef} on light CP-even neutral Higgs
bosons. Direct searches at LEP for light neutral Higgs states produced
in pairs or in association with a $Z$ boson are also
relevant~\cite{Barate:2003sz,Abbiendi:2004gn,Schael:2006cr},
setting relevant limits on the couplings of the light Higgs to SM
gauge bosons. For
the charged Higgs bosons, LEP searches~\cite{Abbiendi:2013hk} and
$B$-physics constraints from $R_b,\,\epsilon_K,\,\Delta m_B,\,B\to
X_s\gamma$, and
$B\to\tau\nu$~\cite{Haisch:2008ar,Mahmoudi:2009zx,Gupta:2009wn,Jung:2010ik,Misiak:2015xwa}
measurements impose the most stringent constraints. These limits apply
to all 2HDMs and impose particularly severe constraints on
\emph{non}-type-I 2HDMs~\cite{Bernon:2015wef} in which there is no
fermiophobic limit.  

As emphasised in numerous studies~\cite{Ilisie:2014hea,Enberg:2016ygw,Delgado:2016arn,Degrande:2017naf,Vega:2018ddp}, the above limits are less stringent (most limits can be rescaled) when the Higgs bosons have highly suppressed couplings to SM fermions as can happen in the type-I 2HDM~\cite{Haber:1978jt} in the large $\tan\beta$ limit~\cite{Akeroyd:1995hg}. For non-doublet extended Higgs sectors one automatically has suppressed couplings to SM fermions when the non-doublet \emph{VEV} is small (or mixing angle in the case of singlets) since they only enter (if at all) through mixing with the SM-like Higgs boson~\cite{Killick:2013mya}.~In the case of fermiophobia, the most robust probes of neutral Higgs bosons are inclusive di-photon~\cite{Delgado:2016arn,Degrande:2017naf,Vega:2018ddp} and multi-photon searches~\cite{Akeroyd:2005pr,Abdallah:2003xf,Aaltonen:2016fnw} which utilise the Drell-Yan pair production channel of a charged and neutral Higgs boson.~Constraints from EW precision data~\cite{Baak:2011ze,ALEPH:2010aa} also apply with the primary effect being that the neutral and charged Higgs bosons are constrained to be not too different in mass.  

\asubsubsubsection{Models with light Higgs bosons}\label{sec:models}

A number of recent studies of the alignment without decoupling limit in
2HDMs have been performed which consider the case where the SM-like
Higgs boson is not the lightest scalar. As shown
in~\cite{Ilisie:2014hea,Bernon:2015wef,Enberg:2016ygw,Arhrib:2017wmo,Arbey:2017gmh,Bhatia:2017ttp,Fox:2017uwr,Haber:2017erd,Haisch:2017gql},
for type-I 2HDMs there are regions of parameter space where, along with
the light CP-even scalar, both the charged and neutral CP-odd Higgs
bosons can be below the SM-like Higgs mass while satisfying the
constraints discussed above. This is in contrast to type-II 2HDM,
where combined constraints from $B$ meson
decays~\cite{Misiak:2015xwa} and EW precision
constraints~\cite{Peskin:1991sw} require the charged and CP-odd neutral
Higgs bosons to be much heavier than the mass range we consider here. Within the MSSM however, the additional particle content results in
  substantially weaker limits from $B$~meson decays and EW data.
In general the allowed regions of parameter space in the type-I 2HDM is
much larger than in other 2HDMs~\cite{Bernon:2015wef,Haber:2017erd},
again due to the presence of a fermiophobic limit at large $\tan\beta$
which opens up more regions of parameter space. 

In the MSSM which is a type-II 2HDM, the alignment without decoupling
limit~\cite{Carena:2013ooa} requires accidental cancellations between
tree level and radiative corrections in the Higgs mass
matrix~\cite{Bechtle:2016kui,Haber:2017erd}. It was shown that a tuning of $\sim 10\%$ is sufficient to find agreement with the Higgs-boson rate measurements~\cite{Bechtle:2016kui}. Depending on the level of alignment required, this can lead to a highly constrained parameter space, especially in the case where the SM-like Higgs is the heavier of the CP-even neutral scalars. In particular, after accounting for all relevant experimental constraints (as well as theoretical uncertainties) recent studies~\cite{Bahl:2018zmf} of the alignment without decoupling limit of the MSSM~\cite{Carena:2013ooa} defined a benchmark plane of allowed parameter space with $\tan\beta \sim 5 - 6$ (and very large values of $\mu$) in which the light CP-even Higgs can be between  $\sim 60 - 100$~\UGeV if the charged Higgs mass is between $\sim 170 - 185$~\UGeV and the neutral CP-odd Higgs is $\sim 130 - 140$~\UGeV. Still larger allowed regions are expected in a global scan, as performed in~\cite{Bechtle:2016kui}.
Recent studies of the NMSSM~\cite{Carena:2015moc,Domingo:2018uim} and
$\mu\nu$SSM~\cite{Biekotter:2017xmf} have also examined the alignment
without decoupling limit finding a larger allowed parameter space than in the
MSSM due to an additional gauge singlet Higgs (or right handed scalar
neutrino). 

For models with non-doublet multiplets the most well known are those involving electroweak triplets. In particular, Higgs triplet models with custodial symmetry~\cite{Sikivie:1980hm}, as in the famous Georgi-Machacek (GM) model~\cite{Georgi:1985nv,Chanowitz:1985ug,Gunion:1989ci,Gunion:1990dt,Hartling:2014zca} or its supersymmetric incarnations~\cite{Cort:2013foa,Garcia-Pepin:2014yfa,Vega:2017gkk}, have been well studied due to their ability to easily satisfy constraints from electroweak precision data. Recent studies~\cite{Davoudiasl:2004aj,Vega:2017gkk,Vega:2018ddp} have shown that GM-like models can allow for light neutral and charged scalars below the SM-like Higgs boson mass. In the alignment limit implied by Higgs coupling measurements, the triplet Higgs \emph{VEV} is constrained to be small though it can still much larger than non-custodial cases~\cite{Tanabashi:2018oca,Haber:1999zh} which are constrained by measurements of the $\rho$ parameter. Custodial symmetry also ensures that the neutral and charged components of the Higgs multiplet have (at least approximately) degenerate masses, making them more difficult to detect due to soft decay products~\cite{Buckley:2009kv,Ismail:2016zby}. For these anti-aligned and fermiophobic Higgs bosons, recent studies have emphasised di- and multi-photon searches~\cite{Aaltonen:2016fnw,Delgado:2016arn,Brooijmans:2016vro,Vega:2018ddp} as robust probes of this scenario.

\asubsubsubsection{Phenomenology of light anti-aligned Higgs bosons}\label{sec:pheno}

In the alignment limit, single electroweak production mechanisms for the
additional `anti-aligned' neutral Higgs bosons (or small \emph{VEV} and Higgs mixing for non-doublets), such as VBF or associated vector boson production, necessarily become suppressed. Thus the dominant production mechanisms become gluon fusion or associated $b\bar{b}$ production when there is a significant coupling to SM quarks. However, these production mechanisms become suppressed when the couplings to fermions are negligible\,\footnote{Of course if they couple to some not too heavy coloured BSM particles, the gluon fusion cross section can be increased.}, as can happen in type-I 2HDM in the large $\tan\beta$ limit~\cite{Akeroyd:2003bt} or non-doublet electroweak sectors which are generically fermiophobic. The same is true for the light charged Higgs bosons production channels $t \to H^\pm b$ and $pp \to H^\pm t b$ which are also obsolete in the fermiophobic limit. Note that for charged scalars coming from larger than doublet representations we can also have $W^\pm Z \to H^\pm$ VBF production, but this is again suppressed in the small non-doublet \emph{VEV} and Higgs mixing limit.

{\bf{Pair Production as a discovery channel}}\label{sec:pair}

A different option that offers new experimental opportunities is
  the Drell-Yan Higgs pair production mechanism. Any extension of the SM Higgs sector by electroweak charged scalars will possess the pair production channels mediated by $W$ and $Z$ bosons and which are not present in the SM. Furthermore, as emphasised in~\cite{Akeroyd:2003bt,Akeroyd:2003xi,Akeroyd:2003jp,Ilisie:2014hea,Delgado:2016arn,Vega:2018ddp}, even in the alignment and fermiophobic limits, this production mechanism is not suppressed and can be as large as $\sim 10\,pb$ at 13~\UTeV and $\sim 50\,pb$ and 27~\UTeV in the mass range we consider (see Sec.~\ref{sec2_HXSWG1}). Thus, Drell-Yan Higgs pair production can be as large or even dominate over single  production mechanisms, for both charged and neutral Higgs bosons. Despite this, the Drell-Yan Higgs pair production mechanism has been largely overlooked in experimental searches with the lone exception being a recent CDF analysis of Tevatron four photon data~\cite{Aaltonen:2016fnw} searching for fermiophobic Higgs bosons.  

The Drell-Yan pair production mechanism is mediated by the vector-Higgs-Higgs coupling. In the alignment limit, this will have vertices that are maximised in this limit and depend only on electroweak couplings and quantum numbers, while some vertices will go to zero depending on which Higgs pairs are being produced~\cite{Akeroyd:2003bt,Akeroyd:2003xi,Akeroyd:2003jp,Ilisie:2014hea,Delgado:2016arn,Vega:2018ddp}. Thus for the non-zero cases the coupling can be written schematically as,
\bea\label{eqn:gvhh}
g_{WH_M^\pm H_N^0} \equiv i g 
\, C_N (p_1 - p_2)^\mu ,~~
g_{ZH_M^0H_N^0} \equiv i \frac{g}{c_W} 
\, C_N (p_1 - p_2)^\mu ,
\eea
where $C_N$ is fixed by the $SU(2)_L\times U(1)_Y$
representation~\cite{Georgi:1985nv,Akeroyd:2003bt,Akeroyd:2010eg,Cort:2013foa,Hartling:2014zca} and $p_{1}, p_{2}$ are the four momenta of the incoming and outgoing scalar momenta. Here $H_N^0$ stands for any neutral Higgs boson and can include CP-even or CP-odd neutral Higgs bosons, as well as $H_M^\pm$ charged Higgs bosons. There is also a photon mediated channel when both Higgs bosons are charged, but we focus on cases where at least one is neutral. In Fig.~\ref{fig:HHprod} we show the leading order $q\bar{q} \to V \to H_M^{\pm,0} H_N^0$ (including \emph{PDFs}) cross section $\times\, C_N^{-2}$ for the $W$ mediated (blue solid) and $Z$ mediated (black dashed) channels at the LHC with $\sqrt{s}=13$~\UTeV (left) and $\sqrt{s}=27$~\UTeV (right) in the mass range $60 - 125$~\UGeV. They are computed with Madgraph~\cite{Alwall:2014hca} using a modified version of the GM model implementation of~\cite{Hartling:2014xma} and rescaling appropriately. There are also NLO contributions which may generate $\gtrsim \mathcal O(1)$ K-factors for Higgs pair production~\cite{Eichten:1984eu,Dawson:1998py,Degrande:2015xnm}. These are not included in our analysis. We show four cases for mass splittings  of $\Delta M \equiv M_{H^{\pm,0}_M} - M_{H_N^0} = 0, 100, 200, 300$~\UGeV as labelled in plot.  
\begin{figure}[tbh]
\begin{center}
\includegraphics[scale=.403]{\main/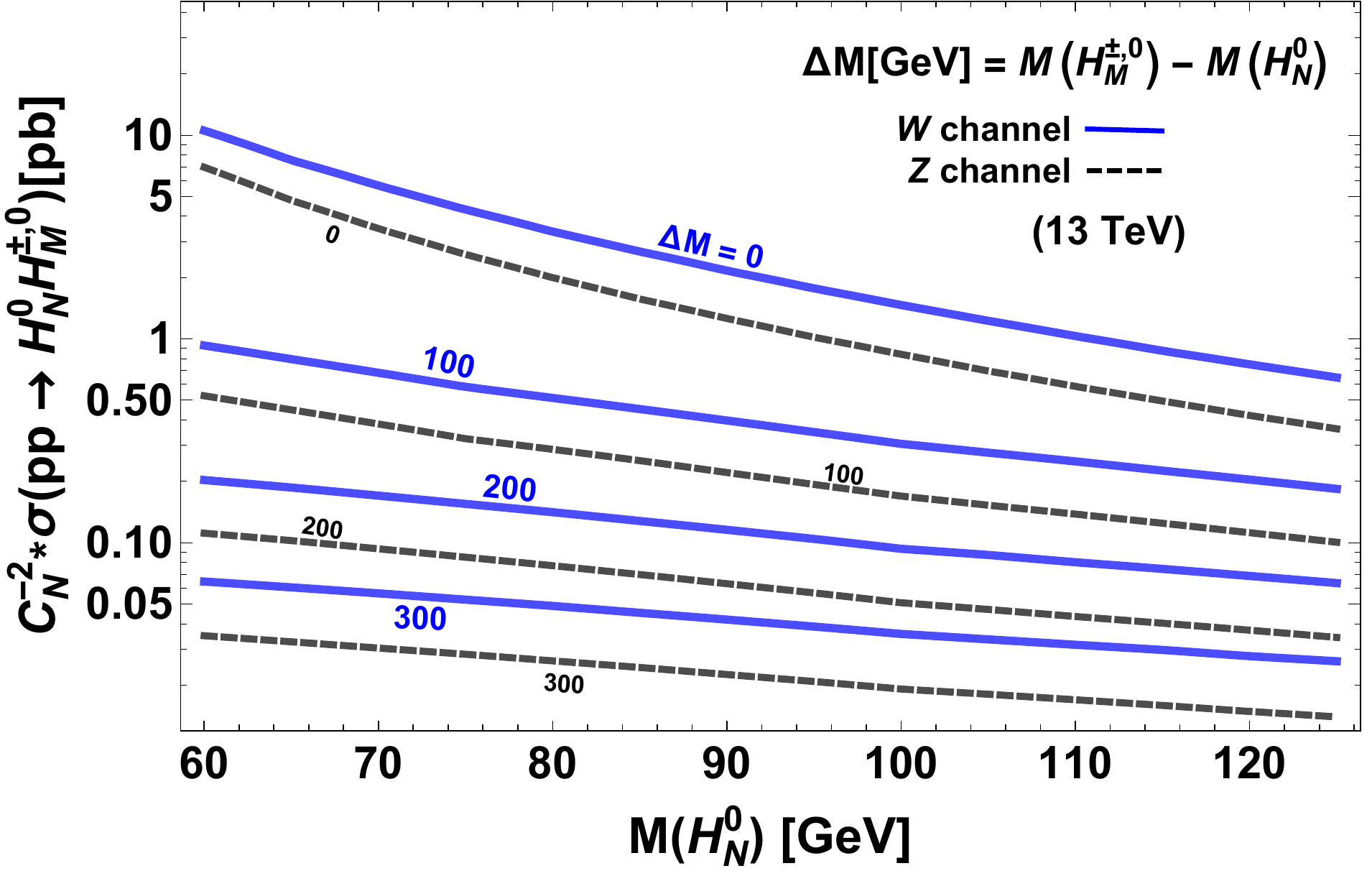}
\includegraphics[scale=.4]{\main/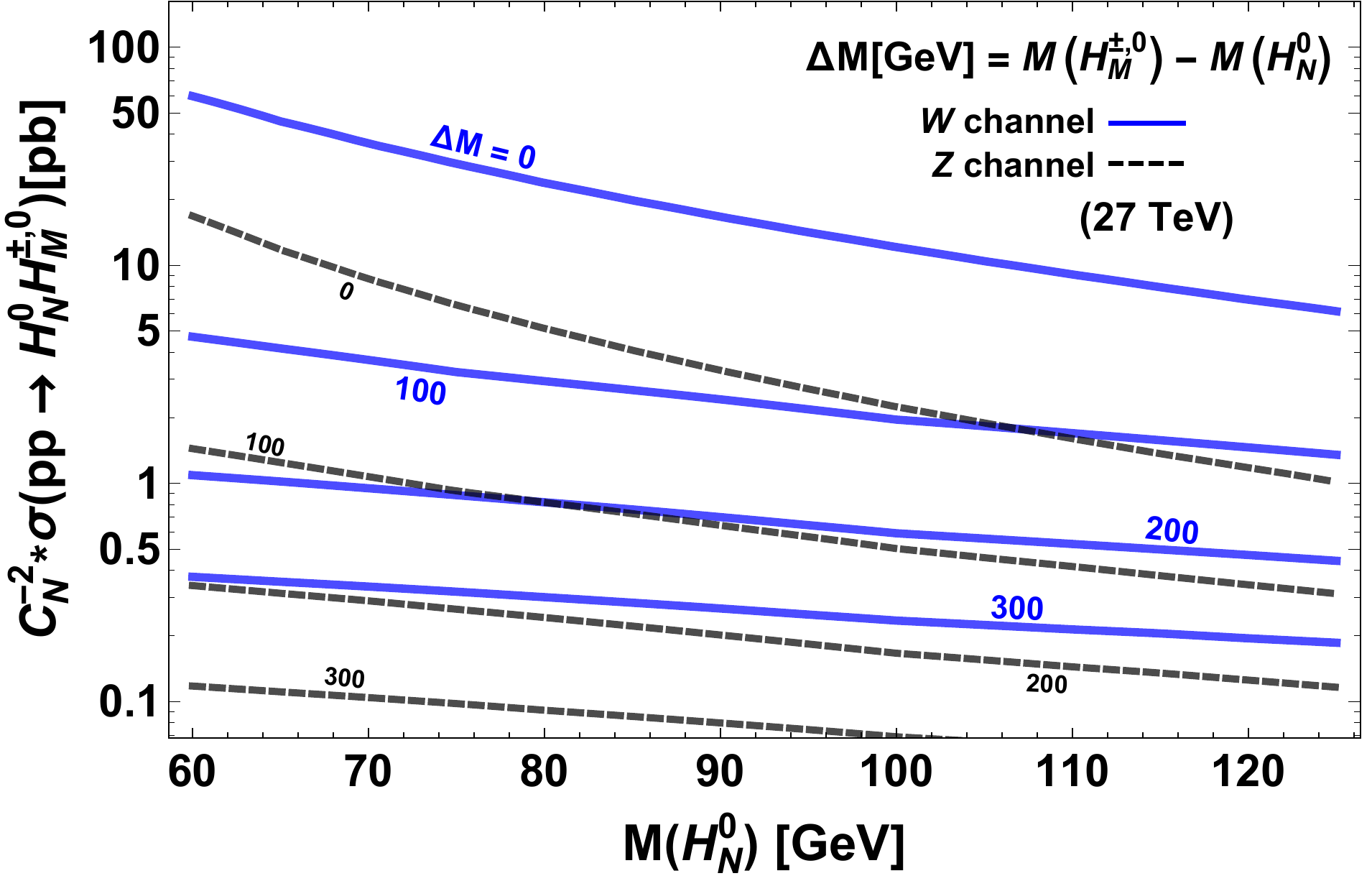}
\end{center}
\caption{Leading order cross sections (with \emph{PDFs}) for the $q\bar{q} \to V \to H_M^{\pm,0} H_N^0$ Higgs pair production mechanism mediated by $W$ (blue solid) and $Z$ (black dashed) bosons at the LHC for $\sqrt{s}=13$~\UTeV (left) and $\sqrt{s}=27$~\UTeV (right) in the mass range $60 - 125$~\UGeV. We show three cases for mass splittings $\Delta M \equiv M_{H^{\pm,0}_M} - M_{H_N^0}= 0, 100, 200, 300$~\UGeV as labelled in plot and have factored out an overall group theory factor $C_N$ (see Eq. (\ref{eqn:gvhh})). The curves for a particular model can be obtained by rescaling with $(C_N)^2$ which is fixed by the $SU(2)_L\times U(1)_Y$ representation.} 
\label{fig:HHprod}
\end{figure}

The dominant decay modes of the neutral Higgs bosons will be to $b\bar{b}$ and $\tau\bar{\tau}$ when there is a significant coupling to SM fermions. In the fermiophobic case, the Higgs bosons can have large branching ratios into EW gauge bosons and in particular photons at low masses. The less emphasised $Z\gamma$ channel may also offer promising opportunities~\cite{Degrande:2017naf}. Inclusive searches for resonances can then be combined with the Drell-Yan production channel to put relatively robust bounds on branching ratios in extended Higgs sectors as done in~\cite{Delgado:2016arn,Vega:2018ddp} for the case of decays into di-photons. For the charged Higgs bosons combing Drell-Yan pair production with decays into $W\gamma$~\cite{Ilisie:2014hea,Degrande:2017naf} or four photon signals~\cite{Aaltonen:2016fnw} offer promising search channels.

{\bf{Suggestions for searches at the HL/HE-LHC}}

We briefly summarise search strategies for (anti)-aligned light
Higgs bosons at the (HL/HE) LHC to be added to the current searches in the mass range we consider, including
$\tau\tau,\,\gamma\gamma,\,b\bar{b}$ searches based on gluon fusion and
$\tau\tau$ searches based on associated $b\bar{b}$
production~\cite{Aad:2014vgg,Khachatryan:2014wca,Khachatryan:2015baw} as
well as recent CMS searches~\cite{CMS:2015mba} for $A \to Zh$ with $Z
\to \ell\bar\ell$ and $h\to \tau\tau, bb$.  


\begin{itemize}
\item Push current conventional Higgs searches in $WW$ and $ZZ$, which
  currently~\cite{Khachatryan:2015cwa,Sirunyan:2018qlb} do not go below
  $\sim 130$~\UGeV, to as low a mass as possible, ideally down to $\sim
  65$~\UGeV. As emphasised in~\cite{Delgado:2016arn,Vega:2018ddp}, this
  can help to rule out cases of a fermiophobic Higgs boson with
  suppressed couplings to photons, which could otherwise escape
  detection. Similarly, heavier Higgs bosons with the ``remaining''
  coupling to SM gauge bosons could be detected.

\item Combine \emph{inclusive} searches for resonances with the
  `universal' Drell-Yan Higgs pair production channel to put robust
  bounds on allowed branching ratios to $\tau\tau$, $b\bar{b}$, $Z\gamma$ and
  $\gamma\gamma$ final states. In the alignment limit,
  these bounds depend only on electroweak couplings and can be applied
  to any extended Higgs boson sector (with appropriate rescaling), in
  some cases providing the strongest limits~\cite{Delgado:2016arn,Vega:2018ddp}.

\item Utilising the Drell-Yan Higgs pair production mechanism, dedicated
  LHC searches for more optimised, but model dependent signals such as
  $4\gamma + V^\ast$~\cite{Akeroyd:2003bt,Aaltonen:2016fnw,Arhrib:2017wmo},
  $4\gamma + V^\ast V^\ast$~\cite{Akeroyd:2003bt}, $3\gamma +
  V^\ast$ where in the last case dedicated phenomenological studies are
  lacking. 

\item Search for $\tau\tau$, $b\bar{b}$, or $\gamma\gamma$ plus missing
  energy as well as mono photon or mono lepton plus missing energy final
  states to cover cases where neutral Higgs may have an invisible
  decay. In particular the $\gamma\gamma$ channel appears to be
  very promising (especially in view of a potential signal at 
  $\sim 96$~\UGeV~\cite{CMS-PAS-HIG-17-013}).
 
\end{itemize}

\asubsubsection[M. Borsato, U. Haisch, J.F. Kamenik, A. Malinauskas, M. Spira]{HL-LHC projections of LHCb searches for 2HDM+S light pseudoscalars}\label{Sec:9.8LHCb}
Several well-motivated extensions of the SM include a new pseudoscalar $a$ with mass below the electroweak scale. A well-known example in the context of supersymmetry is the next-to-minimal supersymmetric SM, where this state can arise as a result of an approximate global $U(1)_R$ symmetry~\cite{Dobrescu:2000yn}. Non-supersymmetric extensions featuring a light pseudoscalar include Little Higgs models, hidden valley scenarios (see~\cite{Curtin:2013fra}, and references therein for details), and simplified models where a complex singlet scalar is coupled to the Higgs potential of the SM or the 2HDM. Light pseudoscalars have been searched via various collider signatures such as exotic decays of the $125 \, {\rm \UGeV}$ scalar $h$ discovered at the LHC (both $h\to aa$ and $h\to aZ$), radiative decays of bottomonium $\Upsilon\to a\gamma$, direct production from $pp$ collisions in association with $b$-jets and also inclusively in $pp\to a+X$, where the main production mode is usually gluon-gluon fusion. The interplay of searches for exotic $h$ decays and direct searches in $pp$ collisions within 2HDM+S models depend on the 2HDM parameters $\alpha, \beta$, on the mixing angle $\theta$, on the physical spin-0 masses, and on the form of the scalar potential (see for instance~\cite{Haisch:2018kqx} for further explanations).

Despite the significantly lower luminosity collected with respect to ATLAS and CMS, LHCb has proven to be capable of placing world-best limits for low-mass pseudoscalars produced in gluon-gluon fusion~\cite{Haisch:2016hzu,Haisch:2018kqx}, by searching simply for resonant pairs of opposite-sign muons~\cite{Aaij:2017rft, Aaij:2018xpt}. Indeed, a large fraction of these light pseudoscalars are produced with large boosts at the LHC and end up in the LHCb acceptance. On top of that, the LHCb detector is capable of triggering on muons with transverse momenta as low as $1.8 \, {\rm \UGeV}$ ($0.5 \, {\rm \UGeV}$) with the current (upgraded) trigger, greatly enhancing its acceptance to $a\to\mu^+\mu^-$ with respect to ATLAS and CMS. A key ingredient of this trigger, is the LHCb capability to efficiently reject  the large background due to pion mis-identification thanks to online availability of offline-quality particle identification based on information from all sub-detectors~\cite{Aaij:2016rxn, Dujany:2015lxd}. On top of that, the large boost of the pseudoscalar $a$  in the forward region allows to separate muons coming from semi-leptonic $B$ decays due to their displacement with respect to the $pp$ collision vertex. 

The HL-LHC sensitivity to prompt di-muon resonances in the context of dark photon searches at LHCb can be found in~\cite{Bediaga:2018lhg}. The kinematic selection used for the projection is inspired by~\cite{Ilten:2016tkc} and rely on the improved performance expected after the upgrade of the LHCb trigger that will be implemented for LHC Run-3. Maintaining this exceptional performance in the HL-LHC era (\ie with 10 times larger instantaneous luminosity) will require a redesign of the muon detector and is briefly discussed in~\cite{Bediaga:2018lhg}.

\begin{figure}[ht!]
\centering
\includegraphics[width=\textwidth]{\main/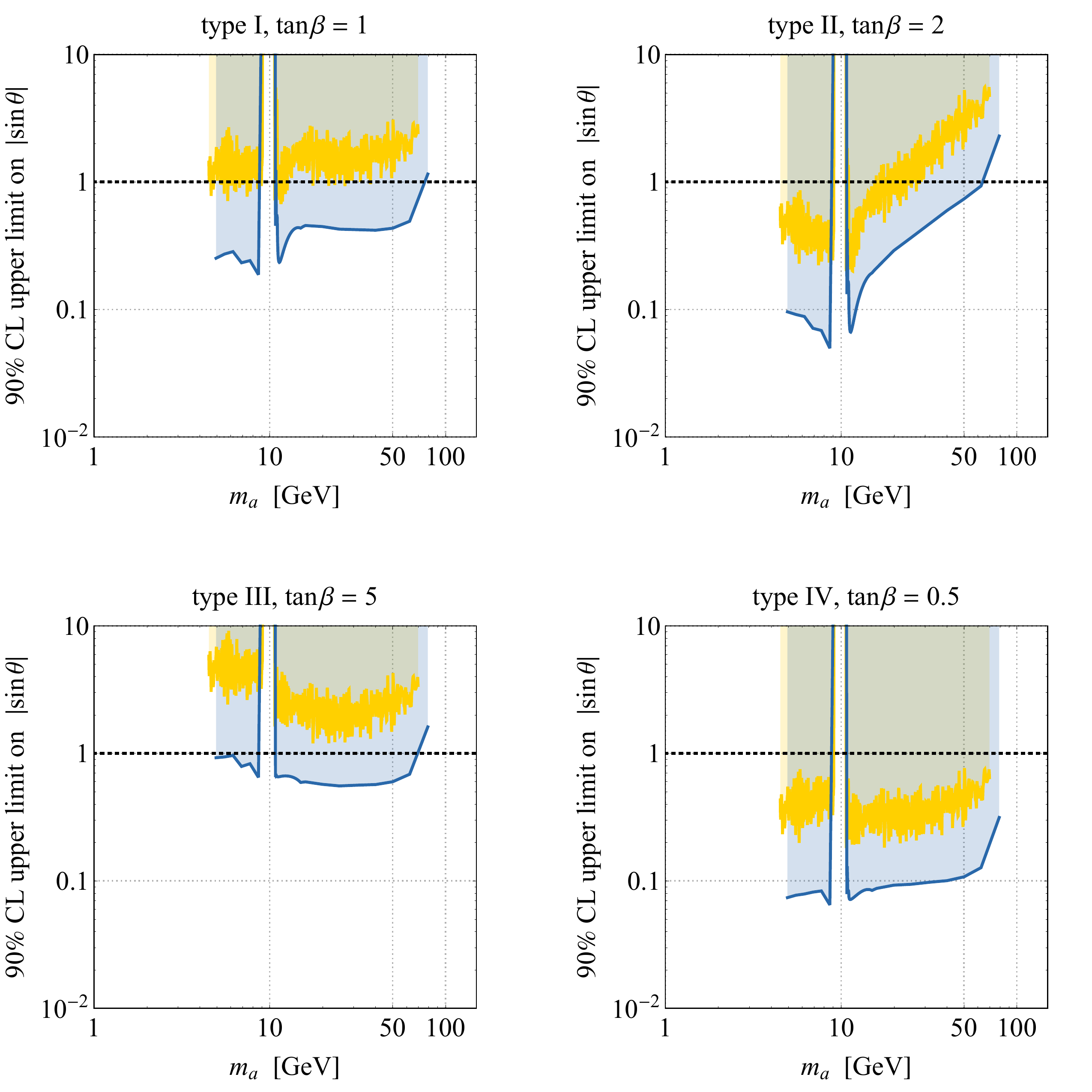}
\vspace{2mm}
\caption{Upper $90\%$ CL limits on $|\!\sin\theta|$ in the 2HDM+S of type~I with $\tan\beta =1$ (top left), type~II with $\tan\beta =2$ (top right), type~III with $\tan\beta = 5$ (bottom left) and type~IV with $\tan\beta = 0.5$ (bottom right). The yellow curves illustrate the results of a recast~\cite{Haisch:2018kqx} of the LHCb search~\cite{Aaij:2017rft} performed with a data set corresponding to $1.6\, {\rm fb}^{-1}$ of $13 \, {\rm \UTeV}$ $pp$ collisions, while the blue contours are our projections to $300 \, {\rm fb}^{-1}$ of $14 \, {\rm \UTeV}$ $pp$ collision data using the expected HL-LHC dark photon limits presented in~\cite{Bediaga:2018lhg}. }\label{fig:lhcb_lightmumu}
\end{figure}

In Figure~\ref{fig:lhcb_lightmumu}, the limits on the dark photon parameter space presented in~\cite{Bediaga:2018lhg} are reinterpreted in the context of the 2HDM+S, following the analysis strategy detailed in~\cite{Haisch:2018kqx}. The production cross section of the pseudoscalar $a$ and its decay rate to muons depend on the mixing angle $\theta$, on the parameter $\tan\beta$ and on the type of the Yukawa sector of the considered 2HDM. Fixing $\tan\beta$ and the type of the 2HDM, upper limits are placed on $|\!\sin\theta|$ as a function of the pseudoscalar mass $m_a$. In all considered cases, LHCb searches in the HL-LHC era~(blue contours) are found to be sensitive to values of $|\!\sin\theta|$ well below 1 for a large range of~$m_a$ values between $5 \, {\rm \UGeV}$ and $70 \, {\rm \UGeV}$. This represents a significant improvement over the LHC Run-2 results~(yellow curves), where only in the 2HDM+S scenario of type IV with $\tan \beta = 0.5$ it was possible to set physical meaningful bounds on the sine of the mixing angle $\theta$, \ie $|\!\sin\theta| < 1$, in the entire range of considered pseudoscalar masses. Notice that spin-0 states with masses around $10 \, {\rm \UGeV}$ can be probed by searches for di-muon resonances in $\Upsilon$ production~\cite{Haisch:2016hzu,Aaij:2018xpt}.


\newpage

\clearpage
\section{Conclusions and Outlook}
\subsection{Higgs properties and EW phenomena at the HL-LHC}
The determination of Higgs boson properties, and their connection to electroweak symmetry breaking (EWSB), is a primary target of the HL-LHC physics programme. Since 2012, the Higgs physics programme has rapidly expanded, with new ideas, more precise predictions and improved analyses, into a major program of precision measurements, as well as searches for rare production and decay processes. Outstanding opportunities have emerged for measurements of fundamental importance, such as the first direct constraints on the Higgs trilinear self-coupling and the natural width. The HL-LHC programme covers also searches for additional Higgs bosons in EWSB scenarios motivated by theories beyond the SM (BSM). Finally, a rigorous effective field theory (EFT) framework allows one to parametrise in a model independent way all EW and Higgs results. 
The studies presented in this report update the key expectations for HL-LHC,  and summarise the interpretation of the future constraints on new physics in terms of EFT couplings. This reappraisal of the future sensitivities relies on the Run~2 analyses improvements and assumes the detector performance targets established in the experiments' upgrade TDRs.
Further improvements should be possible with analyses optimised for the HL-LHC data sets.

The main Higgs boson measurement channels correspond to five production modes (the gluon fusion ggF, the vector boson fusion VBF, the associated production with a vector boson $WH$ and $ZH$, and the associated production with a pair of top quarks $ttH$) and seven decay modes: $H \to \gamma\gamma$, $ZZ^*$, $WW^*$, $\tau^+\tau^-$, $b\overline{b}$, $\mu^+\mu^-$ and $Z\gamma$. The latter two decay channels, as yet unobserved, should become visible during the next two LHC runs. 
The rate measurements in the aforementioned production and decay channels yield measurements of the Higgs couplings in the so-called "$\kappa$-framework". This introduces a set of $\kappa_i$ factors that linearly modify the coupling of the Higgs boson to SM elementary particles, $i$, including the effective couplings to gluons and photons, and assuming no additional BSM contribution to the Higgs total width, $\Gamma_H$. The projected uncertainties, combining ATLAS and CMS, are summarised in Fig.~\ref{fig:summary_K2} of Section~\ref{sec2}. They include today's theory uncertainties reduced by a factor of two, which is close to the uncertainty that would result from using the improved HL-LHC parton distribution functions (PDFs, see Section~\ref{sec2:PDFuncertainties}) and considering signal theory uncertainties as uncorrelated. Except for rare decays, the overall uncertainties will be dominated by the theoretical systematics, with a precision close to percent level.

The main Higgs boson couplings will be measured at HL-LHC with a precision at the percent level. Large statistics will particularly help the study of complex final states, such as those arising from $ttH$ production. 
The constraining power of the current $ttH$ analyses has been limited to plausible improvements in the theory predictions, in particular in the $H\rightarrow b\overline{b}$ channel. The 3.4\% precision on $\kappa_t$ thus obtained is mostly due to the other direct $ttH$ measurement channels.


These coupling measurements assume the absence of sizeable additional contributions to $\Gamma_H$. As recently suggested, the patterns of quantum interference between background and Higgs-mediated production of photon pairs or four leptons are sensitive to $\Gamma_H$. Measuring the off-shell four-fermion final states, and assuming the Higgs couplings to gluons and $ZZ$ evolve off-shell as in the SM, the HL-LHC will extract $\Gamma_H$ with a 20\% precision at 68\%~CL.
Furthermore, combining all Higgs channels, and with the assumption that the couplings to vector bosons are not  larger than the SM ones ($\kappa_V\le 1$), will constrain $\Gamma_H$ with a 5\% precision at 95\%~CL. Invisible Higgs boson decays will be searched for at HL-LHC in all production channels, VBF being the most sensitive. The combination of ATLAS and CMS Higgs boson coupling measurements will set an upper limit on the Higgs invisible branching ratio of 2.5\%, at the 95\% CL.  The precision reach in the measurements of ratios will be at the percent level, with particularly interesting measurements of $\kappa_{\gamma}/\kappa_{\rm Z}$, which serves as a probe of new physics entering the $H \rightarrow \gamma\gamma$ loop, can be measured with an uncertainty of 1.4\%, and $\kappa_{t}/\kappa_{g}$, which serves as probe of new physics entering the $gg \rightarrow H$ loop, with a precision of 3.4\%.

A summary of the limits obtained on  first and second generation quarks from a variety of observables is given in Fig.~\ref{fig:sec7:summary} of section~\ref{sec7}. It includes: (i) HL-LHC projections for {\it exclusive} decays of the Higgs into quarkonia; (ii) constraints from fits to differential cross sections of {\it kinematic} observables (in particular $p_\mathrm{T}$); (iii) constraints on the total {\it width}, $\Gamma_H$, relying on different assumptions (the examples given in  Fig.~\ref{fig:sec7:summary} correspond to a projected limit of 200~\UMeV on the total width from the mass shift from the interference in the di-photon channel between signal and continuous background and the constraint at 68\% CL on the total width from off-shell couplings measurements of 20\%); (iv) a {\it global} fit of Higgs production cross sections (yielding the constraint of 5\% on the width mentioned herein); and (v) the {\it direct search} for Higgs decays to $c\overline{c}$ using inclusive charm tagging techniques. Assuming SM couplings, the latter is expected to lead to the most stringent upper limit of $\kappa_c \lessapprox 2$. A combination of ATLAS, CMS and LHCb results would further improve this constraint to $\kappa_c \lessapprox 1$.



Precision measurements provide an important tool to search for BSM physics associated to mass scales beyond the LHC direct reach. The EFT framework, where the SM Lagrangian is supplemented with dimension-6 operators $\sum_{i} c_i \mathcal{O}^{(6)}_i/\Lambda^2$, allows one to systematically parametrise BSM effects and how they modify SM processes. Figure~\ref{fig:dim6U_HLLHC} of section~\ref{sec8:globalfit} shows the results of a global fit  to observables in Higgs physics, as well as di-boson and Drell-Yan processes at high energy. The fit includes all operators generated by new physics that only couples to SM bosons. These operators can either modify SM amplitudes, or generate new amplitudes. In the former case, the best LHC probes are, for example, precision measurements of Higgs branching ratios. In the case of the operator $\mathcal{O}_H$, 
for example, the constraints in Fig.~\ref{fig:dim6U_HLLHC} translate into a sensitivity to the Higgs compositeness scale $f>1.6$~\UTeV, corresponding to a new physics mass scale of 20~\UTeV for an underlying strongly coupled theory.
The effects associated with some new amplitudes grow quadratically with the energy. For example, Drell-Yan production at large mass can access, via the operators $\mathcal{O}_{2W,2B}$, energy scales of order 12~\UTeV (Fig.~\ref{fig:dim6U_HLLHC}). 

The Run~2 experience in searches for Higgs pair production led to a reappraisal of the HL-LHC sensitivity, including several channels, some of which were not considered in previous projections: $2\textrm{b}2\gamma$, $2\textrm{b}2\tau$, $4\textrm{b}$, $2\textrm{b}\textrm{WW}$, $2\textrm{b}\textrm{ZZ}$.
Assuming the SM Higgs self-coupling $\lambda$, ATLAS and CMS project a sensitivity to the $HH$ signal of approximately 3 $\sigma$. per experiment, leading to a combined observation sensitivity of 4 $\sigma$. These analyses, which make use also of the $HH$ mass spectrum shape, result in the likelihood profile as a function of $\kappa_{\lambda}$ shown in Fig.~\ref{fig:comb_HH} of section~\ref{sec:HH_Combination}. An important feature of these analyses is the presence of the secondary minimum in the likelihood line-shape, due to the degeneracy in the total number of $HH$ signal events for different $\kappa_{\lambda}$ values. We note that at the HL-LHC the secondary minimum can be excluded at 99.4\%~CL, with a constraint on the Higgs self-coupling of $0.5 < \kappa_{\lambda} < 1.5 $ at the 68\%\ CL. The results on HH production studies are statistics limited, therefore a dataset of at least 6\,\iab{}  (ATLAS and CMS combined) is essential to achieve this objective.



Higgs studies at HL-LHC will enhance the sensitivity to BSM physics, exploiting indirect probes via precision measurements, and a multitude of direct search targets, ranging from exotic decays of the 125~\UGeV Higgs boson (e.g. decays including promptly decaying light scalars, light dark photons or axion-like particles, and decays involving long-lived BSM particles) to the production of new Higgs bosons, neutral and charged, at masses above or below 125~\UGeV.
The HL-LHC will be able to probe very rare exotic decay modes of the 125 \UGeV Higgs boson thanks to the huge Higgs data set that will be produced (branching ratios as small as $\mathcal O(10^{-6})$ could be probed for sufficiently clean decay modes). Furthermore, the mass reach for new heavy Higgs bosons can be pushed to few \UTeV. As an example, Fig.~\ref{fig:bench} in section~\ref{Sec:9.5}, shows a summary of the Minimal Supersymmetric SM regions of parameter space that will be probed by ATLAS and CMS either via direct searches of new Higgs bosons decaying to tau lepton pairs, or via indirect 125 \UGeV Higgs coupling measurements. The HL-LHC will have access to new Higgs bosons as heavy as ~2.5 \UTeV at large $\tan\beta$ ($\tan\beta>50$). Complementarily, the interpretation of Higgs precision coupling measurements will exclude Higgs bosons with masses lower than approximately 1~\UTeV over a large range of $\tan\beta$.

\subsection{Potential of the HE-LHC}
With the increase in centre-of-mass energy and luminosity, the Higgs physics programme at HE-LHC will considerably extend the reach of the entire HL-LHC program. Measurements of the Higgs boson trilinear self-coupling, of elusive decay modes (e.g.\,H$\to$c\={c}), of rare (e.g.\,H$\to$Z\textgamma), invisible or exotic decays will become accessible. At the same time, Higgs boson production can be explored at very large transverse momenta.
Projections presented in this section are exploratory and provide qualitative results, due to the absence of clearly defined reference detectors, and in view of the highly challenging pile-up environment. Several approaches have been followed to address this issue, typically assuming experimental performances similar to those currently achieved by LHC detectors. Other studies focused on Higgs bosons produced at finite transverse momentum ($p_T>50$~\UGeV), to reduce the impact of pile-up. The selection of fiducial regions in $p_T$ and rapidity, furthermore, allows measurements of the ratios of rates for different final states, free of uncertainties related to the production dynamics and to luminosity. 

\begin{table}[h]
\centering
  \caption{\label{tab:Hrates}
Higgs production event rates for selected processes
    at 27~\UTeV ($N_{27}$) and
 statistical increase with 
 respect to the statistics of the HL-LHC ($N_{27}=\sigma_{27~\mathrm{\UTeV}} \times 15$~\iab, 
 $N_{14}=\sigma_{14~\mathrm{\UTeV}} \times 3$~\iab).} 
\begin{tabular}{|l|c|c|c|c|c|c|} 
\hline  \hline
&  gg\,$\to$\,H   & VBF &
 WH  &
 ZH  &
 $\rm t\bar{t}H$ &
 HH 
 \\ \hline
$N_{27}$  & $2.2\times 10^9$ & $1.8\times 10^8$ & $5.4\times 10^7$ & $3.7\times
 10^7$ &  $4\times 10^7$ & $2.1 \times 10^6$  \\
$N_{27}/N_{14}$ & 13 & 14 & 12 & 13 &23 & 19
\\ \hline
\hline
\end{tabular}
\end{table}
The statistics expected for some reference production processes, and the increase with respect to the HL-LHC, are shown in
Table~\ref{tab:Hrates}. The Higgs samples will typically increase by a factor between 10 and 25, in part as a result of the 5 times larger luminosity, leading to a potential reduction in the statistical uncertainties by factors of 3 to 5. The biggest improvements arise for the channels favoured by the higher energy, such as ttH and HH.


The potential for the measurement of the Higgs boson trilinear coupling at the HE-LHC has been estimated with methods and in channels similar to those used at the HL-LHC. Extrapolation studies from the current experiments and from phenomenological studies have been carried out in the two most sensitive $HH$ channels at the HL-LHC ($b\overline{b}\gamma\gamma$ and $b\overline{b}\tau^+\tau^-$). 
Several studies were made under different experimental performance and systematic uncertainty assumptions (in some cases neglecting systematic uncertainties), yielding results covering the wide range of precision estimates presented here. At the HE-LHC the $HH$ signal would be observed unambiguously and the combined sensitivity on the trilinear coupling, $\kappa_{\lambda}$ (assuming the SM value), is expected to reach a precision of 10\% to 20\% from the combination of these two channels alone. A comparison of the HE-LHC sensitivity to that of the HL-LHC is displayed in Fig.~\ref{fig:HH_HELHC_comb} of section~\ref{sec:HH_HE}, showing that the secondary minimum still visible in the HL-LHC study is unambiguously excluded at HE-LHC. 
These studies do not include the additional decay channels that have already been studied for HL-LHC, and of others that could become relevant at the HE-LHC. Exclusive production modes are also very interesting to take into consideration for this measurement. The potential improvements from these have not been assessed yet.

The measurement of the couplings of the Higgs boson at HL-LHC relies either on the assumption that no additional undetected contribution to the Higgs boson width is present, or that the couplings of the Higgs boson to vector bosons do not exceed those expected in the SM. In both cases, the foreseen precision in the measurements of most Higgs boson couplings at the HL-LHC is currently limited by the theoretical uncertainty on the signal predictions. The significantly larger dataset and the increase in centre-of-mass energy at HE-LHC would reduce the statistical uncertainty of these measurements to being negligible.
To match the overall precision of the experimental measurements, the extraction of the couplings of the Higgs boson to photons, gluons, W, Z, taus, and b quarks will require significant theoretical improvements in the precision of the theoretical predictions for the signals.

For rare decay processes such as the di-muon channel, from an extrapolation of the HL-LHC projections, a precision of approximately 2\% on the coupling modifier 
should be achievable.
 With the current theoretical systematic uncertainties on the signal and the backgrounds, the direct measurement of the Higgs coupling modifier to top quarks is expected to reach a precision of approximately 3\%. While the substantial additional amount of data at various centre-of-mass energies will undoubtedly be useful to further constrain the systematic modelling uncertainties and further progress in theoretical predictions will be achieved, the potential improvements have not been quantified. Assuming an improvement of the theoretical uncertainties of a factor of 2, the precision on the ttH coupling would reach approximately 2\% (the experimental systematic uncertainty alone is approximately 1\%, assuming performances similar to current LHC experiments). The significant gain in precision will be obtained mostly through ratios of couplings. Studies have shown that the ratio of the $t\overline{t}H$ to $t\overline{t}Z$ ratio could be measured at close to the percent level.

At HE-LHC energies, the $H \to c\bar{c}$ production increases relative to backgrounds, and may be observable with inclusive searches by ATLAS, CMS, and LHCb, depending on $c$-tagging systematic uncertainties. Unfortunately, at the HE-LHC, exclusive searches, kinematic limits, and global fits are not expected to reach the SM level for the $u$, $d$, $s$, and $c$ Yukawas.

Precision measurements provide an important tool to search for BSM physics associated to mass scales beyond the LHC direct reach. The EFT framework, where the SM Lagrangian is supplemented with higher dimension operators 
allows one to systematically parametrise BSM effects and how they modify SM processes. These operators can either modify SM amplitudes, or generate new amplitudes. In the former case, the best LHC probes are, for example, precision measurements of Higgs branching ratios. In the case of the operator $\mathcal{O}_H$, for example, the constraints in Fig.~\ref{fig:dim6U_HELHC} of Section~\ref{sec8:globalfit}, translate into a sensitivity to the Higgs compositeness scale $f>2$~\UTeV, corresponding to a new physics mass scale of 25~\UTeV for an underlying strongly coupled theory.

Effects associated with  new amplitudes grow quadratically (for dimension-6 operators) with the energy. The higher centre-of-mass energy and larger dataset of HE-LHC make it possible to greatly extend the measurable range in the Higgs transverse momenta, providing new opportunities: a 10\% measurement  at 1 \UTeV energy corresponds roughly to a per-mille  precision measurement at the Higgs mass energy. 
In the context of EW physics this will allow  to test, via Drell-Yan processes and  the operators $\mathcal{O}_{2W,2B}$, energy scales of order 25~\UTeV; or, via $WZ$ di-boson processes, mass scales of roughly 6 (100) \UTeV if the underlying new physics is weakly (strongly) coupled.
Figure~\ref{fig:dim6U_HELHC} shows the results of a global fit  to observables in Higgs physics, as well as di-boson and Drell-Yan processes at high energy. 

Another important high-energy measurement concerns the scattering of longitudinally polarised vector bosons: departures from its SM value could betray a composite nature of the Higgs.
The decomposition of measurements of VBS cross-sections into the polarised components based on the decays of the individual vector bosons is experimentally challenging. Preliminary studies show that, thanks to pile-up mitigation techniques that retain Run-2 performance of hadronically decaying W/Z-boson tagging, the precision on the VBS cross section measurement in the semi-leptonic $W V$ + jj $\to \ell\nu$ + jjjj channel can be reduced from 6.5 $\%$ (HL-LHC) to about 2 $\%$ at HE-LHC. From this measurement and from the measurement of the EW production of a Z boson pair, the purely longitudinal final state of the WW and ZZ scattering processes can be extracted with a significance of 5$\sigma$ or more. Similarly, the reach for vector-boson-scattering will be extended by roughly a factor of two in the energy scale of BSM physics, i.e. the sensitivity of the HE-LHC to Wilson coefficients, $f/\Lambda^4$, of dimension eight operators, which describe anomalous quartic gauge couplings, improves by a factor 10-20.

Complementarily, the HE-LHC will offer unprecedented opportunities to directly test light \UTeV-scale new degrees of freedom associated to the Higgs boson and generically arising in models for electroweak symmetry breaking. 
Particularly, due to the large increase in the Higgs data set (see table \ref{tab:Hrates}), very rare exotic Higgs decays could be discovered.
For example, multi-lepton signatures could be produced from Higgs decays to light BSM particles ($X$) as dark photons, or axion-like particles: $h \rightarrow XX \rightarrow b\overline{b}\mu\mu,~4\ell$. The projected HE-LHC reach on the branching ratios of these two decay modes is estimated to be $\sim10^{-5}$ and $\sim10^{-8}$, respectively, extending the HL-LHC reach by a factor of $\sim5$ and $\sim10$, respectively (see Secs. \ref{Sec:2b2muExo}, \ref{Sec:4lExo}). As shown by these numbers, the reach of particularly clean decay modes will see a major gain at the HE-LHC mainly due to the increase in Higgs statistics (from gluon fusion production). 
At the same time, the sample of Higgs bosons produced from sub-leading production
modes in association with other SM particles (e.g. $t\bar th$) will be sizeable, increasing the discovery prospects for rare and more background limited Higgs decay signatures. Therefore, the HL/HE-LHC
Higgs exotic decay program can be uniquely sensitive to the existence of a broad range of new light
weakly coupled particles.

The increase in energy will also open up many opportunities for the direct search of new \UTeV-scale degrees of freedom associated to electroweak symmetry breaking, as new heavy Higgs bosons. In this report, we have studied, for example, the reach for $pp \rightarrow S \rightarrow hh$, with $h$ the 125~\UGeV Higgs boson and $S$ a new Higgs boson, and we have shown that the HE-LHC can extend the reach to $S$ masses that are 1.5-2 times heavier than the masses probed by the HL-LHC (see Sec. \ref{Sec.9.4.2}). Many more studies will be needed to assess the full discovery potential of the HE-LHC to extended Higgs sectors, as arising in many well motivated BSM theories.

\newpage

\section*{Acknowledgements}
We would like to thank the LHC experimental Collaborations and the WLCG for their essential support.
We are especially grateful for the efforts by the computing, generator and validation groups who were
instrumental for the creation of large simulation samples. We thank the detector upgrade groups as well
as the physics and performance groups for their input.  Not least, we thank the many colleagues who
have provided useful comments on the analyses.

\addcontentsline{toc}{chapter}{References}
\bibliographystyle{report}
\bibliography{report}

\end{document}